\documentclass{book}

\usepackage{amssymb,amsmath,amsthm}
\usepackage{makeidx,mathrsfs}         
\usepackage{graphicx}        
\usepackage{upgreek}        

\usepackage{psfrag}
\usepackage{a4wide}
\usepackage{hyperref}

\usepackage{imakeidx}

\makeindex

\theoremstyle{plain}
\newtheorem{thm}{Theorem}[chapter]
\newtheorem{prop}[thm]{Proposition}
\newtheorem{lemma}[thm]{Lemma}
\newtheorem{cor}[thm]{Corollary}
\theoremstyle{definition}
\newtheorem{definition}[thm]{Definition}

\newtheorem{remark}[thm]{Remark}
\newtheorem{example}[thm]{Example}
\newtheorem{remarks}[thm]{Remarks}
\newtheorem{assumption}[thm]{Assumption}
\newtheorem{assumptions}[thm]{Assumptions}

\newcommand{\tn}[1]{\ensuremath{\mathbb{T}^{#1}}}

\newcommand{\rn}[1]{\ensuremath{\mathbb{R}^{#1}}}
\newcommand{\zn}[1]{\ensuremath{\mathbb{Z}^{#1}}}
\newcommand{\sn}[1]{\ensuremath{\mathbb{S}^{#1}}}
\newcommand{\cn}[1]{\ensuremath{\mathbb{C}^{#1}}}

\newcommand{\Mn}[2]{\ensuremath{\mathbf{M}_{#1}(#2)}}
\newcommand{\Gl}[2]{\ensuremath{\mathbf{GL}_{#1}(#2)}}
\newcommand{\bRie}[1]{\ensuremath{\beta_{\mathrm{Ri},#1}}}
\newcommand{\baverage}[1]{\ensuremath{\bar{\beta}_{#1}}}
\newcommand{\nuaverage}[1]{\ensuremath{\bar{\nu}_{#1}}}
\newcommand{\Xaverage}[1]{\ensuremath{\bar{X}_{#1}}}
\newcommand{\Spe}{\mathrm{Sp}}
\newcommand{\mrspan}{\mathrm{span}}
\newcommand{\grad}{\mathrm{grad}}
\newcommand{\Rsp}{\mathrm{Rsp}}
\newcommand{\rsp}{\mathrm{rsp}}
\newcommand{\zo}{\mathbb{Z}}
\newcommand{\ro}{\mathbb{R}}

\newcommand{\rovol}{\mathrm{vol}}

\newcommand{\roS}{\mathrm{S}}

\newcommand{\rosub}{\mathrm{sub}}
\newcommand{\rocsp}{\mathrm{csp}}
\newcommand{\rocr}{\mathrm{cr}}
\newcommand{\roRe}{\mathrm{Re}}
\newcommand{\Rho}{\mathrm{P}}

\newcommand{\co}{\mathbb{C}}
\newcommand{\so}{\mathbb{S}^{1}}

\newcommand{\abs}{\mathrm{abs}}
\newcommand{\rest}{\mathrm{rest}}
\newcommand{\partic}{\mathrm{part}}
\newcommand{\inter}{\mathrm{inter}}

\newcommand{\normSobtsmuzpm}{\|\cdot\|_{t,s,\mu_{0},\pm}}
\newcommand{\normrolma}{\|\cdot\|_{\rolma}}
\newcommand{\normHs}{\|\cdot\|_{(s)}}
\newcommand{\normbkbge}{|\bk|_{\bge}}
\newcommand{\normmcXbh}{|\mcX|_{\bh}}
\newcommand{\normfunAs}{\|\cdot\|_{A,s}}
\newcommand{\normSobts}{\|\cdot\|_{t,s}}
\newcommand{\normSobtrss}{\|\cdot\|_{\trs,s}}
\newcommand{\normRHSA}{\|\cdot\|_{A}}
\newcommand{\diag}{\mathrm{diag}}
\newcommand{\rem}{\mathrm{rem}}
\newcommand{\refer}{\mathrm{ref}}

\newcommand{\aux}{\mathrm{aux}}
\newcommand{\rosh}{\mathrm{sh}}

\newcommand{\rog}{\mathrm{g}}
\newcommand{\rotot}{\mathrm{tot}}
\newcommand{\rosts}{\mathrm{sts}}
\newcommand{\roih}{\mathrm{ih}}
\newcommand{\roh}{\mathrm{h}}
\newcommand{\roim}{\mathrm{im}}
\newcommand{\roi}{\mathrm{i}}
\newcommand{\row}{\mathrm{w}}

\newcommand{\coeff}{\mathrm{coeff}}

\newcommand{\roup}{\mathrm{up}}
\newcommand{\rohom}{\mathrm{hom}}
\newcommand{\roloc}{\mathrm{loc}}
\newcommand{\rolma}{\mathrm{lma}}
\newcommand{\rolts}{\mathrm{lts}}
\newcommand{\ror}{\mathrm{r}}
\newcommand{\rom}{\mathrm{m}}
\newcommand{\roM}{\mathrm{M}}
\newcommand{\roU}{\mathrm{U}}
\newcommand{\ron}{\mathrm{n}}

\newcommand{\rol}{\mathrm{l}}
\newcommand{\ronl}{\mathrm{nl}}
\newcommand{\ros}{\mathrm{s}}
\newcommand{\rot}{\mathrm{t}}
\newcommand{\rod}{\mathrm{d}}
\newcommand{\roder}{\mathrm{der}}
\newcommand{\rodS}{\mathrm{dS}}
\newcommand{\roRie}{\mathrm{Rie}}
\newcommand{\roRi}{\mathrm{Ri}}

\newcommand{\roT}{\mathrm{T}}
\newcommand{\rou}{\mathrm{u}}
\newcommand{\romn}{\mathrm{mn}}
\newcommand{\rolb}{\mathrm{lb}}
\newcommand{\roas}{\mathrm{as}}
\newcommand{\road}{\mathrm{ad}}
\newcommand{\romar}{\mathrm{mar}}
\newcommand{\roini}{\mathrm{ini}}
\newcommand{\romod}{\mathrm{mod}}
\newcommand{\roode}{\mathrm{ode}}
\newcommand{\roopt}{\mathrm{opt}}
\newcommand{\robas}{\mathrm{bas}}
\newcommand{\robal}{\mathrm{bal}}
\newcommand{\ropar}{\mathrm{par}}
\newcommand{\rocu}{\mathrm{cu}}
\newcommand{\roco}{\mathrm{co}}
\newcommand{\rocon}{\mathrm{con}}
\newcommand{\roeff}{\mathrm{eff}}
\newcommand{\romed}{\mathrm{med}}
\newcommand{\roep}{\mathrm{ep}}
\newcommand{\roeep}{\mathrm{eep}}
\newcommand{\rodm}{\mathrm{dm}}

\newcommand{\rosil}{\mathrm{sil}}
\newcommand{\rood}{\mathrm{od}}

\newcommand{\rodiff}{\mathrm{diff}}
\newcommand{\app}{\mathrm{app}}
\newcommand{\wa}{\mathrm{wa}}

\newcommand{\low}{\mathrm{low}}

\newcommand{\sgn}{\mathrm{sgn}}

\newcommand{\bre}{\bar{r}}
\newcommand{\tr}{\mathrm{tr}}
\newcommand{\trs}{\mathrm{ts}}

\newcommand{\per}{\mathrm{per}}
\newcommand{\pre}{\mathrm{pre}}
\newcommand{\spec}{\mathrm{spec}}

\newcommand{\Id}{\mathrm{Id}}
\newcommand{\Ind}{\mathrm{Ind}}
\newcommand{\fin}{\mathrm{fin}}

\newcommand{\mri}{\mathrm{I}}

\newcommand{\bze}{\bar{\zeta}}

\newcommand{\bka}{\bar{\kappa}}

\newcommand{\tildt}{\tilde{t}}
\newcommand{\tx}{\tilde{x}}
\newcommand{\ty}{\tilde{y}}
\newcommand{\tu}{\tilde{u}}
\newcommand{\tf}{\tilde{f}}
\newcommand{\tw}{\tilde{w}}
\newcommand{\tA}{\tilde{A}}
\newcommand{\tB}{\tilde{B}}
\newcommand{\tJ}{\tilde{J}}
\newcommand{\tF}{\tilde{F}}
\newcommand{\tX}{\tilde{X}}
\newcommand{\bA}{\bar{A}}

\newcommand{\bk}{\bar{k}}
\newcommand{\bK}{\bar{K}}
\newcommand{\hK}{\hat{K}}

\newcommand{\bh}{\bar{h}}

\newcommand{\bH}{\bar{H}}
\newcommand{\bge}{\bar{g}}

\newcommand{\bX}{\bar{X}}
\newcommand{\bY}{\bar{Y}}

\newcommand{\bM}{\bar{M}}
\newcommand{\bB}{\bar{B}}

\newcommand{\bJ}{\bar{J}}
\newcommand{\bj}{\bar{j}}
\newcommand{\bt}{\bar{t}}
\newcommand{\bs}{\bar{s}}

\newcommand{\bla}{\bar{\lambda}}
\newcommand{\bu}{\bar{u}}
\newcommand{\bz}{\bar{z}}

\newcommand{\bbe}{\bar{\beta}}
\newcommand{\etab}{\bar{\eta}}

\newcommand{\bga}{\bar{\gamma}}

\renewcommand{\a}{\alpha}
\newcommand{\e}{\epsilon}
\newcommand{\vare}{\varepsilon}

\newcommand{\de}{\delta}
\newcommand{\om}{\omega}
\renewcommand{\b}{\beta}
\newcommand{\g}{\gamma}
\newcommand{\G}{\Gamma}
\renewcommand{\d}{\partial}

\newcommand{\me}{\mathcal{E}}
\newcommand{\mK}{\mathcal{K}}

\newcommand{\ma}{\mathcal{A}}
\newcommand{\mO}{\mathcal{O}}
\newcommand{\mf}{\mathcal{F}}
\newcommand{\mfN}{\mathfrak{N}}
\newcommand{\mfS}{\mathfrak{S}}
\newcommand{\ml}{\mathcal{L}}
\newcommand{\mb}{\mathcal{B}}

\newcommand{\ms}{\mathcal{S}}

\newcommand{\mW}{\mathcal{W}}
\newcommand{\mff}{\mathfrak{f}}
\newcommand{\betafun}{\mathfrak{e}}
\newcommand{\ellderbd}{\mathfrak{b}}
\newcommand{\shiftbd}[1]{\eta_{\rosh,#1}}
\newcommand{\shiftrb}{\eta_{\rosh,0}}

\newcommand{\cruderate}{\eta_{\rocr}}
\newcommand{\nolossrate}{\eta_{\ronl}}
\newcommand{\ellderbdsil}{\mathfrak{b}_{\ros}}
\newcommand{\ellderbdlow}{\mathfrak{b}_{\low}}
\newcommand{\bfcon}{c_{\betafun}}

\newcommand{\mfG}{\mathfrak{G}}
\newcommand{\mfD}{\mathfrak{D}}

\newcommand{\mP}{\mathcal{P}}
\newcommand{\mR}{\mathcal{R}}
\newcommand{\tmR}{\tilde{\mathcal{R}}}
\newcommand{\bmR}{\bar{\mathcal{R}}}

\newcommand{\mcX}{\mathcal{X}}
\newcommand{\mcY}{\mathcal{Y}}
\newcommand{\md}{\mathcal{D}}

\newcommand{\mfr}{\mathfrak{r}}
\newcommand{\mfq}{\mathfrak{q}}
\newcommand{\mfp}{\mathfrak{p}}
\newcommand{\mfg}{\mathfrak{g}}
\newcommand{\mfl}{\mathfrak{l}}
\newcommand{\mfe}{\mathfrak{E}}
\newcommand{\mfh}{\mathfrak{h}}
\newcommand{\mfH}{\mathfrak{H}}

\newcommand{\hF}{\hat{F}}

\newcommand{\hU}{\hat{U}}

\newcommand{\hf}{\hat{f}}
\newcommand{\hk}{\hat{k}}
\newcommand{\he}{\hat{e}}
\newcommand{\hg}{\hat{g}}
\newcommand{\chg}{\check{g}}
\newcommand{\hh}{\hat{h}}

\newcommand{\hrho}{\hat{\rho}}
\newcommand{\hal}{\hat{\a}}

\newcommand{\hA}{\hat{A}}

\newcommand{\chrho}{\check{\rho}}
\newcommand{\chR}{\check{R}}
\newcommand{\chA}{\check{A}}
\newcommand{\chB}{\check{B}}
\newcommand{\chF}{\check{F}}

\newcommand{\chJ}{\check{J}}

\newcommand{\hBl}{\hat{B}}
\newcommand{\hu}{\hat{u}}
\newcommand{\hv}{\hat{v}}

\newcommand{\ha}{\hat{a}}

\newcommand{\hpsi}{\hat{\psi}}
\newcommand{\hchi}{\hat{\chi}}

\newcommand{\is}{\int_{\sn{1}}}
\newcommand{\bx}{\bar{x}}
\newcommand{\bv}{\bar{v}}
\newcommand{\bw}{\bar{w}}

\newcommand{\bp}{\bar{p}}

\newcommand{\trho}{\tilde{\rho}}

\newcommand{\bpsi}{\bar{\psi}}

\newcommand{\bchi}{\bar{\chi}}

\newcommand{\boc}{\mathbf{c}}

\newcommand{\tPhi}{\tilde{\Phi}}

\newcommand{\ldr}[1]{\langle #1\rangle}

\newcommand{\roB}{\mathrm{B}}

\newcommand{\mubox}{\mu_{\roB}}
\newcommand{\lambdabox}{\lambda_{\roB}}
\newcommand{\mutrsbox}{\mu_{\roB,\trs}}
\newcommand{\musilbox}{\mu_{\roB,\rosil}}
\newcommand{\ldrbox}[1]{\langle #1\rangle_{\roB}}
\newcommand{\ldrtrsbox}[1]{\langle #1\rangle_{\roB,\trs}}

\newcommand{\muindexset}{\mathcal{M}}
\newcommand{\indexnot}{\iota}
\newcommand{\indexnottwo}{\varsigma}
\newcommand{\EFindexset}{\mathcal{I}_{\roB}}
\newcommand{\EFindexsetr}[2]{\mathcal{I}_{\roB,#1}^{#2}}
\newcommand{\EFnindexset}{\mathcal{I}_{\roB,\ron}}
\newcommand{\EFnindexsetp}{\mathcal{I}_{\roB,\ron}^{+}}
\newcommand{\EFnindexsetm}{\mathcal{I}_{\roB,\ron}^{-}}
\newcommand{\EFnindexsetpm}{\mathcal{I}_{\roB,\ron}^{\pm}}
\newcommand{\EFtrsindexset}{\mathcal{I}_{\roB,\trs}}
\newcommand{\EFsilindexset}{\mathcal{I}_{\roB,\rosil}}
\newcommand{\projtrs}{\Pi_{\trs}}
\newcommand{\projsil}{\Pi_{\rosil}}

\begin{document}

\author{Hans Ringstr\"{o}m\\ KTH Royal Institute of Technology and Institut Mittag-Leffler}
\title{Linear systems of wave equations on cosmological backgrounds with 
convergent asymptotics}
\maketitle

\frontmatter

\setcounter{tocdepth}{1}
\tableofcontents

\mainmatter

\part{Introductory material, results}\label{part:introduction}

\chapter{Equations and questions}

\section{Introduction}

The subject of these notes is the asymptotic behaviour of solutions to systems of linear wave equations of the following form:
\begin{equation}\label{eq:gensyswaveeq}
\Box_{g}u+Xu+\zeta u=f.
\end{equation}
Here $g$ is a given Lorentz metric on a manifold $M$; $\Box_{g}$ is the wave operator associated with the Lorentz metric $g$, defined by 
$\Box_{g}u:=\mathrm{div}(\grad u)$; $X$ is a smooth $m\times m$ matrix of vector fields on $M$ whose coefficients with respect
to a frame are allowed to be complex (and $1\leq m\in \zo$);  $\zeta$ is a smooth $\Mn{m}{\co}$-valued function on $M$ (where $\Mn{m}{K}$ denotes 
the set of $m\times m$ matrices
with entries in the field $K$); 
\index{$\a$Aa@Notation!Matrix notation!MnmK@$\Mn{m}{K}$}%
and $f$ is a smooth $\cn{m}$-valued function on $M$. We are mainly interested in the real (as opposed to 
the complex) setting, but for technical reasons, it turns out to be convenient to derive results under the above assumptions. The problem is to 
analyse the asymptotic behaviour of solutions $u:M\rightarrow\cn{m}$ to (\ref{eq:gensyswaveeq}). In particular, we are interested in 
the problem of deriving optimal estimates for energies naturally associated with solutions to (\ref{eq:gensyswaveeq}), and in the problem of 
calculating the leading order asymptotics. We are only interested in globally hyperbolic Lorentz manifolds $(M,g)$, and in that setting, there 
is a unique smooth solution to (\ref{eq:gensyswaveeq}), given smooth initial data on a Cauchy hypersurface; a justification for this statement
is provided, e.g., by \cite[Theorem~12.19, p.~144]{minbok} (though we do not need to appeal to this result in these notes). In fact, we here 
think of solutions to (\ref{eq:gensyswaveeq}) as arising in this way. Our main 
motivation for studying (\ref{eq:gensyswaveeq}) is that we are interested in the linearised Einstein equations. Note, however, that 
(\ref{eq:gensyswaveeq}) is not the most general class of systems relevant in that context. For example, in the study of the Einstein--Euler 
equations, it is of interest to consider systems for which the symbol is different for the different components of $u$, a situation not 
covered by (\ref{eq:gensyswaveeq}). Nevertheless, (\ref{eq:gensyswaveeq}) represents quite a general class of equations, and, in order to 
obtain conclusions, we gradually need to impose more and more restrictions on the Lorentz manifolds of interest etc. One assumption we
consistently make is that the Cauchy hypersurfaces of $(M,g)$ are closed manifolds. In fact, $M$ is here always of the form $M:=\bM\times I$,
where $\bM$ is a closed manifold and $I=(t_{-},t_{+})$ is an open interval. In the case of Einstein's equations, this situation is 
of interest in cosmology (i.e., in the study of the universe as a whole), and the asymptotic regimes associated with $t\rightarrow t_{\pm}$ 
should be thought of as representing a cosmological singularity (a big bang or a big crunch) or the expanding direction. The cosmological
setting should be contrasted with the asymptotically Euclidean or asymptotically hyperbolic situation, which is of interest in the study of 
isolated systems (isolated galaxies, stars, black holes etc.). In the isolated systems setting, static and stationary solutions (such as 
the Minkowski and Kerr spacetimes) are the ones of greatest interest. In the cosmological setting, it is more natural to consider solutions 
with contracting directions (toward a big bang/big crunch) and/or expanding directions. In the following section, we take one more step towards 
restricting the class of equations defined by (\ref{eq:gensyswaveeq}) to the one of interest here. 

\section{Equations}\label{section:genReq}

In these notes, we consider equations of the form
\begin{equation}\label{eq:thesystemRge}
\begin{split}
-g^{00}(t)u_{tt}-\textstyle{\sum}_{j,l=1}^{d}g^{jl}(t)\d_{j}\d_{l}u-2\sum_{l=1}^{d}g^{0l}(t)\d_{l}\d_{t}u
-\sum_{r=1}^{R}a^{-2}_{r}(t)\Delta_{g_{r}}u & \\
+\a(t)u_{t}+\textstyle{\sum}_{j=1}^{d}X^{j}(t)\d_{j}u+\zeta(t)u & = f.
\end{split}
\end{equation}
Here $1\leq m\in\zo$, $0\leq d,R\in\zo$ 
\index{$\a$Aa@Notation!Constants!$m$}%
\index{$\a$Aa@Notation!Constants!$d$}%
\index{$\a$Aa@Notation!Constants!$R$}%
and $g^{00},g^{jl},g^{0l},a_{r}\in C^{\infty}(I,\ro)$, 
\index{$\a$Aa@Notation!Metrics!$g^{00}$}%
\index{$\a$Aa@Notation!Metrics!$g^{0l}$}%
\index{$\a$Aa@Notation!Metrics!$g^{jl}$}%
\index{$\a$Aa@Notation!Metrics!$a_{r}$}%
where $I$ is an open interval, $j,l=1,\dots,d$ and 
$r=1,\dots,R$. In case $d=0$, the sum from $1$ to $d$ should be thought of as empty, and similarly for the sum from $1$ to $R$. Moreover, 
$-g^{00}$ and $a_{r}$, $r=1,\dots,R$, take their values in $(0,\infty)$ and for each $t\in I$, $g^{jl}(t)$, $j,l=1,\dots,d$, are the components 
of a positive definite matrix (in (\ref{eq:thesystemRge}) we also abuse notation in that we write $a^{-2}_{r}(t)$ when we, strictly speaking, 
mean $[a_{r}(t)]^{-2}$). In addition, $(M_{r},g_{r})$, $r=1,\dots,R$, 
\index{$\a$Aa@Notation!Manifolds!$M_{r}$}%
\index{$\a$Aa@Notation!Metrics!$g_{r}$}%
are closed Riemannian manifolds and $\Delta_{g_{r}}$ 
\index{$\a$Aa@Notation!Differential operators!$\Delta_{g_{r}}$}%
is the 
Laplace-Beltrami operator on $(M_{r},g_{r})$. The functions $u$ and $f$ should be smooth functions from $M:=\bM\times I$
\index{$\a$Aa@Notation!Manifolds!$M$}%
to $\cn{m}$, where
\begin{equation}\label{eq:bMdef}
\bM:=\tn{d}\times M_{1}\times\cdots\times M_{R};
\end{equation}
\index{$\a$Aa@Notation!Manifolds!$\bM$}%
the differential operators $\d_{j}$
are the standard vector fields on $\tn{d}$; $\d_{t}$ represents standard differentiation with respect to $t$ (for functions on $M$); and 
$\a,X^{j},\zeta\in C^{\infty}[I,\Mn{m}{\co}]$. 
\index{$\a$Aa@Notation!Coefficients, equation!$\a$}%
\index{$\a$Aa@Notation!Coefficients, equation!$X^{j}$}%
\index{$\a$Aa@Notation!Coefficients, equation!$\zeta$}%
Note that (\ref{eq:thesystemRge}) is a special case 
of (\ref{eq:gensyswaveeq}) when the Lorentz metric $g$ on $M$ is of the form
\begin{equation}\label{eq:gfromthesystemRgeintro}
g=g_{00}dt\otimes dt+g_{0i}dt\otimes dx^{i}+g_{i0}dx^{i}\otimes dt+g_{ij}dx^{i}\otimes dx^{j}+\textstyle{\sum}_{r=1}^{R}a_{r}^{2}g_{r},
\end{equation}
\index{$\a$Aa@Notation!Metrics!$g$}%
where we use Einstein's summation convention (to sum over repeated upstairs and downstairs indices) and the coefficients only depend on $t$. 
The $a_{r}$ appearing in this metric can be read off directly from (\ref{eq:thesystemRge}) and the $g_{\g\b}(t)$, $\g,\b\in \{0,\dots,d\}$, are the 
components of the inverse of the matrix with components $g^{\g\b}(t)$, $\g,\b\in \{0,\dots,d\}$. That this inverse exists, and that 
(\ref{eq:gfromthesystemRgeintro}) is a Lorentz metric on $M$, is a consequence of the assumptions; cf. Chapter~\ref{chapter:geometry}, in particular
the beginning of Section~\ref{section:basgeomglobhyp}, for a detailed justification. Moreover, $g_{00}(t)<0$ and $g_{ij}(t)$, $i,j\in \{1,\dots,d\}$, 
are the components of a positive definite matrix for all $t\in I$. In Chapter~\ref{chapter:geometry}, we also demonstrate that 
$(M,g)$ is globally hyperbolic and that each hypersurface
\begin{equation}\label{eq:bMtintro}
\bM_{t}:=\bM\times\{t\},
\end{equation} 
\index{$\a$Aa@Notation!Manifolds!$\bM_{t}$}%
$t\in I$, is a Cauchy hypersurface; cf. Lemma~\ref{lemma:globallyhyperbol}. 

\begin{remark}\label{remark:invofthesystemRgeundconfresc}
There are two important operations that leave the class of equations of the form (\ref{eq:thesystemRge}) invariant. First, multiplying an equation
of the form (\ref{eq:thesystemRge}) by a strictly positive function of $t$ only yields an equation of the same type. Considering only the leading 
order derivatives in (\ref{eq:thesystemRge}), this corresponds to 
a conformal rescaling of the metric (\ref{eq:gfromthesystemRgeintro}) (by a conformal factor depending only on $t$). Secondly, we have the 
freedom of changing the time coordinate. In what follows, we make use of both of these operations in order to reduce the underlying geometry
to a preferred form. 
\end{remark}

Clearly, the step from (\ref{eq:gensyswaveeq}) to (\ref{eq:thesystemRge}) involves significant restrictions. In particular, the fact that the 
coefficients of (\ref{eq:thesystemRge}) only depend on $t$ ensures that (\ref{eq:thesystemRge}) is separable, an important simplification.
Note also that the separability makes it very easy to study solutions under much more general regularity assumptions than smoothness (though
we focus on smooth solutions in these notes). 
On the other hand, our main interest is in cosmological solutions, and in that setting, the standard starting point is the assumption of spatial 
homogeneity. From this point of view, assuming the coefficients to only depend on time is thus natural. 

\subsection{Separable cosmological model manifolds}\label{ssection:sepcosmmodman}

Due to our interest in equations of the form (\ref{eq:thesystemRge}), we restrict our attention to the following class of Lorentz manifolds in 
what follows. 

\begin{definition}\label{def:sepcosmmodmanintro}
A \textit{separable cosmological model manifold} 
\index{Separable cosmological model manifold}%
is a Lorentz manifold $(M,g)$ such that $M=\bM\times I$, where $I=(t_{-},t_{+})$ is an open interval,
$0\leq d,R\in\zo$, $\bM$ is given by (\ref{eq:bMdef}) and
\begin{equation}\label{eq:sepcosmodmeintro}
g = g_{00}dt\otimes dt+g_{0i}dt\otimes dx^{i}+g_{i0}dx^{i}\otimes dt+g_{ij}dx^{i}\otimes dx^{j}+\textstyle{\sum}_{r=1}^{R}a_{r}^{2}g_{r},
\end{equation}
\index{$\a$Aa@Notation!Metrics!$g$}%
where the $g_{\g\b}$ and the $a_{r}$ only depend on $t$; the $(M_{r},g_{r})$ 
are closed Riemannian manifolds; $g_{\g\b}(t)=g_{\b\g}(t)$ for all $t$; $g_{00}(t)<0$ for all $t$; for all $t$, the 
$g_{ij}(t)$ are the components of a positive definite matrix; and $a_{r}(t)>0$ for all $t$.
\end{definition}
\begin{remark}
In the definition, it is always understood that $\g,\b\in\{0,\dots,d\}$; $i,j\in \{1,\dots,d\}$; and $r\in\{1,\dots,R\}$.
\end{remark}
\begin{remark}
The reason for using the word separable is that the wave equation on $(M,g)$ is separable. 
The reason for using the word cosmological is that $\bM$ is closed. 
\end{remark}
\begin{remark}\label{remark:shiftandlapsedef}
Sometimes it is convenient to write $g$ as 
\begin{equation}\label{eq:lapseshiftsepcosmodmeintro}
g=-N^{2}dt\otimes dt+g_{ij}(\chi^{i}dt+dx^{i})\otimes (\chi^{j}dt+dx^{j})+\textstyle{\sum}_{r=1}^{R}a_{r}^{2}g_{r};
\end{equation}
cf. Subsection~\ref{ssection:lapseandshift} below for an explanation of how to calculate $N$ and $\chi^{i}$ in terms of the metric components. 
When using this notation, $N>0$ is called the \textit{lapse function} 
\index{$\a$Aa@Notation!Metrics!$N$, Lapse function}%
\index{Lapse function}%
and $\chi:=\chi^{i}\d_{i}$ the 
\textit{shift vector field}. 
\index{Shift vector field}%
\index{$\a$Aa@Notation!Vector fields!$\chi$}%
Note also that $N$ and $\chi^{i}$ only depend on $t$. 
\end{remark}
In the case that $(M,g)$ is a separable cosmological model manifold, the induced metrics, second fundamental forms and volumes on the 
hypersurfaces $\bM_{t}$ (cf. (\ref{eq:bMtintro})) are of importance. It is therefore convenient to introduce the following terminology.
\begin{definition}\label{def:bkbgeV}
Let $(M,g)$ be a separable cosmological model manifold. Then the metric and second fundamental form of $\bM_{t}$ are denoted
by $\bge_{t}=\bge(t)$ and $\bk_{t}=\bk(t)$ respectively (and they are interpreted as symmetric covariant $2$-tensor fields on $\bM$). 
The trace $\tr_{\bge}\bk$ is referred to as the \textit{mean curvature}
\index{Mean curvature}%
of $\bM_{t}$. Moreover, 
$V(t):=\rovol_{\bge(t)}(\bM)$ denotes the volume of $\bM$ 
\index{Volume of $\bM_{t}$}%
\index{$\a$Aa@Notation!Auxiliary functions!$V$}%
with respect to $\bge(t)$. Finally, $U$ 
\index{$\a$Aa@Notation!Vector fields!$U$}%
denotes the future directed unit timelike vector 
field which is normal to all the hypersurfaces $\bM_{t}$. 
\end{definition}
\begin{remark}\label{remark:UformulaUlnVformintro}
The vector field $U$ can be written 
\begin{equation}\label{eq:UitoNchiintro}
U:=N^{-1}(\d_{t}-\chi^{i}\d_{i}).
\end{equation}
Moreover, $U(\ln V)=\tr_{\bge}\bk$; cf. (\ref{eq:thetaroTrdef})--(\ref{eq:thetadecomp}) below. 
\end{remark}
In the context of general relativity, the metric and second fundamental form induced on $\bM_{t}$ by $g$ are of particular importance, 
since they constitute the geometric part of the initial data for Einstein's equations. 

\subsection{Examples of separable cosmological model manifolds}\label{ssection:exsepcosmoma}

In order to illustrate that there are separable cosmological model manifolds of interest, let us give a few examples. 

\textbf{Minkowski space.}
\index{Minkowski space}%
In the study of hyperbolic PDE's on $\rn{d+1}$, it is natural to begin with equations whose symbols equal that of the standard wave operator 
on $d+1$-dimensional Minkowski space. One reason for this is that physical theories consistent with special relativity give rise to Lorentz 
invariant PDE's. From the point of view of general relativity, Minkowski space also plays a central role. This is partly due to the fact that 
it is a vacuum solution to Einstein's equations describing the geometry in the absence of gravitational fields (and constitutes the weak field 
limit of general relativity), but also because it is a \textit{stable} solution to Einstein's equations, as has been demonstrated in, e.g., 
\cite{cak,lar,lar2,baz}. The Minkowski metric on $\rn{d+1}$ can, of course, also be considered to be a vacuum solution to Einstein's equations on 
$\tn{d}\times\ro$. However, this solution is less natural. The main reason for this is that it is unstable, as we demonstrate below; cf. also
the comments made at the beginning of Section~\ref{section:geoconprefme}.

\textbf{Bianchi type I.} Consider initial data to Einstein's equations on $\tn{d}$. Assume that they are invariant under the natural action of 
$\tn{d}$ on itself. Here we refer to the corresponding maximal Cauchy developments as \textit{Bianchi type I spacetimes}.
\index{Bianchi type I spacetimes}%
In this situation, the spacetime metric can sometimes be written in the form 
\begin{equation}\label{eq:gBianchiI}
g=-dt\otimes dt+\textstyle{\sum}_{i=1}^{d}a_{i}^{2}dx^{i}\otimes dx^{i}
\end{equation}
on $M=\tn{d}\times I$, where $I$ is an open interval; the $a_{i}$ are smooth, strictly positive functions on $I$; $t$ is the coordinate on $I$; and
$dx^{i}$ are the standard one-forms on $\tn{d}$ (considered as forms on $M$).  We are here mainly interested in the situation that the mean curvature 
is non-zero (cf. the discussion at the beginning of Section~\ref{section:geoconprefme}). In the case of the vacuum Einstein equations, the $a_{i}$ 
then take the form $\a_{i}t^{p_{i}}$, where $0<\a_{i}\in\ro$ and $p_{i}\in\ro$ for $i=1,\dots,d$. Moreover, the 
sum of the $p_{i}$ and the sum of the $p_{i}^{2}$ both equal $1$ (these conditions are referred to as the \textit{Kasner relations}). 
\index{Kasner!relations}%
The corresponding
solutions are referred to as the \textit{Kasner solutions}. 
\index{Kasner!solutions}%
Note that when $a_{i}(t)=(t/t_{0})^{p_{i}}$, the initial data induced by
(\ref{eq:gBianchiI}) on the $t=t_{0}$ hypersurface take the form
\[
\bge_{t_{0}}:=\textstyle{\sum}_{j=1}^{d}d\bx^{j}\otimes d\bx^{j},\ \ \
\bk_{t_{0}}:=\sum_{j=1}^{d}\frac{p_{j}}{t_{0}}d\bx^{j}\otimes d\bx^{j},
\]
where $d\bx^{i}$ are the standard one-forms on $\tn{d}$. 
Letting $t_{0}\rightarrow\infty$, it is clear that these initial data converge to those of the Minkowski metric on $\tn{d}\times\ro$. Moreover, this 
convergence is in any $C^{k}$-topology. On the other hand, if none of the $p_{j}$'s equal $1$, then (\ref{eq:gBianchiI}) with $a_{i}(t)=(t/t_{0})^{p_{i}}$ 
is an inextendible solution to Einstein's vacuum equations which is past causally geodesically incomplete and exhibits curvature blow up as 
$t\rightarrow 0+$. In particular, it is thus clear that $d+1$-dimensional Minkowski space, considered as a solution on $\tn{d}\times\ro$, is unstable. 

Concerning the Kasner solutions, let us remark that even though they can be expected to be unstable (both in the expanding and in the contracting
direction),
they have traditionally played an important role in general relativity. The reason for this is that they are the basic building blocks
in the so-called BKL conjecture 
\index{BKL conjecture}%
(due to Belinski\v{\i}, Khalatnikov and Lifschitz) concerning spacelike cosmological (big bang or big crunch) singularities. 
We shall here not discuss this conjecture further, but refer the reader interested
in heuristic and numerical perspectives to \cite{dhan,dah,huar} and \cite{xetal} respectively, as well as references cited therein. 
Nevertheless, due to the central role this conjecture plays in the physics literature, it is of interest to understand the behaviour of 
solutions to the linearised Einstein equations around Kasner solutions. 

\textbf{Bianchi type I, accelerated expansion.} Turning to cosmological solutions that are future stable, it is of interest to consider ones 
with accelerated expansion. The reason for this is partly due to the fact that the currently preferred models of the universe exhibit such
expansion. However, it is also related to the cosmic no-hair conjecture, roughly stating that generic cosmological solutions to Einstein's
equations with a positive cosmological constant generally asymptote to de Sitter space; cf., e.g., \cite{AAR} for a more detailed discussion 
of this topic and a more precise formulation of the cosmic no-hair conjecture. Turning to the results, de Sitter space is stable; cf., 
e.g., \cite{f}. Here de Sitter space with cosmological constant $\Lambda>0$ is the metric 
\[
-dt\otimes dt+\cosh^{2}(Ht)g_{\sn{3}_{H}}
\]
on $\ro\times\sn{3}$, where $H=(\Lambda/3)^{1/2}$ and $g_{\sn{3}_{H}}$ is the standard metric on the $3$-sphere with radius such that 
$\mathrm{Ric}[g_{\sn{3}_{H}}]=2H^{2}g_{\sn{3}_{H}}$. Similarly, 
\begin{equation}\label{eq:flatdS}
g_{\rodS}:=-dt\otimes dt+\textstyle{\sum}_{j=1}^{d}e^{2Ht}dx^{j}\otimes dx^{j},
\end{equation}
where $0<H\in\ro$ and $3\leq d\in\zo$, is a future stable solution to Einstein's vacuum equations with a positive cosmological constant 
$\Lambda=d(d-1)H^{2}/2$. Note that (\ref{eq:flatdS}) is a solution of Bianchi type I. Moreover, the universal covering spacetime 
of $(\tn{d}\times\ro,g_{\rodS})$ is a subset of de Sitter space which, to the future, exhibits the same asymptotics as de Sitter space itself. 
Moreover, future stability holds in the presence of a large class of different matter models; cf., e.g., 
\cite{HRINV,SVE,RAS,SPECK,HAS,stab,AAR,NUN}. In accordance with the cosmic no-hair conjecture, spatially homogeneous but anisotropic solutions 
typically asymptote to de Sitter space; cf., e.g., \cite{wald2,hayounglambda} and \cite[Part~VII]{stab}. In fact, there are even classes of 
solutions that are both highly anisotropic and highly
spatially inhomogeneous that asymptote to the geometry described by (\ref{eq:flatdS}); cf., e.g., \cite{tar,AAR} for two results in the 
Einstein--Vlasov setting. Due to these results, it is clear that the metric (\ref{eq:flatdS}) is of importance. It also describes the asymptotics
of the currently preferred models of the universe. 

The Einstein-non-linear scalar field equations with an exponential potential are considered in, e.g., \cite{HRPL,heinzle,LAI}. The motivation for
studying this setting is that it can be used to model dark energy, the purpose being to explain the accelerated expansion of the universe.
In this setting, the model metrics again take the form (\ref{eq:gBianchiI}), but with $a_{i}(t)=\a_{i}t^{p}$, where $0\neq \a_{i}\in\ro$ and 
$1<p\in\ro$. That these metrics (together with a scalar field corresponding to an appropriate exponential potential) are future stable solutions 
to the Einstein-non-linear scalar field equations is demonstrated in \cite{HRPL}. 

\textbf{Bianchi type I, quiescent singularities.} Concerning the spacetimes mentioned above, we focused on the issue of future stability. However,
it is also of interest to consider big bang/big crunch type singularities. Due to the BKL conjecture,
big bang singularities are typically thought to be 'oscillatory' (in a sense we do not here try to specify). However, there are some special matter
models that are expected to neutralise the oscillations. Two examples of such matter models are stiff fluids and scalar fields. In this setting 
there is support for the BKL picture. The support comes in three main forms. First, there are results in the presence of symmetries (but in the 
absence of small data conditions); cf., e.g., \cite[Section~7]{BianchiIXattr}. Second, there are constructions of large classes of solutions with 
the expected behaviour; cf., e.g., \cite{aarendall,daetal}. Third, there are stability results around symmetric solutions; cf., e.g., 
\cite{rasql,rasq}. The relevant model metrics in this case are, again, of the form (\ref{eq:gBianchiI}), where $a_{i}(t)=\a_{i}t^{p_{i}}$, $0<\a_{i}\in\ro$ 
and $0<p_{i}\in\ro$. Moreover, the sum of the $p_{i}$ equal $1$, but the sum of the $p_{i}^{2}$ is strictly less than $1$. 

\textbf{Higher dimensional analogues of de Sitter space.} Let $0<H\in\ro$ and $(\Sigma,g_{\Sigma})$ be a closed Riemannian manifold of dimension 
$n\geq 3$ such that $\mathrm{Ric}[g_{\Sigma}]=(n-1)H^{2}g_{\Sigma}$. Then the Lorentz metric
\begin{equation}\label{eq:ggendS}
g_{\mathrm{gdS}}:=-dt\otimes dt+\cosh^{2}(Ht)g_{\Sigma}
\end{equation}
on $M_{\mathrm{gdS}}:=\Sigma\times\ro$ is a solution to Einstein's vacuum equations with a positive cosmological constant $\Lambda=n(n-1)H^{2}/2$. 
In the special case that $n=3$ and $\Sigma=\sn{3}$, the Lorentz manifold $(M_{\mathrm{gdS}},g_{\mathrm{gdS}})$ is called \textit{de Sitter space}. 
\index{de Sitter space}%

\textbf{The Nariai spacetimes.} The Nariai spacetimes 
\index{Nariai spacetimes}%
are interesting counterexamples to what is expected to be the generic behaviour of solutions to Einstein's equations with a 
positive cosmological constant (according to the cosmic no-hair conjecture). For $0<H\in\ro$, the Nariai spacetime solving Einstein's 
vacuum equations with a positive cosmological constant 
$\Lambda=H^{2}$ is given by $(M_{\mathrm{N}},g_{\mathrm{N}})$, where $M_{\mathrm{N}}:=\sn{1}\times\sn{2}\times\ro$,
\begin{equation}\label{eq:Nariai}
g_{\mathrm{N}}:=-dt\otimes dt+\cosh^{2}(Ht)dx^{2}+H^{-2}g_{\sn{2}},
\end{equation}
and $g_{\sn{2}}$ is the standard metric on the unit $2$-sphere.

\textbf{The Milne model.} In \cite{aam}, Andersson and Moncrief demonstrate the future stability of the \textit{Milne model} 
\index{Milne model}%
in the class of vacuum
solutions to Einstein's equations. The Milne model is the spacetime $(\Sigma\times(0,\infty),g_{\roM})$, where 
\[
g_{\roM}:=-dt\otimes dt+t^{2}g_{\Sigma}
\]
and $(\Sigma,g_{\Sigma})$ is an appropriate closed hyperbolic manifold. This model plays a central role in a general conjecture
concerning the future asymptotics of cosmological solutions to Einstein's vacuum equations due to Arthur Fischer and Vincent Moncrief on the one hand,
and Michael Anderson on the other; cf., e.g., \cite{anderson,fimon}. 

\textbf{Cosmological solutions with $U(1)$-symmetry.} In \cite{chobmon,chob}, Choquet-Bruhat and Moncrief study the future stability of 
\[
g_{\roU}:= -dt\otimes dt+d\theta\otimes d\theta+t^{2}g_{\Sigma}
\]
on $\sn{1}\times\Sigma\times (0,\infty)$, where $d\theta$ is the standard one-form field on $\sn{1}$ and $(\Sigma,g_{\Sigma})$ is an appropriate 
closed hyperbolic $2$-manifold. In fact, they demonstrated future stability of this solution under $U(1)$-symmetric perturbations.
\index{Uone@$U(1)$-symmetry}%

\textbf{Gowdy and $\tn{2}$-symmetry.} When taking the step from spatially homogeneous solutions to Einstein's equations to spatially inhomogeneous
solutions, the setting which is the easiest to consider is that of polarised $\tn{3}$-Gowdy symmetric solutions
\index{Polarised $\tn{3}$-Gowdy symmetric solutions}%
to Einstein's vacuum equations. 
In that case, the essential part of the equations is given by one linear, scalar wave equation:
\begin{equation}\label{eq:polarisedGowdy}
P_{tt}-e^{2t}P_{\theta\theta}=0
\end{equation}
on $\sn{1}\times\ro$. Here $t\rightarrow\infty$ corresponds to the expanding direction and $t\rightarrow-\infty$ corresponds to the big bang 
singularity. Again, this is an equation of the form (\ref{eq:thesystemRge}). Even though (\ref{eq:polarisedGowdy}) is a simple equation, 
deriving all the asymptotic information needed in order to prove strong cosmic censorship in this setting does require an effort; cf. 
\cite{chisamo,isamo}. Relaxing the symmetry assumptions by considering $\tn{3}$-Gowdy symmetric solutions,
\index{Tthree@$\tn{3}$-Gowdy symmetric solutions}%
the equation (\ref{eq:polarisedGowdy})
is replaced by a system of non-linear wave equations, but with the same symbol. In this case, analysing the asymptotics is more difficult; cf., e.g.,
\cite{expGowdy,asvelGowdy,SCCGowdy}. Moreover, in the expanding direction, the behaviour of solutions can be radically different from that of 
spatially homogeneous solutions; in fact, spatially homogeneous solutions are unstable. This conclusion is strengthened by the analyses carried
out in \cite{LeFaSm} (in the polarised $\tn{2}$-symmetric setting) and in \cite{instabt2sol} (in the $\tn{2}$-symmetric setting). 
\index{Ttwo@$\tn{2}$-symmetric setting}%
Even though we are here interested in the linear setting, a better understanding of linear systems of equations with symbols similar to that of 
(\ref{eq:polarisedGowdy}) can be expected to be useful in interpreting the results of \cite{expGowdy,LeFaSm,instabt2sol}, and possibly in 
extending them to other situations. 

\textbf{Summary.} As is clear from the above, there is a large number of spacetimes such that studying linear systems of equations of the form 
(\ref{eq:thesystemRge}) on these spacetimes is of interest. In several of the examples we have given, stability has already been demonstrated in 
the non-linear setting. It is therefore reasonable to ask if there is any use in considering linear equations on the corresponding backgrounds. 
The reason we do so here is that in several of the cases mentioned, the results are limited in that they do not currently apply to the full range 
of situations one would expect. In fact, in many of the situations discussed above where there are results, it would be desirable to relax the 
symmetry requirements and/or to extend the results to other matter models. When extending the results, a more detailed understanding of the linear 
setting can be expected to be valuable. This is particularly true in the case of Einstein's equations, since, even for a fixed solution to Einstein's
equations coupled to a specific matter model, there are many different ways of formulating the equations and many different ways of breaking the 
gauge invariance. In order to be able to make an informed choice on how to break the gauge invariance, it is very valuable to have tools allowing
a quick determination of how solutions to the corresponding linearised system behave.

\section{Metrics with convergent asymptotics}\label{section:geoconprefme}

In the context of cosmology, the two types of asymptotic regimes that are of greatest interest are expanding directions and big bang/big crunch 
type singularities. One indicator of expansion is that the volume $V$ tends to infinity, and one indicator of a big bang/big crunch is that the 
volume tends to zero. It is also natural to assume that, at least asymptotically, the derivative of the volume in the direction normal to the 
hypersurfaces $\bM_{t}$ is non-zero; this would ensure a strict monotonicity of the volume. Due to Remark~\ref{remark:UformulaUlnVformintro}
this requirement is equivalent to the assumption that the mean curvature is either strictly positive or strictly negative asymptotically. 
In what follows, we impose these conditions. However, before stating them, let us make the following comments concerning static metrics.

\begin{remark}[Static solutions]
\index{Static solutions}%
Due to the importance of static solutions in the asymptotically flat setting, it is perhaps worth commenting on the properties of such solutions in 
the cosmological setting. We already discussed Minkowski space with $\tn{d}$-spatial topology in Subsection~\ref{ssection:exsepcosmoma}. More 
generally, we would expect a static cosmological solution to have the following three properties: it should be a causally geodesically complete 
solution to Einstein's equations; it should be globally hyperbolic with closed spatial topology; and with respect to a suitable foliation by 
spacelike Cauchy hypersurfaces, say $\Sigma_{t}$, the volume of the leaves $\Sigma_{t}$ should be constant. That the volume of the leaves is 
constant implies that the mean curvature of the leaves vanishes (assuming constant mean curvature). Perturbing the initial data induced on 
$\Sigma_{t}$ should thus yield a sequence 
of initial data for Einstein's equations with strictly positive mean curvature and converging to those induced on $\Sigma_{t}$. 
Appealing to Hawking's singularity theorem (assuming the relevant matter fields appearing satisfy the energy conditions needed in order to appeal 
to Hawking's theorem) yields the conclusion that the maximal Cauchy developments of the members of this sequence are all causally geodesically 
incomplete; cf., e.g., \cite[Theorem~55B, p.~432]{oneill}. This means that the solution we started with is unstable. For this reason, static solutions 
are not natural in the context of cosmology. In fact, the asymptotics are typically characterised by expansion and/or contraction in the cosmological 
setting. 
\end{remark}

Due to the above observations, we here focus on separable cosmological model manifolds with one of the following properties. 

\begin{definition}\label{def:futureexpetcintro}
Let $(M,g)$ be a separable cosmological model manifold. 
\begin{itemize}
\item If $V(t)\rightarrow\infty$ as $t\rightarrow t_{+}-$ and there is a $t_{0}\in I$ such that $U(V)>0$ for $t\geq t_{0}$, then $(M,g)$
is said to be \textit{future expanding}. 
\index{Future!expanding separable cosmological model manifold}%
\index{Separable cosmological model manifold!future expanding}%
\item If $V(t)\rightarrow\infty$ as $t\rightarrow t_{-}+$ and there is a $t_{0}\in I$ such that $U(V)<0$ for $t\leq t_{0}$, then $(M,g)$
is said to be \textit{past expanding}. 
\index{Past!expanding separable cosmological model manifold}%
\index{Separable cosmological model manifold!past expanding}%
\item If $V(t)\rightarrow 0$ as $t\rightarrow t_{+}-$ and there is a $t_{0}\in I$ such that $U(V)<0$ for $t\geq t_{0}$, then $(M,g)$
is said to have \textit{big crunch asymptotics}. 
\index{Big crunch asymptotics}%
\item If $V(t)\rightarrow 0$ as $t\rightarrow t_{-}+$ and there is a $t_{0}\in I$ such that $U(V)>0$ for $t\leq t_{0}$, then $(M,g)$
is said to have \textit{big bang asymptotics}. 
\index{Big bang asymptotics}%
\end{itemize}
\end{definition}
\begin{remark}
Let $(M,g)$ be a separable cosmological model manifold. If it is future expanding, then reversing the time coordinate yields a separable 
cosmological model manifold which is past expanding (and vice versa). If $(M,g)$ has big crunch asymptotics, then reversing the time coordinate 
yields a separable cosmological model manifold with big bang asymptotics (and vice versa). For that reason, we here restrict our attention to 
$(M,g)$'s which are future expanding or have big crunch asymptotics. 
\end{remark}
\begin{remark}
Due to Remark~\ref{remark:UformulaUlnVformintro}, $\pm U(V)>0$ is equivalent to $\pm \tr_{\bge}\bk>0$. The conditions involving $U(V)$ could thus
be reformulated in terms of $\tr_{\bge}\bk$. 
\end{remark}
The assumptions introduced in Definition~\ref{def:futureexpetcintro} are not very restrictive. In order to obtain a class
of geometries such that we can draw conclusions concerning solutions to (\ref{eq:thesystemRge}), we need to impose additional conditions. 
One object on which it is natural to impose conditions is the second fundamental form. In the formulation of the conditions, it is convenient to 
consider $\bk$ to be a map from $T\bM$ to itself. This can be achieved by raising one of the indices of $\bk$ by $\bge$, and we denote the 
corresponding object by $\bK$. Note that $\bK_{i}^{\phantom{i}j}=\bk_{i}^{\phantom{i}j}$ with respect to local coordinates. In particular, 
$\tr \bK=\tr_{\bge}\bk$ (note that $\tr\bK$ does not depend on a choice of metric, since for each $p\in\bM$, $(\bK)_{p}\in\mathrm{End}(T_{p}\bM)$).
In many of the situations of interest, $\tr\bK$ either tends to zero or $\pm\infty$ asymptotically. It is therefore natural to normalise
$\bK$. One way of doing so is to focus on $\bK/\tr\bK$. In what follows, we refer to this object as the \textit{expansion normalised second
fundamental form}. 
\index{Expansion normalised!second fundamental form}%
\index{Second fundamental form!expansion normalised}%
If the assumptions of Definition~\ref{def:futureexpetcintro} are satisfied, the expansion normalised second fundamental form
is at least asymptotically well defined. In these notes, we are mainly interested in the case that $\bK/\tr\bK$ converges, a situation we refer 
to as the convergent setting. However, it also turns out to be important to impose conditions on the normal derivative of the mean curvature. 
Imposing such restrictions may seem to be less natural. However, in the context of Einstein's vacuum equations, the relevant conditions often follow 
from a combination of 
the assumption of convergence and the so-called Raychaudhuri equation. In the presence of matter, additional conditions may need to be imposed.
However, the required conditions sometimes follow by combining the assumption of convergence, the Hamiltonian constraint, the Raychaudhuri equation 
and, possibly, energy conditions. Considering, e.g., solutions to Einstein's vacuum equations corresponding to left invariant initial data on 
$3$-dimensional unimodular Lie groups yields solutions such that the following version of the Raychaudhuri equation
\index{Raychaudhuri equation}%
holds:
\[
U(\tr_{\bge}\bk)=-\bk^{ij}\bk_{ij};
\]
cf., e.g., \cite[(20.17), p.~218]{minbok}. This means that if $\bK/\tr\bK$ is convergent (or bounded), then $U[(\tr_{\bge}\bk)^{-1}]$ is 
convergent (or bounded). Nevertheless, in these notes, we do not assume $(M,g)$ to be a solution to Einstein's equations. For this reason, 
we here impose conditions directly on $U[(\tr_{\bge}\bk)^{-1}]$. 

Even though we are mainly interested in the convergent setting, it is sometimes sufficient to assume boundedness. For this reason, we introduce
the following terminology. 
\begin{definition}\label{definition:futbdfutconvgeometryintro}
Let $(M,g)$ be a separable cosmological model manifold and let $\varrho$ be a fixed Riemannian metric on $\bM$. Then $(M,g)$ is said to have 
\textit{future bounded geometry} 
\index{Future!bounded geometry of a spacetime}%
if there is a constant $0<C\in\ro$ and a $t_{0}\in I$ such that $|\tr_{\bge}\bk|>0$ and
\[
|\bk|_{\bge}/|\tr_{\bge}\bk|+|U[(\tr_{\bge}\bk)^{-1}]|\leq C
\]
for all $t\geq t_{0}$. If, in addition, there is a $2$-tensor field $A$ on $\bM$ of mixed type, a constant $a\in\ro$ and constants $0<C,\eta\in\ro$ such
that 
\begin{equation}\label{eq:bkdbtrkconvwithrateintro}
|\bK/\tr\bK-A|_{\varrho}+|U[(\tr_{\bge}\bk)^{-1}]-a|\leq C\exp[-\eta |\ln V(t)|]
\end{equation}
for all $t\geq t_{0}$, then $(M,g)$ is said to be \textit{future convergent}.
\index{Future!convergent spacetime}%
\end{definition}
\begin{remark}
It is of course possible to define the notion of future convergence without a rate. However, in these notes we are only interested in convergence in
the case that there are quantitative bounds of the form (\ref{eq:bkdbtrkconvwithrateintro}). 
\end{remark}
At this stage, it is of interest to make use of the invariance described in Remark~\ref{remark:invofthesystemRgeundconfresc}. By conformally
rescaling the metric and then changing the time coordinate, we are in a position to reduce it to what we refer to as ``canonical form''. 
\begin{lemma}\label{lemma:bdandconvtocanonicalform}
Let $(M,g)$ be a separable cosmological model manifold, cf. Definition~\ref{def:sepcosmmodmanintro}, and let $\varrho$ be a fixed Riemannian
metric on $\bM$. Assume that $(M,g)$ is future expanding and has future bounded geometry; cf. Definitions~\ref{def:futureexpetcintro} and 
\ref{definition:futbdfutconvgeometryintro}. Introduce the metric $\hg:=(\tr_{\bge}\bk)^{2}g$ and the time coordinate
\[
\tau(t):=\ln\frac{V(t)}{V(t_{0})}.
\]
Then the interval $[t_{0},t_{+})$ in $t$-time corresponds to $[0,\infty)$ in $\tau$-time. Moreover, $\hg$ is well defined on 
$[0,\infty)$ in $\tau$-time and can be written
\begin{equation}\label{eq:hgitotauintro}
\hg=-d\tau\otimes d\tau+\hg_{ij}(\hchi^{i}d\tau+dx^{i})\otimes (\hchi^{j}d\tau+dx^{j})+\textstyle{\sum}_{r=1}^{R}\ha_{r}^{2}g_{r}.
\end{equation}
If $\chg$ and $\hk$ are the metric and second fundamental form induced on constant $\tau$-hypersurfaces by $\hg$, then there is a constant 
$0<C\in\ro$ such that  $|\hk|_{\chg}\leq C$ for all $\tau\geq 0$. 

Assuming, in addition to the above, that $(M,g)$ is future convergent, there is a $2$-tensor field $\hA$ on $\bM$ of mixed type such that 
\begin{equation}\label{eq:bkdbtrkconvwithratehverintro}
|\hK-\hA|_{\varrho}\leq Ce^{-\eta\tau}
\end{equation}
for all $\tau\geq 0$, where $0<C,\eta\in\ro$ and $\hK$ is obtained from $\hk$ by raising one of the indices using $\chg$. 
\end{lemma}
\begin{remark}
The proof of this statement is to be found in Section~\ref{section:geometricconvergence}. Moreover, there is a similar result in the 
case of big crunch asymptotics. 
\end{remark}
\begin{remark}
The idea to conformally rescale the metric by multiplying it by the square of the mean curvature is not new; cf., e.g., \cite{huar} in which 
this idea is applied in a much more general (cosmological) setting. 
\end{remark}
\begin{remark}
The functions $\hg_{ij}$, $\ha_{r}$ and $\hchi^{i}$ only depend on $\tau$. Moreover, $\hg(\d_{\tau},\d_{\tau})<0$ by construction. This means that 
$\hg_{ij}\hchi^{i}\hchi^{j}<1$. 
\end{remark}

\subsection{Canonical separable cosmological model manifolds}\label{ssection:cansepcomoma}

With the above observations in mind, it is natural to introduce the following definition.

\begin{definition}\label{def:cansepcosmmodmanintro}
A \textit{canonical separable cosmological model manifold}
\index{Canonical separable cosmological model manifold}%
is a separable cosmological model manifold such that the interval $I=(t_{-},t_{+})$ contains $[0,\infty)$ and such that $N$ appearing in 
(\ref{eq:lapseshiftsepcosmodmeintro}) equals $1$.
\end{definition} 
\begin{remark}\label{remark:guzzcansepcosmmodmet}
Due to the fact that $N=1$, it can be concluded that $g^{00}=-1$; cf. Chapter~\ref{chapter:geometry}, in particular 
Lemma~\ref{lemma:invitoshiftetc}.
\end{remark}
\begin{remark}
In order to obtain a canonical separable cosmological model manifold, it is sufficient to assume that $(M,g)$ is future expanding; to 
conformally rescale $g$ according to $\hg:=(\tr_{\bge}\bk)^{2}g$; and to change the time coordinate as in 
Lemma~\ref{lemma:bdandconvtocanonicalform}. There is a similar statement in the case of big crunch asymptotics.
\end{remark}
\begin{remark}
We are mainly interested in the case that $\bK$ converges to a non-trivial mixed $2$-tensor field on $\bM$; cf. 
Lemma~\ref{lemma:bdandconvtocanonicalform}. However, we prefer to impose specific conditions on $\bK$ in the statements of the results
rather than including it in the definition. 
\end{remark}

It is useful to have some specific examples in mind before proceeding. 

\textbf{Solutions with exponential expansion.} Consider the metrics $g_{\rodS}$, $g_{\mathrm{gdS}}$ and $g_{\mathrm{N}}$, introduced in (\ref{eq:flatdS}),
(\ref{eq:ggendS}) and (\ref{eq:Nariai}) respectively and defined on $\tn{d}\times\ro$, $\Sigma\times\ro$ and $\so\times\sn{2}\times\ro$ 
respectively. They are all canonical separable cosmological model manifolds. Moreover, in the case of $(\tn{d}\times\ro,g_{\rodS})$, 
$\bK=H\Id_{T\tn{d}}$, where we think of $\bK$ as an element of $\mathrm{End}(T\tn{d})$. In the case of $(\Sigma\times\ro,g_{\mathrm{gdS}})$, 
\[
\bK=H\frac{\sinh(Ht)}{\cosh(Ht)}\Id_{T\Sigma}\rightarrow \pm H\Id_{T\Sigma}
\]
as $t\rightarrow\pm\infty$. Finally, in the case of $(\so\times\sn{2}\times\ro,g_{\mathrm{N}})$, 
\[
\bK=H\frac{\sinh(Ht)}{\cosh(Ht)}\Pi_{T\so}\rightarrow \pm H\Pi_{T\so}
\]
as $t\rightarrow\pm\infty$, where $\Pi_{T\so}\in \mathrm{End}[T(\so\times\sn{2})]$ is the projection onto $T\so$. 

\textbf{Examples, Bianchi type I.} Let us consider a metric of the form (\ref{eq:gBianchiI}), where $a_{i}(t)=\a_{i}t^{p_{i}}$, $0<\a_{i}\in\ro$
and $p_{i}\in\ro$. Metrics of this form arise, e.g., in the vacuum setting, in the study of the Einstein-scalar field equations and in the 
study of the Einstein-non-linear scalar field equations with an appropriate exponential potential; cf. Subsection~\ref{ssection:exsepcosmoma}
above. In all of these examples, the sum of the $p_{i}$'s, say $p$, is strictly positive. 

\textit{The expanding direction.} Say now that we want to study the expanding direction of these solutions. According to 
Lemma~\ref{lemma:bdandconvtocanonicalform}, it is then natural to introduce a time coordinate $\tau(t)=\ln (t/t_{0})^{p}$. Choosing $t_{0}=1$, this 
yields $\tau=p\ln t$. On the other hand, 
\[
\bk=\textstyle{\sum}_{i=1}^{d}p_{i}t^{-1}\a_{i}^{2}t^{2p_{i}}dx^{i}\otimes dx^{i},
\]
so that $\tr_{\bge}\bk=pt^{-1}$. Multiplying $g$ (given by (\ref{eq:gBianchiI}) with $a_{i}$ as above) with $(\tr_{\bge}\bk)^{2}$ thus yields
\begin{equation}\label{eq:BianchiIpol}
\begin{split}
(\tr_{\bge}\bk)^{2}g = & -p^{2}t^{-2}dt\otimes dt+\textstyle{\sum}_{i=1}^{d}p^{2}\a_{i}^{2}t^{2(p_{i}-1)}dx^{i}\otimes dx^{i}\\
 = & -d\tau\otimes d\tau+\textstyle{\sum}_{i=1}^{d}\b_{i}^{2}e^{2\g_{i}\tau}dx^{i}\otimes dx^{i},
\end{split}
\end{equation}
where $\b_{i}=p\a_{i}$ and $\g_{i}=(p_{i}-1)/p$. Note that in the case of the Kasner solutions, $p=1$ and the sum of the squares of the $p_{i}$
equals one. There are thus two possibilities: either one $p_{i}$ equals $1$ and all the rest equal zero (the corresponding metrics are referred 
to as \textit{flat Kasner solutions}, 
\index{Flat!Kasner solution}%
\index{Kasner!flat solution}%
since they have a vanishing Riemann curvature tensor - in fact, they are quotients of a part of $d+1$-dimensional Minkowski space); or all the 
$p_{i}<1$. In the first case, there is one $\g_{i}=0$ and all the rest 
equal $-1$. In the second case, all the $\g_{i}<0$. In the second case, it is thus clear that there is contraction in all directions, even though 
the original solution was expanding. Denoting the metric in (\ref{eq:BianchiIpol}) by $\hg$, the associated second fundamental form (of the 
hypersurfaces of constant $\tau$) is given by 
\[
\hk=\textstyle{\sum}_{i=1}^{d}\g_{i}\b_{i}^{2}e^{2\g_{i}\tau}dx^{i}\otimes dx^{i}.
\]
In particular, it is clear that $\hK$ is constant. Moreover, the eigenvalues are given by the $\g_{i}$'s. 

\textit{The contracting direction.} Turning to the contracting direction, it is natural to introduce the time coordinate $\tau=-p\ln t$. 
The computation (\ref{eq:BianchiIpol}) is still valid. The only difference in the contracting direction is that $\g_{i}=(1-p_{i})/p$. In the
case of the Kasner solutions, the typical situation is thus that the metric given by the far right hand side of (\ref{eq:BianchiIpol}) 
exhibits exponential expansion. In the case of a non-vacuum solution to the Einstein-scalar field equations, the sum of the squares of the 
$p_{i}$'s is strictly less than $1$. For that reason the $\g_{i}$'s are always strictly positive, and the metric exhibits exponential expansion. 
In both cases, $\hK$ is constant. 

\textbf{The Milne model.} In this case, the volume is given by $V(t)=t^{d}\rovol_{g_{\Sigma}}(\Sigma)$, where $d$ is the dimension of $\Sigma$. Fixing 
$t_{0}=1$, it is therefore natural to introduce $\tau(t)=d\cdot\ln t$. On the other hand, the second fundamental form is given by 
$\bk=tg_{\Sigma}$, so that $\tr_{\bge}\bk=d\cdot t^{-1}$. Thus
\[
\hg_{\roM}:=(\tr_{\bge}\bk)^{2}g_{\roM}=-d^{2}t^{-2}dt\otimes dt+d^{2}g_{\Sigma}=-d\tau\otimes d\tau+d^{2}g_{\Sigma}.
\]
In particular, the associated second fundamental form is identically zero, even though the Milne model exhibits expansion. 

\textbf{$U(1)$ symmetric solutions.} In this case, $V(t)=2\pi t^{2}\rovol_{g_{\Sigma}}(\Sigma)$. Letting $t_{0}=1$ yields $\tau(t)=2\ln t$.
Moreover, $\tr_{\bge}\bk=2t^{-1}$, so that 
\[
\hg_{\roU}:=-4t^{-2}dt\otimes dt+4t^{-2}d\theta\otimes d\theta+4g_{\Sigma}=-d\tau\otimes d\tau+4e^{-\tau}d\theta\otimes d\theta+4g_{\Sigma}.
\]

\textbf{Model metrics.} Considering the above examples, it is clear that metrics of the following form appear naturally:
\index{Model metric}%
\begin{equation}\label{eq:modelmetricafterresc}
g=-dt\otimes dt+\textstyle{\sum}_{r=1}^{R}\a_{r}^{2}e^{2\b_{r}t}g_{r},
\end{equation}
where $0<\a_{r}\in\ro$, $\b_{r}\in\ro$ and $(M_{r},g_{r})$ are closed Riemannian manifolds (here $(M_{r},g_{r})$ could equal $(\so,d\theta\otimes d\theta)$, 
so that the $d$-torus part appearing in (\ref{eq:sepcosmodmeintro}) should be thought of as being included in (\ref{eq:modelmetricafterresc})). In
fact, we are mainly interested in metrics that, asymptotically, behave as (\ref{eq:modelmetricafterresc}). However, in several of the results of these
notes we impose much weaker conditions. 

\subsection{Silent, transparent and noisy metrics}\label{ssection:siltrsanoimet}

The behaviour of solutions to (\ref{eq:thesystemRge}) is strongly dependent on the asymptotic behaviour of the corresponding metric 
(after a conformal rescaling and change of time coordinate of the type described in Lemma~\ref{lemma:bdandconvtocanonicalform}). In 
Chapters~\ref{chapter:silentequations}, \ref{chapter:transpeq} and \ref{chapter:domnoisspdirintro} we introduce the notions of silent equations; 
transparent equations; and equations with a dominant noisy spatial direction. The purpose of this terminology is to allow us to distinguish between
asymptotics of fundamentally different type. It would be premature to formulate the corresponding definitions
here, but it is useful to define what it means for a model metric (\ref{eq:modelmetricafterresc}) to be silent, transparent or noisy. 

\begin{definition}
Consider a metric of the form (\ref{eq:modelmetricafterresc}). If all the $\b_{r}$ are strictly positive, the metric is said to be \textit{silent}; 
\index{Model metric!silent}%
\index{Silent!model metric}%
if all the $\b_{r}\geq 0$ but there is one $\b_{r}=0$, then the metric is said to be \textit{transparent};
\index{Model metric!transparent}%
\index{Transparent!model metric}%
 and if there is one $\b_{r}$ which is strictly 
negative, then the metric is said to be \textit{noisy}. 
\index{Model metric!noisy}%
\index{Noisy!model metric}%
\end{definition}
In order to motivate the terminology, we consider the three cases separately. 

\textbf{Silent metrics.} Let $g$ be a silent metric of the form (\ref{eq:modelmetricafterresc}) and let $\g$ be a causal curve with respect to this 
metric (so that $g(\dot{\g},\dot{\g})\leq 0$). Note that, in general relativity, light and material objects (or, more generally, information) are assumed
to travel along causal curves, so that causal curves are often spoken of as ``observers''. By a reparametrisation, if necessary, we can assume $\g$ to be of 
the form $\g(t)=(\bga_{1}(t),\dots,\bga_{R}(t),t)$, where $\bga_{r}$ takes its values in $M_{r}$. The assumption of causality is equivalent to 
\[
-1+\textstyle{\sum}_{r=1}^{R}\a_{r}^{2}e^{2\b_{r}t}|\dot{\bga}_{r}(t)|^{2}_{g_{r}}\leq 0.
\]
In particular, $|\dot{\bga}_{r}(t)|_{g_{r}}\leq \a_{r}^{-1}e^{-\b_{r}t}$. This means that $\bga_{r}(t)$ converges to a point $p_{r}\in M_{r}$. If $\lambda$
is another causal curve, we obtain corresponding points $q_{r}\in M_{r}$. For two typical curves, there is one $r\in \{1,\dots,R\}$ such that 
$p_{r}\neq q_{r}$. Given two such curves, there is a $t_{1}$ such that for $t\geq t_{1}$, it is not possible to send information from $\g(t)$ to 
$\lambda$ (in other words, there is no future directed causal curve $\xi$, starting at $\g(t)$ and intersecting $\lambda$ to the future). Due to 
this inability of observers to communicate at late times, the metric (\ref{eq:modelmetricafterresc}) is said to be silent. 

\textbf{Transparent metrics.} Let $g$ be a transparent metric of the form (\ref{eq:modelmetricafterresc}) and assume that $\b_{r}=0$. Given two 
distinct points $p_{r},q_{r}\in M_{r}$, let $\bga_{r}$ be a smooth simple closed curve in $M_{r}$ with constant speed $\a_{r}^{-1}$ (with respect to $g_{r}$) 
such that the image of $\bga_{r}$ contains $p_{r}$ and $q_{r}$. Assume, moreover, that $\bga_{r}(0)=p_{r}$. We think of $\bga_{r}$ as being defined 
on $\ro$ and define $\g(t)=(\bga_{1}(t),\dots,\bga_{R}(t),t)$, 
where $\bga_{s}$ is a constant curve in $M_{s}$ for $s\neq r$. Then $\g$ is a null curve and the $M_{r}$ coordinate of $\g$ equals $p_{r}$ and $q_{r}$ 
infinitely many times. In particular, the spatial coordinate of the curve does not converge. Moreover, if $t_{n}\geq 0$, $1\leq n\in\zo$, are the 
non-negative times at which the $M_{r}$ coordinate of $\g$ equals $p_{r}$, then $t_{n+1}-t_{n}$ is independent of $n$. We think of $\g$ as representing a 
signal propagating in $M$ and let $p\in\bM$ be defined by $(p,0)=\g(0)$. Then the signal $\g$ passes through $(p,t_{n})$ for each $n$. If $\lambda$ is the
observer sitting at $p$, i.e. $\lambda(t)=(p,t)$, then $\lambda$ receives the signal (propagating along $\g$) at times $t_{n}$, and the proper time elapsed 
between received signals (as measured by $\lambda$) equals $t_{n+1}-t_{n}$. In particular, the proper time elapsed is independent of $n$. The word 
transparent is intended to capture the combination of the two properties that information can be sent (along null curves) between points on $M_{r}$ and 
that the time elapsed between received 
signals remains constant. Note, however, that it is not necessarily possible to send information between an arbitrary pair of points on $\bM$. From this 
point of view, it would perhaps be more natural to say that the metric is partially transparent. However, we here use the terminology introduced 
above. 

\textbf{Noisy metrics.} Let $g$ be a noisy metric of the form (\ref{eq:modelmetricafterresc}) and assume that $\b_{r}<0$. Let $p_{r}$, $q_{r}$ and
$\bga_{s}$, $s\in \{1,\dots,R\}$, be as in the transparent setting above, and define
\[
\g(t)=[\bga_{1}(-\b_{r}^{-1}e^{-\b_{r}t}),\dots,\bga_{R}(-\b_{r}^{-1}e^{-\b_{r}t}),t].
\]
Then $\g$ is a null curve and the $M_{r}$ coordinate of $\g$ equals $p_{r}$ and $q_{r}$ infinitely many times. In particular, 
the spatial coordinate of the curve does not converge. Letting $t_{n}$ be defined as in the transparent setting, there is a constant $C$ such that 
\[
t_{n+1}-t_{n}\leq Ce^{\b_{r}t_{n}}.
\]
In particular, the time elapsed between received signals decays exponentially. In short, information can be sent (along null curves) between points on 
$M_{r}$ and the time elapsed between received signals decays exponentially. The word noisy is intended to capture the combination of these two properties. 
Again, it could be argued that the terminology partially noisy would be more justified.

\section{Questions}\label{section:Questions}

\textbf{Energy estimates.}
Our main goal in these notes is to obtain optimal energy estimates for solutions to subclasses of systems of the form (\ref{eq:thesystemRge}). 
In what follows, we are not interested in the most general case, but rather assume the underlying geometry to be a canonical separable cosmological model 
manifold, so that, in particular, $g^{00}=-1$. In other words, we assume that we have already appealed to 
Lemma~\ref{lemma:bdandconvtocanonicalform}, if necessary, and that the relevant metric and time coordinate are the conformally rescaled metric and 
logarithmic ``volume'' time coordinate. In this setting, the main energy of interest is 
\begin{equation}\label{eq:mfebasdef}
\begin{split}
\mfe_{\robas}[u](t) = & \frac{1}{2}\int_{\bM}\left(|u_{t}(\cdot,t)|^{2}+\textstyle{\sum}_{i=1}^{m}[g^{kl}(t)\d_{k}u_{i}(\cdot,t)\d_{l}u_{i}^{*}(\cdot,t)
\phantom{\textstyle{\sum}_{r=1}^{R}}\right.\\
 & \left. \phantom{\textstyle{\frac{1}{2}\sum}_{i}(}+\textstyle{\sum}_{r=1}^{R}
a_{r}^{-2}(t)|\grad_{g_{r}}u_{i}(\cdot,t)|_{g_{r}}^{2}]+|u(\cdot,t)|^{2}\right)\mubox,
\end{split}
\end{equation}
\index{$\a$Aa@Notation!Energies!$\mfe_{\robas}$}%
where $\bM$ is given by (\ref{eq:bMdef}), 
\begin{equation}\label{eq:muboxdef}
\mubox:=dx\wedge \mu_{g_{1}}\wedge\cdots\wedge\mu_{g_{R}},
\end{equation}
\index{$\a$Aa@Notation!Measures!$\mubox$}%
$dx$ is the standard volume form on $\tn{d}$ and $\mu_{g_{r}}$ and $\grad_{g_{r}}$ are the volume form and gradient associated with the Riemannian manifold 
$(M_{r},g_{r})$ respectively; we assume $M_{r}$ to be oriented. Moreover, if $X_{j}$, $j=1,2$, are two real vector fields on $M_{r}$ and $X=X_{1}+iX_{2}$
is a complex vector field, then
\[
|X|_{g_{r}}^{2}:=|X_{1}|_{g_{r}}^{2}+|X_{2}|_{g_{r}}^{2}.
\]
From now on we refer to $\mfe_{\robas}[u]$ as the \textit{basic energy}.
\index{Basic energy}%
There are also associated higher order energies. However, due to the
separability of the equation, estimates for the higher order energies are direct consequences of the estimates for the basic energy. Even though
$\mfe_{\robas}[u]$ is the energy of greatest interest, it is sometimes convenient to consider
\begin{equation}\label{eq:mfehomdefintro}
\begin{split}
\mfe_{\rohom}[u](t) = & \frac{1}{2}\int_{\bM}\left(|u_{t}(\cdot,t)|^{2}+\textstyle{\sum}_{i=1}^{m}[g^{kl}(t)\d_{k}u_{i}(\cdot,t)\d_{l}u_{i}^{*}(\cdot,t)
\phantom{\textstyle{\sum}_{r=1}^{R}}\right.\\
 & \left. \phantom{\textstyle{\frac{1}{2}\sum}_{i}(}+\textstyle{\sum}_{r=1}^{R}a_{r}^{-2}(t)|\grad_{g_{r}}u_{i}(\cdot,t)|_{g_{r}}^{2}]\right)\mubox.
\end{split}
\end{equation}
\index{$\a$Aa@Notation!Energies!$\mfe_{\rohom}$}%

There are several ways of formulating questions related to optimal decay. However, we are here mainly interested in equations such that the basic energy 
grows or decays exponentially (in some degenerate situations, polynomial factors might also occur, but they represent subdominant behaviour). In such
settings it is natural to introduce the following numbers.

\begin{definition}\label{eq:defdecayrates}
Consider (\ref{eq:thesystemRge}). Assume that the associated Lorentz manifold $(M,g)$ is a canonical separable cosmological model manifold. 
Assume, moreover, that $f=0$. Let $\ms$ denote the set of smooth solutions to (\ref{eq:thesystemRge}) (with $f=0$). Assume the set
\[
\ma_{\rocr}:=\{\eta\in\ro\ | \ \forall u\in\ms\ \exists C\in\ro:\mfe_{\robas}[u](t)\leq Ce^{2\eta t}\ \forall t\geq 0\}
\]
to be non-empty and bounded from below. Then $\cruderate$ 
\index{$\a$Aa@Notation!Constants!$\cruderate$}%
is defined to be the infimum of $\ma_{\rocr}$ and is called the \textit{crude exponential bound}
\index{Crude!exponential bound}%
\index{Exponential bound!crude}%
for the growth of the basic energy of solutions to (\ref{eq:thesystemRge}) with $f=0$. Assume the set 
\begin{equation}\label{eq:maronldef}
\ma_{\ronl}:=\{\eta\in\ro\ | \ \exists C\in\ro:\forall u\in\ms,\ \mfe_{\robas}[u](t)\leq C\mfe_{\robas}[u](0)e^{2\eta t}\ \forall t\geq 0\}
\end{equation}
to be non-empty and bounded from below. Then $\nolossrate$ 
\index{$\a$Aa@Notation!Constants!$\nolossrate$}%
is defined to be the infimum of $\ma_{\ronl}$ and is called the \textit{exponential bound}
\index{Exponential bound}%
for the growth of the basic energy of solutions to (\ref{eq:thesystemRge}) with $f=0$. 
\end{definition}
\begin{remark}
Note that $\cruderate\leq\nolossrate$. As we demonstrate below, however, this is the only restriction on the relation between these two numbers. 
\end{remark}
\begin{remark}
Clearly, other definitions are conceivable. For instance, we could replace $\mfe_{\robas}[u](0)$ appearing in (\ref{eq:maronldef}) by a higher order
energy of $u$ at $t=0$, corresponding to, say, the $H^{s_{0}+1}\times H^{s_{0}}$-norm of the initial data. For each $s_{0}\geq 0$, we then obtain a constant, 
say $\a_{\rol}(s_{0})$, describing the behaviour of the energy. Then $\cruderate\leq\a_{\rol}(s_{0})\leq\nolossrate$, and $\a_{\rol}(s_{0})$ is a decreasing 
function of $s_{0}$. Nevertheless, we here focus on the two constants $\cruderate$ and $\nolossrate$. 
\end{remark}

Our main goal in these notes is to calculate $\nolossrate$ for large classes of equations. However, we are also interested in obtaining asymptotic 
information.

\textbf{Asymptotics.} In order to give an idea of the type of asymptotic information we would like to derive, let us first give two 
very simple examples. 

\textit{The contracting direction of polarised $\tn{3}$-Gowdy.} Consider the equation (\ref{eq:polarisedGowdy}) with the time reversed; i.e., the equation 
\begin{equation}\label{eq:polGowdytimereversed}
P_{tt}-e^{-2t}P_{\theta\theta}=0.
\end{equation}
Then the asymptotics in the direction $t\rightarrow\infty$ correspond to the behaviour close to the big bang singularity of polarised 
$\tn{3}$-Gowdy symmetric solutions to Einstein's vacuum equations. Given a smooth solution $P$ to (\ref{eq:polGowdytimereversed}), there are  
$v,\psi\in C^{\infty}(\sn{1},\ro)$ such that 
\begin{equation}\label{eq:polGowdyasympt}
P(\cdot,t)-vt-\psi,\ \ \
P_{t}(\cdot,t)-v
\end{equation}
both decay exponentially in any $C^{k}$-norm as $t\rightarrow\infty$. In fact, one can demonstrate that there is a homeomorphism (in the 
$C^{\infty}$-topology) between initial data to (\ref{eq:polGowdytimereversed}) and asymptotic data (given by $v$ and $\psi$ as described above).
For a justification of these statements, the reader is referred to Example~\ref{example:polt3gowdyderofas} below. 

\textit{The expanding direction of polarised $\tn{3}$-Gowdy.}
Returning to (\ref{eq:polarisedGowdy}), it is clear that the asymptotics as $t\rightarrow\infty$ are quite different from those in the opposite
time direction. However, changing time coordinate to $\tau=e^{t}$, (\ref{eq:polarisedGowdy}) becomes
\begin{equation}\label{eq:polarisedGowdyoriginaltime}
P_{\tau\tau}+\frac{1}{\tau}P_{\tau}-P_{\theta\theta}=0
\end{equation}
on $\sn{1}\times (0,\infty)$. Given a solution $P$ to this equation, there are two constants, say $a$ and $b$, and a solution $\nu$ to the flat 
space wave equation with zero mean value over the circle, such that 
\[
P(\cdot,\tau)-a\ln\tau-b-\tau^{-1/2}\nu(\cdot,\tau),\ \ \
P_{\tau}(\cdot,\tau)-a\tau^{-1}-\tau^{-1/2}\nu_{\tau}(\cdot,\tau)
\]
decay to zero as $\tau^{-3/2}$ with respect to any $C^{k}$-norm. In this case, $a$, $b$ and $\nu$ thus constitute the asymptotic data. In fact, in
this case there is also a homeomorphism (in the $C^{\infty}$-topology) between initial data to (\ref{eq:polarisedGowdyoriginaltime}) and the asymptotic 
data (given by $a$, $b$ and $\nu$ as described above, where $\nu$ is topologised by its initial data). For a justification of these statements, the 
reader is referred to Example~\ref{example:polt3gowdyexpdiras} below.

The equations (\ref{eq:polarisedGowdy}) and (\ref{eq:polGowdytimereversed}) are two very simple examples of systems of the form 
(\ref{eq:thesystemRge}). However, it is of interest to derive analogous asymptotic information, given a solution to (\ref{eq:thesystemRge}). 
Moreover, we would like to demonstrate that the map from initial data to asymptotic data is continuous with respect to a suitable topology. Finally, 
it is of interest to prescribe asymptotic data, and to demonstrate that the map from asymptotic data to initial data is continuous with 
respect to, say, the $C^{\infty}$-topology. These are our goals as far as the asymptotics are concerned.

\section{Outline of the introductory part}

The outline of the remainder of Part~\ref{part:introduction} is the following. 


\textbf{On the notion of balance.}
Considering (\ref{eq:thesystemRge}), it is clear that it is determined by the metric $g$; the coefficients $\a$, $\zeta$ and $X^{j}$; and 
the right hand side $f$. In Section~\ref{section:geoconprefme} above, we introduce conditions on the metric $g$. Even though
most of the results in these notes require stronger assumptions, these conditions give a good idea of the class of metrics we have in mind. 
However, at this stage it is unclear what to assume concerning $\a$, $\zeta$ and the $X^{i}$. In
Chapter~\ref{chapter:onnotofbal}, we therefore describe some of the pathologies that can arise if the assumptions concerning these
functions are too weak. It is natural to start by focusing on $\a$ and $\zeta$. Consider (\ref{eq:thesystemRge}) with $g^{00}=-1$ and $f=0$; cf. 
Remark~\ref{remark:guzzcansepcosmmodmet}. The spatially homogeneous solutions to this equation satisfy
\begin{equation}\label{eq:sphomandhomverbalintro}
u_{tt}+\a(t)u_{t}+\zeta(t)u=0.
\end{equation}
If we allow, e.g., $\|\a\|$ or $\|\zeta\|$ to grow exponentially, there are corresponding equations 
whose solutions grow \textit{super exponentially}
\index{Super exponential!growth}%
\index{Growth!super exponential}%
 (i.e. as $\exp[a\exp(bt)]$ for some $a,b>0$); cf. Section~\ref{section:introductiononthenotionofbal}. There are analogous results if $\|\a\|$ or 
$\|\zeta\|$ grow, say, polynomially. Since we here think of the Lorentz manifolds and time coordinates of interest as arising from an application
of Lemma~\ref{lemma:bdandconvtocanonicalform}, we think of the volume of the time slices in the original manifold as equalling $V(t)=V(t_{0})e^{t-t_{0}}$
(assuming the original Lorentz manifold to be future expanding). From this perspective, the natural length scale is proportional to $e^{t/D}$, where
$D$ is the dimension of $\bM$. For this reason, we here consider growth faster than exponential as 
pathological. As a consequence, we are mainly interested in equations of the form (\ref{eq:thesystemRge}) with a metric satisfying the assumptions of 
Definition~\ref{def:cansepcosmmodmanintro} and matrices $\a$ and $\zeta$ whose norms are bounded to the future. There are similar arguments in the 
case of Lorentz manifolds with big crunch asymptotics. 

Concerning the $X^{i}$, it turns out to be natural to focus on $Y^{i}(t):=X^{i}(t)/[g^{ii}(t)]^{1/2}$ (no summation). If there is a $\b>0$
such that $e^{-\b t}Y^{i}(t)$ converges to a matrix with an eigenvalue with non-zero real part, then the results of Chapter~\ref{chapter:onnotofbal}
demonstrate that the energies of solutions to (\ref{eq:thesystemRge}) typically grow super exponentially (additional technical assumptions are 
needed in order to obtain this conclusion, but we refer the reader to Chapter~\ref{chapter:onnotofbal} for the details). In 
Section~\ref{section:unbaeqfiexsupexpgr}, we state such results for homogeneous equations of the type (\ref{eq:thesystemRge})
with $d=1$ and $R=0$. The solutions with super exponential growth that we construct are smooth. However, if $g^{11}$ decays exponentially, we are 
not able to prove super exponential growth of solutions corresponding to real analytic initial data. In order to determine where the construction
breaks down, it is of interest to consider regularity classes which are intermediate between smoothness and real analyticity. In 
Section~\ref{section:gevreyclassesintro} we therefore determine to which Gevrey classes the initial data of the constructed solutions belong. 

Even though the examples exhibiting super exponential growth are of interest, we would also like to know if they correspond to a generic set
of initial data. We turn to this topic in Section~\ref{section:gensupexpgrointro}, where we consider equations of the form 
(\ref{eq:thesystemRge}) for general $d$ and 
$R$, and with an $f$ which is not necessarily zero. We state two genericity results. First, there is a subset of the set of smooth 
initial data which is generic in the sense of Baire, such that the corresponding solutions exhibit super exponential growth; cf. 
Proposition~\ref{prop:supexpgrowthgencasedenintro}. Second, the set of initial data which do not lead to solutions with super exponential 
growth has infinite co-dimension; cf. Proposition~\ref{prop:supexpgrowthinfcodimintro}. Finally, in Section~\ref{section:mainideaspfsupexpgrowth}
we explain the main ideas behind the proof of the results. 

\textbf{A geometric perspective.} On the basis of the observations made in Chapter~\ref{chapter:onnotofbal}, we are in a position to formulate 
conditions that exclude pathological behaviour (such as super exponential growth). This is the subject of Chapter~\ref{chapter:ageomperintro}.
It turns out that the conditions quite naturally take a geometric form. To begin with, we assume that the metric satisfies the conditions
of Definition~\ref{def:cansepcosmmodmanintro}. Turning to the shift vector field, we already know that $g_{ij}\chi^{i}\chi^{j}<1$. However, it 
is convenient to impose conditions of the type $g_{ij}\chi^{i}\chi^{j}\leq a$ for some $a\in [0,1)$ and all $t\geq 0$; this ensures that 
$-g_{00}$ is uniformly bounded away from zero to the future. In specific results, we impose additional smallness conditions. Concerning the lower 
order terms, we assume $\|\a\|$ and $\|\zeta\|$ to be bounded to the future (though there are some exceptions to this; cf. 
Subsection~\ref{ssection:reintoscODEbeh} below). Turning to the $X^{i}$, it is natural to think of $\mcX=X^{i}\d_{i}$ as a matrix of vector fields on $M$. 
For a fixed $t$, it can also be thought of as a matrix of vector fields on $\tn{d}$. It is of interest to consider the norm of $\mcX$ with respect 
to the metric induced on $\tn{d}\times\{p\}\times \{t\}$ (considered as a metric on $\tn{d}$). It turns out that some of the pathologies described
in Chapter~\ref{chapter:onnotofbal} can be eliminated by assuming the norms of the components of $\mcX$ to be bounded to the future with respect 
to this metric. 

In practice, we also need to impose conditions on the derivatives of the second fundamental form, $\mcX$ etc. One natural way of doing so is
to demand that the Lie derivatives of these objects with respect to $U$ (up to a certain order) are bounded to the future; recall that $U$ is 
defined by (\ref{eq:UitoNchiintro}). We introduce the relevant terminology in Definitions~\ref{definition:bdsonchi}, 
\ref{definition:boundsonbk}, \ref{definition:bdsonmcX} and \ref{definition:Cobal}. In Section~\ref{section:aroughnotofbal}, we then 
state conditions ensuring that the energy of solutions to the homogeneous equation cannot grow faster than exponentially; cf. 
(\ref{eq:mferobasestroughbalcase}). Finally, in Section~\ref{section:geomformunbalsetintro}, we formulate the results of 
Chapter~\ref{chapter:onnotofbal} geometrically. 

\textbf{Silent equations.} 
\index{Silent!equations}%
In Chapter~\ref{chapter:silentequations}, we turn to silent equations. The main condition concerning the metric
in this case is that the second fundamental form has a strictly positive lower bound. This means that the metric is expanding exponentially
in all directions; cf. Subsection~\ref{ssection:siltrsanoimet} above. We also assume the shift vector field to be small. Given that these 
conditions are satisfied and that the norm of $\mcX$ is future bounded, the coefficients of the spatial derivatives of $u$ appearing in 
(\ref{eq:thesystemRge}) decay exponentially. It is therefore 
natural to compare solutions to (\ref{eq:thesystemRge}) with solutions to the equation obtained from (\ref{eq:thesystemRge}) by dropping all 
the terms involving spatial derivatives. Additionally assuming $\a$ and $\zeta$ to converge to constant matrices $\a_{\infty}$ and $\zeta_{\infty}$
respectively, we obtain a ``limit equation'' 
\index{Limit equation}%
which is a constant coefficient ODE for each spatial point. The first result of 
Chapter~\ref{chapter:silentequations} states that the basic energy exhibits the same growth/decay as the basic energy for solutions to the limit 
equation; cf. (\ref{eq:mfeestslossintro}). In particular, the growth/decay is determined by the coefficients of the limit equation 
(i.e. by $\a_{\infty}$ and $\zeta_{\infty}$) and the right hand side. Next, Proposition~\ref{prop:roughas} yields the leading order asymptotics of 
solutions, as well as the conclusion that these asymptotics coincide with those of solutions to the limit equation. Moreover, according to 
Proposition~\ref{prop:spasda}, it is possible to specify the leading order asymptotics. Returning to the questions stated in 
Section~\ref{section:Questions}, we can compute $\cruderate$, derive asymptotics and specify asymptotics. However, the results stated in 
Chapter~\ref{chapter:silentequations} only yield a lower bound on $\nolossrate$. 

\textbf{Transparent equations.} 
\index{Transparent!equations}%
As a next step, it is of interest to consider equations such that the second fundamental form of the 
associated metric is non-negative, but converges to zero in some direction(s). This is the main assumption concerning the metric in the 
definition of transparent equations, the subject of Chapter~\ref{chapter:transpeq}. In this situation, the spatial variables are naturally
divided into ``silent'' and ``transparent'' variables. 
\index{Silent!variables}%
\index{Transparent!variables}%
This division is formalised in Subsection~\ref{ssection:divthevartrssetintro}. 
Additionally, we assume the shift vector field to be small; $\mcX$ to be future bounded; the components of $\mcX$ corresponding to the 
transparent variables to converge; and $\a$ and $\zeta$ to converge. Under these assumptions, it is possible to associate a limit equation
with (\ref{eq:thesystemRge}). The limit equation is a constant coefficient system of linear wave equations in the transparent variables and
the time variable for each silent variable. Similarly to the silent setting, the leading order asymptotics coincide with those of the 
limit equation; cf. Proposition~\ref{prop:trsasymptintro}. We are also in a position to specify the leading order asymptotics; cf. 
Proposition~\ref{prop:spasdatrsintro}.

\textbf{Equations with a dominant noisy spatial direction.} 
\index{Noisy!equations}%
Finally, it is of interest to consider metrics such that the second fundamental
form is negative definite in some directions. In the end, we consider, roughly speaking, the following situation: we assume that there is 
one direction (corresponding to one $\so$-factor in $\tn{d}$ or one of the $M_{r}$) such that $\bk$ behaves as $-\b_{\ron}\bge$ in that 
direction asymptotically, where $\b_{\ron}>0$; and that $\bk\geq (-\b_{\ron}+\eta_{\ron})\bge$ in the remaining directions, where $\eta_{\ron}>0$.
In this sense, there is one ``dominant noisy spatial direction''. In this setting, the spatial variables can be divided into ``noisy'' variables
\index{Noisy!variables}%
and ``subdominant'' variables. 
\index{Subdominant!variables}%
Adding appropriate convergence and boundedness conditions, we again obtain a limit equation. 
\index{Limit equation}%
However, as opposed to the previous cases, it is not a constant coefficient equation. In order to obtain conclusions, we need to limit our attention
to the modes of the equation that correspond to non-trivial dependence on the noisy variables. It turns out that, for a fixed such mode, the 
dominant behaviour is oscillatory, and the frequency of the oscillations grows exponentially. However, it turns out to be possible to average
over the oscillations, and this yields a well-defined overall growth/decay of solutions. We state the corresponding energy estimate in 
Proposition~\ref{prop:roughsobestnoisysetintro}. 

In order to describe the asymptotics, we need to separate the oscillatory part from the overall growth/decay. To this end, it is of interest 
to first focus on solutions to the scalar homogeneous equation obtained from (\ref{eq:thesystemRge}) by setting all the lower order terms 
(i.e., the ones involving $\a$, $\zeta$ and $X^{i}$) as well as the right hand side $f$ to zero. The resulting equation is given by 
(\ref{eq:homtheeqnoiseintro}) below. For (\ref{eq:homtheeqnoiseintro}), we are able not only to derive detailed asymptotics, but in fact to 
demonstrate that there is a homeomorphism (with respect to the $C^{\infty}$-topology) between initial data and asymptotic data. The relevant 
result is stated as Proposition~\ref{prop:asymposccasescalareqnoiseintro}. Given these models for the oscillatory behaviour, we derive the 
leading order asymptotics in Proposition~\ref{prop:asymposccaseitosoltowenoiseintro}. The asymptotic form of the solution is given by a sum of 
terms of the form $e^{At}u_{\row}$, where $A$ is a constant matrix and $u_{\row}$ is a vector valued solution to (\ref{eq:homtheeqnoiseintro}).
Similarly to silent and transparent equations, it is also possible to specify the leading order asymptotics. The relevant result is stated as 
Proposition~\ref{prop:asymposccaseitosoltowenoisespaintro}. Combining the results of Chapter~\ref{chapter:domnoisspdirintro}, we are in 
a position to derive and to specify asymptotics. Moreover, restricting to the parts of the solutions corresponding to modes with non-trivial 
dependence on the noisy variables, we can calculate an analogue of the $\cruderate$ introduced in Definition~\ref{eq:defdecayrates}. 
However, we only obtain a lower bound on $\nolossrate$. 

\textbf{Energy estimates in the asymptotically diagonal setting.} 
\index{Asymptotically diagonal!equations}%
In spite of the detailed asymptotic information obtained in 
Chapters~\ref{chapter:silentequations}--\ref{chapter:domnoisspdirintro}, the corresponding results only yield a lower bound on
$\nolossrate$. In order to be able to calculate $\nolossrate$, we need to make stronger assumptions. In Chapter~\ref{chapter:enestasdiagsett},
we assume, roughly speaking, that the metric becomes diagonal asymptotically; i.e., that the $\so$-factors in $\tn{d}$ become orthogonal asymptotically
(the $M_{r}$-factors are already orthogonal to everything else by construction). Moreover, we assume that the second fundamental form is
convergent and that the shift vector field converges to zero. As a consequence, the metric asymptotically takes the form (\ref{eq:modelmetricafterresc}).
Turning to the lower order terms, we assume that $\a$, $\zeta$ and $X^{j}/(g^{jj})^{1/2}$ converge to $\a_{\infty}$, $\zeta_{\infty}$ and 
$X^{j}_{\infty}$ respectively. Finally, we assume the metric to be non-degenerate in the sense that the $\b_{r}$ appearing in 
(\ref{eq:modelmetricafterresc}) are distinct. Under these circumstances we are able to calculate $\nolossrate$; cf. 
Theorem~\ref{thm:mainoptthmintro}. It is of interest to note that $\nolossrate$ depends on $\a_{\infty}$, $\zeta_{\infty}$, the $\b_{r}$ and 
the $X^{j}_{\infty}$. 

\textbf{Improved asymptotic estimates in the silent setting.} In Chapter~\ref{chapter:silentequations}, we derive asymptotics, given a solution.
Moreover, for given leading order asymptotics, we construct a corresponding solution. Due to the results stated in 
Chapter~\ref{chapter:silentequations}, we know that the map from initial data to asymptotic data is continuous with respect to the 
$C^{\infty}$-topology. In fact, there is an $0\leq s_{0}\in\ro$ such that the $H^{s}$-norm of the asymptotic data is bounded by the $H^{s+s_{0}}$
norm of the initial data. On the other hand, the results stated in Chapter~\ref{chapter:silentequations} do not allow us to determine
$s_{0}$ in terms of the coefficients of the equation. Imposing conditions similar to those of Chapter~\ref{chapter:enestasdiagsett}, we obtain 
a specific value for $s_{0}$; cf. Proposition~\ref{prop:roughasexintro}. In Proposition~\ref{prop:spasdaexintro}, we state similar results 
concerning the map from asymptotic data to initial data. It is of interest to note that the map from initial data to asymptotic data in some 
situations improves the regularity. In other words, there are cases for which the $H^{s}$-norm of the asymptotic data is bounded by the 
$H^{s+s_{0}}$-norm of the initial data, where $s_{0}<0$.

\section{Acknowledgements}

The author would like to acknowledge the support of the G\"{o}ran Gustafsson Foundation for Research in Natural Sciences and Medicine, 
the Swedish Research Council (Reference numbers 621-2012-3103 and 2017-03863) and the Erwin Schr\"{o}dinger Institute. This article was 
in part written during a stay at the Erwin Schr\"{o}dinger Institute in Vienna during the ESI Workshop on ``Geometric Transport Equations 
in General Relativity'' 2017. Finally, the author would like to thank Ellery Ames for comments on an earlier version of this article. 
The author is a professor at KTH Royal Institute of Technology and the Deputy Director of the Institut Mittag-Leffler. 

\chapter{On the notion of balance}\label{chapter:onnotofbal}

\section{Introduction}\label{section:introductiononthenotionofbal}

Consider (\ref{eq:thesystemRge}). The coefficients of this equation can naturally be divided into two groups: those of the highest 
order derivatives and those of the lower order derivatives. The coefficients of the highest order 
derivatives correspond to the Lorentz metric (\ref{eq:sepcosmodmeintro}). Even though we impose additional conditions in particular 
situations, we have already formulated the basic assumptions concerning this metric in Section~\ref{section:geoconprefme}. In particular,
we, from now on, assume the Lorentz manifold corresponding to (\ref{eq:thesystemRge}) to satisfy
Definition~\ref{def:cansepcosmmodmanintro}. Turning to $\a$ and $\zeta$, we can already develop a feeling 
for what types of assumptions are reasonable by considering spatially homogeneous solutions. Let us, to this end, consider 
(\ref{eq:sphomandhomverbalintro}). Assume that there are constants $C_{\rocu},\g,\eta_{\rocu}>0$ and matrices 
$\a_{\infty},\zeta_{\infty}\in\Mn{m}{\co}$ such that 
\[
\|\g^{-1}e^{-\g t}\a(t)-\a_{\infty}\|+
\|\g^{-2}e^{-2\g t}\zeta(t)-\zeta_{\infty}\|\leq C_{\rocu}e^{-\eta_{\rocu} t}
\]
for all $t\geq 0$. Assume, moreover, that one of $\a_{\infty},\zeta_{\infty}$ is non-zero. Define
\begin{equation}\label{eq:Ainfdefintrobal}
A_{\infty}:=\left(\begin{array}{cc} 0 & \Id_{m} \\ -\zeta_{\infty} & -\a_{\infty}\end{array}
\right)
\end{equation}
\index{$\a$Aa@Notation!Matrix notation!Ainfinity@$A_{\infty}$}%
and let $\kappa_{1}$ be the largest real part of an eigenvalue of $A_{\infty}$. Then solutions to (\ref{eq:sphomandhomverbalintro})
behave as $\exp[\kappa_{1}e^{\g t}]$ as $t\rightarrow\infty$; cf. Lemma~\ref{lemma:spsysodes} below for a precise statement. If $\kappa_{1}>0$, then
solutions grow super exponentially. If $\kappa_{1}<0$, then solutions decay super exponentially (meaning that they grow super exponentially to 
the past). We consider both cases to be pathological. The reason for this is that we think of the Lorentz manifolds of interest as arising from 
an application of Lemma~\ref{lemma:bdandconvtocanonicalform}. Such Lorentz manifolds have the property that the volume of the hypersurfaces 
$\bM_{t}$ with respect to the original 
metric is proportional to $e^{\pm t}$ (in the case of big crunch asymptotics, we obtain an exponentially decreasing volume). As a consequence, 
the natural length scale is $e^{\pm t/D}$, where $D:=\dim\bM$. From this point of view, exponential growth/decay is natural, but super exponential
growth/decay is not. In case the eigenvalues of $A_{\infty}$ are purely imaginary, Lemma~\ref{lemma:spsysodes}
does not yield any conclusions. On the other hand, $\kappa_{1}=0$ is a sensitive borderline case, which we choose not to consider further (with
one exception; cf. Subsection~\ref{ssection:reintoscODEbeh} below). For 
these reasons, we wish to avoid $\a$ and $\zeta$ growing exponentially. Another option would be to consider the case that $\a$ and $\zeta$ grow polynomially. 
However, this typically yields faster than exponential growth/decay. Due to these observations, we assume $\|\a\|$ and $\|\zeta\|$ to be future 
bounded in most of the results of these notes. 

\textbf{Imposing conditions on the matrix $\mcX$ of vector fields.} Except for the $X^{j}$'s, we have developed a rough feeling for what conditions to 
impose on the coefficients of (\ref{eq:thesystemRge}). The main purpose of the present chapter is to illustrate some of the pathologies that arise
if we do not impose appropriate bounds on $\mcX$. Before turning to a formal statement of the results, let us consider 
\begin{equation}\label{eq:elexsupexpgrowth}
u_{tt}-e^{-4t}u_{\theta\theta}+e^{-t}u_{\theta}+u_{t}+u=0
\end{equation}
on $\so\times\ro$. One naive way of approaching this equation is to say that the factors in front of the spatial derivatives are exponentially decaying. 
It should thus be possible to ignore the corresponding terms, and the behaviour should be dominated by the ODE part of the equation. Since the solutions
to the corresponding ODE decay exponentially, solutions should in general decay exponentially. Another perspective would be to separate the equation 
(\ref{eq:elexsupexpgrowth}) and to consider the Fourier coefficients. We discuss Fourier decompositions in Section~\ref{section:divintomodes} below (and
we use the notation introduced in Section~\ref{section:divintomodes} without further comment in what follows). However, in the 
particular case of (\ref{eq:elexsupexpgrowth}), we obtain
\begin{equation}\label{eq:elexsupexpgrowthfourier}
\ddot{z}+\dot{z}+(n^{2}e^{-4t}+ine^{-t}+1)z=0,
\end{equation}
where $z$ denotes the $n$'th Fourier coefficient of $u$. We thus expect $z$ to decay exponentially (just as in the case of the equation obtained by dropping the spatial 
derivatives). For each fixed Fourier coefficient, this is indeed what happens; Lemma~\ref{lemma:oderegest} below implies that $z$ decays as $e^{-t/2}$. 
On the other hand, the constant $C$ appearing in the estimate $|z(n,t)|\leq Ce^{-t/2}$ depends on $n$, indicating that there might be a problem when considering, 
e.g., smooth solutions to (\ref{eq:elexsupexpgrowth}). 
Yet another perspective is to study energies. In analogy with the basic energy introduced in (\ref{eq:mfebasdef}), consider
\[
E_{\rou}(t):=\frac{1}{2}\int_{\so}[|u_{t}(\theta,t)|^{2}+e^{-4t}|u_{\theta}(\theta,t)|^{2}+|u(\theta,t)|^{2}]d\theta.
\]
A crude energy estimate yields the conclusion that $\dot{E_{\rou}}(t)\leq e^{t}E_{\rou}(t)$, indicating that the growth might be as bad as $\exp(e^{t})$; i.e., 
super exponential. The different perspectives clearly yield very different conclusions. It is therefore of interest to ask: do solutions decay exponentially 
or grow super exponentially? Below we demonstrate that for generic smooth solutions to (\ref{eq:elexsupexpgrowth}), the crude energy estimate 
$E_{\rou}(t)\leq C\exp(e^{t})$ is essentially optimal; cf. Example~\ref{example:basexsupexpgrowth} and Section~\ref{section:gensupexpgrointro} below. In 
this particular case, the argument that the spatial derivatives in (\ref{eq:elexsupexpgrowth}) can be dropped since they are multiplied by exponentially 
decaying factors is not justified. Later on, we introduce the notion of balance, and it will become clear that the super exponential growth of solutions 
to (\ref{eq:elexsupexpgrowth}) is related to the fact that (\ref{eq:elexsupexpgrowth}) is not balanced. 

In the present chapter, the conditions appearing in the results are largely non-geometric in nature. The reason for this is that we, in the next chapter, 
use the results obtained here to motivate the importance of a geometric perspective. However, in Section~\ref{section:geomformunbalsetintro} below, we 
provide geometric reformulations of the results of the present chapter.

\section[Super exponential growth, first examples]{Unbalanced equations, first examples of super exponential 
growth}\label{section:unbaeqfiexsupexpgr}

To begin with, we state a result concerning equations of the form (\ref{eq:thesystemRge}) with $d=1$, $R=0$ and $f=0$. In order to be able to state
the assumptions in a concise way, it is convenient to introduce the notation
\begin{equation}\label{eq:mflYovarsdefintro}
\mfl(t):=\ln([g^{11}(t)]^{1/2}),\ \ \
Y^{1}(t):=\frac{X^{1}(t)}{[g^{11}(t)]^{1/2}},\ \ \
\varsigma(t):=\frac{g^{01}(t)}{[g^{11}(t)]^{1/2}}.
\end{equation}
\index{$\a$Aa@Notation!Coefficients, equation!$\mfl$}%
\index{$\a$Aa@Notation!Coefficients, equation!$Y^{1}$}%
\index{$\a$Aa@Notation!Coefficients, equation!$\varsigma$}%
It is possible to take a geometric perspective on these objects (and we encourage the reader to do so). However, in the present chapter, we take an analytic 
perspective, and only in the next chapter do we interpret the conclusions geometrically. Given the notation (\ref{eq:mflYovarsdefintro}), the main assumptions 
are the following. 

\begin{assumption}\label{ass:mainassumpubcaseintro}
Consider (\ref{eq:thesystemRge}). Assume the associated metric to be such that $(M,g)$ is a canonical separable cosmological model manifold. 
Assume, moreover, that $d=1$ and that $R=0$. Let $\mfl$, $Y^{1}$ and $\varsigma$ be defined by (\ref{eq:mflYovarsdefintro}). Assume that there
are real numbers $\b_{\rosh}<0$; $\ellderbd,C_{\rosh},C_{X},C_{\roode}>0$; $\b_{X}\geq 0$; and $\b_{\roode}$ such that $\b_{X}>\b_{\roode}$,
\begin{equation}\label{eq:constbdunbalintro}
|\dot{\mfl}(t)|+|\ddot{\mfl}(t)| \leq \ellderbd
\end{equation}
\index{$\a$Aa@Notation!Exponential rates!$\b_{X}$}%
for all $t\geq 0$, and 
\begin{align}
|\varsigma(t)|+|\dot{\varsigma}(t)| \leq & C_{\rosh}e^{\b_{\rosh}t},\label{eq:shiftbdubintro}\\
\|Y^{1}(t)\|+\|\dot{Y}^{1}(t)\| \leq & C_{X}e^{\b_{X}t},\label{eq:Xbdubintro}\\
\|\dot{\a}(t)\|+\|\a(t)\|+\|\dot{\zeta}(t)\|^{1/2}+\|\zeta(t)\|^{1/2} \leq & C_{\roode}e^{\b_{\roode}t}\label{eq:albdubintro}
\end{align}
for all $t\geq 0$. 
\end{assumption}
\begin{remark}
The assumption concerning $\mfl$ guarantees that $g^{11}$ cannot grow (or decay) faster than exponentially. Combining this observation with 
(\ref{eq:shiftbdubintro})--(\ref{eq:albdubintro}) yields the conclusion that none of the coefficients of the equation grows faster than 
exponentially. On the other hand, the assumptions still allow both exponential growth and exponential decay of the coefficients. 
\end{remark}
Assumption~\ref{ass:mainassumpubcaseintro} only involves bounds, and as a consequence there are classes of equations, satisfying the assumptions, whose 
solutions do not exhibit super exponential growth. In order to proceed, we therefore make the following additional assumption.

\begin{assumption}\label{assumption:Yoconvubintro}
Consider (\ref{eq:thesystemRge}). Assume the associated metric to be such that $(M,g)$ is a canonical separable cosmological model manifold. Assume, 
moreover, that $d=1$ and that $R=0$. Let $Y^{1}$ be defined by (\ref{eq:mflYovarsdefintro}) and Assumption~\ref{ass:mainassumpubcaseintro} be 
fulfilled. Assume that there is a $Y^{1}_{\infty}\in\Mn{m}{\co}$ and constants $\eta_{X}>0$ and $K_{X}>0$ such that
\begin{equation}\label{eq:Yoasbehubintro}
\|e^{-\b_{X}t}Y^{1}(t)-Y^{1}_{\infty}\|\leq K_{X}e^{-\eta_{X}t}
\end{equation}
for all $t\geq 0$, where $\b_{X}\geq 0$ is the constant appearing in the statement of Assumption~\ref{ass:mainassumpubcaseintro}.
Finally, assume $Y^{1}_{\infty}$ to have an eigenvalue with a non-zero real part.
\end{assumption}
\begin{remark}
If $\b_{X}>0$, this assumption ensures that $Y^{1}$ grows exponentially. On the other hand, it does not imply that $X^{1}$ grows. In fact, $X^{1}$ could 
decay exponentially. 
\end{remark}

Given that these assumptions hold, we obtain the following result. 
\begin{prop}\label{prop:supexpinstabintro}
Consider (\ref{eq:thesystemRge}). Assume the associated metric to be such that $(M,g)$ is a canonical separable cosmological model manifold. Assume, 
moreover, that $d=1$; that $R=0$; and that $f=0$. Given that Assumptions~\ref{ass:mainassumpubcaseintro} and \ref{assumption:Yoconvubintro} 
hold with $\b_{X}>0$, let $\kappa$ be the largest absolute value of a real part of an eigenvalue of $Y^{1}_{\infty}/2$ (by assumption, $\kappa>0$). 
Let $\e>0$. Then there is a sequence of smooth $\cn{m}$-valued solutions $v_{l}$ to (\ref{eq:thesystemRge}), $1\leq l\in\zo$, and for each $l\geq 1$, 
there is a time sequence $t_{l,k}\rightarrow\infty$ (as $k\rightarrow\infty$) such that for each $s\in\ro$, 
\begin{equation}\label{eq:vlsobconvtozerointro}
\lim_{l\rightarrow\infty}\left(\|v_{l}(\cdot,0)\|_{(s+1)}+\|\d_{t}v_{l}(\cdot,0)\|_{(s)}\right)=0.
\end{equation}
Moreover, 
\begin{equation}\label{eq:mferohomsupexplbintro}
\mfe_{\rohom}[v_{l}](t_{l,k})\geq \exp\left[2\b_{X}^{-1}\kappa(1-\e)e^{\b_{X}t_{l,k}}\right],
\end{equation}
where $\mfe_{\rohom}[u]$ is defined by (\ref{eq:mfehomdefintro}); i.e., in the present setting,
\begin{equation}\label{eq:mferohomdeqodefintro}
\mfe_{\rohom}[u](t):=\frac{1}{2}\is[|u_{t}(x,t)|^{2}+g^{11}(t)|u_{x}(x,t)|^{2}]dx.
\end{equation}
\end{prop}
\begin{remarks}
Here $\|\cdot\|_{(s)}$ denotes the norm introduced in (\ref{eq:HsnormonbM}). The combination of the estimates (\ref{eq:vlsobconvtozerointro}) 
and (\ref{eq:mferohomsupexplbintro}) demonstrates that the solution $u=0$ to (\ref{eq:thesystemRge}) is future unstable. 
\end{remarks}
\begin{remark}
The statement is a consequence of Lemma~\ref{lemma:supexpinstab}; cf. Remark~\ref{remark:supexpgrowthlemmatointroprop}.
\end{remark}
\begin{remark}\label{remark:upbdstenest}
It is of interest to compare the lower bound given by (\ref{eq:mferohomsupexplbintro}) with the upper bound obtained by 
standard energy estimates. Due to Lemma~\ref{lemma:standenestimates}, given a solution $u$ to an equation satisfying the assumptions of 
the proposition, there is an upper bound on the associated energy of the form $C_{\e}\exp[\b_{X}^{-1}\|Y^{1}_{\infty}\|(1+\e)e^{\b_{X}t}]$. 
When $\|Y^{1}_{\infty}\|=2\kappa$ (e.g., when $Y^{1}_{\infty}$ is Hermitian), the estimate (\ref{eq:mferohomsupexplbintro}) is thus essentially
optimal. 
\end{remark}

Let us apply this result to (\ref{eq:elexsupexpgrowth}).

\begin{example}\label{example:basexsupexpgrowth}
Consider (\ref{eq:elexsupexpgrowth}). This equation is such that $d=1$, $R=0$ and $f=0$. Moreover, $g^{11}(t)=e^{-4t}$, $g^{01}(t)=0$, $\a(t)=1$, 
$\zeta(t)=1$ and $X^{1}(t)=e^{-t}$. In particular, the relations $\mfl(t)=-2t$, $Y^{1}(t)=e^{t}$ and $\varsigma(t)=0$ thus hold. As a consequence,
Assumption~\ref{ass:mainassumpubcaseintro} holds with $\b_{\rosh}=-1$, $\ellderbd=2$, $\eta_{\roode}=1$, $C_{\rosh}=1$, $C_{X}=2$, $C_{\roode}=2$,
$\b_{X}=1$ and $\b_{\roode}=0$. Moreover, Assumption~\ref{assumption:Yoconvubintro} holds with $Y^{1}_{\infty}=1$, $\eta_{X}=1$ and $K_{X}=1$; clearly
$Y^{1}_{\infty}/2$ has an eigenvalue with non-zero real part and $\kappa=1/2$. Thus the conclusions of Proposition~\ref{prop:supexpinstabintro} hold, 
so that, given $\e>0$, there is a sequence of smooth solutions $v_{l}$ to (\ref{eq:elexsupexpgrowth}) and time sequences $\{t_{l,k}\}$ such that 
$t_{l,k}\rightarrow\infty$ as $k\rightarrow\infty$ and (\ref{eq:vlsobconvtozerointro}) and (\ref{eq:mferohomsupexplbintro}) hold. In our setting, 
the estimate (\ref{eq:mferohomsupexplbintro}) reads
\[
\mfe_{\rohom}[v_{l}](t_{l,k})\geq \exp\left[(1-\e)e^{t_{l,k}}\right].
\]
This indicates that the crude energy estimate mentioned in connection with (\ref{eq:elexsupexpgrowth}) is essentially optimal. 
\end{example}

The equation in Example~\ref{example:basexsupexpgrowth} is rather simple in that the coefficients of (\ref{eq:elexsupexpgrowth}) are either 
constants or constants times exponential factors. However, similar conclusions hold for, e.g., 
\begin{equation}\label{eq:highosctermssupexpgrowth}
u_{tt}-e^{-4t}u_{\theta\theta}+e^{-t}u_{\theta}+\sin(e^{t/2})u_{t}+e^{t/2}\cos(e^{t/2})u=0. 
\end{equation}
Here one might naively expect the factors $\sin(e^{t/2})$ and $e^{t/2}\cos(e^{t/2})$ to influence the behaviour (due to their growth and
the growth of their derivatives). However, as far as the argument is concerned, the corresponding terms in (\ref{eq:highosctermssupexpgrowth})
are effectively error terms. It is also of interest to consider, e.g.,
\begin{equation}\label{eq:supexpgrowthsupexpdecay}
u_{tt}-e^{-4t}u_{\theta\theta}+e^{-t}u_{\theta}+e^{t/2}u_{t}+e^{t}u=0.
\end{equation}
In the case of (\ref{eq:supexpgrowthsupexpdecay}), initial data with finite frequency content yield solutions with super exponential decay (in fact, 
such solutions roughly behave as $\exp(-e^{t/2})$; cf. Section~\ref{section:ODEunbal}). However, generic smooth solutions exhibit super exponential 
growth; cf. Section~\ref{section:gensupexpgrointro} below. 

Recalling the discussion of the Fourier coefficients of solutions to (\ref{eq:elexsupexpgrowth}), it is clear that solutions to (\ref{eq:elexsupexpgrowth}) 
with finite frequency content decay exponentially. In particular, there is thus a dense set of initial data such that the corresponding solutions exhibit
exponential decay. It is of interest to isolate where the transition between exponential decay and super exponential growth takes place.
To this end, it is useful to introduce function spaces interpolating between initial data with finite frequency content and 
smooth initial data, and to analyse where super exponential growth sets in. In the end, we are not able to demonstrate super exponential growth
for solutions with real analytic initial data (at least not for equations such that the coefficients of the spatial derivatives decay exponentially), 
so that it is sufficient to consider a class of function spaces interpolating between real analyticity and smoothness. This leads us to study Gevrey 
classes.

\section{Gevrey classes}\label{section:gevreyclassesintro}

Let us begin by defining the Gevrey classes in the context of interest here. 
\begin{definition}\label{def:GevreyTdintro}
A real valued function $f$ defined on $\tn{d}$ is said to be a \textit{Gevrey function of order}
\index{Gevrey!function}%
 $0<s\in\ro$ if 
$f\in C^{\infty}(\tn{d},\ro)$ and there are constants $R>0$ and $C>0$ such that 
\begin{equation}\label{eq:GevreycondTdfactintro}
\sup_{x\in \tn{d}}|\d^{\a}f(x)|\leq RC^{|\a|}(\a!)^{s}
\end{equation}
for all $d$-multiindices $\a$. The class of Gevrey functions of order $0<s\in\ro$ on $\tn{d}$ is denoted $G^{s}(\tn{d})$. 
\index{Gevrey!class}%
\index{$\a$Aa@Notation!Function spaces!$G^{s}$}%
\end{definition}
\begin{remarks}
Here we think of functions defined on $\tn{d}$ as being defined on $\rn{d}$ and being $2\pi$-periodic in all of their variables.
The definition can be generalised to functions with values in $\rn{m}$ or $\cn{m}$, and we use the notation 
$G^{s}(\tn{d},\rn{m})$ and $G^{s}(\tn{d},\cn{m})$. Moreover, if $f\in G^{1}(\tn{d})$, 
then $f$ is real analytic. Finally, if $f\in G^{s}(\tn{d})$, where $0<s<1$, then $f$ is entire. 
\end{remarks}
\begin{remark}
There is an alternate characterisation of the Gevrey classes; cf. Lemma~\ref{lemma:fouriercoeffcharofgevreyclass} below.
In fact, let $f\in\ C^{\infty}(\tn{d},\co)$, $0<s\in\ro$ and $a_{n}$, $n\in\zn{d}$, be the complex Fourier coefficients of $f$. Then 
$f\in G^{s}(\tn{d},\co)$ if and only if there are constants $C_{i}>0$, $i=1,2$, 
such that 
\begin{equation}\label{eq:altgevcondintro}
|a_{n}|\leq C_{1}\exp\left(-C_{2}|n|^{1/s}\right)
\end{equation}
for all $n\in\zn{d}$. There is an analogous characterisation in case $f$ takes its values in $\rn{m}$ or $\cn{m}$. 
\end{remark}

When considering solutions to (\ref{eq:thesystemRge}) with $R=0$, it is of interest to verify that Gevrey class regularity is preserved. 
In that context, the following definition is useful. 
\begin{definition}
Let $I\subseteq\ro$ be an open interval, $f\in C^{\infty}(\tn{d}\times I,\cn{m})$ and $0<s\in\ro$. Then $f$ is said to belong to 
$\mfG^{s}_{\roloc,\rou}(I,\tn{d},\cn{m})$
\index{$\a$Aa@Notation!Function spaces!$\mfG^{s}_{\roloc,\rou}(I,\tn{d},\cn{m})$}%
if, for every compact interval $J\subset I$, the following holds: there are constants $C_{1}>0$ and $C_{2}>0$ (depending on $f$ and $J$) such that 
\[
|\hf(n,t)|\leq C_{1}\exp\left(-C_{2}|n|^{1/s}\right)
\]
for all $n\in\zn{d}$ and $t\in J$, where $\hf(n,t)$ is the $n$'th (complex) Fourier coefficient of $f(\cdot,t)$. 
\end{definition}

Consider (\ref{eq:thesystemRge}). Assume the associated metric to be such that $(M,g)$ is a canonical separable cosmological model manifold. Assume, 
moreover, that $R=0$. If the coefficients are smooth, $f\in\mfG^{s}_{\roloc,\rou}(\ro,\tn{d},\cn{m})$, and the initial data at $t=t_{0}$ belong to 
$G^{s}(\tn{d},\cn{m})$, then the corresponding solution to (\ref{eq:thesystemRge}) is smooth and such that $u$ and $u_{t}$ have the same regularity 
as $f$; cf. Lemma~\ref{lemma:gevreypreserv}. 

Given the above terminology, the statement of Proposition~\ref{prop:supexpinstabintro} can be improved to the following result. 

\begin{prop}\label{prop:supexpinstabGevreyregintro}
Assume that the conditions of Proposition~\ref{prop:supexpinstabintro} are fulfilled. Assume, moreover, that there are constants $\b_{1}\in\ro$
and $0<c_{\rom,1}\in\ro$ such that 
\[
c_{\rom,1}e^{2\b_{1}t}\leq g^{11}(t)
\]
for all $t\geq 0$. If $\b_{X}-\b_{1}>0$, then the sequence of functions $v_{l}$ 
constructed in Proposition~\ref{prop:supexpinstabintro} can be assumed to be such that $v_{l},\d_{t}v_{l}\in \mfG_{\roloc,\rou}^{s_{X}}(I,\so,\cn{m})$ for all
$l$, where 
\begin{equation}\label{eq:sXdefintro}
s_{X}:=\frac{\b_{X}-\b_{1}}{\b_{X}}.
\end{equation}
In case $\b_{X}-\b_{1}\leq 0$, one frequency can be chosen, say $n_{a}\in\zo$, such that the 
sequence of functions $v_{l}$ constructed in Proposition~\ref{prop:supexpinstabintro} has the property that if $z_{l}(n,t)$ denotes the 
$n$'th Fourier coefficient of $v_{l}(\cdot,t)$, then $z_{l}(n,t)=0$ for all $t$ unless $n=n_{a}$. 
\end{prop}
\begin{remark}
The statement is an immediate consequence of Proposition~\ref{prop:supexpinstabGevreyreg}.
\end{remark}
\begin{remark}\label{remark:basexsupexpgrowthGevrey}
Returning to Example~\ref{example:basexsupexpgrowth}, it is clear that $\b_{1}=-2$ and $\b_{X}=1$. Thus $s_{X}=3$. 
\end{remark}
\begin{remark}
Consider
\[
u_{tt}-e^{2\b_{1}t}u_{\theta\theta}+e^{(\b_{1}+\b_{X})t}u_{\theta}=0.
\]
If we want the coefficients of the spatial derivatives to decay exponentially, we have to have $\b_{1}<0$ and $\b_{1}+\b_{X}<0$. If we also want
to have super exponential growth, we have to have $\b_{X}>0$. This means that $0<\b_{X}<-\b_{1}$, so that $s_{X}>2$. In this setting, there is thus
a limit on how close we can get to the case of real analytic initial data. If, on the other hand, $0<\b_{X}\leq\b_{1}$, then there are initial
data with finite frequency content such that the corresponding solutions exhibit super exponential growth. 
\end{remark}

It would be of interest to determine if the value of $s_{X}$ appearing in (\ref{eq:sXdefintro}) has any real significance or is just an artefact of 
our argument. However, we do not pursue this topic in these notes. 

\section{Unbalanced equations, generic super exponential growth}\label{section:gensupexpgrointro}

In the previous sections, we demonstrate that there are solutions to equations of the form (\ref{eq:thesystemRge}) that exhibit super exponential 
growth. However, it is not clear if these solutions are very special, or if they represent the typical behaviour. When addressing this question, it 
is of interest to return to general equations of the form (\ref{eq:thesystemRge}). Let us first note that initial data ensuring super exponential 
growth are generic in the Baire sense. 

\begin{prop}\label{prop:supexpgrowthgencasedenintro}
Consider (\ref{eq:thesystemRge}). Assume the associated metric to be such that $(M,g)$ is a canonical separable cosmological model manifold.
Assume, moreover, that there is a $j\in \{1,\dots,d\}$ such that 
\begin{equation}\label{eq:urestrtojdepintro}
u_{tt}-g^{jj}(t)\d_{j}^{2}u-2g^{0j}(t)\d_{t}\d_{j}u+\a(t)u_{t}+X^{j}(t)\d_{j}u+\zeta(t)u=0
\end{equation}
(no summation on $j$) satisfies the conditions of Proposition~\ref{prop:supexpinstabintro}; 
note that (\ref{eq:urestrtojdepintro}) is an equation of the form (\ref{eq:thesystemRge}) with $f=0$, $d=1$ and $R=0$. Let $\e>0$. Then 
there is a set of smooth $\cn{m}$-valued initial data to (\ref{eq:thesystemRge}), say $\ma$, with the following properties. First, $\ma$ 
is the intersection of a countable number of open and dense sets (with respect to the $C^{\infty}$ topology). Second, each element of $\ma$ 
corresponds to a $\cn{m}$-valued solution $u$ to (\ref{eq:thesystemRge}) such that there is a time sequence $\{t_{k}\}$, $1\leq k\in\zo$, 
with the properties that $t_{k}\rightarrow\infty$ and 
\begin{equation}\label{eq:mferohomsupexplbthvintro}
\mfe_{\rohom}[u](t_{k})\geq \exp\left[2\b_{X}^{-1}\kappa(1-\e)e^{\b_{X}t_{k}}\right]
\end{equation}
for all $1\leq k\in\zo$, where $\mfe_{\rohom}[u](t)$ is defined by (\ref{eq:mfehomdefintro}).
\end{prop}
\begin{remark}
In the present proposition, we do not assume that $f=0$, nor do we assume that $R=0$ or that $d=1$. Moreover, we take it to be understood
that the equation (\ref{eq:urestrtojdepintro}) is obtained from (\ref{eq:thesystemRge}) by setting $f$ to zero and dropping all terms involving
spatial derivatives other than $\d_{j}$. 
\end{remark}
\begin{remark}
The result is an immediate consequence of Proposition~\ref{prop:supexpgrowthgencaseden}.
\end{remark}

Even though this result is interesting, we would also like to know that the set of initial data that do not yield solutions with super exponential
growth has positive codimension. The next result states that the codimension is in fact infinite. 

\begin{prop}\label{prop:supexpgrowthinfcodimintro}
Consider (\ref{eq:thesystemRge}). Assume the associated metric to be such that $(M,g)$ is a canonical separable cosmological model manifold.
Assume, moreover, that there is a $j\in \{1,\dots,d\}$ such that (\ref{eq:urestrtojdepintro}) (no summation on $j$) satisfies the 
conditions of Proposition~\ref{prop:supexpinstabintro}. Let $\e>0$. Define
$\mb_{\e}$ by the condition that $\xi\in\mb_{\e}$ iff $\xi$ constitutes smooth $\cn{m}$-valued initial data for (\ref{eq:thesystemRge}) 
such that if $u$ is the solution corresponding to $\xi$, then there is a constant $C_{\e}$ (depending on the solution) such that 
\[
\mfe_{\rohom}[u](t)\leq C_{\e}\exp\left[2\b_{X}^{-1}\kappa(1-\e)e^{\b_{X}t}\right]
\]
for all $t\geq 0$, where $\mfe_{\rohom}$ is defined by (\ref{eq:mfehomdefintro}). Define $\ma_{\e}$ by the condition that $\xi\in\ma_{\e}$ iff $\xi$ 
constitutes smooth $\cn{m}$-valued initial data for (\ref{eq:thesystemRge}) such that if $u$ is the solution corresponding to $\xi$, then there 
is a time sequence $0\leq t_{k}\rightarrow\infty$ such that 
\begin{equation}\label{eq:maeplbdefintro}
\mfe_{\rohom}[u](t_{k})\geq \exp\left[2\b_{X}^{-1}\kappa(1-\e/2)e^{\b_{X}t_{k}}\right]
\end{equation}
for all $k$. Then $\ma_{\e}$ and $\mb_{\e}$ are disjoint. Moreover, there is an infinite dimensional (linear) subspace $P_{\e}$ of the 
vector space of smooth $\cn{m}$-valued initial data for (\ref{eq:thesystemRge}) such that if $\xi\in\mb_{\e}$ and $\xi_{\per}\in P_{\e}$, 
then $\xi+\xi_{\per}\in\ma_{\e}$ unless $\xi_{\per}=0$. 
\end{prop}
\begin{remarks}
The statement of the proposition can be interpreted as saying that $\mb_{\e}$ has infinite codimension in the set of smooth initial data
for (\ref{eq:thesystemRge}). On the other hand, there are examples of equations such that $\mb_{\e}$, even though it has infinite 
codimension, is dense in the set of initial data; this is a consequence of the current proposition and the results of 
Section~\ref{section:modeanalysis}. In fact, there are examples of equations satisfying the conditions of 
Proposition~\ref{prop:supexpgrowthinfcodimintro} such
that all solutions with finite frequency content decay super exponentially. Having one dense set of initial data yielding super exponential
growth and one dense set of initial data yielding super exponential decay may seem to be counter intuitive. However, it should be pointed 
out that the super exponential growth can be detected by only considering solutions with finite frequency content (even when solutions
with finite frequency content exhibit super exponential decay); cf. Subsection~\ref{ssection:unbdgritofinitefre}.
\end{remarks}
\begin{remark}
By the vector space of smooth $\cn{m}$-valued initial data for (\ref{eq:thesystemRge}), we here mean the space
$C^{\infty}(\bM,\cn{m})\times C^{\infty}(\bM,\cn{m})$, where $\bM$ is defined by (\ref{eq:bMdef}).
\end{remark}
\begin{remark}
The statement is an immediate consequence of Proposition~\ref{prop:supexpgrowthinfcodim}.
\end{remark}

To conclude: solutions to equations satisfying the conditions of Propositions~\ref{prop:supexpgrowthgencasedenintro} and 
\ref{prop:supexpgrowthinfcodimintro} typically exhibit super exponential growth. Returning to the conditions, it is thus clear that 
when $X^{j}(t)/[g^{jj}(t)]^{1/2}$ exhibits definite exponential growth, the behaviour of solutions to (\ref{eq:thesystemRge}) might be 
pathological. The next step of our analysis is to introduce the notion of a balanced equation, the purpose being to exclude this 
pathological behaviour. 

\section{A rough idea of the argument in a model case}\label{section:mainideaspfsupexpgrowth}

In order to obtain a complete proof of the statements of this chapter, it is sufficient to read Parts~\ref{part:averaging} and \ref{part:unbadegeq}
of these notes. However, the material in Part~\ref{part:averaging} covers much more general situations than are needed to obtain the results of the
present chapter. Moreover, the level of abstraction is sufficiently high that extracting the essential idea of the argument from the formal proof is 
non-trivial. For that reason, it is of interest to develop some intuition by considering a model case. That is the purpose of the present section. 

\textbf{A model case.} Let us return to (\ref{eq:elexsupexpgrowthfourier}), the equation for the 
Fourier coefficients of (\ref{eq:elexsupexpgrowth}). In the coefficients of this equation, there are terms of three different sizes: $1$, $|n|e^{-t}$ and 
$n^{2}e^{-4t}$. For $|n|\geq 2$, the third term dominates initially. Denote the subinterval of $[0,\infty)$ during which $n^{2}e^{-4t}$ is largest by $I_{1}$; 
it is of the form $I:=[0,t_{\fin}]$. Considering the intervals on which the other terms dominate is also of interest, but in order to demonstrate generic 
super exponential growth, it is sufficient to focus on $I_{1}$. The equation (\ref{eq:elexsupexpgrowthfourier}) can be written $\dot{w}=Aw$, where
\[
w(n,t):=\left(\begin{array}{c} \mfg(n,t)z(n,t) \\ \dot{z}(n,t)\end{array}\right),\ \ \
A(n,t):=\left(\begin{array}{cc} -2 & \mfg(n,t) \\ -\mfg(n,t) -i\frac{ne^{t}}{|n|}-\frac{e^{2t}}{|n|} & -1 \end{array}\right)
\]
and $\mfg(n,t)=|n|e^{-2t}$. The dominant behaviour of solutions in $I_{1}$ is oscillatory with a period roughly given by $T=2\pi/\mfg(n,t)$. 

\textbf{Approximating the oscillations.} Since the dominant behaviour is oscillatory, it is natural to divide the analysis of the evolution into two
steps. First we consider one period of the oscillations; the corresponding analysis yields a matrix describing the evolution over the period. Then we 
draw conclusions concerning the evolution over longer periods of time by multiplying together these matrices. Assuming the oscillations to be sufficiently
rapid, the evolution within one period can be approximated by freezing the coefficients of the equation. Freezing the coefficients of $A$ at $t=t_{0}$, the 
change in the solution from $t_{0}$ to $t_{0}+T_{0}$, where $T_{0}:=2\pi/\mfg(n,t_{0})$, roughly speaking corresponds to multiplication by $\exp[A(n,t_{0})T_{0}]$. 
The matrix $A(n,t_{0})T_{0}$, in its turn, is roughly speaking given by $2\pi B_{0}$, where 
\[
B_{0}:=\left(\begin{array}{cc} 0 & 1 \\ -1 -ir & 0 \end{array}\right)
\]
and $r:=n^{-1}e^{3t_{0}}$. There are various ways of calculating $\exp(2\pi B_{0})$. In Section~\ref{section:calmaexp}, we study the matrix exponentials 
of a general class of matrices that is of interest when approximating oscillations by freezing coefficients. Using these calculations, in particular 
Lemma~\ref{lemma:maexpcomp}, yields
\[
\exp[2\pi B_{0}]\approx \Id_{2}+\pi \left(\begin{array}{cc} 0 & ir \\ -ir & 0 \end{array}\right)
\approx \exp\left[\pi \left(\begin{array}{cc} 0 & ir \\ -ir & 0 \end{array}\right)\right].
\]
The eigenvalues of the matrix inside the exponential on the right hand side are given by $\pm \pi r$. In particular, the largest eigenvalue is given by 
\[
\pi |r|=\pi |n|^{-1}e^{3t_{0}}=\frac{1}{2}e^{t_{0}}T_{0}.
\]
It is thus natural to expect the growth of the solution from $t_{0}$ to $t_{0}+T_{0}$ to be $\exp(e^{t_{0}}T_{0}/2)$. Here $T_{0}$ should be thought of as a 
small increment in $t$, so that summing $e^{t_{0}}T_{0}+e^{t_{1}}T_{1}+\dots$ (where $t_{1}:=t_{0}+T_{0}$, $T_{1}:=2\pi/\mfg(n,t_{1})$ etc.) should roughly
speaking yield the integral of $e^{t}$. The change in absolute value of the solution from $0$ to $t$ should thus, roughly speaking, correspond to 
multiplication by $\exp(e^{t}/2)$. This expectation should be compared with the conclusions of Example~\ref{example:basexsupexpgrowth}. 

\textbf{Gevrey classes.}
The above line of reasoning only concerns one mode. Sooner or later, both $|n|e^{-t}$ and $n^{2}e^{-4t}$ are less than or equal to one. At that point, the
ODE behaviour takes over and yields exponential decay of the Fourier coefficient. However, by combining infinitely many modes,  we obtain super exponential 
growth of smooth solutions. The constructed solutions have the property that the frequency of the dominant Fourier coefficient becomes larger as the time increases.
On the other hand, in order to obtain super exponential growth, the size of the Fourier coefficients of the initial data must not decay too fast with $|n|$; 
otherwise the growth in the intervals of the type $I_{1}$ described above is not sufficient to ensure overall growth. This is why we only obtain super 
exponential growth for the Gevrey indices indicated in Remark~\ref{remark:basexsupexpgrowthGevrey}. 

\textbf{Making the argument rigorous.} The above discussion provides the intuition behind the proof that there are solutions that grow super exponentially.
However, turning the ideas into a rigorous argument requires an effort. The purpose of Part~\ref{part:averaging} of these notes is to calculate the 
evolution over one period to leading order, and to provide an estimate for the error. To this end, we carry out several changes of variables that
serve the purpose of isolating the essential behaviour. Moreover, we calculate the leading order terms in the relevant matrix exponentials (and 
estimate the errors); we estimate the variation of flows associated with the coefficients of (\ref{eq:thesystemRge}) during one period etc. 
Once this has been done, we are interested in 
multiplying together the corresponding matrices in order to derive conclusions concerning the overall behaviour. As $|n|$ grows, so does the number of 
factors. Proving that we can control the error of the product, even though there is no bound on the number of factors, is non-trivial. Moreover, in the 
present setting, we are interested in deriving a lower bound on the size of Fourier coefficients corresponding to appropriately chosen initial
data. The necessary arguments are presented in Sections~\ref{section:PDEunbalterm}--\ref{section:themainestimateseg}.

\chapter{A geometric perspective}\label{chapter:ageomperintro}

\section{Introduction}

Before considering the general case, let us return to (\ref{eq:elexsupexpgrowth}), our basic example of an equation with solutions exhibiting 
super exponential growth. What is the cause of this growth? Considering the assumptions 
required in order for Proposition~\ref{prop:supexpinstabintro} to apply, in particular (\ref{eq:Yoasbehubintro}), it is clear that it is
related to the fact that $Y^{1}$ grows exponentially. In order to obtain a more geometric perspective on the meaning of this object, note 
that the natural Lorentz metric associated with the equation (\ref{eq:elexsupexpgrowth}) is given by 
\begin{equation}\label{eq:gbasexsupexpgrowth}
g=-dt\otimes dt+e^{4t}d\theta\otimes d\theta,
\end{equation}
and defined on $\so\times\ro$. Moreover, for each $t\in\ro$, $g$ induces a metric on $\so$: 
\[
\bge:=e^{4t}d\theta\otimes d\theta.
\]
Turning to the lower order terms, $e^{-t}u_{\theta}$ can be interpreted as $\mcX u$, where $\mcX$ is the vector field $\mcX=e^{-t}\d_{\theta}$. 
Thus, even though the coefficient of $u_{\theta}$ appearing in (\ref{eq:elexsupexpgrowth}) decays exponentially, it is clear that the norm of 
$\mcX$ with respect to $\bge$ does not. In fact, $\bge(\mcX,\mcX)=e^{2t}$. When viewed from a more geometric perspective, it is thus clear that 
$\mcX$ is not small. Given the metric (\ref{eq:gbasexsupexpgrowth}), it is natural to introduce an orthonormal frame $\{e_{0},e_{1}\}$, where 
$e_{0}=\d_{t}$ and $e_{1}=e^{-2t}\d_{\theta}$. With respect to this frame, $\mcX=e^{t}e_{1}$; i.e., $\mcX=Y^{1}e_{1}$. 

\section{Conditions on the metric}

Let us now return to the general case. Consider (\ref{eq:thesystemRge}). Assume the associated metric to be such that $(M,g)$ is a 
canonical separable cosmological model manifold (recall that this implies that $g^{00}=-1$; cf. Remark~\ref{remark:guzzcansepcosmmodmet}). 
Then
\begin{equation}\label{eq:gthesystemRge}
g=-dt\otimes dt+g_{ij}(\chi^{i}dt+dx^{i})\otimes(\chi^{j}dt+dx^{j})+\textstyle{\sum}_{r=1}^{R}a_{r}^{2}g_{r}.
\end{equation}
\index{$\a$Aa@Notation!Metrics!$g$}%
Note that (\ref{eq:gthesystemRge}) induces a family of metrics on $\bM$, given by 
\[
\bge:=g_{ij}dx^{i}\otimes dx^{j}+\textstyle{\sum}_{r=1}^{R}a_{r}^{2}g_{r},
\]
\index{$\a$Aa@Notation!Metrics!$\bge$}%
where we normally do not explicitly state the dependence on $t$. The metric $g$ also induces a family of metrics on $\tn{d}$:
\[
\bh:=g_{ij}dx^{i}\otimes dx^{j}.
\]
\index{$\a$Aa@Notation!Metrics!$\bh$}%
Again, we normally omit explicit reference to the dependence on $t$. 
Next we wish to impose restrictions on the metric (\ref{eq:gthesystemRge}). When studying Einstein's equations, there is a freedom in
choosing the gauge. For this reason there is some freedom in imposing conditions on the metric coefficients. In particular, the lapse
function and the shift vector field, cf. Remark~\ref{remark:shiftandlapsedef}, are related to a choice of foliation of the spacetime. 
In the case of the metric $g$ defined by (\ref{eq:gthesystemRge}), the lapse function equals $1$ and the shift vector field is given by 
$\chi$. Even though one typically has a freedom of imposing 
conditions on the shift vector field, setting it to zero is sometimes problematic. That is the reason why we include it here. On the other hand, it 
is rarely useful to have a shift vector field with a large norm. Moreover, it is often convenient if the foliation is such that $\d_{t}$ is an approximate 
normal to the constant $t$ hypersurfaces. Considering (\ref{eq:gthesystemRge}), the future directed unit normal to $\bM_{t}$ is 
\begin{equation}\label{eq:futdirunitnormU}
U:=\d_{t}-\chi^{i}\d_{i}. 
\end{equation}
To begin with, we therefore impose the following conditions.

\begin{definition}\label{definition:bdsonchi}
Let $(M,g)$ be a canonical separable cosmological model manifold. If there is an $0<\shiftbd{0}\leq 1$ such that 
\begin{equation}\label{eq:ddtuniftimelike}
g(\d_{t},\d_{t})\leq -\shiftbd{0}^{2}
\end{equation}
for all $t\geq 0$, then $\d_{t}$ is said to be \textit{future uniformly timelike}. 
\index{Future uniformly timelike!shift vector field}%
\index{Shift vector field!future uniformly timelike}%
If, in addition, there
is a $1\leq k\in\zo$ and an $0<\shiftbd{k}\in\ro$ such that 
\begin{equation}\label{eq:LieDerchibd}
\textstyle{\sum}_{l=1}^{k}|\ml_{U}^{l}\chi|_{\bge}\leq \shiftbd{k}
\end{equation}
for all $t\geq 0$, then the shift vector field of (\ref{eq:gthesystemRge}) is said to be $C^{k}$-\textit{future bounded}.
\index{$\a$Aa@Notation!Conditions!$C^{k}$-future bounded shift vector field}%
\index{Shift vector field!$C^{k}$-future bounded}%
\end{definition}
\begin{remarks}
The estimate (\ref{eq:ddtuniftimelike}) is equivalent to $g_{ij}\chi^{i}\chi^{j}\leq 1-\shiftrb^{2}$ (note that if $g_{ij}\chi^{i}\chi^{j}\geq 1$, then
$\d_{t}$ is no longer timelike). Moreover, $\ml_{U}\chi$ denotes the Lie derivative of $\chi$ with respect to $U$. 
\end{remarks}
\begin{remarks}
Since $\ml_{\d_{t}}\chi=\ml_{U}\chi$, we could replace $\ml_{U}\chi$ with $\ml_{\d_{t}}\chi$ in (\ref{eq:LieDerchibd}). Note also that since 
$\ml_{U}\chi=\dot{\chi}^{i}\d_{i}$, we sometimes use the notation $\dot{\chi}:=\ml_{U}\chi$. 
\index{$\a$Aa@Notation!Vector fields!$\dot{\chi}$}%
Finally, 
\[
|\ml_{U}^{l}\chi|_{\bge}:=\left[\bge(\ml_{U}^{l}\chi,\ml_{U}^{l}\chi)\right]^{1/2},
\]
where it is understood that $\ml_{U}^{l}\chi$ and $\bge$ are both evaluated at the same time $t$. 
\end{remarks}

Turning to the conditions on $g_{ij}$ and $a_{r}$, it is convenient to express them in terms of the second fundamental form of the hypersurfaces 
$\bM_{t}$; cf. (\ref{eq:bMtintro}). The second fundamental form is given by 
\[
\bk(Z_{1},Z_{2}):=g(\nabla_{Z_{1}}U,Z_{2}),
\]
\index{$\a$Aa@Notation!Second fundamental forms!$\bk$}%
where $\nabla$ is the Levi-Civita connection associated with the Lorentz manifold $(M,g)$, $U$ is given by (\ref{eq:futdirunitnormU}) and 
$Z_{i}$, $i=1,2$, are vector fields tangent to $\bM_{t}$. It can be computed that 
\begin{align*}
\bk = & \frac{1}{2}\sum_{i,j=1}^{d}\d_{t}g_{ij}dx^{i}\otimes dx^{j}+\sum_{r=1}^{R}\frac{\dot{a}_{r}}{a_{r}}a_{r}^{2}g_{r},\\
\ml_{U}\bk = & \frac{1}{2}\sum_{i,j=1}^{d}\d_{t}^{2}g_{ij}dx^{i}\otimes dx^{j}+\sum_{r=1}^{R}
\left[\frac{\ddot{a}_{r}}{a_{r}}+\left(\frac{\dot{a}_{r}}{a_{r}}\right)^{2}\right]a_{r}^{2}g_{r}.
\end{align*}
Again, $\ml_{U}\bk=\ml_{\d_{t}}\bk$, so that in the conditions stated below, we could replace $\ml_{U}\bk$ with $\ml_{\d_{t}}\bk$. 
The basic assumptions we make concerning $\bk$ are the following

\begin{definition}\label{definition:boundsonbk}
Let $(M,g)$ be a canonical separable cosmological model manifold. Let, moreover, $0\leq k\in\zo$. If there is a constant $0<C_{k}\in\ro$ such that 
\[
\textstyle{\sum}_{l=0}^{k}|\ml_{U}^{l}\bk|_{\bge}\leq C_{k}
\]
for all $t\geq 0$, then the second fundamental form is said to be $C^{k}$-\textit{future bounded}. 
\index{$\a$Aa@Notation!Conditions!$C^{k}$-future bounded second fundamental form}%
\index{Second fundamental form!$C^{k}$-future bounded}%
\end{definition}
\begin{remark}
The notation $|\bk|_{\bge}$ is defined by 
\[
|\bk|_{\bge}:=\left(\bge^{ab}\bge^{cd}\bk_{ac}\bk_{bd}\right)^{1/2},
\]
\index{$\a$Aa@Notation!Norms!$\normbkbge$}%
where the lower case Latin indices from the beginning of the alphabet indicate coordinate components on $\bM$ and it is understood that both 
$\bk$ and $\bge$ are evaluated at the same time $t$. In Section~\ref{section:normstensorsapp} we develop different perspectives on this norm. 
\end{remark}
\begin{remark}
The motivation for assuming $\bk$ to be bounded is given in Section~\ref{section:geoconprefme}. However, in some of the arguments, it is not 
sufficient to only assume $\bk$ to be bounded, we also need to impose bounds on $\ml_{U}\bk$. This is the motivation for introducing the 
terminology of the definition. Note also that the basic examples given in Subsection~\ref{ssection:cansepcomoma} are such that the second 
fundamental form is $C^{k}$-future bounded for any $k$.
\end{remark}

\section{The lower order terms}

So far, we introduced bounds on the metric components; cf. Definitions~\ref{definition:bdsonchi} and \ref{definition:boundsonbk}. However,
considering (\ref{eq:thesystemRge}), we also need to impose restrictions on the $X^{j}$, $\a$ and $\zeta$. Concerning the $X^{j}$'s, it is 
convenient to combine them and think of the result as a matrix of vector fields on $M$:
\begin{equation}\label{eq:mcXdef}
\mcX:=X^{j}(t)\d_{j}.
\end{equation}
\index{$\a$Aa@Notation!Vector fields!$\mcX$}%
Due to the results of Chapter~\ref{chapter:onnotofbal}, it is clear that we need to impose bounds on $\mcX$ in order to exclude pathological
behaviour of solutions. When stating the relevant bounds, the following notation is convenient:
\begin{equation}\label{eq:mcXnorm}
|\mcX|_{\bh}:=\left(\textstyle{\sum}_{\varsigma\in \mfS}\sum_{i,j=1}^{d}\bh_{ij}\varsigma_{i}\|X^{i}\|\cdot \varsigma_{j}\|X^{j}\|\right)^{1/2},
\end{equation}
\index{$\a$Aa@Notation!Norms!$\normmcXbh$}%
where it is understood that both $\mcX$ and $\bh$ are evaluated at the same time $t$ and $\mfS$ is the set of elements of $\rn{d}$ whose components 
are plus or minus one; note that $\mfS$ has $2^{d}$ elements. We use similar notation for other matrices of vector fields 
(in the span of the $\{\d_{j}\}$). The reader interested in a motivation for the choice (\ref{eq:mcXnorm}) of norm is referred to the proof of 
Lemma~\ref{lemma:sigmaXbdsandderbds}, in particular (\ref{eq:signsubtlmcXnorm}). 

\begin{definition}\label{definition:bdsonmcX}
Consider (\ref{eq:thesystemRge}). Assume the associated metric to be such that $(M,g)$ is a canonical separable cosmological model manifold.
Define $\mcX$ by (\ref{eq:mcXdef}). If there is a $0\leq k\in\zo$ and a $0<C_{k}\in\ro$ such that 
\begin{equation}\label{eq:LieDermcXbd}
\textstyle{\sum}_{l=0}^{k}|\ml_{U}^{l}\mcX|_{\bh}\leq C_{k}
\end{equation}
for all $t\geq 0$, then $\mcX$ is said to be $C^{k}$-\textit{future bounded}.
\index{$\a$Aa@Notation!Conditions!$C^{k}$-future bounded $\mcX$}%
\index{$\a$Aa@Notation!Conditions!$\mcX$ $C^{k}$-future bounded}%
\end{definition}

Finally, let us turn to the matrix valued functions $\a$ and $\zeta$ appearing in (\ref{eq:thesystemRge}). Due to the results of 
Section~\ref{section:ODEunbal}, we know that if $\a$ and $\zeta$ are allowed to grow exponentially, then spatially homogeneous solutions 
to (\ref{eq:thesystemRge}) (with $f=0$) can exhibit super exponential growth or decay; cf. also the discussion in 
Section~\ref{section:introductiononthenotionofbal}. We consider both of these types of behaviour to be pathological. For these reasons, we 
typically impose the conditions that $\|\a(t)\|$, $\|\dot{\a}(t)\|$, $\|\zeta(t)\|$ and $\|\dot{\zeta}(t)\|$ are bounded to the future.

\section{A rough notion of balance}\label{section:aroughnotofbal}

One basic notion of balance, prohibiting the occurrence of super exponential growth, is the following. 

\begin{definition}\label{definition:Cobal}
Consider (\ref{eq:thesystemRge}). Assume the associated metric to be such that $(M,g)$ is a canonical separable cosmological model manifold.
Let $0\leq k\in\zo$. If $\d_{t}$ is future uniformly timelike; there is a constant $0<C_{k}\in\ro$ such that 
\[
\textstyle{\sum}_{l=0}^{k}[\|\d_{t}^{l}\a(t)\|+\|\d_{t}^{l}\zeta(t)\|]\leq C_{k}
\]
for all $t\geq 0$ (i.e., $\a$ and $\zeta$ are $C^{k}$-future bounded); $\mcX$ is $C^{k}$-future bounded; the shift vector field of (\ref{eq:gthesystemRge}) 
is $C^{k+1}$-future bounded; and the second fundamental form is $C^{k}$-future bounded, then (\ref{eq:thesystemRge}) is said to be 
$C^{k+1}$-\textit{balanced}. 
\index{$\a$Aa@Notation!Conditions!$C^{k}$-balanced equation}%
\index{Equation!$C^{k}$-balanced}%
\end{definition}
\begin{remark}
The bounds on the metric components involve $k+1$ derivatives, but the conditions on the coefficients of the lower order terms in 
(\ref{eq:thesystemRge}) only involve $k$ derivatives. 
\end{remark}

Given a $C^{1}$-balanced equation, the basic energy cannot grow faster than exponentially (except if the right hand side does). In fact, 
there is a constant $0<\eta_{\robal}\in\ro$ (depending only on the constants appearing in the bounds of Definition~\ref{definition:Cobal}) 
such that 
\begin{equation}\label{eq:mferobasestroughbalcase}
\mfe^{1/2}_{\robas}[u](t_{1})\leq e^{\eta_{\robal}|t_{1}-t_{0}|}\mfe^{1/2}_{\robas}[u](t_{0})
+\left|\int_{t_{0}}^{t_{1}}e^{\eta_{\robal}|t_{1}-t|}\|f(\cdot,t)\|_{2}dt\right|
\end{equation}
for all $0\leq t_{0},t_{1}\in\ro$ and all solutions $u$ to (\ref{eq:thesystemRge}), where $\|f(\cdot,t)\|_{2}$ denotes the $L^{2}$-norm of $f(\cdot,t)$. 
This statement follows from Lemma~\ref{lemma:roughenestbalsettingfullsol}; cf. Remark~\ref{remark:roughenestbalsettingfullsol}. In fact, 
estimates of the form (\ref{eq:mferobasestroughbalcase}) hold under slightly weaker conditions; cf. Remark~\ref{remark:roughenestbalsettingfullsol}. 
Even though the estimate (\ref{eq:mferobasestroughbalcase}) is of interest, it only excludes pathological behaviour. We are also interested in 
deriving more detailed information concerning the asymptotics of solutions and concerning the behaviour of the basic energy; the constant $\eta_{\robal}$ 
that results from an application of Lemma~\ref{lemma:roughenestbalsettingfullsol} could very well be quite far from the optimal value. In order to 
obtain more detailed conclusions, we need to make stronger assumptions. 

\section[Geometric formulation, unbalanced setting]{Geometric formulation of the results 
in the unbalanced setting}\label{section:geomformunbalsetintro}

At this stage, it is of interest to return to the unbalanced setting, and to express the conditions geometrically.

\begin{assumptions}\label{assumptions:geomcritunbalsetintro}
Consider (\ref{eq:thesystemRge}). Assume the associated metric to be such that $(M,g)$ is a canonical separable cosmological model 
manifold; cf. Definition~\ref{def:cansepcosmmodmanintro}. Assume, moreover, that $d=1$, that $R=0$ and that there is an 
$0<\shiftbd{0}\leq 1$ such that (\ref{eq:ddtuniftimelike}) holds.  Assume that there are constants $K_{\rosh},K_{\bk},K_{\roode},\bK_{X}>0$, 
$\b_{\rosh}<0$, $\b_{\roode}$ and $\b_{X}>0$ such that $\b_{X}-\b_{\roode}>0$ and 
\begin{align}
|\chi(t)|_{\bge}+|\dot{\chi}(t)|_{\bge} \leq & K_{\rosh}e^{\b_{\rosh}t},\label{eq:shiftexpdecbdsupexpgrintro}\\
|\ddot{\chi}(t)|_{\bge}+|\bk(t)|_{\bge}+|\ml_{U}\bk(t)|_{\bge} \leq & K_{\bk},\label{eq:shiftddbkzadsupexpgrintro}\\
\|\zeta(t)\|^{1/2}+\|\dot{\zeta}(t)\|^{1/2}+\|\a(t)\|+\|\dot{\a}(t)\| \leq & K_{\roode}e^{\b_{\roode}t},\label{eq:zetaalphaanddersupexpgrintro}\\
|\mcX(t)|_{\bh}+|\dot{\mcX}(t)|_{\bh} \leq & \bK_{X}e^{\b_{X}t}\label{eq:mcXamcXdsupexpgrintro}
\end{align}
for all $t\geq 0$. 

Let $e_{1}$ be the unit vector field which is a positive multiple of $\d_{1}$, define 
$\mcY^{1}$ to be the matrix valued function such that $\mcY^{1}e_{1}=\mcX$ and assume that there is a matrix $\mcY^{1}_{\infty}\in \Mn{m}{\co}$ and 
$\etab_{X},L_{X}>0$ such that 
\begin{equation}\label{eq:mcYomcYoinfestintro}
\|e^{-\b_{X}t}\mcY^{1}(t)-\mcY^{1}_{\infty}\|\leq L_{X}e^{-\etab_{X}t}
\end{equation}
for all $t\geq 0$. Assume, moreover, that $\mcY^{1}_{\infty}$ has an eigenvalue with a non-zero real part.

Assume, finally, that there are 
$\bbe_{j}\in\ro$, $j=1,2$, and a continuous function $0\leq\betafun\in L^{1}([0,\infty))$ such that 
\begin{equation}\label{eq:bkbobtwobdellderunbdintro}
(\bbe_{2}-\betafun)\bge\leq \bk\leq(\bbe_{1}+\betafun)\bge
\end{equation}
for all $t\geq 0$.
\end{assumptions}
\begin{remark}
If Assumptions~\ref{assumptions:geomcritunbalsetintro} hold, then Assumptions~\ref{ass:mainassumpubcaseintro} and \ref{assumption:Yoconvubintro} 
hold with $Y^{1}_{\infty}=\mcY^{1}_{\infty}$; cf. Lemma~\ref{lemma:geomcritunbalset} and Remark~\ref{remark:geomcritunbsetintrover}. 
\end{remark}
\begin{remark}
The only quantities we require to converge are the components of $e^{-\b_{X}t}\mcX$ with respect to the unit vector field $e_{1}$. 
Moreover, the condition $\b_{X}-\b_{\roode}>0$ implies that $\mcY^{1}$ is larger in size than $\|\a(t)\|$ and $\|\zeta(t)\|^{1/2}$. 
\end{remark}
\begin{remark}\label{remark:guoolobdgeomunbal}
One consequence of the assumptions is that there is a constant $c_{1}>0$ such that $c_{1}e^{-2\bbe_{1}t}\leq g^{11}(t)$ for 
all $t\geq 0$; this follows from Lemma~\ref{lemma:guaranteeingbdsonguoounbset}. 
\end{remark}

\begin{prop}
Consider (\ref{eq:thesystemRge}). Assume that $d=1$, $R=0$ and $f=0$. Given that Assumptions~\ref{assumptions:geomcritunbalsetintro} hold, let 
$\kappa>0$ be the largest absolute value of a real part of an eigenvalue of $\mcY^{1}_{\infty}/2$. Let $\e>0$. 
Then there is a sequence of smooth $\cn{m}$-valued solutions $v_{l}$ to (\ref{eq:thesystemRge}), $1\leq l\in\zo$, and for each $l\geq 1$, 
there is a time sequence $t_{l,k}\rightarrow\infty$ (as $k\rightarrow\infty$) such that for each $s\in\ro$, 
\begin{equation}\label{eq:vlsobconvtozerointrogeo}
\lim_{l\rightarrow\infty}\left(\|v_{l}(\cdot,0)\|_{(s+1)}+\|\d_{t}v_{l}(\cdot,0)\|_{(s)}\right)=0.
\end{equation}
In addition, 
\begin{equation}\label{eq:mferohomsupexplbintrogeo}
\mfe_{\rohom}[v_{l}](t_{l,k})\geq \exp\left[2\b_{X}^{-1}\kappa(1-\e)e^{\b_{X}t_{l,k}}\right],
\end{equation}
where $\mfe_{\rohom}[u]$ is given by (\ref{eq:mferohomdeqodefintro}). 
Moreover, if $\b_{X}+\bbe_{1}>0$, then the sequence of functions $v_{l}$ can be assumed to be such that $v_{l},\d_{t}v_{l}\in 
\mfG_{\roloc,\rou}^{s_{X}}(I,\so,\cn{m})$ for all $l$, where 
\begin{equation}\label{eq:sXdefintrogeo}
s_{X}:=\frac{\b_{X}+\bbe_{1}}{\b_{X}}.
\end{equation}
In case $\b_{X}+\bbe_{1}\leq 0$, one frequency can be chosen, say $n_{a}\in\zo$, such that the 
sequence of functions $v_{l}$ has the property that if $z_{l}(n,t)$ denotes the 
$n$'th Fourier coefficient of $v_{l}(\cdot,t)$, then $z_{l}(n,t)=0$ for all $t$ unless $n=n_{a}$. 
\end{prop}
\begin{remark}
If $\mcY^{1}_{\infty}$ is Hermitian, (\ref{eq:mferohomsupexplbintrogeo}) is essentially optimal. In fact, in that case there is, given a solution
$u$, a constant $C_{\e}$ such that $\mfe_{\rohom}[u](t)\leq C_{\e}\exp\left[2\b_{X}^{-1}\kappa(1+\e)e^{\b_{X}t}\right]$ for all $t\geq 0$;
cf. Remark~\ref{remark:upbdstenest}. 
\end{remark}
\begin{remark}
The constituents of the right hand side of (\ref{eq:sXdefintrogeo}) have a natural interpretation. The constant $\bbe_{1}$ represents the (average)
maximal expansion of the spacetime, in the sense that it corresponds to the upper bound in (\ref{eq:bkbobtwobdellderunbdintro}). Moreover, 
$\b_{X}$ is the exponential rate of growth of the norm of the vector field $\mcX$ (with respect to the induced metric on the constant-$t$ 
hypersurfaces). The statements concerning the regularity can be summarised as follows. If the average maximal expansion is strictly positive
(i.e., if $\bbe_{1}>0$), then $s_{X}>1$, and the initial data are not analytic. If the average maximal expansion is zero (i.e., if $\bbe_{1}=0$), 
then $s_{X}=1$, and the initial data are analytic. If the average maximal expansion is negative, but strictly greater than minus the growth rate of 
$\mcX$ (i.e., if $-\b_{X}<\bbe_{1}<0$), then $0<s_{X}<1$, and the initial data are entire. If the average maximal expansion is less than or equal to 
minus the growth rate of $\mcX$ (i.e., if $\bbe_{1}\leq -\b_{X}$), then the initial data have only one non-zero Fourier coefficient.
\end{remark}
\begin{proof}
Due to the assumptions of the proposition, the conditions of Proposition~\ref{prop:supexpinstabintro} are fulfilled. Moreover, due to 
Remark~\ref{remark:guoolobdgeomunbal}, the conditions of Proposition~\ref{prop:supexpinstabGevreyregintro} hold with $\b_{1}=-\bbe_{1}$.
Combining the corresponding conclusions yields the statement of the proposition. 
\end{proof}

Next, let us turn to the question of the generic behaviour. 

\begin{prop}\label{prop:gensupexpgrowthgeoform}
Consider (\ref{eq:thesystemRge}). Assume the associated metric to be such that $(M,g)$ is a canonical separable cosmological model manifold.
Assume, moreover, that there is a $j\in \{1,\dots,d\}$ such that 
\begin{equation}\label{eq:urestrtojdepintrogeo}
u_{tt}-g^{jj}(t)\d_{j}^{2}u-2g^{0j}(t)\d_{t}\d_{j}u+\a(t)u_{t}+X^{j}(t)\d_{j}u+\zeta(t)u=0
\end{equation}
(no summation on $j$) satisfies Assumptions~\ref{assumptions:geomcritunbalsetintro}. Then the conclusions of 
Propositions~\ref{prop:supexpgrowthgencasedenintro} and \ref{prop:supexpgrowthinfcodimintro} hold. 
\end{prop}
\begin{remark}
The conclusions of Propositions~\ref{prop:supexpgrowthgencasedenintro} and \ref{prop:supexpgrowthinfcodimintro} can roughly speaking be 
summarised as saying that there is a Baire generic set of initial data corresponding to solutions exhibiting super exponential growth; and 
that the initial data corresponding to solutions that do not exhibit super exponential growth have infinite co-dimension.
\end{remark}
\begin{proof}
The conditions of the proposition ensure that the assumptions required to appeal to Propositions~\ref{prop:supexpgrowthgencasedenintro} and 
\ref{prop:supexpgrowthinfcodimintro} are satisfied. 
\end{proof}

It is of interest to contrast this result with conclusions obtained concerning solutions with finite frequency content. In particular, we here
consider a situation in which the dominant coefficients in (\ref{eq:thesystemRge}) are $\a$ and $\zeta$. Recalling the notation introduced
in Section~\ref{section:divintomodes}, we have the following result. 

\begin{prop}\label{prop:supexpdecfinitefreqcont}
Consider (\ref{eq:thesystemRge}). Assume the associated metric to be such that $(M,g)$ is a canonical separable cosmological model
manifold. Assume, moreover, that $f=0$. Assume that there is a 
$\g>0$; $\a_{\infty}, \zeta_{\infty}\in\Mn{m}{\co}$, one of which is non-zero; an $\eta_{\rocu}>0$; and a constant $C_{\rocu}>0$ such that 
\begin{align*} 
|g^{0l}(t)|+|g^{jl}(t)|^{1/2}+|a^{-1}_{r}(t)|+\|X^{j}(t)\|^{1/2} \leq & C_{\rocu}e^{(\g-\eta_{\rocu})t},\\
\|e^{-\g t}\a(t)-\g\a_{\infty}\|+\|e^{-2\g t}\zeta(t)-\g^{2}\zeta_{\infty}\|^{1/2} \leq & C_{\rocu}e^{-\eta_{\rocu}t}
\end{align*}
for all $t\geq 0$, $j,l=1,\dots,d$, and $r=1,\dots,R$. Let $A_{\infty}$ be defined by (\ref{eq:Ainfdefintrobal}) 
and $\kappa_{1}$ be the largest real part of an eigenvalue of $A_{\infty}$. Consider a solution $u$ to (\ref{eq:thesystemRge}) such that 
if $z$ is defined by (\ref{eq:znutdef}), then $z(\indexnot,\cdot)=0$ for all but a finite number of $\indexnot\in\EFindexset$. 
Fix $\e>0$. Then there is a constant $C_{\e}>0$ such that 
\[
\mfe_{\robas}[u](t)\leq C_{\e}\exp[2(\kappa_{1}+\e)e^{\g t}]\mfe_{\robas}[u](0)
\]
for all $t\geq 0$. Here $\mfe_{\robas}$ is defined by (\ref{eq:mfebasdef}).
\end{prop}
\begin{remark}
The constant $C_{\e}$ only depends on $C_{\rocu}$; the set of $\indexnot\in\EFindexset$ for which $z(\indexnot,\cdot)$ does not vanish; 
the spectrum of the Laplace-Beltrami operator on $(M_{r},g_{r})$ for $r=1,\dots,R$; $A_{\infty}$; $\eta_{\rocu}$; $\g$; and $\e$.
\end{remark}
\begin{remark}
There are large classes of equations satisfying both the assumptions of Proposition~\ref{prop:gensupexpgrowthgeoform} and the assumptions
of Proposition~\ref{prop:supexpdecfinitefreqcont} with $\kappa_{1}<0$. For the corresponding equations, solutions with finite frequency content
decay to zero super exponentially. In other words, there is a dense set of initial data such that the corresponding solutions exhibit
super exponential decay. However, generic solutions grow super exponentially. 
\end{remark}
\begin{remark}
Appealing to Lemma~\ref{lemma:expdeccondmode} instead of to Lemma~\ref{lemma:supexpdeccondmode} yields an analogous result in the case that 
$\a$ and $\zeta$ converge exponentially to elements of $\Mn{m}{\co}$. In that case, assuming the quantity corresponding to $\kappa_{1}$ in 
Proposition~\ref{prop:supexpdecfinitefreqcont} to be negative, solutions with finite frequency content decay exponentially. 
\end{remark}
\begin{proof}
The conclusion is an immediate consequence of Lemma~\ref{lemma:supexpdeccondmode}.
\end{proof}

\chapter{Silent equations}\label{chapter:silentequations}

\section{Introduction}

\textbf{Heuristics.}
Consider (\ref{eq:polGowdytimereversed}). As already mentioned, the direction $t\rightarrow\infty$ corresponds to the big bang singularity of 
polarised $\tn{3}$-Gowdy symmetric solutions to Einstein's equations. In heuristic arguments concerning cosmological singularities, similar equations 
appear. In particular, it is often expected, on heuristic grounds, to be possible to approximate the 
full Einstein equations with equations such that the coefficients multiplying the spatial derivatives of the unknowns decay exponentially. A similar
structure appears in the expanding direction when considering solutions to Einstein's equations with a positive cosmological constant. In the 
physics literature, the exponential decay of the coefficients is often assumed to justify dropping the spatial derivatives. This results in one
ODE for each spatial point, a substantially simplified situation. In order to illustrate this line of reasoning, consider (\ref{eq:polGowdytimereversed})
again. Dropping the term involving the spatial derivatives yields the equation $P_{tt}=0$ with solutions $P(\theta,t)=v(\theta)t+\psi(\theta)$. 
This heuristically motivated conclusion should be compared with the statements made in connection with (\ref{eq:polGowdyasympt}); see also
Example~\ref{example:polt3gowdyderofas} below. 

Due to the results of Chapter~\ref{chapter:onnotofbal}, we know that the above line of reasoning does not always work; cf., in particular, 
Example~\ref{example:basexsupexpgrowth}. On the other hand, perhaps the situation changes if we assume the equation to be balanced in the sense of 
Definition~\ref{definition:Cobal}.  For this reason, it is of interest to consider balanced systems of equations such that the coefficients of the 
spatial derivatives decay exponentially. Moreover, the question we wish to ask is: can the above heuristic picture be given a mathematical 
justification? 

\textbf{Examples.} Considering the examples of geometries given in Section~\ref{section:geoconprefme}, the metrics $g_{\rodS}$ and $g_{\mathrm{gdS}}$ 
yield principal symbols such that the coefficients of the highest order spatial derivatives decay exponentially. Considering the contracting 
direction of Kasner solutions (with the exception of the flat Kasner solutions), or of non-vacuum Bianchi type I solutions to the Einstein-scalar 
field equations, yields the same conclusion. 

\textbf{Geometry.} The above heuristics are formulated in terms of properties of the coefficients of the equations. However, it is desirable
to formulate geometric conditions. Exponential decay of the coefficients of the spatial derivatives is related to exponential expansion. 
Exponential expansion, in its turn, is related to lower bounds on the second fundamental form. In the examples given above, not only do the 
second fundamental forms have a uniform strictly positive lower bound (for late times), they, in fact, converge. However, in the present chapter,
the main assumption is that of a lower bound. It is therefore of interest to introduce the following terminology.

\begin{definition}\label{definition:Cosilenceintro}
Consider (\ref{eq:thesystemRge}). Assume the associated metric to be such that $(M,g)$ is a canonical separable cosmological model manifold.
If $\d_{t}$ is future uniformly timelike and there is a $0<\mu\in\ro$ and a continuous non-negative $\betafun\in L^{1}([0,\infty))$ such that 
\begin{equation}\label{eq:bklowerbdCosilintro}
\bk\geq (\mu-\betafun)\bge,\ \ \
|\chi|_{\bge}\cdot|\dot{\chi}|_{\bge}\leq \betafun
\end{equation}
for all $t\geq 0$; then (\ref{eq:thesystemRge}) is said to be $C^{1}$-\textit{silent}.
\index{Equation!$C^{1}$-silent}%
\index{$\a$Aa@Notation!Conditions!$C^{1}$-silent equation}%
\end{definition}
\begin{remark}
In Subsection~\ref{ssection:siltrsanoimet}, we motivate the terminology ``silent metric'' in the context of the metrics introduced
in (\ref{eq:modelmetricafterresc}). The motivation in the present setting is similar. In fact, let $\g:J\rightarrow M$ be a future directed 
inextendible causal curve in $(M,g)$, where $J=(s_{-},s_{+})$. Then $\g^{0}(s)\rightarrow\infty$ as $s\rightarrow s_{+}$, where $\g^{0}$ denotes 
the $t$-coordinate of $\g$; this is an immediate consequence of the fact that $\bM_{t}$ is a Cauchy hypersurface in $(M,g)$ for each $t\in\ro$
(cf. Lemma~\ref{lemma:globallyhyperbol}) and the tacit assumption that $\d_{t}$ is future oriented. Moreover, given that the 
assumptions of the definition are satisfied, the $\bM$-coordinate of $\g$, say $\bga$, converges to a point, say $\bp[\g]$ 
as $s\rightarrow s_{+}$. Finally, if two future directed inextendible causal curves, say $\g_{i}:J_{i}\rightarrow M$, $i=1,2$, are such that 
$\bp[\g_{1}]\neq \bp[\g_{2}]$, then there are $s_{i}\in J_{i}$ such that 
\[
J^{+}[\g_{1}(s_{1})]\cap J^{+}[\g_{2}(s_{2})]=\varnothing;
\]
the reader interested in an explanation of the terminology $J^{+}(p)$ is referred to \cite{oneill}. 
Thus, sooner or later, the observers $\g_{i}$ lose the ability to communicate to the future. In other words, there is silence. 
For a justification of these statements, the reader is referred to Lemma~\ref{lemma:silenceconsforcauscurves} and 
Remark~\ref{remark:silenceconsforcauscurves}.
\end{remark}

In the present chapter, we are interested in equations (\ref{eq:thesystemRge}) that are $C^{1}$-silent in the sense of 
Definition~\ref{definition:Cosilenceintro}; such that $\mcX$ is $C^{0}$-future bounded; and such that there are 
$\a_{\infty},\zeta_{\infty}\in\Mn{m}{\co}$ and $0<\eta_{\romn},C_{\romn}\in\ro$ with the property that 
\begin{equation}\label{eq:alpahzetaconvest}
\|\a(t)-\a_{\infty}\|+\|\zeta(t)-\zeta_{\infty}\|\leq C_{\romn}e^{-\eta_{\romn}t}
\end{equation}
for all $t\geq 0$. Under these assumptions, all the coefficients in (\ref{eq:thesystemRge}) that multiply spatial derivatives of 
$u$ decay to zero exponentially; cf. Lemma~\ref{lemma:coeffofspderdecexpsil}. Since (\ref{eq:alpahzetaconvest}) is satisfied, it is 
then of interest to compare solutions to (\ref{eq:thesystemRge}) with solutions to the equation
\begin{equation}\label{eq:ODEapproxeqinCosilentsett}
\d_{t}\left(\begin{array}{c} v \\ v_{t}\end{array}\right)=A_{\infty}\left(\begin{array}{c} v \\ v_{t}\end{array}\right)+\left(\begin{array}{c} 0 \\ f\end{array}\right),
\end{equation}
where 
\begin{equation}\label{eq:Asilentdefintro}
A_{\infty}:=\left(\begin{array}{cc} 0 & \Id_{m} \\ -\zeta_{\infty} & -\a_{\infty}\end{array}\right).
\end{equation}
\index{$\a$Aa@Notation!Matrix notation!Ainfinity@$A_{\infty}$}%
Note that if one drops the second term on the right hand side of (\ref{eq:ODEapproxeqinCosilentsett}), then what results is a system
of constant coefficient equations in which no spatial derivatives occur. On the other hand, we have not eliminated the dependence of the 
solution on the spatial 
variables. Thus (\ref{eq:ODEapproxeqinCosilentsett}) should
be interpreted as a system of ODE's for each point in $\bM$. In case $f=0$, one would expect solutions to (\ref{eq:ODEapproxeqinCosilentsett}) 
to grow as $\ldr{t}^{d_{1}-1}e^{\kappa_{1}t}$, where $\kappa_{1}$ is the largest real part of an eigenvalue of $A_{\infty}$ and $d_{1}$ is the largest dimension 
of a Jordan block corresponding to an eigenvalue of $A_{\infty}$ with real part $\kappa_{1}$; cf. Section~\ref{section:jordannormalform} below for a 
justification. Here we use the notation 
\begin{equation}\label{eq:ldrdefinitionintro}
\ldr{\xi}:=(1+|\xi|^{2})^{1/2}
\end{equation}
\index{$\a$Aa@Notation!Auxiliary functions!$\ldr{\cdot}$}%
for $\xi\in\cn{l}$. Due to the importance of $\kappa_{1}$ and $d_{1}$ in the description of the asymptotics, it is convenient to introduce the 
following terminology. 
\begin{definition}\label{def:SpRspdef}
Given $A\in\Mn{k}{\co}$, let $\Spe A$ 
\index{$\a$Aa@Notation!Matrix notation!Spea@$\Spe A$}%
denote the set of eigenvalues of $A$. Moreover, let 
\[
\kappa_{\max}(A):=\sup\{\mathrm{Re}\lambda\ |\ \lambda\in \Spe A\},\ \ \
\kappa_{\min}(A):=\inf\{\mathrm{Re}\lambda\ |\ \lambda\in \Spe A\}.
\]
\index{$\a$Aa@Notation!Matrix notation!kappamax@$\kappa_{\max}$}%
\index{$\a$Aa@Notation!Matrix notation!kappamin@$\kappa_{\min}$}%
Then $\Rsp A$, 
\index{$\a$Aa@Notation!Matrix notation!Rspa@$\Rsp A$}%
the \textit{real eigenvalue spread}
\index{Real eigenvalue spread}%
 of $A$, is defined by $\Rsp A:=\kappa_{\max}(A)-\kappa_{\min}(A)$. In addition, if 
$\kappa\in \{\mathrm{Re}\lambda\ |\ \lambda\in \Spe A\}$, then 
$d_{\max}(A,\kappa)$ 
\index{$\a$Aa@Notation!Matrix notation!dmax@$d_{\max}$}%
is defined to be the largest dimension of a Jordan 
block corresponding to an eigenvalue of $A$ with real part $\kappa$. Finally, if $\kappa\notin \{\mathrm{Re}\lambda\ |\ \lambda\in \Spe A\}$,
then $d_{\max}(A,\kappa):=1$. 
\end{definition}

\section{Results}\label{section:resultssilentsettingintro}

Concerning the energy of solutions to (\ref{eq:thesystemRge}), we obtain the following conclusion. 

\begin{prop}\label{prop:genroughestintro}
Consider (\ref{eq:thesystemRge}). Assume that it is $C^{1}$-silent in the sense of Definition~\ref{definition:Cosilenceintro}; 
that $\mcX$ is $C^{0}$-future bounded; and that there are $\a_{\infty},\zeta_{\infty}\in\Mn{m}{\co}$ and $0<\eta_{\romn},C_{\romn}\in\ro$ 
with the property that (\ref{eq:alpahzetaconvest}) holds for all $t\geq 0$. Then there are constants $C$ and $s_{0}\geq 0$, 
depending only on the coefficients of the equation (\ref{eq:thesystemRge}), such that if $u$ is a smooth solution to (\ref{eq:thesystemRge}), 
then
\begin{equation}\label{eq:mfeestslossintro}
\begin{split}
\mfe_{s}^{1/2}[u](t) \leq & C\ldr{t}^{d_{1}-1}e^{\kappa_{1}t}\mfe_{s+s_{0}}^{1/2}[u](0)\\
 & +C\int_{0}^{t}\ldr{t-\tau}^{d_{1}-1}e^{\kappa_{1}(t-\tau)}\|f(\cdot,\tau)\|_{(s+s_{0})}d\tau
\end{split}
\end{equation}
for all $t\geq 0$ and $s\in\ro$, where $\kappa_{1}:=\kappa_{\max}(A_{\infty})$, $d_{1}:=d_{\max}(A_{\infty},\kappa_{1})$ and $A_{\infty}$ is given by 
(\ref{eq:Asilentdefintro}); cf. Definition~\ref{def:SpRspdef}.
\end{prop}
\begin{remark}
The proposition is a consequence of Lemma~\ref{lemma:genroughest}; cf. Remark~\ref{remark:condequivbasenest}.
\end{remark}
\begin{remarks}
More detailed information concerning the dependence of the constants is to be found in Remark~\ref{remark:deponconstbasenest}. The
energy $\mfe_{s}$ is defined by (\ref{eq:mfedef}). The object $\|f(\cdot,t)\|_{(s)}$ is the $H^{s}$-Sobolev norm of the function $f(\cdot,t)$; 
cf. (\ref{eq:HsnormonbM}). 
\end{remarks}

Next we would like to derive more detailed information concerning the asymptotics. Naively, we would expect a solution to (\ref{eq:thesystemRge})
to be well approximated by a solution to (\ref{eq:ODEapproxeqinCosilentsett}). However, there are obstructions. Say, for the sake
of argument, that 
\begin{equation}\label{eq:fAsnormdefintro}
\|f\|_{A,s}:=\int_{0}^{\infty}e^{-\kappa_{1}\tau}\|f(\cdot,\tau)\|_{(s)}d\tau<\infty
\end{equation}
\index{$\a$Aa@Notation!Norms!$\normfunAs$}%
for $s\in\ro$, where $\kappa_{1}$ is the number defined in the statement of Proposition~\ref{prop:genroughestintro}. Then the energies associated
with solutions $u$ to (\ref{eq:thesystemRge}) are bounded by $C\ldr{t}^{2d_{1}-2}e^{2\kappa_{1}t}$; cf. (\ref{eq:mfeestslossintro}). Moreover, as is 
demonstrated below, this estimate is optimal. On the other hand, when taking the step from (\ref{eq:thesystemRge}) to 
(\ref{eq:ODEapproxeqinCosilentsett}), we have ignored terms of the form
\begin{equation}\label{eq:discardedtermssilentintro}
\chi^{j}\d_{j}\d_{t}u,\ \ \ X^{j}\d_{j}u,\ \ \ (\a-\a_{\infty})u_{t},\ \ \ (\zeta-\zeta_{\infty})u,
\end{equation}
as well as the terms involving second spatial derivatives of $u$. On the other hand, the first two terms appearing in 
(\ref{eq:discardedtermssilentintro})
can be expected to be of the order of magnitude $\ldr{t}^{d_{1}-1}e^{(\kappa_{1}-\mu)t}$, and the last two terms can be expected to be of the order of 
magnitude $\ldr{t}^{d_{1}-1}e^{(\kappa_{1}-\eta_{\romn})t}$; cf. (\ref{eq:alpahzetaconvest}) and Lemma~\ref{lemma:coeffofspderdecexpsil}. Moreover, the 
terms involving second spatial derivatives of $u$ can be expected to be of the order of magnitude $\ldr{t}^{d_{1}-1}e^{(\kappa_{1}-2\mu)t}$; cf. 
Lemma~\ref{lemma:coeffofspderdecexpsil}. For this reason, it would be optimistic to think that the terms appearing in solutions to 
(\ref{eq:ODEapproxeqinCosilentsett}) that are of the order of magnitude $e^{(\kappa_{1}-\b_{\rem})t}$, where
\begin{equation}\label{eq:bremintrodef}
\b_{\rem}:=\min\{\mu,\eta_{\romn}\}
\end{equation}
can be distinguished from the error terms arising when the terms (\ref{eq:discardedtermssilentintro}) are omitted. When considering solutions to 
the homogeneous version of the equation (\ref{eq:ODEapproxeqinCosilentsett}), it therefore seems natural to focus on solutions whose norms 
asymptotically exceed $e^{(\kappa_{1}-\b_{\rem})t}$. For this reason, it is natural to introduce the following terminology.

\begin{definition}\label{def:defofgeneigenspintro}
Let $1\leq n\in\zo$, $B\in\Mn{n}{\co}$ and $P_{B}(X)$ be the characteristic polynomial of $B$. Then
\[
P_{B}(X)=\prod_{\lambda\in\Spe B}(X-\lambda)^{n_{\lambda}},
\] 
where $1\leq n_{\lambda}\in\zo$. Moreover, the \textit{generalised eigenspace of $B$ corresponding
to $\lambda$}, denoted $E_{\lambda}$, 
\index{Generalised eigenspace}%
\index{Generalised eigenspace!corresponding to an eigenvalue}%
is defined by 
\[
E_{\lambda}:=\ker (B-\lambda\Id_{n})^{n_{\lambda}},
\]
where $\Id_{n}$ denotes the $n\times n$-dimensional identity matrix. If $J\subseteq\ro$ is an interval, then the $J$-\textit{generalised eigenspace 
of} $B$, 
\index{Generalised eigenspace!corresponding to an interval}%
denoted $E_{B,J}$, 
\index{$\a$Aa@Notation!Vector spaces!$E_{B,J}$}%
is the subspace of $\cn{n}$ defined to be the direct sum of the generalised eigenspaces of $B$ corresponding to eigenvalues 
with real parts belonging to $J$ (in case there are no eigenvalues with real part belonging to $J$, then $E_{B,J}$ is defined to be $\{0\}$). 
Finally, given $0<\b\in\ro$, the \textit{first generalised eigenspace in the $\b$, $B$-decomposition of $\cn{n}$}, 
\index{First generalised eigenspace}%
\index{Generalised eigenspace!first}%
denoted $E_{B,\b}$, is defined
to be $E_{B,J_{\b}}$, where $J_{\b}:=(\kappa-\b,\kappa]$ and $\kappa:=\kappa_{\max}(B)$; cf. Definition~\ref{def:SpRspdef}.
\end{definition}
\begin{remark}
Our definition of a generalised eigenspace coincides with that of \cite[p.~51]{serre}. 
\end{remark}
\begin{remark}
We discuss the significance of the first generalised eigenspace in greater detail in Section~\ref{section:jordannormalform} below. The reason we use
the terminology \textit{first} generalised eigenspace is that we, in Section~\ref{def:fssubspetc} below, divide $\cn{n}$ into a direct sum of the first 
\textit{and second} generalised eigenspaces.
\end{remark}
\begin{remark}\label{remark:EMbrealifMreal}
If $B\in\Mn{n}{\ro}$, then $\lambda\in\Spe B\Rightarrow\lambda^{*}\in\Spe B$, where $\lambda^{*}$ denotes the conjugate of $\lambda$. 
Moreover, $n_{\lambda}=n_{\lambda^{*}}$, so that $v\in E_{\lambda}$ implies $v^{*}\in E_{\lambda^{*}}$. In particular, given an interval $J\subseteq\ro$,
we can thus choose a basis of $E_{B,J}$ consisting of elements of $\rn{n}$. In short: in the context of real $n\times n$-matrices, we can think of 
$E_{B,J}$ as a subspace of $\rn{n}$. 
\end{remark}

Given this terminology, we have the following result concerning the asymptotics.

\begin{prop}\label{prop:roughas}
Consider (\ref{eq:thesystemRge}). Assume that it is $C^{1}$-silent in the sense of Definition~\ref{definition:Cosilenceintro}; 
that $\mcX$ is $C^{0}$-future bounded; and that there are $\a_{\infty},\zeta_{\infty}\in\Mn{m}{\co}$ and $0<\eta_{\romn},C_{\romn}\in\ro$ 
with the property that (\ref{eq:alpahzetaconvest}) holds for all $t\geq 0$. Assume, moreover, that $f$ is a smooth function such that
for every $s\in\ro$, (\ref{eq:fAsnormdefintro}) holds, where $\kappa_{1}:=\kappa_{\max}(A_{\infty})$ and $A_{\infty}$ is given by (\ref{eq:Asilentdefintro}).
Let $\b_{\rem}$ be given by (\ref{eq:bremintrodef}) and $E_{a}$ be the first generalised eigenspace in the $\b_{\rem}$, $A_{\infty}$-decomposition of $\cn{2m}$. 
Then there are constants $C$, $N$ and 
$s_{\rohom},s_{\roih}\geq 0$, depending only on the coefficients of the equation (\ref{eq:thesystemRge}), such that the following holds. Given a smooth 
solution $u$ to (\ref{eq:thesystemRge}), there is a  unique $V_{\infty}\in C^{\infty}(\bM,E_{a})$ such that 
\begin{equation}\label{eq:uudothsestintro}
\begin{split}
 & \left\|\left(\begin{array}{c} u(\cdot,t) \\ u_{t}(\cdot,t)\end{array}\right)
-e^{A_{\infty}t}V_{\infty}
-\int_{0}^{t}e^{A_{\infty}(t-\tau)}\left(\begin{array}{c} 0 \\ f(\cdot,\tau)\end{array}\right)d\tau\right\|_{(s)} \\
\leq & C\ldr{t}^{N}e^{(\kappa_{1}-\b_{\rem})t}\left(\|u_{t}(\cdot,0)\|_{(s+s_{\rohom})}+\|u(\cdot,0)\|_{(s+s_{\rohom}+1)}
+\|f\|_{A,s+s_{\roih}}\right)
\end{split}
\end{equation}
holds for $t\geq 0$ and all $s\in\ro$. Moreover, 
\begin{equation}\label{eq:uinfudinfHsestintro}
\|V_{\infty}\|_{(s)}\leq C\left(\|u_{t}(\cdot,0)\|_{(s+s_{\rohom})}+\|u(\cdot,0)\|_{(s+s_{\rohom}+1)}+\|f\|_{A,s+s_{\roih}}\right).
\end{equation}
\end{prop}
\begin{remark}
The proposition is a consequence of Lemma~\ref{lemma:roughas}; cf. Remark~\ref{remark:condequivbasassil}.
\end{remark}
\begin{remark}\label{remark:chimcXimprassbetterbrem}
If, in addition to the assumptions of the proposition, there are constants $K_{\roder},\b_{\roder}>0$ such that 
\begin{equation}\label{eq:chimcXexpdecest}
|\chi|_{\bge}+|\mcX|_{\bh}\leq K_{\roder}e^{-\b_{\roder}t}
\end{equation}
for all $t\geq 0$, then $\b_{\rem}$ can be replaced by $\min\{2\mu,\mu+\b_{\roder},\eta_{\romn}\}$. This follows from 
Remark~\ref{remark:improvestchimcXbremgeomset}.
\end{remark}
\begin{remark}
If $\a_{\infty},\zeta_{\infty}\in\Mn{m}{\ro}$, then $E_{a}$ can be thought of as a subspace of $\rn{2m}$ due to 
Remark~\ref{remark:EMbrealifMreal}
\end{remark}
\begin{remarks}
Due to the definition of $E_{a}$, the function $V_{\infty}$ is uniquely determined by the estimate (\ref{eq:uudothsestintro}). 
Viewing $V_{\infty}$ as asymptotic data, the estimate (\ref{eq:uinfudinfHsestintro}) can be interpreted as saying that the map from 
initial data to asymptotic data is continuous with respect to the $C^{\infty}$-topology. Finally, note that 
\[
e^{A_{\infty}t}V_{\infty}+\int_{0}^{t}e^{A_{\infty}(t-\tau)}\left(\begin{array}{c} 0 \\ f(\cdot,\tau)\end{array}\right)d\tau
\]
is a solution to (\ref{eq:ODEapproxeqinCosilentsett}). 
\end{remarks}

Proposition~\ref{prop:roughas} yields asymptotic data, given a solution. However, it is also of interest to specify the asymptotics.
Due to Proposition~\ref{prop:roughas}, it is sufficient to demonstrate that asymptotic data can be specified in the case of homogeneous
equations. For that reason, we focus on the homogeneous setting for the remainder of the section; i.e., we assume $f=0$ in 
(\ref{eq:thesystemRge}). In order to obtain conclusions, we need to strengthen the assumptions of the previous two propositions. In 
particular, we need to assume an upper bound on $\bk$, analogous to the lower bound of Definition~\ref{definition:Cosilenceintro}. The
reason for this is that we need to control how the energy evolves when going backwards in time, and an upper bound on $\bk$ excludes 
pathologically fast growth. 

\begin{prop}\label{prop:spasda}
Consider (\ref{eq:thesystemRge}) with $f=0$. Assume that (\ref{eq:thesystemRge}) is $C^{1}$-silent in the sense of 
Definition~\ref{definition:Cosilenceintro}; that $\mcX$ is $C^{0}$-future bounded; and that there are $\a_{\infty},\zeta_{\infty}\in\Mn{m}{\co}$ 
and $0<\eta_{\romn},C_{\romn}\in\ro$ 
with the property that (\ref{eq:alpahzetaconvest}) holds for all $t\geq 0$. Assume, moreover, that there is a constant 
$0<\mu_{+}\in\ro$ and a non-negative continuous function $\betafun_{+}\in L^{1}([0,\infty))$ such that 
\begin{equation}\label{eq:elldlbintro}
\bk\leq (\mu_{+}+\betafun_{+})\bge
\end{equation}
for all $t\geq 0$. Let $\b_{\rem}$ be given by (\ref{eq:bremintrodef}) and $E_{a}$ be the first generalised eigenspace in the $\b_{\rem}$, 
$A_{\infty}$-decomposition of $\cn{2m}$. Then there is an injective map
\[
\Phi_{\infty}:C^{\infty}(\bM,E_{a})\rightarrow C^{\infty}(\bM,\cn{2m})
\]
such that the following holds. First, 
\begin{equation}\label{eq:Phiinfnobdintro}
\|\Phi_{\infty}(\psi)\|_{(s)}\leq C\|\psi\|_{(s+s_{\infty})}
\end{equation}
for all $s\in\ro$ and all $\psi\in C^{\infty}(\bM,E_{a})$, where the constants $C$ and $s_{\infty}\geq 0$ only depend on the coefficients of the 
equation (\ref{eq:thesystemRge}). Second, if $\psi\in C^{\infty}(\bM,E_{a})$ and $u$ is the solution to (\ref{eq:thesystemRge}) (with $f=0$) such that 
\begin{equation}\label{eq:uuditoPhiinfchiintro}
\left(\begin{array}{c} u(\cdot,0) \\ u_{t}(\cdot,0)\end{array}\right)=\Phi_{\infty}(\psi),
\end{equation}
then 
\begin{equation}\label{eq:estspecasdataintro}
\begin{split}
 & \left\|\left(\begin{array}{c} u(\cdot,t) \\ u_{t}(\cdot,t)\end{array}\right)
-e^{A_{\infty}t}\psi\right\|_{(s)} \\
\leq & C\ldr{t}^{N}e^{(\kappa_{1}-\b_{\rem})t}\left(\|u_{t}(\cdot,0)\|_{(s+s_{\rohom})}+\|u(\cdot,0)\|_{(s+s_{\rohom}+1)}\right)
\end{split}
\end{equation}
for all $t\geq 0$ and $s\in\ro$, where the constants $C$, $N$ and $s_{\rohom}$ have the same dependence as in the case of 
Proposition~\ref{prop:roughas}, $\kappa_{1}:=\kappa_{\max}(A_{\infty})$ and $A_{\infty}$ is given by (\ref{eq:Asilentdefintro}). Finally, if 
$E_{a}=\cn{2m}$, then $\Phi_{\infty}$ is surjective. 
\end{prop}
\begin{remark}
The proposition is a consequence of Lemma~\ref{lemma:spasda}; cf. Remark~\ref{remark:specassilsetpropequivtolem}.
\end{remark}
\begin{remark}\label{remark:chimcXimprassbetterbremspas}
If, in addition to the assumptions of the proposition, there are constants $K_{\roder},\b_{\roder}>0$ such that (\ref{eq:chimcXexpdecest})
holds for all $t\geq 0$, then $\b_{\rem}$ can be replaced by $\min\{2\mu,\mu+\b_{\roder},\eta_{\romn}\}$. This follows from 
Remark~\ref{remark:improvestchimcXbremgeomsetspas}.
\end{remark}
\begin{remarks}
By combining (\ref{eq:Phiinfnobdintro}), (\ref{eq:uuditoPhiinfchiintro}) and (\ref{eq:estspecasdataintro}), the Sobolev norms of $u(\cdot,0)$
and $u_{t}(\cdot,0)$ appearing on the right hand side of (\ref{eq:estspecasdataintro}) can be replaced by a suitable Sobolev norm of $\psi$. The 
estimate (\ref{eq:Phiinfnobdintro}) demonstrates that the map from asymptotic data to initial data is continuous with respect to the 
$C^{\infty}$-topology. Combining Propositions~\ref{prop:roughas} and \ref{prop:spasda} yields the conclusion that when $E_{a}=\cn{2m}$, the map 
from initial data to asymptotic data is a homeomorphism in the $C^{\infty}$-topology. Note also that $\Rsp A_{\infty}<\b_{\rem}$ implies that $E_{a}=\cn{2m}$. 
\end{remarks}

\section{Examples}

In order to illustrate the results, let us consider two simple examples of scalar equations. 

\begin{example}\label{example:polt3gowdyderofas}
Consider (\ref{eq:polGowdytimereversed}), the relevant equation when studying the singularity in polarised $\tn{3}$-Gowdy. In this case,  
\[
g=-dt\otimes dt+e^{2t}d\theta\otimes d\theta,\ \ \
\bge=e^{2t}d\theta\otimes d\theta,\ \ \
\bk=\bge,\ \ \
\chi=0.
\]
Moreover, $\mcX=0$, $\a=0$ and $\zeta=0$. It is thus obvious that the equation (\ref{eq:polGowdytimereversed}) is $C^{1}$-silent with 
$\mu=1$. In addition, (\ref{eq:chimcXexpdecest}) holds with $K_{\roder}=1$ and $\b_{\roder}=1$. Finally, since $\zeta=\a=0$, the matrix 
$A_{\infty}$, introduced in (\ref{eq:Asilentdefintro}), is given by 
\begin{equation}\label{ApolGowdy}
A_{\infty}=\left(\begin{array}{cc} 0 & 1 \\ 0 & 0\end{array}\right).
\end{equation}
Thus $\kappa_{1}=0$. Note that (\ref{eq:alpahzetaconvest}) holds with, e.g., $\eta_{\romn}=2$ and $C_{\romn}=1$. Since $\b_{\roder}=1$ we
can equate $\b_{\rem}$ with $2$; cf. Remarks~\ref{remark:chimcXimprassbetterbrem} and \ref{remark:chimcXimprassbetterbremspas}. 
Finally, since $A_{\infty}$ has only one eigenvalue, the spaces $E_{a}$ appearing in the statements of Propositions~\ref{prop:roughas} and 
\ref{prop:spasda} equal $\cn{2}$. To conclude, the assumptions of all the propositions stated in Section~\ref{section:resultssilentsettingintro} 
are fulfilled. Appealing to Propositions~\ref{prop:roughas} and \ref{prop:spasda} yields a homeomorphism
between initial data to (\ref{eq:polGowdytimereversed}) and asymptotic data, consisting of $\psi\in C^{\infty}(\so,\cn{2})$. If
$\psi=(\psi_{\infty}\ v_{\infty})^{t}$, (\ref{eq:estspecasdataintro}) yields
\[
\left\|\left(\begin{array}{c} P(\cdot,t) \\ P_{t}(\cdot,t)\end{array}\right)
-\left(\begin{array}{c} v_{\infty}t+\psi_{\infty} \\ v_{\infty}\end{array}\right)\right\|_{(s)} \leq C_{s}\ldr{t}^{N}e^{-2t},
\]
where $C_{s}$ in this case is allowed to depend on $s$ and the solution. These asymptotics should be compared with (\ref{eq:polGowdyasympt}).
\end{example}


\begin{example}[The non--flat Kasner solutions]\label{example:nonflatvacuumKasner}
Let us consider the non--flat Kasner solutions in the direction of the singularity. In other words, let us consider a metric $g$ of the 
form (\ref{eq:gBianchiI}), where $a_{i}(t)=t^{p_{i}}$ and the $p_{i}$ are constants satisfying 
\[
\textstyle{\sum}_{i=1}^{d}p_{i}=\textstyle{\sum}_{i=1}^{d}p_{i}^{2}=1,\ \ \
p_{i}<1.
\]
Then 
\[
\Box_{g}u=\frac{1}{\sqrt{-\det g}}\d_{\a}\left(g^{\a\b}\sqrt{-\det g}\ \d_{\b}u\right)=-\frac{1}{t}\d_{t}(tu_{t})+\sum_{i=1}^{d}t^{-2p_{i}}u_{ii}.
\]
Changing time coordinate to $\tau$, where $t=e^{-\tau}$, the Klein-Gordon equation $\Box_{g}u-m^{2}u=0$, where $m$ is a constant, can thus be 
written
\begin{equation}\label{eq:nonflatKasnerKleinGordon}
-u_{\tau\tau}+\textstyle{\sum}_{i=1}^{d}e^{-2\b_{i}\tau}u_{ii}-m^{2}e^{-2\tau}u=0,
\end{equation}
where $\b_{i}=1-p_{i}>0$. The metric associated with (\ref{eq:nonflatKasnerKleinGordon}) is given by 
\begin{equation}\label{eq:gconnonflatKasner}
g_{\rocon}=-d\tau\otimes d\tau+\textstyle{\sum}_{i=1}^{d}e^{2\b_{i}\tau}dx^{i}\otimes dx^{i}
\end{equation}
on $\tn{d}\times\ro$. Let $\bge$ be the metric and $\bk$ be the second fundamental form induced on $\tn{d}\times\{\tau\}$ by $g_{\rocon}$. 
Then, beyond the metric (\ref{eq:gconnonflatKasner}), the functions characterising (\ref{eq:nonflatKasnerKleinGordon}) are given by 
$\alpha(\tau)=0$, $\mcX(\tau)=0$ and $\zeta(\tau)=m^{2}e^{-2\tau}$. Moreover, 
\[
\bk=\textstyle{\sum}_{i=1}^{d}\b_{i}e^{2\b_{i}\tau}dx^{i}\otimes dx^{i};
\]
note that $U=\d_{\tau}$. In particular, 
\[
\bk\geq \min\{\b_{1},\dots,\b_{d}\}\bge.
\]
Since the shift vector field field $\chi$ vanishes in the current setting, it is thus clear that (\ref{eq:nonflatKasnerKleinGordon}) is $C^{1}$-silent, 
with $\mu:=\min\{\b_{1},\dots,\b_{d}\}<1$; cf. Definition~\ref{definition:Cosilenceintro}. Moreover, $\mcX$ is clearly $C^{0}$-future bounded, and 
(\ref{eq:alpahzetaconvest}) holds with $\a_{\infty}=0$, $\zeta_{\infty}=0$, $C_{\romn}=m^{2}$ and $\eta_{\romn}=2$. Finally, (\ref{eq:chimcXexpdecest}) holds 
with $K_{\roder}=1$ and $\b_{\roder}=1$. Thus Proposition~\ref{prop:roughas}
applies, and the relevant matrix $A_{\infty}$ is in the present setting given by (\ref{ApolGowdy}). Moreover, $\kappa_{1}=0$, $\b_{\rem}=2\mu$ and
$E_{a}=\cn{2}$. Given a solution $u$ to (\ref{eq:nonflatKasnerKleinGordon}), appealing to Proposition~\ref{prop:roughas} yields 
$u_{\infty},v_{\infty}\in C^{\infty}(\tn{d},\co)$ such that 
\begin{equation}\label{eq:uudotasnonflatKasner}
\left\|\left(\begin{array}{c} u(\cdot,\tau) \\ u_{\tau}(\cdot,\tau)\end{array}\right)
-\left(\begin{array}{c} v_{\infty}\tau+u_{\infty} \\ v_{\infty}\end{array}\right)\right\|_{(s)} \leq C_{s}\ldr{t}^{N}e^{-2\mu \tau},
\end{equation}
where $C_{s}$ in this case is allowed to depend on $s$ and the solution. In fact, appealing to Propositions~\ref{prop:roughas} and 
\ref{prop:spasda} yields the conclusion that the map from initial data to asymptotic data is a homeomorphism with respect to the 
$C^{\infty}$-topology. 

It is of interest to note that the mass term in the Klein-Gordon equation does not have any influence on the asymptotics. 
\end{example}

\section[Strengths/weaknesses; argument]{Strengths and weaknesses of the results; a rough idea of the argument}\label{section:sileqstweaargintro}

\subsection{Strengths and weaknesses}\label{ssection:strandweaksilsettintro}
One clear strength of the results contained in Propositions~\ref{prop:genroughestintro}, \ref{prop:roughas} and \ref{prop:spasda} is that they
apply to a general class of equations. Moreover, the consequences are quite far reaching; in favourable situations, 
combining Propositions~\ref{prop:roughas} and \ref{prop:spasda} yields a homeomorphism between initial data and asymptotic data. Turning to 
the questions concerning the optimal behaviour of the energy, posed in Section~\ref{section:Questions}, note that combining 
Propositions~\ref{prop:genroughestintro} and \ref{prop:spasda} yields the conclusion that  $\cruderate=\kappa_{1}$, where $\kappa_{1}$ is
introduced in the statement of Proposition~\ref{prop:genroughestintro}. In particular, we can calculate $\cruderate$. How about 
$\nolossrate$? Considering (\ref{eq:mfeestslossintro}), there is clearly one fundamental obstruction to calculating $\nolossrate$: the 
fact that $\mfe_{s+s_{0}}$ appears on the right hand side (where we do not know whether $s_{0}=0$ or not). In other words, the estimate 
(\ref{eq:mfeestslossintro}) potentially involves a loss of derivatives which makes it impossible to deduce an upper bound on $\nolossrate$.
On the other hand, Proposition~\ref{prop:spasda} yields the lower bound $\nolossrate\geq\kappa_{1}$. Below we demonstrate that for any 
$\eta_{1}\geq \eta_{2}$, there are equations satisfying the assumptions of Propositions~\ref{prop:genroughestintro}, \ref{prop:roughas} and 
\ref{prop:spasda} such that $\cruderate\leq \eta_{2}$ and $\nolossrate\geq\eta_{1}$; cf. Example~\ref{example:optimalityone} below. It is therefore 
hard to imagine the perspective taken in the 
present chapter (where we, generally, accept unspecified derivative losses) to be useful in drawing conclusions concerning $\nolossrate$. On the
other hand, when applying the linear theory in the study of non-linear equations, $\nolossrate$ is of much greater interest than 
$\cruderate$. 

\subsection{Outline of the argument}\label{ssection:outloftheargsilsetting}
The proofs of Propositions~\ref{prop:genroughestintro}, \ref{prop:roughas} and \ref{prop:spasda} are not very complicated. They are also quite 
short; the complete arguments are contained in Chapters~\ref{chapter:roughanalysisODEregion} and \ref{chapter:weaksil}. In particular, it is of 
interest to note that an analysis of the type
described in Section~\ref{section:mainideaspfsupexpgrowth}, involving a detailed study of the oscillations, is not necessary. In fact, 
we do not need to appeal to Part~\ref{part:averaging} at all (below we describe the contents of Part~\ref{part:averaging} 
in greater detail, but let us here point out that it is the most important part of these notes in that it allows us to study the oscillations 
in detail and thereby to derive optimal estimates for the energies, our main goal in these notes). 

\textbf{The Fourier coefficients; dividing the time interval.} In order to describe the arguments, it is convenient to use the notation 
introduced in Section~\ref{section:divintomodes}. Consider, in particular, (\ref{eq:fourierthesystemRgesigmaandX}), the equation for the 
Fourier coefficients, where 
$\sigma$ and $X$ are defined by (\ref{eq:sigmaXdefintro}). Due to the assumptions on the shift vector field and on $\mcX$, it can be verified 
that for all $t\geq 0$, $|\sigma(\indexnot,t)|$ and $\|X(\indexnot,t)\|$ are bounded by constants which are independent of $\indexnot$; this is 
a consequence of Lemma~\ref{lemma:sigmaXbdsandderbds} below. In this sense, all the dependence on $\indexnot$ in 
(\ref{eq:fourierthesystemRgesigmaandX}) enters via the function $\mfg(\indexnot,t)$, introduced in (\ref{eq:mfgnutdef}). On the other hand, due to the 
assumptions concerning the second fundamental form, 
\begin{equation}\label{eq:mfgexpdecintro}
\mfg(\indexnot,t)\leq Ce^{-\mu t}\ldr{\nu(\indexnot)}
\end{equation}
for all $t\geq 0$ and $\indexnot\in\EFindexset$, where $C\in\ro$ is independent of $\indexnot$, $\nu(\indexnot)$ is defined by 
(\ref{eq:nugenReqdef}) and we use the notation (\ref{eq:ldrdefinitionintro}); this is a consequence of Lemma~\ref{lemma:condwsil} below. 
Due to (\ref{eq:mfgexpdecintro}), it is natural to divide $[0,\infty)$ into two parts: $[0,T_{\roode}]$ and $[T_{\roode},\infty)$. The 
division is, roughly speaking, defined by the condition that $\mfg(\indexnot,t)\leq 1$ for all $t\geq T_{\roode}$; cf. 
Definition~\ref{def:roughODEtermo} below for the details. Given the definition of $T_{\roode}$, it is natural to divide the analysis into
two parts. First, we analyse the behaviour in $[T_{\roode},\infty)$, which we refer to as the \textit{ODE-regime}. Second, we analyse the behaviour 
in the interval $[0,T_{\roode}]$, which we refer to as the \textit{oscillatory regime}.

\textbf{The ODE regime.} For $t\geq T_{\roode}$, $\mfg(\indexnot,\cdot)$ is not only $\leq 1$, it is also
exponentially decaying. In the interval $[T_{\roode},\infty)$, the equation (\ref{eq:fourierthesystemRgesigmaandX}) can thus be written in the form 
\begin{equation}\label{eq:ODEfmODEregime}
\dot{v}(\indexnot,t)=A_{\infty}v(\indexnot,t)+A_{\rem}(\indexnot,t)v(\indexnot,t)+F(\indexnot,t)
\end{equation}
for an appropriate choice of $v$ and $F$, where $A_{\infty}$ is given by (\ref{eq:Asilentdefintro}). Moreover, $A_{\rem}$ satisfies an estimate of the 
form $\|A_{\rem}(\indexnot,t)\|\leq Ce^{-\b_{\rem} (t-T_{\roode})}$ for all $t\geq T_{\roode}$, where $C\in\ro$ is independent of $t$ and $\indexnot$; and 
$\b_{\rem}$ is given by (\ref{eq:bremintrodef}). In Chapter~\ref{chapter:roughanalysisODEregion}, we develop the tools necessary for analysing
the behaviour of solutions to equations of the form (\ref{eq:ODEfmODEregime}). The discussion is divided into the following steps.

\textit{Algebraic decompositions, energy estimates.} The first step of the analysis is to decompose the equation algebraically in order to isolate 
the leading
order behaviour. This is the subject of Section~\ref{section:ODEappr}. The main idea is very simple, namely to, first, change variables in 
(\ref{eq:ODEfmODEregime}) so that $A_{\infty}$ is transformed into its Jordan normal form; second, to change variables by a time dependent unitary 
matrix in order to eliminate the imaginary parts of the eigenvalues; third, to multiply the variables by a scalar exponential 
function to normalise the eigenvalues so that the largest one is zero; and, fourth, to multiply the variables by a diagonal matrix so that the 
Jordan blocks corresponding to a strictly negative eigenvalue become negative definite (after the final change of variables, the Jordan blocks 
are non-standard in the sense that the non-zero off-diagonal elements all equal a suitably chosen $0<\e\in\ro$, small enough). The end result is 
that (\ref{eq:ODEfmODEregime}) is transformed into 
\begin{equation}\label{eq:ODEfmODEregimereformulation}
\dot{w}(\indexnot,t)=Jw(\indexnot,t)+B_{\rem}(\indexnot,t)w(\indexnot,t)+G(\indexnot,t),
\end{equation}
where $J$ is a real matrix, in non-standard Jordan normal form, such that the Jordan blocks corresponding to negative eigenvalues are negative 
definite; $B_{\rem}$ satisfies the same type of estimate as $A_{\rem}$; and $|G(\indexnot,t)|\leq Ce^{-\kappa_{1}t}|F(\indexnot,t)|$, where 
$\kappa_{1}:=\kappa_{\max}(A_{\infty})$ and $C$ is a constant depending only on $A_{\infty}$. Moreover, $|w|$ is equivalent to $e^{-\kappa_{1}t}|v|$, with constants
of equivalence depending only on $A_{\infty}$. Based on the reformulation (\ref{eq:ODEfmODEregimereformulation}), we derive an energy estimate for 
$t\geq T_{\roode}$ in Section~\ref{section:roughODEest}. 

\textit{Deriving asymptotics.}
Next, we turn to the problem of deriving asymptotics. This is the subject of Section~\ref{section:detailedODEas}. Again, the variables $w$
described above play an important role. However, we can, in general, only expect to obtain partial information concerning the asymptotics of $w$. 
In Section~\ref{section:resultssilentsettingintro} above, we give a heuristic motivation for the analogous statement in the case of 
$C^{1}$-silent equations. We also motivate the necessity of focusing on the 
spaces $E_{a}$ appearing in the statement of Proposition~\ref{prop:roughas}. This issue already appears in the study of systems of the form 
(\ref{eq:ODEfmODEregimereformulation}), and makes it necessary to, beyond the algebraic decompositions already discussed, divide the variables $w$ into 
one group for which we can expect to be able to distinguish the asymptotic behaviour from the error terms arising from $B_{\rem}$, say $w_{a}$, and 
one group for which we cannot, say $w_{b}$. Considering the matrix $J$ appearing in (\ref{eq:ODEfmODEregimereformulation}), $w_{a}$ corresponds
to the Jordan blocks whose eigenvalues are $>-\b_{\rem}$, and $w_{b}$ corresponds to the remaining Jordan blocks. Concerning $w_{a}$, we need to 
calculate an asymptotic expression, and to estimate the error. Concerning $w_{b}$, we only need an estimate. Finally, we need to combine the 
estimates for the two components, and to reinterpret the result in terms of the original variables. Doing so yields Lemma~\ref{lemma:ODEasymp}. 

\textit{Specifying the asymptotics.} Finally, we want to construct solutions with prescribed asymptotics. When doing so, it is sufficient to 
consider the case that $G=0$ in (\ref{eq:ODEfmODEregimereformulation}).
Note also that we can only hope to be able to specify the asymptotics for $w_{a}$. The idea of the argument is to fix a sufficiently late
time, say $t_{0}\geq T_{\roode}$; to specify initial data at $t_{0}$ such that $w_{b}(t_{0})=0$; and to consider the map taking $w_{a}(t_{0})$
to the asymptotic data for $w_{a}$. For sufficiently late $t_{0}$, it can be demonstrated that this map is injective. Since the relevant
vector spaces have the same dimension, this proves that the map is an isomorphism. Inversion yields a map from asymptotic data to 
initial data. In the end, we also have to estimate the norm of this map, and to translate the estimates into the original variables. 
The end result is Lemma~\ref{lemma:spasODEsett}. 

\textbf{The oscillatory regime.} In Chapter~\ref{chapter:weaksil}, we first consider the interval $[0,T_{\roode}]$ for one individual mode. 
Note that as $\ldr{\nu(\indexnot)}\rightarrow\infty$, $T_{\roode}\rightarrow\infty$. Moreover, under the assumptions of the present chapter, 
we do not have any detailed control of the growth of the energy of the modes in $[0,T_{\roode}]$. On the other hand, due to the fact
that the equation is balanced, we know that the growth is not faster than exponential; cf. Lemma~\ref{lemma:oscroughODE}. Moreover, the 
constant appearing in the exponential does not depend on $\indexnot$. Finally, due to (\ref{eq:mfgexpdecintro}), we know that 
$T_{\roode}\leq C\ln\ldr{\nu(\indexnot)}$. Combining these observations, the energy associated with a mode can, in the interval $[0,T_{\roode}]$, 
be estimated by $\ldr{\nu(\indexnot)}^{s_{0}}$ times the initial energy, for a suitable constant $s_{0}$ depending only on the coefficients of 
the equation. From this point of view, the growth in the interval $[0,T_{\roode}]$ simply corresponds to a loss of derivatives. Combining
the rough estimate given by Lemma~\ref{lemma:oscroughODE} with the estimate (\ref{eq:vest}), valid in the ODE-regime, yields 
Proposition~\ref{prop:genroughestintro} after summing over the modes. In order to obtain Proposition~\ref{prop:roughas}, it is essentially sufficient 
to combine the estimates in the oscillatory regime with the ones in the ODE-regime. However, writing down the details turns out to be 
quite technical. Finally, in order to prove Proposition~\ref{prop:spasda}, we need an additional assumption. The reason is that in order
to take the step from initial data at $t=T_{\roode}$ (obtained from asymptotic data by the analysis in the ODE regime) to initial data at 
$t=0$, we need to estimate the energy of a mode at $t=0$ in terms of the energy at $t=T_{\roode}$. In order for the growth (when going backwards 
in time from $T_{\roode}$ to $0$) not to be worse than exponential (so that it does not correspond to more than a loss of derivatives), it is not 
sufficient to have a lower bound on $\bk$, we also need an upper bound. This is the reason for demanding that (\ref{eq:elldlbintro}) hold. 
However, once we have the necessary estimates for the individual modes, it is sufficient to sum them up (even though the details are somewhat 
technical). This yields Proposition~\ref{prop:spasda}.

\chapter{Transparent equations}\label{chapter:transpeq}

\section{Introduction}\label{section:introtransintro}

In Chapter~\ref{chapter:silentequations} above, we present results concerning silent equations. Beyond the bounds on $\mcX$ and the shift vector 
field, and the convergence requirements on $\a$ and $\zeta$ (which we impose in most of these notes), the main assumption in the silent 
setting is that (\ref{eq:bklowerbdCosilintro}) hold. Disregarding the integrable function $\betafun$, the first inequality appearing in 
(\ref{eq:bklowerbdCosilintro}) implies that $\bk$ has a strictly positive lower bound. It is of interest to consider the case that $\bk$ 
(up to an integrable function) has a non-negative lower bound, but that there are some directions in which it degenerates asymptotically. 
This leads to the study of transparent equations. Returning to Subsection~\ref{ssection:cansepcomoma}, the Nariai spacetimes, the flat
Kasner solutions (in the direction of the singularity) and the Milne model (in the expanding direction) are some examples of relevant 
geometries. 

\subsection{Outline}

In the transparent setting, it is necessary to divide the spatial variables into ``silent'' and ``transparent'' variables; cf. 
Subsection~\ref{ssection:divthevartrssetintro} below. Given this division, a formal definition of the above requirements can be given; we demand 
that $\bk$ converges to zero exponentially in the transparent directions, cf. (\ref{eq:bkbdtrssettingintro}) below, and that there is exponential 
expansion in the silent directions, cf. (\ref{eq:bkbdtrssettingsilentpartintro}) below. The conditions imposed on the shift vector field are described 
in Definition~\ref{def:shiftnegligibletrs}. In part, the assumptions are similar to the ones imposed in the silent setting. However, we here 
also demand that the shift vector field be $C^{1}$-future bounded, and that its transparent components converge to zero exponentially. Due to
these assumptions, we can calculate the leading order behaviour of the coefficients of the highest order derivatives in (\ref{eq:thesystemRge});
cf. (\ref{eq:ghotassilparttrseq}) and (\ref{eq:ghotastrsparttrseq}) below. Concerning $\a$ and $\zeta$, we assume that they converge exponentially, 
as in the silent setting. Turning to $\mcX$, the $C^{0}$-future boundedness is sufficient to ensure that the ``silent'' components of $\mcX$
converge to zero exponentially. In order to control the ``transparent'' components, we explicitly assume that they converge exponentially; cf.
(\ref{eq:mcXconvsildir}). Adding up all of the above observations we obtain a limit equation, given by (\ref{eq:thelimitequationinthetransparentcase}) 
below. For fixed silent variables, the limit equation can be thought of as a system of linear wave equations in the transparent variables and the 
time variable. Moreover, this system has constant coefficients. 

\textbf{The Fourier modes.} Turning to the Fourier side, cf. Section~\ref{section:divintomodes}, the situation is very similar to the one 
considered in the previous chapter. However, 
there is one crucial difference: as opposed to the silent setting, the coefficients of the limit equation here depend on $\indexnot$. This leads to 
several complications. First, estimating the distance between two solutions by computing the energy of the difference is not so useful, since 
the energies of different modes typically exhibit different exponential growth/decay. In order to obtain norms such that bounds on these norms
have implications for all the modes, it is necessary to introduce a time and mode dependent normalisation in the definition of the energies/norms. 
Second, when describing the asymptotics, we need to 
introduce spaces analogous to the $E_{a}$ defined in the statement of Proposition~\ref{prop:roughas}. However, in the present context, what $E_{a}$
is depends on the mode. Third, the proof proceeds similarly to the argument described in Section~\ref{section:sileqstweaargintro}. In particular, 
we wish to make an appropriate algebraic decomposition of the matrices appearing in the equation on the Fourier side. The problem that arises in 
the transparent setting is that we, in general, need to decompose infinitely many different matrices. This leads to the problem of obtaining uniform
control of the norms of the matrices used to transform the equations into the preferred form; we return to this topic in greater detail in 
Section~\ref{section:outlinetransintro} below. 

\textbf{Dividing the equation into silent and transparent parts.} Corresponding to the division of the spatial variables into ``silent'' 
\index{Silent!variables}%
\index{Variables!silent}%
and ``transparent'' variables, 
\index{Transparent!variables}%
\index{Variables!transparent}%
there is a division of $\EFindexset$ into $\EFsilindexset$ and $\EFtrsindexset$; cf. Definition~\ref{def:indexnottrssilintro}
below. We can project (\ref{eq:thesystemRge}) to the corresponding subspaces of $L^{2}(\bM,\cn{m})$ (where we tacitly assume the projection to only be in 
the spatial variables). The projection to the subspace corresponding to 
$\EFsilindexset$, say $L^{2}_{\rosil}(\bM,\cn{m})$, yields an equation to which the results of the previous chapter apply. We therefore focus on the 
projection 
to the subspace corresponding to $\EFtrsindexset$, say $L^{2}_{\trs}(\bM,\cn{m})$, in the present chapter. We refer to the latter equation as the 
\textit{transparent part of} (\ref{eq:thesystemRge}). If the (smooth) initial data for (\ref{eq:thesystemRge}) are in 
$L^{2}_{\trs}(\bM,\cn{m})\times L^{2}_{\trs}(\bM,\cn{m})$, then they are said to be \textit{transparent}. 
\index{Initial data!transparent}%
\index{Transparent!initial data}%
Finally, given a solution to (\ref{eq:thesystemRge}),
say $u$, we call the projection of $u$ to $L^{2}_{\trs}(\bM,\cn{m})$ \textit{the transparent part of the solution} $u$, 
\index{Transparent!part of solution}%
\index{Solution!transparent part}%
and we denote it by $u_{\trs}$. 

\textbf{Deriving asymptotics.} Before describing the asymptotics, it is of interest to develop a feeling for the solutions to the limit
equation. In Subsection~\ref{ssection:solvlimeqtrsintro}, we therefore solve the transparent part of the limit equation, given transparent initial data. 
Given a solution $u$ to (\ref{eq:thesystemRge}), our goal is to describe the asymptotics of $u_{\trs}$. We therefore wish to find transparent initial 
data such that the corresponding solution to the transparent part of the limit equation, say $v_{\trs}$, is as ``close'' to $u_{\trs}$ as possible. The 
relevant result is formulated as Proposition~\ref{prop:trsasymptintro} below. Note that it is analogous to Proposition~\ref{prop:roughas} above, with
the following two main differences. First, the vector spaces to which the asymptotic data belong depend on $\indexnot$, for reasons mentioned above. 
Second, in order for estimates of the difference between $u_{\trs}$ and $v_{\trs}$ to uniquely determine the asymptotic data, we, in the definition of the 
norm, need to introduce a time and $\indexnot$ dependent normalising factor for each mode. The definition of the norm is given in 
(\ref{eq:tsnodefintro}) below, and the main estimate of the difference of $u_{\trs}$ and $v_{\trs}$ is given by (\ref{eq:ultmuzpartasfinintro}) below. 
As in the silent setting, the map from initial data to asymptotic data is continuous with respect to the $C^{\infty}$-topology; cf. 
(\ref{eq:Uinfsobnoestfinintro}) below. 

\textbf{Specifying asymptotics, examples and outline.} In Section~\ref{section:specastrsintro} below, we turn to the question of specifying the 
asymptotics. With the exception of the differences described above, the relevant statement is analogous to Proposition~\ref{prop:spasda}, the 
corresponding statement in the silent setting. In particular, we obtain a continuous map from asymptotic data to initial data. In order to illustrate 
the results, we give a simple example in Section~\ref{section:anextheflvaKasol} below. We end the chapter by giving an outline of the argument in 
Section~\ref{section:outlinetransintro}.

\subsection{Dividing the variables}\label{ssection:divthevartrssetintro}

Before stating the requirements on the second fundamental form, we need to specify in which directions there is expansion and in which directions 
there is not. This division can be made in the tangent space, or in the cotangent space. The reason for this is that 
for each hypersurface $\bM_{t}$, the metric $\bge$ induces an isomorphism between $T\bM$ and $T^{*}\bM$. We denote this isomorphism by $\flat$, and 
the inverse by $\sharp$. In particular, $(T^{*}\bM)^{\sharp}=T\bM$. Clearly, the isomorphisms are time dependent, but we omit explicit reference to 
this dependence in the notation. Given a division into directions, either on the tangent or the cotangent side, the relevant isomorphism, i.e. $\flat$ 
or $\sharp$ respectively, yields a division on the other side. In practice, it turns out to be more natural to make the division on the cotangent side. 
Later, when we impose requirements on the second fundamental form, it is convenient to state the conditions in terms of the symmetric contravariant 
$2$-tensor fields $\bge^{\sharp}$ and $\bk^{\sharp}$. 
\index{$\a$Aa@Notation!Metrics!$\bge^{\sharp}$}%
\index{$\a$Aa@Notation!Second fundamental forms!$\bk^{\sharp}$}%
They are defined by the requirement that their components be given by $\bge^{ij}$ and $\bk^{ij}$ respectively. An equivalent, geometric, definition is
\[
\bge^{\sharp}(\eta,\xi):=\bge(\eta^{\sharp},\xi^{\sharp}),\ \ \
\bk^{\sharp}(\eta,\xi):=\bk(\eta^{\sharp},\xi^{\sharp})
\]
for all $\eta,\xi\in T^{*}_{\bp}\bM$ and all $\bp\in\bM$. 

The following definition yields the relevant division both in the tangent and in the cotangent space. 
\begin{definition}\label{def:transpandsilentdivision}
Let $(M,g)$ be a canonical separable cosmological model manifold. Assume that a division of the 
set $\{1,\dots,d\}$ into two disjoint sets $\{j_{1},\dots,j_{d_{\trs}}\}$ and $\{\bj_{1},\dots,\bj_{d_{\rosil}}\}$ is given (if $d_{\trs}=0$ or $d_{\rosil}=0$ the 
corresponding set is empty). These subsets are referred to as the \textit{transparent} and \textit{silent} subsets 
\index{Transparent!subset of $\{1,\dots,d\}$}%
\index{Subset!transparent}%
\index{Silent!subset of $\{1,\dots,d\}$}%
\index{Subset!silent}%
of $\{1,\dots,d\}$ respectively. Assume, moreover, that a 
division of the set $\{1,\dots,R\}$ into the union of two disjoint sets $\{r_{1},\dots,r_{R_{\trs}}\}$ and $\{\bre_{1},\dots,\bre_{R_{\rosil}}\}$ is given 
(if $R_{\trs}=0$ or $R_{\rosil}=0$ the corresponding set is empty). Again, these subsets are referred to as the \textit{transparent} and \textit{silent} 
subsets of $\{1,\dots,R\}$ 
\index{Transparent!subset of $\{1,\dots,R\}$}%
\index{Subset!transparent}%
\index{Silent!subset of $\{1,\dots,R\}$}%
\index{Subset!silent}%
respectively. Finally, assume that $d_{\trs}+R_{\trs}>0$. Given this division, define
\[
\bM_{\trs}:=\tn{d_{\trs}}\times M_{r_{1}}\times\cdots\times M_{r_{R_{\trs}}},\ \ \
\bM_{\rosil}:=\tn{d_{\rosil}}\times M_{\bre_{1}}\times\cdots\times M_{\bre_{R_{\rosil}}}.
\]
\index{$\a$Aa@Notation!Manifolds!$\bM_{\trs}$}%
\index{$\a$Aa@Notation!Manifolds!$\bM_{\rosil}$}%
These manifolds are referred to as the \textit{transparent} and \textit{silent} manifolds 
\index{Transparent!manifold}%
\index{Manifold!transparent}%
\index{Silent!manifold}%
\index{Manifold!silent}%
respectively. 
Given these manifolds, there are associated projections $\pi_{\trs}:\bM\rightarrow\bM_{\trs}$ and $\pi_{\rosil}:\bM\rightarrow\bM_{\rosil}$, given by
\begin{align*}
\pi_{\trs}(x_{1},\dots,x_{d},p_{1},\dots,p_{R}) := & (x_{j_{1}},\dots,x_{j_{d_{\trs}}},p_{r_{1}},\dots,p_{r_{R_{\trs}}}),\\
\pi_{\rosil}(x_{1},\dots,x_{d},p_{1},\dots,p_{R}) := & (x_{\bj_{1}},\dots,x_{\bj_{d_{\rosil}}},p_{\bre_{1}},\dots,p_{\bre_{R_{\rosil}}}),
\end{align*}
where $x_{j}\in\so$, $j\in \{1,\dots,d\}$, and $p_{r}\in M_{r}$, $r\in\{1,\dots,R\}$. These projections induce maps 
$\pi_{\trs}^{*}:T^{*}\bM_{\trs}\rightarrow T^{*}\bM$ and $\pi_{\rosil}^{*}:T^{*}\bM_{\rosil}\rightarrow T^{*}\bM$. Let
\[
T^{*}_{\trs}\bM:=\pi_{\trs}^{*}(T^{*}\bM_{\trs}),\ \ \
T^{*}_{\rosil}\bM:=\pi_{\rosil}^{*}(T^{*}\bM_{\rosil}).
\]
\index{$\a$Aa@Notation!Tangent and cotangent spaces!$T_{\trs}^{*}\bM$}%
\index{$\a$Aa@Notation!Tangent and cotangent spaces!$T_{\rosil}^{*}\bM$}%
These sets are referred to as the \textit{transparent} and \textit{silent} subsets of $T^{*}\bM$ 
\index{Transparent!subset of $T^{*}\bM$}%
\index{Silent!subset of $T^{*}\bM$}%
respectively. Given $t\geq 0$, let 
\[
T_{\trs,t}\bM:=(T^{*}_{\trs}\bM)^{\sharp},\ \ \
T_{\rosil,t}\bM:=(T^{*}_{\rosil}\bM)^{\sharp},
\]
where the operator $\sharp$ is introduced prior to the statement of the definition. 
These sets are referred to as the $t$-\textit{transparent} and $t$-\textit{silent} subsets of $T\bM$ respectively.
\index{Transparent!subset of $T\bM$}%
\index{Silent!subset of $T\bM$}%
\end{definition}

\subsection{Conditions on the second fundamental form and the shift vector field}\label{ssection:condtransparency}

Given the above division of the cotangent space into silent and transparent subsets, we are in a position to introduce the relevant conditions 
on the second fundamental form. 

\begin{definition}\label{def:expandconvdir}
Let $(M,g)$ be a canonical separable cosmological model manifold. Assume that there is a division of $\{1,\dots,d\}$ and $\{1,\dots,R\}$ into transparent 
and silent subsets, as described in Definition~\ref{def:transpandsilentdivision}. Assume, moreover, that there are constants $\eta_{\trs}>0$ and $C_{a}>0$ 
such that
\begin{equation}\label{eq:bkbdtrssettingintrocotan}
|\bk^{\sharp}(\xi,\xi)|\leq C_{a}e^{-\eta_{\trs}t}\bge^{\sharp}(\xi,\xi)
\end{equation}
for all $\xi\in T^{*}_{\trs}\bM$ and $t\geq 0$, where $T^{*}_{\trs}\bM$ is introduced in Definition~\ref{def:transpandsilentdivision}. Then $\bge^{\sharp}$ is 
said to be \textit{convergent on} $T_{\trs}^{*}\bM$. 
\index{$\a$Aa@Notation!Conditions!$\bge^{\sharp}$ convergent}%
Assume that there is a $\b_{\rosil}>0$ and a continuous non-negative function 
$\betafun_{\rosil}\in L^{1}([0,\infty))$ such that 
\begin{equation}\label{eq:bkbdtrssettingsilentpartintrocotan}
\bk^{\sharp}(\xi,\xi)\geq [\b_{\rosil}-\betafun_{\rosil}(t)]\bge^{\sharp}(\xi,\xi)
\end{equation}
for all $\xi\in T^{*}_{\rosil}\bM$ and $t\geq 0$, where $T_{\rosil}^{*}\bM$ is introduced in Definition~\ref{def:transpandsilentdivision}. Then $\bge^{\sharp}$ 
is said to be \textit{expanding on} $T_{\rosil}^{*}\bM$.
\index{$\a$Aa@Notation!Conditions!$\bge^{\sharp}$ expanding}%
\end{definition}
\begin{remark}
The notation $\bge^{\sharp}$ and $\bk^{\sharp}$ is introduced at the beginning of Subsection~\ref{ssection:divthevartrssetintro}.
\end{remark}
\begin{remark}\label{remark:trscondtanspdiv}
The condition (\ref{eq:bkbdtrssettingintrocotan}) is equivalent to the requirement that 
\begin{equation}\label{eq:bkbdtrssettingintro}
|\bk(v,v)|\leq C_{a}e^{-\eta_{\trs}t}\bge(v,v)
\end{equation}
for all $v\in T_{\trs,t}\bM$ and $t\geq 0$, where $T_{\trs,t}\bM$ is introduced in Definition~\ref{def:transpandsilentdivision}. Similarly, the condition
(\ref{eq:bkbdtrssettingsilentpartintrocotan}) is equivalent to the requirement that 
\begin{equation}\label{eq:bkbdtrssettingsilentpartintro}
\bk(v,v)\geq [\b_{\rosil}-\betafun_{\rosil}(t)]\bge(v,v)
\end{equation}
for all $v\in T_{\rosil,t}\bM$ and $t\geq 0$, where $T_{\rosil,t}\bM$ is introduced in Definition~\ref{def:transpandsilentdivision}. 
\end{remark}

In order to be able to draw conclusions, we also need to make assumptions concerning the shift vector field $\chi$.

\begin{definition}\label{def:shiftnegligibletrs}
Let $(M,g)$ be a canonical separable cosmological model manifold.
Assume that there is a division of $\{1,\dots,d\}$ and $\{1,\dots,R\}$ into transparent and silent subsets, as described in 
Definition~\ref{def:transpandsilentdivision}. Assume, in addition, that there is a continuous non-negative function 
$\betafun\in L^{1}([0,\infty))$ such that 
\begin{equation}\label{eq:chichidottrsest}
|\chi|_{\bge}\cdot|\dot{\chi}|_{\bge}\leq \betafun
\end{equation}
holds for all $t\geq 0$, and that the shift vector field is $C^{1}$-future bounded; cf. Definition~\ref{definition:bdsonchi}.
Assume, finally, that there are $C_{b}>0$ and $\eta_{\trs}>0$ such that 
\begin{equation}\label{eq:transpshiftcondintro}
|\xi(\chi)|\leq C_{b}e^{-\eta_{\trs}t}|\xi|_{\bge}
\end{equation}
for all $\xi\in T^{*}_{\trs}\bM$ and $t\geq 0$, where $T^{*}_{\trs}\bM$ is introduced in Definition~\ref{def:transpandsilentdivision}.
Then the shift vector field of $g$ is said to be \textit{asymptotically negligible}. 
\index{Asymptotically negligible!shift vector field}%
\index{Shift vector field!asymptotically negligible}%
\end{definition}
\begin{remark}\label{remark:Xbjkconvtozerogeom}
The estimate (\ref{eq:transpshiftcondintro}) corresponds to imposing exponential decay on the shift vector in the transparent directions. 
This is of course a weaker condition than imposing exponential decay in all directions. 
\end{remark}
\begin{remark}\label{remark:etatsmultoccur}
The constant $\eta_{\trs}$ appears in both (\ref{eq:bkbdtrssettingintrocotan}) and (\ref{eq:transpshiftcondintro}). In situations where we
assume both estimates to hold, we tacitly assume the constants $\eta_{\trs}$ appearing in these estimates to coincide. 
\end{remark}

\subsection{Consequences for the coefficients of the equations}

\textbf{The highest order coefficients.}
Consider (\ref{eq:thesystemRge}). Assume the associated metric to be such that $(M,g)$ is a canonical separable cosmological model manifold.
Assume, moreover, that there is a division of $\{1,\dots,d\}$ and $\{1,\dots,R\}$ into transparent and silent subsets, as described in 
Definition~\ref{def:transpandsilentdivision}. Assume, finally, that Definitions~\ref{def:expandconvdir} and \ref{def:shiftnegligibletrs} are fulfilled. 
Then the leading order coefficients of (\ref{eq:thesystemRge}) have the following properties. First, there is a constant $C$ such that
\begin{equation}\label{eq:ghotassilparttrseq}
|g^{\bj_{k}\bj_{l}}(t)|+a^{-2}_{\bre_{j}}(t)\leq Ce^{-2\b_{\rosil}t},\ \ \
|g^{\bj_{k}j_{i}}(t)|+|g^{0\bj_{k}}(t)|\leq Ce^{-\b_{\rosil}t}
\end{equation}
for all $t\geq 0$, all $k,l\in \{1,\dots,d_{\rosil}\}$, $i\in\{1,\dots,d_{\trs}\}$ and $j\in\{1,\dots,R_{\rosil}\}$. Moreover, for
$k,l\in \{1,\dots,d_{\trs}\}$ and $j\in\{1,\dots,R_{\trs}\}$, there are constants $g^{j_{k}j_{l}}_{\infty}$ and $q_{\infty,r_{j}}$ such that 
\begin{equation}\label{eq:ghotastrsparttrseq}
|g^{j_{k}j_{l}}_{\infty}-g^{j_{k}j_{l}}_{\infty}|+|a^{-2}_{r_{j}}(t)-q_{\infty,r_{j}}|+|g^{0j_{k}}(t)|\leq Ce^{-\eta_{\trs}t}
\end{equation}
for all $t\geq 0$. Finally, $q_{\infty,r_{j}}>0$ and $g^{j_{k}j_{l}}_{\infty}$ are the components of a positive definite matrix. The justification of 
the above statements is to be found in Lemma~\ref{lemma:geometrictoanalytictrans} below. 

\textbf{The lower order coefficients.}
In order to obtain control of the coefficients of the lower order derivatives in (\ref{eq:thesystemRge}), we need to make additional assumptions.
To begin with, we assume that $\mcX$ is $C^{0}$-future bounded and that there are $\a_{\infty},\zeta_{\infty}\in\Mn{m}{\co}$ and $0<\eta_{\romn},C_{\romn}\in\ro$ 
with the property that (\ref{eq:alpahzetaconvest}) holds for all $t\geq 0$. Finally, assume that there are $X^{j_{k}}_{\infty}\in\Mn{m}{\co}$ and
a constant $0<K_{\trs}\in\ro$ such that
\begin{equation}\label{eq:mcXconvsildir}
\|X^{j_{k}}(t)-X^{j_{k}}_{\infty}\|\leq K_{\trs}e^{-\eta_{\trs}t}
\end{equation}
for all $t\geq 0$ and all $k\in \{1,\dots,d_{\trs}\}$ (an observation similar to Remark~\ref{remark:etatsmultoccur} is equally relevant in the case of 
the estimate (\ref{eq:mcXconvsildir})). Under these circumstances, there is a constant $C$ such that 
\[
\|X^{\bj_{k}}(t)\|\leq Ce^{-\b_{\rosil}t}
\]
for all $k\in \{1,\dots,d_{\rosil}\}$ and $t\geq 0$; cf. Remark~\ref{remark:Xnormbdsiltrans}. 

\textbf{The limit equation.}
Combining the above assumptions, it is natural to associate the following \textit{limit equation}
\index{Limit equation!transparent setting}%
\index{Transparent!limit equation}%
 with (\ref{eq:thesystemRge}):
\begin{equation}\label{eq:thelimitequationinthetransparentcase}
\begin{split}
v_{tt}-\textstyle{\sum}_{k,l=1}^{d_{\trs}}g^{j_{k}j_{l}}_{\infty}\d_{j_{k}}\d_{j_{l}}v-\sum_{j=1}^{R_{\trs}}q_{\infty,r_{j}}\Delta_{g_{r_{j}}}v & \\
+\a_{\infty}v_{t}+\textstyle{\sum}_{k=1}^{d_{\trs}}X^{j_{k}}_{\infty}\d_{j_{k}}v+\zeta_{\infty}v & = f.
\end{split}
\end{equation}
Note that this is a constant coefficient equation. Moreover, we can think of it as yielding a system of linear wave equations on $\bM_{\trs}\times I$
for each $\bx\in\bM_{\rosil}$. It is natural to compare the asymptotic behaviour of solutions to (\ref{eq:thesystemRge}) with that of solutions to 
(\ref{eq:thelimitequationinthetransparentcase}). Before doing so, it is of interest to consider the limit equation in its own right. We turn to this
subject next. 

\section{The limit equation}

In the analysis to follow, it is important to keep track of on which variables the solution depends. In fact, if (\ref{eq:thesystemRge}) is an equation 
satisfying the conditions stated in the previous section with $f=0$, and if $u$ is a solution arising from initial data depending only on the 
variables corresponding to $\bM_{\rosil}$, then $u$ is effectively a solution to (\ref{eq:thesystemRge}) where $g^{j_{k}\xi}$, $X^{j_{k}}$ and $a_{r_{j}}^{-2}$ 
have been set to zero for all $k\in \{1,\dots,d_{\trs}\}$, $\xi\in \{0,\dots,d\}$ and $j\in \{1,\dots,R_{\trs}\}$; we say that the solution only depends on 
the silent variables and time. Moreover, the conclusions of Propositions~\ref{prop:genroughestintro} and 
\ref{prop:roughas} apply to the corresponding class of solutions; cf. Remark~\ref{remark:udeponlonsilandt} below. If, on the other hand, the initial data 
depend on some of the variables corresponding to $\bM_{\trs}$, then the asymptotic behaviour can be expected to be different. In order to distinguish the cases 
that have already been handled from what remains to be done, it is therefore convenient to introduce the following terminology.

\begin{definition}\label{def:indexnottrssilintro}
Let $(M,g)$ be a canonical separable cosmological model manifold. Assume that the variables can be divided as described in 
Definition~\ref{def:transpandsilentdivision}. Given $\indexnot\in\EFindexset$, define $\indexnot_{\trs}\in\EFindexset$ as follows: replace 
$n_{\bj_{k}}$ in $\indexnot$ by zero for $k=1,\dots,d_{\rosil}$; and replace $i_{\bre_{k}}$ by $0$ for $k=1,\dots,R_{\rosil}$. Given 
$\indexnot\in\EFindexset$, define $\indexnot_{\rosil}\in\EFindexset$ by $\indexnot_{\rosil}:=\indexnot-\indexnot_{\trs}$. The set 
of $\indexnot\in\EFindexset$ such that $\indexnot_{\trs}=0$ ($\indexnot_{\trs}\neq 0$) is denoted by $\EFsilindexset$ ($\EFtrsindexset$).
\index{$\a$Aa@Notation!Frequency sets!$\EFsilindexset$}%
\index{$\a$Aa@Notation!Frequency sets!$\EFtrsindexset$}%
\end{definition}
\begin{remark}
It is important to keep in mind that the set of $\indexnot_{\trs}$ for $\indexnot\in\EFindexset$ is in general \textit{different} from 
$\EFtrsindexset$, but that the set of $\indexnot_{\rosil}$ for $\indexnot\in\EFindexset$ \textit{equals} 
$\EFsilindexset$. 
\end{remark}
\begin{remark}\label{remark:udeponlonsilandt}
Let (\ref{eq:thesystemRge}) be an equation satisfying the conditions stated in the previous section with $f=0$. Then the conclusions of 
Propositions~\ref{prop:genroughestintro} and \ref{prop:roughas} hold for the class of solutions depending only on the silent variables 
and time. The reason for this is that, due to Lemma~\ref{lemma:geometrictoanalytictrans}, in particular (\ref{eq:dotellsilestfinal});
Lemma~\ref{lemma:sigmaXbdsandderbds}; and (\ref{eq:alpahzetaconvest}); the conditions of Definition~\ref{def:roughODEtermo} are satisfied.
In particular, Lemmas~\ref{lemma:genroughest} and \ref{lemma:roughas} thus apply. Finally, assuming $\bk$ to be $C^{0}$-future bounded and
appealing to Lemma~\ref{lemma:condyieldellderbd}, we conclude that Lemma~\ref{lemma:spasda} applies as well. Thus the conclusions of 
Proposition~\ref{prop:spasda} hold. 
\end{remark}

\subsection{Solving the limit equation}\label{ssection:solvlimeqtrsintro}

Let us consider the limit equation (\ref{eq:thelimitequationinthetransparentcase}). Decomposing it into modes as described in 
Subsection~\ref{ssection:fourdecompmaineq} yields
\begin{equation}\label{eq:fouriertrslimthesystemRgelong}
\begin{split}
\ddot{z}(\indexnot,t)+\mfg^{2}_{\infty}(\indexnot)z(\indexnot,t)
+\a_{\infty}\dot{z}(\indexnot,t) & \\
+\textstyle{\sum}_{l=1}^{d_{\trs}}in_{j_{l}}X^{j_{l}}_{\infty}z(\indexnot,t)+\zeta_{\infty}z(\indexnot,t) & =\hf(\indexnot,t),
\end{split}
\end{equation}
where, applying Einstein's summation convention to $j_{k}$ and $j_{l}$, 
\begin{equation}\label{eq:mfginftytrsdefintro}
\textstyle{\mfg}_{\infty}(\indexnot)=\lim_{t\rightarrow\infty}\mfg(\indexnot,t)=\left(g^{j_{k}j_{l}}_{\infty}n_{j_{k}}n_{j_{l}}
+\sum_{j=1}^{R_{\trs}}q_{\infty,r_{j}}\nu_{r_{j},i_{r_{j}}}^{2}\right)^{1/2}.
\end{equation}
Assuming that $\indexnot\in\EFtrsindexset$ (or, equivalently, that $\mfg_{\infty}(\indexnot)\neq 0$), (\ref{eq:fouriertrslimthesystemRgelong}) can 
be written 
\begin{equation}\label{eq:fourtrslimsyssys}
\d_{t}w(\indexnot,t)=A(\indexnot)w(\indexnot,t)+F(\indexnot,t),
\end{equation}
where 
\begin{equation}\label{eq:Aindexnotdefintro}
A(\indexnot):=\left(\begin{array}{cc} 0 & \mfg_{\infty}(\indexnot)\Id_{m} \\ -\mfg_{\infty}(\indexnot)\Id_{m}
-i[\mfg_{\infty}(\indexnot)]^{-1}n_{j_{l}}X^{j_{l}}_{\infty}-[\mfg_{\infty}(\indexnot)]^{-1}\zeta_{\infty} & 
-\a_{\infty}\end{array}\right).
\end{equation}
\index{$\a$Aa@Notation!Matrix notation!Aiota@$A(\indexnot)$}%
Moreover, 
\begin{equation}\label{eq:vFdeftrsintro}
w(\indexnot,t):=\left(\begin{array}{c} \mfg_{\infty}(\indexnot)z(\indexnot,t) \\ \dot{z}(\indexnot,t)\end{array}\right),\ \ \
F(\indexnot,t):=\left(\begin{array}{c} 0 \\ \hf(\indexnot,t) \end{array}\right).
\end{equation}
The solutions to (\ref{eq:fourtrslimsyssys}) can be written
\begin{equation}\label{eq:vformtrsonemode}
w(\indexnot,t)=e^{A(\indexnot)t}w(\indexnot,0)+\int_{0}^{t}e^{A(\indexnot)(t-\tau)}F(\indexnot,\tau)d\tau.
\end{equation}
Returning to (\ref{eq:thelimitequationinthetransparentcase}), let $f_{\trs}$ denote the function obtained from $f$ by setting $\hf(\indexnot_{\rosil},t)$
to zero for all $t\in I$ and all $\indexnot\in\EFindexset$. 
\index{$\a$Aa@Notation!Projected functions!$f_{\trs}$}%
Define $v_{\trs}$ similarly and introduce 
\begin{equation}\label{eq:Vtrsdefintro}
V_{\trs}(\cdot,t) := \left(\begin{array}{c} (-\Delta_{\infty})^{1/2}v_{\trs}(\cdot,t) \\ \d_{t}v_{\trs}(\cdot,t)\end{array}\right).
\end{equation}
\index{$\a$Aa@Notation!Projected functions!$V_{\trs}$}%
Here $\Delta_{\infty}$ is the differential operator given by 
\begin{equation}\label{eq:Deltainfdefintro}
\Delta_{\infty}:=g^{j_{k}j_{l}}_{\infty}\d_{j_{k}}\d_{j_{l}}+\textstyle{\sum}_{j=1}^{R_{\trs}}q_{\infty,r_{j}}\Delta_{g_{r_{j}}}.
\end{equation}
\index{$\a$Aa@Notation!Differential operators!$\Delta_{\infty}$}%
Moreover, if $\psi\in C^{\infty}(\bM,\cn{k})$ is such that $\hpsi(\indexnot)=0$ for all $\indexnot\in\EFsilindexset$ and $a\in\ro$, then  
$(-\Delta_{\infty})^{a}\psi$ is the function whose $\indexnot$'th Fourier coefficient is given by $0$ in case $\indexnot\in\EFsilindexset$
and by $[\mfg_{\infty}(\indexnot)]^{2a}\hpsi(\indexnot)$ otherwise. Then (\ref{eq:vformtrsonemode}) can formally be summarised into
\begin{equation}\label{eq:capUtrsform}
V_{\trs}(\cdot,t)=e^{\ma_{\trs}t}V_{\trs}(\cdot,0)+\int_{0}^{t}e^{\ma_{\trs}(t-\tau)}F_{\trs}(\cdot,\tau)d\tau,
\end{equation}
where
\begin{align}
F_{\trs}(\cdot,t) := & \left(\begin{array}{c} 0 \\ f_{\trs}(\cdot,t)\end{array}\right),\nonumber\\
\ma_{\trs} := & \left(\begin{array}{cc} 0 & \Id_{m}(-\Delta_{\infty})^{1/2} \\ -\Id_{m}(-\Delta_{\infty})^{1/2}
-(X^{j_{l}}_{\infty}\d_{j_{l}}+\zeta_{\infty})(-\Delta_{\infty})^{-1/2} & 
-\a_{\infty}\end{array}\right).\label{eq:matrsdefintro}
\end{align}
\index{$\a$Aa@Notation!Projected functions!$F_{\trs}$}%
\index{$\a$Aa@Notation!Differential operators!$\ma_{\trs}$}%

\subsection{Describing the asymptotics}\label{ssection:descastrsprel}

Let (\ref{eq:thesystemRge}) be an equation satisfying the assumptions of Section~\ref{section:introtransintro}. If $u$ is a corresponding 
solution, we can divide $u$ into $u=u_{\trs}+u_{\rosil}$, where $u_{\trs}$ is defined as above. Assuming $f$ to satisfy the appropriate 
conditions, $u_{\rosil}$ solves an equation such that Propositions~\ref{prop:genroughestintro} and \ref{prop:roughas} apply; cf. 
Remark~\ref{remark:udeponlonsilandt}. We therefore focus on 
$u_{\trs}$. When describing the asymptotics of $u_{\trs}$, it is convenient to define $U_{\trs}$ to be the right hand side of (\ref{eq:Vtrsdefintro})
with $v_{\trs}$ replaced by $u_{\trs}$. Keeping (\ref{eq:capUtrsform}) in mind, it is then of interest to try to find a $U_{\infty}$ (whose 
$\indexnot$'th Fourier coefficient vanishes for $\indexnot\in\EFsilindexset$) such that 
\begin{equation}\label{eq:Utrsdiffexptobeest}
U_{\trs}(\cdot,t)-e^{\ma_{\trs}t}U_{\infty}-\int_{0}^{t}e^{\ma_{\trs}(t-\tau)}F_{\trs}(\cdot,\tau)d\tau
\end{equation}
is small. In practice, we carry out this comparison on the level of the Fourier coefficients. In other words, we compare solutions to 
(\ref{eq:fourierthesystemRgelong}) with solutions to (\ref{eq:fouriertrslimthesystemRgelong}). This is very similar to the analysis
carried out in the case of silent equations. However, there are two important differences.
\begin{itemize}
\item Consider the equation (\ref{eq:fouriertrslimthesystemRgelong}). For $\indexnot\in\EFsilindexset$, the coefficients of this equation
do not depend on $\indexnot$. However, for $\indexnot\in\EFtrsindexset$, they do. In fact, (\ref{eq:fouriertrslimthesystemRgelong})
corresponds to infinitely many different equations. In particular, as opposed to the silent setting, the asymptotic behaviour depends on 
the mode. This yields complications when making the statement that (\ref{eq:Utrsdiffexptobeest}) is small precise.
\item Just as in the case of silent equations, we can only expect to be able to obtain conclusions concerning the leading order
behaviour of solutions to (\ref{eq:fouriertrslimthesystemRgelong}); cf. the discussion adjacent to (\ref{eq:discardedtermssilentintro}).
Moreover, in order to isolate the leading order behaviour, we need to transform the matrix $A(\indexnot)$ appearing in (\ref{eq:fourtrslimsyssys})
in such a way that the leading order part is separated from the remainder. The problem in the current setting is that there are infinitely
many different matrices of the form $A(\indexnot)$ (as opposed to a single matrix in the silent setting), but we would like to have uniform 
control in the estimates (independent of $\indexnot$). 
\end{itemize}
The first problem can be handled by defining an appropriate norm. To be more precise, assume that the maximal growth of solutions to 
(\ref{eq:fourtrslimsyssys}) with $F(\indexnot,t)=0$, up to polynomial factors, is $e^{\kappa_{\indexnot}t}$ for some $\kappa_{\indexnot}\in\ro$. We then, 
roughly speaking, define a norm in which the 
$\indexnot$'th mode is multiplied with $e^{-\kappa_{\indexnot}t}$. That such a norm decays exponentially then gives relevant information for 
every mode. The second problem is more difficult to deal with. Here we address it by introducing additional restrictions. 

Consider the matrix $A(\indexnot)$ introduced in (\ref{eq:Aindexnotdefintro}). If $X^{j_{l}}_{\infty}=0$ for all $l\in\{1,\dots,d_{\trs}\}$,
then $A(\indexnot)$ is such that as $\mfg_{\infty}(\indexnot)\rightarrow\infty$, there is an asymptotic expansion of the eigenvalues of 
$A(\indexnot)$; cf. Chapter~\ref{chapter:ODEtransp}, in particular Sections~\ref{section:aseigvcalc} and \ref{section:compprojontogensubs}. 
In addition, for a fixed $0<\mu_{0}\in\ro$, the set 
\[
\{A(\indexnot)|\indexnot\in\EFtrsindexset:\mfg_{\infty}(\indexnot)\leq\mu_{0}\}
\]
is finite. Analysing the equation (\ref{eq:fourtrslimsyssys}) for $\indexnot\in\EFtrsindexset$ such that $\mfg_{\infty}(\indexnot)\leq\mu_{0}$
is therefore not a problem; it is essentially sufficient to appeal to the results of Chapter~\ref{chapter:silentequations}. Similarly, 
if we allow $X^{j_{k}}_{\infty}\neq 0$ for some $k\in\{1,\dots,d_{\trs}\}$, but insist that $d_{\trs}=1$ and $R_{\trs}=0$, then the matrices
$A(\indexnot)$ have the same properties. For that reason, we, from now on, restrict to these two situations.

\section{Future asymptotics}\label{section:futastranspintro}

Before stating a result concerning the future asymptotics, we need to give a formal definition of a norm of the type described
in Subsection~\ref{ssection:descastrsprel}. Given $\indexnot\in\EFtrsindexset$, let $\kappa_{\indexnot}:=\kappa_{\max}[A(\indexnot)]$ 
and $\delta_{\indexnot}:=d_{\max}[A(\indexnot),\kappa_{\indexnot}]$; cf. Definition~\ref{def:SpRspdef}. Then we expect generic 
solutions to the homogeneous version of (\ref{eq:fourtrslimsyssys}) 
to behave as $\ldr{t}^{\delta_{\indexnot}-1}e^{\kappa_{\indexnot}t}$. We therefore introduce the norm
\begin{equation}\label{eq:tsnodefintro}
\|\psi\|_{t,s}:=\left(\textstyle{\sum}_{\indexnot\in\EFtrsindexset}\ldr{\nu(\indexnot)}^{2s}\ldr{t}^{-2\de_{\indexnot}+2}e^{-2\kappa_{\indexnot}t}
|\hat{\psi}(\indexnot)|^{2}\right)^{1/2}.
\end{equation}
\index{$\a$Aa@Notation!Norms!$\normSobts$}%
Clearly, $\|\cdot\|_{t,s}$ is not a norm on $C^{\infty}(\bM,\cn{k})$. To justify the terminology, we can consider two functions to be equivalent
if their difference, say $v$, is such that $\hat{v}(\indexnot)=0$ for all $\indexnot\in\EFtrsindexset$. Then $\|\cdot\|_{t,s}$ is a norm
on the corresponding vector space of equivalence classes. In the case of a solution to the homogeneous version of the limit equation 
(\ref{eq:thelimitequationinthetransparentcase}), we expect the norm $\|\cdot\|_{t,s}$ to be bounded for $t\geq 0$, but not better. Next, we are 
interested in situations where the behaviour of solutions is not dominated by the contribution from the inhomogeneity. In the case of silent 
equations, this assumption led us to the introduction of the norm $\|\cdot\|_{A,s}$; cf. (\ref{eq:fAsnormdefintro}). Given $0<\b_{\romar}\in\ro$, 
we introduce an analogous norm in the present setting:
\begin{equation}\label{eq:ftrssnormintro}
\|f\|_{\trs,s}:=\int_{0}^{\infty}\left(\textstyle{\sum}_{\indexnot\in\EFtrsindexset}
\ldr{\nu(\indexnot)}^{2s}e^{-2(\kappa_{\indexnot}-\b_{\romar})t}|\hf(\indexnot,t)|^{2}\right)^{1/2}dt;
\end{equation}
\index{$\a$Aa@Notation!Norms!$\normSobtrss$}%
in order for $\|\cdot\|_{\trs,s}$ to be a norm, we need to give it an interpretation similar to the one given to $\|\cdot\|_{t,s}$ above. 
When using the notation (\ref{eq:ftrssnormintro}), the value of the constant $\b_{\romar}>0$ should be clear from the context. The reason for including 
$\b_{\romar}$ in (\ref{eq:ftrssnormintro}) is related to the complications described in Subsection~\ref{ssection:descastrsprel}.
Due to the fact that there are infinitely many matrices of the form $A(\indexnot)$, infinitely many different matrices are needed
in order to transform all the $A(\indexnot)$ into their preferred form. On the other hand, we need to have uniform bounds on the norms of the 
transformation matrices and their inverses. Under certain circumstances, the latter requirement is not compatible with obtaining as 
detailed an algebraic decomposition of the $A(\indexnot)$'s as the decomposition of the matrix $A$ described in 
Subsection~\ref{ssection:outloftheargsilsetting}. As a consequence, we are not always able to, in the algebraic decompositions, resolve the 
real parts of the eigenvalues exactly. This necessitates the margin represented by $\b_{\romar}$.

\begin{prop}\label{prop:trsasymptintro}
Consider (\ref{eq:thesystemRge}). Assume the associated metric to be such that $(M,g)$ is a canonical separable cosmological 
model manifold. Assume that there is a division of $\{1,\dots,d\}$ and
$\{1,\dots,R\}$ into transparent and silent subsets, as described in Definition~\ref{def:transpandsilentdivision}. Assume, moreover, that 
Definitions~\ref{def:expandconvdir} and \ref{def:shiftnegligibletrs} are fulfilled; that $\mcX$ and the second fundamental form are 
$C^{0}$-future bounded; that there are $\a_{\infty},\zeta_{\infty}\in\Mn{m}{\co}$ and $0<\eta_{\romn},C_{\romn}\in\ro$ with the property that 
(\ref{eq:alpahzetaconvest}) holds for 
all $t\geq 0$; that there are $X^{j_{k}}_{\infty}\in\Mn{m}{\co}$, $k=1,\dots,d_{\trs}$, and a constant $0<K_{\trs}\in\ro$ with the property that
(\ref{eq:mcXconvsildir}) holds
for all $t\geq 0$ and all $k\in \{1,\dots,d_{\trs}\}$; and that if there is an $X^{j_{k}}_{\infty}\neq 0$, then $d_{\trs}=1$ and $R_{\trs}=0$. 
Fix a $0<\b_{\romar}\in\ro$ which is small enough, the bound depending only on the coefficients of the operator $\Delta_{\infty}$ introduced in 
(\ref{eq:Deltainfdefintro}); $X^{j_{k}}_{\infty}$, $k=1,\dots,d_{\trs}$; $\a_{\infty}$, $\zeta_{\infty}$; and the constant 
\[
\b_{\trs}:=\min\{\eta_{\trs},\b_{\rosil},\eta_{\romn}\}.
\]
Assume $f$ to be such that $\|f\|_{\trs,s}<\infty$ for all $s\in\ro$, where $\|\cdot\|_{\trs,s}$ is defined in (\ref{eq:ftrssnormintro}). Then 
there are constants $\sigma_{\rohom},\sigma_{\roih}\in [0,\infty)$, $0\leq N\in \zo$ and $0< C\in\ro$ such that the following holds. Given a 
solution $u$ to (\ref{eq:thesystemRge}), there is a unique $U_{\infty}\in C^{\infty}(\bM,\cn{2m})$ with the following properties
\begin{itemize}
\item the $\indexnot$'th Fourier coefficient of $U_{\infty}$ vanishes unless $\indexnot\in\EFtrsindexset$, 
\item if $\indexnot\in\EFtrsindexset$, then the $\indexnot$'th Fourier coefficient of $U_{\infty}$ belongs to 
$E_{\indexnot}$, where $E_{\indexnot}$ is the first generalised eigenspace in the $(\b_{\trs}-\b_{\romar}),A(\indexnot)$-decomposition 
of $\cn{2m}$, and $A(\indexnot)$ is given by (\ref{eq:Aindexnotdefintro}),
\item the estimate
\begin{equation}\label{eq:ultmuzpartasfinintro}
\begin{split}
 & \left\|\left(\begin{array}{c} (-\Delta_{\infty})^{1/2}u(\cdot,t) \\ \d_{t}u(\cdot,t)\end{array}\right)
-e^{\ma_{\trs}t}U_{\infty}-\int_{0}^{t}e^{\ma_{\trs}(t-\tau)}\left(\begin{array}{c} 0 \\ f(\cdot,\tau)\end{array}\right)d\tau\right\|_{t,s}\\
 \leq & C\ldr{t}^{N}e^{-(\b_{\trs}-\b_{\romar})t}[\|u_{t}(\cdot,0)\|_{(s+\sigma_{\rohom})}+\|u(\cdot,0)\|_{(s+\sigma_{\rohom}+1)}
+\|f\|_{\trs,s+\sigma_{\roih}}]
\end{split}
\end{equation}
holds for all $t\geq 0$ and $s\in\ro$, where $\Delta_{\infty}$ is defined in (\ref{eq:Deltainfdefintro}), $\ma_{\trs}$ is defined in (\ref{eq:matrsdefintro})
and $\|\cdot\|_{t,s}$ is defined in (\ref{eq:tsnodefintro}).
\end{itemize}
Moreover, 
\begin{equation}\label{eq:Uinfsobnoestfinintro}
\|U_{\infty}\|_{(s)}\leq C\|u_{t}(\cdot,0)\|_{(s+\sigma_{\rohom})}+C\|u(\cdot,0)\|_{(s+\sigma_{\rohom}+1)}+
C\|f\|_{\trs,s+\sigma_{\roih}}
\end{equation}
for all $s\in\ro$. 
\end{prop} 
\begin{remark}\label{remark:trsasintro}
The constant $C$ appearing in (\ref{eq:ultmuzpartasfinintro}) and (\ref{eq:Uinfsobnoestfinintro}) only depends on the coefficients of 
(\ref{eq:thesystemRge}), the spectrum of the Riemannian manifolds $(M_{r_{j}},g_{r_{j}})$, $j=1,\dots,R_{\trs}$, and $\b_{\romar}$. The constants
$\sigma_{\rohom}$ and $\sigma_{\roih}$ have the same dependence and $N$ only depends on $m$. 
\end{remark}
\begin{proof}
The proposition is a consequence of Proposition~\ref{prop:trsasympt}; cf. Remark~\ref{remark:geometrictoanalyticweaksil}. 
The statements of Remark~\ref{remark:trsasintro} follow from Remark~\ref{remark:constdepfintrs}. 
\end{proof}

Due to the proposition, we can think of the transparent part of solutions to (\ref{eq:thesystemRge}) as being well approximated by the transparent
part of solutions to the limit equation
(\ref{eq:thelimitequationinthetransparentcase}). Just as in the case of silent equations, we obtain the leading order asymptotics. However,
there is a limit to how detailed asymptotics we can obtain. Moreover, the limit is set by the size of the discrepancy between the actual
equation and the limit equation (quantitatively, the limit is determined by the number $\b_{\trs}$). Finally, (\ref{eq:Uinfsobnoestfinintro})
implies that the map from initial data to asymptotic data is continuous. 

The reader interested in an application of this result is referred to Section~\ref{section:anextheflvaKasol} below.

\section{Specifying the asymptotics}\label{section:specastrsintro}

In the previous section, we derived asymptotics, given a solution. However, it is also of interest to specify the leading order asymptotics, just 
as in the case of silent equations. In analogy with Proposition~\ref{prop:spasda}, it is sufficient to focus on the homogeneous case. 
In this section, we therefore assume that $f=0$ in (\ref{eq:thesystemRge}). When comparing the present context with that of 
Proposition~\ref{prop:spasda}, there is one important difference: we cannot assume the asymptotic data to take its values in one specific 
subspace of $\cn{2m}$; cf. the statement of Proposition~\ref{prop:trsasymptintro}. For this reason, it is convenient to introduce the following
terminology. 

\begin{definition}
Let $(M,g)$ be a canonical separable cosmological model manifold. Assume that the variables can be divided as described in 
Definition~\ref{def:transpandsilentdivision}.
Let $E_{\trs}$ be a function from $\EFtrsindexset$ to the set of vector subspaces of $\cn{2m}$. Then $\psi\in C^{\infty}(\bM,\cn{2m})$ is said to be 
$E_{\trs}$-\textit{adapted} 
\index{$\a$Aa@Notation!Conditions!$E_{\trs}$-adapted function}%
\index{Function!$E_{\trs}$-adapted}%
if the $\indexnot$'th Fourier coefficient of $\psi$ vanishes for $\indexnot\in\EFsilindexset$ and belongs to 
$E_{\trs}(\indexnot)$ for $\indexnot\in\EFtrsindexset$. The set of $E_{\trs}$-adapted elements of $C^{\infty}(\bM,\cn{2m})$ is denoted by 
$C^{\infty}(\bM,\cn{2m};E_{\trs})$. Finally, $C_{\trs}^{\infty}(\bM,\cn{2m})$ denotes the set of elements of $C^{\infty}(\bM,\cn{2m})$ whose 
$\indexnot$'th Fourier coefficients vanish for $\indexnot\in\EFsilindexset$.
\end{definition}

Next, we demonstrate that we can specify the leading order asymptotics. 

\begin{prop}\label{prop:spasdatrsintro}
Assume that the conditions of Proposition~\ref{prop:trsasymptintro} are fulfilled. Let $\b_{\romar}$
and $\b_{\trs}$ be as in the statement of Proposition~\ref{prop:trsasymptintro}. Assume, finally, that $f=0$. Then there is a function 
$E_{\trs}$ from $\EFtrsindexset$ to the set of subspaces of $\cn{2m}$ such that if $E_{\indexnot}:=E_{\trs}(\indexnot)$, then 
the spaces $E_{\indexnot}$ have the properties stated in Proposition~\ref{prop:trsasymptintro}. Moreover, there are constants
$C_{\Phi}>0$ and $s_{\infty}\geq 0$, and an injective linear map $\Phi_{\infty}$ from 
$C^{\infty}(\bM,\cn{2m};E_{\trs})$ to $C_{\trs}^{\infty}(\bM,\cn{2m})$ with the following properties. First, 
\begin{equation}\label{eq:Phiinfnobdtrsintro}
\|\Phi_{\infty}(\psi)\|_{(s)}\leq C_{\Phi}\|\psi\|_{(s+s_{\infty})}
\end{equation}
for all $s\in\ro$ and $\psi\in C^{\infty}(\bM,\cn{2m};E_{\trs})$. Second, if $\psi\in C^{\infty}(\bM,\cn{2m};E_{\trs})$ and $u$ is the solution 
to (\ref{eq:thesystemRge}) (with $f=0$) such that 
\begin{equation}\label{eq:uuditoPhiinfchitrsintro}
\left(\begin{array}{c} u(\cdot,0) \\ u_{t}(\cdot,0)\end{array}\right)=\Phi_{\infty}(\psi),
\end{equation}
then 
\begin{equation}\label{eq:estspecasdatatrsintro}
\begin{split}
 & \left\|\left(\begin{array}{c} (-\Delta_{\infty})^{1/2}u(\cdot,t) \\ u_{t}(\cdot,t)\end{array}\right)
-e^{\ma_{\trs}t}\psi\right\|_{t,s} \\
\leq & C\ldr{t}^{N}e^{-(\b_{\trs}-\b_{\romar})t}\left(\|u_{t}(\cdot,0)\|_{(s+\sigma_{\rohom})}+\|u(\cdot,0)\|_{(s+\sigma_{\rohom}+1)}\right),
\end{split}
\end{equation}
where the constants $C$, $N$ and $\sigma_{\rohom}$ have the same dependence as in the case of Proposition~\ref{prop:trsasymptintro}; cf. 
Remark~\ref{remark:trsasintro}. Finally, if there is an $\e>0$ such that $\Rsp[A(\indexnot)]<\b_{\trs}-\e$ for all $\indexnot\in\EFtrsindexset$, 
cf. Definition~\ref{def:SpRspdef}, then $E_{\trs}$ can be chosen to be such that $E_{\trs}(\indexnot)=\cn{2m}$ for all 
$\indexnot\in\EFtrsindexset$. In that case, $\Phi_{\infty}$ is surjective. 
\end{prop}
\begin{remark}\label{remark:CPhisinftytrsintro}
The constant $C_{\Phi}$ only depends on the coefficients of (\ref{eq:thesystemRge}), the spectrum of the Riemannian manifolds $(M_{r_{j}},g_{r_{j}})$, 
$j=1,\dots,R_{\trs}$, $\b_{\romar}$ and the supremum of $\kappa_{\indexnot}$ for $\indexnot\in\EFtrsindexset$. The constant $s_{\infty}\geq 0$ only depends 
on the coefficients of (\ref{eq:thesystemRge}) and the supremum of $\kappa_{\indexnot}$ for $\indexnot\in\EFtrsindexset$.
\end{remark}
\begin{remark}
By combining (\ref{eq:Phiinfnobdtrsintro}), (\ref{eq:uuditoPhiinfchitrsintro}) and (\ref{eq:estspecasdatatrsintro}), the norms of $u(\cdot,0)$ and $u_{t}(\cdot,0)$ 
appearing on the right hand side of (\ref{eq:estspecasdatatrsintro}) can be replaced by a suitable Sobolev norm of $\psi$. 
\end{remark}
\begin{remark}
In order to obtain a similar result in the case of inhomogeneous equations, it is sufficient to combine Propositions~\ref{prop:trsasymptintro} and
\ref{prop:spasdatrsintro}.
\end{remark}
\begin{remark}
Due to the estimate (\ref{eq:Phiinfnobdtrsintro}), the map from asymptotic data to initial data is continuous. 
\end{remark}
\begin{proof}
The statement follows from Proposition~\ref{prop:spasdatrs} and Remark~\ref{remark:geometrictoanalyticweaksil}. The statements of 
Remark~\ref{remark:CPhisinftytrsintro} follow from Remark~\ref{remark:CPhisinftytrs}.
\end{proof}
Due to Proposition~\ref{prop:spasdatrsintro}, we are in a position to calculate $\cruderate$. The reason for this is the following. First, 
we can divide a solution, say $u$, to the homogeneous version of (\ref{eq:thesystemRge}) into two parts: $u_{\trs}$ and $u_{\rosil}$. Due
to Proposition~\ref{prop:genroughestintro}, the energy of $u_{\rosil}$ does not grow faster than $\ldr{t}^{2d_{1}-2}e^{2\kappa_{1}t}$, where 
$\kappa_{1}$ and $d_{1}$ are defined in the statement of Proposition~\ref{prop:genroughestintro}. Similarly, due to 
Proposition~\ref{prop:trsasymptintro}, for every 
$\e>0$, the energy of $u_{\trs}$ does not grow faster than $e^{2(\kappa+\e)t}$, where $\kappa$ is the supremum of the $\kappa_{\indexnot}$'s introduced 
at the beginning of Section~\ref{section:futastranspintro}. Thus $\cruderate\leq \max\{\kappa_{1},\kappa\}$. Combining this observation with 
Propositions~\ref{prop:spasda} and \ref{prop:spasdatrsintro} yields the conclusion that $\cruderate=\max\{\kappa_{1},\kappa\}$; cf. 
Remark~\ref{remark:udeponlonsilandt} for a justification of the statement that Proposition~\ref{prop:spasda} applies. On the other
hand, we do not obtain any conclusions concerning $\nolossrate$. 

\section{An example: the flat Kasner solution}\label{section:anextheflvaKasol}

In Example~\ref{example:nonflatvacuumKasner}, we consider the Klein-Gordon equation for all the non--flat Kasner solutions.
In the case of the flat Kasner solution, the Klein-Gordon equation also takes the form (\ref{eq:nonflatKasnerKleinGordon}) (though in what
follows, we relabel $\tau$ to $t$). However, all the 
$\b_{i}$ except one equal $1$ and the exceptional exponent, say $\b_{d}$, equals $0$. In the case of solutions that only depend on the first
$d-1$ variables, an analysis similar to the one provided in Example~\ref{example:nonflatvacuumKasner} yields the conclusion that for such solutions, 
there is a homeomorphism from initial data to asymptotic data. Moreover, an estimate of the form (\ref{eq:uudotasnonflatKasner}) holds, where
$\mu=1$. Let us therefore focus on the part of the solution that corresponds to modes with $n_{d}\neq 0$;
\[
u_{\trs}(x,t)=\textstyle{\sum}_{n\in\zn{d},n_{d}\neq 0}z(n,t)(2\pi)^{-d/2}e^{in\cdot x}.
\]
In accordance with Definition~\ref{def:transpandsilentdivision}, there is a division of $\{1,\dots,d\}$ into a silent subset (in our case given 
by $\{1,\dots,d-1\}$) and a transparent subset (in our case given by $\{d\}$). Moreover, Definition~\ref{def:expandconvdir} applies with $C_{a}=1$, 
$\eta_{\trs}=2$, $\b_{\rosil}=1$ and $\betafun_{\rosil}=0$. Since the shift vector field vanishes in the present setting, it is clear that it 
is asymptotically negligible in the sense of Definition~\ref{def:shiftnegligibletrs}. Since $\mcX=0$ and (\ref{eq:alpahzetaconvest}) holds, 
with $C_{\romn}=m^{2}$, $\eta_{\romn}=2$, $\a_{\infty}=0$ and $\zeta_{\infty}=0$, the limit equation (\ref{eq:thelimitequationinthetransparentcase})
takes the form 
\begin{equation}\label{eq:odstlinwaveeq}
v_{tt}-v_{dd}=0.
\end{equation}
Next, note that $\EFindexset=\zn{d}$ in the present setting. For this reason, we here use the notation $n$ instead of $\indexnot$. The 
function $\mfg_{\infty}$, introduced in (\ref{eq:mfginftytrsdefintro}), is here given by $\mfg_{\infty}(n)=|n_{d}|$. 
Turning to $A(\indexnot)$, $w(\indexnot,t)$ and $F(\indexnot,t)$, introduced in (\ref{eq:Aindexnotdefintro}) and 
(\ref{eq:vFdeftrsintro}) , they are given by 
\[
A(n)=\left(\begin{array}{cc} 0 & |n_{d}| \\ -|n_{d}| & 0\end{array}\right),\ \ \
w(n,t)=\left(\begin{array}{c} |n_{d}|z(n,t) \\ \dot{z}(n,t)\end{array}\right)
\]
and $F(n,t)=0$. In particular, 
\[
e^{A(n)t}=\left(\begin{array}{rr} \cos(|n_{d}|t) & \sin(|n_{d}|t) \\ -\sin(|n_{d}|t) & \cos(|n_{d}|t)\end{array}\right).
\]
If $U_{\infty}\in C^{\infty}(\tn{d},\cn{2})$ is a function with Fourier coefficients $\hat{U}_{\infty}$ such that $\hat{U}_{\infty}(n)=0$ if 
$n_{d}=0$, then 
\[
(e^{\ma_{\trs}t}U_{\infty})(x,t)=\textstyle{\sum}_{n\in\zn{d},n_{d}\neq 0}
\left(\begin{array}{rr} \cos(|n_{d}|t) & \sin(|n_{d}|t) \\ -\sin(|n_{d}|t) & \cos(|n_{d}|t)\end{array}\right)\hat{U}_{\infty}(n)
(2\pi)^{-d/2}e^{in\cdot x}.
\]
In practice, the variables $x^{1},\dots,x^{d-1}$ are thus frozen, and we solve the linear wave equation (\ref{eq:odstlinwaveeq}) in the 
$tx^{d}$-directions. 

Let us now return to Proposition~\ref{prop:trsasymptintro}. Note that in the present setting, the assumptions of this proposition are 
satisfied. In addition, $\b_{\trs}=1$. Fix $0<\b_{\romar}\in\ro$ satisfying the restrictions of Proposition~\ref{prop:trsasymptintro}. Given a 
solution $u$ to (\ref{eq:nonflatKasnerKleinGordon}), there is a unique $U_{\infty}\in C^{\infty}(\tn{d},\cn{2})$ with the property that 
$\hat{U}_{\infty}(n)=0$ if $n_{d}=0$, and the property that 
\begin{equation}\label{eq:utrsasflatKasner}
\left\|\left(\begin{array}{c} |\d_{d}|u_{\trs}(\cdot,t) \\ \d_{t}u_{\trs}(\cdot,t)\end{array}\right)
-e^{\ma_{\trs}t}U_{\infty}\right\|_{(s)}
 \leq C_{s}\ldr{t}^{N}e^{-(1-\b_{\romar})t}
\end{equation}
for some constant $0<C_{s}\in\ro$ and all $t\geq 0$. Here the constant $C_{s}$ is allowed to depend on $s$, the solution and the equation, and 
$|\d_{d}|:=(-\d_{d}^{2})^{1/2}$. In fact, appealing
to Proposition~\ref{prop:spasdatrsintro}, the map from initial data to $U_{\infty}$ is a homeomorphism with respect to the $C^{\infty}$ topology.
In order to obtain this conclusion, we use the fact that the eigenvalues of $A(n)$ are purely imaginary. Turning to the function $u_{\trs}$
itself, note that (\ref{eq:utrsasflatKasner}) implies that 
\[
\sum_{n\in\zn{d},n_{d}\neq 0}\ldr{n}^{2s}\left| |n_{d}|z(n,t) -\cos(|n_{d}|t)\hU_{1}(n)-\sin(|n_{d}|t)\hU_{2}(n)\right|^{2}\leq 
C_{s}\ldr{t}^{2N}e^{-2(1-\b_{\romar})t}
\]
for all $t\geq 0$, where $\hU_{i}$, $i=1,2$, are the components of $\hU_{\infty}$. Combining this estimate with (\ref{eq:utrsasflatKasner}) 
yields the conclusion that there is a function $U_{\trs}\in C^{\infty}(\tn{d}\times\ro,\co)$ which solves (\ref{eq:odstlinwaveeq}); has the 
property that $\hat{U}_{\trs}(n,\cdot)=0$ for all $n\in\zn{d}$ such that $n_{d}=0$; and is such that
\[
\|u_{\trs}(\cdot,t)-U_{\trs}(\cdot,t)\|_{(s)}+\|\d_{t}u_{\trs}(\cdot,t)-\d_{t}U_{\trs}(\cdot,t)\|_{(s)}\leq C_{s}\ldr{t}^{N}e^{-(1-\b_{\romar})t}
\]
for all $t\geq 0$. 

\section{Outline of the argument}\label{section:outlinetransintro}

The strengths and weaknesses of the results are quite similar to those in the silent setting; cf. Subsection~\ref{ssection:strandweaksilsettintro}.
Here, we therefore focus on an outline of the proof. The rough structure of the argument in the transparent setting is similar to that of the argument
in the silent setting. In particular, Subsection~\ref{ssection:outloftheargsilsetting} gives a general idea of how to proceed. The main difference
is due to the fact that the set of $A(\indexnot)$ is infinite. This difference gives rise to difficulties that are purely algebraic, but 
somewhat technical to resolve. In order to describe the complications, let us consider the matrices $A(\indexnot)$ in greater detail. 

\textbf{Model matrices.} Consider the case that $X^{j_{k}}_{\infty}\neq 0$ for some $k\in \{1,\dots,d_{\trs}\}$. 
Then, due to the assumptions of Propositions~\ref{prop:trsasymptintro} and \ref{prop:spasdatrsintro}, $d_{\trs}=1$ and $R_{\trs}=0$. Thus 
$\mfg_{\infty}(\indexnot)=(g^{j_{1}j_{1}}_{\infty})^{1/2}|n_{j_{1}}|$, so that $A(\indexnot)$ equals
\begin{equation}\label{eq:Amudeftrsintro}
A_{\mu}:=\left(\begin{array}{cc} 0 & \mu\Id_{m} \\ -\mu\Id_{m}+iV+2\mu^{-1}W & 
U\end{array}\right),
\end{equation}
where $\mu:=\mfg_{\infty}(\indexnot)$; $U:=-\a_{\infty}$; $W:=-\zeta_{\infty}/2$; and
\[
V:=-\frac{n_{j_{1}}}{|n_{j_{1}}|}\frac{X^{j_{1}}_{\infty}}{(g^{j_{1}j_{1}}_{\infty})^{1/2}}.
\]
For $\indexnot\in\EFtrsindexset$, $n_{j_{1}}\neq 0$, so that there are only two possibilities for $V$. In that sense, it is sufficient to 
focus on matrices of the form (\ref{eq:Amudeftrsintro}) for fixed matrices $U,V,W\in\Mn{m}{\co}$. In case all the $X^{j_{k}}_{\infty}$ vanish, 
$A(\indexnot)$ also takes the form (\ref{eq:Amudeftrsintro}), this time with $V=0$. 

\textbf{Asymptotic expansions for the eigenvalues.}
Let us consider $A_{\mu}$ in greater detail. Conjugating $A_{\mu}$ by a matrix independent of $\mu$ yields a matrix $N_{\mu}$ of the form 
\begin{equation}\label{eq:Ninfformulaintrointro}
N_{\mu}=\left(\begin{array}{cc} -i\mu\Id_{m}+R_{\mu}^{-}
& Q_{\mu}^{+}\\
Q_{\mu}^{-} & i\mu\Id_{m}+R_{\mu}^{+}\end{array}\right),
\end{equation}
where
\[
R_{\mu}^{\pm}=\g_{\pm}\mp i\mu^{-1}W_{\pm},\ \ \
Q_{\mu}^{\pm}=\pm i\de_{\pm}+\mu^{-1}V_{\pm}.
\]
Here $\g_{\pm}$, $\de_{\pm}$, $V_{\pm}$ and $W_{\pm}$ are independent of $\mu$. Moreover, the $\g_{\pm}$ are Jordan normal forms of the matrices 
$(U\pm V)/2$. Let $\lambda_{j,\pm}$, $j=1,\dots,p_{\pm}$, be the distinct eigenvalues of $\g_{\pm}$, and $m_{j,\pm}$ be the 
corresponding multiplicities. It can then be demonstrated that there are constants $0<c_{a},\mu_{a}\in\ro$ such that for $\mu\geq\mu_{a}$ and 
$j=1,\dots,p_{\pm}$, there are $m_{j,\pm}$ eigenvalues of $N_{\mu}$ in a ball of radius $c_{a}\mu^{-1/m_{j,\pm}}$ and centre $\pm i\mu+\lambda_{j,\pm}$; 
cf. Lemma~\ref{lemma:Nmueigenvapprox}. In other words, there are asymptotic expansions for the eigenvalues.

\textbf{Asymptotic partial diagonalisation.} Ideally, we would like to diagonalise $N_{\mu}$. However, the potential multiplicities of the 
eigenvalues of $\g_{\pm}$ cause problems. In the end, we therefore only obtain a partial diagonalisation. In fact, for sufficiently large $\mu$, 
there are matrices $T_{\mu}\in\Mn{2m}{\co}$ such that $\|T_{\mu}\|,\|T_{\mu}^{-1}\|\leq 2$ and such that 
\begin{equation}\label{eq:Tmudiagintrointro}
T_{\mu}^{-1}N_{\mu}T_{\mu}=\diag\{N_{\mu,1}^{-},\dots,N_{\mu,p_{-}}^{-},N_{\mu,1}^{+},\dots,N_{\mu,p_{+}}^{+}\},
\end{equation}
where
\[
\|N_{\mu,j}^{\pm}\mp i\mu\Id_{m_{j,\pm}}-\g_{\pm,j}\|\leq C\mu^{-1}.
\]
Here $\g_{\pm,j}$, $j=1,\dots,p_{\pm}$, is the matrix collecting all the Jordan blocks in $\g_{\pm}$ corresponding to the eigenvalue 
$\lambda_{j,\pm}$. The justification for these statements is to be found in Lemma~\ref{lemma:aspardiag}. The proof is quite long, and requires
some background material, which we develop in Section~\ref{section:compprojontogensubs} below.

\textbf{Dividing the matrices according to frequency.}
Fix $0\leq\mu_{0}\in\ro$. Due to the assumptions, the set of $\mfg_{\infty}(\indexnot)$ satisfying $\mfg_{\infty}(\indexnot)\leq \mu_{0}$ is finite. 
Considering the family of matrices $A(\indexnot)$, there is thus a natural way to divide it into two subsets: the matrices corresponding to 
$\mfg_{\infty}(\indexnot)\leq \mu_{0}$ and the matrices corresponding to $\mfg_{\infty}(\indexnot)>\mu_{0}$. The use of this division is due to the fact
that for large $\mu_{0}$, there is a partial diagonalisation of the form (\ref{eq:Tmudiagintrointro}). Choosing $\mu_{0}$ large enough, the behaviour 
of the modes satisfying $\mfg_{\infty}(\indexnot)>\mu_{0}$ can therefore be analysed (up to, possibly, some small error). Since the remaining modes 
only correspond to a finite number of matrices $A(\indexnot)$, their behaviour can be analysed by appealing to the results of the previous chapter.

\textbf{Asymptotic analysis for the modes.} Let us now focus on the Fourier mode corresponding to a $\indexnot\in\EFtrsindexset$. In analogy
with the discussion in Subsection~\ref{ssection:outloftheargsilsetting}, it is natural to divide $[0,\infty)$ into two subsets: 
$[0,T_{\trs}]$ and $(T_{\trs},\infty)$. Here $T_{\trs}$ is defined so that the error terms (i.e., the terms that make up the difference between
the actual equation and the limit equation) are bounded by $Ce^{-\b_{\trs}\bt}$ for $t\geq T_{\trs}$, where $C$ is a constant independent of 
$\indexnot$ and $\bt:=t-T_{\trs}$. Moreover, $T_{\trs}$ is logarithmic in $\ldr{\nu(\indexnot)}$. Just as in the silent setting, the analysis in 
$[0,T_{\trs}]$ consists of a crude energy estimate. Due to the fact that $T_{\trs}$ is logarithmic in $\ldr{\nu(\indexnot)}$, this is sufficient;
our lack of detailed knowledge in $[0,T_{\trs}]$ only corresponds to a finite loss of derivatives. How we analyse the asymptotics in the 
interval $(T_{\trs},\infty)$ depends on the division of the modes into low and high frequencies. The parameter defining the division is $\mu_{0}$. 
A large $\mu_{0}$ yields more detailed control of the eigenvalues (and
corresponds to a smaller $\b_{\romar}$; cf. Section~\ref{section:futastranspintro}). On the other hand, the larger the $\mu_{0}$, the larger
the number of cases to which we need to apply the analysis of the previous chapter (and, thereby, the larger the constants appearing in the
estimates). Given the partial diagonalisation (\ref{eq:Tmudiagintrointro}), we are in a position to carry out an analysis for high
frequencies and $t\geq T_{\trs}$. This is the subject of Section~\ref{section:mbmanhftrs}. In Section~\ref{section:spdatinftrs}, we then turn 
to the problem of specifying the asymptotic data for an individual high frequency mode. 

\textbf{Summing up.} Given the analysis described above and the results of the previous chapter, we are in a position to derive the desired
conclusions concerning transparent equations. This is the subject of Chapter~\ref{chapter:asympttranspcase}. We start, in 
Sections~\ref{section:introweaktrsmainresults} and \ref{section:weasilbalcontrs}, by describing the equations of interest; estimating the 
discrepancy between the actual equation and the limit equation for fixed modes; and defining $T_{\trs}$. The derivation of the asymptotics 
is divided into two parts. In Lemma~\ref{lemma:trsasymptlargefre}, we derive conclusions for the part of the solution corresponding to all 
the high frequency modes. Combining this result with the methods developed in the silent setting yields 
the desired conclusions concerning the asymptotics for the transparent part of the solution; cf. Proposition~\ref{prop:trsasympt}. In 
particular, we obtain a continuous map from initial data to asymptotic data. We end the chapter in Section~\ref{section:specastrs} by 
demonstrating that we can specify the leading order asymptotics; cf. Proposition~\ref{prop:spasdatrs}. Again, the map from asymptotic
data to initial data is continuous.

\chapter{Equations with a dominant noisy spatial direction}\label{chapter:domnoisspdirintro}

\section{Introduction}

In Chapter~\ref{chapter:silentequations} we consider the case that $\bk$ has a strictly positive lower bound and in Chapter~\ref{chapter:transpeq}
we consider the case that $\bk$ asymptotically vanishes in some directions. As a next step, it is natural to allow $\bk$ to be 
negative in some directions. Note that this corresponds to contraction. As far as the asymptotic behaviour is concerned, the direction in which 
the contraction is the strongest is the most important. Moreover, assuming that there is one direction in which the contraction dominates simplifies
the analysis, and we restrict our attention to this case. For reasons mentioned in Subsection~\ref{ssection:siltrsanoimet}, we refer to the
corresponding geometries (and equations) as noisy. Two examples of noisy geometries are given by the Kasner solutions and the $U(1)$-symmetric 
solutions (in the expanding direction); cf. Subsection~\ref{ssection:cansepcomoma}. In the study of the expanding direction of polarised Gowdy
solutions, we also obtain noisy equations. 

\subsection{Outline}\label{ssection:outlinedomnoisspdir}

\textbf{Division of the variables and basic assumptions.}
In order to give a formal definition of what it means for an equation to have a dominant noisy spatial direction, we need to divide the cotangent
space into directions in which the spacetime contracts maximally and the remaining directions. The formal definition of this division is similar to the 
definition in the transparent setting, cf. Definition~\ref{def:transpandsilentdivision}, and is provided in Subsection~\ref{ssection:divtanspnoisyeq}. 
In Subsection~\ref{ssection:divtanspnoisyeq}, we also describe the requirements on the second fundamental form, the shift vector field, the $X^{l}$'s, 
$\a$ and $\zeta$. Concerning the second fundamental form, the main assumption is that there is a $0<\b_{\ron}\in\ro$ such that $\bk^{\sharp}+\b_{\ron}\bge^{\sharp}$
converges to zero exponentially in the dominant noisy spatial direction. Concerning the remaining directions, we only need to assume that the contraction is 
slower, with a margin. Turning to the coefficients of the lower order derivatives, we only need to impose convergence conditions on the $X^{l}$'s 
corresponding to the dominant noisy direction and on $\a$. Beyond these requirements, we demand that (\ref{eq:thesystemRge}) be $C^{2}$-balanced
and that the $C^{1}$-norm of the shift vector field decay exponentially. In other words, the conditions are weaker than the ones imposed in the
silent and transparent settings, in the sense that we do not require $\zeta$ to converge. However, they are stronger in that we require control 
over a higher number of derivatives. In both cases, the reason for the difference is the fact that the solutions exhibit oscillatory behaviour, with
a frequency that grows exponentially. Due to the exponentially growing frequency, $\zeta$ is effectively an error term, as long as it is future
bounded in $C^{1}$. On the other hand, in order to control the evolution in this highly oscillatory setting, we need to approximate the oscillations,
just as for the equations considered in Chapter~\ref{chapter:onnotofbal}; cf. Section~\ref{section:mainideaspfsupexpgrowth}. In order for the 
relevant methods, developed in Part~\ref{part:averaging}, to work, we need to impose bounds on a larger number of derivatives than in the silent and 
transparent settings. 

\textbf{The limit equation.} Just as in the case of silent and transparent equations, there is a limit equation in the noisy setting. 
On the Fourier side, cf. Section~\ref{section:divintomodes}, the homogeneous version of this equation reads
\begin{equation}\label{eq:limiteqnoiseintrointro}
\ddot{z}+\nu_{\ron}^{2}(\indexnot)e^{2\b_{\ron}t}z+i\nu_{\ron}(\indexnot)e^{\b_{\ron}t}\bX_{\ron}(\indexnot)z+\a_{\infty}\dot{z}=0;
\end{equation}
\index{Limit equation!noisy setting}%
cf. (\ref{eq:limiteqinthenoisyset}) below. Here $\nu_{\ron}(\indexnot)$ is the limit of $e^{-\b_{\ron}t}\mfg(\indexnot,t)$ (which exists due to the 
assumptions). Moreover, $\a_{\infty}$ is the limit of $\a$, and $\bX_{\ron}(\indexnot)$ is either zero or, up to a sign, given by a limit that exists
due to the assumptions; cf. (\ref{eq:bXrondefintro})--(\ref{eq:tXinfdefgendirintro}) below. 

At this point, let us remark that the assumptions that 
we make in the present chapter only allow us to draw conclusions for the $\indexnot\in\EFindexset$ such that 
$\nu_{\ron}(\indexnot)\neq 0$. The Fourier modes with $\nu_{\ron}(\indexnot)=0$ correspond to solutions to the so-called \textit{subdominant equation}. 
\index{Equation!subdominant part}%
\index{Subdominant!part of equation}%
This equation is 
obtained by setting all the coefficients in front of the derivatives with respect to the dominant noisy spatial variables to zero. In order 
to obtain asymptotic information concerning solutions to the subdominant equation, we need to make additional assumptions. For instance, if the 
subdominant equation is transparent in the sense of the previous chapter, then the results of the previous two chapters apply; if it is silent, 
the results of Chapter~\ref{chapter:silentequations} apply; and if it has a dominant noisy spatial direction in the sense of the present chapter, 
then we can proceed inductively. 

\textbf{Averaging over the oscillations.} Consider a solution to (\ref{eq:limiteqnoiseintrointro}) corresponding to a $\indexnot\in\EFindexset$ such 
that $\nu_{\ron}(\indexnot)\neq 0$. Since $\nu_{\ron}(\indexnot)e^{\b_{\ron}t}\rightarrow\infty$, it is clear that the dominant behaviour of solutions is 
oscillatory. However, it is also of interest to understand the overall behaviour over longer periods of time. To this end, it is of interest to 
consider one period of the oscillations in detail and then to derive conclusions concerning the overall behaviour by combining the conclusions concerning
the individual periods; cf. the discussion in Section~\ref{section:mainideaspfsupexpgrowth}. In order to develop a feeling for the evolution over
one period, fix a $t_{0}\geq 0$. Then the corresponding period of the oscillations is roughly speaking $T:=2\pi/\mfg_{\infty}(\indexnot,t_{0})$,
where $\mfg_{\infty}(\indexnot,t):=\nu_{\ron}(\indexnot)e^{\b_{\ron}t}$. Moreover, (\ref{eq:limiteqnoiseintrointro}) can be rewritten as
\begin{equation}\label{eq:wdotnoislim}
\dot{w}(\indexnot,t)=A_{\infty}(\indexnot,t)w(\indexnot,t),
\end{equation}
where
\begin{align}
w(\indexnot,t) := & e^{-\b_{\ron}(t-t_{0})/2}\left(\begin{array}{c} e^{\a_{\infty}(t-t_{0})/2}\mfg_{\infty}(\indexnot,t)z(\indexnot,t) \\ 
e^{\a_{\infty}(t-t_{0})/2}\dot{z}(\indexnot,t) \end{array} \right),\nonumber\\
A_{\infty}(\indexnot,t) := & \left(\begin{array}{cc} (\b_{\ron}\Id_{m}+\a_{\infty})/2 & \mfg_{\infty}(\indexnot,t)\Id_{m} \\ 
-\mfg_{\infty}(\indexnot,t)\Id_{m}-i\bY_{\ron}(\indexnot,t)
 & -(\b_{\ron}\Id_{m}+\a_{\infty})/2 \end{array} \right)\label{eq:Ainfindtdefnoislim}
\end{align}
and
\begin{equation}\label{eq:bYrondef}
\bY_{\ron}(\indexnot,t):=e^{\a_{\infty}(t-t_{0})/2}\bX_{\ron}(\indexnot)e^{-\a_{\infty}(t-t_{0})/2}.
\end{equation}
Note that the choice of variables ensures that the sum of the matrices on the diagonal of $A_{\infty}$ equals zero. A naive way of approximating the evolution 
of $w$ from $t_{0}$ to $t_{0}+T$ is to say that it corresponds to multiplication by $\exp[A_{\infty}(\indexnot,t_{0})T]$. However, due to the fact that $T$ decays 
exponentially 
as $t_{0}$ grows, we expect this approximation to gradually become better and better. The matrix $A_{\infty}(\indexnot,t_{0})T$ has a very special form. First, the 
dominant part corresponds to the evolution over exactly one full period of the oscillations. Second, the sum of the matrices on the diagonal vanishes (so that, 
intuitively, the mean contribution from the diagonal matrices over one period should vanish). Third, the only matrix that remains beyond this is 
\[
-i\bY_{\ron}(\indexnot,t_{0})T=-i\bX_{\ron}(\indexnot)T. 
\]
In Section~\ref{section:calmaexp}, we develop methods for calculating exponentials of matrices of exactly this form. The result is 
\[
\exp[A_{\infty}(\indexnot,t_{0})T]\approx \Id_{2m}+\frac{1}{2}\left(\begin{array}{cc} 0 & i\bX_{\ron}(\indexnot) \\ 
-i\bX_{\ron}(\indexnot)
 & 0 \end{array} \right)T.
\]
Combining the above definitions and observations yields the conclusion that if $t_{1}:=t_{0}+T$, then 
\begin{align*}
\left(\begin{array}{c} \mfg_{\infty}(\indexnot,t_{1})z(\indexnot,t_{1}) \\ 
\dot{z}(\indexnot,t_{1}) \end{array} \right) \approx & e^{\b_{\ron}T/2}\left(\begin{array}{cc} e^{-\a_{\infty}T/2} & 0 \\ 
0  & e^{-\a_{\infty}T/2} \end{array} \right)\exp[A_{\infty}(\indexnot,t_{0})T]w(\indexnot,t_{0})\\
 \approx & \left[\Id_{2m}+B_{\infty}(\indexnot)T\right]w(\indexnot,t_{0}),
\end{align*}
where 
\[
B_{\infty}(\indexnot):=\frac{1}{2}\left(\begin{array}{cc} \b_{\ron}\Id_{m}-\a_{\infty} & i\bX_{\ron}(\indexnot) \\ 
-i\bX_{\ron}(\indexnot) & \b_{\ron}\Id_{m}-\a_{\infty} \end{array} \right).
\]
Thus
\begin{align*}
 \left(\begin{array}{c} \mfg_{\infty}(\indexnot,t_{1})z(\indexnot,t_{1}) \\ 
\dot{z}(\indexnot,t_{1}) \end{array} \right)  
\approx & e^{B_{\infty}(\indexnot)T}\left(\begin{array}{c} \mfg_{\infty}(\indexnot,t_{0})z(\indexnot,t_{0}) \\ 
\dot{z}(\indexnot,t_{0}) \end{array} \right).
\end{align*}
Conjugating $B_{\infty}(\indexnot)$ by 
\[
S:=\frac{1}{2}\left(\begin{array}{cc} \Id_{m} & i\Id_{m} \\  i\Id_{m} & \Id_{m} \end{array} \right)
\]
yields
\begin{equation}\label{eq:SBinfSinv}
SB_{\infty}(\indexnot)S^{-1}=\frac{1}{2}\left(\begin{array}{cc} \b_{\ron}\Id_{m}-\a_{\infty}+\bX_{\ron}(\indexnot) & 0 \\ 
0 & \b_{\ron}\Id_{m}-\a_{\infty}-\bX_{\ron}(\indexnot) \end{array} \right).
\end{equation}
Naively, it is thus reasonable to expect the overall evolution to be determined by the diagonal elements of this matrix. Note that 
$\bX_{\ron}(\indexnot)$ can be written as $\pm V$ for a fixed matrix $V$ (independent of $\indexnot$). For this reason, the set consisting
of the matrices on the diagonal of (\ref{eq:SBinfSinv}) is independent of $\indexnot$. In particular, the growth of generic solutions to 
$\dot{v}=B_{\infty}(\indexnot)v$ is independent of $\indexnot$.

\textbf{Outline, results.} After describing the conditions we impose in the present chapter, cf. Subsection~\ref{ssection:divtanspnoisyeq};
and the limit equation, cf. Subsection~\ref{ssection:thelimitequationinthenoisysetting}; we turn to the results in Section~\ref{section:results}. 
First, we derive an energy estimate in Subsection~\ref{ssection:roughsobestintronoise}. The growth corresponds exactly to the generic growth
of solutions to $\dot{v}=B_{\infty}(\indexnot)v$ (which, as we noted above, is independent of $\indexnot$). However, just as in the silent and 
transparent settings, the stated energy estimate involves a loss of derivatives; cf. Proposition~\ref{prop:roughsobestnoisysetintro}. 

\textit{Understanding the oscillations.} Our second goal is to understand the oscillatory character of the solutions. In order to do so, we focus 
on the scalar equation obtained from (\ref{eq:thesystemRge}) by dropping the right hand side and only keeping the terms on the left hand side
that involve second order derivatives of $u$, i.e.,
\begin{equation}\label{eq:homtheeqnoiseintro}
u_{tt}-\textstyle\sum_{j,l=1}^{d}g^{jl}(t)\d_{j}\d_{l}u-2\sum_{l=1}^{d}g^{0l}(t)\d_{l}\d_{t}u
-\sum_{r=1}^{R}a^{-2}_{r}(t)\Delta_{g_{r}}u=0. 
\end{equation}
Considering (\ref{eq:SBinfSinv}), it seems reasonable to expect oscillations with an overall growth of $\mfe[u]$ of the form $e^{\b_{\ron}t}$. This is 
indeed what happens. Moreover, not only are we in a position to derive detailed asymptotics of solutions, we actually obtain a homeomorphism between 
initial data and asymptotic data. Due to this fact, we can consider solutions to (\ref{eq:homtheeqnoiseintro}) as models for the oscillatory behaviour.
The result is stated as Proposition~\ref{prop:asymposccasescalareqnoiseintro}. As an illustration of the result, we consider the future asymptotics of 
polarised vacuum Gowdy solutions. 

\textit{Deriving/specifying the asymptotics. Outline of the proof.} Let us return to (\ref{eq:thesystemRge}). In order to describe the future asymptotics 
of a solution, it turns out to be convenient to compare it with a sum of terms of the form $e^{Rt}u$, where $R$ is a constant matrix and $u$ is a vector 
valued solution to (\ref{eq:homtheeqnoiseintro}). However, describing the detailed combination of matrices and solutions to (\ref{eq:homtheeqnoiseintro}) 
needed is somewhat technical, and is left to Subsection~\ref{ssection:derivasnoiseintr}, in particular 
Proposition~\ref{prop:asymposccaseitosoltowenoiseintro}. Finally, similarly to the silent and transparent settings, we can specify the leading order 
asymptotics. Moreover, in favourable circumstances, we obtain a homeomorphism between initial data and asymptotic data. The relevant result is stated as 
Proposition~\ref{prop:asymposccaseitosoltowenoisespaintro}. In Section~\ref{section:outpfnoiseintro} we end the chapter by giving an outline of 
the proofs of the results.

\subsection{Dividing the cotangent space}\label{ssection:divtanspnoisyeq}

Just as in the transparent setting, we here need to divide the cotangent space into the direct sum of two subspaces. We refer to the relevant subspaces as 
the dominant noisy spatial direction and the subdominant directions. In parallel with the introduction of these subspaces, we state the main convergence 
requirements. Since the formulation of the conditions depends on whether the dominant noisy spatial direction corresponds to an $\so$-factor or an 
$M_{r}$-factor, we state the requirements in two separate definitions. We start by considering the $\so$-case.

\begin{definition}\label{def:domnoisyspasodir}
Let $(M,g)$ be a canonical separable cosmological model manifold. Fix $j\in \{1,\dots,d\}$ and let $\pi_{\ron}$ 
be the map from $\bM$ to $\so$ corresponding to projection onto the $j$'th $\so$-factor in $\tn{d}$. Let $\pi_{\rosub}$ be the map from $\bM$ to 
\[
\bM_{\rosub}:=\tn{d-1}\times M_{1}\times\cdots\times M_{R}
\]
\index{$\a$Aa@Notation!Manifolds!$\bM_{\rosub}$}%
corresponding to the projection onto what remains after removing the $j$'th $\so$-factor. Define
\[
T_{\ron}^{*}\bM:=\pi_{\ron}^{*}(T^{*}\so),\ \ \
T_{\rosub}^{*}\bM:=\pi_{\rosub}^{*}(T^{*}\bM_{\rosub}).
\]
\index{$\a$Aa@Notation!Tangent and cotangent spaces!$T_{\ron}^{*}\bM$}%
\index{$\a$Aa@Notation!Tangent and cotangent spaces!$T_{\rosub}^{*}\bM$}%
The metric $g$ is said to have a \textit{geometric dominant noisy spatial $\so$-direction corresponding to} $j$ 
\index{Geometric!dominant noisy spatial direction}%
\index{Dominant noisy spatial direction!geometric}%
if there are 
$0<\b_{\ron},\eta_{\ron},C_{\ron}\in\ro$ and a continuous non-negative function $\betafun\in L^{1}([0,\infty))$ such that the following holds:
\begin{equation}\label{eq:condonkdomnoisdirintro}
|\bk^{\sharp}(\xi,\xi)+\b_{\ron}\bge^{\sharp}(\xi,\xi)|\leq C_{\ron}e^{-\eta_{\ron}t}\bge^{\sharp}(\xi,\xi)
\end{equation}
for all $\xi\in T_{\ron}^{*}\bM$ and all $t\geq 0$, and 
\begin{equation}\label{eq:bkwwlbsubtmintro}
\bk^{\sharp}(\xi,\xi)\geq [-\b_{\ron}+\eta_{\ron}-\betafun(t)]\bge^{\sharp}(\xi,\xi)
\end{equation}
for all $\xi\in T_{\rosub}^{*}\bM$ and all $t\geq 0$. Consider (\ref{eq:thesystemRge}), where $(M,g)$ is such that the above conditions hold. If
there are constants $\eta_{\romn}>0$, $C_{\romn}>0$ and matrices $X^{j}_{\infty},\a_{\infty}\in\Mn{m}{\co}$ such that 
\begin{equation}\label{eq:Xjalconvnosettingintro}
\|e^{-\b_{\ron}t}X^{j}(t)-X^{j}_{\infty}\|+\|\a(t)-\a_{\infty}\|\leq C_{\romn}e^{-\eta_{\romn}t}
\end{equation}
for all $t\geq 0$; then (\ref{eq:thesystemRge}) is said to be such that \textit{the dominant coefficients are convergent}. 
\index{Convergent!dominant coefficients}%
\index{Dominant coefficients!convergent}%
\end{definition}
\begin{remark}
The definitions of $\bge^{\sharp}$ and $\bk^{\sharp}$ are given at the beginning of Subsection~\ref{ssection:divthevartrssetintro}.
\end{remark}
\begin{remark}
In Remark~\ref{remark:trscondtanspdiv}, we reformulate the conditions (\ref{eq:bkbdtrssettingintrocotan}) and (\ref{eq:bkbdtrssettingsilentpartintrocotan}) 
on the second fundamental form in the transparent setting to (\ref{eq:bkbdtrssettingintro}) and 
(\ref{eq:bkbdtrssettingsilentpartintro}). Using the notation $\sharp$ introduced at the beginning of Subsection~\ref{ssection:divthevartrssetintro}, 
and the notation 
\[
T_{\ron,t}\bM:=[\pi_{\ron}^{*}(T^{*}\so)]^{\sharp},\ \ \
T_{\rosub,t}\bM:=[\pi_{\rosub}^{*}(T^{*}\bM_{\rosub})]^{\sharp},
\]
there is a similar reformulation of the conditions (\ref{eq:condonkdomnoisdirintro}) and (\ref{eq:bkwwlbsubtmintro}).
\end{remark}

We use similar terminology, and identical notation, when the dominant noisy spatial direction corresponds to one of the $M_{r}$. 

\begin{definition}\label{def:domnoisyspaMrdir}
Let $(M,g)$ be a canonical separable cosmological model manifold.
Fix $r\in \{1,\dots,R\}$ and let $\pi_{\ron}$ be the projection from $\bM$ to the $M_{r}$-factor in $\bM$ and let $\pi_{\rosub}$ be the map from 
$\bM$ to 
\[
\bM_{\rosub}:=\tn{d}\times M_{1}\times\cdots \times\hat{M}_{r}\times\cdots\times M_{R}
\]
\index{$\a$Aa@Notation!Manifolds!$\bM_{\rosub}$}%
corresponding to the projection onto what remains after removing $M_{r}$; here a hat signifies omission. Define
\[
T_{\ron}^{*}\bM:=\pi_{\ron}^{*}(T^{*}M_{r}),\ \ \
T_{\rosub}^{*}\bM:=\pi_{\rosub}^{*}(T^{*}\bM_{\rosub}).
\]
\index{$\a$Aa@Notation!Tangent and cotangent spaces!$T_{\ron}^{*}\bM$}%
\index{$\a$Aa@Notation!Tangent and cotangent spaces!$T_{\rosub}^{*}\bM$}%
The metric $g$ is said to have a \textit{geometric dominant noisy spatial direction corresponding to} $M_{r}$ 
\index{Geometric!dominant noisy spatial direction}%
\index{Dominant noisy spatial direction!geometric}%
if there are $0<\b_{\ron},\eta_{\ron},C_{\ron}\in\ro$ and a continuous non-negative function $\betafun\in L^{1}([0,\infty))$ such that 
(\ref{eq:condonkdomnoisdirintro}) holds for all $\xi\in T_{\ron}^{*}\bM$ and all $t\geq 0$, and such that (\ref{eq:bkwwlbsubtmintro}) holds 
for all $\xi\in T_{\rosub}^{*}\bM$ and all $t\geq 0$. Consider (\ref{eq:thesystemRge}), where $(M,g)$ is 
such that the above conditions hold. If there are constants $\eta_{\romn}>0$, $C_{\romn}>0$ and a matrix $\a_{\infty}\in\Mn{m}{\co}$ such that 
\begin{equation}\label{eq:alconvnosettinggeneralisedintro}
\|\a(t)-\a_{\infty}\|\leq C_{\romn}e^{-\eta_{\romn}t}
\end{equation}
for all $t\geq 0$; then (\ref{eq:thesystemRge}) is said to be such that \textit{the dominant coefficients are convergent}. 
\index{Convergent!dominant coefficients}%
\index{Dominant coefficients!convergent}%
\end{definition}

In what follows, it is convenient to make additional assumptions. In order for the statements of the results not to be too
cumbersome, we therefore introduce the following terminology.

\begin{definition}\label{def:noisemainassumptions}
Consider (\ref{eq:thesystemRge}). Assume the associated metric to be such that $(M,g)$ is a canonical separable cosmological model manifold. 
Assume, moreover, the metric $g$ to have a geometric dominant noisy spatial direction
and the dominant coefficients of (\ref{eq:thesystemRge}) to be convergent; i.e., either all the requirements of Definition~\ref{def:domnoisyspasodir}
or all the requirements of Definition~\ref{def:domnoisyspaMrdir} are satisfied. Assume, in addition, that (\ref{eq:thesystemRge}) is $C^{2}$-balanced
and that there are constants $C_{\rosh},\eta_{\rosh}>0$ such that 
\begin{equation}\label{eq:neglshiftvfnoissettintro}
|\chi(t)|_{\bge}+|\dot{\chi}(t)|_{\bge}\leq C_{\rosh}e^{-\eta_{\rosh}t}
\end{equation}
for all $t\geq 0$. Then (\ref{eq:thesystemRge}) is said to be $C^{2}$-\textit{balanced with a geometric dominant noisy spatial direction, convergent
dominant coefficients and a negligible shift vector field}.  
\index{Equation!$C^{2}$-balanced with a geometric dominant noisy spatial direction, convergent
dominant coefficients and a negligible shift vector field}%
\end{definition}

Before turning to the results, note that given assumptions of the form stated in Definitions~\ref{def:domnoisyspasodir} and
\ref{def:domnoisyspaMrdir}, we cannot say much about modes such that there is no spatial variation in the dominant noisy spatial
direction. For that reason, we, from now on, focus on $\indexnot\in\EFnindexset$, where $\EFnindexset$ is defined as follows.

\begin{definition}\label{def:nuronEFnindexsetintro}
Assume that (\ref{eq:thesystemRge}) is $C^{2}$-balanced with a geometric dominant noisy spatial direction, convergent
dominant coefficients and a negligible shift vector field; cf. Definition~\ref{def:noisemainassumptions}. Define
\begin{equation}\label{eq:nurondefintro}
\nu_{\ron}(\indexnot):=\lim_{t\rightarrow\infty}e^{-\b_{\ron}t}\mfg(\indexnot,t),
\end{equation}
\index{$\a$Aa@Notation!Eigenvalues!$\nu_{\ron}(\indexnot)$}%
where $\b_{\ron}$ is given in Definitions~\ref{def:domnoisyspasodir} and \ref{def:domnoisyspaMrdir}. Then the set $\EFnindexset$ is defined 
to consist of the $\indexnot\in\EFindexset$ such that $\nu_{\ron}(\indexnot)\neq 0$. 
\end{definition}
\begin{remark}
The function $\mfg(\indexnot,t)$ is defined by (\ref{eq:mfgnutdef}). Due to Remarks~\ref{remark:geometrictonongeometricnoise} and
\ref{remark:nuronindlimex}, the limit
(\ref{eq:nurondefintro}) exists. Moreover, if $g$ has a geometric dominant noisy spatial $\so$-direction corresponding to $j$, then 
$\indexnot\in\EFindexset$ if and only if $\nu_{\roT,j}(\indexnot)\neq 0$; cf. (\ref{eq:nuroTetcdef}). Similarly, if $g$ has a geometric 
dominant noisy spatial corresponding to $M_{r}$, then $\indexnot\in\EFindexset$ if and only if $\nu_{r,i_{r}}(\indexnot)\neq 0$.
\end{remark}

\subsection{The limit equation}\label{ssection:thelimitequationinthenoisysetting}

Assume that (\ref{eq:thesystemRge}) is $C^{2}$-balanced with a geometric dominant noisy spatial direction, convergent
dominant coefficients and a negligible shift vector field; cf. Definition~\ref{def:noisemainassumptions}.
Consider (\ref{eq:fourierthesystemRge}), the equation for the Fourier coefficients of a solution to (\ref{eq:thesystemRge}), 
and assume that $\indexnot\in\EFnindexset$. In analogy with earlier chapters, it is of interest to isolate a limit equation.
Considering (\ref{eq:nurondefintro}), it seems natural to, as a first approximation, replace $\mfg(\indexnot,t)$ with 
$\nu_{\ron}(\indexnot)e^{\b_{\ron}t}$. Moreover, due to (\ref{eq:Xjalconvnosettingintro}) and (\ref{eq:alconvnosettinggeneralisedintro}), 
it seems reasonable to replace $\a$ with $\a_{\infty}$. Turning to the term involving the $X^{l}$, it can be verified that it, to leading
order, is given by $i\bX_{\ron}(\indexnot)\nu_{\ron}(\indexnot)e^{\b_{\ron}t}z$ (cf. Lemma~\ref{lemma:approxconstofRabprek} for a detailed 
justification). Here
\begin{equation}\label{eq:bXrondefintro}
\bX_{\ron}(\indexnot):=\sgn_{\ron}(\indexnot)\tX_{\ron,\infty}.
\end{equation}
\index{$\a$Aa@Notation!Coefficients, equation!$\bX_{\ron}(\indexnot)$}%
Concerning the definition of the expressions appearing on the right hand side, there are two cases to consider. If (\ref{eq:thesystemRge}) 
has a geometric dominant noisy spatial $\so$-direction corresponding to $j$, and $\indexnot\in\EFnindexset$, then
\begin{equation}\label{eq:tXinfrondefintro}
\tX_{\infty,\ron}:=(g^{jj}_{\infty})^{-1/2}X^{j}_{\infty},\ \ \
\sgn_{\ron}(\indexnot):=\frac{n_{j}}{|n_{j}|}
\end{equation}
\index{$\a$Aa@Notation!Coefficients, equation!$\tX_{\infty,\ron}$}%
\index{$\a$Aa@Notation!Coefficients, equation!$\sgn_{\ron}(\indexnot)$}%
(no summation on $j$); here $g^{jj}_{\infty}:=\lim_{t\rightarrow\infty}e^{-2\b_{\ron}t}g^{jj}(t)$ (that this limit exists and is strictly positive is a 
consequence of Remark~\ref{remark:geometrictonongeometricnoise}). If (\ref{eq:thesystemRge}) has a geometric dominant noisy spatial
direction corresponding to $M_{r}$ and $\indexnot\in\EFnindexset$, then
\begin{equation}\label{eq:tXinfdefgendirintro}
\tX_{\infty,\ron}:=0,\ \ \
\sgn_{\ron}(\indexnot):=1.
\end{equation}
Leaving the significance of $\zeta$ aside for a moment, the preliminary, homogeneous version of the limit equation is given by 
\[
\ddot{z}+\nu_{\ron}^{2}(\indexnot)e^{2\b_{\ron}t}z+i\nu_{\ron}(\indexnot)e^{\b_{\ron}t}\bX_{\ron}(\indexnot)z+\a_{\infty}\dot{z}+\zeta(t)z=0,
\]
where the absence of the term involving the shift vector field is justified by (\ref{eq:neglshiftvfnoissettintro}). 
This equation can be written as a first order system for a vector valued function whose first $m$ components are given by 
$\nu_{\ron}(\indexnot)e^{\b_{\ron}t}z$ and whose last $m$ components are given by $\dot{z}$. If one does so, it becomes clear that $\zeta$ only 
appears divided by $\nu_{\ron}(\indexnot)e^{\b_{\ron}t}$. As a consequence, it seems reasonable to expect $\zeta$ to have a negligible influence
on the leading order asymptotics. In the end, this expectation turns out to be justified. Summing up yields the homogeneous limit equation
\begin{equation}\label{eq:limiteqinthenoisyset}
\ddot{z}+\nu_{\ron}^{2}(\indexnot)e^{2\b_{\ron}t}z+i\nu_{\ron}(\indexnot)e^{\b_{\ron}t}\bX_{\ron}(\indexnot)z+\a_{\infty}\dot{z}=0.
\end{equation}
\index{Limit equation!noisy setting}%
\index{Noisy setting!limit equations}%
Since $\b_{\ron}>0$, it is clear that the leading order behaviour is oscillatory, with a frequency that grows exponentially. However, 
there is also an overall exponential growth/decay. Considering the limit equation, it is natural to expect the growth/decay to be 
determined by $\b_{\ron}$, $\tX_{\infty,\ron}$ and $\a_{\infty}$. In fact, we expect the matrices appearing on the diagonal on the right hand
side of (\ref{eq:SBinfSinv}) to determine the growth/decay of solutions. It is therefore natural to introduce the notation
\begin{equation}\label{eq:Rronpmpmdefintro}
R_{\ron,+}^{\pm}:=\frac{1}{2}(-\a_{\infty}+\b_{\ron}\Id_{m}\pm\tX_{\infty,\ron}).
\end{equation}
\index{$\a$Aa@Notation!Matrix notation!Rnpm@$R_{\ron,+}^{\pm}$}%
Let $R_{\ron,+}:=\diag(R_{\ron,+}^{+},R_{\ron,+}^{-})$, 
\index{$\a$Aa@Notation!Matrix notation!Rnp@$R_{\ron,+}$}%
$\kappa_{\ron,+}:=\kappa_{\max}(R_{\ron,+})$ and 
\index{$\a$Aa@Notation!Matrix notation!kappan@$\kappa_{\ron,+}$}%
$d_{\ron,+}:=d_{\max}(R_{\ron,+},\kappa_{\ron,+})$; 
\index{$\a$Aa@Notation!Matrix notation!dnp@$d_{\ron,+}$}%
cf. Definition~\ref{def:SpRspdef}.

\section{Results}\label{section:results}

Turning to the results, it is natural to begin by stating a rough Sobolev estimate. However, for reasons mentioned above, we can only expect to
be able to estimate the projection of solutions to the subspace corresponding to $\EFnindexset$. This is the subject of 
Subsection~\ref{ssection:roughsobestintronoise}. After that, we 
turn to the problem of deriving asymptotics. As mentioned above, the asymptotic behaviour can roughly speaking be divided into two pieces: 
an oscillatory part and an overall growth/decay. The oscillatory part can be expressed in terms of solutions to (\ref{eq:homtheeqnoiseintro}); 
i.e., the homogeneous wave equation obtained by dropping the right hand side of (\ref{eq:thesystemRge}) as well as all the lower order 
terms on the left hand side. In Subsection~\ref{ssection:modelosc}, we state the asymptotics for solutions to this equation. Moreover, 
we note that there is a homeomorphism between initial data and asymptotic data. An important observation that arises in the 
study of (\ref{eq:homtheeqnoiseintro}) is that solutions (with frequency content contained in $\EFnindexset$) can be divided into
two pieces, one of which is said to be positively oriented, and the other of which is said to be negatively oriented. 

Turning to the asymptotics of solutions to (\ref{eq:thesystemRge}), we restrict our attention to the situation that the right hand
side exhibits slower growth than solutions to the homogeneous equation. Then the solutions can be approximated by a sum of functions of 
the form $e^{At}u_{\row}$, where $A$ is a constant matrix and $u_{\row}$ is a solution to (\ref{eq:homtheeqnoiseintro}). However, there
are two types of orientation that influence what the matrix $A$ is: the orientation of solutions to (\ref{eq:homtheeqnoiseintro})
mentioned above; and the sign of $n_{j}$ (assuming the metric $g$ associated with (\ref{eq:thesystemRge}) has a geometric dominant noisy 
spatial $\so$-direction corresponding to $j$). We describe the relevant results in Subsection~\ref{ssection:derivasnoiseintr}. 
Finally, in Subsection~\ref{ssection:spasnoiseintr}, we turn to the problem of specifying the asymptotics.

\subsection{A rough Sobolev estimate}\label{ssection:roughsobestintronoise}

Next, we formulate a rough estimate of Sobolev type energies. In the statement, we use the following notation. If $f\in C^{\infty}(M,\cn{m})$, 
\begin{equation}\label{eq:frondefintro}
f_{\ron}(p,t):=\textstyle{\sum}_{\indexnot\in\EFnindexset}\hf(\indexnot,t)\varphi_{\indexnot}(p),
\end{equation}
\index{$\a$Aa@Notation!Projected functions!$f_{\ron}$}%
where $\varphi_{\indexnot}$ and $\hf$ are defined by (\ref{eq:varphinudef}) and (\ref{eq:hfnutdef}) respectively. We define
$u_{\ron}$ similarly, and $\mfe_{s}$ by (\ref{eq:mfedef}). 

\begin{prop}\label{prop:roughsobestnoisysetintro}
Assume that (\ref{eq:thesystemRge}) is $C^{2}$-balanced with a geometric dominant noisy spatial direction, convergent dominant 
coefficients and a negligible shift vector field; cf. Definition~\ref{def:noisemainassumptions}. Then there are constants 
$C>0$ and $s_{\ron}\geq 0$ such that if $u$ is a solution to (\ref{eq:thesystemRge}), then
\begin{equation}\label{eq:windtestfinnoisyerapartconvsobintro}
\begin{split}
\mfe_{s}^{1/2}[u_{\ron}](t) \leq &  C\ldr{t}^{d_{\ron,+}-1}e^{\kappa_{\ron,+}t}\mfe_{s+s_{\ron}}^{1/2}[u_{\ron}](0)\\
 & + \int_{0}^{t}C\ldr{t-t'}^{d_{\ron,+}-1}e^{\kappa_{\ron,+}(t-t')}\|f_{\ron}(\cdot,t')\|_{(s+s_{\ron})}dt'
\end{split}
\end{equation}
for all $t\geq 0$ and $s\in\ro$, where $f_{\ron}$ and $u_{\ron}$ are defined by (\ref{eq:frondefintro}) and the adjacent text. Finally, 
$\kappa_{\ron,+}$ and $d_{\ron,+}$ are defined in the text adjacent to (\ref{eq:Rronpmpmdefintro}). 
\end{prop}
\begin{remarks}
The proposition is a consequence of Remark~\ref{remark:roughsobestnoise}. Moreover, $C$ and $s_{\ron}$ only depend on the the Riemannian 
manifolds $(M_{r},g_{r})$, $r=1,\dots,R$, and the coefficients of the equation (\ref{eq:thesystemRge}).
\end{remarks}

\subsection{A model for the oscillations}\label{ssection:modelosc}

As a preparation for the description of the asymptotics of solutions to (\ref{eq:homtheeqnoiseintro}), we need to introduce some terminology.
Assume, first of all, that (\ref{eq:homtheeqnoiseintro}) is $C^{2}$-balanced with a geometric dominant noisy spatial direction, convergent 
dominant coefficients and a negligible shift vector field; cf. Definition~\ref{def:noisemainassumptions}. We then define $\eta_{A}$ by 
\begin{equation}\label{eq:etaAdefinitionintro}
\eta_{A}:=\min\{\eta_{\ron}/2,\eta_{\rosh},\eta_{\romn},\b_{\ron}\},
\end{equation}
where $\b_{\ron}$, $\eta_{\ron}$ and $\eta_{\romn}$ are the constants appearing in Definitions~\ref{def:domnoisyspasodir} and 
\ref{def:domnoisyspaMrdir}; and $\eta_{\rosh}$ is the constant appearing in Definition~\ref{def:noisemainassumptions}. Note that, in the 
present setting, $\eta_{\romn}$ could be removed from the definition of $\eta_{A}$, since $\a=0$ and $X^{l}=0$ for 
(\ref{eq:homtheeqnoiseintro}). However, we also wish to use the definition (\ref{eq:etaAdefinitionintro}) in the study of solutions
to (\ref{eq:thesystemRge}), and in that setting, we need to include $\eta_{\romn}$. In order to describe the oscillations of the solutions, 
the following two functions are of central importance:
\begin{align}
\omega_{\rosh}(\indexnot,t) := & \int_{0}^{t}\sigma(\indexnot,t')\mfg(\indexnot,t')dt',\label{eq:omegashdefnoiseintro}\\
\varphi_{\rotot}(\indexnot,t) := & \int_{0}^{t}[1+\sigma^{2}(\indexnot,t')]^{1/2}\mfg(\indexnot,t')dt',\label{eq:varphirototdefnoiseintro}
\end{align}
\index{$\a$Aa@Notation!Auxiliary functions!$\omega_{\rosh}$}%
\index{$\a$Aa@Notation!Auxiliary functions!$\varphi_{\rotot}$}%
where $\mfg$ is defined by (\ref{eq:mfgnutdef}) and $\sigma$ is defined by (\ref{eq:sigmaXdefintro}). The main result concerning the 
asymptotics of solutions to (\ref{eq:homtheeqnoiseintro}) is the following. 

\begin{prop}\label{prop:asymposccasescalareqnoiseintro}
Consider the equation (\ref{eq:homtheeqnoiseintro}). Assume that it is $C^{2}$-balanced with a geometric dominant noisy spatial direction, 
convergent dominant coefficients and a negligible shift vector field; cf. Definition~\ref{def:noisemainassumptions}. Let $\eta_{A}$ be defined 
by (\ref{eq:etaAdefinitionintro}) and $\chi_{\pm}(\indexnot)\in\co$, $\indexnot\in\EFnindexset$, be such that for every $0\leq k\in\zo$, there 
is a constant $C_{k}<\infty$ with the property that 
\begin{equation}\label{eq:rortbdnoiseintro}
\textstyle{\sum}_{\indexnot\in\EFnindexset}\ldr{\nu(\indexnot)}^{2k}[|\chi_{+}(\indexnot)|^{2}+|\chi_{-}(\indexnot)|^{2}]\leq C_{k}.
\end{equation}
Define $z_{\app}$ and $u_{\app}$ by 
\begin{align}
z_{\app}(\indexnot,t) := & \frac{e^{\b_{\ron}t/2}}{\mfg(\indexnot,t)}\left(\chi_{+}(\indexnot)e^{i[\omega_{\rosh}(\indexnot,t)-\varphi_{\rotot}(\indexnot,t)]}
+\chi_{-}(\indexnot)e^{i[\omega_{\rosh}(\indexnot,t)+\varphi_{\rotot}(\indexnot,t)]}\right)\label{eq:zappdefnoiseintro}\\
u_{\app}(p,t) := & \textstyle{\sum}_{\indexnot\in\EFnindexset}z_{\app}(\indexnot,t)\varphi_{\indexnot}(p),\label{eq:uappdefnoiseintro}
\end{align}
where $\indexnot\in\EFnindexset$ in (\ref{eq:zappdefnoiseintro}); $\omega_{\rosh}$ is defined by (\ref{eq:omegashdefnoiseintro}); $\varphi_{\rotot}$ is 
defined by (\ref{eq:varphirototdefnoiseintro}); and $\mfg(\indexnot,t)$ is defined by (\ref{eq:mfgnutdef}). Then there is a unique smooth 
solution $u$ to (\ref{eq:homtheeqnoiseintro}) such that the following holds: $u=u_{\ron}$, where $u_{\ron}$ is defined in analogy with 
(\ref{eq:frondefintro}); and there are constants $s\in\ro$, $\eta>0$ and $C>0$ such that 
\begin{equation}\label{eq:umiuappuniqueestintro}
\mfe_{s}[u-u_{\app}](t)\leq Ce^{(\b_{\ron}-\eta)t}
\end{equation}
for all $t\geq 0$. Moreover, this unique solution has the property that 
\begin{equation}\label{eq:mfediffhomestnoiseintro}
\mfe_{s}[u-u_{\app}](t)\leq C_{A}e^{(\b_{\ron}-2\eta_{A})t}\textstyle{\sum}_{\indexnot\in\EFnindexset}\ldr{\nu(\indexnot)}^{2(s+s_{\rom})}
[|\chi_{+}(\indexnot)|^{2}+|\chi_{-}(\indexnot)|^{2}]
\end{equation}
for all $t\geq 0$ and $s\in\ro$, where $0\leq s_{\rom}\in\ro$ and $0<C_{A}\in\ro$ only depend on the coefficients of the equation 
(\ref{eq:homtheeqnoiseintro}) and the Riemannian manifolds $(M_{r},g_{r})$, $r=1,\dots,R$; and $\mfe_{s}$ is defined by (\ref{eq:mfedef}). 
Finally,
\begin{equation}\label{eq:mfezeroestnoiseintro}
\mfe^{1/2}_{s}[u](0)\leq 
C_{A}\left(\textstyle{\sum}_{\indexnot\in\EFnindexset}\ldr{\nu(\indexnot)}^{2(s+s_{\rom})}
[|\chi_{+}(\indexnot)|^{2}+|\chi_{-}(\indexnot)|^{2}]\right)^{1/2}
\end{equation}
for all $s\in\ro$, where $C_{A}$ and $s_{\rom}$ have the same dependence as in the case of (\ref{eq:mfediffhomestnoiseintro}).

Conversely, assume that $u$ is a smooth solution to (\ref{eq:homtheeqnoiseintro}) such that $u=u_{\ron}$. Then there are uniquely determined 
$\chi_{\pm}(\indexnot)$, $\indexnot\in\EFnindexset$, such that (\ref{eq:rortbdnoiseintro}) holds, and such that if $u_{\app}$ is defined by 
(\ref{eq:zappdefnoiseintro}) and (\ref{eq:uappdefnoiseintro}), then (\ref{eq:umiuappuniqueestintro}) holds. Moreover, this $u_{\app}$ is 
such that (\ref{eq:mfediffhomestnoiseintro}) holds. Finally,
\begin{equation}\label{eq:chipmmfezeroestnoiseintro}
\left(\textstyle{\sum}_{\indexnot\in\EFnindexset}\ldr{\nu(\indexnot)}^{2s}[|\chi_{+}(\indexnot)|^{2}+|\chi_{-}(\indexnot)|^{2}]\right)^{1/2}\leq 
C_{A}\mfe^{1/2}_{s+s_{\rom}}[u](0)
\end{equation}
for all $s\in\ro$, where $C_{A}$ and $s_{\rom}$ have the same dependence as in the case of (\ref{eq:mfediffhomestnoiseintro}).
\end{prop}
\begin{remark}
The proposition is a consequence of Remark~\ref{remark:homwaveonlyprincsymbnoise}.
\end{remark}
\begin{remark}
Specifying $\chi_{\pm}(\indexnot)\in\co$, $\indexnot\in\EFnindexset$, such that (\ref{eq:rortbdnoiseintro}) holds for all $k\geq 0$ corresponds
to specifying two smooth $\co$-valued functions on $\bM$ with frequency content contained in $\EFnindexset$. These two functions can be 
thought of as the asymptotic data. Moreover, from this point of view, the estimates (\ref{eq:mfezeroestnoiseintro}) and 
(\ref{eq:chipmmfezeroestnoiseintro}) imply that the map from initial data to asymptotic data is a homeomorphism with respect to the 
$C^{\infty}$-topology. 
\end{remark}
\begin{remark}
Since $\mfg(\indexnot,t)\approx \nu_{\ron}(\indexnot)e^{\b_{\ron}t}$, (\ref{eq:zappdefnoiseintro}) implies that $z_{\app}(\indexnot,t)$
decays as $e^{-\b_{\ron}t/2}$. However, the energy of $u_{\app}$ grows as $e^{\b_{\ron}t}$. The reason for this is the following. First of all, 
\begin{align*}
\dot{z}_{\app}(\indexnot,t) \approx & e^{\b_{\ron}t/2}\left(-i\chi_{+}(\indexnot)e^{i[\omega_{\rosh}(\indexnot,t)-\varphi_{\rotot}(\indexnot,t)]}
+i\chi_{-}(\indexnot)e^{i[\omega_{\rosh}(\indexnot,t)+\varphi_{\rotot}(\indexnot,t)]}\right)\\
\mfg(\indexnot,t)z_{\app}(\indexnot,t) = & e^{\b_{\ron}t/2}\left(\chi_{+}(\indexnot)e^{i[\omega_{\rosh}(\indexnot,t)-\varphi_{\rotot}(\indexnot,t)]}
+\chi_{-}(\indexnot)e^{i[\omega_{\rosh}(\indexnot,t)+\varphi_{\rotot}(\indexnot,t)]}\right).
\end{align*}
Thus
\[
|\dot{z}_{\app}(\indexnot,t)|^{2}+\mfg^{2}(\indexnot,t)|z_{\app}(\indexnot,t)|^{2}\approx 2e^{\b_{\ron}t}[|\chi_{+}(\indexnot)|^{2}+|\chi_{-}(\indexnot)|^{2}].
\]
In particular, the contribution of the $\indexnot$'th mode to $\mfe_{s}[u_{\app}]$ grows as $e^{\b_{\ron}t}$ unless the $\chi_{\pm}(\indexnot)$ vanish. 
\end{remark}
\begin{remark}
The estimate (\ref{eq:mfediffhomestnoiseintro}) immediately yields an estimate for the $L^{2}$-norm of the difference between $u$ and $u_{\app}$;
cf. (\ref{eq:mfedef}). However, this estimate is not very good. In order to improve it, assume that there is a non-negative continuous function 
$\betafun_{\low}\in L^{1}([0,\infty))$ such that 
\begin{equation}\label{eq:bklowbddomnoisspdir}
\bk\geq -[\b_{\ron}+\betafun_{\low}(t)]\bge
\end{equation}
for all $t\geq 0$ (in addition to the assumptions of the proposition). Then there is a constant $0<C\in\ro$ (depending only on the spectrum 
of the Riemannian manifold corresponding to the dominant noisy spatial direction and the coefficients of the equation 
(\ref{eq:homtheeqnoiseintro})) such that $e^{\b_{\ron}t}[\mfg(\indexnot,t)]^{-1}\leq C$ for all $t\geq 0$ and all $\indexnot\in \EFnindexset$; cf. 
Remarks~\ref{remark:domnoisspdirlowbdonbk} and \ref{remark:domnoisspdirlowbdonbkgendir}. As a consequence, 
\begin{align*}
 & e^{2\b_{\ron}t}\textstyle{\sum}_{\indexnot\in\EFindexset}\ldr{\nu(\indexnot)}^{2s}|z(\indexnot,t)-z_{\app}(\indexnot,t)|^{2}\\
 = & \textstyle{\sum}_{\indexnot\in\EFindexset}e^{2\b_{\ron}t}[\mfg(\indexnot,t)]^{-2}
\ldr{\nu(\indexnot)}^{2s}\mfg^{2}(\indexnot,t)|z(\indexnot,t)-z_{\app}(\indexnot,t)|^{2}\leq C\mfe_{s}[u-u_{\app}].
\end{align*}
Combining this estimate with (\ref{eq:mfediffhomestnoiseintro}) yields
\begin{equation}\label{eq:umuappsobest}
\begin{split}
 & \textstyle{\sum}_{\indexnot\in\EFindexset}\ldr{\nu(\indexnot)}^{2s}|z(\indexnot,t)-z_{\app}(\indexnot,t)|^{2}\\
 \leq & C_{A}e^{-(\b_{\ron}+2\eta_{A})t}\textstyle{\sum}_{\indexnot\in\EFnindexset}\ldr{\nu(\indexnot)}^{2(s+s_{\rom})}
[|\chi_{+}(\indexnot)|^{2}+|\chi_{-}(\indexnot)|^{2}]
\end{split}
\end{equation}
for all $t\geq 0$ and $s\in\ro$, where $C_{A}$ and $s_{\rom}$ have the same dependence as in the case of (\ref{eq:mfediffhomestnoiseintro}).
\end{remark}

Due to this proposition, we can introduce a notion of orientation of a solution to (\ref{eq:homtheeqnoiseintro}).

\begin{definition}\label{def:orsoltowenoiseintro}
Consider the equation (\ref{eq:homtheeqnoiseintro}). Assume that it is $C^{2}$-balanced with a geometric dominant noisy spatial direction, 
convergent dominant coefficients and a negligible shift vector field; cf. Definition~\ref{def:noisemainassumptions}.
Assume that $u$ is a smooth solution to (\ref{eq:homtheeqnoiseintro}) such that 
$u=u_{\ron}$, where $u_{\ron}$ is defined in analogy with (\ref{eq:frondefintro}). Let $u_{\app}$ be the uniquely associated function of the form 
(\ref{eq:zappdefnoiseintro}) and (\ref{eq:uappdefnoiseintro}), whose existence is guaranteed by
Proposition~\ref{prop:asymposccasescalareqnoiseintro}. Consider (\ref{eq:zappdefnoiseintro}) and (\ref{eq:uappdefnoiseintro}). If 
$\chi_{-}(\indexnot)=0$ for all 
$\indexnot\in\EFnindexset$, then $u$ is said to be \textit{positively oriented}. 
\index{Solution!positively oriented}%
\index{Positively oriented!solution}%
If $\chi_{+}(\indexnot)=0$ for all $\indexnot\in\EFnindexset$, then $u$ 
is said to be \textit{negatively oriented}.
 \index{Solution!negatively oriented}%
\index{Negatively oriented!solution}%
\end{definition}

In order to illustrate the use of Proposition~\ref{prop:asymposccasescalareqnoiseintro}, let us consider a special case. 

\begin{example}\label{example:polt3gowdyexpdiras}
Consider (\ref{eq:polarisedGowdy}). In the case of this equation,
\[
g=-dt\otimes dt+e^{-2t}d\theta\otimes d\theta,\ \ \
\bge=e^{-2t}d\theta\otimes d\theta,\ \ \
\bk=-\bge,\ \ \
\ml_{U}\bk=2\bge.
\]
Moreover, the shift vector field vanishes (as does $\a$, $\zeta$ and $\mcX$). Thus (\ref{eq:polarisedGowdy}) is $C^{2}$-balanced. 
Since $\bk+\bge=0$, it is clear that (\ref{eq:polarisedGowdy}) has a geometric dominant noisy spatial $\so$-direction corresponding to 
$1$. Moreover, $\b_{\ron}=1$ and we can choose $C_{\ron}=1$ and $\eta_{\ron}=2$. That the dominant coefficients are convergent is obvious
(since $\mcX$ and $\a$ vanish), and we can choose $\eta_{\romn}=2$. Finally, (\ref{eq:polarisedGowdy}) has a negligible shift vector field, 
and we can choose $\eta_{\rosh}=2$. Considering (\ref{eq:etaAdefinitionintro}), it is clear that $\eta_{A}=1$. Turning to $\omega_{\rosh}$ 
and $\varphi_{\rotot}$, note that $\omega_{\rosh}=0$, $\mfg(n,t)=e^{t}|n|$ and $\varphi_{\rotot}(n,t)=(e^{t}-1)|n|$. Thus $z_{\app}$ appearing 
in (\ref{eq:zappdefnoiseintro}) can be written
\[
z_{\app}(n,t)=|n|^{-1}e^{-t/2}\left(\chi_{+}(n)\exp[-i(e^{t}-1)|n|]
+\chi_{-}(n)\exp[i(e^{t}-1)|n|]\right).
\]
This means that $u_{\app}$ can be written in the form
\[
u_{\app}(\theta,t)=e^{-t/2}u_{\row}(\theta,e^{t}),
\]
where $u_{\row}$ is a solution to the flat space wave equation $u_{\tau\tau}-u_{\theta\theta}=0$ with mean value zero for all $\tau$ (in the 
present setting, the requirement that $u_{\row}$ have mean value zero corresponds to the frequency content being contained in 
$\EFnindexset$). Moreover, there is a homeomorphism (in the $C^{\infty}$-topology) between the initial data for $P$ (assuming that the mean 
value of $P$ over $\so$ vanishes) and the initial data for $u_{\row}$. Turning to the estimates of the difference between $u_{\app}$ and 
$P$, note that (\ref{eq:umuappsobest}) implies that every Sobolev norm of $P(\cdot,t)-u_{\app}(\cdot,t)$ decays as $e^{-3t/2}$. Consider
\begin{equation}\label{eq:Ptmdtuapp}
P_{t}(\theta,t)-\d_{t}u_{\app}(\theta,t)=P_{t}(\theta,t)-e^{t/2}(\d_{\tau}u_{\row})(\theta,e^{t})+\frac{1}{2}e^{-t/2}u_{\row}(\theta,e^{t}). 
\end{equation}
Due to (\ref{eq:mfediffhomestnoiseintro}), the Sobolev norm of this difference decays as $e^{-t/2}$. However, the Sobolev norm of the 
last term on the right hand side of (\ref{eq:Ptmdtuapp}) also decays as $e^{-t/2}$, so that the sum of the first two terms on the right
hand side of (\ref{eq:Ptmdtuapp}) decays as $e^{-t/2}$. Changing time coordinate to $\tau=e^{t}$, we obtain, by abuse of notation,
\[
\|P(\cdot,\tau)-\tau^{-1/2}u_{\row}(\cdot,\tau)\|_{C^{k}}
+\|P_{\tau}(\cdot,\tau)-\tau^{-1/2}\d_{\tau}u_{\row}(\cdot,\tau)\|_{C^{k}}\leq C_{k}\tau^{-3/2}
\]
for all $0\leq k\in\zo$ and all $\tau\geq 1$. Moreover, there is a homeomorphism between the initial data for $P$ and the initial data
for $u_{\row}$. Adding to these observations the spatially homogeneous solutions to (\ref{eq:polarisedGowdyoriginaltime}) yields the 
conclusions stated in connection with (\ref{eq:polarisedGowdyoriginaltime}). Finally, let us note that the fact that there is a bijection 
follows from \cite{jurke,polgowdaschar}.
\end{example}

\subsection{Deriving asymptotics}\label{ssection:derivasnoiseintr}

Next, we consider the asymptotics of solutions to (\ref{eq:thesystemRge}). As already mentioned, a given solution can be approximated by 
a sum of terms consisting of an oscillatory part and an overall growth/decay. However, what the overall growth/decay is depends on the 
orientation of the oscillatory part, as well as the sign of $n_{j}$ (assuming (\ref{eq:thesystemRge}) has a geometric dominant noisy spatial 
$\so$-direction corresponding to $j$); here the oscillatory part is modelled by a solution to (\ref{eq:homtheeqnoiseintro}) and the notion
of an orientation for a solution to (\ref{eq:homtheeqnoiseintro}) is given by Definition~\ref{def:orsoltowenoiseintro}. 
It is therefore convenient to introduce the following notation.

\begin{definition}\label{definition:EFnindexsetpmintro}
Assume that (\ref{eq:thesystemRge}) is $C^{2}$-balanced with a geometric dominant noisy spatial direction, convergent dominant coefficients and 
a negligible shift vector field; cf. Definition~\ref{def:noisemainassumptions}. In case (\ref{eq:thesystemRge}) has a 
geometric dominant noisy spatial $\so$-direction corresponding to $j$, let
\begin{equation}\label{eq:EFnindpmtorusintro}
\EFnindexsetp:=\{\indexnot\in\EFnindexset : \nu_{\roT,j}(\indexnot)<0\},\ \ \
\EFnindexsetm:=\{\indexnot\in\EFnindexset : \nu_{\roT,j}(\indexnot)>0\};
\end{equation}
\index{$\a$Aa@Notation!Frequency sets!$\EFnindexsetp$}%
\index{$\a$Aa@Notation!Frequency sets!$\EFnindexsetm$}%
cf. (\ref{eq:nuroTetcdef}). In case (\ref{eq:thesystemRge}) has a geometric dominant noisy spatial direction corresponding to, say, $M_{r}$, let 
\begin{equation}\label{eq:EFnindpmgeneralisedintro}
\EFnindexsetp:=\EFnindexset,\ \ \
\EFnindexsetm:=\varnothing.
\end{equation}
\index{$\a$Aa@Notation!Frequency sets!$\EFnindexsetp$}%
\index{$\a$Aa@Notation!Frequency sets!$\EFnindexsetm$}%
\end{definition}

\begin{prop}\label{prop:asymposccaseitosoltowenoiseintro}
Assume that (\ref{eq:thesystemRge}) is $C^{2}$-balanced with a geometric dominant noisy spatial direction, convergent dominant 
coefficients and a negligible shift vector field; cf. Definition~\ref{def:noisemainassumptions}. Let $\eta_{A}$ be defined by 
(\ref{eq:etaAdefinitionintro}) and assume that there is an $0<\eta_{B}\leq \eta_{A}$ such that 
\begin{equation}\label{eq:fronHsbdnoiseintro}
\int_{0}^{\infty}e^{-(\kappa_{\ron,+}-\eta_{B})t}\|f_{\ron}(\cdot,t)\|_{(s)}dt<\infty
\end{equation}
for all $s$, where $\kappa_{\ron,+}$ is defined by the text adjacent to (\ref{eq:Rronpmpmdefintro}); $f_{\ron}$ is defined by 
(\ref{eq:frondefintro}); and $\|\cdot\|_{(s)}$ is defined by (\ref{eq:HsnormonbM}). Let 
\[
A_{\ron,\pm}:=-\frac{1}{2}\left(\a_{\infty}\pm\tX_{\infty,\ron}\right),\ \ \
\kappa_{\ron,\rom}:=\max\{\kappa_{\max}(A_{\ron,+}),\kappa_{\max}(A_{\ron,-})\},
\]
where $\tX_{\infty,\ron}$ is defined by (\ref{eq:tXinfrondefintro}) and (\ref{eq:tXinfdefgendirintro}). Finally, let 
$I_{\ron}:=(\kappa_{\ron,\rom}-\eta_{B},\kappa_{\ron,\rom}]$ and $E_{\ron,\pm}:=E_{A_{\ron,\pm},I_{\ron}}$; cf. Definition~\ref{def:defofgeneigenspintro}. 

If the metric $g$ has a geometric dominant noisy spatial direction corresponding to, say, $M_{r}$,
cf. Definition~\ref{def:domnoisyspaMrdir}, then $E_{\ron,+}=E_{\ron,-}=:E_{\ron}$ and $A_{\ron,+}=A_{\ron,-}=:A_{\ron}$. Moreover, if $u$ is a solution to 
(\ref{eq:thesystemRge}), there is a unique solution $u_{0,\row}\in C^{\infty}(M,E_{\ron})$ to (\ref{eq:homtheeqnoiseintro}) such that 
\begin{itemize}
\item if $z_{0,\row}(\indexnot,t)$ denotes the $\indexnot$'th Fourier coefficient of $u_{0,\row}(\cdot,t)$, then $z_{0,\row}(\indexnot,\cdot)=0$ if 
$\indexnot\notin\EFnindexset$,
\item if $u_{\row}$ is defined by 
\begin{equation}\label{eq:urowdefnoisegenintro}
u_{\row}(p,t) = e^{A_{\ron}t}u_{0,\row}(p,t),
\end{equation}
then there are real constants $C$, $N$ and $s$ such that 
\begin{equation}\label{eq:uuweststatementuniquenoiseintro}
\begin{split}
\mfe_{s}^{1/2}[u_{\ron}-u_{\row}](t) \leq & C\ldr{t}^{N}e^{(\kappa_{\ron,+}-\eta_{B})t}
\end{split}
\end{equation}
for all $t\geq 0$, where $u_{\ron}$ is defined in analogy with (\ref{eq:frondefintro}).
\end{itemize}

If the metric $g$ has a geometric dominant noisy spatial $\so$-direction corresponding to $j$ and $u$ is a solution to 
(\ref{eq:thesystemRge}), then there are unique solutions $u_{\pm,\row}\in C^{\infty}(M,E_{\ron,\pm})$ to (\ref{eq:homtheeqnoiseintro}) 
such that the following holds:
\begin{itemize}
\item $u_{+,\row}=u_{+,\row}^{+}+u_{-,\row}^{-}$ and $u_{-,\row}=u_{+,\row}^{-}+u_{-,\row}^{+}$, where $u_{\tau,\row}^{\upsilon}$, $\tau,\upsilon\in\{+,-\}$, are 
smooth solutions to (\ref{eq:homtheeqnoiseintro}) whose $\indexnot$'th Fourier coefficients are denoted by $z_{\tau,\row}^{\upsilon}(\indexnot,t)$ and
are such that
\begin{itemize}
\item $z_{\pm,\row}^{+}(\indexnot,\cdot)=0$ in case $\indexnot\notin\EFnindexsetp$ and 
$z_{\pm,\row}^{-}(\indexnot,\cdot)=0$ in case $\indexnot\notin\EFnindexsetm$,
\item $u_{+,\row}^{\pm}$ are positively oriented and $u_{-,\row}^{\pm}$ are negatively oriented solutions 
to (\ref{eq:homtheeqnoiseintro}),
\end{itemize}
\item if $u_{\row}$ is defined by 
\begin{equation}\label{eq:urowdefnoisetorusintro}
u_{\row}(p,t) = e^{A_{\ron,+}t}u_{+,\row}(p,t)+e^{A_{\ron,-}t}u_{-,\row}(p,t),
\end{equation}
then there are real constants $C$, $N$ and $s$ such that 
\begin{equation}\label{eq:uuweststatementuniquenoisetorusintro}
\begin{split}
\mfe_{s}^{1/2}[u_{\ron}-u_{\row}](t) \leq & C\ldr{t}^{N}e^{(\kappa_{\ron,+}-\eta_{B})t}
\end{split}
\end{equation}
for all $t\geq 0$.
\end{itemize}
Moreover, in both cases there are constants $C_{B}>0$ and $s_{B}\geq 0$ (depending only on $\eta_{B}$, the Riemannian manifolds $(M_{r},g_{r})$, 
$r=1,\dots,R$, and the coefficients of the equation (\ref{eq:thesystemRge})); and a non-negative integer $N$ (depending only on $m$) such that 
the following holds. If $u$ and $u_{\row}$ are as above, then 
\begin{equation}\label{eq:uuweststatementnoiseintro}
\begin{split}
 & \mfe_{s}^{1/2}[u_{\ron}-u_{\row}](t)\\
 \leq & C_{B}\ldr{t}^{N}e^{(\kappa_{\ron,+}-\eta_{B})t}\left(\mfe_{s+s_{B}}^{1/2}[u_{\ron}](0)
+\int_{0}^{\infty}e^{-(\kappa_{\ron,+}-\eta_{B})t'}\|f_{\ron}(\cdot,t')\|_{(s+s_{B})}dt'\right)
\end{split}
\end{equation}
for all $t\geq 0$ and all $s\in\ro$. In addition, if the metric $g$ has a geometric dominant noisy spatial direction corresponding 
to, say, $M_{r}$, then 
\begin{equation}\label{eq:mfesuzrowestitoidanihintro}
\mfe_{s}^{1/2}[u_{0,\row}](0)\leq C_{B}\left(\mfe_{s+s_{B}}^{1/2}[u_{\ron}](0)
+\int_{0}^{\infty}e^{-(\kappa_{\ron,+}-\eta_{B})t'}\|f_{\ron}(\cdot,t')\|_{(s+s_{B})}dt'\right)
\end{equation}
for all $s\in\ro$, where $C_{B}$ and $s_{B}$ have the same dependence as in the case of (\ref{eq:uuweststatementnoiseintro}). Finally, if the metric $g$ 
has a geometric dominant noisy spatial $\so$-direction corresponding to $j$, then 
\begin{equation}\label{eq:mfesupmrowestitoidanihintro}
\mfe_{s}^{1/2}[u_{\pm,\row}](0)\leq C_{B}\left(\mfe_{s+s_{B}}^{1/2}[u_{\ron}](0)
+\int_{0}^{\infty}e^{-(\kappa_{\ron,+}-\eta_{B})t'}\|f_{\ron}(\cdot,t')\|_{(s+s_{B})}dt'\right)
\end{equation}
for all $s\in\ro$, where $C_{B}$ and $s_{B}$ have the same dependence as in the case of (\ref{eq:uuweststatementnoiseintro}). 
\end{prop}
\begin{remark}
The result is a consequence of Lemma~\ref{lemma:asymposccaseitosoltowenoise}; cf. Remark~\ref{remark:homwaveonlygeneqnoise}.
\end{remark}
\begin{remark}
The asymptotics can be characterised in terms of functions such as $u_{\app}$ introduced in (\ref{eq:zappdefnoiseintro}) and 
(\ref{eq:uappdefnoiseintro}) (as opposed to solutions to (\ref{eq:homtheeqnoiseintro})); cf. Remark~\ref{remark:specupmwanoise}.
\end{remark}
\begin{remark}
Note that $\kappa_{\ron,+}$ and $\kappa_{\ron,\rom}$ are related according to $\kappa_{\ron,+}=\kappa_{\ron,\rom}+\b_{\ron}/2$. 
\end{remark}
\begin{remarks}\label{remarks:Tronbronetcdeponeqintro}
In the statement of the proposition, two equations play an important role: (\ref{eq:thesystemRge}) and (\ref{eq:homtheeqnoiseintro}). Note that it is here 
taken for granted that (\ref{eq:homtheeqnoiseintro}) is the equation obtained from (\ref{eq:thesystemRge}) by setting $\a$, $\zeta$, $X^{j}$ and $f$ to 
zero. In the statement of the present proposition, we speak of $\eta_{A}$, defined by (\ref{eq:etaAdefinitionintro}). Note that this $\eta_{A}$ need
not necessarily coincide with the $\eta_{A}$ associated with the equation (\ref{eq:homtheeqnoiseintro}), say $\eta_{A,\rohom}$, since $\eta_{\romn}$ can 
be omitted from the right hand side of (\ref{eq:etaAdefinitionintro}) in the definition of $\eta_{A,\rohom}$. On the other hand, it is clear that 
$\eta_{A,\rohom}\geq \eta_{A}$, so that when we appeal to Proposition~\ref{prop:asymposccasescalareqnoiseintro}, the discrepancy does not cause complications.
Turning to the constant $\b_{\ron}$, note that it is the same for the two equations; cf. Definitions~\ref{def:domnoisyspasodir} and 
\ref{def:domnoisyspaMrdir}.
\end{remarks}
\begin{remark}
If (\ref{eq:thesystemRge}) has a dominant noisy spatial generalised direction, then $A_{\ron}=-\a_{\infty}/2$, so that (\ref{eq:urowdefnoisegenintro}) can be 
written
\[
u_{\row}(p,t) = e^{-\a_{\infty}t/2}u_{0,\row}(p,t).
\]
\end{remark}
\begin{remark}
The constants $C$, $s$ and $N$ appearing in (\ref{eq:uuweststatementuniquenoiseintro}) and (\ref{eq:uuweststatementuniquenoisetorusintro}) are allowed
to depend on the coefficients of the equation (\ref{eq:thesystemRge}), $\eta_{B}$, the functions $u_{\ron}$, $u_{\row}$, $f_{\ron}$ etc.
\end{remark}
\begin{remark}
If all the Jordan blocks of the matrices $A_{\ron,\pm}$ are trivial, then the $N$ appearing in (\ref{eq:uuweststatementnoiseintro}) can be replaced by $1$.
If $E_{\ron,+}=E_{\ron,-}=\cn{m}$, then $N$ can be replaced by $d_{n,+}-1$, where $d_{\ron,+}$ is defined in connection with (\ref{eq:Rronpmpmdefintro}).
\end{remark}

\subsection{Specifying the asymptotics}\label{ssection:spasnoiseintr}

The result of the previous subsection yields conclusions concerning the asymptotics, given a solution. In fact, 
(\ref{eq:mfesuzrowestitoidanihintro}) and (\ref{eq:mfesupmrowestitoidanihintro}) demonstrate that there is a continuous map from initial data to
asymptotic data. It is then of interest to ask if it is possible to construct a solution, given asymptotic data. If there is a corresponding map,
it is also natural to ask whether it is continuous. In the preparation for the statement of such a result, it is convenient to introduce the following 
terminology. 

\begin{definition}
Consider the equation (\ref{eq:homtheeqnoiseintro}). Assume that it is $C^{2}$-balanced with a geometric dominant noisy spatial direction, 
convergent dominant coefficients and a negligible shift vector field; cf. Definition~\ref{def:noisemainassumptions}. Let $1\leq m\in\zo$ and $V$ be 
a vector subspace of $\cn{m}$. Then
\[
W_{\ron}(\bM,V)\subset C^{\infty}(\bM,V)\times C^{\infty}(\bM,V)
\]
\index{$\a$Aa@Notation!Function spaces!$W_{\ron}(\bM,V)$}%
denotes the set of initial data at $t=0$ for solutions $u$ to (\ref{eq:homtheeqnoiseintro}) such that $u\in C^{\infty}(M,V)$ and the $\indexnot$'th 
Fourier coefficient of $u(\cdot,t)$ vanishes for all $t$ if $\indexnot\notin\EFnindexset$. Moreover, $W_{\ron,+}(\bM,V)$ ($W_{\ron,-}(\bM,V)$) 
\index{$\a$Aa@Notation!Function spaces!$W_{\ron,\pm}(\bM,V)$}%
denotes the subset of $W_{\ron}(\bM,V)$
corresponding to positively (negatively) oriented solutions to (\ref{eq:homtheeqnoiseintro}). In addition, $W_{\ron,\pm}^{+}(\bM,V)$ ($W_{\ron,\pm}^{-}(\bM,V)$) 
\index{$\a$Aa@Notation!Function spaces!$W_{\ron,\pm}^{\pm}(\bM,V)$}%
denotes the subset of $W_{\ron,\pm}(\bM,V)$ consisting of functions whose $\indexnot$'th Fourier coefficients vanish if $\indexnot\notin\EFnindexsetp$ 
($\indexnot\notin\EFnindexsetm$). 
\end{definition}
\begin{remark}
By initial data we here mean $[u(\cdot,0),u_{t}(\cdot,0)]$. 
\end{remark}

\begin{prop}\label{prop:asymposccaseitosoltowenoisespaintro}
Assume that (\ref{eq:thesystemRge}) is $C^{2}$-balanced with a geometric dominant noisy spatial direction, convergent dominant coefficients and a negligible shift 
vector field; cf. Definition~\ref{def:noisemainassumptions}. Assume, moreover, that $f=0$. Let $\eta_{A}$ be defined by (\ref{eq:etaAdefinitionintro}) and 
$0<\eta_{B}\leq \eta_{A}$. Define $A_{\ron,\pm}$ and $E_{\ron,\pm}$ as in the statement of Proposition~\ref{prop:asymposccaseitosoltowenoiseintro}.

If the metric $g$ has a geometric dominant noisy spatial direction corresponding to, say, $M_{r}$, then $E_{\ron,+}=E_{\ron,-}=:E_{\ron}$, and there is an 
injective linear map
\[
\Phi_{\ron,\rog}: W_{\ron}(\bM,E_{\ron})\rightarrow W_{\ron}(\bM,\cn{m})
\]
such that if $\psi\in W_{\ron}(\bM,E_{\ron})$; $u_{0,\row}$ is the solution to (\ref{eq:homtheeqnoiseintro}) corresponding to the initial data $\psi$ at $t=0$; 
$u_{\row}$ is defined by (\ref{eq:urowdefnoisegenintro}); and $u$ is the solution to (\ref{eq:thesystemRge}) (with $f=0$) corresponding to the initial data 
$\Phi_{\ron,\rog}(\psi)$ at $t=0$, then $u=u_{\ron}$ and 
\begin{equation}\label{eq:mfesumuwestasspgenintro}
\mfe_{s}^{1/2}[u-u_{\row}](t)\leq C_{B}\ldr{t}^{N}e^{(\kappa_{\ron,+}-\eta_{B})t}\|\psi\|_{(s+s_{B})}
\end{equation}
for all $t\geq 0$ and $s\in\ro$, where $C_{B}$, $s_{B}$ and $N$ have the same dependence as in the case of (\ref{eq:uuweststatementnoiseintro}).
If the Jordan blocks of the matrix $A_{\ron}$ are trivial, the $N$ appearing in (\ref{eq:mfesumuwestasspgenintro}) can be replaced by $1$.
Moreover, if $E_{\ron}=\cn{m}$, then $N$ can be replaced by $d_{n,+}-1$, where $d_{\ron,+}$ is defined in connection with (\ref{eq:Rronpmpmdefintro}).
If $E_{\ron}=\cn{m}$, then $\Phi_{\ron,\rog}$ is surjective. Finally, there are constants $C_{B}>0$ and $s_{B}\geq 0$ such that 
\begin{equation}\label{eq:Phironrogestnoiseintro}
\|\Phi_{\ron,\rog}(\psi)\|_{(s)}\leq C_{B}\|\psi\|_{(s+s_{B})}
\end{equation}
for all $s\in\ro$ and all $\psi\in W_{\ron}(\bM,E_{\ron})$, where $C_{B}$ and $s_{B}$ have the same dependence as in the case of (\ref{eq:uuweststatementnoiseintro}). 

If the metric $g$ has a geometric dominant noisy spatial $\so$-direction corresponding to $j$, let
\begin{align*}
\mW_{\ron}^{+} := & W_{\ron,+}^{+}(\bM,E_{\ron,+})\oplus W_{\ron,-}^{-}(\bM,E_{\ron,+}),\\
\mW_{\ron}^{-} := & W_{\ron,+}^{-}(\bM,E_{\ron,-})\oplus W_{\ron,-}^{+}(\bM,E_{\ron,-}).
\end{align*}
Then there is an injective linear map
\[
\Phi_{\ron,\roT}: \mW_{\ron}^{+}\times \mW_{\ron}^{-}\rightarrow W_{\ron}(\bM,\cn{m})
\]
such that if $\chi_{\pm}\in \mW_{\ron}^{\pm}$; $u_{\pm,\row}$ is the solution to (\ref{eq:homtheeqnoiseintro}) corresponding to the initial data $\chi_{\pm}$ at $t=0$; 
$u_{\row}$ is defined by (\ref{eq:urowdefnoisetorusintro}); and $u$ is the solution to (\ref{eq:thesystemRge}) (with $f=0$) corresponding to the initial data 
$\Phi_{\ron,\roT}(\chi_{+},\chi_{-})$ at $t=0$, then $u=u_{\ron}$ and 
\begin{equation}\label{eq:mfesumuwestassptorusintro}
\mfe_{s}^{1/2}[u-u_{\row}](t)\leq C_{B}\ldr{t}^{N}e^{(\kappa_{\ron,+}-\eta_{B})t}[\|\chi_{+}\|_{(s+s_{B})}+\|\chi_{-}\|_{(s+s_{B})}]
\end{equation}
for all $t\geq 0$ and $s\in\ro$, where $C_{B}$, $s_{B}$ and $N$ have the same dependence as in the case of (\ref{eq:uuweststatementnoiseintro}). If the Jordan 
blocks of the matrices $A_{\ron,\pm}$ are trivial, the $N$ appearing in (\ref{eq:mfesumuwestassptorusintro}) can be replaced by $1$. Moreover, if 
$E_{\ron,+}=E_{\ron,-}=\cn{m}$, then $N$ can be replaced by $d_{n,+}-1$, where $d_{\ron,+}$ is defined in connection with (\ref{eq:Rronpmpmdefintro}).
If $E_{\ron,+}=E_{\ron,-}=\cn{m}$, then $\Phi_{\ron,\roT}$ is surjective. Finally, there are constants $C_{B}>0$ and $s_{B}\geq 0$ such that 
\begin{equation}\label{eq:PhironroTestnoiseintro}
\|\Phi_{\ron,\roT}(\chi_{+},\chi_{-})\|_{(s)}\leq C_{B}[\|\chi_{+}\|_{(s+s_{B})}+\|\chi_{-}\|_{(s+s_{B})}]
\end{equation}
for all $s\in\ro$ and all $(\chi_{+},\chi_{-})\in \mW_{\ron}^{+}\times \mW_{\ron}^{-}$, where $C_{B}$ and $s_{B}$ have the same dependence as in the 
case of (\ref{eq:uuweststatementnoiseintro}).  
\end{prop}
\begin{remark}
The result is a consequence of Lemma~\ref{lemma:asymposccaseitosoltowenoisespa}; cf. Remark~\ref{remark:homwaveonlygeneqspeasnoise}.
\end{remark}
\begin{remark}
The asymptotic data can be specified in terms of functions such as $u_{\app}$ introduced in (\ref{eq:zappdefnoiseintro}) and 
(\ref{eq:uappdefnoiseintro}) (as opposed to solutions to (\ref{eq:homtheeqnoiseintro})); cf. Remark~\ref{remark:specupmwanoisedatinf}.
\end{remark}
\begin{remark}
Remarks~\ref{remarks:Tronbronetcdeponeqintro} are equally relevant in the present setting. 
\end{remark}

\subsection{Reinterpreting oscillatory ODE behaviour; equations with unbounded $\zeta$}\label{ssection:reintoscODEbeh}

In most of these notes, we assume $\zeta$ to be future bounded. However, there are some situations in which this requirement can be omitted. 
To be more specific, assume that there is a real valued function $\xi_{2}\in C^{\infty}(I,\ro)$ and real numbers $\xi_{2,\infty},\eta_{\aux},C_{\aux}>0$ 
such that 
\begin{equation}\label{eq:xitwoassumptions}
|\xi_{2}(t)-\xi_{2,\infty}|+|\d_{t}\xi_{2}(t)|\leq C_{\aux}e^{-\eta_{\aux}t}
\end{equation}
for all $t\geq 0$. Assume, moreover, $\zeta$ to be of the form
\[
\zeta(t)=\zeta_{1}(t)+\zeta_{2}(t)\Id_{m},\ \ \
\zeta_{2}(t):=\xi_{2}(t)e^{2\b_{\aux}t}
\]
for all $t\geq 0$, where $\b_{\aux}>0$ is a constant. Here $\zeta_{1}$ should be thought of as being such that $\|\zeta_{1}\|$ and 
$\|\d_{t}\zeta_{1}\|$ are bounded (however, $\zeta_{1}$ need not converge). By redefining $\zeta_{1}$ and $\zeta_{2}$ (in such a way
that they remain unchanged in a neighbourhood of $t=\infty$), it can be ensured that $\xi_{2}\geq \xi_{2,\infty}/2$ for all $t\geq 0$. 
For that reason, we from now on assume this inequality to be satisfied. Returning to (\ref{eq:thesystemRge}) with $g^{00}=-1$, 
note that it can be written
\begin{equation}\label{eq:uorigeq}
u_{tt}-g^{jl}\d_{j}\d_{l}u-2g^{0l}\d_{l}\d_{t}u
-\textstyle{\sum}_{r=1}^{R}a^{-2}_{r}(t)\Delta_{g_{r}}u+\a u_{t}+X^{j}\d_{j}u+\zeta_{1}u+\zeta_{2}u = f
.\end{equation}
On the other hand, we can view this equation as being the equation for a Fourier coefficient of a solution to
\begin{equation}\label{eq:Umodeq}
U_{tt}-g^{jl}\d_{j}\d_{l}U-\zeta_{2}\d_{d+1}^{2}U-2g^{0l}\d_{l}\d_{t}U
-\textstyle{\sum}_{r=1}^{R}a^{-2}_{r}(t)\Delta_{g_{r}}U+\a U_{t}+X^{j}\d_{j}U+\zeta_{1}U = F
\end{equation}
on $M_{\aux}:=\bM_{\aux}\times I$, where $\bM_{\aux}:=\tn{d+1}\times M_{1}\times\cdots\times M_{R}$. In fact, given a 
solution $u$ to (\ref{eq:uorigeq}), $U=ue^{ix^{d+1}}$ satisfies (\ref{eq:Umodeq}) with $F=fe^{ix^{d+1}}$ (where $x^{d+1}$
denotes the $(d+1)$'th coordinate on the $(d+1)$-torus). If we can derive asymptotic information concerning solutions to
(\ref{eq:Umodeq}), we can thus do the same for solutions to (\ref{eq:uorigeq}). In fact, it is sufficient if we can 
derive asymptotic information concerning the parts of solutions to (\ref{eq:Umodeq}) corresponding to $\indexnot$'s such 
that $n_{d+1}\neq 0$. There are many situations in which the results of these notes could be applied to 
(\ref{eq:Umodeq}). However, for the sake of definiteness, let us consider one such situation.  

\textbf{Dominant ODE oscillations.} Note that (\ref{eq:Umodeq}) is an equation of the form (\ref{eq:thesystemRge}), where 
the metric is given by 
\[
g_{\aux}:=g+\zeta_{2}^{-1}dx^{d+1}\otimes dx^{d+1}
\]
on $M_{\aux}$. Moreover, $(M_{\aux},g_{\aux})$ is a canonical separable cosmological model manifold, and if $\bk_{\aux}$ is the 
second fundamental form associated with $g_{\aux}$, then 
\[
\bk_{\aux}=\bk-\left(\b_{\aux}+\frac{1}{2}\xi_{2}^{-1}\dot{\xi}_{2}\right)\zeta_{2}^{-1}dx^{d+1}\otimes dx^{d+1},
\]
where $\bk$ is the second fundamental form of $g$. Assume that there is a constant 
$\eta_{\ron}>0$ and a continuous $\betafun\in L^{1}([0,\infty))$ such that 
\begin{equation}\label{eq:bklowbdODEosc}
\bk(v,v)\geq [-\b_{\aux}+\eta_{\ron}-\betafun(t)]\bge(v,v)
\end{equation}
for all $v\in T\bM$ and $t\geq 0$. Define $\bM_{\rosub}:=\bM$; $\pi_{\ron}:\bM_{\aux}\rightarrow\so$ to be the projection onto 
the $(d+1)$'th $\so$-factor in $\bM_{\aux}$; $\pi_{\rosub}$ to be the map from $\bM_{\aux}$ to $\bM_{\rosub}$ corresponding to the projection onto
what remains after removing the $(d+1)$'th $\so$-factor; and let $\b_{\ron}:=\b_{\aux}$. Then (\ref{eq:xitwoassumptions}) ensures that 
(\ref{eq:condonkdomnoisdirintro}) holds. Moreover, (\ref{eq:bklowbdODEosc}) ensures that (\ref{eq:bkwwlbsubtmintro}) holds. Thus
$g_{\aux}$ has a geometric dominant noisy spatial $\so$-direction corresponding to $d+1$. Since $X^{d+1}=0$ in (\ref{eq:Umodeq}), 
the requirement that (\ref{eq:Xjalconvnosettingintro}) hold is equivalent to the requirement that 
\begin{equation}\label{eq:alconvODEosc}
\|\a(t)-\a_{\infty}\|\leq C_{\romn}e^{-\eta_{\romn}t}
\end{equation}
for all $t\geq 0$. From now on, we therefore assume (\ref{eq:alconvODEosc}) to hold, so that the dominant coefficients are convergent;
cf. Definition~\ref{def:domnoisyspasodir}. Turning to the shift vector field, note that it is the same for $g$ and $g_{\aux}$, and assume
that (\ref{eq:neglshiftvfnoissettintro}) holds for all $t\geq 0$. What remains to be verified in order to ensure that the conditions of 
Definition~\ref{def:noisemainassumptions} are satisfied is that (\ref{eq:Umodeq}) is $C^{2}$-balanced. Assuming the shift vector field of 
$g$ to be $C^{2}$-future bounded; $\a$ and $\zeta_{1}$ to be $C^{1}$-future bounded; $\mcX$ to be $C^{1}$-future bounded; $\bk$ to be 
$C^{1}$-future bounded; and (in addition to (\ref{eq:xitwoassumptions})), $\ddot{\xi}_{2}$ to be future bounded, then (\ref{eq:Umodeq}) is 
$C^{2}$-balanced. Making appropriate assumptions concerning $f$, the results of the present section thus apply to (\ref{eq:Umodeq}). 

\begin{example}
Consider the Klein-Gordon equation in the expanding direction of Kasner spacetimes. Due to the calculations carried out
in Example~\ref{example:nonflatvacuumKasner}, the Klein-Gordon equation can be written
\[
t\d_{t}(tu_{t})-\textstyle{\sum}_{i=1}^{d}t^{2\b_{i}}u_{ii}+m^{2}t^{2}u=0,
\]
where $\b_{i}=1-p_{i}$. Letting $\tau=\ln t$, this equation becomes
\[
u_{\tau\tau}-\textstyle{\sum}_{i=1}^{d}e^{2\b_{i}\tau}u_{ii}+m^{2}e^{2\tau}u=0. 
\]
In the discussion above, this equation corresponds to (\ref{eq:uorigeq}). The corresponding version of (\ref{eq:Umodeq}) is
\begin{equation}\label{eq:KlGeK}
U_{\tau\tau}-\textstyle{\sum}_{i=1}^{d}e^{2\b_{i}\tau}U_{ii}-m^{2}e^{2\tau}U_{DD}=0, 
\end{equation}
where $D:=d+1$. If the $\b_{i}$ are all distinct and different from $1$, we can think of (\ref{eq:KlGeK}) as being defined on 
$\tn{D}\times\ro$. If there are several $\b_{i}$'s equalling, say, $\b_{1}$, then we can combine the corresponding circles to a 
$\tn{d_{1}}=:M_{1}$ etc. Due to these observations, we can apply the methods developed so far in these notes to (\ref{eq:KlGeK}). 
Doing so yields a homeomorphism between initial data and asymptotic data (we leave the details of the verification of this statement
to the reader). 
\end{example}

\section{Outline of the proof}\label{section:outpfnoiseintro}

Fix a Fourier mode $\indexnot\in\EFnindexset$. In one respect, the analysis of the asymptotics of a corresponding
Fourier coefficient is similar to the analysis in the silent and transparent cases: 
\begin{itemize}
\item For large $t$, say $t\geq T_{\ron}$, $\mfg(\indexnot,t)$ is well approximated by $\nu_{\ron}(\indexnot)e^{\b_{\ron}t}$, and 
(\ref{eq:fourierthesystemRge}) can effectively be replaced by a limit equation.
\item For small $t$, i.e. $t\in [0,T_{\ron}]$, the solution can be controlled by a crude energy estimate. 
\end{itemize}
The main difference between the noisy setting on the one hand, and the silent and transparent settings on the other, is that the limit equation
in the noisy setting is not a constant coefficient equation, so that quite different methods are needed in order to analyse the behaviour of 
solutions for $t\geq T_{\ron}$. The fact that a crude energy estimate is sufficient to handle the interval $[0,T_{\ron}]$ is a strength in the 
sense that the conditions needed to obtain the results are weak. On the other hand, the crude analysis leads to a loss of derivatives, which makes 
it impossible to calculate $\nolossrate$. 

\textbf{The noisy regime.} Due to the above observations, it is clear that the main difficulty consists in analysing the behaviour of solutions 
in the regime $[T_{\ron},\infty)$. In the introduction to the present chapter, we give a rough idea of how to approximate the corresponding 
oscillatory behaviour in the case of (\ref{eq:limiteqnoiseintrointro}). In Part~\ref{part:averaging}, we develop general methods 
for approximating the evolution over one period of the oscillations. However, the methods developed in Part~\ref{part:averaging} of these notes
apply not only to the equations considered in Chapter~\ref{chapter:onnotofbal} and the equations considered in the present chapter. If we make 
additional assumptions concerning the equations, they also apply in regions of the form $[0,T_{\roode}]$, $[0,T_{\trs}]$ and $[0,T_{\ron}]$ in the 
silent, transparent and noisy settings respectively. Moreover, in Chapter~\ref{chapter:enestasdiagsett} below, we demonstrate that it is possible
to calculate $\nolossrate$ on the basis of such an application. To summarise: since the full range of the applications of the results of 
Part~\ref{part:averaging} only becomes clear at the end of the next chapter, we only give an outline of the methods developed in 
Part~\ref{part:averaging} at the end of Chapter~\ref{chapter:enestasdiagsett}. However, let us describe the main outcome of the analysis. 
The goal is to estimate the evolution of solutions in intervals where the behaviour is oscillatory. To this end, we introduce a time sequence
$\{t_{k}\}$ such that $[t_{k},t_{k+1}]$, roughly speaking, corresponds to one period of the oscillations. The final result of 
Part~\ref{part:averaging} is (\ref{eq:wprekpertransvarpi}), an equation that yields
\begin{equation}\label{eq:psikpoformintronoise}
\psi_{k+1} =A_{k}^{+}\psi_{k}+A_{k}^{+}\int_{t_{k}}^{t_{k+1}}\hF_{k}(t)dt.
\end{equation}
Here $\psi_{k}$ represents the initial data for the Fourier coefficient at $t_{k}$ (the exact relation is somewhat technical, and we refer the
reader interested in the details to Part~\ref{part:averaging}). Moreover, $A_{k}^{+}$ is the matrix taking initial data at $t_{k}$ to initial 
data at $t_{k+1}$ in the homogeneous setting. Finally, $\hF_{k}(t)$ is a reformulated version of the contribution from $\hf(\indexnot,t)$. That
a relation of the form (\ref{eq:psikpoformintronoise}) holds is an immediate consequence of the fact that the underlying equation is a linear
system of ODE's. However, in Part~\ref{part:averaging}, we provide approximations for $A_{k}^{+}$, and estimate the error. Moreover, the 
approximations can be expressed in terms of the coefficients of (\ref{eq:thesystemRge}).

\textbf{Estimates along a time sequence. A rough Sobolev estimate.} The main part of the arguments needed to prove the results of the present 
chapter is contained 
in Parts~\ref{part:averaging} and \ref{part:dominantnoisyspdirection} of these notes. Focusing on Part~\ref{part:dominantnoisyspdirection}, 
we begin, in Sections~\ref{section:termbaassannoise} and \ref{section:oscaddelatiregnoise}, by providing an analytic version of the definition
of a noisy equation (the conditions are such that they are implied by the geometric conditions imposed in the current chapter). Moreover, we 
verify that the conditions needed in order to apply the results of Part~\ref{part:averaging} are satisfied, and we define $T_{\ron}$. In 
Section~\ref{section:detailedasfomaappinit}, we adapt the general framework of Part~\ref{part:averaging} to the context of interest here. 
In particular, we estimate $A_{k}^{+}$ appearing in (\ref{eq:psikpoformintronoise}) in terms of $\a_{\infty}$, $\b_{\ron}$ and 
$\bX_{\ron}(\indexnot)$. Given this approximation, we reformulate the iteration (\ref{eq:psikpoformintronoise}) in order to isolate the 
leading order behaviour; this is the subject of Section~\ref{section:reformeqsnoise}. On the basis of this reformulation, we are then in 
a position to estimate the growth of the individual Fourier coefficients. This is the subject of Section~\ref{section:aroughmodeestnoise}.
Combining this estimate for the Fourier coefficients with a rough energy estimate in the interval $[0,T_{\ron}]$ and summing up over the 
modes yields the Sobolev estimate (\ref{eq:windtestfinnoisyerapartconvsobintro}); cf. Section~\ref{section:roughsobolevestnoise}.

\textbf{Asymptotics along a time sequence.} In Section~\ref{section:asalongtimeseq}, we turn to the problem of deriving the asymptotic
behaviour along a time sequence. The argument is quite long and technical, but in some respects, it is a discrete version of the corresponding
argument in the silent setting. Note, however, that the analysis is only relevant for one mode and along a time sequence contained in 
$[T_{\ron},\infty)$. In Section~\ref{section:spdataatinfnoise}, we turn to the problem of specifying the asymptotic data along the time sequence. 
Again, the argument is long and technical, but roughly speaking a discrete version of the corresponding argument in the silent setting. Deriving 
and specifying asymptotics
along a time sequence is interesting. However, there are two drawbacks: the time sequence is not canonically defined and depends on $\indexnot$;
and the conclusions do not tell us what happens between the times belonging to the sequence. In Section~\ref{section:contsolbeytimeseqnoise}
we begin addressing these issues by estimating the evolution for all $t\geq T_{\ron}$. In order to obtain the desired conclusion, we, again, need
to appeal to the results of Part~\ref{part:averaging}. 

\textbf{The results.} Given the analysis described above, we are in a position to derive the main results of the present chapter; 
Propositions~\ref{prop:asymposccasescalareqnoiseintro}, \ref{prop:asymposccaseitosoltowenoiseintro} and 
\ref{prop:asymposccaseitosoltowenoisespaintro}. In Section~\ref{section:caseofscalhomwaveeqnoise} we demonstrate
Proposition~\ref{prop:asymposccasescalareqnoiseintro}. Even though most of the ingredients are already in place, we need to demonstrate
that we can specify the asymptotics for the individual modes (in the particular case of interest); we need to sum up over the modes; and,
given a solution,  
we need to derive asymptotics. As a next step, we demonstrate a uniqueness result; cf. Section~\ref{section:auniquenresultnoise}. The 
purpose of this result is to ensure that the uniqueness statement made in Proposition~\ref{prop:asymposccaseitosoltowenoiseintro} holds. 
Proving this uniqueness statement is more complicated than proving the corresponding result in the silent setting. The reason for this 
is that in the silent setting, the decay rate on the right hand side of the estimate is strong enough to ensure uniqueness. However, in
the noisy setting, the fact that uniqueness holds is dependent on the orientation of the solutions, and this complicates the proof. 
In Section~\ref{eq:deasinnoissetnoise} we prove Proposition~\ref{prop:asymposccaseitosoltowenoiseintro}: we verify the uniqueness
statement; we derive estimates for the individual modes and late times; we extend the results to early times; we consider
the approximate solutions and use the approximate solutions to construct solutions to the model equation; and finally we derive
the desired estimates. We end by demonstrating that the asymptotic data can be specified in Section~\ref{section:spedataatinf}.

\chapter[The asymptotically diagonal setting]{Energy estimates in the asymptotically 
diagonal setting}\label{chapter:enestasdiagsett}

\section{Introduction}

\textbf{Weaknesses of the results.}
The results of Chapters~\ref{chapter:silentequations}--\ref{chapter:domnoisspdirintro} may, superficially, seem quite satisfactory. First of 
all, they yield optimal energy estimates in the sense that they allow us to calculate $\cruderate$. Second, they yield detailed asymptotics 
and, third, they allow us to specify the leading order asymptotics. In fact, we sometimes obtain a homeomorphism between initial data and 
asymptotic data. In other words, if one is only interested in the linear systems of equations discussed in these notes, then the results are
quite informative. However, our main motivation for studying these equations is that they arise as linearised versions of Einstein's equations. 
Moreover, in the original non-linear setting, the question of stability is of interest. Assume, therefore, that the homogeneous
version of (\ref{eq:thesystemRge}) is the linearised version of a non-linear equation. We would then like to know that if $u$ is a 
solution with initially small energy, then the energy of $u$ remains small. In order to evaluate whether the results of, say, 
Chapter~\ref{chapter:silentequations} allow us to draw such conclusions, say that the conditions of Proposition~\ref{prop:genroughestintro} are 
satisfied and that $f=0$. Assume, moreover, that $d_{1}=1$ and that $\kappa_{1}=-4711$. For a smooth solution $u$, 
Proposition~\ref{prop:genroughestintro} then guarantees that there is a $0<C\in\ro$ such that 
\begin{equation}\label{eq:mfe4711est}
\mfe^{1/2}[u](t)\leq Ce^{-4711t}
\end{equation}
for all $t\geq 0$; recall that $\mfe=\mfe_{0}=\mfe_{\robas}$. 
In particular, the energy is decaying exponentially, so that the solution will eventually be small. Let $0<\e\in\ro$ be given, and assume that 
$\mfe^{1/2}[u](0)\leq \e$. Assuming $\e$ to be small, we know $\mfe^{1/2}[u]$ to be small initially and to decay exponentially in the sense
that there is a constant such that (\ref{eq:mfe4711est}) holds. Does the combination of these two observations ensure that $\mfe$ is always
small? Unfortunately, the answer to this question is no. Given two constants $0<\e,N\in\ro$, there is an equation satisfying the above 
restrictions and a corresponding solution $u$ such that $\mfe^{1/2}[u](0)\leq \e$; such that (\ref{eq:mfe4711est}) holds for some $C$ and all 
$t\geq 0$; and such that there is a 
$0<t_{1}\in\ro$ such that $\mfe^{1/2}[u](t_{1})\geq N$. In this sense, the results of Chapter~\ref{chapter:silentequations} are not very useful
when addressing the issue of stability. On the other hand, assuming $\mfe_{s_{0}}[u](0)\leq C^{-2}\e^{2}$, where $s_{0}$ and $C$ are the constants
appearing in the statement of Proposition~\ref{prop:genroughestintro}, the estimate $\mfe^{1/2}[u](t)\leq \e e^{-4711t}$ holds for all $t\geq 0$. 
However, 
this result involves a loss of derivatives, and, depending on the context, this may substantially reduce its use. For a justification of the above
statements, we refer the reader to Example~\ref{example:optimalityone} below.

\textbf{Calculating $\nolossrate$.} The deficiencies mentioned above are related to our inability to calculate $\nolossrate$ in
the previous chapters. Moreover, this inability is related to the loss of derivatives in, e.g., Propositions~\ref{prop:genroughestintro} and 
\ref{prop:roughsobestnoisysetintro}. Finally, this loss of derivatives is related to the rough energy estimates we carry out in time intervals
of the form $[0,T_{\roode}]$, $[0,T_{\trs}]$ and $[0,T_{\ron}]$ for a fixed mode; cf. the descriptions of the arguments in the previous chapters. 
In order to calculate $\nolossrate$, we need to improve the analysis in these intervals. However, in order to be able to do so, we need to make
stronger assumptions. In Section~\ref{section:asdiageqintronoise} we introduce the relevant conditions. However, it is useful to consider
a few examples before stating the results; this is the subject of Subsection~\ref{ssection:examplesnoiseintro}. We also give an 
outline of the present chapter in Subsection~\ref{ssection:outlineintrooptenest}.

\subsection{Examples}\label{ssection:examplesnoiseintro}

In order to contrast the results of Chapter~\ref{chapter:silentequations} and the results of the present chapter with standard energy estimates, 
let us consider the following two examples. 

\begin{example}
Consider
\begin{equation}\label{eq:elexbal}
u_{tt}-e^{-2t}u_{\theta\theta}+2e^{-t}u_{\theta}+u_{t}+u=0.
\end{equation}
This equation is such that the results of Chapter~\ref{chapter:silentequations} apply. In fact, $\bk=\bge$ and $\chi=0$, so that 
(\ref{eq:elexbal}) is $C^{1}$-future silent in the sense of Definition~\ref{definition:Cosilenceintro}, with $\mu=1$. Moreover, $\mcX=2e^{-t}\d_{\theta}$
is $C^{0}$-future bounded, and since $\a=1$ and $\zeta=1$,  (\ref{eq:alpahzetaconvest}) holds with $C_{\romn}=\eta_{\romn}=1$. In particular, 
the eigenvalues of $A_{\infty}$ introduced in (\ref{eq:Asilentdefintro}) all have real part $\kappa_{1}=-1/2$ and the corresponding Jordan blocks
are trivial. Moreover, Proposition~\ref{prop:genroughestintro} applies and yields the conclusion that there are constants $C$ and $s_{0}$
such that 
\begin{equation}\label{eq:mfeelexbalest}
\mfe^{1/2}_{s}[u](t)\leq Ce^{-t/2}\mfe^{1/2}_{s+s_{0}}[u](0)
\end{equation}
for all $t\geq 0$ and all $s\in\ro$. Since the energy of spatially homogeneous solutions decays as $e^{-t}$, this estimate is optimal
as far as the time dependence is concerned, and $\cruderate=-1/2$. 

On the other hand, considering the problem from the point of view of standard energy estimates, let 
\begin{equation}\label{eq:Erobdef}
\mfe_{\robas}[u](t):=\frac{1}{2}\int_{\so}[|u_{t}(\theta,t)|^{2}+e^{-2t}|u_{\theta}(\theta,t)|^{2}+|u(\theta,t)|^{2}]d\theta.
\end{equation}
It can then be calculated that 
\begin{equation}\label{eq:dErobdetexp}
\frac{d\mfe_{\robas}[u]}{dt}=-\int_{\so}|u_{t}+e^{-t}u_{\theta}|^{2}d\theta.
\end{equation}
Clearly, $\mfe_{\robas}$ is a decreasing function. On the other hand, we can chose initial data so that the right hand side of (\ref{eq:dErobdetexp})
vanishes. In general, we therefore cannot obtain a better estimate than $\d_{t}\mfe_{\robas}\leq 0$. From this point of view, it seems optimistic
to even expect decay. Clearly, there is a tension between this observation and the estimate (\ref{eq:mfeelexbalest}). However, it turns out that 
$\nolossrate=0$; this is a consequence of Theorem~\ref{thm:mainoptthmintro} below. In other words, if $a\in\ro$ and $0<C\in\ro$ are such that
\[
\mfe_{\robas}[u](t)\leq Ce^{2a t}\mfe_{\robas}[u](0)
\]
for all $t\geq 0$ and all solutions $u$ to (\ref{eq:elexbal}), then $a\geq 0$. To conclude, what the decay rate is depends on how many 
derivatives one is prepared to lose. The estimate (\ref{eq:mfeelexbalest}) is the best as far as decay is concerned, but the estimate 
$\mfe_{\robas}[u](t)\leq \mfe_{\robas}[u](0)$ for all $t\geq 0$ is the best estimate if one is not prepared to lose derivatives. 
\end{example}
In the case of (\ref{eq:elexbal}), the equality (\ref{eq:dErobdetexp}) yields a good indication of the value of $\nolossrate$.
In order to illustrate that the situation is sometimes more complicated, we consider the following example. 
\begin{example}
Consider the equation
\begin{equation}\label{eq:optenestabpar}
u_{tt}-e^{-2t}u_{\theta\theta}+ae^{-t}u_{\theta}+bu_{t}=0,
\end{equation}
where $a,b\in\ro$ and $b<0$. In what follows, we focus on real valued solutions. Due to the absence of an undifferentiated $u$, it is convenient to 
consider the energy
\[
\mfe_{\rohom}[u](t):=\frac{1}{2}\int_{\so}[u_{t}^{2}(\theta,t)+e^{-2t}u_{\theta}^{2}(\theta,t)]d\theta.
\]
Due to (\ref{eq:optenestabpar}), 
\[
\frac{d\mfe_{\rohom}[u]}{dt}=\int_{\so}(-ae^{-t}u_{\theta}u_{t}-bu_{t}^{2}-e^{-2t}u_{\theta}^{2})d\theta.
\]
The integrand is a quadratic form in $u_{t}$ and $e^{-t}u_{\theta}$. The largest eigenvalue of the corresponding matrix is 
\[
\lambda_{\max}=\frac{1}{2}\left[-(b+1)+\sqrt{(b-1)^{2}+a^{2}}\right].
\]
Note that if $b<0$ and $a\neq 0$, then $\lambda_{\max}>-b$. For a given $t_{0}$, there are non-vanishing initial data at $t_{0}$ such that 
\[
\d_{t}\mfe_{\rohom}(t_{0})=2\lambda_{\max}\mfe_{\rohom}(t_{0}).
\]
Hoping for a better estimate than $\mfe_{\rohom}(t)\leq e^{2\lambda_{\max}t}\mfe_{\rohom}(0)$ thus seems optimistic. On the other
hand, it can be calculated that for the equation (\ref{eq:optenestabpar}),
\begin{equation}\label{eq:nolossratecalenestprob}
\nolossrate=\frac{1}{2}\max\{-b-1+|a|,-2b\},
\end{equation}
assuming $b<0$; 
cf. Example~\ref{example:optimalitytwo} below. In fact, we demonstrate that there is a constant $C$ such that for every solution $u$ to 
(\ref{eq:optenestabpar}), the energy $\mfe_{\robas}$ introduced in (\ref{eq:Erobdef}) satisfies 
\begin{equation}\label{eq:mfebasenestnolossbaex}
\mfe_{\robas}[u](t)\leq Ce^{2\nolossrate t}\mfe_{\robas}[u](0)
\end{equation}
for all $t\geq 0$. 
Note that, as long as $b<0$ and $a\neq 0$, $\nolossrate<\lambda_{\max}$. Since $\mfe_{\rohom}\leq \mfe_{\robas}$, it follows
that $\mfe_{\rohom}(t)\leq Ce^{2\nolossrate t}\mfe_{\robas}(0)$. This yields a better estimate for the energy $\mfe_{\rohom}$ than we would naively have 
expected. Of course, there is a loss involved in that $\mfe_{\robas}(0)\geq \mfe_{\rohom}(0)$. On the other hand, $\mfe_{\robas}(0)$ involves the same 
number of derivatives of the initial data as $\mfe_{\rohom}(0)$. In this sense, the loss is not severe. Even though this argument does not prove that 
energy estimates cannot be used to obtain optimal conclusions, they do indicate that crude energy estimates sometimes do not suffice. 
\end{example}

\subsection{Outline}\label{ssection:outlineintrooptenest}

As indicated by the title of the chapter, we are here interested in metrics that become diagonal asymptotically. A precise explanation of what this 
means is given in Definition~\ref{def:diagconvgeometricdefintro} below. However, loosely speaking, there are two main conditions. First, in each 
$\so$ and $M_{r}$ direction, the second fundamental form converges; cf. (\ref{eq:diagconvtorusdir}) and (\ref{eq:diagconvgeneraldir}) below. 
Second, the different $\so$-directions are becoming exponentially more orthogonal with time in the sense that (\ref{eq:metricupoffdiaggeomintro}) and 
(\ref{eq:mlUbkvivlestintro}) below hold for some suitable choice of $k$. In the present chapter, we also assume that the shift vector field becomes negligible 
asymptotically, in the sense that (\ref{eq:chidotchinegligibleintro}) below holds. Adding to these conditions on the metric the assumption that the lower 
order coefficients are convergent, in the sense that (\ref{eq:Xjalzeintro}) below holds, and the assumption that (\ref{eq:thesystemRge}) is $C^{2}$-balanced,
we obtain the conclusion that the equation (\ref{eq:thesystemRge}) asymptotically takes the form
\begin{equation}\label{eq:limitequasdiagset}
u_{tt}-\textstyle{\sum}_{j}g^{jj}_{\infty}e^{2\b_{j}t}u_{jj}-\sum_{r}a_{r,\infty}^{-2}e^{2\bRie{r}t}\Delta_{g_{r}}u+\sum_{j}e^{\b_{j}t}X^{j}_{\infty}u_{j}
+\a_{\infty}u_{t}+\zeta_{\infty}u=f,
\end{equation}
\index{Limit equation!asymptotically diagonal setting}%
where $u_{t}=\d_{t}u$, $u_{j}=\d_{j}u$ etc.; cf. Section~\ref{section:asdiageqintronoise} for a justification of this statement. Here 
$0<g^{jj}_{\infty},a_{r,\infty}\in\ro$, $\b_{j},\bRie{r}\in\ro$ and $X^{j}_{\infty},\a_{\infty},\zeta_{\infty}\in\Mn{m}{\co}$. In order to develop some intuition for 
the results, let us consider the limit equation (\ref{eq:limitequasdiagset}). 

\textbf{The Fourier modes.} Introducing
\begin{equation}\label{eq:mfginfasdiagnondeg}
\mfg_{\infty}(\indexnot,t):=\left(\textstyle{\sum}_{j}g^{jj}_{\infty}e^{2\b_{j}t}n_{j}^{2}+\sum_{r}a_{r,\infty}^{-2}e^{2\bRie{r}t}\nu_{r,i_{r}}^{2}(\indexnot)\right)^{1/2},
\end{equation}
\index{$\a$Aa@Notation!Coefficients, Fourier side!$\mfg_{\infty}$}%
the equation for the Fourier coefficients of solutions to the limit equation can be written
\begin{equation}\label{eq:limiteqfourierasdiagset}
\ddot{z}+\mfg^{2}_{\infty}z+\textstyle{\sum}_{j}in_{j}e^{\b_{j}t}X^{j}_{\infty}z+\a_{\infty}\dot{z}+\zeta_{\infty}z=\hf.
\end{equation}
In the present chapter, we focus on the non-degenerate setting, where the set consisting of the union of the $\b_{j}$'s and the $\bRie{r}$'s contains 
$Q:=d+R$ distinct elements. Under this assumption, the interval $[0,\infty)$ can be divided into, at most, $Q+1$ intervals. Each interval corresponds
to one of the functions 
\[
1,\ \ \
(g^{jj}_{\infty})^{1/2}|n_{j}|e^{\b_{j}t},\ \ \
a_{r,\infty}^{-1}\nu_{r,i_{r}}(\indexnot)e^{\bRie{r}t}
\]
being larger than all the others, and we refer to these intervals as \textit{eras}. Let us now assume that $(g^{jj}_{\infty})^{1/2}|n_{j}|e^{\b_{j}t}$ dominates. Naively, it then 
seems reasonable to replace (\ref{eq:limiteqfourierasdiagset}) by 
\begin{equation}\label{eq:limiteqfourierasdiagsetofd}
\ddot{z}+g^{jj}_{\infty}e^{2\b_{j}t}n_{j}^{2}z+in_{j}e^{\b_{j}t}X^{j}_{\infty}z+\a_{\infty}\dot{z}=\hf
\end{equation}
(no summation on $j$). Here we have discarded the term $\zeta_{\infty}z$ since $(g^{jj}_{\infty})^{1/2}|n_{j}|e^{\b_{j}t}\geq 1$ by assumption. When $f=0$,
the equation (\ref{eq:limiteqfourierasdiagsetofd}) is identical to (\ref{eq:limiteqnoiseintrointro}) if we introduce the notation 
\[
\nu_{\ron}(\indexnot):=(g^{jj}_{\infty})^{1/2}|n_{j}|,\ \ \
\b_{\ron}:=\b_{j},\ \ \
\bX_{\ron}(\indexnot):=\frac{n_{j}}{|n_{j}|}(g^{jj}_{\infty})^{-1/2}X^{j}_{\infty}.
\]
If $\b_{j}>0$, we can thus appeal to the line of reasoning presented in Subsection~\ref{ssection:outlinedomnoisspdir} in order to conclude that the 
growth/decay of solutions should be determined by the matrices
\begin{equation}\label{eq:relmatricesnonzerobj}
\frac{1}{2}\left(\b_{j}\Id_{m}-\a_{\infty}\pm(g^{jj}_{\infty})^{-1/2}X^{j}_{\infty}\right);
\end{equation}
cf. (\ref{eq:SBinfSinv}). In fact, the same line of reasoning works in the case that $\b_{j}<0$. In other words, as long as $\b_{j}\neq 0$, the largest 
real part of an eigenvalue of a matrix of the form (\ref{eq:relmatricesnonzerobj}) can naively be expected to determine the overall growth of the Fourier 
coefficient in the era during which $(g^{jj}_{\infty})^{1/2}|n_{j}|e^{\b_{j}t}$ dominates. In the case that $a_{r,\infty}^{-1}\nu_{r,i_{r}}(\indexnot)e^{\bRie{r}t}$ 
dominates with $\bRie{r}\neq 0$, a similar argument applies. In an interval where $1$ dominates, the largest real part of an eigenvalue of the matrix 
$A_{\infty}$ introduced in (\ref{eq:Asilentdefintro}) determines the overall growth/decay. 

\textit{The transparent eras.} Consider an era corresponding to $\b_{j}\neq 0$, $\bRie{r}\neq 0$ or $1$. Then there is one matrix that determines the overall 
growth/decay rate. Now consider an era with $\b_{j}=0$ or $\bRie{r}=0$; from now on, we refer to such eras as \textit{transparent eras}. In the case of a
transparent era, the term $\zeta_{\infty}z$ has to be included in the limit equation. In the interval where, say, $(g^{jj}_{\infty})^{1/2}|n_{j}|e^{\b_{j}t}$ dominates (and $\b_{j}=0$), 
the limit equation thus becomes
\begin{equation}\label{eq:limiteqfourierasdiagsetofdtrs}
\ddot{z}+g^{jj}_{\infty}n_{j}^{2}z+in_{j}X^{j}_{\infty}z+\a_{\infty}\dot{z}+\zeta_{\infty}z=\hf.
\end{equation}
This is a family of infinitely many different systems of constant coefficient ODE's. For each member of the family, there is an associated generic growth
rate for solutions to the corresponding homogeneous equation. On the basis of the results of the previous chapters, we expect the supremum of these growth
rates over all $0\neq n_{j}\in\zo$ to be a lower bound for $\nolossrate$. 

\textbf{Outline of the chapter.} 
In Section~\ref{section:asdiageqintronoise} below, we introduce the conditions that characterise the asymptotically diagonal equations. First, we impose
conditions on the geometry, cf. Definition~\ref{def:diagconvgeometricdefintro}; and then we impose conditions on the lower order coefficients, cf. 
Definition~\ref{def:maincoeffconvasdiagsetintro}. However, in order for the results of the present chapter to apply, we also need to make a non-degeneracy
assumption; cf. Definition~\ref{def:nondegconvabalintro}. Finally, we need to assume that the equation is balanced. For convenience, we summarise the assumptions 
in Definition~\ref{def:geometricstrongassumptions}. 

\textit{The growth/decay rates corresponding to the different eras.}
In Section~\ref{section:theconstgrdecrates}, we introduce the constituent growth/decay rates that determine $\nolossrate$. The constituents are of three 
types. First, there is the ODE part, corresponding to the maximal growth rate of solutions to the homogeneous version of 
(\ref{eq:ODEapproxeqinCosilentsett}), where $A_{\infty}$ is given by (\ref{eq:Asilentdefintro}). Second, there are the growth/decay rates given by the 
greatest real parts of eigenvalues of matrices of the form (\ref{eq:relmatricesnonzerobj}) (for $j$'s such that $\b_{j}\neq 0$). In addition, there are 
similar contributions from the corresponding matrices for $r$'s such that $\bRie{r}\neq 0$. Third, there are the growth/decay rates arising from the 
transparent eras. 
They are given by the supremum of the growth/decay rates of the solutions to the homogeneous versions of (\ref{eq:limiteqfourierasdiagsetofdtrs}) (or the 
relevant analogous equations in case $\bRie{r}=0$). Combining these definitions leads to the introduction of $\kappa_{\rotot,+}$, cf. 
Definition~\ref{def:subdomnondegintro}. In the 
end, we are able to demonstrate that $\nolossrate=\kappa_{\rotot,+}$; cf. Theorem~\ref{thm:mainoptthmintro} below. In the energy estimates, we, in general, have
to count on a loss of the form $e^{\e t}$; cf. (\ref{eq:megenestlspresobsobintro}). However, if (\ref{eq:thesystemRge}) satisfies some additional 
non-degeneracy requirements, this loss can be removed; cf. (\ref{eq:megenestlspresobsilcasesobintro}), (\ref{eq:megenestlspresobstscasesobintro}) and
(\ref{eq:megenestlspresobsncasesobintro}). In Definition~\ref{def:subdomnondegintro}, we introduce not only $\kappa_{\rotot,+}$, but also the relevant 
non-degeneracy conditions. 

\textit{Results/outline of the argument.} Once we have introduced the necessary terminology, we are in a position to formulate the main results. In 
Section~\ref{section:mainenestintro} we state the main energy estimates; cf. Theorem~\ref{thm:sobestndddbaconvintro}. Moreover, we demonstrate
optimality and calculate $\nolossrate$ in Section~\ref{ssection:optimalityintro}; cf. Theorem~\ref{thm:mainoptthmintro}. We end the chapter, in 
Section~\ref{section:outlineofargoptenestintro}, by giving an outline of the argument.

\section{Asymptotically diagonal equations}\label{section:asdiageqintronoise}

In addition to the requirements made in previous chapters, the main assumption concerning the metric in the present chapter is that it becomes 
asymptotically diagonal and is diagonally convergent. These notions are defined as follows.

\begin{definition}\label{def:diagconvgeometricdefintro}
Let $(M,g)$ be a canonical separable cosmological model manifold. For each $l\in \{1,\dots,d\}$ 
let $\pi_{\roT,l}$ be the map from $\bM$ to $\so$ corresponding to projection onto the $l$'th $\so$-factor in $\tn{d}$. For each $r\in \{1,\dots,R\}$,
let $\pi_{\roRi,r}$ be the map from $\bM$ to $M_{r}$ corresponding to projection onto $M_{r}$. 
Then $\bk$ is said to be \textit{diagonally convergent} 
\index{Diagonally convergent!second fundamental form}%
\index{Second fundamental form!diagonally convergent}%
if there are real numbers $\b_{j}$ and $\bRie{r}$, where $j\in \{1,\dots,d\}$ and $r\in \{1,\dots,R\}$,
and constants $C_{\rod},\kappa_{\rod}>0$ such that the following holds:
\begin{align}
|\bk^{\sharp}(v_{j},v_{j})+\b_{j}\bge^{\sharp}(v_{j},v_{j})| \leq & C_{\rod}e^{-\kappa_{\rod}t}\bge^{\sharp}(v_{j},v_{j}),\label{eq:diagconvtorusdir}\\
|\bk^{\sharp}(w_{r},w_{r})+\bRie{r}\bge^{\sharp}(w_{r},w_{r})| \leq & C_{\rod}e^{-\kappa_{\rod}t}\bge^{\sharp}(w_{r},w_{r})\label{eq:diagconvgeneraldir}
\end{align}
for all $t\geq 0$, $j\in \{1,\dots,d\}$, $v_{j}\in \pi_{\roT,j}^{*}(T^{*}\so)$, $r\in \{1,\dots,R\}$, and $w_{r}\in \pi_{\roRi,r}^{*}(T^{*}M_{r})$.

In addition, $\bge$ is said to be \textit{$C^{0}$-asymptotically diagonal} 
\index{$\a$Aa@Notation!Conditions!$C^{0}$-asymptotically diagonal metric}%
if there are constants $C_{\rood},\kappa_{\rood}>0$ such that 
\begin{equation}\label{eq:metricupoffdiaggeomintro}
|\bge^{\sharp}(v_{j},v_{l})|\leq C_{\rood}e^{-\kappa_{\rood}t}|v_{j}|_{\bge}|v_{l}|_{\bge}
\end{equation}
for all $t\geq 0$; $j,l\in \{1,\dots,d\}$ such that $j\neq l$; $v_{j}\in \pi_{\roT,j}^{*}(T^{*}\so)$; and $v_{l}\in \pi_{\roT,l}^{*}(T^{*}\so)$. 
Similarly, $\bge$ is said to be \textit{$C^{k}$-asymptotically diagonal}, 
\index{$\a$Aa@Notation!Conditions!$C^{k}$-asymptotically diagonal!metric}%
$1\leq k\in\zo$, if it is $C^{0}$-asymptotically diagonal and there 
are constants $C_{\rood,k},\kappa_{\rood}>0$ such that 
\begin{equation}\label{eq:mlUbkvivlestintro}
\textstyle{\sum}_{i=0}^{k-1}|(\ml_{U}^{i}\bk)^{\sharp}(v_{j},v_{l})|\leq C_{\rood,k}e^{-\kappa_{\rood}t}|v_{j}|_{\bge}|v_{l}|_{\bge}
\end{equation}
for all $t\geq 0$; $j,l\in \{1,\dots,d\}$ such that $j\neq l$; $v_{j}\in \pi_{\roT,j}^{*}(T^{*}\so)$; and $v_{l}\in \pi_{\roT,j}^{*}(T^{*}\so)$.

Finally, if there are constants $C_{\rosh},\kappa_{\rosh}>0$ such that 
\begin{equation}\label{eq:chidotchinegligibleintro}
|\chi(t)|_{\bge}+|\dot{\chi}(t)|_{\bge}\leq C_{\rosh}e^{-\kappa_{\rosh}t}
\end{equation}
for all $t\geq 0$, then the shift vector field is said to be \textit{negligible}. 
\index{Negligible!shift vector field}%
\index{Shift vector field!negligible}%
\end{definition}
\begin{remark}
In (\ref{eq:diagconvtorusdir})--(\ref{eq:mlUbkvivlestintro}), the symbol $\sharp$ indicates raising all indices using $\bge$; in particular, 
$\bge^{\sharp}$, $\bk^{\sharp}$ and $(\ml_{U}\bk)^{\sharp}$ are the contravariant $2$-tensor fields whose components are given by $\bge^{ij}$, 
$\bk^{ij}$ and $(\ml_{U}\bk)^{ij}$ respectively. Moreover, if $\eta\in T^{*}M$, then $|\eta|_{\bge}:=(\bge^{ij}\eta_{i}\eta_{j})^{1/2}$.  
\end{remark}

Beyond making assumptions concerning the metric appearing in (\ref{eq:thesystemRge}), we also need to make assumptions concerning the lower
order coefficients. 

\begin{definition}\label{def:maincoeffconvasdiagsetintro}
Consider (\ref{eq:thesystemRge}). Assume the associated metric to be such that $(M,g)$ is a canonical separable cosmological model manifold.
Assume, moreover, the associated second fundamental form $\bk$ to be diagonally convergent; cf. 
Definition~\ref{def:diagconvgeometricdefintro}. If there are constants $0<C_{\romn},\kappa_{\romn}\in\ro$ and matrices 
$X_{\infty}^{j},\a_{\infty},\zeta_{\infty}\in \Mn{m}{\co}$, $j\in\{1,\dots,d\}$, such that 
\begin{equation}\label{eq:Xjalzeintro}
\|e^{-\b_{j}t}X^{j}(t)-X^{j}_{\infty}\|+\|\a(t)-\a_{\infty}\|+\|\zeta(t)-\zeta_{\infty}\|\leq C_{\romn}e^{-\kappa_{\romn} t}
\end{equation}
for all $t\geq 0$ and all $j\in\{1,\dots,d\}$, then (\ref{eq:thesystemRge}) is said to be such that the \textit{main coefficients are convergent}; 
\index{Main coefficients!convergent}%
\index{Convergent!main coefficients}%
here the $\b_{j}$ appearing in (\ref{eq:Xjalzeintro}) are the constants whose existence is guaranteed by Definition~\ref{def:diagconvgeometricdefintro}. 
\end{definition}

Consider (\ref{eq:thesystemRge}). Assume the associated metric to be such that $(M,g)$ is a canonical separable cosmological model manifold. 
Assume (\ref{eq:thesystemRge}) to be $C^{2}$-balanced; cf. Definition~\ref{definition:Cobal}. Assume, moreover, the shift vector field to be negligible, 
$\bk$ to be diagonally convergent and $\bge$ to be $C^{2}$-asymptotically diagonal; cf. Definition~\ref{def:diagconvgeometricdefintro}. Finally, assume 
(\ref{eq:thesystemRge}) to be such that the 
main coefficients are convergent; cf. Definition~\ref{def:maincoeffconvasdiagsetintro}. Then Lemma~\ref{lemma:geomecharofdiagdombalaconv} applies.
In particular, $e^{-2\b_{j}t}g^{jj}(t)$ converges to $g^{jj}_{\infty}>0$; $e^{-2\bRie{r}t}a_{r}^{-2}(t)$ converges to $a_{r,\infty}^{-2}>0$; $e^{-\b_{j}t}X^{j}(t)$
converges to $X^{j}_{\infty}$; $\a(t)$ converges to $\a_{\infty}$; and $\zeta(t)$ converges to $\zeta_{\infty}$. Moreover, the shift vector field $g^{0l}$
and the off-diagonal components $g^{jl}$, $j\neq l$, are relatively speaking negligible. This leads to the limit equation 
(\ref{eq:limitequasdiagset}). 

Treating the $\b_{j}$ and the $\bRie{r}$ separately is inconvenient in the discussions to follow. 
We therefore introduce the following terminology. 

\begin{definition}\label{def:nondegconvabalintro}
Consider (\ref{eq:thesystemRge}). Assume the associated metric to be such that $(M,g)$ is a canonical separable cosmological model manifold. 
Assume, moreover, $\bk$ to be diagonally convergent; cf. Definition~\ref{def:diagconvgeometricdefintro}. Define 
$\baverage{1}<\baverage{2}<\dots<\baverage{Q}$ 
\index{$\a$Aa@Notation!Exponential rates!$\baverage{q}$}%
to be the distinct elements of the union of the sets $\{\b_{1},\dots,\b_{d}\}$ and 
$\{\bRie{1},\dots,\bRie{R}\}$, where $\b_{j}$ and $\bRie{r}$ are the numbers whose existence is guaranteed by 
Definition~\ref{def:diagconvgeometricdefintro}. If $Q=R+d$, the equation (\ref{eq:thesystemRge}) is said to be \textit{asymptotically non-degenerate}. 
\index{Equation!asymptotically non-degenerate}%
\index{Asymptotically non-degenerate!equation}%
\end{definition}
\begin{remark}
The numbers $\b_{1},\dots,\b_{d},\bRie{1},\dots,\bRie{R}$ are distinct if and only if the equality $Q=d+R$ holds. 
\end{remark}
In what follows, we are mainly interested in asymptotically non-degenerate equations. In order to simplify the statement of the assumptions, 
we therefore introduce the following terminology. 

\begin{definition}\label{def:geometricstrongassumptions}
The equation (\ref{eq:thesystemRge}) is said to be \textit{geometrically non-degenerate, diagonally dominated, balanced and convergent} 
\index{Equation!geometrically non-degenerate, diagonally dominated, balanced and convergent}%
if the following assumptions hold. First, the associated metric is such that $(M,g)$ is a canonical separable cosmological model manifold. 
Second, (\ref{eq:thesystemRge}) is $C^{2}$-balanced; cf. Definition~\ref{definition:Cobal}. Third, 
the shift vector field is negligible, $\bk$ is diagonally convergent and $\bge$ is $C^{2}$-asymptotically diagonal; cf. 
Definition~\ref{def:diagconvgeometricdefintro}. Fourth, (\ref{eq:thesystemRge}) is such that the main coefficients are convergent; cf. 
Definition~\ref{def:maincoeffconvasdiagsetintro}. Fifth, (\ref{eq:thesystemRge}) is asymptotically non-degenerate; cf. 
Definition~\ref{def:nondegconvabalintro}.
\end{definition}

\section{The constituent growth/decay rates}\label{section:theconstgrdecrates}

Consider (\ref{eq:thesystemRge}). Assume that it is geometrically non-degenerate, diagonally dominated, balanced and convergent; cf. 
Definition~\ref{def:geometricstrongassumptions}. It is then convenient to introduce the following terminology: For $1\leq q\leq Q$ and 
$\indexnot\in\EFindexset$, let 
\begin{equation}\label{eq:nuaveragedefintro}
\nuaverage{q}(\indexnot):=
\left(\sum_{\{j:\b_{j}=\baverage{q}\}}g^{jj}_{\infty}n_{j}^{2}+\sum_{\{r:\bRie{r}=\baverage{q}\}}a_{r,\infty}^{-2}\nu_{r,i_{r}}^{2}(\indexnot)\right)^{1/2}.
\end{equation}
\index{$\a$Aa@Notation!Eigenvalues!$\nuaverage{q}(\indexnot)$}%

The following objects play a central role in the analysis to follow.

\begin{definition}\label{def:tXqinfRqpmkappaqpmintro}
Assume that (\ref{eq:thesystemRge}) is geometrically non-degenerate, diagonally dominated, balanced and convergent; cf. 
Definition~\ref{def:geometricstrongassumptions}. Given $q\in \{1,\dots,Q\}$, define $\tX^{q}_{\infty}$ as follows. If 
$\baverage{q}=\b_{j}$ for some $j\in \{1,\dots,d\}$, then $j$ is uniquely determined by $q$ and
\begin{equation}\label{eq:tXqinfdefintro}
\tX^{q}_{\infty}:=(g^{jj}_{\infty})^{-1/2}X^{j}_{\infty}.
\end{equation}
\index{$\a$Aa@Notation!Coefficients, Fourier side!$\tX^{q}_{\infty}$}%
Moreover, if $\indexnot\in\EFindexset$ is such that $\nu_{\roT,j}(\indexnot)=n_{j}\neq 0$, then 
\[
\Xaverage{\diag}^{q}(\indexnot):=\frac{n_{j}}{|n_{j}|}\tX^{q}_{\infty}.
\]
\index{$\a$Aa@Notation!Coefficients, Fourier side!$\Xaverage{\diag}^{q}(\indexnot)$}%
If $\baverage{q}=\bRie{r}$ for some $r\in \{1,\dots,R\}$, then $r$ is uniquely determined by $q$ and $\tX^{q}_{\infty}:=0$. Moreover, 
$\Xaverage{\diag}^{q}(\indexnot):=0$. For $q\in\{1,\dots,Q\}$, let 
\begin{equation}\label{eq:matrixdetgrowthintro}
R_{q,\pm}^{+}:=\pm\frac{1}{2}(-\a_{\infty}+\baverage{q}\Id_{m}+\tX^{q}_{\infty}),\ \  \
R_{q,\pm}^{-}:=\pm\frac{1}{2}(-\a_{\infty}+\baverage{q}\Id_{m}-\tX^{q}_{\infty}),
\end{equation}
\index{$\a$Aa@Notation!Matrix notation!Rqpm@$R_{q,\pm}^{\pm}$}%
and let
\begin{align}
\kappa_{q,\pm} := & \max\{\kappa_{\max}(R_{q,\pm}^{+}),\kappa_{\max}(R_{q,\pm}^{-})\},\label{eq:kappaqpmdefintro}\\
d_{q,\pm} := & d_{\max}(\diag\{R_{q,\pm}^{+},R_{q,\pm}^{-}\},\kappa_{q,\pm});\label{eq:dqpmdefintro}
\end{align}
\index{$\a$Aa@Notation!Matrix notation!kappaqpm@$\kappa_{q,\pm}$}%
\index{$\a$Aa@Notation!Matrix notation!dqpm@$d_{q,\pm}$}%
cf. Definition~\ref{def:SpRspdef}. Finally, let 
\begin{equation}\label{eq:kappapmdefintro}
\kappa_{\pm}:=\max_{q}\kappa_{q,\pm}.
\end{equation}
\index{$\a$Aa@Notation!Matrix notation!kappapm@$\kappa_{\pm}$}%
\end{definition}
\begin{remark}
It is of interest to interpret the constituents of $R_{q,+}^{\pm}$ geometrically. First of all, the $\baverage{q}$ are the asymptotic
eigenvalues of the second fundamental form. Second, note that
\[
\mcY:=\a\d_{t}+X^{j}\d_{j}
\]
is the matrix of vector fields appearing in (\ref{eq:thesystemRge}). Let $E_{j}$ be the unit one-form field which is a positive multiple
of $dx^{j}$. Then 
\[
\lim_{t\rightarrow\infty}U^{\flat}(\mcY)=-\a_{\infty},\ \ \
\lim_{t\rightarrow\infty}E_{j}(\mcY)=(g^{jj}_{\infty})^{-1/2}X^{j}_{\infty}.
\]
Note that $U^{\flat}$ is metrically equivalent to the future directed unit normal to the hypersurfaces $\bM_{t}$. Moreover, $\pm E_{j}$ are
the two unit one-form fields generating $\pi_{\roT,j}^{*}(T^{*}\so)$. Since the sign corresponds to a choice of orientation of the $j$'th 
$\so$-factor, it is not surprising that $\kappa_{\pm}$ does not depend on it. 
\end{remark}
\begin{remark}
The matrices appearing in (\ref{eq:matrixdetgrowthintro}) should be compared with those appearing in (\ref{eq:relmatricesnonzerobj}).
\end{remark}
\begin{remark}
At this stage, it is not so clear why it would be of interest to consider $R^{\pm}_{q,-}$. However, the argument demonstrating that we can specify asymptotic 
data is based on an analysis of the evolution when going backwards in time. In that context, the matrices $R^{\pm}_{q,-}$ are of importance. 
\end{remark}

The number $\kappa_{+}$ plays an important role in determining the growth rate of the basic energy. However, in the case of equations
with a $q\in \{1,\dots,Q\}$ such that $\baverage{q}=0$, we need to complement $\kappa_{+}$ with additional information; cf. the discussion 
of transparent eras in Subsection~\ref{ssection:outlineintrooptenest}. 

\begin{definition}\label{def:kappaqtrpmetcintro}
Consider the equation (\ref{eq:thesystemRge}). Assume that it is geometrically non-degenerate, diagonally dominated, balanced and convergent; cf. 
Definition~\ref{def:geometricstrongassumptions}. If there is a $q\in \{1,\dots,Q\}$ such that $\baverage{q}=0$, say $q_{\trs}$, define the sets 
$\mK_{q_{\trs},\pm}$ by 
\[
\mK_{q_{\trs},\pm}:=\left\{\kappa_{q_{\trs},\indexnot,\pm}\ |\ \indexnot\in\EFindexset,\ \nuaverage{q_{\trs}}(\indexnot)\neq 0\right\},
\]
\index{$\a$Aa@Notation!Matrix notation!mkqtrs@$\mK_{q_{\trs},\pm}$}%
where $\nuaverage{q}(\indexnot)$ is defined by (\ref{eq:nuaveragedefintro}), $\kappa_{q_{\trs},\indexnot,\pm}$ is defined by
\begin{equation}\label{eq:kapqindnotpmintro}
\kappa_{q_{\trs},\indexnot,\pm}:=\kappa_{\max}[\pm A_{q_{\trs},\infty}(\indexnot)]
\end{equation}
\index{$\a$Aa@Notation!Matrix notation!kappaqtrs@$\kappa_{q_{\trs},\indexnot,\pm}$}%
and 
\begin{equation}\label{eq:Aqtrsinfindnot}
A_{q_{\trs},\infty}(\indexnot) :=  \left(\begin{array}{cc} 0 & \nuaverage{q_{\trs}}(\indexnot)\Id_{m} \\ 
-\nuaverage{q_{\trs}}(\indexnot)\Id_{m}-i\Xaverage{\diag}^{q_{\trs}}(\indexnot)-[\nuaverage{q_{\trs}}(\indexnot)]^{-1}\zeta_{\infty} & -\a_{\infty}\end{array}\right).
\end{equation}
\index{$\a$Aa@Notation!Matrix notation!Aqtrs@$A_{q_{\trs},\infty}(\indexnot)$}%
Define $\bka_{q_{\trs},\pm}$ to be the limit of $\kappa_{q_{\trs},\indexnot,\pm}$ as 
$\nuaverage{q_{\trs}}(\indexnot)\rightarrow\infty$ and define
\[
\kappa_{q_{\trs},\trs,\pm}:=\sup\mK_{q_{\trs},\pm}.
\]
\index{$\a$Aa@Notation!Matrix notation!kappaqtrs@$\kappa_{q_{\trs},\trs,\pm}$}%
It is also convenient to introduce the notation
\begin{equation}\label{eq:kappatrpmdefintro}
\kappa_{\trs,\pm}:=\max\{\kappa_{\pm},\kappa_{q_{\trs},\trs,\pm}\}
\end{equation}
\index{$\a$Aa@Notation!Matrix notation!kappaqtrspm@$\kappa_{\trs,\pm}$}%
\end{definition} 
\begin{remark}
The matrix $A_{q_{\trs},\infty}(\indexnot)$ introduced in (\ref{eq:Aqtrsinfindnot}) should be compared with $A(\indexnot)$ introduced in
(\ref{eq:Aindexnotdefintro}); cf. also (\ref{eq:limiteqfourierasdiagsetofdtrs}).
\end{remark}
\begin{remark}
That the stated limit, $\bka_{q_{\trs},\pm}$, exists is a consequence of Remark~\ref{remark:limitofrealparttranseras}. In this remark, we also 
calculate $\bka_{q_{\trs},\pm}$ in terms of $\a_{\infty}$ and $\tX^{q_{\trs}}_{\infty}$. 
\end{remark}

In Definition~\ref{def:tXqinfRqpmkappaqpmintro}, we introduce the growth/decay rates associated with the eras with $\baverage{q}\neq 0$; and in 
Definition~\ref{def:kappaqtrpmetcintro} we introduce the growth/decay rate associated with the transparent eras. Combining these definitions
with the growth/decay rate corresponding to the spatially homogeneous solutions to (\ref{eq:thesystemRge}) yields a number $\kappa_{\rotot,+}$;
cf. Definition~\ref{def:subdomnondegintro} below. Naively, we would hope $\kappa_{\rotot,+}$ to equal $\nolossrate$. In the end this turns out
to be the case; cf. Theorem~\ref{thm:mainoptthmintro} below. 

In some situations, it is possible to not only calculate $\nolossrate$, but also to obtain the correct polynomial rate; cf. 
(\ref{eq:megenestlspresobsilcasesobintro})--(\ref{eq:megenestlspresobsncasesobintro}) below. However, to obtain such conclusions, we need to 
make suitable non-degeneracy assumptions. The relevant conditions are the following. 

\begin{definition}\label{def:subdomnondegintro}
Consider the equation (\ref{eq:thesystemRge}). Assume that it is geometrically non-degenerate, diagonally dominated, balanced and convergent; cf. 
Definition~\ref{def:geometricstrongassumptions}. If there is a $q_{\trs}\in \{1,\dots,Q\}$ such that $\baverage{q_{\trs}}=0$, let 
$\kappa_{\rotot,+}:=\max\{\kappa_{\trs,+},\kappa_{\rosil,+}\}$, 
\index{$\a$Aa@Notation!Matrix notation!kappatotp@$\kappa_{\rotot,+}$}%
where $\kappa_{\trs,+}$ is defined in (\ref{eq:kappatrpmdefintro}); 
$\kappa_{\rosil,+}:=\kappa_{\max}(A_{\infty})$; 
\index{$\a$Aa@Notation!Matrix notation!kappasilp@$\kappa_{\rosil,+}$}%
and $A_{\infty}$ is defined in (\ref{eq:Asilentdefintro}). If there is no such $q_{\trs}$, let 
$\kappa_{\rotot,+}:=\max\{\kappa_{+},\kappa_{\rosil,+}\}$, 
\index{$\a$Aa@Notation!Matrix notation!kappatotp@$\kappa_{\rotot,+}$}%
where $\kappa_{\rosil,+}$ is defined as before and $\kappa_{+}$ is defined by 
(\ref{eq:kappapmdefintro}). 

The equation (\ref{eq:thesystemRge}) is said to exhibit \textit{subdominant block non-degeneracy} provided that
\index{Subdominant block non-degeneracy}%
\begin{itemize}
\item if there is a $q_{\trs}\in \{1,\dots,Q\}$ such that $\baverage{q_{\trs}}=0$, then $\bka_{q_{\trs},+}<\kappa_{\rotot,+}$, 
\item for $q\in \{1,\dots,Q\}$ such that $\baverage{q}<\max\{0,\baverage{Q}\}$, all the Jordan blocks of $R_{q,+}^{+}$ and $R_{q,+}^{-}$ corresponding 
to the eigenvalues with real part $\kappa_{\rotot,+}$ are trivial, and
\item if $\baverage{Q}>0$ and there is a $q_{\trs}\in \{1,\dots,Q\}$ such that $\baverage{q_{\trs}}=0$, then the Jordan blocks of $A_{q_{\trs},\infty}(\indexnottwo)$ 
corresponding to the eigenvalues with real part $\kappa_{\rotot,+}$ are trivial for all $\indexnottwo\in\EFindexset$ such that 
$\nuaverage{q_{\trs}}(\indexnottwo)\neq 0$.
\end{itemize}
\end{definition}
\begin{remark}
Note that $\baverage{Q}\geq \baverage{q}$ for all $q\in \{1,\dots,Q\}$. 
\end{remark}
\begin{remark}
In Remark~\ref{remark:subdomblockdegbody} below, we give an intuitive motivation for the conditions that characterise subdominant block non-degeneracy. 
\end{remark}

\section{Energy estimates}\label{section:mainenestintro}

We are now in a position to state the main result of these notes. 

\begin{thm}\label{thm:sobestndddbaconvintro}
Consider the equation (\ref{eq:thesystemRge}). Assume that it is geometrically non-degenerate, diagonally dominated, balanced and convergent; cf. 
Definition~\ref{def:geometricstrongassumptions}. Fix $\e>0$. Then there is a constant $0<C_{\e}\in\ro$, depending only on $\e$, the spectra of the Riemannian 
manifolds $(M_{r},g_{r})$, $r\in\{1,\dots,R\}$, and the coefficients of the equation (\ref{eq:thesystemRge}), such that 
\begin{equation}\label{eq:megenestlspresobsobintro}
\mfe_{s}^{1/2}[u](t) \leq  C_{\e}e^{(\kappa_{\rotot,+}+\e)t}\mfe^{1/2}_{s}[u](0)+C_{\e}\int_{0}^{t}e^{(\kappa_{\rotot,+}+\e)(t-t')}\|f(\cdot,t')\|_{(s)}dt'
\end{equation}
for all $t\geq 0$, all solutions $u$ to (\ref{eq:thesystemRge}) and all $s\in\ro$, where $\kappa_{\rotot,+}$ is given by 
Definition~\ref{def:subdomnondegintro}. If, in addition to the above, the equation (\ref{eq:thesystemRge}) exhibits subdominant block non-degeneracy, 
there are the following improvements of (\ref{eq:megenestlspresobsobintro}):
\begin{itemize}
\item If $\baverage{Q}<0$, then the estimate 
\begin{equation}\label{eq:megenestlspresobsilcasesobintro}
\mfe^{1/2}_{s}[u](t) \leq  C\ldr{t}^{d_{\rosil,+}-1}e^{\kappa_{\rotot,+}t}\mfe^{1/2}_{s}[u](0)
+C\int_{0}^{t}\ldr{t-t'}^{d_{\rosil,+}-1}e^{\kappa_{\rotot,+}(t-t')}\|f(\cdot,t')\|_{(s)}dt'
\end{equation}
holds for all $t\geq 0$, all solutions $u$ to (\ref{eq:thesystemRge}) and all $s\in\ro$. Here $d_{\rosil,+}=d_{\max}(A_{\infty},\kappa_{\rotot,+})$; 
$A_{\infty}$ is defined in 
(\ref{eq:Asilentdefintro}); and the constant $C$ only depends on the spectra of the Riemannian manifolds $(M_{r},g_{r})$, $r\in\{1,\dots,R\}$, 
and the coefficients of the equation (\ref{eq:thesystemRge}).
\item If $\baverage{Q}=0$, then 
\begin{equation}\label{eq:megenestlspresobstscasesobintro}
\mfe^{1/2}_{s}[u](t) \leq  C\ldr{t}^{d_{\rosts,+}-1}e^{\kappa_{\rotot,+}t}\mfe^{1/2}_{s}[u](0)
+C\int_{0}^{t}\ldr{t-t'}^{d_{\rosts,+}-1}e^{\kappa_{\rotot,+}(t-t')}\|f(\cdot,t')\|_{(s)}dt'
\end{equation}
holds for all $t\geq 0$, all solutions $u$ to (\ref{eq:thesystemRge}) and all $s\in\ro$. Here $d_{\rosts,+}:=\max\{d_{\rosil,+},d_{\rot,+}\}$; $d_{\rosil,+}$ is defined as 
in the case of (\ref{eq:megenestlspresobsilcasesobintro}); $d_{\rot,+}$ is the largest dimension of a non-trivial Jordan block of 
a matrix of the form $A_{Q,\infty}(\indexnottwo)$ (where $\indexnottwo\in\EFindexset$ is such that $\nuaverage{Q}(\indexnottwo)\neq 0$) corresponding 
to an eigenvalue with real part $\kappa_{\rotot,+}$; and the constant $C$ only depends on the spectra of the Riemannian manifolds $(M_{r},g_{r})$, 
$r\in\{1,\dots,R\}$, and the coefficients of the equation (\ref{eq:thesystemRge}). 
\item If $\baverage{Q}>0$, then 
\begin{equation}\label{eq:megenestlspresobsncasesobintro}
\mfe^{1/2}_{s}[u](t) \leq  C\ldr{t}^{d_{\rotot,+}-1}e^{\kappa_{\rotot,+}t}\mfe^{1/2}_{s}[u](0)
+C\int_{0}^{t}\ldr{t-t'}^{d_{\rotot,+}-1}e^{\kappa_{\rotot,+}(t-t')}\|f(\cdot,t')\|_{(s)}dt'
\end{equation}
holds for all $t\geq 0$, all solutions $u$ to (\ref{eq:thesystemRge}) and all $s\in\ro$. Here $d_{\rotot,+}:=\max\{d_{\rosil,+},d_{\ron,+}\}$; $d_{\rosil,+}$ is defined as 
in the case of (\ref{eq:megenestlspresobsilcasesobintro});
\[
d_{\ron,+}:=\max\{d_{\max}(R_{Q,+}^{+},\kappa_{\rotot,+}),d_{\max}(R_{Q,+}^{-},\kappa_{\rotot,+})\};
\]
and the constant $C$ only depends on the spectra of the Riemannian manifolds $(M_{r},g_{r})$, $r\in\{1,\dots,R\}$, and the coefficients of the equation 
(\ref{eq:thesystemRge}). 
\end{itemize}
\end{thm}
\begin{remark}
The statement follows from Theorem~\ref{thm:sobestndddbaconv}, cf. Remark~\ref{remark:maintheoremgeomcond}. 
\end{remark}

\subsection{Optimality}\label{ssection:optimalityintro}

Even though it is of interest to know that the estimate (\ref{eq:megenestlspresobsobintro}) holds, its value is dependent on whether it is optimal
or not. For this reason, we now address the issue of optimality.

\begin{thm}\label{thm:mainoptthmintro}
Consider the equation (\ref{eq:thesystemRge}) with $f=0$. Assume that it is geometrically non-degenerate, diagonally dominated, balanced and 
convergent; cf. Definition~\ref{def:geometricstrongassumptions}. Then $\nolossrate$, given by Definition~\ref{eq:defdecayrates}, satisfies
$\nolossrate=\kappa_{\rotot,+}$. Assume, in addition to the above, that the equation exhibits subdominant block non-degeneracy.  If 
$\baverage{Q}<0$ and $\kappa_{\rosil,+}=\kappa_{\rotot,+}$, 
then there is a constant $0<C\in\ro$ and a solution $u\neq 0$ to (\ref{eq:thesystemRge}) with $f=0$ such that 
\begin{equation}\label{eq:mfeoptlbsilsettintro}
\mfe^{1/2}[u](t)\geq C\ldr{t}^{d_{\rosil,+}-1}e^{\kappa_{\rotot,+}t}\mfe^{1/2}[u](0)
\end{equation}
for all $t\geq 0$. In this setting, the estimate (\ref{eq:megenestlspresobsilcasesobintro}) is thus optimal for $f=0$. If $\baverage{Q}=0$ and 
either $\kappa_{\rosil,+}=\kappa_{\rotot,+}$ or $\kappa_{Q,\trs,+}=\kappa_{\rotot,+}$ holds, then there is a constant $0<C\in\ro$ and a solution $u\neq 0$ 
to (\ref{eq:thesystemRge}) with $f=0$ such that 
\begin{equation}\label{eq:mfeoptlbtrssettintro}
\mfe^{1/2}[u](t)\geq C\ldr{t}^{d_{\rosts,+}-1}e^{\kappa_{\rotot,+}t}\mfe^{1/2}[u](0)
\end{equation}
for all $t\geq 0$. In this setting, the estimate (\ref{eq:megenestlspresobstscasesobintro}) is thus optimal for $f=0$. Finally, if $\baverage{Q}>0$ 
and either $\kappa_{\rosil,+}=\kappa_{\rotot,+}$ or $\kappa_{Q,+}=\kappa_{\rotot,+}$ holds, then there is a constant $0<C\in\ro$ and a solution $u\neq 0$ 
to (\ref{eq:thesystemRge}) with $f=0$ such that 
\begin{equation}\label{eq:mfeoptlbnsettintro}
\mfe^{1/2}[u](t)\geq C\ldr{t}^{d_{\rotot,+}-1}e^{\kappa_{\rotot,+}t}\mfe^{1/2}[u](0)
\end{equation}
for all $t\geq 0$. In this setting, the estimate (\ref{eq:megenestlspresobsncasesobintro}) is thus optimal for $f=0$.
\end{thm}
\begin{remark}
The statement follows from Theorem~\ref{thm:mainoptthm}, cf. Remark~\ref{remark:maintheoremgeomcond}. 
\end{remark}

\subsection{Examples}

In order to illustrate the results, we give two examples. 

\begin{example}\label{example:optimalityone}
Consider the equation
\begin{equation}\label{eq:modelilnoloss}
u_{tt}-e^{-2t}u_{\theta\theta}+e^{-t}Xu_{\theta}+\a u_{t}+\zeta u=0
\end{equation}
on $\so\times\ro$, where $X,\a,\zeta\in\ro$. The relevant metric in this case is 
\[
g=-dt\otimes dt+e^{2t}d\theta\otimes d\theta.
\]
In particular, $\bge=e^{2t}d\theta\otimes d\theta$ and $\bk=\bge$. Moreover, $\chi=0$. In particular, (\ref{eq:modelilnoloss})
is $C^{1}$-silent in the sense of Definition~\ref{definition:Cosilenceintro}. Note also that $\mcX:=e^{-t}X\d_{\theta}$ is $C^{0}$-future 
bounded and that (\ref{eq:alpahzetaconvest}) holds with $\a_{\infty}=\a$, $\zeta_{\infty}=\zeta$, $C_{\romn}=1$ and $\eta_{\romn}=1$. In
particular, Proposition~\ref{prop:genroughestintro} applies. In order to justify that the results of the present chapter are 
applicable, note that $\bk$ is diagonally convergent (with $\b_{1}=-1$) and that $\bge$ is $C^{2}$-asymptotically diagonal. Moreover, 
the shift vector field is negligible, and the main coefficients are convergent; $X^{1}_{\infty}=X$, $\a_{\infty}=\a$ and $\zeta_{\infty}=\zeta$. 
Finally, (\ref{eq:modelilnoloss}) is $C^{2}$-balanced and asymptotically non-degenerate. To conclude, (\ref{eq:modelilnoloss})
is geometrically non-degenerate, diagonally dominated, balanced and convergent; cf. Definition~\ref{def:geometricstrongassumptions}.
Thus the results of the present chapter apply. 

Assume that $\a>0$, $\zeta=\a^{2}/4+1$ and that $X=2\a+1$. Then Proposition~\ref{prop:genroughestintro} yields the conclusion that there are 
constants $C$ and $s_{0}$ such that 
\[
\mfe_{s}[u](t)\leq C\mfe_{s+s_{0}}[u](0)e^{-\a t}
\]
for all $t\geq 0$. Choosing $\a:=2\cdot 4711$, it is thus clear that (\ref{eq:mfe4711est}) holds. On the other hand, 
\[
R^{\pm}_{1,+}=\frac{1}{2}(-\a+\b_{1}\pm\tX^{1}_{\infty}).
\]
Moreover, $\b_{1}=-1$, $g^{11}_{\infty}=1$ and $X^{1}_{\infty}=X$. In particular, $R^{+}_{1,+}=\a/2$ and $\kappa_{\rotot,+}=\a/2$. Thus, 
Theorem~\ref{thm:mainoptthmintro} yields the conclusion that $\nolossrate=\a/2$. Let $\a:=2\cdot 4711$ and fix $0<\e,N\in\ro$. Then 
$\nolossrate=4711$. Let $\ma_{\ronl}$ be defined by (\ref{eq:maronldef}). Clearly, $4710\notin\ma_{\ronl}$. In other words, regardless of 
the choice of $C>0$, there is a solution $0\neq u\in\ms$ and a $t\geq 0$ such that $\mfe[u](t)\geq C\mfe[u](0)e^{9420t}$. Choosing
$C=\e^{-2}N^{2}$ and normalising $u$ so that $\mfe[u](0)=\e^{2}$, it is clear that 
\[
\mfe^{1/2}[u](0)=\e,\ \ \
\mfe^{1/2}[u](t)\geq N.
\]
This justifies the statements made in the introduction. 
\end{example}

\begin{example}\label{example:optimalitytwo}
Consider the equation (\ref{eq:optenestabpar}). It is of the same type as (\ref{eq:modelilnoloss}), so that the results of the 
present chapter apply. In the case of (\ref{eq:optenestabpar}), 
\[
\kappa_{\rosil,+}=-b,\ \ \
\kappa_{+}=\frac{1}{2}(-b-1+|a|),\ \ \
\kappa_{\rotot,+}=\frac{1}{2}\max\{-b-1+|a|,-2b\},
\]
where we assume that $b<0$. 
Appealing to Theorem~\ref{thm:mainoptthmintro} yields the conclusion that (\ref{eq:nolossratecalenestprob}) holds. Moreover, since 
(\ref{eq:optenestabpar}) exhibits subdominant block non-degeneracy, we know that (\ref{eq:mfebasenestnolossbaex}) holds; cf. 
Theorem~\ref{thm:sobestndddbaconvintro}. 
\end{example}

\section{Outline of the argument, outlook}\label{section:outlineofargoptenestintro}

The proofs of the results of the present chapter are provided in Part~\ref{part:nondegcabeq}, in particular
Chapters~\ref{chapter:diagdomconvbal}--\ref{chapter:estsobnormsolbodoftext}. However, they build upon some of the results of the 
previous chapters, in particular on those of Part~\ref{part:averaging}. We therefore begin by describing the contents of 
Part~\ref{part:averaging}.

\subsection{Averaging over oscillations}

Consider a solution $u$ to (\ref{eq:thesystemRge}). Let $z(\indexnot,t)$ be the $\indexnot$'th Fourier coefficient of $u(\cdot,t)$. 
Then $z$ satisfies (\ref{eq:fourierthesystemRgesigmaandX}). For at least some part of the interval $[0,\infty)$, the behaviour of 
$z$ can be expected to be oscillatory. The purpose of Part~\ref{part:averaging} is to develop methods for estimating how $z$ evolves 
in such oscillatory regimes. Before turning to the formal details, it is useful to develop some intuition. However, a rough idea of 
the perspective developed in Part~\ref{part:averaging} is given by the argument presented in 
Subsection~\ref{ssection:outlinedomnoisspdir}. Turning to the formal details, the first step is to calculate the matrix exponential 
of a special class of matrices; one example of an element of this class is given by $A_{\infty}(\indexnot,t_{0})T$ appearing in 
Subsection~\ref{ssection:outlinedomnoisspdir}.

\textbf{Matrix exponentials and approximations.} In order to estimate the overall evolution of solutions during one period of the
oscillations, it is useful to reformulate the equation; in the case of the limit equation in the noisy setting, the relevant 
reformulation is given by (\ref{eq:wdotnoislim})--(\ref{eq:bYrondef}). In the general case, the 
reformulation is given in Section~\ref{section:chofvariables} below. Freezing the coefficients in the reformulated equation, one is 
led to the problem of calculating the matrix exponential $\exp(2\pi P)$, where $P$ is a matrix of the form 
\begin{equation}\label{eq:Pdecomp}
P=\left(\begin{array}{cc} 0 & \Id_{m} \\ -\Id_{m} & 0 \end{array}\right)
+\left(\begin{array}{cc} i\xi \Id_{m}+R_{11} & 0 \\ R_{21} & -i\xi \Id_{m}-R_{11} \end{array}\right)+E.
\end{equation}
Here $E$ should be thought of as an error term and $\xi$ and $R_{ij}$ should be thought of as being small; cf. Section~\ref{section:calmaexp} 
for a detailed description of the assumptions. Note that $P$ has a very special structure. First of all, the exponential of $2\pi$ times
the first term on the right hand side of (\ref{eq:Pdecomp}) equals $\Id_{2m}$. Second, the second matrix on the right hand side 
of (\ref{eq:Pdecomp}) is trace free, block triangular and of lower order. In Section~\ref{section:calmaexp}, we calculate $\exp(2\pi sP)$ 
up to an error term
for $s\in [-1,1]$. In order to understand the overall evolution over longer periods of time, it is sufficient to focus on the case that
$s=1$. However, we are sometimes interested in going backwards in time (for instance when specifying asymptotics), and in that case, 
it is of interest to know what happens for $s=-1$. Finally, in the noisy setting, we are also interested in knowing what happens during
one period of the oscillations, and this is the motivation for taking an interest in all the $s\in [-1,1]$. 

\textit{Freezing the coefficients.} To estimate the evolution of solutions to non-constant coefficient equations (over one period), we 
freeze the coefficients and calculate the solution to the resulting constant coefficient equation. In order to estimate the error
involved, we need to compare the matrix exponential (which arises when freezing the coefficients) with the appropriate fundamental solution 
of the actual system of equations. This is the topic of Section~\ref{section:freezingcoeffabs}. 

\textbf{Assumptions and preliminary estimates.} Turning to equations of the form (\ref{eq:thesystemRge}), we state the relevant
assumptions in Section~\ref{section:assumponepertermaprel}. The assumptions are restrictive enough that we can analyse the behaviour
over one period of the oscillations. However, they are also general enough that the results apply in the situations considered in 
Chapters~\ref{chapter:onnotofbal} and \ref{chapter:domnoisspdirintro}, as well as in those considered in the present chapter. In
Section~\ref{section:chofvariables} we perform a change of variables that allows us to isolate the leading order behaviour during one
period of the oscillations. It also allows us to appeal to the matrix exponential calculations of Section~\ref{section:calmaexp}. 

\textit{Choosing time intervals.} Above, we speak of ``one period of the oscillations'' etc. In general, this notion is not
uniquely defined. However, for practical purposes, we here take one period of the oscillations to mean an interval $[t_{a},t_{b}]$ such that 
\[
\int_{t_{a}}^{t_{b}}\mfg(\indexnot,t)dt=2\pi.
\]

\textit{Variation within a period; a first approximation of the fundamental solution.} Before appealing to the results of 
Sections~\ref{section:calmaexp} and \ref{section:freezingcoeffabs}, we need to estimate the variation of the coefficients of the equation 
within one period of the oscillations. This is the purpose of Sections~\ref{section:choapptiint} and \ref{section:varcoeffoneper}.
Given the relevant estimates, we are then in a position to give a first approximation of the fundamental solution; this is the purpose
of Section~\ref{section:compappmaexp}. 

\textbf{Evolution over one period; iteration.} In Chapter~\ref{chapter:destopofoscref}, we consider the evolution over one period of the 
oscillations in greater detail. First, we derive a rough estimate in Section~\ref{section:roughestevooneper}. This estimate is then 
refined to a detailed estimate in Section~\ref{section:firstdetailestoneper}. 

\textit{Iteration.} In order to estimate the evolution of the solution over longer periods of time, it is convenient to introduce a 
time sequence $\{t_{k}\}$ such that $[t_{k},t_{k+1}]$ corresponds to one period of the oscillations. Introducing such a time sequence and
changing the variables appropriately yields an iteration relating the solution at $t_{k+1}$ to the solution at $t_{k}$. Moreover, the 
matrices involved in the relation can be well approximated by matrices that can be calculated from the coefficients of the equation. 
The iteration is the basic tool for analysing the behaviour of solutions over longer periods of time. The reader interested in the 
details is referred to Section~\ref{section:iteration}. 

\subsection{The asymptotically diagonal setting}

\textbf{Preliminaries.}
In Part~\ref{part:nondegcabeq} we turn to a study of asymptotically diagonal equations. To begin with, we give an analytical definition
of what this means in Section~\ref{section:diagdomconvbal}; in Part~\ref{part:appendices} we verify that the requirements of the analytical 
definition follow from the geometric conditions introduced in Definition~\ref{def:geometricstrongassumptions}. In Section~\ref{section:diagdomconvbal}, 
we also work out some of the consequences of the definition. 
In particular, we verify that the basic assumptions of Part~\ref{part:averaging} are satisfied. In Section~\ref{section:simplmacoeff}, we 
then turn to the problem of simplifying the matrix coefficients that appear in the iteration mentioned above. Using this simplification,
we are in a position to introduce a notion of ``frequency eras''; cf. Definition~\ref{def:era}. In the non-degenerate setting, a frequency 
era is a time interval in which one of the terms inside the parenthesis on the right hand side of (\ref{eq:mfginfasdiagnondeg}) dominates.
In Section~\ref{section:timeintfreeras}, we simplify the matrix coefficients further in frequency eras. Finally, in 
Section~\ref{section:estsumsfreera}, we estimate sums of the error terms that appear in the iteration. 

\textbf{Analysis for one mode.} Given the preliminaries, we are in a position to analyse how the energy associated with one mode evolves 
over time. This is the subject of Chapter~\ref{chapter:ndddconandbalancedeq}, a chapter in which we restrict our attention to the asymptotically
non-degenerate setting; cf. Definition~\ref{def:nondegconvabalintro}. To begin with, we devote Section~\ref{section:modeonefreqera} to estimating 
the evolution of the energy of one mode in one frequency
era. The basic tool for obtaining the estimates is the iteration. In the end, we therefore need to estimate the norm of a matrix product (where 
the number of factors depends on the frequency). The argument begins by a derivation of as good estimates of the constituent matrices as possible. 
As a second step, we simplify the constituent matrices by conjugating them with appropriate matrices. Finally, we estimate the norm of the 
matrix products; cf. the proof of Lemma~\ref{lemma:maprnoes} for the details. 

\textit{Refined estimates in transparent eras.} Lemma~\ref{lemma:maprnoes} applies both when going forward and when going
backwards in time. In the case that $\baverage{q}\neq 0$, the estimate is essentially optimal (with, possibly, the exception of a loss
$e^{\e t}$ in the decay). However, in the case that $\baverage{q}=0$, there is potentially a substantial loss. We therefore devote 
Section~\ref{section:refmodeanaltranspera} to deriving refined estimates in transparent eras. The main idea in this case is that for 
$\nuaverage{q}(\indexnot)$ large, the results of Lemma~\ref{lemma:maprnoes} yield good estimates. On the other hand, fixing a 
$0< K_{\roco}\in\ro$, there are only finitely many types of frequency eras with $\baverage{q}=0$ and $\nuaverage{q}(\indexnot)\leq K_{\roco}$.
For the latter class of frequency eras, a refined analysis can therefore be carried out, and the loss in the estimates is at worst of the form 
$e^{\e t}$. 

\textit{Estimates for one mode.} Given the analysis described above, we are in a position to derive estimates for one full mode. 
This is the purpose of Section~\ref{section:analysisforonemode}. The desired conclusion is essentially obtained by combining previous 
results. 

\textit{Unbounded frequency eras.} As already mentioned, many of the estimates involve a loss of the type $e^{\e t}$. However, in some cases,
this loss can be avoided. In particular, in the case of a frequency era of the form $[t_{0},\infty)$. In this case the loss $e^{\e t}$ can 
sometimes be replaced by a polynomial loss. Justifying this statement is the subject of Section~\ref{section:thecaseofunbdfreeras}; cf.,
in particular, Lemma~\ref{lemma:westunbfre}.

\textit{Optimality.} Finally, in Section~\ref{ssection:optimality}, we turn to the question of optimality. In order to obtain the desired
conclusions, it is sufficient to consider one frequency era. Moreover, we only need to derive a lower bound on the norm of the matrix
products, say $P_{k}$ appearing in the iteration. However, deriving such a lower bound is somewhat technical. First, we have to conjugate 
the constituent matrices in order to isolate the dominant part; then we have to choose the initial data $\xi_{0}$ appropriately to 
ensure maximal growth of $|P_{k}\xi_{0}|$; and, finally, we have to verify that the error terms do not cause problems. 

\textbf{Optimal energy estimates.} In Chapter~\ref{chapter:estsobnormsolbodoftext}, we are finally in a position to prove the results
stated in the present chapter. The argument essentially consists of summarising the estimates derived in previous chapters. In 
Section~\ref{section:ODEregime}, we begin by dividing $\EFindexset$ into different subsets. Then, in Proposition~\ref{prop:modeestentfuture},
we derive estimates for individual modes, based on this division of $\EFindexset$. Applying Minkowski's inequality to the estimates for 
the individual modes yields Theorem~\ref{thm:sobestndddbaconvintro}; cf. Section~\ref{section:estsobnorm} below. Finally, we prove 
optimality in Section~\ref{section:optsobest}. 

\subsection{Outlook} 

\textbf{Degeneracy.}
There are several ways in which one could think of improving the results of the present chapter. First of all, it would be desirable to 
remove the non-degeneracy condition. It is then to be expected that the one matrix $\tX^{q}_{\infty}$ appearing in (\ref{eq:matrixdetgrowthintro}) 
is replaced by a compact set of 
matrices; cf. $\Xaverage{\diag}^{q}(\indexnot)$ introduced in (\ref{eq:Xavdiagqindexnotdef}) below. This causes complications, but it
should be possible to derive conclusions nonetheless. For a non-degenerate equation with no transparent eras, the growth/decay rate is 
determined by finitely many matrices. In the degenerate setting, it is to be expected that the growth/decay rate is determined by the supremum 
of the real parts of the eigenvalues of matrices belonging to a compact set. Moreover, there would be infinitely many 
different matrices appearing in the estimates for the frequency modes. Nevertheless, it should be possible to carry out a corresponding 
analysis. 

\textbf{Detailed estimates for a single mode.} Considering a fixed mode, the time interval $[0,\infty)$ can be divided into
several parts. First, there is the initial interval $[0,T_{\roini}]$, where $T_{\roini}$ is, roughly speaking, defined by the condition that 
(\ref{eq:limiteqfourierasdiagset}) is a good approximation of the equation for $t\geq T_{\roini}$. The interval $[T_{\roini},\infty)$ is, in its
turn, divided into frequency eras. Due to the methods developed in Part~\ref{part:nondegcabeq} of these notes, we have a good understanding
of how the solutions behave in a frequency era. However, it would be of interest to give a description of how the Fourier coefficient
behaves for all $t\in [T_{\roini},\infty)$. In the present chapter we focus on estimates, but it would be of interest to derive 
more information. A first step is taken in Subsection~\ref{ssection:refinementsenest}, but it should be possible to do better.

\chapter{Improved asymptotic estimates in the silent setting}\label{chapter:impestinsilsetting}

Even though Proposition~\ref{prop:roughas} gives detailed information concerning the asymptotic behaviour of solutions in the silent setting, 
there is one fundamental problem: we have essentially no information concerning the values of $s_{\rohom}$ and $s_{\roih}$. 
Proposition~\ref{prop:spasda} suffers from a similar deficiency. In the present chapter, we state results analogous to these propositions, but
with specific values for the loss of derivatives involved in the estimates. 

\section{Estimating the asymptotic data in terms of the initial data}

Let us start by giving an estimate of the asymptotic data in terms of the initial data, with a loss of derivatives that can be calculated
in terms of the coefficients of the equation. 

\begin{prop}\label{prop:roughasexintro}
Consider the equation (\ref{eq:thesystemRge}). Assume that it is geometrically non-degenerate, diagonally dominated, balanced and convergent; cf. 
Definition~\ref{def:geometricstrongassumptions}. Assume that $\baverage{Q}<0$ and that $f$ is a smooth function such that for every $s\in\ro$, 
\begin{equation}\label{eq:fnosilsdefintro}
\|f\|_{\rosil,s}:=\int_{0}^{\infty}e^{-\kappa_{\rosil,+}t}\|f(\cdot,t)\|_{(s)}dt<\infty,
\end{equation}
where $\kappa_{\rosil,+}:=\kappa_{\max}(A_{\infty})$ and $A_{\infty}$ is defined by (\ref{eq:Asilentdefintro}). Let $\b_{\rem}:=\min\{-\baverage{Q},\kappa_{\romn}\}$, 
where $\kappa_{\romn}$ 
is the constant appearing in (\ref{eq:Xjalzeintro}). Fix $0<\b\leq\b_{\rem}$ and $\e>0$. Let $E_{a}$ be the first generalised eigenspace in the 
$\b,A_{\infty}$-decomposition of $\cn{2m}$; cf. Definition~\ref{def:defofgeneigenspintro}. 
Then there are constants $C_{\e,\b}$ and $N$, where $C_{\e,\b}$ only depends on $\e$, $\b$, the coefficients of the equation (\ref{eq:thesystemRge}) and 
the spectra of the Riemannian manifolds $(M_{r},g_{r})$, $r\in\{1,\dots,R\}$; and $N$ only depends on $m$, such that the following holds. Given a smooth 
solution $u$ to 
(\ref{eq:thesystemRge}), there is a $V_{\infty}\in C^{\infty}(\bM,E_{a})$ such that 
\begin{equation}\label{eq:uudothsestexintro}
\begin{split}
 & \left\|\left(\begin{array}{c} u(\cdot,t) \\ u_{t}(\cdot,t)\end{array}\right)
-e^{A_{\infty}t}V_{\infty}
-\int_{0}^{t}e^{A_{\infty}(t-\tau)}\left(\begin{array}{c} 0 \\ f(\cdot,\tau)\end{array}\right)d\tau\right\|_{(s)} \\
\leq & C_{\e,\b}\ldr{t}^{N}e^{(\kappa_{\rosil,+}-\b)t}\left(\|u_{t}(\cdot,0)\|_{(s+s_{\roh,\b,+}+\e)}+\|u(\cdot,0)\|_{(s+s_{\roh,\b,+}+\e+1)}
+\|f\|_{\rosil,s+s_{\roih,\b}+\e}\right)
\end{split}
\end{equation}
holds for all $t\geq 0$ and all $s\in\ro$. Here
\begin{align}
s_{\roh,\b,+} := & \max_{1\leq q\leq Q}\max\left\{0,-\frac{\b+\kappa_{q,+}-\kappa_{\rosil,+}}{\baverage{q}}\right\},\label{eq:shbpdefintro}\\
s_{\roih,\b} := & \max\left\{-\frac{\kappa_{\roode}}{\baverage{Q}},
-\frac{\b}{\baverage{Q}}+\max_{1\leq q\leq Q}\max\left\{0,-\frac{\kappa_{q,+}-\kappa_{\rosil,+}}{\baverage{q}}\right\}\right\},\label{eq:sihbdefintro}
\end{align}
where $\kappa_{\roode}:=\Rsp A_{\infty}$, cf. Definition~\ref{def:SpRspdef}, and $\kappa_{q,+}$ is defined by (\ref{eq:kappaqpmdefintro}). Moreover, 
\begin{equation}\label{eq:uinfudinfHsestexintro}
\|V_{\infty}\|_{(s)}\leq C_{\e,\b}\left(\|u_{t}(\cdot,0)\|_{(s+s_{\roh,\b}+\e)}+\|u(\cdot,0)\|_{(s+s_{\roh,\b}+\e+1)}+\|f\|_{A,s+s_{\roih,\b}+\e}\right),
\end{equation}
where 
\begin{equation}\label{eq:shbdefintro}
s_{\roh,\b}:=\max_{1\leq q\leq Q}\left(-\frac{\b+\kappa_{q,+}-\kappa_{\rosil,+}}{\baverage{q}}\right).
\end{equation}
\end{prop}
\begin{remark}
The proposition is a consequence of Proposition~\ref{prop:roughasex} and Remark~\ref{remark:optimalasest}. 
\end{remark}
\begin{remark}
The constants $\baverage{q}$, $q\in \{1,\dots,Q\}$, are introduced in Definition~\ref{def:nondegconvabalintro}.
\end{remark}
\begin{remark}
The function $V_{\infty}$ is uniquely determined by the fact that it satisfies the estimate (\ref{eq:uudothsestexintro}). 
\end{remark}
\begin{remark}
Both $s_{\roh,\b,+}$ and $s_{\roih,\b}$ are non-negative. However, $s_{\roh,\b}$ could be strictly negative. In this sense, the function $V_{\infty}$
could be more regular than the initial data. 
\end{remark}
\begin{remark}
As opposed to (\ref{eq:uudothsestintro}), $\b$ need not equal $\b_{\rem}$ in (\ref{eq:uudothsestexintro}). The reason for allowing $\b$ to vary in the 
range $(0,\b_{\rem}]$ is that a higher $\b$ yields more detailed asymptotics, but a lower $\b$ yields a lower loss of derivatives. 
\end{remark}
\begin{remark}
Under some circumstances, the $\e$ appearing in (\ref{eq:uudothsestexintro}) and (\ref{eq:uinfudinfHsestexintro}) can be removed. 
See Remarks~\ref{remark:homimprsilasest} and \ref{remark:homimprsilasfunest} for more details. 
\end{remark}

In order to illustrate the result, let us consider an example.

\begin{example}\label{example:optasestimates}
Consider the equation (\ref{eq:modelilnoloss}). It satisfies the conditions of Proposition~\ref{prop:roughasexintro}. Moreover, $Q=1$; 
$\baverage{1}=-1$; (\ref{eq:Xjalzeintro}) holds with $\kappa_{\romn}=1$; and $\b_{\rem}=1$. Let 
\[
\zeta=\frac{1}{4}\a^{2}\pm \frac{1}{4}Y^{2},
\]
where $Y\in\ro$. There are two cases to consider. 

Assuming, for the sake of argument, that $\zeta\geq \a^{2}/4$, we know that $\kappa_{\rosil,+}=-\a/2$ and that $\kappa_{\roode}=0$. Moreover, 
\[
\kappa_{1,+}=\frac{1}{2}(-\a+\baverage{1}+|X|).
\]
Thus
\[
s_{\roh,\b}=\b-\frac{1}{2}(1-|X|).
\]
In particular, for small $\b$ and $|X|$, we know that $s_{\roh,\b}$ is negative. Note also that, regardless of how small $\b$ is, $E_{a}=\cn{2}$,
and the map from initial data to asymptotic data is a homeomorphism. If $X=0$, (\ref{eq:uinfudinfHsestexintro}) thus implies that for every $\e>0$,
there is a constant $C_{\e}$ such that 
\[
\|V_{\infty}\|_{(s)}\leq C_{\e}\left(\|u_{t}(\cdot,0)\|_{(s-1/2+\e)}+\|u(\cdot,0)\|_{(s+1/2+\e)}\right).
\]
Moreover, for $\b\leq 1/2$, $s_{\roh,\b,+}=0$, so that (\ref{eq:uudothsestexintro}) only involves an $\e$-loss in derivatives. On the other hand, 
the time dependence of the right hand side is then not optimal. In fact, we are allowed to choose $\b=1$, and then the right hand side of
(\ref{eq:uudothsestexintro}) has a time dependence of the form $\ldr{t}^{N}e^{(\kappa_{\rosil,+}-1)t}$ as opposed to $\ldr{t}^{N}e^{(\kappa_{\rosil,+}-1/2)t}$. However, the 
price for the improvement as far as the time dependence is concerned is a loss of derivatives. 

Assume now that $\zeta\leq \a^{2}/4$. Then $\kappa_{\rosil,+}=-\a/2+|Y|/2$. On the other hand, $\kappa_{1,+}$ is the same as before so that 
\[
s_{\roh,\b}=\b-\frac{1}{2}(1+|Y|-|X|).
\]
In this case, it is thus clear that $s_{\roh,\b}$ can be made arbitrarily negative, by making $|Y|$ large enough. On the other hand, 
$\Rsp A_{\infty}=\kappa_{\roode}=|Y|$. Thus, if $\b\leq |Y|$, $E_{a}$ is (complex) one dimensional. If we wish $E_{a}$ to be two dimensional, we have
to have $\b>|Y|$, in which case 
\[
s_{\roh,\b}>\frac{1}{2}(|Y|+|X|-1)\geq -\frac{1}{2}.
\]
If we want complete asymptotic information, we can thus not hope for $s_{\roh,\b}\leq -1/2$. Note also that if $|Y|\geq 1$, we, in the present
context, do not obtain a homeomorphism between initial data and asymptotic data. 
\end{example}

\section{Estimating the initial data in terms of the asymptotic data}

Next, we improve Proposition~\ref{prop:spasda} by expressing the derivative losses in terms of the coefficients of the equation.

\begin{prop}\label{prop:spasdaexintro}
Consider the equation (\ref{eq:thesystemRge}) with $f=0$. Assume that it is geometrically non-degenerate, diagonally dominated, balanced and convergent; 
cf. Definition~\ref{def:geometricstrongassumptions}. Assume that $\baverage{Q}<0$ and let $\b_{\rem}:=\min\{-\baverage{Q},\kappa_{\romn}\}$, where 
$\kappa_{\romn}$ is the constant appearing in (\ref{eq:Xjalzeintro}). Fix $0<\b\leq\b_{\rem}$ and $\e>0$. Let $E_{a}$ be the first generalised eigenspace 
in the $\b,A_{\infty}$-decomposition of $\cn{2m}$; cf. Definition~\ref{def:defofgeneigenspintro}. Then there is a linear injective map
\[
\Phi_{\b,\infty}:C^{\infty}(\bM,E_{a})\rightarrow C^{\infty}(\bM,\cn{2m})
\]
such that the following holds. First, if $\Phi^{j}_{\b,\infty}:C^{\infty}(\bM,E_{a})\rightarrow C^{\infty}(\bM,\cn{m})$, $j=1,2$, are defined by 
the condition that 
\[
\Phi_{\b,\infty}(\psi)=\left(\begin{array}{c} \Phi^{1}_{\b,\infty}(\psi) \\ \Phi^{2}_{\b,\infty}(\psi)\end{array}\right)
\]
for all $\psi\in C^{\infty}(\bM,E_{a})$, then 
\begin{equation}\label{eq:Phiinfnobdexintro}
\|\Phi_{\b,\infty}^{1}(\psi)\|_{(s+1)}+\|\Phi_{\b,\infty}^{2}(\psi)\|_{(s)}\leq C_{\e}\|\psi\|_{(s+s_{\roh,-}+\e)}
\end{equation}
for all $s\in\ro$ and all $\psi\in C^{\infty}(\bM,E_{a})$, where $C_{\e}$ only depends on $\e$, the coefficients of the equation 
(\ref{eq:thesystemRge}) and the spectra of $(M_{r},g_{r})$, $r\in\{1,\dots,R\}$, and 
\begin{equation}\label{eq:srohmdefintro}
 s_{\roh,-}:=\max_{1\leq q\leq Q}\left(-\frac{\kappa_{q,-}+\kappa_{\rosil,+}}{\baverage{q}}\right).
\end{equation}
Second, if $\psi\in C^{\infty}(\bM,E_{a})$ and $u$ is the solution to (\ref{eq:thesystemRge}) (with $f=0$) such that 
\begin{equation}\label{eq:uuditoPhiinfchiexintro}
\left(\begin{array}{c} u(\cdot,0) \\ u_{t}(\cdot,0)\end{array}\right)=\Phi_{\b,\infty}(\psi),
\end{equation}
then 
\begin{equation}\label{eq:estspecasdataexintro}
\begin{split}
 & \left\|\left(\begin{array}{c} u(\cdot,t) \\ u_{t}(\cdot,t)\end{array}\right)
-e^{A_{\infty}t}\psi\right\|_{(s)} \\
\leq & C_{\e,\b}\ldr{t}^{N}e^{(\kappa_{\rosil,+}-\b)t}\left(\|u_{t}(\cdot,0)\|_{(s+s_{\roh,\b,+}+\e)}+\|u(\cdot,0)\|_{(s+s_{\roh,\b,+}+\e+1)}\right)
\end{split}
\end{equation}
holds for all $t\geq 0$ and all $s\in\ro$, 
where $C_{\e,\b}$ only depends on $\e$, $\b$, the coefficients of the equation (\ref{eq:thesystemRge}) and the spectra of the Riemannian manifolds 
$(M_{r},g_{r})$, $r\in\{1,\dots,R\}$, and $N$ only depends on $m$. Finally, if $E_{a}=\cn{2m}$ (i.e., if $\b>\Rsp A_{\infty}$; cf. 
Definition~\ref{def:SpRspdef}), then $\Phi_{\b,\infty}$ is surjective. 
\end{prop}
\begin{remark}
The proposition is a consequence of Proposition~\ref{prop:spasdaex} and Remark~\ref{remark:optimalasspecest}.
\end{remark}
\begin{remark}
Again, the $\e$ appearing in (\ref{eq:Phiinfnobdexintro}) can be removed under appropriate circumstances; cf. Remark~\ref{remark:removingepsilonbackwest}.
\end{remark}

Again, let us illustrate the result with an example. 

\begin{example}
Consider the equation (\ref{eq:modelilnoloss}). It satisfies the conditions of Proposition~\ref{prop:spasdaexintro} and in what follows, we use the 
notation introduced in Example~\ref{example:optasestimates}. As before, there are two cases to consider. 

Assume that $\zeta\geq \a^{2}/4$. Then $\kappa_{\rosil,+}=-\a/2$ and 
\[
\kappa_{1,-}=\frac{1}{2}(\a-\baverage{1}+|X|),
\]
so that 
\[
s_{\roh,-}=\frac{1}{2}(1+|X|).
\]
Note that $|X|$ only leads to a loss. Comparing with Example~\ref{example:optasestimates}, we see that if $X=0$, we can, roughly speaking,
gain $1/2$ derivative when going from initial data to asymptotic data, but we lose roughly $1/2$ derivative when going back. Introducing 
$X\neq 0$ leads to an additional loss in both directions. 

Assume that $\zeta\leq \a^{2}/4$. In this case, $\kappa_{1,-}$ remains the same, but $\kappa_{\rosil,+}$ changes. In fact, 
\[
s_{\roh,-}=\frac{1}{2}(1+|X|+|Y|).
\]
In other words, $X$ and $Y$ only yield a loss of derivatives. Comparing with Example~\ref{example:optasestimates}, we see that the (potential) 
gain caused by $Y$ when going from initial data to asymptotic data is lost when going back. 
\end{example}

\section{Outline of the argument, outlook}

The proofs of Propositions~\ref{prop:roughasexintro} and \ref{prop:spasdaexintro} build on Part~\ref{part:averaging} and 
Chapters~\ref{chapter:diagdomconvbal}--\ref{chapter:estsobnormsolbodoftext}. However, the argument is based on a more refined 
version of the analysis of the Fourier coefficients than the one used to obtain optimal energy estimates. In fact, in order to 
obtain optimal energy estimates, it is sufficient to isolate the frequency era corresponding to the least amount of decay. 
However, in order to calculate the loss of derivatives of interest in the present chapter, we need to consider all the frequency
eras for a given mode. 

\textbf{Refined analysis.} Consider a Fourier coefficient $z$ corresponding to a solution to (\ref{eq:thesystemRge}) and a $\indexnot\in\EFindexset$. 
Under the assumptions of Propositions~\ref{prop:roughasexintro} and \ref{prop:spasdaexintro}, there is a $T_{\roode}$ such that the ODE-behaviour is 
dominant in $[T_{\roode},\infty)$. The interval $[0,T_{\roode}]$ can, in its turn, be divided into 
frequency eras, say $[0,T_{1}],\dots,[T_{l},T_{\roode}]$. In each frequency era, there is a worst case scenario as far as the growth/decay is concerned,
say $\lambda_{j}$ in $[T_{j},T_{j+1}]$. This leads, roughly speaking, to a growth of the form 
\[
\exp[\lambda_{0}T_{1}+\dots+\lambda_{l}(T_{\roode}-T_{l})+\kappa_{\rosil,+}(t-T_{\roode})]
\] 
for $t\geq T_{\roode}$. On the other hand, $T_{l+1}-T_{l}$ can be estimated in terms of $\indexnot$. Moreover, we, in the end, wish to compare this 
growth with $e^{\kappa_{\rosil,+}t}$. This leads to the problem of estimating, e.g.,
\begin{equation}\label{eq:lambdajminkappaonedeltaT}
\exp[(\lambda_{0}-\kappa_{\rosil,+})T_{1}+\dots+(\lambda_{l}-\kappa_{\rosil,+})(T_{\roode}-T_{l})].
\end{equation}
In particular, we are interested in the worst case scenario as we vary $\indexnot$. Carrying out the corresponding analysis leads to the conclusion that 
expressions of the form (\ref{eq:lambdajminkappaonedeltaT}) can be estimated by $C\ldr{\nu(\indexnot)}^{s_{a}}$ for some $s_{a}\in\ro$ depending on the
$\baverage{q}$, the $\kappa_{q,+}$ and $\kappa_{\rosil,+}$. This is, roughly speaking, the argument 
by which one arrives at the number $s_{\roh,\b}$. Note that all the $\lambda_{j}-\kappa_{\rosil,+}$ could be negative; this is the origin of the (potential) gain of 
derivatives in the estimate for $V_{\infty}$. In the case of $s_{\roh,\b,+}$, the estimate has to be valid for all $t\geq 0$, and it is therefore clear that 
$s_{\roh,\b,+}$ has to be bounded from below by zero. 

\textbf{Outline.} The main structure of the proofs of Propositions~\ref{prop:roughasexintro} and \ref{prop:spasdaexintro} is similar to that of the proofs 
of Propositions~\ref{prop:roughas} and \ref{prop:spasda}; cf. Subsection~\ref{ssection:outloftheargsilsetting} for a rough description of the arguments in 
the case of the latter propositions. However, in order to obtain the desired estimates, we need to appeal to the refined analysis described above. 
The details are provided in Sections~\ref{section:roughODEfutasex} and \ref{section:roughODEspecasex}. 

\subsection{Outlook}

In the present chapter, we only consider the silent setting. It would, however, also be of interest to derive similar conclusions in the transparent
and noisy settings. Using the methods developed in these notes, it should be possible to do so. However, the required arguments can be expected to be 
lengthy.

\part{Rough analysis in the silent and in the transparent settings}\label{part:roughansiltrs}

\chapter{ODE's with exponentially decaying error terms}\label{chapter:roughanalysisODEregion}

\textbf{Introduction.} Consider (\ref{eq:fourierthesystemRge}), obtained by decomposing (\ref{eq:thesystemRge}) into Fourier modes. Introducing
additional variables, it can be reformulated as a first order system 
\begin{equation}\label{eq:ODEregmodintro}
\dot{v}(t)=B(t)v(t)+F(t).
\end{equation}
The function $F$ arises from $\hf$ appearing on the right hand side of (\ref{eq:fourierthesystemRge}) and depends on the frequency 
$\indexnot\in\EFindexset$ and $t$. In general, $B$ also depends on both $\indexnot$ and $t$. However, in the silent setting, the matrix 
can be written as a sum $B(t)=A+A_{\rem}(t)$, where $A$ is a constant matrix and $A_{\rem}$ decays exponentially. Moreover, 
$A$ does not depend on $\indexnot$. This is the simplest situation, and the one we consider in the present chapter. The case of transparent equations 
is similar. However, in that case, $A$ depends on $\indexnot$, a more complicated situation which we discuss in Chapter~\ref{chapter:ODEtransp}.
Even though the division of $B$ into $A$ and $A_{\rem}$ corresponds to a division into a dominant part $A$ and an error term $A_{\rem}$, there
is one problem: $A_{\rem}$ depends on $\indexnot$, and $\|A_{\rem}(0)\|$ typically tends on $\infty$ as $|\nu(\indexnot)|\rightarrow\infty$. In 
that sense, $A_{\rem}$
cannot be considered to be negligible. On the other hand, the equations of interest are such that for each $\indexnot$, there is a time, say $T_{\roode}$
(depending on $\indexnot$), such that 
\begin{equation}\label{eq:Aremestintro}
\|A_{\rem}(t)\|\leq C_{\rem}e^{-\b_{\rem}(t-T_{\roode})}
\end{equation}
for $t\geq T_{\roode}$. Moreover, 
$C_{\rem}>0$ and $\b_{\rem}>0$ are independent of $\indexnot$. Here we refer to the interval $[T_{\roode},\infty)$ as the \textit{non-oscillatory} or 
\textit{ODE era} 
\index{ODE!era}%
\index{Era!ODE}%
\index{Non-oscillatory!era}%
\index{Era!non-oscillatory}%
of the relevant Fourier mode, and this is the interval we focus on in the present chapter.

\textbf{Asymptotics.} In this chapter we analyse the asymptotics of solutions to 
\begin{equation}\label{eq:ODEregmod}
\dot{v}(t)=Av(t)+A_{\rem}(t)v(t)+F(t),
\end{equation}
where $A_{\rem}$ satisfies (\ref{eq:Aremestintro}) and $A$ is constant. 
To begin with, it can be demonstrated that, roughly speaking, solutions do not grow faster than solutions to $\dot{v}(t)=Av(t)+F(t)$; cf. 
Lemma~\ref{lemma:oderegest}. In order to obtain detailed asymptotics, it is necessary to make more detailed assumptions concerning $F$. If
$F$ grows more quickly (or decays less slowly) than the fastest growing solution to $\dot{v}=Av$, then $F$ can be expected to have a dominant
influence on the asymptotics. However, there might be cancellations associated with, e.g., oscillatory behaviour in $F$. In order to obtain detailed
asymptotics in such a situation, it can thus be expected to be necessary to make detailed assumptions concerning $F$. This is not our main interest 
here, and we therefore restrict our attention to the case that the fastest growing solutions to $\dot{v}=Av$ grow more quickly than $F$. To be more 
precise, we assume that 
\begin{equation}\label{eq:Fnoone}
\|F\|_{A}:=\int_{0}^{\infty}e^{-\kappa_{1}s}|F(s)|ds<\infty,
\end{equation}
\index{$\a$Aa@Notation!Norms!$\normRHSA$}%
where $\kappa_{1}:=\kappa_{\max}(A)$; cf. Definition~\ref{def:SpRspdef}. Under these circumstances, Lemma~\ref{lemma:oderegest} ensures that 
solutions to (\ref{eq:ODEregmod}) do not grow faster than the fastest growing solutions to $\dot{v}=Av$; i.e., 
\begin{equation}\label{eq:vaprestintro}
|v(t)|\leq C\ldr{t-T_{\roode}}^{d_{1}-1}e^{\kappa_{1}(t-T_{\roode})}
\end{equation}
for $t\geq T_{\roode}$, where we have used the notation (\ref{eq:ldrdefinitionintro}) and $d_{1}:=d_{\max}(A,\kappa_{1})$; cf. 
Definition~\ref{def:SpRspdef}. At this stage, it would seem natural to return to 
(\ref{eq:ODEregmod}) in order to calculate the leading order terms of the asymptotics. Compute, to this end, 
\begin{equation}\label{eq:ODEremainderintro}
\frac{d}{dt}\left(e^{-At}v-\int_{T_{\roode}}^{t}e^{-As}F(s)ds\right)=e^{-At}A_{\rem}(t)v(t).
\end{equation}
Under ideal circumstances, the right hand side decays exponentially, so that we can integrate this equality. This would lead to the conclusion
that the expression inside the parenthesis on the left hand side converges exponentially. Thus $v$ would, to good approximation, behave as a 
solution to (\ref{eq:ODEregmod}) with $A_{\rem}$ set to zero. However, even though $A_{\rem}$ and $v$ satisfy (\ref{eq:Aremestintro}) and 
(\ref{eq:vaprestintro}) respectively, there is no reason why the right hand side of (\ref{eq:ODEremainderintro}) should decay exponentially.
Whether we obtain exponential decay or not is determined by the spread of the real parts of the eigenvalues of $A$; cf. 
Definition~\ref{def:SpRspdef}. If $\Rsp A<\b_{\rem}$, then the right hand side of (\ref{eq:ODEremainderintro}) decays exponentially. However, if 
$\Rsp A\geq\b_{\rem}$ we need to proceed differently. By conjugating $A$ by a suitable $T\in\Gl{k}{\co}$, the resulting matrix
is in block form; $T^{-1}AT=\diag\{A_{a},A_{b}\}$. Moreover, $\kappa_{\max}(A_{a})=\kappa_{\max}(A)$, $\Rsp A_{a}<\b_{\rem}$ and 
$\kappa_{\max}(A_{b})\leq \kappa_{\max}(A)-\b_{\rem}$; cf. Definition~\ref{def:SpRspdef} for an explanation of the notation. This follows from 
transforming $A$ to Jordan normal form and arranging the Jordan blocks in 
a suitable way. Letting $w(t):=T^{-1}v(t)$ and letting $w_{a}$ and $w_{b}$ denote the components of $w$ corresponding to the blocks $A_{a}$ and 
$A_{b}$ respectively, (\ref{eq:ODEregmod}) can be divided into the two equations
\begin{align}
\dot{w}_{a}(t) = & A_{a}w_{a}(t)+[T^{-1}A_{\rem}(t)v(t)]_{a}+[T^{-1}F(t)]_{a},\label{eq:waintro}\\
\dot{w}_{b}(t) = & A_{b}w_{b}(t)+[T^{-1}A_{\rem}(t)v(t)]_{b}+[T^{-1}F(t)]_{b}. \label{eq:wbintro}
\end{align}
Here $\xi_{a}$ and $\xi_{b}$ denote the components of $\xi\in\cn{k}$ corresponding to the blocks $A_{a}$ and $A_{b}$ respectively. 
In the case of $w_{a}$, we can carry out a computation similar to (\ref{eq:ODEremainderintro}) by appealing to (\ref{eq:waintro}). 
In this way we obtain the leading order behaviour of $w_{a}$. However, in the case of $w_{b}$ we only obtain an estimate. This is not 
so surprising, since the second term on the right hand 
side of (\ref{eq:wbintro}) is, on the one hand, an error term (the only information we have concerning $A_{\rem}$ is an estimate). On the
other hand, this term could potentially grow more quickly than the fastest growing solution to $\dot{w}_{b}=A_{b}w_{b}$. For this reason,
it is not realistic to hope to extract asymptotic information concerning $w_{b}$. 

\textbf{Outline.} The outline of the present chapter is as follows. In Section~\ref{section:ODEappr}, we state the equations of interest
and describe in greater detail the algebraic decomposition leading to (\ref{eq:waintro}) and (\ref{eq:wbintro}). Then we proceed to derive 
a rough estimate for solutions to (\ref{eq:ODEregmod}). This is the subject of Section~\ref{section:roughODEest}. Given this information, it 
is then possible to derive detailed asymptotics concerning $w_{a}$ in the above decomposition; cf. Section~\ref{section:detailedODEas}. This
gives a map from initial data for (\ref{eq:ODEregmod}) to asymptotic data for $w_{a}$. However, it is also of interest to go in the other
direction; i.e., to start with asymptotic data for $w_{a}$ and to construct initial data for (\ref{eq:ODEregmod}) such that the corresponding 
solution yields the desired asymptotic data for $w_{a}$. In Section~\ref{section:spdatinfODE} we demonstrate that this is possible. In all of 
the arguments, it is of great importance to keep track of the constants involved, since we, in the end, are interested in an infinite sequence 
of equations and a corresponding infinite sequence of times $T_{\roode}$.

\section{Equations and algebraic decompositions}\label{section:ODEappr}

In this chapter we are concerned with the non-oscillatory, or ODE, era of a Fourier mode of an asymptotically silent equation. However, 
it is convenient to formulate general results that do not refer to the PDE origin of the problem. Let us start by describing the class of 
equations we are interested in. 

\subsection{Class of equations}\label{ssection:ODEreg}
We are interested in equations of the form (\ref{eq:ODEregmod}) for a $\cn{k}$-valued function $v$, $1\leq k\in\zo$, where $F$ is a given smooth
$\cn{k}$-valued function on $\ro$; $A\in\Mn{k}{\co}$; and $A_{\rem}$ is a smooth function from $\ro$ to $\Mn{k}{\co}$. We assume, moreover, that 
there are constants $C_{\rem},T_{\roode}\geq 0$ and $\b_{\rem}>0$ such that 
\begin{equation}\label{eq:Aremass}
\|A_{\rem}(t)\|\leq C_{\rem}e^{-\b_{\rem}\bt}
\end{equation}
for $t\geq T_{\roode}$, where $\bt:=t-T_{\roode}$. We refer to $\b_{\rem}$ as the \textit{ODE margin}. 
\index{ODE!margin}%
\index{Margin!ODE}%

\subsection{Motivation and goal}
The equation (\ref{eq:ODEregmod}) should be thought of as a model equation for the Fourier modes of an asymptotically silent equation. In that setting, 
there are infinitely many different modes (corresponding to infinitely many different matrices $A_{\rem}$). However, there are universal constants 
$\b_{\rem}$ and $C_{\rem}$ such that (\ref{eq:Aremass}) holds for $T_{\roode}$ large enough. Note, however, that $T_{\roode}$ does depend on the mode. Our goal 
is to estimate the difference between $v$ and an appropriate solution to the equation that results by removing the second term on the right hand side from 
(\ref{eq:ODEregmod}). The estimate should be valid for $t\geq T_{\roode}$. Since $A$, $\b_{\rem}$ and $C_{\rem}$ should be thought of as universal 
(independent of mode), it is not of any greater importance to spell out the dependence of constants (appearing in the estimates) on $C_{\rem}$, $\b_{\rem}$ 
and $A$ explicitly. However, we need to keep very careful track of the dependence on $T_{\roode}$.

\subsection{Decomposition of $A$}\label{ssection:decompofA}
The complex $k\times k$-matrix $A$ plays a central role in (\ref{eq:ODEregmod}). Many of the results below will be phrased in terms of different 
algebraic decompositions of $A$. For the sake of clarity, we here collect all the details of these decompositions in one subsection; see also 
Section~\ref{section:jordannormalform} below for a summary of some of the basic facts concerning the Jordan normal form and the matrix exponential. Our motivation
for decomposing $A$ is described in connection with (\ref{eq:waintro}) and (\ref{eq:wbintro}). In particular, the starting point, motivating the 
decomposition, is the fact that the ODE margin might be strictly smaller than the real eigenvalue spread of $A$; i.e., that the inequality
$\b_{\rem}<\Rsp A$ might hold. 

\begin{lemma}\label{lemma:Apredecomp}
Given $1\leq k\in\zo$, $0<\b\in\ro$ and $A\in\Mn{k}{\co}$, there is a $T\in\Gl{k}{\co}$ such that $A_{J}:=T^{-1}AT$ has the following properties.
To begin with,
\[
A_{J}=T^{-1}AT=\diag\{A_{a},A_{b}\},
\]
where $A_{a}\in\Mn{k_{a}}{\co}$; $A_{b}\in\Mn{k_{b}}{\co}$; $1\leq k_{a}\in\zo$ and $0\leq k_{b}\in\zo$ are such that $k_{a}+k_{b}=k$; and $A_{a}$ and 
$A_{b}$ consist of Jordan blocks. Moreover, $\Rsp A_{a}<\b$; $\kappa_{\max}(A_{a})=\kappa_{\max}(A)$; and $\kappa_{\max}(A_{b})\leq \kappa_{\max}(A)-\b$ 
(assuming $k_{b}\geq 1$). 
\end{lemma}
\begin{remark}
See Section~\ref{section:jordannormalform} below for the definition of Jordan blocks and a summary of some of the basic facts concerning the Jordan normal form.
\end{remark}
\begin{remarks}
If $k_{b}=0$, then $\diag\{A_{a},A_{b}\}$ should be interpreted as equalling $A_{a}$. Moreover, the notation $\kappa_{\max}(A)$ is introduced
in Definition~\ref{def:SpRspdef}. The matrices $T$, $A_{a}$ etc. are not unique. However, the dimensions $k_{a}$ and $k_{b}$ are well defined.
\end{remarks}
\begin{proof}
The statement follows by transforming $A$ to Jordan normal form and then arranging the Jordan blocks appropriately. 
\end{proof}

\begin{definition}\label{def:fssubspetc}
Given $1\leq k\in\zo$, $A\in\Mn{k}{\co}$ and $0<\b\in\ro$, let $k_{a}$ and $k_{b}$ be the integers obtained by appealing to 
Lemma~\ref{lemma:Apredecomp}. Below, $k_{a}$ and $k_{b}$ are referred to as the \textit{dimensions of the first and second subspaces} 
(respectively) \textit{of the $\b$, $A$-decomposition of} $\cn{k}$.  
\index{$\a$Aa@Notation!Vector spaces!$\b$, $A$-decomposition of $\cn{k}$}%
\index{$\a$Aa@Notation!Vector spaces!$\b$, $A$-decomposition of $\cn{k}$}%
\index{$\a$Aa@Notation!Vector spaces!$\b$, $A$-decomposition of $\cn{k}$; dimension of first (second) subspace of}%
If $B\in\Mn{k}{\co}$ can be written $B=\diag\{B_{a},B_{b}\}$, where 
$B_{a}\in\Mn{k_{a}}{\co}$ and $B_{b}\in\Mn{k_{b}}{\co}$, then $B_{a}$ ($B_{b}$) is referred to as the \textit{first (second) block of the $\b$, 
$A$-decomposition of} $B$. 
\index{$\a$Aa@Notation!Matrix notation!badec@$\b$, $A$-decomposition of a matrix}%
\index{$\a$Aa@Notation!Matrix notation!badec@$\b$, $A$-decomposition of a matrix; first (second) block of}%
If $\xi\in\cn{k}$ and 
$\xi_{a}\in\cn{k_{a}}$ collects the first $k_{a}$ components of $\xi$ and $\xi_{b}\in\cn{k_{b}}$ collects the last $k_{b}$ components of $\xi$, then 
$\xi_{a}$ ($\xi_{b}$) is referred to as the \textit{first (second) component in the $\b$, $A$-decomposition of $\xi$}. 
\index{$\a$Aa@Notation!Matrix notation!badec@$\b$, $A$-decomposition of a vector}%
\index{$\a$Aa@Notation!Matrix notation!badec@$\b$, $A$-decomposition of a vector; first (second) component of}%
Finally, if $T$ is obtained by 
appealing to Lemma~\ref{lemma:Apredecomp}, $E_{a}:=T(\cn{k_{a}}\times\{0\}^{k_{b}})$ and $E_{b}:=T(\{0\}^{k_{a}}\times\cn{k_{b}})$, then $E_{a}$ 
($E_{b}$) is referred to as the \textit{first (second) generalised eigenspace in the $\b$, $A$-decomposition of $\cn{k}$}. 
\index{$\a$Aa@Notation!Vector spaces!$\b$, $A$-decomposition of $\cn{k}$; first (second) generalised eigenspace}%
The first generalised eigenspace in the $\b$, $A$-decomposition of $\cn{k}$ is also denoted by $E_{A,\b}$. 
\end{definition}
\begin{remark}
The vector space $E_{a}$ ($E_{b}$) is the direct sum of the generalised eigenspaces of $A$ corresponding to eigenvalues in $\Spe A_{a}$ ($\Spe A_{b}$);
here $A_{a}$ and $A_{b}$ are obtained by appealing to Lemma~\ref{lemma:Apredecomp}. In particular, the spaces $E_{a}$ and $E_{b}$ are uniquely
determined by $A$ and $\b>0$. Moreover, $\dim E_{a}=k_{a}$ and $\dim E_{b}=k_{b}$. Finally, note that the present definition is consistent with 
Definition~\ref{def:defofgeneigenspintro}. 
\end{remark}

In our applications, the following estimate will be of interest.

\begin{lemma}\label{lemma:eAtPibnormest}
Given $1\leq k\in\zo$, $A\in\Mn{k}{\co}$ and $0<\b\in\ro$, let $E_{a}$ ($E_{b}$) be the first (second) generalised eigenspace in the $\b$, $A$-decomposition 
of $\cn{k}$. Let $\Pi_{a}:\cn{k}\rightarrow E_{a}$ and $\Pi_{b}:\cn{k}\rightarrow E_{b}$ be defined by the condition that $x=\Pi_{a}x+\Pi_{b}x$ for all 
$x\in\cn{k}$. Let $\kappa_{b}:=\kappa_{\max}(A_{b})$ and $d_{b}:=d_{\max}(A_{b},\kappa_{b})$, where $A_{b}$ is the restriction of $A$ to $E_{b}$; cf. 
Definition~\ref{def:SpRspdef}. Then
\[
\|e^{At}\Pi_{b}\|\leq C_{A}\ldr{t}^{d_{b}-1}e^{\kappa_{b}t}
\]
for all $t\geq 0$, where $C_{A}$ only depends on $A$. 
\end{lemma}
\begin{proof}
Let $T$, $k_{a}$ and $k_{b}$ be obtained by appealing to Lemma~\ref{lemma:Apredecomp} and let $\pi_{b}:\cn{k}\rightarrow \{0\}^{k_{a}}\times\cn{k_{b}}$ be the 
projection onto the last $k_{b}$ elements of $\cn{k}$. Then $\Pi_{b}=T\pi_{b}T^{-1}$, so that 
\[
e^{At}\Pi_{b}=TT^{-1}e^{At}T\pi_{b}T^{-1}=T\diag\{0_{k_{a}},e^{A_{b}t}\}T^{-1},
\]
where $A_{b}$ here denotes the matrix appearing in the statement of Lemma~\ref{lemma:Apredecomp}. In view of the observations concerning the matrix exponential
made in Section~\ref{section:jordannormalform} below, the statement of the lemma follows. 
\end{proof}

Below, the following terminology will be important. 

\begin{definition}\label{def:RAdef}
Given $1\leq k\in\zo$, $A\in\Mn{k}{\co}$ and $0<\b\in\ro$, let $A_{J,\roi}:=i\mathrm{Im}\{A_{J}\}$, where $A_{J}$ is obtained by appealing to
Lemma~\ref{lemma:Apredecomp}. Then $A_{J,\roi}$ is diagonal and purely imaginary, and the smooth function $R_{A}:\ro\rightarrow\Mn{k}{\co}$ defined by 
$R_{A}(t):=\exp(-A_{J,\roi}t)$ is referred to as a \textit{rotation matrix associated with $A$}.
\index{Rotation matrix}%
\index{Rotation matrix!associated with a matrix}%
\end{definition}
\begin{remark}
Note that $\|R_{A}(t)\|=\|[R_{A}(t)]^{-1}\|=1$. 
\end{remark}
In our setting, we are given a matrix $A\in \Mn{k}{\co}$ and a $0<\b_{\rem}\in\ro$. Appealing to Lemma~\ref{lemma:Apredecomp} with $\b$ replaced
by $\b_{\rem}$ then yields a matrix $A_{J}$, which we here refer to as $A_{\pre}$. Let $\kappa_{1}:=\kappa_{\max}(A)$, cf. 
Definition~\ref{def:SpRspdef}, and consider $A_{\inter}:=\mathrm{Re}\{A_{\pre}\}-\kappa_{1}\mathrm{Id}_{k}$. This is a real $k\times k$-matrix consisting of 
Jordan blocks, all of whose diagonal elements are non-positive. It would be convenient if the Jordan blocks in $A_{\inter}$ with strictly negative 
eigenvalues were negative definite. However, this need not be the case. On the other hand, it can be arranged. 

\begin{lemma}\label{lemma:genJordblock}
Let $J\in\Mn{k}{\ro}$ be a Jordan block whose diagonal components equal $\lambda\in\ro$. For every $0<\e\in\ro$, there is then a diagonal matrix 
$D\in\Gl{k}{\ro}$ such that for every $x\in\rn{k}$, 
\[
x^{t}D^{-1}JDx\leq (\lambda+\e)|x|^{2}.
\]
Moreover, $\|D\|=\max\{1,\e^{k-1}\}$ and $\|D^{-1}\|=\max\{1,\e^{-(k-1)}\}$.  
\end{lemma}
\begin{remark}\label{remark:genJordblock}
The difference between $J$ and $D^{-1}JD$ is that the off-diagonal $1$'s in $J$ have been replaced by $\e$. Below, it will be convenient to refer 
to a matrix of the form $D^{-1}JD$ as a \textit{generalised Jordan block}. We also use this terminology when $J$ is complex. 
\index{Jordan block!generalised}%
\index{Generalised!Jordan block}%
\end{remark}
\begin{proof}
The matrix $D$ can be chosen according to $D=\diag\{1,\e,\dots,\e^{k-1}\}$. 
\end{proof}

Applying this to the individual Jordan blocks leads to the following definition. 
\begin{definition}\label{def:JATA}
Given $1\leq k\in\zo$, $A\in\Mn{k}{\co}$ and $0<\b\in\ro$, let $T$ be a matrix of the form obtained in Lemma~\ref{lemma:Apredecomp}. Let, moreover, 
$\kappa_{1}:=\kappa_{\max}(A)$. Then there is a diagonal matrix $D\in\Gl{k}{\ro}$ such that if $T_{A}:=TD$, then $T^{-1}_{A}AT_{A}$ consists of 
generalised Jordan blocks. Moreover, all the generalised Jordan blocks of $J_{A}:=\roRe\{T^{-1}_{A}AT_{A}\}-\kappa_{1}\Id_{k}$ corresponding to
negative eigenvalues are negative definite. A matrix $T_{A}$ obtained in this way is referred to as a 
$\b$-\textit{normalised $A$-conjugation matrix}. 
\index{$\a$Aa@Notation!Matrix notation!bnormalised@$\b$-normalised $A$-conjugation matrix}%
\index{$\a$Aa@Notation!Matrix notation!Aconjugation@$A$-conjugation matrix, $\b$-normalised}%
Moreover, the matrix $J_{A}$ is referred to as a \textit{real $\b$-normalised Jordan decomposition of} $A$.
\index{Real $\b$-normalised Jordan decomposition of a matrix}%
\end{definition}
\begin{remarks}
The notion of a generalised Jordan block is introduced in Remark~\ref{remark:genJordblock}. Note that the definition of $T_{A}$ leads to 
$\|T_{A}\|$ and $\|T_{A}^{-1}\|$ depending on the distance between the largest and the second largest real part of an eigenvalue of $A$. 
On the other hand, $\|T_{A}\|$ and $\|T_{A}^{-1}\|$ depend only on $A$ (partly through this distance). 
\end{remarks}
\begin{remarks}\label{remarks:compandcomm}
Note that the matrices $R_{A}(t)$, $T_{A}$ and $J_{A}$ are dependent on a number of choices. For this reason we speak of a rotation matrix instead of 
the rotation matrix etc. However, we assume the different choices to be consistent with each other in the sense that they are related to each
other as described above. Then $R_{A}(t)$ and $A_{J,\roi}$ (cf. Definition~\ref{def:RAdef}) commute with $J_{A}$, since for each generalised Jordan 
block in $J_{A}$, the corresponding diagonal components in $A_{J,\roi}$ are constant.
\end{remarks}
For future reference, it is of interest to note that $e^{At}$ can be expressed in terms of the above decomposition of $A$. In fact, given 
$A\in\Mn{k}{\co}$ and $0<\b\in\ro$, let $T$ and $J$ be the $T_{A}$ and $J_{A}$ (respectively) constructed in Definition~\ref{def:JATA}.
Let, moreover, $R$ be the matrix valued function $R_{A}$ constructed in Definition~\ref{def:RAdef}. Then 
\begin{equation}\label{TinvATdecomp}
T^{-1}AT=A_{J,\roi}+J+\kappa_{1}\Id_{k},
\end{equation}
so that 
\begin{equation}\label{eq:eAtform}
e^{At}=e^{\kappa_{1}t}T[R(t)]^{-1}e^{Jt}T^{-1};
\end{equation}
note that $R(t)$ commutes with $J$, cf. Remarks~\ref{remarks:compandcomm}. Alternately, since $T^{-1}AT=\diag\{A_{a},A_{b}\}$ (this is true irrespective of 
whether $T$ is the matrix obtained in Lemma~\ref{lemma:Apredecomp} or if $T$ is the matrix $T_{A}$ introduced in Definition~\ref{def:JATA}),
\[
e^{At}=T\diag\{e^{A_{a}t},e^{A_{b}t}\}T^{-1}.
\]

\section{Rough estimate of solutions}\label{section:roughODEest}

The first step in the analysis of the asymptotics is to derive an estimate for $|v(t)|$ for $t\geq T_{\roode}$, where $v$ is a solution to 
(\ref{eq:ODEregmod}). This is the subject of the present section. 

\begin{lemma}\label{lemma:oderegest}
Consider the equation (\ref{eq:ODEregmod}), where $A$, $A_{\rem}$ and $F$ satisfy the assumptions described in Subsection~\ref{ssection:ODEreg}. 
There is a constant $C$, depending only on $C_{\rem}$, $\b_{\rem}$ and the matrix $A$, such that if $v$ is 
a solution to (\ref{eq:ODEregmod}), then 
\begin{equation}\label{eq:vest}
|v(t)|\leq C\ldr{\bt}^{d_{1}-1}e^{\kappa_{1}\bt}|v(T_{\roode})|+C\int_{T_{\roode}}^{t}\ldr{t-s}^{d_{1}-1}e^{\kappa_{1}(t-s)}|F(s)|ds
\end{equation}
for $t\geq T_{\roode}$, where $\bt=t-T_{\roode}$, $\kappa_{1}:=\kappa_{\max}(A)$ and $d_{1}:=d_{\max}(A,\kappa_{1})$; cf. Definition~\ref{def:SpRspdef}.
\end{lemma}
\begin{remark}\label{remark:Cexdep}
The constant $C$ may depend on, e.g., the reciprocal of the difference between the largest and the second largest real part of an eigenvalue of $A$. 
If one is interested in applying the result for an infinite set of matrices $A$, this has to be kept in mind. 
\end{remark}
\begin{proof}
Denoting the sum of $A$ and $A_{\rem}(t)$ by $B(t)$, the equation (\ref{eq:ODEregmod}) takes the same form as (\ref{eq:ODEregmodintro}). Let $\Phi(t;t_{0})$
denote the solution to the initial value problem
\[
\frac{d\Psi}{dt}(t)=B(t)\Psi(t),\ \ \ \Psi(t_{0})=\Id_{k}.
\]
Then the solution to (\ref{eq:ODEregmod}) with initial data $v_{0}$ at $T_{\roode}$ can be written
\begin{equation}\label{eq:vtformtmsver}
v(t)=\Phi(t;T_{\roode})v_{0}+\int_{T_{\roode}}^{t}\Phi(t;s)F(s)ds.
\end{equation}
It is thus of interest to estimate $\|\Phi(t;s)\|$ for $T_{\roode}\leq s\leq t$. Since $|\Phi(t;s)\xi|=|v(t)|$, where $v$ is the solution to the
initial value problem 
\begin{equation}\label{eq:vipB}
\frac{dv}{dt}(t)=B(t)v(t),\ \ \ v(s)=\xi, 
\end{equation}
it is clear that we only need to estimate $|v(t)|/|\xi|$ for solutions $v$ to (\ref{eq:vipB}). 

Given $A$ and $\b_{\rem}$, let $R(t):=R_{A}(t)$, $T:=T_{A}$ and $J:=J_{A}$ be the function/matrices obtained by appealing to 
Definitions~\ref{def:RAdef} and \ref{def:JATA}. Recall that (\ref{TinvATdecomp}) holds and that $R(t)$ and $J$ commute due to 
Remarks~\ref{remarks:compandcomm}. Let $v$ be the solution to (\ref{eq:vipB}) and define $w$ by 
\begin{equation}\label{eq:waGdef}
w(t):=e^{-\kappa_{1}t}R(\bt)T^{-1}v(t).
\end{equation}
Then
\begin{equation}\label{eq:wnoeqge}
\dot{w}=Jw+A_{\rest}w,
\end{equation}
where 
\begin{equation}\label{eq:Arestestge}
\|A_{\rest}(t)\|\leq Ce^{-\b_{\rem}\bt}
\end{equation}
for $t\geq T_{\roode}$, and the constant $C$ only depends on $C_{\rem}$, $\b_{\rem}$, $\|T\|$ and $\|T^{-1}\|$. Let $J_{\tr}$ be the truncated version of $J$, 
obtained from $J$ by setting the Jordan blocks with zero eigenvalue to zero. Let $J_{\rodiff}=J-J_{\tr}$ and define $u$ and $B_{\rest}$ by 
\begin{equation}\label{eq:uBHdef}
u=e^{-J_{\rodiff}(t-s)}w,\ \ \
B_{\rest}(t)=e^{-J_{\rodiff}(t-s)}A_{\rest}(t)e^{J_{\rodiff}(t-s)}.
\end{equation}
Then 
\begin{equation}\label{eq:ubHver}
\dot{u}=J_{\tr}u+B_{\rest}u,
\end{equation}
where
\begin{equation}\label{eq:normBest}
\|B_{\rest}(t)\|\leq C\ldr{t-s}^{2(d_{1}-1)}e^{-\b_{\rem}\bt}
\end{equation}
for $t\geq T_{\roode}$, where the constant $C$ only depends on $C_{\rem}$, $\b_{\rem}$ and $A$, 
and we have used the notation (\ref{eq:ldrdefinitionintro}). The advantage of the version (\ref{eq:ubHver}) of the equation is that $J_{\tr}$ is negative 
semi-definite. Moreover, even in the case that $B_{\rest}$ in (\ref{eq:ubHver}) equals zero, a solution $u$ to (\ref{eq:ubHver}) is typically bounded, 
but no more. Estimate
\[
\frac{d}{dt}|u|^{2}=\ldr{u,\dot{u}}+\ldr{\dot{u},u}\leq 2\|B_{\rest}\||u|^{2}.
\]
Integrating this estimate yields
\begin{equation}\label{eq:utusest}
|u(t)|\leq C|u(s)|,
\end{equation}
for all $T_{\roode}\leq s\leq t$, where $C$ only depends on $C_{\rem}$, $\b_{\rem}$ and $A$. Returning to (\ref{eq:waGdef}) and (\ref{eq:uBHdef}), 
the estimate (\ref{eq:utusest}) yields 
\[
|v(t)|\leq C\ldr{t-s}^{d_{1}-1}e^{\kappa_{1}(t-s)}|v(s)|,
\]
where $C$ has the same dependence as in the case of (\ref{eq:utusest}). In particular, 
\[
\|\Phi(t;s)\|\leq C\ldr{t-s}^{d_{1}-1}e^{\kappa_{1}(t-s)}
\]
for $T_{\roode}\leq s\leq t$, where $C$ has the same dependence as in the case of (\ref{eq:utusest}). Combining this estimate with 
(\ref{eq:vtformtmsver}) yields the conclusion of the lemma. 
\end{proof}

\section{Detailed asymptotics}\label{section:detailedODEas}

Consider (\ref{eq:vest}). If $F=0$, this estimate is optimal (strictly speaking, we only demonstrate that this is true in 
Section~\ref{section:spdatinfODE} below, but in the case that $A_{\rem}=0$, it is obvious). Moreover, the first term on the right hand side 
of (\ref{eq:vest}) exhibits
the asymptotic growth associated with solutions to $\dot{v}=Av$. On the other hand, if $|F|$ grows very quickly, the second term on the right
hand side of (\ref{eq:vest}) might be dominant. Before proceeding to a derivation of more detailed information concerning the asymptotics, it is 
necessary to decide which type of situation is of greatest interest: the case where $F$ dominates the evolution, or the case when the contribution
from $F$ is, at worst, comparable with solutions to the homogeneous system associated with (\ref{eq:ODEregmod}). Here, we are mainly interested in
the latter case, and for that reason, we assume that (\ref{eq:Fnoone}) holds. Then (\ref{eq:vest}) implies that the growth of $v$ is bounded by 
$\ldr{\bt}^{d_{1}-1}e^{\kappa_{1}t}$. This is a natural starting point for deriving more detailed information concerning the asymptotics, the subject of the 
present section. 

\begin{lemma}\label{lemma:ODEasymp}
Consider the equation (\ref{eq:ODEregmod}), where $A$, $A_{\rem}$ and $F$ satisfy (\ref{eq:Fnoone}) and the assumptions described in 
Subsection~\ref{ssection:ODEreg}. Given notation as in Subsection~\ref{ssection:ODEreg}, there is then a constant $C$ and a 
non-negative integer $N$, where $C$ only depends on $C_{\rem}$, $\b_{\rem}$ and $A$, and $N$ only depends on $k$, such that the following 
holds. Fix $0<\b\leq\b_{\rem}$. For every solution $v$ to (\ref{eq:ODEregmod}), there is a unique vector $v_{\infty}\in E_{A,\b}$ such that 
\begin{equation}\label{eq:vinfasres}
\begin{split}
 & \left|v(t)-e^{At}v_{\infty}-\int_{T_{\roode}}^{t}e^{A(t-s)}F(s)ds\right|\\
 \leq & C\ldr{\bt}^{N}e^{(\kappa_{1}-\b)\bt}\left[|v(T_{\roode})|+e^{\kappa_{1}T_{\roode}}\|F\|_{A}\right]
\end{split}
\end{equation}
for all $t\geq T_{\roode}$, where $\kappa_{1}:=\kappa_{\max}(A)$ and the notation $E_{A,\b}$ is introduced in Definition~\ref{def:fssubspetc}.
Moreover, $v_{\infty}$ can be written
\begin{equation}\label{eq:vinfform}
v_{\infty}=e^{-AT_{\roode}}u_{\infty},
\end{equation}
where $u_{\infty}\in E_{A,\b}$,
\begin{equation}\label{eq:winfTodeest}
|u_{\infty}|\leq C\left[|v(T_{\roode})|+e^{\kappa_{1}T_{\roode}}\|F\|_{A}\right],
\end{equation}
and $C$ only depends on $C_{\rem}$, $\b_{\rem}$ and $A$. 
\end{lemma}
\begin{remark}
In the ODE setting, it may seem unnatural not to assume $\b=\b_{\rem}$. The reason for this is that, the larger the $\b$, the more detailed the 
asymptotics. However, we 
wish to apply this lemma in the PDE setting, in which case we are interested in an infinite number of ODE's (one for each mode). In that setting, 
a larger $\b$ is not only associated with more detailed information concerning the asymptotics, it is also associated with a larger loss of 
derivatives. In some applications, it might be important to minimise the loss of derivatives (while still retaining a certain amount of asymptotic
information). This is the reason for not fixing the value of $\b$ to equal $\b_{\rem}$. 
\end{remark}
\begin{remark}
The constants $C_{\rem}$, $\b_{\rem}$ and $T_{\roode}$ are the ones appearing in connection with the assumption (\ref{eq:Aremass}). 
\end{remark}
\begin{remark}
Remark~\ref{remark:Cexdep} is equally relevant here. 
\end{remark}
\begin{remark}
Combining (\ref{eq:vinfform}) and (\ref{eq:winfTodeest}) yields a crude estimate of $|v_{\infty}|$. However, it is of interest to obtain
more detailed information. This is the reason for including (\ref{eq:vinfform}), (\ref{eq:winfTodeest}) and the statement that 
$u_{\infty}\in E_{A,\b}$. 
\end{remark}
\begin{proof}
\textbf{Uniqueness.} Let $v$ be a solution to (\ref{eq:ODEregmod}). Assume that there are $v_{\infty,i}\in E_{A,\b}$, $i=1,2$, such that 
\[
 \left|v(t)-e^{At}v_{\infty,i}-\int_{T_{\roode}}^{t}e^{A(t-s)}F(s)ds\right|\leq C_{i}\ldr{\bt}^{N_{i}}e^{(\kappa_{1}-\b)\bt}
\]
for all $t\geq T_{\roode}$ and some constants $C_{i}$ and $N_{i}\geq 0$ which are allowed to depend on $v$, $F$, $T_{\roode}$ as well as 
$C_{\rem}$, $\b_{\rem}$ and $A$. Then
\[
|e^{At}(v_{\infty,1}-v_{\infty,2})|\leq (C_{1}+C_{2})\ldr{\bt}^{N_{1}+N_{2}}e^{(\kappa_{1}-\b)\bt}
\]
for all $t\geq T_{\roode}$. However, if $v_{\infty,1}\neq v_{\infty,2}$, then the left hand side of this inequality grows more quickly than the right hand 
side, which leads to a contradiction. Therefore $v_{\infty,1}=v_{\infty,2}$, and $v_{\infty}$ is uniquely determined by the estimate (\ref{eq:vinfasres}).

\textbf{Existence.}
To begin with, we introduce notation which is similar to, but not exactly the same as, the one introduced in the proof of 
Lemma~\ref{lemma:oderegest}. Given $A$ and $\b$, let $R(t):=R_{A}(t)$, $T:=T_{A}$ and $J:=J_{A}$ be the function/matrices obtained by appealing to 
Definitions~\ref{def:RAdef} and \ref{def:JATA}. Recall that $\|T\|$ and $\|T^{-1}\|$ only depend on $A$; that (\ref{TinvATdecomp}) holds; and that 
$R(t)$ and $J$ commute due to Remarks~\ref{remarks:compandcomm}. Note, moreover, that $T$, $J$ and $R$ can be chosen independently of $\b$. However,
what $A_{a}$ and $A_{b}$ are depends on $\b$. Define $w$ and $G$ by 
\begin{equation}\label{eq:waGdefex}
w(t):=e^{-\kappa_{1}t}R(\bt)T^{-1}v(t),\ \ \
G(t):=e^{-\kappa_{1}t}R(\bt)T^{-1}F(t),
\end{equation}
where $\bt:=t-T_{\roode}$. Then
\begin{equation}\label{eq:wnoeqgeex}
\dot{w}=Jw+A_{\rest}w+G,
\end{equation}
where 
\begin{equation}\label{eq:Arestestgeex}
\|A_{\rest}\|\leq Ce^{-\b_{\rem}\bt}
\end{equation}
for $t\geq T_{\roode}$, and the constant $C$ only depends on $C_{\rem}$ and $A$. Let $J_{\tr}$ be the truncated version of $J$, 
obtained from $J$ by setting the Jordan blocks with zero eigenvalue to zero. Let $J_{\rodiff}=J-J_{\tr}$ and define $u$, $B$ and $H$ by 
\begin{equation}\label{eq:uBHdefex}
u=e^{-J_{\rodiff}\bt}w,\ \ \
B(t)=e^{-J_{\rodiff}\bt}A_{\rest}(t)e^{J_{\rodiff}\bt},\ \ \
H(t)=e^{-J_{\rodiff}\bt}G(t).
\end{equation}
Then 
\begin{equation}\label{eq:ubHverex}
\dot{u}=J_{\tr}u+Bu+H,
\end{equation}
where
\begin{equation}\label{eq:normBestex}
\|B(t)\|\leq C\ldr{\bt}^{2(d_{1}-1)}e^{-\b_{\rem}\bt}
\end{equation}
for $t\geq T_{\roode}$, where the constant $C$ only depends on $C_{\rem}$ and $A$. For future reference, it is convenient to keep in mind
that (\ref{eq:vest}), (\ref{eq:waGdefex}) and (\ref{eq:uBHdefex}) yield
\begin{equation}\label{eq:uroughestex}
|u(t)|\leq C\ldr{\bt}^{2(d_{1}-1)}|w(T_{\roode})|+C\ldr{\bt}^{2(d_{1}-1)}\|F\|_{A}
\end{equation}
for $t\geq T_{\roode}$, where $C$ only depends on $C_{\rem}$, $\b_{\rem}$ and $A$.

\textbf{Estimating the first component.}
Let $u_{a}(t)$ denote the first component in the $\b,A$-decomposition of $u(t)$; cf. Definition~\ref{def:fssubspetc} for an explanation of the 
terminology. We similarly use the notation $w_{a}$, $(Bu)_{a}$, $G_{a}$, $F_{a}$ etc. 
Let $J_{a}$ and $J_{\tr,a}$ denote the first blocks in the $\b,A$-decomposition of $J$ and $J_{\tr}$ respectively; cf. Definition~\ref{def:fssubspetc}. 
Integrating the $a$'th component of (\ref{eq:ubHverex}) yields
\begin{equation}\label{eq:ubaintform}
e^{-J_{a}\bt}w_{a}(t)=w_{a}(T_{\roode})+\int_{T_{\roode}}^{t}e^{-J_{\tr,a}\bs}(Bu)_{a}(s)ds
+\int_{T_{\roode}}^{t}e^{-J_{a}\bs}G_{a}(s)ds.
\end{equation}
Define
\begin{equation}\label{eq:winfadef}
w_{\infty,a}:=w_{a}(T_{\roode})+\int_{T_{\roode}}^{\infty}e^{-J_{\tr,a}\bs}(Bu)_{a}(s)ds.
\end{equation}
Then (\ref{eq:ubaintform}) implies
\begin{equation}\label{eq:waasestprel}
\left|w_{a}(t)-e^{J_{a}\bt}w_{\infty,a}-\int_{T_{\roode}}^{t}e^{J_{a}(t-s)}G_{a}(s)ds\right|\leq 
\left|\int_{t}^{\infty}e^{-J_{a}(s-t)}e^{J_{\rodiff,a}\bs}(Bu)_{a}(s)ds\right|,
\end{equation}
where $J_{\rodiff,a}$ is defined similarly to $J_{\tr,a}$. Note that (\ref{eq:normBestex}) and (\ref{eq:uroughestex}) yield
\begin{equation}\label{eq:contrtoconstodeapp}
\left|\int_{T_{\roode}}^{\infty}e^{-J_{\tr,a}\bs}(Bu)_{a}(s)ds\right|
\leq C[|w(T_{\roode})|+\|F\|_{A}],
\end{equation}
where $C$ only depends on $C_{\rem}$, $\b_{\rem}$ and $A$. Note, however, that the dependence 
is via, e.g., $(\b_{\rem}+\kappa_{\min})^{-1}$ (where $\kappa_{\min}:=\kappa_{\min}(A_{a})-\kappa_{1}$; cf. Definition~\ref{def:SpRspdef}), a 
quantity which tends to infinity as $|\kappa_{\min}|$ tends to $\b_{\rem}$. Combining this estimate with (\ref{eq:winfadef}) yields
\begin{equation}\label{eq:winfaest}
|w_{\infty,a}|\leq C[|w(T_{\roode})|+\|F\|_{A}],
\end{equation}
where $C$ has the same dependence as in (\ref{eq:contrtoconstodeapp}). Returning to (\ref{eq:waasestprel}), we need to estimate the right hand side.
Due to (\ref{eq:normBestex}) and (\ref{eq:uroughestex}), it can be bounded by 
\begin{equation*}
\begin{split}
 & C\int_{t}^{\infty}e^{-\kappa_{\min}(s-t)}\ldr{\bs}^{N}e^{-\b_{\rem}\bs}[|w(T_{\roode})|+\|F\|_{A}]ds\\
 \leq & C\ldr{\bt}^{N}e^{-\b_{\rem}\bt}[|w(T_{\roode})|+\|F\|_{A}]
\end{split}
\end{equation*}
for $t\geq T_{\roode}$, where $C$ only depends on $C_{\rem}$, $\b_{\rem}$ and $A$, and $N$ is a non-negative integer depending only on $k$. Combining this estimate 
with (\ref{eq:waasestprel}) and (\ref{eq:waGdefex}) yields 
\begin{equation}\label{eq:wawinfaest}
\begin{split}
 & \left|w_{a}(t)-e^{J_{a}\bt}w_{\infty,a}-\int_{T_{\roode}}^{t}e^{J_{a}(t-s)}G_{a}(s)ds\right|\\
 \leq & C\ldr{\bt}^{N}e^{-\b_{\rem}\bt}[e^{-\kappa_{1}T_{\roode}}|v(T_{\roode})|+\|F\|_{A}]
\end{split}
\end{equation}
for $t\geq T_{\roode}$, where $C$ and $N$ have the same dependence as before. 

\textbf{Estimating the second component.}
Let us now turn to $u_{b}(t)$, the second component in the $\b,A$-decomposition of $u(t)$; cf. Definition~\ref{def:fssubspetc} for an explanation of the 
terminology. Similarly to the above, we use the notation $w_{b}$, $(Bu)_{b}$, $G_{b}$, $F_{b}$ etc. In this case, (\ref{eq:ubHverex}) yields
\begin{equation}\label{eq:wbintform}
w_{b}(t)=e^{J_{b}\bt}w_{b}(T_{\roode})+\int_{T_{\roode}}^{t}e^{J_{b}(t-s)}(Bu)_{b}(s)ds
+\int_{T_{\roode}}^{t}e^{J_{b}(t-s)}G_{b}(s)ds;
\end{equation}
note that $J_{\tr,b}=J_{b}$, that $u_{b}=w_{b}$ and that $H_{b}=G_{b}$. To begin with
\begin{equation}\label{eq:wbconsttermest}
|e^{J_{b}\bt}w_{b}(T_{\roode})|\leq C\ldr{\bt}^{N}e^{\kappa_{b}\bt}|w_{b}(T_{\roode})|\leq C\ldr{\bt}^{N}e^{-\b\bt}|w_{b}(T_{\roode})|,
\end{equation}
where $\kappa_{b}:=\kappa_{\max}(A_{b})-\kappa_{1}$, $C$ only depends on $A$ and $N$ only depends on $k$; note that, by definition, $\kappa_{b}\leq -\b$. Next, 
\begin{equation}\label{eq:wbrestintest}
\begin{split}
\left|\int_{T_{\roode}}^{t}e^{J_{b}(t-s)}(Bu)_{b}(s)ds\right| \leq & C\ldr{\bt}^{N}\int_{T_{\roode}}^{t}e^{\kappa_{b}(t-s)}e^{-\b_{\rem}\bs}ds 
[|w(T_{\roode})|+\|F\|_{A}]\\
 \leq & C\ldr{\bt}^{N}e^{-\b\bt}[|w(T_{\roode})|+\|F\|_{A}],
\end{split}
\end{equation}
where $C$ only depends on $C_{\rem}$, $\b_{\rem}$ and $A$; $N$ is a non-negative integer depending only on $k$; and we have appealed to (\ref{eq:normBestex}) 
and (\ref{eq:uroughestex}). Combining (\ref{eq:wbintform}), (\ref{eq:wbconsttermest}) and (\ref{eq:wbrestintest}) yields
\begin{equation}\label{eq:wbwinfbest}
\left|w_{b}(t)-\int_{T_{\roode}}^{t}e^{J_{b}(t-s)}G_{b}(s)ds\right|\leq 
C\ldr{\bt}^{N}e^{-\b\bt}[|w(T_{\roode})|+\|F\|_{A}]
\end{equation}
for $t\geq T_{\roode}$, where $C$ only depends on $C_{\rem}$, $\b_{\rem}$ and $A$, and $N$ is a non-negative integer depending only on $k$. 

\textbf{Summing up.} Let $w_{\infty}\in\cn{k}$ be such that the first component in the $\b$, $A$-decomposition of $w_{\infty}$ is given by $w_{\infty,a}$, 
defined in (\ref{eq:winfadef}), and the second component in the $\b$, $A$-decomposition of $w_{\infty}$ vanishes. Then (\ref{eq:wawinfaest}) and 
(\ref{eq:wbwinfbest}) yield
\begin{equation}\label{eq:wwinfest}
\begin{split}
 & \left|w(t)-e^{J\bt}w_{\infty}-\int_{T_{\roode}}^{t}e^{J(t-s)}G(s)ds\right|\\
 \leq & C\ldr{\bt}^{N}e^{-\b\bt}[e^{-\kappa_{1}T_{\roode}}|v(T_{\roode})|+\|F\|_{A}]
\end{split}
\end{equation}
for $t\geq T_{\roode}$, where $C$ only depends on $C_{\rem}$, $\b_{\rem}$ and $A$; and $N$ is a non-negative 
integer depending only on $k$. Appealing to (\ref{eq:eAtform}) and (\ref{eq:waGdefex}) yields (\ref{eq:vinfasres}), where $v_{\infty}$ is given by 
(\ref{eq:vinfform}) and $u_{\infty}\in E_{A,\b}$ is defined by 
\begin{equation}\label{eq:uinfodedef}
u_{\infty}:=Te^{\kappa_{1}T_{\roode}}w_{\infty}.
\end{equation}
Combining this definition with (\ref{eq:waGdefex}) and (\ref{eq:winfaest}) yields (\ref{eq:winfTodeest}), and the lemma follows. 
\end{proof}

\section{Specifying data at infinity}\label{section:spdatinfODE}

Lemma~\ref{lemma:ODEasymp} contains a statement concerning the leading order behaviour of solutions. However, it is also of interest 
to prove that, given asymptotic data of the form $v_{\infty}$ (described in the statement of Lemma~\ref{lemma:ODEasymp}), there is a solution
with asymptotics as in (\ref{eq:vinfasres}). In order to obtain such a result, it is sufficient to consider the homogeneous version of the 
equation (\ref{eq:ODEregmod}). 

\begin{lemma}\label{lemma:spasODEsett}
Consider the equation
\begin{equation}\label{eq:homveq}
\dot{v}(t)=Av(t)+A_{\rem}(t)v(t),
\end{equation}
where $A\in\Mn{k}{\co}$ and $A_{\rem}\in C^{\infty}[\ro,\Mn{k}{\co}]$ satisfies the estimate (\ref{eq:Aremass}) for some choice of $T_{\roode}\geq 0$, 
$C_{\rem}$ and $\b_{\rem}>0$. Let $0<\b\leq\b_{\rem}$ and $E_{A,\b}$ be the first generalised eigenspace in the $\b,A$-decomposition of $\cn{k}$. Then 
there is a linear injective map $\Psi_{\infty}:E_{A,\b}\rightarrow\cn{k}$ such that if $v_{\infty}\in E_{A,\b}$ and $v$ is the solution to (\ref{eq:homveq}) 
satisfying $v(T_{\roode})=\Psi_{\infty}(v_{\infty})$, then $v$ satisfies (\ref{eq:vinfasres}) with $F=0$. Moreover, for $u_{\infty}\in E_{A,\b}$
\begin{equation}\label{eq:Psiinfnorm}
|\Psi_{\infty}(e^{-AT_{\roode}}u_{\infty})|\leq C|u_{\infty}|,
\end{equation}
where $C$ only depends on $C_{\rem}$, $\b_{\rem}$ and $A$. 
\end{lemma}
\begin{remark}
In the statement of the lemma, as well as the proof, we use the terminology introduced in Subsection~\ref{ssection:decompofA}. 
\end{remark}
\begin{remark}\label{remark:inhomaschar}
The result can be used to derive conclusions concerning the inhomogeneous equation (\ref{eq:ODEregmod}). In order to justify this 
statement, assume that $\|F\|_{A}<\infty$ and let $v_{\mathrm{part}}$ be the solution to (\ref{eq:ODEregmod}) corresponding to the initial data 
$v_{\mathrm{part}}(T_{\roode})=0$. Let $v_{\partic,\infty}\in E_{A,\b}$ be such that (\ref{eq:vinfasres}) holds with $v$ and $v_{\infty}$ replaced by 
$v_{\mathrm{part}}$ and $v_{\partic,\infty}$ respectively. Given $v_{\infty}\in E_{A,\b}$, let $v_{\hom}$ be the solution to (\ref{eq:homveq}) corresponding 
to initial data $v_{\hom}(T_{\roode})=\Psi_{\infty}(v_{\infty}-v_{\partic,\infty})$. Then $v:=v_{\partic}+v_{\hom}$ solves (\ref{eq:ODEregmod}) and satisfies 
(\ref{eq:vinfasres}). In other words, we are also allowed to specify the asymptotic data $v_{\infty}\in E_{A,\b}$ in the case of the inhomogeneous
equation (\ref{eq:ODEregmod}), assuming $\|F\|_{A}<\infty$. In order to obtain estimates, let $u_{\partic,\infty}\in E_{A,\b}$ be such that (\ref{eq:vinfform}) 
and (\ref{eq:winfTodeest}) hold with $v$, $v_{\infty}$ and $u_{\infty}$ replaced by $v_{\mathrm{part}}$, $v_{\partic,\infty}$ and $u_{\partic,\infty}$ respectively. 
Due to (\ref{eq:winfTodeest}), $|u_{\partic,\infty}|\leq Ce^{\kappa_{1}T_{\roode}}\|F\|_{A}$, where $C$ only depends on $C_{\rem}$, $\b_{\rem}$ and $A$. Due to 
(\ref{eq:vinfform}), $v_{\partic,\infty}=e^{-AT_{\roode}}u_{\partic,\infty}$. Representing $v_{\infty}$ according to $v_{\infty}=e^{-A T_{\roode}}u_{\infty}$, it follows 
that 
\begin{equation}\label{eq:vTodeestitoeta}
\begin{split}
|v(T_{\roode})| \leq & |v_{\partic}(T_{\roode})|+|v_{\hom}(T_{\roode})|=|\Psi_{\infty}(v_{\infty}-v_{\partic,\infty})|\\
 = & |\Psi[e^{-AT_{\roode}}(u_{\infty}-u_{\partic,\infty})]|\\
 \leq & C|u_{\infty}-u_{\partic,\infty}|
\leq C\left(|u_{\infty}|+e^{\kappa_{1}T_{\roode}}\|F\|_{A}\right),
\end{split}
\end{equation}
where $C$ is a constant depending only on $C_{\rem}$, $\b_{\rem}$ and $A$. 
\end{remark}
\begin{remark}
The equality $E_{A,\b}=\cn{k}$ holds if $\b>\Rsp A$; cf. 
Definition~\ref{def:SpRspdef}. Moreover, if $E_{A,\b}=\cn{k}$, then the map $\Psi_{\infty}$ is an isomorphism. This means that the solution 
is uniquely specified in terms of the ``data at infinity'', $v_{\infty}$. Due to Remark~\ref{remark:inhomaschar}, the same holds for 
inhomogeneous equations such that $\|F\|_{A}<\infty$. Finally, $|v(T_{\roode})|$ can be estimated in terms of the asymptotic data and $F$; cf. 
(\ref{eq:vTodeestitoeta}). 
\end{remark}
\begin{proof}
Let $T$, $J$ and $R$ be defined as in the proof of Lemma~\ref{lemma:ODEasymp}. Define $w$ as in (\ref{eq:waGdefex}) and let $J_{a}$ be the first 
block in the $\b,A$-decomposition of $J$; cf. Definition~\ref{def:fssubspetc}. Let
\begin{align*}
J_{a,0} = & \diag\{J_{a},0\},\ \ \
\bB(t)=e^{-J_{a,0}\bt}A_{\rest}(t)e^{J_{a,0}\bt},\\
\bJ = & J-J_{a,0},\ \ \
\bu(t)=e^{-J_{a,0}\bt}w(t),
\end{align*}
where $A_{\rest}$ is the matrix appearing in (\ref{eq:wnoeqgeex}). Then (\ref{eq:homveq}) can be written
\begin{equation}\label{eq:budot}
\dot{\bu}=\bJ \bu+\bB\bu,
\end{equation}
where we have appealed to (\ref{eq:wnoeqgeex}) with $G=0$. Due to (\ref{eq:Arestestgeex}), 
\begin{equation}\label{eq:bBnormest}
\|\bB(t)\|\leq C\ldr{\bt}^{N}e^{-\b_{\min}\bt}
\end{equation}
for $t\geq T_{\roode}$, where $\b_{\min}=\b_{\rem}+\kappa_{\min}$; $\kappa_{\min}:=\kappa_{\min}(A_{a})-\kappa_{1}$, where $A_{a}$ is  the first block in the 
$\b,A$-decomposition of $T^{-1}AT$; $C$ is a constant depending only on $C_{\rem}$, $\b_{\rem}$ and $A$; and $N$ is a non-negative integer depending 
only on $k$. Note that $\b_{\min}>0$ and that $\bJ$ is negative semi-definite. Let $t_{0}\geq T_{\roode}$. Due to (\ref{eq:budot}) and the fact that 
$\bJ$ is negative semi-definite, 
\[
\frac{d}{dt}|\bu|^{2}\leq 2\|\bB\|\cdot|\bu|^{2},
\]
so that 
\begin{equation}\label{eq:absuthomest}
|\bu(t)|\leq\exp\left(\int_{t_{0}}^{t}\|\bB(s)\|ds\right)|\bu(t_{0})|
\end{equation}
for $t\geq t_{0}$. Note, in particular, that by choosing $t_{0}$ to be large enough, the first factor on the right hand side of (\ref{eq:absuthomest})
can be assumed to be as close to $1$ as we wish. Let $\bu_{a}(t)$ and $\bu_{b}(t)$ denote the first and second components in the $\b,A$-decomposition
of $\bu(t)$ respectively (we use similar notation for the same decomposition of other vectors and vector valued functions). Then
\begin{equation}\label{eq:emlibui}
\bu_{a}(t)=\bu_{a}(t_{0})+\int_{t_{0}}^{t}(\bB\bu)_{a}(s)ds,
\end{equation}
so that 
\[
\left|\bu_{a}(t)-\bu_{a}(t_{0})\right|\leq C\int_{t_{0}}^{t}\ldr{\bs}^{N}e^{-\b_{\min}\bs}ds
\exp\left(\int_{t_{0}}^{\infty}\|\bB(s)\|ds\right)|\bu(t_{0})|
\]
for $t\geq t_{0}$, where we have appealed to (\ref{eq:bBnormest}) and (\ref{eq:absuthomest}); $C$ only depends on $C_{\rem}$, $\b_{\rem}$ and $A$; and $N$ 
only depends on $k$. For any $\e>0$, it is thus possible to choose $t_{0}$ so that 
\begin{equation}\label{eq:buicontr}
\left|\bu_{a}(t)-\bu_{a}(t_{0})\right|\leq \e |\bu(t_{0})|
\end{equation}
for $t\geq t_{0}$ (where $t_{0}$ only depends on $C_{\rem}$, $\b_{\rem}$, $A$ and $\e$). Given a $t_{0}$ such that (\ref{eq:buicontr}) holds 
with $\e$ replaced by $1/2$, define the map $L_{a}:\cn{k_{a}}\rightarrow \cn{k_{a}}$ as follows (here $k_{a}$ is the dimension of the 
first subspace of the $\b,A$-decomposition of $\cn{k}$; cf. Definition~\ref{def:fssubspetc}). Given $\eta\in\cn{k_{a}}$, let 
$\xi\in\cn{k}$ be such that $\xi_{a}=\eta$ and $\xi_{b}=0$. Let $\bu$ be the solution to (\ref{eq:budot}) with $\bu(t_{0})=\xi$ 
and define $L_{a}\eta$ by 
\begin{equation}\label{eq:Laetadef}
L_{a}\eta:=\lim_{t\rightarrow\infty}\bu_{a}(t).
\end{equation}
Due to (\ref{eq:bBnormest}), (\ref{eq:absuthomest}) and (\ref{eq:emlibui}), it is clear that the limit on the right hand side of (\ref{eq:Laetadef})
exists. Thus $L_{a}\eta$ is well defined. Moreover, it is clear that $L_{a}$ is linear. Due to the fact that (\ref{eq:buicontr}) holds with $\e$ replaced 
by $1/2$ and the fact that $\bu_{b}(t_{0})=0$,
\[
\left|L_{a}\eta-\bu_{a}(t_{0})\right|\leq \frac{1}{2}|\bu_{a}(t_{0})|. 
\]
In particular, $|\eta|/2\leq |L_{a}\eta|$, so that $L_{a}$ is injective. Since $L_{a}$ is a linear map between spaces of the same finite dimension, it 
follows that $L_{a}$ is an isomorphism. Moreover, 
\begin{equation}\label{eq:Lainvest}
|L_{a}^{-1}\eta|\leq 2|\eta|.
\end{equation}
Note that $L_{a}^{-1}$ (essentially) maps data at infinity to data at $t_{0}$. In the end, we wish to translate the data at $t_{0}$ to data at $T_{\roode}$. 
To this end, we wish to solve (\ref{eq:budot}) backwards. Since $t_{0}-T_{\roode}$ can be bounded by a constant depending only on $C_{\rem}$, $\b_{\rem}$ and 
$A$, there is a constant $C$ (with the same dependence) such that for all solutions $\bu$ to (\ref{eq:budot}),
\begin{equation}\label{eq:butodetzest}
|\bu(T_{\roode})|\leq C|\bu(t_{0})|.
\end{equation}
Before defining the map $\Psi_{\infty}$ appearing in the statement of the lemma, we need to clarify how the limit of $\bu_{a}(t)$ is related to the 
$v_{\infty}\in E_{A,\b}$ appearing in (\ref{eq:vinfasres}). However, returning to (\ref{eq:ubaintform}) and (\ref{eq:winfadef}) (and recalling that
$G=0$ in the present setting), it is clear that
\begin{equation}\label{eq:bualimitowinfa}
\lim_{t\rightarrow\infty}\bu_{a}(t)=\lim_{t\rightarrow\infty}e^{-J_{a}\bt}w_{a}(t)=w_{\infty,a}.
\end{equation}
On the other hand, (\ref{eq:vinfform}) and (\ref{eq:uinfodedef}) imply that 
\[
v_{\infty}=e^{-AT_{\roode}}u_{\infty}=e^{-AT_{\roode}}Te^{\kappa_{1}T_{\roode}}\iota_{a}(w_{\infty,a}),
\]
where $\iota_{a}:\cn{k_{a}}\rightarrow\cn{k}$ is defined as follows: if $\eta\in\cn{k_{a}}$, then $\iota_{a}(\eta)$ is the vector $\xi\in\cn{k}$ such
that $\xi_{a}=\eta$ and $\xi_{b}=0$. Note also that for each $v_{\infty}\in E_{A,\b}$, there is a unique $\eta\in\cn{k_{a}}$ such that 
\begin{equation}\label{eq:etavinfrel}
v_{\infty}=e^{-AT_{\roode}}Te^{\kappa_{1}T_{\roode}}\iota_{a}(\eta)
\end{equation}
holds. Define $\Psi_{\infty}(v_{\infty})$ to be the composition of the following maps: first the map which takes $v_{\infty}\in E_{A,\b}$ to the 
$\eta\in\cn{k_{a}}$ such that (\ref{eq:etavinfrel}) holds; second, the map which takes $\eta$ to $\xi:=\iota_{a}\circ L_{a}^{-1}(\eta)$; and, third, 
the map which takes $\xi\in\cn{k}$ to 
\begin{equation}\label{eq:vzdef}
v_{0}:=e^{\kappa_{1}T_{\roode}}T\bu(T_{\roode}),
\end{equation}
where $\bu$ is the solution to (\ref{eq:budot}) with initial data $\bu(t_{0})=\xi$. Note that $\Psi_{\infty}$ is linear and injective, and that 
if (\ref{eq:vinfform}) holds, then
\begin{equation*}
\begin{split}
|\Psi_{\infty}(e^{-AT_{\roode}}u_{\infty})| \leq & Ce^{\kappa_{1}T_{\roode}}|\bu(T_{\roode})|\leq Ce^{\kappa_{1}T_{\roode}}|\bu(t_{0})|\leq Ce^{\kappa_{1}T_{\roode}}|\eta|\\
 \leq & C|u_{\infty}|,
\end{split}
\end{equation*}
where $C$ only depends on $C_{\rem}$, $\b_{\rem}$ and $A$; and we have appealed to (\ref{eq:Lainvest}), (\ref{eq:butodetzest}) and (\ref{eq:vzdef}).
Thus (\ref{eq:Psiinfnorm}) holds. Assume now that $v$ is the solution to (\ref{eq:homveq}) corresponding to $v(T_{\roode})=\Psi_{\infty}(v_{\infty})$.
Then $v$ satisfies (\ref{eq:vinfasres}) with $F=0$ and $v_{\infty}$ replaced by some $\bv_{\infty}\in E_{A,\b}$. Moreover, Lemma~\ref{lemma:ODEasymp}, in 
particular (\ref{eq:vinfform}) and (\ref{eq:uinfodedef}), yields $\bv_{\infty}=e^{-AT_{\roode}}Te^{\kappa_{1}T_{\roode}}\iota_{a}(\bw_{\infty,a})$ for some 
$\bw_{\infty,a}\in \cn{k_{a}}$. In addition, 
\[
\lim_{t\rightarrow\infty}\bu_{a}(t)=\bw_{\infty,a},
\]
where $\bu$ is defined in terms of $v$ in the same way as in the beginning of the present proof; cf. (\ref{eq:bualimitowinfa}). This means that 
\[
\iota_{a}\circ L_{a}^{-1}(\bw_{\infty,a})=\bu(t_{0}),
\]
due to the definition of $L_{a}$. Going through the steps defining $\Psi_{\infty}$, it is thus clear that 
$\Psi_{\infty}(\bv_{\infty})=v(T_{\roode})=\Psi_{\infty}(v_{\infty})$. Due to the injectivity of $\Psi_{\infty}$, it follows that $v_{\infty}=\bv_{\infty}$.
The lemma follows. 
\end{proof}

\chapter[Asymptotics, the silent setting]{Asymptotics for weakly silent, balanced and convergent 
equations}\label{chapter:weaksil}

\section{Introduction}\label{section:introductionweaksil}

In the present chapter, we focus on the simplest setting of interest in these notes, namely that of weakly silent, balanced and convergent equations. 
The goal is to derive asymptotics, as well as to obtain a (partial) characterisation of the solutions in terms of their asymptotics. The formal 
definition of weak silence, balance and convergence is provided in Definition~\ref{def:roughODEtermo} below. Let us briefly justify the 
terminology

\textbf{Weak silence.} Recall that we justified the use of the word silent in, e.g., Chapter~\ref{chapter:silentequations}. The conditions imposed 
in the present chapter are implied by 
the requirements appearing in, e.g., Proposition~\ref{prop:genroughestintro}. In that sense, the conditions appearing here are weaker 
than the conditions imposed in Chapter~\ref{chapter:silentequations}. We therefore speak of weak silence.  

\textbf{Weak balance and weak convergence.} The notion of weak balance involves bounds on $\chi$, $\mcX$, $\a$ and $\zeta$. The bounds on the 
shift vector field and $\mcX$ are formulated in terms of $\sigma$ and $X$ introduced in (\ref{eq:sigmaXdefintro}). Moreover, the bounds on
$\sigma$ and $X$ follow from the requirements that $\chi$ be future uniformly timelike and that $\mcX$ be $C^{0}$-future bounded; cf. 
Lemma~\ref{lemma:sigmaXbdsandderbds} and Definition~\ref{def:roughODEtermo} below. We also demand that the norms of $\a$ and $\zeta$ be future 
bounded. This is to exclude super exponential growth; cf. Section~\ref{section:introductiononthenotionofbal}. Finally, we are mainly interested 
in weakly convergent equations. This means that $\a$ and $\zeta$ converge to limits. The reason we speak of \textit{weak} balance and \textit{weak} 
convergence is that we later impose similar conditions which also involve bounds on the first derivatives of the relevant objects. 

\textbf{Sketch of the analysis.} Consider (\ref{eq:fourierthesystemRge}). Say that $\mfg$, $g^{0l}$ and $X^{l}$ decay to zero exponentially, 
and that $\a$ and $\zeta$ converge exponentially to some limits. Say, moreover, that the decay rate of $\mfg$ is independent of the frequency.
For each frequency $\nu(\indexnot)$, there is then a $T_{\roode}\geq 0$ such that 
(\ref{eq:fourierthesystemRge}) can be reformulated as an equation of the form (\ref{eq:ODEregmod}), where $A_{\rem}$ satisfies an estimate of 
the form (\ref{eq:Aremestintro}); $C_{\rem}$ and $\b_{\rem}$ are independent of the frequency; and $A\in\Mn{2m}{\co}$. In the case of weakly silent, 
balanced and convergent equations, the details of this reformulation is provided in (\ref{eq:dotvroughODE})--(\ref{eq:Aremestgensys}) below.
Due to the results of 
Chapter~\ref{chapter:roughanalysisODEregion}, we can analyse the asymptotics to the future of $T_{\roode}$. However, in order for this analysis to 
be meaningful, we need to control the behaviour in the interval $[0,T_{\roode}]$. How long is this interval? How much can the solution grow in 
it? Can the growth be estimated in a way which does not destroy the asymptotic analysis to the future of $T_{\roode}$? The assumption
of weak silence implies that $\dot{\mfg}/\mfg\leq -\ellderbdsil+\betafun$, where $\ellderbdsil>0$ and $0\leq\betafun\in L^{1}([0,\infty))$ are independent 
of the frequency. Thus $\mfg(t)\leq C\exp[-\ellderbdsil(t-T_{1,\roode})]$ for $t\geq T_{1,\roode}$, where $C$ is a constant independent of the frequency and 
$T_{1,\roode}=\ellderbdsil^{-1}\ln\mfg(0)$. Similar conclusions can be derived for $n_{l}g^{0l}$ and $n_{l}X^{l}$, given that $g^{0l}$ and $X^{l}$ decay 
exponentially. This leads to times $T_{i,\roode}$, $i=2,3$. In comparison with the above, we can then choose $T_{\roode}$ to be the largest of the 
$T_{i,\roode}$, $i=1,2,3$. The next problem is then to estimate how the norm of the solution evolves on the interval 
$[0,T_{\roode}]$. Say, for the sake of argument, 
that $T_{\roode}$ can be chosen to equal $T_{1,\roode}$, and that the norm of the solution cannot grow faster than exponentially on 
$[0,T_{\roode}]$, where the rate is independent of the frequency. Then the norm cannot grow more than by a factor of $e^{CT_{\roode}}$, where the 
constant $C$ is independent of the frequency. However, since $T_{\roode}$ is a constant multiple of $\ln\mfg(0)$, this means that $e^{CT_{\roode}}$ is simply 
a power of $\mfg(0)$ (moreover, the power is independent of the frequency). In other words, we can interpret the growth on the interval $[0,T_{\roode}]$ 
as a loss of regularity. In terms of Sobolev spaces, this loss corresponds to having to estimate the initial data in $H^{s+s_{0}}$ (for some $s_{0}\geq 0$) 
in order to obtain an estimate concerning the asymptotics in $H^{s}$. In short, we can appeal to Chapter~\ref{chapter:roughanalysisODEregion} in order 
to analyse the behaviour to the future of $T_{\roode}$, and we can interpret the possible growth occurring in the interval $[0,T_{\roode}]$ as a loss of 
derivatives. 

\textit{Purpose of the conditions.} The question remains: how do we ensure that $T_{\roode}=T_{1,\roode}$, and that the exponential growth rate
in the interval $[0,T_{\roode}]$ is independent of the frequency? Consider (\ref{eq:fourierthesystemRgesigmaandX}), a reformulation of 
(\ref{eq:fourierthesystemRge}). It is clear that if $|\sigma|$ and $\|X\|$ appearing in (\ref{eq:fourierthesystemRgesigmaandX}) are bounded by 
constants independent of $\indexnot$, then the essential dependence on $\indexnot$ is contained in $\mfg$. Moreover, we can choose $T_{i,\roode}$
to equal $T_{1,\roode}$ for $i=2,3$. This means that $T_{\roode}=T_{1,\roode}$. If we, in addition to the assumption of weak silence, assume 
$|\sigma|$, $\|X\|$, $\|\a\|$ and $\|\zeta\|$ to be bounded by a constant independent of $\indexnot$ (these requirements correspond to demanding 
that (\ref{eq:thesystemRge}) be weakly balanced), then the exponential growth rate of the norm is bounded in $[0,T_{\roode}]$ by a constant 
independent of $\indexnot$. Thus the above line of reasoning can be carried out. 

\textit{Failure of the conditions.} If we would not impose the condition of weak balance, then the norm of solutions to (\ref{eq:fourierthesystemRge}) 
could grow super exponentially on the interval $[0,T_{\roode}]$, even in situations when $\mfg$, $g^{0l}$ and $X^{l}$ all decay to zero exponentially; 
cf. Chapter~\ref{chapter:onnotofbal}. Again, we could interpret this growth as a loss of regularity, but the loss would be of the form 
$\exp[C\mfg(0)^{\eta}]$ for some $C,\eta>0$. Clearly, this corresponds to more than a finite loss of regularity. In particular, note that simply appealing
to Cauchy stability in order to estimate the evolution in the interval $[0,T_{\roode}]$ is not sufficient: Cauchy stability does not exclude the 
possibility of super exponential growth. 

\textbf{Outline.} In Section~\ref{section:weasilbalcon} we give precise definitions of the notions of weak silence, balance and convergence. 
As indicated above, for a given frequency this leads to a division of $[0,\infty)$ into the intervals $[0,T_{\roode}]$ and $[T_{\roode},\infty)$. In 
Chapter~\ref{chapter:roughanalysisODEregion} we already considered the interval $[T_{\roode},\infty)$. The subject of Section~\ref{section:oscreg} is 
to analyse the evolution in the interval $[0,T_{\roode}]$ for one fixed frequency. Combining this analysis with the results of 
Chapter~\ref{chapter:roughanalysisODEregion} yields a basic energy estimate for how the $H^{s}$-norm of the solution evolves. This is the 
subject of Section~\ref{section:roughODEbaenest}; cf., in particular, Lemma~\ref{lemma:genroughest}. Combining the estimates on the interval 
$[0,T_{\roode}]$ with the analysis presented in Section~\ref{section:detailedODEas} leads to detailed asymptotics in the future direction. 
This topic is treated in Section~\ref{section:roughODEfutas}. Finally, appealing to the results of Section~\ref{section:spdatinfODE} leads to 
a partial characterisation of the solution in terms of the asymptotic data (in fact, in some situations there is a homeomorphism between 
initial data and the asymptotic data). This is the subject of Section~\ref{section:roughODEspecas}. 

\section{Weak silence, balance and convergence}\label{section:weasilbalcon}

In order to give a precise meaning to the notion of a weakly silent, balanced and convergent equation, we need to introduce some terminology. Consider 
the equation (\ref{eq:thesystemRge}). Assume the associated metric to be such that $(M,g)$ is a canonical separable cosmological model manifold. 
Given a $0\neq\indexnot\in\EFindexset$, where $\EFindexset$ is defined in connection with (\ref{eq:IgenReqdef}), define 
$\mfg(\indexnot,t)$ by (\ref{eq:mfgnutdef}) and $\ell$, $\sigma$ and $X$ by 
\begin{equation}\label{eq:ellsigmaXgenRdef}
\ell(\indexnot,t):=\ln[\mfg(\indexnot,t)],\ \ \
\sigma(\indexnot,t):=\frac{n_{l}g^{0l}(t)}{\mfg(\indexnot,t)},\ \ \
X(\indexnot,t) :=\frac{n_{l}X^{l}(t)}{\mfg(\indexnot,t)},
\end{equation}
\index{$\a$Aa@Notation!Coefficients, Fourier side!$\ell(\indexnot,t)$}%
\index{$\a$Aa@Notation!Coefficients, Fourier side!$\sigma(\indexnot,t)$}%
\index{$\a$Aa@Notation!Coefficients, Fourier side!$X(\indexnot,t)$}%
where $n_{j}=\nu_{\roT,j}(\indexnot)$; cf. (\ref{eq:nuroTetcdef}). Most of the time, it is convenient to omit the argument $\indexnot$. 
Sometimes we also omit reference to the argument $t$. 

\begin{definition}\label{def:roughODEtermo}
Consider (\ref{eq:thesystemRge}). Assume the associated metric to be such that $(M,g)$ is a canonical separable cosmological model manifold. 
For $0\neq\indexnot\in\EFindexset$, define $\mfg(\indexnot,t)$ by (\ref{eq:mfgnutdef}) and $\ell$, $\sigma$ and $X$ by 
(\ref{eq:ellsigmaXgenRdef}). If there is a constant $C_{\coeff}$ such that
\begin{equation}\label{eq:weakbal}
|\sigma(\indexnot,t)|+\|X(\indexnot,t)\|+\|\a(t)\|+\|\zeta(t)\|\leq C_{\coeff}
\end{equation}
for all $0\neq\indexnot\in\EFindexset$ and all $t\geq 0$, the equation (\ref{eq:thesystemRge}) is said to be 
\textit{weakly balanced}. 
\index{Equation!weakly balanced}%
\index{Weakly balanced!equation}%
If there is a constant $\ellderbdsil>0$ and a continuous non-negative function $\betafun\in L^{1}([0,\infty))$ such that
\begin{equation}\label{eq:weaksil}
\dot{\ell}(\indexnot,t)\leq-\ellderbdsil+\betafun(t)
\end{equation}
for all $0\neq\indexnot\in\EFindexset$ and all $t\geq 0$, the equation (\ref{eq:thesystemRge}) is said to be 
\textit{weakly silent}. 
\index{Equation!weakly silent}%
\index{Weakly silent!equation}%
In case (\ref{eq:thesystemRge}) is weakly silent, $T_{\roode}$ is defined as follows. If $\bfcon:=\|\betafun\|_{1}$ and 
$\mfg(\indexnot,0)\leq e^{-\bfcon}$, then $T_{\roode}:=0$. If 
$\mfg(\indexnot,0)>e^{-\bfcon}$, $T_{\roode}$ is defined as the first $t\geq 0$ such that $\mfg(\indexnot,t)=e^{-\bfcon}$. Finally, if there 
are $\a_{\infty},\zeta_{\infty}\in\Mn{m}{\co}$ and constants $\eta_{\romn}>0$ and $C_{\romn}$ such that 
\begin{equation}\label{eq:alazeas}
\|\a(t)-\a_{\infty}\|+\|\zeta(t)-\zeta_{\infty}\|\leq C_{\romn}e^{-\eta_{\romn}t}
\end{equation}
for all $t\geq 0$, then (\ref{eq:thesystemRge}) is said to be \textit{weakly convergent}.
\index{Equation!weakly convergent}%
\index{Weakly convergent!equation}%
\end{definition}
\begin{remark}
It is to be expected that the requirement that $\betafun$ be continuous can be avoided at the expense of additional technicalities in the 
proofs below. 
\end{remark}
\begin{remark}\label{remark:wbwswcfollowfromgeometry}
If the conditions of one of Propositions~\ref{prop:genroughestintro}, \ref{prop:roughas} and \ref{prop:spasda} are satisfied, then 
(\ref{eq:thesystemRge}) is weakly balanced, weakly silent and 
weakly convergent. The reasons for this are the following. First of all, the norms of $\a$ and $\zeta$ are bounded due to the fact that 
(\ref{eq:alpahzetaconvest}) holds, an estimate which also yields (\ref{eq:alazeas}). That (\ref{eq:thesystemRge}) is weakly silent is a consequence
of Lemma~\ref{lemma:condwsil} and the assumptions of any one of the propositions mentioned at the beginning of the remark. Finally, the desired bounds 
on $|\sigma(\indexnot,t)|$ and $\|X(\indexnot,t)\|$ follow from Lemma~\ref{lemma:sigmaXbdsandderbds}. 
\end{remark}
\begin{remark}
Some of the estimates derived below can be improved if we, in addition to the requirements of Definition~\ref{def:roughODEtermo}, demand that there 
be constants $C_{\roder},\b_{\roder}>0$ such that 
\begin{equation}\label{eq:sigmaXCderbderest}
|\sigma(\indexnot,t)|+\|X(\indexnot,t)\|\leq C_{\roder}e^{-\b_{\roder}t}
\end{equation}
for all $0\neq\indexnot\in\EFindexset$ and all $t\geq 0$. 
\end{remark}

\section{The oscillatory regime}\label{section:oscreg}

For a weakly silent and balanced equation, it is, given a solution to (\ref{eq:fourierthesystemRge}), natural to divide $[0,\infty)$ into the intervals
$[0,T_{\roode}]$ and $[T_{\roode},\infty)$, where $T_{\roode}$ is given by Definition~\ref{def:roughODEtermo}. The methods developed in
Chapter~\ref{chapter:roughanalysisODEregion} can be used to analyse the behaviour in the interval $[T_{\roode},\infty)$. In the present section, we 
estimate how the norm of the solution evolves in the interval $[0,T_{\roode}]$. 

\begin{lemma}\label{lemma:oscroughODE}
Assume that (\ref{eq:thesystemRge}) is weakly balanced and weakly silent. Let $\indexnot\in\EFindexset$, $\mfg$ be given by (\ref{eq:mfgnutdef}) 
and $T_{\roode}$ by Definition~\ref{def:roughODEtermo}. Let $z$ be a solution to (\ref{eq:fourierthesystemRge}) and define $E$ by
\begin{equation}\label{eq:Edefodecase}
E(t):=\frac{1}{2}[|\dot{z}(t)|^{2}+\mfg^{2}(t)|z(t)|^{2}].
\end{equation}
Then
\begin{equation}\label{eq:EroughODEest}
E^{1/2}(t)\leq e^{c_{\betafun}+\eta_{\roode}t}E^{1/2}(0)+e^{c_{\betafun}}\int_{0}^{t}e^{\eta_{\roode}(t-s)}|\hf(s)|ds
\end{equation}
for $t\in [0,T_{\roode}]$, where $c_{\betafun}:=\|\betafun\|_{1}$; $\betafun$ is the continuous non-negative $L^{1}$-function appearing in the definition of 
weak silence; and $\eta_{\roode}>0$ is a constant depending only on $c_{\betafun}$ and the constant $C_{\coeff}$ appearing in the definition of weak balance. 
\end{lemma}
\begin{remark}\label{remark:sigmasizeirr}
The estimate is independent of the size of $|\sigma|$. In this respect, the assumptions can be weakened. However, in later estimates, 
we need control of $\sigma$. Moreover, the estimate (\ref{eq:weakbal}) turns out to be convenient. 
\end{remark}
\begin{proof} 
Estimate
\begin{equation}\label{eq:dEdtoscph}
\begin{split}
\frac{dE}{dt} = & \frac{1}{2}(\dot{z}\cdot\ddot{z}^{*}+\ddot{z}\cdot\dot{z}^{*})+\dot{\ell}\mfg^{2}|z|^{2}+\frac{1}{2}\mfg^{2}
(z\cdot\dot{z}^{*}+\dot{z}\cdot z^{*})\\
 \leq & \frac{1}{2}(\ddot{z}^{*}+\mfg^{2}z^{*})\cdot \dot{z}+
\frac{1}{2}(\ddot{z}+\mfg^{2}z)\cdot \dot{z}^{*}+\betafun \mfg^{2}|z|^{2}\\
 = & \frac{1}{2}(\ddot{z}^{*}+2i\sigma \mfg \dot{z}^{*}+\mfg^{2}z^{*})\cdot \dot{z}+
\frac{1}{2}(\ddot{z}-2i\sigma \mfg \dot{z}+\mfg^{2}z)\cdot \dot{z}^{*}+\betafun \mfg^{2}|z|^{2}.
\end{split}
\end{equation}
Combining this estimate with (\ref{eq:fourierthesystemRgesigmaandX}) yields
\begin{equation}\label{eq:Eprestoscreg}
\frac{dE}{dt}\leq \|\a\| |\dot{z}|^{2}+\|X\|E+\frac{1}{2}\|\zeta\|(|\dot{z}|^{2}+|z|^{2})+|\dot{z}||\hf|+\betafun \mfg^{2}|z|^{2}.
\end{equation}
On the other hand, $\mfg(t)\geq e^{-c_{\betafun}}$ on $[0,T_{\roode}]$, where $c_{\betafun}:=\|\betafun\|_{1}$. Thus (\ref{eq:Eprestoscreg}) yields
\[
\frac{dE}{dt}\leq (C+2\betafun)E+\sqrt{2}E^{1/2}|\hf|
\]
for $t\in [0,T_{\roode}]$, where the constant $C$ only depends on $C_{\coeff}$ and $c_{\betafun}$. As a consequence, it can be 
deduced that (\ref{eq:EroughODEest}) holds. 
\end{proof}

For future reference, it is of interest to note that the following related estimate holds.

\begin{lemma}\label{lemma:roughenestbalsetting}
Assume that (\ref{eq:thesystemRge}) is weakly balanced. Assume, moreover, that there is a constant $C_{\ell}>0$ and a continuous non-negative 
function $\betafun\in L^{1}([0,\infty))$ such that
\begin{equation}\label{eq:elldotbdgeneral}
|\dot{\ell}(\indexnot,t)|\leq C_{\ell}+\betafun(t)
\end{equation}
for all $0\neq \indexnot\in\EFindexset$ and all $t\geq 0$. Let $\indexnot\in\EFindexset$ and $\mfg$ be given by 
(\ref{eq:mfgnutdef}). Let $z$ be a solution to (\ref{eq:fourierthesystemRge}) and define $\me$ by
\begin{equation}\label{eq:medefroughbalset}
\me(\indexnot,t):=\frac{1}{2}[|\dot{z}(\indexnot,t)|^{2}+\mfg^{2}(\indexnot,t)|z(\indexnot,t)|^{2}+|z(\indexnot,t)|^{2}].
\end{equation}
Then
\begin{equation}\label{eq:meestroughbalset}
\me^{1/2}(\indexnot,t_{1})\leq e^{c_{\betafun}+\eta_{\robal}|t_{1}-t_{0}|}\me^{1/2}(\indexnot,t_{0})
+e^{c_{\betafun}}\left|\int_{t_{0}}^{t_{1}}e^{\eta_{\robal}|t_{1}-t|}|\hf(\indexnot,t)|dt\right|
\end{equation}
for $t_{0},t_{1}\geq 0$, where $c_{\betafun}:=\|\betafun\|_{1}$; and $\eta_{\robal}>0$ is a constant depending only on $C_{\ell}$ and 
the constant $C_{\coeff}$ appearing in the definition of weak balance. 
\end{lemma}
\begin{remark}
The reason for including the absolute value sign around the integral on the far right hand side of (\ref{eq:meestroughbalset}) is
that $t_{1}$ might be smaller than $t_{0}$. 
\end{remark}
\begin{proof}
By arguments similar to (\ref{eq:dEdtoscph}) and (\ref{eq:Eprestoscreg}), 
\begin{equation*}
\begin{split}
\left|\frac{d\me}{dt}\right| \leq & (C+2\betafun)\me+\sqrt{2}|\hf|\me^{1/2},
\end{split}
\end{equation*}
where $C$ only depends on $C_{\coeff}$ and $C_{\ell}$. This estimate can be used to deduce the conclusion of the lemma. 
\end{proof}

Consider (\ref{eq:fourierthesystemRgesigmaandX}). Assume that there is a $\mu\in\ro$ such that for each $\indexnot\in\EFindexset$, there 
is a $T\geq 0$ with the property that solutions behave as $e^{\mu t}$ for $t\geq T$. We would then like to derive an estimate corresponding 
to this behaviour which is valid for all $t\geq 0$. This can be done with the help of (\ref{eq:meestroughbalset}), but at the price of 
introducing a $T$-dependent constant.

\begin{cor}\label{cor:roughenestbalsettingztoTest}
Given assumptions and notation as in Lemma~\ref{lemma:roughenestbalsetting}, let $\mu_{l}\in\ro$, $l=1,2$, and $0\leq T\in\ro$. Then
\begin{equation}\label{eq:meestroughbalsetztot}
\me^{1/2}(\indexnot,t)\leq e^{c_{\betafun}+\eta_{1}T}e^{\mu_{1}t}\me^{1/2}(\indexnot,0)
+e^{c_{\betafun}+\eta_{2}T}e^{\mu_{1}t}\int_{0}^{t}e^{\mu_{2}(t-t')}|\hf(\indexnot,t')|dt'
\end{equation}
for $t\in [0,T]$, where $c_{\betafun}$ and $\eta_{\robal}>0$ have the same dependence as in the case of (\ref{eq:meestroughbalset}); and
\[
\eta_{1}:=\max\{0,\eta_{\robal}-\mu_{1}\},\ \ \
\eta_{2}:=\max\{0,-\mu_{1}\}+\max\{0,\eta_{\robal}-\mu_{2}\}.
\]
Similarly,
\begin{equation}\label{eq:meestroughbalsetztoT}
\me^{1/2}(\indexnot,T)\leq e^{c_{\betafun}+\eta_{a}T}e^{\mu_{1}T}\me^{1/2}(\indexnot,0)
+e^{c_{\betafun}+\eta_{b}T}e^{\mu_{1}T}\int_{0}^{T}e^{\mu_{2}(T-t)}|\hf(\indexnot,t)|dt,
\end{equation}
where $c_{\betafun}$ and $\eta_{\robal}>0$ have the same dependence as in the case of (\ref{eq:meestroughbalset}); and
\[
\eta_{a}:=\eta_{\robal}-\mu_{1},\ \ \
\eta_{b}:=-\mu_{1}+\max\{0,\eta_{\robal}-\mu_{2}\}.
\]
\end{cor}
\begin{remark}
Note that the constants $\mu_{l}\in\ro$ are arbitrary. Moreover, the estimates (\ref{eq:meestroughbalsetztot}) and (\ref{eq:meestroughbalsetztoT}) differ 
in the sense that the first estimate applies in the interval $[0,T]$, but the second estimate only applies at $T$. On the other hand, the second estimate 
is better than the estimate obtained by inserting $t=T$ in the first estimate. The $T$-dependent constant should be thought of as a loss of derivatives. 
\end{remark}
\begin{proof}
Note that (\ref{eq:meestroughbalset}) implies that 
\[
\me^{1/2}(\indexnot,t)\leq e^{c_{\betafun}+\eta_{\robal}t}\me^{1/2}(\indexnot,0)
+e^{c_{\betafun}}\int_{0}^{t}e^{\eta_{\robal}(t-t')}|\hf(\indexnot,t')|dt'.
\]
At this stage, we can estimate $\eta_{\robal}t=(\eta_{\robal}-\mu_{1})t+\mu_{1}t\leq \eta_{1}T+\mu_{1}t$ and
\[
\eta_{\robal}(t-t')=-\mu_{1}t+\mu_{1}t+(\eta_{\robal}-\mu_{2})(t-t')+\mu_{2}(t-t')\leq \eta_{2}T+\mu_{1}t+\mu_{2}(t-t')
\]
for $0\leq t'\leq t\leq T$. This yields (\ref{eq:meestroughbalsetztot}) and the proof of (\ref{eq:meestroughbalsetztoT}) is similar. 
\end{proof}

\section{The basic energy estimate}\label{section:roughODEbaenest}

Assume that (\ref{eq:thesystemRge}) is weakly silent, balanced and convergent. In the present section, we then derive energy estimates for solutions 
to (\ref{eq:thesystemRge}) for $t\geq 0$. We do so by carrying out a mode by mode analysis, and then summing up
the resulting estimates. Let, therefore, $z$ be a solution to (\ref{eq:fourierthesystemRge}) corresponding to $\indexnot\in\EFindexset$, and
let $T_{\roode}$ be given by Definition~\ref{def:roughODEtermo}. In order to analyse the behaviour for $t\geq T_{\roode}$, we wish to appeal to the 
results derived in Section~\ref{section:roughODEest}. Before doing so, note that (\ref{eq:alazeas}) holds. Moreover, $\mfg(t)\leq e^{-\ellderbdsil\bt}$ for 
$t\geq T_{\roode}$, where 
\[
\bt:=t-T_{\roode},
\]
so that (\ref{eq:weakbal}) yields
\[
|n_{l}g^{0l}(t)|+\|n_{l}X^{l}(t)\|\leq C_{\coeff}e^{-\ellderbdsil\bt}
\]
for $t\geq T_{\roode}$. When (\ref{eq:thesystemRge}) is weakly silent, balanced and convergent, the equation (\ref{eq:fourierthesystemRge})
can thus be reformulated 
\begin{equation}\label{eq:dotvroughODE}
\dot{v}=A_{\infty}v+A_{\rem}v+F,
\end{equation}
where 
\begin{equation}\label{eq:vAhFdef}
v=\left(\begin{array}{c} z \\ \dot{z}\end{array}\right),\ \ \
A_{\infty}=\left(\begin{array}{cc} 0 & \Id_{m} \\ -\zeta_{\infty} & -\a_{\infty}\end{array}\right),\ \ \
F=\left(\begin{array}{c} 0 \\ \hf \end{array}\right).
\end{equation}
Moreover, $A_{\rem}$ is a smooth function taking its values in $\Mn{2m}{\co}$ and satisfying
\begin{equation}\label{eq:Aremestgensys}
\|A_{\rem}(t)\|\leq C_{\rem}e^{-\b_{\rem}\bt}
\end{equation}
for $t\geq T_{\roode}$, where $\b_{\rem}=\min\{\ellderbdsil,\eta_{\romn}\}$ and the constant $C_{\rem}$ only depends on $C_{\coeff}$ and $C_{\romn}$. 
The equation of interest here thus falls into the category described in Subsection~\ref{ssection:ODEreg}. If we, in addition to the above, 
demand that (\ref{eq:sigmaXCderbderest}) hold, then $\b_{\rem}=\min\{2\ellderbdsil,\ellderbdsil+\b_{\roder},\eta_{\romn}\}$ and $C_{\rem}$ only 
depends on $C_{\coeff}$, $C_{\romn}$ and $C_{\roder}$.

Before stating the results, let us introduce the following terminology. Let $\indexnot\in\EFindexset$. For a solution $z$ to 
(\ref{eq:fourierthesystemRge}) and an $s\in\ro$, define
\[
\me_{s}(\indexnot,t):=\frac{1}{2}\ldr{\nu(\indexnot)}^{2s}[|\dot{z}(\indexnot,t)|^{2}+\mfg^{2}(\indexnot,t)|z(\indexnot,t)|^{2}+|z(\indexnot,t)|^{2}],
\]
\index{$\a$Aa@Notation!Energies!$\me_{s}(\indexnot,t)$}%
where $\nu(\indexnot)$ is given by (\ref{eq:nugenReqdef});
note that $\nu(\indexnot)$ is an element of $\rn{d+R}$, and that $\ldr{\nu(\indexnot)}=[1+|\nu(\indexnot)|^{2}]^{1/2}$. Let, also, $\me:=\me_{0}$,
in accordance with (\ref{eq:medefroughbalset}). 
If $u$ is a solution to (\ref{eq:thesystemRge}), $z$ are the corresponding Fourier modes and $s\in\ro$, consider (\ref{eq:mfedef}):
\begin{equation}\label{eq:mfedefextra}
\mfe_{s}[u](t)=\frac{1}{2}\sum_{\indexnot\in\EFindexset}\ldr{\nu(\indexnot)}^{2s}\left[|\dot{z}(\indexnot,t)|^{2}
+\mfg^{2}(\indexnot,t)|z(\indexnot,t)|^{2}+|z(\indexnot,t)|^{2}\right].
\end{equation}
Finally, in analogy with (\ref{eq:HsnormonbM}), define
\begin{equation}\label{eq:fcdottHsnorm}
\|f(\cdot,t)\|_{(s)}:=\left(\sum_{\indexnot\in\EFindexset}\ldr{\nu(\indexnot)}^{2s}|\hf(\indexnot,t)|^{2}\right)^{1/2},
\end{equation}
where $\hf(\indexnot,t)$ is given by (\ref{eq:hfnutdef}). Let us begin by deriving two simple estimates.

\begin{lemma}\label{lemma:genroughest}
Assume that (\ref{eq:thesystemRge}) is weakly silent, weakly balanced and weakly convergent. Then there are constants $C$ and $s_{0}\geq 0$, 
depending only on the coefficients of the equation (\ref{eq:thesystemRge}), such that if $z$ is a solution to (\ref{eq:fourierthesystemRge})
corresponding to $\indexnot\in\EFindexset$, then
\begin{equation}\label{eq:meestsloss}
\begin{split}
\me_{s}^{1/2}(\indexnot,t) \leq & C\ldr{t}^{d_{1}-1}e^{\kappa_{1}t}\me_{s+s_{0}}^{1/2}(\indexnot,0)\\
 & +C\ldr{\nu(\indexnot)}^{s+s_{0}}\int_{0}^{t}\ldr{t-\tau}^{d_{1}-1}e^{\kappa_{1}(t-\tau)}|\hf(\indexnot,\tau)|d\tau
\end{split}
\end{equation}
for all $t\geq 0$, where $\kappa_{1}:=\kappa_{\max}(A_{\infty})$, $d_{1}:=d_{\max}(A_{\infty},\kappa_{1})$, the matrix $A_{\infty}$ is defined in 
(\ref{eq:vAhFdef}) and $\hf(\indexnot,t)$ is given by (\ref{eq:hfnutdef}). Moreover, 
if $u$ is a smooth solution to (\ref{eq:thesystemRge}), then
\begin{equation}\label{eq:mfeestsloss}
\begin{split}
\mfe_{s}^{1/2}[u](t) \leq & C\ldr{t}^{d_{1}-1}e^{\kappa_{1}t}\mfe_{s+s_{0}}^{1/2}[u](0)\\
 & +C\int_{0}^{t}\ldr{t-\tau}^{d_{1}-1}e^{\kappa_{1}(t-\tau)}\|f(\cdot,\tau)\|_{(s+s_{0})}d\tau
\end{split}
\end{equation}
for all $t\geq 0$ and $s\in\ro$.
\end{lemma}
\begin{remark}
The functions $\kappa_{\max}$ and $d_{\max}$ are introduced in Definition~\ref{def:SpRspdef}.
\end{remark}
\begin{remark}\label{remark:deponconstbasenest}
The constant $C$ appearing in the statement of the lemma only depends on $\ellderbdsil$, $\eta_{\romn}$, $C_{\coeff}$, $C_{\romn}$, $c_{\betafun}$, $A_{\infty}$, 
$g^{ij}(0)$ and $a_{r}(0)$, where $i,j=1,\dots,d$ and $r=0,\dots,R$. Moreover, $s_{0}$ only depends on $c_{\betafun}$, $C_{\coeff}$, $\kappa_{1}$ and 
$\ellderbdsil$. 
\end{remark}
\begin{remark}\label{remark:condequivbasenest}
If the assumptions of Proposition~\ref{prop:genroughestintro} are satisfied, then the assumptions of the present lemma are fulfilled; cf. 
Remark~\ref{remark:wbwswcfollowfromgeometry}.
\end{remark}
\begin{proof}
In the proof of (\ref{eq:meestsloss}), there are two cases to consider: $T_{\roode}=0$ and $T_{\roode}>0$. In the first case, we need only appeal to 
Lemma~\ref{lemma:oderegest}; since $T_{\roode}=0$ implies $\mfg(t)\leq 1$ for all $t\geq 0$, the expressions $|v|$ and $\me^{1/2}$ are equivalent in 
this case. Assume that $T_{\roode}>0$. In the interval $[0,T_{\roode}]$, we can appeal to Lemma~\ref{lemma:oscroughODE} and for 
$t\geq T_{\roode}$, we can appeal to the combination of Lemmas~\ref{lemma:oderegest} and \ref{lemma:oscroughODE}.  When $T_{\roode}>0$, $\me(T_{\roode})$ 
and $E(T_{\roode})$ are equivalent (with a constant of equivalence depending only on the constant $c_{\betafun}$ introduced in Lemma~\ref{lemma:oscroughODE}). 
Combining Lemmas~\ref{lemma:oderegest} and \ref{lemma:oscroughODE} thus yields
\begin{equation*}
\begin{split}
\me^{1/2}(t) \leq & C\ldr{\bt}^{d_{1}-1}e^{\kappa_{1}\bt}E^{1/2}(T_{\roode})
+C\int_{T_{\roode}}^{t}\ldr{t-\tau}^{d_{1}-1}e^{\kappa_{1}(t-\tau)}|\hf(\tau)|d\tau\\
 \leq & C\ldr{t}^{d_{1}-1}e^{\kappa_{1}t}e^{\eta_{+}T_{\roode}}E^{1/2}(0)
+Ce^{\eta_{+}T_{\roode}}\int_{0}^{T_{\roode}}\ldr{t-\tau}^{d_{1}-1}e^{\kappa_{1}(t-\tau)}|\hf(\tau)|d\tau\\
 & +C\int_{T_{\roode}}^{t}\ldr{t-\tau}^{d_{1}-1}e^{\kappa_{1}(t-\tau)}|\hf(\tau)|d\tau
\end{split}
\end{equation*}
for $t\geq T_{\roode}$, where $\eta_{+}=\max\{\eta_{\roode}-\kappa_{1},0\}$. In particular, 
\begin{equation}\label{eq:meestlargt}
\me^{1/2}(t) \leq C\ldr{t}^{d_{1}-1}e^{\kappa_{1}t}e^{\eta_{+}T_{\roode}}\me^{1/2}(0)
+Ce^{\eta_{+}T_{\roode}}\int_{0}^{t}\ldr{t-\tau}^{d_{1}-1}e^{\kappa_{1}(t-\tau)}|\hf(\tau)|d\tau
\end{equation}
for $t\geq T_{\roode}$. Due to (\ref{eq:EroughODEest}), an even better estimate holds for $t\in [0,T_{\roode}]$. In particular, 
(\ref{eq:meestlargt}) thus holds for all $t\geq 0$. However, in order for this estimate to be of any interest, we need to 
estimate $e^{\eta_{+}T_{\roode}}$. Note, to this end, that (\ref{eq:weaksil}) and the definition of $T_{\roode}$ yield
\begin{equation}\label{eq:Todenest}
e^{-c_{\betafun}}=\mfg(T_{\roode})\leq \mfg(0)e^{-\ellderbdsil T_{\roode}+c_{\betafun}}.
\end{equation}
Thus $T_{\roode}\leq \ellderbdsil^{-1}\ln\mfg(0)+2\ellderbdsil^{-1}c_{\betafun}$. In particular, 
\[
e^{\eta_{+}T_{\roode}}\leq C\exp\left[\eta_{+}\ellderbdsil^{-1}\ln\mfg(0)\right]\leq C|\nu(\indexnot)|^{\eta_{+}/\ellderbdsil},
\]
where $C$ only depends on $g^{ij}(0)$, $a_{r}(0)$, $\ellderbdsil$, $c_{\betafun}$, $C_{\coeff}$ and $A_{\infty}$. Thus, letting 
$s_{0}:=\eta_{+}/\ellderbdsil$, the fact that 
(\ref{eq:meestlargt}) holds for all $t\geq 0$ implies that (\ref{eq:meestsloss}) holds. Applying Minkowski's inequality to this estimate 
yields (\ref{eq:mfeestsloss}). 
\end{proof}

For future reference, it is of interest to note that the following related estimate holds.

\begin{lemma}\label{lemma:roughenestbalsettingfullsol}
Assume that (\ref{eq:thesystemRge}) is weakly balanced. Assume, moreover, that there is a constant $C_{\ell}>0$ and a continuous non-negative 
function $\betafun\in L^{1}([0,\infty))$ such that (\ref{eq:elldotbdgeneral}) holds
for all $0\neq \indexnot\in\EFindexset$ and all $t\geq 0$. Let $u$ be a solution to (\ref{eq:thesystemRge}). Then
\begin{equation}\label{eq:meestroughbalsetfullso}
\mfe^{1/2}_{s}[u](t_{1})\leq e^{c_{\betafun}+\eta_{\robal}|t_{1}-t_{0}|}\mfe^{1/2}_{s}[u](t_{0})
+e^{c_{\betafun}}\left|\int_{t_{0}}^{t_{1}}e^{\eta_{\robal}|t_{1}-t|}\|f(\cdot,t)\|_{(s)}dt\right|
\end{equation}
for $t_{0},t_{1}\geq 0$ and $s\in\ro$, where $c_{\betafun}:=\|\betafun\|_{1}$; and $\eta_{\robal}>0$ is a constant depending only on $C_{\ell}$ and 
the constant $C_{\coeff}$ appearing in the definition of weak balance. 
\end{lemma}
\begin{remark}\label{remark:roughenestbalsettingfullsol}
Note that if (\ref{eq:thesystemRge}) is $C^{1}$-balanced in the sense of Definition~\ref{definition:Cobal}, then it satisfies the conditions
of the lemma with $\betafun=0$; this is a consequence of Lemmas~\ref{lemma:condyieldellderbd} and \ref{lemma:sigmaXbdsandderbds}. In fact, 
the conditions of the lemma also hold if we relax the assumptions in Definition~\ref{definition:Cobal} by replacing the conditions that 
\begin{itemize}
\item the shift vector field be $C^{1}$-future bounded, and
\item $|\bk|_{\bge}\leq C_{\bk}$ for $t\geq 0$
\end{itemize}
by the requirements that $\d_{t}$ be future uniformly timelike and
\[
|\bk|_{\bge}+|\chi|_{\bge}\cdot|\dot{\chi}|_{\bge}\leq C_{a}+\betafun(t)
\]  
for all $t\geq 0$, where $0<C_{a}\in\ro$ and $\betafun\in L^{1}([0,\infty))$ is continuous and non-negative. This is a consequence of 
Lemmas~\ref{lemma:condbalweak} and \ref{lemma:sigmaXbdsandderbds}.
\end{remark}
\begin{proof}
The conclusion follows from (\ref{eq:meestroughbalset}) and Minkowski's inequality. 
\end{proof}

\section{Future asymptotics}\label{section:roughODEfutas}

As a next step, we wish to derive asymptotics by appealing to Lemma~\ref{lemma:ODEasymp}; note that the equation (\ref{eq:dotvroughODE})
is of a form such that this lemma is applicable, assuming $f$ to satisfy a suitable integrability condition. 

\begin{lemma}\label{lemma:roughas}
Assume that (\ref{eq:thesystemRge}) is weakly silent, weakly balanced and weakly convergent. Assume, moreover, that $f$ is a smooth function such that
for every $s\in\ro$, 
\begin{equation}\label{eq:fhsintbd}
\|f\|_{A,s}:=\int_{0}^{\infty}e^{-\kappa_{1}\tau}\|f(\cdot,\tau)\|_{(s)}d\tau<\infty
\end{equation}
holds, where $\kappa_{1}:=\kappa_{\max}(A_{\infty})$ and $A_{\infty}$ is defined in (\ref{eq:vAhFdef}); cf. Definition~\ref{def:SpRspdef}. 
Let $\b_{\rem}:=\min\{\ellderbdsil,\eta_{\romn}\}$, where $\ellderbdsil$ and $\eta_{\romn}$ are the constants appearing in the definition of weak silence and weak convergence; 
cf. Definition~\ref{def:roughODEtermo}. Let, moreover, $E_{a}$ be the first generalised eigenspace in the $\b_{\rem},A_{\infty}$-decomposition of $\cn{2m}$; cf. 
Definition~\ref{def:fssubspetc}. Then there are constants $C$, $N$ and $s_{\rohom},s_{\roih}\geq 0$, 
depending only on the coefficients of the equation (\ref{eq:thesystemRge}), such that the following holds. Given a smooth solution $u$ to 
(\ref{eq:thesystemRge}), there is a unique $V_{\infty}\in C^{\infty}(\bM,E_{a})$ such that 
\begin{equation}\label{eq:uudothsest}
\begin{split}
 & \left\|\left(\begin{array}{c} u(\cdot,t) \\ u_{t}(\cdot,t)\end{array}\right)
-e^{A_{\infty}t}V_{\infty}
-\int_{0}^{t}e^{A_{\infty}(t-\tau)}\left(\begin{array}{c} 0 \\ f(\cdot,\tau)\end{array}\right)d\tau\right\|_{(s)} \\
\leq & C\ldr{t}^{N}e^{(\kappa_{1}-\b_{\rem})t}\left(\|u_{t}(\cdot,0)\|_{(s+s_{\rohom})}+\|u(\cdot,0)\|_{(s+s_{\rohom}+1)}
+\|f\|_{A,s+s_{\roih}}\right)
\end{split}
\end{equation}
holds for $t\geq 0$ and all $s\in\ro$; recall that $\bM$ is given by (\ref{eq:bMdef}). Moreover, 
\begin{equation}\label{eq:uinfudinfHsest}
\|V_{\infty}\|_{(s)}\leq C\left(\|u_{t}(\cdot,0)\|_{(s+s_{\rohom})}+\|u(\cdot,0)\|_{(s+s_{\rohom}+1)}+\|f\|_{A,s+s_{\roih}}\right).
\end{equation}
\end{lemma}
\begin{remark}\label{remark:condequivbasassil}
If the assumptions of Proposition~\ref{prop:roughas} are satisfied, then the assumptions of the present lemma are fulfilled; cf. 
Remark~\ref{remark:wbwswcfollowfromgeometry}. Note also that the fact that the $\b_{\rem}$ appearing in the statement of the present lemma
coincides with the $\b_{\rem}$ appearing in the statement of Proposition~\ref{prop:roughas} is a consequence of the fact that, in the 
context of Proposition~\ref{prop:roughas}, $\ellderbd_{\ros}=\mu$; cf. (\ref{eq:mfgdotestbetafun}).
\end{remark}
\begin{remark}\label{remark:depofconroughas}
Recall that the notation $C_{\coeff}$, $C_{\romn}$, $\ellderbdsil$, $\eta_{\romn}$ and $\betafun$ is introduced in Definition~\ref{def:roughODEtermo}. Let 
$c_{\betafun}:=\|\betafun\|_{1}$. Then the constant $C$ only depends on $c_{\betafun}$, $C_{\coeff}$, $C_{\romn}$, $\ellderbdsil$, $\eta_{\romn}$, 
$\eta_{\roode}$, $A_{\infty}$, $g^{ij}(0)$ and $a_{r}(0)$, where $\eta_{\roode}$ is the constant appearing in Lemma~\ref{lemma:oscroughODE}. The
constant $N$ only depends on $m$. Finally, $s_{\rohom}=(\b_{\rem}+\eta_{+})/\ellderbdsil$ where $\eta_{+}:=\max\{\eta_{\roode}-\kappa_{1},0\}$; 
and $s_{\roih}:=\max\{\b_{\rem}+\eta_{+},\kappa+1\}/\ellderbdsil$, where $\kappa:=\Rsp A_{\infty}$ and the notation $\Rsp A_{\infty}$ is introduced 
in Definition~\ref{def:SpRspdef}.
\end{remark}
\begin{remark}\label{remark:improvestchimcXbrem}
If, in addition to the assumptions of the lemma, the estimate (\ref{eq:sigmaXCderbderest}) holds, then $\b_{\rem}$ can be replaced by 
$\min\{2\ellderbdsil,\ellderbdsil+\b_{\roder},\eta_{\romn}\}$; cf. the comments made in connection with (\ref{eq:Aremestgensys}). The dependence 
of the constants then changes in that they also depend on $C_{\roder}$ and $\b_{\roder}$. 
\end{remark}
\begin{remark}\label{remark:improvestchimcXbremgeomset}
If the assumptions of Proposition~\ref{prop:roughas} are satisfied and there are constants $K_{\roder}$ and $\b_{\roder}$ such that 
(\ref{eq:chimcXexpdecest}) holds for all $t\geq 0$, then the conclusions of Lemma~\ref{lemma:roughas} hold with $\b_{\rem}$ replaced by
$\min\{2\mu,\mu+\b_{\roder},\eta_{\romn}\}$. The reason for this is that (\ref{eq:chimcXexpdecest}) and Lemma~\ref{lemma:sigmaXbdsandderbds} 
imply that (\ref{eq:sigmaXCderbderest}) holds so that Remark~\ref{remark:improvestchimcXbrem} applies. 
\end{remark}
\begin{proof}
\textbf{Uniqueness.} The proof of uniqueness is essentially identical to the argument presented at the beginning of the proof of 
Lemma~\ref{lemma:ODEasymp}. We therefore omit the details here. 

\textbf{Strategy, existence.}
In order to prove the statement, we derive estimates for each mode and then sum up the result. For a given $\indexnot\in\EFindexset$, it is natural to 
divide the interval $[0,\infty)$ into two parts: $[0,T_{\roode}]$ and $[T_{\roode},\infty)$. On the latter interval, we can appeal to 
Lemma~\ref{lemma:ODEasymp}. However, we have no similar estimate on the interval $[0,T_{\roode}]$. In fact, there is no reason to expect an expression 
such as the left hand side of (\ref{eq:vinfasres}) to be small on $[0,T_{\roode}]$. The way to overcome this problem is to 
\begin{itemize}
\item observe that the solution cannot grow more than exponentially on $[0,T_{\roode}]$; cf. Lemma~\ref{lemma:oscroughODE};
\item note that this exponential growth cannot exceed a constant multiple of $\ldr{\nu(\indexnot)}^{\kappa}$ for some $\kappa$, so that the growth can be 
reinterpreted in terms of a loss of regularity. 
\end{itemize}
\textbf{Preliminary observations; the objects that need to be estimated.}
Consider (\ref{eq:vinfasres}). We wish to reformulate this estimate to one in which $T_{\roode}$ does not appear, and we wish to prove that it is valid 
for all $t\geq 0$ (note that $v$, $v_{\infty}$ and $F$ appearing in this estimate are all dependent on $\indexnot$ even though this is not explicitly 
stated). Introducing
\begin{equation}\label{eq:vinfmoddef}
v_{\infty,\romod}:=v_{\infty}-\int_{0}^{T_{\roode}}e^{-A_{\infty}\tau}F(\tau)d\tau,
\end{equation}
(\ref{eq:vinfasres}) can be written
\begin{equation}\label{eq:vasadallt}
\begin{split}
 & \left|v(t)-e^{A_{\infty}t}v_{\infty,\romod}-\int_{0}^{t}e^{A_{\infty}(t-\tau)}F(\tau)d\tau\right|\\
 \leq & C\ldr{t}^{N}e^{(\kappa_{1}-\b_{\rem})t}e^{\b_{\rem}T_{\roode}}\left[e^{-\kappa_{1}T_{\roode}}|v(T_{\roode})|+\|F\|_{A}\right];
\end{split}
\end{equation}
recall that this estimate holds on $[T_{\roode},\infty)$ and that the notation $\|F\|_{A}$ is introduced in (\ref{eq:Fnoone}). Note also that the constant
$C$ only depends on $C_{\rem}$, $\b_{\rem}$ and $A_{\infty}$; due to the comments made in connection with (\ref{eq:Aremestgensys}), this means that the constant
only depends on $C_{\coeff}$, $C_{\romn}$, $\ellderbdsil$, $\eta_{\romn}$ and $A_{\infty}$. In accordance with the above strategy, 
we wish to estimate each term appearing on the left hand side separately on the interval $[0,T_{\roode}]$. Moreover, we wish to estimate the right hand 
side in terms of initial data and $\indexnot$, but we do not wish to have any dependence on $|v(T_{\roode})|$ or $T_{\roode}$. In practice, it turns out to be 
convenient to estimate
\begin{equation}\label{eq:zTodeterms}
e^{(\b_{\rem}-\kappa_{1})t}|v(t)|,\ \ \
e^{(\b_{\rem}-\kappa_{1})t}|e^{A_{\infty}t}v_{\infty}|,\ \ \
\left|e^{(\b_{\rem}-\kappa_{1})t}\int_{t}^{T_{\roode}}e^{A_{\infty}(t-\tau)}F(\tau)d\tau\right|
\end{equation}
for $t\in [0,T_{\roode}]$; this yields the desired estimate on the interval $[0,T_{\roode}]$. In order to obtain a good estimate for $t\geq T_{\roode}$, we need
to estimate 
\begin{equation}\label{eq:Todeinftyterms}
e^{(\b_{\rem}-\kappa_{1})T_{\roode}}|v(T_{\roode})|,\ \ \
e^{\b_{\rem}T_{\roode}}\|F\|_{A}.
\end{equation}
Finally, we wish to estimate $v_{\infty,\romod}$ introduced in (\ref{eq:vinfmoddef}). Note that if we have estimated the first expression appearing in 
(\ref{eq:zTodeterms}) on $[0,T_{\roode}]$, then we have also estimated the first expression appearing in (\ref{eq:Todeinftyterms}). 

\textbf{Estimating the size of the solution.} Assume that $T_{\roode}>0$. Then we can appeal to Lemma~\ref{lemma:oscroughODE} in order to obtain
\begin{equation*}
\begin{split}
e^{(\b_{\rem}-\kappa_{1})t}|v(t)|  \leq & \sqrt{2}e^{2c_{\betafun}}e^{\eta_{\rohom,1}t}E^{1/2}(0)+\sqrt{2}e^{2c_{\betafun}}e^{\eta_{\roih,1}t}
\int_{0}^{t}e^{-\kappa_{1}\tau}|\hf(\tau)|d\tau
\end{split}
\end{equation*}
for $t\in [0,T_{\roode}]$, where $\eta_{\rohom,1}:=\max\{\b_{\rem}+\eta_{\roode}-\kappa_{1},0\}$, $\eta_{\roih,1}:=\b_{\rem}+\max\{\eta_{\roode}-\kappa_{1},0\}$, 
$\betafun$ is the function appearing in the definition of weak silence and $c_{\betafun}:=\|\betafun\|_{1}$ (for future reference, it is convenient to 
keep in mind that the same estimate holds with $\b_{\rem}$ replaced by $0$). Due to (\ref{eq:Todenest}), it then follows that 
\begin{equation}\label{eq:vsupesztode}
\begin{split}
 & \ldr{\nu(\indexnot)}^{s}\sup_{t\in [0,T_{\roode}]}e^{(\b_{\rem}-\kappa_{1})t}|v(t)| \\
\leq & C\me_{s+s_{\rohom,1}}^{1/2}(0)+C\ldr{\nu(\indexnot)}^{s+s_{\roih,1}}\int_{0}^{T_{\roode}}e^{-\kappa_{1}\tau}|\hf(\tau)|d\tau,
\end{split}
\end{equation}
where $s_{\rohom,1}:=\eta_{\rohom,1}/\ellderbdsil$, $s_{\roih,1}:=\eta_{\roih,1}/\ellderbdsil$ and the constant $C$ only depends on $c_{\betafun}$, $\ellderbdsil$, $\eta_{\romn}$, 
$\eta_{\roode}$, $\kappa_{1}$,
$g^{ij}(0)$ and $a_{r}(0)$ (and, again, the analogous estimate with $\b_{\rem}$ replaced by $0$ holds). One particular consequence of this estimate is 
\begin{equation}\label{eq:vFATodeest}
\begin{split}
 & \ldr{\nu(\indexnot)}^{s}e^{(\b_{\rem}-\kappa_{1})T_{\roode}}|v(T_{\roode})|+\ldr{\nu(\indexnot)}^{s}e^{\b_{\rem}T_{\roode}}\|F\|_{A}\\
 \leq & C\me_{s+s_{\rohom,1}}^{1/2}(0)+C\ldr{\nu(\indexnot)}^{s+s_{\roih,1}}\|\hf\|_{A},
\end{split}
\end{equation}
where the constant $C$ has the same dependence as in the case of (\ref{eq:vsupesztode}) and we have used the notation (\ref{eq:Fnoone}). Moreover, 
the analogous estimate with $\b_{\rem}$ replaced by $0$ holds.  Clearly, there are various ways in which (\ref{eq:vFATodeest})
could be improved in terms of the power of $\ldr{\nu(\indexnot)}$. However, we do not aim for an optimal estimate in that regard here. 

\textbf{Estimating the limit.} Let us consider the 
vector $v_{\infty}$ appearing in (\ref{eq:vinfasres}) in greater detail. To begin with, it is given by (\ref{eq:vinfform}), where 
$u_{\infty}$ satisfies the estimate (\ref{eq:winfTodeest}), in which the constant $C$ only depends on $C_{\coeff}$, 
$C_{\romn}$, $\ellderbdsil$, $\eta_{\romn}$ and $A_{\infty}$. Note that due to (\ref{eq:winfTodeest}) and estimates of the above type (in fact, the estimate 
(\ref{eq:vFATodeest}) where $\b_{\rem}$ has been set to zero), 
\begin{equation}\label{eq:winfestitoEFA}
\ldr{\nu(\indexnot)}^{s}|u_{\infty}|\leq Ce^{\kappa_{1}T_{\roode}}\me_{s+s_{0}}^{1/2}(0)+Ce^{\kappa_{1}T_{\roode}}\ldr{\nu(\indexnot)}^{s+s_{0}}\|\hf\|_{A},
\end{equation}
where $s_{0}:=\eta_{+}/\ellderbdsil$, $\eta_{+}:=\max\{\eta_{\roode}-\kappa_{1},0\}$ and $C$ only depends on $c_{\betafun}$, $C_{\coeff}$, $C_{\romn}$, $\ellderbdsil$, 
$\eta_{\romn}$, 
$\eta_{\roode}$, $A_{\infty}$, $g^{ij}(0)$ and $a_{r}(0)$ (since we appeal to (\ref{eq:winfTodeest}), the constant depends on $C_{\rem}$). Since $v_{\infty}$ is given by 
(\ref{eq:vinfform}) and since $u_{\infty}\in E_{A_{\infty},\b_{\rem}}$, it follows that
\begin{equation}\label{eq:vinfestitoidaf}
\ldr{\nu(\indexnot)}^{s}|v_{\infty}|\leq C\me_{s+s_{\rohom,2}}^{1/2}(0)+C\ldr{\nu(\indexnot)}^{s+s_{\roih,2}}\|\hf\|_{A},
\end{equation}
where $s_{\rohom,2}:=s_{\roih,1}$, $s_{\roih,2}:=s_{\roih,1}$ and the constant $C$ has the same dependence as in the case of (\ref{eq:winfestitoEFA}). In 
order to estimate the second expression in (\ref{eq:zTodeterms}), note that (\ref{eq:vinfform}) yields
\begin{equation}\label{eq:vinfztodeest}
e^{(\b_{\rem}-\kappa_{1})t}e^{A_{\infty}t}v_{\infty}=e^{(\b_{\rem}-\kappa_{1})t}e^{A_{\infty}\bt}u_{\infty}.
\end{equation}
This equation, the fact that $u_{\infty}\in E_{A_{\infty},\b_{\rem}}$ and the fact that $t\leq T_{\roode}$ lead to the conclusion that 
\[
|e^{(\b_{\rem}-\kappa_{1})t}e^{A_{\infty}t}v_{\infty}|\leq Ce^{(\b_{\rem}-\kappa_{1})T_{\roode}}|u_{\infty}|
\]
for $t\in [0,T_{\roode}]$, where $C$ only depends on $\b_{\rem}$ and $A_{\infty}$. Combining this estimate with (\ref{eq:winfestitoEFA}) yields
\begin{equation}\label{eq:vinfztodeestfin}
\ldr{\nu(\indexnot)}^{s}\sup_{t\in [0,T_{\roode}]}|e^{(\b_{\rem}-\kappa_{1})t}e^{A_{\infty}t}v_{\infty}|
\leq C\me_{s+s_{\rohom,2}}^{1/2}(0)+C\ldr{\nu(\indexnot)}^{s+s_{\roih,2}}\|\hf\|_{A},
\end{equation}
where the constant $C$ has the same dependence as in the case of (\ref{eq:winfestitoEFA}). 

\textbf{Estimating the contribution of the right hand side of (\ref{eq:thesystemRge}).} Let us turn to 
\begin{equation}\label{eq:FconzTodepart}
\begin{split}
 & e^{(\b_{\rem}-\kappa_{1})t}\int_{t}^{T_{\roode}}e^{A_{\infty}(t-\tau)}F(\tau)d\tau\\
 = & \int_{t}^{T_{\roode}}e^{\b_{\rem}\tau}e^{(A_{\infty}-\kappa_{1}\Id_{2m}+\b_{\rem}\Id_{2m})(t-\tau)}e^{-\kappa_{1}\tau}F(\tau)d\tau.
\end{split}
\end{equation}
The situation is similar to (\ref{eq:vinfztodeest}). However, there is a fundamental difference in that we do not know that $F$ takes its values in 
$E_{A_{\infty},\b_{\rem}}$. In order to estimate the right hand side of (\ref{eq:FconzTodepart}), it is convenient to 
consider the cases $\b_{\rem}-\kappa>0$ and $\b_{\rem}-\kappa\leq 0$ separately, where $\kappa:=\Rsp A_{\infty}$ and the notation $\Rsp A_{\infty}$ is introduced in 
Definition~\ref{def:SpRspdef}. In the second case, there is potentially a polynomial loss, which we estimate crudely by an exponentially growing 
factor. This yields
\begin{equation}\label{eq:Fintfinest}
\ldr{\nu(\indexnot)}^{s}\sup_{t\in [0,T_{\roode}]}\left|e^{(\b_{\rem}-\kappa_{1})t}\int_{t}^{T_{\roode}}e^{A_{\infty}(t-\tau)}F(\tau)d\tau\right|
\leq C\ldr{\nu(\indexnot)}^{s+s_{\roih,3}}\|\hf\|_{A},
\end{equation}
where $s_{\roih,3}=\max\{\b_{\rem},\kappa+1\}/\ellderbdsil$ and $C$ only depends on $\ellderbdsil$, $\eta_{\romn}$, $c_{\betafun}$, $A_{\infty}$, $g^{ij}(0)$ and $a_{r}(0)$. 

\textbf{Combining the estimates.} Since the 
second term on the right hand side of (\ref{eq:vinfmoddef}) (multiplied by $\ldr{\nu(\indexnot)}^{s}$) can be estimated by 
the left hand side of (\ref{eq:Fintfinest}), we have all the estimates we need. To begin with, (\ref{eq:vinfmoddef}), (\ref{eq:vinfestitoidaf})
and (\ref{eq:Fintfinest}) yield
\begin{equation}\label{eq:vinfromodest}
\ldr{\nu(\indexnot)}^{s}|v_{\infty,\romod}|\leq C\me_{s+s_{\infty,\rohom}}^{1/2}(0)+C\ldr{\nu(\indexnot)}^{s+s_{\infty,\roih}}\|\hf\|_{A},
\end{equation}
where $s_{\infty,\rohom}:=(\b_{\rem}+\eta_{+})/\ellderbdsil$, $s_{\infty,\roih}:=\max\{\b_{\rem}+\eta_{+},\kappa+1\}/\ellderbdsil$ and $C$ has the same dependence as in the case of 
(\ref{eq:winfestitoEFA}). Moreover, combining (\ref{eq:vsupesztode}), (\ref{eq:vinfztodeestfin}) and (\ref{eq:Fintfinest}) yields
\begin{equation}\label{eq:vasadalltmod}
\begin{split}
 & \ldr{\nu(\indexnot)}^{s}\left|v(t)-e^{A_{\infty}t}v_{\infty,\romod}-\int_{0}^{t}e^{A_{\infty}(t-\tau)}F(\tau)d\tau\right|\\
 \leq & Ce^{(\kappa_{1}-\b_{\rem})t}\left(\me_{s+s_{\rohom,4}}^{1/2}(0)+\ldr{\nu(\indexnot)}^{s+s_{\roih,4}}\|\hf\|_{A}\right)
\end{split}
\end{equation}
for $0\leq t\leq T_{\roode}$, where $s_{\rohom,4}:=(\b_{\rem}+\eta_{+})/\ellderbdsil$, $s_{\roih,4}:=\max\{\b_{\rem}+\eta_{+},\kappa+1\}/\ellderbdsil$ and $C$ has the same 
dependence as in the case of (\ref{eq:winfestitoEFA}). Combining (\ref{eq:vasadallt}) and (\ref{eq:vFATodeest}) yields an estimate similar to 
(\ref{eq:vasadalltmod}) which is valid for $t\geq T_{\roode}$. Combining the two yields
\begin{equation}\label{eq:vastgtTode}
\begin{split}
 & \ldr{\nu(\indexnot)}^{s}\left|v(t)-e^{A_{\infty}t}v_{\infty,\romod}-\int_{0}^{t}e^{A_{\infty}(t-\tau)}F(\tau)d\tau\right|\\
 \leq & C\ldr{t}^{N}e^{(\kappa_{1}-\b_{\rem})t}\left(\me_{s+s_{\rohom}}^{1/2}(0)+\ldr{\nu(\indexnot)}^{s+s_{\roih}}\|\hf\|_{A}\right)
\end{split}
\end{equation}
for $t\geq 0$, where $s_{\rohom}:=(\b_{\rem}+\eta_{+})/\ellderbdsil$; $s_{\roih}:=\max\{\b_{\rem}+\eta_{+},\kappa+1\}/\ellderbdsil$; $C$ has the same dependence as 
in the case of (\ref{eq:winfestitoEFA}); and $N$ is a non-negative integer depending only on $m$. Moreover, in case $T_{\roode}=0$, (\ref{eq:vastgtTode})
is an immediate consequence of (\ref{eq:vinfasres}).

\textbf{Projecting the data at infinity.} At this stage, we would like to sum up the estimate (\ref{eq:vastgtTode}) over the different modes in order to 
obtain (\ref{eq:uudothsest}). Comparing (\ref{eq:vastgtTode}) and (\ref{eq:uudothsest}), it seems reasonable to choose $V_{\infty}$ so that its 
$\indexnot$'th mode is given by $v_{\infty,\romod}(\indexnot)$. However, there is one problem with doing so: we do not know that $v_{\infty,\romod}(\indexnot)$
belongs to $E_{a}$, since it is not clear that the second term on the right hand side of (\ref{eq:vinfmoddef}) belongs to this space. In fact, we first
need to project $v_{\infty,\romod}(\indexnot)$ to $E_{a}$. This leaves an error term we need to estimate. It is given by
\begin{equation}\label{eq:projectiontobeestimated}
e^{A_{\infty}t}\Pi_{b}\left(\int_{0}^{T_{\roode}}e^{-A_{\infty}\tau}F(\tau)d\tau\right),
\end{equation}
where $\Pi_{b}$ denotes the projection to the second generalised eigenspace in the $\b_{\rem}$, $A_{\infty}$-decomposition of $\cn{2m}$. Appealing to 
Lemma~\ref{lemma:eAtPibnormest} and (\ref{eq:Fintfinest}) yields 
\begin{equation*}
\begin{split}
\left|e^{A_{\infty}t}\Pi_{b}\left(\int_{0}^{T_{\roode}}e^{-A_{\infty}\tau}F(\tau)d\tau\right)\right| \leq & C_{A}\ldr{t}^{d_{b}-1}e^{\kappa_{b}t}
\left|\int_{0}^{T_{\roode}}e^{-A_{\infty}\tau}F(\tau)d\tau\right|\\
 \leq & C\ldr{t}^{d_{b}-1}e^{\kappa_{b}t}\ldr{\nu(\indexnot)}^{s_{\roih,3}}\|\hf\|_{A},
\end{split}
\end{equation*}
where $d_{b}$ and $\kappa_{b}$ are given in the statement of Lemma~\ref{lemma:eAtPibnormest} and $C$ has the same dependence as in the case of 
(\ref{eq:Fintfinest}). Since $\kappa_{b}\leq \kappa_{1}-\b_{\rem}$, this estimate yields
\[
\ldr{\nu(\indexnot)}^{s}|e^{A_{\infty}t}\Pi_{b}v_{\infty,\romod}|\leq C\ldr{t}^{N}e^{(\kappa_{1}-\b_{\rem})t}\ldr{\nu(\indexnot)}^{s+s_{\roih}}\|\hf\|_{A},
\]
where $C$ and $N$ have the same dependence as in the case of (\ref{eq:vastgtTode}). Thus (\ref{eq:vastgtTode}) holds with $v_{\infty,\romod}$ 
replaced by $\Pi_{a}v_{\infty,\romod}$.

\textbf{Summing up.}
Given $p=(x,p_{1},\dots,p_{R})\in\bM$, where $\bM$ is given by (\ref{eq:bMdef}), define $V_{\infty}(p)$ by
\[
V_{\infty}(p):=\textstyle{\sum}_{\indexnot\in\EFindexset}\Pi_{a}[v_{\infty,\romod}(\indexnot)]\varphi_{\indexnot}(p),
\]
where $\varphi_{\indexnot}$ is given by (\ref{eq:varphinudef}). In order to justify that the sum on the right hand side makes sense and yields a smooth 
function, it is sufficient to appeal to the fact that $u$ is smooth; the assumption that (\ref{eq:fhsintbd}) holds for all $s$; 
(\ref{eq:vinfromodest}); and the Minkowski inequality. Then (\ref{eq:vastgtTode}) (with $v_{\infty,\romod}$ replaced by $\Pi_{a}v_{\infty,\romod}$) implies that
\begin{equation*}
\begin{split}
 & \left\|\left(\begin{array}{c} u(\cdot,t) \\ u_{t}(\cdot,t)\end{array}\right)
-e^{A_{\infty}t}V_{\infty}
-\int_{0}^{t}e^{A_{\infty}(t-\tau)}\left(\begin{array}{c} 0 \\ f(\cdot,\tau)\end{array}\right)d\tau\right\|_{(s)} \\
\leq & C\ldr{t}^{N}e^{(\kappa_{1}-\b_{\rem})t}\left(\mfe_{s+s_{\rohom}}^{1/2}[u](0)+\|f\|_{A,s+s_{\roih}}\right)
\end{split}
\end{equation*}
for $t\geq 0$, where we have appealed to the Minkowski inequality and $C$ has the same dependence as in the case of (\ref{eq:winfestitoEFA}). 
Thus (\ref{eq:uudothsest}) holds. Note also that, due to (\ref{eq:vinfromodest}), 
\[
\|V_{\infty}\|_{(s)}\leq C\left(\mfe_{s+s_{\rohom}}^{1/2}[u](0)
+\|f\|_{A,s+s_{\roih}}\right),
\]
where $C$ has the same dependence as in the case of (\ref{eq:winfestitoEFA}). Thus (\ref{eq:uinfudinfHsest}) holds. 
\end{proof}

\section{Specifying the asymptotics}\label{section:roughODEspecas}

As a next step, it is of interest to go in the direction opposite to that of Lemma~\ref{lemma:roughas}; i.e., to specify $V_{\infty}$ and to find a 
solution such that (\ref{eq:uudothsest}) holds. Since it is sufficient to consider the case of homogeneous equations, let us assume that $f=0$.

\begin{lemma}\label{lemma:spasda}
Assume that (\ref{eq:thesystemRge}) is weakly balanced, weakly silent and weakly convergent. Assume, moreover, that $f=0$ and that there is a constant 
$\ellderbdlow>0$ and a non-negative continuous function $\betafun_{\low}\in L^{1}([0,\infty))$ such that 
\begin{equation}\label{eq:elldlb}
\dot{\ell}(\indexnot,t)\geq -\ellderbdlow-\betafun_{\low}(t)
\end{equation}
for all $t\geq 0$ and all $0\neq \indexnot\in\EFindexset$; recall that $\ell$ is defined in (\ref{eq:ellsigmaXgenRdef}).
Let $A_{\infty}$ be defined by (\ref{eq:vAhFdef}) and let $\b_{\rem}=\min\{\eta_{\romn},\ellderbdsil\}$, where $\ellderbdsil$ and $\eta_{\romn}$ are the constants 
appearing in 
(\ref{eq:weaksil}) and (\ref{eq:alazeas}) respectively. Finally, let $E_{a}$ be the first generalised eigenspace in the $\b_{\rem},A_{\infty}$-decomposition of 
$\cn{2m}$; cf. Definition~\ref{def:fssubspetc}. Then there is an injective map
\[
\Phi_{\infty}:C^{\infty}(\bM,E_{a})\rightarrow C^{\infty}(\bM,\cn{2m})
\]
such that the following holds. First, 
\begin{equation}\label{eq:Phiinfnobd}
\|\Phi_{\infty}(\chi)\|_{(s)}\leq C\|\chi\|_{(s+s_{\infty})}
\end{equation}
for all $s\in\ro$ and all $\chi\in C^{\infty}(\bM,E_{a})$, where the constants $C$ and $s_{\infty}\geq 0$ only depend on $C_{\coeff}$, $C_{\romn}$, $\ellderbdsil$, 
$\ellderbdlow$, $\eta_{\romn}$, $A_{\infty}$, $\|\betafun\|_{1}$, $\|\betafun_{\low}\|_{1}$, $g^{ij}(0)$, $i,j=1,\dots,d$, and $a_{r}(0)$, $r=1,\dots,R$. Secondly, 
if $\chi\in C^{\infty}(\bM,E_{a})$ and $u$ is the solution to (\ref{eq:thesystemRge}) (with $f=0$) such that 
\begin{equation}\label{eq:uuditoPhiinfchi}
\left(\begin{array}{c} u(\cdot,0) \\ u_{t}(\cdot,0)\end{array}\right)=\Phi_{\infty}(\chi),
\end{equation}
then 
\begin{equation}\label{eq:estspecasdata}
\begin{split}
 & \left\|\left(\begin{array}{c} u(\cdot,t) \\ u_{t}(\cdot,t)\end{array}\right)
-e^{A_{\infty}t}\chi\right\|_{(s)} \\
\leq & C\ldr{t}^{N}e^{(\kappa_{1}-\b_{\rem})t}\left(\|u_{t}(\cdot,0)\|_{(s+s_{\rohom})}+\|u(\cdot,0)\|_{(s+s_{\rohom}+1)}\right)
\end{split}
\end{equation}
for all $t\geq 0$ and all $s\in\ro$, 
where the constants $C$, $N$ and $s_{\rohom}$ have the same dependence as in the case of Lemma~\ref{lemma:roughas}; cf. Remark~\ref{remark:depofconroughas}.
Finally, if $E_{a}=\cn{2m}$ (i.e., if $\b_{\rem}>\Rsp A_{\infty}$; cf. Definition~\ref{def:SpRspdef}), then $\Phi_{\infty}$ is surjective. 
\end{lemma}
\begin{remark}
By combining (\ref{eq:Phiinfnobd}), (\ref{eq:uuditoPhiinfchi}) and (\ref{eq:estspecasdata}), the norms of $u(\cdot,0)$ and $u_{t}(\cdot,0)$ appearing on 
the right hand side of (\ref{eq:estspecasdata}) can be replaced by a suitable Sobolev norm of $\chi$. 
\end{remark}
\begin{remark}
In order to obtain a similar result in the case of inhomogeneous equations, it is sufficient to combine Lemmas~\ref{lemma:roughas} and
\ref{lemma:spasda}; cf. Remark~\ref{remark:inhomaschar}. 
\end{remark}
\begin{remark}\label{remark:specassilsetpropequivtolem}
If the assumptions of Proposition~\ref{prop:spasda} hold, then the conditions of the lemma are fulfilled. The reasons for this are the 
following. First of all, (\ref{eq:thesystemRge}) is weakly balanced, weakly silent and weakly convergent due to 
Remark~\ref{remark:wbwswcfollowfromgeometry}. The lower bound on $\dot{\ell}$ is a consequence of the assumptions of Proposition~\ref{prop:spasda}
and Remark~\ref{remark:bkesttoelldotestuppbd}. The fact that the $\b_{\rem}$ appearing in the statement of the present lemma
coincides with the $\b_{\rem}$ appearing in the statement of Proposition~\ref{prop:spasda} is a consequence of the fact that, in the 
context of Proposition~\ref{prop:spasda}, $\ellderbd_{\ros}=\mu$; cf. (\ref{eq:mfgdotestbetafun}).
\end{remark}
\begin{remark}\label{remark:improvestchimcXbremspas}
If, in addition to the assumptions of the lemma, the estimate (\ref{eq:sigmaXCderbderest}) holds, then $\b_{\rem}$ can be replaced by 
$\min\{2\ellderbdsil,\ellderbdsil+\b_{\roder},\eta_{\romn}\}$; cf. the comments made in connection with (\ref{eq:Aremestgensys}). The dependence 
of the constants then changes in that they also depend on $C_{\roder}$ and $\b_{\roder}$. 
\end{remark}
\begin{remark}\label{remark:improvestchimcXbremgeomsetspas}
If the assumptions of Proposition~\ref{prop:spasda} are satisfied and there are constants $K_{\roder}$ and $\b_{\roder}$ such that 
(\ref{eq:chimcXexpdecest}) holds for all $t\geq 0$, then the conclusions of Lemma~\ref{lemma:spasda} hold with $\b_{\rem}$ replaced by
$\min\{2\mu,\mu+\b_{\roder},\eta_{\romn}\}$. The reason for this is that (\ref{eq:chimcXexpdecest}) and Lemma~\ref{lemma:sigmaXbdsandderbds} 
imply that (\ref{eq:sigmaXCderbderest}) holds so that Remark~\ref{remark:improvestchimcXbremspas} applies. 
\end{remark}
\begin{proof}
Let $\chi\in C^{\infty}(\bM,E_{a})$, $\indexnot\in \EFindexset$ and $\hchi(\indexnot):=\ldrbox{\chi,\varphi_{\indexnot}}\in E_{a}$; cf. the notation introduced
in Subsection~\ref{ssection:specprodset}. Let $u$ be a solution to (\ref{eq:thesystemRge}) and let $z(\indexnot,t)$ be given by (\ref{eq:znutdef}). 
Note that the $\indexnot$'th mode of the expression inside the norm on the left hand side of (\ref{eq:estspecasdata}) can be written
\begin{equation}\label{eq:vasitochih}
v-e^{A_{\infty}t}\hchi,
\end{equation}
where we use the notation (\ref{eq:vAhFdef}). On the other hand, $v$ satisfies (\ref{eq:dotvroughODE}) with $F=0$, and $A_{\rem}$ satisfies 
(\ref{eq:Aremestgensys}) for $t\geq T_{\roode}$. Thus the conditions needed to apply Lemma~\ref{lemma:spasODEsett} are satisfied. In fact, specifying
$v$ by imposing the initial condition $v(T_{\roode}):=\Psi_{\infty}[\hchi(\indexnot)]$ yields the conclusion that (\ref{eq:vasitochih}) is small asymptotically; 
here $\Psi_{\infty}$ is the map constructed in Lemma~\ref{lemma:spasODEsett}. Note that $\Psi_{\infty}$ depends on $\indexnot$. However, the estimates
involving $\Psi_{\infty}$ that we need do not. In order to estimate $|v(T_{\roode})|$ in terms of $|\hchi(\indexnot)|$, define $u_{\infty}\in E_{a}$ by 
\[
e^{-A_{\infty}T_{\roode}}u_{\infty}=\hchi(\indexnot).
\]
Then $|u_{\infty}|=|e^{A_{\infty}T_{\roode}}\hchi(\indexnot)|$, so that (\ref{eq:Psiinfnorm}) yields the estimate
\begin{equation}\label{eq:vTodeabsestasdata}
\begin{split}
|v(T_{\roode})| = & |\Psi_{\infty}(e^{-A_{\infty}T_{\roode}}u_{\infty})|\leq C|u_{\infty}|=C|e^{A_{\infty}T_{\roode}}\hchi(\indexnot)|\\
 \leq & Ce^{(\kappa_{1}+1)T_{\roode}}|\hchi(\indexnot)|
\end{split}
\end{equation}
where $C$ only depends on $C_{\rem}$, $\b_{\rem}$ and $A_{\infty}$. 
In order to estimate $\me_{s}(0)$, it is natural to divide the analysis into two cases. To begin with, assume that $T_{\roode}=0$. Then 
\begin{equation}\label{eq:mesestTodeez}
\begin{split}
\me_{s}^{1/2}(0)\leq & \ldr{\nu(\indexnot)}^{s}|v(0)| = \ldr{\nu(\indexnot)}^{s}|v(T_{\roode})|\leq C\ldr{\nu(\indexnot)}^{s}|\hchi(\indexnot)|
\end{split}
\end{equation}
for all $s$, where $C$ only depends on $C_{\rem}$, $\b_{\rem}$ and $A_{\infty}$ and we have appealed to (\ref{eq:vTodeabsestasdata}) and the fact that $T_{\roode}=0$. 
In case $T_{\roode}>0$, it is natural to proceed as in the case of (\ref{eq:dEdtoscph}). Since we wish to estimate $dE/dt$ from below, we need a lower bound 
on $\dot{\ell}$, which is provided by (\ref{eq:elldlb}). An estimate similar to (\ref{eq:dEdtoscph}) yields the conclusion that 
$dE/dt\geq -2(\eta_{\low}+\betafun_{\low})E$ on $[0,T_{\roode}]$, where the constant $\eta_{\low}$ only depends on $C_{\coeff}$, $c_{\betafun}$ and $\ellderbdlow$, 
where $c_{\betafun}:=\|\betafun\|_{1}$, $\betafun$ is the function appearing in (\ref{eq:weaksil}) and we have used the fact that $\mfg(t)\geq e^{-c_{\betafun}}$ on 
$[0,T_{\roode}]$. Using the notation $c_{\low}:=\|\betafun_{\low}\|_{1}$ and appealing to 
(\ref{eq:vTodeabsestasdata}) yields
\begin{equation}\label{eq:daatinftodaatzest}
\begin{split}
E^{1/2}(0) \leq & e^{\eta_{\low}T_{\roode}+c_{\low}}E^{1/2}(T_{\roode})=2^{-1/2}e^{\eta_{\low}T_{\roode}+c_{\low}}|v(T_{\roode})|\\
 \leq & Ce^{(\eta_{\low}+\kappa_{1}+1)T_{\roode}}|\hchi(\indexnot)|,
\end{split}
\end{equation}
where $C$ only depends on $C_{\rem}$, $\b_{\rem}$, $c_{\low}$ and $A_{\infty}$. Thus
\begin{equation}\label{eq:mesestTodenz}
\me_{s}^{1/2}(0)\leq C\ldr{\nu(\indexnot)}^{s+s_{\infty}}|\hchi(\indexnot)|
\end{equation}
for all $s\in\ro$, where $C$ only depends on $C_{\coeff}$, $C_{\romn}$, $\ellderbdlow$, $\ellderbdsil$, $\eta_{\romn}$, $c_{\low}$, $c_{\betafun}$, $A_{\infty}$, $g^{ij}(0)$ ($i,j=1,\dots,d$) 
and $a_{r}(0)$ ($r=0,\dots,R$); $s_{\infty}\geq 0$ only depends on $C_{\coeff}$, $\ellderbdlow$, $\ellderbdsil$, $c_{\betafun}$ and $A_{\infty}$ (it can be chosen to equal 
$\max\{\eta_{\low}+\kappa_{1}+1,0\}/\ellderbdsil$); and we have appealed to (\ref{eq:Todenest}). Combining (\ref{eq:mesestTodeez}) and (\ref{eq:mesestTodenz}) yields 
the conclusion that (\ref{eq:mesestTodenz}) holds regardless of whether $T_{\roode}=0$ or not. 

Let $\Phi_{\infty}$ be the map taking $\chi\in C^{\infty}(\bM,E_{a})$
to the element of $C^{\infty}(\bM,\cn{2m})$ whose $\indexnot$'th mode is given by $v(0)$, where $v$ is constructed as above. Then 
(\ref{eq:mesestTodenz}) implies that (\ref{eq:Phiinfnobd}) holds; note that this implies that $\Phi_{\infty}$ maps $\chi\in C^{\infty}(\bM,E_{a})$ to an element
of $C^{\infty}(\bM,\cn{2m})$. Since $\Phi_{\infty}$ is injective on the level of Fourier coefficients (this is a consequence of Lemma~\ref{lemma:spasODEsett}), 
it follows that $\Phi_{\infty}$ is injective. Next, let $\chi\in C^{\infty}(\bM,E_{a})$ and let $u$ be the solution to (\ref{eq:thesystemRge}) (with $f=0$) satisfying 
(\ref{eq:uuditoPhiinfchi}). Due to Lemma~\ref{lemma:roughas}, we know that (\ref{eq:estspecasdata}) holds with $\chi$ replaced by some 
$V_{\infty}\in C^{\infty}(\bM,E_{a})$. Moreover, the constants $C$, $N$ and $s_{\hom}$ have the dependence stated in Lemma~\ref{lemma:roughas}. In order to determine 
$V_{\infty}$ in terms of $\chi$, note that (\ref{eq:estspecasdata}) (with $\chi$ replaced by $V_{\infty}$) implies that 
\begin{equation}\label{eq:vprelaspf}
\left|v(t)-e^{A_{\infty}t}v_{\infty}\right| \leq C\ldr{t}^{N}e^{(\kappa_{1}-\b_{\rem})t}
\end{equation}
where $C$ depends on the initial data and $\indexnot$; $v$ is defined as in (\ref{eq:vAhFdef}) where $z$ is the $\indexnot$'th mode of $u$; and $v_{\infty}$ is 
the $\indexnot$'th mode of $V_{\infty}$. Due to the construction of $\Phi_{\infty}$, we know that the estimate (\ref{eq:vprelaspf}) holds 
with $v_{\infty}$ replaced by $\hchi(\indexnot)$. Thus
\[
\left|e^{A_{\infty}t}[\hchi(\indexnot)-v_{\infty}]\right| \leq C\ldr{t}^{N}e^{(\kappa_{1}-\b_{\rem})t}
\]
for $t\geq 0$. Since $\hchi(\indexnot)-v_{\infty}\in E_{a}$, this estimate implies that $\hchi(\indexnot)-v_{\infty}=0$. Thus $V_{\infty}=\chi$, and 
the lemma follows, except for the statement concerning surjectivity. 

Assume that $E_{a}=\cn{2m}$ and let $\psi\in C^{\infty}(\bM,\cn{2m})$. We wish to demonstrate that $\psi$ is in the image of $\Phi_{\infty}$. Let 
$u$ be the solution to (\ref{eq:thesystemRge}) with initial data given by (\ref{eq:uuditoPhiinfchi}), where the right hand side has been replaced by 
$\psi$. Appealing to Lemma~\ref{lemma:roughas} yields a $V_{\infty}\in C^{\infty}(\bM,\cn{2m})$ such that 
(\ref{eq:uudothsest}) holds (with $f=0$). Let $\bu$ be the solution to (\ref{eq:thesystemRge}) with initial data given by (\ref{eq:uuditoPhiinfchi}),
where the right hand side has been replaced by $\Phi_{\infty}(V_{\infty})$. Then, by the above arguments, (\ref{eq:uudothsest}) holds (with $f=0$)
and $u$ replaced by $\bu$. In particular, there is thus a constant $C$, depending on $u$, $\bu$ etc., such that 
\begin{equation}\label{eq:uidminusbuidrough}
\left\|\left(\begin{array}{c} u(\cdot,t) \\ u_{t}(\cdot,t)\end{array}\right)
-\left(\begin{array}{c} \bu(\cdot,t) \\ \bu_{t}(\cdot,t)\end{array}\right)\right\|_{(s)} 
\leq C\ldr{t}^{N}e^{(\kappa_{1}-\b_{\rem})t}
\end{equation}
for all $t\geq 0$ and all $s\in\ro$. Let $z$ and $\bz$ be the Fourier coefficients of $u$ and $\bu$ respectively, let $v$ and $\bv$ be defined in
analogy with (\ref{eq:vAhFdef}), starting with $z$ and $\bz$ respectively. Finally, let $V=v-\bv$. If $V(\indexnot,0)=0$ for all $\indexnot\in\EFindexset$, 
then $u=\bu$ (since $V(\indexnot,\cdot)$ solves a homogeneous equation, $V(\indexnot,0)=0$ for all $\indexnot\in\EFindexset$ implies that $V(\indexnot,t)=0$ for all 
$\indexnot\in\EFindexset$ and all $t\in I$), so that $\Phi_{\infty}(V_{\infty})=\psi$. Thus $\psi$ is in the image of $\Phi_{\infty}$. Assume now that there is 
a $\indexnot\in\EFindexset$ such that $V(\indexnot,0)\neq 0$. Note that $V(\indexnot,\cdot)$ is a solution to an equation to which 
Lemma~\ref{lemma:spasODEsett} applies. Moreover, applying Lemma~\ref{lemma:spasODEsett} yields a $\Psi_{\infty}$ which is linear and bijective. Since
$V(\indexnot,0)\neq 0$, there is a $0\neq\chi\in\cn{2m}$ such that $\Psi_{\infty}(\chi)=V(\indexnot,T_{\roode})$, where $T_{\roode}$ is given by 
Definition~\ref{def:roughODEtermo} (for the relevant $\indexnot$ under consideration). This means that 
\[
|V(\indexnot,t)-e^{A_{\infty}t}\chi|\leq C\ldr{t}^{N}e^{(\kappa_{1}-\b_{\rem})t}
\]
for all $t\geq T_{\roode}$ and a constant $C$ depending on $V$, $\indexnot$ etc. On the other hand, (\ref{eq:uidminusbuidrough}) implies that the same estimate holds
with $\chi$ set to zero. Due to the fact that $\Rsp A_{\infty}<\b_{\rem}$, these estimates are contradictory. To conclude, every element of $C^{\infty}(\bM,\cn{2m})$ is 
in the image of $\Phi_{\infty}$. The lemma follows. 
\end{proof}

\chapter{ODE analysis in the transparent setting}\label{chapter:ODEtransp}

\section{Introduction}

Consider (\ref{eq:thesystemRge}). In the previous chapter we consider the silent setting. Then $g^{ij}$, $X^{i}$, $g^{0i}$, 
$i,j=1,\dots,d$, and $a_{r}^{-2}$, $r=1,\dots,R$, converge to zero exponentially. Next, we wish to consider the transparent setting. In that case, 
these functions are all bounded. Moreover, some of them converge to zero exponentially. However, some converge to non-zero functions. 
Let
\begin{equation}\label{eq:gijinfetcdef}
g^{ij}_{\infty}:=\lim_{t\rightarrow\infty}g^{ij}(t),\ \ \
g^{0i}_{\infty}:=\lim_{t\rightarrow\infty}g^{0i}(t),\ \ \
q_{\infty,r}:=\lim_{t\rightarrow\infty}a_{r}^{-2}(t),\ \ \
X^{i}_{\infty}:=\lim_{t\rightarrow\infty}X^{i}(t).
\end{equation}
We here assume that these limits exist. We also assume the convergence to be exponential. Note that $q_{r,\infty}\geq 0$, and that $g^{ij}_{\infty}$ are 
the components of a positive semi-definite matrix. Given $\indexnot\in\EFindexset$, let 
\begin{equation}\label{eq:mfginftydef}
\mfg_{\infty}(\indexnot):=\lim_{t\rightarrow\infty}\mfg(\indexnot,t)=
\left(\textstyle{\sum}_{j,l=1}^{d}g^{jl}_{\infty}n_{j}n_{l}+\sum_{r=1}^{R}q_{\infty,r}\nu_{r,i_{r}}^{2}(\indexnot)\right)^{1/2}.
\end{equation}
\index{$\a$Aa@Notation!Coefficients, Fourier side!$\mfg_{\infty}(\indexnot)$}%
Assume that $\a$ and $\zeta$ converge, exponentially, to $\a_{\infty}$ and $\zeta_{\infty}$ respectively and that $\mfg_{\infty}(\indexnot)\neq 0$.
Then (\ref{eq:fourierthesystemRge}) can be written
\begin{equation}\label{eq:vdottrscase}
\dot{v}=A(\indexnot)v+A_{\indexnot,\rem}v+F_{\indexnot},
\end{equation}
where 
\begin{equation}\label{eq:Aindexnotdef}
A(\indexnot):=\left(\begin{array}{cc} 0 & \mfg_{\infty}\Id_{m} \\ -\mfg_{\infty}\Id_{m}-i\mfg_{\infty}^{-1}n_{l}X^{l}_{\infty}-\mfg_{\infty}^{-1}\zeta_{\infty} & 
2in_{l}g^{0l}_{\infty}\Id_{m}-\a_{\infty}\end{array}\right).
\end{equation}
Moreover, 
\begin{equation}\label{eq:vFdeftrs}
v(\indexnot,t):=\left(\begin{array}{c} \mfg_{\infty}(\indexnot)z(\indexnot,t) \\ \dot{z}(\indexnot,t)\end{array}\right),\ \ \
F_{\indexnot}(t):=\left(\begin{array}{c} 0 \\ \hf(\indexnot,t) \end{array}\right).
\end{equation}
Finally, 
\begin{equation}\label{eq:Aremtrsest}
\|A_{\indexnot,\rem}(t)\|\leq C_{\trs}e^{-\b_{\trs}\bt}
\end{equation}
for $t\geq T_{\trs}$. Here $0<C_{\trs},\b_{\trs}\in\ro$ are independent of $\indexnot$, but $T_{\trs}$ does depend on $\indexnot$. Moreover, 
$\bt:=t-T_{\trs}$. Formally, the equation (\ref{eq:vdottrscase}) is identical to (\ref{eq:ODEregmod}). However, there is a fundamental difference.  
In (\ref{eq:ODEregmod}), the matrix $A$ is fixed once and for all, but in (\ref{eq:vdottrscase}) it depends on $\indexnot$. In the present chapter, 
we would, essentially, like to repeat the 
analysis presented in Chapter~\ref{chapter:roughanalysisODEregion}. However, this involves dividing $\cn{2m}$ into the first and second generalised 
eigenspace in the $\b_{\trs},A(\indexnot)$-decomposition of $\cn{2m}$. As opposed to the case studied in Chapter~\ref{chapter:roughanalysisODEregion}, 
there are potentially infinitely many different such generalised eigenspaces in the present setting. Similarly, there are potentially infinitely many 
different matrices $T_{A(\indexnot)}$ with the properties stated in Definition~\ref{def:JATA}; one for each $\indexnot$. This is a problem, since if we 
choose the matrices $T_{A(\indexnot)}$ so that their norms are uniformly bounded, there is no reason why $\|T_{A(\indexnot)}^{-1}\|$ should be uniformly bounded 
for $\indexnot\in\EFindexset$. Since $\|T_{A(\indexnot)}^{-1}\|$ appears in the estimates, this could potentially cause a problem. In some of the estimates, 
an even more subtle dependence on $A(\indexnot)$ appears. As a consequence, it is clear that we, in the present setting, need to keep very detailed track 
of how the algebraic constructions carried out in Section~\ref{section:ODEappr} depend on $A(\indexnot)$. 

Consider $A(\indexnot)$ defined by (\ref{eq:Aindexnotdef}). The bottom left component of this matrix contains the expression 
$-i\mfg_{\infty}^{-1}n_{l}X^{l}_{\infty}$. By varying $\indexnot$, this expression gives rise to a large number of linear combinations of the matrices 
$X^{l}_{\infty}$. In fact, one would typically expect the closure of the set of $-i\mfg_{\infty}^{-1}n_{l}X^{l}_{\infty}$, $\indexnot\in\EFindexset$, to be 
$d$-dimensional. Dealing with such a situation can be expected to be quite difficult, and we do not attempt to do so here. In fact, we consider, 
exclusively, the following two restricted situations. 

\begin{definition}\label{def:Xnontriv}
Consider (\ref{eq:thesystemRge}). Assume the associated metric to be such that $(M,g)$ is a canonical separable cosmological model manifold. 
Then (\ref{eq:thesystemRge}) is said to be an $X$-\textit{non-trivial weakly transparent equation} 
\index{Weakly transparent equation!$X$-non-trivial}%
\index{$\a$Aa@Notation!Conditions!$X$-non-trivial weakly transparent equation}%
if the following holds. The limits 
(\ref{eq:gijinfetcdef}) exist and there is a $j\in\{1,\dots,d\}$ such that $X^{j}_{\infty}\neq 0$. Moreover,
$X^{l}_{\infty}=0$ for $l\neq j$; $q_{\infty,r}=0$ for $r=1,\dots,R$; $g^{jj}_{\infty}>0$; $g^{kl}_{\infty}=0$ unless $k=l=j$; and 
$g^{0l}_{\infty}=0$ for $l=1,\dots,d$. Given an $X$-non-trivial weakly transparent equation (\ref{eq:thesystemRge}), 
let $d_{\trs}:=1$ and $j_{1}:=j$. 
\end{definition}
\begin{remarks}\label{remarks:Xnontriv}
If (\ref{eq:thesystemRge}) is an $X$-non-trivial weakly transparent equation, then $\mfg_{\infty}=(g^{jj}_{\infty})^{1/2}|n_{j}|$. Thus, assuming 
$n_{j}\neq 0$, 
\[
-i\mfg_{\infty}^{-1}n_{l}X^{l}_{\infty}=-i(g^{jj}_{\infty})^{-1/2}\sgn(n_{j})X^{j}_{\infty}. 
\]
In other words, the left hand side can only be one of two possible matrices. Note also that if we would allow $q_{\infty,r}>0$, then 
the set of $-i\mfg_{\infty}^{-1}n_{l}X^{l}_{\infty}$ could potentially have every point on the line between $-i(g^{jj}_{\infty})^{-1/2}X^{j}_{\infty}$
and $i(g^{jj}_{\infty})^{-1/2}X^{j}_{\infty}$ as an accumulation point. This leads to complications we wish to avoid. 
\end{remarks}

\begin{definition}\label{def:Xdegenerate}
Consider (\ref{eq:thesystemRge}). Assume the associated metric to be such that $(M,g)$ is a canonical separable cosmological model manifold. 
Then (\ref{eq:thesystemRge}) is said to be an $X$-\textit{degenerate weakly transparent equation} 
\index{Weakly transparent equation!$X$-degenerate}%
\index{$\a$Aa@Notation!Conditions!$X$-degenerate weakly transparent equation}%
if the following holds. The limits 
(\ref{eq:gijinfetcdef}) exist, $X^{j}_{\infty}=0$ and $g^{0j}_{\infty}=0$ for $j=1,\dots,d$. Moreover, $\{1,\dots,d\}$ 
can be written as the union of two disjoint sets $\{j_{1},\dots,j_{d_{\trs}}\}$ and $\{\bj_{1},\dots,\bj_{d_{\rosil}}\}$ (if $d_{\trs}=0$ or $d_{\rosil}=0$ 
the corresponding set is empty)
such that the spatial limit metric is positive definite on $P_{\infty}:=\mrspan\{e_{j_{1}},\dots,e_{j_{d_{\trs}}}\}$ and vanishes on 
$N_{\infty}:=\mrspan\{e_{\bj_{1}},\dots,e_{\bj_{d_{\rosil}}}\}$. Finally, if $d_{\trs}=0$, there is an $r\in \{1,\dots,R\}$ such that $q_{\infty,r}>0$. 
\end{definition}
\begin{remarks}
In the definition, the spatial limit metric corresponds to the matrix with components $g^{jk}_{\infty}$. Moreover, the statement that the spatial 
limit metric vanishes on $N_{\infty}$ means that if $v\in\rn{d}$ and $w\in N_{\infty}$, then $g^{jk}_{\infty}v_{j}w_{k}=0$. 
\end{remarks}
\begin{remarks}
One purpose of the restriction on the spatial limit metric is to ensure that the only accumulation point of 
$\{\mfg_{\infty}(\indexnot):\indexnot\in\EFindexset\}$ is $\infty$. To see that this is a consequence of the restriction,
fix $R>0$ and assume that $\mfg_{\infty}(\indexnot)\leq R$. If $q_{\infty,r}>0$, this means that the $\nu_{r,i}$ component of $\nu(\indexnot)$ is bounded;
cf. (\ref{eq:nugenReqdef}) for an explanation of the terminology. Thus $\nu_{r,i}$ takes one of a finite number of values. 
Similarly, $n_{j_{l}}$, $l=1,\dots,d_{\trs}$, can only take one of a finite number of values due to the positive definiteness of the limit metric on 
$P_{\infty}$. In other words, the only elements of $\nu(\indexnot)$ that contribute to $\mfg_{\infty}(\indexnot)$ are restricted to belong to a 
finite set if $\mfg_{\infty}(\indexnot)\leq R$.  
\end{remarks}

Returning to (\ref{eq:Aindexnotdef}), it is clear that if one of the Definitions~\ref{def:Xnontriv} and \ref{def:Xdegenerate} is satisfied, then
the matrices $A(\indexnot)$ are of the form 
\begin{equation}\label{eq:Amudef}
A_{\mu}:=\left(\begin{array}{cc} 0 & \mu\Id_{m} \\ -\mu\Id_{m}+iV+2\mu^{-1}W & 
U\end{array}\right),
\end{equation}
where $U,V,W\in\Mn{m}{\co}$. Moreover, $\mu\in\muindexset$, where $\muindexset$ is a countable subset of $(0,\infty)$ with $\infty$ as its only 
accumulation point. The factor $2$ is included for future convenience. 

\section{Asymptotic eigenvalue calculation}\label{section:aseigvcalc}

The first step of our analysis is to calculate the asymptotic values of the eigenvalues of the matrix $A_{\mu}$ introduced in (\ref{eq:Amudef}). Let, to this
end,  
\begin{equation}\label{eq:UpmSodef}
U_{\pm}:=\frac{1}{2}(U\pm V),\ \ \
S_{1}:=\left(\begin{array}{cc} i\Id_{m} & \Id_{m}\\ \Id_{m} & i\Id_{m}\end{array}\right). 
\end{equation}
\index{$\a$Aa@Notation!Matrix notation!Upm@$U_{\pm}$}%
Then
\[
S_{1}^{-1}A_{\mu}S_{1}=\left(\begin{array}{cc} -i\mu\Id_{m}+U_{-}+i\mu^{-1}W
& iU_{+}+\mu^{-1}W\\
-iU_{-}+\mu^{-1}W & i\mu\Id_{m}+U_{+}-i\mu^{-1}W\end{array}\right).
\]
Let $S_{\pm}\in\Gl{m}{\co}$ be such that 
\begin{equation}\label{eq:gapmdefintro}
\g_{\pm}:=(S_{\pm})^{-1}U_{\pm}S_{\pm}
\end{equation}
\index{$\a$Aa@Notation!Matrix notation!gammapm@$\g_{\pm}$}%
is of Jordan normal form. Letting $S_{2}:=\diag\{S_{-},S_{+}\}$ and $S_{12}:=S_{1}S_{2}$ then yields
\begin{equation}\label{eq:Nmudef}
N_{\mu}:=S_{12}^{-1}A_{\mu}S_{12}=\left(\begin{array}{cc} -i\mu\Id_{m}+\g_{-}+i\mu^{-1}W_{-} 
& i\de_{+}+\mu^{-1}W_{-+}\\
-i\de_{-}+\mu^{-1}W_{+-} & i\mu\Id_{m}+\g_{+}-i\mu^{-1}W_{+}\end{array}\right),
\end{equation}
\index{$\a$Aa@Notation!Matrix notation!Nmu@$N_{\mu}$}%
where
\[
W_{\pm}:=(S_{\pm})^{-1}WS_{\pm},\ \ \
W_{+-}:=(S_{+})^{-1}WS_{-},\ \ \
W_{-+}:=(S_{-})^{-1}WS_{+}.
\]
\index{$\a$Aa@Notation!Matrix notation!Wpmmp@$W_{\pm,\mp}$}%
\index{$\a$Aa@Notation!Matrix notation!Wpm@$W_{\pm}$}%
Moreover, $\de_{\pm}$ is defined by 
\[
\de_{+}:=(S_{-})^{-1}U_{+}S_{+},\ \ \
\de_{-}:=(S_{+})^{-1}U_{-}S_{-}
\]
\index{$\a$Aa@Notation!Matrix notation!deltapm@$\de_{\pm}$}%
and the $\g_{\pm}$ are defined in (\ref{eq:gapmdefintro}). For the sake of brevity, it is convenient to introduce
\begin{align}
R_{\mu}^{-} := & \g_{-}+i\mu^{-1}W_{-},\ \ \
R_{\mu}^{+}:=\g_{+}-i\mu^{-1}W_{+},\label{eq:Rinfpmdefintro}\\
Q_{\mu}^{+} := & i\de_{+}+\mu^{-1}W_{-+},\ \ \
Q_{\mu}^{-}:=-i\de_{-}+\mu^{-1}W_{+-}.\label{eq:Qinfpmdefintro}
\end{align}
\index{$\a$Aa@Notation!Matrix notation!Rmupm@$R_{\mu}^{\pm}$}%
\index{$\a$Aa@Notation!Matrix notation!Qmupm@$Q_{\mu}^{\pm}$}%
Then
\begin{equation}\label{eq:Ninfformulaintro}
N_{\mu}=\left(\begin{array}{cc} -i\mu\Id_{m}+R_{\mu}^{-}
& Q_{\mu}^{+}\\
Q_{\mu}^{-} & i\mu\Id_{m}+R_{\mu}^{+}\end{array}\right).
\end{equation}

\begin{lemma}\label{lemma:Nmueigenvapprox}
Consider the matrix $N_{\mu}$ given by (\ref{eq:Ninfformulaintro}). Let $\lambda_{j,\pm}$, $j=1,\dots,p_{\pm}$, be the distinct eigenvalues of 
$\g_{\pm}$, and $m_{j,\pm}$ be the corresponding algebraic multiplicities, where the matrices $\g_{\pm}$ are introduced in (\ref{eq:gapmdefintro}). 
There are then constants $0<c_{a},\mu_{a}\in\ro$ such that for $\mu\geq\mu_{a}$ and $j=1,\dots,p_{\pm}$, there are $m_{j,\pm}$ eigenvalues of $N_{\mu}$ 
in a ball of radius $c_{a}\mu^{-1/m_{j,\pm}}$ and centre $\pm i\mu+\lambda_{j,\pm}$. 
\end{lemma}
\begin{remark}
Since the sum of the $m_{j,+}$ equal $m$, and similarly for the $m_{j,-}$, each eigenvalue of $N_{\mu}$ has to belong to a ball of radius 
$c_{a}\mu^{-1/m_{j,\pm}}$ and centre $\pm i\mu+\lambda_{j,\pm}$ for some choice of $j$ and $\pm$. 
\end{remark}
\begin{remark}
The constants $c_{a}$ and $\mu_{a}$ only depend on $U$, $V$ and $W$. 
\end{remark}
\begin{proof}
In order to obtain estimates for the eigenvalues, we proceed step by step. Moreover, it is convenient to consider 
\begin{equation}\label{eq:Ninfformulaintrorver}
N_{\mu,r}=\left(\begin{array}{cc} -i\mu\Id_{m}+R_{\mu,r}^{-}
& Q_{\mu,r}^{+}\\
Q_{\mu,r}^{-} & i\mu\Id_{m}+R_{\mu,r}^{+}\end{array}\right)
\end{equation}
for $r\in [0,1]$. Here 
\[
R_{\mu,r}^{-} := \g_{-}+ir\mu^{-1}W_{-},\ \ \
R_{\mu,r}^{+}:=\g_{+}-ir\mu^{-1}W_{+},\ \ \
Q_{\mu,r}^{+}:=rQ_{\mu}^{+},\ \ \
Q_{\mu,r}^{-}:=rQ_{\mu}^{-}.
\]
Note that $N_{\mu,1}=N_{\mu}$. To begin with, we appeal to Gershgorin's Theorem; cf., e.g., \cite[Proposition~5.12, p.~102]{serre}. For 
$j=1,\dots,2m$, let
\[
\rho_{j}:=\sum_{l\neq j}|N_{\mu,r,lj}|\leq \varrho_{0}+\mu^{-1}\varrho_{1},
\]
where $0\leq \varrho_{l}\in\ro$, $l=0,1$, are independent of $\mu$ and $r\in [0,1]$, and $N_{\mu,r,lj}$ is the $lj$'th component of $N_{\mu,r}$. Then 
the eigenvalues of $N_{\mu,r}$ are contained in the discs $B_{\rho_{j}}(N_{\mu,r,jj})$ (no summation on $j$). In particular, there are thus constants 
$0<\mu_{0},R\in\ro$ such that for $\mu\geq \mu_{0}$ and $r\in [0,1]$, the two balls $B_{R}(\pm i\mu)$ are disjoint and contain the eigenvalues of 
$N_{\mu,r}$. Assume that $\mu\geq\mu_{0}$, that $r\in [0,1]$ and that $\lambda$ is an eigenvalue belonging to $B_{R}(-i\mu)$. Consider
\begin{equation}\label{eq:Nnumlidwaintro}
N_{\mu,r}-\lambda\Id_{2m}=\left(\begin{array}{cc} N^{-}_{\mu,r,\lambda} & Q_{\mu,r}^{+}\\ Q_{\mu,r}^{-} & 
N^{+}_{\mu,r,\lambda}\end{array}\right),
\end{equation}
where
\[
N^{\pm}_{\mu,r,\lambda}:=-(\lambda\mp i\mu)\Id_{m}+R_{\mu,r}^{\pm}. 
\]
Due to the assumptions, the diagonal elements of $N^{+}_{\mu,r,\lambda}$ are contained in a ball centred at $2i\mu$ and with radius bounded 
by a constant independent of $\mu$ and $r\in [0,1]$ (for $\mu$ large enough). There is thus a $\mu_{1}\geq\mu_{0}$ such that for 
$\mu\geq\mu_{1}$ and $r\in [0,1]$, $N^{+}_{\mu,r,\lambda}$ is 
invertible and
\begin{equation}\label{eq:Nmlanuinvestwaintro}
\| (N^{+}_{\mu,r,\lambda})^{-1}\|\leq c_{0}\mu^{-1}
\end{equation}
for some constant $c_{0}$ independent of $\mu$ and $r\in [0,1]$. In particular, 
\[
0=\det(N_{\mu,r}-\lambda\Id_{2m})=\det\left(\begin{array}{cc} N^{-}_{\mu,r,\lambda}-Q_{\mu,r}^{+}(N^{+}_{\mu,r,\lambda})^{-1}Q_{\mu,r}^{-} & Q_{\mu,r}^{+}\\ 0 & 
N^{+}_{\mu,r,\lambda}\end{array}\right).
\]
Since $N^{+}_{\mu,r,\lambda}$ is invertible, this implies 
\[
\det\left[N^{-}_{\mu,r,\lambda}-Q_{\mu,r}^{+}(N^{+}_{\mu,r,\lambda})^{-1}Q_{\mu,r}^{-}\right]=0.
\]
On the other hand, 
\begin{equation}\label{eq:essdetmatwaintro}
N^{-}_{\mu,r,\lambda}-Q_{\mu,r}^{+}(N^{+}_{\mu,r,\lambda})^{-1}Q_{\mu,r}^{-}=-(\lambda+i\mu)\Id_{m}+\g_{-}+\Rho_{\mu,r},
\end{equation}
where $\Rho_{\mu,r}$ is a matrix satisfying 
\begin{equation}\label{eq:Rhobdnuwaintro}
\|\Rho_{\mu,r}\|\leq c_{0}\mu^{-1}
\end{equation}
for $\mu\geq\mu_{1}$ and $r\in [0,1]$, where $c_{0}$, again, is independent of $\mu$ and $r\in [0,1]$. Note that the norms of the first two terms 
on the right hand side of (\ref{eq:essdetmatwaintro}) are (individually) bounded by a constant independent of $\mu$ and $r$. Moreover, the third term 
on the 
right hand side of (\ref{eq:essdetmatwaintro}) satisfies the bound (\ref{eq:Rhobdnuwaintro}). Note that we can consider the determinant of the right 
hand side of (\ref{eq:essdetmatwaintro}) as a multilinear function, say $\mathrm{Det}$, of the corresponding columns:
\begin{equation}\label{eq:detDetmultlinewaintro}
\begin{split}
 & \det[-(\lambda+i\mu)\Id_{m}+\g_{-}+\Rho_{\mu,r}] = \mathrm{Det}(K_{\mu,1}+\Rho_{\mu,r,1},\dots,K_{\mu,m}+\Rho_{\mu,r,m})\\ 
 = & \mathrm{Det}(K_{\mu,1},\dots,K_{\mu,m})+\mathrm{Det}(K_{\mu,1},\dots,K_{\mu,m-1},\Rho_{\mu,r,m})\\
 & +\mathrm{Det}(K_{\mu,1},\dots,\Rho_{\mu,r,m-1},K_{\mu,m}+\Rho_{\mu,r,m})+\dots+\mathrm{Det}(\Rho_{\mu,r,1},\dots,K_{\mu,m}+\Rho_{\mu,r,m}),
\end{split}
\end{equation}
where $K_{\mu,l}$ denotes the $l$'th column of the sum of the first two matrices on the right hand side of (\ref{eq:essdetmatwaintro})
and $\Rho_{\mu,r,l}$ denotes the $l$'th column of $\Rho_{\mu,r}$. We know that the left hand side of (\ref{eq:detDetmultlinewaintro}) is zero.
Moreover, due to (\ref{eq:Rhobdnuwaintro}) and the above observations, all the terms on the far right hand side of (\ref{eq:detDetmultlinewaintro}), 
except for the first one, can be bounded by a $c_{0}\mu^{-1}$ for some constant $c_{0}$ independent of $\mu$ and $r\in [0,1]$. Thus
\[
|\det[-(\lambda+i\mu)\Id_{m}+\g_{-}]|\leq c_{0}\mu^{-1}
\]
for $\mu\geq\mu_{1}$ and some constant $c_{0}$ independent of $\mu$ and $r\in [0,1]$. With notation as in the statement of the lemma, it follows that 
\begin{equation}\label{eq:detbdlajswaintro}
|(\lambda-\lambda_{1,-}+i\mu)^{m_{1,-}}\cdots (\lambda-\lambda_{p_{-},-}+i\mu)^{m_{p_{-},-}}|\leq c_{0}\mu^{-1}
\end{equation}
for $\mu\geq\mu_{1}$;
recall that $\g_{-}$ is of Jordan normal form. Due to this estimate, it is clear that the smallest of the $|\lambda-\lambda_{s,-}+i\mu|$ has 
to be bounded by $c_{0}\mu^{-1/m}$. In particular, there are constants $\mu_{2}\geq\mu_{1}$ and $0<c_{0}\in\ro$ such that for $\mu\geq\mu_{2}$
and $r\in [0,1]$, there is one and only one $j\in \{1,\dots,p_{-}\}$ such that $|\lambda-\lambda_{j,-}+i\mu|\leq c_{0}\mu^{-1/m}$. In order to 
justify the uniqueness of $j$, we use the fact that if $|\lambda-\lambda_{j,-}+i\mu|\leq c_{0}\mu^{-1/m}$, then, for $l\neq j$, 
\[
|\lambda-\lambda_{l,-}+i\mu|\geq |\lambda_{l,-}-\lambda_{j,-}|-|\lambda-\lambda_{j,-}+i\mu|\geq |\lambda_{l,-}-\lambda_{j,-}|-c_{0}\mu^{-1/m}.
\]
Since the first term on the far right hand side has a strictly positive lower bound for all $l\neq j$ (independent of $\mu$ and $r$), and since 
the second term converges to zero as $\mu\rightarrow\infty$, it is clear that the desired statement holds for $\mu$ large enough. We also obtain
the conclusion that for $\mu\geq\mu_{2}$, there is a $c_{1}>0$, independent of $\mu$ and $r\in [0,1]$, such that for $l\in \{1,\dots,p_{-}\}$ 
with $l\neq j$, $|\lambda-\lambda_{l,-}+i\mu|\geq c_{1}$. Combining this observation with (\ref{eq:detbdlajswaintro}) yields
\begin{equation}\label{eq:eigvalnujpwaintro}
|\lambda-\lambda_{j,-}+i\mu|\leq c_{2}\mu^{-1/m_{j,-}}
\end{equation}
for $\mu\geq\mu_{2}$ and some constant $c_{2}$ independent of $\mu$ and $r\in [0,1]$. 

In order to prove that there are $m_{j,-}$ eigenvalues of $N_{\mu}$ in a ball of radius $c_{2}\mu^{-1/m_{j,-}}$ and centre $-i\mu+\lambda_{j,-}$, we 
proceed as in the proof of \cite[Theorem~5.7, p. 102]{serre}. Let 
\[
\mP_{\mu,r}(z)=\det(N_{\mu,r}-z\Id_{2m}).
\]
By the above, we, first of all, know that the zeros of this polynomial are contained in the disjoint balls $B_{R}(\pm i\mu)$. Focusing on the 
zeros contained in $B_{R}(-i\mu)$, they are contained in the balls of radius $c_{2}\mu^{-1/m_{j,-}}$ and centre $-i\mu+\lambda_{j,-}$ (and for 
$\mu$ large enough, we can assume the balls with the same centre and twice the radius to be disjoint). Let $\Gamma$ be a curve parametrising the 
boundary circle of the ball of radius $c_{2}\mu^{-1/m_{j,-}}$ and centre $-i\mu+\lambda_{j,-}$. Define
\[
Z_{\mu}(r):=\frac{1}{2\pi i}\int_{\Gamma}\frac{\mP_{\mu,r}'(z)}{\mP_{\mu,r}(z)}dz.
\]
Note that $\mP_{\mu,r}(z)$ is continuous (in $r$ and $z$) and non-vanishing on $\Gamma$ for all $r\in [0,1]$, assuming $\mu$ to be large enough. 
Moreover, $Z_{\mu}(r)$ is a continuous function of $r$. Finally, $Z_{\mu}(r)$ counts the number of zeros of the polynomial $\mP_{\mu,r}(z)$ inside 
$\Gamma$. Since $Z_{\mu}(r)$ is integer valued and continuous, $Z_{\mu}(1)=Z_{\mu}(0)$. On the other hand, by the definition of $N_{\mu,r}$,
it is clear that $Z_{\mu}(0)=m_{j,-}$. To conclude, there are $m_{j,-}$ eigenvalues of $N_{\mu}$ in a ball of radius 
$c_{a}\mu^{-1/m_{j,-}}$ and centre $-i\mu+\lambda_{j,-}$. By a similar argument, there are $m_{j,+}$ eigenvalues of $N_{\mu}$ in a ball of radius 
$c_{a}\mu^{-1/m_{j,+}}$ and centre $i\mu+\lambda_{j,+}$. The lemma follows. 
\end{proof}

\section{Computing projections onto generalised eigenspaces}\label{section:compprojontogensubs}

In the previous section, we obtained asymptotic estimates for the eigenvalues of the matrix $A_{\mu}$ introduced in (\ref{eq:Amudef}). The purpose of the 
present section is to take the first step towards obtaining asymptotic expressions for the generalised eigenspaces of interest when carrying out 
an analysis similar to that described in Chapter~\ref{chapter:roughanalysisODEregion}. To begin with, we need to recall how to evaluate holomorphic
functions on complex square matrices. 

\subsection{Evaluating holomorphic functions on complex square matrices} 
Let $A\in\Mn{m}{\co}$ and $f$ be a holomorphic function on an open set $U\subseteq\co$ (which need not be connected). 
Assume that $\Spe A\subset U$ and let $P$ be a polynomial such that 
$P^{(r)}(\lambda)=f^{(r)}(\lambda)$ for every $\lambda\in\Spe A$ and every $0\leq r\leq m-1$; that such a polynomial exists
follows, e.g., from \cite[Theorem~1, p.~42]{spitzb}. Given this polynomial, $f(A)$ is defined by $f(A):=P(A)$; that 
$f(A)$ does not depend on the particular polynomial chosen is a consequence of the Cayley-Hamilton theorem. The above
discussion is merely a summary of the beginning of \cite[Section~5.5]{serre}, and we refer the reader interested in more 
details to \cite{serre}. Due to \cite[Proposition~5.7, pp.~94--95]{serre}, this definition leads, e.g., to the following natural 
properties: if $M$ is non-singular, then $f(M^{-1}AM)=M^{-1}f(A)M$;
$(fg)(A)=f(A)g(A)$; $\Spe f(A)=f(\Spe A)$; $(f\circ g)(A)=f(g(A))$ etc. In all of these equalities, we assume $f$ and
$g$ to be holomorphic functions defined on open sets $U\subseteq\co$ and $V\subseteq\co$ respectively. Moreover, we assume $U$
and $V$ to be such that the expressions formulated above make sense. 

Next we want to describe a different way of defining $f(A)$. The corresponding perspective is based on results from complex analysis, 
and we therefore briefly recall the relevant results and notation. 

\begin{thm}[Cauchy's Theorem]\label{thm:Cauchy}
Let $U\subset\co$ be an open set and $f:U\rightarrow\co$ be a holomorphic function. If $0\leq m\in\zo$ and $\G$ is a cycle in $U$ such that 
$\Ind_{\G}(z)=0$ for all $z\notin U$, then 
\[
\frac{1}{m!}f^{(m)}(z)\cdot \Ind_{\G}(z)=\frac{1}{2\pi i}\int_{\G}\frac{f(w)}{(w-z)^{m+1}}dw
\]
for all $z\in U-\G^{*}$. 
\end{thm}
\begin{remarks}
Here a \textit{cycle} 
\index{Cycle}%
$\G$ is a collection of closed paths, say $\g_{j}$ ($j=1,\dots,l$), where a \textit{path} 
\index{Path}%
is a piecewise continuously differentiable curve defined on a compact interval (such that the left- and right-hand derivatives always exist, and 
the number of derivative discontinuities is finite). Moreover, $\G^{*}$ is the union of the ranges of the $\g_{j}$'s and if $h$ is a continuous 
function on $\G^{*}$, then
\[
\int_{\G}h(z)dz:=\sum_{j=1}^{l}\int_{\g_{j}}h(z)dz.
\]
Finally, if $z\notin\G^{*}$, 
\[
\Ind_{\G}(z):=\frac{1}{2\pi i}\int_{\G}\frac{d\zeta}{\zeta-z}.
\]
\index{$\a$Aa@Notation!Index!$\Ind_{\G}(z)$}%
\end{remarks}
The theorem is essentially a consequence of \cite[Theorem~10.35, pp.~218--219]{rudin}; cf. also \cite[Exercise~7, p.~228]{rudin}. 

Combining Cauchy's Theorem and the above definition of $f(A)$ yields the following result (which is merely a slight reformulation of 
\cite[Proposition~5.8, p.~95]{serre}). The reader interested in a proof is referred to \cite[p.~96]{serre}. 
\begin{thm}\label{thm:DunfoTaylor}
Let $U\subset\co$ be an open set and $f:U\rightarrow\co$ be a holomorphic function. Assume that $A\in\Mn{m}{\co}$ is such that 
$\Spe A\subset U$. If $\G$ is a cycle in $U$ such that $\lambda\notin\G^{*}$ for all $\lambda\in\Spe A$; $\Ind_{\G}(\lambda)=1$ for all 
$\lambda\in\Spe A$; and $\Ind_{\G}(z)=0$ for all $z\notin U$, then 
\[
f(A)=\frac{1}{2\pi i}\int_{\G}f(z)(z\Id_{m}-A)^{-1}dz.
\]
\end{thm}
Next we use these observations to derive formulae for the projections onto the generalised eigenspaces of 
$A_{\mu}$ that we are interested in. 

\subsection{Formulae for projections onto generalised eigenspaces}\label{ssection:formproj}
Due to Lemma~\ref{lemma:Nmueigenvapprox}, we know that the eigenvalues of $N_{\mu}$ are contained in balls centred at
\[
\mu_{j,+}:=i\mu+\lambda_{j,+},\ \ \
\mu_{l,-}:=-i\mu+\lambda_{l,-},
\]
where $j=1,\dots,p_{+}$ and $l=1,\dots,p_{-}$. Let $r>0$ be small enough that the balls $B_{2r}(\lambda_{j,+})$ are disjoint for $j=1,\dots,p_{+}$, 
and the balls $B_{2r}(\lambda_{l,-})$ are disjoint for $l=1,\dots,p_{-}$. Then, there is a $\mu_{0}>0$ such that for $\mu\geq\mu_{0}$, the balls 
$B_{2r}(\mu_{j,+})$ and $B_{2r}(\mu_{l,-})$, $j=1,\dots,p_{+}$ and $l=1,\dots,p_{-}$, are all disjoint. Moreover, for $\mu\geq\mu_{0}$, 
all the eigenvalues of $N_{\mu}$ belong to $U_{\mu,0}$, defined to be the union of the $B_{r}(\mu_{j,+})$ and $B_{r}(\mu_{l,-})$ for $j=1,\dots,p_{+}$ 
and $l=1,\dots,p_{-}$; this is due to Lemma~\ref{lemma:Nmueigenvapprox}. A final application of Lemma~\ref{lemma:Nmueigenvapprox} allows us to assume
that there are $m_{j,+}$ and $m_{l,-}$ eigenvalues in $B_{r}(\mu_{j,+})$ and $B_{r}(\mu_{l,-})$ respectively. Define
\[
\g_{\mu,j,+}(t):=\mu_{j,+}+re^{2\pi i t},\ \ \
\g_{\mu,l,-}(t):=\mu_{l,-}+re^{2\pi i t}
\]
for $j=1,\dots,p_{+}$, $l=1,\dots,p_{-}$. Let $\G_{\mu}$ be the cycle consisting of all the $\g_{\mu,j,+}$ and $\g_{\mu,l,-}$. Finally, let $U_{\mu}$ 
be the union of all the balls $B_{3r/2}(\mu_{j,+})$, $j=1,\dots,p_{+}$, and $B_{3r/2}(\mu_{l,-})$, $l=1,\dots,p_{-}$; $f_{\mu,j,+}:U_{\mu}\rightarrow\co$ 
be defined by $f_{\mu,j,+}(z)=1$ for $z\in B_{3r/2}(\mu_{j,+})$ and $f_{\mu,j,+}(z)=0$ for $z\notin B_{3r/2}(\mu_{j,+})$; and let $f_{\mu,l,-}$ be defined 
analogously. Let $\Omega_{\mu,j,\pm}$ be the complement of $\g_{\mu,j,\pm}^{*}$, where $\g_{\mu,j,\pm}^{*}$ denotes the range 
of $\g_{\mu,j,\pm}$. Then $\Ind_{\g_{\mu,j,\pm}}$ is an integer valued function on $\Omega_{\mu,j,\pm}$ which is constant on each component of $\Omega_{\mu,j,\pm}$ and 
vanishes on the unbounded component of $\Omega_{\mu,j,\pm}$; cf., e.g., \cite[Theorem~10.10, p.~203]{rudin}. Moreover, it is obvious that 
$\Ind_{\g_{\mu,j,\pm}}(\mu_{j,\pm})=1$. Thus it is clear that $\Ind_{\g_{\mu,j,\pm}}(z)=0$ if $z\notin\bar{B}_{r}(\mu_{j,\pm})$ and 
$\Ind_{\g_{\mu,j,\pm}}(z)=1$ if $z\in B_{r}(\mu_{j,\pm})$. In particular, $\Ind_{\G_{\mu}}(z)=1$ if $z\in U_{\mu,0}$ and $\Ind_{\G_{\mu}}(z)=0$ if $z\notin\bar{U}_{\mu,0}$. 
Thus $U_{\mu}$ and $\Gamma_{\mu}$ fulfil the conditions of Theorem~\ref{thm:DunfoTaylor}. Moreover, the $f_{\mu,j,\pm}$ are holomorphic on $U_{\mu}$;
$f_{\mu,j,+}f_{\mu,l,-}=0$ for all $j=1,\dots,p_{+}$ and $l=1,\dots,p_{-}$; $f_{\mu,j,+}f_{\mu,l,+}=0$ for all $j,l=1,\dots,p_{+}$ such that $j\neq l$;
$f_{\mu,j,-}f_{\mu,l,-}=0$ for all $j,l=1,\dots,p_{-}$ such that $j\neq l$; and $f_{\mu,j,\pm}^{2}=f_{\mu,j,\pm}$ for all $j=1,\dots,p_{\pm}$. Define 
$\pi_{\mu,j,\pm}:=f_{\mu,j,\pm}(N_{\mu})$. Then the $\pi_{\mu,j,\pm}$'s have properties similar to the $f_{\mu,j,\pm}$. Moreover,
\[
\sum_{j=1}^{p_{+}}\pi_{\mu,j,+}+\sum_{j=1}^{p_{-}}\pi_{\mu,j,-}=\sum_{j=1}^{p_{+}}f_{\mu,j,+}(N_{\mu})+\sum_{j=1}^{p_{-}}f_{\mu,j,-}(N_{\mu})=\Id_{2m},
\]
where the last equality is a consequence of the fact that the sum of the $f_{\mu,j,\pm}$ equals $1$ on $\Spe N_{\mu}$. Letting 
$E_{\mu,j,\pm}:=\pi_{\mu,j,\pm}(\cn{2m})$, it is thus clear that $\cn{2m}$ is the direct sum of the $E_{\mu,j,\pm}$. Moreover, since $\pi_{\mu,j,\pm}$ and $N_{\mu}$ 
commute, it is clear that $N_{\mu}(E_{\mu,j,\pm})\subseteq E_{\mu,j,\pm}$. Finally, it is of interest to relate the $E_{\mu,j,\pm}$'s
to the generalised eigenspaces associated with the eigenvalues of $N_{\mu}$. To that end, we appeal to the following lemma. 

\begin{lemma}
Let each of the sets $U\subset\co$ and $V\subset\co$ be a union of a finite number of disjoint open balls with finite radius. Assume, moreover, that for 
each component of $V$, its closure is contained in a component of $U$, and that each component of $U$ contains a unique component of $V$.  
Assume that $A\in\Mn{k}{\co}$ is such that $\Spe A\subset V$. Let $V_{a}\subset U_{a}$, be components of 
$V$ and $U$ respectively. Let $f:U\rightarrow\co$ be defined by the requirement that $f(z)=1$ for $z\in U_{a}$ and $f(z)=0$ for $z\notin U_{a}$. Then 
$f(A)$ is the projection onto the direct sum of the generalised eigenspaces corresponding to the eigenvalues in $\Spe A\cap V_{a}$. 
\end{lemma}
\begin{remark}
In particular, given the terminology introduced prior to the statement of the lemma, $\pi_{\mu,j,\pm}$ is the projection onto the direct sum of the 
generalised eigenspaces of $N_{\mu}$ corresponding to the eigenvalues in $B_{r}(\mu_{j,\pm})$. 
\end{remark}
\begin{proof}
Let $\G$ consist of the boundary circles of the components of $V$. Then $U$, $f$, $A$ and $\Gamma$ are such that the conditions of 
Theorem~\ref{thm:DunfoTaylor} are satisfied. Let $\g_{a}$ be a curve parametrising the boundary circle $\d V_{a}$. Then Theorem~\ref{thm:DunfoTaylor} 
yields
\[
\pi:=f(A)=\frac{1}{2\pi i}\int_{\g_{a}}(z\Id_{k}-A)^{-1}dz.
\]
Let $M$ be a non-singular matrix such that 
\begin{equation}\label{eq:psJorddecomp}
M^{-1}AM=\diag\{\lambda_{1}\Id_{l_{1}}+N_{1},\dots,\lambda_{r}\Id_{l_{r}}+N_{r}\},
\end{equation}
where $\lambda_{1},\dots,\lambda_{r}$ are the (distinct) eigenvalues of $A$, $l_{1},\dots,l_{r}$ are the corresponding algebraic multiplicities,
and $N_{l}$, $l=1,\dots,r$, are upper triangular nilpotent matrices. That such a matrix $M$ exists follows from the Jordan normal decomposition 
of $A$. Fixing $j\in\{1,\dots,r\}$, it is clear that 
\[
M^{-1}(A-\lambda_{j}\Id_{k})^{l_{j}}M
=\diag\{[(\lambda_{1}-\lambda_{j})\Id_{l_{1}}+N_{1}]^{l_{j}},\dots,[(\lambda_{r}-\lambda_{j})\Id_{l_{r}}+N_{r}]^{l_{j}}\}.
\]
Note that the $j$'th block vanishes and that all the other blocks are invertible. Thus
\begin{equation}\label{eq:geneigspljitoMim}
\mathrm{ker}(A-\lambda_{j}\Id_{k})^{l_{j}}=M(\{0\}^{l_{1}+\dots+l_{j-1}}\times\cn{l_{j}}\times \{0\}^{l_{j+1}+\dots+l_{r}}).
\end{equation}
Thus the generalised eigenspace associated with $\lambda_{j}$ is given by the right hand side of (\ref{eq:geneigspljitoMim}). Let us relate this
to $\pi$. Due to the fact that $M^{-1}f(A)M=f(M^{-1}AM)$ for invertible matrices $M$, it is sufficient to consider the diagonal blocks on
the right hand side of (\ref{eq:psJorddecomp}) separately. In fact,  
\begin{equation*}
\begin{split}
 & M^{-1}\pi M=f(M^{-1}AM)\\
 = & \frac{1}{2\pi i}\int_{\g_{a}}\diag\{[(z-\lambda_{1})\Id_{l_{1}}-N_{1}]^{-1},\dots,[(z-\lambda_{r})\Id_{l_{r}}-N_{r}]^{-1}\}dz.
\end{split}
\end{equation*}
On the other hand, 
\[
[(z-\lambda_{j})\Id_{l_{j}}-N_{j}]^{-1}=(z-\lambda_{j})^{-1}\Id_{l_{j}}+(z-\lambda_{j})^{-2}N_{j}+\dots +(z-\lambda_{j})^{-l_{j}}N_{j}^{l_{j}-1}.
\]
Combining these observations with Theorem~\ref{thm:Cauchy} yields 
\begin{equation}\label{eq:pijconjindliidli}
M^{-1}\pi M=\diag\{\Ind_{\g_{a}}(\lambda_{1})\Id_{l_{1}},\dots,\Ind_{\g_{a}}(\lambda_{r})\Id_{l_{r}}\}.
\end{equation}
Assume now that $\lambda_{j}\in V_{a}$ and let $x\in \mathrm{ker}(A-\lambda_{j}\Id_{k})^{l_{j}}$. Then $\Ind_{\g_{a}}(\lambda_{j})=1$.
Moreover (\ref{eq:geneigspljitoMim}) implies that $x=My$, where $y\in \{0\}^{l_{1}+\dots+l_{j-1}}\times\cn{l_{j}}\times \{0\}^{l_{j+1}+\dots+l_{r}}$.
Combining these observations with (\ref{eq:pijconjindliidli}) yields
\[
\pi(x)=\pi (My)=My=x.
\]
Similarly, if $\lambda_{j}\notin V_{a}$ and $x\in \mathrm{ker}(A-\lambda_{j}\Id_{k})^{l_{j}}$, then $\pi(x)=0$. Thus $\pi$
is the identity on the generalised eigenspaces corresponding to eigenvalues belonging to $V_{a}$. Moreover, it annihilates the
generalised eigenspaces corresponding to eigenvalues not belonging to $V_{a}$. The lemma follows. 
\end{proof}

\subsection{Asymptotic partial diagonalisation}\label{ssection:aspardiag}

Given the above background, we are in a position to prove that there is, asymptotically, a partial diagonalisation of matrices of the 
form (\ref{eq:Amudef}). 

\begin{lemma}\label{lemma:aspardiag}
Let $1\leq m\in\zo$, $U,V,W\in\Mn{m}{\co}$ and define $A_{\mu}$ by (\ref{eq:Amudef}), where $0<\mu\in\ro$. Define $\g_{\pm}$ by 
(\ref{eq:gapmdefintro}) and $N_{\mu}$ by (\ref{eq:Nmudef}). Let $\lambda_{j,\pm}$, $j=1,\dots,p_{\pm}$, denote the distinct eigenvalues of 
$\g_{\pm}$, and let $m_{j,\pm}$ denote their respective algebraic multiplicities. Finally, let $\g_{\pm,j}$, $j=1,\dots,p_{\pm}$, be the matrix collecting
all the Jordan blocks in $\g_{\pm}$ corresponding to the eigenvalue $\lambda_{j,\pm}$. Then there is a $0<\mu_{0}\in\ro$ and, for each $\mu\geq\mu_{0}$, 
a $T_{\mu}\in\Gl{2m}{\co}$ such that $\|T_{\mu}\|,\|T_{\mu}^{-1}\|\leq 2$ and such that 
\begin{equation}\label{eq:Tmudiag}
T_{\mu}^{-1}N_{\mu}T_{\mu}=\diag\{N_{\mu,1}^{-},\dots,N_{\mu,p_{-}}^{-},N_{\mu,1}^{+},\dots,N_{\mu,p_{+}}^{+}\}.
\end{equation}
Moreover, 
\[
\|N_{\mu,j}^{\pm}\mp i\mu\Id_{m_{j,\pm}}-\g_{\pm,j}\|\leq C\mu^{-1}.
\]
Finally, the constants $\mu_{0}$ and $C$ only depend on $U$, $V$ and $W$.
\end{lemma}
\begin{remark}
Assuming that 
\[
\g_{\pm}=\diag\{\g_{\pm,1},\dots,\g_{\pm,p_{\pm}}\},
\]
$T_{\mu}$ can be chosen so that $T_{\mu}\rightarrow \Id_{2m}$ as $\mu\rightarrow\infty$. 
\end{remark}
\begin{proof}
Given the notation introduced in Subsection~\ref{ssection:formproj}, we are interested in the projections $\pi_{\mu,j,\pm}$, in particular
their limits as $\mu\rightarrow\infty$. The reason is that these projections can be used to calculate the matrix $T_{\mu}$, the existence
of which is the main conclusion of the lemma. Note that
\begin{equation*}
\begin{split}
\pi_{\mu,j,\pm} = & \frac{1}{2\pi i}\int_{\g_{\mu,j,\pm}}(z\Id_{2m}-N_{\mu})^{-1}dz\\
= & \frac{1}{2\pi i}\int_{\g}[(z+\mu_{j,\pm})\Id_{2m}-N_{\mu}]^{-1}dz,
\end{split}
\end{equation*}
where $\g(t):=re^{2\pi i t}$ and the range of $j$ depends on the sign $\pm$. Let us introduce the notation 
\[
N_{\mu,j,\pm}:=N_{\mu}-\mu_{j,\pm}\Id_{2m}. 
\]
Then 
\begin{align}
N_{\mu,j,+} = & \left(\begin{array}{cc} -(2i\mu+\lambda_{j,+})\Id_{m}+R_{\mu}^{-} & Q_{\mu}^{+}\\ Q_{\mu}^{-} & -\lambda_{j,+}\Id_{m}+R_{\mu}^{+}
\end{array}\right),\label{eq:Ninfjpform}\\
N_{\mu,l,-} = & \left(\begin{array}{cc} -\lambda_{l,-}\Id_{m}+R_{\mu}^{-} & Q_{\mu}^{+}\\ Q_{\mu}^{-} & (2i\mu-\lambda_{l,-})\Id_{m}+R_{\mu}^{+}
\end{array}\right).\label{eq:Ninflmform}
\end{align}
Moreover, 
\begin{equation}\label{eq:pinujpmsforwa}
\pi_{\mu,j,\pm} = \frac{1}{2\pi i}\int_{\g}(z\Id_{2m}-N_{\mu,j,\pm})^{-1}dz,
\end{equation}
where the range of $j$ is different depending on the sign $\pm$. Considering $N_{\mu,l,-}$, it is clear that the diagonal components of 
the bottom right block tend to infinity along the imaginary axis as $\mu\rightarrow\infty$. Moreover, the remaining components of the bottom 
right block remain bounded, as well as the remaining blocks. In the case of $N_{\mu,j,+}$ the situation is similar. In this case it is the diagonal components 
of the top left block which tend to infinity along the imaginary axis. The same statements are true for $z\Id_{2m}-N_{\mu,l,-}$ and $z\Id_{2m}-N_{\mu,j,+}$ 
respectively. Let us introduce $A_{\mu,j}^{\pm}(z)$ and $D_{\mu,j}^{\pm}(z)$ by
\begin{equation}\label{eq:zIdmNnujpmwa}
z\Id_{2m}-N_{\mu,j,\pm}=\left(\begin{array}{cc} A_{\mu,j}^{\pm}(z) & -Q_{\mu}^{+}\\ -Q_{\mu}^{-} & D_{\mu,j}^{\pm}(z)\end{array}\right),
\end{equation}
where the range of $j$ depends on the sign $\pm$. Since $|z|=r$, it is clear that there is a $0<\mu_{0}\in\ro$ such that for $\mu\geq\mu_{0}$, 
$D_{\mu,l}^{-}(z)$ and $A_{\mu,j}^{+}(z)$ are invertible and 
\begin{equation}\label{eq:DmulmAmujpinvest}
\|[D_{\mu,l}^{-}(z)]^{-1}\|+\|[A_{\mu,j}^{+}(z)]^{-1}\|\leq c_{0}\mu^{-1},
\end{equation}
where $c_{0}$ is independent of $\mu$ but depends on $r$ (though the dependence on $r$ can be eliminated by demanding that $r\leq 1$). In the process 
of calculating the right hand side of (\ref{eq:pinujpmsforwa}), it is useful to note that 
\begin{equation}\label{eq:invintprojwa}
\begin{split}
 & \left(\begin{array}{cc} A_{\mu,l}^{-}(z) & -Q_{\mu}^{+}\\ -Q_{\mu}^{-} & D_{\mu,l}^{-}(z)\end{array}\right)\\
 = & \left(\begin{array}{cc} B_{\mu,l}^{-}(z) & -Q_{\mu}^{+}\\ 0 & D_{\mu,l}^{-}(z)\end{array}\right)
\left(\begin{array}{cc} \Id_{m} & 0\\ -[D_{\mu,l}^{-}(z)]^{-1}Q_{\mu}^{-} & \Id_{m}\end{array}\right),
\end{split}
\end{equation}
where 
\begin{equation}\label{eq:Bnujpzdefwa}
B_{\mu,l}^{-}(z):=A_{\mu,l}^{-}(z)-Q_{\mu}^{+}[D_{\mu,l}^{-}(z)]^{-1}Q_{\mu}^{-}.
\end{equation}
Note that, by construction, the left (and thereby the right) hand side of (\ref{eq:zIdmNnujpmwa}) is invertible. Thus the left (and thereby the right) hand side of 
(\ref{eq:invintprojwa}) is invertible. In particular, the first factor on the right hand side of (\ref{eq:invintprojwa}) is invertible.
Since this factor is a block diagonal matrix, the individual blocks are invertible. In particular, $B_{\mu,l}^{-}(z)$ is invertible for the 
relevant range of $z$. In the end, we wish to compute the integral (along $\g$) of the inverse of the left hand side of (\ref{eq:invintprojwa}). 
Moreover, we wish to compute the matrix up to an error of the form $O(\mu^{-2})$. Introducing the notation
\begin{equation}\label{eq:EFGHdefwa}
\left(\begin{array}{cc} A_{\mu,l}^{-}(z) & -Q_{\mu}^{+}\\ -Q_{\mu}^{-} & D_{\mu,l}^{-}(z)\end{array}\right)^{-1}
=:\left(\begin{array}{cc} E_{\mu,l}^{-}(z) & F_{\mu,l}^{-}(z)\\ G_{\mu,l}^{-}(z) & H_{\mu,l}^{-}(z)\end{array}\right),
\end{equation}
it can be computed that 
\begin{align}
E_{\mu,l}^{-}(z) = & [B_{\mu,l}^{-}(z)]^{-1},\label{eq:Enujpzdefwa}\\
F_{\mu,l}^{-}(z) = & [B_{\mu,l}^{-}(z)]^{-1}Q_{\mu}^{+}[D_{\mu,l}^{-}(z)]^{-1},\label{eq:Fnujpzdefwa}\\
G_{\mu,l}^{-}(z) = & [D_{\mu,l}^{-}(z)]^{-1}Q_{\mu}^{-}[B_{\mu,l}^{-}(z)]^{-1},\label{eq:Gnujpzdefwa}\\
H_{\mu,l}^{-}(z) = & [D_{\mu,l}^{-}(z)]^{-1}Q_{\mu}^{-}[B_{\mu,l}^{-}(z)]^{-1}Q_{\mu}^{+}[D_{\mu,l}^{-}(z)]^{-1}
+[D_{\mu,l}^{-}(z)]^{-1}.\label{eq:Hnujpzdefwa}
\end{align}
In what follows, we need a bound on $\|[B_{\mu,l}^{-}(z)]^{-1}\|$. Note, to this end, that 
\[
A_{\mu,l}^{-}(z)=(z+\lambda_{l,-})\Id_{m}-\g_{-}-i\mu^{-1}W_{-}.
\]
Combining this observation with (\ref{eq:DmulmAmujpinvest}) and (\ref{eq:Bnujpzdefwa}) leads to the conclusion that 
\[
\|B_{\mu,l}^{-}(z)-(z+\lambda_{l,-})\Id_{m}+\g_{-}\|\leq C\mu^{-1},
\]
where $C$ is independent of $\mu$ but depends on $r$. Since $(z+\lambda_{l,-})\Id_{m}-\g_{-}$ is an upper triangular matrix, we only need to focus
on its diagonal elements. They are of the form $z+\lambda_{l,-}-\lambda_{j,-}$. By construction, this expression is, in absolute value, bounded from
below by a strictly positive constant. Note, however, that the lower bound depends on $r$. In particular, it depends on a lower bound on $r$ (when 
taking $l=j$). To conclude, there is thus a uniform bound on $\|[B_{\mu,l}^{-}(z)]^{-1}\|$ which holds for all $\mu$ large enough. 

Next, note that 
\begin{equation}\label{eq:Dnujpzinvwa}
[D_{\mu,l}^{-}(z)]^{-1}=(z-2i\mu+\lambda_{l,-})^{-1}[\Id_{m}-(z-2i\mu+\lambda_{l,-})^{-1}R_{\mu}^{+}]^{-1},
\end{equation}
so that 
\[
\frac{1}{2\pi i}\int_{\g}[D_{\mu,l}^{-}(z)]^{-1}dz=0
\]
for large $\mu$. On the other hand, the norm of the first term on the right hand side of (\ref{eq:Hnujpzdefwa}) is $O(\mu^{-2})$.
Note, however, that the corresponding estimate depends on a lower bound on $r$ (this is due to the fact that the upper bound on 
$\|[B_{\mu,l}^{-}(z)]^{-1}\|$ depends on a lower bound on $r$). Thus
\begin{equation}\label{eq:intgHnujpzwa}
\left\|\frac{1}{2\pi i}\int_{\g}H_{\mu,l}^{-}(z)dz\right\|\leq c_{0}\mu^{-2}.
\end{equation}
Before turning to (\ref{eq:Fnujpzdefwa}) and (\ref{eq:Gnujpzdefwa}), note that (\ref{eq:Dnujpzinvwa}) yields
\begin{equation}\label{eq:Dnujpzestwa}
\left\|[D_{\mu,l}^{-}(z)]^{-1}-\frac{i}{2\mu}\Id_{m}\right\|\leq c_{0}\mu^{-2}.
\end{equation}
Thus
\begin{equation}\label{eq:Fnujpzestwa}
\left\|F_{\mu,l}^{-}(z)-\frac{i}{2\mu}[B_{\mu,l}^{-}(z)]^{-1}Q_{\mu}^{+}\right\|\leq c_{0}\mu^{-2}.
\end{equation}
There is a similar estimate concerning $G_{\mu,l}^{-}(z)$, so that it is sufficient to focus on the integral of $[B_{\mu,l}^{-}(z)]^{-1}$ 
over $\g$. Due to (\ref{eq:Rinfpmdefintro}), (\ref{eq:Ninflmform}), (\ref{eq:zIdmNnujpmwa}), (\ref{eq:Bnujpzdefwa}) and (\ref{eq:Dnujpzestwa}) 
\[
\left\|B_{\mu,l}^{-}(z)-(z+\lambda_{l,-})\Id_{m}+\g_{-}+\mu^{-1}L_{\mu}^{-}\right\|\leq c_{0}\mu^{-2},
\]
where 
\[
L_{\mu}^{-}:=iW_{-}+\frac{i}{2}Q_{\mu}^{+}Q_{\mu}^{-}.
\]
Up to an error of order $O(\mu^{-2})$, the inverse of $B_{\mu,l}^{-}(z)$ is thus given by 
\[
[\Id_{m}-[(z+\lambda_{l,-})\Id_{m}-\g_{-}]^{-1}\mu^{-1}L_{\mu}^{-}]^{-1}[(z+\lambda_{l,-})\Id_{m}-\g_{-}]^{-1}.
\]
Again, since we are only interested in computing $[B_{\mu,l}^{-}(z)]^{-1}$ to order $O(\mu^{-2})$, it is 
sufficient to focus on 
\begin{equation}\label{eq:invBinflmzlot}
[(z+\lambda_{l,-})\Id_{m}-\g_{-}]^{-1}+[(z+\lambda_{l,-})\Id_{m}-\g_{-}]^{-1}\mu^{-1}L_{\mu}^{-}[(z+\lambda_{l,-})\Id_{m}-\g_{-}]^{-1}.
\end{equation}
Recall that $\g_{-}$ is in Jordan normal form and compute
\begin{equation}\label{eq:Klambdadefwa}
\begin{split}
K_{\lambda}(z) := & 
\left(\begin{array}{ccccc} z-\lambda & -1 & 0 & \cdots & 0\\ 0 & z-\lambda & -1 & \cdots & 0\\ \vdots & \vdots & \vdots & \vdots & \vdots \\
0 & 0 & 0 & \cdots & z-\lambda\end{array}\right)^{-1}\\
 = & \left(\begin{array}{ccccc} (z-\lambda)^{-1} & (z-\lambda)^{-2} & (z-\lambda)^{-3} & \cdots & (z-\lambda)^{-l}\\ 
0 & (z-\lambda)^{-1} & (z-\lambda)^{-2} & \cdots & (z-\lambda)^{-l+1}\\ \vdots & \vdots & \vdots & \vdots & \vdots \\
0 & 0 & 0 & \cdots & (z-\lambda)^{-1}\end{array}\right),
\end{split}
\end{equation}
where $l$ is the dimension of the Jordan block. Assuming $\lambda\notin\g^{*}$, there are thus two possibilities: If $\lambda\in B_{r}(0)$,
\begin{equation}\label{eq:intKlambda}
\frac{1}{2\pi i}\int_{\g}K_{\lambda}(z)dz=\Id_{l}.
\end{equation}
If $\lambda\notin\bar{B}_{r}(0)$, then the integral vanishes. As a consequence, 
\[
\frac{1}{2\pi i}\int_{\g}[(z+\lambda_{l,-})\Id_{m}-\g_{-}]^{-1}dz=\Pi_{l,-},
\]
where $\Pi_{l,-}$ is the projection to the subspace of $\cn{m}$ corresponding to the diagonal components of $\g_{-}$ that equal $\lambda_{l,-}$. 
What we need to focus on is thus the second term on the right hand side of (\ref{eq:invBinflmzlot}). Before proceeding, it is useful to introduce 
some notation. Recall that $\lambda_{l,-}$, $l=1,\dots,p_{-}$, are the distinct eigenvalues of $\g_{-}$, and that $m_{l,-}$ is the algebraic multiplicity 
of $\lambda_{l,-}$. We may thus assume, without loss of generality, that $\g_{-}=\diag\{\g_{-,11},\dots,\g_{-,p_{-}p_{-}}\}$, where $\g_{-,jj}\in\Mn{m_{j,-}}{\co}$, 
$j=1,\dots,p_{-}$. Moreover, $\g_{-,jj}$ consists of all the Jordan blocks corresponding to the eigenvalue $\lambda_{j,-}$. If $M\in\Mn{m}{\co}$ and 
$r,s\in\{1,\dots,p_{-}\}$, we define $M_{rs}\in\Mn{m_{r,-}\times m_{s,-}}{\co}$ by 
\[
M=:\left(\begin{array}{cccc} M_{11} & M_{12} & \cdots & M_{1p_{-}} \\ M_{21} & M_{22} & \cdots & M_{2p_{-}} \\ \vdots & \vdots & \vdots & \vdots \\
M_{p_{-}1} & M_{p_{-}2} & \cdots & M_{p_{-}p_{-}}\end{array}\right).
\]
This notation is consistent with the notation $\g_{-,jj}$. Moreover, $\g_{-,rs}=0$ if $r\neq s$. Let $\mK$ be defined by 
\[
\mK:=\frac{1}{2\pi i}\int_{\g}[(z+\lambda_{l,-})\Id_{m}-\g_{-}]^{-1}L_{\mu}^{-}[(z+\lambda_{l,-})\Id_{m}-\g_{-}]^{-1}dz.
\]
Then 
\begin{equation}\label{eq:mKrlformulawa}
\mK_{rs}=\frac{1}{2\pi i}\int_{\g}[(z+\lambda_{l,-})\Id_{m_{r,-}}-\g_{-,rr}]^{-1}L_{\mu,rs}^{-}[(z+\lambda_{l,-})\Id_{m_{s,-}}-\g_{-,ss}]^{-1}dz.
\end{equation}
If $r\neq l$ and $s\neq l$, it is clear that the integrand appearing on the right hand side of (\ref{eq:mKrlformulawa}) is analytic
in $B_{r}(0)$, so that $\mK_{rs}=0$. If $r=s=l$, it is clear (keeping (\ref{eq:Klambdadefwa}) in mind) that the components of the integrand
on the right hand side of (\ref{eq:mKrlformulawa}) are (finite) power series in $z^{-1}$, with the lowest power being $2$. Combining
this observation with Theorem~\ref{thm:Cauchy} yields $\mK_{ll}=0$. The only non-zero components of $\mK$ are thus $\mK_{rl}$ and $\mK_{lr}$
for $r\neq l$. Summing up the above observations, it is clear that 
\begin{equation}\label{eq:intgBinvnujpzwa}
\left\|\frac{1}{2\pi i}\int_{\g}[B_{\mu,l}^{-}(z)]^{-1}dz-\Pi_{l,-}-\mu^{-1}\mK\right\|\leq c_{0}\mu^{-2}.
\end{equation}
Combining this estimate with (\ref{eq:Fnujpzestwa}) and the analogous estimate for $G_{\mu,l}^{-}(z)$ yields
\begin{align}
\left\|\frac{1}{2\pi i}\int_{\g}F_{\mu,l}^{-}(z)dz-\frac{i}{2\mu}\Pi_{l,-}Q_{\mu}^{+}\right\| \leq & c_{0}\mu^{-2},\label{eq:intgFnujpzwa}\\
\left\|\frac{1}{2\pi i}\int_{\g}G_{\mu,l}^{-}(z)dz-\frac{i}{2\mu}Q_{\mu}^{-}\Pi_{l,-}\right\| \leq & c_{0}\mu^{-2}.\label{eq:intgGnujpzwa}
\end{align}
Combining (\ref{eq:pinujpmsforwa}), (\ref{eq:zIdmNnujpmwa}), (\ref{eq:EFGHdefwa}), (\ref{eq:Enujpzdefwa}), (\ref{eq:intgHnujpzwa}),
(\ref{eq:intgBinvnujpzwa}), (\ref{eq:intgFnujpzwa}) and (\ref{eq:intgGnujpzwa})
\[
\left\|\pi_{\mu,l,-}-\left(\begin{array}{cc} \Pi_{l,-}+\mu^{-1}\mK & \frac{i}{2\mu}\Pi_{l,-}Q_{\mu}^{+} \\
\frac{i}{2\mu}Q_{\mu}^{-}\Pi_{l,-} & 0 \end{array}\right)\right\|\leq c_{0}\mu^{-2}.
\]
Let $e_{l,k}^{-}$, $k=1,\dots,m_{l,-}$, be the standard basis for the image of $\Pi_{l,-}$. Then
\[
\left\|\pi_{\mu,l,-}\left(\begin{array}{c} e_{l,k}^{-} \\ 0 \end{array}\right)-\left(\begin{array}{c} e_{l,k}^{-}+\mu^{-1}\mK e_{l,k}^{-} \\
\frac{i}{2\mu}Q_{\mu}^{-} e_{l,k}^{-}\end{array}\right)\right\|\leq c_{0}\mu^{-2}. 
\]
Thus
\[
\he_{l,k}^{-}:=\pi_{\mu,l,-}\left(\begin{array}{c} e_{l,k}^{-} \\ 0 \end{array}\right)
\]
is a basis for $E_{\mu,l,-}$ for $\mu$ large enough. Since we know that $N_{\mu}$ leaves $E_{\mu,l,-}$ invariant, there is a matrix
$N_{\mu,l}^{-}\in\Mn{m_{l,-}}{\co}$ such that 
\begin{equation}\label{eq:hNnujlkdefwa}
N_{\mu}\he_{l,r}^{-}=\sum_{s=1}^{m_{l,-}}N_{\mu,l,sr}^{-}\he_{l,s}^{-}.
\end{equation}
We wish to compute $N_{\mu,l}^{-}$. Consider, to this end, 
\begin{equation*}
\begin{split}
N_{\mu}\he_{l,k}^{-} = & \left(\begin{array}{cc} -i\mu\Id_{m}+R_{\mu}^{-} & Q_{\mu}^{+}\\ Q_{\mu}^{-} & i\mu\Id_{m}+R_{\mu}^{+}\end{array}\right)
\left(\begin{array}{c} e_{l,k}^{-}+\mu^{-1}\mK e_{l,k}^{-} \\
\frac{i}{2\mu}Q_{\mu}^{-} e_{l,k}^{-}\end{array}\right)+O(\mu^{-1})\\
 = & \left(\begin{array}{c} -i\mu e_{l,k}^{-}+\g_{-}e_{l,k}^{-}-i\mK e_{l,k}^{-} \\ \frac{1}{2}Q_{\mu}^{-}e_{l,k}^{-} \end{array}\right)+O(\mu^{-1}).
\end{split}
\end{equation*}
Let us take the scalar product of this equality with $\he_{l,r}^{-}$ (when we say scalar product, we here mean the ordinary scalar product
and not the Hermitian inner product). This yields
\begin{equation}\label{eq:Nnuhebasformwa}
\begin{split}
(N_{\mu}\he_{l,k}^{-})\cdot \he_{l,r}^{-} = & (-i\mu e_{l,k}^{-}+\g_{-}e_{l,k}^{-}-i\mK e_{l,k}^{-})\cdot (e_{l,r}^{-}+\mu^{-1}\mK e_{l,r}^{-})+O(\mu^{-1})\\
 = & -i\mu\de_{kr}+(\g_{-} e_{l,k}^{-})\cdot e_{l,r}^{-}+O(\mu^{-1}),
\end{split}
\end{equation}
where we have used the fact that $e_{l,r}^{-}\cdot (\mK e_{l,k}^{-})=0$. On the other hand, 
\begin{equation}\label{eq:hejlpchofbaswa}
\he_{l,r}^{-}\cdot \he_{l,s}^{-}=\de_{rs}+O(\mu^{-2}),
\end{equation}
where we have, again, used the fact that $e_{l,r}^{-}\cdot (\mK e_{l,k}^{-})=0$. Combining (\ref{eq:hNnujlkdefwa}), (\ref{eq:Nnuhebasformwa}) and 
(\ref{eq:hejlpchofbaswa}) yields 
\[
N_{\mu,l,rs}^{-}=-i\mu\de_{rs}+(\g_{-} e_{l,s}^{-})\cdot e_{l,r}^{-}+O(\mu^{-1}). 
\]
There is a similar analysis concerning $E_{\mu,j,+}$ (which we provide below), and as a consequence, there is $0<\mu_{0}\in\ro$, and, for $\mu\geq\mu_{0}$, 
a matrix $T_{\mu}$ (which converges to $\Id_{2m}$ as $\mu\rightarrow\infty$) such that (\ref{eq:Tmudiag}) holds. The lemma follows.

\textbf{Analysis in the case of $E_{\mu,j,+}$.} The analysis in the case of $E_{\mu,j,+}$ is similar, but we include it for the sake of completeness. 
In the process of calculating the right hand side of (\ref{eq:pinujpmsforwa}), it is useful to note that 
\begin{equation}\label{eq:invintprojwaplus}
\begin{split}
 & \left(\begin{array}{cc} A_{\mu,l}^{+}(z) & -Q_{\mu}^{+}\\ -Q_{\mu}^{-} & D_{\mu,l}^{+}(z)\end{array}\right)\\
 = & \left(\begin{array}{cc} A_{\mu,l}^{+}(z) & 0\\ -Q_{\mu}^{-} & B_{\mu,l}^{+}(z)\end{array}\right)
\left(\begin{array}{cc} \Id_{m} & -[A_{\mu,l}^{+}(z)]^{-1}Q_{\mu}^{+}\\ 0 & \Id_{m}\end{array}\right),
\end{split}
\end{equation}
where 
\begin{equation}\label{eq:Bnujpzdefwaplus}
B_{\mu,l}^{+}(z):=D_{\mu,l}^{+}(z)-Q_{\mu}^{-}[A_{\mu,l}^{+}(z)]^{-1}Q_{\mu}^{+}.
\end{equation}
Introducing the notation
\begin{equation}\label{eq:EFGHdefwaplus}
\left(\begin{array}{cc} A_{\mu,l}^{+}(z) & -Q_{\mu}^{+}\\ -Q_{\mu}^{-} & D_{\mu,l}^{+}(z)\end{array}\right)^{-1}
=:\left(\begin{array}{cc} E_{\mu,l}^{+}(z) & F_{\mu,l}^{+}(z)\\ G_{\mu,l}^{+}(z) & H_{\mu,l}^{+}(z)\end{array}\right),
\end{equation}
it can be computed that 
\begin{align}
E_{\mu,l}^{+}(z) = & [A_{\mu,l}^{+}(z)]^{-1}Q_{\mu}^{+}[B_{\mu,l}^{+}(z)]^{-1}Q_{\mu}^{-}[A_{\mu,l}^{+}(z)]^{-1}
+[A_{\mu,l}^{+}(z)]^{-1},\label{eq:Enujpzdefwaplus}\\
F_{\mu,l}^{+}(z) = & [A_{\mu,l}^{+}(z)]^{-1}Q_{\mu}^{+}[B_{\mu,l}^{+}(z)]^{-1},\label{eq:Fnujpzdefwaplus}\\
G_{\mu,l}^{+}(z) = & [B_{\mu,l}^{+}(z)]^{-1}Q_{\mu}^{-}[A_{\mu,l}^{+}(z)]^{-1},\label{eq:Gnujpzdefwaplus}\\
H_{\mu,l}^{+}(z) = & [B_{\mu,l}^{+}(z)]^{-1}.\label{eq:Hnujpzdefwaplus}
\end{align}
Note that 
\begin{equation}\label{eq:Dnujpzinvwaplus}
[A_{\mu,l}^{+}(z)]^{-1}=(z+2i\mu+\lambda_{l,+})^{-1}[\Id_{m}-(z+2i\mu+\lambda_{l,+})^{-1}R_{\mu}^{-}]^{-1},
\end{equation}
so that 
\[
\frac{1}{2\pi i}\int_{\g}[A_{\mu,l}^{+}(z)]^{-1}dz=0
\]
for large $\mu$. On the other hand, the norm of the first term on the right hand side of (\ref{eq:Enujpzdefwaplus}) is $O(\mu^{-2})$. 
Thus
\begin{equation}\label{eq:intgHnujpzwaplus}
\left\|\frac{1}{2\pi i}\int_{\g}E_{\mu,l}^{+}(z)dz\right\|\leq c_{0}\mu^{-2}.
\end{equation}
Before turning to (\ref{eq:Fnujpzdefwaplus}) and (\ref{eq:Gnujpzdefwaplus}), note that (\ref{eq:Dnujpzinvwaplus}) yields
\begin{equation}\label{eq:Dnujpzestwaplus}
\left\|[A_{\mu,l}^{+}(z)]^{-1}+\frac{i}{2\mu}\Id_{m}\right\|\leq c_{0}\mu^{-2}.
\end{equation}
Thus
\begin{equation}\label{eq:Fnujpzestwaplus}
\left\|F_{\mu,l}^{+}(z)+\frac{i}{2\mu}Q_{\mu}^{+}[B_{\mu,l}^{+}(z)]^{-1}\right\|\leq c_{0}\mu^{-2}.
\end{equation}
There is a similar estimate concerning $G_{\mu,l}^{+}(z)$, so that it is sufficient to focus on the integral of $[B_{\mu,l}^{+}(z)]^{-1}$ 
over $\g$. Due to (\ref{eq:Rinfpmdefintro}), (\ref{eq:Ninfjpform}), (\ref{eq:zIdmNnujpmwa}), (\ref{eq:Bnujpzdefwaplus}) and (\ref{eq:Dnujpzestwaplus}) 
\[
\left\|B_{\mu,l}^{+}(z)-(z+\lambda_{l,+})\Id_{m}+\g_{+}-\mu^{-1}L_{\mu}^{+}\right\|\leq c_{0}\mu^{-2},
\]
where 
\[
L_{\mu}^{+}:=iW_{+}+\frac{i}{2}Q_{\mu}^{-}Q_{\mu}^{+}.
\]
Up to an error of order $O(\mu^{-2})$, the inverse of $B_{\mu,l}^{+}(z)$ is thus given by 
\[
[\Id_{m}+[(z+\lambda_{l,+})\Id_{m}-\g_{+}]^{-1}\mu^{-1}L_{\mu}^{+}]^{-1}[(z+\lambda_{l,+})\Id_{m}-\g_{+}]^{-1}.
\]
Again, since we are only interested in computing $[B_{\mu,l}^{+}(z)]^{-1}$ to order $O(\mu^{-2})$, it is 
sufficient to focus on 
\begin{equation}\label{eq:invBinflmzlotplus}
[(z+\lambda_{l,+})\Id_{m}-\g_{+}]^{-1}-[(z+\lambda_{l,+})\Id_{m}-\g_{+}]^{-1}\mu^{-1}L_{\mu}^{+}[(z+\lambda_{l,+})\Id_{m}-\g_{+}]^{-1}.
\end{equation}
Due to (\ref{eq:Klambdadefwa}) and (\ref{eq:intKlambda}),
\[
\frac{1}{2\pi i}\int_{\g}[(z+\lambda_{l,+})\Id_{m}-\g_{+}]^{-1}dz=\Pi_{l,+},
\]
where $\Pi_{l,+}$ is the projection to the subspace of $\cn{m}$ corresponding to the diagonal components of $\g_{+}$ that equal $\lambda_{l,+}$.
What we need to focus on is thus the second term appearing in (\ref{eq:invBinflmzlotplus}). Before proceeding, it is useful to introduce 
some notation. Recall that $\lambda_{l,+}$, $l=1,\dots,p_{+}$, are the distinct eigenvalues of $\g_{+}$, and that $m_{l,+}$ is the algebraic multiplicity 
of $\lambda_{l,+}$. We may thus assume, without loss of generality, that $\g_{+}=\diag\{\g_{+,11},\dots,\g_{+,p_{+}p_{+}}\}$, where $\g_{+,jj}\in\Mn{m_{j,+}}{\co}$, 
$j=1,\dots,p_{+}$. Moreover, $\g_{+,jj}$ consists of all the Jordan blocks corresponding to the eigenvalue $\lambda_{j,+}$. If $M\in\Mn{m}{\co}$ and 
$r,s\in\{1,\dots,p_{+}\}$, we define $M_{rs}\in\Mn{m_{r,+}\times m_{s,+}}{\co}$ by 
\[
M=:\left(\begin{array}{cccc} M_{11} & M_{12} & \cdots & M_{1p_{+}} \\ M_{21} & M_{22} & \cdots & M_{2p_{+}} \\ \vdots & \vdots & \vdots & \vdots \\
M_{p_{+}1} & M_{p_{+}2} & \cdots & M_{p_{+}p_{+}}\end{array}\right).
\]
This notation is consistent with the notation $\g_{+,jj}$. Moreover, $\g_{+,rs}=0$ if $r\neq s$. Let $\ml$ be defined by 
\[
\ml:=\frac{1}{2\pi i}\int_{\g}[(z+\lambda_{l,+})\Id_{m}-\g_{+}]^{-1}L_{\mu}^{+}[(z+\lambda_{l,+})\Id_{m}-\g_{+}]^{-1}dz.
\]
Then 
\begin{equation}\label{eq:mKrlformulawaplus}
\ml_{rs}=\frac{1}{2\pi i}\int_{\g}[(z+\lambda_{l,+})\Id_{m_{r,+}}-\g_{+,rr}]^{-1}L_{\mu,rs}^{+}[(z+\lambda_{l,+})\Id_{m_{s,+}}-\g_{+,ss}]^{-1}dz.
\end{equation}
If $r\neq l$ and $s\neq l$, it is clear that the integrand appearing on the right hand side of (\ref{eq:mKrlformulawa}) is analytic
in $B_{r}(0)$, so that $\ml_{rs}=0$. If $r=s=l$, it is clear (keeping (\ref{eq:Klambdadefwa}) in mind) that the components of the integrand
on the right hand side of (\ref{eq:mKrlformulawa}) are (finite) power series in $z^{-1}$, with the lowest power being $2$. Combining
this observation with Theorem~\ref{thm:Cauchy} yields $\ml_{ll}=0$. The only non-zero components of $\ml$ are thus $\ml_{rl}$ and $\ml_{lr}$
for $r\neq l$. Summing up the above observations, it is clear that 
\begin{equation}\label{eq:intgBinvnujpzwaplus}
\left\|\frac{1}{2\pi i}\int_{\g}[B_{\mu,l}^{+}(z)]^{-1}dz-\Pi_{l,+}+\mu^{-1}\ml\right\|\leq c_{0}\mu^{-2}.
\end{equation}
Combining this estimate with (\ref{eq:Fnujpzestwaplus}) and the analogous estimate for $G_{\mu,l}^{+}(z)$ yields
\begin{align}
\left\|\frac{1}{2\pi i}\int_{\g}F_{\mu,l}^{+}(z)dz+\frac{i}{2\mu}Q_{\mu}^{+}\Pi_{l,+}\right\| \leq & c_{0}\mu^{-2},\label{eq:intgFnujpzwaplus}\\
\left\|\frac{1}{2\pi i}\int_{\g}G_{\mu,l}^{+}(z)dz+\frac{i}{2\mu}\Pi_{l,+}Q_{\mu}^{-}\right\| \leq & c_{0}\mu^{-2}.\label{eq:intgGnujpzwaplus}
\end{align}
Combining (\ref{eq:pinujpmsforwa}), (\ref{eq:zIdmNnujpmwa}), (\ref{eq:EFGHdefwaplus}), (\ref{eq:Hnujpzdefwaplus}), (\ref{eq:intgHnujpzwaplus}),
(\ref{eq:intgBinvnujpzwaplus}), (\ref{eq:intgFnujpzwaplus}) and (\ref{eq:intgGnujpzwaplus})
\[
\left\|\pi_{\mu,l,+}-\left(\begin{array}{cc} 0 & -\frac{i}{2\mu}Q_{\mu}^{+}\Pi_{l,+} \\
-\frac{i}{2\mu}\Pi_{l,+}Q_{\mu}^{-} & \Pi_{l,+}-\mu^{-1}\ml \end{array}\right)\right\|\leq c_{0}\mu^{-2}.
\]
Let $e_{l,k}^{+}$, $k=1,\dots,m_{l,+}$, be the standard basis for the image of $\Pi_{l,+}$. Then
\[
\left\|\pi_{\mu,l,+}\left(\begin{array}{c} 0 \\ e_{l,k}^{+} \end{array}\right)-\left(\begin{array}{c}  -\frac{i}{2\mu}Q_{\mu}^{+} e_{l,k}^{+} \\
e_{l,k}^{+}-\mu^{-1}\ml e_{l,k}^{+} \end{array}\right)\right\|\leq c_{0}\mu^{-2}. 
\]
Thus
\[
\he_{l,k}^{+}:=\pi_{\mu,l,+}\left(\begin{array}{c} e_{l,k}^{+} \\ 0 \end{array}\right)
\]
is a basis for $E_{\mu,l,+}$ for $\mu$ large enough. Since we know that $N_{\mu}$ leaves $E_{\mu,l,+}$ invariant, there is a matrix
$N_{\mu,l}^{+}\in\Mn{m_{l,+}}{\co}$ such that 
\begin{equation}\label{eq:hNnujlkdefwaplus}
N_{\mu}\he_{l,r}^{+}=\sum_{s=1}^{m_{l,+}}N_{\mu,l,sr}^{+}\he_{l,s}^{+}.
\end{equation}
We wish to compute $N_{\mu,l}^{+}$. Consider, to this end, 
\begin{equation*}
\begin{split}
N_{\mu}\he_{l,k}^{+} = & \left(\begin{array}{cc} -i\mu\Id_{m}+R_{\mu}^{-} & Q_{\mu}^{+}\\ Q_{\mu}^{-} & i\mu\Id_{m}+R_{\mu}^{+}\end{array}\right)
\left(\begin{array}{c}  -\frac{i}{2\mu}Q_{\mu}^{+} e_{l,k}^{+} \\
e_{l,k}^{+}-\mu^{-1}\ml e_{l,k}^{+} \end{array}\right)+O(\mu^{-1})\\
 = & \left(\begin{array}{c} \frac{1}{2}Q_{\mu}^{+}e_{l,k}^{+} \\ i\mu e_{l,k}^{+}+\g_{+}e_{l,k}^{+}-i\ml e_{l,k}^{+} \end{array}\right)+O(\mu^{-1}).
\end{split}
\end{equation*}
Let us take the scalar product of this equality with $\he_{l,r}^{+}$. This yields
\begin{equation}\label{eq:Nnuhebasformwaplus}
\begin{split}
(N_{\mu}\he_{l,k}^{+})\cdot \he_{l,r}^{+} = & (i\mu e_{l,k}^{+}+\g_{+}e_{l,k}^{+}-i\ml e_{l,k}^{+})\cdot (e_{l,r}^{+}-\mu^{-1}\ml e_{l,r}^{+})+O(\mu^{-1})\\
 = & i\mu\de_{kr}+(\g_{+} e_{l,k}^{+})\cdot e_{l,r}^{+}+O(\mu^{-1}),
\end{split}
\end{equation}
where we have used the fact that $e_{l,r}^{+}\cdot (\ml e_{l,k}^{+})=0$. On the other hand, 
\begin{equation}\label{eq:hejlpchofbaswaplus}
\he_{l,r}^{+}\cdot \he_{l,s}^{+}=\de_{rs}+O(\mu^{-2}),
\end{equation}
where we have, again, used the fact that $e_{l,r}^{+}\cdot (\ml e_{l,k}^{+})=0$. Combining (\ref{eq:hNnujlkdefwaplus}), (\ref{eq:Nnuhebasformwaplus}) and 
(\ref{eq:hejlpchofbaswaplus}) yields 
\[
N_{\mu,l,rs}^{+}=i\mu\de_{rs}+(\g_{+} e_{l,s}^{+})\cdot e_{l,r}^{+}+O(\mu^{-1}). 
\]
This completes the analysis. 
\end{proof}

\section{Mode by mode analysis}\label{section:mbmanhftrs}

Consider the index set $\EFindexset$ introduced in connection with (\ref{eq:IgenReqdef}) and fix $1\leq m\in\zo$. Assume that for each 
$\indexnot\in\EFindexset$, there is a matrix $A(\indexnot)\in\Mn{2m}{\co}$ and functions $A_{\indexnot,\rem}\in C^{\infty}(\ro,\Mn{2m}{\co})$
and $F_{\indexnot}\in C^{\infty}(\ro,\cn{2m})$. We are then interested in the equation
\begin{equation}\label{eq:dotvAmu}
\dot{v}=A(\indexnot)v+A_{\indexnot,\rem}v+F_{\indexnot}.
\end{equation}
Here the matrix valued function $A_{\indexnot,\rem}$ is assumed to satisfy the estimate 
\begin{equation}\label{eq:Amuremesttrscase}
\|A_{\indexnot,\rem}(t)\|\leq C_{\trs}e^{-\b_{\trs}\bt}
\end{equation}
for $t\geq T_{\trs}$, where $0<C_{\trs},\b_{\trs}\in\ro$ are independent of $\indexnot$; $T_{\trs}$ depends on $\indexnot$; and $\bt:=t-T_{\trs}$.
In order to restrict the possible matrices $A(\indexnot)$, we think of (\ref{eq:dotvAmu}) as arising from an equation of the form 
(\ref{eq:thesystemRge}) satisfying one of Definitions~\ref{def:Xnontriv} and \ref{def:Xdegenerate}. In the present setting, it is convenient
to make assumptions that follow from these definitions without referring to them or to (\ref{eq:thesystemRge}). One way of doing this is by 
introducing a function $\mu:\EFindexset\rightarrow [0,\infty)$ with the property that if $\muindexset:=\mu(\EFindexset)$, then the only 
accumulation point of $\muindexset$ is $\infty$. Note that if (\ref{eq:thesystemRge}) satisfies one of Definitions~\ref{def:Xnontriv} and 
\ref{def:Xdegenerate}, and if $\mu(\indexnot):=\mfg_{\infty}(\indexnot)$, then this requirement is satisfied. Given the function $\mu$, we 
restrict our attention to matrices $A(\indexnot)$ of the form 
\begin{equation}\label{eq:Aindexnotitomuindexnot}
A(\indexnot):= 
A_{\mu(\indexnot)}=\left(\begin{array}{cc} 0 & \mu(\indexnot)\Id_{m} \\ -\mu(\indexnot)\Id_{m}+iV+2[\mu(\indexnot)]^{-1}W & 
U\end{array}\right),
\end{equation}
where $V,W,U\in\Mn{m}{\co}$ and we have used the notation (\ref{eq:Amudef}). Below we frequently omit reference to the argument $\indexnot$ 
and simply write $\mu$ instead of $\mu(\indexnot)$. When we do, $\mu^{-1}$ should be interpreted as $[\mu(\indexnot)]^{-1}$. For a fixed 
$\mu_{a}\in\muindexset$, we can analyse the asymptotics of solutions to (\ref{eq:dotvAmu}) for $\indexnot\in\mu^{-1}(\mu_{a})$. In fact, we
can appeal to the results obtained in Sections~\ref{section:detailedODEas} and \ref{section:spdatinfODE}. However, the constants appearing
in the results depend on $A(\indexnot)=A_{\mu_{a}}$. Appealing to the analysis in these sections is thus acceptable for a finite number of 
$\mu$-values. In the present setting, this is not sufficient, and we need to proceed differently. Considering the analysis carried out in 
Sections~\ref{section:aseigvcalc} and \ref{section:compprojontogensubs}, it is clear that there is an asymptotic 
partial diagonalisation of the matrices $A_{\mu}$ as $\mu\rightarrow\infty$. Moreover, the real parts of the eigenvalues of $A_{\mu}$ converge 
to specific values as $\mu\rightarrow\infty$. Since the only accumulation point of $\muindexset$ is $\infty$, this leads to the following 
approach to obtaining estimates for all $\indexnot\in\EFindexset$. First, we appeal to the asymptotic partial diagonalisation in order to 
obtain estimates for $\mu$ large enough, say $\mu\geq\mu_{0}$. Second, we note that the remaining $\mu$-values are finite in number. Thus we 
can handle them by appealing to Sections~\ref{section:detailedODEas} and \ref{section:spdatinfODE}. By increasing $\mu_{0}$, the asymptotic 
analysis becomes more
and more refined. However, there is a price to pay for this refinement: the supremum of the constants involved in the estimates for the 
remaining $\mu$ values can be expected to tend to infinity as $\mu_{0}$ tends to infinity. What is preferable might depend on the context. For 
that reason, it is natural to introduce a parameter $0<\b_{\romar}\in\ro$ which quantifies the quality of the asymptotic analysis; a small 
$\b_{\romar}$-value corresponds to good control in the asymptotic region where $\mu$ is large. However, the constants appearing in the estimates
can be expected to tend to infinity as $\b_{\romar}\rightarrow 0+$. 

Just as in the case of Lemma~\ref{lemma:ODEasymp}, we need to make assumptions concerning the integrability properties of $F_{\indexnot}$ in order 
to obtain detailed information concerning the asymptotics of solutions to (\ref{eq:dotvAmu}). In the next lemma, we therefore assume that the
following norm of $F_{\indexnot}$ is bounded:
\begin{equation}\label{eq:Findexnotrolmadef}
\|F_{\indexnot}\|_{\rolma}:=\int_{0}^{\infty}e^{-(\kappa_{\indexnot}-\b_{\romar})s}|F_{\indexnot}(s)|ds,
\end{equation}
\index{$\a$Aa@Notation!Norms!$\normrolma$}%
where $\kappa_{\indexnot}:=\kappa_{\max}[A(\indexnot)]$ 
\index{$\a$Aa@Notation!Matrix notation!kappaiota@$\kappa_{\indexnot}$}%
and the choice of $\b_{\romar}>0$ should be clear from the context. Note that we abuse notation
slightly here, in that the norm depends on $\indexnot$ without this being explicitly stated. 

In Lemma~\ref{lemma:ODEasymp}, the first generalised eigenspace in the $\b_{\rem},A$-decomposition of $\cn{2m}$ played an important role. Similar
algebraic constructions can be expected to be important in the present setting as well. However, since we need to have uniform (i.e., independent
of $\indexnot$) bounds on the norms 
of the matrices, say $T_{\indexnot}$ and $T_{\indexnot}^{-1}$, putting $A(\indexnot)$ into Jordan normal form, in order for the 
$\b_{\rem},A(\indexnot)$-decomposition of $\cn{2m}$ to be of use, and since such such uniform bounds are not to be expected in the present setting,
the algebraic constructions we use here have to be 
somewhat more rough. On the other hand, we can appeal to the partial diagonalisation derived in Subsection~\ref{ssection:aspardiag}. This leads
to the following asymptotic substitute for the notion of a first generalised eigenspace in the $\b_{\rem},A$-decomposition of $\cn{2m}$.
\begin{definition}\label{def:Einfindexnotb}
Let $1\leq m\in\zo$, $\EFindexset$ be the set introduced in connection with (\ref{eq:IgenReqdef}) and $\mu$ be a function 
$\mu:\EFindexset\rightarrow [0,\infty)$. Let $U,V,W\in\Mn{m}{\co}$ and define $A(\indexnot)$ by (\ref{eq:Aindexnotitomuindexnot}). Let $\lambda_{j,\pm}$, 
$j=1,\dots,p_{\pm}$, be the eigenvalues of $U_{\pm}=(U\pm V)/2$, and let $\kappa_{\infty}$ be the largest real part of an element of 
$\Spe (U_{+})\cup\Spe(U_{-})$. Let $0<r_{a},\mu_{a}\in\ro$ (depending only on $U$, $V$ and $W$) be such that if $\mu\geq\mu_{a}$, then the closed balls 
with radius $r_{a}$ and centre $\pm i\mu+\lambda_{j,\pm}$ are disjoint and a distance at least $r_{a}$ apart. Due to Lemma~\ref{lemma:Nmueigenvapprox}, 
there is a $\mu_{0}\geq\mu_{a}$ such that if $\mu(\indexnot)\geq\mu_{0}$, then the set $\Spe[A(\indexnot)]$ is contained in the disjoint union of the open 
balls of radius $r_{a}$ and centre $\pm i\mu(\indexnot)+\lambda_{j,\pm}$. Given a $0<\b\in\ro$ and a $\indexnot\in\EFindexset$ such that 
$\mu(\indexnot)\geq\mu_{0}$, the vector space $E^{\infty}_{\indexnot,\b}$ 
\index{$\a$Aa@Notation!Vector spaces!$E^{\infty}_{\indexnot,\b}$}%
is defined to be the direct sum of the generalised eigenspaces of $A(\indexnot)$ corresponding to eigenvalues in the balls of 
radius $r_{a}$ and centre $\pm i\mu(\indexnot)+\lambda_{j,\pm}$ for $j$ such that $\kappa_{\infty}-\mathrm{Re}(\lambda_{j,\pm})<\b$. 
\end{definition}
\begin{lemma}\label{lemma:modebymodetrs}
Let $1\leq m\in\zo$, $\EFindexset$ be the set introduced in connection with (\ref{eq:IgenReqdef}) and $\mu$ be a function
$\mu:\EFindexset\rightarrow [0,\infty)$. Let $\EFtrsindexset\subseteq\EFindexset$ be the set of $\indexnot\in\EFindexset$ such that 
$\mu(\indexnot)>0$. Let $U,V,W\in\Mn{m}{\co}$ and define $A(\indexnot)$ by (\ref{eq:Aindexnotitomuindexnot}) for $\indexnot\in\EFtrsindexset$.
Assume that for each $\indexnot\in\EFtrsindexset$, there are functions $A_{\indexnot,\rem}\in C^{\infty}(\ro,\Mn{2m}{\co})$
and $F_{\indexnot}\in C^{\infty}(\ro,\cn{2m})$. The function $A_{\indexnot,\rem}$ is assumed to be such that there are constants 
$0<\b_{\trs},C_{\trs}\in\ro$ (independent of $\indexnot$) and a constant $T_{\trs}$ (which is allowed to depend on $\indexnot$) such 
that (\ref{eq:Amuremesttrscase}) holds for $t\geq T_{\trs}$. Fix a $0<\b_{\romar}\in\ro$ which is small enough, the bound depending only on 
$U,V$ and $\b_{\trs}$. Assume the functions $F_{\indexnot}$ to be such that $\|F_{\indexnot}\|_{\rolma}<\infty$ for $\indexnot\in\EFtrsindexset$;
cf. (\ref{eq:Findexnotrolmadef}). Then there are constants $0<\mu_{0}\in\ro$ and $0< C\in\ro$, where 
$\mu_{0}$ only depends on $U,V,W$ and $\b_{\romar}$, and $C$ only depends on $U,V,\b_{\trs},\b_{\romar}$ and $C_{\trs}$, such that the following holds. If 
$\indexnot\in\mu^{-1}([\mu_{0},\infty))$ and $v$ is a solution to (\ref{eq:dotvAmu}), then there is a $v_{\infty}\in E^{\infty}_{\indexnot,\b_{\trs}}$ such that
\begin{equation}\label{eq:vvinfesttrsstmt}
\begin{split}
 & \left|v(t)-e^{A(\indexnot)t}v_{\infty}-\int_{T_{\trs}}^{t}e^{A(\indexnot)(t-s)}F_{\indexnot}(s)ds\right|\\
 \leq & Ce^{(\kappa_{\indexnot} -\b_{\trs}+\b_{\romar})\bt}[|v(T_{\trs})|+e^{\kappa_{\indexnot} T_{\trs}}\|F_{\indexnot}\|_{\rolma}]
\end{split}
\end{equation}
for $t\geq T_{\trs}$, where $E^{\infty}_{\indexnot,\b_{\trs}}$ is given by Definition~\ref{def:Einfindexnotb} and $\kappa_{\indexnot}:=\kappa_{\max}[A(\indexnot)]$; 
cf. Definition~\ref{def:SpRspdef}. Moreover, 
\begin{equation}\label{eq:vinftyitouinfty}
v_{\infty}:=e^{-A(\indexnot)T_{\trs}}u_{\infty}
\end{equation}
for a $u_{\infty}\in E^{\infty}_{\indexnot,\b_{\trs}}$ satisfying the estimate
\begin{equation}\label{eq:uinfest}
|u_{\infty}|\leq C[|v(T_{\trs})|+e^{\kappa_{\indexnot}T_{\trs}}\|F_{\indexnot}\|_{\rolma}].
\end{equation}
Here $C$ has the same dependence as in the case of (\ref{eq:vvinfesttrsstmt}). 
\end{lemma}
\begin{remark}
Note that if $A_{\mu}$ is given by (\ref{eq:Amudef}), then $\kappa_{\max}(A_{\mu})\rightarrow\kappa_{\infty}$ as $\mu\rightarrow\infty$, where
$\kappa_{\infty}$ is introduced in Definition~\ref{def:Einfindexnotb}; this is a consequence of Lemma~\ref{lemma:Nmueigenvapprox}. In fact, we could 
replace $\kappa_{\indexnot}$ by $\kappa_{\infty}$ in the statement of the lemma. 
\end{remark}
\begin{proof}
It is convenient to divide the proof into several steps. To begin with we derive several reformulations of the equations.

\textbf{First reformulation of the equations; partial diagonalisation.}
Due to (\ref{eq:Nmudef}) and Lemma~\ref{lemma:aspardiag}, we know that there is a $0<\mu_{0}\in\ro$ such that if $\mu(\indexnot)\geq\mu_{0}$,
then  (\ref{eq:dotvAmu}) can be reformulated to 
\[
\dot{v}_{\pre}=\bar{A}(\indexnot)v_{\pre}+\bar{A}_{\indexnot,\rem}v_{\pre}+\bar{F}_{\indexnot},
\]
where $T_{\indexnot}:=S_{12}T_{\mu(\indexnot)}$ and 
\begin{equation}\label{eq:vpreetcdef}
v_{\pre}:=T_{\indexnot}^{-1}v,\ \ \
\bar{A}(\indexnot):=T_{\indexnot}^{-1}A(\indexnot)T_{\indexnot},\ \ \
\bar{A}_{\indexnot,\rem}:=T_{\indexnot}^{-1}A_{\indexnot,\rem}T_{\indexnot},\ \ \
\bar{F}_{\indexnot}:=T_{\indexnot}^{-1}F_{\indexnot}.
\end{equation}
Note that $\|T_{\indexnot}\|$ and $\|T_{\indexnot}^{-1}\|$ are bounded by numerical constants. In addition, $\bar{A}(\indexnot)$ is given by the right 
hand side of (\ref{eq:Tmudiag}) with $\mu$ replaced by $\mu(\indexnot)$. Moreover, 
\begin{equation}\label{eq:Nmujpmexpr}
N_{\mu(\indexnot),j}^{\pm}=\pm i\mu(\indexnot) \Id_{m_{j,\pm}}+\g_{\pm,j}+\mR_{\indexnot,j}^{\pm},
\end{equation}
where 
\begin{equation}\label{eq:mRmujpmest}
\|\mR_{\indexnot,j}^{\pm}\|\leq c_{0}[\mu(\indexnot)]^{-1}
\end{equation}
for $\mu(\indexnot)\geq\mu_{0}$, where $c_{0}$ and $\mu_{0}$ only depend on $U,V$ and $W$. Letting $M_{\indexnot,j}^{\pm}$ denote the sum of the first 
two terms on the right hand side of (\ref{eq:Nmujpmexpr}), let 
\[
M_{\indexnot}:=\diag\{M_{\indexnot,1}^{-},\dots,M_{\indexnot,p_{-}}^{-},M_{\indexnot,1}^{+},\dots,M_{\indexnot,p_{+}}^{+}\}.
\]

\textbf{Second reformulation of the equations; eliminating the imaginary parts.}
Note that the imaginary part of $M_{\indexnot,j}^{\pm}$ is a multiple of the identity, so that the imaginary part of $M_{\indexnot}$ commutes with 
$\bar{A}(\indexnot)$. Introducing $M_{\indexnot,\roim}:=\mathrm{Im}(M_{\indexnot})$ and 
\[
v_{\fin}:=\exp\left(-iM_{\indexnot,\roim}\bt\right)v_{\pre}
\]
thus yields
\[
\dot{v}_{\fin}=\tA(\indexnot)v_{\fin}+\tA_{\indexnot,\rem}v_{\fin}+\tF_{\indexnot},
\]
where
\[
\tA_{\indexnot,\rem}:=\exp\left(-iM_{\indexnot,\roim}\bt\right)\bar{A}_{\indexnot,\rem}\exp\left(iM_{\indexnot,\roim}\bt\right),\ \ \
\tF_{\indexnot}:=\exp\left(-iM_{\indexnot,\roim}\bt\right)\bar{F}_{\indexnot}.
\]
Moreover, 
\[
\tA(\indexnot)=\tJ+\tmR_{\indexnot},\ \ \
\tJ:=\diag\{\tJ_{1},\dots,\tJ_{K}\},\ \ \
\tmR_{\indexnot}:=\diag\{\tmR_{\indexnot,1},\dots,\tmR_{\indexnot,K}\},
\]
where $\dim \tmR_{\indexnot,k}=\dim \tJ_{k}$ and the $\tmR_{\indexnot,k}$ satisfy estimates similar to (\ref{eq:mRmujpmest}). Finally, the $\tJ_{k}$ are the 
real parts of the $\g_{\pm,j}$. 

\textbf{Third reformulation of the equations; normalising the real parts.}
Let $\kappa_{\infty}$ be the constant introduced in Definition~\ref{def:Einfindexnotb}. Note that this means that $\kappa_{\infty}$ is the 
largest real part of an eigenvalue of $\g_{\pm}$. Define
\[
w:=e^{-\kappa_{\infty} t}Dv_{\fin},
\] 
where $D$ is a real non-degenerate diagonal matrix that remains to be determined. Then
\begin{equation}\label{eq:dotwBmurem}
\dot{w}=B(\indexnot)w+B_{\indexnot,\rem}w+G_{\indexnot},
\end{equation}
where
\[
B(\indexnot):=D\tA(\indexnot)D^{-1}-\kappa_{\infty}\Id_{2m},\ \ \
B_{\indexnot,\rem}:=D\tA_{\indexnot,\rem}D^{-1},\ \ \
G_{\indexnot}:=e^{-\kappa_{\infty} t}D\tF_{\indexnot}.
\]
Moreover, 
\[
B(\indexnot)=J+\mR_{\indexnot},\ \ \
J:=\diag\{J_{1},\dots,J_{K}\},\ \ \
\mR_{\indexnot}:=\diag\{\mR_{\indexnot,1},\dots,\mR_{\indexnot,K}\},
\]
where $\dim \mR_{\indexnot,k}=\dim J_{k}$ and the $\mR_{\indexnot,k}$ satisfy estimates similar to (\ref{eq:mRmujpmest}) (however, in this case the estimate
depends on $D$). Here the $J_{k}$ consist of generalised Jordan blocks (their non-zero off-diagonal components need not equal $1$) and their diagonal 
components are non-positive. Moreover, we choose $D$ so that if the diagonal components of $J_{k}$ are strictly negative, then $J_{k}$
is negative definite; cf. Lemma~\ref{lemma:genJordblock} and Remark~\ref{remark:genJordblock}. The price we need to pay in order to achieve this
is that $D$ depends on $\g_{\pm}$. For the resulting $J_{k}$, we can then ensure that the symmetrisation of $J_{k}+\mR_{\indexnot,k}$ is negative definite 
by demanding that $\mu(\indexnot)$ be large enough; the symmetrisation of $A\in\Mn{k}{\co}$ is given by $(A+A^{\dagger})/2$, where $A^{\dagger}$ is the 
transpose of the complex conjugate of $A$. The lower bound on $\mu(\indexnot)$ required to achieve this, say $\mu_{0}$ depends only on $U,V$ and $W$. 
Finally, we can ensure that if the diagonal component of $J_{k}$ is $\kappa_{k}$, then 
\begin{equation}\label{eq:Jkkappakest}
\|J_{k}+\mR_{\indexnot,k}-\kappa_{k}\Id_{d_{k}}\|+\|J_{k}-\kappa_{k}\Id_{d_{k}}\|\leq \b_{\romar}/4,
\end{equation}
where $d_{k}$ is the dimension of $J_{k}$. In order to ensure that this estimate holds, we need to choose a $D$ depending only on $\g_{\pm}$ and 
$\b_{\romar}$. Moreover, we need to demand that $\mu(\indexnot)\geq\mu_{0}$, where $\mu_{0}$ only depends on $U,V,W$ and $\b_{\romar}$. 
At this stage, it is of interest to note that there is a $\mu_{0}>0$ (depending only on $U,V$ and $W$) such that for $\mu(\indexnot)\geq\mu_{0}$, 
\begin{equation}\label{eq:Bmuremest}
\|B_{\indexnot,\rem}(t)\|\leq Ce^{-\b_{\trs}\bt}
\end{equation}
for $t\geq T_{\trs}$, where $C$ only depends on $U,V$, $\b_{\romar}$ and $C_{\trs}$; here we use the fact that $\|T_{\indexnot}^{-1}\|$ and 
$\|T_{\indexnot}\|$ are bounded by numerical constants for $\mu(\indexnot)\geq\mu_{0}$. Similarly, for $\mu(\indexnot)\geq\mu_{0}$ and $t\geq 0$, 
\begin{equation}\label{eq:Gindexnotest}
|G_{\indexnot}|\leq Ce^{-\kappa_{\infty} t}|F_{\indexnot}|,
\end{equation}
where $C$ only depends on $U,V$ and $\b_{\romar}$, and $\mu_{0}$ only depends on $U,V$ and $W$. 

\textbf{Fourth reformulation of the equations; obtaining negative semi-definiteness.}
If the diagonal components of $J_{k}$ are zero, let 
$\bJ_{k}=\bmR_{\indexnot,k}=0$. If the diagonal components of $J_{k}$ are non-zero, let $\bJ_{k}=J_{k}$ and $\bmR_{\indexnot,k}=\mR_{\indexnot,k}$. Finally, let
\[
\bJ:=\diag\{\bJ_{1},\dots,\bJ_{K}\},\ \ \
\bmR_{\indexnot}:=\diag\{\bmR_{\indexnot,1},\dots,\bmR_{\indexnot,K}\},\ \ \
\bB(\indexnot)=\bJ+\bmR_{\indexnot}
\]
and $B_{\indexnot,\rodiff}:=B(\indexnot)-\bB(\indexnot)$; note that $B_{\indexnot,\rodiff}$, $B(\indexnot)$ and $\bB(\indexnot)$ all commute. By construction 
there is a $0<\mu_{0}\in\ro$ (depending only on $U,V,W$ and $\b_{\romar}$) such that for $\mu(\indexnot)\geq\mu_{0}$, 
\[
\|B_{\indexnot,\rodiff}\|\leq\b_{\romar}/4. 
\]
Let 
\[
u(t):=\exp\left(-B_{\indexnot,\rodiff}\bt\right)w(t). 
\]
Then
\begin{equation}\label{eq:dotumu}
\dot{u}=\bB(\indexnot)u+\bB_{\indexnot,\rem}u+H_{\indexnot},
\end{equation}
where
\[
\bB_{\indexnot,\rem}(t):=\exp\left(-B_{\indexnot,\rodiff}\bt\right)B_{\indexnot,\rem}(t)\exp\left(B_{\indexnot,\rodiff}\bt\right),\ \ \
H_{\indexnot}(t):=\exp\left(-B_{\indexnot,\rodiff}\bt\right)G_{\indexnot}(t).
\]
Note that for $\mu(\indexnot)\geq\mu_{0}$, 
\[
\|\bB_{\indexnot,\rem}(t)\|\leq Ce^{-(\b_{\trs}-\b_{\romar}/2)\bt}
\]
for $t\geq T_{\trs}$, where $C$ has the same dependence as in the case of (\ref{eq:Bmuremest}) and $\mu_{0}$ only depends on $U$, $V$, $W$ and 
$\b_{\romar}$. Similarly, if $\mu(\indexnot)\geq\mu_{0}$,
\begin{equation}\label{eq:absHmuest}
|H_{\indexnot}|\leq Ce^{-\kappa_{\infty} t+\b_{\romar}\bt/4}|F_{\indexnot}|,
\end{equation}
where $C$ has the same dependence as in the case of (\ref{eq:Gindexnotest}) and $\mu_{0}$ depends only on $U,V,W$ and $\b_{\romar}$. 

\textbf{Rough estimate of the solutions.}
The main advantage of (\ref{eq:dotumu}) is that the symmetrisation of $\bB(\indexnot)$ is negative semi-definite. Thus
\begin{equation*}
\begin{split}
\frac{d}{dt}|u|^{2} = & \ldr{\dot{u},u}+\ldr{u,\dot{u}}
\leq \ldr{[\bB(\indexnot)+\bB(\indexnot)^{\dagger}]u,u}+2\|\bB_{\indexnot,\rem}\||u|^{2}+2|H_{\indexnot}||u|\\
 \leq & 2\|\bB_{\indexnot,\rem}\||u|^{2}+2|H_{\indexnot}||u|.
\end{split}
\end{equation*}
Define
\[
\sigma_{\indexnot}(t):=-\int_{T_{\trs}}^{t}\|\bB_{\indexnot,\rem}(s)\|ds.
\]
Then
\[
\frac{d}{dt}(e^{2\sigma_{\indexnot}}|u|^{2})\leq 2e^{2\sigma_{\indexnot}}|H_{\indexnot}||u|.
\]
This estimate can be used to deduce that 
\[
e^{\sigma_{\indexnot}(t)}|u(t)|\leq |u(T_{\trs})|+\int_{T_{\trs}}^{t}e^{\sigma_{\indexnot}(s)}|H_{\indexnot}(s)|ds
\]
for $t\geq T_{\trs}$. Assuming $\b_{\romar}\leq\b_{\trs}/2$, there is thus a constant $C$, depending only on $U$, $V$, $\b_{\romar}$, $\b_{\trs}$ 
and $C_{\trs}$, such that 
\[
|u(t)|\leq C|u(T_{\trs})|+C\int_{T_{\trs}}^{t}e^{-\kappa_{\infty} s+\b_{\romar}\bs/4}|F_{\indexnot}(s)|ds
\]
for $t\geq T_{\trs}$ and $\indexnot$ such that $\mu(\indexnot)\geq\mu_{0}$. Here $\mu_{0}$ has the same dependence as in the case of 
(\ref{eq:absHmuest}) and we have appealed to (\ref{eq:absHmuest}). Combining the above definitions, this leads to the estimate
\begin{equation}\label{eq:vroughest}
|v(t)|\leq Ce^{(\kappa_{\infty}+\b_{\romar}/4)\bt}|v(T_{\trs})|+Ce^{\b_{\romar}\bt/4}\int_{T_{\trs}}^{t}e^{\kappa_{\infty}(t-s)+\b_{\romar}\bs/4}|F_{\indexnot}(s)|ds
\end{equation}
for $t\geq T_{\trs}$ and $\mu(\indexnot)\geq\mu_{0}$, where $C$ depends only on $U$, $V$, $\b_{\romar}$, $\b_{\trs}$ and $C_{\trs}$, and $\mu_{0}$
depends only on $U,V,W$ and $\b_{\romar}$. It is of interest to note that this estimate implies
\begin{equation}\label{eq:vkappaindexnotroughest}
|v(t)|\leq Ce^{(\kappa_{\indexnot}+\b_{\romar}/2)\bt}|v(T_{\trs})|+Ce^{\b_{\romar}\bt/2}\int_{T_{\trs}}^{t}e^{\kappa_{\indexnot}(t-s)+\b_{\romar}\bs/2}|F_{\indexnot}(s)|ds
\end{equation}
for $t\geq T_{\trs}$ and $\mu(\indexnot)\geq\mu_{0}$, where $\kappa_{\indexnot}=\kappa_{\max}[A(\indexnot)]$ and $C$ and $\mu_{0}$ have the 
same dependence as in the case of (\ref{eq:vroughest}). For future reference, it is also of interest to keep in mind that 
\begin{equation}\label{eq:wroughest}
|w(t)|\leq Ce^{\b_{\romar}\bt/4}\left[|w(T_{\trs})|+\int_{T_{\trs}}^{t}e^{-\kappa_{\infty} s+\b_{\romar}\bs/4}|F_{\indexnot}(s)|ds\right]
\end{equation}
for $t\geq T_{\trs}$ and $\mu(\indexnot)\geq\mu_{0}$, where $C$ and $\mu_{0}$ have the same dependence as in the case of (\ref{eq:vroughest}).

\textbf{Splitting the equations.}
Let $\mK_{\pm}$ denote the set of real parts of elements of $\Spe (\g_{\pm})$. Then $\kappa_{\infty}$ is the largest element of 
$\mK:=\mK_{+}\cup\mK_{-}$. Given $\b_{\trs}$, there are two possibilities. Either $\kappa_{\infty}-\b_{\trs}\in\mK$ or $\kappa_{\infty}-\b_{\trs}\notin\mK$.
In either case, we let $\kappa_{\min}$ be the smallest element of $\mK$ which is strictly larger than $\kappa_{\infty}-\b_{\trs}$. Note that the
distance between $\kappa_{\min}$ and $\kappa_{\infty}-\b_{\trs}$ has a positive lower bound which only depends on $U,V$ and $\b_{\trs}$. 
Let $\kappa_{\romar}:=\kappa_{\infty}-\kappa_{\min}$. For $k=1,\dots,K$, consider $J_{k}$. If the diagonal component of $J_{k}$ is in the interval 
$[-\kappa_{\romar},0]$, let $J_{k,a}:=J_{k}$, $J_{k,b}:=0$, $\mR_{k,a}:=\mR_{\indexnot,k}$, $\mR_{k,b}:=0$, $\Pi_{k,a}:=\Id_{d_{k}}$ and $\Pi_{k,b}:=0$. Similarly, 
if the diagonal component 
of $J_{k}$ is in the interval $(-\infty,-\b_{\trs}]$, let $J_{k,a}:=0$, $J_{k,b}:=J_{k}$, $\mR_{k,a}:=0$, $\mR_{k,b}:=\mR_{\indexnot,k}$, $\Pi_{k,a}:=0$ 
and $\Pi_{k,b}:=\Id_{d_{k}}$. Let, moreover, 
\begin{equation}\label{eq:Jabtrsdef}
J_{a}:=\diag\{J_{1,a},\dots,J_{K,a}\},\ \ \
J_{b}:=\diag\{J_{1,b},\dots,J_{K,b}\}.
\end{equation}
Define $\mR_{a}$, $\mR_{b}$, $\Pi_{a}$ and $\Pi_{b}$ similarly, and let $B_{\indexnot,a}:=J_{a}+\mR_{a}$ and $B_{\indexnot,b}:=J_{b}+\mR_{b}$. For a vector 
$\xi$ with values in $\cn{2m}$ we similarly divide its components into $\xi_{a},\xi_{b}\in\cn{2m}$ such that $\xi=\xi_{a}+\xi_{b}$; in fact, 
$\xi_{a}=\Pi_{a}(\xi)$ and $\xi_{b}=\Pi_{b}(\xi)$. For future reference, it is of interest to keep in mind that if $t\geq 0$, then 
\begin{equation}\label{eq:emBmuatnormest}
\|e^{-B_{\indexnot,a}t}\|\leq e^{(\kappa_{\romar}+\b_{\romar}/4)t}\leq e^{(\b_{\trs}-\b_{\romar}/2)t},
\end{equation}
where we have appealed to (\ref{eq:Jkkappakest}), and the last inequality is based on the assumption that $\b_{\romar}$ is small enough, 
the bound depending only on $\b_{\trs}$ and $\g_{\pm}$. Similarly, for $t\geq 0$, 
\begin{equation}\label{eq:eBmubtnormest}
\|e^{B_{\indexnot,b}t}\Pi_{b}\|\leq e^{-(\b_{\trs}-\b_{\romar}/2)t}.
\end{equation}
Let us return to (\ref{eq:dotwBmurem}). Given the above terminology, it can be divided into two components
\begin{align}
\dot{w}_{a} = & B_{\indexnot,a}w_{a}+(B_{\indexnot,\rem}w)_{a}+G_{\indexnot,a},\label{eq:wacomp}\\
\dot{w}_{b} = & B_{\indexnot,b}w_{b}+(B_{\indexnot,\rem}w)_{b}+G_{\indexnot,b};\label{eq:wbcomp}
\end{align}
note that while $B_{\indexnot}$ respects the division into components, the matrix $B_{\indexnot,\rem}$ cannot be expected to do so.

\textbf{Analysis of the first equation.} Integrating (\ref{eq:wacomp}) yields 
\begin{equation}\label{eq:ubaintformtrs}
e^{-B_{\indexnot,a}\bt}w_{a}(t)=w_{a}(T_{\trs})+\int_{T_{\trs}}^{t}e^{-B_{\indexnot,a}\bs}(B_{\indexnot,\rem}w)_{a}(s)ds
+\int_{T_{\trs}}^{t}e^{-B_{\indexnot,a}\bs}G_{\indexnot,a}(s)ds.
\end{equation}
Define
\begin{equation}\label{eq:winfadeftrs}
w_{\infty,a}:=w_{a}(T_{\trs})+\int_{T_{\trs}}^{\infty}e^{-B_{\indexnot,a}\bs}(B_{\indexnot,\rem}w)_{a}(s)ds.
\end{equation}
Then (\ref{eq:ubaintformtrs}) implies
\begin{equation}\label{eq:waasestpreltrs}
\begin{split}
 & \left|w_{a}(t)-e^{B_{\indexnot,a}\bt}w_{\infty,a}-\int_{T_{\trs}}^{t}e^{B_{\indexnot,a}(t-s)}G_{\indexnot,a}(s)ds\right|\\
 \leq & \left|\int_{t}^{\infty}e^{-B_{\indexnot,a}(s-t)}(B_{\indexnot,\rem}w)_{a}(s)ds\right|.
\end{split}
\end{equation}
Appealing to (\ref{eq:Bmuremest}), (\ref{eq:wroughest}) and (\ref{eq:emBmuatnormest}) yields
\begin{equation}\label{eq:contrtoconstodeapptrs}
\left|\int_{T_{\trs}}^{\infty}e^{-B_{\indexnot,a}\bs}(B_{\indexnot,\rem}w)_{a}(s)ds\right|
\leq C\left[|w(T_{\trs})|+e^{-\b_{\romar}T_{\trs}/4}\int_{T_{\trs}}^{\infty}e^{-(\kappa_{\infty}-\b_{\romar}/4)s}|F_{\indexnot}(s)|ds\right].
\end{equation}
This estimate holds assuming $\mu(\indexnot)\geq\mu_{0}$, where $\mu_{0}$ only depends on $U,V,W$ and $\b_{\romar}$. Moreover, the constant 
$C$ only depends on $U,V$, $\b_{\romar}$, $\b_{\trs}$ and $C_{\trs}$. Finally, $\b_{\romar}$ has to be small enough, the bound depending
only on $\g_{\pm}$ and $\b_{\trs}$. In the end, we wish to replace $\kappa_{\infty}$ by $\kappa_{\indexnot}$ in (\ref{eq:contrtoconstodeapptrs}).
This can be achieved by replacing $\b_{\romar}/4$ by $\b_{\romar}$ in $e^{-(\kappa_{\infty}-\b_{\romar}/4)s}$. Moreover, it is convenient to introduce the notation
(\ref{eq:Findexnotrolmadef}). Combining (\ref{eq:winfadeftrs}) and (\ref{eq:contrtoconstodeapptrs}) then yields
\begin{equation}\label{eq:winfaesttrs}
|w_{\infty,a}|\leq C[|w(T_{\trs})|+e^{-\b_{\romar}T_{\trs}/4}\|F_{\indexnot}\|_{\rolma}]
\end{equation}
for $\mu(\indexnot)\geq\mu_{0}$, where $C$ and $\mu_{0}$ have the same dependence as in (\ref{eq:contrtoconstodeapptrs}), and 
$\b_{\romar}$ has to satisfy the same bound as in the case of (\ref{eq:contrtoconstodeapptrs}). Returning to (\ref{eq:waasestpreltrs}), we 
need to estimate the right hand side. Appealing to (\ref{eq:Bmuremest}), (\ref{eq:wroughest}) and (\ref{eq:emBmuatnormest}) again yields 
the conclusion that it can be bounded by 
\begin{equation*}
\begin{split}
Ce^{-(\b_{\trs}-\b_{\romar}/4)\bt}[|w(T_{\trs})|+e^{-\b_{\romar}T_{\trs}/4}\|F_{\indexnot}\|_{\rolma}]
\end{split}
\end{equation*}
for $t\geq T_{\trs}$ and $\mu(\indexnot)\geq\mu_{0}$, where $C$ and $\mu_{0}$ have the same dependence as in (\ref{eq:contrtoconstodeapptrs}), and 
$\b_{\romar}$ has to satisfy the same bound as in the case of (\ref{eq:contrtoconstodeapptrs}). Combining this estimate 
with (\ref{eq:waasestpreltrs}) yields 
\begin{equation}\label{eq:wawinfaesttrs}
\begin{split}
 & \left|w_{a}(t)-e^{B_{\indexnot,a}\bt}w_{\infty,a}-\int_{T_{\trs}}^{t}e^{B_{\indexnot,a}(t-s)}G_{\indexnot,a}(s)ds\right|\\
 \leq & Ce^{-(\b_{\trs}-\b_{\romar}/4)\bt}[e^{-\kappa_{\infty} T_{\trs}}|v(T_{\trs})|+e^{-\b_{\romar}T_{\trs}/4}\|F_{\indexnot}\|_{\rolma}]
\end{split}
\end{equation}
for $t\geq T_{\trs}$ and $\mu(\indexnot)\geq\mu_{0}$, where $C$ and $\mu_{0}$ have the same dependence as in (\ref{eq:contrtoconstodeapptrs}), and 
$\b_{\romar}$ has to satisfy the same bound as in the case of (\ref{eq:contrtoconstodeapptrs}).

\textbf{Analysis of the second equation.}
Let us now turn to (\ref{eq:wbcomp}). Integrating this equation yields
\begin{equation}\label{eq:wbintformtrs}
w_{b}(t)=e^{B_{\indexnot,b}\bt}w_{b}(T_{\trs})+\int_{T_{\trs}}^{t}e^{B_{\indexnot,b}(t-s)}(B_{\indexnot,\rem}w)_{b}(s)ds
+\int_{T_{\trs}}^{t}e^{B_{\indexnot,b}(t-s)}G_{\indexnot,b}(s)ds.
\end{equation}
Appealing to (\ref{eq:eBmubtnormest}) yields
\begin{equation}\label{eq:wbconsttermesttrs}
|e^{B_{\indexnot,b}\bt}w_{b}(T_{\trs})|\leq e^{-(\b_{\trs}-\b_{\romar}/2)\bt}|w_{b}(T_{\trs})|.
\end{equation}
Next, (\ref{eq:Bmuremest}), (\ref{eq:wroughest}) and (\ref{eq:eBmubtnormest}) yield
\begin{equation}\label{eq:wbrestintesttrs}
\begin{split}
\left|\int_{T_{\trs}}^{t}e^{B_{\indexnot,b}(t-s)}(B_{\indexnot,\rem}w)_{b}(s)ds\right| 
 \leq & Ce^{-(\b_{\trs}-\b_{\romar}/2)\bt}[|w(T_{\trs})|+e^{-\b_{\romar}T_{\trs}/4}\|F_{\indexnot}\|_{\rolma}]
\end{split}
\end{equation}
for $t\geq T_{\trs}$ and $\mu(\indexnot)\geq\mu_{0}$, where $C$ and $\mu_{0}$ have the same dependence as in (\ref{eq:contrtoconstodeapptrs}), and 
$\b_{\romar}$ has to satisfy the same bound as in the case of (\ref{eq:contrtoconstodeapptrs}). Combining (\ref{eq:wbintformtrs}), 
(\ref{eq:wbconsttermesttrs}) and (\ref{eq:wbrestintesttrs}) yields
\begin{equation}\label{eq:wbwinfbesttrs}
\left|w_{b}(t)-\int_{T_{\trs}}^{t}e^{B_{\indexnot,b}(t-s)}G_{\indexnot,b}(s)ds\right|\leq 
Ce^{-(\b_{\trs}-\b_{\romar}/2)\bt}[|w(T_{\trs})|+e^{-\b_{\romar}T_{\trs}/4}\|F_{\indexnot}\|_{\rolma}]
\end{equation}
for $t\geq T_{\trs}$ and $\mu(\indexnot)\geq\mu_{0}$, where $C$ and $\mu_{0}$ have the same dependence as in (\ref{eq:contrtoconstodeapptrs}), and 
$\b_{\romar}$ has to satisfy the same bound as in the case of (\ref{eq:contrtoconstodeapptrs}).

\textbf{Combining the estimates.} Let $w_{\infty}\in\cn{2m}$ be defined by $w_{\infty}:=w_{\infty,a}+w_{\infty,b}$, where $w_{\infty,b}=0$. Then 
(\ref{eq:wawinfaesttrs}) and (\ref{eq:wbwinfbesttrs}) yield
\begin{equation}\label{eq:wwinfesttrs}
\begin{split}
 & \left|w(t)-e^{B(\indexnot)\bt}w_{\infty}-\int_{T_{\trs}}^{t}e^{B(\indexnot)(t-s)}G_{\indexnot}(s)ds\right|\\
 \leq & Ce^{-(\b_{\trs}-\b_{\romar}/2)\bt}[e^{-\kappa_{\infty} T_{\trs}}|v(T_{\trs})|+e^{-\b_{\romar}T_{\trs}/4}\|F_{\indexnot}\|_{\rolma}]
\end{split}
\end{equation}
for $t\geq T_{\trs}$ and $\mu(\indexnot)\geq\mu_{0}$, where $C$ and $\mu_{0}$ have the same dependence as in (\ref{eq:contrtoconstodeapptrs}), and 
$\b_{\romar}$ has to satisfy the same bound as in the case of (\ref{eq:contrtoconstodeapptrs}). This estimate can be reformulated as
\begin{equation}\label{eq:vvinfesttrs}
\begin{split}
 & \left|v(t)-e^{A(\indexnot)t}v_{\infty}-\int_{T_{\trs}}^{t}e^{A(\indexnot)(t-s)}F_{\indexnot}(s)ds\right|\\
 \leq & Ce^{\kappa_{\infty} t-(\b_{\trs}-\b_{\romar}/2)\bt}[e^{-\kappa_{\infty} T_{\trs}}|v(T_{\trs})|+e^{-\b_{\romar}T_{\trs}/4}\|F_{\indexnot}\|_{\rolma}]
\end{split}
\end{equation}
for $t\geq T_{\trs}$ and $\mu(\indexnot)\geq\mu_{0}$, where $C$ and $\mu_{0}$ have the same dependence as in (\ref{eq:contrtoconstodeapptrs}), and 
$\b_{\romar}$ has to satisfy the same bound as in the case of (\ref{eq:contrtoconstodeapptrs}). Moreover, 
\begin{equation}\label{eq:vinftrspfdef}
v_{\infty}:=e^{\kappa_{\infty} T_{\trs}}e^{-A(\indexnot)T_{\trs}}T_{\indexnot}D^{-1}w_{\infty}. 
\end{equation}
Sacrificing a part of the margin yields (\ref{eq:vvinfesttrsstmt})
for $t\geq T_{\trs}$ and $\mu(\indexnot)\geq\mu_{0}$, where $C$ and $\mu_{0}$ have the same dependence as in (\ref{eq:contrtoconstodeapptrs}), and 
$\b_{\romar}$ has to satisfy the same bound as in the case of (\ref{eq:contrtoconstodeapptrs}). Finally, let 
$u_{\infty}:=e^{\kappa_{\infty} T_{\trs}}T_{\indexnot}D^{-1}w_{\infty}$. Then $u_{\infty}\in E^{\infty}_{\indexnot,\b_{\trs}}$ due to the definition of $w_{\infty}$, $D$ and 
$T_{\indexnot}$. Moreover, the estimate (\ref{eq:winfaesttrs}) yields the conclusion that (\ref{eq:uinfest}) holds. The lemma follows. 
\end{proof}

\section{Specifying data at infinity}\label{section:spdatinftrs}

The present section is quite analogous to Section~\ref{section:spdatinfODE}; the goal is to construct a solution to (\ref{eq:dotvAmu}) corresponding
to the asymptotic data $v_{\infty}$ appearing in (\ref{eq:vvinfesttrsstmt}). However, in the present setting the fact that the matrix
$A(\indexnot)$ appearing in (\ref{eq:dotvAmu}) depends on the mode number $\indexnot$ causes additional complications. Fortunately, it turns out
that these complications can be handled using the methods developed in the previous section. Just as in the case of Section~\ref{section:spdatinfODE},
it is, moreover, sufficient to consider homogeneous equations. 

\begin{lemma}\label{lemma:spastrssett}
Let $1\leq m\in\zo$, $\EFindexset$ be the set introduced in connection with (\ref{eq:IgenReqdef}) and $\mu$ be a function
$\mu:\EFindexset\rightarrow [0,\infty)$. Let $\EFtrsindexset\subseteq\EFindexset$ be the set of $\indexnot\in\EFindexset$ such that 
$\mu(\indexnot)>0$. Let $U,V,W\in\Mn{m}{\co}$ and define $A(\indexnot)$ by (\ref{eq:Aindexnotitomuindexnot}) for $\indexnot\in\EFtrsindexset$.
Assume that for each $\indexnot\in\EFtrsindexset$, there is a function $A_{\indexnot,\rem}\in C^{\infty}(\ro,\Mn{2m}{\co})$ with the property that there 
are constants $0<\b_{\trs},C_{\trs}\in\ro$ (independent of $\indexnot$) and a constant $T_{\trs}$ (which is allowed to depend on $\indexnot$) such 
that (\ref{eq:Amuremesttrscase}) holds for $t\geq T_{\trs}$. Fix a $0<\b_{\romar}\in\ro$ which is small enough, the bound depending only on 
$U,V$ and $\b_{\trs}$. Then there are constants $0<\mu_{0}\in\ro$ and $0< C\in\ro$, where 
$\mu_{0}$ only depends on $U,V,W$ and $\b_{\romar}$, and $C$ only depends on $U,V,\b_{\trs},\b_{\romar}$ and $C_{\trs}$, such that the following holds. If 
$\indexnot\in\mu^{-1}([\mu_{0},\infty))$, there is a linear injective map $\Psi_{\indexnot,\infty}:E^{\infty}_{\indexnot,\b_{\trs}}\rightarrow\cn{2m}$ such that if 
$v_{\infty}\in E^{\infty}_{\indexnot,\b_{\trs}}$ and $v$ is the solution to 
\begin{equation}\label{eq:dotvAmuhom}
\dot{v}=A(\indexnot)v+A_{\indexnot,\rem}v
\end{equation}
with initial data $v(T_{\trs})=\Psi_{\indexnot,\infty}(v_{\infty})$, then $v$ satisfies (\ref{eq:vvinfesttrsstmt})  (with $F_{\indexnot}=0$) for $t\geq T_{\trs}$.
Moreover,
\begin{equation}\label{eq:Psiinfnormtrs}
|\Psi_{\indexnot,\infty}[e^{-A(\indexnot)T_{\trs}}\eta]|\leq C|\eta|
\end{equation}
for $\eta\in E^{\infty}_{\indexnot,\b_{\trs}}$, where $C$ only depends on $U,V,\b_{\romar},\b_{\trs}$ and $C_{\trs}$.
\end{lemma}
\begin{remark}\label{remark:inhomaschartrs}
The result can be used to derive conclusions concerning the inhomogeneous equation (\ref{eq:dotvAmu}). In order to justify this statement, 
assume that $\|F_{\indexnot}\|_{\rolma}<\infty$, where $\b_{\romar}$ and $\indexnot$ are such that the assumptions of Lemmas~\ref{lemma:modebymodetrs} 
and \ref{lemma:spastrssett} are satisfied. Let $v_{\mathrm{part}}$ be the solution to (\ref{eq:dotvAmu}) corresponding to the initial data 
$v_{\mathrm{part}}(T_{\trs})=0$. Let $v_{\partic,\infty}\in E^{\infty}_{\indexnot,\b_{\trs}}$ be such that (\ref{eq:vvinfesttrsstmt}) holds with $v$ and $v_{\infty}$ 
replaced by 
$v_{\mathrm{part}}$ and $v_{\partic,\infty}$ respectively. Given $v_{\infty}\in E^{\infty}_{\indexnot,\b_{\trs}}$, let $v_{\hom}$ be the solution to (\ref{eq:dotvAmuhom}) 
corresponding to initial data $v_{\hom}(T_{\trs})=\Psi_{\indexnot,\infty}(v_{\infty}-v_{\partic,\infty})$. Then $v:=v_{\partic}+v_{\hom}$ solves (\ref{eq:dotvAmu}) and 
satisfies (\ref{eq:vvinfesttrsstmt}). In other words, we are also allowed to specify the asymptotic data $v_{\infty}\in E^{\infty}_{\indexnot,\b_{\trs}}$ in 
the case of the inhomogeneous equation (\ref{eq:dotvAmu}), assuming $\|F_{\indexnot}\|_{\rolma}<\infty$. In order to obtain estimates, note first of 
all that (\ref{eq:vinftyitouinfty}) and (\ref{eq:uinfest}) hold due to  Lemma~\ref{lemma:modebymodetrs}, where $u_{\infty}\in E^{\infty}_{\indexnot,\b_{\trs}}$
is defined by the first equation. There is a similar statement  with $v_{\infty}$ and $u_{\infty}$ replaced by $v_{\mathrm{part},\infty}$ and 
$u_{\mathrm{part},\infty}$. Keeping in mind that $v_{\mathrm{part}}(T_{\trs})=0$, this yields
\begin{equation}\label{eq:vTodeestitoetatrs}
\begin{split}
|v(T_{\trs})| = & |v_{\hom}(T_{\trs})|=|\Psi_{\indexnot,\infty}(v_{\infty}-v_{\partic,\infty})|
=|\Psi_{\indexnot,\infty}[e^{-A(\indexnot)T_{\trs}}(u_{\infty}-u_{\partic,\infty})]|\\
 \leq & C|u_{\infty}-u_{\partic,\infty}|\leq C[|u_{\infty}|+e^{\kappa_{\indexnot}T_{\trs}}\|F_{\indexnot}\|_{\rolma}],
\end{split}
\end{equation}
where we have appealed to (\ref{eq:Psiinfnormtrs}) and the constant $C$ only depends on $U,V,\b_{\romar},\b_{\trs}$ and $C_{\trs}$.
\end{remark}
\begin{remark}
The equality $E^{\infty}_{\indexnot,\b_{\trs}}=\cn{2m}$ holds if $\b_{\trs}$ is strictly larger than the difference between the largest and the 
smallest real parts of elements of $\Spe (U_{+})\cup\Spe(U_{-})$; cf. Definition~\ref{def:Einfindexnotb}.
According to the statement, the map $\Psi_{\indexnot,\infty}$ is then an isomorphism. This means that the solution is uniquely 
specified in terms of the ``data at infinity'', $v_{\infty}$. Due to Remark~\ref{remark:inhomaschartrs}, the same holds for 
inhomogeneous equations such that $\|F_{\indexnot}\|_{\rolma}<\infty$. Finally, $|v(T_{\trs})|$ can be estimated in terms of the asymptotic data 
and $F_{\indexnot}$; cf. (\ref{eq:vTodeestitoetatrs}). 
\end{remark}
\begin{proof}
The argument is based on the proof of Lemma~\ref{lemma:modebymodetrs}. We therefore use the notation introduced in that proof without further comment.
A natural starting point is the third reformulation of the equations, as given in the proof of Lemma~\ref{lemma:modebymodetrs}. Let
\[
\tw(t):=e^{-B_{\indexnot,a}\bt}w(t).
\]
Then (\ref{eq:dotwBmurem}) with $G_{\indexnot}=0$ implies
\begin{equation}\label{eq:dottwtrs}
\dot{\tw}=B_{\indexnot,b}\tw+\tB_{\indexnot,\rem}\tw,
\end{equation}
where
\[
\tB_{\indexnot,\rem}(t):=e^{-B_{\indexnot,a}\bt}B_{\indexnot,\rem}(t)e^{B_{\indexnot,a}\bt}.
\]
Due to (\ref{eq:Jkkappakest}), (\ref{eq:Bmuremest}) and the definition of $B_{\indexnot,a}$, 
\[
\|\tB_{\indexnot,\rem}(t)\|\leq Ce^{-(\b_{\trs}-\kappa_{\romar}-\b_{\romar}/2)\bt},
\]
where $C$ only depends on $U,V,\b_{\romar}$ and $C_{\trs}$. However, to obtain this estimate, we need to demand that $\mu(\indexnot)\geq\mu_{0}$, 
where $\mu_{0}$ only depends on $U,V,W$ and $\b_{\romar}$. Note that $\b_{\trs}-\kappa_{\romar}>0$ and that $\kappa_{\romar}$ is determined by the
eigenvalues of $U_{\pm}$ (or, equivalently, $\g_{\pm}$) and $\b_{\trs}$. Assuming $\b_{\romar}$ to be small enough, the bound depending only on $U,V$ 
and $\b_{\trs}$, we thus have
\begin{equation}\label{eq:tBindexnotest}
\|\tB_{\indexnot,\rem}(t)\|\leq Ce^{-(\b_{\trs}-\kappa_{\romar})\bt/2},
\end{equation}
where $C$ only depends on $U,V,\b_{\romar}$ and $C_{\trs}$. Note also that the symmetrisation of $B_{\indexnot,b}$ is negative semi-definite. 
Let $t_{0}\geq T_{\trs}$. By an argument similar to the derivation of (\ref{eq:absuthomest}), 
\begin{equation}\label{eq:absuthomesttrs}
|\tw(t)|\leq\exp\left(\int_{t_{0}}^{t}\|\tB_{\indexnot,\rem}(s)\|ds\right)|\tw(t_{0})|
\end{equation}
for $t\geq t_{0}$. Note, in particular, that by choosing $t_{0}-T_{\trs}$ to be large enough (the bound depending only on $U,V,\b_{\romar},\b_{\trs}$ and 
$C_{\trs}$), the first factor on the right hand side of (\ref{eq:absuthomesttrs}) can be assumed to be as close to $1$ as we wish. Let $w_{a}$ and 
$w_{b}$ be defined as in the proof of Lemma~\ref{lemma:modebymodetrs} and define 
\begin{equation}\label{eq:twatwbdef}
\tw_{a}(t):=e^{-B_{\indexnot,a}\bt}w_{a}(t),\ \ \
\tw_{b}(t):=e^{-B_{\indexnot,a}\bt}w_{b}(t).
\end{equation}
Note that $\tw_{b}=w_{b}$. Then appealing to (\ref{eq:wacomp}) with $G_{\indexnot,a}=0$ yields
\[
\dot{\tw}_{a}=e^{-B_{\indexnot,a}\bt}(B_{\indexnot,\rem}w)_{a}=(e^{-B_{\indexnot,a}\bt}B_{\indexnot,\rem}w)_{a}
=(\tB_{\indexnot,\rem}\tw)_{a}.
\]
Thus
\begin{equation}\label{eq:emlibuitrs}
\tw_{a}(t)=\tw_{a}(t_{0})+\int_{t_{0}}^{t}(\tB_{\indexnot,\rem}\tw)_{a}(s)ds,
\end{equation}
so that 
\[
\left|\tw_{a}(t)-\tw_{a}(t_{0})\right|\leq C\int_{t_{0}}^{t}e^{-(\b_{\trs}-\kappa_{\romar})\bs/2}ds
\exp\left(\int_{t_{0}}^{\infty}\|\tB_{\indexnot,\rem}(s)\|ds\right)|\tw(t_{0})|
\]
for $t\geq t_{0}$, where we have appealed to (\ref{eq:tBindexnotest}) and (\ref{eq:absuthomesttrs}) and $C$ only depends on $U,V,\b_{\romar}$
and $C_{\trs}$. For any $\e>0$, it is thus possible to choose $t_{0}$ so that 
\begin{equation}\label{eq:buicontrtrs}
\left|\tw_{a}(t)-\tw_{a}(t_{0})\right|\leq \e |\tw(t_{0})|
\end{equation}
for $t\geq t_{0}$ (where $\bt_{0}:=t_{0}-T_{\trs}$ only depends on $U,V,\b_{\romar},\b_{\trs},C_{\trs}$ and $\e$). Before proceeding, let 
$V_{a}:=\Pi_{a}(\cn{2m})$ and $V_{b}:=\Pi_{b}(\cn{2m})$, where $\Pi_{a}$ and $\Pi_{b}$ are introduced in connection with (\ref{eq:Jabtrsdef}). 
Given a $t_{0}$ such that (\ref{eq:buicontrtrs}) holds with $\e$ replaced by $1/2$, define the map $L_{a}:V_{a}\rightarrow V_{a}$ as follows. 
Given $\eta\in V_{a}$, let $\tw$ be the solution to (\ref{eq:dottwtrs}) with $\tw(t_{0})=\eta$ and define $L_{a}\eta$ by 
\[
L_{a}\eta:=\lim_{t\rightarrow\infty}\tw_{a}(t).
\]
Due to (\ref{eq:tBindexnotest}), (\ref{eq:absuthomesttrs}) and (\ref{eq:emlibuitrs}), it is clear that the limit on the right hand side exists. 
Thus $L_{a}\eta$ is well defined. Moreover, it is clear that $L_{a}$ is linear. Due to the fact that (\ref{eq:buicontrtrs}) holds with $\e$ replaced 
by $1/2$ and the fact that $\tw_{b}(t_{0})=0$,
\[
\left|L_{a}\eta-\tw_{a}(t_{0})\right|\leq \frac{1}{2}|\tw_{a}(t_{0})|. 
\]
In particular, $|\eta|/2\leq |L_{a}\eta|$, so that $L_{a}$ is injective. Since $L_{a}$ is a linear map between spaces of the same finite dimension, it 
follows that $L_{a}$ is an isomorphism. Moreover, 
\begin{equation}\label{eq:Lainvesttrs}
|L_{a}^{-1}\eta|\leq 2|\eta|.
\end{equation}
Note that $L_{a}^{-1}$ (essentially) maps data at infinity to data at $t_{0}$. In the end, we wish to translate the data at $t_{0}$ to data at $T_{\trs}$. 
To this end, we wish to solve (\ref{eq:dottwtrs}) backwards. Since $\bt_{0}$ can be bounded by a constant depending only on 
$U,V,\b_{\romar},\b_{\trs}$ and $C_{\trs}$, there is a constant $C$ (with the same dependence) such that for all solutions $\tw$ to 
(\ref{eq:dottwtrs})
\begin{equation}\label{eq:butodetzesttrs}
|\tw(T_{\trs})|\leq C|\tw(t_{0})|.
\end{equation}
Before defining the map $\Psi_{\indexnot,\infty}$ appearing in the statement of the lemma, we need to clarify how the limit of $\tw_{a}(t)$ is related to the 
$v_{\infty}\in E^{\infty}_{\indexnot,\b_{\trs}}$ appearing in (\ref{eq:vvinfesttrsstmt}). However, returning to (\ref{eq:ubaintformtrs}) and (\ref{eq:winfadeftrs}) 
(and recalling that $G_{\indexnot,a}=0$ in the present setting), it is clear that
\begin{equation}\label{eq:bualimitowinfatrs}
\lim_{t\rightarrow\infty}\tw_{a}(t)=\lim_{t\rightarrow\infty}e^{-B_{\indexnot,a}\bt}w_{a}(t)=w_{\infty,a}.
\end{equation}
On the other hand, (\ref{eq:vinftrspfdef}) and the fact that $w_{\infty}=w_{\infty,a}$ imply that
\[
v_{\infty}=e^{\kappa_{\infty} T_{\trs}}e^{-A(\indexnot)T_{\trs}}T_{\indexnot}D^{-1}w_{\infty,a}. 
\]
Note also that for each $v_{\infty}\in E^{\infty}_{\indexnot,\b_{\trs}}$, there is a unique $\eta\in V_{a}$ such that 
\begin{equation}\label{eq:etavinfreltrs}
v_{\infty}=e^{\kappa_{\infty} T_{\trs}}e^{-A(\indexnot)T_{\trs}}T_{\indexnot}D^{-1}\eta
\end{equation}
holds. Define $\Psi_{\indexnot,\infty}(v_{\infty})$ to be the composition of the following maps: first the map which takes 
$v_{\infty}\in E^{\infty}_{\indexnot,\b_{\trs}}$ to the $\eta\in V_{a}$ such that (\ref{eq:etavinfreltrs}) holds; 
second, the map which takes $\eta$ to $\xi:=L_{a}^{-1}(\eta)$; and, third, the map which takes $\xi\in V_{a}$ to 
\begin{equation}\label{eq:vzdeftrs}
v_{0}:=e^{\kappa_{\infty} T_{\trs}}T_{\indexnot}D^{-1}\tw(T_{\trs}),
\end{equation}
where $\tw$ is the solution to (\ref{eq:dottwtrs}) with initial data $\tw(t_{0})=\xi$. Note that $\Psi_{\indexnot,\infty}$ is linear and injective, and that 
for $\eta\in E^{\infty}_{\indexnot,\b_{\trs}}$
\[
|\Psi_{\indexnot,\infty}[e^{-A(\indexnot)T_{\trs}}\eta]|\leq C|\eta|,
\]
where $C$ only depends on $U,V,\b_{\romar},\b_{\trs}$ and $C_{\trs}$, and we have appealed to (\ref{eq:Lainvesttrs}), (\ref{eq:butodetzesttrs}),
(\ref{eq:etavinfreltrs}) and (\ref{eq:vzdeftrs}). Thus (\ref{eq:Psiinfnormtrs}) holds. 

Assume now that $v$ is the solution to (\ref{eq:dotvAmuhom}) corresponding to $v(T_{\trs})=\Psi_{\indexnot,\infty}(v_{\infty})$ for some 
$v_{\infty}\in E^{\infty}_{\indexnot,\b_{\trs}}$. Then $v$ satisfies (\ref{eq:vvinfesttrsstmt}) with $F_{\indexnot}=0$ and $v_{\infty}$ replaced by some 
$\bv_{\infty}\in E^{\infty}_{\indexnot,\b_{\trs}}$. Moreover, the proof of Lemma~\ref{lemma:modebymodetrs}, in  particular (\ref{eq:vinftrspfdef}), yields
\[
\bv_{\infty}:=e^{\kappa_{\infty} T_{\trs}}e^{-A(\indexnot)T_{\trs}}T_{\indexnot}D^{-1}\bw_{\infty},
\]
where $\bw_{\infty}\in V_{a}$. In addition, 
\[
\lim_{t\rightarrow\infty}\tw_{a}(t)=\bw_{\infty},
\]
where $\tw_{a}$ is defined by (\ref{eq:twatwbdef}). This means that 
\[
L_{a}^{-1}(\bw_{\infty})=\tw(t_{0}),
\]
due to the definition of $L_{a}$ and the fact that $\tw(t_{0})=\tw_{a}(t_{0})$. Going through the steps defining $\Psi_{\indexnot,\infty}$, it is thus 
clear that $\Psi_{\indexnot,\infty}(\bv_{\infty})=v(T_{\trs})=\Psi_{\indexnot,\infty}(v_{\infty})$. Due to the injectivity of $\Psi_{\indexnot,\infty}$, it follows 
that $v_{\infty}=\bv_{\infty}$. The lemma follows. 
\end{proof}

\section{Algebraic properties of $A(\indexnot)$}

As a final section of the present chapter, it is convenient to collect some of the algebraic properties of $A(\indexnot)$ that we need
in what follows. 

\begin{lemma}\label{lemma:expAintrestetaest}
Let $1\leq m\in\zo$, $\EFindexset$ be the set introduced in connection with (\ref{eq:IgenReqdef}) and $\mu$ be a function
$\mu:\EFindexset\rightarrow [0,\infty)$. Let $\EFtrsindexset\subseteq\EFindexset$ be the set of $\indexnot\in\EFindexset$ such that 
$\mu(\indexnot)>0$. Let $U,V,W\in\Mn{m}{\co}$ and define $A(\indexnot)$ by (\ref{eq:Aindexnotitomuindexnot}) for $\indexnot\in\EFtrsindexset$.
Let $\b>0$. Then there are constants $C$ and $\mu_{0}>0$ (depending only on $U$, $V$, $W$ and $\b$) such that 
\[
|e^{-A(\indexnot)t}\eta|\leq Ce^{-(\kappa_{\indexnot}-\b)t}|\eta|
\] 
for all $\indexnot$ such that $\mu(\indexnot)\geq\mu_{0}$; all $\eta\in E^{\infty}_{\indexnot,\b}$; and all $t\geq 0$. Using the notation introduced in
Definition~\ref{def:Einfindexnotb}, let $e^{\infty}_{\indexnot,\b}$ be the direct sum of the generalised eigenspaces of $A(\indexnot)$ corresponding to 
eigenvalues in the balls of radius $r_{a}$ and centre $\pm i\mu(\indexnot)+\lambda_{j,\pm}$ for $j$ such that $\kappa_{\infty}-\mathrm{Re}(\lambda_{j,\pm})\geq \b$.
Given $\e>0$, there are constants $C$ and $\mu_{0}>0$ (depending only on $U$, $V$, $W$ and $\e$) such that 
\[
|e^{A(\indexnot)t}\eta|\leq Ce^{(\kappa_{b}+\e)t}|\eta|
\]
for all $\indexnot$ such that $\mu(\indexnot)\geq\mu_{0}$; all $\eta\in e^{\infty}_{\indexnot,\b}$; and all $t\geq 0$, where $\kappa_{b}$ is the largest 
element of $\{\mathrm{Re}(\lambda_{j,\pm})\}$ which is $\leq \kappa_{\infty}-\b$. 
\end{lemma}
\begin{proof}
Let $\mu_{0}>0$ be larger than the $\mu_{0}$ appearing in Definition~\ref{def:Einfindexnotb} and let $\indexnot$ be such that $\mu(\indexnot)\geq\mu_{0}$,
$\eta\in E^{\infty}_{\indexnot,\b}$ and $t\geq 0$. Keeping the change of variables introduced at the beginning of the proof of 
Lemma~\ref{lemma:modebymodetrs} in mind, 
\begin{equation}\label{eq:vinftybAexprpre}
e^{-A(\indexnot)t}\eta=T_{\indexnot}T_{\indexnot}^{-1}e^{-A(\indexnot)t}T_{\indexnot}T_{\indexnot}^{-1}\eta
=T_{\indexnot}e^{-\bA(\indexnot)t}T_{\indexnot}^{-1}\eta,
\end{equation}
where 
\begin{equation}\label{eq:barAindexnotdiagpre}
\bar{A}(\indexnot)=\diag\{N_{\indexnot,1}^{-},\dots,N_{\indexnot,p_{-}}^{-},N_{\indexnot,1}^{+},\dots,N_{\indexnot,p_{+}}^{+}\};
\end{equation}
cf. (\ref{eq:Tmudiag}). The blocks appearing on the right hand side of (\ref{eq:barAindexnotdiagpre}) are given by 
\begin{equation}\label{eq:Nmujpmexprinpre}
N_{\indexnot,j}^{\pm}=\pm i\mu(\indexnot) \Id_{m_{j,\pm}}+\g_{\pm,j}+\mR_{\indexnot,j}^{\pm},
\end{equation}
where 
\begin{equation}\label{eq:mRmujpmestinpre}
\|\mR_{\indexnot,j}^{\pm}\|\leq c_{0}[\mu(\indexnot)]^{-1}
\end{equation}
for $\mu(\indexnot)\geq\mu_{0}$, where $c_{0}$ and $\mu_{0}$ only depend on $U$, $V$ and $W$.
Recall that the notation $\g_{\pm,j}$ is introduced in Lemma~\ref{lemma:aspardiag}. The matrices $\g_{\pm,j}$ are defined in terms of 
$\g_{\pm}$, which are defined in terms of $U$ and $V$. Recall, finally, that $\g_{\pm,j}$
consists of the Jordan blocks with diagonal components $\lambda_{j,\pm}$. The blocks appearing on the right hand side of (\ref{eq:barAindexnotdiagpre}) 
correspond to a division of the generalised eigenspaces of $A(\indexnot)$; cf. Definition~\ref{def:Einfindexnotb}. The blocks of 
greatest interest are the ones corresponding to eigenvalues that belong to the sets described in Definition~\ref{def:Einfindexnotb}, 
since they correspond to the only non-zero components of $T_{\indexnot}^{-1}\eta$. Considering (\ref{eq:vinftybAexprpre}), it is thus 
clear that we only need to estimate 
\[
\|T_{\indexnot}\|\cdot\|T_{\indexnot}^{-1}\|\cdot\|\exp(-N_{\indexnot,j}^{\pm}t)\|\cdot|\eta|
\]
in the case that the real parts of the diagonal components of $\g_{\pm,j}$ are strictly larger than $\kappa_{\infty}-\b$, where the notation is 
taken from Definition~\ref{def:Einfindexnotb}. Note that 
\begin{equation}\label{eq:expmNinjpmTtrsnestpre}
\begin{split}
\exp(-N_{\indexnot,j}^{\pm}t) = & \exp[\mp i\mu(\indexnot)t-i\mathrm{Im}(\lambda_{j,\pm})t]
\exp[-\mathrm{Re}(\lambda_{j,\pm})t]\\
 & \cdot\exp[-(\g_{\pm,j}-\lambda_{j,\pm}\Id_{m_{j,\pm}}+\mR_{\indexnot,j}^{\pm})t].
\end{split}
\end{equation}
Note that the first factor on the right hand side has absolute value one, and that 
\begin{equation}\label{eq:ReljpmTtrsestpre}
\exp[-\mathrm{Re}(\lambda_{j,\pm})t]\leq e^{-(\kappa_{\infty}-\b)t}.
\end{equation}
In fact, the latter estimate can be improved; since there are only finitely many eigenvalues $\lambda_{j,\pm}$ and 
$\mathrm{Re}(\lambda_{j,\pm})>\kappa_{\infty}-\b$ for all the eigenvalues of relevance here, there is a gap between the smallest
$\mathrm{Re}(\lambda_{j,\pm})$ satisfying $\mathrm{Re}(\lambda_{j,\pm})>\kappa_{\infty}-\b$, and $\kappa_{\infty}-\b$ (and the gap only depends on 
$\b$, $U$ and $V$). 
What remains to estimate is the norm of the last factor on the right hand side of (\ref{eq:expmNinjpmTtrsnestpre}). In order to 
simplify the notation, note that this factor can be written $\exp[-(N+R)t]$, where $N$ consists of Jordan blocks corresponding
to the eigenvalue zero (in other words, the only potentially non-zero elements in $N$ are ones immediately above the diagonal) and 
$\|R\|$ is bounded by the right hand side of (\ref{eq:mRmujpmestinpre}). Given $\eta_{\romar}>0$ it is possible to conjugate $N$ by 
a diagonal matrix, say $S_{\romar}\in\Gl{m_{j,\pm}}{\ro}$, such that $\|S_{\romar}NS_{\romar}^{-1}\|\leq\eta_{\romar}/2$; cf. Lemma~\ref{lemma:genJordblock}
and its proof. The maximal element of 
$\{\|S_{\romar}\|,\|S_{\romar}^{-1}\|\}$ then depends only on $\eta_{\romar}$. Given such an $\eta_{\romar}$, there is then a $\mu_{0}>0$,
depending only on $\eta_{\romar}$, $U$, $V$ and $W$, such that if $\mu(\indexnot)\geq\mu_{0}$, then 
$\|S_{\romar}RS_{\romar}^{-1}\|\leq\eta_{\romar}/2$. Thus
\[
\|\exp[-(N+R)t]\|\leq \|S_{\romar}^{-1}\exp[-S_{\romar}(N+R)S_{\romar}^{-1}t]S_{\romar}\|\leq 
Ce^{\eta_{\romar}t},
\]
where the constant only depends on $\eta_{\romar}$. Combining this estimate with (\ref{eq:expmNinjpmTtrsnestpre}), (\ref{eq:ReljpmTtrsestpre})
and the comments made in connection with (\ref{eq:ReljpmTtrsestpre}), it is clear that by first fixing $\eta_{\romar}>0$ small enough
(in fact, we specify $\eta_{\romar}$ in terms of $U$, $V$ and $\b$) and then requiring
$\mu_{0}$ to be large enough, the bound depending only on $U$, $V$, $W$ and $\b$, we can ensure that
\[
\|\exp(-N_{\indexnot,j}^{\pm}t)\|\leq Ce^{-(\kappa_{\indexnot}-\b)t},
\]
where $C$ only depends on $U$, $V$, $W$ and $\b$. In order to obtain this conclusion, 
we have also appealed to the fact that $|\kappa_{\infty}-\kappa_{\indexnot}|\leq c_{0}[\mu(\indexnot)]^{1/m}$, where $c_{0}$ only
depends on $U$, $V$ and $W$; cf. Lemma~\ref{lemma:Nmueigenvapprox}. Combining this estimate with previous observations yields
the first statement of the lemma. The proof of the second statement is similar.
\end{proof}

\begin{lemma}\label{lemma:genepmaintest}
Let $1\leq m\in\zo$, $\EFindexset$ be the set introduced in connection with (\ref{eq:IgenReqdef}) and $\mu$ be a function
$\mu:\EFindexset\rightarrow [0,\infty)$. Let $\EFtrsindexset\subseteq\EFindexset$ be the set of $\indexnot\in\EFindexset$ such that 
$\mu(\indexnot)>0$. Let $U,V,W\in\Mn{m}{\co}$ and define $A(\indexnot)$ by (\ref{eq:Aindexnotitomuindexnot}) for $\indexnot\in\EFtrsindexset$.
Let $\e>0$. Then there are constants $C$ and $\mu_{0}>0$ (depending only on $U$, $V$, $W$ and $\e$) such that 
\begin{align}
\|e^{-A(\indexnot)t}\| \leq & Ce^{-(\kappa_{\min,\indexnot}-\e)t},\label{eq:expmaintnoest}\\
\|e^{A(\indexnot)t}\| \leq & Ce^{(\kappa_{\indexnot}+\e)t}\label{eq:exppaintnoest}
\end{align}
for all $\indexnot$ such that $\mu(\indexnot)\geq\mu_{0}$; all $\eta\in \cn{2m}$; and all $t\geq 0$. Here 
$\kappa_{\indexnot}:=\kappa_{\max}[A(\indexnot)]$ and $\kappa_{\min,\indexnot}:=\kappa_{\min}[A(\indexnot)]$. 
\end{lemma}
\begin{remark}\label{remark:Upmsuffinest}
In (\ref{eq:expmaintnoest}) and (\ref{eq:exppaintnoest}), $\kappa_{\min,\indexnot}$ and $\kappa_{\indexnot}$ can be replaced by 
$\kappa_{b}$ and $\kappa_{a}$ respectively, where $\kappa_{a}$ ($\kappa_{b}$) is the largest (smallest) real part of an element of 
$\Spe (U_{+})\cup\Spe(U_{-})$; cf. the notation introduced in Definition~\ref{def:Einfindexnotb}.
\end{remark}
\begin{proof}
The proof is based on arguments similar to those of the proof of Lemma~\ref{lemma:expAintrestetaest}. We therefore 
leave the details to the reader. 
\end{proof}

\chapter[Asymptotics, transparent case]{Asymptotics for weakly transparent, balanced and convergent 
equations}\label{chapter:asympttranspcase}

\section{Introduction}\label{section:introweaktrsmainresults}

In the present chapter, we consider the asymptotics of solutions to so-called weakly transparent, balanced and convergent equations;
cf. Definition~\ref{def:weaktrans} below for a clarification of the terminology. 
The arguments and results are quite analogous to those of Chapter~\ref{chapter:weaksil}; in particular, we derive asymptotics and 
provide a partial characterisation of solutions in terms of asymptotic data. The intuitive reason why we refer to the equations studied 
here as transparent is that the spatial coordinate of a causal curve no-longer needs to converge; cf. the discussion in 
Subsection~\ref{ssection:siltrsanoimet}. The relevance of the notions of balance and convergence is the same in the present setting
as in the case of Chapter~\ref{chapter:weaksil}. The main steps of the analysis are the following. 

\textbf{Silent and transparent variables.} One important difference when comparing the present setting with silent equations
is that for transparent equations, it is natural to divide the spatial variables into silent and transparent variables; cf. 
Definitions~\ref{def:Xnontrtrsdivvar} and \ref{def:degtrsdivvar} below. Roughly speaking, the silent variables correspond to the 
subspace (of the cotangent space of $\bM$) on which the spatial part of the inverse metric degenerates as $t\rightarrow\infty$, and 
the transparent variables correspond to the subspace on which the spatial part of the inverse metric converges to a positive definite 
bilinear form as $t\rightarrow\infty$. Projecting a solution onto the modes that correspond to dependence only on the silent variables 
leads to a situation analogous to that considered in Chapter~\ref{chapter:weaksil}. For that reason we here focus on the situation that 
results when projecting onto the modes that correspond to dependence on at least one of the transparent variables. 

\textbf{Defining the notion of weak transparency.} Before proceeding to the analysis of the asymptotics, we have to give a precise meaning 
to the notion of weak transparency, and this is done in Definition~\ref{def:weaktrans} below. The first criterion is that one of 
Definitions~\ref{def:Xnontriv} and \ref{def:Xdegenerate} hold. The reason for making this assumption is discussed in the introduction of
Chapter~\ref{chapter:ODEtransp}. The function $\mfg^{2}(\indexnot,t)$ can then be divided into three parts; one corresponding to the silent
variables, one corresponding to the transparent variables and one corresponding to a mixture. The one corresponding to the silent variables 
should satisfy a condition similar to (\ref{eq:weaksil}). In fact, we demand that (\ref{eq:weaksiltrs}) hold. Concerning the part of 
$\mfg^{2}(\indexnot,t)$ corresponding to the transparent variables we demand that it converge exponentially; this is one consequence of 
(\ref{eq:weaktrs}). However, the estimate (\ref{eq:weaktrs}) also implies that $g^{0j}$ converges to zero exponentially if $j$ corresponds 
to a transparent variable. Concerning the mixture, it turns out that it can be considered to be an error term. Finally, we require the $X^{j}$ to 
converge exponentially when $j$ corresponds to a transparent variable; cf. (\ref{eq:Xjtrsrocon}). To summarise, it is clear that due
to the fact that there is a division of the variables and that the asymptotic behaviour of the inverse metric is qualitatively different on 
the corresponding subspaces of the cotangent space, the definition of weak transparency is somewhat more involved than the definition of 
weak silence. It also requires more of an effort to arrive at (\ref{eq:vdottrscase})--(\ref{eq:Aremtrsest}), in particular to calculate the 
exact dependence of the constants $C_{\trs}$ and $\b_{\trs}$ on the constants and functions appearing in the definition of weak transparency,
balance and convergence. We derive the estimates we need in Section~\ref{section:weasilbalcontrs}. 

\textbf{Analysis prior to the transparent era.} In Chapter~\ref{chapter:ODEtransp} we obtain conclusions starting from the assumption that 
(\ref{eq:vdottrscase})--(\ref{eq:Aremtrsest}) hold. However, this analysis is only valid at late times; i.e., for $t\geq T_{\trs}$, where 
$T_{\trs}$ depends on the mode under consideration. Here we need estimates for all $t\geq 0$. We therefore have to take the step from $t=0$ to 
$t=T_{\trs}$. This is achieved by appealing to Lemma~\ref{lemma:roughenestbalsetting}. The proof of this lemma is very simple. On the other hand, 
the corresponding part of the argument is the source of the greatest weakness of the results. The reason for this is that we only obtain a very 
rough estimate of the number of derivatives lost in the estimates of the asymptotics. 

\textbf{Deriving and specifying asymptotics.} Finally, we both derive and specify asymptotics; cf. Sections~\ref{section:derastrs} and 
\ref{section:specastrs}. Partly, the presentation parallels that of Sections~\ref{section:roughODEfutas} and \ref{section:roughODEspecas}.
However, there are two fundamental differences. One is that in the case of silent equations, all the modes have the same asymptotic 
behaviour. This is not true for transparent equations. In fact, if $v$ is a solution to (\ref{eq:vdottrscase}) and $F_{\indexnot}$ satisfies
appropriate bounds, we expect $v$ to behave asymptotically as $\ldr{t}^{\de_{\indexnot}-1}e^{\kappa_{\indexnot}t}$, where 
$\kappa_{\indexnot}:=\kappa_{\max}[A(\indexnot)]$ and $\de_{\indexnot}:=d_{\max}[A(\indexnot),\kappa_{\indexnot}]$; cf. Definition~\ref{def:SpRspdef}. 
This leads to a problem when estimating the difference between a solution to (\ref{eq:thesystemRge}) and a solution to the limit equation 
(obtained by replacing the coefficients (but not the right hand side) of (\ref{eq:thesystemRge}) by their limits as $t\rightarrow\infty$). In 
fact, to say that the difference satisfies a given bound says more for some modes than for others. It is therefore natural to apply an operator 
which multiplies the Fourier coefficient corresponding to a mode $\indexnot$ with $\ldr{t}^{-\de_{\indexnot}+1}e^{-\kappa_{\indexnot}t}$ before estimating 
the difference with respect to a Sobolev norm. A related issue is that of the optimal size of the difference. In 
Section~\ref{section:roughODEfutas} we obtain (\ref{eq:uudothsest}). Normalising this estimate as above corresponds to multiplying both sides
with $\ldr{t}^{-d_{1}+1}e^{-\kappa_{1}t}$, using the terminology introduced in Lemma~\ref{lemma:roughas}. The time dependent part of the right hand 
side then takes the form of a polynomial times $e^{-\b_{\rem}t}$. It would be optimistic to hope for more than this (though it is of course possible
to discuss the optimal degree of the polynomial). In the transparent setting, it is even optimistic to hope for a time dependence on the right 
hand side of the form $\ldr{t}^{N}e^{-\b_{\trs}t}$, where $\b_{\trs}$ is the constant appearing in (\ref{eq:Aremtrsest}); for weakly transparent, 
balanced and convergent equations, $\b_{\trs}$ is defined by (\ref{eq:btrsdef}) below. The reason for this is that we have to handle an infinite number 
of matrices $A(\indexnot)$. This leads to complications that make it natural to replace the polynomial by a factor of the form $e^{\b_{\romar}t}$. On the 
other hand, $\b_{\romar}>0$ can be chosen freely. However, the constants appearing in the estimates depend on $\b_{\romar}$, and they can be 
expected to blow up as $\b_{\romar}\rightarrow 0$. A second difference between the silent and transparent settings is that we here have
to divide the analysis into two parts. First we need to consider the modes for which $\mfg_{\infty}(\indexnot)$, introduced in (\ref{eq:mfginftydef}), 
is large. 
The analysis in that case is based on Sections~\ref{section:mbmanhftrs} and \ref{section:spdatinftrs}. How large $\mfg_{\infty}(\indexnot)$ has
to be depends, among other things, on $\b_{\romar}$. Once the lower bound, say $\mu_{0}$, on $\mfg_{\infty}(\indexnot)$ has been fixed, it remains
to consider the modes satisfying $0<\mfg_{\infty}(\indexnot)<\mu_{0}$ (the modes satisfying $\mfg_{\infty}(\indexnot)=0$ correspond to the 
silent part of the equation and can be analysed as in Chapter~\ref{chapter:weaksil}). However, for these modes, we can appeal to the 
analysis in Chapter~\ref{chapter:roughanalysisODEregion}. The reason this is possible is that our definition of weak transparency is such that 
there are only a finite number of matrices $A(\indexnot)$ for $\indexnot$ satisfying $0<\mfg_{\infty}(\indexnot)<\mu_{0}$. 

\textbf{Geometric perspective.} In the present chapter, we derive conclusions based on assumptions that are purely analytic in nature. 
However, the relevant assumptions follow from quite natural geometric conditions. In Chapter~\ref{chapter:transpeq}, we describe the corresponding
geometric perspective, and the reader is encouraged to read that chapter in parallel with the present one.

\section{Weak transparency, balance and convergence}\label{section:weasilbalcontrs}

As indicated in the previous section, several of the arguments and definitions of the present chapter are based on a division of the 
spatial variables into two groups. We therefore begin by defining this division. In the present chapter we only do so
in case (\ref{eq:thesystemRge}) is an $X$-non-trivial weakly transparent equation
or an $X$-degenerate weakly transparent equation; cf. Definitions~\ref{def:Xnontriv} and \ref{def:Xdegenerate}. 

\begin{definition}\label{def:Xnontrtrsdivvar}
Assume that (\ref{eq:thesystemRge}) is an $X$-non-trivial weakly transparent equation.
Assume, moreover, that the integer $j\in \{1,\dots,d\}$ has the properties stated in Definition~\ref{def:Xnontriv} and that $\bM$ is 
given by (\ref{eq:bMdef}). Then the \textit{silent variables} 
\index{Silent!variables}%
\index{Variables!silent}%
are all the variables in $\bM$ except the one corresponding to 
the $j$'th $\sn{1}$-factor in $\tn{d}$. The \textit{transparent variables} 
\index{Transparent!variables}%
\index{Variables!transparent}%
are the variable corresponding to the $j$'th $\sn{1}$-factor 
in $\tn{d}$. Finally, define $d_{\rosil}:=d-1$; $d_{\trs}:=1$; $R_{\rosil}:=R$; $R_{\trs}:=0$; $j_{1}:=j$; $\bj_{i}:=i$ for $i<j$; $\bj_{i}=i+1$
for $i\geq j$; and $\bre_{j}=j$ for $j=1,\dots,R$. 
\end{definition}
\begin{remark}
The reasons for introducing the notation $d_{\rosil}$, $d_{\trs}$ etc. becomes clear in Definition~\ref{def:degtrsdivvar} below.
\end{remark}

\begin{definition}\label{def:degtrsdivvar}
Assume that (\ref{eq:thesystemRge}) is an $X$-degenerate weakly transparent equation.
Assume, moreover, that the integers $j_{i}$, $i=1,\dots,d_{\trs}$, and $\bj_{i}$, $i=1,\dots,d_{\rosil}$, have the properties stated in 
Definition~\ref{def:Xdegenerate}, and that $\bM$ is given by (\ref{eq:bMdef}). Then the \textit{silent variables} 
\index{Silent!variables}%
\index{Variables!silent}%
are all the variables 
in $\bM$ corresponding to the $\bj_{i}$'th $\sn{1}$-factors in $\tn{d}$, $i=1,\dots,d_{\rosil}$, together with all the variables corresponding to 
$M_{r}$ for $r$'s such that $q_{\infty,r}=0$; the corresponding $r$'s are denoted $\bre_{j}$ for $j=1,\dots,R_{\rosil}$.  Moreover, the 
\textit{transparent variables} 
\index{Transparent!variables}%
\index{Variables!transparent}%
are all the variables in $\bM$ corresponding to the 
$j_{i}$'th $\sn{1}$-factors in $\tn{d}$, $i=1,\dots,d_{\trs}$, together with all the variables corresponding to $M_{r}$ for $r$'s such that 
$q_{\infty,r}>0$; the corresponding $r$'s are denoted $r_{j}$ for $j=1,\dots,R_{\trs}$. 
\end{definition}

Given that (\ref{eq:thesystemRge}) is an $X$-non-trivial or an $X$-degenerate weakly transparent equation, we can introduce the 
following terminology:
\begin{align}
\mfg_{\trs}(\indexnot,t) := & 
\left(\textstyle{\sum}_{k,l=1}^{d_{\trs}}g^{j_{k}j_{l}}(t)n_{j_{k}}n_{j_{l}}
+\sum_{j=1}^{R_{\trs}}a_{r_{j}}^{-2}(t)\nu_{r_{j},i_{r_{j}}}^{2}(\indexnot)\right)^{1/2},\label{eq:mfgtrsdef}\\
\mfg_{\rosil}(\indexnot,t) := & 
\left(\textstyle{\sum}_{k,l=1}^{d_{\rosil}}g^{\bj_{k}\bj_{l}}(t)n_{\bj_{k}}n_{\bj_{l}}
+\sum_{j=1}^{R_{\rosil}}a_{\bre_{j}}^{-2}(t)\nu_{\bre_{j},i_{\bre_{j}}}^{2}(\indexnot)\right)^{1/2}.\label{eq:mfgsildef}
\end{align}
\index{$\a$Aa@Notation!Coefficients, Fourier side!$\mfg_{\trs}(\indexnot,t)$}%
\index{$\a$Aa@Notation!Coefficients, Fourier side!$\mfg_{\rosil}(\indexnot,t)$}%
In case the upper limit of a sum is zero, it should be understood as being empty. Note that, by assumption, 
\[
\lim_{t\rightarrow\infty}\mfg_{\trs}(\indexnot,t)=
\left(\textstyle{\sum}_{k,l=1}^{d_{\trs}}g^{j_{k}j_{l}}_{\infty}n_{j_{k}}n_{j_{l}}+\sum_{j=1}^{R_{\trs}}q_{\infty,r_{j}}\nu_{r_{j},i_{r_{j}}}^{2}(\indexnot)\right)^{1/2}=
\mfg_{\infty}(\indexnot),
\]
where $\mfg_{\infty}(\indexnot)$ is given by (\ref{eq:mfginftydef}). Moreover, $g^{j_{k}j_{l}}_{\infty}$, $k,l=1,\dots,d_{\trs}$, are the components 
of a positive definite matrix, and $q_{\infty,r_{j}}>0$ for $j=1,\dots,R_{\trs}$. It is also of interest to note that 
$\mfg_{\rosil}(\indexnot,t)\rightarrow 0$ as $t\rightarrow \infty$. It would be very convenient if the quantity
\[
\mfg(\indexnot,t)-[\mfg_{\trs}^{2}(\indexnot,t)+\mfg_{\rosil}^{2}(\indexnot,t)]^{1/2}
\]
were zero. However, this is unfortunately not the case, due to the occurrence of cross terms of the form $g^{j_{i}\bj_{l}}(t)n_{j_{i}}n_{\bj_{l}}$ in
$\mfg^{2}(\indexnot,t)$. Such terms have to be dealt with separately. 

It is convenient to divide $\indexnot$ into a transparent and a silent part. 

\begin{definition}\label{def:indexnottrssil}
Assume that (\ref{eq:thesystemRge}) is an $X$-non-trivial or an $X$-degenerate weakly transparent equation, and divide the variables
of $\bM$ as described in Definitions~\ref{def:Xnontrtrsdivvar} and \ref{def:degtrsdivvar} respectively. 
Given $\indexnot\in\EFindexset$, define $\indexnot_{\trs}\in\EFindexset$ as follows: replace $n_{\bj_{k}}$ in $\indexnot$ by zero for 
$k=1,\dots,d_{\rosil}$; and replace $i_{\bre_{k}}$ by $0$ for $k=1,\dots,R_{\rosil}$. Given $\indexnot\in\EFindexset$,
define $\indexnot_{\rosil}\in\EFindexset$ by $\indexnot_{\rosil}:=\indexnot-\indexnot_{\trs}$. The set of $\indexnot\in\EFindexset$ such that 
$\indexnot_{\trs}=0$ ($\indexnot_{\trs}\neq 0$) is denoted by $\EFsilindexset$ ($\EFtrsindexset$).
\index{$\a$Aa@Notation!Frequency sets!$\EFsilindexset$}%
\index{$\a$Aa@Notation!Frequency sets!$\EFtrsindexset$}%
\end{definition}
\begin{remark}
It is important to keep in mind that the set of $\indexnot_{\trs}$ for $\indexnot\in\EFindexset$ is in general \textit{different} from 
$\EFtrsindexset$, but that the set of $\indexnot_{\rosil}$ for $\indexnot\in\EFindexset$ \textit{equals} 
$\EFsilindexset$. 
\end{remark}
With this notation,
\[
\mfg_{\trs}(\indexnot,t)=\mfg(\indexnot_{\trs},t),\ \ \
\mfg_{\rosil}(\indexnot,t)=\mfg(\indexnot_{\rosil},t).
\]
It is convenient to introduce projections associated with this division of the variables. 
\begin{definition}\label{def:projtrssil}
Assume that (\ref{eq:thesystemRge}) is an $X$-non-trivial or an $X$-degenerate weakly transparent equation and define the sets $\EFsilindexset$ 
and $\EFtrsindexset$ as in Definition~\ref{def:indexnottrssil}. Let $s\in\ro$. Then $\projtrs,\projsil:H^{s}(\bM)\rightarrow H^{s}(\bM)$ 
\index{$\a$Aa@Notation!Projections!$\projtrs$}%
\index{$\a$Aa@Notation!Projections!$\projsil$}%
are defined as follows. Given $u\in H^{s}(\bM)$ with Fourier coefficients $\hu(\indexnot)$ determined by 
(\ref{eq:huindexnotdef}), $\projtrs u$ is the function whose $\indexnot$'th Fourier coefficient equals zero if $\indexnot\in\EFsilindexset$ 
and equals $\hu(\indexnot)$ if $\indexnot\in\EFtrsindexset$. Similarly, $\projsil u$ is the function whose $\indexnot$'th Fourier coefficient 
equals $\hu(\indexnot)$ if $\indexnot\in\EFsilindexset$ and equals zero if $\indexnot\in\EFtrsindexset$. 
\end{definition}
\begin{remarks}\label{remarks:usilutrs}
Clearly, the norms of $\projtrs$ and $\projsil$ are bounded by $1$. Moreover, these operators map smooth functions to smooth functions
and we can apply them to functions with values in $\cn{k}$. If $u\in C^{\infty}(\bM\times I,\cn{k})$, then 
$u_{\trs},u_{\rosil}$ are defined by $u_{\trs}(\cdot,t)=\projtrs[u(\cdot,t)]$ and $u_{\rosil}(\cdot,t)=\projsil[u(\cdot,t)]$. It can then be verified that 
$u_{\trs},u_{\rosil}\in C^{\infty}(\bM\times I,\cn{k})$. 
\end{remarks}

Finally, we are in a position to define what is meant by a weakly transparent, balanced and convergent equation. 

\begin{definition}\label{def:weaktrans}
Consider (\ref{eq:thesystemRge}). Assume the associated metric $g$ to be such that $(M,g)$ is a canonical separable cosmological 
model manifold. Then (\ref{eq:thesystemRge}) is said to be \textit{weakly transparent} 
\index{Weakly transparent!equation}%
\index{Equation!weakly transparent}%
if the 
following conditions hold. First, (\ref{eq:thesystemRge}) is either an $X$-non-trivial or an $X$-degenerate weakly transparent equation;
cf. Definitions~\ref{def:Xnontriv} and \ref{def:Xdegenerate}. Second, there is a constant $\b_{\rosil}>0$ and a continuous non-negative 
function $\betafun_{\rosil}\in L^{1}([0,\infty))$ such that
\begin{equation}\label{eq:weaksiltrs}
\dot{\ell}(\indexnot_{\rosil},t)\leq-\b_{\rosil}+\betafun_{\rosil}(t)
\end{equation}
for all $\indexnot\in\EFindexset$ such that $\indexnot_{\rosil}\neq 0$ and all $t\geq 0$; here the notation $\ell$ and $\indexnot_{\rosil}$ 
is introduced in (\ref{eq:ellsigmaXgenRdef}) and Definition~\ref{def:indexnottrssil} respectively. Third, there are constants $K_{\trs},\eta_{\trs}>0$ 
such that 
\begin{equation}\label{eq:weaktrs}
|\dot{\ell}(\indexnot_{\trs},t)|+|\sigma(\indexnot_{\trs},t)|\leq K_{\trs}e^{-\eta_{\trs}t}
\end{equation}
for all $\indexnot\in\EFindexset$ such that $\indexnot_{\trs}\neq 0$ and all $t\geq 0$. Fourth, 
\begin{equation}\label{eq:Xjtrsrocon}
\|X^{j_{k}}(t)-X^{j_{k}}_{\infty}\|\leq K_{\trs}e^{-\eta_{\trs}t}
\end{equation}
for all $t\geq 0$ and all $k=1,\dots,d_{\trs}$, where the notation $j_{k}$ and $d_{\trs}$ is introduced in Definitions~\ref{def:Xnontrtrsdivvar} 
\ref{def:degtrsdivvar}. 
\end{definition}
\begin{remark}
As in the case of Definition~\ref{def:roughODEtermo}, it is to be expected that the requirement that $\betafun_{\rosil}$ be continuous can be 
avoided at the expense of additional technicalities in the proofs below. For future reference, we use the notation
$c_{\rosil}:=\|\betafun_{\rosil}\|_{1}$.
\index{$\a$Aa@Notation!Constants!$c_{\rosil}$}%
\end{remark}
\begin{remark}
The equation (\ref{eq:thesystemRge}) is said to be weakly transparent, balanced and convergent, if it is weakly transparent in the sense of 
Definition~\ref{def:weaktrans}, weakly balanced in the sense of Definition~\ref{def:roughODEtermo} and weakly convergent in the sense of 
Definition~\ref{def:roughODEtermo}. 
\end{remark}

It is of interest to note that if (\ref{eq:thesystemRge}) is weakly transparent, balanced and convergent, then restricting
(\ref{eq:thesystemRge}) to solutions which only depend on the silent variables (cf. Definitions~\ref{def:Xnontrtrsdivvar} and 
\ref{def:degtrsdivvar}) yields an equation which is weakly silent, balanced and convergent. In other words, the results of 
Chapter~\ref{chapter:weaksil} apply. What we need to do in the present chapter is thus to analyse the modes such that 
$\indexnot_{\trs}\neq 0$. Comparing Definition~\ref{def:roughODEtermo} with Definition~\ref{def:weaktrans}, it is clear that we 
still need to define the time corresponding to $T_{\roode}$ appearing in Definition~\ref{def:roughODEtermo}. Before doing so, let us
introduce some terminology. 
\begin{definition}\label{def:CiniStrsdef}
Assume that (\ref{eq:thesystemRge}) is weakly transparent in the sense of Definition~\ref{def:weaktrans}. Let $\EFindexset$ be the set 
introduced in connection with (\ref{eq:IgenReqdef}). Define $C_{\roini}\geq 1$ to be the smallest constant such that
\begin{equation}\label{eq:Cinidef}
C_{\roini}^{-1}|\nu(\indexnot)|\leq \mfg(\indexnot,0)\leq C_{\roini}|\nu(\indexnot)|
\end{equation}
\index{$\a$Aa@Notation!Constants!$C_{\roini}$}%
for all $\indexnot\in\EFindexset$; here $\nu(\indexnot)$ is defined by (\ref{eq:nugenReqdef}). Define $\roS_{\trs}$ by 
\begin{equation}\label{eq:roSdef}
\roS_{\trs}:=\left[\inf\{\mfg(\indexnot,0):\indexnot\in\EFtrsindexset\}\right]^{-1},
\end{equation}
\index{$\a$Aa@Notation!Constants!$\roS_{\trs}$}%
where the notation $\EFtrsindexset$ is introduced in Definition~\ref{def:indexnottrssil}.
\end{definition}
\begin{remark}
The constant $C_{\roini}$ only depends on $g^{ij}(0)$ and $a_{r}(0)$. However, the constant $\roS_{\trs}$ depends not only on these quantities, but
also on the spectrum of the Laplace-Beltrami operator on $(M_{r_{k}},g_{r_{k}})$ for $k=1,\dots,R_{\trs}$. 
\end{remark}

Before defining the constant $T_{\trs}$, which here plays a role analogous to that of $T_{\roode}$ in Chapter~\ref{chapter:weaksil}, it is 
convenient to rewrite (\ref{eq:fourierthesystemRge}) in the form (\ref{eq:vdottrscase}) for suitable choices of $A(\indexnot)$, $A_{\indexnot,\rem}$ 
and $F_{\indexnot}$, and to derive a general estimate for $\|A_{\indexnot,\rem}(t)\|$.

\begin{lemma}\label{lemma:Aremtrsest}
Assume that (\ref{eq:thesystemRge}) is weakly transparent in the sense of Definition~\ref{def:weaktrans}, weakly balanced in the sense of 
Definition~\ref{def:roughODEtermo} and weakly convergent in the sense of Definition~\ref{def:roughODEtermo}. Fix a $\indexnot\in\EFtrsindexset$. 
Then (\ref{eq:fourierthesystemRge}) can be formulated as (\ref{eq:vdottrscase}), where $A(\indexnot)$ is given by 
\begin{equation}\label{eq:Aindexnotdeftrs}
A(\indexnot):=\left(\begin{array}{cc} 0 & \mfg_{\infty}\Id_{m} \\ -\mfg_{\infty}\Id_{m}-i\mfg_{\infty}^{-1}n_{l}X^{l}_{\infty}-\mfg_{\infty}^{-1}\zeta_{\infty} & 
-\a_{\infty}\end{array}\right);
\end{equation}
$v$ and $F_{\indexnot}$ are given by (\ref{eq:vFdeftrs}); $\mfg_{\infty}$ is given by (\ref{eq:mfginftydef}); $X^{l}_{\infty}$ is defined by 
(\ref{eq:gijinfetcdef});
and $\a_{\infty}$ and $\zeta_{\infty}$ are defined by (\ref{eq:alazeas}). Moreover, $A_{\indexnot,\rem}$ satisfies the estimate
\begin{equation}\label{eq:Aremtrsgenest}
\begin{split}
\|A_{\indexnot,\rem}(t)\| \leq & Ce^{-\min\{\eta_{\trs},\eta_{\romn}\}t}+Ce^{-\min\{\b_{\rosil},\eta_{\trs}\}t}\mfg(\indexnot_{\trs},0)+Ce^{-\b_{\rosil}t}\mfg(\indexnot_{\rosil},0)\\
 & +Ce^{-\b_{\rosil}t}\mfg^{2}(\indexnot_{\rosil},0)
\end{split}
\end{equation}
for $t\geq 0$, where the constant $C$ only depends on $K_{\trs},C_{\romn},C_{\coeff},C_{\roini},c_{\rosil},\roS_{\trs}$ and $\eta_{\trs}$.
\end{lemma}
\begin{remark}
Here $K_{\trs}$ and $\eta_{\trs}$ are the constants appearing in (\ref{eq:weaktrs}) and (\ref{eq:Xjtrsrocon}); $C_{\romn}$ and $\eta_{\romn}$ are the 
constants appearing in (\ref{eq:alazeas}); $C_{\coeff}$ is the constant appearing in (\ref{eq:weakbal}); the notation $C_{\roini}$ and $\roS_{\trs}$ is 
introduced in Definition~\ref{def:CiniStrsdef}; $\b_{\rosil}$ is the constant appearing in (\ref{eq:weaksiltrs});
and $c_{\rosil}:=\|\betafun_{\rosil}\|_{1}$, 
\index{$\a$Aa@Notation!Constants!$c_{\rosil}$}%
where $\betafun_{\rosil}$ is the function appearing in (\ref{eq:weaksiltrs}).
\end{remark}
\begin{remarks}\label{remarks:Aindexnotproperties}
Given that the assumptions of Lemma~\ref{lemma:Aremtrsest} are satisfied, the only possibility for $X^{l}_{\infty}$ to be non-zero
is that (\ref{eq:thesystemRge}) is an $X$-non-trivial weakly transparent equation. Then there is one (and only one) $j\in\{1,\dots,d\}$ such 
that $X^{j}_{\infty}\neq 0$, and $-i\mfg_{\infty}^{-1}n_{l}X^{l}_{\infty}$ is one of two possible matrices, irrespective of the value of 
$\indexnot\in\EFtrsindexset$; cf. Remarks~\ref{remarks:Xnontriv}. Thus, if the assumptions of Lemma~\ref{lemma:Aremtrsest} are satisfied,
$A(\indexnot)$ is of the form (\ref{eq:Amudef}), where $\mu=\mfg_{\infty}(\indexnot)$; $V$ is one of the three matrices $0$, 
$(g^{j_{1}j_{1}}_{\infty})^{-1/2}X^{j_{1}}_{\infty}$ or $-(g^{j_{1}j_{1}}_{\infty})^{-1/2}X^{j_{1}}_{\infty}$ (the first possibility only occurs if 
(\ref{eq:thesystemRge}) is an $X$-degenerate weakly transparent equation, and the latter two possibilities only occur when
(\ref{eq:thesystemRge}) is an $X$-non-trivial weakly transparent equation, in which case $j_{1}$ is the number appearing in 
Definition~\ref{def:Xnontrtrsdivvar}); $W:=-\zeta_{\infty}/2$; and $U:=-\a_{\infty}$.  
In addition, due to the requirements appearing in Definitions~\ref{def:Xnontriv} and \ref{def:Xdegenerate}, the set 
$\muindexset:=\{\mfg_{\infty}(\indexnot):\indexnot\in\EFindexset\}$ has $\infty$ as its only accumulation point. Finally, the real parts of the 
eigenvalues of $A(\indexnot)$ converge as $\mfg_{\infty}(\indexnot)\rightarrow\infty$; cf. Lemma~\ref{lemma:Nmueigenvapprox}. 
\end{remarks}
\begin{proof}
Due to (\ref{eq:weaktrs}), 
\begin{equation}\label{eq:mfgintrsest}
C^{-1}\mfg(\indexnot_{\trs},0)\leq\mfg(\indexnot_{\trs},t)\leq C\mfg(\indexnot_{\trs},0)
\end{equation}
for $t\geq 0$, where $C\geq 1$ only depends on $\eta_{\trs}$ and $K_{\trs}$. Taking the limit as $t\rightarrow\infty$ and keeping in mind
that $\mfg(\indexnot_{\trs},t)$ converges to $\mfg_{\infty}(\indexnot_{\trs})$ yields
\begin{equation}\label{eq:mfginftyintrsest}
C^{-1}\mfg(\indexnot_{\trs},0)\leq\mfg_{\infty}(\indexnot_{\trs})\leq C\mfg(\indexnot_{\trs},0),
\end{equation}
where $C\geq 1$ only depends on $\eta_{\trs}$ and $K_{\trs}$. Appealing to (\ref{eq:weaksiltrs}) yields
\begin{equation}\label{eq:mfginsilest}
\mfg(\indexnot_{\rosil},t)\leq e^{-\b_{\rosil}t+c_{\rosil}}\mfg(\indexnot_{\rosil},0)
\end{equation}
for $t\geq 0$, where $c_{\rosil}:=\|\betafun_{\rosil}\|_{1}$. Before proceeding, note that, using the notation introduced in (\ref{eq:nuroTetcdef})
and Definition~\ref{def:indexnottrssil}, 
\[
n=\nu_{\roT}(\indexnot)=\nu_{\roT}(\indexnot_{\trs})+\nu_{\roT}(\indexnot_{\rosil}),\ \ \
n_{l}g^{0l}(t)=[\nu_{\roT}(\indexnot_{\trs})]_{l}g^{0l}(t)+[\nu_{\roT}(\indexnot_{\rosil})]g^{0l}(t).
\]
On the other hand (\ref{eq:sigmaXdefintro}) yields 
\[
\sigma(\indexnot_{\trs},t)=\frac{[\nu_{\roT}(\indexnot_{\trs})]_{l}g^{0l}(t)}{\mfg(\indexnot_{\trs},t)},\ \ \
\sigma(\indexnot_{\rosil},t)=\frac{[\nu_{\roT}(\indexnot_{\rosil})]_{l}g^{0l}(t)}{\mfg(\indexnot_{\rosil},t)}.
\]
Combining the last four equalities yields
\[
n_{l}g^{0l}(t)=\sigma(\indexnot_{\trs},t)\mfg(\indexnot_{\trs},t)+\sigma(\indexnot_{\rosil},t)\mfg(\indexnot_{\rosil},t).
\]
Due to this observation, (\ref{eq:weaktrs}), (\ref{eq:mfgintrsest}), (\ref{eq:mfginsilest}) and the fact that the equation is weakly balanced, it follows
that
\begin{equation}\label{eq:nlgzltrsest}
\begin{split}
|n_{l}g^{0l}(t)| \leq & |\sigma(\indexnot_{\trs},t)|\cdot \mfg(\indexnot_{\trs},t)
+|\sigma(\indexnot_{\rosil},t)|\cdot\mfg(\indexnot_{\rosil},t)\\
 \leq & Ce^{-\eta_{\trs}t}\mfg(\indexnot_{\trs},0)+Ce^{-\b_{\rosil}t}\mfg(\indexnot_{\rosil},0)
\end{split}
\end{equation}
for $t\geq 0$, where $C$ depends only on $K_{\trs},C_{\coeff},c_{\rosil}$ and $\eta_{\trs}$, and $C_{\coeff}$ is the constant appearing in 
(\ref{eq:weakbal}); note that the fact that the equation is weakly balanced implies that $|\sigma(\indexnot_{\rosil},t)|\leq C_{\coeff}$. 
In case $g^{0j_{k}}\equiv 0$ for $k=1,\dots,d_{\trs}$, the first term on the far right hand side of (\ref{eq:nlgzltrsest}) can be removed. 
Similarly, we can estimate
\begin{equation}\label{eq:nlXldifftrsest}
\begin{split}
\|n_{l}X^{l}(t)-n_{l}X_{\infty}^{l}\| \leq & \textstyle{\sum}_{k=1}^{d_{\trs}}\|n_{j_{k}}X^{j_{k}}(t)-n_{j_{k}}X^{j_{k}}_{\infty}\|
+\|X(\indexnot_{\rosil},t)\|\cdot \mfg(\indexnot_{\rosil},t)\\
 \leq &  K_{\trs}e^{-\eta_{\trs}t}\textstyle{\sum}_{k=1}^{d_{\trs}}|n_{j_{k}}|+Ce^{-\b_{\rosil}t}\mfg(\indexnot_{\rosil},0)
\end{split}
\end{equation}
where $C$ only depends on $C_{\coeff}$ and $c_{\rosil}$; we have appealed to (\ref{eq:Xjtrsrocon}) and the fact that 
$\|X(\indexnot_{\rosil},t)\|\leq C_{\coeff}$, an estimate which holds due to the fact that the equation is weakly balanced. In the final estimate,
we need to divide both sides of (\ref{eq:nlXldifftrsest}) by $\mfg_{\infty}(\indexnot_{\trs})$. Note, to this end, that 
\[
[\mfg_{\infty}(\indexnot_{\trs})]^{-1}\textstyle{\sum}_{k=1}^{d_{\trs}}|n_{j_{k}}|
\leq C[\mfg(\indexnot_{\trs},0)]^{-1}\textstyle{\sum}_{k=1}^{d_{\trs}}|n_{j_{k}}|\leq C,
\]
where we have appealed to (\ref{eq:Cinidef}) and (\ref{eq:mfginftyintrsest}) and the constant $C$ only depends on $K_{\trs},C_{\roini}$ and 
$\eta_{\trs}$. When dividing the second term on the far right hand side of (\ref{eq:nlXldifftrsest}) by $\mfg_{\infty}(\indexnot_{\trs})$,
we can similarly appeal to (\ref{eq:roSdef}) and (\ref{eq:mfginftyintrsest}). Summing up yields
\begin{equation}\label{eq:nlXldifftrsestgn}
[\mfg_{\infty}(\indexnot_{\trs})]^{-1}\|n_{l}X^{l}(t)-n_{l}X_{\infty}^{l}\|\leq C[e^{-\eta_{\trs}t}+e^{-\b_{\rosil}t}\mfg(\indexnot_{\rosil},0)]
\end{equation}
for $t\geq 0$, where $C$ only depends on $K_{\trs},C_{\roini},\roS_{\trs},c_{\rosil},C_{\coeff}$ and $\eta_{\trs}$. In this estimate we tacitly assume
$\indexnot$ to be such that $\indexnot_{\trs}\neq 0$, since it does not make sense otherwise. An alternate version of (\ref{eq:nlXldifftrsestgn})
is obtained by replacing $\mfg(\indexnot_{\rosil},0)$ appearing on the right hand side by $[\mfg(\indexnot_{\trs},0)]^{-1}\mfg(\indexnot_{\rosil},0)$.
In that case the constant $C$ appearing in (\ref{eq:nlXldifftrsestgn}) does not depend on $\roS_{\trs}$. Finally, let us consider 
\begin{equation}\label{eq:mfgsdiffest}
\mfg^{2}(\indexnot,t)-\mfg_{\infty}^{2}(\indexnot)=\mfg^{2}(\indexnot_{\rosil},t)+\mfg^{2}(\indexnot_{\trs},t)
+\mfG_{\rem}(\indexnot,t)-\mfg_{\infty}^{2}(\indexnot),
\end{equation}
where $\mfg_{\infty}(\indexnot)$ is defined in (\ref{eq:mfginftydef}) and 
\[
\mfG_{\rem}(\indexnot,t):=2\textstyle{\sum}_{k=1}^{d_{\trs}}\sum_{l=1}^{d_{\rosil}}g^{j_{k}\bj_{l}}n_{j_{k}}n_{\bj_{l}}.
\] 
The first term on the right hand side of (\ref{eq:mfgsdiffest}) can be estimated using (\ref{eq:mfginsilest}). 
In order to estimate $\mfG_{\rem}$, note that the Cauchy-Schwarz inequality yields
\begin{equation}\label{eq:mfGremCSest}
|\mfG_{\rem}(\indexnot,t)|\leq 2\mfg(\indexnot_{\trs},t)\mfg(\indexnot_{\rosil},t)
\end{equation}
for all $t\geq 0$. In particular
\begin{equation}\label{eq:mfGremest}
|\mfG_{\rem}(\indexnot,t)|\leq e^{\b_{\rosil}t}\mfg^{2}(\indexnot_{\rosil},t)+e^{-\b_{\rosil}t}\mfg^{2}(\indexnot_{\trs},t)
\end{equation}
for all $t\geq 0$. Combining this observation with (\ref{eq:mfgsdiffest}) yields
\begin{equation}\label{eq:mfgsqminusmfginfsqprelest}
|\mfg^{2}(\indexnot,t)-\mfg_{\infty}^{2}(\indexnot)|\leq |\mfg^{2}(\indexnot_{\trs},t)-\mfg_{\infty}^{2}(\indexnot)|
+Ce^{-\b_{\rosil}t}\mfg^{2}(\indexnot_{\rosil},0)+Ce^{-\b_{\rosil}t}\mfg^{2}(\indexnot_{\trs},0)
\end{equation}
for $t\geq 0$, where $C$ only depends on $\eta_{\trs}$, $K_{\trs}$ and $c_{\rosil}$ and we have appealed to (\ref{eq:mfgintrsest})
and (\ref{eq:mfginsilest}). The first term on the right hand side of (\ref{eq:mfgsqminusmfginfsqprelest}) can be estimated using 
(\ref{eq:weaktrs}). In fact, if $\indexnot_{\trs}\in\EFindexset$ is such that $\indexnot_{\trs}\neq 0$, then (\ref{eq:weaktrs}) can be used 
to deduce that 
\[
\left|\frac{\mfg_{\infty}(\indexnot_{\trs})}{\mfg(\indexnot_{\trs},t)}-1\right|\leq Ce^{-\eta_{\trs}t},
\]
where $C$ only depends on $K_{\trs}$ and $\eta_{\trs}$. Combining this estimate with (\ref{eq:mfginftyintrsest}) yields
\begin{equation}\label{eq:mfgsqdifftrsest}
|\mfg^{2}(\indexnot_{\trs},t)-\mfg_{\infty}^{2}(\indexnot_{\trs})|\leq Ce^{-\eta_{\trs}t}\mfg^{2}(\indexnot_{\trs},0)
\end{equation}
for $t\geq 0$, where $C$ only depends on $K_{\trs}$ and $\eta_{\trs}$. Clearly, this estimate holds even if $\indexnot_{\trs}=0$. To conclude,
\[
|\mfg^{2}(\indexnot,t)-\mfg_{\infty}^{2}(\indexnot)|\leq Ce^{-\b_{\rosil}t}\mfg^{2}(\indexnot_{\rosil},0)
+Ce^{-\min\{\b_{\rosil},\eta_{\trs}\}t}\mfg^{2}(\indexnot_{\trs},0)
\]
for $t\geq 0$, where we have appealed to (\ref{eq:mfgsqminusmfginfsqprelest}) and (\ref{eq:mfgsqdifftrsest}); the constant only depends 
on $K_{\trs},c_{\rosil}$ and $\eta_{\trs}$; and we have used the fact that $\mfg_{\infty}(\indexnot)=\mfg_{\infty}(\indexnot_{\trs})$. Dividing this 
estimate by $\mfg_{\infty}(\indexnot_{\trs})$ and appealing to (\ref{eq:roSdef}) and (\ref{eq:mfginftyintrsest}) yields
\begin{equation}\label{eq:gsqdiffginfnormal}
[\mfg_{\infty}(\indexnot_{\trs})]^{-1}|\mfg^{2}(\indexnot,t)-\mfg_{\infty}^{2}(\indexnot)|\leq Ce^{-\b_{\rosil}t}\mfg^{2}(\indexnot_{\rosil},0)
+Ce^{-\min\{\b_{\rosil},\eta_{\trs}\}t}\mfg(\indexnot_{\trs},0)
\end{equation}
for $t\geq 0$, where the constant $C$ only depends on $K_{\trs},c_{\rosil},\roS_{\trs}$ and $\eta_{\trs}$ and we, again, assume $\indexnot_{\trs}\neq 0$.
An alternate version of (\ref{eq:gsqdiffginfnormal}) is obtained by replacing $\mfg^{2}(\indexnot_{\rosil},0)$ appearing on the right hand side by 
$[\mfg(\indexnot_{\trs},0)]^{-1}\mfg^{2}(\indexnot_{\rosil},0)$. In that case the constant $C$ appearing in (\ref{eq:gsqdiffginfnormal}) does not depend 
on $\roS_{\trs}$. Finally, note that 
\begin{equation}\label{eq:zetadiffgnorm}
[\mfg_{\infty}(\indexnot_{\trs})]^{-1}\|\zeta(t)-\zeta_{\infty}\|\leq Ce^{-\eta_{\romn}t}
\end{equation}
for $t\geq 0$, where $C$ only depends on $K_{\trs}$, $\eta_{\trs}$, $C_{\romn}$ and $\roS_{\trs}$. 

Next, the equation (\ref{eq:fourierthesystemRge}) can be formulated as (\ref{eq:vdottrscase}), where $A$ is given by (\ref{eq:Aindexnotdeftrs})
and $v$ and $F_{\indexnot}$ are given by (\ref{eq:vFdeftrs}). Due to (\ref{eq:alazeas}), (\ref{eq:nlgzltrsest}), 
(\ref{eq:nlXldifftrsestgn}), (\ref{eq:gsqdiffginfnormal}) and (\ref{eq:zetadiffgnorm}), $A_{\indexnot,\rem}$ satisfies the estimate 
(\ref{eq:Aremtrsgenest}). The lemma follows.
\end{proof}

Finally, we are in a position to define constants $C_{\trs}>0$ and $\b_{\trs}>0$ such that an estimate of the form (\ref{eq:Aremtrsest}) holds.

\begin{definition}\label{def:btrsTtrs}
Assume that (\ref{eq:thesystemRge}) is weakly transparent in the sense of Definition~\ref{def:weaktrans}, weakly balanced in the sense of 
Definition~\ref{def:roughODEtermo} and weakly convergent in the sense of Definition~\ref{def:roughODEtermo}. Fix a $\indexnot\in\EFtrsindexset$.
Then $\b_{\trs}$ and $T_{\trs}$ are defined by 
\begin{align}
\b_{\trs} := & \min\{\eta_{\trs},\b_{\rosil},\eta_{\romn}\},\label{eq:btrsdef}\\
T_{\trs} := & \frac{1}{\min\{\b_{\rosil},\eta_{\trs}\}}\ln [1+|\nu(\indexnot_{\trs})|]+\frac{2}{\b_{\rosil}}\ln [1+|\nu(\indexnot_{\rosil})|].\label{eq:Ttrsdef}
\end{align}
\end{definition}
\begin{remark}\label{remark:Ctrsdef}
Combining Lemma~\ref{lemma:Aremtrsest} and Definition~\ref{def:btrsTtrs}, it is clear that 
\begin{equation}\label{eq:Aremtrsderest}
\|A_{\indexnot,\rem}(t)\|\leq C_{\trs}e^{-\b_{\trs}\bt}
\end{equation}
for all $t\geq T_{\trs}$, where $\bt:=t-T_{\trs}$ and $C_{\trs}$ only depends on $K_{\trs},C_{\romn},C_{\coeff},C_{\roini},c_{\rosil},\roS_{\trs}$ and $\eta_{\trs}$.
\end{remark}

Given a solution $u$ to (\ref{eq:thesystemRge}), we can divide it into two parts, $u=u_{\rosil}+u_{\trs}$; cf. Remarks~\ref{remarks:usilutrs}.
The results of Chapter~\ref{chapter:weaksil} apply to 
$u_{\rosil}$. We can therefore focus on $u_{\trs}$. Let $\indexnot\in\EFtrsindexset$. Then, due to Lemma~\ref{lemma:Aremtrsest} and 
Definition~\ref{def:btrsTtrs}, we can rewrite (\ref{eq:fourierthesystemRge}) as (\ref{eq:vdottrscase}), where $A(\indexnot)$ 
is given by (\ref{eq:Aindexnotdeftrs}), $v$ and $F_{\indexnot}$ are given by (\ref{eq:vFdeftrs}), and $A_{\indexnot,\rem}$ satisfies the estimate 
(\ref{eq:Aremtrsderest}).
It is thus natural to divide the interval $[0,\infty)$ into $[0,T_{\trs}]$ and $[T_{\trs},\infty)$. On the interval $[T_{\trs},\infty)$, we can appeal
to the results of Chapter~\ref{chapter:ODEtransp}. What remains to do is thus, first, to analyse how the solution to 
(\ref{eq:fourierthesystemRge}) develops in the time period $[0,T_{\trs}]$; and, second, to sum up the estimates over all the 
$\indexnot\in\EFtrsindexset$. On the other hand, in order to be able to estimate the solution in the interval $[0,T_{\trs}]$, 
we need to make stronger assumptions than those contained in Definition~\ref{def:weaktrans}. The reason for this is that the conditions
(\ref{eq:weaksiltrs}) and (\ref{eq:weaktrs}) do not impose any restrictions on, e.g., $\dot{g}^{j_{k}\bj_{l}}$. On the other hand, the assumptions
made in the statement of Lemma~\ref{lemma:roughenestbalsetting} are sufficient, since they allow us to apply this lemma.

\section{Deriving asymptotics}\label{section:derastrs}

As a next step, we wish to derive asymptotics for solutions to (\ref{eq:thesystemRge}) for all $t\geq 0$. Doing so is more complicated 
in the present setting than in Section~\ref{section:roughODEfutas}. The reason for this is that in Section~\ref{section:roughODEfutas}, the equations 
of interest are such that all modes have the same asymptotic behaviour. For weakly transparent equations, the asymptotic behaviour is mode dependent, 
and how to formulate estimates is therefore less obvious; cf. the discussion in the introduction to the present chapter. 
In order to develop some intuition, it is useful to consider the limit equation defined by (\ref{eq:thesystemRgeinfty}) below.
Moreover, it is useful to divide solutions $u$ to (\ref{eq:thesystemRge}) according to $u=u_{\rosil}+u_{\trs}$, where $u_{\rosil}$ and $u_{\trs}$ are defined 
as in Remarks~\ref{remarks:usilutrs}. 

\textbf{The limit equation.}
If $u$ is a solution to (\ref{eq:thesystemRge}), then it is of interest to compare the asymptotics of $u_{\trs}$ with the asymptotics of 
an appropriately chosen solution to the limit equation. By the \textit{limit equation}, 
\index{Limit equation!transparent setting}%
we here mean the equation obtained by letting the 
coefficients appearing in (\ref{eq:thesystemRge}) tend to infinity. It can be written
\begin{equation}\label{eq:thesystemRgeinfty}
\begin{split}
u_{tt}-\textstyle{\sum}_{j,l=1}^{d}g^{jl}_{\infty}\d_{j}\d_{l}u
-\sum_{r=1}^{R}q_{\infty,r}\Delta_{g_{r}}u 
+\a_{\infty}u_{t}+\sum_{j=1}^{d}X^{j}_{\infty}\d_{j}u+\zeta_{\infty}u = f.
\end{split}
\end{equation}
In order to arrive at this equation, we assume the metric associated with (\ref{eq:thesystemRge}) to be such that $(M,g)$ is a canonical 
separable cosmological model manifold; the limits (\ref{eq:gijinfetcdef}) to exist; (\ref{eq:alazeas}) to hold; and $g^{0j}_{\infty}$ to equal
zero for $j=1,\dots,d$. Note that 
$g^{0j}_{\infty}=0$, $j=1,\dots,d$, holds if one of Definitions~\ref{def:Xnontriv} and \ref{def:Xdegenerate} is satisfied; i.e., if, e.g.,
(\ref{eq:thesystemRge}) is a weakly transparent equation. In this context, it is also convenient to introduce a \textit{limiting
Laplace-Beltrami operator}:
\index{Limiting Laplace-Beltrami operator}%
\begin{equation}\label{eq:Deltainftydef}
\Delta_{\infty}:=\textstyle{\sum}_{j,l=1}^{d}g^{jl}_{\infty}\d_{j}\d_{l}+\sum_{r=1}^{R}q_{\infty,r}\Delta_{g_{r}}
\end{equation}
on $\bM$. Needless to say, this operator is typically not elliptic on $\bM$, since it has degeneracies associated with, e.g., coefficients
of the form $q_{\infty,r}$ that vanish. When considering a solution $u$ to (\ref{eq:thesystemRgeinfty}), it is convenient to divide it into
$u=u_{\rosil}+u_{\trs}$ (as described in Remark~\ref{remarks:usilutrs}), just as in the case of solutions to (\ref{eq:thesystemRge}). On the 
Fourier side, $\Delta_{\infty}$ corresponds to multiplication with $-\mfg_{\infty}^{2}(\indexnot)$. For $\indexnot\in\EFtrsindexset$, this 
expression is strictly negative. For that reason, it makes sense to apply $(-\Delta_{\infty})^{-1/2}$ to $u_{\trs}$ (on the Fourier side, 
this corresponds to multiplication by $[\mfg_{\infty}(\indexnot)]^{-1}$). 

\textbf{Solving the limit equation.} Assume the conditions of Lemma~\ref{lemma:Aremtrsest} to be fulfilled and let (\ref{eq:thesystemRgeinfty})
be the associated limit equation. Let $u$ be a solution to (\ref{eq:thesystemRgeinfty}). Then 
\begin{equation}\label{eq:Usil}
\dot{U}_{\rosil}=A_{\infty}U_{\rosil}+F_{\rosil},
\end{equation}
where $A_{\infty}$ is the matrix given by (\ref{eq:vAhFdef}), $U_{\rosil}$ is the vector valued function with components $u_{\rosil}$ and $\d_{t}u_{\rosil}$, 
and $F_{\rosil}$ is the vector valued function with components $0$ and $f_{\rosil}$, where $f_{\rosil}$ is obtained from $f$ as described in 
Remark~\ref{remarks:usilutrs}. Solutions to (\ref{eq:Usil}) can be written
\[
U_{\rosil}(p,t)=e^{A_{\infty}t}U_{\rosil}(p,0)+\int_{0}^{t}e^{A_{\infty}(t-\tau)}F_{\rosil}(p,\tau)d\tau,
\]
where $p\in\bM$. In particular, the behaviour of $U_{\rosil}$ is independent of the Fourier mode. In the case of $u_{\trs}$, the situation is 
more complicated. Given $\indexnot\in\EFtrsindexset$, let 
\[
v_{\trs}(\indexnot,t):=\left(\begin{array}{c} \mfg_{\infty}(\indexnot)\hu_{\trs}(\indexnot,t)\\ \d_{t}\hu_{\trs}(\indexnot,t)\end{array}\right),
\ \ \
F_{\trs,\indexnot}(t):=\left(\begin{array}{c} 0 \\ \hf_{\trs}(\indexnot,t)\end{array}\right)
\]
and let $A(\indexnot)$ be given by (\ref{eq:Aindexnotdeftrs}) (keeping Remarks~\ref{remarks:Aindexnotproperties} in mind). Here
$f_{\trs}$ is obtained from $f$ as described in Remark~\ref{remarks:usilutrs} and we use the notation introduced in (\ref{eq:huindexnotdef}).
Then
\[
v_{\trs}(\indexnot,t)=e^{A(\indexnot)t}v_{\trs}(\indexnot,0)+\int_{0}^{t}e^{A(\indexnot)(t-\tau)}F_{\trs,\indexnot}(\tau)d\tau.
\]
A formal way of writing this equality so that it is valid for all $\indexnot\in\EFtrsindexset$ at once is
\[
U_{\trs}(\cdot,t)=e^{\ma_{\trs}t}U_{\trs}(\cdot,0)+\int_{0}^{t}e^{\ma_{\trs}(t-\tau)}F_{\trs}(\cdot,\tau)d\tau,
\]
where
\begin{align}
F_{\trs}(\cdot,t) := & \left(\begin{array}{c} 0 \\ f_{\trs}(\cdot,t)\end{array}\right),\nonumber\\
U_{\trs}(\cdot,t) := & \left(\begin{array}{c} (-\Delta_{\infty})^{1/2}u_{\trs}(\cdot,t) \\ \d_{t}u_{\trs}(\cdot,t)\end{array}\right),\nonumber\\
\ma_{\trs} := & \left(\begin{array}{cc} 0 & \Id_{m}(-\Delta_{\infty})^{1/2} \\ -\Id_{m}(-\Delta_{\infty})^{1/2}
-(X^{l}_{\infty}\d_{l}+\zeta_{\infty})(-\Delta_{\infty})^{-1/2} & 
-\a_{\infty}\end{array}\right).\label{eq:matrsdef}
\end{align}

\textbf{Norms.} When measuring the norm of, e.g., $U_{\trs}$, it is convenient to rescale the individual Fourier coefficients before 
applying Sobolev norms. The purpose of this rescaling is to normalise the size of the coefficients so that they are asymptotically
comparable. Since the analysis of the asymptotics is divided into two parts according to the size of $\mfg_{\infty}(\indexnot)$, 
cf. the introduction, it is also convenient to restrict attention to certain ranges of the frequencies. Given $1\leq k\in\zo$, 
$\psi\in C^{\infty}(\bM,\cn{k})$, $s\in\ro$, $0\leq t\in\ro$ and $0<\mu_{0}\in\ro$, we therefore introduce the semi-norm 
\begin{equation}\label{eq:tsmuznodef}
\|\psi\|_{t,s,\mu_{0},\pm}:=\left(\textstyle{\sum}_{\indexnot\in\EFindexsetr{\mu_{0}}{\pm}}\ldr{\nu(\indexnot)}^{2s}\ldr{t}^{-2\de_{\indexnot}+2}e^{-2\kappa_{\indexnot}t}
|\hat{\psi}(\indexnot)|^{2}\right)^{1/2},
\end{equation}
\index{$\a$Aa@Notation!Norms!$\normSobtsmuzpm$}%
where $\hat{\psi}$ is defined in analogy with (\ref{eq:huindexnotdef}); $\kappa_{\indexnot}:=\kappa_{\max}[A(\indexnot)]$ and 
$\delta_{\indexnot}:=d_{\max}[A(\indexnot),\kappa_{\indexnot}]$, cf. Definition~\ref{def:SpRspdef}; $\EFindexsetr{\mu_{0}}{+}$ is the set 
of $\indexnot\in\EFindexset$ such that $\mfg_{\infty}(\indexnot)\geq\mu_{0}$; and $\EFindexsetr{\mu_{0}}{-}$ is the set of 
$\indexnot\in\EFtrsindexset$ such that $\mfg_{\infty}(\indexnot)\leq\mu_{0}$. Similarly, 
\begin{equation}\label{eq:tsnodef}
\|\psi\|_{t,s}:=\left(\textstyle{\sum}_{\indexnot\in\EFtrsindexset}\ldr{\nu(\indexnot)}^{2s}\ldr{t}^{-2\de_{\indexnot}+2}e^{-2\kappa_{\indexnot}t}
|\hat{\psi}(\indexnot)|^{2}\right)^{1/2}.
\end{equation}
\index{$\a$Aa@Notation!Norms!$\normSobts$}%
In the case of the right hand side of the equation, we also use the semi-norm
\begin{equation}\label{eq:ftrssnorm}
\|f\|_{\trs,s}:=\int_{0}^{\infty}\left(\textstyle{\sum}_{\indexnot\in\EFtrsindexset}
\ldr{\nu(\indexnot)}^{2s}e^{-2(\kappa_{\indexnot}-\b_{\romar})t}|\hf(\indexnot,t)|^{2}\right)^{1/2}dt,
\end{equation}
\index{$\a$Aa@Notation!Norms!$\normSobtrss$}%
where the value of the constant $\b_{\romar}>0$ should be clear from the context. 

When estimating $u_{\trs}$, it is convenient to divide the elements of $\EFtrsindexset$ into the two sets $\EFindexsetr{\mu_{0}}{\pm}$.
For appropriately chosen $\mu_{0}$ we can appeal to Lemma~\ref{lemma:modebymodetrs} for $\mfg_{\infty}(\indexnot)\geq\mu_{0}$. For the 
remaining $\indexnot$ there are only a finite number of matrices $A(\indexnot)$. To derive corresponding estimates, it is thus sufficient 
to appeal to the results of Chapter~\ref{chapter:weaksil} a finite number of 
times. We start by combining Lemmas~\ref{lemma:modebymodetrs} and \ref{lemma:roughenestbalsetting}.

\begin{lemma}\label{lemma:trsasymptlargefre}
Assume that (\ref{eq:thesystemRge}) is weakly transparent in the sense of Definition~\ref{def:weaktrans}, weakly balanced in the sense of 
Definition~\ref{def:roughODEtermo} and weakly convergent in the sense of Definition~\ref{def:roughODEtermo}. Assume, finally, that 
there is a constant $0<C_{\ell}\in\ro$ and a continuous non-negative function $\betafun\in L^{1}([0,\infty))$ 
such that (\ref{eq:elldotbdgeneral}) holds for all $0\neq\indexnot\in\EFindexset$ and all $t\geq 0$. Fix a $0<\b_{\romar}\in\ro$ which is small 
enough, the bound depending only on $\a_{\infty}$, $\zeta_{\infty}$, the limits appearing in (\ref{eq:gijinfetcdef}) and the constant
$\b_{\trs}$ introduced in Definition~\ref{def:btrsTtrs}. Assume $f$ to be such that $\|f\|_{\trs,s}<\infty$ for all $s\in\ro$, where 
$\|\cdot\|_{\trs,s}$ is defined in (\ref{eq:ftrssnorm}). Then there are constants $s_{\infty,\rohom},s_{\infty,\roih}\in [0,\infty)$, 
$0<\mu_{0}\in\ro$ and $0< C\in\ro$ such that the following holds. 
Given a solution $u$ to (\ref{eq:thesystemRge}), there is a unique $U_{\infty}\in C^{\infty}(\bM,\cn{2m})$ with the following properties
\begin{itemize}
\item the $\indexnot$'th Fourier coefficient of $U_{\infty}$ vanishes unless $\indexnot\in\EFindexsetr{\mu_{0}}{+}$, where the notation 
$\EFindexsetr{\mu_{0}}{+}$ is introduced in connection with (\ref{eq:tsmuznodef}),
\item if $\indexnot\in\EFindexsetr{\mu_{0}}{+}$, then the $\indexnot$'th Fourier coefficient of $U_{\infty}$ belongs to $E_{\indexnot}$, where
$E_{\indexnot}$ is the first generalised eigenspace in the $(\b_{\trs}-\b_{\romar}),A(\indexnot)$-decomposition of $\cn{2m}$,
\item the estimate
\begin{equation}\label{eq:ultmuzpartas}
\begin{split}
 & \left\|\left(\begin{array}{c} (-\Delta_{\infty})^{1/2}u(\cdot,t) \\ \d_{t}u(\cdot,t)\end{array}\right)
-e^{\ma_{\trs}t}U_{\infty}-\int_{0}^{t}e^{\ma_{\trs}(t-\tau)}\left(\begin{array}{c} 0 \\ f(\cdot,\tau)\end{array}\right)d\tau\right\|_{t,s,\mu_{0},+}\\
 \leq & Ce^{-(\b_{\trs}-\b_{\romar})t}[\|u_{t}(\cdot,0)\|_{(s+s_{\infty,\rohom})}+\|u(\cdot,0)\|_{(s+s_{\infty,\rohom}+1)}
+\|f\|_{\trs,s+s_{\infty,\roih}}]
\end{split}
\end{equation}
holds for $t\geq 0$ and $s\in\ro$, where $\Delta_{\infty}$ is defined in (\ref{eq:Deltainftydef}), $\ma_{\trs}$ is defined in (\ref{eq:matrsdef})
and $\|\cdot\|_{t,s,\mu_{0},+}$ is defined in (\ref{eq:tsmuznodef}). 
\end{itemize}
Moreover, 
\begin{equation}\label{eq:Uinfsobnoest}
\|U_{\infty}\|_{(s)}\leq C\|u_{t}(\cdot,0)\|_{(s+s_{\infty,\rohom})}+C\|u(\cdot,0)\|_{(s+s_{\infty,\rohom}+1)}+
C\|f\|_{\trs,s+s_{\infty,\roih}}
\end{equation}
for all $s\in\ro$. 
\end{lemma}
\begin{remark}\label{remark:Cmuzdephighfre}
The constant $C$ only depends on $C_{\roini}$, $C_{\romn}$, $C_{\coeff}$, $\roS_{\trs}$, $K_{\trs}$, $\eta_{\trs}$, $c_{\rosil}$, $c_{\betafun}$, 
$C_{\ell}$, $\b_{\rosil}$, 
the limits appearing in (\ref{eq:gijinfetcdef}), $\a_{\infty}$, $\zeta_{\infty}$, $\b_{\trs}$ and $\b_{\romar}$; $\mu_{0}$ only depends on 
$\b_{\trs}$, $\b_{\romar}$, $\a_{\infty}$, $\zeta_{\infty}$ and the limits appearing in (\ref{eq:gijinfetcdef}); and $s_{\infty,\rohom}$ and 
$s_{\infty,\roih}$ only depend on $C_{\coeff}$, $C_{\ell}$, $\eta_{\trs}$, $\eta_{\romn}$, $\b_{\rosil}$, $\a_{\infty}$, 
$\zeta_{\infty}$ and the limits appearing in (\ref{eq:gijinfetcdef}). Here $c_{\betafun}:=\|\betafun\|_{1}$ and $c_{\rosil}:=\|\betafun_{\rosil}\|_{1}$. 
\end{remark}
\begin{remark}
The terms
\[
e^{\ma_{\trs}t}U_{\infty}+\int_{0}^{t}e^{\ma_{\trs}(t-\tau)}\left(\begin{array}{c} 0 \\ f(\cdot,\tau)\end{array}\right)d\tau
\]
appearing on the left hand side of (\ref{eq:ultmuzpartas}) correspond to a solution to the limit equation (\ref{eq:thesystemRgeinfty}).
\end{remark}
\begin{remark}
Due to the statement of the lemma, $U_{\infty}$ is uniquely determined by the solution. Moreover, the map from the initial data of $u$ to 
$U_{\infty}$ is continuous in the $C^{\infty}$-topology. 
\end{remark}
\begin{proof}
The proof is based on a combination of Lemmas~\ref{lemma:roughenestbalsetting} and \ref{lemma:modebymodetrs}. However, the justification that we 
are allowed to appeal to Lemma~\ref{lemma:modebymodetrs} depends on which of Definitions~\ref{def:Xnontriv} and 
\ref{def:Xdegenerate} is satisfied. If (\ref{eq:thesystemRge}) is an $X$-non-trivial weakly transparent equation, then there are only two 
possibilities for $-\mfg_{\infty}^{-1}n_{l}X^{l}_{\infty}$ appearing in the bottom left component of the right hand side of (\ref{eq:Aindexnotdeftrs}),
namely $\pm(g^{j_{1}j_{1}}_{\infty})^{-1/2}X^{j_{1}}_{\infty}$; cf. Remark~\ref{remarks:Aindexnotproperties}. When appealing to Lemma~\ref{lemma:modebymodetrs},
the matrices $U,V,W$ are thus given by $U:=-\a_{\infty}$, $V:=\mp (g^{j_{1}j_{1}}_{\infty})^{-1/2}X^{j_{1}}_{\infty}$ and $W:=-\zeta_{\infty}/2$. Since there are two 
possibilities for $V$, we, strictly speaking, need to apply Lemma~\ref{lemma:modebymodetrs} twice. If (\ref{eq:thesystemRge}) is an $X$-degenerate 
weakly transparent equation, then $U:=-\a_{\infty}$, $V:=0$ and $W:=-\zeta_{\infty}/2$, and we only need to appeal to Lemma~\ref{lemma:modebymodetrs}
once. However, it is of interest to note that, irrespective of whether (\ref{eq:thesystemRge}) is an $X$-non-trivial weakly transparent equation or 
an $X$-degenerate weakly transparent equation, the matrices $U,V$ and $W$ only depend on $\a_{\infty}$, $\zeta_{\infty}$ and the limits appearing in 
(\ref{eq:gijinfetcdef}). 

When appealing to Lemma~\ref{lemma:modebymodetrs}, we let $\mu(\indexnot):=\mfg_{\infty}(\indexnot)$ for $\indexnot\in\EFindexset$. Note that 
then $\mu(\indexnot)$ is strictly positive if and only if $\indexnot\in\EFtrsindexset$. Thus the notation $\EFtrsindexset$ used in the present
chapter coincides with the notation $\EFtrsindexset$ used in the statement of Lemma~\ref{lemma:modebymodetrs}. We have already clarified what 
the matrices $U,V$ and $W$ are, and due to Remark~\ref{remarks:Aindexnotproperties}, it is clear that $A(\indexnot)$ is of the form 
(\ref{eq:Aindexnotitomuindexnot}) for $\indexnot\in\EFtrsindexset$. Turning to the $A_{\indexnot,\rem}$ and $F_{\indexnot}$ appearing in 
Lemma~\ref{lemma:modebymodetrs}, we have already defined them in the case of $\indexnot\in\EFtrsindexset$; cf. Lemma~\ref{lemma:Aremtrsest}. 
The constants $\b_{\trs}$, $C_{\trs}$ and $T_{\trs}$ are all introduced in Definition~\ref{def:btrsTtrs} and Remark~\ref{remark:Ctrsdef}. The upper 
bound on the constant $\b_{\romar}$ appearing in Lemma~\ref{lemma:modebymodetrs} only depends on $U,V$ and $\b_{\trs}$. In the present setting, that 
means that it only depends on $\a_{\infty}$, $\zeta_{\infty}$, the limits appearing in (\ref{eq:gijinfetcdef}) and $\b_{\trs}$, as stated in the 
present lemma. Finally, the condition
$\|F_{\indexnot}\|_{\rolma}<\infty$ clearly holds for all $\indexnot\in\EFtrsindexset$. Thus Lemma~\ref{lemma:modebymodetrs} applies and we can 
appeal to the conclusions of that lemma. If (\ref{eq:thesystemRge}) is an $X$-non-trivial weakly transparent equation, we apply 
Lemma~\ref{lemma:modebymodetrs} twice, and there are two corresponding $\mu_{0}$ values, and two corresponding constants $C$. In that case, the 
constant $\mu_{0}$ appearing in the statement of the present lemma is the largest of these two $\mu_{0}$-values, and similarly for $C$. 

\textbf{Rough estimate.} Let us start by obtaining a rough estimate. Due to (\ref{eq:vkappaindexnotroughest}),
\begin{equation}\label{eq:vkappaindexnotroughesttrs}
|v(\indexnot,t)|\leq Ce^{(\kappa_{\indexnot}+\b_{\romar}/2)\bt}|v(\indexnot,T_{\trs})|
+Ce^{\b_{\romar}\bt/2}\int_{T_{\trs}}^{t}e^{\kappa_{\indexnot}(t-s)+\b_{\romar}\bs/2}|F_{\indexnot}(s)|ds
\end{equation}
for $t\geq T_{\trs}$ and $\mfg_{\infty}(\indexnot)\geq\mu_{0}$, where $C$ and $\mu_{0}$ have the same dependence as in 
Remark~\ref{remark:Cmuzdephighfre}. Note that it is not immediately obvious that $|v|$ and $\me^{1/2}$ are equivalent on $[T_{\trs},\infty)$, 
where the notation $\me$ is introduced in (\ref{eq:medefroughbalset}). In order to see that this is the case, note first that 
(\ref{eq:mfgintrsest}) and (\ref{eq:mfginftyintrsest}) hold, where the constant $C$ only depends on $K_{\trs}$ and $\eta_{\trs}$. Thus
there is a constant $C\geq 1$, depending only on $\roS_{\trs}$, $K_{\trs}$ and $\eta_{\trs}$, such that
\begin{equation}\label{eq:vEprelequiv}
C^{-1}[|\dot{z}|^{2}+\mfg^{2}_{\trs}|z|^{2}+|z|^{2}]\leq |v|^{2}\leq C[|\dot{z}|^{2}+\mfg^{2}_{\trs}|z|^{2}+|z|^{2}]
\end{equation}
on $[T_{\trs},\infty)$. Moreover, for $t\geq T_{\trs}$, (\ref{eq:Cinidef}), (\ref{eq:roSdef}), (\ref{eq:mfgintrsest}), (\ref{eq:mfginsilest})
and (\ref{eq:Ttrsdef}) yield
\begin{equation}\label{eq:mfgsilmfgtrslateequiv}
\begin{split}
\mfg(\indexnot_{\rosil},t) \leq & e^{-\b_{\rosil}t+c_{\rosil}}\mfg(\indexnot_{\rosil},0)\leq e^{-\b_{\rosil}t+c_{\rosil}}C_{\roini}|\nu(\indexnot_{\rosil})|
\leq C_{\roini}e^{c_{\rosil}}\\
 \leq & C_{\roini}e^{c_{\rosil}}\roS_{\trs}\mfg(\indexnot_{\trs},0)\leq C\mfg(\indexnot_{\trs},t),
\end{split}
\end{equation}
where $C$ only depends on $C_{\roini},c_{\rosil},K_{\trs},\roS_{\trs}$ and $\eta_{\trs}$. Appealing to 
(\ref{eq:mfGremCSest}) and (\ref{eq:mfgsilmfgtrslateequiv}) yields
\begin{equation*}
\begin{split}
\mfg^{2}(\indexnot,t) = & \mfg^{2}(\indexnot_{\trs},t)+\mfg^{2}(\indexnot_{\rosil},t)+\mfG_{\rem}(\indexnot,t)\\
 \geq & \frac{1}{2}\mfg^{2}(\indexnot_{\trs},t)-\mfg^{2}(\indexnot_{\rosil},t)
\geq \frac{1}{2}\mfg^{2}(\indexnot_{\trs},t)-C
\end{split}
\end{equation*}
for all $t\geq T_{\trs}$, where $C$ only depends on $C_{\roini}$ and $c_{\rosil}$. In particular, 
\[
\mfg^{2}(\indexnot_{\trs},t)+1\leq C[\mfg^{2}(\indexnot,t)+1]
\]
for all $t\geq T_{\trs}$, where $C$ only depends on $C_{\roini}$ and $c_{\rosil}$. Combining this estimate with 
(\ref{eq:vEprelequiv}) and (\ref{eq:mfgsilmfgtrslateequiv}) yields a constant $C\geq 1$ such that 
\begin{equation}\label{eq:mevequivtrans}
C^{-2}\me(t)\leq |v(t)|^{2}\leq C^{2}\me(t) 
\end{equation}
for $t\geq T_{\trs}$, where $C$ has the same dependence as in the case of (\ref{eq:mfgsilmfgtrslateequiv}). Combining this estimate with 
(\ref{eq:vkappaindexnotroughesttrs}) yields
\begin{equation}\label{eq:Etrslateest}
\me^{1/2}(t)\leq Ce^{(\kappa_{\indexnot}+\b_{\romar}/2)\bt}\me^{1/2}(T_{\trs})
+Ce^{\b_{\romar}\bt/2}\int_{T_{\trs}}^{t}e^{\kappa_{\indexnot}(t-s)+\b_{\romar}\bs/2}|F_{\indexnot}(s)|ds
\end{equation}
for $t\geq T_{\trs}$ and $\mfg_{\infty}(\indexnot)\geq\mu_{0}$, where $C$ and $\mu_{0}$ have the same dependence as in Remark~\ref{remark:Cmuzdephighfre}. 
Appealing to (\ref{eq:meestroughbalset}) yields
\begin{equation}\label{eq:ETtrstermest}
\begin{split}
Ce^{(\kappa_{\indexnot}+\b_{\romar}/2)\bt}\me^{1/2}(T_{\trs}) \leq & Ce^{c_{\betafun}}e^{(\kappa_{\indexnot}+\b_{\romar}/2)t}e^{\delta_{\trs}T_{\trs}}\me^{1/2}(0)\\
 & +Ce^{c_{\betafun}}e^{\delta_{\trs}T_{\trs}}\int_{0}^{T_{\trs}}e^{(\kappa_{\indexnot}+\b_{\romar}/2)(t-s)}|F_{\indexnot}(s)|ds,
\end{split}
\end{equation}
where $c_{\betafun}:=\|\betafun\|_{1}$, $C$ has the same dependence as in the case of (\ref{eq:Etrslateest}), and 
\[
\delta_{\indexnot,\trs}:=\max\{\eta_{\robal}-\kappa_{\indexnot},0\},\ \ \
\delta_{\trs}:=\sup_{\indexnot\in\EFtrsindexset}\delta_{\indexnot,\trs}.
\]
Note also that $\eta_{\robal}$ depends only on $C_{\ell}$ and $C_{\coeff}$. 
In order to estimate $e^{\de_{\trs}T_{\trs}}$ from above, note that (\ref{eq:Ttrsdef}) implies that
\begin{equation}\label{eq:edetrsTtrsest}
e^{\de_{\trs}T_{\trs}}=[1+|\nu(\indexnot_{\trs})|]^{\de_{\trs}/\min\{\eta_{\trs},\b_{\rosil}\}}[1+|\nu(\indexnot_{\rosil})|]^{2\de_{\trs}/\b_{\rosil}}
\leq C\ldr{\nu(\indexnot)}^{s_{0}},
\end{equation}
where $C$ only depends on $C_{\ell}$, $C_{\coeff}$, $\a_{\infty}$, $\zeta_{\infty}$, the limits appearing in (\ref{eq:gijinfetcdef}), $\b_{\rosil}$
and $\eta_{\trs}$; and 
\[
s_{0}:=[(\min\{\eta_{\trs},\b_{\rosil}\})^{-1}+2\b_{\rosil}^{-1}]\de_{\trs}.
\]
In particular, $s_{0}$ only depends on $\eta_{\trs}$, $\b_{\rosil}$, $C_{\ell}$, $C_{\coeff}$, the limits appearing in (\ref{eq:gijinfetcdef}), $\a_{\infty}$
and $\zeta_{\infty}$. Combining (\ref{eq:Etrslateest}), (\ref{eq:ETtrstermest}) and (\ref{eq:edetrsTtrsest}) yields
\begin{equation}\label{eq:EesttltTtrs}
\begin{split}
\me^{1/2}(t)\leq  & Ce^{(\kappa_{\indexnot}+\b_{\romar}/2)t}\ldr{\nu(\indexnot)}^{s_{0}}\me^{1/2}(0)\\
 & +Ce^{(\kappa_{\indexnot}+\b_{\romar}/2)t}\ldr{\nu(\indexnot)}^{s_{0}}\int_{0}^{t}e^{-(\kappa_{\indexnot}-\b_{\romar}/2)s}|F_{\indexnot}(s)|ds
\end{split}
\end{equation}
for $t\geq T_{\trs}$ and $\mfg_{\infty}(\indexnot)\geq\mu_{0}$, where $C$ and $\mu_{0}$ have the same dependence as in Remark~\ref{remark:Cmuzdephighfre}.
In order to obtain an estimate for $0\leq t\leq T_{\trs}$, we appeal to (\ref{eq:meestroughbalset}) again. This yields
\begin{align*}
\me^{1/2}(t) \leq & e^{c_{\betafun}+\eta_{\robal}t}\me^{1/2}(0)+e^{c_{\betafun}}\int_{0}^{t}e^{\eta_{\robal}(t-u)}|F_{\indexnot}(u)|du\\
 \leq & e^{c_{\betafun}}e^{\de_{\trs}T_{\trs}}e^{(\kappa_{\indexnot}+\b_{\romar}/2)t}\me^{1/2}(0)
+e^{c_{\betafun}}e^{\de_{\trs}T_{\trs}}\int_{0}^{t}e^{(\kappa_{\indexnot}+\b_{\romar}/2)(t-u)}|F_{\indexnot}(u)|du
\end{align*}
for $0\leq t\leq T_{\trs}$. Keeping (\ref{eq:edetrsTtrsest}) in mind, this estimate yields the conclusion that (\ref{eq:EesttltTtrs}) holds for 
$0\leq t\leq T_{\trs}$. Thus (\ref{eq:EesttltTtrs}) holds for all $t\geq 0$.

\textbf{Asymptotics, introduction.} In order to obtain asymptotics, we wish to appeal to Lemma~\ref{lemma:modebymodetrs}. However, before we do so, 
we need to impose restrictions on $\b_{\romar}$. Let, to this end, $\lambda_{j,\pm}$ be given by Definition~\ref{def:Einfindexnotb}, where
$U$ and $V$ are constructed from $\a_{\infty}$, $\zeta_{\infty}$ and the limits appearing in (\ref{eq:gijinfetcdef}) as described in the beginning
of the proof. Let $\kappa_{j,\pm}$ denote the real parts of $\lambda_{j,\pm}$ and let $\kappa_{\infty}$ denote the
largest of the $\kappa_{j,\pm}$. Given $\b_{\trs}>0$, there are two possibilities. Either there is a $\kappa_{j,\pm}$, say $\kappa_{c}$, such that 
$\kappa_{c}=\kappa_{\infty}-\b_{\trs}$, or there is no such $\kappa_{c}$. Regardless, there is a $\b_{\mathrm{spe}}>0$ (depending only on $\a_{\infty}$, 
$\zeta_{\infty}$, the limits appearing in (\ref{eq:gijinfetcdef}) and $\b_{\trs}$) such that there is no $\kappa_{j,\pm}$ in the interval 
$(\kappa_{\infty}-\b_{\trs},\kappa_{\infty}-\b_{\trs}+2\b_{\mathrm{spe}})$. Fix $0<\b_{\romar}<\b_{\mathrm{spe}}$ (we here also assume $\b_{\romar}$ to be 
small enough that Lemma~\ref{lemma:modebymodetrs} can be applied). Then, for $\mu_{0}$ large enough (the bound depending 
only on $\a_{\infty}$, $\zeta_{\infty}$, the limits appearing in (\ref{eq:gijinfetcdef}), $\b_{\trs}$ and $\b_{\romar}$) there is a constant $C$ (with the 
same dependence) such that if $\mfg_{\infty}(\indexnot)\geq\mu_{0}$, then 
\begin{itemize}
\item the generalised eigenspaces of $A(\indexnot)$ of which $E^{\infty}_{\indexnot,\b_{\trs}}$ consists are all the generalised eigenspaces such that 
the corresponding eigenvalues have real parts $>\kappa_{\indexnot}-\b_{\trs}+\b_{\romar}$,
\item if $e^{\infty}_{\indexnot,\b_{\trs}}$ is the space introduced in the statement of Lemma~\ref{lemma:expAintrestetaest}, then 
\begin{equation}\label{eq:eAteinfindnotbest}
|e^{A(\indexnot)t}\eta|\leq Ce^{(\kappa_{\indexnot}-\b_{\trs}+\b_{\romar}/2)t}|\eta|
\end{equation}
for all $t\geq 0$ and all $\eta\in e^{\infty}_{\indexnot,\b_{\trs}}$.
\end{itemize}
Appealing to Lemma~\ref{lemma:modebymodetrs} yields a $v_{\infty}\in E^{\infty}_{\indexnot,\b_{\trs}}$ such
that (\ref{eq:vvinfesttrsstmt}) holds for $t\geq T_{\trs}$, where $E^{\infty}_{\indexnot,\b_{\trs}}$ is given by Definition~\ref{def:Einfindexnotb}
(even though $v_{\infty}$ depends on $\indexnot$, we omit reference to this dependence for the sake of brevity). Introducing 
\begin{equation}\label{eq:vinfmoddeftrs}
v_{\infty,\romod}:=v_{\infty}-\int_{0}^{T_{\trs}}e^{-A(\indexnot)\tau}F_{\indexnot}(\tau)d\tau,
\end{equation}
the estimate (\ref{eq:vvinfesttrsstmt}) can be reformulated as 
\begin{equation}\label{eq:vvinfesttrsstmtextra}
\begin{split}
 & \left|v(t)-e^{A(\indexnot)t}v_{\infty,\romod}-\int_{0}^{t}e^{A(\indexnot)(t-s)}F_{\indexnot}(s)ds\right|\\
 \leq & Ce^{\kappa_{\indexnot} t-(\b_{\trs}-\b_{\romar})\bt}[e^{-\kappa_{\indexnot} T_{\trs}}|v(T_{\trs})|+\|F_{\indexnot}\|_{\rolma}]
\end{split}
\end{equation}
for $t\geq T_{\trs}$ and $\mfg_{\infty}(\indexnot)\geq\mu_{0}$, where $\mu_{0}$ and $C$ have the same dependence as in Remark~\ref{remark:Cmuzdephighfre}.
As in the proof of Lemma~\ref{lemma:roughas}, we need to project $v_{\infty,\romod}$ onto the correct subspace. However, we only
write down the corresponding estimates at the end of the argument.  

\textbf{The objects to be estimated.} In order to obtain the desired conclusion, we proceed as in the proof of Lemma~\ref{lemma:roughas}. 
As in that proof, we thus need to estimate
\begin{equation}\label{eq:zTodetermstrs}
e^{(\b_{\trs}-\kappa_{\indexnot})t}|v(t)|,\ \ \
e^{(\b_{\trs}-\kappa_{\indexnot})t}|e^{A(\indexnot)t}v_{\infty}|,\ \ \
\left|e^{(\b_{\trs}-\kappa_{\indexnot})t}\int_{t}^{T_{\trs}}e^{A(\indexnot)(t-\tau)}F_{\indexnot}(\tau)d\tau\right|
\end{equation}
for $t\in [0,T_{\trs}]$; this yields the desired estimate on the interval $[0,T_{\trs}]$. In order to obtain a good estimate for $t\geq T_{\trs}$, we need
to estimate 
\begin{equation}\label{eq:Todeinftytermstrs}
e^{(\b_{\trs}-\kappa_{\indexnot})T_{\trs}}|v(T_{\trs})|,\ \ \
e^{\b_{\trs}T_{\trs}}\|F_{\indexnot}\|_{\rolma}.
\end{equation}
Finally, we wish to estimate $v_{\infty,\romod}$ introduced in (\ref{eq:vinfmoddeftrs}). Note that if we have estimated the first expression appearing in 
(\ref{eq:zTodetermstrs}) on $[0,T_{\trs}]$, then we have also estimated the first expression appearing in (\ref{eq:Todeinftytermstrs}). 

\textbf{Estimating the solution.} Note, to begin with, that
\begin{equation}\label{eq:mfginfsplitestimatewithloss}
\mfg_{\infty}^{2}(\indexnot)\leq C\mfg^{2}(\indexnot_{\trs},t)\leq C[\mfg^{2}(\indexnot,t)+\mfg^{2}(\indexnot_{\rosil},t)]
\end{equation}
for $t\geq 0$, where $C$ only depends on $K_{\trs}$ and $\eta_{\trs}$ and we have appealed to (\ref{eq:mfgintrsest}), (\ref{eq:mfginftyintrsest}) 
and the fact that $\mfg_{\infty}(\indexnot)=\mfg_{\infty}(\indexnot_{\trs})$. We have also used the fact that (\ref{eq:mfGremCSest}) yields
\begin{equation}\label{eq:mfGremestdommfgsqindnottrs}
|\mfG_{\rem}(\indexnot,t)|\leq \frac{1}{2}\mfg^{2}(\indexnot_{\trs},t)+2\mfg^{2}(\indexnot_{\rosil},t)
\end{equation}
for $t\geq 0$; note that 
(\ref{eq:mfGremestdommfgsqindnottrs}) implies that 
\[
\mfg^{2}(\indexnot,t)\geq \frac{1}{2}\mfg^{2}(\indexnot_{\trs},t)-\mfg^{2}(\indexnot_{\rosil},t)
\]
for $t\geq 0$. The estimate (\ref{eq:mfginfsplitestimatewithloss}) yields
\[
|v(\indexnot,t)|\leq C[1+\mfg^{2}(\indexnot_{\rosil},t)]^{1/2}\me^{1/2}(\indexnot,t)\leq C\ldr{\nu(\indexnot)}\me^{1/2}(\indexnot,t)
\]
for $t\geq 0$, where $C$ only depends on $K_{\trs}$, $C_{\roini}$, $c_{\rosil}$ and $\eta_{\trs}$. Combining this estimate with 
Lemma~\ref{lemma:roughenestbalsetting} yields
\begin{equation}\label{eq:ebtrskaindexnottvprelest}
\begin{split}
e^{(\b_{\trs}-\kappa_{\indexnot})t}|v(\indexnot,t)|  \leq & Ce^{c_{\betafun}}\ldr{\nu(\indexnot)}[e^{\g_{\rohom,1}(\indexnot)t}\me^{1/2}(\indexnot,0)
+e^{\g_{\roih,1}(\indexnot)t}\int_{0}^{t}e^{-\kappa_{\indexnot}\tau}|\hf(\indexnot,\tau)|d\tau]
\end{split}
\end{equation}
for $t\in [0,T_{\trs}]$, where $\g_{\rohom,1}(\indexnot):=\max\{\b_{\trs}+\eta_{\robal}-\kappa_{\indexnot},0\}$; 
$\g_{\roih,1}(\indexnot):=\b_{\trs}+\max\{\eta_{\robal}-\kappa_{\indexnot},0\}$; and the constant $C$ only depends on $K_{\trs}$, $\eta_{\trs}$, $C_{\roini}$ and
$c_{\rosil}$. For future reference, it is useful to keep in mind that the same estimate holds with $\b_{\trs}$ 
replaced by $0$. Due to estimates of the form (\ref{eq:edetrsTtrsest}), it then follows that 
\begin{equation}\label{eq:vsupesztodetrs}
\begin{split}
 & \ldr{\nu(\indexnot)}^{s}\sup_{t\in [0,T_{\trs}]}e^{(\b_{\trs}-\kappa_{\indexnot})t}|v(\indexnot,t)| \\
\leq & C\ldr{\nu(\indexnot)}^{s+s_{\rohom,1}}\me^{1/2}(\indexnot,0)
+C\ldr{\nu(\indexnot)}^{s+s_{\roih,1}}\int_{0}^{T_{\trs}}e^{-\kappa_{\indexnot}\tau}|\hf(\indexnot,\tau)|d\tau
\end{split}
\end{equation}
for $\indexnot\in \EFindexsetr{\mu_{0}}{+}$, where 
\begin{align*}
s_{\rohom,1} := & 1+[(\min\{\eta_{\trs},\b_{\rosil}\})^{-1}+2\b_{\rosil}^{-1}]\textstyle{\sup}_{\indexnot\in \EFindexsetr{\mu_{0}}{+}}\g_{\rohom,1}(\indexnot),\\
s_{\roih,1} := & 1+[(\min\{\eta_{\trs},\b_{\rosil}\})^{-1}+2\b_{\rosil}^{-1}]\textstyle{\sup}_{\indexnot\in \EFindexsetr{\mu_{0}}{+}}\g_{\roih,1}(\indexnot)
\end{align*}
and the constant $C$ has the same dependence as in Remark~\ref{remark:Cmuzdephighfre}. Again, this estimate also holds when 
$\b_{\trs}$ is replaced by zero. One particular consequence of this estimate is 
\begin{equation}\label{eq:vFATodeesttrs}
\begin{split}
 & \ldr{\nu(\indexnot)}^{s}e^{(\b_{\trs}-\kappa_{\indexnot})T_{\trs}}|v(\indexnot,T_{\trs})|+\ldr{\nu(\indexnot)}^{s}e^{\b_{\trs}T_{\trs}}\|F_{\indexnot}\|_{\rolma}\\
 \leq & C\ldr{\nu(\indexnot)}^{s+s_{\rohom,1}}\me^{1/2}(\indexnot,0)+C\ldr{\nu(\indexnot)}^{s+s_{\roih,1}}\|\hf\|_{\rolma}
\end{split}
\end{equation}
for $\indexnot\in \EFindexsetr{\mu_{0}}{+}$, where the constant $C$ has the same dependence as in Remark~\ref{remark:Cmuzdephighfre} and we have 
used the notation (\ref{eq:Findexnotrolmadef}). Moreover, the analogous estimate with $\b_{\trs}$ replaced by $0$ holds. 

\textbf{Estimating the limiting value.} Next, let us turn to the problem
of estimating the second expression appearing in (\ref{eq:zTodetermstrs}). To begin with, $v_{\infty}$ is such that (\ref{eq:vinftyitouinfty}) 
and (\ref{eq:uinfest}) are satisfied, where $u_{\infty}\in E^{\infty}_{\indexnot,\b_{\trs}}$. Appealing to Lemma~\ref{lemma:expAintrestetaest},
it is clear that for $\mu_{0}$ large enough, the bound depending only on the limits appearing in (\ref{eq:gijinfetcdef}), $\a_{\infty}$,
$\zeta_{\infty}$ and $\b_{\trs}$, 
\begin{equation}\label{eq:vinftytrsfinest}
|v_{\infty}|\leq  Ce^{-(\kappa_{\indexnot}-\b_{\trs})T_{\trs}}|u_{\infty}|
\end{equation}
for $\indexnot\in\EFindexsetr{\mu_{0}}{+}$, where $C$ only depends on the limits appearing in (\ref{eq:gijinfetcdef}), $\a_{\infty}$,
$\zeta_{\infty}$ and $\b_{\trs}$. On the other hand, $u_{\infty}\in E^{\infty}_{\indexnot,\b_{\trs}}$ satisfies the estimate
\begin{equation}\label{eq:uinfesttrs}
|u_{\infty}|\leq C[|v(T_{\trs})|+e^{\kappa_{\indexnot}T_{\trs}}\|F_{\indexnot}\|_{\rolma}],
\end{equation}
where $C$ has the same dependence as in Remark~\ref{remark:Cmuzdephighfre}.
Combining (\ref{eq:vFATodeesttrs}), (\ref{eq:vinftytrsfinest}) and (\ref{eq:uinfesttrs}) yields
\begin{equation}\label{eq:vinftytrsfinsobest}
\ldr{\nu(\indexnot)}^{s}|v_{\infty}|\leq C\ldr{\nu(\indexnot)}^{s+s_{\rohom,1}}\me^{1/2}(0)+C\ldr{\nu(\indexnot)}^{s+s_{\roih,1}}\|\hf\|_{\rolma}
\end{equation}
for $\indexnot\in \EFindexsetr{\mu_{0}}{+}$, where $C$ has the same dependence as in Remark~\ref{remark:Cmuzdephighfre}. Moreover, the 
constants $s_{\rohom,1}$ and $s_{\roih,1}$ are defined in connection with (\ref{eq:vsupesztodetrs}).

In order to estimate the second expression in (\ref{eq:zTodetermstrs}), it is of interest to note that
\begin{equation}\label{eq:vinfztodeesttrs}
e^{A(\indexnot)t}v_{\infty}=e^{-A(\indexnot)(T_{\trs}-t)}u_{\infty}.
\end{equation}
Here we are only interested in the case that $t\in [0,T_{\trs}]$. The argument which led to the estimate (\ref{eq:vinftytrsfinest})
thus yields
\[
e^{(\b_{\trs}-\kappa_{\indexnot})t}|e^{A(\indexnot)t}v_{\infty}|\leq e^{(\b_{\trs}-\kappa_{\indexnot})t}Ce^{-(\kappa_{\indexnot}-\b_{\trs})(T_{\trs}-t)}|u_{\infty}|
\]
for $\indexnot\in \EFindexsetr{\mu_{0}}{+}$ and $t\in [0,T_{\trs}]$, where $C$ and the lower bound on $\mu_{0}$ only depend on the limits 
appearing in (\ref{eq:gijinfetcdef}), $\a_{\infty}$, $\zeta_{\infty}$ and $\b_{\trs}$. Combining this estimate with (\ref{eq:vFATodeesttrs}) 
and (\ref{eq:uinfesttrs}) yields
\begin{equation}\label{eq:vinfztodeestfintrs}
\ldr{\nu(\indexnot)}^{s}\sup_{t\in [0,T_{\trs}]}|e^{(\b_{\trs}-\kappa_{\indexnot})t}e^{A(\indexnot)t}v_{\infty}|
\leq C\ldr{\nu(\indexnot)}^{s+s_{\rohom,1}}\me^{1/2}(0)+C\ldr{\nu(\indexnot)}^{s+s_{\roih,1}}\|\hf\|_{\rolma}
\end{equation}
for $\indexnot\in \EFindexsetr{\mu_{0}}{+}$, where $C$ and $\mu_{0}$ have the same dependence as in Remark~\ref{remark:Cmuzdephighfre}.

\textbf{Estimating the right hand side.} Let us turn to 
\[
\int_{t}^{T_{\trs}}e^{A(\indexnot)(t-\tau)}F_{\indexnot}(\tau)d\tau.
\]
Due to Lemma~\ref{lemma:genepmaintest} and Remark~\ref{remark:Upmsuffinest}, we can, for a fixed $\eta_{\romar}>0$, estimate 
\[
\|e^{A(\indexnot)(t-\tau)}\|\leq Ce^{(\kappa_{b}-\eta_{\romar})(t-\tau)}
\]
for $t\leq\tau$ and $\mfg_{\infty}(\indexnot)\geq\mu_{0}$, where $\kappa_{b}$ ($\kappa_{a}$) is the smallest (largest) real part of an element 
in the set of eigenvalues $\lambda_{j,\pm}$, and $C$ and $\mu_{0}$ only depend on $\eta_{\romar}$, the limits appearing in (\ref{eq:gijinfetcdef}),
$\a_{\infty}$ and $\zeta_{\infty}$.

When estimating the last expression appearing in (\ref{eq:zTodetermstrs}), it is convenient to 
divide the situation into two cases; $\b_{\trs}>\kappa_{a}-\kappa_{b}$ and $\b_{\trs}\leq\kappa_{a}-\kappa_{b}$. In the first case, we choose
$\eta_{\romar}$ small enough, the bound depending only on $\b_{\trs}$, $\a_{\infty}$, $\zeta_{\infty}$ and the limits appearing in (\ref{eq:gijinfetcdef}), 
and then a $\mu_{0}$ large enough, with the same dependence, such that $\kappa_{b}-\kappa_{\indexnot}+\b_{\trs}-\eta_{\romar}>0$ for 
$\indexnot\in \EFindexsetr{\mu_{0}}{+}$. In the second case, we choose $\eta_{\romar}=\b_{\romar}/2$ and then 
$\mu_{0}$ large enough, the bound depending only on $\b_{\romar}$, $\a_{\infty}$, $\zeta_{\infty}$ and the limits appearing in (\ref{eq:gijinfetcdef}).
Summing up yields
\begin{equation}\label{eq:Fintfinesttrs}
\ldr{\nu(\indexnot)}^{s}\sup_{t\in [0,T_{\trs}]}\left|e^{(\b_{\trs}-\kappa_{\indexnot})t}\int_{t}^{T_{\trs}}e^{A(\indexnot)(t-\tau)}F_{\indexnot}(\tau)d\tau\right|
\leq C\ldr{\nu(\indexnot)}^{s+s_{\roih,2}}\|\hf\|_{\rolma}
\end{equation}
for $\mfg_{\infty}(\indexnot)\geq\mu_{0}$, where 
\[
s_{\roih,2}=\max\{\b_{\trs},\kappa_{a}-\kappa_{b}\}[(\min\{\eta_{\trs},\b_{\rosil}\})^{-1}+2\b_{\rosil}^{-1}]
\]
and $C$ and $\mu_{0}$ have the same dependence as in Remark~\ref{remark:Cmuzdephighfre}.

\textbf{Combining the estimates.} 
Since the the absolute value of the second term on the right hand side of (\ref{eq:vinfmoddeftrs}) (multiplied by $\ldr{\nu(\indexnot)}^{s}$) 
can be estimated by the left hand side of (\ref{eq:Fintfinesttrs}), we have all the estimates we need. To begin with, (\ref{eq:vinfmoddeftrs}), 
(\ref{eq:vinftytrsfinsobest}) and (\ref{eq:Fintfinesttrs}) yield
\begin{equation}\label{eq:vinfromodesttrs}
\ldr{\nu(\indexnot)}^{s}|v_{\infty,\romod}|\leq 
C\ldr{\nu(\indexnot)}^{s+s_{\infty,\rohom}}\me^{1/2}(0)+C\ldr{\nu(\indexnot)}^{s+s_{\infty,\roih}}\|\hf\|_{\rolma}
\end{equation}
for $\mfg_{\infty}(\indexnot)\geq\mu_{0}$, where $s_{\infty,\rohom}:=s_{\rohom,1}$; $s_{\infty,\roih}:=\max\{s_{\roih,1},s_{\roih,2}\}$; 
and $C$ and $\mu_{0}$ have the same dependence as in Remark~\ref{remark:Cmuzdephighfre}.
Moreover, combining (\ref{eq:vsupesztodetrs}), (\ref{eq:vinfztodeestfintrs}) and (\ref{eq:Fintfinesttrs}) yields
\begin{equation}\label{eq:vasadalltmodtrs}
\begin{split}
 & \ldr{\nu(\indexnot)}^{s}\left|v(t)-e^{A(\indexnot)t}v_{\infty,\romod}-\int_{0}^{t}e^{A(\indexnot)(t-\tau)}F_{\indexnot}(\tau)d\tau\right|\\
 \leq & Ce^{(\kappa_{\indexnot}-\b_{\trs})t}\left(\ldr{\nu(\indexnot)}^{s+s_{\infty,\rohom}}\me^{1/2}(0)+\ldr{\nu(\indexnot)}^{s+s_{\infty,\roih}}\|\hf\|_{\rolma}\right)
\end{split}
\end{equation}
for $0\leq t\leq T_{\trs}$ and $\mfg_{\infty}(\indexnot)\geq\mu_{0}$, where $C$ and $\mu_{0}$ have the same dependence as in 
Remark~\ref{remark:Cmuzdephighfre}. Combining (\ref{eq:vvinfesttrsstmtextra}) and (\ref{eq:vFATodeesttrs}) yields 
\begin{equation}\label{eq:vasadalltmodtrslat}
\begin{split}
 & \ldr{\nu(\indexnot)}^{s}\left|v(t)-e^{A(\indexnot)t}v_{\infty,\romod}-\int_{0}^{t}e^{A(\indexnot)(t-\tau)}F_{\indexnot}(\tau)d\tau\right|\\
 \leq & Ce^{(\kappa_{\indexnot}-\b_{\trs}+\b_{\romar})t}\left(\ldr{\nu(\indexnot)}^{s+s_{\infty,\rohom}}\me^{1/2}(0)
+\ldr{\nu(\indexnot)}^{s+s_{\infty,\roih}}\|\hf\|_{\rolma}\right)
\end{split}
\end{equation}
for $\mfg_{\infty}(\indexnot)\geq\mu_{0}$ and $t\geq T_{\trs}$, where $C$ and $\mu_{0}$ have the same dependence as in 
Remark~\ref{remark:Cmuzdephighfre}. Combining (\ref{eq:vasadalltmodtrs}) and (\ref{eq:vasadalltmodtrslat}), 
it is clear that the latter estimate holds for all $t\geq 0$. 

\textbf{Projection onto the right subspace.} Next we would like to define a function by the condition that its Fourier coefficients are given by 
$v_{\infty,\romod}$ and then sum up the estimates (\ref{eq:vasadalltmodtrslat}) in order to obtain (\ref{eq:ultmuzpartas}). However, we do not 
yet know that $v_{\infty,\romod}$ belongs to the correct space. We therefore have to project it to $E^{\infty}_{\indexnot,\b_{\trs}}$. Let $\Pi_{a,\indexnot}$
denote projection onto $E^{\infty}_{\indexnot,\b_{\trs}}$ and $\Pi_{b,\indexnot}$ denote projection onto $e^{\infty}_{\indexnot,\b_{\trs}}$. Then we need to 
estimate
\begin{equation*}
\begin{split}
 & \left|e^{A(\indexnot)t}\Pi_{b,\indexnot}\left(\int_{0}^{T_{\trs}}e^{-A(\indexnot)\tau}F_{\indexnot}(\tau)d\tau\right)\right| \\
 \leq & Ce^{(\kappa_{\indexnot}-\b_{\trs}+\b_{\romar}/2)t}\left|\int_{0}^{T_{\trs}}e^{-A(\indexnot)\tau}F_{\indexnot}(\tau)d\tau\right|
\end{split}
\end{equation*}
where we have appealed to (\ref{eq:eAteinfindnotbest}) and the constant only depends on $\a_{\infty}$, $\zeta_{\infty}$, the limits appearing in 
(\ref{eq:gijinfetcdef}) and $\b_{\romar}$; note that the norm of $\Pi_{b,\indexnot}$ is bounded by a constant depending only on $\a_{\infty}$, 
$\zeta_{\infty}$ and the limits appearing in (\ref{eq:gijinfetcdef}). Combining this estimate with (\ref{eq:Fintfinesttrs}) yields
\[
\ldr{\nu(\indexnot)}^{s}\left|e^{A(\indexnot)t}\Pi_{b,\indexnot}\left(\int_{0}^{T_{\trs}}e^{-A(\indexnot)\tau}F_{\indexnot}(\tau)d\tau\right)\right|
\leq C\ldr{\nu(\indexnot)}^{s+s_{\roih,2}}e^{(\kappa_{\indexnot}-\b_{\trs}+\b_{\romar}/2)t}\|\hf\|_{\rolma}
\]
for all $t\geq 0$ and all $s$, where $C$ has the same dependence as in Remark~\ref{remark:Cmuzdephighfre}. Combining this estimate with 
(\ref{eq:vasadalltmodtrslat}), it is clear that we can replace $v_{\infty,\romod}$ with $\Pi_{a,\indexnot}[v_{\infty,\romod}]$ on the left hand side of 
(\ref{eq:vasadalltmodtrslat}).

\textbf{Summing up.}
As indicated previously, the derivation of the asymptotics proceeds in several steps. In the present proof, we derive asymptotics for the part 
of the solution that corresponds to $\indexnot\in\EFindexset$ such that $\mfg_{\infty}(\indexnot)$ is large. Due to the above arguments, there are 
$\mu_{0}$ and $C$, with dependence as in Remark~\ref{remark:Cmuzdephighfre}, such that if $u$ is a solution to (\ref{eq:thesystemRge}), then the modes of 
$u$ corresponding to $\indexnot\in\EFindexsetr{\mu_{0}}{+}$ satisfy (\ref{eq:vasadalltmodtrslat}) for $t\geq 0$, where $v_{\infty,\romod}(\indexnot)$ 
can be replaced by $\Pi_{a,\indexnot}[v_{\infty,\romod}(\indexnot)]$. In particular, for each such mode, we have constructed an amplitude 
$\Pi_{a,\indexnot}[v_{\infty,\romod}(\indexnot)]$. Combining these amplitudes then yields a function, defined as follows. Given 
$p\in\bM$, define $U_{\infty}(p)$ by
\[
U_{\infty}(p):=\textstyle{\sum}_{\indexnot\in\EFindexsetr{\mu_{0}}{+}}\Pi_{a,\indexnot}[v_{\infty,\romod}(\indexnot)]\varphi_{\indexnot}(p),
\]
where $\varphi_{\indexnot}$ is given by (\ref{eq:varphinudef}). In order to justify that the sum on the right hand side makes sense and yields a 
smooth function, it is sufficient to appeal to the fact that $u$ is smooth; the assumption that $\|f\|_{\trs,s}<\infty$ for all $s\in\ro$; 
(\ref{eq:vinfromodesttrs}); and the Minkowski inequality. In fact, (\ref{eq:Uinfsobnoest}) holds, where $C$ and $\mu_{0}$ have the same dependence 
as in Remark~\ref{remark:Cmuzdephighfre}. Similarly, (\ref{eq:vasadalltmodtrslat}) implies that (\ref{eq:ultmuzpartas}) holds, where $C$ and $\mu_{0}$ 
have the same dependence as in Remark~\ref{remark:Cmuzdephighfre}. That the $\indexnot$'th Fourier coefficient belongs to a space of the stated type 
follows from the observations made above concerning $E^{\infty}_{\indexnot,\b_{\trs}}$.

\textbf{Uniqueness.} 
In order to justify the uniqueness, let us assume that there are two functions $U_{\infty,i}$, $i=1,2$, satisfying the assumptions of the lemma. 
Assume, moreover, that there is a $\indexnot\in\EFindexsetr{\mu_{0}}{+}$ such that the corresponding Fourier coefficients, say $v_{\infty,i}(\indexnot)$,
$i=1,2$, differ. Then (\ref{eq:ultmuzpartas}) implies that 
\begin{equation}\label{eq:Uinfuniqpf}
|e^{A(\indexnot)t}[v_{\infty,1}(\indexnot)-v_{\infty,2}(\indexnot)]|\leq C\ldr{t}^{\de_{\indexnot}-1}e^{(\kappa_{\indexnot}-\b_{\trs}+\b_{\romar})t}
\end{equation}
for $t\geq 0$, where the constant $C$ depends on the solution and $f$. On the other hand, the left hand side grows faster than the right hand side
due to the assumption that $v_{\infty,i}(\indexnot)\in E_{\indexnot}$, where $E_{\indexnot}$ has properties stated in the lemma. This yields the desired 
contradiction and the lemma follows. 
\end{proof}

Finally, we are in a position to consider all the $\indexnot\in\EFtrsindexset$ simultaneously. We do so by combining 
the results of Chapter~\ref{chapter:weaksil} with Lemma~\ref{lemma:trsasymptlargefre}.

\begin{prop}\label{prop:trsasympt}
Assume that (\ref{eq:thesystemRge}) is weakly transparent in the sense of Definition~\ref{def:weaktrans}, weakly balanced in the sense of 
Definition~\ref{def:roughODEtermo} and weakly convergent in the sense of Definition~\ref{def:roughODEtermo}. Assume, finally, that 
there is a constant $0<C_{\ell}\in\ro$ and a continuous non-negative function $\betafun\in L^{1}([0,\infty))$ 
such that (\ref{eq:elldotbdgeneral}) holds for all $0\neq\indexnot\in\EFindexset$ and all $t\geq 0$. Fix a $0<\b_{\romar}\in\ro$ which is small 
enough, the bound depending only on $\a_{\infty}$, $\zeta_{\infty}$, the limits appearing in (\ref{eq:gijinfetcdef}) and the constant
$\b_{\trs}$ introduced in Definition~\ref{def:btrsTtrs}. Assume $f$ to be such that $\|f\|_{\trs,s}<\infty$ for all $s\in\ro$, where 
$\|\cdot\|_{\trs,s}$ is defined in (\ref{eq:ftrssnorm}). Then there are constants $\sigma_{\rohom},\sigma_{\roih}\in [0,\infty)$, $0\leq N\in \zo$ and 
$0< C\in\ro$ such that the following holds. Given a solution $u$ to (\ref{eq:thesystemRge}), there is a unique 
$U_{\infty}\in C^{\infty}(\bM,\cn{2m})$ with the following properties
\begin{itemize}
\item the $\indexnot$'th Fourier coefficient of $U_{\infty}$ vanishes unless $\indexnot\in\EFtrsindexset$, 
\item if $\indexnot\in\EFtrsindexset$, then the $\indexnot$'th Fourier coefficient of $U_{\infty}$ belongs to 
$E_{\indexnot}$, where $E_{\indexnot}$ is the first generalised eigenspace in the $(\b_{\trs}-\b_{\romar}),A(\indexnot)$-decomposition 
of $\cn{2m}$, and $A(\indexnot)$ is given by (\ref{eq:Aindexnotdeftrs}),
\item the estimate
\begin{equation}\label{eq:ultmuzpartasfin}
\begin{split}
 & \left\|\left(\begin{array}{c} (-\Delta_{\infty})^{1/2}u(\cdot,t) \\ \d_{t}u(\cdot,t)\end{array}\right)
-e^{\ma_{\trs}t}U_{\infty}-\int_{0}^{t}e^{\ma_{\trs}(t-\tau)}\left(\begin{array}{c} 0 \\ f(\cdot,\tau)\end{array}\right)d\tau\right\|_{t,s}\\
 \leq & C\ldr{t}^{N}e^{-(\b_{\trs}-\b_{\romar})t}[\|u_{t}(\cdot,0)\|_{(s+\sigma_{\rohom})}+\|u(\cdot,0)\|_{(s+\sigma_{\rohom}+1)}
+\|f\|_{\trs,s+\sigma_{\roih}}]
\end{split}
\end{equation}
holds for $t\geq 0$ and $s\in\ro$, where $\Delta_{\infty}$ is defined in (\ref{eq:Deltainftydef}), $\ma_{\trs}$ is defined in (\ref{eq:matrsdef})
and $\|\cdot\|_{t,s}$ is defined in (\ref{eq:tsnodef}).
\end{itemize}
Moreover, 
\begin{equation}\label{eq:Uinfsobnoestfin}
\|U_{\infty}\|_{(s)}\leq C\|u_{t}(\cdot,0)\|_{(s+\sigma_{\rohom})}+C\|u(\cdot,0)\|_{(s+\sigma_{\rohom}+1)}+
C\|f\|_{\trs,s+\sigma_{\roih}}
\end{equation}
for all $s\in\ro$. 
\end{prop} 
\begin{remark}\label{remark:constdepfintrs}
The constants $C$, $N$, $\sigma_{\rohom}$ and $\sigma_{\roih}$ have the following dependence: $C$ only depends on $C_{\roini}$, $g^{ij}(0)$, $a_{r}(0)$, 
$\a_{\infty}$, $\zeta_{\infty}$, the limits appearing in (\ref{eq:gijinfetcdef}), $\roS_{\trs}$, $\b_{\rosil}$, $\b_{\romar}$, $\eta_{\romn}$, $\eta_{\trs}$, 
$c_{\rosil}$, $K_{\trs}$, $C_{\ell}$, $c_{\betafun}$, $C_{\coeff}$, $C_{\romn}$ and the spectra of the Laplace-Beltrami operators on the Riemannian manifolds 
$(M_{r_{j}},g_{r_{j}})$, $j=1,\dots,R_{\trs}$; $N$ only depends on $m$; and $\sigma_{\rohom}$ and $\sigma_{\roih}$ only depend on $\eta_{\romn}$, 
$\eta_{\trs}$, $\b_{\rosil}$, $\b_{\romar}$, $c_{\rosil}$, $C_{\coeff}$, $C_{\ell}$, $K_{\trs}$, $\a_{\infty}$, $\zeta_{\infty}$, the limits appearing 
in (\ref{eq:gijinfetcdef}) and the spectra of the Laplace-Beltrami operators on the Riemannian manifolds $(M_{r_{j}},g_{r_{j}})$, $j=1,\dots,R_{\trs}$.
\end{remark}
\begin{remark}\label{remark:geometrictoanalyticweaksil}
If the conditions of Proposition~\ref{prop:trsasymptintro} are satisfied, then the assumptions of the present proposition are satisfied. In order 
to justify this statement, note first that since Definitions~\ref{def:expandconvdir} and \ref{def:shiftnegligibletrs} are fulfilled, the conclusions
of Lemma~\ref{lemma:geometrictoanalytictrans} hold. Thus the first three limits appearing in (\ref{eq:gijinfetcdef}) exist. Turning to the fourth
limit, it also exists due to (\ref{eq:mcXconvsildir}), Remark~\ref{remark:Xnormbdsiltrans} and the fact that $\mcX$ is $C^{0}$-future bounded. 
Moreover, it is clear that if there is an $X^{j_{k}}_{\infty}\neq 0$, then Definition~\ref{def:Xnontriv} is satisfied. If, on the other hand, 
$X^{j_{k}}_{\infty}=0$ for all the $j_{k}$, then Definition~\ref{def:Xdegenerate} is satisfied; cf. the conclusions of 
Lemma~\ref{lemma:geometrictoanalytictrans}. Thus the first condition of Definition~\ref{def:weaktrans} is satisfied. That the second and third
conditions are satisfied follows from Lemma~\ref{lemma:geometrictoanalytictrans}. The fourth condition follows from (\ref{eq:mcXconvsildir}).
Thus (\ref{eq:thesystemRge}) is weakly transparent in the sense of Definition~\ref{def:weaktrans}. Combining the fact that (\ref{eq:alpahzetaconvest})
holds; the fact that $\mcX$ is $C^{0}$-future bounded; the fact that Definition~\ref{def:shiftnegligibletrs} is satisfied; and 
Lemma~\ref{lemma:sigmaXbdsandderbds}, yields the conclusion that (\ref{eq:thesystemRge}) is weakly balanced and weakly convergent; cf. 
Definition~\ref{def:roughODEtermo}. Finally, there is a constant $0<C_{\ell}\in\ro$ 
such that (\ref{eq:elldotbdgeneral}) (with $\betafun=0$) holds for all $0\neq\indexnot\in\EFindexset$ and all $t\geq 0$; this is a consequence of
the fact that the second fundamental form is $C^{0}$-future bounded and Definition~\ref{def:shiftnegligibletrs}; cf. 
Lemma~\ref{lemma:condyieldellderbd}. In addition to the above, it is of interest to note that the constants $\b_{\rosil}$, $\eta_{\trs}$ and 
$\eta_{\romn}$ appearing in Proposition~\ref{prop:trsasymptintro} are the same as the $\b_{\rosil}$, $\eta_{\trs}$ and $\eta_{\romn}$ appearing in the 
statement of Proposition~\ref{prop:trsasympt}.
\end{remark}
\begin{proof}
Consider a solution $u$ to (\ref{eq:thesystemRge}). In Lemma~\ref{lemma:trsasymptlargefre}, we derive asymptotics for the Fourier modes
such that $\mfg_{\infty}(\indexnot)\geq\mu_{0}$ for a suitably chosen $0<\mu_{0}\in\ro$. What remains is to analyse the asymptotics for Fourier 
modes such that $\mfg_{\infty}(\indexnot)\leq\mu_{0}$. To this end, it is convenient to partially Fourier transform
(\ref{eq:thesystemRge}). In fact, in analogy with the terminology introduced in (\ref{eq:varphinudef})--(\ref{eq:ldrboxdef}), we here
introduce the following notation. First, 
\begin{equation}\label{eq:bMdeftrs}
\bM_{\trs}:=\tn{d_{\trs}}\times \textstyle{\prod}_{k=1}^{R_{\trs}}M_{r_{k}},
\end{equation} 
where $d_{\trs}$, $R_{\trs}$ are introduced in Definitions~\ref{def:Xnontrtrsdivvar} and \ref{def:degtrsdivvar} (below we use additional 
notation introduced in these definitions without further comment). Note that the manifold $\bM_{\trs}$ corresponds to the transparent
variables. Analogously, one can define $\bM_{\rosil}$ corresponding to the silent variables. In analogy with (\ref{eq:varphinudef}), we
introduce
\begin{equation}\label{eq:varphinudeftrs}
\varphi_{\indexnot,\trs}(p):=(2\pi)^{-d_{\trs}/2}\textstyle{\prod}_{k=1}^{d_{\trs}}
e^{in_{j_{k}}\cdot x^{j_{k}}}\cdot\prod_{k=1}^{R_{\trs}}\varphi_{r_{k},i_{r_{k}}}(p_{r_{k}})
\end{equation}
for $p\in\bM$ and $\indexnot\in\EFindexset$ (note, however, that the right hand side only depends on $\indexnot_{\trs}$). Note that we can 
consider $\varphi_{\indexnot,\trs}$ to be a function on $\bM_{\trs}$. Introduce the $(d_{\rosil}+R_{\rosil})$-form $\musilbox$ on $\bM$ by 
\begin{equation}\label{eq:muboxdefsil}
\musilbox:=\textstyle{\bigwedge}_{k=1}^{d_{\rosil}}dx^{\bj_{k}}\wedge \bigwedge_{k=1}^{R_{\rosil}}\mu_{g_{\bre_{k}}}
\end{equation}
and the $(d_{\trs}+R_{\trs})$-form $\mutrsbox$ on $\bM$ by 
\begin{equation}\label{eq:muboxdeftrs}
\mutrsbox:=\pm \textstyle{\bigwedge}_{k=1}^{d_{\trs}}dx^{j_{k}}\wedge \bigwedge_{k=1}^{R_{\trs}}\mu_{g_{r_{k}}},
\end{equation}
where the sign is chosen so that 
\[
\mutrsbox\wedge\musilbox=\mubox.
\]
It is convenient to introduce 
\begin{equation}\label{eq:ldrboxdeftrs}
\ldrtrsbox{u,v}:=\int_{\bM_{\trs}}uv^{*}\mutrsbox
\end{equation}
for $u\in C^{\infty}(\bM,\cn{k})$ and $v\in C^{\infty}(\bM,\co)$, where the star denotes complex conjugation. Note that 
$\ldrtrsbox{u,v}$ is a function on $\bM_{\rosil}$. In particular, given a solution $u$ to (\ref{eq:thesystemRge}), we 
define (with slight abuse of notation)
\[
\tu(\indexnot_{\trs},p_{\rosil},t):=\ldrtrsbox{u(\cdot,t),\varphi_{\indexnot_{\trs},\trs}}.
\]
This function satisfies the equation
\begin{equation}\label{eq:thesystemRgepafo}
\begin{split}
\tu_{tt}-\textstyle{\sum}_{k,l=1}^{d_{\rosil}}g^{\bj_{k}\bj_{l}}(t)\d_{\bj_{k}}\d_{\bj_{l}}\tu-2\sum_{l=1}^{d_{\rosil}}g^{0\bj_{l}}(t)\d_{\bj_{l}}\d_{t}\tu
-\sum_{k=1}^{R_{\rosil}}a^{-2}_{\bre_{k}}(t)\Delta_{g_{\bre_{k}}}\tu & \\
+\a_{\romod}(t)\tu_{t}+\textstyle{\sum}_{k=1}^{d_{\rosil}}X_{\romod}^{\bj_{k}}(t)\d_{\bj_{k}}\tu+\zeta_{\romod}(t)\tu & = \tf,
\end{split}
\end{equation}
where $\tf$ is defined in the same way as $\tu$ and 
\begin{align*}
\a_{\romod}(t) := & \a(t)-2\textstyle{\sum}_{k=1}^{d_{\trs}}in_{j_{k}}g^{0j_{k}}(t)\Id_{m},\\
\zeta_{\romod}(t) := & \zeta(t)+\mfg^{2}(\indexnot_{\trs},t)\Id_{m}+i\textstyle{\sum}_{k=1}^{d_{\trs}}n_{j_{k}}X^{j_{k}}(t),\\
X^{\bj_{k}}_{\romod}(t) := & X^{\bj_{k}}(t)-2\textstyle{\sum}_{l=1}^{d_{\trs}}in_{j_{l}}g^{j_{l}\bj_{k}}(t)\Id_{m}.
\end{align*}
Next, we wish to verify that (\ref{eq:thesystemRgepafo}) is an equation such that the results of Chapter~\ref{chapter:weaksil} apply. 
We do so in several steps.

\textbf{Properties of the metric.} As a first step, we verify that the metric on $M_{\rosil}:=\bM_{\rosil}\times I$, say $g_{\rosil}$, 
associated with the equation (\ref{eq:thesystemRgepafo}) is such that $(M_{\rosil},g_{\rosil})$ is a canonical separable cosmological 
model manifold. Note that $g^{00}_{\rosil}(t)=-1$ and that $g^{kl}_{\rosil}(t)=g^{\bj_{k}\bj_{l}}(t)$ are the components of a positive definite
matrix for all $t\in I$. Finally, $g^{0l}_{\rosil}(t)=g^{0\bj_{l}}(t)$. As a consequence, there are smooth functions $g_{\rosil,\a\b}$, 
$\a,\b\in\{0,\dots,d_{\rosil}\}$, from $I$ to $\ro$ such that $g_{\rosil,\a\b}(t)$ are the components of the inverse of the matrix with
components $g^{\a\b}_{\rosil}(t)$. 
Moreover, $g_{\rosil,00}(t)<0$ and $g_{\rosil,ij}(t)$ are the components of a positive definite matrix for all $t$. These statements are a 
consequence of \cite[Lemma~8.5, p.~72]{minbok}. To conclude, $g_{\rosil}$ can be written
\[
g_{\rosil}=g_{\rosil,00}dt\otimes dt+g_{\rosil,0i}dt\otimes dx^{i}+g_{\rosil,i0}dx^{i}\otimes dt+g_{\rosil,ij}dx^{i}\otimes dx^{j}
+\textstyle{\sum}_{j=1}^{R_{\rosil}}a_{\bre_{j}}^{2}g_{\bre_{j}}
\]
Moreover, it is clear that $(M_{\rosil},g_{\rosil})$ is a separable cosmological model manifold; that $g_{\rosil}$ is a Lorentz metric follows
from \cite[Lemma~8.5, p.~72]{minbok}. Due to Remark~\ref{remark:huzzformgenerallapse}, it is clear that the lapse function equals one.
Finally, $I$ contains $[0,\infty)$, so that $(M_{\rosil},g_{\rosil})$ is a canonical separable cosmological model manifold.

\textbf{Estimating the coefficients of the equation.}
Let us start by considering $\a_{\romod}$ and $\zeta_{\romod}$. To begin with, 
\begin{equation}\label{eq:amodroest}
\|\a_{\romod}(t)\|\leq \|\a(t)\|+2|\sigma(\indexnot_{\trs},t)|\cdot\mfg(\indexnot_{\trs},t)\leq C(1+\mu_{0})
\end{equation}
for all $t\geq 0$ and all $\indexnot\in \EFindexsetr{\mu_{0}}{-}$, where the notation $\EFindexsetr{\mu_{0}}{-}$ is introduced in 
connection with (\ref{eq:tsmuznodef}). Here $C$ only depends on $C_{\coeff}$ appearing in (\ref{eq:weakbal}) and $K_{\trs}$ and $\eta_{\trs}$ 
appearing in (\ref{eq:weaktrs}). Moreover, we have appealed to the estimates (\ref{eq:mfgintrsest}), (\ref{eq:mfginftyintrsest}) and the 
bound $\mfg_{\infty}(\indexnot)\leq\mu_{0}$. Similarly, 
\begin{equation}\label{eq:zetamodroest}
\|\zeta_{\romod}(t)\|\leq \|\zeta(t)\|+\mfg^{2}(\indexnot_{\trs},t)+\|X(\indexnot_{\trs},t)\|\cdot\mfg(\indexnot_{\trs},t)\leq 
C(1+\mu_{0}^{2})
\end{equation}
for all $t\geq 0$ and all $\indexnot\in \EFindexsetr{\mu_{0}}{-}$,
where $C$ only depends on $C_{\coeff}$, $K_{\trs}$ and $\eta_{\trs}$, and we have appealed to the estimates (\ref{eq:mfgintrsest}), 
(\ref{eq:mfginftyintrsest}) and the bound $\mfg_{\infty}(\indexnot)\leq\mu_{0}$. Finally, for $\indexnot\in \EFindexsetr{\mu_{0}}{-}$
such that $\indexnot_{\rosil}\neq 0$, 
\[
\|X_{\romod}(\indexnot_{\rosil},t)\|=\frac{\|n_{\bj_{k}}X^{\bj_{k}}_{\romod}(t)\|}{\mfg(\indexnot_{\rosil},t)}\leq 
\frac{\|n_{\bj_{k}}X^{\bj_{k}}(t)\|}{\mfg(\indexnot_{\rosil},t)}
+2\frac{|n_{\bj_{k}}n_{j_{l}}g^{j_{l}\bj_{k}}(t)|}{\mfg(\indexnot_{\rosil},t)}.
\]
The first term on the far right hand side is bounded by $C_{\coeff}$ for $t\geq 0$. Due to (\ref{eq:mfGremCSest}), the second term on the 
far right hand side is bounded by $2\mfg(\indexnot_{\trs},t)$. Thus
\begin{equation}\label{eq:Xmodesttrs}
\|X_{\romod}(\indexnot_{\rosil},t)\|\leq C_{\coeff}+2\mfg(\indexnot_{\trs},t)\leq C(1+\mu_{0})
\end{equation}
for all $t\geq 0$ and all $\indexnot\in \EFindexsetr{\mu_{0}}{-}$ such that $\indexnot_{\rosil}\neq 0$. Here $C$ only depends on $C_{\coeff}$, 
$K_{\trs}$ and $\eta_{\trs}$. Introducing 
\begin{align*}
\a_{\romod,\infty} := & \a_{\infty},\\
\zeta_{\romod,\infty} := & \zeta_{\infty}+\mfg_{\infty}^{2}(\indexnot_{\trs})\Id_{m}+i\textstyle{\sum}_{k=1}^{d_{\trs}}n_{j_{k}}X^{j_{k}}_{\infty},
\end{align*}
it can similarly be verified that 
\begin{align}
\|\a_{\romod}(t)-\a_{\romod,\infty}\| \leq & C(1+\mu_{0})e^{-\eta_{\romod,\romn}t},\label{eq:amodas}\\
\|\zeta_{\romod}(t)-\zeta_{\romod,\infty}\| \leq & C(1+\mu_{0}^{2})e^{-\eta_{\romod,\romn}t}\label{eq:zetamodas}
\end{align}
for all $t\geq 0$ and all $\indexnot\in \EFindexsetr{\mu_{0}}{-}$,
where $C$ only depends on $C_{\romn}$, $C_{\roini}$, $K_{\trs}$ and $\eta_{\trs}$; $\eta_{\romod,\romn}:=\min\{\eta_{\romn},\eta_{\trs}\}$; and we
have appealed to (\ref{eq:alazeas}), (\ref{eq:weaktrs}), (\ref{eq:Xjtrsrocon}), (\ref{eq:Cinidef}), (\ref{eq:mfgintrsest}), 
(\ref{eq:mfginftyintrsest}) and (\ref{eq:mfgsqdifftrsest}). It is thus clear that (\ref{eq:thesystemRgepafo}) is weakly 
convergent in the sense of Definition~\ref{def:roughODEtermo}; in (\ref{eq:alazeas}) we simply have to replace $\eta_{\romn}$
by $\eta_{\romod,\romn}$ and $C_{\romn}$ by $C_{\romod,\romn}:=C(2+\mu_{0}+\mu_{0}^{2})$, where $C$ is the constant appearing in 
(\ref{eq:amodas}) and (\ref{eq:zetamodas}). 

\textbf{Weak balance; weak silence; and generalised eigenspaces.}
Before turning to the topic of weak balance, it is important to note that there is one $\mfg$, $\sigma$ and $X$ associated with the 
equation (\ref{eq:thesystemRge}) and one $\mfg$, $\sigma$ and $X$ associated with the equation (\ref{eq:thesystemRgepafo}) (say $\mfg_{\romod}$, 
$\sigma_{\romod}$ and $X_{\romod}$), and it is important not to confuse them with one another. Moreover, the set $\EFindexset$ is different for 
the two equations. However, we can think of $\EFsilindexset$ as the set $\EFindexset$ corresponding to the equation (\ref{eq:thesystemRgepafo}).
Moreover, in practice, $\mfg_{\romod}=\mfg$ and $\sigma_{\romod}=\sigma$, assuming that we only evaluate $\mfg$ and $\sigma$ at 
$\indexnot\in\EFsilindexset$.  Finally, we should consider 
$\indexnot_{\trs}\neq 0$ to be fixed once and for all. Combining (\ref{eq:weakbal}) (which holds for (\ref{eq:thesystemRge}) by assumption)
with (\ref{eq:amodroest}), (\ref{eq:zetamodroest}) and (\ref{eq:Xmodesttrs}), it follows that (\ref{eq:thesystemRgepafo}) is weakly balanced 
(and the constant $C_{\coeff}$ is replaced by a constant depending only on $C_{\coeff}$, $K_{\trs}$, $\eta_{\trs}$ and $\mu_{0}$). Finally, that 
(\ref{eq:thesystemRgepafo}) is weakly silent is an immediate consequence of (\ref{eq:weaksiltrs}). When appealing to 
Lemma~\ref{lemma:roughas}, it is important to keep track of what $\kappa_{1}$, $\b_{\rem}$ and $E_{a}$ are. In the present
context, $\kappa_{1}=\kappa_{\indexnot}$, where the notation $\kappa_{\indexnot}$ is 
introduced in connection with (\ref{eq:tsmuznodef}); this is justified by (\ref{eq:Aromoddef})--(\ref{eq:Dinfconj}) below. Moreover,
$\b_{\rem}=\min\{\eta_{\romod,\romn},\b_{\rosil}\}=\min\{\eta_{\romn},\eta_{\trs},\b_{\rosil}\}$, so that $\b_{\rem}=\b_{\trs}$; cf. 
(\ref{eq:btrsdef}). Comparing the statement of the present proposition with the statement of Lemma~\ref{lemma:roughas}, it is clear that 
we would like to appeal to Lemma~\ref{lemma:roughas} with $\b_{\rem}$ replaced by $\b_{\trs}-\b_{\romar}$. Formally, this can be achieved by, 
in the application of Lemma~\ref{lemma:roughas}, replacing $\eta_{\romod,\romn}$ by $\b_{\trs}-\b_{\romar}$. Appealing to Lemma~\ref{lemma:roughas}
with this modification, $E_{a}$ is the direct sum of the generalised eigenspaces of $A_{\romod}$ (defined below) corresponding to eigenvalues
with real part $>\kappa_{\indexnot}-(\b_{\trs}-\b_{\romar})$. 

\textbf{Applying the results of Chapter~\ref{chapter:weaksil}.}
When comparing with the results of Chapter~\ref{chapter:weaksil}, the matrix $A_{\infty}$ is replaced by 
\begin{equation}\label{eq:Aromoddef}
A_{\romod}:=\left(\begin{array}{cc} 0 & \Id_{m} \\ -\zeta_{\infty}-\mfg_{\infty}^{2}\Id_{m}-i\sum_{k=1}^{d_{\trs}}n_{j_{k}}X^{j_{k}}_{\infty} & -\a_{\infty}
\end{array}\right). 
\end{equation}
Introducing the matrix 
\begin{equation}\label{eq:Dinfdef}
D_{\infty}:=\diag(\mfg_{\infty}^{-1}\Id_{m},\Id_{m}),
\end{equation}
it can be verified that 
\begin{equation}\label{eq:Dinfconj}
D_{\infty}^{-1}A_{\romod}D_{\infty}=A(\indexnot),
\end{equation}
where $A(\indexnot)$ is given by (\ref{eq:Aindexnotdeftrs}). Due to this observation, it is clear that the requirement (\ref{eq:fhsintbd})
can be reformulated to 
\begin{equation}\label{eq:odetrtotrcondonf}
\int_{0}^{\infty}e^{-\kappa_{\indexnot}\tau}\|\tf(\indexnot_{\trs},\cdot,\tau)\|_{(s)}d\tau<\infty.
\end{equation}
Moreover, due to the requirement $\|f\|_{\trs,s}<\infty$ for all $s\in\ro$ and the definition (\ref{eq:ftrssnorm}), it is clear 
that (\ref{eq:odetrtotrcondonf}) holds. Let $\b_{\rem}=\b_{\trs}-\b_{\romar}$ be defined as above and $E_{a}$ be the first generalised eigenspace
in the $\b_{\rem},A_{\romod}$-decomposition of $\cn{2m}$ (note that $D_{\infty}^{-1}E_{a}$ is the first generalised eigenspace in the 
$(\b_{\trs}-\b_{\romar}),A(\indexnot)$-decomposition of $\cn{2m}$). Then Lemma~\ref{lemma:roughas} is applicable, and there are constants 
$C$, $N$, $s_{\rohom}$
and $s_{\roih}$ such that, given a solution $\tu$ to (\ref{eq:thesystemRgepafo}), there is a $V_{\infty}\in C^{\infty}(\bM_{\rosil},E_{a})$ such 
that (\ref{eq:uudothsest}) and (\ref{eq:uinfudinfHsest}) hold with $u$, $f$ and $A_{\infty}$ replaced by $\tu$, $\tf$ and $A_{\romod}$ respectively.
Moreover, the constants have the following dependence: the constant $C$ only depends on $c_{\rosil}$, $C_{\coeff}$, $K_{\trs}$, $\eta_{\trs}$, $\mu_{0}$,
$\b_{\romar}$, $C_{\romn}$, $C_{\roini}$, $\b_{\rosil}$, $\eta_{\romn}$, $A_{\romod}$, $g^{ij}(0)$ and $a_{r}(0)$. Since there are only finitely many 
$\indexnot_{\trs}$ satisfying $\mfg_{\infty}(\indexnot_{\trs})\leq\mu_{0}$, we only need to consider a finite number of matrices of the form 
(\ref{eq:Aromoddef}). The dependence on $A_{\romod}$ can thus be replaced by a dependence on $\a_{\infty}$, $\zeta_{\infty}$, the limits appearing in 
(\ref{eq:gijinfetcdef}), $\mu_{0}$ and the spectra of the Laplace-Beltrami operators on the Riemannian manifolds $(M_{r_{j}},g_{r_{j}})$, 
$j=1,\dots,R_{\trs}$. This yields
a constant for which Lemma~\ref{lemma:roughas} is applicable irrespective of $\indexnot\in\EFindexsetr{\mu_{0}}{-}$. The constant $N$ only depends 
on $m$. Finally, by a similar argument, $s_{\rohom}\geq 0$ and $s_{\roih}\geq 0$ can be chosen to depend only $\eta_{\romn}$, $\eta_{\trs}$, $\b_{\rosil}$, 
$\b_{\romar}$, $C_{\coeff}$, $K_{\trs}$, $c_{\rosil}$, $\a_{\infty}$, $\zeta_{\infty}$, the limits appearing in (\ref{eq:gijinfetcdef}), 
the spectra of the Laplace-Beltrami operators on the Riemannian manifolds $(M_{r_{j}},g_{r_{j}})$, $j=1,\dots,R_{\trs}$, and $\mu_{0}$ (in 
particular, they are thus independent of $\indexnot\in\EFindexsetr{\mu_{0}}{-}$). Thus, given a solution $\tu$ to (\ref{eq:thesystemRgepafo}), there 
is a $V_{\infty}(\indexnot_{\trs},\cdot)\in C^{\infty}(\bM_{\rosil},E_{a})$ such that 
\begin{equation}\label{eq:tuudothsest}
\begin{split}
 & \left\|\left(\begin{array}{c} \tu(\indexnot_{\trs},\cdot,t) \\ \tu_{t}(\indexnot_{\trs},\cdot,t)\end{array}\right)
-e^{A_{\romod}t}V_{\infty}(\indexnot_{\trs},\cdot)
-\int_{0}^{t}e^{A_{\romod}(t-\tau)}\left(\begin{array}{c} 0 \\ \tf(\indexnot_{\trs},\cdot,\tau)\end{array}\right)d\tau\right\|_{(s)} \\
\leq & C\ldr{t}^{N}e^{(\kappa_{\indexnot}-\b_{\trs}+\b_{\romar})t}\left(\|\tu_{t}(\indexnot_{\trs},\cdot,0)\|_{(s+s_{\rohom})}+\|\tu(\indexnot_{\trs},\cdot,0)\|_{(s+s_{\rohom}+1)}
\phantom{\tf_{s_{\roih}}}\right.\\
 & \left. +\|\tf(\indexnot_{\trs},\cdot)\|_{A,s+s_{\roih}}\right)
\end{split}
\end{equation}
holds for $t\geq 0$; recall that $\bM_{\rosil}$ is defined in connection with (\ref{eq:bMdeftrs}). Moreover, 
\begin{equation}\label{eq:tuinfudinfHsest}
\begin{split}
\|V_{\infty}(\indexnot_{\trs},\cdot)\|_{(s)} \leq & C\left(\|\tu_{t}(\indexnot_{\trs},\cdot,0)\|_{(s+s_{\rohom})}+\|\tu(\indexnot_{\trs},\cdot,0)\|_{(s+s_{\rohom}+1)}
\phantom{\tf_{s_{\roih}}}\right.\\
 & \left.+\|\tf(\indexnot_{\trs},\cdot)\|_{A,s+s_{\roih}}\right).
\end{split}
\end{equation}
Appealing to (\ref{eq:Dinfdef}) and (\ref{eq:Dinfconj}), the estimate (\ref{eq:tuudothsest}) can be reformulated to 
\begin{equation}\label{eq:tuudothsestref}
\begin{split}
 & \left\|\left(\begin{array}{c} \mfg_{\infty}(\indexnot)\tu(\indexnot_{\trs},\cdot,t) \\ \tu_{t}(\indexnot_{\trs},\cdot,t)\end{array}\right)
-e^{A(\indexnot)t}W_{\infty}(\indexnot_{\trs},\cdot)
-\int_{0}^{t}e^{A(\indexnot)(t-\tau)}\left(\begin{array}{c} 0 \\ \tf(\indexnot_{\trs},\cdot,\tau)\end{array}\right)d\tau\right\|_{(s)} \\
\leq & C\ldr{t}^{N}e^{(\kappa_{\indexnot}-\b_{\trs}+\b_{\romar})t}\left(\|\tu_{t}(\indexnot_{\trs},\cdot,0)\|_{(s+s_{\rohom})}+\|\tu(\indexnot_{\trs},\cdot,0)\|_{(s+s_{\rohom}+1)}
\phantom{\tf_{s_{\roih}}}\right.\\
 & \left. +\|\tf(\indexnot_{\trs},\cdot)\|_{A,s+s_{\roih}}\right),
\end{split}
\end{equation}
where $C$, $N$, $s_{\rohom}$ and $s_{\roih}$ have the same dependence as in the case of (\ref{eq:tuudothsest}). Moreover, $W_{\infty}$ satisfies
an estimate of the form (\ref{eq:tuinfudinfHsest}) and takes its values in the first generalised eigenspace of the 
$(\b_{\trs}-\b_{\romar}),A(\indexnot)$-decomposition of $\cn{2m}$, a space having the properties required of $E_{\indexnot}$ in the statement of the 
proposition. Note that (\ref{eq:tuinfudinfHsest}) and (\ref{eq:tuudothsestref}) hold for the individual
modes. One way to see this is to replace $f$ on the right hand side of (\ref{eq:thesystemRge}) by a function, say $f_{1}$, whose $\indexnot$'th 
mode is that of $f$ and all of whose remaining modes vanish. Replace $u$ by the solution, say $u_{1}$, to (\ref{eq:thesystemRge}) with $f$ replaced by 
$f_{1}$ and initial data given by the functions whose $\indexnot$'th modes are those of $u(\cdot,0)$ and $u_{t}(\cdot,0)$ respectively (and all of 
whose remaining modes vanish). Then it is 
clear that (\ref{eq:tuinfudinfHsest}) and (\ref{eq:tuudothsestref}) hold for the $\indexnot$'th mode. In particular, (\ref{eq:tuudothsestref})
implies that 
\begin{equation}\label{eq:fotuudothsestref}
\begin{split}
 & \left|\left(\begin{array}{c} \mfg_{\infty}(\indexnot)\hu(\indexnot,t) \\ \hu_{t}(\indexnot,t)\end{array}\right)
-e^{A(\indexnot)t}\hU_{\infty}(\indexnot)
-\int_{0}^{t}e^{A(\indexnot)(t-\tau)}\left(\begin{array}{c} 0 \\ \hf(\indexnot,\tau)\end{array}\right)d\tau\right| \\
\leq & C\ldr{t}^{N}e^{(\kappa_{\indexnot}-\b_{\trs}+\b_{\romar})t}\left(\ldr{\nu(\indexnot_{\rosil})}^{s_{\rohom}}|\hu_{t}(\indexnot,0)|
+\ldr{\nu(\indexnot_{\rosil})}^{s_{\rohom}+1}|\hu(\indexnot,0)|
\phantom{\tf_{s_{\roih}}}\right.\\
 & \left. +\ldr{\nu(\indexnot_{\rosil})}^{s_{\roih}}\|\hf(\indexnot,\cdot)\|_{A}\right)
\end{split}
\end{equation}
for $\indexnot\in\EFindexsetr{\mu_{0}}{-}$, where $\hU_{\infty}(\indexnot)$ is the $\indexnot_{\rosil}$'th mode of $W_{\infty}(\indexnot_{\trs},\cdot)$. 
Thus
\begin{equation}\label{eq:ultmuzpartaslowfr}
\begin{split}
 & \left\|\left(\begin{array}{c} (-\Delta_{\infty})^{1/2}u(\cdot,t) \\ \d_{t}u(\cdot,t)\end{array}\right)
-e^{\ma_{\trs}t}U_{\infty}-\int_{0}^{t}e^{\ma_{\trs}(t-\tau)}\left(\begin{array}{c} 0 \\ f(\cdot,\tau)\end{array}\right)d\tau\right\|_{t,s,\mu_{0},-}\\
 \leq & C\ldr{t}^{N}e^{-(\b_{\trs}-\b_{\romar})t}[\|u_{t}(\cdot,0)\|_{(s+s_{\rohom},\mu_{0},-)}+\|u(\cdot,0)\|_{(s+s_{\rohom}+1,\mu_{0},-)}
+\|f\|_{\trs,s+s_{\roih},\mu_{0},-}],
\end{split}
\end{equation}
where $C$, $N$, $s_{\rohom}$ and $s_{\roih}$ have the same dependence as in the case of (\ref{eq:tuudothsest}). Moreover, 
\begin{equation}\label{eq:tuinfudinfHsestlowfr}
\begin{split}
\|U_{\infty}\|_{(s,\mu_{0},-)} \leq & C\left(\|u_{t}(\cdot,0)\|_{(s+s_{\rohom},\mu_{0},-)}+\|u(\cdot,0)\|_{(s+s_{\rohom}+1,\mu_{0},-)}
\phantom{\tf_{s_{\roih}}}\right.\\
 & \left.+\|f\|_{\trs,s+s_{\roih},\mu_{0},-}\right),
\end{split}
\end{equation}
and
\begin{align*}
\|\psi\|_{(s,\mu_{0},-)} := & \left(\textstyle{\sum}_{\indexnot\in\EFindexsetr{\mu_{0}}{-}}\ldr{\nu(\indexnot)}^{2s}|\hpsi(\indexnot)|^{2}\right)^{1/2},\\
\|f\|_{\trs,s,\mu_{0},-} := & \int_{0}^{\infty}\left(\textstyle{\sum}_{\indexnot\in\EFindexsetr{\mu_{0}}{-}}\ldr{\nu(\indexnot)}^{2s}
\ldr{t}^{2\de_{\indexnot}-2}e^{-2\kappa_{\indexnot}t}|\hf(\indexnot,t)|^{2}\right)^{1/2}dt.
\end{align*}
\textbf{Summarising.} To summarise, the conclusions of the proposition are a consequence of the following argument.
First, we fix a $\b_{\romar}$ as in the statement of the proposition. In particular, $\b_{\romar}>0$ should be small enough that 
Lemma~\ref{lemma:trsasymptlargefre} is applicable. Second, we choose a $\mu_{0}$ as in the statement of Lemma~\ref{lemma:trsasymptlargefre}. 
Assume, moreover, $\mu_{0}$ to have the dependence stated in Remark~\ref{remark:Cmuzdephighfre}. Third, for the given $\b_{\romar}$ and $\mu_{0}$, 
we appeal to Lemma~\ref{lemma:trsasymptlargefre} and the above arguments of the present proof. Fourth, we define $U_{\infty}$ by combining the 
function constructed in Lemma~\ref{lemma:trsasymptlargefre} with the function constructed above. As a consequence of the construction, 
the $\indexnot$'th Fourier coefficient of $U_{\infty}$ vanishes if $\indexnot\notin\EFtrsindexset$ and takes its values in $E_{\indexnot}$ for 
$\indexnot\in\EFtrsindexset$. Moreover, $E_{\indexnot}$ is the first generalised eigenspace in the $(\b_{\trs}-\b_{\romar}),A(\indexnot)$-decomposition 
of $\cn{2m}$. Combining (\ref{eq:ultmuzpartas}) and (\ref{eq:ultmuzpartaslowfr}) yields the conclusion that (\ref{eq:ultmuzpartasfin}) holds.
Combining (\ref{eq:Uinfsobnoest}) and (\ref{eq:tuinfudinfHsestlowfr}) yields the conclusion that (\ref{eq:Uinfsobnoestfin}) holds. Moreover, the 
constants have the dependence stated in Remark~\ref{remark:constdepfintrs}. Finally, the proof of uniqueness is in practice identical to the proof 
of uniqueness in Lemma~\ref{lemma:trsasymptlargefre}. The proposition follows. 
\end{proof}

\section{Specifying the asymptotics}\label{section:specastrs}

In the present section, we derive results similar to those of Section~\ref{section:roughODEspecas}, but for weakly transparent equations. Considering
the statement of Proposition~\ref{prop:trsasympt}, it is clear that there are some differences between the present setting and that considered
in Section~\ref{section:roughODEspecas}. In particular, we cannot expect the limiting function $U_{\infty}$ appearing in (\ref{eq:ultmuzpartasfin})
to take its values in one specific subspace of $\cn{2m}$. In fact, each mode of $U_{\infty}$ can be expected to belong to a different subspace. For
that reason, it is convenient to introduce the following terminology. Let $E_{\trs}$ be a map from $\EFtrsindexset$ to the set of vector subspaces
of $\cn{2m}$. Then we say that $\chi\in C^{\infty}(\bM,\cn{2m})$ is $E_{\trs}$-\textit{adapted} if the $\indexnot$'th Fourier coefficient of $\chi$ 
vanishes for $\indexnot\in\EFsilindexset$ and belongs to $E_{\trs}(\indexnot)$ for $\indexnot\in\EFtrsindexset$. We denote the set of 
$E_{\trs}$-adapted elements of $C^{\infty}(\bM,\cn{2m})$ by $C^{\infty}(\bM,\cn{2m};E_{\trs})$. It is also convenient to introduce the notation 
$C_{\trs}^{\infty}(\bM,\cn{2m})$ to denote the set of elements of $C^{\infty}(\bM,\cn{2m})$ whose $\indexnot$'th Fourier coefficients
vanish for $\indexnot\in\EFsilindexset$.

\begin{prop}\label{prop:spasdatrs}
Assume that (\ref{eq:thesystemRge}) is weakly transparent in the sense of Definition~\ref{def:weaktrans}, weakly balanced in the sense of 
Definition~\ref{def:roughODEtermo} and weakly convergent in the sense of Definition~\ref{def:roughODEtermo}. Assume, finally, that 
there is a constant $0<C_{\ell}\in\ro$ and a continuous non-negative function $\betafun\in L^{1}([0,\infty))$ 
such that (\ref{eq:elldotbdgeneral}) holds for all $0\neq\indexnot\in\EFindexset$ and all $t\geq 0$. Fix a $0<\b_{\romar}\in\ro$ which is small 
enough, the bound depending only on $\a_{\infty}$, $\zeta_{\infty}$, the limits appearing in (\ref{eq:gijinfetcdef}) and the constant
$\b_{\trs}$ introduced in Definition~\ref{def:btrsTtrs}. Assume, moreover, that $f=0$.
Then there is a function $E_{\trs}$ from $\EFtrsindexset$ to the set of subspaces of $\cn{2m}$ such that if $E_{\indexnot}:=E_{\trs}(\indexnot)$, then 
the spaces $E_{\indexnot}$ have the properties stated in Proposition~\ref{prop:trsasympt}. Moreover, there are constants $C_{\Phi}>0$ and 
$s_{\infty}\geq 0$, and an injective linear map $\Phi_{\infty}$ from 
$C^{\infty}(\bM,\cn{2m};E_{\trs})$ to $C_{\trs}^{\infty}(\bM,\cn{2m})$ with the following properties. First, 
\begin{equation}\label{eq:Phiinfnobdtrs}
\|\Phi_{\infty}(\chi)\|_{(s)}\leq C_{\Phi}\|\chi\|_{(s+s_{\infty})}
\end{equation}
for all $s\in\ro$ and all $\chi\in C^{\infty}(\bM,\cn{2m};E_{\trs})$. Second, if $\chi\in C^{\infty}(\bM,\cn{2m};E_{\trs})$ and $u$ is the solution to 
(\ref{eq:thesystemRge}) (with $f=0$) such that 
\begin{equation}\label{eq:uuditoPhiinfchitrs}
\left(\begin{array}{c} u(\cdot,0) \\ u_{t}(\cdot,0)\end{array}\right)=\Phi_{\infty}(\chi),
\end{equation}
then 
\begin{equation}\label{eq:estspecasdatatrs}
\begin{split}
 & \left\|\left(\begin{array}{c} (-\Delta_{\infty})^{1/2}u(\cdot,t) \\ u_{t}(\cdot,t)\end{array}\right)
-e^{\ma_{\trs}t}\chi\right\|_{t,s} \\
\leq & C\ldr{t}^{N}e^{-(\b_{\trs}-\b_{\romar})t}\left(\|u_{t}(\cdot,0)\|_{(s+\sigma_{\rohom})}+\|u(\cdot,0)\|_{(s+\sigma_{\rohom}+1)}\right)
\end{split}
\end{equation}
for all $t\geq 0$ and all $s\in\ro$, where the constants $C$, $N$ and $\sigma_{\rohom}$ have the same dependence as in the case of 
Proposition~\ref{prop:trsasympt}; cf. 
Remark~\ref{remark:constdepfintrs}. Finally, if there is an $\e>0$ such that $\Rsp[A(\indexnot)]<\b_{\trs}-\e$ for all $\indexnot\in\EFtrsindexset$, 
then $E_{\trs}$ can be chosen to be such that $E_{\trs}(\indexnot)=\cn{2m}$ for all $\indexnot\in\EFtrsindexset$. Then $\Phi_{\infty}$ is surjective. 
\end{prop}
\begin{remark}\label{remark:CPhisinftytrs}
The constant $C_{\Phi}$ only depends on $\b_{\romar}$, $\b_{\rosil}$, $\a_{\infty}$, $\zeta_{\infty}$, the limits appearing in (\ref{eq:gijinfetcdef}), 
$C_{\roini}$, $C_{\ell}$, $C_{\romn}$, $C_{\coeff}$, $c_{\rosil}:=\|\betafun_{\rosil}\|_{1}$, $c_{\betafun}:=\|\betafun\|_{1}$, $\roS_{\trs}$, $\eta_{\trs}$, 
$\eta_{\romn}$, $K_{\trs}$, the supremum of $\kappa_{\indexnot}$ for $\indexnot\in\EFtrsindexset$ and the spectra of the Laplace-Beltrami operators 
on the Riemannian manifolds $(M_{r_{j}},g_{r_{j}})$, 
$j=1,\dots,R_{\trs}$; and $s_{\infty}\geq 0$ only depends on $C_{\coeff}$, $C_{\ell}$, $\b_{\rosil}$, $\eta_{\trs}$ and the supremum of 
$\kappa_{\indexnot}$ for $\indexnot\in\EFtrsindexset$.
\end{remark}
\begin{remark}
By combining (\ref{eq:Phiinfnobdtrs}), (\ref{eq:uuditoPhiinfchitrs}) and (\ref{eq:estspecasdatatrs}), the norms of $u(\cdot,0)$ and $u_{t}(\cdot,0)$ 
appearing on the right hand side of (\ref{eq:estspecasdatatrs}) can be replaced by a suitable Sobolev norm of $\chi$. 
\end{remark}
\begin{remark}
In order to obtain a similar result in the case of inhomogeneous equations, it is sufficient to combine Propositions~\ref{prop:trsasympt} and
\ref{prop:spasdatrs}; cf. Remark~\ref{remark:inhomaschar}. 
\end{remark}
\begin{proof}
Let $\chi\in C^{\infty}(\bM,\cn{2m})$, $\indexnot\in \EFtrsindexset$ and $\hchi(\indexnot):=\ldrbox{\chi,\varphi_{\indexnot}}$; cf. the notation introduced
in Subsection~\ref{ssection:specprodset}. Let $u$ be a solution to (\ref{eq:thesystemRge}) (with $f=0$) and let $z(\indexnot,t)$ be given by 
(\ref{eq:znutdef}). Note that the $\indexnot$'th Fourier coefficient of the expression inside the norm on the left hand side of 
(\ref{eq:estspecasdatatrs}) can be written
\begin{equation}\label{eq:vasitochihtrs}
v(\indexnot,t)-e^{A(\indexnot)t}\hchi(\indexnot), 
\end{equation}
where we use the notation (\ref{eq:vFdeftrs}) and (\ref{eq:Aindexnotdeftrs}). Clearly, we want to construct a $v$ such that this object, 
after multiplication with $e^{-\kappa_{\indexnot}t}$, decays exponentially. On the other hand, $v$ satisfies (\ref{eq:vdottrscase}) with $F_{\indexnot}=0$, 
where $A_{\indexnot,\rem}$ satisfies (\ref{eq:Aremtrsderest}) for $t\geq T_{\trs}$; cf. Definition~\ref{def:btrsTtrs}
for an explanation of the terminology used. In order to construct $v$, it thus seems natural to appeal to Lemmas~\ref{lemma:spasODEsett} and
\ref{lemma:spastrssett}.

Due to the above observations, the strategy of the proof is the following. In order to construct the
initial data corresponding to $\chi$, we consider the different Fourier modes, as above, and appeal to  Lemmas~\ref{lemma:spasODEsett} and
\ref{lemma:spastrssett}. There are three things to keep in mind when appealing to these lemmas. First, Lemma~\ref{lemma:spastrssett} only 
applies for $\mfg_{\infty}(\indexnot)$ sufficiently large; second, when appealing to Lemma~\ref{lemma:spasODEsett}, the constants appearing
depend on $A(\indexnot)$; and, third, in order to be allowed to appeal to Lemmas~\ref{lemma:spasODEsett} and \ref{lemma:spastrssett}, 
$\hchi(\indexnot)$ has to belong to an appropriate vector subspace of $\cn{2m}$, say $E_{\trs}(\indexnot)$ (what space this is depends on the 
lemma to which we appeal). It is therefore natural to begin by appealing to Lemma~\ref{lemma:spastrssett}, and, in the process, to fix a choice
of $\mu_{0}>0$ and to make a choice of $E_{\trs}(\indexnot)$ for $\mfg_{\infty}(\indexnot)\geq\mu_{0}$. For the $\indexnot\in\EFtrsindexset$ to 
which Lemma~\ref{lemma:spastrssett} does not apply, we then appeal to Lemma~\ref{lemma:spasODEsett} and make a corresponding choice of 
$E_{\trs}(\indexnot)$. Since the number of matrices $A(\indexnot)$ for $0<\mfg_{\infty}(\indexnot)<\mu_{0}$ is finite, the dependence of the 
constants on $A(\indexnot)$ is manageable. The outcome of the analysis is a construction of a map $E_{\trs}$ from $\EFtrsindexset$ to the 
set of subspaces of $\cn{2m}$; given $\indexnot\in\EFtrsindexset$ and $\hchi(\indexnot)\in E_{\trs}(\indexnot)$, a construction of 
$v(\indexnot,T_{\trs})$; estimates of 
$|v(\indexnot,T_{\trs})|$ in terms of the asymptotic data $\hchi(\indexnot)$; and an asymptotic estimate of (\ref{eq:vasitochihtrs}).
Once $v(\indexnot,T_{\trs})$ has been constructed, the next step is to estimate $v(\indexnot,0)$ in terms of $\hchi(\indexnot)$. We do so
by appealing to Lemma~\ref{lemma:roughenestbalsetting}. Finally, some additional arguments are required to verify that all the conclusions 
of the proposition hold. 

\textbf{From asymptotic data to $T_{\trs}$.}
Note, to begin with, that the terminology of the present setting is related to that of Lemma~\ref{lemma:spastrssett} as described 
at the beginning of the proof of Lemma~\ref{lemma:trsasymptlargefre}. Fixing $\b_{\romar}>0$ small enough, the bound depending only on 
$\b_{\trs}$, $\a_{\infty}$, $\zeta_{\infty}$ and the limits appearing in (\ref{eq:gijinfetcdef}), there are then constants $\mu_{0}>0$ and 
$C>0$ such that the conclusions of Lemma~\ref{lemma:spastrssett} hold. Moreover, $\mu_{0}$ only depends on $\b_{\romar}$, $\a_{\infty}$, $\zeta_{\infty}$ 
and the limits appearing in (\ref{eq:gijinfetcdef}), and $C$ only depends on $C_{\trs}$, $\b_{\trs}$, $\b_{\romar}$, $\a_{\infty}$, 
$\zeta_{\infty}$ and the limits appearing in (\ref{eq:gijinfetcdef}). Let $\Psi_{\indexnot,\infty}$
be the map constructed in Lemma~\ref{lemma:spastrssett}. Note that this map depends on $\indexnot$, but that the constants appearing in the 
estimates involving $\Psi_{\indexnot,\infty}$ do not. Note, moreover, that $\Psi_{\indexnot,\infty}$ maps $E_{\indexnot,\b_{\trs}}^{\infty}$ to $\cn{2m}$. It is 
thus clear that we have 
to choose $E_{\trs}(\indexnot)=E_{\indexnot,\b_{\trs}}^{\infty}$ for $\mfg_{\infty}(\indexnot)\geq\mu_{0}$. Assuming $\b_{\romar}$ to be small enough, the
bound depending only on $\b_{\trs}$, $\a_{\infty}$, $\zeta_{\infty}$ and the limits appearing in (\ref{eq:gijinfetcdef}), and then $\mu_{0}$ large 
enough, the bound depending only on $\b_{\romar}$, $\b_{\trs}$, $\a_{\infty}$, $\zeta_{\infty}$ and the limits appearing in (\ref{eq:gijinfetcdef}), it can be 
ensured that the spaces $E_{\indexnot}:=E_{\trs}(\indexnot)$ have the 
properties stated in Proposition~\ref{prop:trsasympt} for $\mfg_{\infty}(\indexnot)\geq\mu_{0}$. 
This is justified by the argument presented in the proof of Lemma~\ref{lemma:trsasymptlargefre}. 
Given $\indexnot\in\EFindexsetr{\mu_{0}}{+}$, let 
us specify $v$ by imposing the initial condition $v(\indexnot,T_{\trs}):=\Psi_{\indexnot,\infty}[\hchi(\indexnot)]$. Then (\ref{eq:vasitochihtrs}) is 
small asymptotically. In order to estimate $|v(\indexnot,T_{\trs})|$ in terms of $|\hchi(\indexnot)|$, define 
$u_{\infty}(\indexnot)\in E_{\indexnot,\b_{\trs}}^{\infty}$ by 
\begin{equation}\label{eq:uinftyindexnotdef}
e^{-A(\indexnot)T_{\trs}}u_{\infty}(\indexnot)=\hchi(\indexnot).
\end{equation}
Then $|u_{\infty}|=|e^{A(\indexnot)T_{\trs}}\hchi(\indexnot)|$, so that (\ref{eq:Psiinfnormtrs}) yields the estimate
\begin{equation}\label{eq:vTodeabsestasdatatrs}
\begin{split}
|v(\indexnot,T_{\trs})| = & |\Psi_{\indexnot,\infty}(e^{-A(\indexnot)T_{\trs}}u_{\infty})|\leq C|u_{\infty}(\indexnot)|=C|e^{A(\indexnot)T_{\trs}}\hchi(\indexnot)|
\end{split}
\end{equation}
where $C$ only depends on $\b_{\romar}$, $\b_{\trs}$, $C_{\trs}$, $\a_{\infty}$, $\zeta_{\infty}$ and the limits appearing in (\ref{eq:gijinfetcdef}). In order 
to estimate the far right hand side, note that Lemma~\ref{lemma:genepmaintest} implies that there are $\mu_{0}$ and $C$, depending only on $\a_{\infty}$, 
$\zeta_{\infty}$ and the limits appearing in (\ref{eq:gijinfetcdef}), such that 
\[
\|\exp[A(\indexnot)T_{\trs}]\|\leq C\exp[(\kappa_{\indexnot}+2)T_{\trs}]
\]
for all $\indexnot\in\EFindexsetr{\mu_{0}}{+}$. Combining this estimate with (\ref{eq:vTodeabsestasdatatrs}) yields
\begin{equation}\label{eq:absvinttrsest}
|v(\indexnot,T_{\trs})|\leq C\exp[(\kappa_{\indexnot}+2)T_{\trs}]|\hchi(\indexnot)|,
\end{equation}
where $C$ only depends on $\b_{\romar}$, $\b_{\trs}$, $C_{\trs}$, $\a_{\infty}$, $\zeta_{\infty}$ and the limits appearing in (\ref{eq:gijinfetcdef}). 

For $\indexnot\in\EFtrsindexset$
such that $\mfg_{\infty}(\indexnot)<\mu_{0}$, we appeal to Lemma~\ref{lemma:spasODEsett}. When doing so, we replace $A$, $A_{\rem}$ and $E_{A,\b}$ by 
$A(\indexnot)$, $A_{\indexnot,\rem}$ and $E_{A(\indexnot),\b_{\trs}-\b_{\romar}}$ respectively. Moreover, we replace $k$, $T_{\roode}$, $\b_{\rem}$ and $C_{\rem}$ by $2m$, 
$T_{\trs}$, $\b_{\trs}$ and $C_{\trs}$ respectively. Finally, the map $\Psi_{\infty}$ appearing in Lemma~\ref{lemma:spasODEsett} we here refer to as 
$\Psi_{\indexnot,\infty}:E_{A(\indexnot),\b_{\trs}-\b_{\romar}}\rightarrow\cn{2m}$. In particular, it is thus clear that for $\indexnot\in\EFtrsindexset$ such that 
$\mfg_{\infty}(\indexnot)<\mu_{0}$, we let $E_{\trs}(\indexnot):=E_{A(\indexnot),\b_{\trs}-\b_{\romar}}$; note that this space has the desired properties. 
For $\indexnot\in\EFtrsindexset$ such that $\mfg_{\infty}(\indexnot)<\mu_{0}$, let $v(\indexnot,T_{\trs}):=\Psi_{\indexnot,\infty}[\hchi(\indexnot)]$
as before and define $u_{\infty}(\indexnot)$ by (\ref{eq:uinftyindexnotdef}). Then (\ref{eq:Psiinfnorm}) yields
\begin{equation}\label{eq:absvinttrsestlf}
|v(\indexnot,T_{\trs})|\leq C\exp[(\kappa_{\indexnot}+2)T_{\trs}]|\hchi(\indexnot)|,
\end{equation}
where $C$ only depends on $C_{\trs}$, $\b_{\trs}$ and $A(\indexnot)$. However, since there are only finitely many matrices $A(\indexnot)$
for $\mfg_{\infty}(\indexnot)<\mu_{0}$, the dependence on $A(\indexnot)$ can be replaced by a dependence on $\mu_{0}$, $\a_{\infty}$, $\zeta_{\infty}$, 
the limits appearing in (\ref{eq:gijinfetcdef}) and the spectra of the Laplace-Beltrami operators on the Riemannian manifolds $(M_{r_{j}},g_{r_{j}})$, 
$j=1,\dots,R_{\trs}$. Defining $\kappa$ to be the supremum of the set of $\kappa_{\indexnot}$ for $\indexnot\in\EFtrsindexset$ (note that this 
supremum is finite), (\ref{eq:absvinttrsest}) and (\ref{eq:absvinttrsestlf}) yield
\begin{equation}\label{eq:vinTtrsfinalest}
|v(\indexnot,T_{\trs})|\leq C\exp[(\kappa+2)T_{\trs}]|\hchi(\indexnot)|
\end{equation}
for all $\indexnot\in\EFtrsindexset$, where $C$ only depends on $\b_{\romar}$, $\b_{\trs}$, $C_{\trs}$, $\mu_{0}$, $\a_{\infty}$, $\zeta_{\infty}$,
the limits appearing in (\ref{eq:gijinfetcdef}) and the spectra of the Laplace-Beltrami operators on the Riemannian manifolds $(M_{r_{j}},g_{r_{j}})$, 
$j=1,\dots,R_{\trs}$. In practice we fix $\mu_{0}$ to a value depending only on $\b_{\romar}$, $\b_{\trs}$, $\a_{\infty}$, 
$\zeta_{\infty}$ and the limits appearing in (\ref{eq:gijinfetcdef}). Thus the dependence of $C$ on $\mu_{0}$ can be eliminated. 
To conclude, we have defined $E_{\trs}(\indexnot)$ for $\indexnot\in\EFtrsindexset$ and this set has the desired properties;
we have defined $v(\indexnot,T_{\trs})$ in terms of $\hchi(\indexnot)$; we know that (\ref{eq:vinTtrsfinalest}) holds; and, finally, we 
know that if $v(\indexnot,T_{\trs}):=\Psi_{\indexnot,\infty}[\hchi(\indexnot)]$, then 
\begin{equation}\label{eq:vintasest}
|v(\indexnot,t)-e^{A(\indexnot)t}\hchi(\indexnot)|\leq C\ldr{t}^{N}e^{(\kappa_{\indexnot}-\b_{\trs}+\b_{\romar})\bt}|v(\indexnot,T_{\trs})|
\end{equation}
for $t\geq T_{\trs}$. If $\mfg_{\infty}(\indexnot)\geq\mu_{0}$, then (\ref{eq:vintasest}) is a consequence of (\ref{eq:vvinfesttrsstmt}) with 
$F_{\indexnot}=0$; recall that (\ref{eq:vvinfesttrsstmt}) holds as a consequence of Lemma~\ref{lemma:spastrssett}. If 
$\mfg_{\infty}(\indexnot)<\mu_{0}$, then 
(\ref{eq:vintasest}) is a consequence of (\ref{eq:vinfasres}) with $F=0$; recall that (\ref{eq:vinfasres}) holds as a consequence of 
Lemma~\ref{lemma:spasODEsett}. Finally, the constant $C$ appearing in (\ref{eq:vintasest}) only depends on $C_{\trs}$, $\b_{\trs}$, $\b_{\romar}$,
$\mu_{0}$, $\a_{\infty}$, $\zeta_{\infty}$, the limits appearing in (\ref{eq:gijinfetcdef}) and the spectra of the Laplace-Beltrami operators on 
the Riemannian manifolds $(M_{r_{j}},g_{r_{j}})$, $j=1,\dots,R_{\trs}$. In practice we fix $\mu_{0}$ to a value depending 
only on $\b_{\romar}$, $\b_{\trs}$, $\a_{\infty}$, $\zeta_{\infty}$ and the limits appearing in (\ref{eq:gijinfetcdef}). Thus the dependence of $C$ on $\mu_{0}$ 
can be eliminated. From now on, we assume $\mu_{0}$ to have been fixed in this way, so that the constants appearing in (\ref{eq:vinTtrsfinalest}) 
and (\ref{eq:vintasest}) are independent of $\mu_{0}$. 

\textbf{From $T_{\trs}$ to $t=0$.} Next we wish to estimate $|v(\indexnot,0)|$. Due to the definition of $T_{\trs}$, we know that $T_{\trs}>0$; 
cf. (\ref{eq:Ttrsdef}). Note also that (\ref{eq:mevequivtrans}) holds, where $C$ only depends on $C_{\roini},c_{\rosil},K_{\trs},\roS_{\trs}$,
$\eta_{\trs}$, and $c_{\rosil}:=\|\betafun_{\rosil}\|_{1}$. Thus, appealing to (\ref{eq:meestroughbalset}) and (\ref{eq:mevequivtrans}),
\[
\me^{1/2}(\indexnot,0)\leq e^{c_{\betafun}+\eta_{\robal}T_{\trs}}\me^{1/2}(\indexnot,T_{\trs})\leq Ce^{\eta_{\robal}T_{\trs}}|v(\indexnot,T_{\trs})|,
\]
where $C$ only depends on $C_{\roini},c_{\rosil},K_{\trs},\roS_{\trs}$, $\eta_{\trs}$ and $c_{\betafun}:=\|\betafun\|_{1}$; and 
$\eta_{\robal}$ only depends on $C_{\coeff}$ and $C_{\ell}$. Combining this estimate with (\ref{eq:vinTtrsfinalest}) yields
\begin{equation}\label{eq:Esrfinestz}
\me^{1/2}(\indexnot,0)\leq C\exp[(\kappa+\eta_{\robal}+2)T_{\trs}]|\hchi(\indexnot)|,
\end{equation}
where $C$ only depends on $C_{\roini},c_{\rosil},K_{\trs},\roS_{\trs}$, $\eta_{\trs}$, $c_{\betafun}$, $\b_{\romar}$, $\b_{\trs}$, $C_{\trs}$, 
$\mu_{0}$, $\a_{\infty}$, $\zeta_{\infty}$, the limits appearing in (\ref{eq:gijinfetcdef}) and the spectra of the Laplace-Beltrami operators on 
the Riemannian manifolds $(M_{r_{j}},g_{r_{j}})$, $j=1,\dots,R_{\trs}$. Thus
\begin{equation}\label{eq:Phiinfprelnoest}
\ldr{\nu(\indexnot)}^{s}(|z(\indexnot,0)|^{2}+|\dot{z}(\indexnot,0)|^{2})^{1/2}\leq C\ldr{\nu(\indexnot)}^{s+s_{\infty}}|\hchi(\indexnot)|
\end{equation}
for $\indexnot\in\EFtrsindexset$, where $s_{\infty}:=(\kappa+2+\eta_{\robal})[2\b_{\rosil}^{-1}+(\min\{\b_{\rosil},\eta_{\trs}\})^{-1}]$ and 
$C$ only depends on $C_{\roini},c_{\rosil},K_{\trs},\roS_{\trs}$, $\eta_{\trs}$, $c_{\betafun}$, $\b_{\romar}$, $\b_{\trs}$, $\b_{\rosil}$, $\kappa$,
$C_{\ell}$, $C_{\coeff}$, $C_{\trs}$, 
$\mu_{0}$, $\a_{\infty}$, $\zeta_{\infty}$, the limits appearing in (\ref{eq:gijinfetcdef}) and the spectra of the Laplace-Beltrami operators on 
the Riemannian manifolds $(M_{r_{j}},g_{r_{j}})$, $j=1,\dots,R_{\trs}$. Let $\Phi_{\infty}$ be the map taking $\chi\in C^{\infty}(\bM,\cn{2m};E_{\trs})$
to the distribution whose $\indexnot$'th mode is given by $v_{0}(\indexnot,0)$, where
\[
v_{0}(\indexnot,t):=\left(\begin{array}{c} z(\indexnot,t) \\ \dot{z}(\indexnot,t)\end{array}\right)
\]
and $z$ is constructed as above. Then (\ref{eq:Phiinfprelnoest}) implies that (\ref{eq:Phiinfnobdtrs}) holds; note that this implies that 
$\Phi_{\infty}$ maps $\chi\in C^{\infty}(\bM,\cn{2m};E_{\trs})$ to an element of $C^{\infty}_{\trs}(\bM,\cn{2m})$. Since $\Phi_{\infty}$ is injective on 
the level of Fourier coefficients (this is a consequence of Lemmas~\ref{lemma:spasODEsett} and \ref{lemma:spastrssett}), it follows that 
$\Phi_{\infty}$ is injective. 

\textbf{Verification of the properties of $\Phi_{\infty}$.} Fix $\b_{\romar}>0$ such that Proposition~\ref{prop:trsasympt} and the above arguments apply.
Let, moreover $\chi\in C^{\infty}(\bM,\cn{2m};E_{\trs})$ and let $u$ be the solution to (\ref{eq:thesystemRge}) (with $f=0$) satisfying 
(\ref{eq:uuditoPhiinfchitrs}). Due to Proposition~\ref{prop:trsasympt}, we know that (\ref{eq:estspecasdatatrs}) holds with $\chi$ replaced by some 
$U_{\infty}\in C^{\infty}(\bM,\cn{2m};E_{\trs})$. Moreover, the constants $C$, $N$ and $\sigma_{\hom}$ have the dependence stated in 
Proposition~\ref{prop:trsasympt}. In order to determine $U_{\infty}$ in terms of $\chi$, note that (\ref{eq:estspecasdatatrs}) (with $\chi$ replaced 
by $U_{\infty}$) implies that 
\begin{equation}\label{eq:vprelaspftrs}
\left|v(\indexnot,t)-e^{A(\indexnot)t}u_{\infty}(\indexnot)\right| \leq C\ldr{t}^{N}e^{(\kappa_{\indexnot}-\b_{\trs}+\b_{\romar})t}
\end{equation}
where $C$ depends on the initial data and $\indexnot$; $v$ is defined as in (\ref{eq:vFdeftrs}) where $z(\indexnot,\cdot)$ is the $\indexnot$'th 
mode of $u$; and $u_{\infty}(\indexnot)$ is the $\indexnot$'th mode of $U_{\infty}$. Due to the construction of $\Phi_{\infty}$, we know that the 
estimate (\ref{eq:vintasest}) holds. Thus
\[
\left|e^{A(\indexnot)t}[\hchi(\indexnot)-u_{\infty}(\indexnot)]\right| \leq C\ldr{t}^{N}e^{(\kappa_{\indexnot}-\b_{\trs}+\b_{\romar})t}
\]
for $t\geq T_{\trs}$. Since $\hchi(\indexnot)-u_{\infty}(\indexnot)\in E_{\trs}(\indexnot)=E_{\indexnot}$, and since $E_{\indexnot}$ has the properties stated in 
Proposition~\ref{prop:trsasympt}, this estimate implies that $\hchi(\indexnot)-u_{\infty}(\indexnot)=0$. Thus $U_{\infty}=\chi$, and 
the lemma follows, except for the statement concerning surjectivity. 

If there is an $\e>0$ such that $\Rsp[A(\indexnot)]<\b_{\trs}-\e$ for all $\indexnot\in\EFtrsindexset$, then, choosing $\b_{\romar}$ to be small enough, 
$E_{\trs}(\indexnot)=\cn{2m}$ for all $\indexnot\in\EFtrsindexset$. Thus $\Phi_{\infty}$ is a map from $C^{\infty}_{\trs}(\bM,\cn{2m})$ to itself. Let 
$\psi\in C^{\infty}_{\trs}(\bM,\cn{2m})$. We wish to demonstrate that $\psi$ is in the image of $\Phi_{\infty}$. Let $u$ be the solution to (\ref{eq:thesystemRge}) 
(with $f=0$) with initial data given by (\ref{eq:uuditoPhiinfchitrs}), where the right hand side has been replaced by $\psi$. Appealing to 
Proposition~\ref{prop:trsasympt} yields a $U_{\infty}\in C_{\trs}^{\infty}(\bM,\cn{2m})$ such that (\ref{eq:ultmuzpartasfin}) holds (with $f=0$). Let $\bu$ be the 
solution to (\ref{eq:thesystemRge}) with initial data given by (\ref{eq:uuditoPhiinfchitrs}),
where the right hand side has been replaced by $\Phi_{\infty}(U_{\infty})$. Then, by the above arguments, (\ref{eq:ultmuzpartasfin}) holds (with $f=0$)
and $u$ replaced by $\bu$. In particular, there is thus a constant $C$, depending on $u$, $\bu$ etc., such that 
\begin{equation}\label{eq:uidminusbuidtransparent}
\left\|\left(\begin{array}{c} (-\Delta_{\infty})^{1/2}u(\cdot,t) \\ u_{t}(\cdot,t)\end{array}\right)
-\left(\begin{array}{c} (-\Delta_{\infty})^{1/2}\bu(\cdot,t) \\ \bu_{t}(\cdot,t)\end{array}\right)\right\|_{t,s} 
\leq C\ldr{t}^{N}e^{-(\b_{\trs}-\b_{\romar})t}
\end{equation}
for all $t\geq 0$ and all $s\in\ro$. Let $z$ and $\bz$ be the Fourier coefficients of $u$ and $\bu$ respectively, let $v$ and $\bv$ be defined in
analogy with (\ref{eq:vFdeftrs}), starting with $z$ and $\bz$ respectively. Finally, let $V=v-\bv$. If $V(\indexnot,0)=0$ for all $\indexnot\in\EFtrsindexset$, 
then $u=\bu$ (since $V(\indexnot,\cdot)$ solves a homogeneous equation, $V(\indexnot,0)=0$ for all $\indexnot\in\EFtrsindexset$ implies that $V(\indexnot,t)=0$ 
for all $\indexnot\in\EFtrsindexset$ and all $t\in I$), so that $\Phi_{\infty}(U_{\infty})=\psi$. Thus $\psi$ is in the image of $\Phi_{\infty}$. Assume now that 
there is a $\indexnot\in\EFtrsindexset$ such that $V(\indexnot,0)\neq 0$. Note that $V(\indexnot,\cdot)$ is a solution to an equation to which 
either Lemma~\ref{lemma:spasODEsett} or Lemma~\ref{lemma:spastrssett} applies; cf. the above. In fact, due to the above arguments, for each 
$\indexnot\in\EFtrsindexset$,
there is a map $\Psi_{\indexnot,\infty}$ which is linear and bijective. Since $V(\indexnot,0)\neq 0$, there is a $0\neq\chi\in\cn{2m}$ such that 
$\Psi_{\indexnot,\infty}(\chi)=V(\indexnot,T_{\trs})$, where $T_{\trs}$ is given by (\ref{eq:Ttrsdef}) (for the relevant $\indexnot$ under consideration). 
Due to the above arguments, 
\[
|V(\indexnot,t)-e^{A(\indexnot)t}\chi|\leq C\ldr{t}^{N}e^{[\kappa_{\indexnot}-(\b_{\trs}-\b_{\romar})]t}
\]
for all $t\geq T_{\trs}$ and a constant $C$ depending on $V$, $\indexnot$ etc. Here $\kappa_{\indexnot}:=\kappa_{\max}[A(\indexnot)]$. On the other hand, 
(\ref{eq:uidminusbuidtransparent}) implies that the same estimate holds with $\chi$ set to zero. Due to the fact that $\Rsp A(\indexnot)<\b_{\trs}-\b_{\romar}$, 
these estimates are contradictory. To conclude, every element of $C^{\infty}_{\trs}(\bM,\cn{2m})$ is in the image of $\Phi_{\infty}$. The lemma follows. 
\end{proof}

\part{Averaging over oscillations}\label{part:averaging}

\chapter{Introduction and technical preliminaries}\label{chapter:avoscintroatechnprel}

\section{Introduction}

When (\ref{eq:thesystemRge}) is weakly silent or weakly transparent, the limit equation (obtained by replacing the coefficients in 
(\ref{eq:thesystemRge}) by their limits as $t\rightarrow\infty$) is either a system of second order ODE's or a system of linear PDE's
with constant coefficients. The solutions to the limit equation can then be calculated by solving linear algebra problems. However, it 
is also of interest to consider situations in which some of the coefficients of the equation exhibit exponential growth asymptotically. 
In that setting, there is no constant coefficient limit equation with which to compare, and it is less obvious how to proceed. Moreover,
if some of the $g^{ij}$ asymptotically grow exponentially, then the Fourier coefficients of a solution can typically be expected to 
oscillate with a frequency that grows exponentially. This may seem to be an unpleasant situation to handle. However, when considering
such equations in greater detail, it becomes clear that an exponentially growing frequency in some respects simplifies the analysis. 
The reason for this is the following. Consider a Fourier coefficient of a solution to the homogeneous version of (\ref{eq:thesystemRge}). 
This coefficient solves 
a (typically non-constant coefficient) system of homogeneous ODE's. In the case that the frequency of the oscillations grows exponentially, 
it is natural to try to approximate the evolution by considering one period of the oscillations at a time. If $[t_{a},t_{b}]$ is one 
period of the oscillations, one would like to calculate the matrix, say $\Phi(t_{b};t_{a})$, that maps initial data at $t_{a}$ to the 
corresponding solution at $t_{b}$. Here we are not able to do so. On the other hand, we can approximate $\Phi(t_{b};t_{a})$ by calculating
the corresponding matrix obtained by freezing the coefficients at $t_{a}$ in the original equation. In order for this calculation to be 
of interest, we need to estimate the error. However, due to the fact that the frequency of the oscillations increases exponentially,
the size of the interval $[t_{a},t_{b}]$ decreases exponentially. Under rather mild conditions on the coefficients of (\ref{eq:thesystemRge}), 
the approximation of $\Phi(t_{b};t_{a})$ therefore becomes better and better. In the end, we are interested in the asymptotic behaviour as 
$t\rightarrow\infty$. We then need to calculate the matrix product of a sequence of matrices of the form $\Phi(t_{b};t_{a})$ (and the number 
of factors in the product tends to infinity asymptotically). In practice we only calculate $\Phi(t_{b};t_{a})$ up to some error term, 
and it is thus necessary to verify that this error term is manageable even when the number of factors in the matrix product tends to infinity. 
Again, due to the exponential decay of the length of the periods of the oscillations, this can be done. 

In the present part of these notes, we develop methods that are suitable for dealing with situations of the above type. However, as we argue
in Subsection~\ref{ssection:reformeq} below, the arguments are also useful in other contexts. The analysis is quite technical, but we give 
an overview of the arguments involved in Subsection~\ref{ssection:mastofanal} below.

\subsection{Reformulating the equation}\label{ssection:reformeq}

The above description gives a rough idea of the argument in the case of an exponentially growing frequency. However, the methods described 
above are also useful in the case of, e.g., weakly silent equations. In particular, under suitable assumptions they can be used to obtain 
better (in fact, optimal) estimates for how the energy $E$ of a Fourier coefficient develops over a period of the form $[0,T_{\roode}]$; cf. 
Section~\ref{section:oscreg}. In order to understand why this is the case, as well as why the methodology described above is useful even more 
generally, let us reformulate (\ref{eq:fourierthesystemRge}). Introducing the notation $X$ and $\sigma$ as in (\ref{eq:ellsigmaXgenRdef}), the 
equation (\ref{eq:fourierthesystemRge}) can be written
\begin{equation}\label{eq:fourierthesystemRgereform}
\ddot{z}+\mfg^{2}z-2i\sigma\mfg\dot{z} +\a\dot{z}+iX\mfg z+\zeta z=\hf.
\end{equation}
Assuming $\hf=0$, (\ref{eq:fourierthesystemRgereform}) can be written
\begin{equation}\label{eq:dotveqAv}
\dot{v}=Av,
\end{equation}
where
\begin{equation}\label{eq:vadefmodeqintroavosc}
v:=\left(\begin{array}{c} \mfg z\\ \dot{z}\end{array}\right),\ \ \
A:=\left(\begin{array}{cc} \dot{\ell}\Id_{m} & \mfg\Id_{m}\\ -\mfg\Id_{m}-iX-\zeta/\mfg & -\a+2i\sigma\mfg\Id_{m}\end{array}\right)
\end{equation}
and $\ell:=\ln\mfg$. Say now that the equation is weakly balanced in the sense that (\ref{eq:weakbal}) holds for $t\geq 0$, where $C_{\coeff}$ 
is independent of $\indexnot$. Assume, moreover, that $|\sigma|$ decays exponentially and that $\dot{\ell}$ is bounded. Then it is clear that when 
$\mfg$ is large, the dominant feature of the evolution associated with (\ref{eq:dotveqAv}) is oscillations with a period of (roughly) $2\pi/\mfg$. 
In fact, it is not even necessary to assume that $\|\a\|$, $\|\zeta\|$ and $\|X\|$ are bounded; it is sufficient that they are small in comparison 
with $\mfg$. Thus considering one period at a time can be expected to be of interest for unbalanced equations. Moreover, 
even in the case of a weakly silent equation (for which $\mfg$ converges to zero exponentially), there are time intervals 
during which $\mfg$ is large, and the behaviour is oscillatory. In such situations it therefore also makes sense to consider one period
of the oscillations at a time. 

\subsection{Main steps of the analysis}\label{ssection:mastofanal}

\textbf{Determining a setting in which it is meaningful to average over one period of the oscillations.}
Due to the discussion of the previous subsection, it is clear that in regions where $\mfg$ is dominant, the evolution associated with 
(\ref{eq:dotveqAv}) is essentially oscillatory. It is therefore of interest to consider the evolution over one period. Moreover, this
perspective is of interest even for equations that violate (\ref{eq:weakbal}); i.e., for equations that are not weakly balanced. This
heuristic insight is the basis for the present part of these notes. Turning to the actual analysis, it is natural to start by stating 
the assumptions needed. We do so at the beginning of Chapter~\ref{chapter:destopofosc} below. It is of interest to note that the 
assumptions are stronger than those made in, e.g., Chapter~\ref{chapter:weaksil} in some respects, but weaker in other respects. In 
particular, in Chapter~\ref{chapter:destopofosc} we make assumptions concerning the time derivatives of $\dot{\ell}$, $X$, $\a$ and $\zeta$,
even though there are no corresponding assumptions in Definition~\ref{def:roughODEtermo}. On the other hand, the assumptions stated in 
Chapter~\ref{chapter:destopofosc} do not exclude the possibility that $X$, $\a$ and $\zeta$ might grow exponentially. The reason for 
making assumptions concerning the time derivatives of, e.g., $X$, $\a$ and $\zeta$ is that, in the present setting, we do not want these 
quantities to vary too much during one period of the oscillations. In Chapter~\ref{chapter:weaksil}, on the other hand, we do not consider
the oscillations in detail, and therefore we do not need to control the variation of $X$, $\a$ and $\zeta$ during a period of the oscillations.

\textbf{Reformulating the equations and deriving rough estimates for one period of the oscillations.}
Once suitable assumptions have been stated, it is convenient to reformulate the equations. The purpose of the reformulation is twofold. 
First, it is convenient to extract the overall growth (or decay) during one period. This is achieved by reformulating the equation to, say,
$\dot{w}=Bw+F$, where the real part of the trace of $B$ is (essentially) zero. Second, it is convenient to change the time coordinate so 
that the dominant coefficients of the equation become time independent. Deriving such a reformulation is the subject of 
Section~\ref{section:chofvariables} below. Once the reformulation is derived, we define what is meant by one period of the oscillations.
In fact, we define one period to be an interval $[t_{a},t_{b}]$ such that 
\[
\int_{t_{a}}^{t_{b}}\mfg(\indexnot,t)dt=2\pi.
\]
In preparation for the analysis of the evolution over one period, we need to estimate how the coefficients of the equation vary in an interval
of the form $[t_{a},t_{b}]$; this is the subject of Sections~\ref{section:choapptiint} and \ref{section:varcoeffoneper} below. Once we have gotten 
this far, it is natural to proceed as described at the beginning of the present introduction. In other words, we, on the one hand, freeze the 
coefficients and calculate the evolution over one period.
Moreover, we, on the other hand, estimate the error associated with this procedure. In the situation of interest in 
Chapter~\ref{chapter:destopofosc}, the formulae for the coefficients of the equation are quite involved; cf. 
(\ref{eq:Aoodefge})--(\ref{eq:Attdefge}) below. Carrying out the calculations using these specific coefficients is not very illuminating;
it is not so clear what the relevant structure is. Moreover, the calculations involved are quite cumbersome. In order to estimate the 
evolution associated with freezing the coefficients in the equation, we therefore present an abstract argument in 
Section~\ref{section:calmaexp} below. The disadvantage of this procedure is of course that the motivation for the assumptions
made in Section~\ref{section:calmaexp} only becomes clear in Chapter~\ref{chapter:destopofosc}. A second important technical step 
is to compare the evolution associated with freezing the coefficients with the actual
evolution. Again, doing this in the particular case of interest makes the analysis obscure and cumbersome. We therefore present an 
abstract argument in Section~\ref{section:freezingcoeffabs} below. Combining the above ingredients yields rough estimates of the 
evolution over a period. We write down the result in Section~\ref{section:compappmaexp}. 

\textbf{Variables adapted to the evolution over long periods of time. Setting up an iteration.} The above analysis is adapted to one particular
period of the oscillations. In fact, the variables we use depend on the starting point, say $t_{a}$, of the period $[t_{a},t_{b}]$ of interest.
However, our main interest is in the long term behaviour of solutions. In that setting, it is important to derive estimates for variables which 
do not depend on the particular period. In Section~\ref{section:roughestevooneper} we introduce suitable variables and derive a rough estimate
for how they evolve during one period. In fact, the variables are 
\[
w(\indexnot,t):=\exp\left(-i\int_{0}^{t}\sigma(\indexnot,s)\mfg(\indexnot,s)ds\right)
\left(\begin{array}{c} \mfg(\indexnot,t) z(\indexnot,t) \\ \dot{z}(\indexnot,t)\end{array}\right).
\]
In particular, $|w(\indexnot,t)|^{2}/2$ equals the energy $E$ introduced in (\ref{eq:Edefodecase}). To first approximation, the evolution over
one period is simply a rotation; cf. Lemma~\ref{lemma:Xiroughappr}. Even though this observation is of interest, we need more detailed 
information. We therefore proceed, in Section~\ref{section:firstdetailestoneper}, by deriving a more detailed estimate. Once this main 
estimate has been derived, it is natural to define an iteration. The iteration corresponds to a time sequence $\{t_{k}\}$ such that 
$[t_{k},t_{k+1}]$ is the interval of one period of the oscillations. The purpose of the iteration is to enable us to estimate the evolution 
over long periods of time. In connection with the introduction of the iteration, we also change the variables slightly, the goal being to 
make the change in norm over one period of the oscillations as clear as possible. The resulting variables are equivalent to $w$ at the times 
$t_{k}$. The relevant statements are given in Lemmas~\ref{lemma:wprekit} and \ref{lemma:wkfinlemma}. The final result of the arguments is
an iteration on which a more detailed analysis can be based. However, in order to be able to draw more detailed conclusions, we need to 
estimate matrix products in situations where the number of factors tends to infinity. In order to be able to estimate such products, we need
to make additional assumptions. Since there are different types of assumptions that lead to interesting conclusions, the presentation, after
Section~\ref{section:iteration}, branches off into Parts~\ref{part:unbadegeq}, \ref{part:dominantnoisyspdirection} and \ref{part:nondegcabeq}.

\section{On a class of matrix exponentials}\label{section:calmaexp}

As mentioned in the introduction, we are here interested in solutions to ODE's that exhibit oscillatory behaviour. Moreover, we wish 
to estimate the evolution over one period of the oscillations. The analysis is divided into two steps. First, we estimate the evolution
associated with freezing the coefficients of the equations. Second, we estimate the error associated with freezing the 
coefficients. In the present section, we consider the case of frozen coefficients. However, the statements of the 
relevant results are abstract in nature; there is no reference to the equation in which we are interested in in the end. 
The reader interested in a motivation for the particular assumptions we make here is therefore referred to Chapter~\ref{chapter:destopofosc}.

We begin the present section with an elementary lemma. The result is not very deep and could be derived under much more general 
circumstances. However, the precise statement is adapted to the applications of it in the proof of the main result of this 
section, Lemma~\ref{lemma:maexpcomp}. After the statement and proof of the preliminary Lemma~\ref{lemma:diffmaexp}, we introduce the
terminology specifying the matrices we are interested in here; cf. (\ref{eq:Pdef})--(\ref{eq:medef}) below. Given this terminology, 
we are in a position to state and prove the main result, Lemma~\ref{lemma:maexpcomp}. 

\begin{lemma}\label{lemma:diffmaexp}
Let $a_{k}\in\co$, $0\leq k\in\zo$ be a sequence of complex numbers with the following property: there is a constant $C>0$ 
and an integer $0\leq l\in\zo$ such that $|a_{k}|\leq C\ldr{k}^{l}$
for all $k\geq 0$. Given $1\leq p\in\zo$, $A\in\Mn{p}{\co}$ and a real number $a$, define $f(A)$ and $f_{\abs}(a)$ by 
\[
f(A):=\sum_{k=0}^{\infty}\frac{a_{k}}{k!}A^{k},\ \ \
f_{\abs}(a):=\sum_{k=0}^{\infty}\frac{|a_{k+1}|}{k!}a^{k}.
\]
Let $A,R\in\Mn{p}{\co}$. Then 
\[
\|f(A+R)-f(A)\|\leq f_{\abs}(\|A\|+\|R\|)\cdot \|R\|.
\]
\end{lemma}
\begin{remark}
Note that if $a_{k}=1$, then $f=\exp$ and $f_{\abs}=\exp$, so that 
\[
\|\exp(A+R)-\exp(A)\|\leq \exp(\|A\|+\|R\|)\cdot \|R\|.
\]
\end{remark}
\begin{proof}
Note that 
\[
f(A+R)-f(A)=\sum_{k=1}^{\infty}\frac{a_{k}}{k!}\int_{0}^{1}\frac{d}{ds}(A+sR)^{k}ds.
\]
Since
\[
\left\|\frac{d}{ds}(A+sR)^{k}\right\|=\|R(A+sR)^{k-1}+\dots+(A+sR)^{k-1}R\|\leq k\|R\|(\|A\|+\|R\|)^{k-1},
\]
we conclude that 
\begin{equation*}
\begin{split}
\|f(A+R)-f(A)\| \leq & \sum_{k=1}^{\infty}\frac{|a_{k}|}{k!}k\|R\|(\|A\|+\|R\|)^{k-1}\\
 = & \|R\|\sum_{k=1}^{\infty}\frac{|a_{k}|}{(k-1)!}(\|A\|+\|R\|)^{k-1}=\|R\|f_{\abs}(\|A\|+\|R\|).
\end{split}
\end{equation*}
The lemma follows. 
\end{proof}

Next, we wish to calculate the matrix exponential of a complex $2m\times 2m$-matrix $P$. However, we are interested in matrices
of a special form. To begin with
\begin{equation}\label{eq:Pdef}
P:=\left(\begin{array}{cc} P_{11} & P_{12} \\ P_{21} & P_{22}\end{array}
\right),
\end{equation}
where
\begin{align}
P_{11} := & i\xi\Id_{m}+R_{11}+E_{11},\ \ \
P_{12} := \Id_{m}+E_{12},\label{eq:Poootdef}\\
P_{21} := & -\Id_{m}+R_{21}+E_{21},\ \ \
P_{22} :=  -i\xi\Id_{m}-R_{11}+E_{22}.\label{eq:Ptottdef}
\end{align}
In these expressions, $\xi\in\ro$ and $R_{ij},E_{ij}\in\Mn{m}{\co}$, $i,j\in \{1,2\}$. Intuitively, $R_{ij}$ and $\xi$ should be thought of as 
``first order terms'' that we need to keep track of. However, the $E_{ij}$ should be thought of as ``second order terms'' that we don't need 
to keep track of. As a first, basic, assumption, we require a universal bound on the first and second order terms: 
\begin{equation}\label{eq:Cobd}
|\xi|+\|R_{11}\|+\|R_{21}\|+\textstyle{\sum}_{i,j=1}^{2}\|E_{ij}\|\leq C_{1}
\end{equation}
for some constant $C_{1}$. We here consider the $E_{ij}$ to be small; they should be thought of as error terms. Similarly, 
$\|R_{ij}\|^{2}$ should be thought of as small. Finally, $|\xi|(\|R_{11}\|+\|R_{21}\|)$ should be thought of as small. Define $\me$ by 
\begin{equation}\label{eq:medef}
\me:=\|R_{11}\|^{2}+\|R_{21}\|^{2}+|\xi|(\|R_{11}\|+\|R_{21}\|)+\textstyle{\sum}_{i,j=1}^{2}\|E_{ij}\|.
\end{equation}
Any term which can be bounded by a constant times $\me$ should be thought of as an error term. Given the above terminology
and assumptions, the following lemma holds.

\begin{lemma}\label{lemma:maexpcomp}
Let $P$, $P_{ij}$, $\xi$, $R_{ij}$, $E_{ij}$ and $\me$ be defined as above. Assume, moreover, that (\ref{eq:Cobd}) is satisfied for 
some $0<C_{1}\in\ro$. Then
\begin{equation}\label{eq:exptpisP}
\begin{split}
& \left\|\exp(2\pi sP)-\cos(s\om)\Id_{2m}
-\nu^{-1}\sin(s\om)\left(\begin{array}{cc} i\xi\Id_{m} & \Id_{m} \\ -\Id_{m} & -i\xi\Id_{m}\end{array}\right)\right.\\
 & -\pi s\sin(s\om)\left(\begin{array}{cc} R_{21} & 0 \\ 0 & R_{21}\end{array}\right)
-\frac{1}{2}\sin(s\om)\left(\begin{array}{cc} 2R_{11} & R_{21} \\ R_{21} & -2R_{11}\end{array}\right)\\
 & \left.+\pi s\cos(s\om)\left(\begin{array}{cc} 0 & R_{21} \\ -R_{21} & 0\end{array}\right)\right\|\leq C\me
\end{split}
\end{equation}
for all $s\in [-1,1]$, where $\nu:=(1+\xi^{2})^{1/2}$, $\om:=2\pi\nu$ and the constant $C$ only depends on $C_{1}$. In particular,
\begin{equation}\label{eq:exptpiP}
\begin{split}
& \left\|\exp(\pm 2\pi P)-\cos(\om)\Id_{2m}
\mp \nu^{-1}\sin(\om)\left(\begin{array}{cc} i\xi\Id_{m} & \Id_{m} \\ -\Id_{m} & -i\xi\Id_{m}\end{array}\right)\right.\\
 & \left.\pm\pi \left(\begin{array}{cc} 0 & R_{21} \\ -R_{21} & 0\end{array}\right)\right\|\leq C\me,
\end{split}
\end{equation}
where the constant $C$ only depends on $C_{1}$.
\end{lemma}
\begin{remark}
In case $\xi=0$, (\ref{eq:exptpiP}) can be simplified to 
\[
\left\|\exp(\pm 2\pi P)-\Id_{2m}
\pm\pi \left(\begin{array}{cc} 0 & R_{21} \\ -R_{21} & 0\end{array}\right)\right\|\leq C\me,
\]
where the constant $C$ only depends on $C_{1}$.
\end{remark}
\begin{remark}\label{remark:Qsqspecstru}
In the course of the proof below, it becomes clear that the structure of $P$ is advantageous
when it comes to computing matrix exponentials. In fact, ignoring the error terms $E_{ij}$ in $P$ yields a matrix $Q$
such that $Q^{2}$ is block triangular; cf. (\ref{eq:Qsqform}) below. Moreover, ignoring error terms in $Q^{2}$, it is block diagonal.
This fact significantly simplifies the calculation of $Q^{2k}$, and thereby $Q^{2k+1}$. On the other hand, these are the
matrices we need to calculate in order to compute the matrix exponential. The main reason why $Q^{2}$ has this nice structure
is that the sum of the matrices on the diagonal of $Q$ equal zero, the matrix in the top right corner of $Q$ (i.e., $Q_{12}$,
using notation analogous to the above) is $\Id_{m}$, and the matrix on the bottom left is $-\Id_{m}$ plus lower order terms.
It is also of interest to note that $R_{11}$ does not contribute to the relevant terms in $Q^{2}$. 
\end{remark}
\begin{proof}
Note that $P=Q+E$, where
$E$ is defined in terms of the $E_{ij}$ in the same way as $P$ is defined in terms of the $P_{ij}$; cf. (\ref{eq:Pdef}). Due 
to Lemma~\ref{lemma:diffmaexp}, 
\begin{equation}\label{eq:exptpisPmQ}
\|\exp(2\pi sP)-\exp(2\pi s Q)\|\leq 2\pi \exp(2\pi\|Q\|+2\pi\|E\|)\|E\|\leq C\me
\end{equation}
for $s\in [-1,1]$, where the constant $C$ only depends on $C_{1}$. Here we are only interested in calculating $\exp(2\pi sP)$ up to an 
error term. For that reason, it is sufficient to calculate $\exp(2\pi sQ)$. Letting
\[
Q_{11}=i\xi\Id_{m}+R_{11},\ \ \
Q_{12}=\Id_{m},\ \ \
Q_{21}=-\Id_{m}+R_{21},\ \ \
Q_{22}=-Q_{11},
\]
$Q$ is defined in terms of the $Q_{ij}$ in the same way as $P$ is defined in terms of the $P_{ij}$. Note also that 
\begin{equation}\label{eq:Qsqform}
Q^{2}=\left(\begin{array}{cc} Q_{11}^{2}+Q_{21} & 0 \\ \left[Q_{21},Q_{11}\right] & Q_{11}^{2}+Q_{21}\end{array}\right).
\end{equation}
Let us calculate
\[
Q_{11}^{2}+Q_{21}=-(1+\xi^{2})\Id_{m}+R_{21}
+2i\xi R_{11}+R_{11}^{2}.
\]
The last two terms on the right hand side should be thought of as error terms. Moreover, $[Q_{21},Q_{11}]=[R_{21},R_{11}]$, which is 
an error term. Thus
\[
Q^{2}=\left(\begin{array}{cc} L & 0 \\ 0 & L\end{array}\right)+F,
\]
where 
\[
L:=-(1+\xi^{2})\Id_{m}+R_{21},\ \ \
F:=\left(\begin{array}{cc} 2i\xi R_{11}+R_{11}^{2} & 0 \\ \left[R_{21},R_{11}\right] & 2i\xi R_{11}+R_{11}^{2}\end{array}\right).
\]
By an argument similar to the proof of Lemma~\ref{lemma:diffmaexp}, 
\begin{equation}\label{eq:Qeventermssum}
\sum_{k=0}^{\infty}\frac{1}{(2k)!}(2\pi s Q)^{2k}=\sum_{k=0}^{\infty}\frac{(2\pi s)^{2k}}{(2k)!}
\left(\begin{array}{cc} L^{k} & 0 \\ 0 & L^{k}\end{array}\right)+O(\me)
\end{equation}
for $s\in [-1,1]$, where the constant implicit in the ordo expression only depends on $C_{1}$. Similarly, 
\begin{equation}\label{eq:Qoddtermssum}
\sum_{k=0}^{\infty}\frac{1}{(2k+1)!}(2\pi sQ)^{2k+1}=\sum_{k=0}^{\infty}\frac{(2\pi s)^{2k+1}}{(2k+1)!}
\left(\begin{array}{cc} L^{k} & 0 \\ 0 & L^{k}\end{array}\right)Q+O(\me)
\end{equation}
for $s\in [-1,1]$, where the constant implicit in the ordo expression only depends on $C_{1}$. Before calculating $L^{k}$, let us 
note that, using the notation $\nu$ introduced in the statement of the lemma, $L=-\nu^{2}\Id_{m}+R_{21}$. Moreover, 
\[
L^{k}=(-1)^{k}\nu^{2k}\Id_{m}+k(-1)^{k-1}\nu^{2(k-1)}R_{21}+
\sum_{l=2}^{k}\left(\begin{array}{c} k \\ l\end{array}\right)(-1)^{k-l}\nu^{2(k-l)}R_{21}^{l}.
\]
Let us denote the terms on the right hand side by $L_{i,k}$, $i=0,1,2$, where $L_{1,0}=L_{2,0}=L_{2,1}=0$. For $k\geq 2$, 
\begin{equation*}
\begin{split}
\|L_{2,k}\| \leq & \sum_{l=2}^{k}\left(\begin{array}{c} k \\ l\end{array}\right)\nu^{2(k-l)}\|R_{21}\|^{l}
\leq \|R_{21}\|^{2}\sum_{l=2}^{k}\left(\begin{array}{c} k \\ l\end{array}\right)\nu^{2(k-l)}(\max\{1,\|R_{21}\|\})^{l}\\
 \leq & \|R_{21}\|^{2}(\nu^{2}+\max\{1,\|R_{21}\|\})^{k}.
\end{split}
\end{equation*}
Combining this estimate with (\ref{eq:Qeventermssum}) and (\ref{eq:Qoddtermssum}) yields
\begin{align*}
\sum_{k=0}^{\infty}\frac{1}{(2k)!}(2\pi s Q)^{2k} = & \sum_{k=0}^{\infty}\frac{(2\pi s)^{2k}}{(2k)!}
\left(\begin{array}{cc} L_{0,k}+L_{1,k} & 0 \\ 0 & L_{0,k}+L_{1,k}\end{array}\right)+O(\me),\\
\sum_{k=0}^{\infty}\frac{1}{(2k+1)!}(2\pi sQ)^{2k+1} = & \sum_{k=0}^{\infty}\frac{(2\pi s)^{2k+1}}{(2k+1)!}
\left(\begin{array}{cc} L_{0,k}+L_{1,k} & 0 \\ 0 & L_{0,k}+L_{1,k}\end{array}\right)Q+O(\me).
\end{align*}
Let us calculate
\begin{align*}
\sum_{k=0}^{\infty}\frac{(2\pi s)^{2k}}{(2k)!}L_{0,k} = & \sum_{k=0}^{\infty}\frac{(-1)^{k}}{(2k)!}(s\om)^{2k}\Id_{m}=\cos(s\om)\Id_{m},\\
\sum_{k=0}^{\infty}\frac{(2\pi s)^{2k+1}}{(2k+1)!}L_{0,k} = & \frac{1}{\nu}
\sum_{k=0}^{\infty}\frac{(-1)^{k}}{(2k+1)!}(s\om)^{2k+1}\Id_{m}=\nu^{-1}\sin(s\om)\Id_{m}.
\end{align*}
Next, let us calculate 
\begin{equation*}
\begin{split}
\sum_{k=0}^{\infty}\frac{(2\pi s)^{2k}}{(2k)!}L_{1,k} = & \sum_{k=1}^{\infty}\frac{(2\pi s)^{2k}}{(2k)!}k(-1)^{k-1}\nu^{2(k-1)}R_{21}\\
 = & \frac{\pi s}{\nu}\sum_{k=1}^{\infty}\frac{(-1)^{k-1}}{(2k-1)!}(s\om)^{2k-1}R_{21}
=\frac{\pi s}{\nu}\sin(s\om)R_{21}.
\end{split}
\end{equation*}
Moreover,
\begin{equation*}
\begin{split}
\sum_{k=0}^{\infty}\frac{(2\pi s)^{2k+1}}{(2k+1)!}L_{1,k} = & 
\sum_{k=0}^{\infty}\frac{(2\pi s)^{2k+1}}{(2k+1)!}k(-1)^{k-1}\nu^{2(k-1)}R_{21}\\
 = & \frac{1}{2}\sum_{k=0}^{\infty}\frac{(2\pi s)^{2k+1}}{(2k+1)!}(2k+1)(-1)^{k-1}\nu^{2(k-1)}R_{21}\\
 & -\frac{1}{2}\sum_{k=0}^{\infty}\frac{(2\pi s)^{2k+1}}{(2k+1)!}(-1)^{k-1}\nu^{2(k-1)}R_{21}.
\end{split}
\end{equation*}
Similarly to earlier computations, this can be simplified to 
\[
\sum_{k=0}^{\infty}\frac{(2\pi s)^{2k+1}}{(2k+1)!}L_{1,k}=-\frac{\pi s}{\nu^{2}}\cos(s\om)R_{21}+\frac{1}{2\nu^{3}}\sin(s\om)R_{21}.
\]
Adding up the above computations yields
\begin{equation}\label{eq:exptpisQ}
\begin{split}
\exp(2\pi sQ) = & \cos(s\om)\Id_{2m}+\frac{\pi s}{\nu}\sin(s\om)
\left(\begin{array}{cc} R_{21} & 0 \\ 0 & R_{21}\end{array}\right)
 +\nu^{-1}\sin(s\om)Q\\
 & +\left[-\frac{\pi s}{\nu^{2}}\cos(s\om)+\frac{1}{2\nu^{3}}\sin(s\om)\right]
\left(\begin{array}{cc} R_{21} & 0 \\ 0 & R_{21}\end{array}\right)Q+O(\me).
\end{split}
\end{equation}
Note that 
\[
\frac{1}{\nu^{l}}=1+O(\xi^{2})
\]
for $l=1,2,3$ and the constant implicit in the ordo-expression is numerical. Note also that 
\[
\left(\begin{array}{cc} R_{21} & 0 \\ 0 & R_{21}\end{array}\right)Q
=\left(\begin{array}{cc} 0 & R_{21} \\ -R_{21} & 0\end{array}\right)+O(\me).
\]
Inserting this information into (\ref{eq:exptpisQ}) yields
\begin{equation*}
\begin{split}
\exp(2\pi sQ) = & \cos(s\om)\Id_{2m}
+\nu^{-1}\sin(s\om)\left(\begin{array}{cc} i\xi\Id_{m} & \Id_{m} \\ -\Id_{m} & -i\xi\Id_{m}\end{array}\right)\\
 & +\pi s\sin(s\om)\left(\begin{array}{cc} R_{21} & 0 \\ 0 & R_{21}\end{array}\right)
+\frac{1}{2}\sin(s\om)\left(\begin{array}{cc} 2R_{11} & R_{21} \\ R_{21} & -2R_{11}\end{array}\right)\\
 & -\pi s\cos(s\om)\left(\begin{array}{cc} 0 & R_{21} \\ -R_{21} & 0\end{array}\right)
+O(\me).
\end{split}
\end{equation*}
Combining this calculation with (\ref{eq:exptpisPmQ}) yields (\ref{eq:exptpisP}). In order to derive (\ref{eq:exptpiP}), note that 
\begin{equation}\label{eq:sinomcosom}
\sin(\pm\om)=\pm\sin[2\pi (\nu-1)]=O(\xi^{2}),\ \ \
\cos(\pm\om)=\cos[2\pi (\nu-1)]=1+O(\xi^{4}).
\end{equation}
In case $s=\pm 1$, several of the terms in (\ref{eq:exptpisP}) can thus be dropped (or simplified). This yields the estimate
(\ref{eq:exptpiP}).
\end{proof}

\section{Approximating a fundamental solution by a matrix exponential}\label{section:freezingcoeffabs}

As described in Subsection~\ref{ssection:mastofanal}, the strategy for estimating the evolution over one period is to first 
estimate the evolution for the equation with frozen coefficients; this we did in the previous section. The second 
step is then to estimate the error associated with freezing the coefficients. This is the subject of the present section. 

Consider the initial value problem
\begin{align}
\frac{d\Phi}{ds}(s) = & M(s)\Phi(s),\label{eq:ivpfseq}\\
\Phi(s_{a}) = & \Id_{k},\label{eq:ivpfsid}
\end{align}
where $M\in C^{\infty}[I,\Mn{k}{\co}]$, $1\leq k\in\zo$ and $I$ is an open interval containing $s_{a}$. We denote the 
solution (when evaluated at $s$) by $\Phi(s;s_{a})$. It is also of interest to consider
\begin{align}
\frac{d\Psi}{ds}(s) = & \Psi(s)M(s),\label{eq:reivpfseq}\\
\Psi(s_{a}) = & \Id_{k},\label{eq:reivpfsid}
\end{align}
with $M$ and $s_{a}$ as above. In this case, we denote the solution (when evaluated at $s$) by $\Psi(s;s_{a})$. Fix a compact 
interval $J$ such that $s_{a}\in J\subset I$ (the length of $J$ should be thought of as being small; in practice it will correspond
to one period of the oscillations). It is then of interest to estimate the differences
\[
\|\exp[M(s_{a})(s-s_{a})]-\Phi(s;s_{a})\|,\ \ \
\|\exp[M(s_{a})(s-s_{a})]-\Psi(s;s_{a})\|
\]
for $s\in J$, and to obtain similar estimates for the inverses. 
\begin{lemma}\label{lemma:apprfunsol}
Let $s_{a}\in\ro$, $1\leq k\in\zo$, $I$ be an open interval containing $s_{a}$ and $M\in C^{\infty}[I,\Mn{k}{\co}]$. 
Let $\Phi(s;s_{a})$ denote the solution to (\ref{eq:ivpfseq}) and (\ref{eq:ivpfsid}), evaluated at $s$, and let 
$\Psi(s;s_{a})$ denote the solution to (\ref{eq:reivpfseq}) and (\ref{eq:reivpfsid}), evaluated at $s$. Let $J\subset I$ be a 
compact interval such that $s_{a}\in J$ and assume that there is a constant $0<C_{1}\in\ro$ such that 
\begin{equation}\label{eq:ACoest}
\|M\|_{C(J)}\cdot |J|\leq C_{1},
\end{equation}
where $|J|$ denotes the length of the interval $J$. Then
\begin{align}
\|\exp[M(s_{a})(\cdot-s_{a})]-\Phi(\cdot;s_{a})\|_{C(J)}
 \leq  & C\|M-M(s_{a})\|_{C(J)}\cdot |J|,\label{eq:Phiest}\\
\|\exp[M(s_{a})(\cdot-s_{a})]-\Psi(\cdot;s_{a})\|_{C(J)}
 \leq  & C\|M-M(s_{a})\|_{C(J)}\cdot |J|,\label{eq:Psiest}
\end{align}
where the constant $C$ only depends on an upper bound on $C_{1}$. Moreover, 
\begin{align}
\|\exp[-M(s_{a})(\cdot-s_{a})]-[\Phi(\cdot;s_{a})]^{-1}\|_{C(J)}
 \leq  & C\|M-M(s_{a})\|_{C(J)}\cdot |J|,\label{eq:invPhiest}\\
\|\exp[-M(s_{a})(\cdot-s_{a})]-[\Psi(\cdot;s_{a})]^{-1}\|_{C(J)}
 \leq  & C\|M-M(s_{a})\|_{C(J)}\cdot |J|,\label{eq:invPsiest}
\end{align}
where the constant $C$ only depends on an upper bound on $C_{1}$.
\end{lemma}
\begin{remark}
In this lemma, we use the notation
\[
\|M\|_{C(J)}=\sup_{s\in J}\|M(s)\|.
\]
\end{remark}
\begin{proof}
Note that if we define $\tPhi$ by 
\[
\tPhi(s;s_{a}):=\exp[-M(s_{a})(s-s_{a})]\Phi(s;s_{a}),
\]
then 
\begin{equation}\label{eq:PhitPhi}
\Phi(s;s_{a})=\exp[M(s_{a})(s-s_{a})]+\exp[M(s_{a})(s-s_{a})][\tPhi(s;s_{a})-\Id_{k}].
\end{equation}
It is therefore of interest to estimate $\|\tPhi(s;s_{a})-\Id_{k}\|$. Note, to this end, that 
\[
\frac{d\tPhi}{ds}(s;s_{a})=\exp[-M(s_{a})(s-s_{a})][M(s)-M(s_{a})]\exp[M(s_{a})(s-s_{a})]\tPhi(s;s_{a}).
\]
Due to (\ref{eq:ACoest}), 
\begin{equation}\label{eq:preGronwall}
\|\tPhi(s;s_{a})-\Id_{k}\|\leq \left|\int_{s_{a}}^{s}\left\|\frac{d\tPhi}{dt}(t;s_{a})\right\|dt\right|
\leq C\left|\int_{s_{a}}^{s}\|M(t)-M(s_{a})\|\cdot\|\tPhi(t;s_{a})\|dt\right|
\end{equation}
for $s\in J$, where the constant $C$ only depends on an upper bound on $C_{1}$; note that 
\begin{equation}\label{eq:Asmtaest}
\|\exp[M(s_{a})(s-s_{a})]\|\leq \exp[\|M(s_{a})\|\cdot |J|]\leq e^{C_{1}}
\end{equation}
for $s\in J$
etc. due to (\ref{eq:ACoest}), and that $s$ could be strictly less than $s_{a}$ (this is the reason we include the absolute value
signs outside the integrals even though the integrands are non-negative). The estimate (\ref{eq:preGronwall}) implies
\[
\|\tPhi(s;s_{a})\|\leq 1+ C\left|\int_{s_{a}}^{s}\|M(t)-M(s_{a})\|\cdot\|\tPhi(t;s_{a})\|dt\right|.
\]
Appealing to Gr\"{o}nwall's lemma then yields
\[
\|\tPhi(s;s_{a})\|\leq \exp\left(C\left|\int_{s_{a}}^{s}\|M(t)-M(s_{a})\|dt\right|\right)
\leq \exp(2CC_{1})
\]
for $s\in J$, where $C$ is 
a constant which only depends on an upper bound on $C_{1}$. Returning to (\ref{eq:preGronwall}), we conclude that 
\[
\sup_{s\in J}\|\tPhi(s;s_{a})-\Id_{k}\|\leq C\sup_{s\in J}\|M(s)-M(s_{a})\|\cdot |J|,
\]
where the constant $C$ only depends on an upper bound on $C_{1}$. Combining this estimate with (\ref{eq:PhitPhi}) and (\ref{eq:Asmtaest}) 
yields (\ref{eq:Phiest}). Turning to (\ref{eq:Psiest}), note that if we use the symbol $\dagger$ to indicate the conjugate transpose of 
a square complex matrix, then $\Psi^{\dagger}$ satisfies (\ref{eq:ivpfseq}) and (\ref{eq:ivpfsid}) with $M$ replaced by $M^{\dagger}$. 
Moreover, since $\|L^{\dagger}\|=\|L\|$ for any complex square matrix $L$, it is clear that (\ref{eq:ACoest}) is satisfied with $M$ 
replaced by $M^{\dagger}$. We are thus allowed to appeal to (\ref{eq:Phiest}) in order to obtain 
\[
\sup_{s\in J}\|\exp[M^{\dagger}(s_{a})(s-s_{a})]-\Psi^{\dagger}(s;s_{a})\|
 \leq C\sup_{s\in J}\|M^{\dagger}(s)-M^{\dagger}(s_{a})\|\cdot |J|.
\]
Since
\[
\exp[M^{\dagger}(s_{a})(s-s_{a})]-\Psi^{\dagger}(s;s_{a})
=\left\{\exp[M(s_{a})(s-s_{a})]-\Psi(s;s_{a})\right\}^{\dagger}
\]
and since the conjugate transpose is an isometry, we conclude that (\ref{eq:Psiest}) holds. In order to prove (\ref{eq:invPhiest})
and (\ref{eq:invPsiest}), note that 
\[
\frac{d\Phi^{-1}}{ds}=-\Phi^{-1}\frac{d\Phi}{ds}\Phi^{-1}=-\Phi^{-1} M,\ \ \
\frac{d\Psi^{-1}}{ds}=-\Psi^{-1}\frac{d\Psi}{ds}\Psi^{-1}=-M\Psi^{-1}.
\]
Appealing to (\ref{eq:Phiest}) and (\ref{eq:Psiest}) with $M$ replaced by $-M$ yields (\ref{eq:invPhiest})
and (\ref{eq:invPsiest}).
\end{proof}

\section{Inhomogeneous equations}

In the following chapters, we are interested in estimating the evolution of solutions to inhomogeneous equations. In the present 
section, we indicate how the above analysis can be used to that end. 

Consider an equation of the form 
\begin{equation}\label{eq:inhab}
\frac{dv}{ds}=M(s)v+F(s).
\end{equation}
Here $v$ is $\cn{2m}$ valued, and $v$ and $M$ can be written
\[
v=\left(\begin{array}{c} x\\ y \end{array}\right),\ \ \
M=\left(\begin{array}{cc} M_{11} & M_{12} \\ M_{21} & M_{22}\end{array}
\right)
\]
where $x$ and $y$ each take their values in $\cn{m}$ and the $M_{ij}$ take their values in $\Mn{m}{\co}$. Fix $s_{a}$ in the 
domain of definition, say $I$, of $M$ and let $J\subset I$ be a compact interval. The dominant term in $M(s_{a})$ should be 
thought of as being
\[
\left(\begin{array}{cc} 0 & \nu_{0}\Id_{m} \\ -\nu_{0}\Id_{m} & 0\end{array}
\right).
\]
where $\nu_{0}$ is a large positive number, and the interval $J$ should be thought of as roughly corresponding to one period of the 
associated oscillation (with $s_{a}$ being either the left or the right endpoint of $J$). Let $\Phi(s;s_{a})$ denote the solution to 
(\ref{eq:ivpfseq}) and (\ref{eq:ivpfsid}), evaluated at $s$. Then the solution to (\ref{eq:inhab}) can be written
\[
v(s)=\Phi(s;s_{a})v(s_{a})+\Phi(s;s_{a})\int_{s_{a}}^{s}[\Phi(u;s_{a})]^{-1}F(u)du.
\]
We are mainly interested in the value of $v$ at $s_{b}$, where $s_{b}$ is the endpoint of $J$ which is different from $s_{a}$. We are 
therefore particularly interested in estimating $\Phi(s_{b};s_{a})$ and $[\Phi(u;s_{a})]^{-1}$ for $u\in J$. However, we can appeal to 
Lemma~\ref{lemma:apprfunsol} in order to approximate these matrices using matrix exponentials. Moreover, the matrix exponentials 
of interest are of the form considered in Lemma~\ref{lemma:maexpcomp}.

\chapter[One period; terminology and preliminaries]{One period of the oscillations; terminology and preliminary estimates}\label{chapter:destopofosc}

In the present part of these notes, we are interested in solutions to (\ref{eq:fourierthesystemRge}) in situations where the behaviour is 
oscillatory. Moreover, due to the applications we have in mind, the arguments should apply even when estimates of the form (\ref{eq:weakbal})
are violated. It is therefore necessary to formulate assumptions that are different from the ones we have made previously. We do so in 
Section~\ref{section:assumponepertermaprel} below. Once the appropriate assumptions have been made, we need to reformulate 
(\ref{eq:fourierthesystemRge}) to a form suitable for appealing to the results of Section~\ref{section:calmaexp}. We take a first step in 
this direction in Section~\ref{section:chofvariables}. Most of the remaining part of the chapter is devoted to 
verifying that not only can the equation be reformulated to an appropriate form, its coefficients also satisfy bounds such that the results of 
Section~\ref{section:freezingcoeffabs} apply. 

The assumptions stated in Section~\ref{section:assumponepertermaprel} below involve bounds on, 
e.g., $\dot{\ell}$ and $\ddot{\ell}$, where $\ell$ is introduced in (\ref{eq:ellsigmaXgenRdef}). Such bounds are of central importance, 
since we are considering situations in which the leading order behaviour is oscillatory, and the frequency of the oscillations is roughly
speaking given by $\mfg$. As an initial step in the analysis, we therefore
first define what we mean by one period of the oscillations, and then analyse the variation of expressions involving $\mfg$ over one such period. 
This is the subject of Section~\ref{section:choapptiint}. Given this information, we are then in a position to analyse the variation of the 
remaining coefficients of the equation; cf. Section~\ref{section:varcoeffoneper}. To end the chapter, we, in Section~\ref{section:compappmaexp}, 
obtain a first rough estimate of the evolution over one period.

\section{Assumptions}\label{section:assumponepertermaprel}

Definitions~\ref{def:roughODEtermo} and \ref{def:weaktrans} are two examples of assumptions that we have made concerning the coefficients of 
(\ref{eq:thesystemRge}) so far. However, they are not appropriate in the present context. The reason for this is that, on the one hand, we 
are here interested in unbalanced equations, which are excluded by Definition~\ref{def:roughODEtermo}; in that sense the assumptions in 
Definition~\ref{def:roughODEtermo} are too strong. On the other hand, we also need to control the variation of the
coefficients over one period of the oscillations. We therefore need to bound, e.g., $\|\dot{\a}\|$. In that sense, 
Definition~\ref{def:roughODEtermo} is too weak. Even though we here admit unbalanced equations, we do not want, e.g., 
$\|\a\|$ to grow too quickly. For that reason, we impose restrictions that imply that $\|\a\|$ etc. cannot grow faster than exponentially. 

\begin{definition}\label{def:oscad}
Consider (\ref{eq:thesystemRge}). Assume the associated metric to be such that $(M,g)$ is a canonical separable cosmological model manifold. 
Define $\mfg$ by (\ref{eq:mfgnutdef}) and $\ell$, $\sigma$ and $X$ by (\ref{eq:ellsigmaXgenRdef}). Then (\ref{eq:thesystemRge}) is said to be 
\textit{oscillation adapted} 
\index{Oscillation adapted!equation}%
\index{Equation!oscillation adapted}%
if there are functions $\mff_{\rosh},\mff_{X},\mff_{\roode}\in C^{\infty}(\ro,\ro_{+})$, where 
$\ro_{+}:=(0,\infty)$, and a constant $0<\ellderbd\in\ro$ such that the following holds. First,
\begin{equation}\label{eq:estdtlnge}
|\dot{\ell}(\indexnot,t)|+|\ddot{\ell}(\indexnot,t)|\leq \ellderbd
\end{equation}
for all $t\geq 0$ and all $0\neq\indexnot\in\EFindexset$. Second,
\begin{align}
|\sigma(\indexnot,t)|+|\dot{\sigma}(\indexnot,t) \leq & \mff_{\rosh}(t),\label{eq:shiftbd}\\
\|X(\indexnot,t)\|+\|\dot{X}(\indexnot,t)\| \leq & \mff_{X}(t),\label{eq:Xbd}\\
\|\dot{\a}(t)\|+\|\a(t)\| \leq & \mff_{\roode}(t),\label{eq:albd}\\
\|\dot{\zeta}(t)\|+\|\zeta(t)\| \leq & \mff_{\roode}^{2}(t)\label{eq:zetabd}
\end{align}
for all $t\geq 0$ and all $0\neq\indexnot\in\EFindexset$. Third,
\begin{equation}\label{eq:mffXdotbdetc}
\frac{|\dot{\mff}_{\rosh}(t)|}{\mff_{\rosh}(t)}+\frac{|\dot{\mff}_{X}(t)|}{\mff_{X}(t)}+\frac{|\dot{\mff}_{\roode}(t)|}{\mff_{\roode}(t)}\leq \ellderbd
\end{equation}
for all $t\geq 0$.
\end{definition}
\begin{remark}
The constant $\ellderbd$ and the functions $\mff_{\rosh},\mff_{X},\mff_{\roode}$ are all independent of $\indexnot\in\EFindexset$.
\end{remark}
\begin{remarks}
In order to motivate the requirements, note that we do not want the coefficients $g^{jl}$ and $a^{-2}_{r}$ to vary too quickly over time. This is
the reason for the bound on $\dot{\ell}$; it ensures that $\mfg$ does not grow or decay faster than exponentially. The bounds concerning
$\sigma$, $X$, $\a$ and $\zeta$ are still quite general. However, it is clear from (\ref{eq:mffXdotbdetc}) that the right hand sides of 
(\ref{eq:shiftbd})--(\ref{eq:zetabd}) are not allowed to grow faster than exponentially. Moreover, the bound (\ref{eq:mffXdotbdetc}), in
the end, ensures that the variation of the functions $\mff_{\rosh},\mff_{X},\mff_{\roode}$ over one period of the oscillations is negligible. 
Nevertheless, we impose more detailed restrictions on these functions below. 
\end{remarks}


\section{Change of variables}\label{section:chofvariables}

In order to analyse how the solution evolves over one period of the oscillations, it is 
convenient to change variables (so that we can apply, e.g., Lemma~\ref{lemma:maexpcomp}).
It would be preferable if it were possible to change the variables once and for all. 
However, that turns out to lead to technical difficulties. For that reason we here fix one
time, say $0\leq t_{a}\in\ro$, and then adapt the variables so that we are in a position
to analyse the variation of solutions in a small neighbourhood of $t_{a}$.

\textbf{Change of variables.}
Fix $t_{a}\geq 0$, and let $U(t;t_{a})$ denote the matrix valued solution (evaluated at $t$)
to the initial value problem
\[
U_{t}(t;t_{a})=\frac{1}{2}\a(t)U(t;t_{a}),\ \ \
U(t_{a};t_{a})=\Id_{m}.
\]
Let 
\begin{equation}\label{eq:xyFodefshiftge}
\left(\begin{array}{c} x_{1} \\ y_{1}\\ F_{1} \end{array}\right) := 
\exp\left(-i\int_{0}^{t}\sigma(s)\mfg(s)ds\right)
\left(\begin{array}{c} \mfg z \\ \dot{z}\\ \hf \end{array}\right),
\end{equation}
where $z$ is a solution to (\ref{eq:fourierthesystemRge}) and we suppress all the arguments but the variable of integration. Define $x$, $y$ and $F_{2}$ by
\begin{equation}\label{eq:xyFdefge}
\left(\begin{array}{c} x(\indexnot,t) \\ y(\indexnot,t)\\ F_{2}(\indexnot,t) \end{array}\right) := 
\exp\left[-\frac{1}{2}\int_{t_{a}}^{t}\dot{\ell}(\indexnot,s)ds\right]
\left(\begin{array}{c} U(t;t_{a})x_{1}(\indexnot,t) \\ U(t;t_{a})y_{1}(\indexnot,t)\\ U(t;t_{a})F_{1}(\indexnot,t) \end{array}\right).
\end{equation}
Given $x$ and $y$ defined by (\ref{eq:xyFdefge}), let $v$ be defined by 
\begin{equation}\label{eq:vdefge}
v:=\left(\begin{array}{c} x\\ y \end{array}\right).
\end{equation}
Appealing to (\ref{eq:fourierthesystemRge}), it can be computed that 
\begin{align}
\dot{x} = & \frac{1}{2}(\a+\dot{\ell}\Id_{m}-2i\sigma\mfg\Id_{m})x+\mfg y,\label{eq:dotxprelprel}\\
\dot{y} = & -\mfg x-\frac{1}{2}(\a+\dot{\ell}\Id_{m}
-2i\sigma\mfg\Id_{m})y
+\left[\a,U(\cdot;t_{a})\right]\left[U(\cdot;t_{a})\right]^{-1}y\label{eq:dotyprelprel}\\
 & -U(\cdot;t_{a})\left(iX+Z\right)
\left[U(\cdot;t_{a})\right]^{-1}x+F_{2},\nonumber
\end{align}
where $Z(\indexnot,t):=\zeta(t)/\mfg(\indexnot,t)$. At this stage, it is natural to pause to consider what has been gained by
this reformulation. Freezing the coefficients of $x$ and $y$ in (\ref{eq:dotxprelprel}) and (\ref{eq:dotyprelprel}) at $t=t_{a}$, 
interpreting the frozen coefficients as the components of a matrix in $\Mn{2m}{\co}$, and dividing the matrix by $\mfg(t_{a})$ 
yields 
\begin{equation}\label{eq:Padef}
P_{a}:=\left(\begin{array}{cc} R_{11}+i\xi\Id_{m} & \Id_{m} \\ -\Id_{m}+R_{21}+E_{21} & -R_{11}-i\xi\Id_{m}\end{array}
\right),
\end{equation}
where 
\[
R_{11}:=\frac{1}{2\mfg(\indexnot,t_{a})}[\a(t_{a})+\dot{\ell}(\indexnot,t_{a})\Id_{m}],\ \ \
R_{21}:=-i\frac{X(\indexnot,t_{a})}{\mfg(\indexnot,t_{a})},\ \ \
E_{21}:=-\frac{Z(\indexnot,t_{a})}{\mfg(\indexnot,t_{a})}
\]
and $\xi:=-\sigma(\indexnot,t_{a})$. Thus $P_{a}$ has the structure considered in Section~\ref{section:calmaexp}. The advantage of this 
structure is discussed in Remark~\ref{remark:Qsqspecstru}; cf. also the statement of Lemma~\ref{lemma:maexpcomp}.

\textbf{Change of time coordinate.} The above discussion demonstrates that freezing the coefficients of 
(\ref{eq:dotxprelprel}) and (\ref{eq:dotyprelprel}) yields a matrix of a form appropriate for appealing to the results of 
Section~\ref{section:calmaexp}. However, when estimating the error associated with freezing the coefficients, it is convenient 
to change the time coordinate so that the largest coefficients appearing in the equations for $x$ and $y$ are constant. To that 
end, we introduce a new time coordinate
\begin{equation}\label{eq:taudefge}
\tau(\indexnot,t):=\int_{t_{0}}^{t}\frac{\mfg(\indexnot,s)}{\mfg(\indexnot,t_{0})}ds.
\end{equation}
Here $0\leq t_{0}\in\ro$ is a parameter which is typically different from $t_{a}$. The purpose of introducing such a parameter is
to indicate a starting point at which $\mfg$ is sufficiently large for the behaviour to be oscillatory. On the other hand, we wish
to use the same coordinate $\tau$ for many periods of the oscillations close to $t_{0}$. That is the reason we do not necessarily
require $t_{0}$ to equal $t_{a}$. Roughly speaking, $t_{0}$ fixes an oscillatory ``era'', whereas $t_{a}$ fixes the starting point 
of one period of the oscillations. In what follows, it is convenient to suppress the dependence of $\tau$ on $\indexnot$, and we 
often do so in what follows. Moreover, we abuse notation in the way it is normally done when changing coordinates; we think of 
$\tau$ as being a function and a real number interchangeably; and we use the notation $\a$, e.g., for both the function mapping 
$t$ to $\a(t)$ and the function mapping $\tau$ to $\a[t(\tau)]$ etc. Changing time coordinate according to (\ref{eq:taudefge})
yields
\begin{align*}
\frac{dx}{d\tau} = & \frac{\mfg_{0}}{2\mfg}(\a+\dot{\ell}\Id_{m}
-2i\sigma\mfg\Id_{m})x+\mfg_{0}y,\\
\frac{dy}{d\tau} = & -\mfg_{0}x-\frac{\mfg_{0}}{2\mfg}(\a+\dot{\ell}\Id_{m}
-2i\sigma\mfg\Id_{m})y+\frac{\mfg_{0}}{\mfg}
\left[\a,U\right]U^{-1}y\\
 & -U\frac{\mfg_{0}}{\mfg}(iX+Z)U^{-1}x+F_{3},
\end{align*}
where $\mfg_{0}(\indexnot):=\mfg(\indexnot,t_{0})$, 
\begin{equation}\label{eq:Fthdef}
F_{3}:=\frac{\mfg_{0}}{\mfg}F_{2}
\end{equation}
and we, in this case, have omitted all the arguments. Moreover, we think of $\a$ as the function
of $\tau$ mapping $\tau$ to $\a[t(\tau)]$ etc. Finally, the dots appearing still refer to the time 
coordinate $t$.  

\textbf{Matrix formulation of the equations.} Define
\begin{align}
A_{11} := & \frac{1}{2}\frac{\a\mfg_{0}}{\mfg}+\frac{1}{2}\frac{\dot{\ell}\mfg_{0}}{\mfg}\Id_{m}
-i\sigma\mfg_{0}\Id_{m},\label{eq:Aoodefge}\\
A_{12} := & \mfg_{0}\Id_{m},\label{eq:Aotdefge}\\
A_{21} := & -\mfg_{0}\Id_{m}-U\frac{\mfg_{0}}{\mfg}(iX+Z)U^{-1},
\label{eq:Atodefge}\\
A_{22} := & -A_{11}+\frac{\mfg_{0}}{\mfg}
\left[\a,U\right]U^{-1}.\label{eq:Attdefge}
\end{align}
Let $A$ and $F$ be defined by 
\begin{equation}\label{eq:AFdef}
A:=\left(\begin{array}{cc} A_{11} & A_{12} \\ A_{21} & A_{22}\end{array}
\right),\ \ \
F:=\left(\begin{array}{c} 0\\ F_{3} \end{array}\right)
\end{equation}
With this notation, (\ref{eq:fourierthesystemRge}) yields
\begin{equation}\label{eq:fttsysreform}
\frac{dv}{d\tau}= Av+F.
\end{equation}

\section{Choosing appropriate time intervals}\label{section:choapptiint}

Our long term goal is to estimate how the energy of solutions to (\ref{eq:fttsysreform}) evolve over time. To this end, 
it is natural to divide the interval $[0,\infty)$ into different regions. Consider the matrix $A$ defined by (\ref{eq:AFdef}) 
and (\ref{eq:Aoodefge})--(\ref{eq:Attdefge}). In intervals where the expression $\mfg_{0}\Id_{m}$, appearing in $A_{12}$ and 
$A_{21}$, dominates, the behaviour can be expected to be oscillatory; below, we loosely speak of such intervals as ``oscillatory 
regions'' or ``oscillatory eras'' (though we do not provide a formal definition of the meaning of this terminology). Considering 
the homogeneous equation associated with (\ref{eq:fttsysreform}), we then, roughly speaking, expect oscillations with a period of 
$2\pi/\mfg_{0}$. It is therefore of interest to analyse how solutions to (\ref{eq:fttsysreform}) 
evolve over one such period. Keeping (\ref{eq:taudefge}) in mind, a period of $2\pi/\mfg_{0}$ in $\tau$-time corresponds to an 
interval $[t_{a},t_{b}]$ (or $[t_{b},t_{a}]$) in $t$-time such that $t_{a},t_{b}\geq 0$ and such that 
\begin{equation}\label{eq:percondge}
\left|\int_{t_{a}}^{t_{b}}\mfg(\indexnot,t)dt\right|=2\pi,
\end{equation}
where we include the absolute value sign in order to allow for the possibility that $t_{b}$ might be strictly smaller than $t_{a}$. 
To begin with, it is thus natural to verify that there are time intervals with this property 
and to analyse the variation of quantities such as $\mfg$ in such intervals. 

\begin{lemma}\label{lemma:gaintvarge}
Assume that (\ref{eq:thesystemRge}) is oscillation adapted, let $0\neq\indexnot\in\EFindexset$
and define $\mfg$ by (\ref{eq:mfgnutdef}). Assume that there is a $t_{a}\geq 0$ such that 
\begin{equation}\label{eq:gtalbge}
\mfg(\indexnot,t_{a})\geq \max\{4\pi\ellderbd,2\},
\end{equation}
where $\ellderbd$ is the constant appearing in (\ref{eq:estdtlnge}). Then there is a $t_{b}\geq t_{a}$ 
such that (\ref{eq:percondge}) holds. If, in addition to (\ref{eq:gtalbge}), $t_{a}\geq 2\pi$, then there is a $0\leq t_{b}<t_{a}$
such that (\ref{eq:percondge}) holds. Fix a $t_{b}\geq 0$ such that (\ref{eq:percondge}) holds and let $I_{a}$ be the interval with 
endpoints $t_{a}$ and $t_{b}$. Then $\mfg(\indexnot,t) \geq 1$ and 
\begin{align}
\frac{1}{2} \leq & \frac{\mfg(\indexnot,t)}{\mfg(\indexnot,t_{a})}\leq 2,\label{eq:gttaroughge}\\
\frac{\pi}{\mfg(\indexnot,t_{a})} \leq & |I_{a}| \leq \frac{4\pi}{\mfg(\indexnot,t_{a})}\label{eq:tatbroughge}
\end{align}
for all $t\in I_{a}$, where $|I_{a}|$ denotes the length of $I_{a}$. Finally, 
\begin{align}
\left|\frac{1}{\mfg(\indexnot,t)}-\frac{1}{\mfg(\indexnot,t_{a})}
\right| \leq & \frac{C}{\mfg^{2}(\indexnot,t_{a})},\label{eq:oogdiffge}\\
\left|\frac{\dot{\mfg}(\indexnot,t)}{\mfg^{2}(\indexnot,t)}-\frac{\dot{\mfg}(\indexnot,t_{a})}{\mfg^{2}(\indexnot,t_{a})}\right|
 \leq & \frac{C}{\mfg^{2}(\indexnot,t_{a})},\label{eq:gdogsdiffge}\\
\left||I_{a}|-\frac{2\pi}{\mfg(\indexnot,t_{a})}\right| \leq & \frac{C}{\mfg^{2}(\indexnot,t_{a})},\label{eq:tbmtagestge}\\
\left|\int_{I_{a}}\frac{\dot{\mfg}(\indexnot,s)}{\mfg(\indexnot,s)}ds
-\frac{2\pi\dot{\mfg}(\indexnot,t_{a})}{\mfg^{2}(\indexnot,t_{a})}\right| \leq & \frac{C}{\mfg^{2}(\indexnot,t_{a})}\label{eq:Fbpoestge}
\end{align}
for all $t\in I_{a}$, 
where the constant $C$ only depends on the parameter $\ellderbd$ appearing in the estimate
(\ref{eq:estdtlnge}).
\end{lemma}
\begin{remark}
In the statement of the lemma, we use the convention that 
\[
\int_{t_{a}}^{t_{b}}f(s)ds=\sgn(t_{b}-t_{a}) \int_{I_{a}}f(s)ds,
\]
where $\sgn(t)$ denotes the sign of $t$. 
\end{remark}
\begin{remark}
In the proof, we omit reference to the argument $\indexnot$ for the sake of brevity. 
\end{remark}
\begin{proof}
To begin with, (\ref{eq:estdtlnge}) implies that $\mfg(t)\geq\mfg(t_{a})\exp[-\ellderbd|t-t_{a}|]$. Thus
\begin{equation}\label{eq:intmfgfies}
\left|\int_{t_{a}}^{t}\mfg(s)ds\right|\geq \left|\int_{t_{a}}^{t}\mfg(t_{a})\exp[-\ellderbd|s-t_{a}|]ds\right|
=\ellderbd^{-1}\mfg(t_{a})[1-e^{-\ellderbd|t-t_{a}|}]
\end{equation}
for $t\geq 0$. Assuming $\mfg(t_{a})\geq 4\pi\ellderbd$ and $t\geq 0$ to be such that $\ellderbd|t-t_{a}|\geq\ln 2$, it is 
clear that the right hand side is bounded from below by $2\pi$. In particular, there is thus a $t_{b}\geq t_{a}$ such that 
(\ref{eq:percondge}) holds. A more crude way to estimate the left hand side of (\ref{eq:intmfgfies}) is the following:
\begin{equation}\label{eq:intmfgsees}
\left|\int_{t_{a}}^{t}\mfg(s)ds\right|\geq |t-t_{a}|\mfg(t_{a})e^{-\ellderbd|t-t_{a}|}.
\end{equation}
Combining (\ref{eq:intmfgfies}) and (\ref{eq:intmfgsees}) yields the following conclusion. Assume that (\ref{eq:gtalbge}) holds 
and that $t_{a}\geq 2\pi$. Then there are two possibilities. Either $\ellderbd t_{a}\geq\ln 2$, in which case (\ref{eq:intmfgfies}) 
implies that there is a $0\leq t_{b}<t_{a}$ such that (\ref{eq:percondge}) holds, or $\ellderbd t_{a}\leq\ln 2$, in which case 
(\ref{eq:intmfgsees}) yields the same conclusion. The statements concerning the existence of $t_{b}$ follow. 

Fix a $t_{b}\geq 0$ such that (\ref{eq:percondge}) holds and define $I_{a}$ as in the statement of the lemma. Then (\ref{eq:intmfgfies})
with $t=t_{b}$ yields
\[
2\pi=\left|\int_{t_{a}}^{t_{b}}\mfg(t)dt\right|\geq \ellderbd^{-1}\mfg(t_{a})[1-e^{-\ellderbd l_{a}}]\geq 4\pi
[1-e^{-\ellderbd l_{a}}],
\]
where $l_{a}:=|I_{a}|$. Thus $e^{-\ellderbd l_{a}}\geq 1/2$. Combining this estimate with (\ref{eq:estdtlnge}) yields
\begin{equation}\label{eq:gvariationge}
\frac{1}{2}\leq e^{-\ellderbd l_{a}}\leq\frac{\mfg(t)}{\mfg(t_{a})}\leq e^{\ellderbd l_{a}}\leq 2
\end{equation}
for $t\in I_{a}$. This estimate implies (\ref{eq:gttaroughge}). Moreover, combining (\ref{eq:gtalbge}) with (\ref{eq:gttaroughge}) yields 
the conclusion that $\mfg(t)\geq 1$ for all $t\in I_{a}$. The estimate (\ref{eq:gvariationge}) also yields
\[
\frac{1}{2}\mfg(t_{a})l_{a}\leq\left|\int_{t_{a}}^{t_{b}}\mfg(t)dt\right|\leq 2\mfg(t_{a})l_{a}.
\]
Combining this estimate with (\ref{eq:percondge}) yields (\ref{eq:tatbroughge}). 

We are now in a position to turn to the variation of $\mfg$ and $\dot{\mfg}$ etc. during one
period. To begin with, note that 
\begin{equation*}
\begin{split}
\left|\frac{1}{\mfg(t)}-\frac{1}{\mfg(t_{a})}\right|
 = & \frac{1}{\mfg(t_{a})}\left|\frac{\mfg(t_{a})}{\mfg(t)}-1\right|
\leq \frac{1}{\mfg(t_{a})}\left(e^{\ellderbd |t-t_{a}|}-1\right)
\leq \frac{1}{\mfg(t_{a})}\left(e^{\ellderbd l_{a}}-1\right)\\
 \leq & \frac{1}{\mfg(t_{a})}\left(\exp\left[\frac{4\pi\ellderbd}{\mfg(t_{a})}\right]-1\right)
\leq \frac{C}{\mfg^{2}(t_{a})}
\end{split}
\end{equation*}
for $t\in I_{a}$, where the constant $C$ only depends on the parameter $\ellderbd$ appearing in the assumption
(\ref{eq:estdtlnge}). In particular, (\ref{eq:oogdiffge}) holds. Let us turn to the 
variation of $\dot{\mfg}/\mfg^{2}$. Note, to begin with, that
\begin{equation}\label{eq:vardtlngge}
\begin{split}
\left|\frac{\dot{\mfg}(t)}{\mfg(t)}-\frac{\dot{\mfg}(t_{a})}{\mfg(t_{a})}\right|
 \leq & \left|\int_{t_{a}}^{t}\ddot{\ell}(s)ds\right|
\leq \ellderbd l_{a}\leq\frac{C}{\mfg(t_{a})}
\end{split}
\end{equation}
for $t\in I_{a}$, where the constant $C$ only depends on the parameter $\ellderbd$ appearing in the assumption
(\ref{eq:estdtlnge}). Due to this estimate, 
\[
\left|\frac{\dot{\mfg}(t)}{\mfg^{2}(t)}-\frac{\dot{\mfg}(t_{a})}{\mfg^{2}(t_{a})}\right|
\leq\left|\frac{1}{\mfg(t)}\left(\frac{\dot{\mfg}(t)}{\mfg(t)}
-\frac{\dot{\mfg}(t_{a})}{\mfg(t_{a})}\right)\right|+
\left|\left(\frac{1}{\mfg(t)}-\frac{1}{\mfg(t_{a})}\right)\frac{\dot{\mfg}(t_{a})}{\mfg(t_{a})}\right|
\leq\frac{C}{\mfg^{2}(t_{a})}
\]
for $t\in I_{a}$, where the constant $C$ only depends on the parameter $\ellderbd$ appearing in the assumption
(\ref{eq:estdtlnge}), and we have used (\ref{eq:oogdiffge}) and (\ref{eq:gttaroughge}).
In particular, (\ref{eq:gdogsdiffge}) holds. Next, let us derive a better estimate for
$l_{a}$. To begin with, 
\[
2\pi=\int_{I_{a}}\mfg(t)dt=\int_{I_{a}}\mfg(t)\mfg(t_{a})\left[\frac{1}{\mfg(t_{a})}
-\frac{1}{\mfg(t)}\right]dt+l_{a}\mfg(t_{a}).
\]
Thus
\[
\left|l_{a}-\frac{2\pi}{\mfg(t_{a})}\right|\leq
\int_{I_{a}}\mfg(t)\left|\frac{1}{\mfg(t_{a})}
-\frac{1}{\mfg(t)}\right|dt\leq\frac{C}{\mfg^{2}(t_{a})},
\]
where the constant $C$ only depends on the parameter $\ellderbd$ appearing in the assumption
(\ref{eq:estdtlnge}) and we have appealed to (\ref{eq:gttaroughge})--(\ref{eq:oogdiffge}).
Thus (\ref{eq:tbmtagestge}) holds. Next, note that 
\[
\left|\int_{I_{a}}\frac{\dot{\mfg}(s)}{\mfg(s)}ds
-\frac{\dot{\mfg}(t_{a})}{\mfg(t_{a})}l_{a}\right|\leq 
\int_{I_{a}}\left|\frac{\dot{\mfg}(s)}{\mfg(s)}-\frac{\dot{\mfg}(t_{a})}{\mfg(t_{a})}\right|ds
\leq \frac{C}{\mfg^{2}(t_{a})},
\]
where the constant $C$ only depends on the parameter $\ellderbd$ appearing in the assumption
(\ref{eq:estdtlnge}) and we have appealed to (\ref{eq:tatbroughge}) and (\ref{eq:vardtlngge}).
Combining this estimate with (\ref{eq:estdtlnge}) and (\ref{eq:tbmtagestge}) yields
(\ref{eq:Fbpoestge}). The lemma follows. 
\end{proof}

\section{Variation of the coefficients during one period}\label{section:varcoeffoneper}

In order to estimate the behaviour of solutions to (\ref{eq:fttsysreform}), it is convenient to appeal to 
Lemma~\ref{lemma:apprfunsol}. In order to be allowed to do so, we need to verify two things. First, we need
an estimate of the form (\ref{eq:ACoest}) and second, we need an estimate of the variation of $A$ in the 
interval of interest. To obtain such estimates is the purpose of the next lemma. 

\begin{lemma}\label{lemma:varofA}
Assume that (\ref{eq:thesystemRge}) is oscillation adapted and let $0\neq\indexnot\in\EFindexset$.
Given that, in addition, the assumptions of Lemma~\ref{lemma:gaintvarge} are satisfied, let $t_{a}$, $t_{b}$ 
and $I_{a}$ be defined as in the statement of Lemma~\ref{lemma:gaintvarge}. Finally, let $\tau_{a}:=\tau(t_{a})$, 
$\tau_{b}:=\tau(t_{b})$ and 
\begin{equation}\label{eq:vareadef}
\vare_{a}:=\frac{\mff_{\rosh}(t_{a})}{\mfg(t_{a})}[1+\mff_{\roode}(t_{a})+\mff_{X}(t_{a})]
+\frac{1+\mff_{X}^{2}(t_{a})+\mff_{\roode}^{2}(t_{a})}{\mfg^{2}(t_{a})}.
\end{equation}
\index{$\a$Aa@Notation!Constants!$\vare_{a}$}%
Then if $A$ is defined as in Section~\ref{section:chofvariables}, 
\begin{equation}\label{eqAtintest}
\|A\|_{C(J_{a})}\cdot |J_{a}|\leq C_{a,\ellderbd}\vare_{a}^{1/2}+C_{\ellderbd}\mff_{\rosh}(t_{a})+C,
\end{equation}
where $J_{a}$ is the interval in $\tau$-time corresponding to $I_{a}$; $C$ is a numerical constant; $C_{\ellderbd}$ only depends 
on an upper bound on $\ellderbd$; and $C_{a,\ellderbd}$ only depends on an upper bound on $\vare_{a}$ and $\ellderbd$. Moreover, 
\begin{equation}\label{eq:Avarbd}
\|A-A(\tau_{a})\|_{C(J_{a})}\cdot |J_{a}|\leq C_{a,\ellderbd}\vare_{a},
\end{equation}
where $C_{a,\ellderbd}$ only depends on an upper bound on $\vare_{a}$ and the parameter $\ellderbd$ appearing 
in (\ref{eq:estdtlnge}).
\end{lemma}
\begin{proof}
Let $I_{a}$ and $J_{a}$ be as in the statement of the lemma. We consider the different variables to be functions of $t$ or 
functions of $\tau$ as convenient. In 
particular, instead of computing the supremum of $\|A\|$ over $J_{a}$, we can consider $\|A\|$ to be a function of $t$ and 
compute the supremum over $I_{a}$. In what follows, we wish to express all the estimates in terms of $\mfg(t_{a})$,
$\mff_{\rosh}(t_{a})$, $\mff_{X}(t_{a})$ and $\mff_{\roode}(t_{a})$. In order to be allowed to do so, we need to estimate the 
variation of $\mff_{\rosh}$, $\mff_{X}$ and $\mff_{\roode}$ in $I_{a}$. Due to (\ref{eq:mffXdotbdetc}), (\ref{eq:gtalbge})
and (\ref{eq:tatbroughge}), it is clear that 
\begin{equation}\label{eq:mffvindvar}
\mff_{\rosh}(t)\leq e^{2\pi\ellderbd}\mff_{\rosh}(t_{a}),\ \ \
\mff_{X}(t)\leq e^{2\pi\ellderbd}\mff_{X}(t_{a}),\ \ \
\mff_{\roode}(t)\leq e^{2\pi\ellderbd}\mff_{\roode}(t_{a})
\end{equation}
for all $t\in I_{a}$. 

\textbf{Estimating the norm of $A$.} Most of the functions of which $A$ is constructed can immediately be estimated using 
(\ref{eq:estdtlnge})--(\ref{eq:mffXdotbdetc}). However, this is not true of $\|U\|$ and $\|U^{-1}\|$.  In order 
to estimate these objects, we appeal to Lemma~\ref{lemma:apprfunsol} with $M$, $s_{a}$ and $J$ replaced by $\a/2$, 
$t_{a}$ and $I_{a}$ respectively. Then 
\begin{equation}\label{eq:alCobd}
\frac{1}{2}\|\a\|_{C(I_{a})}|I_{a}|\leq 2\pi e^{2\pi\ellderbd}\frac{\mff_{\roode}(t_{a})}{\mfg(t_{a})}
\leq 2\pi e^{2\pi\ellderbd}\vare_{a}^{1/2},
\end{equation}
where we have used (\ref{eq:albd}), (\ref{eq:tatbroughge}) and (\ref{eq:mffvindvar}). Thus 
Lemma~\ref{lemma:apprfunsol} applies with $C_{1}$ replaced by the right hand side of (\ref{eq:alCobd}), so that 
\begin{equation}\label{eq:Uestdetailed}
\begin{split}
\|\exp[\a(t_{a})(\cdot-t_{a})/2]-U(\cdot;t_{a})\|_{C(I_{a})}
 \leq & C\|\a-\a(t_{a})\|_{C(I_{a})}|I_{a}|\\
 \leq & Ce^{2\pi\ellderbd}\mff_{\roode}(t_{a})|I_{a}|^{2}\leq C_{a,\ellderbd}\frac{\mff_{\roode}(t_{a})}{\mfg^{2}(t_{a})}
\end{split}
\end{equation}
where the constant $C$ only depends on an upper bound on $C_{1}$ (and thus only on an upper bound on $\ellderbd$ and $\vare_{a}$);
$C_{a,\ellderbd}$ only depends on an upper bound on $\ellderbd$ and $\vare_{a}$; and we 
have used (\ref{eq:albd}), (\ref{eq:tatbroughge}) and (\ref{eq:mffvindvar}). The estimate for $U^{-1}$ is analogous. 
Note, in particular, that
\begin{equation}\label{eq:UUinvmidest}
\|U(\cdot;t_{a})-\Id_{m}\|_{C(I_{a})}+\|[U(\cdot;t_{a})]^{-1}-\Id_{m}\|_{C(I_{a})}\leq C_{a,\ellderbd}\frac{\mff_{\roode}(t_{a})}{\mfg(t_{a})},
\end{equation}
where the constant $C_{a,\ellderbd}$ only depends on an upper bound on $\ellderbd$ and $\vare_{a}$. 

Let us now estimate the norm of the components of $A$ (after multiplication by $|J_{a}|$). Note, to begin with, 
that $|J_{a}|=2\pi/\mfg_{0}$; cf. (\ref{eq:taudefge}). Thus
\begin{equation}\label{eq:Aootintest}
\|A_{11}(\tau)\|\cdot |J_{a}|\leq \pi\left[\frac{\mff_{\roode}(t)}{\mfg(t)}+\frac{\ellderbd}{\mfg(t)}\right]+2\pi\mff_{\rosh}(t)
\leq C_{\ellderbd}[\vare_{a}^{1/2}+\mff_{\rosh}(t_{a})]
\end{equation}
for $\tau\in J_{a}$, where $C_{\ellderbd}$ only depends on an upper bound on $\ellderbd$ and we have used
(\ref{eq:estdtlnge}), (\ref{eq:shiftbd}), (\ref{eq:albd}), (\ref{eq:gttaroughge}) and (\ref{eq:mffvindvar}). Next, 
\begin{equation}\label{eq:Aottintest}
\|A_{12}(\tau)\|\cdot |J_{a}|=2\pi
\end{equation}
for all $\tau$. Turning to $A_{21}$, 
\begin{equation}\label{eq:Atotintest}
\|A_{21}(\tau)\|\cdot |J_{a}|\leq 2\pi\left[1+C_{a,\ellderbd}\left(\frac{\mff_{X}(t)}{\mfg(t)}+\frac{\mff_{\roode}^{2}(t)}{\mfg^{2}(t)}\right)\right]
\leq 2\pi+C_{a,\ellderbd}\vare_{a}^{1/2}
\end{equation}
for $\tau\in J_{a}$, where $C_{a,\ellderbd}$ only depends on an upper bound on $\vare_{a}$ and $\ellderbd$, and we have used
(\ref{eq:Xbd}), (\ref{eq:zetabd}), (\ref{eq:gttaroughge}), (\ref{eq:mffvindvar}) and (\ref{eq:UUinvmidest}). Before turning to
$A_{22}$, we need to estimate the last term on the right hand side of (\ref{eq:Attdefge}). Estimate
\begin{equation*}
\begin{split}
 & \left\|\frac{\mfg_{0}}{\mfg(t)}\left[\a(t),U(t;t_{a})\right][U(t;t_{a})]^{-1}\right\|\\
 \leq & \left\|\frac{\mfg_{0}}{\mfg(t)}\left[\a(t),U(t;t_{a})-\exp(\a(t_{a})(t-t_{a})/2)\right][U(t;t_{a})]^{-1}\right\|\\
 & +\left\|\frac{\mfg_{0}}{\mfg(t)}\left[\a(t)-\a(t_{a}),\exp(\a(t_{a})(t-t_{a})/2)\right][U(t;t_{a})]^{-1}\right\|.
\end{split}
\end{equation*}
Appealing to (\ref{eq:albd}), (\ref{eq:gttaroughge}), (\ref{eq:tatbroughge}), (\ref{eq:mffvindvar}), (\ref{eq:Uestdetailed}) and 
(\ref{eq:UUinvmidest}) yields
\begin{equation}\label{eq:commerrterm}
\left\|\frac{\mfg_{0}}{\mfg(t)}\left[\a(t),U(t;t_{a})\right][U(t;t_{a})]^{-1}\right\|\leq C_{a,\ellderbd}\mfg_{0}\frac{\mff_{\roode}(t_{a})}{\mfg^{2}(t_{a})}
\leq C_{a,\ellderbd}\mfg_{0}\vare_{a}
\end{equation}
for all $t\in I_{a}$, where the constant $C_{a,\ellderbd}$ only depends on an upper bound on $\ellderbd$ and $\vare_{a}$. Combining this estimate with 
(\ref{eq:Aootintest}) yields 
\begin{equation}\label{eq:Atttintest}
\|A_{22}(\tau)\|\cdot |J_{a}|\leq C_{a,\ellderbd}\vare_{a}^{1/2}+C_{\ellderbd}\mff_{\rosh}(t_{a})
\end{equation}
for $\tau\in J_{a}$, where $C_{a,\ellderbd}$ only depends on an upper bound on $\ellderbd$ and $\vare_{a}$, and $C_{\ellderbd}$ only depends on an upper bound
on $\ellderbd$. Combining (\ref{eq:Aootintest}), (\ref{eq:Aottintest}), (\ref{eq:Atotintest}) and (\ref{eq:Atttintest}) yields (\ref{eqAtintest}).

\textbf{The variation of $A$.} Let us now turn to the variation of the $A_{ij}$. Since the $A_{ij}$ are given by 
(\ref{eq:Aoodefge})--(\ref{eq:Attdefge}), we need to estimate the variation of 
\[
\frac{\a\mfg_{0}}{\mfg},\ \ \
\frac{\dot{\ell}\mfg_{0}}{\mfg},\ \ \
\sigma \mfg_{0},\ \ \
\frac{\mfg_{0}}{\mfg}UXU^{-1},\ \ \
\frac{\mfg_{0}}{\mfg}UZU^{-1},\ \ \
\frac{\mfg_{0}}{\mfg}\left[\a,U\right]U^{-1}
\]
in the interval $I_{a}$. Estimate
\begin{equation}\label{eq:firsting}
\left\|\frac{\a(t)\mfg_{0}}{\mfg(t)}-\frac{\a(t_{a})\mfg_{0}}{\mfg(t_{a})}\right\|\leq C_{\ellderbd}\frac{\mff_{\roode}(t_{a})\mfg_{0}}{\mfg^{2}(t_{a})}
\leq C_{\ellderbd}\mfg_{0}\vare_{a}
\end{equation}
for all $t\in I_{a}$,
where we have used (\ref{eq:albd}), (\ref{eq:gttaroughge}), (\ref{eq:tatbroughge}), (\ref{eq:oogdiffge}) and (\ref{eq:mffvindvar}) and the constant 
$C_{\ellderbd}$ only depends on an upper bound on the parameter $\ellderbd$ appearing in (\ref{eq:estdtlnge}). Moreover, 
\begin{equation}\label{eq:seconding}
\left\|\frac{\dot{\ell}(t)\mfg_{0}}{\mfg(t)}-\frac{\dot{\ell}(t_{a})\mfg_{0}}{\mfg(t_{a})}\right\|\leq C_{\ellderbd}\frac{\mfg_{0}}{\mfg^{2}(t_{a})}
\leq C_{\ellderbd}\mfg_{0}\vare_{a}
\end{equation}
for all $t\in I_{a}$,
where the constant $C_{\ellderbd}$ only depends on an upper bound on $\ellderbd$, and we have used the estimate (\ref{eq:gdogsdiffge}). 
Appealing to (\ref{eq:shiftbd}), (\ref{eq:tatbroughge}) and (\ref{eq:mffvindvar}), 
\begin{equation}\label{eq:thirding}
\left\|\sigma(t)\mfg_{0}-\sigma(t_{a})\mfg_{0}\right\|\leq C_{\ellderbd}\mfg_{0}\frac{\mff_{\rosh}(t_{a})}{\mfg(t_{a})}
\leq C_{\ellderbd}\mfg_{0}\vare_{a}
\end{equation}
for all $t\in I_{a}$,
where the constant $C_{\ellderbd}$ only depends on an upper bound on $\ellderbd$. Next, let us estimate
\begin{equation}\label{eq:fourthing}
\begin{split}
 & \left\|\frac{\mfg_{0}}{\mfg(t)}U(t;t_{a})X(t)[U(t;t_{a})]^{-1}-\frac{\mfg_{0}}{\mfg(t_{a})}X(t_{a})\right\|\\
 \leq & C_{a,\ellderbd}\mfg_{0}\frac{1+\mff_{\roode}^{2}(t_{a})+\mff_{X}^{2}(t_{a})}{\mfg^{2}(t_{a})}\leq C_{a,\ellderbd}\mfg_{0}\vare_{a}
\end{split}
\end{equation}
for all $t\in I_{a}$,
where the constant only depends on an upper bound on $\ellderbd$ and $\vare_{a}$ and we have used (\ref{eq:Xbd}),
(\ref{eq:gttaroughge}), (\ref{eq:tatbroughge}), (\ref{eq:oogdiffge}), (\ref{eq:mffvindvar}) and (\ref{eq:UUinvmidest}).
Similarly, 
\begin{equation}\label{eq:fifthing}
\left\|\frac{\mfg_{0}}{\mfg(t)}U(t;t_{a})Z(t)[U(t;t_{a})]^{-1}-\frac{\mfg_{0}}{\mfg(t_{a})}Z(t_{a})\right\|\leq
C_{a,\ellderbd}\mfg_{0}\frac{\mff_{\roode}^{2}(t_{a})}{\mfg^{2}(t_{a})}\leq C_{a,\ellderbd}\mfg_{0}\vare_{a}
\end{equation}
for all $t\in I_{a}$,
where the constant only depends on an upper bound on $\ellderbd$ and $\vare_{a}$ and we have used (\ref{eq:zetabd}),
(\ref{eq:gttaroughge}), (\ref{eq:tatbroughge}), (\ref{eq:oogdiffge}), (\ref{eq:mffvindvar}) and (\ref{eq:UUinvmidest}).
Combining (\ref{eq:firsting}), (\ref{eq:seconding}), (\ref{eq:thirding}), (\ref{eq:fourthing}), (\ref{eq:fifthing}) with 
(\ref{eq:commerrterm}) yields (\ref{eq:Avarbd}). The lemma follows. 
\end{proof}

\section{Computing the approximate fundamental solution}\label{section:compappmaexp}

Due to Lemma~\ref{lemma:varofA}, we have the estimates needed for appealing to Lemma~\ref{lemma:apprfunsol}. In particular, 
\begin{align}
\|\exp[A(\tau_{a})(\cdot-\tau_{a})]-\Phi(\cdot;\tau_{a})\|_{C(J_{a})}
 \leq & C_{a,\ellderbd,\rosh}\vare_{a},\label{eq:Phiappr}\\
\|\exp[-A(\tau_{a})(\cdot-\tau_{a})]-[\Phi(\cdot;\tau_{a})]^{-1}\|_{C(J_{a})}
 \leq & C_{a,\ellderbd,\rosh}\vare_{a},\label{eq:Phiinvappr}
\end{align}
where $J_{a}$, $\vare_{a}$ and $\tau_{a}$ are defined as in the statement of Lemma~\ref{lemma:varofA}; and 
the constant $C_{a,\ellderbd,\rosh}$ only depends on an upper bound on $\vare_{a}$, the parameter $\ellderbd$ appearing in (\ref{eq:estdtlnge})
and $\mff_{\rosh}(t_{a})$. Moreover, $\Phi(\cdot;\tau_{a})$ is the solution to the initial value problem
\begin{equation}\label{eq:ivpPhi}
\frac{d\Phi}{d\tau}(\tau)=A(\tau)\Phi(\tau),\ \ \ \Phi(\tau_{a})=\Id_{2m},
\end{equation}
where $A$ is the matrix defined in Section~\ref{section:chofvariables}. Since the evolution of a solution to 
(\ref{eq:fttsysreform}) over a period can be written
\begin{equation}\label{eq:vintform}
v(\tau_{b})=\Phi(\tau_{b};\tau_{a})v(\tau_{a})+\Phi(\tau_{b};\tau_{a})\int_{\tau_{a}}^{\tau_{b}}[\Phi(\tau;\tau_{a})]^{-1}F(\tau)d\tau,
\end{equation}
where $\tau_{b}$ is defined as in the statement of Lemma~\ref{lemma:varofA}, we are mainly interested in approximating $\Phi(\tau_{b};\tau_{a})$ 
and $[\Phi(\tau';\tau_{a})]^{-1}$ for $\tau'\in J_{a}$. This is the purpose of the next lemma. 

\begin{lemma}\label{lemma:phiaphiiappr}
Assume that (\ref{eq:thesystemRge}) is oscillation adapted and let $0\neq\indexnot\in\EFindexset$.
Given that, in addition, the assumptions of Lemma~\ref{lemma:gaintvarge} are satisfied, let $t_{a}$, $t_{b}$, $\tau_{a}$, $\tau_{b}$,
$I_{a}$ and $J_{a}$ be defined as in the statement of Lemma~\ref{lemma:varofA}. Let $A$ and $\tau$ be defined as in 
Section~\ref{section:chofvariables} and let $\Phi(\tau;\tau_{a})$ be the solution to (\ref{eq:ivpPhi}). Then
\begin{equation}\label{eq:PIPappdef}
\|\Phi_{\app}(\tau_{b};\tau_{a})-\Phi(\tau_{b};\tau_{a})\|
+\|\Phi^{\mri}_{\app}(\cdot;\tau_{a})-[\Phi(\cdot;\tau_{a})]^{-1}\|_{C(J_{a})}
 \leq C_{a,\ellderbd,\rosh}\vare_{a},
\end{equation}
where the constant $C_{a,\ellderbd,\rosh}$ only depends on an upper bound on $\vare_{a}$ defined in 
(\ref{eq:vareadef}); the parameter $\ellderbd$ appearing in (\ref{eq:estdtlnge}); and $\mff_{\rosh}(t_{a})$. In this estimate, 
\begin{equation}\label{eq:Pappdef}
\begin{split}
\Phi_{\app}(\tau_{b};\tau_{a}) := & \cos(\om)\Id_{2m}
+\sgn(t_{b}-t_{a})\cdot\nu^{-1}\sin(\om)\left(\begin{array}{cc} i\xi\Id_{m} & \Id_{m} \\ -\Id_{m} & -i\xi\Id_{m}\end{array}\right)\\
 & +\sgn(t_{b}-t_{a})\cdot\frac{i\pi}{\mfg(t_{a})} \left(\begin{array}{cc} 0 & X(t_{a}) \\ -X(t_{a}) & 0\end{array}\right)
\end{split}
\end{equation}
where $\xi:=-\sigma(t_{a})$; $\nu:=(1+\xi^{2})^{1/2}$; and $\om:=2\pi\nu$. Moreover, 
\begin{equation}\label{eq:PIappdef}
\begin{split}
\Phi^{\mri}_{\app}(\tau;\tau_{a}) := & \cos(s\om)\Id_{2m}
+\nu^{-1}\sin(s\om)\left(\begin{array}{cc} i\xi\Id_{m} & \Id_{m} \\ -\Id_{m} & -i\xi\Id_{m}\end{array}\right)\\
 & +\pi s\sin(s\om)\left(\begin{array}{cc} R_{21} & 0 \\ 0 & R_{21}\end{array}\right)
+\frac{1}{2}\sin(s\om)\left(\begin{array}{cc} 2R_{11} & R_{21} \\ R_{21} & -2R_{11}\end{array}\right)\\
 & -\pi s\cos(s\om)\left(\begin{array}{cc} 0 & R_{21} \\ -R_{21} & 0\end{array}\right),
\end{split}
\end{equation}
where $\xi$, $\nu$ and $\omega$ are defined as above, $R_{11}$ and $R_{21}$ are defined in (\ref{eq:Rijetcdef}) below, and 
\begin{equation}\label{eq:sdef}
s:=-\frac{\mfg_{0}}{2\pi}(\tau-\tau_{a}),
\end{equation}
where $\mfg_{0}:=\mfg(t_{0})$ and $t_{0}$ is introduced in connection with the definition of $\tau$; cf. (\ref{eq:taudefge}).
\end{lemma}
\begin{remark}
When convenient, we write 
\[
\Phi_{\app}=\Phi_{\app}^{0}+\Phi_{\app}^{1},\ \ \ 
\Phi^{\mri}_{\app}=\Phi^{\mri,0}_{\app}+\Phi^{\mri,1}_{\app} 
\]
where $\Phi_{\app}^{0}$ and $\Phi^{\mri,0}_{\app}$ are the sum of the first two terms on the right hand sides of (\ref{eq:Pappdef})
and (\ref{eq:PIappdef}) respectively. 
\end{remark}
\begin{remark}
It is important to keep in mind that, regardless of the choice of $s,\xi\in\ro$, the matrix $\Phi^{\mri,0}_{\app}$
is an isometry of $\cn{2m}$ (with respect to the standard inner product); i.e., $\Phi^{\mri,0}_{\app}$ is unitary. 
\end{remark}
\begin{proof}
Due to 
(\ref{eq:Phiappr}) and (\ref{eq:Phiinvappr}), it is sufficient to calculate $\exp\left[A(\tau_{a})T_{0}s\right]$, for $s\in [-1,1]$,
where $T_{0}:=2\pi/\mfg_{0}$. Returning to Section~\ref{section:calmaexp}, in particular Lemma~\ref{lemma:maexpcomp}, we are thus interested
in calculating $\exp(2\pi sP)$, where $P=A(\tau_{a})/\mfg_{0}$. Note that the $P$ defined in this way is also given by (\ref{eq:Pdef}), 
(\ref{eq:Poootdef}) and (\ref{eq:Ptottdef}), where $E_{11}$, $E_{12}$ and $E_{22}$ are defined to be zero, and 
\begin{equation}\label{eq:Rijetcdef}
\xi:=-\sigma(t_{a}),\ \ \ R_{11}:=\frac{1}{2}\frac{\a(t_{a})}{\mfg(t_{a})}+\frac{1}{2}\frac{\dot{\ell}(t_{a})}{\mfg(t_{a})}\Id_{m},\ \ \
R_{21}:=-i\frac{X(t_{a})}{\mfg(t_{a})},\ \ \
E_{21}:=-\frac{Z(t_{a})}{\mfg(t_{a})}.
\end{equation}
The next step is to apply Lemma~\ref{lemma:maexpcomp}. To be allowed to do so, we need to verify that the bound (\ref{eq:Cobd})
holds. Estimate
\begin{equation}\label{eq:Cosubst}
|\xi|+\|R_{11}\|+\|R_{21}\|+\textstyle{\sum}_{i,j}\|E_{ij}\|\leq \mff_{\rosh}(t_{a})+C_{a,\ellderbd}\vare_{a}^{1/2},
\end{equation}
where the constant $C_{a,\ellderbd}$ only depends on an upper bound on $\ellderbd$ and $\vare_{a}$. Thus Lemma~\ref{lemma:maexpcomp} is applicable 
with $C_{1}$ replaced by the right hand side of (\ref{eq:Cosubst}). It is of interest to express $\me$, defined in 
(\ref{eq:medef}), in terms of $\mfg(t_{a})$ etc. Note, to this end, that in our situation
\[
\|R_{11}\|^{2}\leq\frac{1}{2}\frac{\mff_{\roode}^{2}(t_{a})+\ellderbd^{2}}{\mfg^{2}(t_{a})},\ \ \
\|R_{21}\|^{2}\leq\frac{\mff_{X}^{2}(t_{a})}{\mfg^{2}(t_{a})},\ \ \
\|E_{21}\|\leq\frac{\mff_{\roode}^{2}(t_{a})}{\mfg^{2}(t_{a})}.
\]
Moreover, 
\[
|\xi|(\|R_{11}\|+\|R_{21}\|)\leq \frac{\mff_{\rosh}(t_{a})}{\mfg(t_{a})}[\ellderbd+\mff_{\roode}(t_{a})+\mff_{X}(t_{a})].
\]
Thus $\me\leq C_{\ellderbd}\vare_{a}$ for some constant $C_{\ellderbd}$ only depending on an upper bound on $\ellderbd$. Summing up the above observations, 
we conclude that (\ref{eq:exptpisP}) and (\ref{eq:exptpiP}) are applicable, and that the constant $C$ appearing in these estimates
only depends on an upper bound on $\vare_{a}$, the parameter $\ellderbd$ appearing in (\ref{eq:estdtlnge}) and $\mff_{\rosh}(t_{a})$. Moreover, 
$\me$ can be replaced by $\vare_{a}$. Thus (\ref{eq:PIPappdef}) holds, where $\Phi_{\app}$ and $\Phi^{\mri}_{\app}$ are given by 
(\ref{eq:Pappdef}) and (\ref{eq:PIappdef}) respectively. Moreover, the constant $C_{a,\ellderbd,\rosh}$ has the properties stated in the lemma. 
\end{proof}

\chapter[One period; detailed estimates]{One period of the oscillations; reformulation and 
detailed estimates}\label{chapter:destopofoscref}

The variables introduced in Section~\ref{section:chofvariables} are adapted to one period of the oscillations. 
However, we are interested in drawing conclusions concerning the evolution over longer periods of time. This requires, first of all,
a change of variables, and it turns out to be convenient to work with $w$ introduced in (\ref{eq:wdefshiftge}) below. Using 
(\ref{eq:vintform}), we can derive an expression for how $w$ evolves over one period of the oscillations; cf. (\ref{eq:wofitoXietc})
below. The main point of the initial part of the present chapter is to estimate the matrices appearing in this expression; in 
Section~\ref{section:roughestevooneper} we derive a rough, and in Section~\ref{section:firstdetailestoneper} a
detailed, estimate. With these estimates as a starting point, we are in a position to set up an iteration; cf. Section~\ref{section:iteration}.
The purpose of the iteration is to make it possible to estimate how $w$ develops over longer periods of time. As a starting point, it
is natural to consider how $w(t_{k})$ develops, where $\{t_{k}\}$ is the time sequence associated with the iteration. However, in order
to simplify the formulae as much as possible, it is convenient to modify $w(t_{k})$ somewhat. To this end, we introduce $w_{\pre,k}$ and 
$w_{\fin,k}$ in (\ref{eq:wkpredef}) and (\ref{eq:wfinkdef}) respectively. The norms of these variables are equivalent to that of $w(t_{k})$.
In Lemmas~\ref{lemma:wprekit} and \ref{lemma:wkfinlemma}, we derive formulae for how $w_{\pre,k}$ and $w_{\fin,k}$ evolve. In the end, 
we obtain (\ref{eq:wprekpertransvarpi}) and (\ref{eq:differenceeqrepres}), which are the basis for the analysis to follow.

\section{Rough estimate of the evolution during one period}\label{section:roughestevooneper}

The motivation for considering the function $v$ introduced in (\ref{eq:vdefge}) is the desire to simplify the analysis of how the solution 
evolves over one period. By a period, we here mean an interval $I_{a}$ of the form described in 
Lemma~\ref{lemma:gaintvarge}. In order to consider the evolution of the solution over longer time intervals, we, however, need to consider
other functions. In what follows, we focus our attention on $w$, defined by
\begin{equation}\label{eq:wdefshiftge}
w(t):=\left(\begin{array}{c} x_{1}(t) \\ y_{1}(t)\end{array}\right)
=\exp\left[\frac{1}{2}\int_{t_{a}}^{t}\dot{\ell}(t')dt'\right]
\left(\begin{array}{cc} [U(t;t_{a})]^{-1} & 0 \\ 0 & [U(t;t_{a})]^{-1}\end{array}
\right)v(t).
\end{equation}
Here the first equality is a definition and the second is a consequence of (\ref{eq:xyFdefge}). Due to (\ref{eq:wdefshiftge}) it is perhaps not 
surprising that the following matrix will play an important role in the analysis: 
\begin{equation}\label{eq:Xidef}
\Xi(\tau;\tau_{a})=\exp\left[\frac{1}{2}\int_{t_{a}}^{t}\dot{\ell}(t')dt'\right]
\left(\begin{array}{cc} [U(t;t_{a})]^{-1} & 0 \\ 0 & [U(t;t_{a})]^{-1}\end{array}
\right)\Phi(\tau;\tau_{a}),
\end{equation}
where $\Phi(\tau;\tau_{a})$ is the solution to (\ref{eq:ivpPhi});
in this formula, we take it for granted that $t$ and $\tau$ are related according to $\tau(t)=\tau$. Since (\ref{eq:vintform}) holds with 
$\tau_{b}$ replaced by $\tau$, 
\begin{equation}\label{eq:wofitoXietc}
w(t)=\Xi(\tau;\tau_{a})w(t_{a})+\Xi(\tau;\tau_{a})\int_{\tau_{a}}^{\tau}[\Phi(\tau';\tau_{a})]^{-1}F(\tau')d\tau',
\end{equation}
where we have used the fact that $w(t_{a})=v(\tau_{a})$. Here, and in the formulae below, $\tau_{a}$, $\tau_{b}$ and $J_{a}$ are defined as in the 
statement of Lemma~\ref{lemma:varofA}. We are mainly interested in calculating the right hand side of (\ref{eq:wofitoXietc})
when $\tau=\tau_{b}$. However, in some situations, it is of interest to obtain a rough approximation of the right hand side when 
$\tau\in J_{a}$. For that reason, we here derive a rough estimate for $\Xi(\tau;\tau_{a})$. 

\begin{lemma}\label{lemma:Xiroughappr}
Assume that (\ref{eq:thesystemRge}) is oscillation adapted and let $0\neq\indexnot\in\EFindexset$.
Given that, in addition, the assumptions of Lemma~\ref{lemma:gaintvarge} are satisfied, let $t_{a}$, $t_{b}$, $\tau_{a}$, $\tau_{b}$,
$I_{a}$ and $J_{a}$ be defined as in the statement of Lemma~\ref{lemma:varofA}. Let $A$ and $\tau$ be defined as in 
Section~\ref{section:chofvariables} and let $\Phi(\tau;\tau_{a})$ be the solution to (\ref{eq:ivpPhi}). Finally, let $\Xi(\tau;\tau_{a})$ 
be defined by (\ref{eq:Xidef}). Then, for $\tau\in J_{a}$, 
\begin{equation}\label{eq:Xiroughest}
\begin{split}
 & \left\|\Xi(\tau;\tau_{a})-\mfD_{a}(t)\right\|+\left\|[\Xi(\tau;\tau_{a})]^{-1}-[\mfD_{a}(t)]^{T}\right\| \\
 \leq & C_{a,\ellderbd,\rosh}\frac{1+\mff_{\roode}(t_{a})+\mff_{X}(t_{a})}{\mfg(t_{a})}+C_{a,\ellderbd,\rosh}\vare_{a}+C_{a,\ellderbd,\rosh}\mff_{\rosh}(t_{a}),
\end{split}
\end{equation}
where $C_{a,\ellderbd,\rosh}$ only depends on an upper bound on $\vare_{a}$ defined in (\ref{eq:vareadef}); the parameter $\ellderbd$ appearing in 
(\ref{eq:estdtlnge}); and $\mff_{\rosh}(t_{a})$. Moreover, $\mfD_{a}(t)$ is defined by 
\begin{equation}\label{eq:varphiamdadef}
\varphi_{a}(t):=\int_{t_{a}}^{t}\mfg(t')dt',\ \ \
\mfD_{a}(t):=\left(\begin{array}{rr} \cos\varphi_{a}(t)\mathrm{Id}_{m}
& \sin\varphi_{a}(t)\mathrm{Id}_{m} \\  -\sin\varphi_{a}(t)\mathrm{Id}_{m} & \cos\varphi_{a}(t)\mathrm{Id}_{m}\end{array}
\right)
\end{equation}
and $[\mfD_{a}(t)]^{T}$ denotes the transpose of the matrix $\mfD_{a}(t)$. 
\end{lemma}
\begin{remark}
In (\ref{eq:Xiroughest}) we take it for granted that $t$ and $\tau$ are related according to $\tau(t)=\tau$, where
$\tau(t)$ is defined in (\ref{eq:taudefge}).
\end{remark}
\begin{remark}
Due to (\ref{eq:varphiamdadef}), it is clear that the leading order behaviour of $\Xi(\tau;\tau_{a})$ over one period of the oscillations
is just a rotation. 
\end{remark}
\begin{proof}
Consider $\Xi(\tau;\tau_{a})$, defined by  (\ref{eq:Xidef}). The first factor on the right hand side of (\ref{eq:Xidef}) is given by 
\[
\exp\left[\frac{1}{2}\int_{t_{a}}^{t}\dot{\ell}(t')dt'\right]
=\frac{\mfg^{1/2}(t)}{\mfg^{1/2}(t_{a})}.
\]
Appealing to (\ref{eq:gttaroughge}) and (\ref{eq:oogdiffge}) thus yields
\begin{equation}\label{eq:intexpelldpre}
\left|\exp\left[\frac{1}{2}\int_{t_{a}}^{t}\dot{\ell}(t')dt'\right]-1\right|\leq \frac{C_{\ellderbd}}{\mfg(t_{a})}
\end{equation}
for $t\in I_{a}$, where $C_{\ellderbd}$ only depends on the parameter $\ellderbd$ appearing in (\ref{eq:estdtlnge}). Turning to $[U(t;t_{a})]^{-1}$, 
note that (\ref{eq:UUinvmidest}) holds. Combining this estimate with (\ref{eq:intexpelldpre}) yields
\begin{equation}\label{eq:ffestpre}
\left\|\exp\left[\frac{1}{2}\int_{t_{a}}^{t}\dot{\ell}(t')dt'\right]
\left(\begin{array}{cc} [U(t;t_{a})]^{-1} & 0 \\ 0 & [U(t;t_{a})]^{-1}\end{array}
\right)-\Id_{2m}\right\|\leq \frac{C_{\ellderbd}}{\mfg(t_{a})}+C_{a,\ellderbd}\frac{\mff_{\roode}(t_{a})}{\mfg(t_{a})}
\end{equation}
for $t\in I_{a}$, 
where the constant $C_{a,\ellderbd}$ only depends on an upper bound on $\ellderbd$ and $\vare_{a}$; here $\vare_{a}$ is given by (\ref{eq:vareadef}). Similarly,
\begin{equation}\label{eq:invffestpre}
\left\|\exp\left[-\frac{1}{2}\int_{t_{a}}^{t}\dot{\ell}(t')dt'\right]
\left(\begin{array}{cc} U(t;t_{a}) & 0 \\ 0 & U(t;t_{a})\end{array}
\right)-\Id_{2m}\right\|\leq \frac{C_{\ellderbd}}{\mfg(t_{a})}+C_{a,\ellderbd}\frac{\mff_{\roode}(t_{a})}{\mfg(t_{a})}.
\end{equation}
In the present setting, we consider the right hand side of (\ref{eq:ffestpre}) to be an error term. As a consequence, the first two factors 
appearing in the definition of $\Xi(\tau;\tau_{a})$ can be replaced by $\Id_{2m}$. Next, let us approximate $\Phi(\tau;\tau_{a})$. In order 
to do so, it is convenient to return to Section~\ref{section:compappmaexp}, in particular (\ref{eq:Phiappr}) and the proof of 
Lemma~\ref{lemma:phiaphiiappr}. Due to the proof of Lemma~\ref{lemma:phiaphiiappr}, we know that (\ref{eq:exptpisP}) holds with $P=A(\tau_{a})/\mfg_{0}$.
Moreover, the constant $C$ appearing in this estimate only depends on an upper bound on $\vare_{a}$, the parameter $\ellderbd$ appearing in (\ref{eq:estdtlnge}) 
and $\mff_{\rosh}(t_{a})$. In addition, $\me$ can be replaced by $\vare_{a}$. Finally, since $R_{11}$ and $R_{21}$ are given by (\ref{eq:Rijetcdef}), we know 
that 
\[
\|R_{11}\|+\|R_{21}\|\leq \frac{\ellderbd+\mff_{\roode}(t_{a})+\mff_{X}(t_{a})}{\mfg(t_{a})}.
\]
Appealing to (\ref{eq:exptpisP}) with $P$ replaced by $A(\tau_{a})/\mfg_{0}$ yields
\begin{equation}\label{eq:exptpisPexpre}
\begin{split}
& \left\|\exp(2\pi sP)-\cos(s\om_{a})\Id_{2m}
-\nu_{a}^{-1}\sin(s\om_{a})\left(\begin{array}{cc} i\xi_{a}\Id_{m} & \Id_{m} \\ -\Id_{m} & -i\xi_{a}\Id_{m}\end{array}\right)\right\|\\
 \leq & C_{\ellderbd}\frac{1+\mff_{\roode}(t_{a})+\mff_{X}(t_{a})}{\mfg(t_{a})}+C_{a,\ellderbd,\rosh}\vare_{a},
\end{split}
\end{equation}
where $\xi_{a}=-\sigma(t_{a})$; $\nu_{a}=(1+\xi_{a}^{2})^{1/2}$; $\om_{a}=2\pi\nu_{a}$; $C_{\ellderbd}$ only depends on $\ellderbd$; and $C_{a,\ellderbd,\rosh}$ only 
depends on an 
upper bound on $\vare_{a}$, the parameter $\ellderbd$ appearing in (\ref{eq:estdtlnge}), and $\mff_{\rosh}(t_{a})$. In (\ref{eq:exptpisPexpre}), $s$ takes its 
values in $[-1,1]$, so that 
\[
|\cos(s\om_{a})-\cos(2\pi s)|\leq |s|\cdot|\om_{a}-2\pi|\leq 2\pi |\nu_{a}-1|\leq \pi\mff_{\rosh}^{2}(t_{a}). 
\]
Similarly, 
\[
|\sin(s\om_{a})-\sin(2\pi s)|+|\nu_{a}^{-1}-1|\leq (\pi+1)\mff_{\rosh}^{2}(t_{a}). 
\]
Finally, $|\xi_{a}|\leq \mff_{\rosh}(t_{a})$, so that (\ref{eq:exptpisPexpre}) yields
\begin{equation}\label{eq:exptpisPexsvpre}
\begin{split}
& \left\|\exp(2\pi sP)-\cos(2\pi s)\Id_{2m}
-\sin(2\pi s)\left(\begin{array}{cc} 0 & \Id_{m} \\ -\Id_{m} & 0\end{array}\right)\right\|\\
 \leq & C_{\ellderbd}\frac{1+\mff_{\roode}(t_{a})+\mff_{X}(t_{a})}{\mfg(t_{a})}+C_{a,\ellderbd,\rosh}\vare_{a}+C_{\rosh}\mff_{\rosh}(t_{a}),
\end{split}
\end{equation}
where $C_{\ellderbd}$ and $C_{a,\ellderbd,\rosh}$ have the same dependence as before and $C_{\rosh}$ only depends on an upper bound on $\mff_{\rosh}(t_{a})$. 
In this estimate, $s=s_{\pm}$, where
\begin{equation}\label{eq:tpisrewrpre}
2\pi s_{\pm}=\pm 2\pi\frac{\tau-\tau_{a}}{T_{0}}=\pm\mfg_{0}\int_{t_{a}}^{t}\frac{\mfg(t')}{\mfg_{0}}dt'=\pm\int_{t_{a}}^{t}\mfg(t')dt',
\end{equation}
and $T_{0}=2\pi/\mfg_{0}$. Here the plus sign is the relevant sign when estimating $\Xi(\tau;\tau_{a})$ and the minus sign is the relevant sign when 
estimating $[\Xi(\tau;\tau_{a})]^{-1}$. Define $\varphi_{a}$ and $\mfD_{a}$ by (\ref{eq:varphiamdadef}). Then (\ref{eq:exptpisPexsvpre}) yields
\begin{equation}\label{eq:exptpisPextvpre}
\begin{split}
& \left\|\exp(2\pi s_{+}P)-\mfD_{a}(t)\right\|\leq 
C_{\ellderbd}\frac{1+\mff_{\roode}(t_{a})+\mff_{X}(t_{a})}{\mfg(t_{a})}+C_{a,\ellderbd,\rosh}\vare_{a}+C_{\rosh}\mff_{\rosh}(t_{a})
\end{split}
\end{equation}
for $t\in I_{a}$. An analogous estimate holds for $s_{-}$. Combining these estimates with (\ref{eq:Phiappr}) and 
(\ref{eq:Phiinvappr}) yields
\begin{equation}\label{eq:Phitautaukestpre}
\begin{split}
 & \left\|\Phi(\tau;\tau_{a})-\mfD_{a}(t)\right\|+\left\|[\Phi(\tau;\tau_{a})]^{-1}-[\mfD_{a}(t)]^{T}\right\|\\
 \leq & C_{\ellderbd}\frac{1+\mff_{\roode}(t_{a})+\mff_{X}(t_{a})}{\mfg(t_{a})}+C_{a,\ellderbd,\rosh}\vare_{a}+C_{\rosh}\mff_{\rosh}(t_{a})
\end{split}
\end{equation}
for $t\in I_{a}$, 
where $\tau$ and $t$ are related according to (\ref{eq:taudefge}); the constants have the same dependence as before; and $[\mfD_{a}(t)]^{T}$ 
denotes the transpose of the matrix $\mfD_{a}(t)$. Finally, combining (\ref{eq:ffestpre}), (\ref{eq:invffestpre}) and (\ref{eq:Phitautaukestpre}) 
yields (\ref{eq:Xiroughest}). The lemma follows. 
\end{proof}

\subsection{Estimating solutions to inhomogeneous equations}\label{ssection:estsolinhomeqprelobs}

For future reference, it is of interest to make the following observation concerning the problem of estimating solutions to inhomogeneous equations. 
To begin with, (\ref{eq:fourierthesystemRge}), (\ref{eq:xyFodefshiftge}) and (\ref{eq:wdefshiftge}) yield
\begin{equation}\label{eq:wmamfprel}
\dot{w}(\indexnot,t)=\ma(\indexnot,t)w(\indexnot,t)+\mf(\indexnot,t),
\end{equation}
where
\begin{align*}
\ma(\indexnot,t) := & \left(\begin{array}{cc} \dot{\ell}(\indexnot,t)\Id_{m}-i\sigma(\indexnot,t)\mfg(\indexnot,t)\Id_{m} & \mfg(\indexnot,t)\Id_{m} \\
-\mfg(\indexnot,t)\Id_{m}-iX(\indexnot,t)-[\mfg(\indexnot,t)]^{-1}\zeta(t) & i\sigma(\indexnot,t)\mfg(\indexnot,t)\Id_{m}-\a(t) \end{array}\right),\\
\mf(\indexnot,t) := & \exp\left(-i\int_{0}^{t}\sigma(\indexnot,t')\mfg(\indexnot,t')dt'\right)\left(\begin{array}{c} 0 \\ \hf(\indexnot,t) \end{array}\right).
\end{align*}
Given $t',t\in I$, let $\Phi_{\indexnot}(t;t')$ be the solution to the initial value problem 
\[
\frac{d\Phi}{dt}=\ma\Phi,\ \ \ \Phi(t')=\Id_{2m},
\]
evaluated at $t$. Then solutions to (\ref{eq:wmamfprel}) satisfy
\begin{equation}\label{eq:winhomformredtoPhiest}
w(\indexnot,t)=\Phi_{\indexnot}(t;t_{a})w(\indexnot,t_{a})+\int_{t_{a}}^{t}\Phi_{\indexnot}(t;t')\mf(\indexnot,t')dt'.
\end{equation}
In order to estimate $|w(\indexnot,t)|$ on, say, $[t_{a},t_{b}]$ in terms of $|w(\indexnot,t_{a})|$ and an integral involving $|\hf|$, it is thus sufficient to 
estimate $\|\Phi_{\indexnot}(t;t')\|$ for $t_{a}\leq t'\leq t\leq t_{b}$. There is a similar statement if one wants the estimate to be in terms of 
$|w(\indexnot,t_{b})|$ and an integral involving $|\hf|$.

\section{First detailed estimate of the evolution over one period}\label{section:firstdetailestoneper}

In the previous section, we derived a rough estimate for $\Xi(\tau;\tau_{a})$. However, in order to obtain information concerning the 
long time behaviour of solutions, we need to derive detailed estimates. The purpose of the present section is to take a first step in 
that direction.

\begin{lemma}\label{lemma:oneperiod}
Assume that (\ref{eq:thesystemRge}) is oscillation adapted and let $0\neq\indexnot\in\EFindexset$.
Given that, in addition, the assumptions of Lemma~\ref{lemma:gaintvarge} are satisfied, let $t_{a}$, $t_{b}$, $\tau_{a}$, $\tau_{b}$,
$I_{a}$ and $J_{a}$ be defined as in the statement of Lemma~\ref{lemma:varofA}. Let $A$ and $\tau$ be defined as in 
Section~\ref{section:chofvariables} and let $\Xi(\tau;\tau_{a})$ be defined by (\ref{eq:Xidef}). Then
\begin{equation}\label{eq:wtbtatrans}
\begin{split}
w(t_{b}) = & \Xi(\tau_{b};\tau_{a})w(t_{a})\\
 & +\Xi(\tau_{b};\tau_{a})
\int_{t_{a}}^{t_{b}}[\Xi(\tau(t);\tau_{a})]^{-1}\exp\left(-i\int_{0}^{t}\sigma(t')\mfg(t')dt'\right)
\left(\begin{array}{c} 0 \\ \hf(t)\end{array}\right)dt,
\end{split}
\end{equation}
where $w$ is defined in (\ref{eq:wdefshiftge}) and $z$ is a solution to (\ref{eq:fourierthesystemRge}). Moreover, 
\begin{equation}\label{eq:Xiappest}
\|\Xi(\tau_{b};\tau_{a})-\Xi_{\app}(\tau_{b};\tau_{a})\| 
+\|[\Xi(\cdot;\tau_{a})]^{-1}-\Xi^{\mri}_{\app}(\cdot;\tau_{a})\|_{C(J_{a})}\leq C_{a,\ellderbd,\rosh}\vare_{a},
\end{equation}
where the constant $C_{a,\ellderbd,\rosh}$ only depends on an upper bound on $\vare_{a}$; the parameter $\ellderbd$ 
appearing in (\ref{eq:estdtlnge}); and $\mff_{\rosh}(t_{a})$. In (\ref{eq:Xiappest}), $\Xi_{\app}$ is given by 
\begin{equation}\label{eq:Xiappdef}
\begin{split}
\Xi_{\app}(\tau_{b};\tau_{a}) = & \cos(\om)\Id_{2m}
+\sgn(t_{b}-t_{a})\cdot \nu^{-1}\sin(\om)\left(\begin{array}{cc} i\xi\Id_{m} & \Id_{m} \\ -\Id_{m} & -i\xi\Id_{m}\end{array}\right)\\
 & +\sgn(t_{b}-t_{a})\cdot \frac{\pi}{\mfg(t_{a})} \left(\begin{array}{cc} \dot{\ell}(t_{a})\Id_{m}-\a(t_{a}) & iX(t_{a}) \\ -iX(t_{a}) & 
\dot{\ell}(t_{a})\Id_{m}-\a(t_{a})
\end{array}\right),
\end{split}
\end{equation}
where $\xi=-\sigma(t_{a})$, $\nu=(1+\xi^{2})^{1/2}$ and $\om=2\pi\nu$. Finally, the $\Xi^{\mri}_{\app}(\tau;\tau_{a})$ appearing 
in (\ref{eq:Xiappest}) is given by 
\begin{equation}\label{eq:XiIappdef}
\begin{split}
\Xi^{\mri}_{\app}(\tau;\tau_{a}) = & \cos(s\om)\Id_{2m}
+\nu^{-1}\sin(s\om)\left(\begin{array}{cc} i\xi\Id_{m} & \Id_{m} \\ -\Id_{m} & -i\xi\Id_{m}\end{array}\right)\\
 & +\pi s\sin(s\om)\left(\begin{array}{cc} R_{21} & R_{\aux} \\ -R_{\aux} & R_{21}\end{array}\right)
+\frac{1}{2}\sin(s\om)\left(\begin{array}{cc} 2R_{11} & R_{21} \\ R_{21} & -2R_{11}\end{array}\right)\\
 & +\pi s\cos(s\om)\left(\begin{array}{cc} R_{\aux} & -R_{21} \\ R_{21} & R_{\aux}\end{array}\right),
\end{split}
\end{equation}
where $R_{11}$ and $R_{21}$ are defined in (\ref{eq:Rijetcdef}); $R_{\aux}$ is defined in (\ref{eq:Rauxdef}) below; and $s$ is defined 
by (\ref{eq:sdef}). 
\end{lemma}
\begin{proof}
Let us start by deriving (\ref{eq:wtbtatrans}). Note that (\ref{eq:wofitoXietc}) yields
\begin{equation}\label{eq:wtbtatransprel}
w(t_{b})=\Xi(\tau_{b};\tau_{a})w(t_{a})+\Xi(\tau_{b};\tau_{a})\int_{\tau_{a}}^{\tau_{b}}[\Phi(\tau;\tau_{a})]^{-1}F(\tau)d\tau.
\end{equation}
Let us consider the second factor in the last term on the right hand side. It can be written
\begin{equation}\label{eq:wtbintreform}
\begin{split}
\int_{\tau_{a}}^{\tau_{b}}[\Phi(\tau;\tau_{a})]^{-1}F(\tau)d\tau
 = & \int_{t_{a}}^{t_{b}}[\Phi(\tau(t);\tau_{a})]^{-1}F[\tau(t)]\frac{\mfg(t)}{\mfg_{0}}dt\\
 = & \int_{t_{a}}^{t_{b}}[\Phi(\tau(t);\tau_{a})]^{-1}\left(\begin{array}{c} 0 \\ F_{2}(t)\end{array}\right)dt\\
 = & \int_{t_{a}}^{t_{b}}[\Xi(\tau(t);\tau_{a})]^{-1}\exp\left(-i\int_{0}^{t}\sigma(t')\mfg(t')dt'\right)
\left(\begin{array}{c} 0 \\ \hf(t)\end{array}\right)dt.
\end{split}
\end{equation}
Combining (\ref{eq:wtbtatransprel}) and (\ref{eq:wtbintreform}) yields (\ref{eq:wtbtatrans}).

\textbf{Calculating $\Xi_{\app}(\tau_{b};\tau_{a})$.} The next step is to calculate $\Xi_{\app}(\tau_{b};\tau_{a})$. 
Considering (\ref{eq:Xidef}), it is clear that we need to estimate $[U(t_{b};t_{a})]^{-1}$. By an estimate similar to 
(\ref{eq:Uestdetailed}), 
\begin{equation}\label{eq:Uestdetailedaux}
\begin{split}
\|\exp[-\a(t_{a})(t_{b}-t_{a})/2]-[U(t_{b};t_{a})]^{-1}\|\leq C_{a,\ellderbd}\frac{\mff_{\roode}(t_{a})}{\mfg^{2}(t_{a})},
\end{split}
\end{equation}
where $C_{a,\ellderbd}$ only depends on an upper bound on $\ellderbd$ and $\vare_{a}$. In particular, 
\begin{equation}\label{eq:Uinvperest}
\left\|[U(t_{b};t_{a})]^{-1}-\Id_{m}+\sgn(t_{b}-t_{a})\cdot
\frac{\pi\a(t_{a})}{\mfg(t_{a})}\right\|\leq C_{a,\ellderbd}\frac{\mff_{\roode}^{2}(t_{a})+1}{\mfg^{2}(t_{a})}
\leq C_{a,\ellderbd}\vare_{a},
\end{equation}
where $C_{a,\ellderbd}$ only depends on an upper bound on $\ellderbd$ and $\vare_{a}$, and we have used (\ref{eq:tbmtagestge})
and (\ref{eq:alCobd}). Appealing to (\ref{eq:Fbpoestge}), 
\begin{equation}\label{eq:elldperint}
\left|\exp\left[\frac{1}{2}\int_{t_{a}}^{t_{b}}\dot{\ell}(t')dt'\right]-1-\sgn(t_{b}-t_{a})\cdot\frac{\pi\dot{\ell}(t_{a})}{\mfg(t_{a})}\right|\leq 
\frac{C_{\ellderbd}}{\mfg^{2}(t_{a})}\leq C_{\ellderbd}\vare_{a},
\end{equation}
where the constant only depends on an upper bound on $\ellderbd$. Combining (\ref{eq:Uinvperest}) and (\ref{eq:elldperint})
with (\ref{eq:PIPappdef}), (\ref{eq:Pappdef}) and (\ref{eq:Xidef}) yields
\[
\|\Xi(\tau_{b};\tau_{a})-\Xi_{\app}(\tau_{b};\tau_{a})\|\leq C_{a,\ellderbd,\rosh}\vare_{a},
\]
where $\Xi_{\app}(\tau_{b};\tau_{a})$ is given by (\ref{eq:Xiappdef}) and
and the constant $C_{a,\ellderbd,\rosh}$ only depends on an upper bound on $\vare_{a}$; the parameter $\ellderbd$ 
appearing in (\ref{eq:estdtlnge}); and $\mff_{\rosh}(t_{a})$. Note, however, that in order to obtain this estimate, we have 
used the fact that 
\[
|\sin(2\pi\nu)|=|\sin[2\pi(\nu-1)]|\leq C\mff_{\rosh}^{2}(t_{a}),\ \ \
|\cos(2\pi\nu)-1|\leq C\mff_{\rosh}^{4}(t_{a}),
\]
where $C$ is a numerical constant; cf. (\ref{eq:sinomcosom}). 

\textbf{Calculating $\Xi^{\mri}_{\app}(\tau;\tau_{a})$.} Next we need to calculate $\Xi^{\mri}_{\app}(\tau;\tau_{a})$. 
Note to this end that (\ref{eq:Uestdetailed}) yields
\begin{equation}\label{eq:Uestdetailedauxnp}
\begin{split}
\|\exp[\a(t_{a})(t-t_{a})/2]-U(t;t_{a})\|\leq C_{a,\ellderbd}\frac{\mff_{\roode}(t_{a})}{\mfg^{2}(t_{a})}
\end{split}
\end{equation}
for $t\in I_{a}$, where $C_{a,\ellderbd}$ only depends on an upper bound on $\ellderbd$ and $\vare_{a}$. Thus
\begin{equation}\label{eq:Uestnp}
\left\|U(t;t_{a})-\Id_{m}-\frac{1}{2}\a(t_{a})(t-t_{a})\right\|\leq C_{a,\ellderbd}\frac{\mff_{\roode}^{2}(t_{a})+1}{\mfg^{2}(t_{a})}
\leq C_{a,\ellderbd}\vare_{a}
\end{equation}
for $t\in I_{a}$, 
where $C_{a,\ellderbd}$ only depends on an upper bound on $\ellderbd$ and $\vare_{a}$. By an argument which is essentially identical to the first step of
the proof of (\ref{eq:Fbpoestge}), 
\begin{equation}\label{eq:ellestnp}
\left|\int_{t_{a}}^{t}\dot{\ell}(t')dt'-\dot{\ell}(t_{a})(t-t_{a})\right|\leq \frac{C_{\ellderbd}}{\mfg^{2}(t_{a})}
\end{equation}
for $t\in I_{a}$, 
where $C_{\ellderbd}$ only depends on an upper bound on $\ellderbd$. It is of interest to express the $t-t_{a}$ appearing in this formula in terms
of the $s$ defined by (\ref{eq:sdef}). Note, to this end, that (\ref{eq:taudefge}) and (\ref{eq:sdef}) yield
\[
s=-\frac{1}{2\pi}\int_{t_{a}}^{t}\mfg(t')dt'=-\frac{1}{2\pi}\int_{t_{a}}^{t}\mfg(t')\mfg(t_{a})\left(\frac{1}{\mfg(t_{a})}-\frac{1}{\mfg(t')}\right)dt'
-\frac{1}{2\pi}\mfg(t_{a})(t-t_{a}).
\]
Using the results of Lemma~\ref{lemma:gaintvarge} to estimate the first term on the right hand side, it can then be verified
that 
\[
\left|t-t_{a}+\frac{2\pi s}{\mfg(t_{a})}\right|\leq \frac{C_{\ellderbd}}{\mfg^{2}(t_{a})}
\]
for $t\in I_{a}$, 
where $C_{\ellderbd}$ only depends on an upper bound on $\ellderbd$. Combining this estimate with (\ref{eq:Uestnp}) and (\ref{eq:ellestnp}) yields
\begin{equation}\label{eq:Psifac}
\left\|\exp\left[-\frac{1}{2}\int_{t_{a}}^{t}\dot{\ell}(t')dt'\right]U(t;t_{a})-\Id_{m}
-\frac{\pi s\dot{\ell}(t_{a})}{\mfg(t_{a})}\Id_{m}+\frac{\pi s}{\mfg(t_{a})}\a(t_{a})\right\|\leq C_{a,\ellderbd}\vare_{a}
\end{equation}
for $t\in I_{a}$, where $C_{a,\ellderbd}$ only depends on an upper bound on $\ellderbd$ and $\vare_{a}$. On the other hand, (\ref{eq:Xidef}) implies 
that 
\[
[\Xi(\tau;\tau_{a})]^{-1}=[\Phi(\tau;\tau_{a})]^{-1}\exp\left[-\frac{1}{2}\int_{t_{a}}^{t}\dot{\ell}(t')dt'\right]
\left(\begin{array}{cc} U(t;t_{a}) &  \\ 0 & U(t;t_{a})\end{array}\right).
\]
Combining this formula with (\ref{eq:PIPappdef}) and (\ref{eq:Psifac}) yields
\begin{equation}\label{eq:Xiinterest}
\|\Xi^{\mri}_{\inter}(\tau;\tau_{a})-[\Xi(\tau;\tau_{a})]^{-1}\|\leq C_{a,\ellderbd,\rosh}\vare_{a}
\end{equation}
for $\tau\in J_{a}$, 
where the constant $C_{a,\ellderbd,\rosh}$ only depends on an upper bound on $\ellderbd$, $\vare_{a}$ and $\mff_{\rosh}(t_{a})$. Moreover, 
\begin{equation}\label{eq:XiIinterm}
\begin{split}
\Xi^{\mri}_{\inter}(\tau;\tau_{a}) := & \Phi^{\mri}_{\app}(\tau;\tau_{a})\left[\Id_{2m}+
\pi s\left(\begin{array}{cc} R_{\aux} & 0 \\ 0 & R_{\aux}\end{array}\right)\right]\\
 = & \Phi^{\mri}_{\app}(\tau;\tau_{a})+
\pi s\left(\begin{array}{rr} \cos(s\om)R_{\aux} & \sin(s\om)R_{\aux} \\ -\sin(s\om)R_{\aux} & 
\cos(s\om)R_{\aux}\end{array}\right)+\Xi^{\mri}_{\rem}(\tau;\tau_{a})
\end{split}
\end{equation}
for $\tau\in J_{a}$, where
\begin{equation}\label{eq:Rauxdef}
R_{\aux}:=\frac{1}{\mfg(t_{a})}\left[\dot{\ell}(t_{a})\Id_{m}-\a(t_{a})\right]
\end{equation}
and
\begin{equation}\label{eq:Xiremest}
\|\Xi^{\mri}_{\rem}(\tau;\tau_{a})\|\leq C_{a,\ellderbd,\rosh}\vare_{a}
\end{equation}
for $\tau\in J_{a}$, 
where the constant $C_{a,\ellderbd,\rosh}$ only depends on an upper bound on $\ellderbd$, $\vare_{a}$ and $\mff_{\rosh}(t_{a})$.
In order to justify the second equality in (\ref{eq:XiIinterm}), note that 
\[
\left\|\Phi^{\mri,1}_{\app}(\tau;\tau_{a})
\left(\begin{array}{cc} R_{\aux} & 0 \\ 0 & R_{\aux}\end{array}\right)\right\|\leq C\|R_{\aux}\|(\|R_{11}\|+\|R_{21}\|)
\leq C_{\ellderbd}\vare_{a}
\]
for $\tau\in J_{a}$, where the constant $C$ is numerical and the constant $C_{\ellderbd}$ only depends on an upper bound on $\ellderbd$. Moreover, 
\[
|\nu^{-1}-1|\|R_{\aux}\|+|\xi|\|R_{\aux}\|\leq C_{\ellderbd}\vare_{a}
\]
for $\tau\in J_{a}$, where the constant $C_{\ellderbd}$ only depends on an upper bound on $\ellderbd$. 

Combining (\ref{eq:Xiinterest}), (\ref{eq:XiIinterm}) and (\ref{eq:Xiremest}) with (\ref{eq:PIappdef}) yields
\[
\|[\Xi(\cdot;\tau_{a})]^{-1}-\Xi^{\mri}_{\app}(\cdot;\tau_{a})\|_{C(J_{a})}\leq C_{a,\ellderbd,\rosh}\vare_{a},
\]
where $\Xi^{\mri}_{\app}(\tau;\tau_{a})$ is given by (\ref{eq:XiIappdef}) and the constant $C_{a,\ellderbd,\rosh}$ only 
depends on an upper bound on $\vare_{a}$; the parameter $\ellderbd$ appearing in (\ref{eq:estdtlnge}); and $\mff_{\rosh}(t_{a})$.
\end{proof}

\section{Iteration}\label{section:iteration}

Previous sections contain a description of how solutions evolve over one period of the oscillations; cf., in particular, 
Lemma~\ref{lemma:oneperiod}. In what follows, we wish to iterate the corresponding estimate. To begin with, it is therefore
natural to define a time sequence. Since we are interested in going both forward and backward in time, we in fact define
two sequences. 

\begin{definition}\label{def:tkdefge}
Assume that (\ref{eq:thesystemRge}) is oscillation adapted and let $0\neq\indexnot\in\EFindexset$. Assume, moreover, $t_{0}\in [0,\infty)$ to
be such that (\ref{eq:gtalbge}) holds with $t_{a}$ replaced by $t_{0}$. Assume that $t_{k}\geq t_{0}$ has been defined for some 
$0\leq k\in\zo$ and that (\ref{eq:gtalbge}) holds with $t_{a}$ replaced by $t_{k}$. Then $t_{k+1}$ is defined by the condition that 
\begin{equation}\label{eq:inttktkpomfgcond}
\int_{t_{k}}^{t_{k+1}}\mfg(\indexnot,t)dt=2\pi.
\end{equation}
If there is a final $k$ such that (\ref{eq:gtalbge}) holds with $t_{a}$ replaced by $t_{k}$, then this $k$ is denoted by $k_{\fin}$. Moreover,
$t_{\fin}:=t_{k_{\fin}}$. 
If there is no such final $k$, then $k_{\fin}:=\infty$ and $t_{\fin}:=\infty$. If, in addition to the above, $t_{0}\geq 2\pi$, then 
$t_{-1}$ is defined by the condition that (\ref{eq:inttktkpomfgcond}) holds with $k=-1$. Assume $t_{k}\leq t_{0}$ has been defined for some 
$0\geq k\in\zo$; that (\ref{eq:gtalbge}) holds with $t_{a}$ replaced by $t_{k}$; and that $t_{k}\geq 2\pi$. Then $t_{k-1}$ is defined by the 
condition that (\ref{eq:inttktkpomfgcond}) holds with $k$ replaced by $k-1$. Let $k_{\roini}$ be the first $0\geq k\in\zo$ such that either 
$0\leq t_{k}<2\pi$ or $\mfg(t_{k})<\max\{4\pi\ellderbd,2\}$. Finally, $t_{\roini}:=t_{k_{\roini}}$. 
\end{definition}
\begin{remark}
That the definition of $t_{k}$ for $k_{\roini}\leq k\leq k_{\fin}$ makes sense is ensured by Lemma~\ref{lemma:gaintvarge}. 
\end{remark}

In order to obtain a clearer picture of the development of the solution, it is convenient to 
introduce the transformation
\begin{align}
T_{\pre,k} := & \frac{1}{2}\left(\begin{array}{cc} (\nu_{k}-\xi_{k})\Id_{m} & i\Id_{m} \\ 
i(\nu_{k}+\xi_{k})\Id_{m} & \Id_{m}\end{array}\right),\label{eq:Tprekage}\\
T_{\pre,k}^{-1} = & \frac{1}{\nu_{k}}\left(\begin{array}{cc} \Id_{m} & -i\Id_{m} \\ 
-i(\xi_{k}+\nu_{k})\Id_{m} & (\nu_{k}-\xi_{k})\Id_{m}\end{array}\right).\label{eq:Tprekainvge}
\end{align}
Here and below we use the notation $\xi_{k}:=-\sigma(t_{k})$, $\nu_{k}:=(1+\xi_{k}^{2})^{1/2}$ and $\om_{k}:=2\pi\nu_{k}$. Note that if $\mff_{\rosh}(t)$
is bounded by a constant independent of $0\leq t\in\ro$, then the norms of both of the matrices appearing in (\ref{eq:Tprekage}) and 
(\ref{eq:Tprekainvge}) are bounded by constants independent of $k$ and $0\neq\indexnot\in\EFindexset$. It turns out that the analysis is simplified 
by considering the variable
\begin{equation}\label{eq:wkpredef}
w_{\pre,k}:=T_{\pre,k}w(t_{k})
\end{equation}
instead of $w(t_{k})$. 

\begin{lemma}\label{lemma:wprekit}
Assume that (\ref{eq:thesystemRge}) is oscillation adapted and let $0\neq\indexnot\in\EFindexset$. Assume $t_{0}\in [0,\infty)$ to 
be such that (\ref{eq:gtalbge}) holds with $t_{a}$ replaced by $t_{0}$, and define $t_{k}$ and $w_{\pre,k}$ as in 
Definition~\ref{def:tkdefge} and (\ref{eq:wkpredef}) respectively. Then there is a matrix $\Xi^{k,+}_{\pre}$ for $0\leq k<k_{\fin}$
and, in case $t_{0}\geq 2\pi$, a matrix $\Xi^{k,-}_{\pre}$ for $k_{\roini}<k\leq 0$. Moreover, there is a matrix valued function 
$\Xi_{\pre}(\cdot;\tau_{k})$ such that
\begin{equation}\label{eq:wprekpertrans}
\begin{split}
w_{\pre,k\pm 1} = & \Xi^{k,\pm}_{\pre}w_{\pre,k}\\
 & +\Xi^{k,\pm}_{\pre}\int_{t_{k}}^{t_{k\pm 1}}[\Xi_{\pre}(\tau(t);\tau_{k})]^{-1}
\exp\left(-i\int_{0}^{t}\sigma(t')\mfg(t')dt'\right)
T_{\pre,k}\left(\begin{array}{c} 0 \\ \hf(t)\end{array}\right)dt
\end{split}
\end{equation}
where $0\leq k<k_{\fin}$ for the equality with a plus sign and $k_{\roini}<k\leq 0$ for the equality with a minus sign (in the case of 
a minus sign, it is also assumed that $t_{0}\geq 2\pi$). Moreover, there is a matrix $\Xi^{k,\pm}_{\pre,\app}$ and a matrix valued function 
$\Xi_{\pre,\app}^{\mri}(\cdot;\tau_{k})$ such that 
\begin{equation}\label{eq:XikXikIpreest}
\|\Xi^{k,\pm}_{\pre}-\Xi^{k,\pm}_{\pre,\app}\|+\|[\Xi_{\pre}(\cdot;\tau_{k})]^{-1}-\Xi_{\pre,\app}^{\mri}(\cdot;\tau_{k})\|_{C(J_{k,\pm})}\leq C_{\vare,\ellderbd,\rosh}\vare_{k},
\end{equation}
where $J_{k,\pm}$ is the interval in $\tau$-time corresponding to the interval with endpoints $t_{k}$ and $t_{k\pm 1}$; and the constant $C_{\vare,\ellderbd,\rosh}$ 
only depends on an upper bound on $\ellderbd$, $\vare_{k}$ and 
$\mff_{\rosh}(t_{k})$ (here $\vare_{k}$ is defined to be the right hand side of (\ref{eq:vareadef}) with $t_{a}$ replaced by $t_{k}$). Finally, in
(\ref{eq:XikXikIpreest}), $\Xi^{k,\pm}_{\pre,\app}$ is defined by 
\begin{equation}\label{eq:Xikpreappdefst}
\begin{split}
\Xi^{k,\pm}_{\pre,\app} := & 
\left(\begin{array}{cc} e^{\mp i\om_{k}}\Id_{m} & 0 \\ 0 & e^{\pm i\om_{k}}\Id_{m} \end{array}\right)
 \pm\pi\left(\begin{array}{cc} R_{\pre,k}^{10}
& 0 \\  0 & R_{\pre,k}^{11}\end{array}
\right)
\end{split}
\end{equation}
and $\Xi_{\pre,\app}^{\mri}$ is defined by 
\begin{equation}\label{eq:XiIappdefst}
\begin{split}
\Xi^{\mri}_{\pre,\app}(\tau;\tau_{k}) := & \left(\begin{array}{cc} e^{-is\om_{k}}\Id_{m} & 0 \\ 0 & e^{is\om_{k}}\Id_{m} \end{array}\right)
 +\pi s\left(\begin{array}{cc} e^{-is\om_{k}}R_{\pre,k}^{10} & 0 \\ 0 & e^{is\om_{k}}R_{\pre,k}^{11}\end{array}\right)\\
 & -\frac{i}{2}\sin(s\om_{k})\left(\begin{array}{cc} 0 & R_{\pre,k}^{00} \\ -R_{\pre,k}^{01} & 0\end{array}\right),
\end{split}
\end{equation}
where $s$ is defined in (\ref{eq:sdef}) (with $\tau_{a}$ replaced by $\tau_{k}$) and
\begin{equation}\label{eq:Rabprekdef}
R^{ab}_{\pre,k} :=  \frac{1}{\mfg(t_{k})}[(-1)^{a}\a(t_{k})+\dot{\ell}(t_{k})\Id_{m}+(-1)^{b}X(t_{k})]
\end{equation}
for $a,b=0,1$. 
\end{lemma}
\begin{remark}\label{remark:kpm}
All the equalities and the estimates in the lemma which refer to the intervals $[t_{k-1},t_{k}]$, $k_{\roini}<k\leq 0$, are based on the assumption that 
$t_{0}\geq 2\pi$. 
\end{remark}
\begin{remark}
The $t_{0}$ appearing in (\ref{eq:taudefge}) can be assumed to equal the $t_{0}$ appearing in the assumptions of the lemma. 
\end{remark}
\begin{remark}
In case $k_{\fin}\geq 1$, $\Xi^{k,-}_{\pre}$ can be defined for finite $k$ such that $1\leq k\leq k_{\fin}$. Similarly, in case 
$k_{\roini}\leq -1$, $\Xi^{k,+}_{\pre}$ can
be defined for $k_{\roini}+1\leq k\leq 0$. The remaining statements of the lemma are then valid for the corresponding $k$ and $\mp$. 
\end{remark}
\begin{proof}
The main purpose of introducing $T_{\pre,k}$ is that it has the property that 
\begin{equation}\label{eq:Tkdiagzc}
\begin{split}
 & T_{\pre,k}\left[\cos(s\om_{k})\Id_{2m}
+\nu_{k}^{-1}\sin(s\om_{k})\left(\begin{array}{cc} i\xi_{k}\Id_{m} & \Id_{m} \\ -\Id_{m} & -i\xi_{k}\Id_{m}\end{array}\right)\right]
T_{\pre,k}^{-1}\\
 = & \left(\begin{array}{cc} e^{-is\om_{k}}\Id_{m} & 0 \\ 0 & e^{is\om_{k}}\Id_{m} \end{array}\right).
\end{split}
\end{equation}
Note also that 
\begin{equation}\label{eq:Tprekmbgge}
\begin{split}
 & \left\|T_{\pre,k}-\frac{1}{2}\left(\begin{array}{rr} \Id_{m} & i\Id_{m}\\ 
i\Id_{m} & \Id_{m}\end{array}\right)\right\|
+\left\|T_{\pre,k}^{-1}-\left(\begin{array}{rr} \Id_{m} & -i\Id_{m}\\ 
-i\Id_{m} & \Id_{m}\end{array}\right)\right\|\\
 \leq & C(|\nu_{k}-1|+|\xi_{k}|)\leq C\mff_{\rosh}(t_{k}),
\end{split}
\end{equation}
where the constant $C$ is numerical. Thus
\begin{equation}\label{eq:Tkdiagfc}
\begin{split}
 & \left\|\frac{\pi}{\mfg(t_{k})}T_{\pre,k}\left(\begin{array}{cc} \dot{\ell}(t_{k})\Id_{m}-\a(t_{k})
& iX(t_{k}) \\  -iX(t_{k}) & \dot{\ell}(t_{k})\Id_{m}-\a(t_{k})\end{array}
\right)T_{\pre,k}^{-1}\right.\\
& \left.\phantom{\|}-\frac{\pi}{\mfg(t_{k})}
\left(\begin{array}{cc} \dot{\ell}(t_{k})\Id_{m}-\a(t_{k})+X(t_{k})
& 0 \\  0 & \dot{\ell}(t_{k})\Id_{m}-\a(t_{k})-X(t_{k})\end{array}
\right)\right\|\leq C_{\ellderbd,\rosh}\vare_{k},
\end{split}
\end{equation}
where $\vare_{k}$ is defined by (\ref{eq:vareadef}) with $t_{a}$ replaced by $t_{k}$ and $C_{\ellderbd,\rosh}$ only depends on an 
upper bound on $\ellderbd$ and $\mff_{\rosh}(t_{k})$. The relations (\ref{eq:Tkdiagzc}) and (\ref{eq:Tkdiagfc}) demonstrate
that $T_{\pre,k}$ diagonalises the matrices of interest. However, there is one problem. In practice, the first factor on the 
left hand side of (\ref{eq:Tkdiagzc}) should be $T_{\pre,k\pm 1}$, not $T_{\pre,k}$. For that reason, it is of interest to 
estimate $T_{\pre,k\pm 1}T_{\pre,k}^{-1}-\Id_{2m}$. Note that 
\begin{equation}\label{eq:errordiffinvge}
T_{\pre,k\pm 1}T_{\pre,k}^{-1}-\Id_{2m}=
\frac{1}{2\nu_{k}}\left(\begin{array}{rr} \mfq_{k,\pm}\Id_{m} & 
-i\mfq_{k,\pm}\Id_{m} \\ 
i\mfr_{k,\pm}\Id_{m} & \mfr_{k,\pm}\Id_{m}\end{array}\right),
\end{equation}
where 
\[
\mfq_{k,\pm}=\nu_{k\pm 1}-\nu_{k}+\xi_{k}-\xi_{k\pm 1},\ \ \
\mfr_{k,\pm}=\nu_{k\pm 1}-\nu_{k}+\xi_{k\pm 1}-\xi_{k}.
\]
It is thus sufficient to estimate $|\nu_{k\pm 1}-\nu_{k}|$ and $|\xi_{k\pm 1}-\xi_{k}|$. In fact, since 
\begin{equation}\label{eq:nukdiffest}
|\nu_{k\pm 1}-\nu_{k}|=\frac{|\nu_{k\pm 1}^{2}-\nu_{k}^{2}|}{\nu_{k\pm 1}+\nu_{k}}
=\frac{|\xi_{k\pm 1}-\xi_{k}||\xi_{k\pm 1}+\xi_{k}|}{\nu_{k\pm 1}+\nu_{k}}\leq |\xi_{k\pm 1}-\xi_{k}|,
\end{equation}
it is sufficient to estimate $|\xi_{k\pm 1}-\xi_{k}|$. On the other hand 
\begin{equation}\label{eq:xikdiffest}
|\xi_{k\pm 1}-\xi_{k}|=\left|\int_{t_{k}}^{t_{k\pm 1}}\dot{\sigma}(t')dt'\right|\leq e^{2\pi\ellderbd}\mff_{\rosh}(t_{k})|t_{k\pm 1}-t_{k}|
\leq 4\pi e^{2\pi\ellderbd}\frac{\mff_{\rosh}(t_{k})}{\mfg(t_{k})},
\end{equation}
where we have used (\ref{eq:shiftbd}), (\ref{eq:tatbroughge}) and (\ref{eq:mffvindvar}); note that the last two estimates
are applicable in the present context with $t_{a}$ replaced by $t_{k}$. Thus
\begin{equation}\label{eq:TprekpmoTinvmo}
\|T_{\pre,k\pm 1}T_{\pre,k}^{-1}-\Id_{2m}\|\leq C_{\ellderbd}\frac{\mff_{\rosh}(t_{k})}{\mfg(t_{k})},
\end{equation}
where $C_{\ellderbd}$ is a constant only depending on an upper bound on $\ellderbd$. 

Next we wish to compute the relation between $w_{\pre,k}$ and $w_{\pre,k\pm 1}$. Note, to this end, that 
\begin{equation*}
\begin{split}
w_{\pre,k\pm 1} = & \Xi^{k,\pm}_{\pre}w_{\pre,k}\\
 & +\Xi^{k,\pm}_{\pre}\int_{t_{k}}^{t_{k\pm 1}}[\Xi_{\pre}(\tau(t);\tau_{k})]^{-1}
\exp\left(-i\int_{0}^{t}\sigma(t')\mfg(t')dt'\right)
T_{\pre,k}\left(\begin{array}{c} 0 \\ \hf(t)\end{array}\right)dt,
\end{split}
\end{equation*}
where we have appealed to (\ref{eq:wtbtatrans}) and
\[
\Xi^{k,\pm}_{\pre}:=T_{\pre,k\pm 1}\Xi(\tau_{k\pm 1};\tau_{k})T_{\pre,k}^{-1},\ \ \
\Xi_{\pre}(\tau;\tau_{k}):=T_{\pre,k}\Xi(\tau;\tau_{k})T_{\pre,k}^{-1}.
\]
To begin with, we wish to compute an approximation for the matrix $\Xi^{k,\pm}_{\pre}$. Due to (\ref{eq:Xiappest}),
(\ref{eq:Xiappdef}), (\ref{eq:Tprekmbgge}) and (\ref{eq:TprekpmoTinvmo}), 
\[
\|\Xi^{k,\pm}_{\pre}-T_{\pre,k}\Xi_{\app}(\tau_{k\pm 1};\tau_{k})T_{\pre,k}^{-1}\|\leq C_{\vare,\ellderbd,\rosh}\vare_{k},
\]
where the constant $C_{\vare,\ellderbd,\rosh}$ only depends on an upper bound on $\ellderbd$, $\vare_{k}$ and $\mff_{\rosh}(t_{k})$. Combining this 
estimate with (\ref{eq:Tkdiagzc}) and (\ref{eq:Tkdiagfc}) yields
\begin{equation}\label{eq:Xikpreest}
\|\Xi^{k,\pm}_{\pre}-\Xi^{k,\pm}_{\pre,\app}\|\leq C_{\vare,\ellderbd,\rosh}\vare_{k},
\end{equation}
where the constant $C_{\vare,\ellderbd,\rosh}$ only depends on an upper bound on $\ellderbd$, $\vare_{k}$ and $\mff_{\rosh}(t_{k})$.
In (\ref{eq:Xikpreest}), $\Xi^{k,\pm}_{\pre,\app}$ is defined by (\ref{eq:Xikpreappdefst}). 
Similarly, 
\begin{equation}\label{eq:Xipreest}
\|[\Xi_{\pre}(\cdot;\tau_{k})]^{-1}-\Xi_{\pre,\app}^{\mri}(\cdot;\tau_{k})\|\leq C_{\vare,\ellderbd,\rosh}\vare_{k},
\end{equation}
where the constant $C_{\vare,\ellderbd,\rosh}$ only depends on an upper bound on $\ellderbd$, $\vare_{k}$ and $\mff_{\rosh}(t_{k})$.
In (\ref{eq:Xipreest}), $\Xi_{\pre,\app}^{\mri}$ is defined by (\ref{eq:XiIappdefst}). The lemma follows. 
\end{proof}

Let 
\begin{align}
\md_{k}(\indexnot) := & 
\exp\left[i\int_{0}^{t_{k}}[1+\sigma^{2}(\indexnot,t)]^{1/2}\mfg(\indexnot,t)dt\right],\label{eq:mdkdef}\\
D_{k}(\indexnot) := & \left(\begin{array}{cc} \md_{k}(\indexnot)\Id_{m}
& 0 \\  0 & [\md_{k}(\indexnot)]^{-1}\Id_{m}\end{array}\right),\label{eq:mDkdef}\\
w_{\fin,k} := & D_{k}w_{\pre,k}.\label{eq:wfinkdef}
\end{align}
With these definitions, we obtain the following simplification. 

\begin{lemma}\label{lemma:wkfinlemma}
Assume that (\ref{eq:thesystemRge}) is oscillation adapted and let $0\neq\indexnot\in\EFindexset$. Assume $t_{0}\in [0,\infty)$ to
be such that (\ref{eq:gtalbge}) holds with $t_{a}$ replaced by $t_{0}$, and define $t_{k}$ and $w_{\fin,k}$ as in 
Definition~\ref{def:tkdefge} and (\ref{eq:wfinkdef}) respectively. Then there is a matrix $\Xi^{k,+}_{\fin}$ for $0\leq k<k_{\fin}$
and, in case $t_{0}\geq 2\pi$, a matrix $\Xi^{k,-}_{\fin}$ for $k_{\roini}<k\leq 0$. Moreover, there is a matrix valued function 
$\Xi_{\fin}(\cdot;\tau_{k})$ such that
\begin{equation}\label{eq:wprekpertransfin}
\begin{split}
w_{\fin,k\pm 1} = & \Xi^{k,\pm}_{\fin}w_{\fin,k}+\Xi^{k,\pm}_{\fin}\int_{t_{k}}^{t_{k\pm 1}}\hF_{k}(t)dt,
\end{split}
\end{equation}
where $0\leq k<k_{\fin}$ for the equality with a plus sign; $k_{\roini}<k\leq 0$ for the equality with a minus sign (in the case of 
a minus sign, it is also assumed that $t_{0}\geq 2\pi$);
\begin{equation}\label{eq:Fkfindef}
\hF_{k}(t):=[\Xi_{\fin}(\tau(t);\tau_{k})]^{-1}
\exp\left(-i\int_{0}^{t}\sigma(t')\mfg(t')dt'\right)
T_{\fin,k}\left(\begin{array}{c} 0 \\ \hf(t)\end{array}\right);
\end{equation}
and $T_{\fin,k}:=D_{k}T_{\pre,k}$. Moreover, there is a matrix $\Xi^{k,\pm}_{\fin,\app}$ and a matrix valued function 
$\Xi_{\fin,\app}^{\mri}(\cdot;\tau_{k})$ such that 
\begin{equation}\label{eq:XikXikIpreestfin}
\|\Xi^{k,\pm}_{\fin}-\Xi^{k,\pm}_{\fin,\app}\|+\|[\Xi_{\fin}(\cdot;\tau_{k})]^{-1}-\Xi_{\fin,\app}^{\mri}(\cdot;\tau_{k})\|_{C(J_{k,\pm})}\leq 
C_{\vare,\ellderbd,\rosh}\vare_{k},
\end{equation}
where $J_{k,\pm}$ is the interval in $\tau$-time corresponding to the interval with endpoints $t_{k}$ and $t_{k\pm 1}$;
and the constant $C_{\vare,\ellderbd,\rosh}$ only depends on an upper bound on $\ellderbd$, $\vare_{k}$ and 
$\mff_{\rosh}(t_{k})$ (here $\vare_{k}$ is defined to be the right hand side of (\ref{eq:vareadef}) with $t_{a}$ replaced by $t_{k}$). Finally, in
(\ref{eq:XikXikIpreestfin}),  $\Xi^{k,\pm}_{\fin,\app}$ is defined by 
\begin{equation}\label{eq:Xikpreappdefstfin}
\begin{split}
\Xi^{k,\pm}_{\fin,\app} := & \Id_{2m} \pm\pi\left(\begin{array}{cc} R_{\pre,k}^{10}
& 0 \\  0 & R_{\pre,k}^{11}\end{array}
\right),
\end{split}
\end{equation}
where $R^{ab}_{\pre,k}$ is defined in (\ref{eq:Rabprekdef}). Moreover, $\Xi_{\fin,\app}^{\mri}$ is defined by 
\begin{equation}\label{eq:XiIappdefstfin}
\begin{split}
\Xi^{\mri}_{\fin,\app}(\tau;\tau_{k}) := & \left(\begin{array}{cc} e^{-is\om_{k}}\Id_{m} & 0 \\ 0 & e^{is\om_{k}}\Id_{m} \end{array}\right)
+\pi s\left(\begin{array}{cc} e^{-is\om_{k}}R_{\pre,k}^{10} & 0 \\ 0 & e^{is\om_{k}}R_{\pre,k}^{11}\end{array}\right)\\
 & -\frac{i}{2}\sin(s\om_{k})\left(\begin{array}{cc} 0 & \md_{k}^{2}R_{\pre,k}^{00} \\ -\md_{k}^{-2}R_{\pre,k}^{01} & 0\end{array}\right),
\end{split}
\end{equation}
where $s$ is defined in (\ref{eq:sdef}) (with $\tau_{a}$ replaced by $\tau_{k}$), $\md_{k}$ is defined in (\ref{eq:mdkdef}) and
$R^{ab}_{\pre,k}$ is defined in (\ref{eq:Rabprekdef}).
\end{lemma}
\begin{remark}
It is of interest to note that $\Xi^{k,\pm}_{\fin,\app}$ is independent of the functions $g^{0l}$ and $\zeta$ appearing in the 
equation (\ref{eq:thesystemRge}). 
\end{remark}
\begin{remark}
Remark~\ref{remark:kpm} is equally relevant in the present setting. 
\end{remark}
\begin{remark}\label{remark:extvalXipmfin}
In case $k_{\fin}\geq 1$, $\Xi^{k,-}_{\fin}$ can be defined for finite $k$ such that $1\leq k\leq k_{\fin}$. Similarly, in case 
$k_{\roini}\leq -1$, $\Xi^{k,+}_{\fin}$ can
be defined for $k_{\roini}+1\leq k\leq 0$. The remaining statements of the lemma are then valid for the corresponding $k$ and $\mp$. 
\end{remark}
\begin{proof}
Note that $D_{k}$, defined in (\ref{eq:mDkdef}), and its inverse both have norm $1$. Note also that 
\[
\pm\om_{k}=\pm 2\pi [1+\sigma^{2}(t_{k})]^{1/2}=\int_{t_{k}}^{t_{k\pm 1}}[1+\sigma^{2}(t_{k})]^{1/2}\mfg(t)dt;
\]
recall that (\ref{eq:percondge}) holds with $t_{a}$ and $t_{b}$ replaced by $t_{k}$ and $t_{k\pm 1}$
respectively. Thus
\begin{equation}\label{eq:mdkpomdkinv}
\begin{split}
\md_{k\pm 1}\md_{k}^{-1} = &
\exp\left[i\int_{t_{k}}^{t_{k\pm 1}}[1+\sigma^{2}(t)]^{1/2}\mfg(t)dt\right]\\
 = & e^{\pm i\om_{k}}
\exp\left[i\int_{t_{k}}^{t_{k\pm 1}}\left([1+\sigma^{2}(t)]^{1/2}
-[1+\sigma^{2}(t_{k})]^{1/2}\right)\mfg(t)dt\right].
\end{split}
\end{equation}
However, by estimates similar to (\ref{eq:nukdiffest}) and (\ref{eq:xikdiffest}),
\[
\left|[1+\sigma^{2}(t)]^{1/2}-[1+\sigma^{2}(t_{k})]^{1/2}\right|\leq C_{\ellderbd}\frac{\mff_{\rosh}(t_{k})}{\mfg(t_{k})}
\]
for $t\in I_{k,\pm}$, where $I_{k,\pm}$ is the interval with endpoints $t_{k}$ and $t_{k\pm 1}$, and the constant $C_{\ellderbd}$ only depends 
on an upper bound on $\ellderbd$. Combining this estimate with (\ref{eq:mdkpomdkinv}) yields
\begin{equation}\label{eq:mdkpotinvk}
\left|\md_{k\pm 1}\md_{k}^{-1}-e^{\pm i\om_{k}}\right|+
\left|\md_{k\pm 1}^{-1}\md_{k}-e^{\mp i\om_{k}}\right|\leq C_{\ellderbd}\frac{\mff_{\rosh}(t_{k})}{\mfg(t_{k})},
\end{equation}
where the constant $C_{\ellderbd}$ only depends on an upper bound on $\ellderbd$. Similarly,
\begin{equation}\label{eq:eiomkest}
|\exp\left(\pm i\om_{k}\right)-1|=|\exp\left(\pm 2\pi i(\nu_{k}-1)\right)-1|\leq 2\pi |\nu_{k}-1|\leq 2\pi\mff_{\rosh}(t_{k}).
\end{equation}
Defining
\begin{equation}\label{eq:Xikfindef}
\Xi^{k,\pm}_{\fin}:=D_{k\pm 1}\Xi^{k,\pm}_{\pre}D_{k}^{-1},\ \ \ 
\Xi_{\fin}(\tau;\tau_{k}):=D_{k}\Xi_{\pre}(\tau;\tau_{k})D_{k}^{-1},\ \ \
T_{\fin,k}:=D_{k}T_{\pre,k}
\end{equation}
yields (\ref{eq:wprekpertransfin}). Combining these definitions with Lemma~\ref{lemma:wprekit}, (\ref{eq:mdkpotinvk})
and (\ref{eq:eiomkest}) yields the conclusions of the lemma. 
\end{proof}

Since we appeal to (\ref{eq:wprekpertransfin}) repeatedly in applications, it is convenient to simplify the notation. We therefore
write (\ref{eq:wprekpertransfin}) as
\begin{equation}\label{eq:wprekpertransvarpi}
\begin{split}
\psi_{k\pm 1} = & A_{k}^{\pm}\psi_{k}+A_{k}^{\pm}\int_{t_{k}}^{t_{k\pm 1}}\hF_{k}(t)dt,
\end{split}
\end{equation}
where $\psi_{k}:=w_{\fin,k}$ and $A_{k}^{\pm}:=\Xi^{k,\pm}_{\fin}$. Applying this equality iteratively yields, e.g.,
\begin{equation}\label{eq:differenceeqrepres}
\psi_{k+1}=A_{k}^{+}\cdots A_{0}^{+}\psi_{0}+A_{k}^{+}\int_{t_{k}}^{t_{k+1}}\hF_{k}(t)dt+\cdots+
A_{k}^{+}\cdots A_{0}^{+}\int_{t_{0}}^{t_{1}}\hF_{0}(t)dt.
\end{equation}
Due to equalities of this type, it is clearly of interest to calculate matrix products such as $A_{k}^{+}\cdots A_{k-l}^{+}$. 
This is an important part of the analysis in the next two parts of these notes.

\part{Unbalanced equations}\label{part:unbadegeq}

\chapter{Unbalanced equations}

\section{Introduction}

Consider the requirement (\ref{eq:weakbal}) for an equation to be weakly balanced. There are various ways of violating this 
condition. However, we here focus on two situations. First, there is the possibility that the norms of $\a(t)$ and $\zeta(t)$ are
unbounded. Then pathological behaviour might arise even on the level of spatially homogeneous solutions. Another
option is that the norm of $X(\indexnot,t)$ is unbounded. In that case, we clearly need to consider spatially inhomogeneous solutions in order
to see something interesting. Finally, there is the possibility that $\sigma(\indexnot,t)$ is not bounded. This is not a situation we
consider in greater detail here. One reason for this is that the main motivation for studying (\ref{eq:thesystemRge}) is that
it arises when linearising Einstein's equations around certain cosmological solutions. In that context, the size of $\sigma(\indexnot,t)$ 
can be influenced by the choice of gauge. In fact, we expect it to be possible to ensure that $\sigma(\indexnot,t)$ is small by an 
appropriate choice of gauge. For this reason, we do not consider the case that $\sigma(\indexnot,t)$ is large here. 

In order to be able to obtain results, we need to assume more than that, say, the norms of $\a(t)$ and $\zeta(t)$ are unbounded. 
In fact, if these quantities are unbounded and strongly oscillatory, it can be expected to be hard to draw conclusions. In
the present chapter we therefore consider situations in which at least one of the following two assumptions holds:
\begin{itemize}
\item $e^{-\b_{\a}t}\a(t)$ and $e^{-2\b_{\zeta}t}\zeta(t)$ converge exponentially for some real numbers $\b_{\a}$ and $\b_{\zeta}$, where 
at least one of $\b_{\a}$ and $\b_{\zeta}$ is strictly positive and the corresponding limit is non-zero,
\item $e^{-\b_{X}t}X^{j}(t)/[g^{jj}(t)]^{1/2}$ converges to a matrix $Y^{j}_{\infty}$ for some $j\in\{1,\dots,d\}$ and some $\b_{X}>0$, where 
$Y^{j}_{\infty}$ has an eigenvalue with non-zero real part.
\end{itemize}
In practice, we need to make additional assumptions, but these requirements are the essential ones. Below, we informally refer to the first 
situation as the ODE-unbalanced setting, and the second situation as the PDE-unbalanced setting. 

\textbf{The ODE-unbalanced setting.} Dealing with the ODE-unbalanced setting
is fairly straightforward, and we record the corresponding results in Section~\ref{section:ODEunbal}. In practice, we consider equations
of the form 
\begin{equation}\label{eq:zazeroeff}
\ddot{z}+\a(t) \dot{z}+\zeta(t) z=0,
\end{equation}
where $\a,\zeta\in C^{\infty}[\ro,\Mn{m}{\co}]$, and $\a$ and $\zeta$ converge after rescaling by an appropriate
exponential factor. The conclusion we obtain in the ODE-unbalanced setting is that solutions typically
either grow or decay super exponentially; cf. Lemma~\ref{lemma:spsysodes}. This is one reason for considering such equations to be pathological. 
At the end of the chapter, in Section~\ref{section:modeanalysis}, we consider the behaviour of the individual modes, and provide conditions 
ensuring super exponential decay; cf. Lemma~\ref{lemma:supexpdeccondmode}. This result is based on the analysis in Section~\ref{section:ODEunbal}.

\textbf{The PDE-unbalanced setting.} The PDE-unbalanced setting is substantially more complicated than the ODE-unbalanced setting. The reason for
this is that in order to be able to derive conclusions in the former case, we need to appeal to the results of Part~\ref{part:averaging} of these 
notes. On the other hand, one simplifying aspect is the fact that in order to demonstrate super exponential growth of the energy, it is 
sufficient to consider (\ref{eq:thesystemRge}) in the case that $d=1$, $R=0$ and $f=0$. The reason for this is that if we consider the natural energy 
associated with (\ref{eq:thesystemRge}) for general $d$ and $R$, then this energy is bounded from below by the energy of the part of the 
solution which only depends on $t$ and, say, $x^{j}$, the $j$'th ``coordinate'' on the $d$-torus. If we can demonstrate super exponential growth
of such solutions, we are thus done. The steps of the analysis can then roughly be divided as follows. 

\textit{Specifying the class of equations, preliminary steps.} In Section~\ref{section:PDEunbalterm} we introduce conditions such that the results 
of Part~\ref{part:averaging} apply, and such that we, in the end, are able to obtain super exponential growth. Under these assumptions, we then 
consider the modes of the solutions. Since the solutions of interest are defined on $I\times\so$, we only need to keep track of one frequency, 
say $n_{1}$. For a given equation and a given $n_{1}$, we gradually focus our attention on a more and more restricted time interval.
The reason for doing so is that we want to construct a time sequence of the form introduced in Definition~\ref{def:tkdefge} and to appeal to
Lemma~\ref{lemma:wkfinlemma}. In fact, by appealing to Lemma~\ref{lemma:wkfinlemma}, we can estimate how the solution evolves along the time 
sequence. For this to be useful we, however, need to restrict the time interval so that the time sequence is well defined and so that the 
the error terms arising in Lemma~\ref{lemma:wkfinlemma} are small. Moreover, even though Lemma~\ref{lemma:wkfinlemma} yields interesting information
concerning the evolution of the solution, we, e.g., need to know that the constituents of the matrix $R^{ab}_{\pre,k}$ appearing on the right hand
side of (\ref{eq:Xikpreappdefstfin}) have some definite behaviour. Due to the form of our assumptions, we can only expect this to be the case in 
an asymptotic regime, where $t$ is large. Summing up, it is thus necessary to restrict the value of $n_{1}$ in such a way that there is a time 
interval in which the behaviour is oscillatory and so that the definition of the time sequence makes sense. Second, it is necessary to restrict 
the time sequence from above so that the error terms appearing in Lemma~\ref{lemma:wkfinlemma} are small. Third, $t$ has to be large enough that
we are in the asymptotic regime. Since there are bounds on $t$ both from above and below, there is a risk that the different conditions contradict
each other. However, this can be avoided by simply increasing the frequency $n_{1}$. In fact, it turns out that the upper bounds (when they exist) 
increase at least logarithmically with $n_{1}$. On the other hand, the lower bounds are independent of $n_{1}$; they only depend on the coefficients 
of (\ref{eq:thesystemRge}). In Section~\ref{section:subdivinterv} we derive upper bounds on $t$ ensuring that the error terms appearing in 
Lemma~\ref{lemma:wkfinlemma} are small and that the sequence introduced in Definition~\ref{def:tkdefge} is well defined. In 
Section~\ref{section:bapropofiterseg}, we then introduce the condition that the quotient $e^{-\b_{X}t}X^{j}(t)/[g^{jj}(t)]^{1/2}$ should converge to a 
matrix $Y^{j}_{\infty}$ for some $j\in\{1,\dots,d\}$ and some $\b_{X}>0$, where $Y^{j}_{\infty}$ has an eigenvalue with non-zero real part. Moreover, 
we reformulate the iteration derived in Lemma~\ref{lemma:wkfinlemma} so that the source of the super exponential growth becomes more apparent. 
Finally, we derive an estimate for the matrix appearing in the new iteration; cf. Lemma~\ref{lemma:hBkapprfirstimpro}. This result is the 
basis for the proof of super exponential growth. 

\textit{The main estimate.} Once we have taken the preliminary steps, we are in a position to prove super exponential growth of the individual
modes. This is the subject of Section~\ref{section:themainestimateseg}. The main tool is the iteration derived previously. In practice we 
would thus like to estimate a matrix product. On the other hand, we are interested in situations where the number of factors tends to infinity. 
Moreover, we do not know the individual factors exactly; we only have an estimate. Finally, the product can be expected to contract some directions 
and expand others. On the other hand, we only need to find some appropriate initial datum for the iteration that leads to super exponential 
growth. 

The first step of the analysis is to demonstrate that the sums of the error terms (that appear when estimating how solutions to the iteration 
grow) can be approximated by integrals which, in their turn, can be computed. The relevant statement is given in Lemma~\ref{lemma:roughintestub}.
In Lemma~\ref{lemma:supexpgrowthmodeest} we then state the main estimate. The proof of this lemma is based, first, on a restriction of the time 
interval as described above. Second, given an $\e>0$, we reformulate the iteration so as to ensure that the influence of the off-diagonal parts 
of the Jordan blocks
on the growth of the solution is bounded by an error term whose size is quantified by $\e$. Once these steps have been taken, we divide 
the components, say $\chrho_{k}$, of the iteration, into two parts, say $\chrho_{k,a}$ and $\chrho_{k,b}$. The idea is then that $\chrho_{k,a}$ 
corresponds to the components of $\chrho_{k}$ with maximal growth. However, we modify the iteration so that the maximal growth corresponds
to no growth at all. In other words, we wish to demonstrate that 
$\chrho_{k,a}$ does not decay too much, whereas $\chrho_{k,b}$ is much smaller in size than $|\chrho_{k,a}|$. In the end, it turns out to be possible
to prove that if $|\chrho_{k,b}|$ is much smaller than $|\chrho_{k,a}|$ initially (where initially here means late enough that we are in the 
asymptotic regime), then it remains much smaller for a long period of time. Moreover, during this period, $\chrho_{k,a}$ does not decay too much. 
Returning to the original variables, this then yields super exponential growth. 

\textit{Super exponential instability.} Given the main estimate, we are in a position to prove a super exponential instability result. In fact, 
given assumptions of the type described above, the solution $u=0$ to (\ref{eq:thesystemRge}) (where $d=1$, $R=0$ and $f=0$) is unstable in the following 
sense. Fix $\e>0$. Then there is a sequence $v_{l}$, $1\leq l\in\zo$, of solutions to (\ref{eq:thesystemRge}) such that the corresponding initial 
data at $t=0$ converge to zero with respect to every Sobolev norm. Moreover, for each $1\leq l\in\zo$, there is a sequence $t_{l,k}\rightarrow
\infty$ (as $k\rightarrow\infty$) with the property that 
\[
\mfe_{\rohom}[v_{l}](t_{l,k})\geq \exp\left[2\b_{X}^{-1}\kappa(1-\e)e^{\b_{X}t_{l,k}}\right],
\]
where $\b_{X}>0$ is the constant appearing above; $\kappa$ is the largest absolute value of a real part of an eigenvalue of the matrix 
$Y^{1}_{\infty}/2$ (cf. the above); and 
\[
\mfe_{\rohom}[v](t):=\frac{1}{2}\is[|v_{t}(x,t)|^{2}+g^{11}(t)|v_{x}(x,t)|^{2}]dx.
\]
The relevant result is stated in Lemma~\ref{lemma:supexpinstab}. 

\textit{Generic super exponential growth, general equations.} As a next step, it is natural to consider general equations of the form 
(\ref{eq:thesystemRge}). The only assumption we need to make is that there is one $j\in\{1,\dots,d\}$ such that 
\begin{equation}\label{eq:jversupexpgrowth}
u_{tt}-g^{jj}(t)\d_{j}^{2}u-2g^{0j}(t)\d_{t}\d_{j}u+\a(t)u_{t}+X^{j}(t)\d_{j}u+\zeta(t)u=0
\end{equation}
satisfies the assumptions guaranteeing super exponential instability; cf. the above. Proving density of initial data yielding super 
exponential growth is then fairly straightforward. However, as we demonstrate in Section~\ref{section:modeanalysis}, the set of initial 
data yielding super exponential decay is sometimes also dense. It is therefore of interest to prove that super exponential growth 
represents the generic behaviour. In the end we prove two corresponding results; Propositions~\ref{prop:supexpgrowthgencaseden}
and \ref{prop:supexpgrowthinfcodim}. Proposition~\ref{prop:supexpgrowthgencaseden} states that there is a set of initial data, say
$\ma$, which is a countable intersection of open and dense sets (with respect to the $C^{\infty}$-topology) and such that solutions
corresponding to initial data in $\ma$ exhibit super exponential growth. Proposition~\ref{prop:supexpgrowthinfcodim} states, roughly
speaking, that the set of initial data whose corresponding solutions do not exhibit super exponential growth has infinite codimension.
Finally, in Subsection~\ref{ssection:unbdgritofinitefre} we also demonstrate that the super exponential growth of solutions can be
detected by solutions with finite frequency content; cf. Proposition~\ref{prop:supexpgrowthfinitefreq}. This is of interest due to the 
fact that solutions with finite frequency content sometime exhibit super exponential decay. 

\textit{Energy estimates.} In Section~\ref{section:standenestimates} we derive upper bounds on the growth of the energy under the
same assumptions that we make when proving the super exponential instability. In particular, we consider (\ref{eq:jversupexpgrowth})
with $j=1$. In the end we prove that if $Y^{1}_{\infty}$ is Hermitian, then the energy of solutions
does not grow faster than $\exp\left[2\b_{X}^{-1}\kappa(1+\e)e^{\b_{X}t}\right]$; cf. Lemma~\ref{lemma:standenestimates} and 
Remark~\ref{remark:standenestimates}. In that setting, the previously derived lower bounds on the growth of the energy are thus
essentially optimal. 

\textit{Gevrey classes.} In Section~\ref{section:gevreyclasses}, we consider solutions corresponding to initial data in Gevrey classes. 
The reason for taking an interest in this topic is the following. There are equations of the form (\ref{eq:jversupexpgrowth}) (with $j=1$) 
such that there are smooth 
solutions exhibiting super exponential growth and such that solutions with finite frequency content exhibit super exponential decay. It 
is therefore of interest to consider function classes interpolating between smooth and real analytic solutions and to maximise the degree 
of regularity of the solutions exhibiting super exponential growth. This is the topic of Section~\ref{section:gevreyclasses}. To begin 
with, we provide two different characterisations of Gevrey classes; a standard characterisation and one more adapted to the problem we
are interested in here. We also demonstrate that Gevrey type regularity is preserved under the evolution, given appropriate assumptions.
Finally, we then calculate a Gevrey index characterising the best regularity (we can demonstrate) of the solutions exhibiting super 
exponential growth; cf. Proposition~\ref{prop:supexpinstabGevreyreg}.

\textit{Mode analysis.} Finally, in Section~\ref{section:modeanalysis}, we analyse how the modes of the solution develop over time.

\section{Unbalanced equations, the ODE setting}\label{section:ODEunbal}

Consider the homogeneous version of (\ref{eq:thesystemRge}) when $g^{00}=-1$. The simplest special case of this equation is obtained 
by considering spatially homogeneous solutions. The relevant equation is then given by 
\begin{equation}\label{eq:ODEconvcoeff}
\ddot{x}+\a(t) \dot{x}+\zeta(t) x=0,
\end{equation}
where $\a,\zeta\in C^{\infty}[I,\Mn{m}{\co}]$. 

\begin{definition}\label{def:convandunbalsphom}
Let $\a,\zeta\in C^{\infty}[I,\Mn{m}{\co}]$, where $I$ is an open interval and $[0,\infty)\subset I$. Then (\ref{eq:ODEconvcoeff}) is said 
to be \textit{convergent and unbalanced} 
\index{ODE!convergent and unbalanced}%
\index{Convergent and unbalanced!ODE}%
if the following holds:
\begin{itemize}
\item there are constants $C_{\rocu},\g,\eta_{\rocu}>0$ and matrices $\a_{\infty},\zeta_{\infty}\in\Mn{m}{\co}$ such that 
\[
\|\g^{-1}e^{-\g t}\a(t)-\a_{\infty}\|+
\|\g^{-2}e^{-2\g t}\zeta(t)-\zeta_{\infty}\|\leq C_{\rocu}e^{-\eta_{\rocu} t}
\]
for $t\geq 0$,
\item one of $\a_{\infty},\zeta_{\infty}$ is non-zero. 
\end{itemize}
\end{definition}
\begin{remark}
Due to the requirements, the norm of at least one of the matrices $\a(t)$ and $\zeta(t)$ is unbounded as $t\rightarrow\infty$. 
\end{remark}
\begin{remark}
The case that $e^{-\g t}\a(t)$ and $e^{-2\g t}\zeta(t)$ converge exponentially for some real number $\g\leq 0$ can be dealt 
with as in Chapter~\ref{chapter:roughanalysisODEregion}. We therefore do not consider this case here. 
\end{remark}
Given that (\ref{eq:ODEconvcoeff}) is convergent and unbalanced, the matrix
\begin{equation}\label{eq:Ainfdef}
A_{\infty}:=\left(\begin{array}{cc} 0 & \Id_{m} \\ -\zeta_{\infty} & -\a_{\infty}\end{array}
\right)
\end{equation}
determines the dominant part of the asymptotic behaviour. 

\begin{lemma}\label{lemma:spsysodes}
Consider the equation (\ref{eq:ODEconvcoeff}). Assume that it is convergent and unbalanced, cf. Definition~\ref{def:convandunbalsphom},
and define $A_{\infty}$ by (\ref{eq:Ainfdef}). Let $\kappa_{1}:=\kappa_{\max}(A_{\infty})$; cf. Definition~\ref{def:SpRspdef}.  For every $\e>0$, 
there is a constant $C_{+,\e}>0$ such that for every solution $x$ to (\ref{eq:ODEconvcoeff})
\begin{equation}\label{eq:xpxdub}
|x(t)|+|\dot{x}(t)|\leq C_{+,\e}\exp[(\kappa_{1}+\e)e^{\g t}][|x(0)|+|\dot{x}(0)|]
\end{equation}
for all $t\geq 0$. Moreover, for every $\e>0$, there is a solution $x$ to (\ref{eq:ODEconvcoeff}) and a constant $C_{-,\e}>0$ such that 
\begin{equation}\label{eq:xpxdlb}
C_{-,\e}\exp[(\kappa_{1}-\e)e^{\g t}]\leq |x(t)|+|\dot{x}(t)|
\end{equation}
for all $t\geq 0$. 
\end{lemma}
\begin{remark}
The lemma illustrates that convergent and unbalanced equations typically have solutions with pathological properties. In fact, if 
$\kappa_{1}>0$, there are solutions that grow super exponentially. If $\kappa_{1}<0$, there are solutions that decay super exponentially
(meaning that they grow super exponentially towards the past). Both of these situations we consider to be pathological. 
\end{remark}
\begin{remark}
The constants $C_{+,\e}$ and $C_{-,\e}$ only depend on $C_{\rocu}$, $A_{\infty}$, $\eta_{\rocu}$, $\g$ and $\e$
\end{remark}
\begin{remarks}
If $\kappa_{1}$ is non-zero, (\ref{eq:xpxdub}) and (\ref{eq:xpxdlb}) yield a rough idea of the behaviour of solutions. However, this 
is not true if $\kappa_{1}=0$. In that case, a further analysis is necessary. The solutions with a lower bound of the form (\ref{eq:xpxdlb}) 
(that we construct in the proof) are such that they are stable under 
perturbations in the sense that a small perturbation of the corresponding initial data yields solutions satisfying a bound of the form  
(\ref{eq:xpxdlb}). Moreover, given a solution to (\ref{eq:ODEconvcoeff}) satisfying 
\[
|x(t)|+|\dot{x}(t)|\leq C\exp[(\kappa_{1}-\eta)e^{\g t}]
\]
for some constants $C,\eta>0$ and all $t\geq 0$, it can be perturbed to a solution satisfying (\ref{eq:xpxdlb}); this is an immediate
consequence of the existence of solutions satisfying (\ref{eq:xpxdlb}).  
\end{remarks}
\begin{proof}
Letting $x(t)=y(e^{\g t})$, (\ref{eq:ODEconvcoeff}) can be written
\begin{equation}\label{eq:transf1}
y''(e^{\g t})+e^{-\g t}y'(e^{\g t})+\g^{-1}e^{-\g t}\a(t)y'(e^{\g t})+\g^{-2}e^{-2\g t}\zeta(t)y(e^{\g t})=0.
\end{equation}
Introducing the time coordinate $\tau=e^{\g t}$, (\ref{eq:transf1}) can be written
\[
y''+\tau^{-1}y'+\a_{\roeff}(\tau)y'+\zeta_{\roeff}(\tau)y=0,
\]
where 
\[
\a_{\roeff}(\tau):=\g^{-1}\tau^{-1}\a(\g^{-1}\ln\tau),\ \ \
\zeta_{\roeff}(\tau):=\g^{-2}\tau^{-2}\zeta(\g^{-1}\ln\tau).
\]
Moreover, 
\[
\|\a_{\roeff}(\tau)-\a_{\infty}\|+\|\zeta_{\roeff}(\tau)-\zeta_{\infty}\|\leq C_{\rocu}\tau^{-\eta_{\rocu}/\g}
\]
for $\tau\geq 1$, where $\a_{\infty}$ and $\zeta_{\infty}$ are the matrices given in Definition~\ref{def:convandunbalsphom}. Introduce
\[
v:=\left(\begin{array}{c} y\\ y' \end{array}\right),\ \ \
A_{\rem}(\tau):=\left(\begin{array}{cc} 0 & 0 \\ -\zeta_{\roeff}(\tau)+\zeta_{\infty} & -\tau^{-1}\Id_{m}-\a_{\roeff}(\tau)+\a_{\infty}\end{array}
\right)
\]
and let $A_{\infty}$ be as in the statement of the lemma. Then 
\begin{equation}\label{eq:vAinfpArem}
v'=A_{\infty}v+A_{\rem}(\tau)v.
\end{equation}

\textbf{Upper bound on the solution.}
Let $T$ be a complex matrix such that $B_{\infty}:=TA_{\infty}T^{-1}$ consists of generalised Jordan blocks, where the blocks are ordered in
such a way that the ones corresponding to eigenvalues with maximal real part are on the top left; say that the first $m_{\ros}$ elements on the
diagonal of $B_{\infty}$ have maximal real part and define $m_{\rou}$ so that $m_{\ros}+m_{\rou}=2m$. Note that, given $\e>0$, $T$ can be chosen so 
that the generalised Jordan blocks are such that the non-zero off-diagonal terms  equal $\e/2$; cf. Lemma~\ref{lemma:genJordblock} and 
its proof. Note, moreover, that $T$ only depends on $A_{\infty}$ and $\e$. The equation (\ref{eq:vAinfpArem}) can then be rewritten
\begin{equation}\label{eq:diagode}
w'=B_{\infty}w+B_{\rem}(\tau)w,
\end{equation}
where $B_{\rem}(\tau):=TA_{\rem}(\tau)T^{-1}$ and $w:=Tv$. Note that the norm of $B_{\rem}(\tau)$ is bounded by $C_{\rocu,\e}\ldr{\tau}^{-\eta_{\min}}$
for all $\tau\geq 1$, where $\eta_{\min}=\min\{1,\eta_{\rocu}/\g\}$ and the constant $C_{\rocu,\e}$ only depends on $C_{\rocu}$, $A_{\infty}$ 
and $\e$. Due to (\ref{eq:diagode}), 
\[
\frac{d}{d\tau}|w|^{2}\leq 2\left(\kappa_{1}+\e/2+C_{\rocu,\e}\ldr{\tau}^{-\eta_{\min}}\right)|w|^{2}.
\]
Given $\e>0$ and a solution $w$, there is thus a constant $C_{\roup,\e}$ such that 
\[
|w(\tau)|\leq C_{\roup,\e}\exp[(\kappa_{1}+3\e/4)\tau]|w(1)|
\]
for $\tau\geq 1$, where $C_{\roup,\e}$ only depends on $C_{\rocu}$, $A_{\infty}$, $\eta_{\rocu}$, $\g$ and $\e$. Since
\begin{equation}\label{eq:xxdotwequiv}
|x(t)|+|\dot{x}(t)|=|y(e^{\g t})|+\g e^{\g t}|y'(e^{\g t})|\leq C_{\fin,\e}e^{\g t}|w(e^{\g t})|,
\end{equation}
(\ref{eq:xpxdub}) holds, where $C_{\fin,\e}$ and $C_{+,\e}$ only depend on $C_{\rocu}$, $A_{\infty}$, $\eta_{\rocu}$, $\g$ and $\e$.

\textbf{Lower bound on solutions.}
In order to obtain a lower bound, it is convenient to divide the components of $w$ into two parts, say $w_{\ros}$ and $w_{\rou}$, 
where $w_{\ros}$ and $w_{\rou}$ take their values in $\cn{m_{\ros}}$ and $\cn{m_{\rou}}$ respectively (in case $m_{\rou}=0$, it is 
trivial to prove the stated lower bound, so we here assume that $m_{\rou}>0$). It is convenient to rewrite
(\ref{eq:diagode}) as 
\begin{align}
w_{\ros}' = & B_{\infty,11}w_{\ros}+B_{\rem,11}(\tau)w_{\ros}+B_{\rem,12}(\tau)w_{\rou},\\
w_{\rou}' = & B_{\infty,22}w_{\rou}+B_{\rem,21}(\tau)w_{\ros}+B_{\rem,22}(\tau)w_{\rou}.
\end{align}
Let  $\kappa_{2}$ be the second largest real part of an eigenvalue of $A_{\infty}$. Then $\kappa_{2}<\kappa_{1}$. Note 
that $\|B_{\rem,ij}(\tau)\|$ satisfies the same bound as $\|B_{\rem}(\tau)\|$. Thus
\begin{align}
\frac{d}{d\tau}|w_{\ros}|^{2} \geq & 2\kappa_{1}|w_{\ros}|^{2}-\e|w_{\ros}|^{2}
-C_{\rocu,\e}\ldr{\tau}^{-\eta_{\min}}(3|w_{\ros}|^{2}+|w_{\rou}|^{2}),\label{eq:wrossqpr}\\
\frac{d}{d\tau}|w_{\rou}|^{2} \leq & 2\kappa_{2}|w_{\rou}|^{2}+\e|w_{\rou}|^{2}
+C_{\rocu,\e}\ldr{\tau}^{-\eta_{\min}}(|w_{\ros}|^{2}+3|w_{\rou}|^{2}).
\end{align}
Assume that $|w_{\ros}(\tau)|\neq 0$ and compute
\begin{equation}\label{eq:quoprime}
\frac{d}{d\tau}\frac{|w_{\rou}|^{2}}{|w_{\ros}|^{2}}\leq 2(\kappa_{2}-\kappa_{1}+\e)\frac{|w_{\rou}|^{2}}{|w_{\ros}|^{2}}
+K_{\rocu,\e}\ldr{\tau}^{-\eta_{\min}}\left(1+\frac{|w_{\rou}|^{2}}{|w_{\ros}|^{2}}+\frac{|w_{\rou}|^{4}}{|w_{\ros}|^{4}}\right),
\end{equation}
where $K_{\rocu,\e}$ has the same dependence as $C_{\rocu,\e}$. 
Without loss of generality, we may assume $\e$ to be such that $\e\leq (\kappa_{1}-\kappa_{2})/2$. Given that such an $\e>0$ has been 
fixed, fix $\tau_{0}$ so that if $K_{\rocu,\e}$ is the constant appearing on the right hand side of (\ref{eq:quoprime}), then 
$K_{\rocu,\e}\ldr{\tau}^{-\eta_{\min}}\leq (\kappa_{1}-\kappa_{2})/8$ for $\tau\geq \tau_{0}$. Fix initial data for $w_{\ros}$ and $w_{\rou}$ at 
$\tau_{0}$ in such a way that $|w_{\rou}(\tau_{0})|^{2}/|w_{\ros}(\tau_{0})|^{2}\leq 1/4$. Assume, in order to obtain a contradiction, that 
there is a $\tau_{1}\geq \tau_{0}$ such that $|w_{\rou}(\tau_{1})|^{2}/|w_{\ros}(\tau_{1})|^{2}=1/2$. Then, at $\tau=\tau_{1}$, 
\[
\frac{d}{d\tau}\frac{|w_{\rou}|^{2}}{|w_{\ros}|^{2}} \leq\frac{1}{4}(\kappa_{2}-\kappa_{1})<0.
\]
Thus $|w_{\rou}|^{2}/|w_{\ros}|^{2}$ is strictly decreasing at $\tau=\tau_{1}$, which means that $|w_{\rou}(\tau)|^{2}/|w_{\ros}(\tau)|^{2}\leq 1/2$
for all $\tau\geq \tau_{0}$. Choose initial data at $\tau_{0}$ so that $w_{\rou}(\tau_{0})=0$ and $|w_{\ros}(\tau_{0})|=1$. Returning to 
(\ref{eq:wrossqpr}), it is then clear that for any $\e>0$, there is a constant $C_{\low,\e}>0$ such that 
\[ 
C_{\low,\e}e^{(\kappa_{1}-\e)\tau}\leq |w(\tau)|
\]
for all $\tau\geq \tau_{0}$, where $C_{\low,\e}$ only depends on $C_{\rocu}$, $A_{\infty}$, $\eta_{\rocu}$, $\g$ and $\e$. Since we can change the 
upper bound in (\ref{eq:xxdotwequiv}) to a lower bound by simply removing the 
factor $e^{\g t}$ on the far right hand side, and changing the value of $C_{\fin,\e}$, it is clear that the constructed solution satisfies
(\ref{eq:xpxdlb}) for $t\geq t_{0}:=\g^{-1}\ln\tau_{0}$. Note also that the solution is stable under perturbations of the initial data. 
In order to ensure that the estimate holds in $[0,t_{0}]$, note that $t_{0}$ only depends on $C_{\rocu}$, $A_{\infty}$, $\e$, $\eta_{\rocu}$
and $\g$. Combining this observation with Definition~\ref{def:convandunbalsphom} yields the conclusion that 
\[
\|\a(t)\|+\|\zeta(t)\|\leq C_{0,\e}
\] 
for $0\leq t\leq t_{0}$, where $C_{0,\e}$ only depends on $C_{\rocu}$, $A_{\infty}$, $\e$, $\eta_{\rocu}$ and $\g$. Using this estimate, it can be
verified that (\ref{eq:xpxdlb}) is valid for all $t\geq 0$. 
\end{proof}

\section{Unbalanced equations, the PDE setting}\label{section:PDEunbalterm}

Let us turn to the case that the equation (\ref{eq:thesystemRge}) is not balanced due to a lack of boundedness of the norm of 
$X(\indexnot,t)$. Since we are only interested in producing counterexamples, 
it turns out to be sufficient to consider solutions that depend on only one of the spatial coordinates. In fact, we here assume that 
$d=1$ and that $R=0$. Then (\ref{eq:thesystemRge}) is an equation on $\so\times I$. When considering the corresponding equation for the Fourier 
coefficients, (\ref{eq:fourierthesystemRge}), it turns out to be sufficient to assume that $n_{1}>0$ (and we therefore typically make this 
assumption from now on). In this context, the definitions (\ref{eq:mfgnutdef}) and (\ref{eq:ellsigmaXgenRdef}) become
\[
\mfg(n_{1},t)=[g^{11}(t)]^{1/2}|n_{1}|,\ \ \
\ell(n_{1},t)=\ln\mfg(n_{1},t)
\]
and
\begin{equation}\label{eq:Xsigodnp}
X(n_{1},t)=\frac{n_{1}X^{1}(t)}{[g^{11}(t)]^{1/2}|n_{1}|}=\frac{X^{1}(t)}{[g^{11}(t)]^{1/2}},\ \ \
\sigma(n_{1},t)=\frac{g^{01}(t)}{[g^{11}(t)]^{1/2}}.
\end{equation}
In what follows, it will also be convenient to introduce the notation
\begin{equation}\label{eq:mflYovarsdef}
\mfl(t):=\ln[g^{11}(t)]^{1/2},\ \ \
Y^{1}(t):=\frac{X^{1}(t)}{[g^{11}(t)]^{1/2}},\ \ \
\varsigma(t):=\frac{g^{01}(t)}{[g^{11}(t)]^{1/2}}.
\end{equation}
The reason we introduce this notation is that the relations (\ref{eq:Xsigodnp}) do not hold in general; they are based on the assumption
that $n_{1}>0$. Note also that $Y^{1}$, $\varsigma$ and $\mfl$ are independent of $n_{1}$ and that $\dot{\mfl}=\dot{\ell}$. Next, we want to 
state conditions that ensure that (\ref{eq:thesystemRge}) is oscillation adapted; cf. Definition~\ref{def:oscad}.

\begin{lemma}\label{lemma:mainassumpubcase}
Consider (\ref{eq:thesystemRge}). Assume the associated metric to be such that $(M,g)$ is a canonical separable cosmological model
manifold. Assume, moreover, that $d=1$ and that $R=0$. Let $\mfl$, $Y^{1}$ and $\varsigma$ be defined by (\ref{eq:mflYovarsdef}). Assume that there
are real numbers $\b_{\rosh}<0$, $\ellderbd>0$, $\eta_{\roode}>0$, $C_{\rosh}>0$, $C_{X}>0$, $C_{\roode}>0$, $\b_{X}\geq 0$ and $\b_{\roode}$ such that 
\begin{align}
\b_{X}-\b_{\roode} = & \eta_{\roode},\label{eq:etaroodedef}\\
|\b_{\rosh}|+|\b_{X}|+|\b_{\roode}| \leq & \ellderbd,\label{eq:betalowerbd}\\
|\dot{\mfl}(t)|+|\ddot{\mfl}(t)| \leq & \ellderbd\label{eq:mfldaddbd}
\end{align}
for all $t\geq 0$, and 
\begin{align}
|\varsigma(t)|+|\dot{\varsigma}(t)| \leq & C_{\rosh}e^{\b_{\rosh}t},\label{eq:shiftbdub}\\
\|Y^{1}(t)\|+\|\dot{Y}^{1}(t)\| \leq & C_{X}e^{\b_{X}t},\label{eq:Xbdub}\\
\|\dot{\a}(t)\|+\|\a(t)\| \leq & C_{\roode}e^{\b_{\roode}t},\label{eq:albdub}\\
\|\dot{\zeta}(t)\|+\|\zeta(t)\| \leq & C_{\roode}^{2}e^{2\b_{\roode}t}\label{eq:zetabdub}
\end{align}
for all $t\geq 0$. Then (\ref{eq:thesystemRge}) is oscillation adapted, with $\mff_{\rosh}$, $\mff_{X}$ and $\mff_{\roode}$ replaced by 
$C_{\rosh}e^{\b_{\rosh}t}$, $C_{X}e^{\b_{X}t}$ and  $C_{\roode}e^{\b_{\roode}t}$ respectively. 
\end{lemma}
\begin{remark} 
The statement is of course true regardless of the sign of $\b_{\rosh}$ and $\b_{X}$, and regardless of whether the left hand side of 
(\ref{eq:etaroodedef}) is positive or not. However, we are here only interested in the case $\b_{\rosh}<0$, $\b_{X}\geq 0$ and 
$\eta_{\roode}>0$, which is why we immediately restrict to that case.  
\end{remark}
\begin{proof}
The statement follows from the facts that $\dot{\mfl}=\dot{\ell}$, $\varsigma=\pm\sigma$ and $Y^{1}=\pm X$ (depending on whether $n_{1}$ is 
positive or negative). 
\end{proof}

\section{Subdividing the time interval}\label{section:subdivinterv}

Before turning to the actual estimates, it is natural to restrict the time interval so that the results of 
Chapter~\ref{chapter:destopofoscref} can be applied. This is the purpose of the present section.
In what follows, we are only interested in the case that $n_{1}$ is very large. In that context, there is a time interval 
$[0,t_{\max}]$ such that the conditions of Lemma~\ref{lemma:gaintvarge} are satisfied. In order to obtain a more quantitative control 
over the size of this interval, it is convenient to make the following assumption.

\begin{assumption}\label{assumption:guooass}
Consider (\ref{eq:thesystemRge}). Assume the associated metric to be such that $(M,g)$ is a canonical separable cosmological model
manifold. Assume, moreover, that $d=1$ and that $R=0$. Assume, finally, that there are $\b_{1},\b_{2}\in\ro$ and constants $c_{\rom,l}>0$, 
$l=1,2$, such that 
\begin{equation}\label{eq:cmebgoobd}
c_{\rom,1}e^{2\b_{1}t}\leq g^{11}(t) \leq c_{\rom,2}e^{2\b_{2}t}
\end{equation}
for all $t\geq 0$. 
\end{assumption}
\begin{remark}\label{remark:gupoolowandupbdsupexpgrowth}
The estimate (\ref{eq:cmebgoobd}) is an immediate consequence of (\ref{eq:mfldaddbd}). Nevertheless, we state it explicitly here
in order to introduce the terminology $\b_{l}$ and $c_{\rom,l}$. 
\end{remark}
 
When considering an interval $I_{a}$ of the form constructed in Lemma~\ref{lemma:gaintvarge}, it is of interest to bound the 
quantity $\vare_{a}$ appearing in (\ref{eq:vareadef}):
\begin{equation}\label{eq:vareadefub}
\vare_{a}:=\frac{\mff_{\rosh}(t_{a})}{\mfg(t_{a})}[1+\mff_{\roode}(t_{a})+\mff_{X}(t_{a})]
+\frac{1+\mff_{X}^{2}(t_{a})+\mff_{\roode}^{2}(t_{a})}{\mfg^{2}(t_{a})}.
\end{equation}
Given that the assumptions of Lemma~\ref{lemma:mainassumpubcase} hold and that Assumption~\ref{assumption:guooass} holds, 
\begin{equation}\label{eq:vareaest}
\vare_{a}\leq C_{\ropar}\frac{e^{(\b_{X}+\b_{\rosh})t_{a}}}{|n_{1}|e^{\b_{1}t_{a}}}
+C_{\ropar}\frac{e^{2\b_{X}t_{a}}}{n_{1}^{2}e^{2\b_{1}t_{a}}},
\end{equation}
where $C_{\ropar}$ is a constant depending only on $\boc_{\ropar}$ defined by 
\begin{equation}\label{eq:bocropardef}
\boc_{\ropar}:=(\b_{\rosh},\ellderbd,\eta_{\roode},C_{\rosh},C_{X},C_{\roode},\b_{X},\b_{\roode},\b_{1},\b_{2},c_{\rom,1},c_{\rom,2}).
\end{equation}
In the analysis to follow, there are two things we want to be fulfilled. First of all, we need to know that the conditions of 
Lemma~\ref{lemma:gaintvarge} are fulfilled. Second, we wish to ensure that the right hand side of (\ref{eq:vareaest}) is small.
The following lemma provides assumptions that ensure that these conditions hold. 

\begin{lemma}\label{lemma:condvaresmallub}
Consider (\ref{eq:thesystemRge}). Assume the associated metric to be such that $(M,g)$ is a canonical separable cosmological model
manifold. Assume, moreover, that $d=1$ and that $R=0$. 
Assume, finally, that the conditions of Lemma~\ref{lemma:mainassumpubcase} are fulfilled and that Assumption~\ref{assumption:guooass} holds. 
Let $0<\e\leq 1$. If $\b_{X}-\b_{1}>0$, there are constants $\nu_{\ropar,\e}$ and $C_{\ropar,\e}$, depending only on $\boc_{\ropar}$, defined in 
(\ref{eq:bocropardef}), and $\e$, such that if $|n_{1}|\geq\nu_{\ropar,\e}$, then 
\begin{equation}\label{eq:Troparedef}
T_{\ropar,\e}:=\frac{1}{\b_{X}-\b_{1}}\ln|n_{1}|+C_{\ropar,\e}>0
\end{equation}
and
\begin{equation}\label{eq:varealtemfglb}
\vare_{a}\leq\e,\ \ \
\mfg(n_{1},t)\geq \max\{4\pi\ellderbd,2\},\ \ \
\frac{e^{\b_{X}t}}{\mfg(n_{1},t)}\leq\e
\end{equation}
for $0\leq t,t_{a}\leq T_{\ropar,\e}$, where $\vare_{a}$ is defined in (\ref{eq:vareadefub}). If $\b_{X}-\b_{1}\leq 0$, there is a constant 
$\nu_{\ropar,\e}$, depending only on $\boc_{\ropar}$ and $\e$, such that if $|n_{1}|\geq\nu_{\ropar,\e}$, then (\ref{eq:varealtemfglb}) holds
for all $t,t_{a}\geq 0$. 
\end{lemma}
\begin{remark}\label{remark:notbrestrub}
In what follows, we assume $n_{1}$ to be such that (\ref{eq:varealtemfglb}) holds with $\e=1$ for $t\in [0,t_{\max})$, where 
$t_{\max}=\infty$ in case $\b_{X}-\b_{1}\leq 0$ and $t_{\max}=T_{\ropar,1}$ in case $\b_{X}-\b_{1}>0$; here $T_{\ropar,\e}$ is defined in 
(\ref{eq:Troparedef}). Note that it is sufficient to assume that $|n_{1}|\geq \nu_{\ropar}$ (where $\nu_{\ropar}$ only depends on 
$\boc_{\ropar}$) in order to obtain this conclusion. 
\end{remark}
\begin{remark}\label{remark:sufconnesequb}
Given that the assumptions of Remark~\ref{remark:notbrestrub} are satisfied, it is clear that we can construct a sequence $\{t_{k}\}$ as in 
Definition~\ref{def:tkdefge} with $0\leq t_{0}<t_{\max}$. Then $t_{\fin}\geq t_{\max}-2\pi$, where $t_{\max}$ is introduced in 
Remark~\ref{remark:notbrestrub}.
\end{remark}
\begin{proof}
The three conditions we want to hold are (\ref{eq:varealtemfglb}). Let us, to begin with, focus on the second condition. Due to 
Assumption~\ref{assumption:guooass}, it is ensured by requiring 
\begin{equation}\label{eq:osccondub}
c_{\rom,1}e^{2\b_{1}t}|n_{1}|^{2}\geq (\max\{4\pi\ellderbd,2\})^{2}.
\end{equation}
As a consequence, there is a $0<\nu_{\ropar}\in\ro$, only depending on $\boc_{\ropar}$, such that the following holds. 
 If $\b_{1}\geq 0$ and $n_{1}$ satisfies $|n_{1}|\geq\nu_{\ropar}$, then (\ref{eq:osccondub}) holds for all $t\geq 0$. 
If $\b_{1}<0$ and $n_{1}$ satisfies $|n_{1}|\geq\nu_{\ropar}$, then (\ref{eq:osccondub}) holds for all $t\geq 0$ such that 
\begin{equation}\label{eq:tbdficonub}
t\leq -\frac{1}{\b_{1}}\ln|n_{1}|+C_{\ropar},
\end{equation}
where $C_{\ropar}$ only depends on $\boc_{\ropar}$, and the right hand side is strictly positive if $|n_{1}|\geq\nu_{\ropar}$. 

As a second step, we wish to ensure that the first and third estimates in (\ref{eq:varealtemfglb}) hold. To prove that the first 
estimate holds, it is, due to (\ref{eq:vareaest}), sufficient to assume that right hand side of (\ref{eq:vareaest}) is bounded by $\e$. 
Given $0<\e\leq 1$, there is a $C_{\ropar}$ (depending only on $\boc_{\ropar}$) such that if
\begin{equation}\label{eq:oscapprub}
\frac{e^{\b_{X}t_{a}}}{|n_{1}|e^{\b_{1}t_{a}}}\leq C_{\ropar}\e,
\end{equation}
then the right hand side of (\ref{eq:vareaest}) is bounded from above by $\e$ (note that a condition of this type is also sufficient 
in order to ensure that the third estimate in (\ref{eq:varealtemfglb}) holds). In this case, given $0<\e\leq 1$, there is a $\nu_{\ropar,\e}>0$,
depending only on $\boc_{\ropar}$ and $\e$, such that the following holds. If $\b_{X}-\b_{1}\leq 0$ and $|n_{1}|\geq \nu_{\ropar,\e}$, then 
(\ref{eq:oscapprub}) holds for all $t_{a}\geq 0$. If $\b_{X}-\b_{1}>0$ and $|n_{1}|\geq \nu_{\ropar,\e}$, then (\ref{eq:oscapprub}) holds for 
all $t_{a}\geq 0$ such that 
\begin{equation}\label{eq:tabdficonub}
t_{a}\leq\frac{1}{\b_{X}-\b_{1}}\ln|n_{1}|+C_{\ropar,\e},
\end{equation}
where $C_{\ropar,\e}$ only depends on $\boc_{\ropar}$ and $\e$, and the right hand side is strictly positive if $|n_{1}|\geq\nu_{\ropar,\e}$.

In order to prove the first statement of the lemma, assume that $\b_{X}-\b_{1}>0$. Then there are two possibilities: either $\b_{1}<0$ or 
$\b_{1}\geq 0$. If $\b_{1}<0$ and $|n_{1}|\geq\max\{\nu_{\ropar},\nu_{\ropar,\e}\}$, then it is sufficient to assume that $t$ and $t_{a}$ satisfy
(\ref{eq:tbdficonub}) and (\ref{eq:tabdficonub}) in order for (\ref{eq:varealtemfglb}) to be satisfied. However, these restrictions on $t$ and
$t_{a}$ can be translated into $0\leq t,t_{a}\leq T_{\ropar,\e}$. In case $\b_{1}\geq 0$ and $|n_{1}|\geq\nu_{\ropar}$, the second estimate in 
(\ref{eq:varealtemfglb}) holds for all $t\geq 0$. Regardless of whether $\b_{1}<0$ or $\b_{1}\geq 0$, the first statement of the lemma thus
holds. 

In order to prove the second statement, assume that $\b_{X}-\b_{1}\leq 0$. Since $\b_{X}\geq 0$, this implies that $\b_{1}\geq 0$. Due to 
the above arguments, it is thus sufficient to assume that $|n_{1}|\geq\max\{\nu_{\ropar},\nu_{\ropar,\e}\}$ in order to ensure that 
(\ref{eq:varealtemfglb}) holds for all $t\geq 0$.
\end{proof}

\section{Basic properties of the iteration}\label{section:bapropofiterseg}

As a next step, we wish to apply Lemma~\ref{lemma:wkfinlemma}, in particular (\ref{eq:wprekpertransfin}); note that if the assumptions
of Lemma~\ref{lemma:condvaresmallub} and Remark~\ref{remark:notbrestrub} are satisfied, there is a non-empty time sequence of the form 
given in Definition~\ref{def:tkdefge}, starting at $0\leq t_{0}<t_{\max}$. 
We are here mainly interested in homogeneous equations, and in that case, (\ref{eq:wprekpertransfin}) implies that 
$w_{\fin,k+1} =\Xi^{k,+}_{\fin}w_{\fin,k}$. However, before proceeding, it turns out to be convenient to transform $w_{\fin,k}$ slightly.
Moreover, since the transformation depends on more detailed knowledge concerning the coefficients of (\ref{eq:thesystemRge}) than
we have thus far assumed, we need to start by making an additional assumption. 

\begin{assumption}\label{assumption:Yoconvub}
Consider (\ref{eq:thesystemRge}). Assume the associated metric to be such that $(M,g)$ is a canonical separable cosmological model
manifold. Assume, moreover, that $d=1$ and that $R=0$.
Let $Y^{1}$ be defined by (\ref{eq:mflYovarsdef}) and the assumptions of Lemma~\ref{lemma:mainassumpubcase} be fulfilled. 
Assume that there is a $Y^{1}_{\infty}\in\Mn{m}{\co}$ and constants $\eta_{X}>0$ and $K_{X}>0$ such that
\begin{equation}\label{eq:Yoasbehub}
\|e^{-\b_{X}t}Y^{1}(t)-Y^{1}_{\infty}\|\leq K_{X}e^{-\eta_{X}t}
\end{equation}
for all $t\geq 0$, where $\b_{X}\geq 0$ is the constant appearing in the statement of Lemma~\ref{lemma:mainassumpubcase}.
\end{assumption}
\begin{remark}
In what follows, we are mainly interested in the case that $Y^{1}_{\infty}$ has an eigenvalue with a non-zero real part. However,
we state this assumption separately in the lemmas to follow. 
\end{remark}
Let $R_{\infty}$ be defined by 
\begin{equation}\label{eq:Rinftydef}
R_{\infty}:=\frac{1}{2}\left(\begin{array}{cc} Y^{1}_{\infty}
& 0 \\  0 & -Y^{1}_{\infty}\end{array}
\right).
\end{equation}
Let $T$ be a complex $2m\times 2m$-matrix such that 
\begin{equation}\label{eq:JTRinfdefub}
J:=TR_{\infty}T^{-1}
\end{equation}
consists of Jordan blocks $J_{1},\dots,J_{r}$; $J=\mathrm{diag}\{J_{1},\dots,J_{r}\}$. Denote the eigenvalue corresponding to 
$J_{s}$ by $\lambda_{s}=\kappa_{s}+i\zeta_{s}$, where $\kappa_{s}$ and $\zeta_{s}$ are real. Let $l_{s}$ be the dimension of the Jordan block 
$J_{s}$ and define
\begin{equation}\label{eq:Skdefub}
S_{k}:=\mathrm{diag}\{\exp(-i\zeta_{1}\b_{X}^{-1}e^{\b_{X}t_{k}})\Id_{l_{1}},\dots,\exp(-i\zeta_{r}\b_{X}^{-1}e^{\b_{X}t_{k}})\Id_{l_{r}}\}.
\end{equation}
Here $t_{k}$ is the time sequence given by Definition~\ref{def:tkdefge}, starting at $0\leq t_{0}<t_{\max}$; note that the sequence is non-empty if 
the assumptions of Lemma~\ref{lemma:condvaresmallub} and Remark~\ref{remark:notbrestrub} are satisfied (cf. Remark~\ref{remark:sufconnesequb}). 
Moreover, we here assume that $\b_{X}>0$, where $\b_{X}$ is the 
constant introduced in Lemma~\ref{lemma:mainassumpubcase} and Assumption~\ref{assumption:Yoconvub}. Defining $\rho_{k}$ and $B_{k}$ according to 
\begin{equation}\label{eq:rhokBkdefub}
\rho_{k}:=S_{k}Tw_{\fin,k},\ \ \
B_{k}:=S_{k+1}T\Xi^{k,+}_{\fin}T^{-1}S_{k}^{-1},
\end{equation}
the equation (\ref{eq:wprekpertransfin}) becomes $\rho_{k+1}=B_{k}\rho_{k}$ (assuming $f=0$). Recall here that $w_{\fin,k}$ is defined in 
(\ref{eq:wfinkdef}) and that $\Xi^{k,+}_{\fin}$ is introduced in the statement of Lemma~\ref{lemma:wkfinlemma}; cf. also (\ref{eq:Xikfindef}). 
Let $\kappa:=\kappa_{\max}(R_{\infty})$; cf. Definition~\ref{def:SpRspdef}. Introducing
\begin{equation}\label{eq:hrhokhBkdefub}
\hrho_{k}:=\exp\left(-\b_{X}^{-1}e^{\b_{X}t_{k}}\kappa\right)\rho_{k},\ \ \
\hBl_{k}:=\exp\left(-\b_{X}^{-1}e^{\b_{X}t_{k+1}}\kappa+\b_{X}^{-1}e^{\b_{X}t_{k}}\kappa\right)B_{k},
\end{equation}
(\ref{eq:wprekpertransfin}) can then be written
\begin{equation}\label{eq:hrhokit}
\hrho_{k+1}=\hBl_{k}\hrho_{k}.
\end{equation}
The natural next step is to approximate $\hBl_{k}$ in order to estimate the product of the matrices. 

\begin{lemma}\label{lemma:hBkapprfirstimpro}
Consider (\ref{eq:thesystemRge}). Assume the associated metric to be such that $(M,g)$ is a canonical separable cosmological model
manifold. Assume, moreover, that $d=1$ and that $R=0$.
Assume that the conditions of Lemma~\ref{lemma:mainassumpubcase} are fulfilled with $\b_{X}>0$ and that Assumptions~\ref{assumption:guooass} and
\ref{assumption:Yoconvub} hold. If $\b_{X}-\b_{1}>0$, there are constants $C_{\ropar}$ and $\nu_{\ropar}>0$ (depending only on $\boc_{\ropar}$ introduced 
in (\ref{eq:bocropardef})), and if $\b_{X}-\b_{1}\leq 0$, there is a constant $\nu_{\ropar}$ (depending only on $\boc_{\ropar}$) such that the 
following holds. If $|n_{1}|\geq \nu_{\ropar}$, then (\ref{eq:varealtemfglb}), with $\e$ replaced by $1$, holds for $0\leq t,t_{a}<t_{\max}$, where
\begin{equation}\label{eq:tmaxdefftry}
t_{\max}:=\left\{\begin{array}{cl} T_{\ropar,1}, & \b_{X}-\b_{1}>0\\ \infty, & \b_{X}-\b_{1}\leq 0\end{array}\right.
\end{equation}
and 
\[
T_{\ropar,1}:=\frac{1}{\b_{X}-\b_{1}}\ln|n_{1}|+C_{\ropar,1}>0;
\]
cf. (\ref{eq:Troparedef}). Let $\{t_{k}\}$ be a time sequence of the form constructed in Definition~\ref{def:tkdefge}, where $0\leq t_{0}<t_{\max}$.
Then $t_{\fin}$, given in Definition~\ref{def:tkdefge}, satisfies $t_{\fin}\geq \max\{t_{0},t_{\max}-2\pi\}$. Assume that $n_{1}\geq\nu_{\ropar}$. 
Then, if $k\geq 0$ and $t_{k}<t_{\max}$, 
\begin{equation}\label{eq:hBlkhBlkapp}
\left\|\hBl_{k}-\hBl_{k,\app}\right\|\leq C_{\roeep}\left(
\frac{e^{(\b_{X}-\eta_{\romn})t_{k}}}{\mfg(t_{k})}+\frac{e^{2\b_{X}t_{k}}}{\mfg^{2}(t_{k})}\right), 
\end{equation}
where $\eta_{\romn}>0$ is defined by
\begin{equation}\label{eq:etaromnub}
\eta_{\romn}:=\min\{\eta_{X},\eta_{\roode},\b_{X},-\b_{\rosh}\};
\end{equation}
$\hBl_{k,\app}$ is given by 
\begin{equation}\label{eq:hBlkappJrondef}
\hBl_{k,\app}:=\Id_{2m}+\frac{2\pi e^{\b_{X}t_{k}}}{\mfg(t_{k})}J_{\ron},\ \ \
J_{\ron}:=\mathrm{Re}\{J\}-\kappa\Id_{2m};
\end{equation}
$J$ is the matrix introduced in (\ref{eq:JTRinfdefub}); $\kappa:=\kappa_{\max}(R_{\infty})$; and $C_{\roeep}$ only depends on $\boc_{\roeep}$ defined by
\begin{equation}\label{eq:bcroeepdef}
\boc_{\roeep}:=(\boc_{\ropar},\eta_{X},K_{X},J,T),
\end{equation}
where the $J$ is defined by (\ref{eq:JTRinfdefub}) and $T$ is the matrix defined in connection with (\ref{eq:JTRinfdefub}) 
(in (\ref{eq:bcroeepdef}), $J$ and $T$ should be thought of as vectors whose components are the entries of the matrices $J$ and $T$
respectively).
\end{lemma}
\begin{proof}
Since the conditions of Lemma~\ref{lemma:mainassumpubcase} are fulfilled, we know that (\ref{eq:thesystemRge}) is oscillation 
adapted, with $\mff_{\rosh}$ etc. as in the statement of Lemma~\ref{lemma:mainassumpubcase}. Moreover, $\indexnot=n_{1}\neq 0$.
Finally, $t_{0}$ is such that (\ref{eq:gtalbge}) holds with $t_{a}$ replaced by $t_{0}$. Thus the conditions of Lemma~\ref{lemma:wkfinlemma}
are satisfied. 

Let us begin by considering $\Xi^{k,+}_{\fin}$. Due to Lemma~\ref{lemma:wkfinlemma}, we know that the estimate (\ref{eq:XikXikIpreestfin})
holds. Consider the right hand side. Note that $\vare_{k}$ is given by (\ref{eq:vareadefub}) with $t_{a}$ replaced by $t_{k}$, and that
$\vare_{k}\leq 1$ due to the definition of $t_{\max}$ and the fact that $t_{k}<t_{\max}$. Keeping the assumptions of 
Lemma~\ref{lemma:mainassumpubcase} in mind, (\ref{eq:XikXikIpreestfin}) yields
\begin{equation}\label{eq:Xikfinprelestub}
\|\Xi^{k,+}_{\fin}-\Xi^{k,+}_{\fin,\app}\|\leq C_{\ropar} \left(
\frac{e^{(\b_{X}+\b_{\rosh})t_{k}}}{\mfg(t_{k})}+\frac{e^{2\b_{X}t_{k}}}{\mfg^{2}(t_{k})}\right),
\end{equation}
where $C_{\ropar}$ only depends on $\boc_{\ropar}$; recall that 
$\mff_{\rosh}(t)\leq C_{\rosh}$ for all $t\geq 0$. Next, let us turn to $\Xi^{k,+}_{\fin,\app}$. Due to (\ref{eq:Xikpreappdefstfin}), it is 
of interest to approximate $R_{\pre,k}^{ab}$, defined in (\ref{eq:Rabprekdef}). Due to (\ref{eq:etaroodedef}), (\ref{eq:mfldaddbd}),
(\ref{eq:albdub}) and (\ref{eq:Yoasbehub}),
\begin{equation}\label{eq:eq:Rprekprelestub}
\left\|R_{\pre,k}^{ab}-(-1)^{b}\frac{e^{\b_{X}t_{k}}}{\mfg(t_{k})}Y^{1}_{\infty}\right\|\leq C_{\roep}\frac{e^{(\b_{X}-\eta_{\romn})t_{k}}}{\mfg(t_{k})}
\end{equation}
where we have used the fact that $X=Y^{1}$ when $n_{1}>0$, $\eta_{\romn}$ is defined by (\ref{eq:etaromnub}) and $C_{\roep}$ only depends on 
\begin{equation}\label{eq:bocroepdef}
\boc_{\roep}:=(\boc_{\ropar},\eta_{X},K_{X}\}. 
\end{equation}
In the definition (\ref{eq:etaromnub}) of $\eta_{\romn}$, we could have excluded $-\beta_{\rosh}$; the estimate (\ref{eq:eq:Rprekprelestub})
would still hold. However, in order not to have to introduce additional notation later, we have included $-\b_{\rosh}$ already here. 
Combining (\ref{eq:Xikfinprelestub}) and (\ref{eq:eq:Rprekprelestub}) with (\ref{eq:Xikpreappdefstfin}) and (\ref{eq:Rinftydef})
yields
\begin{equation}\label{eq:Xikestfinaux}
\|\Xi^{k,+}_{\fin}-\Xi^{k,+}_{\app}\|\leq C_{\roep} \left(
\frac{e^{(\b_{X}-\eta_{\romn})t_{k}}}{\mfg(t_{k})}+\frac{e^{2\b_{X}t_{k}}}{\mfg^{2}(t_{k})}\right),
\end{equation}
where the constant $C_{\roep}$ only depends on $\boc_{\roep}$ and 
\[
\Xi^{k,+}_{\app}:=\Id_{2m}+\frac{2\pi e^{\b_{X}t_{k}}}{\mfg(t_{k})}R_{\infty}.
\]
Due to the definition of $T$ and $J$, cf. (\ref{eq:JTRinfdefub}), it is clear that 
\begin{equation}\label{eq:JordecXikappub}
T\Xi^{k,+}_{\app}T^{-1}=\Id_{2m}+\frac{2\pi e^{\b_{X}t_{k}}}{\mfg(t_{k})}J.
\end{equation}
Just as in the second reformulation of the equations in the proof of Lemma~\ref{lemma:modebymodetrs}, it is of 
interest to eliminate the imaginary parts; this is the reason for introducing $\rho_{k}$ defined in (\ref{eq:rhokBkdefub}). It is also of 
interest to subtract the largest real part of the eigenvalues of $J$. This is the reason for introducing $\hrho_{k}$ defined in (\ref{eq:hrhokhBkdefub}). 
In order to estimate the effect of the transformation involved in defining, e.g., $\rho_{k}$, let $\lambda\in\co$ and consider
\[
\exp\left(\b_{X}^{-1}e^{\b_{X}t_{k+1}}\lambda-\b_{X}^{-1}e^{\b_{X}t_{k}}\lambda\right)
=\exp\left[\b_{X}^{-1}e^{\b_{X}t_{k}}\lambda(e^{\b_{X}(t_{k+1}-t_{k})}-1)\right].
\]
Note that 
\[
\left|e^{\b_{X}(t_{k+1}-t_{k})}-1-\frac{2\pi\b_{X}}{\mfg(t_{k})}\right|\leq \frac{C}{\mfg^{2}(t_{k})},
\]
where the constant $C$ only depends on $\b_{X}$ and the constant $\ellderbd$ appearing in the estimate (\ref{eq:estdtlnge}); cf. 
Lemma~\ref{lemma:gaintvarge}, in particular (\ref{eq:tatbroughge}) and (\ref{eq:tbmtagestge}). Thus
\[
\left|\b_{X}^{-1}e^{\b_{X}t_{k+1}}\lambda-\b_{X}^{-1}e^{\b_{X}t_{k}}\lambda-\frac{2\pi\lambda}{\mfg(t_{k})}e^{\b_{X}t_{k}}\right|
\leq C\frac{e^{\b_{X}t_{k}}}{\mfg^{2}(t_{k})},
\]
where $C$ only depends on $\lambda$, $\b_{X}$ and the constant $\ellderbd$ appearing in the estimate (\ref{eq:estdtlnge}). To
conclude, 
\begin{equation}\label{eq:explambdabXub}
\left|\exp\left(\b_{X}^{-1}e^{\b_{X}t_{k+1}}\lambda-\b_{X}^{-1}e^{\b_{X}t_{k}}\lambda\right)-1-\frac{2\pi\lambda}{\mfg(t_{k})}e^{\b_{X}t_{k}}\right|
\leq C\frac{e^{2\b_{X}t_{k}}}{\mfg^{2}(t_{k})},
\end{equation}
where $C$ only depends on $\lambda$, $\b_{X}$, the constant $\ellderbd$ appearing in the estimate (\ref{eq:estdtlnge}), and an upper bound on 
$e^{\b_{X}t_{k}}/\mfg(t_{k})$. On the other hand, due to (\ref{eq:varealtemfglb}), $e^{\b_{X}t_{k}}/\mfg(t_{k})\leq 1$. Thus 
$C$ only depends on $\lambda$, $\b_{X}$ and $\ellderbd$. Defining $S_{k}$ as in (\ref{eq:Skdefub}), the estimate (\ref{eq:explambdabXub}) thus 
yields 
\begin{equation}\label{eq:skposkinv}
\left\|S_{k+1}S_{k}^{-1}-\Id_{2m}+\frac{2\pi}{\mfg(t_{k})}e^{\b_{X}t_{k}}\diag\{i\zeta_{1}\Id_{l_{1}},\dots,i\zeta_{r}\Id_{l_{r}}\}\right\|
\leq C_{\roeep}\frac{e^{2\b_{X}t_{k}}}{\mfg^{2}(t_{k})},
\end{equation}
where $C_{\roeep}$ only depends on $\boc_{\roeep}$ introduced in (\ref{eq:bcroeepdef}). Let
\[
B_{k,\app}:=S_{k+1}T\Xi^{k,+}_{\app}T^{-1}S_{k}^{-1}=S_{k+1}\left(\Id_{2m}+\frac{2\pi e^{\b_{X}t_{k}}}{\mfg(t_{k})}J\right)S_{k}^{-1},
\]
where the last equality is due to (\ref{eq:JordecXikappub}). Due to (\ref{eq:skposkinv}) and the fact that $S_{k}$ and $J$ commute, 
\[
\left\|B_{k,\app}-\Id_{2m}-\frac{2\pi e^{\b_{X}t_{k}}}{\mfg(t_{k})}\mathrm{Re}\{J\}\right\|\leq C_{\roeep}\frac{e^{2\b_{X}t_{k}}}{\mfg^{2}(t_{k})},
\]
where $C_{\roeep}$ only depends on $\boc_{\roeep}$. Combining this estimate with (\ref{eq:Xikestfinaux}) yields
\[
\left\|B_{k}-\Id_{2m}-\frac{2\pi e^{\b_{X}t_{k}}}{\mfg(t_{k})}\mathrm{Re}\{J\}\right\|\leq C_{\roeep}\left(
\frac{e^{(\b_{X}-\eta_{\romn})t_{k}}}{\mfg(t_{k})}+\frac{e^{2\b_{X}t_{k}}}{\mfg^{2}(t_{k})}\right), 
\]
where $C_{\roeep}$ only depends on $\boc_{\roeep}$. Applying a similar argument a second time yields the conclusion of the lemma. 
\end{proof}

\section{The main estimate}\label{section:themainestimateseg}

On a technical level, the main result of this chapter is Lemma~\ref{lemma:supexpgrowthmodeest}; the proof of the existence of solutions
exhibiting super exponential growth is based on it. The purpose of the present section is to prove this lemma. The main tool we use
to obtain the desired result is the iteration (\ref{eq:hrhokit}) and the estimate (\ref{eq:hBlkhBlkapp}). As a consequence, it is not 
surprising that the matrix $J_{\ron}$, introduced in (\ref{eq:hBlkappJrondef}), plays an important role. In the analysis to follow, it 
turns out to be convenient to divide $J_{\ron}$ into a part consisting of the Jordan blocks of $J_{\ron}$ with zero eigenvalue, say $J_{a}$, 
and a part consisting of the Jordan blocks of $J_{\ron}$ with strictly negative eigenvalue, say $J_{b}$. We can then, without loss of generality, 
assume that $J_{\ron}=\mathrm{diag}\{J_{a},J_{b}\}$. Let $l_{a}$ and $l_{b}$ be the dimensions of $J_{a}$ and $J_{b}$ respectively (note that $l_{b}$ 
could equal zero, in which case $J_{\ron}=J_{a}$). Before proceeding, it is convenient to introduce the following terminology. 
\begin{definition}\label{def:doddecofma}
If $1\leq l\in\zo$ and $M\in\Mn{l}{\co}$, then the matrices $M_{\rod},M_{\rood}\in\Mn{l}{\co}$ are defined
by the following conditions: $M_{\rod}$ is a diagonal matrix; the diagonal components of $M_{\rood}$ are zero; and $M=M_{\rod}+M_{\rood}$.
\end{definition}
\begin{remark}\label{remark:matriceswithind}
If the matrix $M$ of interest already has one index, say that it is given by $M=J_{\ron}$, then we write $J_{\ron,\rod}:=M_{\rod}$ and 
$J_{\ron,\rood}:=M_{\rood}$ e.g. 
\end{remark}
It is also useful to divide matrices into blocks according to the following definition. 
\begin{definition}\label{def:mataaabetcdecomp}
Let $0\leq j_{a},j_{b}\in\zo$ and $1\leq j\in\zo$ be such that $j=j_{a}+j_{b}$. If $M\in\Mn{j}{\co}$, it is divided into 
blocks $M_{aa}$, $M_{ab}$, $M_{ba}$ and $M_{bb}$ according to the following conditions: $M_{aa}\in\Mn{j_{a}}{\co}$; $M_{ab}$ 
is a complex $j_{a}\times j_{b}$ matrix; $M_{ba}$ is a complex $j_{b}\times j_{a}$ matrix; $M_{bb}\in\Mn{j_{b}}{\co}$; and
\[
M=\left(\begin{array}{cc} M_{aa} & M_{ab} \\ M_{ba} & M_{bb}\end{array}\right).
\]
Moreover, if $\xi\in\cn{j}$, then we denote the vector consisting of the first $j_{a}$ and the vector consisting of the last $j_{b}$ components 
of $\xi$ by $\xi_{a}\in\cn{j_{a}}$ and $\xi_{b}\in\cn{j_{b}}$ respectively.
\end{definition}
\begin{remarks}
When writing $M_{aa}$ etc., the integers $j$, $j_{a}$, $j_{b}$ and the matrix $M$ should be clear from the context. In practice, we are, in the 
present section, interested in the case that $j=2m$, $j_{a}=l_{a}$ and $j_{b}=l_{b}$, where $l_{a}$ and $l_{b}$ are defined immediately prior to
Definition~\ref{def:doddecofma}. Concerning vectors and matrices that already have an index, we use a convention similar to that introduced 
in Remark~\ref{remark:matriceswithind}. 
\end{remarks}

In the analysis below, we need to estimate the following type of sums. 

\begin{lemma}\label{lemma:roughintestub}
Assume that (\ref{eq:thesystemRge}) is oscillation adapted, let $0\neq\indexnot\in\EFindexset$ and define $\mfg$ by (\ref{eq:mfgnutdef}). 
Assume that there is a $t_{0}\geq 0$ such that (\ref{eq:gtalbge}) holds with $t_{a}$ replaced by $t_{0}$. Let the sequence $\{t_{k}\}$ and 
$k_{\fin}$ be given by Definition~\ref{def:tkdefge}. If $0\leq k_{0}\leq k_{1}\leq k_{\fin}$ and $\lambda\in\ro$, it then follows that 
\[
\sum_{k=k_{0}}^{k_{1}}\frac{e^{\lambda t_{k}}}{\mfg(t_{k})}\leq C\int_{t_{k_{0}}}^{t_{k_{1}+1}}e^{\lambda t}dt,
\]
where $C$ only depends on $\lambda$. 
\end{lemma}
\begin{proof}
Due to (\ref{eq:tatbroughge}),
\[
\sum_{k=k_{0}}^{k_{1}}\frac{e^{\lambda t_{k}}}{\mfg(t_{k})}\leq \frac{1}{\pi}\sum_{k=k_{0}}^{k_{1}}\int_{t_{k}}^{t_{k+1}}e^{\lambda t_{k}}dt.
\]
On the other hand, for $t\in [t_{k},t_{k+1}]$, 
\[
e^{\lambda t_{k}}\leq e^{\lambda t}e^{|\lambda (t_{k}-t)|}\leq \exp\left(\frac{4\pi|\lambda|}{\max\{2,4\pi\ellderbd\}}\right)e^{\lambda t}
\leq e^{2\pi|\lambda|}e^{\lambda t},
\]
where we have used (\ref{eq:tatbroughge}) and the fact that $\mfg(t_{k})\geq \max\{2,4\pi\ellderbd\}$. The lemma follows. 
\end{proof}

We are now in a position to prove the main estimate of this section.

\begin{lemma}\label{lemma:supexpgrowthmodeest}
Consider (\ref{eq:thesystemRge}). Assume the associated metric to be such that $(M,g)$ is a canonical separable cosmological model
manifold. Assume, moreover, that $f=0$, $d=1$ and $R=0$. 
Assume that the conditions of Lemma~\ref{lemma:mainassumpubcase} are fulfilled with $\b_{X}>0$ and that Assumptions~\ref{assumption:guooass} and
\ref{assumption:Yoconvub} hold. Finally, assume that $\kappa>0$, where $\kappa$ is the largest real part of an eigenvalue of the 
matrix $R_{\infty}$ defined by (\ref{eq:Rinftydef}). Let $\e>0$. If $\b_{X}-\b_{1}>0$, there are constants
$K_{\roeep,\e}$, $N_{\roeep,\e}$, $T_{0,\e}\geq 0$ and $T_{\romed,\e}\geq T_{0,\e}$, depending only on $\e$ and $\boc_{\roeep}$, where $\boc_{\roeep}$ is
introduced in (\ref{eq:bcroeepdef}), and if $\b_{X}-\b_{1}\leq 0$, there are constants $N_{\roeep,\e}$, $T_{0,\e}\geq 0$ and $T_{\romed,\e}\geq T_{0,\e}$, 
depending only on $\e$ and $\boc_{\roeep}$, such that the following holds. Let $n_{1}\geq N_{\roeep,\e}$ and define $T_{\max,\e}$ by 
\begin{equation}\label{eq:tmaxeeepver}
T_{\max,\e}:=\left\{\begin{array}{cl} T_{\roeep,\e}, & \b_{X}-\b_{1}>0\\ \infty, & \b_{X}-\b_{1}\leq 0\end{array}\right.,
\end{equation}
where 
\begin{equation}\label{eq:Teepedefml}
T_{\roeep,\e}:=\frac{1}{\b_{X}-\b_{1}}\ln|n_{1}|+K_{\roeep,\e}.
\end{equation}
Then $T_{\max,\e}>T_{\romed,\e}$; (\ref{eq:gtalbge}) holds for $0\leq t_{a}<T_{\max,\e}$; and if $\{t_{k}\}$ is the time sequence given by 
Definition~\ref{def:tkdefge} with $t_{0}$ replaced by $T_{0,\e}$, then there are solutions to (\ref{eq:fourierthesystemRge}) (corresponding 
to $f=0$ and frequency $n_{1}$) such that $|w(T_{0,\e})|=1$ and such that
\begin{equation}\label{eq:wsupexpgrowth}
|w(t_{k+1})|\geq \exp\left[\b_{X}^{-1}\kappa(1-\e)e^{\b_{X}t_{k+1}}\right]
\end{equation}
for all $k$ such that $T_{\romed,\e}\leq t_{k}<T_{\max,\e}$; recall that if $z$ is a solution to (\ref{eq:fourierthesystemRge}),
then $w$ is defined by (\ref{eq:xyFodefshiftge}) and (\ref{eq:wdefshiftge}). 
\end{lemma}
\begin{remark}
The constant $K_{\roeep,\e}$ can be chosen in such a way that $t_{\fin}$, given in Definition~\ref{def:tkdefge}, satisfies 
$t_{\fin}\geq T_{\max,\e}$; cf. Remark~\ref{remark:sufconnesequb}. Moreover, since $T_{\romed,\e}$ is fixed (depending only on 
$\e$ and $\boc_{\roeep}$), it is clear that by choosing $n_{1}$ to be large enough, the interval $[T_{\romed,\e},T_{\max,\e}]$
can be assumed to contain a large number of elements of the sequence $\{t_{k}\}$. 
\end{remark}
\begin{remark}
The condition $t\geq T_{0,\e}$ ensures that we are in the asymptotic regime. The time $T_{\romed,\e}$ is chosen so that 
significant growth can take place in the interval $[T_{0,\e},T_{\romed,\e}]$. 
\end{remark}
\begin{remark}
In terms of the solution $z$ to (\ref{eq:fourierthesystemRge}), the estimate (\ref{eq:wsupexpgrowth}) can be written
\begin{equation*}
\begin{split}
 & [|\dot{z}(t_{k+1})|^{2}+\mfg^{2}(t_{k+1})|z(t_{k+1})|^{2}]^{1/2}\\
 \geq & \exp\left[\b_{X}^{-1}\kappa(1-\e)e^{\b_{X}t_{k+1}}\right]
[|\dot{z}(T_{0,\e})|^{2}+\mfg^{2}(T_{0,\e})|z(T_{0,\e})|^{2}]^{1/2},
\end{split}
\end{equation*}
where we have used the fact that $|w(T_{0,\e})|=1$ as well as (\ref{eq:xyFodefshiftge}) and (\ref{eq:wdefshiftge}).
\end{remark}
\begin{proof}
Before turning to the actual proof, it is useful to make a few remarks concerning the constants and parameters appearing in it. 
In the course of the proof, we bound $K_{\roeep,\e}$ from above by conditions that are gradually more restrictive. Similarly, we 
bound $N_{\roeep,\e}$, $T_{0,\e}\geq 0$ and $T_{\romed,\e}\geq T_{0,\e}$ from below by conditions that are gradually more restrictive.
On the other hand, all of these constants depend only on $\e$ and $\boc_{\roeep}$. By demanding that $n_{1}$ be large enough, it 
can thus be ensured that $T_{\max,\e}$ is as much larger than $T_{0,\e}$ and $T_{\romed,\e}$ as we wish. 

Turning to the parameter $\e$, it is convenient to introduce it only at the end of the proof. To begin with, we instead introduce
two parameters $0<\e_{j}\leq 1$, $j=1,2$. In the end, we describe how the $\e_{j}$ are determined by $\e$ and $\boc_{\roeep}$. 

In order to obtain a clearer picture of the restrictions on the constants $K_{\roeep,\e}$ etc., we proceed step by step, and 
conclude each step with a description of the restrictions that need to be imposed. 

\textbf{Step 1, dealing with the presence of non-trivial Jordan blocks.}
If one of the Jordan blocks in $R_{\infty}$ corresponding to an eigenvalue with real part $\kappa$ is non-trivial, it is convenient to 
reformulate the relation (\ref{eq:hrhokit}). The reason for this is that we wish to control the influence of the non-trivial nature
of such Jordan blocks on the evolution. In order to quantify this control, we use the parameter $\e_{1}$. 

Let $0<\e_{1}\leq 1$ and recall that $J_{\ron}$ is defined in (\ref{eq:hBlkappJrondef}). By conjugating $J_{\ron}$ by a suitable 
diagonal matrix, say $T_{1}$, it can be ensured that 
\[
\chJ_{\ron}:=T_{1}J_{\ron}T_{1}^{-1}
\]
is such that 
\begin{equation}\label{eq:chJronroodest}
\|\chJ_{\ron,\rood}\|\leq\e_{1},
\end{equation}
where the terminology $\chJ_{\ron,\rood}$ is introduced in Definition~\ref{def:doddecofma}; cf. Lemma~\ref{lemma:genJordblock} and its 
proof. Note, however, that the norms of $T_{1}$ and 
$T_{1}^{-1}$ only depend on $\e_{1}$. Define the matrices $\chJ_{a}$ and $\chJ_{b}$ by $\chJ_{a}:=\chJ_{\ron,aa}$ and $\chJ_{b}:=\chJ_{\ron,bb}$,
where we use the notation introduced in Definition~\ref{def:mataaabetcdecomp} (with $j_{a}=l_{a}$, $j_{b}=l_{b}$ and $j=2m$). 
Then (\ref{eq:chJronroodest}) yields
\begin{equation}\label{eq:chJaroodchJbroodest}
\|\chJ_{a,\rood}\|\leq \|\chJ_{\ron,\rood}\|\leq\e_{1},\ \ \
\|\chJ_{b,\rood}\|\leq \|\chJ_{\ron,\rood}\|\leq\e_{1}.
\end{equation}
Let $\nu_{\ropar}$ and $C_{\ropar}$ be defined as in Lemma~\ref{lemma:hBkapprfirstimpro}. For $|n_{1}|\geq \nu_{\ropar}$, let $t_{\max}$ be 
defined by (\ref{eq:tmaxdefftry}). Assume that $0\leq t_{0}<t_{\max}$ has been specified (in the present context, $t_{0}$ remains to be determined) 
and let the sequence $\{t_{k}\}$ be given as in Definition~\ref{def:tkdefge}. Then, for $k\geq 0$ and $t_{k}<t_{\max}$ we can appeal to 
(\ref{eq:hBlkhBlkapp}). Moreover, introducing
\begin{equation}\label{eq:chrhokchBkdef}
\chrho_{k}:=T_{1}\hrho_{k},\ \ \
\chB_{k}:=T_{1}\hBl_{k}T_{1}^{-1},\ \ \
\chB_{k,\app}:=T_{1}\hBl_{k,\app}T_{1}^{-1},
\end{equation}
(\ref{eq:hBlkhBlkapp}) yields
\begin{equation}\label{eq:chBkchBkapp}
\left\|\chB_{k}-\chB_{k,\app}\right\|\leq C_{\roeep,\e_{1}}\left(
\frac{e^{(\b_{X}-\eta_{\romn})t_{k}}}{\mfg(t_{k})}+\frac{e^{2\b_{X}t_{k}}}{\mfg^{2}(t_{k})}\right), 
\end{equation}
where $C_{\roeep,\e_{1}}$ only depends on $\boc_{\roeep}$ defined in (\ref{eq:bcroeepdef}) and $\e_{1}$. In addition, 
\begin{equation}\label{eq:chBkappdef}
\chB_{k,\app}=\Id_{2m}+\frac{2\pi e^{\b_{X}t_{k}}}{\mfg(t_{k})}\chJ_{\ron}.
\end{equation}
Finally, (\ref{eq:hrhokit}) yields
\begin{equation}\label{eq:chrhokit}
\chrho_{k+1}=\chB_{k}\chrho_{k}.
\end{equation}
Define $\chR_{k}$ by 
\[
\chB_{k}=\chB_{k,\app}+\chR_{k}.
\]
Then $\|\chR_{k}\|$ is bounded by the right hand side of (\ref{eq:chBkchBkapp}). Using the notation introduced in 
Definition~\ref{def:mataaabetcdecomp} (with $j_{a}=l_{a}$, $j_{b}=l_{b}$ and $j=2m$), (\ref{eq:chrhokit}) can then be written 
\begin{align}
\chrho_{k+1,a} = & \chB_{k,\app,aa}\chrho_{k,a}+(\chR_{k}\chrho_{k})_{a},\label{eq:chrhokait}\\
\chrho_{k+1,b} = & \chB_{k,\app,bb}\chrho_{k,b}+(\chR_{k}\chrho_{k})_{b}.\label{eq:chrhokbit}
\end{align}
\textbf{Restrictions, step 1.} Assume that $T_{0,\e}\geq 0$, $K_{\roeep,\e}\leq C_{\ropar}$ (in case $\b_{X}-\b_{1}>0$) and
$N_{\roeep,\e}\geq\nu_{\ropar}$. Fix $n_{1}$ satisfying $|n_{1}|\geq N_{\roeep,\e}$ and define $T_{\max,\e}$ by (\ref{eq:tmaxeeepver}), 
where $T_{\roeep,\e}$ is given by (\ref{eq:Teepedefml})
in case $\b_{X}-\b_{1}>0$. Construct the sequence $\{t_{k}\}$ as in Definition~\ref{def:tkdefge}, where the $t_{0}$ satisfying 
$0\leq t_{0}<T_{\max,\e}$ is arbitrary. Then the conclusions of step 1 hold for $|n_{1}|\geq N_{\roeep,\e}$ and $t_{0}\leq t_{k}<T_{\max,\e}$. 
Finally, let us remark that if $K_{\roeep,\e}$ and $N_{\roeep,\e}$ are chosen inappropriately, then $T_{\max,\e}$ might be negative. 
However, for a given choice of $K_{\roeep,\e}$ (depending only on the $\e_{j}$'s, $\e$ and $\boc_{\roeep}$), there is a choice of 
$N_{\roeep,\e}\geq\nu_{\ropar}$ with the same dependence, such that $T_{\max,\e}>0$ if $|n_{1}|\geq N_{\roeep,\e}$.

\textbf{Step 2, limiting the time sequence.} In order to be able to proceed, we need to obtain better control of the right
hand side of (\ref{eq:chBkchBkapp}). There are two aspects to this. First, we want $e^{\b_{X}t_{k}}/\mfg(t_{k})$ to be small. 
This can be ensured by appealing to Lemma~\ref{lemma:condvaresmallub}. Second, we want $e^{-\eta_{\romn}t_{k}}$ to be small. This
can be ensured by assuming that $t_{k}$ is sufficiently large. In order to quantify the control, we use the parameter $\e_{2}$.

Let $0<\e_{2}\leq 1$. Then Lemma~\ref{lemma:condvaresmallub} ensures the following. If $\b_{X}-\b_{1}>0$, then there are constants 
$\nu_{\ropar,\e_{2}}$ and $C_{\ropar,\e_{2}}$, depending only on $\boc_{\ropar}$ and $\e_{2}$, such that if $|n_{1}|\geq \nu_{\ropar,\e_{2}}$, then 
(\ref{eq:Troparedef}) holds (with $\e$ replaced by $\e_{2}$), and (\ref{eq:varealtemfglb}) holds for $0\leq t,t_{a}\leq T_{\ropar,\e_{2}}$
(with $\e$ replaced by $\e_{2}$). Similarly, if $\b_{X}-\b_{1}\leq 0$, 
there is a constant $\nu_{\ropar,\e_{2}}$, depending only on $\boc_{\ropar}$ and $\e_{2}$, such that if $|n_{1}|\geq \nu_{\ropar,\e_{2}}$, then 
(\ref{eq:varealtemfglb}) holds for $0\leq t,t_{a}< \infty$ (with $\e$ replaced by $\e_{2}$). Defining 
\begin{equation}\label{eq:tmaxetwo}
t_{\max,\e_{2}}:=\left\{\begin{array}{cl} T_{\ropar,\e_{2}}, & \b_{X}-\b_{1}>0\\ \infty, & \b_{X}-\b_{1}\leq 0\end{array}\right.
\end{equation}
yields the conclusion that (\ref{eq:varealtemfglb}) (with $\e$ replaced by $\e_{2}$) holds for $0\leq t,t_{a}<t_{\max,\e_{2}}$. 
Here we take for granted that $\nu_{\ropar,\e_{2}}\geq \nu_{\ropar}$ and $C_{\ropar,\e_{2}}\leq C_{\ropar}$, where $\nu_{\ropar}$ and $C_{\ropar}$ are
the constants introduced in Lemma~\ref{lemma:hBkapprfirstimpro}. As a consequence, $t_{\max,\e_{2}}\leq t_{\max}$. 
Before turning to specific bounds we wish to impose on $\e_{2}$, let us 
introduce the notation $\kappa_{\ron,\max}$ for the largest 
eigenvalue of $-J_{\ron}$. Let, moreover, $\kappa_{\ron}$ be the smallest strictly positive eigenvalue of $-J_{\ron}$. It is of course 
possible that there is no strictly positive eigenvalue of $-J_{\ron}$. However, this case corresponds to $l_{b}=0$, and only simplifies 
the analysis; we leave it to the reader to modify the analysis presented below in order to deal with the case $l_{b}=0$. In what follows, 
we always assume that 
\begin{equation}\label{eq:etwofirstbd}
\e_{2},\e_{1}\leq \frac{1}{2\pi\max\{\kappa_{\ron,\max},1\}}.
\end{equation}
One advantage of making this assumption is that, if it holds, then
\begin{equation}\label{eq:chJbnormest}
\left\|\Id_{l_{b}}+\frac{2\pi e^{\b_{X}t_{k}}}{\mfg(t_{k})}\chJ_{b}\right\|\leq 1-\frac{2\pi e^{\b_{X}t_{k}}}{\mfg(t_{k})}(\kappa_{\ron}-\e_{1})
\end{equation}
for $t_{0}\leq t_{k}<t_{\max,\e_{2}}$, where we have appealed to (\ref{eq:varealtemfglb}) (with $\e$ replaced by $\e_{2}$) and 
(\ref{eq:chJaroodchJbroodest}). For future reference, it is also of interest to note that 
\begin{equation}\label{eq:chJanormest}
1-\frac{2\pi e^{\b_{X}t_{k}}}{\mfg(t_{k})}\e_{1}\leq
\left|\left(\Id_{l_{a}}+\frac{2\pi e^{\b_{X}t_{k}}}{\mfg(t_{k})}\chJ_{a}\right)\xi\right|\leq 
1+\frac{2\pi e^{\b_{X}t_{k}}}{\mfg(t_{k})}\e_{1}
\end{equation}
for $\xi\in\cn{l_{a}}$ such that $|\xi|=1$. 

Fix $0<\e_{1}\leq 1$  satisfying (\ref{eq:etwofirstbd}), and define $T_{1}$, $\chrho_{k}$ etc. as above. Assume that $0<\e_{2}\leq 1$
satisfies (\ref{eq:etwofirstbd}) and
\begin{equation}\label{eq:eptwosecdbd}
C_{\roeep,\e_{1}}\e_{2}\leq \pi\e_{1},
\end{equation}
where $C_{\roeep,\e_{1}}$ is the constant appearing on the right hand side of (\ref{eq:chBkchBkapp}). Then 
\begin{equation}\label{eq:chRkbdstone}
\|\chR_{k}\|\leq C_{\roeep,\e_{1}}e^{-\eta_{\romn}t_{k}}\frac{e^{\b_{X}t_{k}}}{\mfg(t_{k})}+\pi\e_{1}\frac{e^{\b_{X}t_{k}}}{\mfg(t_{k})}
\end{equation}
for all $k$ such that $t_{0}\leq t_{k}< t_{\max,\e_{2}}$, where we have appealed to (\ref{eq:varealtemfglb}) (with $\e$ replaced by $\e_{2}$)
and the fact that $\|\chR_{k}\|$ is bounded 
by the right hand side of (\ref{eq:chBkchBkapp}). In (\ref{eq:chRkbdstone}), the factor $\pi$ in the second term on the right hand 
side has been inserted for convenience. In order to improve the first 
term on the right hand side of (\ref{eq:chRkbdstone}), it is necessary to impose a lower bound on $t_{k}$. To this end, let 
$t_{\min,\e_{1}}\geq 0$, depending only on $\boc_{\roeep}$ and $\e_{1}$, be such that 
\begin{equation}\label{eq:tminecond}
C_{\roeep,\e_{1}}e^{-\eta_{\romn}t}\leq \pi\e_{1}
\end{equation}
for $t\geq t_{\min,\e_{1}}$, where $C_{\roeep,\e_{1}}$ is the specific constant appearing on the right hand side of (\ref{eq:chRkbdstone}). Construct the 
sequence $\{t_{k}\}$ as in Definition~\ref{def:tkdefge} but with $t_{\min,\e_{1}}\leq t_{0}<t_{\max,\e_{2}}$. Then, for $t_{0}\leq t_{k}<t_{\max,\e_{2}}$, 
\begin{equation}\label{eq:chRkbdsttwo}
\|\chR_{k}\|\leq 2\pi\e_{1}\frac{e^{\b_{X}t_{k}}}{\mfg(t_{k})}.
\end{equation}
\textbf{Restrictions, step 2.} Fix $0<\e_{1}\leq 1$ satisfying (\ref{eq:etwofirstbd}), and fix $0<\e_{2}\leq 1$ satisfying (\ref{eq:etwofirstbd})
and (\ref{eq:eptwosecdbd}). Let $\nu_{\ropar,\e_{2}}\geq \nu_{\ropar}$ and $C_{\ropar,\e_{2}}\leq C_{\ropar}$ be defined as above; cf. the text adjacent
to (\ref{eq:tmaxetwo}). Finally, let $t_{\min,\e_{1}}\geq 0$ be such that 
(\ref{eq:tminecond}) holds for $t\geq t_{\min,\e_{1}}$. Assume that $T_{0,\e}\geq t_{\min,\e_{1}}$, $K_{\roeep,\e}\leq C_{\ropar,\e_{2}}$ (in case 
$\b_{X}-\b_{1}>0$) and $N_{\roeep,\e}\geq\nu_{\ropar,\e_{2}}$.  Fix $n_{1}$ satisfying $|n_{1}|\geq N_{\roeep,\e}$ and define $T_{\max,\e}$ by 
(\ref{eq:tmaxeeepver}), where $T_{\roeep,\e}$ is given by (\ref{eq:Teepedefml}) in case $\b_{X}-\b_{1}>0$. Construct the sequence $\{t_{k}\}$ as 
in Definition~\ref{def:tkdefge}, where $T_{0,\e}\leq t_{0}<T_{\max,\e}$ is arbitrary. Then the conclusions of steps 1 and 2 hold for 
$t_{0}\leq t_{k}<T_{\max,\e}$. Finally, note that if $K_{\roeep,\e}$ and $N_{\roeep,\e}$ are chosen inappropriately, then $T_{\max,\e}$ might 
be less than or equal to $T_{0,\e}$. However, for a given choice of $K_{\roeep,\e}$ and $T_{0,\e}$ (depending only on the $\e_{j}$'s, $\e$ and 
$\boc_{\roeep}$), there is a choice of $N_{\roeep,\e}\geq\nu_{\ropar}$ with the same dependence, such that 
$T_{\max,\e}>T_{0,\e}$ if $|n_{1}|\geq N_{\roeep,\e}$.

\textbf{Step 3, controlling the components of the solutions to the difference equation.} Assume that the restrictions arising
in step 2 are satisfied and let $t_{0}$ etc. satisfy the corresponding conditions. Let us return to 
(\ref{eq:chrhokait}) and (\ref{eq:chrhokbit}). Due to (\ref{eq:chrhokait}),
\begin{equation}\label{eq:chrhokpoalb}
\begin{split}
|\chrho_{k+1,a}| \geq & |\chB_{k,\app,aa}\chrho_{k,a}|-|(\chR_{k}\chrho_{k})_{a}|\\
 \geq & |\chB_{k,\app,aa}\chrho_{k,a}|-\e_{1}\frac{2\pi e^{\b_{X}t_{k}}}{\mfg(t_{k})}|\chrho_{k}|\\
 \geq & |\chrho_{k,a}|-\e_{1}\frac{4\pi e^{\b_{X}t_{k}}}{\mfg(t_{k})}(|\chrho_{k,a}|
+|\chrho_{k,b}|),
\end{split}
\end{equation}
where we have appealed to (\ref{eq:chJanormest}) and (\ref{eq:chRkbdsttwo}); from now on we always tacitly assume that $k$ is such
that $T_{0,\e}\leq t_{0}\leq t_{k}<T_{\max,\e}$. Similarly, (\ref{eq:chrhokbit}) yields
\begin{equation}\label{eq:chrhokpoalbb}
\begin{split}
|\chrho_{k+1,b}| \leq & |\chrho_{k,b}|-\frac{2\pi e^{\b_{X}t_{k}}}{\mfg(t_{k})}(\kappa_{\ron}-\e_{1})|\chrho_{k,b}|
+\e_{1}\frac{2\pi e^{\b_{X}t_{k}}}{\mfg(t_{k})}|\chrho_{k}|\\
 \leq & \left(1-\kappa_{\ron}\frac{2\pi e^{\b_{X}t_{k}}}{\mfg(t_{k})}\right) |\chrho_{k,b}|
+\e_{1}\frac{4\pi e^{\b_{X}t_{k}}}{\mfg(t_{k})}(|\chrho_{k,b}|+|\chrho_{k,a}|),
\end{split}
\end{equation}
where we have appealed to (\ref{eq:chJbnormest}) and (\ref{eq:chRkbdsttwo}). In particular, (\ref{eq:chrhokpoalb}) and (\ref{eq:chrhokpoalbb})  
yield
\begin{align}
\frac{|\chrho_{k+1,a}|}{|\chrho_{k,a}|}  \geq & 1-\e_{1}\frac{4\pi e^{\b_{X}t_{k}}}{\mfg(t_{k})}
\left(1+\frac{|\chrho_{k,b}|}{|\chrho_{k,a}|}\right),\label{eq:chrhokaqit}\\
\frac{|\chrho_{k+1,b}|}{|\chrho_{k,a}|}
 \leq & \left(1-\kappa_{\ron}\frac{2\pi e^{\b_{X}t_{k}}}{\mfg(t_{k})}\right)\frac{|\chrho_{k,b}|}{|\chrho_{k,a}|}
+\e_{1}\frac{4\pi e^{\b_{X}t_{k}}}{\mfg(t_{k})}\left(1+\frac{|\chrho_{k,b}|}{|\chrho_{k,a}|}\right).\label{eq:chrhokbqit}
\end{align}
Assuming $\e_{2}$ to be small enough ($\e_{2}\leq c_{0,1}$, where $c_{0,1}>0$ is a numerical constant) and the inequality
\begin{equation}\label{eq:bootstrapchorhokit}
\frac{|\chrho_{k,b}|}{|\chrho_{k,a}|}\leq 1
\end{equation}
to hold, the absolute value of the second term on the right hand side of (\ref{eq:chrhokaqit}) is bounded from above by $1/4$. 
Under these assumptions, (\ref{eq:chrhokaqit}) yields
\begin{equation}\label{eq:chrhokaqitinv}
\frac{|\chrho_{k,a}|}{|\chrho_{k+1,a}|}  \leq  1+\e_{1}\frac{8\pi e^{\b_{X}t_{k}}}{\mfg(t_{k})}
\left(1+\frac{|\chrho_{k,b}|}{|\chrho_{k,a}|}\right).
\end{equation}
Thus, if (\ref{eq:bootstrapchorhokit}) holds and $\e_{2}\leq c_{0,1}$, then (\ref{eq:chrhokbqit}) and (\ref{eq:chrhokaqitinv}) imply that 
\begin{equation}\label{eq:chorhokpoabitfin}
\frac{|\chrho_{k+1,b}|}{|\chrho_{k+1,a}|}
 \leq \left(1-\kappa_{\ron}\frac{2\pi e^{\b_{X}t_{k}}}{\mfg(t_{k})}\right)\frac{|\chrho_{k,b}|}{|\chrho_{k,a}|}
+c_{0,2}\e_{1}\frac{2\pi e^{\b_{X}t_{k}}}{\mfg(t_{k})},
\end{equation}
where $c_{0,2}\geq 1$ is a numerical constant. Let us now assume that $\e_{1}$ has been chosen so that 
\begin{equation}\label{eq:eopetbd}
\e_{1}\leq \frac{1}{c_{0,2}}\min\left\{\frac{\kappa_{\ron}}{4},\frac{1}{2}\right\}.
\end{equation}
It is of interest to draw the following two conclusions from (\ref{eq:chorhokpoabitfin}), given that (\ref{eq:eopetbd}) is fulfilled. 
To begin with, 
\[
\frac{|\chrho_{k+1,b}|}{|\chrho_{k+1,a}|}\leq \frac{|\chrho_{k,b}|}{|\chrho_{k,a}|}+\frac{1}{2},
\]
where we have used the fact that the first factor in the first term on the right hand side of (\ref{eq:chorhokpoabitfin}) is $\leq 1$;
the fact that (\ref{eq:etwofirstbd}) holds; and the fact that (\ref{eq:varealtemfglb}) (with $\e$ replaced by $\e_{2}$) holds for 
$t=t_{k}$. In particular, we obtain the following two conclusions. 

\textit{Small quotient initially.} If $\e_{2}\leq c_{0,1}$, $|\chrho_{k,b}|/|\chrho_{k,a}|\leq 1/2$ and 
(\ref{eq:eopetbd}) holds, then $|\chrho_{k+1,b}|/|\chrho_{k+1,a}|\leq 1$. In other words, $|\chrho_{k,b}|/|\chrho_{k,a}|$ cannot jump
from being less than or equal to $1/2$ to being strictly larger than $1$. 

\textit{Intermediate quotient initially.} Assume now that $\e_{2}\leq c_{0,1}$, $1/2\leq |\chrho_{k,b}|/|\chrho_{k,a}|\leq 1$ and 
(\ref{eq:eopetbd}) hold. Then 
\begin{equation*}
\begin{split}
\frac{|\chrho_{k+1,b}|}{|\chrho_{k+1,a}|}
 \leq & \left(1-\kappa_{\ron}\frac{2\pi e^{\b_{X}t_{k}}}{\mfg(t_{k})}\right)\frac{|\chrho_{k,b}|}{|\chrho_{k,a}|}
+\frac{\kappa_{\ron}}{4}\frac{2\pi e^{\b_{X}t_{k}}}{\mfg(t_{k})}\\
 \leq & \left(1-\kappa_{\ron}\frac{2\pi e^{\b_{X}t_{k}}}{\mfg(t_{k})}\right)\frac{|\chrho_{k,b}|}{|\chrho_{k,a}|}
+\frac{\kappa_{\ron}}{2}\frac{2\pi e^{\b_{X}t_{k}}}{\mfg(t_{k})}\frac{|\chrho_{k,b}|}{|\chrho_{k,a}|}<\frac{|\chrho_{k,b}|}{|\chrho_{k,a}|},
\end{split}
\end{equation*}
where we used (\ref{eq:chorhokpoabitfin}) and (\ref{eq:eopetbd}) in the first step and $|\chrho_{k,b}|/|\chrho_{k,a}|\geq 1/2$ in the 
second step. In other words, if $1/2\leq |\chrho_{k,b}|/|\chrho_{k,a}|\leq 1$, then $|\chrho_{k+1,b}|/|\chrho_{k+1,a}|<|\chrho_{k,b}|/|\chrho_{k,a}|$.

Combining these two observations, we obtain the following conclusion. If $\e_{2}\leq c_{0,1}$, (\ref{eq:eopetbd}) and
$|\chrho_{0,b}|/|\chrho_{0,a}|\leq 1/2$ hold, then $|\chrho_{k,b}|/|\chrho_{k,a}|\leq 1$ for all $k\geq 0$ such that $t_{k}<T_{\max,\e}$. 

\textbf{Restrictions, step 3.} Fix $0<\e_{1}\leq 1$ satisfying (\ref{eq:etwofirstbd}) and (\ref{eq:eopetbd}), and fix $0<\e_{2}\leq 1$ 
satisfying $\e_{2}\leq c_{0,1}$, (\ref{eq:etwofirstbd}) and (\ref{eq:eptwosecdbd}). Given these parameters, impose restrictions as in 
step 2. Then the conclusions of steps 1, 2 and 3 hold. Moreover, if $|\chrho_{0,b}|/|\chrho_{0,a}|\leq 1/2$ holds, then 
$|\chrho_{k,b}|/|\chrho_{k,a}|\leq 1$ for all $k\geq 0$ such that $t_{k}<T_{\max,\e}$. 

\textbf{Step 4, growth estimate.}
Assume that the restrictions arising in step 3 are satisfied and let $t_{0}$ etc. satisfy the corresponding conditions. 
Let us return to (\ref{eq:chrhokpoalb}). Given that $|\chrho_{k,b}|/|\chrho_{k,a}|\leq 1$, this estimate yields
\begin{equation}\label{eq:chrhokpoaabsitfinal}
|\chrho_{k+1,a}| \geq \left(1-\e_{1}\frac{8\pi e^{\b_{X}t_{k}}}{\mfg(t_{k})}\right)|\chrho_{k,a}|
\end{equation}
for $k\geq 0$ such that $t_{k}<T_{\max,\e}$. Note that when $l_{b}=0$, the estimate (\ref{eq:chrhokpoaabsitfinal}) is an immediate
consequence of (\ref{eq:chrhokpoalb}); the issue of the relative sizes of $|\chrho_{k,b}|$ and $|\chrho_{k,a}|$ does not arise in 
that case (needless to say). 

Before proceeding, note that $\ln(1-x)+2x\geq 0$ for $0\leq x\leq 1/2$. Assuming 
$\e_{1}$ to be such that $\e_{1}\leq 1/8$, (\ref{eq:etwofirstbd}) yields
\begin{equation}\label{eq:assumpeonefin}
\e_{1}\frac{8\pi e^{\b_{X}t_{k}}}{\mfg(t_{k})}\leq \frac{1}{2}.
\end{equation}
Combining these observations yields
\begin{equation*}
\begin{split}
 & \ln\left[\left(1-\e_{1}\frac{8\pi e^{\b_{X}t_{k}}}{\mfg(t_{k})}\right)\cdots
\left(1-\e_{1}\frac{8\pi e^{\b_{X}t_{0}}}{\mfg(t_{0})}\right)\right]\\
 \geq & -\e_{1}\sum_{l=0}^{k}\frac{16\pi e^{\b_{X}t_{l}}}{\mfg(t_{l})}
 \geq -C_{X}\e_{1}e^{\b_{X}t_{k+1}},
\end{split}
\end{equation*}
where we have appealed to Lemma~\ref{lemma:roughintestub}, and the constant $C_{X}$ only depends on $\b_{X}$. Combining this
estimate with (\ref{eq:chrhokpoaabsitfinal}) yields
\[
|\chrho_{k+1,a}|\geq \exp\left[-C_{X}\e_{1}e^{\b_{X}t_{k+1}}\right]|\chrho_{0,a}|,
\]
assuming that $|\chrho_{l,b}|/|\chrho_{l,a}|\leq 1$ for $l=0,\dots,k$. 
Choosing $\chrho_{0,a}$ and $\chrho_{0,b}$ so that $|\chrho_{0,a}|>0$ and $\chrho_{0,b}=0$, it is then clear that 
$|\chrho_{k,b}|/|\chrho_{k,a}|\leq 1$ for all $k\geq 0$ such that $t_{k}<T_{\max,\e}$. Then
\[
|\chrho_{k+1}|\geq \exp\left[-C_{X}\e_{1}e^{\b_{X}t_{k+1}}\right]|\chrho_{0}|
\]
for all $k\geq 0$ such that $t_{k}<T_{\max,\e}$. Returning to the definition of $\chrho_{k}$ in terms of $w$ (cf., in particular,
(\ref{eq:wkpredef}), (\ref{eq:wfinkdef}), (\ref{eq:rhokBkdefub}), (\ref{eq:hrhokhBkdefub})
and (\ref{eq:chrhokchBkdef})), this estimate yields
\begin{equation}\label{eq:wtkpogrowth}
\begin{split}
|w(t_{k+1})|\geq C_{\roeep,\e_{1}}\exp\left[\b_{X}^{-1}\kappa (e^{\b_{X}t_{k+1}}
-e^{\b_{X}t_{0}})-C_{X}\e_{1}e^{\b_{X}t_{k+1}}\right]|w(t_{0})|
\end{split}
\end{equation}
for all $k\geq 0$ such that $t_{k}<T_{\max,\e}$, where $C_{\roeep,\e_{1}}>0$ only depends on $\e_{1}$ and $\boc_{\roeep}$. Note, moreover, that
\[
|w(t_{k})|=[|\dot{z}(t_{k})|^{2}+\mfg^{2}(t_{k})|z(t_{k})|^{2}]^{1/2}.
\]
\textbf{Restrictions, step 4.} Fix $0<\e_{1}\leq 1/8$ satisfying (\ref{eq:etwofirstbd}) and (\ref{eq:eopetbd}), and 
fix $0<\e_{2}\leq 1$ satisfying $\e_{2}\leq c_{0,1}$, (\ref{eq:etwofirstbd}) and (\ref{eq:eptwosecdbd}). Given these parameters, impose restrictions 
as in step 2. Assume, moreover, that $|\chrho_{0,b}|=0$ and $|\chrho_{0,a}|>0$. Then the conclusions of steps 1-4 hold. 

\textbf{Step 5, summing up.} Let $0<\e\leq 1$. Let, moreover, $C_{X}$ be the constant appearing in (\ref{eq:wtkpogrowth}) (depending only on 
$\b_{X}$). Let $\e_{1}>0$ be the largest constant satisfying $\e_{1}\leq 1/8$, (\ref{eq:etwofirstbd}), (\ref{eq:eopetbd}) and 
\begin{equation}\label{eq:eoecond}
C_{X}\e_{1}\leq \frac{\e}{2}\b_{X}^{-1}\kappa.
\end{equation}
Recall that $c_{0,2}$ appearing in (\ref{eq:eopetbd}) is a numerical constant and that $\e_{1}\leq 1/8$ ensures that (\ref{eq:assumpeonefin})
holds. In particular, it is thus clear that $\e_{1}>0$ only depends on $\e$ and $\boc_{\roeep}$. Next, fix $0<\e_{2}\leq 1$ to be the largest 
constant such that (\ref{eq:etwofirstbd}) and (\ref{eq:eptwosecdbd}) hold, and such that $\e\leq c_{0,1}$, where $c_{0,1}$ is the 
numerical constant introduced in connection with (\ref{eq:bootstrapchorhokit}). Again, $\e_{2}>0$ only 
depends on $\e$ and $\boc_{\roeep}$. Let $\nu_{\ropar,\e_{2}}\geq\nu_{\ropar}$ and $C_{\ropar,\e_{2}}\leq C_{\ropar}$ be defined as in the text 
adjacent to (\ref{eq:tmaxetwo}),
and define $t_{\min,\e_{1}}$ to be the smallest non-negative real number such that (\ref{eq:tminecond}) holds for $t\geq t_{\min,\e_{1}}$. 
Then $\nu_{\ropar,\e_{2}}$, $C_{\ropar,\e_{2}}$ and $t_{\min,\e_{1}}$ only depend on $\e$ and $\boc_{\roeep}$. Let $K_{\roeep,\e}:=C_{\ropar,\e_{2}}$ (in case 
$\b_{X}-\b_{1}>0$) and $T_{0,\e}:=t_{\min,\e_{1}}$. Choose, moreover, $T_{\romed,\e}$ to be the smallest number such that $T_{\romed,\e}\geq T_{0,\e}$
and such that 
\begin{equation}\label{eq:Tmededefcond}
-\ln C_{\roeep,\e_{1}}+\b_{X}^{-1}\kappa e^{\b_{X}T_{0,\e}}\leq \frac{\e}{2}\b_{X}^{-1}\kappa e^{\b_{X}t}
\end{equation}
for $t\geq T_{\romed,\e}$, where $C_{\roeep,\e_{1}}$ is the constant appearing on the right hand side of (\ref{eq:wtkpogrowth}). Note that 
$T_{\romed,\e}$ only depends on $\e$ and $\boc_{\roeep}$. Finally, let $N_{\roeep,\e}$ be the smallest 
number which is larger than $\nu_{\ropar,\e_{2}}$ and such that if $|n_{1}|\geq N_{\roeep,\e}$, then $T_{\max,\e}$, defined by (\ref{eq:tmaxeeepver}) 
(where $T_{\roeep,\e}$ is given by (\ref{eq:Teepedefml}) in case $\b_{X}-\b_{1}>0$), is such that $T_{\romed,\e}+2\pi\leq T_{\max,\e}$. Then
$N_{\roeep,\e}$ only depends on $\e$ and $\boc_{\roeep}$. Finally, let $t_{0}:=T_{0,\e}$ and construct the sequence $\{t_{k}\}$ as in 
Definition~\ref{def:tkdefge}. 

Consider a solution $z$ to (\ref{eq:fourierthesystemRge}) such that $|\chrho_{0,b}|=0$ and $|\chrho_{0,a}|>0$. Then, since $\e_{1}$ and
$\e_{2}$ satisfy the restrictions imposed in step 4 and $T_{0,\e}$, $K_{\roeep}$ and $N_{\roeep,\e}$ satisfy the restrictions imposed
in step 2, the conclusions of steps 1-4 hold. 
If $T_{\romed,\e}\leq t_{k}<T_{\max,\e}$, then (\ref{eq:wtkpogrowth}), (\ref{eq:eoecond}) and (\ref{eq:Tmededefcond}) yield
\[
|w(t_{k+1})|\geq \exp\left[\b_{X}^{-1}\kappa(1-\e)e^{\b_{X}t_{k+1}}\right]|w(t_{0})|.
\]
The lemma follows.
\end{proof}

\section{Lower bounds on the growth of the energy}

In the present section, we construct solutions to (\ref{eq:thesystemRge}) exhibiting super exponential growth, given that the conditions 
stated in Lemma~\ref{lemma:mainassumpubcase} and Assumptions~\ref{assumption:guooass} and \ref{assumption:Yoconvub} are satisfied. 

\begin{lemma}\label{lemma:supexpinstab}
Consider (\ref{eq:thesystemRge}). Assume the associated metric to be such that $(M,g)$ is a canonical separable cosmological model
manifold. Assume, moreover, that $f=0$, $d=1$ and $R=0$. 
Assume that the conditions of Lemma~\ref{lemma:mainassumpubcase} are fulfilled with $\b_{X}>0$ and that Assumptions~\ref{assumption:guooass} and
\ref{assumption:Yoconvub} hold. Finally, assume that $\kappa>0$, where $\kappa$ is the largest real part of an eigenvalue of the matrix 
$R_{\infty}$ defined by (\ref{eq:Rinftydef}). Let $\e>0$. Then there is a sequence of smooth $\cn{m}$-valued solutions $v_{l}$ to 
(\ref{eq:thesystemRge}) (with $d=1$, $R=0$ and $f=0$), $1\leq l\in\zo$, and for each $l\geq 1$, there is a time sequence 
$t_{l,k}\rightarrow\infty$ (as $k\rightarrow\infty$) such that for each $s\in\ro$, 
\begin{equation}\label{eq:vlsobconvtozero}
\lim_{l\rightarrow\infty}\left(\|v_{l}(\cdot,0)\|_{(s+1)}+\|\d_{t}v_{l}(\cdot,0)\|_{(s)}\right)=0.
\end{equation}
Moreover, 
\begin{equation}\label{eq:mferohomsupexplb}
\mfe_{\rohom}[v_{l}](t_{l,k})\geq \exp\left[2\b_{X}^{-1}\kappa(1-\e)e^{\b_{X}t_{l,k}}\right],
\end{equation}
where 
\begin{equation}\label{eq:mferohomdeqodef}
\mfe_{\rohom}[u](t):=\frac{1}{2}\is[|u_{t}(x,t)|^{2}+g^{11}(t)|u_{x}(x,t)|^{2}]dx.
\end{equation}
\end{lemma}
\begin{remark}
In (\ref{eq:mferohomdeqodef}), we denote the coordinate on $\so$ by $x$ instead of by $x^{1}$ for the sake of brevity.
\end{remark}
\begin{remark}\label{remark:supexpgrowthlemmatointroprop}
If the assumptions of Proposition~\ref{prop:supexpinstabintro} are satisfied, then the conditions of Lemma~\ref{lemma:mainassumpubcase} are 
fulfilled with $\b_{X}>0$ (even though the requirement (\ref{eq:betalowerbd}) does not appear explicitly in Assumption~\ref{ass:mainassumpubcaseintro}, 
it follows from the remaining assumptions (by increasing $\ellderbd$, if necessary)). That Assumption~\ref{assumption:guooass} holds is an immediate
consequence of this; cf. Remark~\ref{remark:gupoolowandupbdsupexpgrowth}. That Assumption~\ref{assumption:Yoconvub} holds is an immediate consequence
of Assumption~\ref{assumption:Yoconvubintro}. Finally, considering the definition of $R_{\infty}$ (given in (\ref{eq:Rinftydef})), it is clear that 
the $\kappa$ introduced in the statement of Lemma~\ref{lemma:supexpinstab} coincides with the $\kappa$ introduced in the statement of 
Proposition~\ref{prop:supexpinstabintro}. Moreover, $\kappa>0$ due to Assumption~\ref{assumption:Yoconvubintro}. To conclude, if the 
assumptions of Proposition~\ref{prop:supexpinstabintro} are satisfied, then the assumptions of Lemma~\ref{lemma:supexpinstab} are satisfied, so 
that the conclusions of Lemma~\ref{lemma:supexpinstab} hold. 
\end{remark}
\begin{proof}
Considering Lemma~\ref{lemma:supexpgrowthmodeest}, it is clear that there are two cases to consider: $\b_{X}-\b_{1}\leq 0$ and 
$\b_{X}-\b_{1}>0$. 

\textbf{The case $\b_{X}-\b_{1}\leq 0$.} Fix $\e>0$ and apply Lemma~\ref{lemma:supexpgrowthmodeest} with $\e$ replaced by $\e/2$. 
Then there are corresponding constants $T_{0,\e}$, $T_{\romed,\e}$ and $N_{\roeep,\e}$ (depending only on $\boc_{\roeep}$ and $\e$) such
that the following holds. For $n_{1}\geq N_{\roeep,\e}$, there is a solution $z$ to (\ref{eq:fourierthesystemRge}) (corresponding to 
$f=0$ and frequency $n_{1}$) such that 
\begin{equation}\label{eq:wsupexpgrowthconofsolfv}
|w(t_{k+1})|\geq \exp\left[\b_{X}^{-1}\kappa(1-\e/2)e^{\b_{X}t_{k+1}}\right]
\end{equation}
for all $k\geq 0$ such that $t_{k}\geq T_{\romed,\e}$, where $\{t_{k}\}$ is the sequence constructed in Definition~\ref{def:tkdefge} and 
starting at $t_{0}:=T_{0,\e}$. Moreover, $|w(t_{0})|=1$. Consider the solution $u$ to (\ref{eq:thesystemRge}) (where $d=1$, $R=0$ and 
$f=0$) defined by 
\begin{equation}\label{eq:uxotonefre}
u(x,t):=(2\pi)^{-1/2}z(n_{1},t)e^{in_{1}x};
\end{equation}
for the sake of brevity, we here use $x$ instead of $x^{1}$ to denote the variable on $\so$. The reason for including the factor 
$(2\pi)^{-1/2}$ is that we wish the relation (\ref{eq:znutdef}) to hold. Compute
\begin{equation}\label{eq:mferohomutkpolb}
\mfe_{\rohom}[u](t_{k+1})=\frac{1}{2}|w(t_{k+1})|^{2}\geq \frac{1}{2}\exp\left[2\b_{X}^{-1}\kappa(1-\e/2)e^{\b_{X}t_{k+1}}\right]
\end{equation}
for $k\geq 0$ such that $t_{k}\geq T_{\romed,\e}$, where $\mfe_{\rohom}[u]$ is defined in the statement of the lemma. Letting $s_{k}=t_{k+1}$ and 
$v_{l}$ be defined by $v_{l}(x,t):=u(x,t)/l$ yields
\[
\mfe_{\rohom}[v_{l}](s_{k})\geq \frac{1}{2l^{2}}\exp\left[\b_{X}^{-1}\kappa\e e^{\b_{X}s_{k}}\right]
\exp\left[2\b_{X}^{-1}\kappa(1-\e)e^{\b_{X}s_{k}}\right].
\]
For $s_{k}$ large enough (the bound depending only on $\e$, $l$, $\b_{X}$ and $\kappa$), it is thus clear that 
\[
\mfe_{\rohom}[v_{l}](s_{k})\geq \exp\left[2\b_{X}^{-1}\kappa(1-\e)e^{\b_{X}s_{k}}\right]. 
\]
On the other hand, it is clear that $\d_{t}v_{l}(\cdot,t_{0})$ and $v_{l}(\cdot,t_{0})$ converge to zero in any $H^{s}$-norm as 
$l\rightarrow\infty$. By Cauchy stability, the same is true of $\d_{t}v_{l}(\cdot,0)$ and $v_{l}(\cdot,0)$. The lemma follows in
the case that $\b_{X}-\b_{1}\leq 0$. 

\textbf{The case $\b_{X}-\b_{1}> 0$.} Fix $\e>0$ and apply Lemma~\ref{lemma:supexpgrowthmodeest} with 
$\e$ replaced by $\e/2$. Then there are corresponding constants $T_{0,\e}$, $T_{\romed,\e}$, $K_{\roeep,\e}$ and $N_{\roeep,\e}$ 
(depending only on $\boc_{\roeep}$ and $\e$) such that the following holds. For $n_{1}\geq N_{\roeep,\e}$, there is a solution 
$z$ to (\ref{eq:fourierthesystemRge}) (corresponding to $f=0$ and frequency $n_{1}$) such that 
\begin{equation}\label{eq:wsupexpgrowthconofsol}
|w(t_{k+1})|\geq \exp\left[\b_{X}^{-1}\kappa(1-\e/2)e^{\b_{X}t_{k+1}}\right]
\end{equation}
for all $k\geq 0$ such that $T_{\romed,\e}\leq t_{k}<T_{\max,\e}$, where $\{t_{k}\}$ is the sequence constructed in 
Definition~\ref{def:tkdefge} and starting at $t_{0}:=T_{0,\e}$. Moreover, $|w(t_{0})|=1$. Note that 
\[
T_{\max,\e}=\frac{1}{\b_{X}-\b_{1}}\ln n_{1}+K_{\roeep,\e}. 
\]
Note that if $t_{k_{\max}}$ is the largest admissible $t_{k}$, then, by definition, $t_{k_{\max}+1}\geq T_{\max,\e}$.
Thus (\ref{eq:wsupexpgrowthconofsol})  yields 
\[
|w(t_{k_{\max}+1})|\geq \exp\left[\mu_{\roeep,\e}n_{1}^{\frac{\b_{X}}{\b_{X}-\b_{1}}}\right]
\exp\left[\b_{X}^{-1}\kappa(1-\e)e^{\b_{X}t_{k_{\max}+1}}\right],
\]
where 
\begin{equation}\label{eq:muroeepedef}
\mu_{\roeep,\e}:=\frac{\e\kappa}{2\b_{X}}e^{\b_{X}K_{\roeep,\e}}>0. 
\end{equation}
Let $u_{n_{1}}$ be the $\cn{m}$-valued solution to (\ref{eq:thesystemRge}) defined by (\ref{eq:uxotonefre}) and note that
\begin{equation}\label{eq:mfevnolb}
\mfe_{\rohom}[u_{n_{1}}](t_{k_{\max}+1})\geq \frac{1}{2}\exp\left[2\mu_{\roeep,\e}n_{1}^{\frac{\b_{X}}{\b_{X}-\b_{1}}}\right]
\exp\left[2\b_{X}^{-1}\kappa(1-\e)e^{\b_{X}t_{k_{\max}+1}}\right]
\end{equation}
and that $\mfe_{\rohom}[u_{n_{1}}](t_{0})=1/2$. Note that $k_{\max}$ depends on $n_{1}$. For the sake of clarity, we therefore write 
$k_{\max}(n_{1})$ in what follows. Define $v$ by 
\begin{equation}\label{eq:vsuexpgrowth}
v(x,t):=\sum_{n_{1}\geq N_{\roeep,\e}}\sqrt{2}\exp\left[-\mu_{\roeep,\e}n_{1}^{\frac{\b_{X}}{\b_{X}-\b_{1}}}\right]u_{n_{1}}(x,t).
\end{equation}
Then $v$ is a smooth $\cn{m}$-valued solution to (\ref{eq:thesystemRge}). In fact, it is more than smooth, as we demonstrate below.  
Due to (\ref{eq:mfevnolb}) it is clear that for each $n_{1}\geq N_{\roeep,\e}$, 
\begin{equation}\label{eq:mfehomvtkmaxnolb}
\mfe_{\rohom}[v](t_{k_{\max}(n_{1})+1})\geq \exp\left[2\b_{X}^{-1}\kappa(1-\e)e^{\b_{X}t_{k_{\max}(n_{1})+1}}\right].
\end{equation}
Since $t_{k_{\max}(n_{1})+1}\rightarrow\infty$ as $n_{1}\rightarrow\infty$, it is clear that there is a sequence $s_{k}\rightarrow\infty$
such that 
\[
\mfe_{\rohom}[v](s_{k})\geq \exp\left[2\b_{X}^{-1}\kappa(1-\e)e^{\b_{X}s_{k}}\right].
\]
Let 
\begin{equation}\label{eq:vldefgevcase}
v_{l}(x,t):=\sum_{n_{1}\geq l,N_{\roeep,\e}}\sqrt{2}\exp\left[-\mu_{\roeep,\e}n_{1}^{\frac{\b_{X}}{\b_{X}-\b_{1}}}\right]u_{n_{1}}(x,t).
\end{equation}
Then $\d_{t}v_{l}(\cdot,t_{0})$ and $v_{l}(\cdot,t_{0})$ converge to zero in any $H^{s}$-norm as $l\rightarrow\infty$. By Cauchy
stability, the same is true of $\d_{t}v_{l}(\cdot,0)$ and $v_{l}(\cdot,0)$. Finally, there is a time sequence $t_{l,k}$ with the 
desired properties. The lemma follows. 
\end{proof}

\section{Super exponential growth of solutions}

In this chapter we have, up till now, focused on the equation (\ref{eq:thesystemRge}) in the case that $d=1$, $R=0$ 
and $f=0$. However, the results we have obtained yield conclusions for more general equations. We turn to this topic
next. 

\subsection{Density of solutions yielding super exponential growth}

Let us start by proving that the initial data that yield solutions with super exponential growth are dense. 

\begin{prop}\label{prop:supexpgrowthgencase}
Consider (\ref{eq:thesystemRge}). Assume the associated metric to be such that $(M,g)$ is a canonical separable cosmological model
manifold. Assume, moreover, that there is a $j\in \{1,\dots,d\}$ such that 
\begin{equation}\label{eq:urestrtojdep}
u_{tt}-g^{jj}(t)\d_{j}^{2}u-2g^{0j}(t)\d_{t}\d_{j}u+\a(t)u_{t}+X^{j}(t)\d_{j}u+\zeta(t)u=0
\end{equation}
(no summation on $j$) satisfies the conditions of Lemma~\ref{lemma:supexpinstab}. Let $\e>0$. For every $\cn{m}$-valued solution $u$ to 
(\ref{eq:thesystemRge}), there is then a sequence of $\cn{m}$-valued solutions $u_{l}$, $1\leq l\in\zo$, such that the initial data of 
$u_{l}-u$ at $t=0$ converge to zero with respect to every Sobolev norm and such that for every $1\leq l\in\zo$ there is 
a sequence $t_{l,k}\rightarrow\infty$ (as $k\rightarrow\infty$) such that 
\begin{equation}\label{eq:mferohomsupexplbsv}
\mfe_{\rohom}[u_{l}](t_{l,k})\geq \exp\left[2\b_{X}^{-1}\kappa(1-\e)e^{\b_{X}t_{l,k}}\right],
\end{equation}
where $\mfe_{\rohom}$ is given by (\ref{eq:mfehomdefintro}).
\end{prop}
\begin{remark}
Since the initial data of $u_{l}-u$ at $t=0$ converge to zero with respect to every Sobolev norm, it is clear that they also converge
to zero in the $C^{\infty}$ topology. 
\end{remark}
\begin{remark}
Due to the fact that $(M,g)$ is a canonical separable cosmological model manifold, the metric arising from (\ref{eq:urestrtojdep})
also corresponds to a canonical separable cosmological model manifold. The reason for this is the following. Let $h^{00}(t)=-1$, 
$h^{01}(t)=g^{0j}(t)$ and $h^{11}(t)=g^{jj}(t)$ (no summation on $j$). Then $h^{11}$ is strictly positive by assumption, so that the 
matrix with components $h^{\a\b}(t)$ has an inverse with components $h_{\a\b}(t)$. Moreover, $h_{00}<0$, $h_{11}>0$ and $h_{\a\b}$
defines a Lorentz metric $h$ on $M_{j}:=\so\times I$. The justification of these statements is given at the beginning of 
Section~\ref{section:basgeomglobhyp}. Finally, that $(M_{j},h)$ is a canonical separable cosmological model manifold follows from 
the fact that $I$ contains $[0,\infty)$ and the fact that $N=1$; the latter statement is a consequence of Lemma~\ref{lemma:invitoshiftetc}, 
Remark~\ref{remark:huzzformgenerallapse} and the fact that $h^{00}=-1$.
\end{remark}
\begin{remark}
Using (\ref{eq:ldruvparsival}), it can be verified that 
\begin{equation}\label{eq:mfehomfourchar}
\mfe_{\rohom}[u](t)=\frac{1}{2}\sum_{\indexnot\in\EFindexset}\left(|\dot{z}(\indexnot,t)|^{2}+\mfg^{2}(\indexnot,t)|z(\indexnot,t)|^{2}\right).
\end{equation}
\end{remark}
\begin{proof}
Let $u$ be a $\cn{m}$-valued solution to (\ref{eq:thesystemRge}). There are two possibilities. Either there is a time sequence 
$t_{k}\rightarrow\infty$ such that (\ref{eq:mferohomsupexplbsv}) holds with $u_{l}$ replaced by $u$ and $t_{l,k}$ replaced by $t_{k}$.
In that case we can choose $u_{l}=u$ for all $1\leq l\in\zo$. The second possibility is that there is a $T>0$ such that 
\begin{equation}\label{eq:mfehomuuppbd}
\mfe_{\rohom}[u](t)\leq \exp\left[2\b_{X}^{-1}\kappa(1-\e)e^{\b_{X}t}\right]
\end{equation}
for all $t\geq T$. In this case we apply Lemma~\ref{lemma:supexpinstab}, with $\e$ replaced by $\e/2$, to the equation (\ref{eq:urestrtojdep}). 
This leads to a sequence of solutions $v_{l}$ to (\ref{eq:urestrtojdep}) such that the initial data of $v_{l}$ at $t=0$ converge to zero with
respect to any Sobolev norm and such that (\ref{eq:mferohomsupexplb}) holds with $\e$ replaced by $\e/2$. We can also, without loss of generality,
assume that the sequence $t_{l,k}$ to be such that $t_{l,k}\geq T$ for all $1\leq l,k\in\zo$. Then
\begin{equation*}
\begin{split}
\mfe_{\rohom}^{1/2}[u+v_{l}](t_{l,k}) \geq &  \mfe_{\rohom}^{1/2}[v_{l}](t_{l,k})-\mfe_{\rohom}^{1/2}[u](t_{l,k})\\
 \geq & \exp\left[\b_{X}^{-1}\kappa(1-\e/2)e^{\b_{X}t_{l,k}}\right]-\exp\left[\b_{X}^{-1}\kappa(1-\e)e^{\b_{X}t_{l,k}}\right].
\end{split}
\end{equation*}
Thus $u_{l}:=u+v_{l}$ has the desired properties, and the proposition follows. 
\end{proof}

\subsection{Genericity of solutions yielding super exponential growth}\label{ssection:genexpgrowth}

Consider an equation satisfying the conditions of Proposition~\ref{prop:supexpgrowthgencase}. Then the set of initial data leading to solutions 
that exhibit super exponential growth is dense. However, there are also classes of equations satisfying the conditions of 
Proposition~\ref{prop:supexpgrowthgencase} such that the set of initial data leading to solutions that exhibit super exponential \textit{decay} 
is dense; cf. Section~\ref{section:modeanalysis} below. It is therefore of interest to ask what the generic behaviour of solutions is. To address 
this question, we first have to decide on a
notion of genericity. One way is to give a measure theoretic definition. However, that is not so natural in the present situation, since the 
space of initial data is infinite dimensional. Another option would be to define a set to be generic if it is open and dense. However, in the 
present setting we are interested in distinguishing between two sets which are both (potentially) dense. For these reasons, we here use the 
following definition of genericity: a set of initial data is \textit{generic} if it is a countable
intersection of open and dense sets with respect to the $C^{\infty}$ topology. Clearly, one could also consider other function spaces, for instance
Sobolev spaces. However, we here only state the results in the smooth setting.

\begin{prop}\label{prop:supexpgrowthgencaseden}
Consider (\ref{eq:thesystemRge}). Assume the associated metric to be such that $(M,g)$ is a canonical separable cosmological model
manifold. Assume, moreover, that there is a $j\in \{1,\dots,d\}$ such that (\ref{eq:urestrtojdep}) (no summation on $j$) satisfies the 
conditions of Lemma~\ref{lemma:supexpinstab}. Let $\e>0$. Then 
there is a set of smooth $\cn{m}$-valued initial data to (\ref{eq:thesystemRge}), say $\ma$, with the following properties. First, $\ma$ 
is the intersection of a countable number of open and dense sets (with respect to the $C^{\infty}$ topology). Second, each element of $\ma$ 
corresponds to a $\cn{m}$-valued solution $u$ to (\ref{eq:thesystemRge}) such that there is a time sequence $\{t_{k}\}$, $1\leq k\in\zo$, 
with the properties that $t_{k}\rightarrow\infty$ and 
\begin{equation}\label{eq:mferohomsupexplbthv}
\mfe_{\rohom}[u](t_{k})\geq \exp\left[2\b_{X}^{-1}\kappa(1-\e)e^{\b_{X}t_{k}}\right]
\end{equation}
for all $1\leq k\in\zo$, where $\mfe_{\rohom}[u](t)$ is defined by (\ref{eq:mfehomdefintro}). 
\end{prop}
\begin{remark}
Due to this statement, it is clear that the generic behaviour of solutions is to exhibit super exponential growth. 
\end{remark}
\begin{remark}
Due to the proof, the sets $\ma_{N}$ whose intersection yield $\ma$ are open with respect to any Sobolev space topology on the 
set of initial data. Moreover, the density is demonstrated by perturbing the initial data by a sequence of functions converging to 
zero in the $C^{\infty}$-topology. It should thus not be a problem to prove genericity in Sobolev-type regularity classes of initial
data. 
\end{remark}
\begin{proof}
The goal of the proof is to construct the set $\ma$. We do so by defining sets $\ma_{N}$, $N_{0}\leq N\in\zo$, with the following properties. 
First, $\ma$ is the intersection of the $\ma_{N}$. Second, the $\ma_{N}$ are open and dense (with respect to the $C^{\infty}$ topology). 
Third, to each solution $u$ corresponding to initial data in $\ma$ there is a corresponding time sequence $t_{k}\rightarrow\infty$ such
that (\ref{eq:mferohomsupexplbthv}) holds. The definition of $\ma_{N}$ depends on two parameters $\e$ and $N_{0}$ and a sequence $\{T_{k}\}$. 
The parameter $\e$ is fixed in the statement of the proposition. However, the definitions of $N_{0}$ and the sequence $T_{k}$ depend on 
whether $\b_{X}-\b_{1}\leq 0$ or $\b_{X}-\b_{1}>0$. On the other hand, it is convenient to first give a general definition of $\ma_{N}$ and 
only later define $N_{0}$ and the sequence $T_{k}$. 

\textbf{Definition of $\ma_{N}$.}
Fix $0<\e\leq 1$ and $1\leq N_{0}\in\zo$. Let $0<T_{k}\in\ro$, $N_{0}\leq k\in\zo$, be a sequence tending to $\infty$ as $k\rightarrow\infty$. 
For $N_{0}\leq N\in\zo$, then
$\xi\in\ma_{N}$ if and only if the following holds. First $\xi$ are smooth $\cn{m}$-valued initial data to (\ref{eq:thesystemRge}).
Second, if $u$ is the solution corresponding to $\xi$ and $z(\indexnot,t)$ is defined by (\ref{eq:znutdef}), then there is a
$k\geq N$ such that 
\begin{equation}\label{eq:zinotlbTk}
|\dot{z}(\indexnot_{k},T_{k})|^{2}+\mfg^{2}(\indexnot_{k},T_{k})|z(\indexnot_{k},T_{k})|^{2}>
2\exp\left[2\b_{X}^{-1}\kappa(1-\e)e^{\b_{X}T_{k}}\right].
\end{equation}
Here $\indexnot_{k}$ is defined by the conditions that the $j$'th component of $\indexnot_{k}$ equals $k$ and all the remaining components 
equal zero.

Let us first prove that the $\ma_{N}$ are open sets. Let $u$ be a solution corresponding to initial data, say $\xi$, in $\ma_{N}$. 
Then there is a $N\leq k\in\zo$ as above. In particular, (\ref{eq:zinotlbTk}) holds. Since the condition (\ref{eq:zinotlbTk}) is 
open, there is an open neighbourhood, say $U$, of $\xi$ such that solutions corresponding to initial data in $U$ also have the property that 
(\ref{eq:zinotlbTk}) holds. Note that $U$ is open with respect to any choice of Sobolev regularity on the initial data, as well
as with respect to the $C^{\infty}$-topology. Assume now that $\xi$ is in the intersection of all the $\ma_{N}$ for $N_{0}\leq N\in\zo$. For each
such $N$, let $k=k(N)$ be as above and let $t_{N}:=T_{k(N)}$. Since $k(N)\geq N$, it is clear that $t_{N}\rightarrow\infty$. Moreover, due
to (\ref{eq:zinotlbTk}) and the fact that (\ref{eq:mfehomfourchar}) holds, (\ref{eq:mferohomsupexplbthv}) is satisfied for all $k$. What remains
to be done is to prove that the $\ma_{N}$ are dense for a suitable choice of $N_{0}$ and $\{T_{k}\}$. However, to achieve this goal, we need to 
divide the analysis into the two cases $\b_{X}-\b_{1}>0$ and $\b_{X}-\b_{1}\leq 0$, just as in the proof of Lemma~\ref{lemma:supexpinstab}.

\textbf{The case $\b_{X}-\b_{1}\leq 0$.} Fix $\e>0$ and apply Lemma~\ref{lemma:supexpgrowthmodeest} with $\e$ replaced by $\e/2$ to the 
equation (\ref{eq:urestrtojdep}). Then there are corresponding constants $T_{0,\e}$, $T_{\romed,\e}$ and $N_{\roeep,\e}$ (depending only on 
$\boc_{\roeep}$ and $\e$) such that the conclusions stated in the proof of Lemma~\ref{lemma:supexpinstab} (concerning the case $\b_{X}-\b_{1}\leq 0$)
hold. Let $N_{0}$ be the smallest strictly positive integer which is larger than $N_{\roeep,\e}$. 
Define $T_{l}$, $N_{0}\leq l\in\zo$, such that $T_{l}\geq \ln l+1$ and that $T_{l}$ equals $t_{k+1}$ for some $k$ with the property that 
$t_{k}\geq T_{\romed,\e}$ (where $t_{k}$ is defined as in the proof of Lemma~\ref{lemma:supexpinstab} in the case that $n_{1}=l$). Then 
$T_{l}\rightarrow\infty$ as $l\rightarrow\infty$. Defining $\ma_{N}$ as above, with this choice of $\{T_{l}\}$, we need to prove that $\ma_{N}$ is 
dense. Define $u_{n_{1}}$ to be the right hand side of (\ref{eq:uxotonefre}). Then 
\[
\mfe_{\rohom}[u_{l}](T_{l})\geq \frac{1}{2}\exp\left[2\b_{X}^{-1}\kappa(1-\e/2)e^{\b_{X}T_{l}}\right]
\]
and $\mfe_{\rohom}[u_{l}](t_{0})=1/2$. Define $v_{l}$ by 
\[
v_{l}:=\sqrt{2}\exp\left[-\b_{X}^{-1}\kappa\e e^{\b_{X}T_{l}}/4\right]u_{l}.
\]
Then
\[
\mfe_{\rohom}[v_{l}](t_{0})\leq \exp\left[-\b_{X}^{-1}\kappa\e e^{\b_{X}T_{l}}/2\right].
\]
Since $T_{l}\geq\ln l$, this implies that every Sobolev norm of the initial data of $v_{l}$ induced at $t_{0}$ converges to zero as $l\rightarrow
\infty$ (and, due to Cauchy stability, the same is true with $t_{0}$ replaced by $0$). Fix $N\geq N_{0}$ and let $u$ be a solution to 
(\ref{eq:thesystemRge}). Let $z$ be defined by (\ref{eq:znutdef}). We need to prove that the initial data of $u$ at $t=0$, say $\xi$, are in the 
closure of $\ma_{N}$. If $\xi\in\ma_{N}$, there is nothing to prove, so let us assume that $\xi\notin\ma_{N}$. Then for all $k\geq N$, 
\[
|\dot{z}(\indexnot_{k},T_{k})|^{2}+\mfg^{2}(\indexnot_{k},T_{k})|z(\indexnot_{k},T_{k})|^{2}
\leq 2\exp\left[2\b_{X}^{-1}\kappa(1-\e)e^{\b_{X}T_{k}}\right]. 
\]
If $z_{k}(\indexnot,t)$ is defined by (\ref{eq:znutdef}) with $u$ replaced by $u+v_{k}$, we thus have 
\begin{equation}\label{eq:upvNlb}
\begin{split}
 & [|\dot{z}_{k}(\indexnot_{k},T_{k})|^{2}+\mfg^{2}(\indexnot_{k},T_{k})|z_{k}(\indexnot_{k},T_{k})|^{2}]^{1/2} \\
\geq & \sqrt{2}\mfe_{\rohom}^{1/2}[v_{k}](T_{k})-[|\dot{z}(\indexnot_{k},T_{k})|^{2}+\mfg^{2}(\indexnot_{k},T_{k})|z(\indexnot_{k},T_{k})|^{2}]^{1/2}\\
 \geq &  
\sqrt{2}\exp\left[\b_{X}^{-1}\kappa(1-3\e/4)e^{\b_{X}T_{k}}\right]-\sqrt{2}\exp\left[\b_{X}^{-1}\kappa(1-\e)e^{\b_{X}T_{k}}\right].
\end{split}
\end{equation}
Then (\ref{eq:upvNlb}) implies that 
the inequality (\ref{eq:zinotlbTk}) (with $z$ replaced by $z_{k}$) is satisfied 
for $T_{k}$ large enough. Thus the initial data of $u+v_{k}$ at $t=0$, say $\xi_{k}$, belong to $\ma_{N}$ for $k$ large enough. On the other 
hand, $\xi_{k}$ converges to $\xi$ with respect to any Sobolev space topology. This proves that 
$\ma_{N}$ is not only open, but also dense. The proposition follows in case $\b_{X}-\b_{1}\leq 0$.

\textbf{The case $\b_{X}-\b_{1}>0$.} Fix $\e>0$ and apply Lemma~\ref{lemma:supexpgrowthmodeest} with 
$\e$ replaced by $\e/2$. Then there are corresponding constants $T_{0,\e}$, $T_{\romed,\e}$, $K_{\roeep,\e}$ and $N_{\roeep,\e}$ 
(depending only on $\boc_{\roeep}$ and $\e$) such that the conclusions stated in the proof of Lemma~\ref{lemma:supexpinstab} (concerning the 
case $\b_{X}-\b_{1}>0$) hold. Let $N_{0}$ be the smallest strictly positive integer which is larger than $N_{\roeep,\e}$. For
$N_{0}\leq l\in\zo$, define $T_{l}:=t_{k_{\max}(l)+1}$, where the sequence $t_{k}$ as well as $k_{\max}(l)$ are defined as in the proof of 
Lemma~\ref{lemma:supexpinstab} (in the case that $n_{1}=l$). Finally, define $\ma_{N}$ as above and let $u_{n_{1}}$ denote the right hand side of 
(\ref{eq:uxotonefre}). Then the function
\[
v_{l}:=\sqrt{2}\exp(-\mu_{\roeep,\e}l^{\b_{X}/(\b_{X}-\b_{1})}/2)u_{l},
\]
where $\mu_{\roeep,\e}$ is given by (\ref{eq:muroeepedef}), is such that we can finish the proof of density as in the case of $\b_{X}-\b_{1}\leq 0$.
The proposition follows. 
\end{proof}

Even though Proposition~\ref{prop:supexpgrowthgencaseden} does demonstrate the genericity of initial data yielding solutions with 
super exponential growth, the notion of genericity can be questioned; in the finite dimensional setting, a dense $G_{\delta}$ set
need not have full measure. We therefore here complement Proposition~\ref{prop:supexpgrowthgencaseden} with an additional result. 

\begin{prop}\label{prop:supexpgrowthinfcodim}
Consider (\ref{eq:thesystemRge}). Assume the associated metric to be such that $(M,g)$ is a canonical separable cosmological model
manifold. Assume, moreover, that there is a $j\in \{1,\dots,d\}$ such that (\ref{eq:urestrtojdep}) (no summation on $j$) satisfies the 
conditions of Lemma~\ref{lemma:supexpinstab}. Let $\e>0$. Define
$\mb_{\e}$ by the condition that $\xi\in\mb_{\e}$ iff $\xi$ constitutes smooth $\cn{m}$-valued initial data for (\ref{eq:thesystemRge}) 
such that if $u$ is the solution corresponding to $\xi$, then there is a constant $C_{\e}$ (depending on the solution) such that 
\[
\mfe_{\rohom}[u](t)\leq C_{\e}\exp\left[2\b_{X}^{-1}\kappa(1-\e)e^{\b_{X}t}\right]
\]
for all $t\geq 0$. Define $\ma_{\e}$ by the condition that $\xi\in\ma_{\e}$ iff $\xi$ constitutes smooth $\cn{m}$-valued initial data 
for (\ref{eq:thesystemRge}) such that if $u$ is the solution corresponding to $\xi$, then there is a time sequence 
$0\leq t_{k}\rightarrow\infty$ such that 
\begin{equation}\label{eq:maeplbdef}
\mfe_{\rohom}[u](t_{k})\geq \exp\left[2\b_{X}^{-1}\kappa(1-\e/2)e^{\b_{X}t_{k}}\right]
\end{equation}
for all $k$. Then $\ma_{\e}$ and $\mb_{\e}$ are disjoint. Moreover, there is an infinite dimensional (linear) subspace $P_{\e}$ of the 
vector space of smooth $\cn{m}$-valued initial data for (\ref{eq:thesystemRge}) such that if $\xi\in\mb_{\e}$ and $\xi_{\per}\in P_{\e}$, 
then $\xi+\xi_{\per}\in\ma_{\e}$ unless $\xi_{\per}=0$. 
\end{prop}
\begin{remark}
The statement of the proposition can be interpreted as saying that $\mb_{\e}$ has infinite codimension in the set of smooth initial data
for (\ref{eq:thesystemRge}). 
\end{remark}
\begin{remark}
By the vector space of smooth $\cn{m}$-valued initial data for (\ref{eq:thesystemRge}), we here mean the space
$C^{\infty}(\bM,\cn{m})\times C^{\infty}(\bM,\cn{m})$, where $\bM$ is defined by (\ref{eq:bMdef}).
\end{remark}
\begin{proof}
That $\ma_{\e}$ and $\mb_{\e}$ are disjoint is an immediate consequence of the definitions. In order to construct the
subspace $P_{\e}$, let us return to the proof of Lemma~\ref{lemma:supexpinstab}. As in the case of that proof, we need to divide the
analysis into the two cases $\b_{X}-\b_{1}\leq 0$ and $\b_{X}-\b_{1}>0$. 

\textbf{The case $\b_{X}-\b_{1}\leq 0$.} When $\b_{X}-\b_{1}\leq 0$, we proceed as in the proof of Lemma~\ref{lemma:supexpinstab}, but 
with $\e$ replaced by $\e/2$ (in other words, we, to start with, apply Lemma~\ref{lemma:supexpgrowthmodeest} with $\e$ replaced by $\e/4$). 
This yields, in particular, numbers $T_{0,\e}$, $T_{\romed,\e}$, $N_{\roeep,\e}$ etc., depending only on $\boc_{\roeep}$ and $\e$. For $n_{1}\geq N_{\roeep,\e}$, we 
can then define $u_{n_{1}}$ by the right hand side of (\ref{eq:uxotonefre}) (where $x$ should be interpreted as $x^{j}$ in the present
setting). Moreover, (\ref{eq:mferohomutkpolb}) with $\e$ replaced by $\e/2$ holds and yields
\begin{equation}\label{eq:mfehomyetanolb}
\mfe_{\rohom}[u_{n_{1}}](t_{k+1})\geq \frac{1}{2}\exp\left[2\b_{X}^{-1}\kappa(1-\e/4)e^{\b_{X}t_{k+1}}\right]
\end{equation}
for all $k\geq 0$ such that $t_{k}\geq T_{\romed,\e}$ (where the sequence $t_{k}$ is constructed as in the proof of Lemma~\ref{lemma:supexpinstab}). Note, 
however, that the sequence $t_{k}$ depends on $n_{1}$. The initial data corresponding to $u_{n_{1}}$ for different $n_{1}$'s are linearly independent. 
Moreover, if 
\begin{equation}\label{eq:usumexp}
u=\textstyle{\sum}_{n_{1}\geq N_{\roeep,\e}}a_{n_{1}}u_{n_{1}}
\end{equation}
for some complex numbers $a_{n_{1}}$ (which we assume to be such that the sum converges to a smooth function), then 
\begin{equation}\label{eq:mferohomsumform}
\mfe_{\rohom}[u]=\textstyle{\sum}_{n_{1}\geq N_{\roeep,\e}}|a_{n_{1}}|^{2}\mfe_{\rohom}[u_{n_{1}}].
\end{equation}
Let $P_{\e}$ be the closure, in the $C^{\infty}$-topology, of the space of smooth $\cn{m}$-valued initial data spanned by the 
initial data of the $u_{n_{1}}$'s for $n_{1}\geq N_{\roeep,\e}$.
For future reference, it is of interest to note that elements of $P_{\e}$ can be considered to be initial data to 
(\ref{eq:urestrtojdep}). Note also that $P_{\e}$ is infinite dimensional. Let now $\xi\in\mb_{\e}$, $\xi_{\per}\in P_{\e}$,
$v$ be the solution to (\ref{eq:thesystemRge}) corresponding to $\xi$, and let $u$ be the solution to (\ref{eq:urestrtojdep})
corresponding to $\xi_{\per}$. Then $v+u$ is the solution to (\ref{eq:thesystemRge}) corresponding to $\xi+\xi_{\per}$. Moreover,
\begin{equation}\label{eq:mfeupvlb}
\mfe_{\rohom}^{1/2}[v+u]\geq \mfe_{\rohom}^{1/2}[u]-\mfe_{\rohom}^{1/2}[v]
\geq\left(\textstyle{\sum}_{n_{1}\geq N_{\roeep,\e}} |a_{n_{1}}|^{2}\mfe_{\rohom}[u_{n_{1}}]\right)^{1/2}-\mfe_{\rohom}^{1/2}[v],
\end{equation}
where we have expressed $u$ as in (\ref{eq:usumexp}). There are two possibilities. Either all the $a_{n_{1}}$'s vanish, in which case 
there is nothing to prove, or there is one $n_{1}\geq N_{\roeep,\e}$ such that $a_{n_{1}}\neq 0$. Assume that the latter occurs, and let 
$t_{k}$ be the time sequence associated with $n_{1}$ as in the proof of Lemma~\ref{lemma:supexpinstab}. Then (\ref{eq:mfehomyetanolb}) and 
(\ref{eq:mfeupvlb}) imply that if $k\geq 0$ and $t_{k}\geq T_{\romed,\e}$, 
\begin{equation*}
\begin{split}
\mfe_{\rohom}^{1/2}[v+u](t_{k+1}) \geq & |a_{n_{1}}|\mfe_{\rohom}^{1/2}[u_{n_{1}}](t_{k+1})-\mfe_{\rohom}^{1/2}[v](t_{k+1})\\
 \geq & 2^{-1/2}|a_{n_{1}}|\exp\left[\b_{X}^{-1}\kappa(1-\e/4)e^{\b_{X}t_{k+1}}\right]\\
 & -C_{\e}^{1/2}\exp\left[\b_{X}^{-1}\kappa(1-\e)e^{\b_{X}t_{k+1}}\right].
\end{split}
\end{equation*}
For $k$ large enough, the estimate (\ref{eq:maeplbdef}) thus holds with $u$ replaced by $u+v$. Thus $\xi+\xi_{\per}\in\ma_{\e}$.
This finishes the proof in case $\b_{X}-\b_{1}\leq 0$. 

\textbf{The case $\b_{X}-\b_{1}>0$.} When $\b_{X}-\b_{1}>0$, we again proceed as in the proof of Lemma~\ref{lemma:supexpinstab}, but with 
$\e$ replaced by $\e/4$ (in other words, we, to start with, apply Lemma~\ref{lemma:supexpgrowthmodeest} with $\e$ replaced by $\e/8$). 
This yields, in particular, numbers $N_{\roeep,\e}$, $K_{\roeep,\e}$ etc., depending only on $\boc_{\roeep}$ and $\e$. Define, moreover, 
$\mu_{\roeep,\e}$ by (\ref{eq:muroeepedef}) (with $\e$ replaced by $\e/4$). Let $p>1$ be a prime number, and define $k_{p,\roeep,\e}$ by 
\[
k_{p,\roeep,\e}:=\frac{\ln N_{\roeep,\e}}{\ln p}.
\]
We then define $v_{p}$ by 
\[
v_{p}:=\sum_{k\geq k_{p,\roeep,\e}}\sqrt{2}\exp\left[-\mu_{\roeep,\e}p^{k\frac{\b_{X}}{\b_{X}-\b_{1}}}\right]u_{p^{k}},
\]
where $u_{n_{1}}$ is defined by the right hand side of (\ref{eq:uxotonefre}); cf. (\ref{eq:vsuexpgrowth}). Due to an estimate similar to 
(\ref{eq:mfehomvtkmaxnolb}), there is, for each $p$, a sequence $t_{p,l}\rightarrow\infty$ as $k\rightarrow\infty$ such that
\[
\mfe_{\rohom}[v_{p}](t_{p,l})\geq \exp\left[2\b_{X}^{-1}\kappa(1-\e/4)e^{\b_{X}t_{p,l}}\right].
\]
Moreover, all the $v_{p}$ are linearly independent. In fact, formulae analogous to (\ref{eq:usumexp}) and (\ref{eq:mferohomsumform})
hold. We can now finish the proof as in the case $\b_{X}-\b_{1}\leq 0$. The proposition follows. 
\end{proof}

\subsection{Universal bounds on the growth of the energy}\label{ssection:unbdgritofinitefre}

Consider the equation (\ref{eq:thesystemRge}) in case $f=0$. Let $\ms$ be the set of smooth solutions to (\ref{eq:thesystemRge}),
and let $\ms_{\fin}$ be the set of smooth solutions to (\ref{eq:thesystemRge}) with finite frequency content. In other words, 
$u\in\ms_{\fin}$ iff $z(\indexnot,\cdot)$ defined by (\ref{eq:znutdef}) is only non-zero for a finite number of $\indexnot\in\EFindexset$.
Define $\mfG,\mfG_{\fin}:I\rightarrow (0,\infty)$ by
\begin{equation}\label{eq:mfgmfgfindef}
\mfG(t):=\sup_{0\neq u\in\ms}\frac{\mfe[u](t)}{\mfe[u](0)},\ \ \
\mfG_{\fin}(t):=\sup_{0\neq u\in\ms_{\fin}}\frac{\mfe[u](t)}{\mfe[u](0)},
\end{equation}
where
\[
\mfe[u](t):=\mfe_{\rohom}[u](t)+\frac{1}{2}\int_{\bM}|u(\cdot,t)|^{2}\mubox.
\]
Clearly, an argument is required to justify that $\mfG$ and $\mfG_{\fin}$ take their values in $(0,\infty)$. However, since it is sufficient
to appeal to standard energy estimates, we omit the justification here. The main point of introducing, e.g., $\mfG$ is of course that 
\[
\mfe[u](t)\leq \mfG(t)\mfe[u](0)
\]
for all $t\in I$ and all smooth solutions $u$. 

\begin{prop}\label{prop:supexpgrowthfinitefreq}
Consider (\ref{eq:thesystemRge}). Assume the associated metric to be such that $(M,g)$ is a canonical separable cosmological model
manifold. Assume, moreover, that $f=0$ and that there is a $j\in \{1,\dots,d\}$ such that (\ref{eq:urestrtojdep}) (no summation on $j$) 
satisfies the conditions of Lemma~\ref{lemma:supexpinstab}. Define $\mfG$ and $\mfG_{\fin}$ by (\ref{eq:mfgmfgfindef}). Then $\mfG=\mfG_{\fin}$ 
and for every $\e>0$, there is a time sequence $t_{k}\rightarrow\infty$ such that 
$\mfG(t_{k})\geq \exp\left[2\b_{X}^{-1}\kappa(1-\e)e^{\b_{X}t_{k}}\right]$.
\end{prop}
\begin{remark}
As we remarked at the beginning of Subsection~\ref{ssection:genexpgrowth}, there are equations of the form considered in the proposition such
that there is a dense subset of initial data yielding super exponential growth, and a dense subset of initial data yielding super exponential
decay. Moreover, the latter set contains the initial data with finite frequency content. Due to the statement of the proposition, it is, however,
clear that the super exponential growth can be detected by only considering solutions that decay super exponentially. 
\end{remark}
\begin{proof}
That $\mfG=\mfG_{\fin}$ is an immediate consequence of Cauchy stability combined with the fact that the set of initial data with finite frequency 
content is dense in the set of all smooth initial data. Fix $\e>0$. Let $v_{l}$ be the sequence of functions obtained by appealing to 
Lemma~\ref{lemma:supexpinstab}; by assumption, (\ref{eq:urestrtojdep}) satisfies the required conditions. Then we can consider the $v_{l}$ to 
be solutions to (\ref{eq:thesystemRge}) with $f=0$. Fix an $l$ large enough that $\mfe[v_{l}](0)\leq 1$; that this inequality holds for $l$ large 
enough is a consequence of (\ref{eq:vlsobconvtozero}). For this fixed $l$, (\ref{eq:mferohomsupexplb}) then implies that 
\[
\mfG(t_{l,k})\geq \mfe_{\rohom}[v_{l}](t_{l,k})\geq \exp\left[2\b_{X}^{-1}\kappa(1-\e)e^{\b_{X}t_{l,k}}\right].
\]
The proposition follows. 
\end{proof}

\section{Standard energy estimates}\label{section:standenestimates}

Consider the equation (\ref{eq:thesystemRge}) in case $g^{00}=-1$, $d=1$, $R=0$ and $f=0$; i.e.,
\begin{equation}\label{eq:odthesystemgehom}
u_{tt}-g^{11}(t)\d_{1}^{2}u-2g^{01}(t)\d_{1}\d_{t}u+X^{1}(t)\d_{1}u+\a(t)u_{t}+\zeta(t)u=0.
\end{equation}
The purpose of the present section is to derive energy estimates for solutions to this equation, given that the conditions stated in 
Lemma~\ref{lemma:mainassumpubcase} are fulfilled with $\b_{X}>0$ and that Assumptions~\ref{assumption:guooass} and \ref{assumption:Yoconvub}
hold. Consider the energy
\begin{equation}\label{eq:Erobasdef}
E_{\robas}(t):=\frac{1}{2}\is [|u_{t}(x,t)|^{2}+g^{11}(t)|\d_{1}u(x,t)|^{2}+e^{2\b_{\roode}t}|u(x,t)|^{2}]dx.
\end{equation}
\begin{lemma}\label{lemma:standenestimates}
Consider (\ref{eq:odthesystemgehom}). Assume the associated metric to be such that $(M,g)$ is a canonical separable cosmological model
manifold. Assume that the conditions stated in Lemma~\ref{lemma:mainassumpubcase} are fulfilled with $\b_{X}>0$ and that
Assumptions~\ref{assumption:guooass} and \ref{assumption:Yoconvub} hold. Then there is a constant $C_{\roep}$, depending only 
on $\boc_{\roep}$ introduced in 
(\ref{eq:bocroepdef}), such that the following holds. If $u$ is a $\cn{m}$-valued $C^{2}$-solution to (\ref{eq:odthesystemgehom})
and $E_{\robas}$ is defined by (\ref{eq:Erobasdef}), then 
\[
E_{\robas}(t)\leq \exp\left[\b_{X}^{-1}\|Y^{1}_{\infty}\|e^{\b_{X}t}+
C_{\roep}\left(\ldr{t}+e^{\b_{\roode}t}+e^{(\b_{X}-\eta_{X})t}\right)\right]E_{\robas}(0)
\]
for all $t\geq 0$. 
\end{lemma}
\begin{remark}\label{remark:standenestimates}
If $Y^{1}_{\infty}$ is Hermitian, then $\|Y^{1}_{\infty}\|/2$ equals $\kappa$, defined to be the largest absolute value of an eigenvalue 
of $Y^{1}_{\infty}/2$. In that case it thus follows that, given $\e>0$, there is a $T_{\roeep,\e}$, depending only $\e$ and $\boc_{\roeep}$ such that 
\[
E_{\robas}(t)\leq \exp\left[2\b_{X}^{-1}\kappa(1+\e)e^{\b_{X}t}\right]E_{\robas}(0)
\]
for all $t\geq T_{\roeep,\e}$. This estimate should be compared with (\ref{eq:mferohomsupexplb}); note that the $\kappa$ appearing in 
(\ref{eq:mferohomsupexplb}) equals the $\kappa$ defined in the present remark. 
\end{remark}
\begin{proof}
Time differentiation of $E_{\robas}$ yields
\begin{equation*}
\begin{split}
\d_{t}E_{\robas} = & \frac{1}{2}\is[\ldr{u_{t},u_{tt}}+\ldr{u_{tt},u_{t}}+(\d_{t}g^{11})|\d_{1}u|^{2}+g^{11}\ldr{\d_{1}u,\d_{1}\d_{t}u}\\
 & +g^{11}\ldr{\d_{1}\d_{t}u,\d_{1}u}+2\b_{\roode}e^{2\b_{\roode}t}|u|^{2}+e^{2\b_{\roode}t}\ldr{u,u_{t}}+e^{2\b_{\roode}t}\ldr{u_{t},u}]dx\\
 \leq & \is [\ellderbd g^{11}|\d_{1}u|^{2}+\b_{\roode}e^{2\b_{\roode}t}|u|^{2}]dx
+\frac{1}{2}e^{\b_{\roode}t}\is[|u_{t}|^{2}+e^{2\b_{\roode}t}|u|^{2}]dx\\
 & +\frac{1}{2}\is [\ldr{u_{tt}-g^{11}\d_{1}^{2}u,u_{t}}+\ldr{u_{t},u_{tt}-g^{11}\d_{1}^{2}u}]dx,
\end{split}
\end{equation*}
where we have used (\ref{eq:mflYovarsdef}) and (\ref{eq:mfldaddbd}). The first two terms on the far right hand side of this inequality
can easily be estimated in terms of the basic energy $E_{\robas}$. Let us therefore focus on the last term. It can be written
\begin{equation*}
\begin{split}
 & \frac{1}{2}\is [\ldr{u_{tt}-g^{11}\d_{1}^{2}u,u_{t}}+\ldr{u_{t},u_{tt}-g^{11}\d_{1}^{2}u}]dx \\
 = & \frac{1}{2}\is [\ldr{-X^{1}\d_{1}u-\a u_{t}-\zeta u,u_{t}}
+\ldr{u_{t},-X^{1}\d_{1}u-\a u_{t}-\zeta u}]dx\\
 & + \is [\ldr{g^{01}\d_{1}\d_{t}u,u_{t}}+\ldr{u_{t},g^{01}\d_{1}\d_{t}u}]dx
\end{split}
\end{equation*}
Note that the last term on the right hand side can be written
\[
\is \d_{1}[g^{01}|u_{t}|^{2}]dx=0,
\]
since $g^{01}$ is independent of the spatial variable. Moreover, 
\[
-\frac{1}{2}\is [\ldr{u_{t},\a u_{t}}+\ldr{\a u_{t},u_{t}}]dx \leq \|\a\|\is |u_{t}|^{2}dx\leq C_{\roode}e^{\b_{\roode}t}\is |u_{t}|^{2}dx
\]
for $t\geq 0$, where we have appealed to (\ref{eq:albdub}). Similarly, (\ref{eq:zetabdub}) yields 
\[
-\frac{1}{2}\is [\ldr{u_{t},\zeta u}+\ldr{\zeta u,u_{t}}]dx
\leq \frac{1}{2}C_{\roode}^{2}e^{\b_{\roode}t}\is [|u_{t}|^{2}+e^{2\b_{\roode}t}|u|^{2}]dx
\]
for $t\geq 0$. Let us focus on 
\begin{equation}\label{eq:Xoneterminenest}
\begin{split}
 & -\frac{1}{2}\is [\ldr{X^{1}\d_{1}u,u_{t}}+\ldr{u_{t},X^{1}\d_{1}u}]dx\\
 = & -\frac{1}{2}\is [\ldr{(Y^{1}-e^{\b_{X}t}Y^{1}_{\infty})e_{1}u,u_{t}}+\ldr{u_{t},(Y^{1}-e^{\b_{X}t}Y^{1}_{\infty})e_{1}u}]dx\\
 & -\frac{1}{2}\is[\ldr{e^{\b_{X}t}Y^{1}_{\infty}e_{1}u,u_{t}}+\ldr{u_{t},e^{\b_{X}t}Y^{1}_{\infty}e_{1}u}]dx,
\end{split}
\end{equation}
where we have used the notation (\ref{eq:mflYovarsdef}) and $e_{1}:=(g^{11})^{1/2}\d_{1}$.
The first term on the right hand side of (\ref{eq:Xoneterminenest}) can be estimated by 
\[
\frac{1}{2}K_{X}e^{(\b_{X}-\eta_{X})t}\is (|u_{t}|^{2}+g^{11}|\d_{1}u|^{2})dx
\]
for $t\geq 0$, where we have appealed to (\ref{eq:Yoasbehub}). Let us focus on the second term on the right hand side of 
(\ref{eq:Xoneterminenest}). It can be estimated by 
\[
\frac{1}{2}\|Y^{1}_{\infty}\|e^{\b_{X}t}\is[|u_{t}|^{2}+g^{11}|\d_{1}u|^{2}]dx.
\]
Adding up all the above observations yields
\[
\d_{t}E_{\robas}\leq \|Y^{1}_{\infty}\|e^{\b_{X}t}E_{\robas}+
C_{\roep}(1+e^{\b_{\roode}t}+e^{(\b_{X}-\eta_{X})t})E_{\robas},
\]
where the constant $C_{\roep}$ only depends on $\boc_{\roep}$ introduced in (\ref{eq:bocroepdef}). The lemma follows. 
\end{proof}

\section{Gevrey classes}\label{section:gevreyclasses}

Consider the proof of Lemma~\ref{lemma:supexpinstab}, in particular the definition of the sequence $v_{l}$; cf. (\ref{eq:vldefgevcase}).
The fall off of the Fourier coefficients of $v_{l}$ is faster than what is required in order for the solutions $v_{l}$ to be smooth.
It is of interest to introduce terminology quantifying this added fall off. This naturally leads us to the Gevrey classes. If 
$\Omega\subseteq\rn{d}$ is an open set and $s\geq 1$ is a real number, one way to define the class of Gevrey functions of order $s$
on $\Omega$ is the following; cf. \cite[Definition~1.4.1, p. 19]{LR}.

\begin{definition}\label{def:GevreyOmega}
A real valued function $f$ defined on $\Omega$ is said to be a \textit{Gevrey function of order} $1\leq s\in\ro$ on $\Omega$ if 
$f\in C^{\infty}(\Omega,\ro)$ and for each compact set $K\subset\Omega$, there are constants $R>0$ and $C>0$ such that 
\[
\sup_{x\in K}|\d^{\a}f(x)|\leq RC^{|\a|}(\a!)^{s}
\]
for all $d$-multiindices $\a$. The class of Gevrey functions of order $1\leq s\in\ro$ on $\Omega$ is denoted $G^{s}(\Omega)$. 
\end{definition}
\begin{remarks}
If $f\in G^{1}(\Omega)$, then $f$ is real analytic. Needless to say, the definition can be generalised to functions with values in 
$\rn{m}$ or $\cn{m}$. In this case, we use the notation $G^{s}(\Omega,\rn{m})$ and $G^{s}(\Omega,\cn{m})$ respectively. 
\end{remarks}

Here we are interested in functions on $\tn{d}$. We can think of such functions as being defined on $\rn{d}$ and being $2\pi$-periodic
in all of their variables. In that setting, referring to a compact set $K$, as in Definition~\ref{def:GevreyOmega}, becomes superfluous. 
Moreover, in what follows, it turns out to be of interest to generalise the notion of a Gevrey function to the case of $0<s<1$. This leads us to 
the following definition.

\begin{definition}\label{def:GevreyTd}
A real valued function $f$ defined on $\tn{d}$ is said to be a \textit{Gevrey function of order} $0<s\in\ro$ on $\tn{d}$ if 
$f\in C^{\infty}(\tn{d},\ro)$ and there are constants $R>0$ and $C>0$ such that 
\begin{equation}\label{eq:GevreycondTdfact}
\sup_{x\in \tn{d}}|\d^{\a}f(x)|\leq RC^{|\a|}(\a!)^{s}
\end{equation}
for all $d$-multiindices $\a$. The class of Gevrey functions of order $0<s\in\ro$ on $\tn{d}$ is denoted $G^{s}(\tn{d})$. 
\end{definition}
\begin{remarks}
Here we think of functions defined on $\tn{d}$ as being defined on $\rn{d}$ and being $2\pi$-periodic in all of their variables.
As before, the definition can be generalised to functions with values in $\rn{m}$ or $\cn{m}$, and we use the notation 
$G^{s}(\tn{d},\rn{m})$ and $G^{s}(\tn{d},\cn{m})$. Moreover, if $f\in G^{1}(\tn{d})$, 
then $f$ is real analytic. Finally, if $f\in G^{s}(\tn{d})$, where $0<s<1$, then $f$ is entire. 
\end{remarks}

It is sometimes of interest to reformulate the condition (\ref{eq:GevreycondTdfact}). Clearly, if there are constants $R>0$ and $C>0$ such that 
(\ref{eq:GevreycondTdfact}) holds for all $d$-multiindices $\a$, then 
\begin{equation}\label{eq:GevreycondTdnfact}
\sup_{x\in \tn{d}}|\d^{\a}f(x)|\leq RC^{|\a|}|\a|^{s|\a|}
\end{equation}
for all $d$-multiindices $\a$. However, this implication can also be turned around. More precisely, if there are constants $R>0$ and $C>0$ such 
that (\ref{eq:GevreycondTdnfact}) holds for all $d$-multiindices $\a$, then there are constants $R_{1}>0$ and $C_{1}>0$ such that 
(\ref{eq:GevreycondTdfact}), with $R$ and $C$ replaced by $R_{1}$ and $C_{1}$ respectively, holds for all $d$-multiindices $\a$. A proof of
this statement is to be found, e.g., in the proof of \cite[Proposition~1.4.2, pp. 19--20]{LR}. For completeness, let us, however, note that 
it follows from Stirling's formula (cf., e.g., \cite[(103), p.~194]{babyrudin}) and the estimate $|\a|!\leq d^{|\a|}\a!$. Note, moreover, that 
the latter estimate is a consequence of 
\[
d^{N}=(1+\dots+1)^{N}=\sum_{|\a|=N}\frac{N!}{\a!}.
\]
Even though the characterisations (\ref{eq:GevreycondTdfact}) and (\ref{eq:GevreycondTdnfact}) are of interest, we are here mainly interested
in characterisations expressed in terms of the Fourier coefficients of the function $f$. 

\begin{lemma}\label{lemma:fouriercoeffcharofgevreyclass}
Let $f\in\ C^{\infty}(\tn{d},\co)$, $0<s\in\ro$ and $a_{n}$, $n\in\zn{d}$, be the complex Fourier coefficients of $f$. Then 
$f\in G^{s}(\tn{d},\co)$ if and only if there are constants $C_{i}>0$, $i=1,2$, 
such that 
\begin{equation}\label{eq:altgevcond}
|a_{n}|\leq C_{1}\exp\left(-C_{2}|n|^{1/s}\right)
\end{equation}
for all $n\in\zn{d}$. 
\end{lemma}
\begin{proof}
To begin with, assume that $f\in G^{s}(\tn{d},\co)$ for some $s>0$. If $0\leq k\in\zo$ and $0\neq n\in\zn{d}$, then
\begin{equation*}
\begin{split}
|n|^{k}|a_{n}| \leq & d^{k/2}(|n_{1}|^{k}+\dots+|n_{d}|^{k})\frac{1}{(2\pi)^{d}}\left|\int_{\tn{d}}f(x)e^{-in\cdot x}dx\right|\\
  = & \frac{d^{k/2}}{(2\pi)^{d}}\sum_{j=1}^{d}\left|\int_{\tn{d}}(-in_{j})^{k}f(x)e^{-in\cdot x}dx\right|\\
 \leq & d^{k/2}\sum_{j=1}^{d}\sup_{x\in\tn{d}}|\d_{j}^{k}f(x)|\leq SB^{k}k^{sk},
\end{split}
\end{equation*}
where we have appealed to (\ref{eq:GevreycondTdnfact}), $B=\sqrt{d}C$ and $S=d\cdot R$; here $C$ and $R$ are the constants
appearing in (\ref{eq:GevreycondTdnfact}). For a given $n$, it is of interest to choose $k$ so that 
\[
h(k):=B^{k}k^{sk}|n|^{-k}=\exp\left(s\cdot(B^{-1}|n|)^{1/s}\frac{k}{(B^{-1}|n|)^{1/s}}\ln\left[\frac{k}{(B^{-1}|n|)^{1/s}}\right]\right)
\]
is minimal. Letting $\xi=k(B^{-1}|n|)^{-1/s}$, the relevant question is that of minimising $r(\xi):=\xi\ln \xi$. Thus the minimum
is obtained when $\xi=\xi_{0}:=e^{-1}$, i.e., when $k=k_{0}:=(B^{-1}|n|)^{1/s}e^{-1}$. Of course, $k_{0}$ need not be a non-negative 
integer. Let $k_{1}$ be the smallest integer satisfying $k_{1}\geq k_{0}$ and let $\xi_{1}=k_{1}(B^{-1}|n|)^{-1/s}$. Then 
$0\leq \xi_{1}-\xi_{0}\leq (B^{-1}|n|)^{-1/s}$, so that 
\[
|r(\xi_{1})-r(\xi_{0})|\leq \frac{e}{2}(B^{-1}|n|)^{-2/s}
\]
for $|n|\geq 1$, where we have used the fact that $r'(\xi_{0})=0$ and the fact that $|r''(\xi)|\leq e$ for $\xi\geq \xi_{0}$. Thus 
\[
h(k_{1})=\exp\left[s\cdot(B^{-1}|n|)^{1/s}r(\xi_{1})\right]
=h(k_{0})\exp\left(\frac{es}{2}(B^{-1}|n|)^{-1/s}\right)\leq Kh(k_{0}),
\]
where $K$ is a constant which is independent of $n\neq 0$. Thus
\[
|a_{n}|\leq SKh(k_{0})=SK\exp\left(-\frac{s}{eB^{1/s}}|n|^{1/s}\right).
\]
Assume now that there are constants $C_{i}>0$, $i=1,2$, such that (\ref{eq:altgevcond}) holds for all $n\in\zn{d}$. 
Let $\a\neq 0$ be a $d$-multiindex. Then
\begin{equation}\label{eq:ukprelest}
\begin{split}
|\d^{\a}f(x)| = & \left|\textstyle{\sum}_{n\in\zn{d}}(in)^{\a}a_{n}e^{in\cdot x}\right|\\
 \leq & \textstyle{\sum}_{n\in\zn{d}}C_{1}\exp\left(-C_{2}|n|^{1/s}/2\right)
|n|^{|\a|}\exp\left(-C_{2}|n|^{1/s}/2\right).
\end{split}
\end{equation}
At this stage, it is natural to maximise $\psi(\xi)=-c_{0}\xi^{1/s}+k\ln \xi$ for $\xi>0$, where $c_{0}=C_{2}/2$ and $k=|\a|$. The maximum is 
obtained when $\xi^{1/s}=sk/c_{0}$. Thus
\[
|n|^{k}\exp\left(-C_{2}|n|^{1/s}/2\right)=\exp[\psi(|n|)]\leq \left(\frac{s}{e c_{0}}\right)^{sk}k^{sk}.
\]
Combining this estimate with (\ref{eq:ukprelest}) yields
\[
|\d^{\a}f(x)|\leq \sum_{n\in\zn{d}}C_{1}\exp\left(-C_{2}|n|^{1/s}/2\right)\left(\frac{s}{e c_{0}}\right)^{sk}k^{sk}
\leq RC^{|\a|}|\a|^{s|\a|},
\]
for all $x\in\tn{d}$, $0\neq\a\in\zn{d}$ and a suitable choice of constants $R$ and $C$.
\end{proof}

\subsection{Preservation of Gevrey class regularity}

Next we want to verify that it is meaningful to speak of solutions to (\ref{eq:thesystemRge}) with Gevrey class regularity
(in the case that $R=0$). In order for this to be possible, it is, however, necessary to impose conditions on $f$. This leads 
us to the following definition. 

\begin{definition}
Let $I\subseteq\ro$ be an open interval, $f\in C^{\infty}(\tn{d}\times I,\cn{m})$ and $0<s\in\ro$. Then $f$ is said to belong to 
$\mfG^{s}_{\roloc,\rou}(I,\tn{d},\cn{m})$ if, for every compact interval $J\subset I$, the following holds: there are constants $C_{1}>0$ 
and $C_{2}>0$ (depending on $f$ and $J$) such that 
\[
|\hf(n,t)|\leq C_{1}\exp\left(-C_{2}|n|^{1/s}\right)
\]
for all $n\in\zn{d}$ and $t\in J$, where $\hf(n,t)$ is the $n$'th (complex) Fourier coefficient of $f(\cdot,t)$. 
\end{definition}
\begin{remark}
Note that if $f\in\mfG^{s}_{\roloc,\rou}(I,\tn{d},\cn{m})$, then $f(\cdot,t)\in G^{s}(\tn{d},\cn{m})$ for every $t\in I$. 
\end{remark}

\begin{lemma}\label{lemma:gevreypreserv}
Let $1\leq d,m\in\zo$, $R=0$, $0<s\in\ro$ and $I\subseteq\ro$ be an open interval. Assume that $g^{00},g^{0l},g^{lj}\in C^{\infty}(I,\ro)$ are such 
that for every $t\in I$, $g^{lj}(t)$ are the components of a positive definite matrix and $g^{00}(t)=-1$. Assume, moreover, that $X^{j}$, $\a$, 
$\zeta\in C^{\infty}[I,\Mn{m}{\co}]$. Finally, assume that $f\in\mfG^{s}_{\roloc,\rou}(I,\tn{d},\cn{m})$.
Let $t_{0}\in I$ and $u$ be the solution to (\ref{eq:thesystemRge}) (with $R=0$) on $\tn{d}\times I$ corresponding to initial data 
$u(\cdot,t_{0}),u_{t}(\cdot,t_{0})\in G^{s}(\tn{d},\cn{m})$. Then $u,u_{t}\in\mfG^{s}_{\roloc,\rou}(I,\tn{d},\cn{m})$.
\end{lemma}
\begin{proof}
Since the initial data are in $G^{s}(\tn{d},\cn{m})$, it is clear that the corresponding solution to (\ref{eq:thesystemRge})
is smooth. Let $z(n,t)$ denote the $n$'th Fourier coefficient of $u(\cdot,t)$ and consider (\ref{eq:fourierthesystemRge}); note that 
$\indexnot=n$ in the present context. Define
\begin{align*}
\mfH_{n}(t) := & |\dot{z}(n,t)|^{2}+[\mfg^{2}(n,t)+1]|z(n,t)|^{2},\\
H_{n}(t) := & |\dot{z}(n,t)|^{2}+\ldr{n}^{2}|z(n,t)|^{2},
\end{align*}
where $\mfg(n,t)$ is given by (\ref{eq:mfgnutdef}). In order to prove the statement of 
the lemma, it is sufficient to estimate 
$H_{n}(t)$ for $t$ in a compact interval $J$ contained in $I$. In practice we can, without loss of generality, assume that 
$J=[t_{0},t_{1}]$ for some $t_{1}>t_{0}$ (the proof in the case $J=[t_{1},t_{0}]$ is similar and in the case of a general $J$, 
$J\subseteq [t_{a},t_{0}]\cup [t_{0},t_{b}]$ for suitably chosen $t_{a}$ and $t_{b}$). It is of interest to note that there 
are constants $K_{j}>0$, $j=1,2$, depending only on $g^{ij}(t)$ for $t\in [t_{0},t_{1}]$, such that 
\[
K_{1}H_{n}(t)\leq \mfH_{n}(t)\leq K_{2}H_{n}(t)
\]
for $t\in [t_{0},t_{1}]$. It can then be estimated that 
\begin{equation}\label{eq:dotmfHntest}
\begin{split}
\dot{\mfH}_{n}(t) \leq &  2\|\a(t)\| |\dot{z}(n,t)|^{2}+2\textstyle{\sum}_{j}|n_{j}|\|X^{j}(t)\| |z(n,t)||\dot{z}(n,t)|\\
 & +2(\|\zeta(t)\|+1) |z(n,t)||\dot{z}(n,t)|+2|\dot{z}(n,t)||\hf(n,t)|\\
 & +\dot{g}^{lj}(t)n_{j}n_{l}|z(n,t)|^{2} \leq C\mfH_{n}(t)+|\hf(n,t)|^{2}
\end{split}
\end{equation}
for $t\in [t_{0},t_{1}]$, where the constant $C>0$ only depends on the coefficients of the equation (\ref{eq:thesystemRge}) (and not on $f$). 
In order to obtain this estimate, we have appealed to the fact that the $g^{jl}$ are continuously differentiable on $I$, and that 
$\a$, $X$ and $\zeta$ are continuous. Due to (\ref{eq:dotmfHntest}), 
\[
\mfH_{n}(t)\leq e^{C(t_{1}-t_{0})}\mfH_{n}(t_{0})+\int_{t_{0}}^{t_{1}}e^{C(t_{1}-\tau)}|\hf(n,\tau)|^{2}d\tau
\]
for $t\in [t_{0},t_{1}]$. Combining this estimate with the assumptions yields the desired conclusion. 
\end{proof}

\subsection{Solutions with super exponential growth and Gevrey regularity}

The purpose of the present subsection is to make more precise statements concerning the regularity of the solutions constructed in 
Lemma~\ref{lemma:supexpinstab}. 

\begin{prop}\label{prop:supexpinstabGevreyreg}
Assume that the conditions of Lemma~\ref{lemma:supexpinstab} are fulfilled. Let $\e>0$. If $\b_{X}-\b_{1}>0$, then the sequence of functions $v_{l}$ 
constructed in Lemma~\ref{lemma:supexpinstab} can be assumed to such that $v_{l},\d_{t}v_{l}\in \mfG_{\roloc,\rou}^{s_{X}}(I,\so,\cn{m})$ for all
$l$, where 
\begin{equation}\label{eq:sXdef}
s_{X}:=\frac{\b_{X}-\b_{1}}{\b_{X}}.
\end{equation}
In case $\b_{X}-\b_{1}\leq 0$, one frequency can be chosen, say $n_{a}\in\zo$, such that the 
sequence of functions $v_{l}$ constructed in Lemma~\ref{lemma:supexpinstab} has the property that if $z_{l}(n,t)$ denotes the 
$n$'th Fourier coefficient of $v_{l}(\cdot,t)$, then $z_{l}(n,t)=0$ for all $t$ unless $n=n_{a}$.
\end{prop}
\begin{remarks}
If $\b_{1}=0$, the functions $v_{l}$ and $\d_{t}v_{l}$ are real analytic on constant $t$ hypersurfaces. If $\b_{1}>0$, they are entire on constant $t$ 
hypersurfaces. In general, the larger the $\b_{1}$, the better the regularity of the functions $v_{l}$. Considering (\ref{eq:cmebgoobd}),
it is clear that improving the lower bound in (\ref{eq:cmebgoobd}) corresponds to improving the regularity of the $v_{l}$. It is of 
interest to ask if regularity of the form $\mfG_{\roloc,\rou}^{s_{X}}(I,\so,\cn{m})$, where $s_{X}$ is given by (\ref{eq:sXdef}), is 
optimal if the parameter $\b_{1}$ appearing in the lower bound in (\ref{eq:cmebgoobd}) is optimal. However, we do not address this question here.
\end{remarks}
\begin{proof}
The statement is a consequence of Lemma~\ref{lemma:gevreypreserv} and the proof of Lemma~\ref{lemma:supexpinstab}.
\end{proof}

\section{Mode analysis}\label{section:modeanalysis}

It is of interest to note that in some cases, a mode by mode analysis gives a picture of the future asymptotics which is radically different from 
that of Lemma~\ref{lemma:supexpinstab} and Propositions~\ref{prop:supexpgrowthgencase}, \ref{prop:supexpgrowthgencaseden} and 
\ref{prop:supexpgrowthinfcodim}. In the present section, we illustrate this by a sequence of lemmas. 

\begin{lemma}\label{lemma:roughmodenest}
Consider (\ref{eq:thesystemRge}). Assume the associated metric to be such that $(M,g)$ is a canonical separable cosmological model
manifold. Assume, moreover, that $f=0$. Assume that there is a $\g\geq 0$ and a constant $C>0$ such that 
\[
|g^{0l}(t)|+|g^{jl}(t)|^{1/2}+|a^{-1}_{r}(t)|+\|X^{j}(t)\|^{1/2}+\|\a(t)\|+\|\zeta(t)\|^{1/2} \leq Ce^{\g t}
\]
for all $t\geq 0$, $j,l=1,\dots,d$, and $r=1,\dots,R$. Fix $\indexnot\in\EFindexset$, let $z(\indexnot,t)$ be defined by (\ref{eq:znutdef}),
where $u$ is a solution to (\ref{eq:thesystemRge}) with $f=0$, and define
\[
E_{\indexnot}(t):=|\dot{z}(\indexnot,t)|^{2}+e^{2\g t}|z(\indexnot,t)|^{2}.
\]
Then there is a constant $C_{\indexnot}$, depending on $\indexnot$, such that 
\[
E_{\indexnot}(t)\leq \exp\left[C_{\indexnot}(e^{\g t}+t)\right]E_{\indexnot}(0)
\]
for $t\geq 0$. 
\end{lemma}
\begin{remark}
There is a large class of equations that satisfy both the conditions of Proposition~\ref{prop:supexpgrowthgencaseden} and the present 
lemma with $\g=\b_{\roode}$. This illustrates that the individual modes can behave quite differently from generic smooth solutions.
\end{remark}
\begin{proof}
Under the assumptions of the lemma, (\ref{eq:fourierthesystemRge}) with $\hf=0$ can be written
\begin{equation}\label{eq:roeffeqhomcase}
\ddot{z}+\a_{\roeff}(t)\dot{z}+\zeta_{\roeff}(t) z=0,
\end{equation}
where 
\[
\|\a_{\roeff}(t)\|+\|\zeta_{\roeff}(t)\|^{1/2}\leq C_{\indexnot}e^{\g t}
\]
for all $t\geq 0$. Note, however, that $C_{\indexnot}$ depends on the frequency $\indexnot$. It can then be estimated that 
\[
\d_{t}E_{\indexnot}(t)\leq C_{\indexnot}e^{\g t}E_{\indexnot}(t)
\]
for all $t\geq 0$. The lemma follows.
\end{proof}

In order to obtain more detailed conclusions, we need to make more detailed assumptions. 

\begin{lemma}\label{lemma:supexpdeccondmode}
Consider (\ref{eq:thesystemRge}). Assume the associated metric to be such that $(M,g)$ is a canonical separable cosmological model
manifold. Assume, moreover, that $f=0$. Assume that there is a 
$\g>0$; $\a_{\infty}, \zeta_{\infty}\in\Mn{m}{\co}$, one of which is non-zero; an $\eta_{\rocu}>0$; and a constant $C_{\rocu}>0$ such that 
\begin{align*} 
|g^{0l}(t)|+|g^{jl}(t)|^{1/2}+|a^{-1}_{r}(t)|+\|X^{j}(t)\|^{1/2} \leq & C_{\rocu}e^{(\g-\eta_{\rocu})t},\\
\|e^{-\g t}\a(t)-\g\a_{\infty}\|+\|e^{-2\g t}\zeta(t)-\g^{2}\zeta_{\infty}\|^{1/2} \leq & C_{\rocu}e^{-\eta_{\rocu}t}
\end{align*}
for all $t\geq 0$, $j,l=1,\dots,d$, and $r=1,\dots,R$. Let $A_{\infty}$ be defined by (\ref{eq:Ainfdef}) 
and $\kappa_{1}:=\kappa_{\max}(A_{\infty})$; cf. Definition~\ref{def:SpRspdef}. Fix $\indexnot\in\EFindexset$ and $\e>0$. Then there is a constant 
$C_{+,\e}>0$ such that the following holds. If $z(\indexnot,t)$ is a solution to (\ref{eq:fourierthesystemRge}) with $\hf=0$, then
\[
|\dot{z}(\indexnot,t)|+|z(\indexnot,t)|\leq C_{+,\e}\exp[(\kappa_{1}+\e)e^{\g t}][|\dot{z}(\indexnot,0)|+|z(\indexnot,0)|]
\]
for all $t\geq 0$. 
\end{lemma}
\begin{remark}
The constant $C_{+,\e}$ only depends on $C_{\rocu}$, $\indexnot$, the spectrum of the Laplace-Beltrami operator on $(M_{r},g_{r})$ for $r=1,\dots,R$,
$A_{\infty}$, $\eta_{\rocu}$, $\g$ and $\e$.
\end{remark}
\begin{remark}
There is a large class of equations satisfying the conditions of Propositions~\ref{prop:supexpgrowthgencaseden},
\ref{prop:supexpgrowthinfcodim} and the present lemma with 
$\g=\b_{\roode}$ and $\kappa_{1}<0$. For such equations, the energy of generic smooth solutions grows super exponentially. However, the energies of 
the individual modes decay to zero super exponentially. In fact, there is a set of initial data, say $\mathcal{B}$, such that the corresponding
solutions decay super exponentially and $\mathcal{B}$ is dense in the set of initial data. In fact, we can take $\mathcal{B}$ to be the set of initial 
data with finite frequency content. In that setting, it is thus clear that we cannot improve the conclusions of 
Proposition~\ref{prop:supexpgrowthgencaseden} to say that $\ma$ is open and dense. 
\end{remark}
\begin{proof}
As in the proof of Lemma~\ref{lemma:roughmodenest}, the equation (\ref{eq:fourierthesystemRge}) with $\hf=0$ is equivalent to (\ref{eq:roeffeqhomcase}).
However, under the assumptions of the present lemma, 
\[
\|e^{-\g t}\a_{\roeff}(t)-\g\a_{\infty}\|+\|e^{-2\g t}\zeta_{\roeff}(t)-\g^{2}\zeta_{\infty}\|^{1/2} \leq C_{\indexnot,\rocu,\spec}e^{-\eta_{\rocu} t}
\]
for all $t\geq 0$, where $C_{\indexnot,\rocu,\spec}$ only depends on $\indexnot$, $C_{\rocu}$ and the spectrum of the Laplace-Beltrami operator on the Riemannian
manifolds $(M_{r},g_{r})$, $r=1,\dots,R$. Given this information, the statement follows from Lemma~\ref{lemma:spsysodes}.  
\end{proof}

The case where $\g=0$ is also of interest, but it requires a slightly different formulation. 

\begin{lemma}\label{lemma:expdeccondmode}
Consider (\ref{eq:thesystemRge}). Assume the associated metric to be such that $(M,g)$ is a canonical separable cosmological model
manifold. Assume, moreover, that $f=0$. Assume that there are $\a_{\infty}, \zeta_{\infty}\in\Mn{m}{\co}$; an $\eta_{\rocu}>0$; and a constant 
$C_{\rocu}>0$ such that 
\begin{align*} 
|g^{0l}(t)|+|g^{jl}(t)|+|a_{r}^{-2}(t)|+\|X^{j}(t)\| \leq & C_{\rocu}e^{ -\eta_{\rocu} t},\\
\|\a(t)-\a_{\infty}\|+\|\zeta(t)-\zeta_{\infty}\| \leq & C_{\rocu}e^{-\eta_{\rocu} t}
\end{align*}
for all $t\geq 0$, $j,l=1,\dots,d$, and $r=1,\dots,R$. Let $A_{\infty}$ be defined by (\ref{eq:Ainfdef}), $\kappa_{1}:=\kappa_{\max}(A_{\infty})$
and $d_{1}:=d_{\max}(A_{\infty},\kappa_{1})$; cf. Definition~\ref{def:SpRspdef}. Fix $\indexnot\in\EFindexset$. Then there is a constant $C_{+}>0$ 
such that the following holds. If $z(\indexnot,t)$ is a solution to (\ref{eq:fourierthesystemRge}) with $\hf=0$, then
\[
|\dot{z}(\indexnot,t)|+|z(\indexnot,t)|\leq C_{+}\ldr{t}^{d_{1}-1}e^{\kappa_{1}t}(|\dot{z}(\indexnot,0)|+|z(\indexnot,0)|)
\]
for all $t\geq 0$. 
\end{lemma}
\begin{remark}
The constant $C_{+}$ only depends on $C_{\rocu}$, $\indexnot$, the spectrum of the Laplace-Beltrami operator on $(M_{r},g_{r})$ for $r=1,\dots,R$,
$A_{\infty}$ and $\eta_{\rocu}$.
\end{remark}
\begin{proof}
As before, (\ref{eq:fourierthesystemRge}) with $\hf=0$ is equivalent to (\ref{eq:roeffeqhomcase}). However, here
\[
\|\a_{\roeff}(t)-\a_{\infty}\|+\|\zeta_{\roeff}(t)-\zeta_{\infty}\| \leq C_{\rocu,\indexnot,\spec}e^{-\eta_{\rocu} t}
\]
for all $t\geq 0$, where $C_{\indexnot,\rocu,\spec}$ only depends on $\indexnot$, $C_{\rocu}$ and the spectrum of the Laplace-Beltrami operator on the Riemannian
manifolds $(M_{r},g_{r})$, $r=1,\dots,R$. Thus (\ref{eq:fourierthesystemRge}) can be reformulated to 
\begin{equation}\label{eq:veqcasegaeqzero}
\dot{v}(t)=A_{\infty}v(t)+A_{\rem}(t)v(t),
\end{equation}
where
\[
v(t):=\left(\begin{array}{c} z(\indexnot,t) \\ \dot{z}(\indexnot,t)\end{array}\right),\ \ \
\|A_{\rem}(t)\|\leq C_{\rocu,\indexnot,\spec}e^{-\eta_{\rocu} t}
\]
for all $t\geq 0$. Thus (\ref{eq:veqcasegaeqzero}) is an equation satisfying the conditions described in Subsection~\ref{ssection:ODEreg}
with $T_{\roode}=0$. The desired conclusion is a consequence of Lemma~\ref{lemma:oderegest}.
\end{proof}

\part{Equations with a dominant noisy spatial direction}\label{part:dominantnoisyspdirection}

\chapter{Terminology and basic estimates}

\section{Introduction}

\textbf{The notion of a dominant noisy spatial direction.}
In Chapters~\ref{chapter:weaksil} and \ref{chapter:asympttranspcase}, we consider silent and transparent equations. In the present part of 
these notes, we turn to so-called noisy equations. We give a formal definition of what we mean by noisy equations in Definition~\ref{def:domnospdir}
below. However, considering the model equation
\[
u_{tt}-\textstyle{\sum}_{j=1}^{d}g_{\infty}^{jj}e^{2\b_{j}t}\d_{j}^{2}u-\sum_{r=1}^{R}a_{r,\infty}^{-2}e^{2\bRie{r}t}\Delta_{g_{r}}u+\a_{\infty}u_{t}
+\sum_{j=1}^{d}X^{j}_{\infty}e^{\b_{j}t}\d_{j}u+\zeta_{\infty}u=f
\]
(where $g_{\infty}^{jj}>0$ and $a_{r,\infty}>0$), the conditions in Definition~\ref{def:domnospdir} amount to requiring one of the exponents, 
i.e. one of the elements of 
\[
\{\b_{1},\dots,\b_{d},\bRie{1},\dots,\bRie{R}\},
\]
to be strictly positive and strictly larger than all the other elements. In practice, it turns out that when there is one spatial direction which is noisy 
(meaning that the corresponding exponent is strictly positive) and dominant (meaning that the corresponding exponent is strictly larger than all the other 
exponents), then the detailed behaviour of the remaining coefficients of the highest order spatial derivatives is not so important; cf. 
Definition~\ref{def:domnospdir} below. Turning to the coefficients of the lower order derivatives, the coefficient in front 
of $u$ does not have to converge, as long as it and its time derivative are bounded. The coefficient in front of $u_{t}$ does, however, need to converge 
exponentially. In case $g^{jj}$ is the dominant coefficient in front of the second order spatial derivatives, and $e^{-2\b_{j}t}g^{jj}(t)$ converges to a 
strictly positive number, then we also require $e^{-\b_{j}t}X^{j}(t)$ to converge. Again, we refer to Definition~\ref{def:domnospdir} below for the precise 
list of requirements. 

\textbf{Averaging over oscillations.}
The main tools we use in order to analyse the asymptotic behaviour of solutions to noisy equations are the ones developed in Part~\ref{part:averaging} of 
these notes. However, in order to be allowed to appeal to the relevant results, we need to make additional assumptions. In particular, we
require the equation to be strongly balanced and to have a negligible shift vector field; cf. Definition~\ref{def:strongbalneglshift} below. Given that the 
requirements of both Definition~\ref{def:domnospdir} and Definition~\ref{def:strongbalneglshift} below are satisfied, it can be verified that the equation 
is oscillation adapted; cf. Lemma~\ref{lemma:strongbalneglshiftimploscad}. Thus the results of Part~\ref{part:averaging} apply. 

\textbf{Defining the asymptotic regime.}
In order to be able to draw conclusions, we need to focus on Fourier modes such that the coefficients corresponding to the dominant noisy spatial direction
appear in (\ref{eq:fourierthesystemRge}). This corresponds to focusing on a subset, say $\EFnindexset$, of $\EFindexset$; cf. Definition~\ref{def:nuronEFnindexset}
below for a formal definition. However, even for $\indexnot\in\EFnindexset$, the behaviour of solutions to (\ref{eq:fourierthesystemRge}) need not initially
be determined by the coefficients corresponding to the dominant noisy spatial direction. For that reason, it is necessary to, given 
$\indexnot\in\EFnindexset$, determine a late time regime in which these coefficients have a dominant influence. After some preliminary observations concerning
the asymptotic behaviour of $\mfg(\indexnot,t)$, we, given $\indexnot\in\EFnindexset$, define a number $T_{\ron}\geq 0$, cf. Definition~\ref{def:Tron} below, 
characterising the late time regime. 

\textbf{Behaviour along a time sequence, rough estimates.}
Once the notion of a late time regime has been defined, we introduce a time sequence $\{t_{k}\}$, given by Definition~\ref{def:tkdefge} and the requirement
that $t_{0}=T_{\ron}$. In order to analyse the behaviour along this time sequence, we appeal to the results of Part~\ref{part:averaging}, in particular
Lemma~\ref{lemma:wkfinlemma}. Considering Lemma~\ref{lemma:wkfinlemma}, it is clear that it is of interest to derive detailed estimates for the right 
hand side of (\ref{eq:Xikpreappdefstfin}).
This is the purpose of Section~\ref{section:detailedasfomaappinit}, and leads to an asymptotic expansion for the matrix $A_{k}^{\pm}$ appearing in 
(\ref{eq:wprekpertransvarpi}); cf. Lemma~\ref{lemma:Akpmasexpnoise}. Given this asymptotic expansion, we are in a position to analyse the asymptotics of 
solutions. However, the analysis proceeds in several steps. First, it is convenient to reformulate the equations. This is the subject of 
Section~\ref{section:reformeqsnoise}, and the goal is to isolate the leading order behaviour. Given this reformulation, we are in a position to 
derive an estimate for how the energy associated with a mode develops in the late time regime; cf. Section~\ref{section:aroughmodeestnoise}. 
Combining this estimate with Lemma~\ref{lemma:roughenestbalsetting} then yields an estimate for the Sobolev norm of the part of the solution associated
with modes $\indexnot\in\EFnindexset$; cf. Section~\ref{section:roughsobolevestnoise}.

\section{Terminology and basic assumptions}\label{section:termbaassannoise}

We begin by defining the notion of a dominant noisy spatial direction. 

\begin{definition}\label{def:domnospdir}
Consider (\ref{eq:thesystemRge}). Assume the associated metric to be such that $(M,g)$ is a canonical separable cosmological model
manifold. Assume, moreover, that there are constants $\b_{\ron}>0$, $\eta_{\romar}>0$ and $C_{\ron}>0$ 
such that one of the following conditions holds:
\begin{itemize}
\item there is one $j\in \{1,\dots,d\}$ and a $g^{jj}_{\infty}\in (0,\infty)$ such that 
\begin{equation*}
\begin{split}
 & |e^{-2\b_{\ron}t}g^{jj}(t)-g^{jj}_{\infty}|+|e^{-2\b_{\ron}t}\d_{t}g^{jj}(t)-2\b_{\ron}g^{jj}_{\infty}| \\
 & +e^{-2\b_{\ron}t}\left(\textstyle{\sum}_{k,l\neq j}|g^{kl}(t)| +\sum_{l\neq j}|g^{jl}(t)|+\sum_{r=1}^{R}a_{r}^{-2}(t)\right)\\
 & +e^{-2\b_{\ron}t}\left(\textstyle{\sum}_{k,l\neq j}|\d_{t}g^{kl}(t)| +\sum_{l\neq j}|\d_{t}g^{jl}(t)|+\sum_{r=1}^{R}|\d_{t}(a_{r}^{-2})(t)|\right)
\leq C_{\ron}e^{-\eta_{\romar}t}
\end{split}
\end{equation*}
for all $t\geq 0$, or
\item there is one $r_{\ron}\in\{1,\dots,R\}$ and a $q_{\infty,\ron}\in (0,\infty)$ such that 
\begin{equation*}
\begin{split}
 & |e^{-2\b_{\ron}t}a^{-2}_{r_{\ron}}(t)-q_{\infty,\ron}|+|e^{-2\b_{\ron}t}\d_{t}(a^{-2}_{r_{\ron}})(t)-2\b_{\ron}q_{\infty,\ron}|\\
 & +e^{-2\b_{\ron}t}\left(\textstyle{\sum}_{k,l=1}^{d}|g^{kl}(t)|+\sum_{r\neq r_{\ron}}a_{r}^{-2}(t)\right)\\
 & +e^{-2\b_{\ron}t}\left(\textstyle{\sum}_{k,l=1}^{d}|\d_{t}g^{kl}(t)|+\sum_{r\neq r_{\ron}}|\d_{t}(a_{r}^{-2})(t)|\right)\leq C_{\ron}e^{-\eta_{\romar}t}
\end{split}
\end{equation*}
for all $t\geq 0$. 
\end{itemize}
Then (\ref{eq:thesystemRge}) is said to have a \textit{dominant noisy spatial direction}. 
\index{Dominant noisy spatial direction}%
In the first case, (\ref{eq:thesystemRge}) is said to 
have a \textit{dominant noisy spatial torus direction} 
\index{Dominant noisy spatial torus direction}%
and $j$ is said to correspond to the \textit{torus direction}. Moreover, the spectrum
of the Laplace-Beltrami operator of the flat metric on $\so$ is referred to as the \textit{spectrum of the Riemannian manifold corresponding to 
the dominant noisy spatial direction}. 
\index{Spectrum of the Riemannian manifold!corresponding to the dominant noisy spatial direction}%
In the second case, (\ref{eq:thesystemRge}) is said to have a \textit{dominant noisy spatial generalised direction} 
\index{Dominant noisy spatial generalised direction}%
and $r_{\ron}$ is said to correspond to the \textit{generalised direction}. 
\index{Generalised!direction}%
\index{Direction!generalised}%
Moreover, the spectrum of the Laplace-Beltrami operator of $(M_{r_{\ron}},g_{r_{\ron}})$ is referred to as the 
\textit{spectrum of the Riemannian manifold corresponding to the dominant noisy spatial direction}.
\index{Spectrum of the Riemannian manifold!corresponding to the dominant noisy spatial direction}%
If (\ref{eq:thesystemRge}) has a dominant noisy spatial torus direction; $j$ corresponds to the 
torus direction; and there are constants $\eta_{\romn}>0$, $C_{\romn}>0$ and matrices $X^{j}_{\infty},\a_{\infty}\in\Mn{m}{\co}$ such that 
\begin{equation}\label{eq:Xjalconvnosetting}
\|e^{-\b_{\ron}t}X^{j}(t)-X^{j}_{\infty}\|+\|\a(t)-\a_{\infty}\|\leq C_{\romn}e^{-\eta_{\romn}t}
\end{equation}
for all $t\geq 0$; then (\ref{eq:thesystemRge}) is said to be such that \textit{the dominant coefficients are convergent}. 
\index{Dominant coefficients!convergent}%
\index{Convergent!dominant coefficients}%
If (\ref{eq:thesystemRge}) 
has a dominant noisy spatial generalised direction; and there are constants $\eta_{\romn}>0$, 
$C_{\romn}>0$ and a matrix $\a_{\infty}\in\Mn{m}{\co}$ such that 
\begin{equation}\label{eq:alconvnosettinggeneralised}
\|\a(t)-\a_{\infty}\|\leq C_{\romn}e^{-\eta_{\romn}t}
\end{equation}
for all $t\geq 0$; then (\ref{eq:thesystemRge}) is said to be such that \textit{the dominant coefficients are convergent}. 
\index{Dominant coefficients!convergent}%
\index{Convergent!dominant coefficients}%
\end{definition}
\begin{remark}\label{remark:geometrictonongeometricnoise}
Assume that (\ref{eq:thesystemRge}) is $C^{2}$-balanced with a geometric dominant noisy spatial direction, convergent
dominant coefficients and a negligible shift vector field; cf. Definition~\ref{def:noisemainassumptions}. Then the conditions of 
Definition~\ref{def:domnospdir} are satisfied. This is a consequence of Lemmas~\ref{lemma:geometrictoanalyticnoise}
and \ref{lemma:geometrictoanalyticnoiseRie}. 
\end{remark}

In order not to have to discuss the two cases described in Definition~\ref{def:domnospdir} separately, it is convenient to introduce
the following terminology. 

\begin{definition}\label{def:nuronEFnindexset}
Assume that (\ref{eq:thesystemRge}) has a dominant noisy spatial direction. Define
\begin{equation}\label{eq:nurondef}
\nu_{\ron}(\indexnot):=\lim_{t\rightarrow\infty}e^{-\b_{\ron}t}\mfg(\indexnot,t),
\end{equation}
where $\b_{\ron}$ is given by Definition~\ref{def:domnospdir}. Then the set $\EFnindexset$ is defined to consist of the 
$\indexnot\in\EFindexset$ such that $\nu_{\ron}(\indexnot)\neq 0$. 
\end{definition}
\begin{remark}\label{remark:nuronindlimex}
Due to Definition~\ref{def:domnospdir}, the limit on the right hand side of (\ref{eq:nurondef}) exists. 
\end{remark}

In what follows, we restrict our attention to balanced equations such that the shift components are negligible. 

\begin{definition}\label{def:strongbalneglshift}
Consider (\ref{eq:thesystemRge}). Assume the associated metric to be such that $(M,g)$ is a canonical separable cosmological model
manifold. Then (\ref{eq:thesystemRge}) is said to be \textit{strongly balanced} 
\index{Strongly!balanced}%
\index{Balanced!strongly}%
if there is a constant $C$ such that
\begin{equation*}
\begin{split}
 & |\dot{\ell}(\indexnot,t)|+|\ddot{\ell}(\indexnot,t)|+|\sigma(\indexnot,t)|+|\dot{\sigma}(\indexnot,t)|
+\|X(\indexnot,t)\|\\
 & +\|\dot{X}(\indexnot,t)\|+\|\dot{\a}(t)\|+\|\a(t)\|+\|\dot{\zeta}(t)\|+\|\zeta(t)\| \leq C
\end{split}
\end{equation*}
for all $t\geq 0$ and all $0\neq\indexnot\in\EFindexset$, where $\ell$, $\sigma$ and $X$ are defined by (\ref{eq:ellsigmaXgenRdef}). 
If there are constants $C$ and $\eta_{\rosh}>0$ such that 
\begin{equation}\label{eq:sigmasigmadotbdneglshiftvectornoise}
|\sigma(\indexnot,t)|+|\dot{\sigma}(\indexnot,t)|\leq Ce^{-\eta_{\rosh}t}
\end{equation}
for all $t\geq 0$ and all $0\neq\indexnot\in\EFindexset$, then (\ref{eq:thesystemRge}) is said to have a \textit{negligible shift vector field}.
\index{Negligible!shift vector field}%
\index{Shift vector field!negligible}%
\end{definition}
\begin{remark}\label{remark:strongbalanceneglshiftnoisegeometric}
Consider (\ref{eq:thesystemRge}). Assume the associated metric to be such that $(M,g)$ is a canonical separable cosmological model
manifold. Assume, moreover, that (\ref{eq:thesystemRge}) is $C^{2}$-balanced and that there are constants $C_{\rosh},\eta_{\rosh}>0$ such that 
\[
|\chi(t)|_{\bge}+|\dot{\chi}(t)|_{\bge}\leq C_{\rosh}e^{-\eta_{\rosh}t}
\] 
for all $t\geq 0$. Then the conditions of Definition~\ref{def:strongbalneglshift} are satisfied. This is a consequence of 
Lemmas~\ref{lemma:condyieldellderbd} and \ref{lemma:sigmaXbdsandderbds}.
\end{remark}

The following objects will play a central role in the analysis

\begin{definition}\label{def:tXinfrondefetc}
Assume that (\ref{eq:thesystemRge}) has a dominant noisy spatial direction and is such that the dominant coefficients are convergent. 
If (\ref{eq:thesystemRge}) has a dominant noisy spatial torus direction, say $j$, and $\indexnot\in\EFnindexset$, define
\[
\tX_{\infty,\ron}:=(g^{jj}_{\infty})^{-1/2}X^{j}_{\infty},\ \ \
\sgn_{\ron}(\indexnot):=\frac{n_{j}}{|n_{j}|}.
\]
If (\ref{eq:thesystemRge}) has a dominant noisy spatial generalised direction and $\indexnot\in\EFnindexset$, define
\[
\tX_{\infty,\ron}:=0,\ \ \
\sgn_{\ron}(\indexnot):=1.
\]
Define $\bX_{\ron}(\indexnot):=\sgn_{\ron}(\indexnot)\tX_{\infty,\ron}$. Moreover, 
\begin{equation}\label{eq:Rronpmpmdef}
R_{\ron,\pm}^{+}:=\pm\frac{1}{2}(-\a_{\infty}+\b_{\ron}\Id_{m}+\tX_{\infty,\ron}),\ \  \
R_{\ron,\pm}^{-}:=\pm\frac{1}{2}(-\a_{\infty}+\b_{\ron}\Id_{m}-\tX_{\infty,\ron}).
\end{equation}
Finally, 
\begin{align}
\kappa_{\ron,\pm} := & \max\{\kappa_{\max}(R_{\ron,\pm}^{+}),\kappa_{\max}(R_{\ron,\pm}^{-})\},\label{eq:kapparonpmdef}\\
\kappa_{\ron} := & \kappa_{\ron,+}+\kappa_{\ron,-}.\label{eq:kapparondef}
\end{align}
\end{definition}
\begin{remark}\label{remark:kapparonetcdep}
The quantities $R_{\ron,\pm}^{+}$, $R_{\ron,\pm}^{-}$ and $\kappa_{\ron,\pm}$ are all independent of $\indexnot$. 
\end{remark}

\section{Oscillation adaptedness and definition of the late time regime}\label{section:oscaddelatiregnoise}

If the conditions of Definition~\ref{def:strongbalneglshift} are satisfied, then (\ref{eq:thesystemRge})
is oscillation adapted; cf. Definition~\ref{def:oscad}. In fact, the following holds.

\begin{lemma}\label{lemma:strongbalneglshiftimploscad}
Assume that (\ref{eq:thesystemRge}) is strongly balanced and has a negligible shift vector field. Then (\ref{eq:thesystemRge})
is oscillation adapted; cf. Definition~\ref{def:oscad}. Moreover, 
\begin{equation}\label{eq:mffXodeshdefnoise}
\mff_{X}(t):=C,\ \ \ \mff_{\roode}(t):=C,\ \ \ \mff_{\rosh}(t):=Ce^{-\eta_{\rosh}t},
\end{equation}
where $C>0$ only depends on the coefficients of the equation (\ref{eq:thesystemRge}). 
\end{lemma}
\begin{remark}\label{remark:deponcoefincldeponRmfds}
Dependence on the coefficients includes dependence on the Riemannian manifolds $(M_{r},g_{r})$, $r=1,\dots,R$. For future reference, 
it is of interest to keep in mind that one particular consequence of this observation is that dependence on the minimum of 
$\nu_{\ron}(\indexnot)$ for $\indexnot\in\EFnindexset$ is included in dependence on the coefficients of the equation (\ref{eq:thesystemRge}). 
\end{remark}

The following observation is also of interest in what follows. 

\begin{lemma}\label{lemma:mfgnuronindnotrelerror}
Assume that (\ref{eq:thesystemRge}) has a dominant noisy spatial direction. Then, if 
$\indexnot\in\EFnindexset$,
\begin{equation}\label{eq:mfgnuronebntrel}
\left|\frac{\mfg(\indexnot,t)}{\nu_{\ron}(\indexnot)e^{\b_{\ron}t}}-1\right|\leq K_{\ron}|\nu(\indexnot)|^{2}e^{-\eta_{\romar}t}
\end{equation}
for $t\geq 0$, where $K_{\ron}\geq 1$ only depends on the coefficients of the equation (\ref{eq:thesystemRge}). 
\end{lemma}
\begin{proof}
Since (\ref{eq:thesystemRge}) has a dominant noisy spatial direction,
\[
|\mfg^{2}(\indexnot,t)-\nu_{\ron}^{2}(\indexnot)e^{2\b_{\ron}t}|\leq C|\nu(\indexnot)|^{2}e^{(2\b_{\ron}-\eta_{\romar})t},
\]
where $C$ only depends on the coefficients of the equation (\ref{eq:thesystemRge}). For $\indexnot\in\EFnindexset$, this estimate implies
that (\ref{eq:mfgnuronebntrel}) holds. 
\end{proof}

In order to be able to draw conclusions, given assumptions of the type introduced in Definition~\ref{def:domnospdir}, we need to restrict our
attention to modes such that $\nu_{\ron}(\indexnot)\neq 0$. Moreover, for a given $\indexnot\in\EFnindexset$, we need 
to focus on a time interval, say $[T_{\ron},\infty)$ in which $\nu_{\ron}^{2}(\indexnot)e^{2\b_{\ron}t}$ is the dominant part of $\mfg^{2}(\indexnot,t)$. 
Finally, we wish $T_{\ron}$ to be such that a time sequence of the type introduced in Definition~\ref{def:tkdefge} can be defined on $[T_{\ron},\infty)$;
the reason for this is that we wish to use the results of Chapter~\ref{chapter:destopofoscref} in order to analyse the asymptotics. Next we 
demonstrate that such a $T_{\ron}$ can be defined.

\begin{definition}\label{def:Tron}
Assume that (\ref{eq:thesystemRge}) is strongly balanced, has a negligible shift vector field, a dominant noisy spatial direction and is such
that the dominant coefficients are convergent. Let $K_{\ron}$ be the constant appearing in 
the estimate (\ref{eq:mfgnuronebntrel}). Given $\indexnot\in\EFnindexset$, define $T_{\ron,1}$ by 
\begin{equation}\label{eq:Trononedef}
K_{\ron}|\nu(\indexnot)|^{2}e^{-\eta_{\romar}T_{\ron,1}}=\frac{1}{2}.
\end{equation}
Let $\ellderbd$ be the constant appearing in (\ref{eq:estdtlnge}) and (\ref{eq:mffXdotbdetc}) (that these estimates hold is a consequence of 
Lemma~\ref{lemma:strongbalneglshiftimploscad}). Given $\indexnot\in\EFnindexset$, define $T_{\ron,2}$ by 
\begin{equation}\label{eq:Trontwodef}
\nu_{\ron}(\indexnot)e^{\b_{\ron}T_{\ron,2}}=2\max\{4\pi\ellderbd,2,\pi(\kappa_{\ron}+1)\}.
\end{equation}
Define
\begin{equation}\label{eq:Trondef}
T_{\ron}:=\max\{T_{\ron,1},T_{\ron,2},0\}. 
\end{equation}
\end{definition}
Given this terminology, the following statements hold. 
\begin{lemma}\label{lemma:mfgnuronebrontequiv}
Assume that (\ref{eq:thesystemRge}) is strongly balanced, has a negligible shift vector field, a dominant noisy spatial direction
and is such that the dominant coefficients are convergent. Given $\indexnot\in\EFnindexset$, 
define $T_{\ron}$ by (\ref{eq:Trondef}). Then 
\begin{equation}\label{eq:condensurtimeseqwd}
\mfg(\indexnot,t)\geq \max\{4\pi\ellderbd,2\}
\end{equation}
for all $t\geq T_{\ron}$. Moreover, 
\begin{equation}\label{eq:mfgnunebnexpdec}
\left|\frac{\mfg(\indexnot,t)}{\nu_{\ron}(\indexnot)e^{\b_{\ron}t}}-1\right|\leq\frac{1}{2}e^{-\eta_{\romar}\bt},\ \ \
\left|\frac{\nu_{\ron}(\indexnot)e^{\b_{\ron}t}}{\mfg(\indexnot,t)}-1\right|\leq e^{-\eta_{\romar}\bt}
\end{equation}
for all $t\geq T_{\ron}$, where $\bt:=t-T_{\ron}$. 
\end{lemma}
\begin{remark}
Due to (\ref{eq:condensurtimeseqwd}), it is clear that we can start a sequence $\{t_{k}\}$ of the type introduced in Definition~\ref{def:tkdefge} 
at $t_{0}:=T_{\ron}$. Moreover, $t_{\fin}=\infty$, where $t_{\fin}$ is given by Definition~\ref{def:tkdefge}. 
\end{remark}
\begin{proof}
Due to (\ref{eq:mfgnuronebntrel}), (\ref{eq:Trononedef}) and (\ref{eq:Trondef}), 
\begin{equation}\label{eq:mfgnunebneq}
\frac{1}{2}\leq \frac{\mfg(\indexnot,t)}{\nu_{\ron}(\indexnot)e^{\b_{\ron}t}}\leq 2
\end{equation}
for all $t\geq T_{\ron}$. Combining this estimate with (\ref{eq:Trontwodef}) and (\ref{eq:Trondef}) yields the conclusion that 
(\ref{eq:condensurtimeseqwd}) holds for all $t\geq T_{\ron}$. Combining (\ref{eq:mfgnuronebntrel}),
(\ref{eq:Trononedef}), (\ref{eq:Trondef}) and (\ref{eq:mfgnunebneq}) yields (\ref{eq:mfgnunebnexpdec}). 
\end{proof}

\section[Detailed asymptotics of the matrices]{Detailed asymptotics of the matrices 
appearing in the iteration}\label{section:detailedasfomaappinit}

Next, let us estimate the constituents of $R^{ab}_{\pre,k}$; cf. (\ref{eq:Rabprekdef}) and (\ref{eq:Xikpreappdefstfin}). 

\begin{lemma}\label{lemma:approxconstofRabprek}
Assume that (\ref{eq:thesystemRge}) is strongly balanced, has a negligible shift vector field, a dominant noisy spatial direction, and 
is such that the dominant coefficients are convergent. Let $\indexnot\in\EFnindexset$ and define $T_{\ron}$ by (\ref{eq:Trondef}). Then 
\begin{equation}\label{eq:alphadividedbymfgasnossett}
\left\|\frac{\a(t)}{\mfg(\indexnot,t)}-\frac{\a_{\infty}}{\nu_{\ron}(\indexnot)e^{\b_{\ron}t}}\right\|
\leq \frac{Ce^{-\eta_{1}\bt}}{\nu_{\ron}(\indexnot)e^{\b_{\ron}t}}
\end{equation}
for all $t\geq T_{\ron}$, where $C$ only depends on the coefficients of the equation (\ref{eq:thesystemRge}) and 
$\eta_{1}:=\min\{\eta_{\romar},\eta_{\romn}\}$. Moreover, 
\begin{equation}\label{eq:elldotdividedbymfgasnossett}
\left|\frac{\dot{\ell}(\indexnot,t)}{\mfg(\indexnot,t)}-\frac{\b_{\ron}}{\nu_{\ron}(\indexnot)e^{\b_{\ron}t}}\right|
\leq  \frac{Ce^{-\eta_{\romar}\bt}}{\nu_{\ron}(\indexnot)e^{\b_{\ron}t}}
\end{equation}
for all $t\geq T_{\ron}$, where $C$ only depends on the coefficients of the equation (\ref{eq:thesystemRge}). Finally
\begin{equation}\label{eq:nkXkbygsqminlimesttorusstmt}
\left\|\frac{X(\indexnot,t)}{\mfg(\indexnot,t)}-\frac{\bX_{\ron}(\indexnot)}{\nu_{\ron}(\indexnot)e^{\b_{\ron}t}}\right\|
\leq C\frac{e^{-\eta_{2}\bt}}{\nu_{\ron}(\indexnot)e^{\b_{\ron}t}}
\end{equation}
for all $t\geq T_{\ron}$, where $C$ only depends on the coefficients of the equation (\ref{eq:thesystemRge}) and 
\begin{equation}\label{eq:etatwodefnoise}
\eta_{2}:=\min\{\eta_{\romn},\eta_{\romar}/2\}.
\end{equation}
\end{lemma}
\begin{proof}
That (\ref{eq:alphadividedbymfgasnossett}) holds is a consequence of (\ref{eq:Xjalconvnosetting}), (\ref{eq:alconvnosettinggeneralised}), 
(\ref{eq:mfgnunebnexpdec}) and the fact that the equation is strongly balanced. 

\textbf{Estimating $\dot{\ell}/\mfg$.} Note that 
\begin{equation*}
\begin{split}
\left|\d_{t}\mfg^{2}-2\b_{\ron}\nu_{\ron}^{2}e^{2\b_{\ron}t}\right|
 = & \left|\textstyle{\sum}_{j,l=1}^{d}\d_{t}g^{jl}(t)n_{j}n_{l}+\sum_{r=1}^{R}\d_{t}(a_{r}^{-2})(t)\nu_{r,i_{r}}^{2}-2\b_{\ron}\nu_{\ron}^{2}e^{2\b_{\ron}t}\right|\\
 \leq & C|\nu|^{2}e^{(2\b_{\ron}-\eta_{\romar})t}
\end{split}
\end{equation*}
for all $t\geq 0$, where $C$ only depends on the coefficients of the equation (\ref{eq:thesystemRge}) and we have appealed to the 
assumptions of Definition~\ref{def:domnospdir}. Thus, 
\[
\left|\frac{\d_{t}\mfg^{2}}{\nu_{\ron}^{3}e^{3\b_{\ron}t}}-\frac{2\b_{\ron}}{\nu_{\ron}e^{\b_{\ron}t}}\right|\leq \frac{C}{\nu_{\ron}e^{\b_{\ron}t}}
|\nu|^{2}e^{-\eta_{\romar}t}
\]
for all $t\geq 0$, 
where $C$ only depends on the coefficients of the equation (\ref{eq:thesystemRge}); recall Remark~\ref{remark:deponcoefincldeponRmfds}. 
Keeping (\ref{eq:Trononedef}) and (\ref{eq:Trondef}) in mind, this estimate yields
\begin{equation}\label{eq:prelldotasnoise}
\left|\frac{\d_{t}\mfg^{2}}{\nu_{\ron}^{3}e^{3\b_{\ron}t}}-\frac{2\b_{\ron}}{\nu_{\ron}e^{\b_{\ron}t}}\right|\leq 
\frac{Ce^{-\eta_{\romar}\bt}}{\nu_{\ron}e^{\b_{\ron}t}},
\end{equation}
for all $t\geq T_{\ron}$, where $C$ only depends on the coefficients of the equation (\ref{eq:thesystemRge}). Estimate
\begin{equation*}
\begin{split}
\left|\frac{\dot{\ell}}{\mfg}-\frac{\b_{\ron}}{\nu_{\ron}e^{\b_{\ron}t}}\right|
 = & \left|\frac{1}{2\mfg^{3}}\d_{t}\mfg^{2}-\frac{\b_{\ron}}{\nu_{\ron}e^{\b_{\ron}t}}\right|\\
 \leq & \left|\frac{\nu_{\ron}^{3}e^{3\b_{\ron}t}}{2\mfg^{3}}
\left(\frac{\d_{t}\mfg^{2}}{\nu_{\ron}^{3}e^{3\b_{\ron}t}}-\frac{2\b_{\ron}}{\nu_{\ron}e^{\b_{\ron}t}}\right)\right|+
\left|\frac{\nu_{\ron}^{3}e^{3\b_{\ron}t}}{2\mfg^{3}}\frac{2\b_{\ron}}{\nu_{\ron}e^{\b_{\ron}t}}-\frac{\b_{\ron}}{\nu_{\ron}e^{\b_{\ron}t}}\right|.
\end{split}
\end{equation*}
Due to (\ref{eq:mfgnunebneq}) and (\ref{eq:prelldotasnoise}), we can estimate the first term on the right hand side. Due to 
(\ref{eq:mfgnunebnexpdec}) and (\ref{eq:mfgnunebneq}) we can estimate the second term on the right hand side. Summing up yields
(\ref{eq:elldotdividedbymfgasnossett}). 

\textbf{Estimating $X/\mfg$.} Finally, let us turn to the expression $X/\mfg$ appearing in (\ref{eq:Rabprekdef}). Assume first that the equation 
(\ref{eq:thesystemRge}) has a dominant noisy spatial generalised direction. Fix a $k\in \{1,\dots,d\}$, assume that $n_{k}\neq 0$ and estimate 
\[
\left\|n_{k}X^{k}(t)\right\|\leq \left\|[g^{kk}(t)n_{k}^{2}]^{-1/2}n_{k}X^{k}(t)\right\|[g^{kk}(t)n_{k}^{2}]^{1/2}\leq C [g^{kk}(t)n_{k}^{2}]^{1/2}
\]
(no summation on $k$) for all $t\geq 0$, where we have appealed to the fact that equation is strongly balanced. Here $C$ only depends on the 
coefficients of the equation (\ref{eq:thesystemRge}). This estimate yields
\begin{equation}\label{eq:Xdivbymfgestgensett}
\begin{split}
\left\|\frac{X(t)}{\mfg(t)}\right\| \leq & C\sum_{k=1}^{d}\frac{[g^{kk}(t)n_{k}^{2}]^{1/2}}{\nu_{\ron}e^{\b_{\ron}t}}\frac{1}{\nu_{\ron}e^{\b_{\ron}t}}\\
 \leq & Ce^{-\eta_{\romar}t/2}|\nu|\frac{1}{\nu_{\ron}e^{\b_{\ron}t}}
\leq C\frac{e^{-\eta_{\romar}\bt/2}}{\nu_{\ron}e^{\b_{\ron}t}}
\end{split}
\end{equation}
for all $t\geq T_{\ron}$, where we have appealed to (\ref{eq:Trononedef}), (\ref{eq:Trondef}), (\ref{eq:mfgnunebneq}) and the assumptions of 
Definition~\ref{def:domnospdir}. Moreover, $C$ only depends on the coefficients of the equation (\ref{eq:thesystemRge}). Thus
(\ref{eq:nkXkbygsqminlimesttorusstmt}) holds in this setting. 

Assume now that (\ref{eq:thesystemRge}) has a dominant noisy spatial torus direction, say $j$. Then
\[
\tX_{\infty,\ron}=(g^{jj}_{\infty})^{-1/2}X^{j}_{\infty},\ \ \
\sgn_{\ron}(\indexnot)=\frac{n_{j}}{|n_{j}|}.
\]
Estimate
\[
\left\|\frac{n_{k}X^{k}(t)}{\mfg^{2}(t)}-\frac{\bX_{\ron}}{\nu_{\ron}e^{\b_{\ron}t}}\right\|
\leq \sum_{k\neq j}\frac{\| n_{k}X^{k}(t)\|}{\mfg^{2}(t)}
+\left\|\frac{n_{j}X^{j}(t)}{\mfg^{2}(t)}-\frac{n_{j}X^{j}_{\infty}e^{\b_{\ron}t}}{\nu_{\ron}^{2}e^{2\b_{\ron}t}}\right\|
\] 
(no summation on $j$). The first term on the right hand side we can estimate by an argument similar to the one leading to 
(\ref{eq:Xdivbymfgestgensett}). The second term we can estimate by appealing to (\ref{eq:Xjalconvnosetting}) and (\ref{eq:mfgnunebnexpdec}).
Summing up yields
\begin{equation}\label{eq:nkXkbygsqminlimesttorus}
\left\|\frac{n_{k}X^{k}(t)}{\mfg^{2}(t)}-\frac{\bX_{\ron}}{\nu_{\ron}e^{\b_{\ron}t}}\right\|
\leq C\frac{e^{-\eta_{2}\bt}}{\nu_{\ron}e^{\b_{\ron}t}}
\end{equation}
for all $t\geq T_{\ron}$, where $C$ only depends on the coefficients of the equation (\ref{eq:thesystemRge}) and $\eta_{2}$ is given by 
(\ref{eq:etatwodefnoise}). Thus (\ref{eq:nkXkbygsqminlimesttorusstmt}) holds in this setting, and the lemma follows. 
\end{proof}

Next we wish to approximate the matrix $\Xi^{k,\pm}_{\fin}$ appearing in the statement of Lemma~\ref{lemma:wkfinlemma}. Before doing so,
recall (\ref{eq:Rronpmpmdef}) and define, for $\indexnot\in\EFnindexset$, 
\begin{equation}\label{eq:Rronpmdef}
R_{\ron}(\indexnot):=\left(\begin{array}{cc} R_{\ron,+}^{\upsilon(\indexnot)} & 0 \\ 0 & R_{\ron,+}^{-\upsilon(\indexnot)} \end{array}\right),
\end{equation}
where $\upsilon(\indexnot)=+$ if $\sgn_{\ron}(\indexnot)=1$ and $\upsilon(\indexnot)=-$ if $\sgn_{\ron}(\indexnot)=-1$. Similarly, 
$-\upsilon(\indexnot)=-$ if $\sgn_{\ron}(\indexnot)=1$ and $-\upsilon(\indexnot)=+$ if $\sgn_{\ron}(\indexnot)=-1$. Note that the 
only effect of varying $\indexnot$ in (\ref{eq:Rronpmdef}) is that the blocks appearing on the right hand side might change place. 
For future reference, it is also of interest to note that
\[
R_{\ron}(\indexnot)=\frac{1}{2}\left(\begin{array}{cc} -\a_{\infty}+\b_{\ron}\Id_{m}+\bX_{\ron}(\indexnot) & 0 \\ 
0 & -\a_{\infty}+\b_{\ron}\Id_{m}-\bX_{\ron}(\indexnot) \end{array}\right).
\]

\begin{lemma}\label{lemma:Akpmasexpnoise}
Assume that (\ref{eq:thesystemRge}) is strongly balanced, has a negligible shift vector field, a dominant noisy spatial direction, and is 
such that the dominant coefficients are convergent. Let $\indexnot\in\EFnindexset$ and define $T_{\ron}$ by (\ref{eq:Trondef}). Let $\{t_{k}\}$ 
be the time sequence given by Definition~\ref{def:tkdefge} and the condition that $t_{0}:=T_{\ron}$. Then
\begin{equation}\label{eq:Akpmestnoise}
\left\|A_{k}^{\pm}-\Id_{2m}\mp\frac{2\pi}{\nu_{\ron}(\indexnot)e^{\b_{\ron}t_{k}}}R_{\ron}(\indexnot)\right\|
\leq \frac{Ce^{-\eta_{A}\bt_{k}}}{\nu_{\ron}(\indexnot)e^{\b_{\ron}t_{k}}}
\end{equation}
for $k\geq 0$ in the case of a plus sign and $k\geq 1$ in the case of a minus sign, where $A_{k}^{\pm}:=\Xi^{k,\pm}_{\fin}$; $\Xi^{k,\pm}_{\fin}$ 
is introduced in the statement of Lemma~\ref{lemma:wkfinlemma}; $\bt_{k}:=t_{k}-T_{\ron}$;
\begin{equation}\label{eq:etaAdefinition}
\eta_{A}:=\min\{\eta_{\romar}/2,\eta_{\rosh},\eta_{\romn},\b_{\ron}\};
\end{equation}
and $C$ only depends on the coefficients of the equation (\ref{eq:thesystemRge}); cf. Remark~\ref{remark:deponcoefincldeponRmfds}.
\end{lemma}
\begin{remark}
In order to obtain the conclusions for $A^{-}_{k}$ for $k\geq 1$, we have appealed to Remark~\ref{remark:extvalXipmfin}. 
\end{remark}
\begin{proof}
Appealing to Lemma~\ref{lemma:approxconstofRabprek}, 
\[
\left\|R^{ab}_{\pre,k}-\frac{1}{\nu_{\ron}(\indexnot)e^{\b_{\ron}t_{k}}}\left[(-1)^{a}\a_{\infty}+\b_{\ron}\Id_{m}
+(-1)^{b}\sgn_{\ron}(\indexnot)\tX_{\infty,\ron}\right]\right\|\leq \frac{Ce^{-\eta_{2}\bt_{k}}}{\nu_{\ron}(\indexnot)e^{\b_{\ron}t_{k}}}
\]
for $t_{k}\geq T_{\ron}$, where $C$ only depends on the coefficients of the equation (\ref{eq:thesystemRge}) and $\eta_{2}$ is given
by (\ref{eq:etatwodefnoise}). Combining this observation with (\ref{eq:Xikpreappdefstfin}) and (\ref{eq:Rronpmdef}) yields
\begin{equation}\label{eq:Xikpmfinapp}
\left\|\Xi^{k,\pm}_{\fin,\app}-\Id_{2m}\mp\frac{2\pi}{\nu_{\ron}(\indexnot)e^{\b_{\ron}t_{k}}}R_{\ron}(\indexnot)\right\|
\leq \frac{Ce^{-\eta_{2}\bt_{k}}}{\nu_{\ron}(\indexnot)e^{\b_{\ron}t_{k}}}
\end{equation}
for $t_{k}\geq T_{\ron}$, where $C$ only depends on the coefficients of the equation (\ref{eq:thesystemRge}). As a final step, we 
need to appeal to (\ref{eq:XikXikIpreestfin}). We therefore need to estimate the right hand side of this estimate. Note, to begin with, 
that 
\begin{equation}\label{eq:varekestnoise}
\vare_{k}\leq C\frac{e^{-\eta_{\rosh}t_{k}}}{\mfg(t_{k})}+\frac{C}{\mfg^{2}(t_{k})},
\end{equation}
where $C$ only depends on the coefficients of the equation (\ref{eq:thesystemRge}). This is due to the fact that $\vare_{k}$ is given by
the right hand side of (\ref{eq:vareadef}), with $t_{a}$ replaced by $t_{k}$, and Lemma~\ref{lemma:strongbalneglshiftimploscad}. In particular,
it is thus clear that $\ellderbd$, $\vare_{k}$ and $\mff_{\rosh}(t_{k})$ are bounded by constants depending only on the coefficients of the 
equation (\ref{eq:thesystemRge}). Combining this observation with (\ref{eq:XikXikIpreestfin}), (\ref{eq:Xikpmfinapp}) and 
(\ref{eq:varekestnoise}) yields the conclusion of the lemma. 
\end{proof}

\section{Reformulating the iteration}\label{section:reformeqsnoise}

The purpose of the present section is to reformulate the iteration given by (\ref{eq:wprekpertransvarpi}) in such a way that it is easier 
to derive estimates. 

\textbf{Transforming to generalised Jordan normal form.} Fix $\e>0$ and let $T_{\ron,\e}$ be such that 
\[
J_{\ron,\e}:=T_{\ron,\e}^{-1}R_{\ron}(\indexnot)T_{\ron,\e}
\]
is in generalised Jordan normal form with the non-zero off-diagonal components equal to $\e$; cf. Remark~\ref{remark:genJordblock}. Due to the 
remarks made in connection with (\ref{eq:Rronpmdef}), 
it is clear that $T_{\ron,\e}$ can be chosen so that $J_{\ron,\e}$ is independent of $\indexnot$. Moreover, even though $T_{\ron,\e}$ depends on 
$\indexnot$, the difference between different $T_{\ron,\e}$'s corresponds to a permutation of the blocks of which $R_{\ron}(\indexnot)$ consists. 
Therefore the norms of $T_{\ron,\e}$ and $T_{\ron,\e}^{-1}$ are independent of $\indexnot$. Introducing
\begin{equation}\label{eq:Akjendefnoise}
A_{k,\ron,\e}^{\pm}:=T_{\ron,\e}^{-1}A_{k}^{\pm}T_{\ron,\e},
\end{equation}
the following relations hold:
\begin{align}
T_{\ron,\e}^{-1}A_{k_{1}}^{+}\cdots A_{k_{0}}^{+}T_{\ron,\e} = & A_{k_{1},\ron,\e}^{+}\cdots A_{k_{0},\ron,\e}^{+},\label{eq:Akdiagprodplusnoise}\\
T_{\ron,\e}^{-1}A_{k_{0}}^{-}\cdots A_{k_{1}}^{-}T_{\ron,\e} = & A_{k_{0},\ron,\e}^{-}\cdots A_{k_{1},\ron,\e}^{-}.\label{eq:Akdiagprodminusnoise}
\end{align}
Moreover, (\ref{eq:Akpmestnoise}) yields
\[
\|A_{k,\ron,\e}^{\pm}-A^{\app,\pm}_{k,\ron,\e}\|\leq \frac{C_{\e}e^{-\eta_{A}\bt_{k}}}{\nu_{\ron}(\indexnot)e^{\b_{\ron}t_{k}}},
\]
where $C_{\e}$ only depends on $\e$ and the coefficients of (\ref{eq:thesystemRge}). Finally,  
\[
A^{\app,\pm}_{k,\ron,\e}:=\Id_{2m}\pm\frac{2\pi}{\nu_{\ron}(\indexnot)e^{\b_{\ron}t_{k}}}J_{\ron,\e}.
\]

\textbf{Adapting to the leading order exponential behaviour.} Note that 
\begin{align}
e^{-\lambda (t_{1,+}-t_{0,+})}A_{k_{1},\ron,\e}^{+}\cdots A_{k_{0},\ron,\e}^{+} = & 
\hA_{k_{1},\ron,\e}^{+}\cdots \hA_{k_{0},\ron,\e}^{+},\label{eq:Akprodscaleredplusnoise}\\
e^{-\lambda (t_{0,-}-t_{1,-})}A_{k_{0},\ron,\e}^{-}\cdots A_{k_{1},\ron,\e}^{-} = & 
\hA_{k_{0},\ron,\e}^{-}\cdots \hA_{k_{1},\ron,\e}^{-},\label{eq:Akprodscaleredminusnoise}
\end{align}
where 
\[
t_{1,+}:=t_{k_{1}+1},\ \ \
t_{0,+}:=t_{k_{0}},\ \ \
t_{1,-}:=t_{k_{1}},\ \ \
t_{0,-}:=t_{k_{0}-1},
\]
$\lambda$ is a real number and 
\begin{equation}\label{eq:hAkjendefnoise}
\hA_{k,\ron,\e}^{\pm}:=e^{-\lambda (t_{k\pm 1}-t_{k})}A_{k,\ron,\e}^{\pm}.
\end{equation}
Moreover,
\[
\left|e^{-\lambda (t_{k\pm 1}-t_{k})}-1+\lambda (t_{k\pm 1}-t_{k})\right|\leq\frac{C_{\lambda}}{\mfg^{2}(\indexnot,t_{k})}
\]
where the constant $C_{\lambda}$ only depends on $\lambda$ and we have appealed to (\ref{eq:tatbroughge}); for future reference, 
it is of interest to keep in mind that a similar estimate holds when $\lambda$ is complex. Combining this estimate with 
(\ref{eq:tbmtagestge}) and (\ref{eq:mfgnunebnexpdec}) yields
\begin{equation}\label{eq:elambdadelttestnoise}
\left|e^{-\lambda (t_{k\pm 1}-t_{k})}-1\pm\frac{2\pi\lambda}{\nu_{\ron}(\indexnot)e^{\b_{\ron}t_{k}}}\right|\leq
\frac{C_{\lambda}e^{-\eta_{\romar}\bt_{k}}}{\nu_{\ron}(\indexnot)e^{\b_{\ron}t_{k}}}+\frac{C_{\lambda}}{\mfg^{2}(\indexnot,t_{k})},
\end{equation}
where $C_{\lambda}$ only depends on $\lambda$ and the coefficients of (\ref{eq:thesystemRge}); again, a similar estimate
holds when $\lambda$ is complex. Thus
\[
\|\hA_{k,\ron,\e}^{\pm}-\hA^{\app,\pm}_{k,\ron,\e}\|\leq \frac{C_{\e,\lambda}e^{-\eta_{A}\bt_{k}}}{\nu_{\ron}(\indexnot)e^{\b_{\ron}t_{k}}},
\]
where
\[
\hA^{\app,\pm}_{k,\ron,\e}:=\Id_{2m}\pm\frac{2\pi}{\nu_{\ron}(\indexnot)e^{\b_{\ron}t_{k}}}[J_{\ron,\e}-\lambda\Id_{2m}].
\]
Moreover, $C_{\e,\lambda}$ only depends on the coefficients of (\ref{eq:thesystemRge}), $\e$ and $\lambda$.

\textbf{Eliminating the imaginary parts.}
Before proceeding, it is of interest to eliminate the imaginary parts of the eigenvalues of $J_{\ron,\e}$. Note that $J_{\ron,\e}-\lambda\Id_{2m}$ consists 
of generalised Jordan blocks. Assume that the blocks have dimension $l_{1},\dots,l_{D}$. Let $\zeta_{r}$ be the imaginary part 
of the eigenvalue corresponding to the $r$'th generalised Jordan block. Define $S_{k}$ and $\chA_{k,\ron,\e}^{\pm}$ via
\begin{equation}\label{eq:ChAkjedefnoise}
S_{k}:=\mathrm{diag}\{e^{-i\zeta_{1}t_{k}}\Id_{l_{1}},\dots,e^{-i\zeta_{D}t_{k}}\Id_{l_{D}}\},\ \ \
\chA_{k,\ron,\e}^{\pm}:=S_{k\pm 1}\hA_{k,\ron,\e}^{\pm}S_{k}^{-1}.
\end{equation}
Note that $\|S_{k}\|=\|S_{k}^{-1}\|=1$ and that 
\begin{align}
S_{k_{1}+1}\hA_{k_{1},\ron,\e}^{+}\cdots \hA_{k_{0},\ron,\e}^{+}S_{k_{0}}^{-1} = & \chA_{k_{1},\ron,\e}^{+}\cdots \chA_{k_{0},\ron,\e}^{+},\label{eq:AkprodSkredplusnoise}\\
S_{k_{0}-1}\hA_{k_{0},\ron,\e}^{-}\cdots \hA_{k_{1},\ron,\e}^{-}S_{k_{1}}^{-1} = & \chA_{k_{0},\ron,\e}^{-}\cdots \chA_{k_{1},\ron,\e}^{-}.\label{eq:AkprodSkredminusnoise}
\end{align}
Moreover, 
\begin{equation}\label{eq:chAkmchAkappnoise}
\|\chA_{k,\ron,\e}^{\pm}-\chA^{\app,\pm}_{k,\ron,\e}\|\leq \frac{C_{\e,\lambda}e^{-\eta_{A}\bt_{k}}}{\nu_{\ron}(\indexnot)e^{\b_{\ron}t_{k}}}
\end{equation}
for all $k\geq 0$, where
\begin{equation}\label{eq:chAkjedefnoise}
\chA^{\app,\pm}_{k,\ron,\e}:=\Id_{2m}\pm\frac{2\pi}{\nu_{\ron}(\indexnot)e^{\b_{\ron}t_{k}}}[\mathrm{Re}\{J_{\ron,\e}\}-\lambda\Id_{2m}];
\end{equation}
this is a consequence of an estimate similar to (\ref{eq:elambdadelttestnoise}). Finally, the constant $C_{\e,\lambda}$ appearing in 
(\ref{eq:chAkmchAkappnoise}) only depends on $\e$, $\lambda$ and the coefficients of the equation (\ref{eq:thesystemRge}).

\section{A rough mode estimate}\label{section:aroughmodeestnoise}

As a first application of the framework developed above, we derive an estimate for the growth of the energy of solutions in the late time regime.

\begin{lemma}\label{lemma:windestnoise}
Assume that (\ref{eq:thesystemRge}) is strongly balanced, has a negligible shift vector field, a dominant noisy spatial direction, and is such that 
the dominant coefficients are convergent. Then there is a constant $C$, depending only on the coefficients of the equation (\ref{eq:thesystemRge}), 
such that the following holds. Let $\indexnot\in\EFnindexset$ and define $T_{\ron}$ by (\ref{eq:Trondef}). If $z$ is a solution to 
(\ref{eq:fourierthesystemRge}) and $w$ is defined by (\ref{eq:wdefshiftge}) and (\ref{eq:xyFodefshiftge}), then
\begin{equation}\label{eq:windtestfinnoisyerapartconv}
\begin{split}
|w(\indexnot,t)| \leq &  C\ldr{t-t_{a}}^{d_{\ron,+}-1}e^{\kappa_{\ron,+}(t-t_{a})}|w(\indexnot,t_{a})|\\
 & + \int_{t_{a}}^{t}C\ldr{t-t'}^{d_{\ron,+}-1}e^{\kappa_{\ron,+}(t-t')}|\hf(\indexnot,t')|dt'
\end{split}
\end{equation}
for all $T_{\ron}\leq t_{a}\leq t$. Here $\kappa_{\ron,+}$ is defined by (\ref{eq:kapparonpmdef}) and 
$d_{\ron,+}:=d_{\max}(R_{\ron}(\indexnot),\kappa_{\ron,+})$, where $R_{\ron}(\indexnot)$ is defined by (\ref{eq:Rronpmdef}); cf. 
Definition~\ref{def:SpRspdef}.
\end{lemma}
\begin{remark}
The constant $d_{\ron,+}$ is independent of $\indexnot\in\EFnindexset$; cf. the comments made in connection with (\ref{eq:Rronpmdef}).
\end{remark}
\begin{remark}
Dependence on the coefficients includes dependence on the Riemannian manifolds $(M_{r},g_{r})$, $r=1,\dots,R$; cf. 
Remark~\ref{remark:deponcoefincldeponRmfds}. 
\end{remark}

\begin{proof} 
\textbf{Reducing the problem.}
Recall the observations made in Subsection~\ref{ssection:estsolinhomeqprelobs}. Due to these observations, (\ref{eq:windtestfinnoisyerapartconv})
is a consequence of the estimate
\begin{equation}\label{eq:Phiindnottsnormestnoise}
\|\Phi_{\indexnot}(t;t')\|\leq C\ldr{t-t'}^{d_{\ron,+}-1}e^{\kappa_{\ron,+}(t-t')}
\end{equation}
for all $T_{\ron}\leq t'\leq t$, where $C$ only depends on the coefficients of the equation (\ref{eq:thesystemRge}). The remainder of the proof is 
devoted to demonstrating (\ref{eq:Phiindnottsnormestnoise}). In order to prove this estimate, it is sufficient to prove that 
\begin{equation}\label{eq:wPhiindnottsnormestnoise}
|w(\indexnot,t)|\leq C\ldr{t-t'}^{d_{\ron,+}-1}e^{\kappa_{\ron,+}(t-t')}|w(\indexnot,t')|
\end{equation}
for every solution to (\ref{eq:fourierthesystemRge}) with $\hf=0$ and all $T_{\ron}\leq t'\leq t$, where $C$ has the same dependence as in the case of 
(\ref{eq:Phiindnottsnormestnoise}).

\textbf{Notation.}
In the remainder of the proof, we use the notation introduced in Section~\ref{section:reformeqsnoise}. Recall (\ref{eq:kapparonpmdef}) and note that
$\kappa_{\ron,+}=\kappa_{\max}[R_{\ron}(\indexnot)]$, where $\kappa_{\ron,+}$ is independent of $\indexnot$; cf. Remark~\ref{remark:kapparonetcdep}. Let, 
moreover, 
\begin{equation}\label{eq:JRronepsilondef}
J_{R,\ron,\e}:=\mathrm{Re}\{J_{\ron,\e}\}-\kappa_{\ron,+}\Id_{2m},
\end{equation}
$J_{\tr}$ be the matrix obtained from $J_{R,\ron,\e}$ by setting the generalised Jordan blocks with zero eigenvalue to zero, and $J_{\rodiff}:=J_{R,\ron,\e}-J_{\tr}$.
Note that the diagonal components of $J_{R,\ron,\e}$ are elements of $[-\kappa_{\ron},0]$, where $\kappa_{\ron}$ is introduced in (\ref{eq:kapparondef}). 
Finally, comparing with Section~\ref{section:reformeqsnoise}, we take $\lambda$ to equal $\kappa_{\ron,+}$ from now on. 

\textbf{Estimates for times belonging to the sequence, preliminaries.} Fix an integer $k_{0}\geq 0$. To begin with, we are interested in proving that 
(\ref{eq:wPhiindnottsnormestnoise}) holds for $t'=t_{k_{0}}$ and $t=t_{k+1}$, where $0\leq k_{0}\leq k$. Consider (\ref{eq:wprekpertransvarpi}), where 
$\psi_{k}=w_{\fin,k}$, $A_{k}^{\pm}=\Xi^{k,\pm}_{\fin}$ and $\hF_{k}$ is replaced by $0$. This equality implies that
\begin{equation}\label{eq:rhokpoitnoise}
\rho_{k+ 1}=\bA_{k,\ron,\e}^{+}\rho_{k},
\end{equation}
where 
\begin{align}
\rho_{k} := & e^{-\kappa_{\ron,+}t_{k}}e^{-J_{\rodiff}\tildt_{k}}S_{k}T_{\ron,\e}^{-1}\psi_{k},\label{eq:rhokbAkdefnoise}\\
\bA_{k,\ron,\e}^{+} := & e^{-J_{\rodiff}\tildt_{k+1}}\chA_{k,\ron,\e}^{+}e^{J_{\rodiff}\tildt_{k}},\nonumber
\end{align}
$\tildt_{k}:=t_{k}-t_{k_{0}}$, $\lambda=\kappa_{\ron,+}$ and $S_{k}$ and $T_{\ron,\e}$ are defined as above. Thus, if $k\geq k_{0}$, 
\begin{equation}\label{eq:rhokpointfornoise}
\begin{split}
\rho_{k+1} = & \bA_{k,\ron,\e}^{+}\cdots \bA_{k_{0},\ron,\e}^{+}\rho_{k_{0}}.
\end{split}
\end{equation}
Compute
\begin{equation}\label{eq:bAauxexpnoise}
\bA_{k,\ron,\e}^{+}=e^{-J_{\rodiff}\tildt_{k+1}}(\chA_{k,\ron,\e}^{+}-\chA^{\app,+}_{k,\ron,\e})e^{J_{\rodiff}\tildt_{k}}
+e^{-J_{\rodiff}(t_{k+1}-t_{k})}\chA^{\app,+}_{k,\ron,\e};
\end{equation}
note that $J_{\rodiff}$ and $\chA^{\app,+}_{k,\ron,\e}$ commute. On the other hand, 
\begin{equation*}
\begin{split}
e^{-J_{\rodiff}(t_{k+1}-t_{k})}\chA^{\app,+}_{k,\ron,\e}
 = & \chA^{\app,+}_{k,\ron,\e}-J_{\rodiff}(t_{k+1}-t_{k})+\frac{2\pi}{\nu_{\ron}(\indexnot)e^{\b_{\ron}t_{k}}}J_{R,\ron,\e}[e^{-J_{\rodiff}(t_{k+1}-t_{k})}-\Id_{2m}]\\
 & +[e^{-J_{\rodiff}(t_{k+1}-t_{k})}-\Id_{2m}+J_{\rodiff}(t_{k+1}-t_{k})].
\end{split}
\end{equation*}
Introducing 
\begin{equation}\label{eq:bAkjappdefnoise}
\bA^{\app,+}_{k,\ron,\e}:=\Id_{2m}+\frac{2\pi}{\nu_{\ron}(\indexnot)e^{\b_{\ron}t_{k}}}J_{\tr}
\end{equation}
and appealing to (\ref{eq:tbmtagestge}) and (\ref{eq:mfgnunebnexpdec}) yields
\[
\|e^{-J_{\rodiff}(t_{k+1}-t_{k})}\chA^{\app,+}_{k,\ron,\e}-\bA^{\app,+}_{k,\ron,\e}\|\leq \frac{C_{\e}e^{-\eta_{A}\bt_{k}}}{\nu_{\ron}(\indexnot)e^{\b_{\ron}t_{k}}}
\]
for all $k\geq k_{0}$, where $C_{\e}$ only depends on $\e$ and the coefficients of the equation (\ref{eq:thesystemRge}). Combining this estimate 
with (\ref{eq:chAkmchAkappnoise}) and (\ref{eq:bAauxexpnoise}) yields
\begin{equation}\label{eq:bAkronepapprest}
\|\bA_{k,\ron,\e}^{+}-\bA^{\app,+}_{k,\ron,\e}\|\leq \frac{C_{\e}e^{-\eta_{A}\bt_{k}}}{\nu_{\ron}(\indexnot)e^{\b_{\ron}t_{k}}}\ldr{\tildt_{k}}^{N}
\end{equation}
for all $k\geq k_{0}$, where $C_{\e}$ only depends on $\e$ and the coefficients of the equation (\ref{eq:thesystemRge}), and $N$ is a non-negative 
integer depending only on $m$. 

\textbf{Estimating the matrix products.}
Consider (\ref{eq:bAkjappdefnoise}). By construction, all the diagonal elements of the second term on the right hand side are non-positive. 
Moreover, they are strictly smaller than $1$ in absolute value due to (\ref{eq:Trontwodef}). Finally, all non-trivial generalised Jordan 
blocks in $J_{\tr}$ have the property that the corresponding diagonal components are strictly negative. By choosing $\e$ sufficiently small,
the bound depending only on the coefficients of the equation, it can be ensured that 
\begin{equation}\label{eq:bappplkroneest}
\|\bA^{\app,+}_{k,\ron,\e}\|\leq 1
\end{equation}
for all $k\geq k_{0}$. Fixing such an $\e$, the estimates below are independent of $\e$. One particular consequence of (\ref{eq:bappplkroneest}) is that
for $0\leq k_{0}\leq k_{1}$,
\[
\|\bA_{k_{1},\ron,\e}^{+}\cdots\bA_{k_{0},\ron,\e}^{+}\|\leq (1+\|\bA_{k_{1},\ron,\e}^{+}-\bA^{\app,+}_{k_{1},\ron,\e}\|)
\cdots(1+\|\bA_{k_{0},\ron,\e}^{+}-\bA^{\app,+}_{k_{0},\ron,\e}\|).
\]
Since $\ln(1+x)\leq x$ for $x\geq 0$, the logarithm of the right hand side is bounded by 
\begin{equation*}
\begin{split}
\sum_{k=k_{0}}^{k_{1}}\|\bA_{k,\ron,\e}^{+}-\bA^{\app,+}_{k,\ron,\e}\| \leq & 
\sum_{k=k_{0}}^{k_{1}}\frac{Ce^{-\eta_{A}\bt_{k}}}{\nu_{\ron}(\indexnot)e^{\b_{\ron}t_{k}}}\ldr{\tildt_{k}}^{N}
\leq \int_{t_{k_{0}}}^{t_{k_{1}+1}}C\ldr{t-t_{k_{0}}}^{N}e^{-\eta_{A}\bt}dt\\
 \leq & Ce^{-\eta_{A}(t_{k_{0}}-T_{\ron})},
\end{split}
\end{equation*}
where $C$ only depends on the coefficients of the equation (\ref{eq:thesystemRge}). In order to obtain this estimate, we appealed to 
Lemma~\ref{lemma:gaintvarge}, (\ref{eq:mfgnunebneq}) and (\ref{eq:bAkronepapprest}). In particular, for $0\leq k_{0}\leq k$, 
\begin{equation}\label{eq:bAkprodnoise}
\|\bA_{k,\ron,\e}^{+}\cdots\bA_{k_{0},\ron,\e}^{+}\|\leq C,
\end{equation}
where $C$ only depends on the coefficients of the equation (\ref{eq:thesystemRge}). Combining this estimate with (\ref{eq:rhokpointfornoise}) yields
\begin{equation}\label{eq:rhokpobdnoise}
|\rho_{k+1}|\leq C|\rho_{k_{0}}|
\end{equation}
for all $k\geq k_{0}$, where the constant $C$ has the same dependence as in the case of (\ref{eq:bAkprodnoise}). Returning to the definition of $\rho_{k}$ in 
terms of $\psi_{k}$, cf. (\ref{eq:rhokbAkdefnoise}), and the definition of $\psi_{k}=w_{\fin,k}$, cf. (\ref{eq:wfinkdef}), this estimate yields
\begin{equation}\label{eq:wkposharpestimatenoise}
\begin{split}
|w(\indexnot,t_{k+1})| \leq & Ce^{\kappa_{\ron,+}\tildt_{k+1}}\ldr{\tildt_{k+1}}^{d_{\ron,+}-1}|w(\indexnot,t_{k_{0}})|,
\end{split}
\end{equation}
where the constant $C$ has the same dependence as in the case of (\ref{eq:bAkprodnoise}). Here $d_{\ron,+}$ is the largest dimension of a 
Jordan block of $R_{\ron}(\indexnot)$ corresponding to an eigenvalue with real part $\kappa_{\ron,+}$. Thus (\ref{eq:wPhiindnottsnormestnoise}) holds if 
$t'=t_{k_{0}}$ and $t=t_{k+1}$. 

\textbf{General $t'$ and $t$.} In general, $t'\in [t_{k_{0}},t_{k_{0}+1}]$ and $t\in [t_{k},t_{k+1}]$ for some $0\leq k_{0}\leq k$. Appealing to 
Lemma~\ref{lemma:roughenestbalsetting} in order to estimate the evolution between $t'$ and $t_{k_{0}}$ and between $t_{k}$ and $t$ yields
the conclusion that (\ref{eq:wPhiindnottsnormestnoise}) holds for all $t_{0}\leq t'\leq t$. Moreover, $C$ has the same dependence as in the case of 
(\ref{eq:bAkprodnoise}).
\end{proof}

\section{A rough Sobolev estimate}\label{section:roughsobolevestnoise}

Assume that (\ref{eq:thesystemRge}) is strongly balanced, has a negligible shift vector field, a dominant noisy spatial direction, and is 
such that the dominant coefficients are convergent. Due to, e.g., Lemma~\ref{lemma:windestnoise} above, it is then clear that we 
can draw conclusions concerning the behaviour for $\indexnot\in\EFnindexset$. What happens for $\indexnot\notin\EFnindexset$ is, however,
less clear. For this reason, it is natural to divide (\ref{eq:thesystemRge}) into two parts, one corresponding to $\EFnindexset$, and one 
corresponding to the complement of $\EFnindexset$. If $f\in C^{\infty}(M,\cn{m})$, it is therefore convenient to introduce the notation
\begin{equation}\label{eq:frondef}
f_{\ron}(p,t):=\textstyle{\sum}_{\indexnot\in\EFnindexset}\hf(\indexnot,t)\varphi_{\indexnot}(p),
\end{equation}
where $\varphi_{\indexnot}$ and $\hf$ are defined by (\ref{eq:varphinudef}) and (\ref{eq:hfnutdef}) respectively. Noting that $\mfe_{s}$ is defined by 
(\ref{eq:mfedef}), we have the following result.

\begin{lemma}
Assume that (\ref{eq:thesystemRge}) is strongly balanced, has a negligible shift vector field, a dominant noisy spatial direction, and is such 
that the dominant coefficients are convergent. Then there are constants $C$ and $s_{\ron}$, depending only on the coefficients of the equation 
(\ref{eq:thesystemRge}), such that if $u$ is a solution to (\ref{eq:thesystemRge}), then
\begin{equation}\label{eq:windtestfinnoisyerapartconvsob}
\begin{split}
\mfe_{s}^{1/2}[u_{\ron}](t) \leq &  C\ldr{t}^{d_{\ron,+}-1}e^{\kappa_{\ron,+}t}\mfe_{s+s_{\ron}}^{1/2}[u_{\ron}](0)\\
 & + \int_{0}^{t}C\ldr{t-t'}^{d_{\ron,+}-1}e^{\kappa_{\ron,+}(t-t')}\|f_{\ron}(\cdot,t')\|_{(s+s_{\ron})}dt'
\end{split}
\end{equation}
for all $t\geq 0$ and $s\in\ro$, where $f_{\ron}$ and $u_{\ron}$ are defined by (\ref{eq:frondef}) and the adjacent text. Finally, 
$\kappa_{\ron,+}$ and $d_{\ron,+}$ are defined as in the statement of Lemma~\ref{lemma:windestnoise}.
\end{lemma}
\begin{remark}
Dependence on the coefficients includes dependence on the Riemannian manifolds $(M_{r},g_{r})$, $r=1,\dots,R$; cf. 
Remark~\ref{remark:deponcoefincldeponRmfds}. 
\end{remark}
\begin{remark}\label{remark:roughsobestnoise}
Assume that (\ref{eq:thesystemRge}) is $C^{2}$-balanced with a geometric dominant noisy spatial direction, convergent dominant 
coefficients and a negligible shift vector field; cf. Definition~\ref{def:noisemainassumptions}. Then the assumptions of the 
lemma are satisfied; cf. Remarks~\ref{remark:geometrictonongeometricnoise} and \ref{remark:strongbalanceneglshiftnoisegeometric}.
\end{remark}
\begin{proof}
Due to Lemma~\ref{lemma:windestnoise}, we know that 
\begin{equation}\label{eq:windtestfinnoisyerapartconvsobwn}
\begin{split}
|w(\indexnot,t)| \leq &  C\ldr{\bt}^{d_{\ron,+}-1}e^{\kappa_{\ron,+}\bt}|w(\indexnot,T_{\ron})|
+\int_{T_{\ron}}^{t}C\ldr{t-t'}^{d_{\ron,+}-1}e^{\kappa_{\ron,+}(t-t')}|\hf(\indexnot,t')|dt'
\end{split}
\end{equation}
for all $\indexnot\in\EFnindexset$ and $t\geq T_{\ron}$, where $\bt:=t-T_{\ron}$ and $C$ only depends on the coefficients of the equation 
(\ref{eq:thesystemRge}). Since $\mfg(\indexnot,t)\geq 2$ for $t\geq T_{\ron}$, cf. (\ref{eq:condensurtimeseqwd}), it is clear that 
$\me^{1/2}(\indexnot,t)$ and $|w(\indexnot,t)|$ are equivalent for $t\geq T_{\ron}$, the constants of equivalence being numerical. 
One particular consequence of (\ref{eq:windtestfinnoisyerapartconvsobwn}) is thus that
\begin{equation}\label{eq:mestnoisycasetgeqTron}
\me^{1/2}(\indexnot,t)\leq C\ldr{\bt}^{d_{\ron,+}-1}e^{\kappa_{\ron,+}\bt}\me^{1/2}(\indexnot,T_{\ron})
+\int_{T_{\ron}}^{t}C\ldr{t-t'}^{d_{\ron,+}-1}e^{\kappa_{\ron,+}(t-t')}|\hf(\indexnot,t')|dt'
\end{equation}
for all $\indexnot\in\EFnindexset$ and $t\geq T_{\ron}$, where $C$ only depends on the coefficients of the equation 
(\ref{eq:thesystemRge}). On the other hand, Lemma~\ref{lemma:roughenestbalsetting} yields 
\begin{equation}\label{eq:westinitnoisyinterv}
\me^{1/2}(\indexnot,t)\leq e^{\eta_{\robal}t}\me^{1/2}(\indexnot,0)+\int_{0}^{t}e^{\eta_{\robal}(t-t')}|\hf(\indexnot,t')|dt'
\end{equation}
for all $t\geq 0$, where $\eta_{\robal}\geq 0$ only depends on the coefficients of the equation (\ref{eq:thesystemRge}). Letting 
\[
\eta_{\romod}:=\max\{\eta_{\robal}-\kappa_{\ron,+},0\},
\]
it is of interest to estimate
\[
\eta_{\romod}T_{\ron}\leq \frac{2\eta_{\romod}}{\eta_{\romar}}\ln |\nu(\indexnot)|+C,
\]
where $C$ only depends on the coefficients of the equation (\ref{eq:thesystemRge}); cf. Definition~\ref{def:Tron}. Combining this estimate with 
(\ref{eq:westinitnoisyinterv}) yields 
\begin{equation}\label{eq:westinitnoisyintervTronint}
\me^{1/2}(\indexnot,t)\leq Ce^{\kappa_{\ron,+}t}\ldr{\nu(\indexnot)}^{s_{\ron}}\me^{1/2}(\indexnot,0)
+C\ldr{\nu(\indexnot)}^{s_{\ron}}\int_{0}^{t}e^{\kappa_{\ron,+}(t-t')}|\hf(\indexnot,t')|dt'
\end{equation}
for all $0\leq t\leq T_{\ron}$, where $s_{\ron}:=2\eta_{\romod}/\eta_{\romar}$ and $C$ only depends on the coefficients of the equation 
(\ref{eq:thesystemRge}). Combining (\ref{eq:mestnoisycasetgeqTron}) and (\ref{eq:westinitnoisyintervTronint}) yields
\begin{equation*}
\begin{split}
\me_{s}^{1/2}(\indexnot,t) \leq &  C\ldr{t}^{d_{\ron,+}-1}e^{\kappa_{\ron,+}t}\me_{s+s_{\ron}}^{1/2}(\indexnot,0)
 + C\ldr{\nu(\indexnot)}^{s+s_{\ron}}\int_{0}^{t}\ldr{t-t'}^{d_{\ron,+}-1}e^{\kappa_{\ron,+}(t-t')}|\hf(\indexnot,t')|dt'
\end{split}
\end{equation*}
for all $t\geq 0$, $s\in\ro$ and $\indexnot\in\EFnindexset$, where $C$ only depends on the coefficients of the equation (\ref{eq:thesystemRge}). 
Appealing to Minkowski's inequality yields the conclusion of the lemma. 
\end{proof}

\chapter{Asymptotics for one mode and late times}

Given the framework developed in the previous chapter, we are in a position to derive information concerning the asymptotics. In the present chapter,
we focus on one single mode $\indexnot\in\EFnindexset$. Moreover, we only derive conclusions in the late time regime $[T_{\ron},\infty)$, where $T_{\ron}$ 
is introduced in Definition~\ref{def:Tron}. To begin with, we derive asymptotic information by analysing how the solution behaves along the time sequence 
$\{t_{k}\}$; this is the subject of Section~\ref{section:asalongtimeseq}. In Section~\ref{section:spdataatinfnoise} we then turn to the problem of 
specifying the asymptotic data. However, the considerations are, again, restricted to a time sequence in the interval $[T_{\ron},\infty)$. Finally, in
Section~\ref{section:contsolbeytimeseqnoise}, we extend the estimates derived along the time sequence to asymptotic estimates valid for all 
$t\geq T_{\ron}$. 

\section{Asymptotics along a time sequence}\label{section:asalongtimeseq}

Lemma~\ref{lemma:windestnoise} yields an estimate for the growth of the energy of one mode. In what follows, we want to derive more detailed information. 
On the interval $[T_{\ron},\infty)$, it is natural to carry out the analysis in two steps. First, we consider how the solution
evolves along the sequence $\{t_{k}\}$. Second, we analyse the development during one period. In this section begin by analysing the 
asymptotics along the time sequence. 

\begin{lemma}\label{lemma:asymposccasenoise}
Assume that (\ref{eq:thesystemRge}) is strongly balanced, has a negligible shift vector field, a dominant noisy spatial direction, and is such that the 
dominant coefficients are convergent. Let $\indexnot\in\EFnindexset$ and define $T_{\ron}$ by (\ref{eq:Trondef}). Let $\{t_{k}\}$ be the time sequence 
given by Definition~\ref{def:tkdefge} and the condition that $t_{0}:=T_{\ron}$. Let $\eta_{A}$ be defined by (\ref{eq:etaAdefinition}) and assume that 
there is an $0<\eta_{B}\leq \eta_{A}$ such that 
\[
\int_{0}^{\infty}e^{-(\kappa_{\ron,+}-\eta_{B})t}|\hf(\indexnot,t)|dt<\infty,
\]
where $\kappa_{\ron,+}$ is defined by (\ref{eq:kapparonpmdef}). Then there is a constant $C_{B}$, depending only on $\eta_{B}$ and the coefficients of 
the equation (\ref{eq:thesystemRge}); and a non-negative
integer $N$, depending only on $m$, such that the following holds. If $z$ is a solution to (\ref{eq:fourierthesystemRge}), then there is a 
$\psi_{\infty}$ belonging to the first generalised eigenspace of the $\eta_{B}$, $R_{\ron}(\indexnot)$-decomposition of $\cn{2m}$ (cf. 
Definition~\ref{def:fssubspetc}) and satisfying
\begin{equation}\label{eq:Psiinfestnoise}
|e^{R_{\ron}(\indexnot)T_{\ron}}\psi_{\infty}|\leq 
C_{B}\left(|w(\indexnot,T_{\ron})|+e^{\kappa_{\ron,+}T_{\ron}}\int_{T_{\ron}}^{\infty}e^{-(\kappa_{\ron,+}-\eta_{B})t}|\hf(\indexnot,t)|dt\right),
\end{equation}
such that 
\begin{equation}\label{eq:zzdottkestnoisestmt}
\begin{split}
 & \left|\left(\begin{array}{c} \mfg(\indexnot,t_{k})z(\indexnot,t_{k}) \\ \dot{z}(\indexnot,t_{k})\end{array}\right)-
\exp\left(i\int_{0}^{t_{k}}\sigma(\indexnot,t')\mfg(\indexnot,t')dt'\right)T_{\pre,k}^{-1}[D_{k}(\indexnot)]^{-1}e^{R_{\ron}(\indexnot)t_{k}}\psi_{\infty}\right|\\
 \leq & C_{B}\ldr{\bt_{k}}^{N}e^{(\kappa_{\ron,+}-\eta_{B})\bt_{k}}\left(|w(\indexnot,T_{\ron})|+e^{\kappa_{\ron,+}T_{\ron}}
\int_{0}^{\infty}e^{-(\kappa_{\ron,+}-\eta_{B})t}|\hf(\indexnot,t)|dt\right)
\end{split}
\end{equation}
holds for all $k\geq 0$, where $\bt_{k}:=t_{k}-t_{0}$. In these estimates, $w$ is given by (\ref{eq:wdefshiftge}) and (\ref{eq:xyFodefshiftge}); 
$\mfg$ is given by (\ref{eq:mfgnutdef});
$\sigma$ is given by (\ref{eq:ellsigmaXgenRdef}); $T_{\pre,k}$ is given by (\ref{eq:Tprekage}); $D_{k}(\indexnot)$ is given by (\ref{eq:mDkdef}); and
$R_{\ron}(\indexnot)$ is given by (\ref{eq:Rronpmdef}).  
\end{lemma}
\begin{remark}
The norms of the matrix 
\[
\exp\left(i\int_{0}^{t_{k}}\sigma(t')\mfg(t')dt'\right)T_{\pre,k}^{-1}D_{k}^{-1}
\]
and its inverse are bounded by constants depending only on the coefficients of the equation (\ref{eq:thesystemRge}). In particular, $\psi_{\infty}$ is 
thus uniquely determined by the estimate (\ref{eq:zzdottkestnoisestmt}) and the condition that it belong to the first generalised eigenspace of the 
$\eta_{B}$, $R_{\ron}(\indexnot)$-decomposition of $\cn{2m}$.
\end{remark}
\begin{remark}
In case all the Jordan blocks of $R_{\ron}(\indexnot)$ are trivial, the $N$ appearing in (\ref{eq:zzdottkestnoisestmt}) can be replaced by 
$1$. Moreover, if the first generalised eigenspace of the $\eta_{B}$, $R_{\ron}(\indexnot)$-decomposition of $\cn{2m}$ equals $\cn{2m}$, then 
the $N$ appearing in (\ref{eq:zzdottkestnoisestmt}) can be replaced by $d_{\ron,+}-1$, where $d_{\ron,+}:=d_{\max}(R_{\ron}(\indexnot),\kappa_{\ron,+})$.
\end{remark}
\begin{remark}
Dependence on the coefficients includes dependence on the Riemannian manifolds $(M_{r},g_{r})$, $r=1,\dots,R$; cf. 
Remark~\ref{remark:deponcoefincldeponRmfds}. 
\end{remark}
\begin{proof}
The proof is based on the reformulation of the iteration introduced in Section~\ref{section:reformeqsnoise}. Moreover, we here let $\lambda=\kappa_{\ron,+}$. 
Correspondingly, we introduce 
\begin{align}
\zeta_{k} := & e^{-\kappa_{\ron,+}t_{k}}S_{k}T_{\ron,\e}^{-1}\psi_{k},\label{eq:zetakdefnoise}\\
\chF_{k}(t) := & e^{-\kappa_{\ron,+}t_{k}}S_{k}T_{\ron,\e}^{-1}\hF_{k}(t),\label{eq:chFkdefnoise}
\end{align}
where $\psi_{k}=w_{\fin,k}$ and $\hF_{k}$ is given by (\ref{eq:Fkfindef}). As opposed to the proof of Lemma~\ref{lemma:windestnoise}, the value of $\e>0$ is
not important in the present proof. We therefore set it to $1$ here. Nevertheless, we keep the $\e$'s that appear in the notation of
Section~\ref{section:reformeqsnoise} in order not to have to introduce additional symbols. Given the definitions (\ref{eq:zetakdefnoise}) and 
(\ref{eq:chFkdefnoise}), equation (\ref{eq:wprekpertransvarpi}) yields
\begin{equation}\label{eq:zetakpoitnoise}
\zeta_{k+1}=\chA_{k,\ron,\e}^{+}\zeta_{k}+\chA_{k,\ron,\e}^{+}\int_{t_{k}}^{t_{k+1}}\chF_{k}(t)dt. 
\end{equation}
Due to (\ref{eq:chAkmchAkappnoise}) and (\ref{eq:chAkjedefnoise}), it is clear that 
\begin{equation}\label{eq:chAkronepminapproxest}
\|\chA_{k,\ron,\e}^{+}-\chA^{\app,+}_{k,\ron,\e}\|\leq \frac{Ce^{-\eta_{A}\bt_{k}}}{\nu_{\ron}(\indexnot)e^{\b_{\ron}t_{k}}},
\end{equation}
where $C$ only depends on the coefficients of the equation (\ref{eq:thesystemRge}); $\eta_{A}$ is given by (\ref{eq:etaAdefinition}); 
$\bt_{k}:=t_{k}-T_{\ron}=t_{k}-t_{0}$; and 
\[
\chA^{\app,+}_{k,\ron,\e}:=\Id_{2m}+\frac{2\pi}{\nu_{\ron}(\indexnot)e^{\b_{\ron}t_{k}}}J_{R,\ron,\e},
\]
where $J_{R,\ron,\e}$ is given by  (\ref{eq:JRronepsilondef}). Moreover, the estimate (\ref{eq:windtestfinnoisyerapartconv}) implies 
\begin{equation}\label{eq:zetakpobdnoise}
|\zeta_{k}|\leq C\ldr{\bt_{k}}^{d_{\ron,+}-1}|\zeta_{0}|+\int_{t_{0}}^{t_{k}}C\ldr{t_{k}-t'}^{d_{\ron,+}-1}e^{-\kappa_{\ron,+}t'}|\hf(\indexnot,t')|dt'
\end{equation}
for all $k\geq 0$, where $C$ only depends on the coefficients of the equation (\ref{eq:thesystemRge}). Let $\eta_{B}$ be as in the statement of the lemma. 
In the analysis of the future asymptotics 
of $\zeta_{k}$, it is convenient to divide the components of $\zeta_{k}$ into $\zeta_{k,a}$ and $\zeta_{k,b}$. Here $\zeta_{k,a}$ and $\zeta_{k,b}$ denote the 
first and second component in the $\eta_{B}$, $J_{R,\ron,\e}$-decomposition of $\zeta_{k}$ respectively; cf. Definition~\ref{def:fssubspetc}. Similarly, we 
denote the first and second blocks of the $\eta_{B}$, $J_{R,\ron,\e}$-decomposition of $J_{R,\ron,\e}$ by $J_{R,a}$ and $J_{R,b}$ respectively. Let us start by 
considering $\zeta_{k,a}$. 

\textbf{Estimating the first component.} With notation as described above, (\ref{eq:zetakpoitnoise}) yields
\begin{equation}\label{eq:zkpoemJritnoise}
e^{-J_{R,a}\bt_{k+1}}\zeta_{k+1,a}=e^{-J_{R,a}\bt_{k+1}}[\chA_{k,\ron,\e}^{+}\zeta_{k}]_{a}
+e^{-J_{R,a}\bt_{k+1}}\int_{t_{k}}^{t_{k+1}}[\chA_{k,\ron,\e}^{+}\chF_{k}(t)]_{a}dt,
\end{equation}
where $\chi_{a}$ is the first component in the $\eta_{B}$, $J_{R,\ron,\e}$-decomposition of the vector $\chi\in\cn{2m}$. Let us rewrite the first term 
on the right hand side according to 
\begin{equation}\label{eq:zkitftrefnoise}
\begin{split}
e^{-J_{R,a}\bt_{k+1}}[\chA_{k,\ron,\e}^{+}\zeta_{k}]_{a} = & e^{-J_{R,a}\bt_{k+1}}[\chA_{k,\ron,\e}^{\app,+}]_{a}\zeta_{k,a}\\
 & +e^{-J_{R,a}\bt_{k+1}}[(\chA_{k,\ron,\e}^{+}-\chA_{k,\ron,\e}^{\app,+})\zeta_{k}]_{a}\\
 = & e^{-J_{R,a}\bt_{k}}\zeta_{k,a}+e^{-J_{R,a}\bt_{k}}[B_{k}\zeta_{k}]_{a},
\end{split}
\end{equation}
where $B_{k}$ is a matrix valued function such that 
\begin{equation}\label{eq:Bkestnoise}
\|B_{k}\|\leq \frac{Ce^{-\eta_{A}\bt_{k}}}{\nu_{\ron}(\indexnot)e^{\b_{\ron}t_{k}}},
\end{equation}
where $C$ only depends on the coefficients of the equation (\ref{eq:thesystemRge}). Letting 
\begin{equation}\label{eq:chikdefdetas}
\chi_{k}:=e^{-J_{R,a}\bt_{k}}\zeta_{k,a},
\end{equation}
the equalities (\ref{eq:zkpoemJritnoise}) and (\ref{eq:zkitftrefnoise}) yield
\begin{equation}\label{eq:chikpoformnoise}
\begin{split}
\chi_{k+1} = & \chi_{k}+e^{-J_{R,a}\bt_{k}}[B_{k}\zeta_{k}]_{a}
+\int_{t_{k}}^{t_{k+1}}e^{-J_{R,a}\bt_{k+1}}[\chA_{k,\ron,\e}^{+}\chF_{k}(t)]_{a}dt\\
 = & \chi_{0}+\sum_{l=0}^{k}e^{-J_{R,a}\bt_{l}}[B_{l}\zeta_{l}]_{a}
+\sum_{l=0}^{k}\int_{t_{l}}^{t_{l+1}}e^{-J_{R,a}\bt_{l+1}}[\chA_{l,\ron,\e}^{+}\chF_{l}(t)]_{a}dt.
\end{split}
\end{equation}
At this stage it is of interest to divide the right hand side into a limiting value and an error term. Let us begin by defining the limiting value:
\begin{equation}\label{eq:zetainftyadefnoise}
\zeta_{\infty,a}:=\zeta_{0,a}+\sum_{l=0}^{\infty}e^{-J_{R,a}\bt_{l}}[B_{l}\zeta_{l}]_{a}
+\sum_{l=0}^{\infty}\int_{t_{l}}^{t_{l+1}}e^{-J_{R,a}\bt_{l+1}}[\chA_{l,\ron,\e}^{+}\chF_{l}(t)]_{a}dt;
\end{equation}
note that the sums appearing on the right hand side are convergent (we justify this statement in (\ref{eq:zetainfaeststnoise})
and (\ref{eq:zetainfaestttnoise}) below). Given the definition (\ref{eq:zetainftyadefnoise}), the equality (\ref{eq:chikpoformnoise}) can be written
\begin{equation}\label{eq:zetakpoformnoise}
\begin{split}
\zeta_{k+1,a} = & e^{J_{R,a}\bt_{k+1}}\zeta_{\infty,a}-\sum_{l=k+1}^{\infty}e^{-J_{R,a}(t_{l}-t_{k+1})}[B_{l}\zeta_{l}]_{a}\\
 & -\sum_{l=k+1}^{\infty}\int_{t_{l}}^{t_{l+1}}e^{-J_{R,a}(t_{l+1}-t_{k+1})}[\chA_{l,\ron,\e}^{+}\chF_{l}(t)]_{a}dt.
\end{split}
\end{equation}
The remainder of the analysis concerning the first component is naturally divided into two parts. First, we need to estimate the limiting value
$\zeta_{\infty,a}$. Second, we need to estimate the error terms appearing in (\ref{eq:zetakpoformnoise}); i.e., the last two terms on the right hand 
side of (\ref{eq:zetakpoformnoise}). 

\textit{Estimating the limiting value.} Let us estimate the terms appearing on the right hand side of (\ref{eq:zetainftyadefnoise}) one by one. 
To begin with,
\begin{equation}\label{eq:zetainfaestftnoise}
|\zeta_{0,a}|\leq |\zeta_{0}|\leq Ce^{-\kappa_{\ron,+}t_{0}}|\psi_{0}|\leq Ce^{-\kappa_{\ron,+}t_{0}}|w(\indexnot,t_{0})|;
\end{equation}
cf. (\ref{eq:zetakdefnoise}), (\ref{eq:wkpredef}), (\ref{eq:wfinkdef}) and recall that $\psi_{k}=w_{\fin,k}$. Here, $C$ only depends on the coefficients 
of the equation (\ref{eq:thesystemRge}). Next, 
\begin{equation}\label{eq:zetainfaeststnoise}
\begin{split}
\left|\sum_{l=0}^{\infty}e^{-J_{R,a}\bt_{l}}[B_{l}\zeta_{l}]_{a}\right|
 \leq & C\sum_{l=0}^{\infty}\ldr{\bt_{l}}^{N}e^{-\eta_{\min,a}\bt_{l}}\frac{e^{-\eta_{A}\bt_{l}}}{\nu_{\ron}(\indexnot)e^{\b_{\ron}t_{l}}}\\
 & \cdot\left[\ldr{\bt_{l}}^{d_{\ron,+}-1}|\zeta_{0}|+\int_{t_{0}}^{t_{l}}\ldr{t_{l}-t'}^{d_{\ron,+}-1}e^{-\kappa_{\ron,+}t'}|\hf(\indexnot,t')|dt'\right]\\
 \leq & C_{B}\left(|\zeta_{0}|+\int_{t_{0}}^{\infty}e^{-\kappa_{\ron,+}t}|\hf(\indexnot,t)|dt\right),
\end{split}
\end{equation}
where we have appealed to (\ref{eq:zetakpobdnoise}) and (\ref{eq:Bkestnoise}); $C_{B}$ only depends on $\eta_{B}$ and the coefficients of the equation 
(\ref{eq:thesystemRge});
and $N$ is a non-negative integer depending only on $m$. Moreover, $\eta_{\min,a}$ is the smallest eigenvalue of $J_{R,a}$ and we have used the fact
that $-\eta_{\min,a}<\eta_{B}\leq \eta_{A}$. Finally,
\begin{equation}\label{eq:zetainfaestttnoise}
\begin{split}
\left|\sum_{l=0}^{\infty}\int_{t_{l}}^{t_{l+1}}e^{-J_{R,a}\bt_{l+1}}[\chA_{l,\ron,\e}^{+}\chF_{l}(t)]_{a}dt\right|
\leq &  C\int_{t_{0}}^{\infty}\ldr{\bt}^{N}e^{-\eta_{\min,a}\bt}e^{-\kappa_{\ron,+}t}|\hf(\indexnot,t)|dt\\
\leq &  C_{B}\int_{t_{0}}^{\infty}e^{-(\kappa_{\ron,+}-\eta_{B})t}|\hf(\indexnot,t)|dt,
\end{split}
\end{equation}
where $C_{B}$ only depends on $\eta_{B}$ and the coefficients of the equation (\ref{eq:thesystemRge}). Combining 
(\ref{eq:zetainfaestftnoise})--(\ref{eq:zetainfaestttnoise}) yields the conclusion
\begin{equation}\label{eq:zetainftyaestnoise}
|\zeta_{\infty,a}|\leq C_{B}\left(e^{-\kappa_{\ron,+}t_{0}}|w(\indexnot,t_{0})|+\int_{t_{0}}^{\infty}e^{-(\kappa_{\ron,+}-\eta_{B})t}|\hf(\indexnot,t)|dt\right),
\end{equation}
where $C_{B}$ has the same dependence as in the case of (\ref{eq:zetainfaestttnoise}). 

\textit{Estimating the error terms.} Let us estimate the last two terms on the right hand side of (\ref{eq:zetakpoformnoise}). The second
term on the right hand side of (\ref{eq:zetakpoformnoise}) can be estimated as follows:
\begin{equation}\label{eq:seterestnoise}
\begin{split}
 & \left|\sum_{l=k+1}^{\infty}e^{-J_{R,a}(t_{l}-t_{k+1})}[B_{l}\zeta_{l}]_{a}\right| \\
\leq &  \sum_{l=k+1}^{\infty}C\ldr{t_{l}-t_{k+1}}^{N}e^{-\eta_{\min,a}(t_{l}-t_{k+1})}
\frac{e^{-\eta_{A}\bt_{l}}}{\nu_{\ron}(\indexnot)e^{\b_{\ron}t_{l}}}\\
 & \cdot \left(
\ldr{\bt_{l}}^{d_{\ron,+}-1}|\zeta_{0}|+\int_{t_{0}}^{t_{l}}\ldr{t_{l}-t'}^{d_{\ron,+}-1}e^{-\kappa_{\ron,+}t'}|\hf(\indexnot,t')|dt'
\right)\\
 \leq & \int_{t_{k+1}}^{\infty}C\ldr{t-t_{k+1}}^{N}e^{-\eta_{\min,a}(t-t_{k+1})}e^{-\eta_{A}\bt}\ldr{\bt}^{d_{\ron,+}-1}dt\\
 & \cdot \left(|\zeta_{0}|+\int_{t_{0}}^{\infty}e^{-\kappa_{\ron,+}t'}|\hf(\indexnot,t')|dt'
\right)\\
 \leq & C_{B}\ldr{\bt_{k+1}}^{d_{\ron,+}-1}e^{-\eta_{A}\bt_{k+1}}\left(|\zeta_{0}|+\int_{t_{0}}^{\infty}e^{-\kappa_{\ron,+}t'}|\hf(\indexnot,t')|dt'
\right),
\end{split}
\end{equation}
where $\eta_{\min,a}$ is the smallest eigenvalue of $J_{R,a}$ and we have appealed to (\ref{eq:zetakpobdnoise}) and (\ref{eq:Bkestnoise}); note 
that $-\eta_{\min,a}<\eta_{A}$. Moreover, $C_{B}$ only depends on $\eta_{B}$ and the coefficients of the equation (\ref{eq:thesystemRge}). Let us 
estimate the third term on the right hand side of 
(\ref{eq:zetakpoformnoise}):
\begin{equation}\label{eq:thtermestnoise}
\begin{split}
 & \left|\sum_{l=k+1}^{\infty}\int_{t_{l}}^{t_{l+1}}e^{-J_{R,a}(t_{l+1}-t_{k+1})}[\chA_{l,\ron,\e}^{+}\chF_{l}(t)]_{a}dt\right|\\
 \leq & C\int_{t_{k+1}}^{\infty}\ldr{t-t_{k+1}}^{N}e^{-\eta_{\min,a}(t-t_{k+1})}e^{-\kappa_{\ron,+}t}|\hf(\indexnot,t)|dt\\
 \leq & C_{B}e^{-\eta_{B}t_{k+1}}\int_{0}^{\infty}e^{-(\kappa_{\ron,+}-\eta_{B})t}|\hf(\indexnot,t)|dt,
\end{split}
\end{equation}
where $C_{B}$ only depends on $\eta_{B}$ and the coefficients of the equation (\ref{eq:thesystemRge}). Combining (\ref{eq:zetakpoformnoise}), 
(\ref{eq:seterestnoise}) and (\ref{eq:thtermestnoise}) yields
\begin{equation}\label{eq:zetakpoalonoise}
\begin{split}
|\zeta_{k+1,a}-e^{J_{R,a}\bt_{k+1}}\zeta_{\infty,a}| \leq &
C_{B}\ldr{\bt_{k+1}}^{d_{\ron,+}-1}e^{-\eta_{B}\bt_{k+1}}\\
 & \cdot \left(|\zeta_{0}|+\int_{0}^{\infty}e^{-(\kappa_{\ron,+}-\eta_{B})t}|\hf(\indexnot,t)|dt\right),
\end{split}
\end{equation}
where $C_{B}$ has the same dependence as in (\ref{eq:thtermestnoise}). Moreover, in case $\eta_{A}>\eta_{B}$, the polynomial factor can be removed. 
Before reformulating this estimate, note that if 
\[
S(t):=\mathrm{diag}\{e^{-i\zeta_{1}t}\Id_{l_{1}},\dots,e^{-i\zeta_{D}t}\Id_{l_{D}}\},
\]
cf. (\ref{eq:ChAkjedefnoise}), then
\[
e^{J_{R,\ron,\e}t}=e^{-\kappa_{\ron,+}t}S(t)T_{\ron,\e}^{-1}\exp[R_{\ron}(\indexnot)t]T_{\ron,\e},
\]
where $R_{\ron}(\indexnot)$ is defined in (\ref{eq:Rronpmdef}). Thus
\begin{equation}\label{eq:expRinfjtnoise}
\exp[R_{\ron}(\indexnot)t]=e^{\kappa_{\ron,+}t}T_{\ron,\e}[S(t)]^{-1}e^{J_{R,\ron,\e}t}T_{\ron,\e}^{-1}.
\end{equation}
With these observations in mind, it can be verified that the estimate (\ref{eq:zetakpoalonoise}) implies that
\begin{equation}\label{eq:psikaestnoise}
\begin{split}
 & |[T_{\ron,\e}^{-1}\psi_{k+1}]_{a}-[T_{\ron,\e}^{-1}\exp[R_{\ron}(\indexnot)\bt_{k+1}]T_{\ron,\e}]_{a}\chi_{\infty,a}|\\
 \leq & C_{B}\ldr{\bt_{k+1}}^{d_{\ron,+}-1}e^{\kappa_{\ron,+}t_{k+1}}e^{-\eta_{B}\bt_{k+1}}
\left(|\zeta_{0}|+\int_{0}^{\infty}e^{-(\kappa_{\ron,+}-\eta_{B})t}|\hf(\indexnot,t)|dt\right),
\end{split}
\end{equation}
where $C_{B}$ has the same dependence as in (\ref{eq:thtermestnoise}),
\begin{equation}\label{eq:psiinfadefnoise}
\chi_{\infty,a}:=e^{\kappa_{\ron,+}t_{0}}S_{0,a}^{-1}\zeta_{\infty,a}
\end{equation}
and $S_{k,a}$ denotes the first block in the $\eta_{B},J_{R,\ron,\e}$-decomposition of $S_{k}$. Finally, if $\eta_{A}>\eta_{B}$, then the polynomial
factor can be removed. Note that, due to (\ref{eq:zetainftyaestnoise}), 
\begin{equation}\label{eq:psiinfaestnoise}
|\chi_{\infty,a}|\leq C_{B}\left(|w(\indexnot,t_{0})|+e^{\kappa_{\ron,+}t_{0}}\int_{t_{0}}^{\infty}e^{-(\kappa_{\ron,+}-\eta_{B})t}|\hf(\indexnot,t)|dt\right),
\end{equation}
where $C_{B}$ has the same dependence as in (\ref{eq:thtermestnoise}).

\textbf{Estimating the second component.}
Let us turn to the second component in the $\eta_{B},J_{R,\ron,\e}$-decomposition of $\zeta_{k}$. To begin with, (\ref{eq:zetakpoitnoise}) yields
\begin{equation}\label{eq:zkbpoemJritnoise}
e^{-J_{R,b}\bt_{k+1}}\zeta_{k+1,b}=e^{-J_{R,b}\bt_{k+1}}[\chA_{k,\ron,\e}^{+}\zeta_{k}]_{b}
+e^{-J_{R,b}\bt_{k+1}}\int_{t_{k}}^{t_{k+1}}[\chA_{k,\ron,\e}^{+}\chF_{k}(t)]_{b}dt,
\end{equation}
where $\zeta_{k,b}$ denotes the second component in the $\eta_{B},J_{R,\ron,\e}$-decomposition of $\zeta_{k}$. Let us rewrite the first term on the 
right hand side according to 
\begin{equation}\label{eq:zkbitftrefnoise}
\begin{split}
e^{-J_{R,b}\bt_{k+1}}[\chA_{k,\ron,\e}^{+}\zeta_{k}]_{b} = & e^{-J_{R,b}\bt_{k+1}}[\chA_{k,\ron,\e}^{\app,+}]_{b}\zeta_{k,b}\\
 & +e^{-J_{R,b}\bt_{k+1}}[(\chA_{k,\ron,\e}^{+}-\chA_{k,\ron,\e}^{\app,+})\zeta_{k}]_{b}\\
 = & e^{-J_{R,b}\bt_{k}}\zeta_{k,b}+e^{-J_{R,b}\bt_{k}}[B_{k}\zeta_{k}]_{b},
\end{split}
\end{equation}
where $B_{k}$ is a matrix valued function satisfying (\ref{eq:Bkestnoise}). Letting $\phi_{k}:=e^{-J_{R,b}\bt_{k}}\zeta_{k,b}$,
the equalities (\ref{eq:zkbpoemJritnoise}) and (\ref{eq:zkbitftrefnoise}) yield
\begin{equation}\label{eq:chikbpoformnoise}
\begin{split}
\phi_{k+1} = & \phi_{k}+e^{-J_{R,b}\bt_{k}}[B_{k}\zeta_{k}]_{b}
+\int_{t_{k}}^{t_{k+1}}e^{-J_{R,b}\bt_{k+1}}[\chA_{k,\ron,\e}^{+}\chF_{k}(t)]_{b}dt\\
 = & \phi_{0}+\sum_{l=0}^{k}e^{-J_{R,b}\bt_{l}}[B_{l}\zeta_{l}]_{b}
+\sum_{l=0}^{k}\int_{t_{l}}^{t_{l+1}}e^{-J_{R,b}\bt_{l+1}}[\chA_{l,\ron,\e}^{+}\chF_{l}(t)]_{b}dt.
\end{split}
\end{equation}
Thus
\begin{equation}\label{eq:zetakpobitnoise}
\begin{split}
\zeta_{k+1,b} = & e^{J_{R,b}\bt_{k+1}}\zeta_{0,b}+\sum_{l=0}^{k}e^{J_{R,b}(t_{k+1}-t_{l})}[B_{l}\zeta_{l}]_{b}\\
 & +\sum_{l=0}^{k}\int_{t_{l}}^{t_{l+1}}e^{J_{R,b}(t_{k+1}-t_{l+1})}[\chA_{l,\ron,\e}^{+}\chF_{l}(t)]_{b}dt.
\end{split}
\end{equation}
Let us estimate the terms on the right hand side one by one. To begin with, 
\begin{equation}\label{eq:ftbcasenoise}
|e^{J_{R,b}\bt_{k+1}}\zeta_{0,b}|\leq C\ldr{\bt_{k+1}}^{N}e^{\eta_{\max,b}\bt_{k+1}}|\zeta_{0,b}|\leq C\ldr{\bt_{k+1}}^{N}e^{-\eta_{B}\bt_{k+1}}|\zeta_{0}|,
\end{equation}
where $\eta_{\max,b}$ is the largest eigenvalue of $J_{R,b}$ (note that $\eta_{\max,b}\leq-\eta_{B}$ due to the definition of $J_{R,b}$), $C$
only depends on the coefficients of the equation (\ref{eq:thesystemRge}), and $N$ is a non-negative integer depending only on $m$. 
Note that if all the Jordan blocks of $R_{\ron}(\indexnot)$ are trivial, then $N=0$. Turning to the second term on the right hand side of 
(\ref{eq:zetakpobitnoise}),
\begin{equation}\label{eq:stbcasenoise}
\begin{split}
\left|\sum_{l=0}^{k}e^{J_{R,b}(t_{k+1}-t_{l})}[B_{l}\zeta_{l}]_{b}\right| \leq &
C\int_{t_{0}}^{t_{k+1}}\ldr{t_{k+1}-t}^{N}e^{\eta_{\max,b}(t_{k+1}-t)}e^{-\eta_{A}\bt}
\ldr{\bt}^{d_{\ron,+}-1}dt\\
 & \cdot \left(|\zeta_{0}|+\int_{t_{0}}^{t_{k+1}}e^{-\kappa_{\ron,+}t}|\hf(\indexnot,t)|dt\right)\\
 \leq & C\ldr{\bt_{k+1}}^{N}e^{-\eta_{B}\bt_{k+1}}\left(|\zeta_{0}|+\int_{t_{0}}^{t_{k+1}}e^{-\kappa_{\ron,+}t}|\hf(\indexnot,t)|dt\right),
\end{split}
\end{equation}
where we have appealed to (\ref{eq:zetakpobdnoise}) and (\ref{eq:Bkestnoise}). Moreover, $C$ only depends on the coefficients of the 
equation (\ref{eq:thesystemRge}), and $N$ is a non-negative integer depending only on $m$. In addition, $N=1$ if all the Jordan blocks of 
$R_{\ron}(\indexnot)$ are trivial. Finally, let us estimate the third term on the right hand side of (\ref{eq:zetakpobitnoise}):
\begin{equation}\label{eq:ttbcasenoise}
\begin{split}
 & \left|\sum_{l=0}^{k}\int_{t_{l}}^{t_{l+1}}e^{J_{R,b}(t_{k+1}-t_{l+1})}[\chA_{l,\ron,\e}^{+}\chF_{l}(t)]_{b}dt\right|\\
 \leq & C\int_{t_{0}}^{t_{k+1}}\ldr{t_{k+1}-t}^{N}e^{\eta_{\max,b}(t_{k+1}-t)}e^{-\kappa_{\ron,+}t}|\hf(\indexnot,t)|dt\\
 \leq & C\ldr{\bt_{k+1}}^{N}e^{-\eta_{B}t_{k+1}}\int_{t_{0}}^{t_{k+1}}e^{-(\kappa_{\ron,+}-\eta_{B})t}|\hf(\indexnot,t)|dt,
\end{split}
\end{equation} 
where $C$ and $N$ have the same dependence as in the case of (\ref{eq:stbcasenoise}). Moreover, $N=0$ if all the Jordan blocks of 
$R_{\ron}(\indexnot)$ are trivial. Combining (\ref{eq:zetakpobitnoise}), (\ref{eq:ftbcasenoise}), (\ref{eq:stbcasenoise}) and 
(\ref{eq:ttbcasenoise}) yields
\begin{equation}\label{eq:psikbestnoise}
|[T_{\ron,\e}^{-1}\psi_{k+1}]_{b}|\leq C\ldr{\bt_{k+1}}^{N}e^{\kappa_{\ron,+}t_{k+1}}e^{-\eta_{B}\bt_{k+1}}\left(|\zeta_{0}|
+\int_{0}^{\infty}e^{-(\kappa_{\ron,+}-\eta_{B})t}|\hf(\indexnot,t)|dt\right),
\end{equation}
where $C$ and $N$ have the same dependence as in the case of (\ref{eq:stbcasenoise}). Moreover, $N=1$ if all the Jordan blocks of 
$R_{\ron}(\indexnot)$ are trivial.

\textbf{Combining the estimates.}
Let $\chi_{\infty}$ be the vector such that the first component in the $\eta_{B},J_{R,\ron,\e}$-decomposition of $\chi_{\infty}$ is given by 
$\chi_{\infty,a}$, cf. (\ref{eq:psiinfadefnoise}), and the the second component is zero. Define
\begin{equation}\label{eq:bpsipsidefnoise}
\bpsi_{\infty}:=T_{\ron,\e}\chi_{\infty},\ \ \
\psi_{\infty}:=e^{-R_{\ron}(\indexnot)T_{\ron}}\bpsi_{\infty}.
\end{equation}
Then (\ref{eq:psiinfaestnoise}) yields (\ref{eq:Psiinfestnoise}), and it is clear that $\bpsi_{\infty}$ and $\psi_{\infty}$ belong to the first 
generalised eigenspace of the $\eta_{B}$, $R_{\ron}(\indexnot)$-decomposition of $\cn{2m}$. 

In order to prove (\ref{eq:zzdottkestnoisestmt}), note that (\ref{eq:psikaestnoise}) and (\ref{eq:psikbestnoise}) can be combined to yield
\begin{equation}\label{eq:psikestnoise}
\begin{split}
 & |\psi_{k+1}-e^{R_{\ron}(\indexnot)\bt_{k+1}}\bpsi_{\infty}|=|\psi_{k+1}-e^{R_{\ron}(\indexnot)t_{k+1}}\psi_{\infty}|\\
 \leq & C_{B}\ldr{\bt_{k+1}}^{N}e^{(\kappa_{\ron,+}-\eta_{B})\bt_{k+1}}\left(|w(\indexnot,t_{0})|
+e^{\kappa_{\ron,+}t_{0}}\int_{0}^{\infty}e^{-(\kappa_{\ron,+}-\eta_{B})t}|\hf(\indexnot,t)|dt\right)
\end{split}
\end{equation}
for $k\geq 0$, where $C_{B}$ has the same dependence as in the case of (\ref{eq:zetainfaestttnoise}) and $N$ is a non-negative integer depending 
only on $m$. Moreover, $N=1$ if all the Jordan blocks of $R_{\ron}(\indexnot)$ are trivial. In addition, if the first generalised eigenspace of the 
$\eta_{B}$, $R_{\ron}(\indexnot)$-decomposition of $\cn{2m}$ equals $\cn{2m}$, then the $N$ appearing in (\ref{eq:psikestnoise}) can be replaced 
by $d_{\ron,+}-1$, where $d_{\ron,+}:=d_{\max}(R_{\ron}(\indexnot),\kappa_{\ron,+})$. It can also be verified that (\ref{eq:psikestnoise}) holds for $k=-1$. 
Retracing the definitions, 
\begin{equation}\label{eq:psikitofznoise}
\psi_{k}=D_{k}(\indexnot)T_{\pre,k}\exp\left(-i\int_{0}^{t_{k}}\sigma(\indexnot,t')\mfg(\indexnot,t')dt'\right)
\left(\begin{array}{c} \mfg(\indexnot,t_{k})z(\indexnot,t_{k}) \\ \dot{z}(\indexnot,t_{k})\end{array}\right).
\end{equation}
Thus (\ref{eq:psikestnoise}) yields (\ref{eq:zzdottkestnoisestmt}). The lemma follows. 
\end{proof}

\section{Specifying data at infinity}\label{section:spdataatinfnoise}

As a next step, we wish to address the problem of specifying data at infinity, proceeding along an appropriate time sequence. 
We here restrict our attention to the case of homogeneous equations. 

\begin{lemma}\label{lemma:asdataosccasenoise}
Assume that (\ref{eq:thesystemRge}) is strongly balanced, has a negligible shift vector field, a dominant noisy spatial direction, and is such that the dominant 
coefficients are convergent. Assume, moreover, that $f=0$. Let $\indexnot\in\EFnindexset$ and define $T_{\ron}$ by (\ref{eq:Trondef}). Let $\{t_{k}\}$ be the 
time sequence given by Definition~\ref{def:tkdefge} and the condition that $t_{0}:=T_{\ron}$. Let $\eta_{A}$ be defined by (\ref{eq:etaAdefinition}) and 
fix an $0<\eta_{B}\leq \eta_{A}$. Let $E_{a}$ denote the first generalised eigenspace of the $\eta_{B}$, $R_{\ron}(\indexnot)$-decomposition of $\cn{2m}$,
where $R_{\ron}(\indexnot)$ is given by (\ref{eq:Rronpmdef}); cf. Definition~\ref{def:fssubspetc}. Then there is an injective linear map 
$\Psi_{\infty}:E_{a}\rightarrow\cn{2m}$ such that if $\chi\in E_{a}$ and $z$ is the solution to the homogeneous version of 
(\ref{eq:fourierthesystemRge}) with initial data at $T_{\ron}$ determined by
\begin{equation}\label{eq:Psiinfchitoidnoise}
\Psi_{\infty}(\chi)=\left(\begin{array}{c} \mfg(\indexnot,T_{\ron})z(\indexnot,T_{\ron}) \\ \dot{z}(\indexnot,T_{\ron})\end{array}\right),
\end{equation}
then (\ref{eq:zzdottkestnoisestmt}) holds with $\psi_{\infty}$ replaced by $\chi$ and $\hf(\indexnot,\cdot)$ replaced by $0$. Moreover,
\begin{equation}\label{eq:Psiinfchitoidnoiseest}
|\Psi_{\infty}(e^{-R_{\ron}(\indexnot)T_{\ron}}\chi)|\leq C|\chi|
\end{equation}
for all $\chi\in E_{a}$, where $C$ only depends on the coefficients of the equation (\ref{eq:thesystemRge}).
\end{lemma}
\begin{remark}
In analogy with Remark~\ref{remark:inhomaschar}, it is possible to combine Lemmas~\ref{lemma:asymposccasenoise} and \ref{lemma:asdataosccasenoise} in
order to obtain analogous conclusions in the inhomogeneous setting. 
\end{remark}
\begin{remark}
Dependence on the coefficients includes dependence on the Riemannian manifolds $(M_{r},g_{r})$, $r=1,\dots,R$; cf. 
Remark~\ref{remark:deponcoefincldeponRmfds}. 
\end{remark}
\begin{proof}
\textbf{Notation and preliminary estimates.} In what follows, we use the notation introduced in Section~\ref{section:reformeqsnoise} without 
further comment. Moreover, we let $\lambda=\kappa_{\ron,+}$ and define $J_{R,\ron,\e}$ by (\ref{eq:JRronepsilondef}). Denote the first and second 
blocks of the $\eta_{B}$, $J_{R,\ron,\e}$-decomposition of $J_{R,\ron,\e}$ by $J_{R,a}$ and $J_{R,b}$ respectively, and let
\[
J_{a}:=\diag\{J_{R,a},0\},\ \ \
J_{b}:=\diag\{0,J_{R,b}\},\ \ \
\tA_{k,\ron,\e}^{+}:=e^{-J_{a}\bt_{k+1}}\chA_{k,\ron,\e}^{+}e^{J_{a}\bt_{k}}
\]
be $2m\times 2m$-matrices, where $\bt_{k}:=t_{k}-t_{0}$. Compute
\begin{equation}\label{eq:tAauxexpnoise}
\tA_{k,\ron,\e}^{+}=e^{-J_{a}\bt_{k+1}}[\chA_{k,\ron,\e}^{+}-\chA^{\app,+}_{k,\ron,\e}]e^{J_{a}\bt_{k}}
+e^{-J_{a}(t_{k+1}-t_{k})}\chA^{\app,+}_{k,\ron,\e};
\end{equation}
note that $J_{a}$ and $\chA^{\app,+}_{k,\ron,\e}$ commute. On the other hand, 
\begin{equation*}
\begin{split}
e^{-J_{a}(t_{k+1}-t_{k})}\chA^{\app,+}_{k,\ron,\e}
 = & \chA^{\app,+}_{k,\ron,\e}-J_{a}(t_{k+1}-t_{k})+\frac{2\pi}{\nu_{\ron}(\indexnot)e^{\b_{\ron}t_{k}}}J_{R,\ron,\e}[e^{-J_{a}(t_{k+1}-t_{k})}-\Id_{2m}]\\
 & +[e^{-J_{a}(t_{k+1}-t_{k})}-\Id_{2m}+J_{a}(t_{k+1}-t_{k})].
\end{split}
\end{equation*}
Introducing 
\begin{equation}\label{eq:tAkjappdefnoise}
\tA^{\app,+}_{k,\ron,\e}:=\Id_{2m}+\frac{2\pi}{\nu_{\ron}(\indexnot)e^{\b_{\ron}t_{k}}}J_{b}
\end{equation}
and appealing to (\ref{eq:tbmtagestge}) and (\ref{eq:mfgnunebnexpdec}) yields 
\[
\|e^{-J_{a}(t_{k+1}-t_{k})}\chA^{\app,+}_{k,\ron,\e}-\tA^{\app,+}_{k,\ron,\e}\|\leq \frac{C_{\e}e^{-\eta_{A}\bt_{k}}}{\nu_{\ron}(\indexnot)e^{\b_{\ron}t_{k}}}
\]
for all $k\geq 0$, where $C_{\e}$ only depends on $\e$ and the coefficients of the equation (\ref{eq:thesystemRge}). Combining this estimate with 
(\ref{eq:chAkmchAkappnoise}) and (\ref{eq:tAauxexpnoise}) yields
\begin{equation}\label{eq:tAkjenestnoise}
\|\tA_{k,\ron,\e}^{+}-\tA^{\app,+}_{k,\ron,\e}\|\leq \frac{C_{\e}}{\nu_{\ron}(\indexnot)e^{\b_{\ron}t_{k}}}\ldr{\bt_{k}}^{N}e^{-\eta_{A,a}\bt_{k}}
\end{equation}
for all $k\geq 0$, where $C_{\e}$ only depends on $\e$ and the coefficients of the equation (\ref{eq:thesystemRge}); $N$ is a non-negative integer depending 
only on $m$; and $\eta_{A,a}$ is the difference 
between $\eta_{A}$ and the largest eigenvalue of the matrix $-J_{a}$ (note that this difference is strictly positive due to the definition of $J_{a}$). Note 
also that, by the definition of $J_{b}$, $\|\tA^{\app,+}_{k,\ron,\e}\|\leq 1$ (in order to obtain this estimate, we, however, need to assume $\e$ to be small
enough, the bound mentioned in connection with (\ref{eq:bappplkroneest}) being sufficient; from now on we fix such an $\e$, depending only on the 
coefficients of the equation (\ref{eq:thesystemRge}), so that dependence on $\e$ from now on can be replaced by dependence on the coefficients 
of the equation (\ref{eq:thesystemRge})). In particular, if $0\leq k_{a}\leq k_{b}$ are integers, then 
\begin{equation}\label{eq:tAkprodnoise}
\|\tA_{k_{b},\ron,\e}^{+}\cdots\tA_{k_{a},\ron,\e}^{+}\|\leq C
\end{equation}
for a constant $C$ depending only on the coefficients of the equation (\ref{eq:thesystemRge}); the proof of this statement is similar to the proof 
of (\ref{eq:bAkprodnoise}).

\textbf{Asymptotics of the leading order component.}
Let us return to (\ref{eq:wprekpertransvarpi}) in case $\hf=0$. Then $\psi_{k+1} = A_{k}^{+}\psi_{k}$. Introducing 
\begin{equation}\label{eq:trhokdefnoise}
\trho_{k}:=e^{-\kappa_{\ron,+}t_{k}}e^{-J_{a}\bt_{k}}S_{k}T_{\ron,\e}^{-1}\psi_{k}, 
\end{equation}
we thus have 
\begin{equation}\label{eq:trhoitnoise}
\trho_{k+1}=\tA_{k,\ron,\e}^{+}\trho_{k}.
\end{equation}
Due to (\ref{eq:tAkprodnoise}), 
\begin{equation}\label{eq:trhokbdnoise}
|\trho_{k}|\leq C|\trho_{k_{0}}|
\end{equation}
for $0\leq k_{0}\leq k$, where $C$ has the same dependence as in the case of (\ref{eq:tAkprodnoise}). Letting $\trho_{k,a}$ denote
the first component in the $\eta_{B},J_{R,\ron,\e}$-decomposition of $\trho_{k}$, (\ref{eq:trhoitnoise}) implies
\[
\trho_{k+1,a}=\trho_{k,a}+[(\tA_{k,\ron,\e}^{+}-\tA^{\app,+}_{k,\ron,\e})\trho_{k}]_{a}
=\trho_{k_{0},a}+\textstyle{\sum}_{l=k_{0}}^{k}[(\tA_{l,\ron,\e}^{+}-\tA^{\app,+}_{l,\ron,\e})\trho_{l}]_{a}
\]
for every $0\leq k_{0}\leq k$. Combining this equality with (\ref{eq:tAkjenestnoise}) and (\ref{eq:trhokbdnoise}) yields
\begin{equation}\label{eq:trhokacauchynoise}
\begin{split}
|\trho_{k+1,a}-\trho_{k_{0},a}| \leq & \sum_{l=k_{0}}^{k}\frac{C}{\nu_{\ron}(\indexnot)e^{\b_{\ron}t_{l}}}\ldr{\bt_{l}}^{N}e^{-\eta_{A,a}\bt_{l}}|\trho_{k_{0}}|\\
 \leq & C\int_{t_{k_{0}}}^{t_{k+1}}\ldr{\bt}^{N}e^{-\eta_{A,a}\bt}dt|\trho_{k_{0}}|\leq
C\ldr{\bt_{k_{0}}}^{N}e^{-\eta_{A,a}\bt_{k_{0}}}|\trho_{k_{0}}|
\end{split}
\end{equation}
for all $0\leq k_{0}\leq k$, where $C$ has the same dependence as in the case of (\ref{eq:tAkprodnoise}). In particular, given an $\vare>0$, we can 
choose a $\bt_{k_{0}}$ (depending only on $\vare$ and the coefficients of the equation (\ref{eq:thesystemRge})) such that 
\begin{equation}\label{eq:trhoconvnoise}
|\trho_{k+1,a}-\trho_{k_{0},a}|\leq \vare |\trho_{k_{0}}|
\end{equation}
for all $0\leq k_{0}\leq k$. 

\textbf{From asymptotic data to a finite time hypersurface.}
From now on, the structure of the argument is quite similar to the end of the proof of Lemma~\ref{lemma:spasODEsett}, though
some of the details are different. Let $k_{0}$ be such
that (\ref{eq:trhoconvnoise}) holds with $\vare$ replaced by $1/2$ and let $l_{a}$ denote the dimension of the first subspace of the 
$\eta_{B},J_{R,\ron,\e}$-decomposition of $\cn{2m}$. Define the map $L_{a}:\cn{l_{a}}\rightarrow \cn{l_{a}}$ as follows. Given $\eta\in\cn{l_{a}}$, 
let $\xi\in\cn{2m}$ be such that $\xi_{a}=\eta$ and $\xi_{b}=0$ (where we divide $\xi$ into components $\xi_{a}$ and $\xi_{b}$ in the same way 
as $\trho_{k}$). Let $\trho_{k}$ be the solution to (\ref{eq:trhoitnoise}) such that $\trho_{k_{0}}=\xi$ and define $L_{a}\eta$ by 
\begin{equation}\label{eq:Ladefnoise}
L_{a}\eta:=\lim_{k\rightarrow\infty}\trho_{k,a}.
\end{equation}
Due to estimates such as (\ref{eq:trhokacauchynoise}), it is clear that $\trho_{k,a}$ is a Cauchy sequence, so that the limit on the right hand side of 
(\ref{eq:Ladefnoise}) exists. Thus $L_{a}\eta$ is well defined. Moreover, it is clear
that $L_{a}$ is linear. Due to the fact that (\ref{eq:trhoconvnoise}) holds with $\varepsilon$ replaced by $1/2$ and the fact that $\trho_{k_{0},b}=0$,
\[
\left|L_{a}\eta-\trho_{k_{0},a}\right|\leq \frac{1}{2}|\trho_{k_{0},a}|.
\]
In particular, $|\eta|/2\leq |L_{a}\eta|$, so that $L_{a}$ is injective. Since $L_{a}$ is a linear map between spaces of the same finite dimension, it 
follows that $L_{a}$ is an isomorphism. Moreover, 
\begin{equation}\label{eq:Lainvestoscnoise}
|L_{a}^{-1}\eta|\leq 2|\eta|.
\end{equation}
Note that $L_{a}^{-1}$ (essentially) maps data at infinity to data at $t_{k_{0}}$. In the end, we wish to translate the data at $t_{k_{0}}$ to data at 
$t_{0}$. To this end, we wish to solve (\ref{eq:trhoitnoise}) backwards. 

\textbf{From asymptotic data to $t=t_{0}$.} In order to solve (\ref{eq:trhoitnoise}) backwards, we can proceed similarly to the above; i.e., study the 
backwards iteration. However, this involves unnecessary complications. In fact, since $\bt_{k_{0}}$ is bounded by a constant depending only on the 
coefficients of the equation (\ref{eq:thesystemRge}), it is sufficient to appeal to Lemma~\ref{lemma:roughenestbalsetting}. In order to justify this
statement, note, to begin with, that (\ref{eq:trhokdefnoise}) yields
\begin{equation}\label{eq:trhozeroitowindextzero}
|\trho_{0}|\leq Ce^{-\kappa_{\ron,+}t_{0}}|w(\indexnot,t_{0})|,
\end{equation}
where $C$ only depends on the coefficients of the equation (\ref{eq:thesystemRge}). On the other hand, $\me^{1/2}(\indexnot,t)$ appearing in 
(\ref{eq:meestroughbalset}) is equivalent to $|w(\indexnot,t)|$ for $t\geq T_{\ron}$, the constant of equivalence being numerical. Combining
(\ref{eq:meestroughbalset}) with (\ref{eq:trhozeroitowindextzero}) yields
\[
|\trho_{0}|\leq Ce^{-\kappa_{\ron,+}t_{0}}e^{\eta_{\robal}\bt_{k_{0}}}|w(\indexnot,t_{k_{0}})|,
\]
where $C$ only depends on the coefficients of the equation (\ref{eq:thesystemRge}). On the other hand, combining (\ref{eq:trhokdefnoise}) with the 
fact that $\bt_{k_{0}}$ is bounded by a constant depending only on the coefficients of the equation (\ref{eq:thesystemRge}), 
\[
|w(\indexnot,t_{k_{0}})|\leq Ce^{\kappa_{\ron,+}t_{k_{0}}}|\trho_{k_{0}}|,
\]
where $C$ only depends on the coefficients of the equation (\ref{eq:thesystemRge}). Combining the last two estimates yields
\[
|\trho_{0}|\leq Ce^{\kappa_{\ron,+}\bt_{k_{0}}}e^{\eta_{\robal}\bt_{k_{0}}}|\trho_{k_{0}}|,
\]
where $C$ only depends on the coefficients of the equation (\ref{eq:thesystemRge}). To conclude, 
\begin{equation}\label{eq:trhoztrhokzest}
|\trho_{0}|\leq C|\trho_{k_{0}}|,
\end{equation}
where $C$ only depends on the coefficients of the equation (\ref{eq:thesystemRge}).

\textbf{Defining the map $\Psi_{\infty}$.} Given the above information, we are in a position to define the map taking data ``at infinity''
to data at the finite time $t_{0}=T_{\ron}$. Let $\Psi_{\infty}$ be the composition of the following maps: first the map which takes 
$\chi\in E_{a}$ to 
\begin{equation}\label{eq:varsigmaadef}
\varsigma_{a}:=L_{a}^{-1}\left(e^{-\kappa_{\ron,+}T_{\ron}}S_{0,a}[T_{\ron,\e}^{-1}e^{R_{\ron}(\indexnot)T_{\ron}}\chi]_{a}\right);
\end{equation}
second, the map which with $\varsigma_{a}$ associates the vector $\xi\in\cn{2m}$ such that $\xi_{a}=\varsigma_{a}$ and $\xi_{b}=0$; third, 
the map which with $\xi$ associates $\trho_{0}$, where $\trho_{k}$ is the solution to (\ref{eq:trhoitnoise}) such that $\trho_{k_{0}}=\xi$;
and fourth, the map which with $\trho_{0}$ associates the element
\begin{equation}\label{eq:Phiinfdefnoise}
\exp\left(i\int_{0}^{t_{0}}\sigma(\indexnot,t')\mfg(\indexnot,t')dt'\right)
T_{\pre,0}^{-1}D_{0}^{-1}e^{\kappa_{\ron,+}t_{0}}T_{\ron,\e}S_{0}^{-1}\trho_{0}.
\end{equation}
Returning to the definitions (\ref{eq:xyFodefshiftge}), (\ref{eq:wdefshiftge}), (\ref{eq:wkpredef}), (\ref{eq:wfinkdef}), (\ref{eq:trhokdefnoise}) 
and recalling that $\psi_{k}=w_{\fin,k}$, it is clear that if, given $\chi\in E_{a}$, we let $z$ be the solution to (\ref{eq:fourierthesystemRge})
such that 
\begin{equation}\label{eq:Phiinfetaitoznoise}
\Psi_{\infty}(\chi)=\left(\begin{array}{c} \mfg(\indexnot,t_{0})z(\indexnot,t_{0}) \\ \dot{z}(\indexnot,t_{0})\end{array}\right)
\end{equation}
holds, then the $\trho_{k}$ associated with the solution according to (\ref{eq:trhokdefnoise}) satisfies $\trho_{k_{0}}=\xi$, where $\xi_{a}=\varsigma_{a}$,
$\xi_{b}=0$ and $\varsigma_{a}$ is given by (\ref{eq:varsigmaadef}). Note that $\Psi_{\infty}$ is linear and injective, and that 
\[
|\Psi_{\infty}(e^{-R_{\ron}(\indexnot)T_{\ron}}\bchi)|\leq C|\bchi|
\]
for all $\bchi\in E_{a}$, where $C$ only depends on the coefficients of the equation (\ref{eq:thesystemRge}). In order to obtain this estimate we appeal 
to (\ref{eq:trhoztrhokzest}) and the definition of $\Psi_{\infty}$. 

Let now $\chi\in E_{a}$ and $z$ be the solution to (\ref{eq:fourierthesystemRge}) such that (\ref{eq:Phiinfetaitoznoise}) holds. Then, by 
construction,
\begin{equation}\label{eq:limktrhokafirstver}
\lim_{k\rightarrow\infty}\trho_{k,a}=L_{a}(\varsigma_{a})=e^{-\kappa_{\ron,+}T_{\ron}}S_{0,a}[T_{\ron,\e}^{-1}e^{R_{\ron}(\indexnot)T_{\ron}}\chi]_{a}.
\end{equation}
On the other hand, keeping (\ref{eq:zetakdefnoise}), (\ref{eq:chikdefdetas}), (\ref{eq:chikpoformnoise}), (\ref{eq:zetainftyadefnoise}),
(\ref{eq:psiinfadefnoise}) and (\ref{eq:trhokdefnoise}) in mind (in the case that $\hf(\indexnot,\cdot)=0$) yields
\begin{equation}\label{eq:limktrhokasecondver}
\lim_{k\rightarrow\infty}\trho_{k,a}=e^{-\kappa_{\ron,+}T_{\ron}}S_{0,a}\chi_{\infty,a}.
\end{equation}
Letting $\chi_{\infty}$ be such that the first component in the $\eta_{B}$, $J_{R,\ron,\e}$-decomposition of $\chi_{\infty}$ is given by $\chi_{\infty,a}$
and the second component vanishes, the equalities (\ref{eq:limktrhokafirstver}) and (\ref{eq:limktrhokasecondver}) yield
\[
\chi_{\infty}=T_{\ron,\e}^{-1}e^{R_{\ron}(\indexnot)T_{\ron}}\chi,
\]
where we have used the fact that $\chi\in E_{a}$. Recalling (\ref{eq:bpsipsidefnoise}), we conclude that the $\psi_{\infty}$ appearing in
the statement of Lemma~\ref{lemma:asymposccasenoise} equals $\chi$. The lemma follows.
\end{proof}

\section{Controlling the solution beyond the time sequence}\label{section:contsolbeytimeseqnoise}

The next step is to analyse how the solution behaves in between the times belonging to the time sequence $\{t_{k}\}$. Before doing so, it is convenient
to introduce the following terminology:
\begin{equation}\label{eq:Sinfjndefnoise}
S_{\ron}(\indexnot):=\frac{1}{2}\left(\begin{array}{cc} -\a_{\infty}+\b_{\ron}\Id_{m} & i\bX_{\ron}(\indexnot)\\
-i\bX_{\ron}(\indexnot) & -\a_{\infty}+\b_{\ron}\Id_{m} \end{array}\right),
\end{equation}
cf. the notation introduced in Definition~\ref{def:tXinfrondefetc}, and 
\begin{equation}\label{eq:omegashdefnoise}
\omega_{\rosh}(\indexnot,t):=\int_{0}^{t}\sigma(\indexnot,t')\mfg(\indexnot,t')dt'.
\end{equation}
Moreover, 
\begin{equation}\label{eq:mfDrototdefnoise}
\mfD_{\rotot}(\indexnot,t):=\left(\begin{array}{rr} \cos\varphi_{\rotot}(\indexnot,t)\mathrm{Id}_{m}
& \sin\varphi_{\rotot}(\indexnot,t)\mathrm{Id}_{m} \\  -\sin\varphi_{\rotot}(\indexnot,t)\mathrm{Id}_{m} & 
\cos\varphi_{\rotot}(\indexnot,t)\mathrm{Id}_{m}\end{array}\right),
\end{equation}
where
\begin{equation}\label{eq:varphirototdefnoise}
\varphi_{\rotot}(\indexnot,t):=\int_{0}^{t}[1+\sigma^{2}(\indexnot,t')]^{1/2}\mfg(\indexnot,t')dt'.
\end{equation}

\begin{lemma}\label{lemma:asymposccasefinnoise}
Assume that (\ref{eq:thesystemRge}) is strongly balanced, has a negligible shift vector field, a dominant noisy spatial direction, and is such that the dominant 
coefficients are convergent. Let $\indexnot\in\EFnindexset$ and define $T_{\ron}$ by (\ref{eq:Trondef}). Let $\eta_{A}$ be defined by (\ref{eq:etaAdefinition}) 
and assume that there is an $0<\eta_{B}\leq \eta_{A}$ such that 
\[
\int_{0}^{\infty}e^{-(\kappa_{\ron,+}-\eta_{B})t}|\hf(\indexnot,t)|dt<\infty,
\]
where $\kappa_{\ron,+}$ is defined by (\ref{eq:kapparonpmdef}). Then there is a constant $C_{B}$, depending only on $\eta_{B}$ and the coefficients of the 
equation (\ref{eq:thesystemRge}); and a non-negative
integer $N$, depending only on $m$, such that the following holds. If $z$ is a solution to (\ref{eq:fourierthesystemRge}), then there is a 
$z_{\infty}$ belonging to the first generalised eigenspace of the $\eta_{B}$, $S_{\ron}(\indexnot)$-decomposition of $\cn{2m}$ (cf. 
Definition~\ref{def:fssubspetc}) and satisfying
\begin{equation}\label{eq:zinfestnoise}
|e^{S_{\ron}(\indexnot)T_{\ron}}z_{\infty}|\leq 
C_{B}\left(|w(\indexnot,T_{\ron})|+e^{\kappa_{\ron,+}T_{\ron}}\int_{T_{\ron}}^{\infty}e^{-(\kappa_{\ron,+}-\eta_{B})t}|\hf(\indexnot,t)|dt\right),
\end{equation}
such that 
\begin{equation}\label{eq:zzdottestfinnoise}
\begin{split}
 & \left|\left(\begin{array}{c} \mfg(\indexnot,t)z(\indexnot,t) \\ \dot{z}(\indexnot,t)\end{array}\right)-e^{i\omega_{\rosh}(\indexnot,t)}
\mfD_{\rotot}(\indexnot,t)e^{S_{\ron}(\indexnot)t}z_{\infty}\right|\\
 \leq & C_{B}\ldr{\bt}^{N}e^{(\kappa_{\ron,+}-\eta_{B})\bt}\left(|w(\indexnot,T_{\ron})|+e^{\kappa_{\ron,+}T_{\ron}}
\int_{0}^{\infty}e^{-(\kappa_{\ron,+}-\eta_{B})t}|\hf(\indexnot,t)|dt\right)
\end{split}
\end{equation}
holds for all $t\geq T_{\ron}$, where $\bt:=t-T_{\ron}$.
\end{lemma}
\begin{remark}
The estimate (\ref{eq:zzdottestfinnoise}) can also be formulated as (\ref{eq:wtestfinnoise}).
\end{remark}
\begin{remark}\label{remark:SrontrivJordbl}
In case all the Jordan blocks of $S_{\ron}(\indexnot)$ are trivial, the $N$ appearing in (\ref{eq:zzdottestfinnoise}) can be replaced by 
$1$. Moreover, if the first generalised eigenspace of the $\eta_{B}$, $S_{\ron}(\indexnot)$-decomposition of $\cn{2m}$ equals $\cn{2m}$, then 
the $N$ appearing in (\ref{eq:zzdottestfinnoise}) can be replaced by $d_{\ron,+}-1$, where $d_{\ron,+}:=d_{\max}(S_{\ron}(\indexnot),\kappa_{\ron,+})$.
\end{remark}
\begin{remark}
Dependence on the coefficients includes dependence on the Riemannian manifolds $(M_{r},g_{r})$, $r=1,\dots,R$; cf. 
Remark~\ref{remark:deponcoefincldeponRmfds}. 
\end{remark}

\begin{proof}
Consider (\ref{eq:zzdottkestnoisestmt}). The first step of the analysis is to simplify the left hand side. 

\textbf{Simplifying (\ref{eq:zzdottkestnoisestmt}).} To begin with, we replace $T_{\pre,k}^{-1}$ with $T_{\pre}^{-1}$, where
\begin{equation}\label{eq:Tpredefnoise}
T_{\pre}:=\frac{1}{2}\left(\begin{array}{rr} \mathrm{Id}_{m}
& i\mathrm{Id}_{m} \\  i\mathrm{Id}_{m} & \mathrm{Id}_{m}\end{array}
\right)
\end{equation}
In order to justify that this is allowed, note that 
\begin{equation*}
\begin{split}
T_{\pre,k}^{-1}-T_{\pre}^{-1}
 = & \frac{1}{\nu_{k}}\left(\begin{array}{cc} \Id_{m} & -i\Id_{m} \\ 
-i(\xi_{k}+\nu_{k})\Id_{m} & (\nu_{k}-\xi_{k})\Id_{m}\end{array}\right)-
\frac{1}{\nu_{k}}\left(\begin{array}{rr} \Id_{m} & -i\Id_{m} \\ 
-i\Id_{m} & \Id_{m}\end{array}\right)\\
 & +\left(\frac{1}{\nu_{k}}-1\right)\left(\begin{array}{rr} \Id_{m} & -i\Id_{m} \\ 
-i\Id_{m} & \Id_{m}\end{array}\right);
\end{split}
\end{equation*}
cf. (\ref{eq:Tprekage}) and (\ref{eq:Tprekainvge}) and the notation introduced in the adjacent text. Thus
\begin{equation}\label{eq:TprekinvminusTpreinv}
\|T_{\pre,k}^{-1}-T_{\pre}^{-1}\|\leq Ce^{-\eta_{\rosh}t_{k}}
\end{equation}
for all $k\geq 0$, where $C$ only depends on the coefficients of the equation (\ref{eq:thesystemRge}). As a consequence of this estimate, 
(\ref{eq:Psiinfestnoise}) and the fact that the largest real part of an eigenvalue of $R_{\ron}(\indexnot)$ is $\kappa_{\ron,+}$, 
\begin{equation}\label{eq:estasymptreplTprekbyTpre}
\begin{split}
 & \left|\exp\left(i\int_{0}^{t_{k}}\sigma(\indexnot,t')\mfg(\indexnot,t')dt'\right)(T_{\pre,k}^{-1}-T_{\pre}^{-1})[D_{k}(\indexnot)]^{-1}
e^{R_{\ron}(\indexnot)t_{k}}\psi_{\infty}\right|\\
 \leq & C_{B}e^{-\eta_{\rosh}t_{k}}\ldr{\bt_{k}}^{d_{\ron,+}-1}
e^{\kappa_{\ron,+}\bt_{k}}\left(|w(\indexnot,T_{\ron})|+e^{\kappa_{\ron,+}T_{\ron}}\int_{0}^{\infty}e^{-(\kappa_{\ron,+}-\eta_{B})t}|\hf(\indexnot,t)|dt\right),
\end{split}
\end{equation}
where $C_{B}$ depends only on $\eta_{B}$ and the coefficients of the equation (\ref{eq:thesystemRge}). Combining this estimate with 
(\ref{eq:zzdottkestnoisestmt}) yields
\begin{equation}\label{eq:zzdottkeststnoise}
\begin{split}
 & \left|\left(\begin{array}{c} \mfg(\indexnot,t_{k})z(\indexnot,t_{k}) \\ \dot{z}(\indexnot,t_{k})\end{array}\right)-
\exp\left(i\int_{0}^{t_{k}}\sigma(\indexnot,t')\mfg(\indexnot,t')dt'\right)T_{\pre}^{-1}[D_{k}(\indexnot)]^{-1}e^{R_{\ron}(\indexnot)t_{k}}\psi_{\infty}\right|\\
 \leq & C_{B}\ldr{\bt_{k}}^{N}e^{(\kappa_{\ron,+}-\eta_{B})\bt_{k}}
\left(|w(\indexnot,T_{\ron})|+e^{\kappa_{\ron,+}T_{\ron}}\int_{0}^{\infty}e^{-(\kappa_{\ron,+}-\eta_{B})t}|\hf(\indexnot,t)|dt\right),
\end{split}
\end{equation}
where $C_{B}$ has the same dependence as in the case of (\ref{eq:estasymptreplTprekbyTpre}) and $N$ only depends on $m$. Moreover, in case all the 
Jordan blocks of $R_{\ron}(\indexnot)$ are trivial, the $N$ appearing in (\ref{eq:zzdottkeststnoise}) can be replaced by $1$. In addition, if the first 
generalised eigenspace of the $\eta_{B}$, $R_{\ron}(\indexnot)$-decomposition of $\cn{2m}$ equals $\cn{2m}$, then the $N$ appearing in 
(\ref{eq:zzdottkeststnoise}) can be replaced by $d_{\ron,+}-1$. Let
\begin{align}
\varphi_{\rosh}(\indexnot,t) := & \int_{0}^{t}\left([1+\sigma^{2}(\indexnot,t')]^{1/2}-1\right)\mfg(\indexnot,t')dt'+
\int_{0}^{t_{0}}\mfg(\indexnot,t')dt'.\label{eq:varshdefnoise}
\end{align}
Then
\[
\varphi_{\rosh}(\indexnot,t_{k})=\int_{0}^{t_{k}}[1+\sigma^{2}(\indexnot,t')]^{1/2}\mfg(\indexnot,t')dt'-2\pi k;
\]
note that by the definition of the time sequence $\{t_{k}\}$,
\[
\int_{t_{0}}^{t_{k}}\mfg(\indexnot,t')dt'=2\pi k.
\] 
Define
\[
\mfD_{\rosh}(\indexnot,t):=\left(\begin{array}{rr} \cos\varphi_{\rosh}(\indexnot,t)\mathrm{Id}_{m}
& \sin\varphi_{\rosh}(\indexnot,t)\mathrm{Id}_{m} \\  -\sin\varphi_{\rosh}(\indexnot,t)\mathrm{Id}_{m} & \cos\varphi_{\rosh}(\indexnot,t)\mathrm{Id}_{m}\end{array}
\right).
\]
Then the equality
\[
T_{\pre}^{-1}[D_{k}(\indexnot)]^{-1}T_{\pre}=\mfD_{\rosh}(\indexnot,t_{k})
\]
holds, and (\ref{eq:zzdottkeststnoise}) can be rewritten
\begin{equation}\label{eq:zzdottkestttnoise}
\begin{split}
 & \left|\left(\begin{array}{c} \mfg(\indexnot,t_{k})z(\indexnot,t_{k}) \\ \dot{z}(\indexnot,t_{k})\end{array}\right)
-e^{i\omega_{\rosh}(\indexnot,t_{k})}\mfD_{\rosh}(\indexnot,t_{k})e^{S_{\ron}(\indexnot)t_{k}}z_{\infty}\right|\\
 \leq & C_{B}\ldr{\bt_{k}}^{N}e^{(\kappa_{\ron,+}-\eta_{B})\bt_{k}}\left(|w(\indexnot,T_{\ron})|+e^{\kappa_{\ron,+}T_{\ron}}
\int_{0}^{\infty}e^{-(\kappa_{\ron,+}-\eta_{B})t}|\hf(\indexnot,t)|dt\right)
\end{split}
\end{equation}
for all $k\geq 0$, where $z_{\infty}:=T_{\pre}^{-1}\psi_{\infty}$; $S_{\ron}(\indexnot)$ is defined by (\ref{eq:Sinfjndefnoise}); and $\omega_{\rosh}$ is defined by 
(\ref{eq:omegashdefnoise}). Moreover, $C_{B}$ and $N$ have the same dependence as in the case of (\ref{eq:zzdottkeststnoise}). In addition, in case all the 
Jordan blocks of $S_{\ron}(\indexnot)$ are trivial, the $N$ appearing in (\ref{eq:zzdottkestttnoise}) can be replaced by $1$. Finally, if the first 
generalised eigenspace of the $\eta_{B}$, $S_{\ron}(\indexnot)$-decomposition of $\cn{2m}$ equals $\cn{2m}$, then the $N$ appearing in 
(\ref{eq:zzdottkestttnoise}) can be replaced by $d_{\ron,+}-1$. Note that $d_{\ron,+}=d_{\max}(S_{\ron}(\indexnot),\kappa_{\ron,+})$ and that $z_{\infty}$
belongs to the first generalised eigenspace of the $\eta_{B}$, $S_{\ron}(\indexnot)$-decomposition of $\cn{2m}$. Moreover, given the definition
of $z_{\infty}$, (\ref{eq:zinfestnoise}) is an immediate consequence of (\ref{eq:Psiinfestnoise}). Keeping
(\ref{eq:xyFodefshiftge}), (\ref{eq:wdefshiftge}) and (\ref{eq:omegashdefnoise}) in mind, the estimate (\ref{eq:zzdottkestttnoise}) can 
be written
\begin{equation}\label{eq:zzdottkestttnoisewver}
\begin{split}
 & \left|w(\indexnot,t_{k})-\mfD_{\rosh}(\indexnot,t_{k})e^{S_{\ron}(\indexnot)t_{k}}z_{\infty}\right|\\
 \leq & C_{B}\ldr{\bt_{k}}^{N}e^{(\kappa_{\ron,+}-\eta_{B})\bt_{k}}\left(|w(\indexnot,T_{\ron})|+e^{\kappa_{\ron,+}T_{\ron}}
\int_{0}^{\infty}e^{-(\kappa_{\ron,+}-\eta_{B})t}|\hf(\indexnot,t)|dt\right)
\end{split}
\end{equation}
for all $k\geq 0$, where $C_{B}$ and $N$ have the same dependence as in the case of (\ref{eq:zzdottkeststnoise}). Next we consider 
the behaviour of the solution between the $t_{k}$'s.

\textbf{Estimating the solution for all $t$'s.} In order to estimate the solution in the intervals between the $t_{k}$'s, let us recall that for 
$t\in[t_{k},t_{k+1}]$
\begin{equation}\label{eq:wofitoXietcknoise}
w(\indexnot,t)=\Xi(\tau;\tau_{k})w(\indexnot,t_{k})+\Xi(\tau;\tau_{k})\int_{\tau_{k}}^{\tau}[\Phi(\tau';\tau_{k})]^{-1}F(\indexnot,\tau')d\tau',
\end{equation}
where we have appealed to (\ref{eq:wofitoXietc}) and $\tau$ and $\tau_{k}$ are assumed to correspond to $t$ and $t_{k}$ according to (\ref{eq:taudefge}).
At this stage we would like to appeal to Lemma~\ref{lemma:Xiroughappr}. Due to Lemma~\ref{lemma:strongbalneglshiftimploscad} and the definition
of the time sequence $\{t_{k}\}$, it is clear that this is allowed, with $t_{a}$ replaced by $t_{k}$, $t_{b}$ replaced by $t_{k+1}$, where $k\geq 0$, etc.
Moreover, since $\varepsilon_{a}$ is given by (\ref{eq:vareadef}) and Lemmas~\ref{lemma:strongbalneglshiftimploscad} and \ref{lemma:mfgnuronebrontequiv} 
apply, we can replace $\varepsilon_{a}$ in the estimates with 
\[
\varepsilon_{k}\leq\frac{Ce^{-\eta_{\rosh}t_{k}}}{\nu_{\ron}(\indexnot)e^{\b_{\ron}t_{k}}}+\frac{C}{\nu_{\ron}^{2}(\indexnot)e^{2\b_{\ron}t_{k}}},
\]
where $C$ only depends on the coefficients of the equation (\ref{eq:thesystemRge}). Keeping in mind that $\mff_{\roode}$, $\mff_{X}$ and $\mff_{\rosh}$ are given 
by (\ref{eq:mffXodeshdefnoise}), (\ref{eq:Xiroughest}) yields
\begin{equation}\label{eq:Xiroughestkcasenoise}
\begin{split}
 \left\|\Xi(\tau;\tau_{k})-\mfD_{\rom}(\indexnot,t)\right\|
 \leq & \frac{C}{\nu_{\ron}(\indexnot)e^{\b_{\ron}t_{k}}}+Ce^{-\eta_{\rosh}t_{k}}
\end{split}
\end{equation}
for all $t\in [t_{k},t_{k+1}]$, 
where $\mfD_{\rom}$ equals $\mfD_{a}$, defined in (\ref{eq:varphiamdadef}), with $a$ replaced by $k$ and $C$ only depends on the coefficients of the 
equation (\ref{eq:thesystemRge}). Define
\[
\varphi_{\rom}(\indexnot,t):=\int_{0}^{t}\mfg(\indexnot,t')dt'+\varphi_{\rom,0}(\indexnot),\ \ \
\varphi_{\rom,0}(\indexnot):=-\int_{0}^{t_{0}}\mfg(\indexnot,t')dt'.
\]
Then
\[
\int_{t_{k}}^{t}\mfg(\indexnot,t')dt'=\varphi_{\rom}(\indexnot,t)-2\pi k,
\]
so that 
\[
\mfD_{\rom}(\indexnot,t)=\left(\begin{array}{rr} \cos\varphi_{\rom}(\indexnot,t)\mathrm{Id}_{m}
& \sin\varphi_{\rom}(\indexnot,t)\mathrm{Id}_{m} \\  -\sin\varphi_{\rom}(\indexnot,t)\mathrm{Id}_{m} & \cos\varphi_{\rom}(\indexnot,t)\mathrm{Id}_{m}\end{array}
\right).
\]
Combining this observation with (\ref{eq:zinfestnoise}), (\ref{eq:zzdottkestttnoisewver}) and (\ref{eq:Xiroughestkcasenoise}) and the fact that the largest 
real part of an eigenvalue of $S_{\ron}(\indexnot)$ is $\kappa_{\ron,+}$, 
\begin{equation}\label{eq:wktmixestnoise}
\begin{split}
 & \left|\Xi(\tau;\tau_{k})w(\indexnot,t_{k})-\mfD_{\rom}(\indexnot,t)\mfD_{\rosh}(\indexnot,t_{k})
e^{S_{\ron}(\indexnot)t_{k}}z_{\infty}\right|\\
 \leq & C_{B}\ldr{\bt_{k}}^{N}e^{(\kappa_{\ron,+}-\eta_{B})\bt_{k}}\left(|w(\indexnot,T_{\ron})|
+e^{\kappa_{\ron,+}T_{\ron}}\int_{0}^{\infty}e^{-(\kappa_{\ron,+}-\eta_{B})t}|\hf(\indexnot,t)|dt\right)
\end{split}
\end{equation}
for all $k\geq 0$, where $C_{B}$ only depends on $\eta_{B}$ and the coefficients of the equation (\ref{eq:thesystemRge}) and $N$ only depends on $m$. 
Moreover, in case all the Jordan blocks of $S_{\ron}(\indexnot)$ are trivial, the $N$ appearing in (\ref{eq:wktmixestnoise}) can be replaced by $1$. 
In addition, if the first generalised eigenspace of the $\eta_{B}$, $S_{\ron}(\indexnot)$-decomposition of $\cn{2m}$ equals $\cn{2m}$, then the $N$ 
appearing in (\ref{eq:wktmixestnoise}) can be replaced by $d_{\ron,+}-1$, where $d_{\ron,+}=d_{\max}(S_{\ron}(\indexnot),\kappa_{\ron,+})$.

Next we wish to estimate
\[
\Xi(\tau;\tau_{k})\int_{\tau_{k}}^{\tau}[\Phi(\tau';\tau_{k})]^{-1}F(\indexnot,\tau')d\tau'.
\]
Note, to this end, that $\|\Xi(\tau;\tau_{k})\|$ is bounded by a constant on the interval of interest, where the constant depends only on the coefficients of 
the equation (\ref{eq:thesystemRge}); this is 
a consequence of (\ref{eq:Xiroughestkcasenoise}). That $\|[\Phi(\tau;\tau_{k})]^{-1}\|$ can be bounded in the same way on the interval of interest is a 
consequence of (\ref{eqAtintest}) and (\ref{eq:Phiinvappr}). Moreover, (\ref{eq:xyFodefshiftge}), (\ref{eq:xyFdefge}), (\ref{eq:taudefge}),
(\ref{eq:Fthdef}) and (\ref{eq:AFdef}) yield the conclusion that 
\begin{equation*}
\begin{split}
\int_{\tau_{k}}^{\tau}[\Phi(\tau';\tau_{k})]^{-1}F(\indexnot,\tau')d\tau' = & \int_{t_{k}}^{t}[\Phi(\tau(t');\tau_{k})]^{-1}
\exp\left[-\frac{1}{2}\int_{t_{k}}^{t'}\dot{\ell}(\indexnot,t'')dt''\right]\\
 & \cdot \exp\left(-i\int_{0}^{t'}\sigma(\indexnot,t'')\mfg(\indexnot,t'')dt''\right)\left(\begin{array}{c} 0 \\ U(t';t_{k})\hf(\indexnot,t')\end{array}\right)dt'.
\end{split}
\end{equation*}
That the integral involving $\dot{\ell}$ is bounded on the interval of interest is a consequence of (\ref{eq:gttaroughge}) and that $\|U(t';t_{k})\|$
is bounded on the interval of interest by a constant depending only on the coefficients of the equation (\ref{eq:thesystemRge}) is a consequence of 
(\ref{eq:UUinvmidest}). Combining these observations yields the conclusion that 
\begin{equation*}
\begin{split}
 & \left|\Xi(\tau;\tau_{k})\int_{\tau_{k}}^{\tau}[\Phi(\tau';\tau_{k})]^{-1}F(\indexnot,\tau')d\tau'\right| \leq  C\int_{t_{k}}^{t}|\hf(\indexnot,t')|dt'\\
 \leq & Ce^{(\kappa_{\ron,+}-\eta_{B})\bt_{k}}e^{\kappa_{\ron,+}t_{0}}\int_{t_{k}}^{t}e^{-(\kappa_{\ron,+}-\eta_{B})t'}|\hf(\indexnot,t')|dt',
\end{split}
\end{equation*}
where $C$ is a constant depending only on the coefficients of the equation (\ref{eq:thesystemRge}). Combining this estimate with 
(\ref{eq:wofitoXietcknoise}) and (\ref{eq:wktmixestnoise}) yields
\begin{equation}\label{eq:wtestnoisefinal}
\begin{split}
 & \left|w(\indexnot,t)-\mfD_{\rom}(\indexnot,t)\mfD_{\rosh}(\indexnot,t_{k})
e^{S_{\ron}(\indexnot)t_{k}}z_{\infty}\right|\\
 \leq & C_{B}\ldr{\bt_{k}}^{N}e^{(\kappa_{\ron,+}-\eta_{B})\bt_{k}}\left(|w(\indexnot,T_{\ron})|
+e^{\kappa_{\ron,+}T_{\ron}}\int_{0}^{\infty}e^{-(\kappa_{\ron,+}-\eta_{B})t}|\hf(\indexnot,t)|dt\right)
\end{split}
\end{equation}
for all $k\geq 0$ and $t\in [t_{k},t_{k+1}]$, where $C_{B}$ only depends on $\eta_{B}$ and the coefficients of the equation (\ref{eq:thesystemRge}) and $N$ only 
depends on $m$. Moreover, in case all the Jordan blocks of $S_{\ron}(\indexnot)$ are trivial, the $N$ appearing in (\ref{eq:wtestnoisefinal}) can be replaced 
by $1$. In addition, if the first generalised eigenspace of the $\eta_{B}$, $S_{\ron}(\indexnot)$-decomposition of $\cn{2m}$ equals $\cn{2m}$, then the $N$ 
appearing in (\ref{eq:wtestnoisefinal}) can be replaced by $d_{\ron,+}-1$, where $d_{\ron,+}=d_{\max}(S_{\ron}(\indexnot),\kappa_{\ron,+})$.

We would like to change the remaining $t_{k}$'s appearing in (\ref{eq:wtestnoisefinal}) to $t$'s. Changing $\bt_{k}$ to $\bt$ on the right hand side is clearly 
allowed. Changing $t_{k}$ appearing in $e^{S_{\ron}(\indexnot)t_{k}}$ to $t$ is also allowed due to (\ref{eq:tatbroughge}), (\ref{eq:zinfestnoise}) and the fact that the 
largest real part of an eigenvalue of $S_{\ron}(\indexnot)$ is $\kappa_{\ron,+}$. In order to justify that we are allowed to change the argument in $\mfD_{\rosh}$, 
note that, due to (\ref{eq:varshdefnoise}), 
\[
|\varphi_{\rosh}(\indexnot,t)-\varphi_{\rosh}(\indexnot,t_{k})|\leq \frac{1}{2}\int_{t_{k}}^{t}\sigma^{2}(\indexnot,t')\mfg(\indexnot,t')dt'\leq Ce^{-2\eta_{\rosh}t}
\]
for all $t\in [t_{k},t_{k+1}]$, where $C$ only depends on the coefficients of the equation (\ref{eq:thesystemRge}). Thus (\ref{eq:wtestnoisefinal}) can be rewritten
\begin{equation}\label{eq:wtestfinnoise}
\begin{split}
 & \left|w(\indexnot,t)-\mfD_{\rotot}(\indexnot,t)e^{S_{\ron}(\indexnot)t}z_{\infty}\right|\\
 \leq & C_{B}\ldr{\bt}^{N}e^{(\kappa_{\ron,+}-\eta_{B})\bt}\left(|w(\indexnot,T_{\ron})|
+e^{\kappa_{\ron,+}T_{\ron}}\int_{0}^{\infty}e^{-(\kappa_{\ron,+}-\eta_{B})t}|\hf(\indexnot,t)|dt\right)
\end{split}
\end{equation}
for all $k\geq 0$ and $t\in [t_{k},t_{k+1}]$, where $C_{B}$ only depends on $\eta_{B}$ and the coefficients of the equation (\ref{eq:thesystemRge}) and $N$ only 
depends on $m$. Moreover, in case all the Jordan blocks of $S_{\ron}(\indexnot)$ are trivial, the $N$ appearing in (\ref{eq:wtestfinnoise}) can be replaced 
by $1$. In addition, if the first generalised eigenspace of the $\eta_{B}$, $S_{\ron}(\indexnot)$-decomposition of $\cn{2m}$ equals $\cn{2m}$, then the $N$ 
appearing in (\ref{eq:wtestfinnoise}) can be replaced by $d_{\ron,+}-1$, where $d_{\ron,+}=d_{\max}(S_{\ron}(\indexnot),\kappa_{\ron,+})$. Finally, 
$\mfD_{\rotot}$ is given by (\ref{eq:mfDrototdefnoise}), where $\varphi_{\rotot}$ is defined by (\ref{eq:varphirototdefnoise}). The lemma follows.
\end{proof}

\chapter{Asymptotics for all future times}

\section{Introduction}

\textbf{Asymptotics and asymptotic characterisation of solutions for a class of scalar equations.} In the previous chapter we fixed one mode 
$\indexnot\in\EFnindexset$ 
and analysed the asymptotic behaviour of solutions to (\ref{eq:fourierthesystemRge}) in the corresponding late time regime $[T_{\ron},\infty)$.
In the present chapter, we derive estimates valid for all $t\geq 0$ and all $\indexnot\in\EFnindexset$. Similarly to the case of silent and transparent
equations, our goal is to derive asymptotics, given a solution, and to construct solutions with given asymptotics. However, in the present setting, it 
is natural to view the asymptotics as composed of two parts. On the one hand, there is the oscillatory behaviour which is already apparent when considering
scalar equations such as (\ref{eq:homtheeqnoise}) below. Second, there is the overall growth and decay of solutions to systems such as (\ref{eq:thesystemRge}).
In order to analyse the oscillatory part of the behaviour, we consider the equation (\ref{eq:homtheeqnoise}) separately in 
Section~\ref{section:caseofscalhomwaveeqnoise}. In particular, we prove that solutions can, in this case, be characterised by asymptotic data, so that there
is a bijection between asymptotic data and initial data. In fact, the bijection is even a homeomorphism with respect to the $C^{\infty}$ topology. In the
course of the analysis, it turns out to be natural to define two subclasses of solutions to (\ref{eq:homtheeqnoise}). The subclasses appear naturally
when considering the asymptotics, and lead to the definitions of positively and negatively oriented solutions in Definition~\ref{def:orsoltowenoise} below.

\textbf{Uniqueness of asymptotic representations.}
Before turning to the asymptotics of solutions to general equations, it is useful to prove a uniqueness result concerning the type of functions 
we use to represent the asymptotics later on. This is the subject of Section~\ref{section:auniquenresultnoise}. It is of interest to note that the notion 
of an orientation, introduced in Definition~\ref{def:orsoltowenoise}, plays an important role in the formulation and proof of uniqueness. Moreover, the 
uniqueness result demonstrates that the function representing the asymptotic behaviour of a solution $u$ to (\ref{eq:thesystemRge}), say 
$u_{\row}$, is uniquely determined by the estimate we derive for the difference between $u$ and $u_{\row}$. 

\textbf{Asymptotics for solutions to general equations.}
Given the above preliminaries, we describe the asymptotics of solutions to general equations in Section~\ref{eq:deasinnoissetnoise}. Interestingly, it 
turns out that the functions representing the asymptotics can be written $e^{A_{\ron,+}t}u_{+,\row}+e^{A_{\ron,-}t}u_{-,\row}$, where the $A_{\ron,\pm}$ 
are constant matrices determined by the coefficients of (\ref{eq:thesystemRge}), and the $u_{\pm,\row}$ are vector valued solutions to (\ref{eq:homtheeqnoise}).
In fact, the $u_{\pm,\row}$ take their values in appropriate generalised eigenspaces of $A_{\ron,\pm}$, say $E_{\ron,\pm}$, respectively. In this representation of 
the asymptotics, the functions $u_{\pm,\row}$ should be thought of as constituting the asymptotic data. However, since there is a bijection between solutions to 
(\ref{eq:homtheeqnoise}) and asymptotic data for solutions to (\ref{eq:homtheeqnoise}), we could represent the asymptotic data of solutions to 
(\ref{eq:thesystemRge}) in terms of the asymptotic data for the $u_{\pm,\row}$. 

\textbf{Specifying the asymptotics of solutions to general equations.}
Finally, in Section~\ref{section:spedataatinf} we demonstrate that, given solutions $u_{\pm,\row}$ to (\ref{eq:homtheeqnoise}) taking their values in 
$E_{\ron,\pm}$ respectively, there is a solution $u$ to (\ref{eq:thesystemRge}) (with $f=0$) with the corresponding asymptotic behaviour. Moreover, the corresponding 
map from asymptotic data (represented, e.g., by the initial data for the functions $u_{\pm,\row}$) to the initial data for $u$ is continuous in the 
$C^{\infty}$-topology. Finally, under favourable circumstances, this map is surjective, giving a homeomorphism from asymptotic data for solutions to
(\ref{eq:thesystemRge}) (with $f=0$) to the corresponding initial data.

\section{The case of scalar homogeneous wave equations}\label{section:caseofscalhomwaveeqnoise}

Next we turn to the problem of specifying the asymptotic behaviour of solutions to the scalar homogeneous equation
\begin{equation}\label{eq:homtheeqnoise}
u_{tt}-\textstyle\sum_{j,l=1}^{d}g^{jl}(t)\d_{j}\d_{l}u-2\sum_{l=1}^{d}g^{0l}(t)\d_{l}\d_{t}u
-\sum_{r=1}^{R}a^{-2}_{r}(t)\Delta_{g_{r}}u  = 0.
\end{equation}
As in Section~\ref{section:roughsobolevestnoise}, we focus on $u_{\ron}$, defined in analogy with (\ref{eq:frondef}).

\begin{lemma}\label{lemma:asymposccasescalareqnoise}
Consider the equation (\ref{eq:homtheeqnoise}).  Assume that it is strongly balanced, has a negligible shift vector field, a dominant noisy spatial direction, 
and is such that the dominant coefficients are convergent. Let $\eta_{A}$ be defined by (\ref{eq:etaAdefinition}) and let $0<\eta_{B}\leq \eta_{A}$.
Let $\chi_{\pm}(\indexnot)\in\co$, $\indexnot\in\EFnindexset$, be such that for every $0\leq k\in\zo$, there is a constant $C_{k}<\infty$ with the 
property that 
\begin{equation}\label{eq:rortbdnoise}
\textstyle{\sum}_{\indexnot\in\EFnindexset}\ldr{\nu(\indexnot)}^{2k}[|\chi_{+}(\indexnot)|^{2}+|\chi_{-}(\indexnot)|^{2}]\leq C_{k}.
\end{equation}
Define $z_{\app}$ and $u_{\app}$ by 
\begin{align}
z_{\app}(\indexnot,t) := & \frac{e^{\b_{\ron}t/2}}{\mfg(\indexnot,t)}\left(\chi_{+}(\indexnot)e^{i[\omega_{\rosh}(\indexnot,t)-\varphi_{\rotot}(\indexnot,t)]}
+\chi_{-}(\indexnot)e^{i[\omega_{\rosh}(\indexnot,t)+\varphi_{\rotot}(\indexnot,t)]}\right)\label{eq:zappdefnoise}\\
u_{\app}(p,t) := & \sum_{\indexnot\in\EFnindexset}z_{\app}(\indexnot,t)\varphi_{\indexnot}(p),\label{eq:uappdefnoise}
\end{align}
where $\indexnot\in\EFnindexset$ in (\ref{eq:zappdefnoise}); $\omega_{\rosh}$ is defined by (\ref{eq:omegashdefnoise}); $\varphi_{\rotot}$ is defined by
(\ref{eq:varphirototdefnoise}); and $\mfg(\indexnot,t)$ is defined by (\ref{eq:mfgnutdef}). Then there is a unique smooth solution $u$ to 
(\ref{eq:homtheeqnoise}) such that the following holds: $u=u_{\ron}$, where $u_{\ron}$ is defined in analogy with (\ref{eq:frondef}); and there are
constants $s\in\ro$, $\eta>0$ and $C>0$ such that 
\begin{equation}\label{eq:umiuappuniqueest}
\mfe_{s}[u-u_{\app}](t)\leq Ce^{(\b_{\ron}-\eta)t}
\end{equation}
for all $t\geq 0$. Moreover, this unique solution has the property that 
\begin{equation}\label{eq:mfediffhomestnoise}
\mfe_{s}[u-u_{\app}](t)\leq C_{B}e^{(\b_{\ron}-2\eta_{B})t}\textstyle{\sum}_{\indexnot\in\EFnindexset}\ldr{\nu(\indexnot)}^{2(s+s_{\rom})}[r_{+}^{2}(\indexnot)+r_{-}^{2}(\indexnot)]
\end{equation}
for all $t\geq 0$ and $s\in\ro$, where $C_{B}$ and $s_{\rom}$ only depend on $\eta_{B}$ and the coefficients of the equation (\ref{eq:homtheeqnoise}); 
$\mfe_{s}$ is defined by (\ref{eq:mfedef}); and $r_{\pm}(\indexnot)=|\chi_{\pm}(\indexnot)|$. Finally,
\begin{equation}\label{eq:mfezeroestnoise}
\mfe^{1/2}_{s}[u](0)\leq 
C_{B}\left(\textstyle{\sum}_{\indexnot\in\EFnindexset}\ldr{\nu(\indexnot)}^{2(s+s_{\rom})}[r_{+}^{2}(\indexnot)+r_{-}^{2}(\indexnot)]\right)^{1/2}
\end{equation}
for all $s\in\ro$, where $C_{B}$ and $s_{\rom}$ have the same dependence as in the case of (\ref{eq:mfediffhomestnoise}).

Conversely, assume that $u$ is a smooth solution to (\ref{eq:homtheeqnoise}) such that $u=u_{\ron}$. Then there are uniquely determined 
$\chi_{\pm}(\indexnot)$, $\indexnot\in\EFnindexset$, such that (\ref{eq:rortbdnoise}) holds, and such that if $u_{\app}$ is defined by 
(\ref{eq:zappdefnoise}) and (\ref{eq:uappdefnoise}), then (\ref{eq:umiuappuniqueest}) holds. Moreover, this $u_{\app}$ is such that 
(\ref{eq:mfediffhomestnoise}) holds. Finally, 
\begin{equation}\label{eq:chipmmfezeroestnoise}
\left(\textstyle{\sum}_{\indexnot\in\EFnindexset}\ldr{\nu(\indexnot)}^{2s}[|\chi_{+}(\indexnot)|^{2}+|\chi_{-}(\indexnot)|^{2}]\right)^{1/2}\leq 
C_{B}\mfe^{1/2}_{s+s_{\rom}}[u](0)
\end{equation}
for all $s\in\ro$, where $C_{B}$ and $s_{\rom}$ have the same dependence as in the case of (\ref{eq:mfediffhomestnoise}).
\end{lemma}
\begin{remark}\label{remark:homwaveonlyprincsymbnoise}
Assume that (\ref{eq:homtheeqnoise}) is $C^{2}$-balanced with a geometric dominant noisy spatial direction, convergent dominant 
coefficients and a negligible shift vector field; cf. Definition~\ref{def:noisemainassumptions}. Then the assumptions of the 
lemma are satisfied; cf. Remarks~\ref{remark:geometrictonongeometricnoise} and \ref{remark:strongbalanceneglshiftnoisegeometric}.
Moreover, if $\eta_{\rosh}$, $\eta_{\ron}$, $\b_{\ron}$ and $\eta_{\romn}$ are the constants arising in Definition~\ref{def:noisemainassumptions},
then $\eta_{\rosh}$, $\b_{\ron}$ and $\eta_{\romn}$ are the same as in the present context. Moreover, $\eta_{\romar}=\min\{2\eta_{\rosh},\eta_{\ron}\}$
in the case of a geometric dominant noisy spatial $\so$-direction. In addition, $\eta_{\romar}=\eta_{\ron}$ in the case of a geometric dominant 
noisy spatial direction corresponding to, say, $M_{r}$. In both cases, $\eta_{A}=\min\{\eta_{\ron}/2,\eta_{\rosh},\eta_{\romn},\b_{\ron}\}$. 
\end{remark}
\begin{remark}
The function $u_{\app}$ is smooth and complex valued; the smoothness is an immediate consequence of the definition and 
(\ref{eq:rortbdnoise}). 
\end{remark}
\begin{remark}
The way the lemma is formulated, there is no apparent gain in letting $\eta_{B}$ be strictly smaller than $\eta_{A}$. However, the value of $s_{\rom}$ depends 
on $\eta_{B}$; the smaller the $\eta_{B}$, the smaller the $s_{\rom}$.
\end{remark}
\begin{remark}
Dependence on the coefficients includes dependence on the Riemannian manifolds $(M_{r},g_{r})$, $r=1,\dots,R$; cf. 
Remark~\ref{remark:deponcoefincldeponRmfds}. 
\end{remark}
\begin{remark}
The constants $C$, $s$ and $\eta$ appearing in (\ref{eq:umiuappuniqueest}) are allowed to depend on $u$, $u_{\app}$ etc. 
\end{remark}
\begin{proof}
The proof is naturally divided into four parts. First, we demonstrate uniqueness. Second, we consider one mode, constructing a solution corresponding 
to given asymptotic data. Third, we sum up the estimates for the different modes. Finally, we arrive at asymptotics starting with a solution to 
(\ref{eq:homtheeqnoise}) such that $u=u_{\ron}$. 

\textbf{Uniqueness.} Assume that there are two solutions, say $u_{l}$, $l=1,2$, to (\ref{eq:homtheeqnoise}) such that $u_{l}=(u_{l})_{\ron}$ and 
such that (\ref{eq:umiuappuniqueest}) holds with $u$ replaced by $u_{l}$. Then the difference, say $u:=u_{1}-u_{2}$, satisfies 
\[
\mfe_{s}[u](t)\leq Ce^{(\b_{\ron}-\eta)t}
\]
for all $t\geq 0$. Defining $z$ by (\ref{eq:znutdef}) and letting $\indexnot\in\EFnindexset$, this implies that 
\[
|\dot{z}(\indexnot,t)|+\mfg(\indexnot,t)|z(\indexnot,t)|\leq Ce^{(\b_{\ron}-\eta)t/2}
\]
for all $t\geq 0$, where $C$ is allowed to depend on $u_{l}$, $l=1,2$, $\indexnot$, etc. We wish to compare this estimate with the conclusions of 
Lemma~\ref{lemma:asymposccasenoise}. Note, to this end, that $R_{\ron}(\indexnot)=\b_{\ron}\Id_{2}/2$ in the present context. Moreover, 
$\kappa_{\ron,+}=\b_{\ron}/2$. Combining these observations, it is clear that the only way for (\ref{eq:zzdottkestnoisestmt}) to be satisfied is 
that $\psi_{\infty}=0$. In fact, $\psi_{\infty}(\indexnot)=0$ for all $\indexnot\in\EFnindexset$. 

Let us now assume that $u\neq 0$. Then there is a $\indexnot\in\EFnindexset$ such that $z(\indexnot,\cdot)\neq 0$.
In particular, the initial data for $z(\indexnot,\cdot)$ at $T_{\ron}$ are non-zero. In order to obtain uniqueness, we demonstrate that this 
statement contradicts the conclusions of Lemma~\ref{lemma:asdataosccasenoise}. Using the notation of Lemma~\ref{lemma:asdataosccasenoise},
note that $m=1$ and that $E_{a}=\cn{2}$ in the present setting. Since $\Psi_{\infty}$ is injective, we conclude that it is surjective. 
There is thus a $\chi\in\cn{2}$ such that (\ref{eq:Psiinfchitoidnoise}) holds. Since the right hand side of this equality is non-zero, it is 
clear that $\chi$ is non-zero. Moreover, Lemma~\ref{lemma:asdataosccasenoise} implies that (\ref{eq:zzdottkestnoisestmt}) holds with 
$\psi_{\infty}$ replaced by $\chi$. By the above argument, we conclude that $\chi=0$, a contradiction. Thus $u=0$, and uniqueness follows. 

\textit{Uniqueness of the approximation.} Let $u=u_{\ron}$ be a solution to (\ref{eq:homtheeqnoise}). Assume that there are $\chi_{\pm,l}(\indexnot)$,
$\indexnot\in\EFnindexset$, $l=1,2$,  such that (\ref{eq:rortbdnoise}) holds with $\chi_{\pm}(\indexnot)$ replaced by $\chi_{\pm,l}(\indexnot)$, and such 
that if $u_{\app,l}$ is defined by (\ref{eq:zappdefnoise}) and (\ref{eq:uappdefnoise}) with $\chi_{\pm}(\indexnot)$ replaced by $\chi_{\pm,l}(\indexnot)$, 
then (\ref{eq:umiuappuniqueest}) holds with $u_{\app}$ replaced by $u_{\app,l}$, $l=1,2$. Then
\begin{equation}\label{eq:uappestgivingunique}
\mfe_{s}[u_{\app}](t)\leq Ce^{(\b_{\ron}-\eta)t}
\end{equation}
for all $t\geq 0$, where $u_{\app}:=u_{\app,1}-u_{\app,2}$. Note that $u_{\app}$ is defined by (\ref{eq:zappdefnoise}) and (\ref{eq:uappdefnoise}), where 
$\chi_{\pm}=\chi_{\pm,1}-\chi_{\pm,2}$. Fix a $\indexnot\in\EFnindexset$ and consider $z_{\app}(\indexnot,\cdot)$ given by (\ref{eq:zappdefnoise}). Due to 
(\ref{eq:uappestgivingunique}), we know that 
\begin{equation}\label{eq:limitszappzappdot}
\lim_{t\rightarrow\infty}e^{-\b_{\ron}t/2}\mfg(\indexnot,t)z_{\app}(\indexnot,t)=0,\ \ \
\lim_{t\rightarrow\infty}e^{-\b_{\ron}t/2}\dot{z}_{\app}(\indexnot,t)=0.
\end{equation}
The first equality implies that 
\begin{equation}\label{eq:chipchimfirstlimit}
\lim_{t\rightarrow\infty}\left(\chi_{+}(\indexnot)e^{i[\omega_{\rosh}(\indexnot,t)-\varphi_{\rotot}(\indexnot,t)]}
+\chi_{-}(\indexnot)e^{i[\omega_{\rosh}(\indexnot,t)+\varphi_{\rotot}(\indexnot,t)]}\right)=0.
\end{equation}
Appealing to the definitions of $\omega_{\rosh}$ and $\varphi_{\rotot}$; the assumed bounds on $\sigma$ and $\dot{\ell}$; and the fact that 
$e^{-\b_{\ron}t}\mfg(\indexnot,t)$ converges to a non-zero number (cf. Lemma~\ref{lemma:mfgnuronebrontequiv}), it can be verified that the 
second equality in (\ref{eq:limitszappzappdot}) implies that 
\begin{equation}\label{eq:chipchimsecondlimit}
\lim_{t\rightarrow\infty}\left(-\chi_{+}(\indexnot)e^{i[\omega_{\rosh}(\indexnot,t)-\varphi_{\rotot}(\indexnot,t)]}
+\chi_{-}(\indexnot)e^{i[\omega_{\rosh}(\indexnot,t)+\varphi_{\rotot}(\indexnot,t)]}\right)=0.
\end{equation}
Combining (\ref{eq:chipchimfirstlimit}) and (\ref{eq:chipchimsecondlimit}) yields the conclusion that $\chi_{+}(\indexnot)=\chi_{-}(\indexnot)=0$
for all $\indexnot\in\EFnindexset$. Thus $\chi_{\pm,1}(\indexnot)=\chi_{\pm,2}(\indexnot)$ for all $\indexnot\in\EFnindexset$, and the desired 
uniqueness follows. 

\textbf{One mode.}
We wish to appeal to Lemmas~\ref{lemma:asdataosccasenoise} and \ref{lemma:asymposccasefinnoise}. Let, to this end, $z$ denote the $\indexnot$'th Fourier 
coefficient of a solution $u$ to (\ref{eq:homtheeqnoise}). Then $z$ satisfies
\begin{equation}\label{eq:fourierhomtheeqnoise}
\ddot{z}(\indexnot,t)-2i\sigma(\indexnot,t)\mfg(\indexnot,t)\dot{z}(\indexnot,t)+\mfg^{2}(\indexnot,t)z(\indexnot,t)=0.
\end{equation}
As is clear from the statement of the lemma, we are here only interested in $\indexnot\in\EFnindexset$. We therefore assume $\indexnot\in\EFnindexset$
from now on. When comparing the present setting with Lemma~\ref{lemma:asymposccasefinnoise}, it is clear that $\a=0$ and $X=0$, 
so that $S_{\ron}(\indexnot)=\b_{\ron}\Id_{2}/2$. Thus (\ref{eq:zzdottestfinnoise}) with $0<\eta_{B}\leq\eta_{A}$ yields
\begin{equation}\label{eq:zzdottestfinscnoise}
\begin{split}
 & \left|\left(\begin{array}{c} \mfg(\indexnot,t)z(\indexnot,t) \\ \dot{z}(\indexnot,t)\end{array}\right)-e^{i\omega_{\rosh}(\indexnot,t)}e^{\b_{\ron}t/2}
\mfD_{\rotot}(\indexnot,t)z_{\infty}\right|\\
 \leq & C_{B}e^{(\b_{\ron}/2-\eta_{B})\bt}\left(|\dot{z}(\indexnot,T_{\ron})|+\mfg(\indexnot,T_{\ron})|z(\indexnot,T_{\ron})|\right)
\end{split}
\end{equation}
for all $t\geq T_{\ron}$, where $\bt:=t-T_{\ron}$; $z_{\infty}\in\cn{2}$; $T_{\ron}$ is defined by (\ref{eq:Trondef}); and $C_{B}$ only depends on $\eta_{B}$ and 
the coefficients of the equation (\ref{eq:homtheeqnoise}). In order to justify the absence of polynomial powers in the estimate 
(\ref{eq:zzdottestfinscnoise}), recall Remark~\ref{remark:SrontrivJordbl}. Note, in particular, that since the first generalised eigenspace of the 
$\eta_{B}$, $S_{\ron}(\indexnot)$-decomposition of $\cn{2}$ equals $\cn{2}$, $N$ appearing in (\ref{eq:zzdottestfinnoise}) can be replaced by $d_{\ron,+}-1$.
Moreover, since $d_{\ron,+}=1$ in the present setting, $N$ can be replaced by $0$. In what follows, it is useful to keep in mind that, due to 
Lemma~\ref{lemma:asdataosccasenoise}, $z_{\infty}$ is freely specifiable. The reason for this is that $E_{a}$ appearing in the statement of 
Lemma~\ref{lemma:asdataosccasenoise} equals $\cn{2}$. Moreover, the reason why $E_{a}=\cn{2}$ is that $\Rsp [R_{\ron}(\indexnot)]=0$; cf. 
Definition~\ref{def:SpRspdef}. Recall, finally, that the $\psi_{\infty}$ appearing in (\ref{eq:zzdottkestnoisestmt}) is related to the 
$z_{\infty}$ appearing in (\ref{eq:zzdottestfinnoise}) according to $z_{\infty}=T_{\pre}^{-1}\psi_{\infty}$. Let now $z$ be the solution to 
(\ref{eq:fourierhomtheeqnoise}) corresponding to asymptotic data $z_{\infty}$. Then Lemma~\ref{lemma:asdataosccasenoise} yields
\begin{equation}\label{eq:windexnotTronestitoasdata}
|w(\indexnot,T_{\ron})|\leq |\dot{z}(\indexnot,T_{\ron})|+\mfg(\indexnot,T_{\ron})|z(\indexnot,T_{\ron})|\leq Ce^{\b_{\ron}T_{\ron}/2}|z_{\infty}|,
\end{equation}
where the constant $C$ only depends on the coefficients of the equation (\ref{eq:homtheeqnoise}). Denote the components of $z_{\infty}$ by $z_{1}$ and 
$z_{2}$. Then (\ref{eq:zzdottestfinscnoise}) can be written 
\begin{equation}\label{eq:zzdottestfinscexpnoise}
\begin{split}
 & \left|\left(\begin{array}{c} \mfg(\indexnot,t)z(\indexnot,t) \\ \dot{z}(\indexnot,t)\end{array}\right)-e^{i\omega_{\rosh}(\indexnot,t)}e^{\b_{\ron}t/2}
\left(\begin{array}{c} z_{1}\cos\varphi_{\rotot}(\indexnot,t)+z_{2}\sin\varphi_{\rotot}(\indexnot,t) \\ 
-z_{1}\sin\varphi_{\rotot}(\indexnot,t)+z_{2}\cos\varphi_{\rotot}(\indexnot,t)\end{array}\right)\right|\\
 \leq & C_{B}e^{(\b_{\ron}/2-\eta_{B})\bt}e^{\b_{\ron}T_{\ron}/2}|z_{\infty}|
\end{split}
\end{equation}
for all $t\geq T_{\ron}$, where the constant $C_{B}$ has the same dependence as in the case of (\ref{eq:zzdottestfinscnoise}). Let 
\begin{equation}\label{eq:zapppreldefnoise}
z_{\app}(\indexnot,t):=\frac{1}{\mfg(\indexnot,t)}e^{i\omega_{\rosh}(\indexnot,t)}e^{\b_{\ron}t/2}[z_{1}\cos\varphi_{\rotot}(\indexnot,t)+z_{2}\sin\varphi_{\rotot}(\indexnot,t)].
\end{equation}
Then
\begin{equation}\label{eq:zapprdotprelnoise}
\begin{split}
\dot{z}_{\app}(\indexnot,t) = & \left[-\dot{\ell}(\indexnot,t)+\frac{\b_{\ron}}{2}+i\dot{\omega}_{\rosh}(\indexnot,t)\right]z_{\app}(\indexnot,t)\\
 & +\frac{1}{\mfg(\indexnot,t)}\dot{\varphi}_{\rotot}(\indexnot,t)e^{i\omega_{\rosh}(\indexnot,t)}e^{\b_{\ron}t/2}[-z_{1}\sin\varphi_{\rotot}(\indexnot,t)+z_{2}\cos\varphi_{\rotot}(\indexnot,t)].
\end{split}
\end{equation}
Due to (\ref{eq:omegashdefnoise}) and (\ref{eq:varphirototdefnoise}), 
\[
\dot{\varphi}_{\rotot}(\indexnot,t)=[1+\sigma^{2}(\indexnot,t)]^{1/2}\mfg(\indexnot,t),\ \ \
\dot{\omega}_{\rosh}(\indexnot,t)=\sigma(\indexnot,t)\mfg(\indexnot,t).
\]
For this reason, the absolute value of the first term on the right hand side of (\ref{eq:zapprdotprelnoise}) can be bounded by the 
right hand side of (\ref{eq:zzdottestfinscexpnoise}). Moreover, in the second term on the right hand side of (\ref{eq:zapprdotprelnoise}),
we can replace $\dot{\varphi}_{\rotot}(\indexnot,t)$ by $\mfg(\indexnot,t)$ at the price of an error term whose absolute value can be estimated by the 
right hand side of (\ref{eq:zzdottestfinscexpnoise}). To conclude, (\ref{eq:zzdottestfinscexpnoise}) can be rewritten
\begin{equation}\label{eq:zzdottestfinscexpappnoise}
\begin{split}
 & \left|\left(\begin{array}{c} \mfg(\indexnot,t)z(\indexnot,t) \\ \dot{z}(\indexnot,t)\end{array}\right)
-\left(\begin{array}{c} \mfg(\indexnot,t)z_{\app}(\indexnot,t) \\ \dot{z}_{\app}(\indexnot,t)\end{array}\right)\right|
 \leq C_{B}e^{(\b_{\ron}/2-\eta_{B})\bt}e^{\b_{\ron}T_{\ron}/2}|z_{\infty}|
\end{split}
\end{equation}
for all $t\geq T_{\ron}$, where the constant $C_{B}$ has the same dependence as in the case of (\ref{eq:zzdottestfinscnoise}). It is convenient to rewrite
(\ref{eq:zapppreldefnoise}) by introducing $\chi_{\pm}$ according to 
\begin{equation}\label{eq:zetaldefnoise}
\chi_{\pm}=\frac{1}{2}(z_{1}\pm iz_{2}).
\end{equation}
Then (\ref{eq:zapppreldefnoise}) becomes
\[
z_{\app}(\indexnot,t)=\frac{e^{\b_{\ron}t/2}}{\mfg(\indexnot,t)}\left(\chi_{+}e^{i[\omega_{\rosh}(\indexnot,t)-\varphi_{\rotot}(\indexnot,t)]}
+\chi_{-}e^{i[\omega_{\rosh}(\indexnot,t)+\varphi_{\rotot}(\indexnot,t)]}\right).
\]
In order for (\ref{eq:zzdottestfinscexpappnoise}) to be of use when summing up the modes, we need to know that it holds for all $t\geq 0$. 
We therefore need to consider the interval $[0,T_{\ron}]$ in greater detail. The fact that the equation is strongly balanced implies that 
Lemma~\ref{lemma:roughenestbalsetting} is applicable (with $\betafun=0$). Thus, we are allowed to combine (\ref{eq:meestroughbalset}) and 
(\ref{eq:windexnotTronestitoasdata}) in order to conclude
\begin{equation}\label{eq:mesqrtestztoTron}
\begin{split}
\me^{1/2}(\indexnot,t) \leq & e^{\eta_{\robal}|t-T_{\ron}|}\me^{1/2}(\indexnot,T_{\ron})\leq Ce^{\eta_{\robal}T_{\ron}}e^{\b_{\ron}T_{\ron}/2}|z_{\infty}|\\
 \leq & Ce^{(\b_{\ron}/2-\eta_{B})t}e^{\eta_{\rom}T_{\ron}}|z_{\infty}|
\end{split}
\end{equation}
for all $t\in [0,T_{\ron}]$, where $\eta_{\robal}$ and $C$ only depend on the coefficients of the equation (\ref{eq:homtheeqnoise}). Moreover, 
\[
\eta_{\rom}:=\max\{\eta_{B}-\b_{\ron}/2,0\}+\eta_{\robal}+\beta_{\ron}/2.
\]
On the other hand, the definition of $z_{\app}$ implies that 
\[
\mfg(\indexnot,t)|z_{\app}(\indexnot,t)|+|\dot{z}_{\app}(\indexnot,t)|\leq Ce^{(\b_{\ron}/2-\eta_{B})t}e^{\eta_{B}T_{\ron}}|z_{\infty}|
\]
for all $t\in [0,T_{\ron}]$, where $C$ only depends on the coefficients of the equation (\ref{eq:homtheeqnoise}). Combining this estimate with 
(\ref{eq:zzdottestfinscexpappnoise}) and (\ref{eq:mesqrtestztoTron}) yields 
\begin{equation}\label{eq:zzdottestfinscexpappnoiseallfutt}
\begin{split}
 & \left|\left(\begin{array}{c} \mfg(\indexnot,t)z(\indexnot,t) \\ \dot{z}(\indexnot,t)\end{array}\right)
-\left(\begin{array}{c} \mfg(\indexnot,t)z_{\app}(\indexnot,t) \\ \dot{z}_{\app}(\indexnot,t)\end{array}\right)\right|
 \leq C_{B}e^{(\b_{\ron}/2-\eta_{B})t}e^{\eta_{\rom}T_{\ron}}|z_{\infty}|
\end{split}
\end{equation}
for all $t\geq 0$, where the constant $C_{B}$ has the same dependence as in the case of (\ref{eq:zzdottestfinscnoise}).

\textbf{Summing up the estimates for the individual modes.}
Let $\chi_{\pm}(\indexnot)$ be as in the statement of the lemma, and define $z_{\app}$ and $u_{\app}$ according to (\ref{eq:zappdefnoise}) 
and (\ref{eq:uappdefnoise}) respectively. In order to construct the solution $u$ with the desired properties, let $z(\indexnot,t)$ 
be the solution to (\ref{eq:fourierhomtheeqnoise}) corresponding to the asymptotic data specified by $\chi_{\pm}(\indexnot)$. In order to 
translate (\ref{eq:zzdottestfinscexpappnoiseallfutt}) into an estimate for $u$, we need to estimate $e^{\eta_{\rom}T_{\ron}}$. Note, to this end, that
\begin{equation}\label{eq:Tronuppbdnoise}
T_{\ron}\leq \frac{2}{\eta_{\romar}}\ln\ldr{\nu(\indexnot)}+C,
\end{equation}
where $C$ only depends on the coefficients of the equation (\ref{eq:homtheeqnoise}); cf. Definition~\ref{def:Tron}. Thus
\[
e^{\eta_{\rom}T_{\ron}}\leq C_{B}\ldr{\nu(\indexnot)}^{s_{\rom,1}},
\]
where $s_{\rom,1}=2\eta_{\rom}/\eta_{\romar}$ and the constant $C_{B}$ only depends on $\eta_{B}$ and the coefficients of the equation (\ref{eq:homtheeqnoise}). 
Thus (\ref{eq:zzdottestfinscexpappnoiseallfutt}) implies 
\begin{equation}\label{eq:zzdottestpresumtnoise}
\begin{split}
 & \left|\left(\begin{array}{c} \mfg(\indexnot,t)z(\indexnot,t) \\ \dot{z}(\indexnot,t)\end{array}\right)
-\left(\begin{array}{c} \mfg(\indexnot,t)z_{\app}(\indexnot,t) \\ \dot{z}_{\app}(\indexnot,t)\end{array}\right)\right|\\
 \leq & C_{B}e^{(\b_{\ron}/2-\eta_{B})t}\ldr{\nu(\indexnot)}^{s_{\rom,1}}[r_{+}(\indexnot)+r_{-}(\indexnot)]
\end{split}
\end{equation}
for all $t\geq 0$, where the constant $C_{B}$ has the same dependence as in the case of (\ref{eq:zzdottestfinscnoise}) and $r_{\pm}(\indexnot):=|\chi_{\pm}(\indexnot)|$.
Similarly, (\ref{eq:mesqrtestztoTron}) implies that 
\begin{equation}\label{eq:mesqrtevatzero}
\me^{1/2}(\indexnot,0) \leq C_{B}\ldr{\nu(\indexnot)}^{s_{\rom,1}}[r_{+}(\indexnot)+r_{-}(\indexnot)],
\end{equation}
where the constant $C_{B}$ has the same dependence as in the case of (\ref{eq:zzdottestfinscnoise}). 

In order to prove (\ref{eq:mfediffhomestnoise}), 
we need to estimate the energy $\me_{s}$ of $z-z_{\app}$. Such an estimate almost follows from (\ref{eq:zzdottestpresumtnoise}). However, there is 
one piece missing: we need to estimate $\mfg(\indexnot,t)$ from below. The estimate $|\dot{\ell}(\indexnot,t)|\leq C_{\ell}$ for
all $t\geq 0$ and $0\neq\indexnot\in\EFindexset$, where $C_{\ell}$ only depends on the coefficients of the equation, implies that 
\[
\left|\ln\frac{\mfg(\indexnot,T_{\ron})}{\mfg(\indexnot,t)}\right|\leq C_{\ell}T_{\ron}
\]
for all $t\in [0,T_{\ron}]$. In particular, 
\begin{equation}\label{eq:mfglowbdztTron}
e^{\b_{\ron}t}[\mfg(\indexnot,t)]^{-1}\leq e^{(C_{\ell}+\b_{\ron})T_{\ron}}\leq C\ldr{\nu(\indexnot)}^{s_{E}}
\end{equation}
for all $t\in [0,T_{\ron}]$, where $C$ and $s_{E}$ only depend on the coefficients of the equation (\ref{eq:homtheeqnoise}) and we appealed to 
(\ref{eq:condensurtimeseqwd}) and (\ref{eq:Tronuppbdnoise}). On the other hand, (\ref{eq:mfgnunebnexpdec}) yields the conclusion that 
$2\mfg(\indexnot,t)\geq \nu_{\ron}(\indexnot)e^{\b_{\ron}t}$ for all $t\geq T_{\ron}$. Combining this estimate with (\ref{eq:mfglowbdztTron}) yields the 
conclusion that 
\begin{equation}\label{eq:mfglowbdallfuttimenoise}
[\mfg(\indexnot,t)]^{-1}\leq C\ldr{\nu(\indexnot)}^{s_{E}}e^{-\b_{\ron}t}
\end{equation}
for all $t\geq 0$, where $C$ and $s_{E}$ only depend on the coefficients of the equation (\ref{eq:homtheeqnoise}). Combining this estimate with 
(\ref{eq:zzdottestpresumtnoise}) yields
\begin{equation}\label{eq:mezzdottestpresumtnoise}
\me_{s}^{1/2}[z-z_{\app}](\indexnot,t)\leq C_{B}e^{(\b_{\ron}/2-\eta_{B})t}\ldr{\nu(\indexnot)}^{s+s_{\rom}}[r_{+}(\indexnot)+r_{-}(\indexnot)]
\end{equation}
for all $t\geq 0$, where $C_{B}$ only depends on $\eta_{B}$ and the coefficients of the equation (\ref{eq:homtheeqnoise}) and $s_{\rom}=s_{\rom,1}+s_{E}$. 
Define $u$ by
\[
u(p,t):=\textstyle{\sum}_{\indexnot\in\EFnindexset}z(\indexnot,t)\varphi_{\indexnot}(p).
\]
Due to the estimate (\ref{eq:mezzdottestpresumtnoise}), the definition of $z_{\app}$, (\ref{eq:rortbdnoise}) and the fact that $z(\indexnot,t)$ solves
(\ref{eq:fourierhomtheeqnoise}), it is clear that $u$ is a smooth solution to (\ref{eq:homtheeqnoise}). Due to (\ref{eq:mezzdottestpresumtnoise}) it 
is clear that (\ref{eq:mfediffhomestnoise}) holds. Similarly, (\ref{eq:mesqrtevatzero}) yields (\ref{eq:mfezeroestnoise}).

\textbf{Deriving asymptotics for a given solution.}
Finally, let us turn to the converse statement. Let $u$ be a smooth solution to (\ref{eq:homtheeqnoise}) such that $u=u_{\ron}$. Then we can appeal to 
Lemma~\ref{lemma:asymposccasefinnoise}. In particular, if $z$ is the $\indexnot$'th mode of $u$, where $\indexnot\in\EFnindexset$,
the estimate (\ref{eq:zzdottestfinnoise}) yields
\begin{equation}\label{eq:zzdottestfinauxnoise}
\begin{split}
 & \left|\left(\begin{array}{c} \mfg(\indexnot,t)z(\indexnot,t) \\ \dot{z}(\indexnot,t)\end{array}\right)-e^{i\omega_{\rosh}(\indexnot,t)}
\mfD_{\rotot}(\indexnot,t)e^{\b_{\ron}t/2}z_{\infty}\right|\\
 \leq & C_{B}e^{(\b_{\ron}/2-\eta_{B})\bt}\left(|\dot{z}(\indexnot,T_{\ron})|+\mfg(\indexnot,T_{\ron})|z(\indexnot,T_{\ron})|\right),
\end{split}
\end{equation}
where $C_{B}$ only depends on $\eta_{B}$ and the coefficients of the equation (\ref{eq:thesystemRge}). Moreover, combining 
Lemma~\ref{lemma:roughenestbalsetting} (with $\betafun=0$) and (\ref{eq:zinfestnoise}) yields
\begin{equation}\label{eq:zetainfestauxnoise}
\begin{split}
|z_{\infty}| \leq & C_{B}e^{-\b_{\ron}T_{\ron}/2}\left(|\dot{z}(\indexnot,T_{\ron})|+\mfg(\indexnot,T_{\ron})|z(\indexnot,T_{\ron})|\right)\\
 \leq & C_{B}e^{\eta_{\robal}T_{\ron}-\b_{\ron}T_{\ron}/2}\me^{1/2}(\indexnot,0),
\end{split}
\end{equation}
where $C_{B}$ has the same dependence as in the case of (\ref{eq:zzdottestfinauxnoise}). Define $\chi_{\pm}(\indexnot)$ according to 
(\ref{eq:zetaldefnoise}), where $z_{1}$, $z_{2}$ are the components of $z_{\infty}$. Then (\ref{eq:Tronuppbdnoise}) and (\ref{eq:zetainfestauxnoise})
yield (\ref{eq:chipmmfezeroestnoise}). In particular, the $\chi_{\pm}(\indexnot)$ satisfy (\ref{eq:rortbdnoise}). Thus we can associate
a smooth solution, say $\bu$, with the asymptotic data $\chi_{\pm}(\indexnot)$ as in the first part of the statement of the lemma. However, since 
the Fourier coefficients of $u$ satisfy (\ref{eq:zzdottestfinauxnoise}) and $\bu$ satisfies (\ref{eq:mfediffhomestnoise}) (with $u$ replaced by $\bu$), 
the Fourier coefficients of $u$ and $\bu$ have to coincide, so that $u=\bu$; the reason the Fourier coefficients have to coincide is that they
have the same asymptotic data, and that, due to Lemma~\ref{lemma:asdataosccasenoise}, the asymptotic data uniquely determine the solution in the 
present setting. The lemma follows. 
\end{proof}

Due to Lemma~\ref{lemma:asymposccasescalareqnoise}, we can assign a meaning to the notion of orientation of a complex valued solution
to (\ref{eq:homtheeqnoise}). 

\begin{definition}\label{def:orsoltowenoise}
Consider the equation (\ref{eq:homtheeqnoise}).  Assume that it is strongly balanced, has a negligible shift vector field, a dominant noisy spatial direction, 
and is such that the dominant coefficients are convergent. Assume that $u$ is a smooth solution to (\ref{eq:homtheeqnoise}) such that 
$u=u_{\ron}$, where $u_{\ron}$ is defined in analogy with (\ref{eq:frondef}). Let $u_{\app}$ be the uniquely associated function of the form 
(\ref{eq:zappdefnoise}) and (\ref{eq:uappdefnoise}), whose existence is guaranteed by
Lemma~\ref{lemma:asymposccasescalareqnoise}. Consider (\ref{eq:zappdefnoise}) and (\ref{eq:uappdefnoise}). If $\chi_{-}(\indexnot)=0$ for all 
$\indexnot\in\EFnindexset$, then $u$ is said to be \textit{positively oriented}. If $\chi_{+}(\indexnot)=0$ for all $\indexnot\in\EFnindexset$, then $u$ 
is said to be \textit{negatively oriented}. 
\end{definition}

\section{A uniqueness result}\label{section:auniquenresultnoise}

Next, we demonstrate a uniqueness result for vector valued solutions to (\ref{eq:homtheeqnoise}). 

\begin{lemma}\label{lemma:orientationuniquenessnoise}
Consider the equation (\ref{eq:homtheeqnoise}).  Assume that it is strongly balanced, has a negligible shift vector field, a dominant noisy spatial direction, 
and is such that the dominant coefficients are convergent. Let $1\leq k\in\zo$; $A_{\pm}\in\Mn{k}{\co}$; and $0<\b\in\ro$. Using the notation of 
Definitions~\ref{def:SpRspdef} and \ref{def:defofgeneigenspintro}, let
\[
\kappa_{\rom}=\max\{\kappa_{\max}(A_{+}),\kappa_{\max}(A_{-})\};
\]
$J:=(\kappa_{\rom}-\b,\kappa_{\rom}]$; and 
$E_{\pm}:=E_{A_{\pm},J}$. Let $u_{\pm}\in C^{\infty}(M,E_{\pm})$ be two solutions to (\ref{eq:homtheeqnoise}) and denote the $\indexnot$'th Fourier coefficient of 
$u_{\pm}(\cdot,t)$ by $z_{\pm}(\indexnot,t)$ for $\indexnot\in\EFindexset$. Assume that the following holds:
\begin{itemize}
\item $z_{\pm}(\indexnot,t)=0$ for all $\indexnot\notin\EFnindexset$ and all $t\in\ro$,
\item the components of $u_{+}$ are positively oriented and the components of $u_{-}$ are negatively oriented.
\end{itemize}
Define $u$ by 
\begin{equation}\label{eq:uAouoAtutdefnoise}
u(p,t):=e^{A_{+}t}u_{+}(p,t)+e^{A_{-}t}u_{-}(p,t).
\end{equation}
Finally, assume that there is an $s\in\ro$ and constants $C$ and $N$, which are allowed to depend on the solutions $u_{\pm}$, such that 
\begin{equation}\label{eq:mfeshomeaouoeatutestnoise}
\mfe_{s}^{1/2}[u](t)\leq C\ldr{t}^{N}e^{(\kappa_{\rom}+\b_{\ron}/2-\b)t}
\end{equation}
for all $t\geq 0$. Then $u_{+}=u_{-}=0$. 
\end{lemma}
\begin{remarks}
The notion of positive and negative orientation is introduced in Definition~\ref{def:orsoltowenoise}. The notation $\mfe_{s}[u](t)$ is 
introduced in (\ref{eq:mfedef}).
\end{remarks}
\begin{proof}
Note, to begin with, that if $z(\indexnot,t)$ is the $\indexnot$'th Fourier coefficient of the function $u(\cdot,t)$, defined by
(\ref{eq:uAouoAtutdefnoise}), then
\[
z(\indexnot,t)=e^{A_{+}t}z_{+}(\indexnot,t)+e^{A_{-}t}z_{-}(\indexnot,t),
\]
where the $z_{\pm}$ are defined in the statement of the lemma. Note also that, due to the assumptions of the lemma, there is no restriction
in assuming $\indexnot\in\EFnindexset$. We therefore do so here and in what follows. Due to (\ref{eq:mfeshomeaouoeatutestnoise}),
\begin{equation}\label{eq:zzdotuniqargnoise}
|\dot{z}(\indexnot,t)|+\mfg(\indexnot,t)|z(\indexnot,t)|\leq C\ldr{t}^{N}e^{(\kappa_{\rom}+\b_{\ron}/2-\b)t}
\end{equation}
for all $t\geq 0$, 
where $C$ depends on $\indexnot$, $s$ and the solutions $u_{\pm}$. Note that $e^{A_{\pm}t}z_{\pm}(\indexnot,t)$ can be written
\begin{equation}\label{eq:eAltzldecompnoise}
e^{A_{\pm}t}z_{\pm}(\indexnot,t)=\textstyle{\sum}_{r=1}^{k_{\pm}}\phi_{\pm,r}(t)z_{\pm,r}(\indexnot,t),
\end{equation}
where the $\phi_{\pm,r}$ are linearly independent $E_{\pm}$-valued solutions to the equation $\dot{x}=A_{\pm}x$ and $z_{\pm,r}$ are $\co$-valued solutions to 
(\ref{eq:fourierhomtheeqnoise}). Moreover, the $\phi_{\pm,r}$ are of the form
\begin{equation}\label{eq:philrdecompnoise}
\phi_{\pm,r}(t)=e^{\lambda_{\pm,r}t}\psi_{\pm,r}(t),
\end{equation}
where $\lambda_{\pm,r}$ is an eigenvalue of $A_{\pm}$ and $\psi_{\pm,r}$ is an $E_{\pm}$-valued function, each of whose components are polynomials (in fact,
$\psi_{\pm,r}$ (and thereby $\phi_{\pm,r}$) takes its values in the generalised eigenspace corresponding to $\lambda_{\pm,r}$). In what
follows, we denote the real part of $\lambda_{\pm,r}$ by $\kappa_{\pm,r}$. The ordering of the $\phi_{\pm,r}$ is such that if $r_{1}<r_{2}$, then 
$\kappa_{\pm,r_{1}}\geq\kappa_{\pm,r_{2}}$. In addition, the $z_{+,r}$'s correspond to positively oriented solutions to (\ref{eq:homtheeqnoise}) and the 
$z_{-,r}$'s correspond to negatively oriented solutions to (\ref{eq:homtheeqnoise}). Finally, note that we, without loss of generality, can assume
that $\kappa_{\pm,r}>\kappa_{\rom}-\b$ for all $r$.  Turning to the $z_{\pm,r}$'s, note that, due to Lemma~\ref{lemma:asymposccasescalareqnoise}, they can 
be approximated by 
\begin{align}
z_{\pm,r,\app}(\indexnot,t) := & \frac{e^{\b_{\ron}t/2}}{\mfg(\indexnot,t)}\chi_{\pm,r}(\indexnot)e^{i[\omega_{\rosh}(\indexnot,t)\mp\varphi_{\rotot}(\indexnot,t)]}.\label{eq:zorappnoisepm}
\end{align}
In fact, there is a constant $C_{\pm,r}$, depending on $z_{\pm,r}$ and $\indexnot$, such that 
\begin{equation}\label{eq:zzdotlruniqargnoise}
|\dot{z}_{\pm,r}(\indexnot,t)-\dot{z}_{\pm,r,\app}(\indexnot,t)|+\mfg(\indexnot,t)|z_{\pm,r}(\indexnot,t)-z_{\pm,r,\app}(\indexnot,t)|\leq C_{\pm,r}e^{(\b_{\ron}/2-\eta_{B})t}
\end{equation}
for all $t\geq 0$; cf. Lemma~\ref{lemma:asymposccasescalareqnoise}, in particular (\ref{eq:mfediffhomestnoise}). Thus
\begin{equation}\label{eq:mfgzlrrbdnoise}
|\dot{z}_{\pm,r}(\indexnot,t)|+\mfg(\indexnot,t)|z_{\pm,r}(\indexnot,t)|\leq C_{\pm,r}e^{\b_{\ron}t/2}
\end{equation}
for all $t\geq 0$. The argument proceeds by a consideration of several different cases.

\textbf{Different $\kappa_{\pm,1}$'s.} Say now, for the sake of argument, that $\kappa_{+,1}>\kappa_{-,1}$. Assume, moreover, that $\kappa_{+,r}=\kappa_{+,1}$ 
for $r=1,\dots,r_{a}$, and that $\kappa_{+,r}<\kappa_{+,1}$ for $r>r_{a}$. Then
\begin{equation}\label{eq:eAtztlimznoise}
\lim_{t\rightarrow\infty}e^{-(\kappa_{+,1}+\b_{\ron}/2)t}\mfg(\indexnot,t)e^{A_{-}t}z_{-}(\indexnot,t)=0,
\end{equation}
where we have appealed to (\ref{eq:eAltzldecompnoise}), (\ref{eq:philrdecompnoise}), (\ref{eq:mfgzlrrbdnoise}) and the fact that $\kappa_{-,1}<\kappa_{+,1}$. 
Combining (\ref{eq:eAtztlimznoise}) with a similar argument concerning the sum on the right hand side of (\ref{eq:eAltzldecompnoise}) in the case of the 
plus sign and for $r>r_{a}$ yields
\[
\lim_{t\rightarrow\infty}e^{-(\kappa_{+,1}+\b_{\ron}/2)t}\textstyle{\sum}_{r=1}^{r_{a}}\mfg(\indexnot,t)e^{\lambda_{+,r}t}\psi_{+,r}(t)z_{+,r}(\indexnot,t)=0,
\]
where we have also appealed to (\ref{eq:zzdotuniqargnoise}), (\ref{eq:philrdecompnoise}) and the fact that $\kappa_{\pm,r}>\kappa_{\rom}-\b$. Combining this limit 
with (\ref{eq:zorappnoisepm}) and (\ref{eq:zzdotlruniqargnoise}) yields
\[
\lim_{t\rightarrow\infty}\textstyle{\sum}_{r=1}^{r_{a}}e^{(\lambda_{+,r}-\kappa_{+,1})t}\psi_{+,r}(t)\chi_{+,r}(\indexnot)=0.
\]
Since $\psi_{+,r}$ takes its values in the generalised eigenspace corresponding to $\lambda_{+,r}$, we may focus on each distinct $\lambda_{+,r}$ separately. 
In other words, we may, in practice, assume that $\lambda_{+,r}$ is independent of $r$. Thus
\begin{equation}\label{eq:psizetazlimnoise}
\lim_{t\rightarrow\infty}\textstyle{\sum}_{r=1}^{r_{a}}\psi_{+,r}(t)\chi_{+,r}(\indexnot)=0.
\end{equation}
Let the non-negative integers $m_{1}>m_{2}>\cdots>m_{p}$ be such that all the polynomial powers occurring in the functions $\psi_{+,r}$ belong
to $\{m_{1},\dots,m_{p}\}$. Then $\psi_{+,r}(t)$ can be written
\[
\psi_{+,r}(t)=\textstyle{\sum}_{j=1}^{p}t^{m_{j}}v_{r,j},
\]
where $v_{r,j}\in\cn{k}$. Due to (\ref{eq:psizetazlimnoise}) this observation yields
\[
0=\lim_{t\rightarrow\infty}t^{-m_{1}}\textstyle{\sum}_{r=1}^{r_{a}}\psi_{+,r}(t)\chi_{+,r}(\indexnot)=\sum_{r=1}^{r_{a}}v_{r,1}\chi_{+,r}(\indexnot).
\]
Proceeding power by power yields
\[
\textstyle{\sum}_{r=1}^{r_{a}}v_{r,j}\chi_{+,r}(\indexnot)=0.
\]
However, this implies that 
\[
\textstyle{\sum}_{r=1}^{r_{a}}\phi_{+,r}(t)\chi_{+,r}(\indexnot)=0
\]
for all $t$. Since the functions $\phi_{+,r}$ are linearly independent, it follows that $\chi_{+,r}(\indexnot)=0$ for $1\leq r\leq r_{a}$. Thus 
$z_{+,r,\app}(\indexnot,t)=0$, so that $z_{+,r}(\indexnot,t)=0$ for $1\leq r\leq r_{a}$. 

In case $\kappa_{+,1}<\kappa_{-,1}$, the argument is similar. In short, if the $\kappa_{\pm,1}$'s are different, we can eliminate the corresponding 
components of $z_{+}$ or $z_{-}$. Proceeding step by step, we are then either finished or left with the situation that the real part of the eigenvalues 
equal. Without loss of generality, we can assume that the $\kappa_{\pm,1}$'s equal. 

\textbf{Equal maximal real parts of the eigenvalues.} Assume that the $\kappa_{\pm,1}$'s equal. In fact, let us assume that 
$\kappa_{\pm,1}=\dots=\kappa_{\pm,r_{\pm}}$, but that $\kappa_{\pm,r_{\pm}+1}<\kappa_{\pm,1}$ (in case $r_{\pm}<k_{\pm}$). Let us introduce the notation
\[
\xi_{\pm}(\indexnot,t):=\textstyle{\sum}_{r=1}^{r_{\pm}}\phi_{\pm,r}(t)z_{\pm,r,\app}(\indexnot,t),
\]
where the $z_{\pm,r,\app}$ are defined by (\ref{eq:zorappnoisepm}). Then, due to (\ref{eq:zzdotuniqargnoise}), 
(\ref{eq:mfgzlrrbdnoise}) and an argument similar to the derivation of (\ref{eq:eAtztlimznoise}), 
\begin{equation}\label{eq:xidaxiefdecaynoise}
\lim_{t\rightarrow\infty}e^{-(\kappa_{+,1}+\b_{\ron}/2)t}\left(|\dot{\xi}_{-}(\indexnot,t)+\dot{\xi}_{+}(\indexnot,t)|
+\mfg(\indexnot,t)|\xi_{-}(\indexnot,t)+\xi_{+}(\indexnot,t)|\right)=0.
\end{equation}
On the other hand, 
\begin{equation}\label{eq:xidotorientationnoise}
\lim_{t\rightarrow\infty}e^{-(\kappa_{+,1}+\b_{\ron}/2)t}\left(|\dot{\xi}_{+}(\indexnot,t)+i\mfg(\indexnot,t)\xi_{+}(\indexnot,t)|
+|\dot{\xi}_{-}(\indexnot,t)-i\mfg(\indexnot,t)\xi_{-}(\indexnot,t)|\right)=0.
\end{equation}
Combining (\ref{eq:xidaxiefdecaynoise}) and (\ref{eq:xidotorientationnoise}) yields
\[
\lim_{t\rightarrow\infty}e^{-(\kappa_{+,1}+\b_{\ron}/2)t}\mfg(\indexnot,t)\left(|\xi_{-}(\indexnot,t)+\xi_{+}(\indexnot,t)|
+|\xi_{-}(\indexnot,t)-\xi_{+}(\indexnot,t)|\right)=0.
\]
As a consequence, the arguments described to deal with the case of different $\kappa_{\pm,1}$'s apply. Thus $z_{\pm,r}(\indexnot,t)=0$ for all $t\in I$
and $r=1,\dots,r_{\pm}$. Proceeding step by step, it follows that $z_{\pm,r}(\indexnot,t)=0$ for all $t\in I$ and $r=1,\dots,k_{\pm}$. Applying the 
argument for all $\indexnot\in\EFnindexset$ yields $u_{+}=u_{-}=0$, and the lemma follows. 
\end{proof}

\section{Detailed asymptotics in the noisy setting}\label{eq:deasinnoissetnoise}

Next, we turn to the problem of expressing the asymptotics of solutions to (\ref{eq:thesystemRge}) in terms of solutions to 
(\ref{eq:homtheeqnoise}). Let us begin by introducing the following terminology. 

\begin{definition}\label{definition:EFnindexsetpm}
Assume that (\ref{eq:thesystemRge}) is strongly balanced, has a negligible shift vector field, a dominant noisy spatial direction, and is such that the dominant 
coefficients are convergent. In case (\ref{eq:thesystemRge}) has a dominant noisy spatial torus direction, say $j$, let
\begin{equation}\label{eq:EFnindpmtorus}
\EFnindexsetp:=\{\indexnot\in\EFnindexset : \nu_{\roT,j}(\indexnot)<0\},\ \ \
\EFnindexsetm:=\{\indexnot\in\EFnindexset : \nu_{\roT,j}(\indexnot)>0\};
\end{equation}
cf. (\ref{eq:nuroTetcdef}). In case (\ref{eq:thesystemRge}) has a dominant noisy spatial generalised direction, let 
\begin{equation}\label{eq:EFnindpmgeneralised}
\EFnindexsetp:=\EFnindexset,\ \ \
\EFnindexsetm:=\varnothing.
\end{equation}
\end{definition}

\begin{lemma}\label{lemma:asymposccaseitosoltowenoise}
Assume that (\ref{eq:thesystemRge}) is strongly balanced, has a negligible shift vector field, a dominant noisy spatial direction, and is such that the dominant 
coefficients are convergent. Let $\eta_{A}$ be defined by (\ref{eq:etaAdefinition}) and assume that there is an 
$0<\eta_{B}\leq \eta_{A}$ such that 
\begin{equation}\label{eq:fronHsbdnoise}
\int_{0}^{\infty}e^{-(\kappa_{\ron,+}-\eta_{B})t}\|f_{\ron}(\cdot,t)\|_{(s)}dt<\infty
\end{equation}
for all $s\in\ro$, where $\kappa_{\ron,+}$ is defined by (\ref{eq:kapparonpmdef}); $f_{\ron}$ is defined by (\ref{eq:frondef}); and $\|\cdot\|_{(s)}$ is 
defined by (\ref{eq:HsnormonbM}). Let 
\[
A_{\ron,\pm}:=-\frac{1}{2}\left(\a_{\infty}\pm\tX_{\infty,\ron}\right),\ \ \
\kappa_{\ron,\rom}:=\max\{\kappa_{\max}(A_{\ron,+}),\kappa_{\max}(A_{\ron,-})\},
\]
where $\tX_{\infty,\ron}$ is introduced in Definition~\ref{def:tXinfrondefetc}. Finally, let $J:=(\kappa_{\ron,\rom}-\eta_{B},\kappa_{\ron,\rom}]$ and 
$E_{\ron,\pm}:=E_{A_{\ron,\pm},J}$; cf. Definition~\ref{def:defofgeneigenspintro}. 

If (\ref{eq:thesystemRge}) has a dominant noisy spatial generalised direction, 
cf. Definition~\ref{def:domnospdir}, then $E_{\ron,+}=E_{\ron,-}=:E_{\ron}$ and $A_{\ron,+}=A_{\ron,-}=:A_{\ron}$. Moreover, if $u$ is a solution to 
(\ref{eq:thesystemRge}), there is a unique solution $u_{0,\row}\in C^{\infty}(M,E_{\ron})$ to (\ref{eq:homtheeqnoise}) such that 
\begin{itemize}
\item if $z_{0,\row}(\indexnot,t)$ denotes the $\indexnot$'th Fourier coefficient of $u_{0,\row}(\cdot,t)$, then $z_{0,\row}(\indexnot,t)=0$ for all
$t\in I$ and $\indexnot\notin\EFnindexset$,
\item if $u_{\row}$ is defined by 
\begin{equation}\label{eq:urowdefnoisegen}
u_{\row}(p,t) = e^{A_{\ron}t}u_{0,\row}(p,t),
\end{equation}
then there are real constants $C$, $N$ and $s$ such that 
\begin{equation}\label{eq:uuweststatementuniquenoise}
\begin{split}
\mfe_{s}^{1/2}[u_{\ron}-u_{\row}](t) \leq & C\ldr{t}^{N}e^{(\kappa_{\ron,+}-\eta_{B})t}
\end{split}
\end{equation}
for all $t\geq 0$, where $u_{\ron}$ is defined in analogy with (\ref{eq:frondef}).
\end{itemize}

If (\ref{eq:thesystemRge}) has a dominant noisy spatial torus direction, say $j$, and $u$ is a solution to (\ref{eq:thesystemRge}), then there 
are unique solutions $u_{\pm,\row}\in C^{\infty}(M,E_{\ron,\pm})$ to (\ref{eq:homtheeqnoise}) such that the following holds:
\begin{itemize}
\item $u_{+,\row}=u_{+,\row}^{+}+u_{-,\row}^{-}$ and $u_{-,\row}=u_{+,\row}^{-}+u_{-,\row}^{+}$, where $u_{\tau,\row}^{\upsilon}$, $\tau,\upsilon\in\{+,-\}$, are 
smooth solutions to (\ref{eq:homtheeqnoise}) whose $\indexnot$'th Fourier coefficients are denoted by $z_{\tau,\row}^{\upsilon}(\indexnot,t)$ and
are such that
\begin{itemize}
\item $z_{\pm,\row}^{+}(\indexnot,\cdot)=0$ in case $\indexnot\notin\EFnindexsetp$ and 
$z_{\pm,\row}^{-}(\indexnot,\cdot)=0$ in case $\indexnot\notin\EFnindexsetm$,
\item $u_{+,\row}^{\pm}$ are positively oriented and $u_{-,\row}^{\pm}$ are negatively oriented solutions 
to (\ref{eq:homtheeqnoise}),
\end{itemize}
\item if $u_{\row}$ is defined by 
\begin{equation}\label{eq:urowdefnoisetorus}
u_{\row}(p,t) = e^{A_{\ron,+}t}u_{+,\row}(p,t)+e^{A_{\ron,-}t}u_{-,\row}(p,t),
\end{equation}
then there are real constants $C$, $N$ and $s$ such that 
\begin{equation}\label{eq:uuweststatementuniquenoisetorus}
\begin{split}
\mfe_{s}^{1/2}[u_{\ron}-u_{\row}](t) \leq & C\ldr{t}^{N}e^{(\kappa_{\ron,+}-\eta_{B})t}
\end{split}
\end{equation}
for all $t\geq 0$.
\end{itemize}
Moreover, in both cases there are constants $C_{B}>0$ and $s_{B}\geq 0$ (depending only on $\eta_{B}$ and the coefficients of the equation 
(\ref{eq:thesystemRge})); and a non-negative integer $N$ (depending only on $m$) such that the following holds. If $u$ and $u_{\row}$ are as 
above, then 
\begin{equation}\label{eq:uuweststatementnoise}
\begin{split}
 & \mfe_{s}^{1/2}[u_{\ron}-u_{\row}](t)\\
 \leq & C_{B}\ldr{t}^{N}e^{(\kappa_{\ron,+}-\eta_{B})t}\left(\mfe_{s+s_{B}}^{1/2}[u_{\ron}](0)
+\int_{0}^{\infty}e^{-(\kappa_{\ron,+}-\eta_{B})t'}\|f_{\ron}(\cdot,t')\|_{(s+s_{B})}dt'\right)
\end{split}
\end{equation}
for all $t\geq 0$ and all $s\in\ro$. In addition, if (\ref{eq:thesystemRge}) has a dominant noisy spatial generalised direction, then 
\begin{equation}\label{eq:mfesuzrowestitoidanih}
\mfe_{s}^{1/2}[u_{0,\row}](0)\leq C_{B}\left(\mfe_{s+s_{B}}^{1/2}[u_{\ron}](0)
+\int_{0}^{\infty}e^{-(\kappa_{\ron,+}-\eta_{B})t'}\|f_{\ron}(\cdot,t')\|_{(s+s_{B})}dt'\right)
\end{equation}
for all $s\in\ro$, where $C_{B}$ and $s_{B}$ have the same dependence as in the case of (\ref{eq:uuweststatementnoise}). Finally, if (\ref{eq:thesystemRge}) 
has a dominant noisy spatial torus direction, then 
\begin{equation}\label{eq:mfesupmrowestitoidanih}
\mfe_{s}^{1/2}[u_{\pm,\row}](0)\leq C_{B}\left(\mfe_{s+s_{B}}^{1/2}[u_{\ron}](0)
+\int_{0}^{\infty}e^{-(\kappa_{\ron,+}-\eta_{B})t'}\|f_{\ron}(\cdot,t')\|_{(s+s_{B})}dt'\right)
\end{equation}
for all $s\in\ro$, where $C_{B}$ and $s_{B}$ have the same dependence as in the case of (\ref{eq:uuweststatementnoise}). 
\end{lemma}
\begin{remark}\label{remark:homwaveonlygeneqnoise}
Assume that (\ref{eq:thesystemRge}) is $C^{2}$-balanced with a geometric dominant noisy spatial direction, convergent dominant 
coefficients and a negligible shift vector field; cf. Definition~\ref{def:noisemainassumptions}. Assume, moreover, that 
(\ref{eq:fronHsbdnoise}) holds for all $s\in\ro$. Then the assumptions of the lemma are satisfied; cf. 
Remarks~\ref{remark:geometrictonongeometricnoise} and \ref{remark:strongbalanceneglshiftnoisegeometric}. How to calculate
$\eta_{A}$ etc. is clarified in Remark~\ref{remark:homwaveonlyprincsymbnoise}. 
\end{remark}
\begin{remark}
Note that $\kappa_{\ron,+}$ and $\kappa_{\ron,\rom}$ are related according to $\kappa_{\ron,+}=\kappa_{\ron,\rom}+\b_{\ron}/2$. 
\end{remark}
\begin{remarks}\label{remarks:Tronbronetcdeponeq}
In the statement of the lemma, two equations play an important role: (\ref{eq:thesystemRge}) and (\ref{eq:homtheeqnoise}). Note that it is here taken 
for granted that (\ref{eq:homtheeqnoise}) is the equation obtained from (\ref{eq:thesystemRge}) by setting $\a$, $\zeta$, $X^{j}$
and $f$ to zero. In the statement of the present lemma, we speak of $\eta_{A}$, defined by (\ref{eq:etaAdefinition}). Note that this $\eta_{A}$ need
not necessarily coincide with the $\eta_{A}$ associated with the equation (\ref{eq:homtheeqnoise}), say $\eta_{A,\rohom}$, since $\eta_{\romn}$ can 
be omitted from the right hand side of (\ref{eq:etaAdefinition}) in the definition of $\eta_{A,\rohom}$. On the other hand, it is clear that 
$\eta_{A,\rohom}\geq \eta_{A}$, so that when we appeal to Lemma~\ref{lemma:asymposccasescalareqnoise}, the discrepancy does not cause complications.
Turning to the constant $\b_{\ron}$, note that it is the same for the two equations; cf. Definition~\ref{def:domnospdir}. Finally, given 
$\indexnot\in\EFnindexset$, the $T_{\ron}$ given by Definition~\ref{def:Tron} need not be the same for the two equations (\ref{eq:thesystemRge}) and 
(\ref{eq:homtheeqnoise}); we obtain, say, $T_{\ron}$ and $T_{\ron,\rohom}$ respectively. However, $\eta_{\romar}$ appearing in Definition~\ref{def:domnospdir}
and $K_{\ron}$ appearing in the statement of Lemma~\ref{lemma:mfgnuronindnotrelerror} can be assumed to be the same for the two equations. Moreover, by 
deteriorating the value of the constant $\ellderbd$ appearing in (\ref{def:oscad}) and (\ref{eq:mffXdotbdetc}) for one of the equations, it can be 
assumed that $T_{\ron}=T_{\ron,\rohom}$. 
\end{remarks}
\begin{remark}
If (\ref{eq:thesystemRge}) has a dominant noisy spatial generalised direction, then $A_{\ron}=-\a_{\infty}/2$, so that (\ref{eq:urowdefnoisegen}) can be 
written
\[
u_{\row}(p,t) = e^{-\a_{\infty}t/2}u_{0,\row}(p,t).
\]
\end{remark}
\begin{remark}
The constants $C$, $s$ and $N$ appearing in (\ref{eq:uuweststatementuniquenoise}) and (\ref{eq:uuweststatementuniquenoisetorus}) are allowed
to depend on the coefficients of the equation (\ref{eq:thesystemRge}), $\eta_{B}$, the functions $u_{\ron}$, $u_{\row}$, $f_{\ron}$ etc.
\end{remark}
\begin{remark}
Dependence on the coefficients includes dependence on the Riemannian manifolds $(M_{r},g_{r})$, $r=1,\dots,R$; cf. 
Remark~\ref{remark:deponcoefincldeponRmfds}. 
\end{remark}
\begin{remark}\label{remark:Ngennoisas}
If all the Jordan blocks of the matrices $A_{\ron,\pm}$ are trivial, then the $N$ appearing in (\ref{eq:uuweststatementnoise}) can be replaced by $1$.
If $E_{\ron,+}=E_{\ron,-}=\cn{m}$, then $N$ can be replaced by $d_{n,+}-1$, where $d_{\ron,+}$ is defined as in the statement of 
Lemma~\ref{lemma:windestnoise}.
\end{remark}
\begin{remark}\label{remark:specupmwanoise}
Due to the proof, the solutions $u_{\pm,\row}$ and $u_{0,\row}$ to (\ref{eq:homtheeqnoise}) can be replaced by functions of the form described in 
the statement of Lemma~\ref{lemma:asymposccasescalareqnoise}. To be more precise, let us first consider the case that (\ref{eq:thesystemRge}) 
has a dominant noisy spatial generalised direction. If $u$ is a solution to (\ref{eq:thesystemRge}), there are then uniquely determined 
$\chi_{\pm}(\indexnot)\in E_{\ron}$, $\indexnot\in\EFnindexset$, with the following properties: for every $0\leq k\in\zo$, there is a constant $C_{k}$ 
such that (\ref{eq:rortbdnoise}) holds; and if $u_{0,\wa}$ is given by the right hand side of (\ref{eq:uappdefnoise}), where $z_{\app}$ is given by 
(\ref{eq:zappdefnoise}), then $u_{\wa}=e^{A_{\ron}t}u_{0,\wa}$ is such that there are constants $C$, $N$ and $s$ with the property that 
\begin{equation}\label{eq:uronminuwauniquenessestimate}
\mfe_{s}^{1/2}[u_{\ron}-u_{\wa}](t)\leq C\ldr{t}^{N}e^{(\kappa_{\ron,+}-\eta_{B})t}
\end{equation}
for all $t\geq 0$. 

Next, assume that (\ref{eq:thesystemRge}) has a dominant noisy spatial torus direction. If $u$ is a solution to (\ref{eq:thesystemRge}), there 
are then uniquely determined $\zeta_{\pm}^{+}(\indexnot)$ and $\zeta_{\pm}^{-}(\indexnot)$, $\indexnot\in\EFindexset$, with the following properties: 
$\zeta_{\pm}^{-}(\indexnot)=0$ for all $\indexnot\notin\EFnindexsetm$ and $\zeta_{\pm}^{+}(\indexnot)=0$ for all $\indexnot\notin\EFnindexsetp$;
$\zeta_{+}^{+}(\indexnot),\zeta_{-}^{-}(\indexnot)\in E_{\ron,+}$ and $\zeta_{+}^{-}(\indexnot),\zeta_{-}^{+}(\indexnot)\in E_{\ron,-}$ for all 
$\indexnot\in\EFindexset$; for every $0\leq k\in\zo$, there is a constant $C_{k}$ such that 
\begin{equation}\label{eq:zetapmpmestnoiseapprwa}
\textstyle{\sum}_{\indexnot\in\EFnindexset}\sum_{\pm}\ldr{\nu(\indexnot)}^{2k}\left[|\zeta_{+}^{\pm}(\indexnot)|^{2}+|\zeta_{-}^{\pm}(\indexnot)|^{2}\right]\leq C_{k};
\end{equation}
and if $u_{+,\wa}^{\pm}$ and $u_{-,\wa}^{\pm}$ are the functions whose $\indexnot$'th Fourier coefficients are given by 
\begin{align}
z_{+,\wa}^{\pm}(\indexnot,t) = & \frac{e^{\b_{\ron} t/2}}{\mfg(\indexnot,t)}e^{i[\omega_{\rosh}(\indexnot,t)-\varphi_{\rotot}(\indexnot,t)]}\zeta_{+}^{\pm}(\indexnot),
\label{eq:zpwapmdef}\\
z_{-,\wa}^{\pm}(\indexnot,t) = & \frac{e^{\b_{\ron} t/2}}{\mfg(\indexnot,t)}e^{i[\omega_{\rosh}(\indexnot,t)+\varphi_{\rotot}(\indexnot,t)]}\zeta_{-}^{\pm}(\indexnot)
\label{eq:zmwapmdef}
\end{align}
for $\indexnot\in\EFindexset$ (when $\indexnot\notin\EFnindexset$, the right hand sides of (\ref{eq:zpwapmdef}) and (\ref{eq:zmwapmdef}) should be
interpreted as equalling zero), if $u_{\pm,\wa}$ are given by 
\begin{equation}\label{eq:upmwadef}
u_{+,\wa}=u_{+,\wa}^{+}+u_{-,\wa}^{-},\ \ \
u_{-,\wa}=u_{+,\wa}^{-}+u_{-,\wa}^{+},
\end{equation}
and if $u_{\wa}=e^{A_{\ron,+}t}u_{+,\wa}+e^{A_{\ron,-}t}u_{-,\wa}$, then there are constants $C$, $N$ and $s$ with the property that 
(\ref{eq:uronminuwauniquenessestimate}) holds for all $t\geq 0$. Note that, by definition, $u_{\pm,\wa}$ takes its values in $E_{\ron,\pm}$. 

In both cases there are constants $C_{B}>0$ and $s_{B}\geq 0$ (depending only on $\eta_{B}$ and the coefficients of the equation 
(\ref{eq:thesystemRge})) and a non-negative integer $N$ (depending only on $m$) such that if $u$ and $u_{\wa}$ are as above, then 
\begin{equation}\label{eq:uuweststatementnoisewa}
\begin{split}
 & \mfe_{s}^{1/2}[u_{\ron}-u_{\wa}](t)\\
 \leq & C_{B}\ldr{t}^{N}e^{(\kappa_{\ron,+}-\eta_{B})t}\left(\mfe_{s+s_{B}}^{1/2}[u_{\ron}](0)
+\int_{0}^{\infty}e^{-(\kappa_{\ron,+}-\eta_{B})t'}\|f_{\ron}(\cdot,t')\|_{(s+s_{B})}dt'\right)
\end{split}
\end{equation}
for all $t\geq 0$ and all $s\in\ro$. In addition, if all the Jordan blocks of the matrices $A_{\ron,\pm}$ are trivial, then the $N$ appearing in 
(\ref{eq:uuweststatementnoisewa}) can be replaced by $1$, and if $E_{\ron,+}=E_{\ron,-}=\cn{m}$, then $N$ can be replaced by $d_{n,+}-1$. Turning to 
the $\zeta_{\pm}^{+}$ and $\zeta_{\pm}^{-}$, there are constants $C_{B}>0$, $s_{a}$ and $s_{b}$ (depending only on $\eta_{B}$ and the coefficients of 
the equation (\ref{eq:thesystemRge})), such that 
\begin{equation}\label{eq:zetalpmHsnormnoise}
\begin{split}
 & \left(\textstyle{\sum}_{\indexnot\in\EFnindexset}\sum_{\pm}\ldr{\nu(\indexnot)}^{2s}
\left[|\zeta_{+}^{\pm}(\indexnot)|^{2}+|\zeta_{-}^{\pm}(\indexnot)|^{2}\right]\right)^{1/2}\\
 \leq & C_{B}\left(\mfe_{s+s_{a}}^{1/2}[u_{\ron}](0)
+\int_{0}^{\infty}e^{-(\kappa_{\ron,+}-\eta_{B})t}\|f_{\ron}(\cdot,t)\|_{(s+s_{b})}dt\right)
\end{split}
\end{equation}
for all $s\in\ro$. In the case of a dominant noisy spatial generalised direction, the same estimate holds, assuming we introduce the 
notation $\zeta_{\pm}^{-}(\indexnot)=0$, $\zeta_{+}^{+}(\indexnot)=\chi_{+}(\indexnot)$ and $\zeta_{-}^{+}(\indexnot)=\chi_{-}(\indexnot)$ for all 
$\indexnot\in\EFindexset$; note that with this notation, $u_{0,\wa}:=u_{+,\wa}+u_{-,\wa}$. 

Finally, the functions $u_{0,\row}$ and $u_{0,\wa}$ (appearing in the statement of the lemma and the present remark, respectively) are uniquely
associated as described in the statement of Lemma~\ref{lemma:asymposccasescalareqnoise}. Similarly, the functions 
$u_{+,\row}^{\pm}$ and $u_{+,\wa}^{\pm}$ (as well as the functions $u_{-,\row}^{\pm}$ and $u_{-,\wa}^{\pm}$) are uniquely
associated as described in the statement of Lemma~\ref{lemma:asymposccasescalareqnoise}.
\end{remark}
\begin{proof}
In order to prove uniqueness, it is convenient to divide the analysis into two different cases.

\textbf{Uniqueness, dominant noisy spatial generalised direction.} Assume that there are two functions, say $v_{l}$, $l=1,2$, with the same
properties as the function $u_{0,\row}$ appearing in the statement of the lemma. Letting $v:=v_{1}-v_{2}$ and $u_{\rodiff}:=e^{A_{\ron}t}v$, we know 
that 
\begin{equation}\label{eq:mfesudiffestnoisspgen}
\mfe_{s}^{1/2}[u_{\rodiff}](t)\leq C\ldr{t}^{N}e^{(\kappa_{\ron,\rom}+\b_{\ron}/2-\eta_{B})t}
\end{equation}
for all $t\geq 0$. Since $u_{\rodiff}$ can be written
\[
u_{\rodiff}=e^{A_{\ron,+}t}v_{+}+e^{A_{\ron,-}t}v_{-},
\]
where the components of $v_{+}$ are positively oriented and the components of $v_{-}$ are negatively oriented; the $\indexnot$'th Fourier coefficient 
of $v_{\pm}(\cdot,t)$ vanishes if $\indexnot\notin\EFnindexset$; $v_{\pm}$ takes its values in $E_{\ron,\pm}$; and since (\ref{eq:mfesudiffestnoisspgen}) holds, 
it is clear that Lemma~\ref{lemma:orientationuniquenessnoise} is applicable and yields the conclusion that $v=0$. The desired uniqueness follows.

\textbf{Uniqueness, dominant noisy spatial torus direction.} Assume now that there are two pairs of functions with the same properties as 
$u_{\pm,\row}$, say $v_{\pm,l}$, $l=1,2$. Define $v_{\pm}:=v_{\pm,1}-v_{\pm,2}$ and
\[
u_{\rodiff}:=e^{A_{\ron,+}t}v_{+}+e^{A_{\ron,-}t}v_{-}.
\]
Define $u_{\rodiff,\pm}$ to be the projection of $u_{\rodiff}$ defined by setting the $\indexnot$'th Fourier coefficient of $u_{\rodiff}$ to zero for
every $\indexnot\notin\EFnindexsetpm$. We define $v_{+,\pm}$ and $v_{-,\pm}$ similarly. Then 
(\ref{eq:uuweststatementuniquenoisetorus}) implies that 
\[
\mfe_{s}^{1/2}[u_{\rodiff,\pm}](t)\leq C\ldr{t}^{N}e^{(\kappa_{\ron,\rom}+\b_{\ron}/2-\eta_{B})t}
\]
for all $t\geq 0$. Moreover, $v_{+,+}$ and $v_{-,-}$ are positively oriented and $v_{+,-}$ and $v_{-,+}$ are negatively oriented. Combining these
observations with the fact that $v_{+,\pm}$ takes its values in $E_{\ron,+}$, the fact that $v_{-,\pm}$ takes its values in $E_{\ron,-}$ and 
Lemma~\ref{lemma:orientationuniquenessnoise} yields the conclusion that $v_{\pm}=0$. The desired uniqueness statement thus holds.

\textbf{Existence: preliminary estimates for late times.} In order to demonstrate existence, let us return to Lemma~\ref{lemma:asymposccasefinnoise}, 
in particular (\ref{eq:zzdottestfinnoise}). Note that 
\begin{align}
T_{\pre}\mfD_{\rotot}(\indexnot,t)T_{\pre}^{-1} = & 
\left(\begin{array}{cc} e^{-i\varphi_{\rotot}(\indexnot,t)}\Id_{m} & 0 \\ 0 & e^{i\varphi_{\rotot}(\indexnot,t)}\Id_{m}\end{array}\right),\label{eq:TpreconjmfDrototnoise}\\
T_{\pre}e^{S_{\ron}(\indexnot)t}T_{\pre}^{-1} = & e^{\b_{\ron}t/2}
\left(\begin{array}{cc} e^{-[\a_{\infty}-\bX_{\ron}(\indexnot)]t/2} & 0 \\ 0 & e^{-[\a_{\infty}+\bX_{\ron}(\indexnot)]t/2}\end{array}\right),\label{eq:TpreconjexpSjintnoise}
\end{align}
where $\bX_{\ron}(\indexnot)$ is introduced in Definition~\ref{def:tXinfrondefetc} and $T_{\pre}$ is defined in (\ref{eq:Tpredefnoise}). 
Introduce $\psi_{\pm}\in\cn{m}$ according to 
\begin{equation}\label{eq:poptTprezinfdefnoise}
\psi_{\infty}=\left(\begin{array}{c} \psi_{+} \\ \psi_{-}\end{array}\right):=T_{\pre}z_{\infty}.
\end{equation}
Note that $\psi_{\infty}$ is the vector whose existence is guaranteed in the statement of Lemma~\ref{lemma:asymposccasenoise}; in particular,
$\psi_{\infty}$ belongs to the first generalised eigenspace of the $\eta_{B}$, $R_{\ron}(\indexnot)$-decomposition of $\cn{2m}$, where 
$R_{\ron}(\indexnot)$ is defined by (\ref{eq:Rronpmdef}) and satisfies $R_{\ron}(\indexnot)=T_{\pre}S_{\ron}(\indexnot)T_{\pre}^{-1}$. In other words, if 
$J_{R}:=(\kappa_{\ron,+}-\eta_{B},\kappa_{\ron,+}]$, then $\psi_{\infty}$ takes its values in the $J_{R}$-generalised eigenspace of 
$R_{\ron}(\indexnot)$. However, due to the form of the matrix $R_{\ron}(\indexnot)$, cf. (\ref{eq:Rronpmdef}), 
this is equivalent to the statement that $\psi_{+}$ takes its values in the $J$-generalised eigenspace of $[-\a_{\infty}+\bX_{\ron}(\indexnot)]/2$ and 
$\psi_{-}$ takes its values in the $J$-generalised eigenspace of $[-\a_{\infty}-\bX_{\ron}(\indexnot)]/2$. Combining (\ref{eq:TpreconjmfDrototnoise}), 
(\ref{eq:TpreconjexpSjintnoise}) and (\ref{eq:poptTprezinfdefnoise}) yields 
\[
\mfD_{\rotot}(\indexnot,t)e^{S_{\ron}(\indexnot)t}z_{\infty}=e^{\b_{\ron}t/2}T_{\pre}^{-1}
\left(\begin{array}{c} e^{-[\a_{\infty}-\bX_{\ron}(\indexnot)]t/2-i\varphi_{\rotot}(\indexnot,t)}\psi_{+} \\ 
e^{-[\a_{\infty}+\bX_{\ron}(\indexnot)]t/2+i\varphi_{\rotot}(\indexnot,t)}\psi_{-}\end{array}\right).
\]
Introduce
\begin{align*}
z_{\app}(\indexnot,t) := & \frac{e^{\b_{\ron} t/2}}{\mfg(\indexnot,t)}e^{i\omega_{\rosh}(\indexnot,t)}\left(e^{-[\a_{\infty}-\bX_{\ron}(\indexnot)]t/2-i\varphi_{\rotot}(\indexnot,t)}
\psi_{+}\right.\\
 & \phantom{\frac{e^{\b_{\ron} t/2}}{\mfg(\indexnot,t)}e^{i\omega_{\rosh}(\indexnot,t)}((}\left.-ie^{-[\a_{\infty}+\bX_{\ron}(\indexnot)]t/2+i\varphi_{\rotot}(\indexnot,t)}\psi_{-}\right),\\
p_{\app}(\indexnot,t) := & e^{\b_{\ron} t/2+i\omega_{\rosh}(\indexnot,t)}\left(-ie^{-[\a_{\infty}-\bX_{\ron}(\indexnot)]t/2-i\varphi_{\rotot}(\indexnot,t)}\psi_{+}\right.\\
 & \phantom{e^{\b_{\ron} t/2+i\omega_{\rosh}(\indexnot,t)}))}\left.+e^{-[\a_{\infty}+\bX_{\ron}(\indexnot)]t/2+i\varphi_{\rotot}(\indexnot,t)}\psi_{-}\right).
\end{align*}
Then (\ref{eq:zzdottestfinnoise}) implies that 
\begin{equation}\label{eq:estzzappzdpappnoise}
\begin{split}
 & |\dot{z}(\indexnot,t)-p_{\app}(\indexnot,t)|+\mfg(\indexnot,t)|z(\indexnot,t)-z_{\app}(\indexnot,t)|\\
 \leq & C_{B}\ldr{\bt}^{N}e^{(\kappa_{\ron,+}-\eta_{B})\bt}\left(|w(\indexnot,T_{\ron})|+e^{\kappa_{\ron,+}T_{\ron}}
\int_{0}^{\infty}e^{-(\kappa_{\ron,+}-\eta_{B})t'}|\hf(\indexnot,t')|dt'\right)
\end{split}
\end{equation}
holds for all $t\geq T_{\ron}$, where $\bt:=t-T_{\ron}$; $C_{B}$ only depends on $\eta_{B}$ and the coefficients of the equation (\ref{eq:thesystemRge}); 
and $N$ is a non-negative integer depending only on $m$. Moreover, $N=1$ in case the Jordan blocks of the matrices $A_{\ron,\pm}$ are trivial. In 
addition, if $E_{\ron,+}=E_{\ron,-}=\cn{m}$, then $N$ can be replaced by $d_{n,+}-1$, where $d_{\ron,+}$ is defined as in the statement of 
Lemma~\ref{lemma:windestnoise}. The justification for the latter two statements is to be found in Remark~\ref{remark:SrontrivJordbl}; recall that 
$S_{\ron}(\indexnot)$ and $R_{\ron}(\indexnot)$ are related by a similarity transformation. 

\textbf{Reformulating the preliminary estimate.} It would be convenient to replace $p_{\app}$ with $\dot{z}_{\app}$ in (\ref{eq:estzzappzdpappnoise}). 
In order to justify that this can be done, let 
\begin{equation}\label{eq:Zdefnoise}
Z(t):=\frac{1}{\mfg(t)}e^{i\omega_{\rosh}(t)}\mfD_{\rotot}(t)e^{S_{\ron}t}z_{\infty}
=\frac{1}{\mfg(t)}\left(\begin{array}{c} \mfg(t)z_{\app}(t) \\ p_{\app}(t)\end{array}\right),
\end{equation}
where we have omitted the argument $\indexnot$ for the sake of brevity. Due to (\ref{eq:zinfestnoise}), 
\begin{equation}\label{eq:eSinfbtzinfestnoise}
|e^{S_{\ron}t}z_{\infty}|
\leq C_{B}\ldr{\bt}^{d_{\ron,+}-1}e^{\kappa_{\ron,+}\bt}\left(|w(T_{\ron})|+e^{\kappa_{\ron,+}T_{\ron}}\int_{T_{\ron}}^{\infty}e^{-(\kappa_{\ron,+}-\eta_{B})t}|\hf(t)|dt\right)
\end{equation}
for all $t\geq T_{\ron}$, where $C_{B}$ has the same dependence as in the case of (\ref{eq:estzzappzdpappnoise}). In particular, $\mfg |Z|$ is thus 
bounded by the right hand side of (\ref{eq:eSinfbtzinfestnoise}). Compute
\begin{equation}\label{eq:Zdotcompnoise}
\begin{split}
\dot{Z} = & -\dot{\ell}Z+i\sigma\mfg Z+\frac{1}{\mfg}e^{i\omega_{\rosh}}\dot{\mfD}_{\rotot}e^{S_{\ron}t}z_{\infty}
 +\frac{1}{\mfg}e^{i\omega_{\rosh}}\mfD_{\rotot}S_{\ron}e^{S_{\ron}t}z_{\infty}.
\end{split}
\end{equation}
Combining (\ref{eq:eSinfbtzinfestnoise}); (\ref{eq:Zdotcompnoise}); (\ref{eq:estdtlnge}); the fact that $\mfg(t)\geq c_{0}e^{\b_{\ron}t}$ for some 
$c_{0}>0$ depending only on the coefficients of the equation (\ref{eq:thesystemRge}) and all $t\geq T_{\ron}$; and the fact that 
$|\sigma(t)|\leq Ce^{-\eta_{\rosh}t}$ for some constant $C>0$ depending only on the coefficients of the equation (\ref{eq:thesystemRge}), it follows that 
\begin{equation*}
\begin{split}
 & \left|\dot{Z}(t)-\dot{\mfD}_{\rotot}(t)[\mfD_{\rotot}(t)]^{-1}Z(t)\right|\\
 \leq & C_{B}\ldr{\bt}^{d_{\ron,+}-1}e^{(\kappa_{\ron,+}-\eta_{B})\bt}\left(|w(T_{\ron})|+e^{\kappa_{\ron,+}T_{\ron}}\int_{T_{\ron}}^{\infty}e^{-(\kappa_{\ron,+}-\eta_{B})t}|\hf(t)|dt\right)
\end{split}
\end{equation*}
for all $t\geq T_{\ron}$, where $C_{B}$ has the same dependence as in the case of (\ref{eq:estzzappzdpappnoise}). Since 
$\dot{\varphi}_{\rotot}(t)=[1+\sigma^{2}(t)]^{1/2}\mfg(t)$ and $|\sigma(t)|\leq Ce^{-\eta_{\rosh}t}$, this estimate implies that 
\begin{equation*}
\begin{split}
 & \left|\dot{Z}(t)-\left(\begin{array}{cc} 0 & \Id_{m} \\ -\Id_{m} & 0\end{array}\right)\mfg(t)Z(t)\right|\\
 \leq & C_{B}\ldr{\bt}^{d_{\ron,+}-1}e^{(\kappa_{\ron,+}-\eta_{B})\bt}\left(|w(T_{\ron})|+e^{\kappa_{\ron,+}T_{\ron}}\int_{T_{\ron}}^{\infty}e^{-(\kappa_{\ron,+}-\eta_{B})t}|\hf(t)|dt\right)
\end{split}
\end{equation*}
for all $t\geq T_{\ron}$, where $C_{B}$ has the same dependence as in the case of (\ref{eq:estzzappzdpappnoise}). In particular, this estimate 
implies that (\ref{eq:estzzappzdpappnoise}) with $p_{\app}$ replaced by $\dot{z}_{\app}$ holds.

\textbf{Eliminating $|w(T_{\ron})|$.} In order to proceed, it is of interest to estimate the right hand side of 
(\ref{eq:estzzappzdpappnoise}). Appealing to Lemma~\ref{lemma:roughenestbalsetting},
\begin{equation}\label{eq:wtkzestnoise}
\me^{1/2}(t)\leq e^{\eta_{\robal}t}\me^{1/2}(0)+\int_{0}^{t}e^{\eta_{\robal}(t-t')}|\hf(t')|dt'
\end{equation}
for all $t\geq 0$. Note that $\me^{1/2}$ and $|w|$ are equivalent for all $t\geq T_{\ron}$, the constant of equivalence being numerical; 
this is due to (\ref{eq:condensurtimeseqwd}). Combining (\ref{eq:wtkzestnoise}) (with $t=T_{\ron}$) with (\ref{eq:estzzappzdpappnoise}) 
(with $p_{\app}$ replaced by $\dot{z}_{\app}$) yields
\begin{equation}\label{eq:estzzappzdpappnoisesv}
\begin{split}
 & |\dot{z}(t)-\dot{z}_{\app}(t)|+\mfg(t)|z(t)-z_{\app}(t)|\\
 \leq & C_{B}\ldr{t}^{N}e^{(\kappa_{\ron,+}-\eta_{B})t}\left(e^{\eta_{1,+}T_{\ron}}\me^{1/2}(0)+e^{\eta_{2,+}T_{\ron}}
\int_{0}^{\infty}e^{-(\kappa_{\ron,+}-\eta_{B})t'}|\hf(t')|dt'\right)
\end{split}
\end{equation}
for all $t\geq T_{\ron}$, where $C_{B}$ and $N$ have the same dependence as in the case of (\ref{eq:estzzappzdpappnoise}), and
\[
\eta_{1,+}:=\max\{\eta_{\robal}-\kappa_{\ron,+}+\eta_{B},0\},\ \ \
\eta_{2,+}:=\max\{\eta_{\robal}-\kappa_{\ron,+}+\eta_{B},\eta_{B}\}.
\] 
If the Jordan blocks of the matrices $A_{\ron,\pm}$ are trivial, the $N$ appearing in (\ref{eq:estzzappzdpappnoisesv}) can be replaced by
$1$. Moreover, if $E_{\ron,+}=E_{\ron,-}=\cn{m}$, then $N$ can be replaced by $d_{n,+}-1$.

\textbf{Deriving an estimate for early times.}
Next we need to verify that an estimate such as (\ref{eq:estzzappzdpappnoisesv}) holds on $[0,T_{\ron}]$. To this end, we need to estimate
$z$ and $z_{\app}$ separately. Due to (\ref{eq:wtkzestnoise}), 
\begin{equation}\label{eq:zdotgzestztoTron}
\me^{1/2}(t) \leq Ce^{(\kappa_{\ron,+}-\eta_{B})t}\left(e^{\eta_{1,+}T_{\ron}}\me^{1/2}(0)+e^{\eta_{1,+}T_{\ron}}
\int_{0}^{t}e^{-(\kappa_{\ron,+}-\eta_{B})t'}|\hf(t')|dt'\right)
\end{equation}
for all $t\in [0,T_{\ron}]$, where $C$ is numerical. In order to estimate $\dot{z}_{\app}$ and $\mfg z_{\app}$ for $t\in [0,T_{\ron}]$, note that 
\begin{equation}\label{eq:mfgZestztoTron}
\begin{split}
\mfg(t)|Z(t)| \leq & \|e^{S_{\ron}(t-T_{\ron})}\|\cdot |e^{S_{\ron}T_{\ron}}z_{\infty}|\\
 \leq & C_{B}e^{\eta_{\ros}T_{\ron}}\left(|w(T_{\ron})|+e^{\kappa_{\ron,+}T_{\ron}}\int_{T_{\ron}}^{\infty}e^{-(\kappa_{\ron,+}-\eta_{B})t'}|\hf(t')|dt'\right)\\
 \leq & C_{B}e^{(\kappa_{\ron,+}-\eta_{B})t}\left(e^{\eta_{3,+}T_{\ron}}\me^{1/2}(0)+e^{\eta_{4,+}T_{\ron}}
\int_{0}^{\infty}e^{-(\kappa_{\ron,+}-\eta_{B})t'}|\hf(t')|dt'\right)
\end{split}
\end{equation}
for all $t\in [0,T_{\ron}]$, where $C_{B}$ has the same dependence as in the case of (\ref{eq:estzzappzdpappnoise}); $\eta_{\ros}\geq 0$ only 
depends on the coefficients of the equation (\ref{eq:thesystemRge}); 
\begin{align*}
\eta_{\rod,\pm} := & \max\{0,\pm(\kappa_{\ron,+}-\eta_{B})\},\\
\eta_{3,+} := & \eta_{\ros}+\eta_{\robal}+\eta_{\rod,-},\\
\eta_{4,+} := & \max\{\max\{\eta_{\ros}+\kappa_{\ron,+},0\}+\eta_{\rod,-},\eta_{\ros}+\eta_{1,+}+\eta_{\rod,+}\};
\end{align*}
and we appealed to (\ref{eq:zinfestnoise}), (\ref{eq:Zdefnoise}) and (\ref{eq:wtkzestnoise}). In order to estimate $\dot{Z}$, let us 
return to (\ref{eq:Zdotcompnoise}). This equality can be rewritten 
\[
\dot{Z} = -\dot{\ell}Z+i\sigma\mfg Z+\dot{\mfD}_{\rotot}\mfD_{\rotot}^{-1}Z+\mfD_{\rotot}S_{\ron}\mfD_{\rotot}^{-1}Z.
\]
Since $|\dot{\ell}|$, $|\sigma|$ and $\mfg^{-1}\|\dot{\mfD}_{\rotot}\|$ are bounded for all $t\geq 0$ by constants depending only on the coefficients
of the equation (\ref{eq:thesystemRge}), this equality yields
\begin{equation}\label{eq:Zdotintermnoise}
|\dot{Z}|\leq C(1+\mfg^{-1})\mfg|Z|
\end{equation}
for all $t\geq 0$, where $C$ only depends on the coefficients of the equation (\ref{eq:thesystemRge}) and we used the fact that $\mfD_{\rotot}$
is unitary. We would like to combine this estimate with (\ref{eq:mfgZestztoTron}) in order to estimate $|\dot{Z}|$. To be allowed to do so, we need
to estimate $\mfg$ from below. We already derived such an estimate in the proof of Lemma~\ref{lemma:asymposccasescalareqnoise}; cf. 
(\ref{eq:mfglowbdallfuttimenoise}). The derivation is equally valid in the present setting, and we conclude that (\ref{eq:mfglowbdallfuttimenoise})
holds for all $t\geq 0$, where $C$ and $s_{E}$ only depend on the coefficients of the equation (\ref{eq:thesystemRge}). Combining 
(\ref{eq:mfglowbdallfuttimenoise}), (\ref{eq:mfgZestztoTron}) and (\ref{eq:Zdotintermnoise}) yields 
\begin{equation*}
\begin{split}
 & |\dot{Z}(\indexnot,t)|+\mfg(\indexnot,t)|Z(\indexnot,t)|+|Z(\indexnot,t)|\\
 \leq & C_{B}e^{(\kappa_{\ron,+}-\eta_{B})t}\ldr{\nu(\indexnot)}^{s_{E}}\left(e^{\eta_{3,+}T_{\ron}}\me^{1/2}(\indexnot,0)+e^{\eta_{4,+}T_{\ron}}
\int_{0}^{\infty}e^{-(\kappa_{\ron,+}-\eta_{B})t'}|\hf(\indexnot,t')|dt'\right)
\end{split}
\end{equation*} 
for all $t\in [0,T_{\ron}]$, where $C_{B}$ has the same dependence as in the case of (\ref{eq:estzzappzdpappnoise}). Combining this estimate with 
(\ref{eq:estzzappzdpappnoisesv}) and (\ref{eq:zdotgzestztoTron}) yields (recall that $\mfg(t)\geq 2$ for $t\geq T_{\ron}$)
\begin{equation}\label{eq:estzzappzdpappnoisesvitonus}
\begin{split}
 & |\dot{z}(t)-\dot{z}_{\app}(t)|+\mfg(t)|z(t)-z_{\app}(t)|+|z(t)-z_{\app}(t)|\\
 \leq & C_{B}\ldr{t}^{N}e^{(\kappa_{\ron,+}-\eta_{B})t}\ldr{\nu}^{s_{E}}\left(e^{\eta_{\roh}T_{\ron}}\me^{1/2}(0)+e^{\eta_{\roih}T_{\ron}}
\int_{0}^{\infty}e^{-(\kappa_{\ron,+}-\eta_{B})t'}|\hf(t')|dt'\right)\\
 \leq & C_{B}\ldr{t}^{N}e^{(\kappa_{\ron,+}-\eta_{B})t}\left(\ldr{\nu}^{s_{\roh}}\me^{1/2}(0)+\ldr{\nu}^{s_{\roih}}
\int_{0}^{\infty}e^{-(\kappa_{\ron,+}-\eta_{B})t'}|\hf(t')|dt'\right)
\end{split}
\end{equation}
for all $t\geq 0$, where $C_{B}$ and $N$ have the same dependence as in the case of (\ref{eq:estzzappzdpappnoise}) and we appealed to 
(\ref{eq:Tronuppbdnoise}). If the Jordan blocks of the matrices $A_{\ron,\pm}$ are trivial, the $N$ appearing in 
(\ref{eq:estzzappzdpappnoisesvitonus}) can be replaced by $1$. Moreover, if $E_{\ron,+}=E_{\ron,-}=\cn{m}$, then $N$ can be replaced by $d_{n,+}-1$. 
Finally, 
\begin{align*}
\eta_{\roh} := & \max\{\eta_{1,+},\eta_{3,+}\},\ \ \
 \eta_{\roih}:=\max\{\eta_{2,+},\eta_{4,+}\},\\
s_{\roh} := & \frac{2\eta_{\roh}}{\eta_{\romar}}+s_{E},\ \ \
 s_{\roih} := \frac{2\eta_{\roih}}{\eta_{\romar}}+s_{E}.
\end{align*}

\textbf{Summing up; the approximate solution.} Let 
\begin{equation}\label{eq:uappdefsvnoise}
u_{\ron}(p,t):=\textstyle{\sum}_{\indexnot\in\EFnindexset}z(\indexnot,t)\varphi_{\indexnot}(p),\ \ \
u_{\app}(p,t):=\sum_{\indexnot\in\EFnindexset}z_{\app}(\indexnot,t)\varphi_{\indexnot}(p).
\end{equation}
Then (\ref{eq:estzzappzdpappnoisesvitonus}) yields
\begin{equation}\label{eq:umuappenestnoise}
\begin{split}
 & \mfe_{s}^{1/2}[u_{\ron}-u_{\app}](t)\\
 \leq & C_{B}\ldr{t}^{N}e^{(\kappa_{\ron,+}-\eta_{B})t}\left(\mfe_{s+s_{\roh}}^{1/2}[u_{\ron}](0)
+\int_{0}^{\infty}e^{-(\kappa_{\ron,+}-\eta_{B})t'}\|f_{\ron}(\cdot,t')\|_{(s+s_{\roih})}dt'\right)
\end{split}
\end{equation}
for all $s\in\ro$ and all $t\geq 0$, where $C_{B}$ and $N$ have the same dependence as in the case of (\ref{eq:estzzappzdpappnoise}). Note also
that $s_{\roh}$ and $s_{\roih}$ only depend on $\eta_{B}$ and the coefficients of the equation (\ref{eq:thesystemRge}). Moreover, if the Jordan blocks 
of the matrices $A_{\ron,\pm}$ are trivial, the $N$ appearing in (\ref{eq:umuappenestnoise}) can be replaced by $1$. Finally, if 
$E_{\ron,+}=E_{\ron,-}=\cn{m}$, then $N$ can be replaced by $d_{n,+}-1$. 

\textbf{Reformulating and estimating the approximate solution.} Next, we wish to rewrite $u_{\app}$. Introduce, to this end, 
\begin{align*}
z_{+,\wa}(\indexnot,t) := & \frac{e^{\b_{\ron} t/2}}{\mfg(\indexnot,t)}e^{i[\omega_{\rosh}(\indexnot,t)-\varphi_{\rotot}(\indexnot,t)]}\psi_{+}(\indexnot),\\
z_{-,\wa}(\indexnot,t) := & -\frac{e^{\b_{\ron} t/2}}{\mfg(\indexnot,t)}e^{i[\omega_{\rosh}(\indexnot,t)+\varphi_{\rotot}(\indexnot,t)]}i\psi_{-}(\indexnot).
\end{align*}
Then $z_{\app}(\indexnot,t)$ can be written 
\begin{equation}\label{eq:zappitozwnoise}
z_{\app}(\indexnot,t)=e^{-[\a_{\infty}-\bX_{\ron}(\indexnot)]t/2}z_{+,\wa}(\indexnot,t)+e^{-[\a_{\infty}+\bX_{\ron}(\indexnot)]t/2}z_{-,\wa}(\indexnot,t).
\end{equation}
Let us estimate the $\psi_{\pm}$. Keeping (\ref{eq:poptTprezinfdefnoise}) in mind, it is sufficient to estimate $z_{\infty}$. Moreover, 
$z_{\infty}$ can be estimated by appealing to (\ref{eq:zinfestnoise}):
\begin{equation*}
\begin{split}
|\psi_{+}|+|\psi_{-}| \leq & C|z_{\infty}|\leq C\| e^{-S_{\ron}T_{\ron}}\|\cdot |e^{S_{\ron}T_{\ron}}z_{\infty}|\\
 \leq & C_{B}\| e^{-S_{\ron}(\indexnot)T_{\ron}}\|
\left(|w(T_{\ron})|+e^{\kappa_{\ron,+}T_{\ron}}\int_{T_{\ron}}^{\infty}e^{-(\kappa_{\ron,+}-\eta_{B})t'}|\hf(t')|dt'\right)\\
 \leq & C_{B}e^{\eta_{\ros}T_{\ron}}\left(e^{\eta_{\robal}T_{\ron}}\me^{1/2}(0)+e^{\eta_{a}T_{\ron}}\int_{0}^{T_{\ron}}e^{-(\kappa_{\ron,+}-\eta_{B})t'}|\hf(t')|dt'\right.\\
 & \left. +e^{\kappa_{\ron,+}T_{\ron}}\int_{T_{\ron}}^{\infty}e^{-(\kappa_{\ron,+}-\eta_{B})t'}|\hf(t')|dt'\right),
\end{split}
\end{equation*}
where we appealed to (\ref{eq:wtkzestnoise}); $C_{B}$ has the same dependence as in the case of (\ref{eq:estzzappzdpappnoise}); 
$\eta_{\ros}\geq 0$ only depends on the coefficients of the equation (\ref{eq:thesystemRge}); and 
\[
\eta_{a}:=\eta_{\robal}+\max\{0,\kappa_{\ron,+}-\eta_{\robal}-\eta_{B}\}.
\]
Due to (\ref{eq:Tronuppbdnoise}), this estimate yields
\begin{equation}\label{eq:zoztinfestnoise}
|\psi_{+}(\indexnot)|+|\psi_{-}(\indexnot)|\leq C_{B}\left(\me_{s_{a}}^{1/2}(\indexnot,0)
+\ldr{\nu(\indexnot)}^{s_{b}}\int_{0}^{\infty}e^{-(\kappa_{\ron,+}-\eta_{B})t'}|\hf(\indexnot,t')|dt'\right),
\end{equation}
where $C_{B}$ has the same dependence as in the case of (\ref{eq:estzzappzdpappnoise}) and
\[
s_{a}:=\frac{2}{\eta_{\romar}}(\eta_{\ros}+\eta_{\robal}),\ \ \
s_{b}:=\frac{2}{\eta_{\romar}}(\eta_{\ros}+\max\{\eta_{a},\kappa_{\ron,+}\});
\]
note that $s_{a}$ and $s_{b}$ only depend on $\eta_{B}$ and the coefficients of the equation (\ref{eq:thesystemRge}). Thus
\begin{equation}\label{eq:zetalHsnormnoise}
\begin{split}
 & \left(\textstyle{\sum}_{\indexnot\in\EFnindexset}\ldr{\nu(\indexnot)}^{2s}\left[|\psi_{+}(\indexnot)|^{2}+|\psi_{-}(\indexnot)|^{2}\right]\right)^{1/2}\\
 \leq & C_{B}\left(\mfe_{s+s_{a}}^{1/2}[u_{\ron}](0)
+\int_{0}^{\infty}e^{-(\kappa_{\ron,+}-\eta_{B})t'}\|f_{\ron}(\cdot,t')\|_{(s+s_{b})}dt'\right)
\end{split}
\end{equation}
for all $s\in\ro$, where $C_{B}$ has the same dependence as in the case of (\ref{eq:estzzappzdpappnoise}). 

Due to the assumptions of the lemma, the right hand side of (\ref{eq:zetalHsnormnoise}) is bounded for all $s$. In particular, the components of 
$\psi_{\pm}(\indexnot)$ satisfy estimates such that 
Lemma~\ref{lemma:asymposccasescalareqnoise} can be applied with $z_{\app}$ replaced with the components of $z_{\pm,\wa}$. However, there is one 
problem associated with doing so: in (\ref{eq:zappitozwnoise}), there are $\indexnot$-dependent factors multiplying $z_{\pm,\wa}$. On the other hand, 
this problem only occurs if the equation has a dominant noisy spatial torus direction and is related to the sign of $\nu_{\roT,j}(\indexnot)$. It is 
therefore convenient to use the notation introduced in Definition~\ref{definition:EFnindexsetpm}; recall also that $\bX_{\ron}(\indexnot)=0$ for all 
$\indexnot\in\EFnindexset$ in case the equation has a dominant noisy spatial generalised direction. Define
\begin{equation}\label{eq:ulwapmdefnoise}
u_{\pm,\wa}^{+}(p,t):=\textstyle{\sum}_{\indexnot\in\EFnindexsetp}z_{\pm,\wa}(\indexnot,t)\varphi_{\indexnot}(p),\ \ \
u_{\pm,\wa}^{-}(p,t):=\sum_{\indexnot\in\EFnindexsetm}z_{\pm,\wa}(\indexnot,t)\varphi_{\indexnot}(p).
\end{equation}
Due to the definition of $z_{\pm,\wa}(\indexnot,t)$ and the properties of the $\psi_{\pm}$'s, it is clear that $u_{+,\wa}^{+}$ and $u_{-,\wa}^{-}$ take their 
values in $E_{\ron,+}$ and that $u_{+,\wa}^{-}$ and $u_{-,\wa}^{+}$ take their values in $E_{\ron,-}$, where the $E_{\ron,\pm}$ are defined in the statement of the lemma. 

\textit{Comparing with Remark~\ref{remark:specupmwanoise}.}
It is of interest to compare the above conclusions and notation with Remark~\ref{remark:specupmwanoise}. Let us first consider the case of a dominant 
noisy spatial torus direction. Define $\zeta_{+}^{\pm}(\indexnot)=\psi_{+}(\indexnot)$ for all $\indexnot\in\EFnindexsetpm$ and $\zeta_{+}^{\pm}(\indexnot)=0$ 
for all $\indexnot\notin\EFnindexsetpm$. Define, similarly, $\zeta_{-}^{\pm}(\indexnot)=-i\psi_{-}(\indexnot)$ for all $\indexnot\in\EFnindexsetpm$ and 
$\zeta_{-}^{\pm}(\indexnot)=0$ for all $\indexnot\notin\EFnindexsetpm$. Then $\zeta_{+}^{+}(\indexnot),\zeta_{-}^{-}(\indexnot)\in E_{\ron,+}$ and 
$\zeta_{+}^{-}(\indexnot),\zeta_{-}^{+}(\indexnot)\in E_{\ron,-}$ for all $\indexnot\in\EFindexset$. Note also that the estimates 
(\ref{eq:zetapmpmestnoiseapprwa}) and (\ref{eq:zetalpmHsnormnoise}) follow from (\ref{eq:zetalHsnormnoise}). Defining $z_{+,\wa}^{\pm}$ and 
$z_{-,\wa}^{\pm}$ by (\ref{eq:zpwapmdef}) and (\ref{eq:zmwapmdef}) respectively, it is clear that $z_{+,\wa}^{\pm}$ and $z_{-,\wa}^{\pm}$ are the Fourier
coefficients of $u_{+,\wa}^{\pm}$ and $u_{-,\wa}^{\pm}$, introduced in (\ref{eq:ulwapmdefnoise}), respectively. In short, the $u_{+,\wa}^{\pm}$ and 
$u_{-,\wa}^{\pm}$ introduced in (\ref{eq:ulwapmdefnoise}) coincide with the $u_{+,\wa}^{\pm}$ and $u_{-,\wa}^{\pm}$ introduced in 
Remark~\ref{remark:specupmwanoise}. Defining $u_{\pm,\wa}$ by (\ref{eq:upmwadef}), the function $u_{\app}$, introduced in 
(\ref{eq:uappdefsvnoise}), can be written
\begin{equation}\label{eq:uappwadefnoise}
\begin{split}
u_{\app}(p,t) = & e^{A_{\ron,+}t}u_{+,\wa}(p,t)+e^{A_{\ron,-}t}u_{-,\wa}(p,t),
\end{split}
\end{equation}
where $A_{\ron,\pm}$ is introduced in the statement of the lemma. In particular, $u_{\app}$ equals the function $u_{\wa}$ introduced in 
Remark~\ref{remark:specupmwanoise}. Due to (\ref{eq:umuappenestnoise}), the estimates (\ref{eq:uronminuwauniquenessestimate}) and
(\ref{eq:uuweststatementnoisewa}) thus hold. To conclude, $\zeta_{+}^{\pm}$ and $\zeta_{+}^{\pm}$ have the properties stated in 
Remark~\ref{remark:specupmwanoise}, except, possibly, for the uniqueness (which remains to be verified). 

Next, let us turn to the case of a dominant noisy spatial generalised direction. Let $\chi_{+}(\indexnot)=\psi_{+}(\indexnot)$ and 
$\chi_{-}(\indexnot)=-i\psi_{-}(\indexnot)$ for all $\indexnot\in\EFnindexset$, and let $\chi_{\pm}(\indexnot)=0$ for all $\indexnot\notin\EFnindexset$. 
Then $\chi_{\pm}(\indexnot)\in E_{\ron}$ for all $\indexnot\in\EFindexset$. Moreover, (\ref{eq:rortbdnoise}) and (\ref{eq:zetalpmHsnormnoise}) follow 
from (\ref{eq:zetalHsnormnoise}) (assuming (\ref{eq:zetalpmHsnormnoise}) is interpreted as stated immediately below this estimate). If $u_{0,\wa}$ 
is given by the right hand side of (\ref{eq:uappdefnoise}), where $z_{\app}$ is given by (\ref{eq:zappdefnoise}), then the function $u_{\app}$, 
introduced in (\ref{eq:uappdefsvnoise}), can be written $u_{\app}=e^{A_{\ron}t}u_{0,\wa}$. In other words, the function $u_{\wa}$, introduced in 
Remark~\ref{remark:specupmwanoise}, equals $u_{\app}$. Due to (\ref{eq:umuappenestnoise}), the estimates (\ref{eq:uronminuwauniquenessestimate}) and
(\ref{eq:uuweststatementnoisewa}) thus hold. To conclude, $\chi_{\pm}$ have the properties stated in Remark~\ref{remark:specupmwanoise}, except, 
possibly, for the uniqueness (which remains to be verified).

\textbf{Constructing solutions to (\ref{eq:homtheeqnoise}).} At this stage, we can appeal to Lemma~\ref{lemma:asymposccasescalareqnoise}. 
This yields four solutions $u_{\pm,\row}^{+}$ and $u_{\pm,\row}^{-}$ to (\ref{eq:homtheeqnoise}) in the case of a dominant noisy spatial torus 
direction. In what follows, we also use the notation
\[
u_{+,\row}=u_{+,\row}^{+}+u_{-,\row}^{-},\ \ \
u_{-,\row}=u_{+,\row}^{-}+u_{-,\row}^{+}. 
\]
In the case of a dominant noisy spatial generalised direction, the $\chi_{\pm}$ give rise to functions $u_{\pm,\row}$. In order to be able to discuss
the torus setting and the generalised setting at the same time, it is convenient to introduce the notation $u_{\pm,\row}^{-}:=0$,
$u_{\pm,\row}^{+}:=u_{\pm,\row}$ and $u_{0,\row}:=u_{+,\row}+u_{-,\row}$ in the generalised setting. Before proceeding, note that the following holds:
\begin{itemize}
\item if $z_{0,\row}(\indexnot,t)$ denotes the $\indexnot$'th Fourier coefficient of $u_{0,\row}(\cdot,t)$, then $z_{0,\row}(\indexnot,\cdot)=0$ if 
$\indexnot\notin\EFnindexset$,
\item $z_{0,\row}$, and thereby $u_{0,\row}$, takes its values in $E_{\ron}$,
\item if the $\indexnot$'th Fourier coefficients of $u_{\tau,\row}^{\upsilon}$, $\tau,\upsilon\in\{+,-\}$, are denoted by 
$z_{\tau,\row}^{\upsilon}(\indexnot,\cdot)$, then 
\begin{itemize}
\item $z_{\pm,\row}^{+}(\indexnot,\cdot)=0$ in case $\indexnot\notin\EFnindexsetp$ and 
$z_{\pm,\row}^{-}(\indexnot,\cdot)=0$ in case $\indexnot\notin\EFnindexsetm$,
\item $u_{+,\row}^{\pm}$ are positively oriented and $u_{-,\row}^{\pm}$ are negatively oriented solutions 
to (\ref{eq:homtheeqnoise}),
\end{itemize}
\item $u_{+,\row}^{+}$ and $u_{-,\row}^{-}$ take their values in $E_{\ron,+}$ and $u_{-,\row}^{+}$ and $u_{-,\row}^{+}$ take their values in $E_{\ron,-}$,
so that $u_{\pm,\row}$ takes its values in $E_{\ron,\pm}$. 
\end{itemize}
With the above notation, the conclusions of Lemma~\ref{lemma:asymposccasescalareqnoise} yield
\begin{equation}\label{eq:mfediffhomestlvernoise}
\mfe_{s}[u_{\pm,\row}-u_{\pm,\wa}](t)\leq C_{B}e^{(\b_{\ron}-2\eta_{B})t}
\textstyle{\sum}_{\indexnot\in\EFnindexset}\ldr{\nu(\indexnot)}^{2(s+s_{\rom})}[r_{+}^{2}(\indexnot)+r_{-}^{2}(\indexnot)]
\end{equation}
for all $t\geq 0$, where the constants have the dependence stated in Lemma~\ref{lemma:asymposccasescalareqnoise} and 
$r_{\pm}(\indexnot)=|\psi_{\pm}(\indexnot)|$. Combining (\ref{eq:zetalHsnormnoise}) with (\ref{eq:mfediffhomestlvernoise}) yields
\begin{equation}\label{eq:mfediffhomestsvnoise}
\begin{split}
 & \mfe_{s}^{1/2}[u_{\pm,\row}-u_{\pm,\wa}](t)\\
 \leq & C_{B}e^{(\b_{\ron}/2-\eta_{B})t}
\left(\mfe_{s+s_{1}}^{1/2}[u_{\ron}](0)
+\int_{0}^{\infty}e^{-(\kappa_{\ron,+}-\eta_{B})t'}\|f_{\ron}(\cdot,t')\|_{(s+s_{2})}dt'\right)
\end{split}
\end{equation}
for all $s\in\ro$ and all $t\geq 0$, where $s_{1}:=s_{a}+s_{\rom}$, $s_{2}:=s_{b}+s_{\rom}$, and $C_{B}$ has the same dependence as in 
the case of (\ref{eq:estzzappzdpappnoise}). In addition combining (\ref{eq:mfezeroestnoise}) and (\ref{eq:zetalHsnormnoise}) yields
(\ref{eq:mfesuzrowestitoidanih}) and (\ref{eq:mfesupmrowestitoidanih}). 

\textbf{The final estimate.} Define $u_{\row}$ by (\ref{eq:urowdefnoisetorus}) in the case of a dominant noisy spatial torus
direction. Define $u_{\row}$ by (\ref{eq:urowdefnoisegen}) in the case of a dominant noisy spatial generalised direction. Combining 
these definitions with (\ref{eq:uappwadefnoise}) and (\ref{eq:mfediffhomestsvnoise}) yields the conclusion 
\begin{equation}\label{eq:urowminusuappintermediateest}
\begin{split}
 & \mfe_{s}^{1/2}[u_{\row}-u_{\app}](t)\\
 \leq & C_{B}\ldr{t}^{d_{n,+}-1}e^{(\kappa_{\ron,+}-\eta_{B})t}
\left(\mfe_{s+s_{1}}^{1/2}[u_{\ron}](0)
+\int_{0}^{\infty}e^{-(\kappa_{\ron,+}-\eta_{B})t'}\|f_{\ron}(\cdot,t')\|_{(s+s_{2})}dt'\right)
\end{split}
\end{equation}
for all $s\in\ro$ and all $t\geq 0$, where $C_{B}$ has the same dependence as in the case of (\ref{eq:estzzappzdpappnoise}).
Finally, combining this estimate 
with (\ref{eq:umuappenestnoise}) yields (\ref{eq:uuweststatementnoise}). Comparing the statements of the lemma with the above constructions, 
it is clear that the lemma holds.

\textbf{Uniqueness of the approximation.} Finally, let us turn to the uniqueness statements contained in Remark~\ref{remark:specupmwanoise}.
Starting with the case of a dominant noisy spatial generalised direction, assume that there are $\chi_{\pm,l}$, $l=1,2$, with the properties
stated in the remark. This gives rise to two functions $u_{0,\wa,l}$ with values in $E_{\ron}$. Appealing to Lemma~\ref{lemma:asymposccasescalareqnoise},
we also obtain functions $u_{0,\row,l}$, $l=1,2$, with values in $E_{\ron}$. Defining $u_{\wa,l}:=e^{A_{\ron}t}u_{0,\wa,l}$ and $u_{\row,l}:=e^{A_{\ron}t}u_{0,\row,l}$, 
(\ref{eq:uronminuwauniquenessestimate}) (with $u_{\wa}$ replaced by $u_{\wa,l}$) and Lemma~\ref{lemma:asymposccasescalareqnoise} imply
\[
\mfe_{s}^{1/2}[u_{\ron}-u_{\row,l}](t)\leq C\ldr{t}^{N}e^{(\kappa_{\ron,+}-\eta_{B})t}
\]
for all $t\geq 0$. Due to the uniqueness part of the lemma, this estimate implies that $u_{0,\row,1}=u_{0,\row,2}$. By the uniqueness part of 
Lemma~\ref{lemma:asymposccasescalareqnoise}, we conclude that $\chi_{\pm,1}=\chi_{\pm,2}$. The proof of uniqueness in the torus setting is 
similar. The same is true of the final statement in Remark~\ref{remark:specupmwanoise}.
\end{proof}

\section{Specifying data at infinity}\label{section:spedataatinf}

If the assumptions of Lemma~\ref{lemma:asymposccaseitosoltowenoise} are satisfied, we know that, given a smooth solution $u$ to 
(\ref{eq:thesystemRge}), there are uniquely associated solutions $u_{\pm,\row}$ (or $u_{0,\row}$) to (\ref{eq:homtheeqnoise}) such that 
$u_{\ron}$ is well approximated by $u_{\row}$ defined by (\ref{eq:urowdefnoisetorus}) (or (\ref{eq:urowdefnoisegen})); cf. (\ref{eq:uuweststatementnoise}). 
Even though this result provides some information concerning the asymptotics, it does leave several questions open. The first 
question is: given that $u_{\pm,\row}$ (or $u_{0,\row}$) have the properties stated in Lemma~\ref{lemma:asymposccaseitosoltowenoise}, except for 
the estimates concerning the difference between $u_{\ron}$ and $u_{\row}$, is there a solution $u$ to (\ref{eq:thesystemRge}) such that 
(\ref{eq:uuweststatementnoise}) holds? The estimates (\ref{eq:mfesuzrowestitoidanih}), (\ref{eq:mfesupmrowestitoidanih}) and 
(\ref{eq:zetalpmHsnormnoise}) demonstrate that the map from solutions to (\ref{eq:thesystemRge}) to the asymptotic data is continuous. 
Once we have constructed a map from asymptotic data to initial data, we here also wish to demonstrate that it is continuous. As in the case 
of Lemma~\ref{lemma:spasda} and  Proposition~\ref{prop:spasdatrs}, we focus our attention on the case of homogeneous equations. 
Before stating the lemma, it is convenient to introduce the following terminology. 

\begin{definition}
Let $1\leq m\in\zo$ and $V$ be a vector subspace of $\cn{m}$. Then
\[
W_{\ron}(\bM,V)\subset C^{\infty}(\bM,V)\times C^{\infty}(\bM,V)
\]
denotes the set of initial data at $t=0$ for solutions $u$ to (\ref{eq:homtheeqnoise}) such that $u\in C^{\infty}(M,V)$ and the $\indexnot$'th Fourier 
coefficient of $u(\cdot,t)$ vanishes for all $t$ if $\indexnot\notin\EFnindexset$. Moreover, $W_{\ron,+}(\bM,V)$ ($W_{\ron,-}(\bM,V)$) denotes the subset of 
$W_{\ron}(\bM,V)$
corresponding to positively (negatively) oriented solutions to (\ref{eq:homtheeqnoise}). In addition, $W_{\ron,\pm}^{+}(\bM,V)$ ($W_{\ron,\pm}^{-}(\bM,V)$) denotes 
the subset of $W_{\ron,\pm}(\bM,V)$ consisting of functions whose $\indexnot$'th Fourier coefficients vanish if $\indexnot\notin\EFnindexsetp$ 
($\indexnot\notin\EFnindexsetm$). 
\end{definition}
\begin{remark}
By initial data we here mean $[u(\cdot,0),u_{t}(\cdot,0)]$. 
\end{remark}

\begin{lemma}\label{lemma:asymposccaseitosoltowenoisespa}
Assume that (\ref{eq:thesystemRge}) is strongly balanced, has a negligible shift vector field, a dominant noisy spatial direction, and is such that the dominant 
coefficients are convergent. Assume, moreover, that $f=0$. Let $\eta_{A}$ be defined by (\ref{eq:etaAdefinition}) and $0<\eta_{B}\leq \eta_{A}$. Define 
$A_{\ron,\pm}$ and $E_{\ron,\pm}$ as in the statement of Lemma~\ref{lemma:asymposccaseitosoltowenoise}.

If (\ref{eq:thesystemRge}) has a dominant noisy spatial generalised direction, then $E_{\ron,+}=E_{\ron,-}=:E_{\ron}$ and $A_{\ron,+}=A_{\ron,-}=:A_{\ron}$.
Moreover, there is an injective linear map
\[
\Phi_{\ron,\rog}: W_{\ron}(\bM,E_{\ron})\rightarrow W_{\ron}(\bM,\cn{m})
\]
such that if $\chi\in W_{\ron}(\bM,E_{\ron})$; $u_{0,\row}$ is the solution to (\ref{eq:homtheeqnoise}) corresponding to the initial data $\chi$ at $t=0$; 
$u_{\row}$ is defined by (\ref{eq:urowdefnoisegen}); and $u$ is the solution to (\ref{eq:thesystemRge}) (with $f=0$) corresponding to the initial data 
$\Phi_{\ron,\rog}(\chi)$ at $t=0$, then $u=u_{\ron}$ and 
\begin{equation}\label{eq:mfesumuwestasspgen}
\mfe_{s}^{1/2}[u-u_{\row}](t)\leq C_{B}\ldr{t}^{N}e^{(\kappa_{\ron,+}-\eta_{B})t}\|\chi\|_{(s+s_{B})}
\end{equation}
for all $t\geq 0$ and $s\in\ro$, where $C_{B}$, $s_{B}$ and $N$ have the same dependence as in the case of (\ref{eq:uuweststatementnoise}). If the Jordan 
blocks of the matrix $A_{\ron}$ are trivial, the $N$ appearing in (\ref{eq:mfesumuwestasspgen}) can be replaced by $1$. Moreover, if 
$E_{\ron}=\cn{m}$, then $N$ can be replaced by $d_{n,+}-1$, where $d_{\ron,+}$ is defined as in the statement of 
Lemma~\ref{lemma:windestnoise}. If $E_{\ron}=\cn{m}$, then $\Phi_{\ron,\rog}$ is surjective. Finally, there are constants $C_{B}>0$ 
and $s_{B}\geq 0$ (depending only on $\eta_{B}$ and the coefficients of the equation (\ref{eq:thesystemRge})) such that 
\begin{equation}\label{eq:Phironrogestnoise}
\|\Phi_{\ron,\rog}(\chi)\|_{(s)}\leq C_{B}\|\chi\|_{(s+s_{B})}
\end{equation}
for all $s\in\ro$ and all $\chi\in W_{\ron}(\bM,E_{\ron})$. 

If (\ref{eq:thesystemRge}) has a dominant noisy spatial torus direction, let
\begin{align*}
\mW_{\ron}^{+} := & W_{\ron,+}^{+}(\bM,E_{\ron,+})\oplus W_{\ron,-}^{-}(\bM,E_{\ron,+}),\\
\mW_{\ron}^{-} := & W_{\ron,+}^{-}(\bM,E_{\ron,-})\oplus W_{\ron,-}^{+}(\bM,E_{\ron,-}).
\end{align*}
Then there is an injective linear map
\[
\Phi_{\ron,\roT}: \mW_{\ron}^{+}\times \mW_{\ron}^{-}\rightarrow W_{\ron}(\bM,\cn{m})
\]
such that if $\chi_{\pm}\in \mW_{\ron}^{\pm}$; $u_{\pm,\row}$ is the solution to (\ref{eq:homtheeqnoise}) corresponding to the initial data $\chi_{\pm}$ at $t=0$; 
$u_{\row}$ is defined by (\ref{eq:urowdefnoisetorus}); and $u$ is the solution to (\ref{eq:thesystemRge}) (with $f=0$) corresponding to the initial data 
$\Phi_{\ron,\roT}(\chi_{+},\chi_{-})$ at $t=0$, then $u=u_{\ron}$ and 
\begin{equation}\label{eq:mfesumuwestassptorus}
\mfe_{s}^{1/2}[u-u_{\row}](t)\leq C_{B}\ldr{t}^{N}e^{(\kappa_{\ron,+}-\eta_{B})t}[\|\chi_{+}\|_{(s+s_{B})}+\|\chi_{-}\|_{(s+s_{B})}]
\end{equation}
for all $t\geq 0$ and $s\in\ro$, where $C_{B}$, $s_{B}$ and $N$ have the same dependence as in the case of (\ref{eq:uuweststatementnoise}). Moreover, $N=1$ 
in case the Jordan blocks of the matrices $A_{\ron,\pm}$ are trivial. In addition, if $E_{\ron,+}=E_{\ron,-}=\cn{m}$, then $N$ can be replaced by $d_{n,+}-1$, 
where $d_{\ron,+}$ is defined as in the statement of Lemma~\ref{lemma:windestnoise}.
If $E_{\ron,+}=E_{\ron,-}=\cn{m}$, then $\Phi_{\ron,\roT}$ is surjective. Finally, there are constants $C_{B}>0$ and $s_{B}\geq 0$ (depending only on $\eta_{B}$ 
and the coefficients of the equation (\ref{eq:thesystemRge})) such that 
\begin{equation}\label{eq:PhironroTestnoise}
\|\Phi_{\ron,\roT}(\chi_{+},\chi_{-})\|_{(s)}\leq C_{B}[\|\chi_{+}\|_{(s+s_{B})}+\|\chi_{-}\|_{(s+s_{B})}]
\end{equation}
for all $s\in\ro$ and all $(\chi_{+},\chi_{-})\in \mW_{\ron}^{+}\times \mW_{\ron}^{-}$. 
\end{lemma}
\begin{remark}
Dependence on the coefficients includes dependence on the Riemannian manifolds $(M_{r},g_{r})$, $r=1,\dots,R$; cf. 
Remark~\ref{remark:deponcoefincldeponRmfds}. 
\end{remark}
\begin{remark}\label{remark:homwaveonlygeneqspeasnoise}
Assume that (\ref{eq:thesystemRge}) is $C^{2}$-balanced with a geometric dominant noisy spatial direction, convergent dominant coefficients and a 
negligible shift vector field; cf. Definition~\ref{def:noisemainassumptions}. Assume, moreover, that $f=0$. Then the assumptions of the 
lemma are satisfied; cf. Remarks~\ref{remark:geometrictonongeometricnoise} and \ref{remark:strongbalanceneglshiftnoisegeometric}. How to calculate
$\eta_{A}$ etc. is clarified in Remark~\ref{remark:homwaveonlyprincsymbnoise}. 
\end{remark}
\begin{remark}
Remarks~\ref{remarks:Tronbronetcdeponeq} are equally relevant in the present setting. 
\end{remark}
\begin{remark}\label{remark:specupmwanoisedatinf}
It is also possible to specify functions $u_{\pm,\wa}$ and $u_{0,\wa}$ as in Remark~\ref{remark:specupmwanoise}, and then to construct a 
solution $u$ to (\ref{eq:thesystemRge}) (with $f=0$) such that $u=u_{\ron}$ and such that 
\begin{equation}\label{eq:uuappeststatementnoise}
\begin{split}
\mfe_{s}^{1/2}[u-u_{\wa}](t) \leq & C_{B}\ldr{t}^{N}e^{(\kappa_{\ron,+}-\eta_{B})t}\mfe_{s+s_{B}}^{1/2}[u](0)
\end{split}
\end{equation}
for all $t\geq 0$ and all $s\in\ro$, where the constants $C_{B}$, $s_{B}$ and $N$ have the same dependence as in the case of (\ref{eq:uuweststatementnoise});
\[
u_{\wa}(p,t):=e^{A_{\ron}t}u_{0,\wa}(p,t)
\]
in case (\ref{eq:thesystemRge}) has a dominant noisy spatial generalised direction; and
\begin{equation}\label{eq:uwadefnoiseremark}
u_{\wa}(p,t):=e^{A_{\ron,+}t}u_{+,\wa}(p,t)+e^{A_{\ron,-}t}u_{-,\wa}(p,t)
\end{equation}
in case (\ref{eq:thesystemRge}) has a dominant noisy spatial torus direction. Moreover, if the $u_{\pm,\wa}$ and $u_{0,\wa}$ are defined in 
terms of the $\zeta_{\pm}^{+}$ and $\zeta_{\pm}^{-}$ as described in Remark~\ref{remark:specupmwanoise}, then 
\begin{equation}\label{eq:contdeponasdatanoisewacase}
\mfe_{s}^{1/2}[u](0)\leq C_{B}\left(\textstyle{\sum}_{\indexnot\in\EFnindexset}\sum_{\pm}\ldr{\nu(\indexnot)}^{2(s+s_{B})}[|\zeta_{\pm}^{+}(\indexnot)|^{2}+
|\zeta_{\pm}^{-}(\indexnot)|^{2}]\right)^{1/2}
\end{equation}
for all $s\in\ro$, where $C_{B}$ and $s_{B}$ have the same dependence as in the case of (\ref{eq:uuweststatementnoise}). Finally, note that by 
combining (\ref{eq:uuappeststatementnoise}) and (\ref{eq:contdeponasdatanoisewacase}), the last factor on the right hand side of 
(\ref{eq:uuappeststatementnoise}) can be replaced by the last factor on the right hand side of (\ref{eq:contdeponasdatanoisewacase}). 
\end{remark}
\begin{proof}
Note that when (\ref{eq:thesystemRge}) has a dominant noisy spatial generalised direction, then 
\[
\mW_{\ron}^{+}=W_{\ron,+}^{+}(\bM,E_{\ron,+}),\ \ \
\mW_{\ron}^{-}=W_{\ron,-}^{+}(\bM,E_{\ron,-}),
\]
so that $\mW_{\ron}^{+}\times \mW_{\ron}^{-}\cong W_{\ron}(\bM,E_{\ron})$ in that case. For that reason, we below assume that (\ref{eq:thesystemRge}) 
has a dominant noisy spatial torus direction and leave it to the reader to verify (in the course of the argument), that the case of a 
dominant noisy spatial generalised direction can be considered to be a special case of this. 

\textbf{Constructing $u_{\pm,\row}$ from $\chi_{\pm}$.}
Let $\chi_{\pm}\in \mW_{\ron}^{\pm}$ be given. Then, by definition, $\chi_{+}$ corresponds to two solutions $u^{+}_{+,\row}$
and $u^{-}_{-,\row}$ to (\ref{eq:homtheeqnoise}); and $\chi_{-}$ corresponds to two solutions $u^{-}_{+,\row}$ and $u^{+}_{-,\row}$ to (\ref{eq:homtheeqnoise}).
Moreover, $u_{\pm,\row}^{+}$ and $u_{\pm,\row}^{-}$ have the properties stated in Lemma~\ref{lemma:asymposccaseitosoltowenoise}. Appealing to 
Lemma~\ref{lemma:asymposccasescalareqnoise} yields functions $u_{\pm,\wa}^{+}$ and $u_{\pm,\wa}^{-}$ characterised by $\zeta_{\pm}^{+}$ and $\zeta_{\pm}^{-}$ 
as described in Remark~\ref{remark:specupmwanoise}. Moreover,
\begin{equation}\label{eq:zetapmpmbdspasnoise}
\left(\textstyle{\sum}_{\indexnot\in\EFnindexset}\sum_{\pm}\ldr{\nu(\indexnot)}^{2s}[|\zeta_{\pm}^{+}(\indexnot)|^{2}+
|\zeta_{\pm}^{-}(\indexnot)|^{2}]\right)^{1/2}\leq C_{B}[\|\chi_{+}\|_{(s+s_{B})}+\|\chi_{-}\|_{(s+s_{B})}],
\end{equation}
where $C_{B}$ and $s_{B}$ have the same dependence as in the case of (\ref{eq:uuweststatementnoise}); cf. (\ref{eq:chipmmfezeroestnoise}). In addition, 
\begin{equation}\label{eq:mfesbdupmrowmupmwa}
\mfe_{s}^{1/2}[u_{\pm,\row}-u_{\pm,\wa}](t)\leq C_{B}e^{(\b_{\ron}/2-\eta_{B})t}[\|\chi_{+}\|_{(s+s_{B})}+\|\chi_{-}\|_{(s+s_{B})}]
\end{equation}
for all $t\geq 0$ and all $s\in\ro$, where $C_{B}$ and $s_{B}$ have the same dependence as in the case of (\ref{eq:uuweststatementnoise}), the $u_{\pm,\wa}$ 
are defined by (\ref{eq:upmwadef}) and the $u_{\pm,\row}$ are defined analogously; cf. the statement of Lemma~\ref{lemma:asymposccaseitosoltowenoise}. In 
order to obtain (\ref{eq:mfesbdupmrowmupmwa}), we appealed to (\ref{eq:mfediffhomestnoise}) and (\ref{eq:zetapmpmbdspasnoise}). Defining $u_{\row}$ by 
(\ref{eq:urowdefnoisetorus}) and $u_{\wa}$ by (\ref{eq:uwadefnoiseremark}), the estimate (\ref{eq:mfesbdupmrowmupmwa}) yields
\begin{equation}\label{eq:mfesbdurowmuwa}
\mfe_{s}^{1/2}[u_{\row}-u_{\wa}](t)\leq C_{B}\ldr{t}^{d_{\ron,+}-1}e^{(\kappa_{\ron,+}-\eta_{B})t}[\|\chi_{+}\|_{(s+s_{B})}+\|\chi_{-}\|_{(s+s_{B})}]
\end{equation}
for all $t\geq 0$ and all $s\in\ro$, where $C_{B}$ and $s_{B}$ have the same dependence as in the case of (\ref{eq:uuweststatementnoise}). Comparing this 
estimate with the statement of the lemma, it is of interest to construct a solution $u$ to (\ref{eq:thesystemRge}) (with $f=0$) and to estimate the 
difference between $u$ and $u_{\wa}$ by the right hand side of (\ref{eq:mfesumuwestassptorus}), as well as to estimate the 
initial data for $u$ in terms of $\chi_{\pm}$; cf. also Remark~\ref{remark:specupmwanoisedatinf}. 

\textbf{Asymptotics along a time sequence.} 
In order to specify the initial data for $u$, we wish to appeal to Lemma~\ref{lemma:asdataosccasenoise}. Note that the $\chi$ appearing in the statement of 
Lemma~\ref{lemma:asdataosccasenoise} corresponds to the $\psi_{\infty}$ appearing in (\ref{eq:zzdottkestnoisestmt}). However, since $\psi_{\infty}$ depends on 
$\indexnot$, we here sometimes write $\psi_{\infty}(\indexnot)$. Due to (\ref{eq:zzdottkestnoisestmt}), it is of interest to compare
\begin{equation}\label{eq:asymptnoisealongtkprelexpression}
\exp\left(i\int_{0}^{t_{k}}\sigma(\indexnot,t')\mfg(\indexnot,t')dt'\right)T_{\pre,k}^{-1}[D_{k}(\indexnot)]^{-1}e^{R_{\ron}(\indexnot)t_{k}}\psi_{\infty}(\indexnot)
\end{equation}
with the $\indexnot$'th Fourier coefficient of $u_{\wa}$, say $z_{\wa}(\indexnot,\cdot)$. Note that the first factor of the expression 
(\ref{eq:asymptnoisealongtkprelexpression}) can be written $e^{i\omega_{\rosh}(\indexnot,t_{k})}$, and that, due to (\ref{eq:TprekinvminusTpreinv}), 
\begin{equation}\label{eq:TprekinvreplbyTpreinvnoisespeas}
\left|e^{i\omega_{\rosh}(t_{k})}(T_{\pre,k}^{-1}-T_{\pre}^{-1})D_{k}^{-1}e^{R_{\ron}t_{k}}\psi_{\infty}\right|\leq C\ldr{t_{k}}^{d_{\ron,+}-1}e^{(\kappa_{\ron,+}-\eta_{B})t_{k}}|\psi_{\infty}|
\end{equation}
for all $k\geq 0$, where $C$ only depends on the coefficients of the equation (\ref{eq:thesystemRge}). On the other hand, if 
\begin{equation}\label{eq:psiinfdecompnoise}
\psi_{\infty}(\indexnot)=\left(\begin{array}{c} \psi_{+}(\indexnot) \\ \psi_{-}(\indexnot)\end{array}\right)
=\left(\begin{array}{c} \psi_{+}^{+}(\indexnot)+\psi_{+}^{-}(\indexnot) \\ \psi_{-}^{+}(\indexnot)+\psi_{-}^{-}(\indexnot)\end{array}\right),
\end{equation}
where $\psi_{\pm}^{+}(\indexnot)=0$ for $\indexnot\notin\EFnindexsetp$ and $\psi_{\pm}^{-}(\indexnot)=0$ for $\indexnot\notin\EFnindexsetm$, then
\begin{equation*}
\begin{split}
e^{i\omega_{\rosh}(\indexnot,t_{k})}[D_{k}(\indexnot)]^{-1}e^{R_{\ron}(\indexnot)t_{k}}\psi_{\infty}(\indexnot) = & 
\left(\begin{array}{c} e^{i(\omega_{\rosh}-\varphi_{\rotot})(\indexnot,t_{k})}e^{[-\a_{\infty}+\b_{\ron}\Id_{m}+\bX_{\ron}(\indexnot)]t_{k}/2}\psi_{+}(\indexnot) \\ 
e^{i(\omega_{\rosh}+\varphi_{\rotot})(\indexnot,t_{k})}e^{[-\a_{\infty}+\b_{\ron}\Id_{m}-\bX_{\ron}(\indexnot)]t_{k}/2}\psi_{-}(\indexnot)\end{array}\right)\\
 = & 
\left(\begin{array}{c} e^{i(\omega_{\rosh}-\varphi_{\rotot})(\indexnot,t_{k})}[e^{R_{\ron,+}^{-}t_{k}}\psi_{+}^{+}(\indexnot)+e^{R_{\ron,+}^{+}t_{k}}\psi_{+}^{-}(\indexnot)] \\ 
e^{i(\omega_{\rosh}+\varphi_{\rotot})(\indexnot,t_{k})}[e^{R_{\ron,+}^{+}t_{k}}\psi_{-}^{+}(\indexnot)+e^{R_{\ron,+}^{-}t_{k}}\psi_{-}^{-}(\indexnot)]\end{array}\right).
\end{split}
\end{equation*}
Thus
\begin{equation}\label{eq:expasalongatimesequence}
\begin{split}
 & e^{i\omega_{\rosh}(\indexnot,t_{k})}T_{\pre}^{-1}[D_{k}(\indexnot)]^{-1}e^{R_{\ron}(\indexnot)t_{k}}\psi_{\infty}(\indexnot)\\
 = & e^{i(\omega_{\rosh}-\varphi_{\rotot})(\indexnot,t_{k})}\left(\begin{array}{c} e^{R_{\ron,+}^{-}t_{k}}\psi_{+}^{+}(\indexnot)+e^{R_{\ron,+}^{+}t_{k}}\psi_{+}^{-}(\indexnot) \\ 
-ie^{R_{\ron,+}^{-}t_{k}}\psi_{+}^{+}(\indexnot)-ie^{R_{\ron,+}^{+}t_{k}}\psi_{+}^{-}(\indexnot)\end{array}\right)\\
 & +e^{i(\omega_{\rosh}+\varphi_{\rotot})(\indexnot,t_{k})}\left(\begin{array}{c} 
-ie^{R_{\ron,+}^{+}t_{k}}\psi_{-}^{+}(\indexnot)-ie^{R_{\ron,+}^{-}t_{k}}\psi_{-}^{-}(\indexnot) \\ 
e^{R_{\ron,+}^{+}t_{k}}\psi_{-}^{+}(\indexnot)+e^{R_{\ron,+}^{-}t_{k}}\psi_{-}^{-}(\indexnot)\end{array}\right).
\end{split}
\end{equation}

\textbf{Expanding $z_{\wa}$.} Note that
\begin{equation*}
\begin{split}
z_{\wa}(\indexnot,t) = & e^{A_{\ron,+}t}z_{+,\wa}(\indexnot,t)+e^{A_{\ron,-}t}z_{-,\wa}(\indexnot,t)\\
 = & e^{A_{\ron,+}t}[z_{+,\wa}^{+}(\indexnot,t)+z_{-,\wa}^{-}(\indexnot,t)]
+e^{A_{\ron,-}t}[z_{+,\wa}^{-}(\indexnot,t)+z_{-,\wa}^{+}(\indexnot,t)],
\end{split}
\end{equation*}
where the $z_{\pm,\wa}$, $z_{\pm,\wa}^{+}$ and $z_{\pm,\wa}^{-}$ have the properties stated in Remark~\ref{remark:specupmwanoise}. Let us estimate
\begin{equation}\label{eq:gzwazwadotfirstapp}
\begin{split}
 & \left|\left(\begin{array}{c} \mfg z_{\wa} \\ \dot{z}_{\wa}\end{array}\right)-
\left(\begin{array}{c} e^{A_{\ron,+}t}\mfg z_{+,\wa}^{+} \\ e^{A_{\ron,+}t}\dot{z}_{+,\wa}^{+}\end{array}\right)
-\left(\begin{array}{c} e^{A_{\ron,+}t}\mfg z_{-,\wa}^{-} \\ e^{A_{\ron,+}t}\dot{z}_{-,\wa}^{-}\end{array}\right)
 -\left(\begin{array}{c} e^{A_{\ron,-}t}\mfg z_{+,\wa}^{-} \\ e^{A_{\ron,-}t}\dot{z}_{+,\wa}^{-}\end{array}\right)\right.\\
 & \left.\phantom{\|}-\left(\begin{array}{c} e^{A_{\ron,-}t}\mfg z_{-,\wa}^{+} \\ e^{A_{\ron,-}t}\dot{z}_{-,\wa}^{+}\end{array}\right)\right|
\leq C_{B}\ldr{t}^{d_{\ron,+}-1}e^{(\kappa_{\ron,+}-\eta_{B})t}\ldr{\nu}^{s_{E}}\sum_{\pm}(|\zeta_{\pm}^{+}|+|\zeta_{\pm}^{-}|)
\end{split}
\end{equation}
for all $t\geq 0$, where $C_{B}$ has the same dependence as in the case of (\ref{eq:uuweststatementnoise}) and we omitted the
arguments $\indexnot$ and $t$ for the sake of brevity. Moreover, we appealed to (\ref{eq:mfglowbdallfuttimenoise}), which holds
in the present setting. By arguments similar to those presented in connection with (\ref{eq:Zdotcompnoise}), we can replace $\dot{z}_{+,\wa}^{\pm}$ and 
$\dot{z}_{-,\wa}^{\pm}$ appearing in (\ref{eq:gzwazwadotfirstapp}) by $-i\mfg z_{+,\wa}^{\pm}$ and $i\mfg z_{-,\wa}^{\pm}$ respectively; cf. 
(\ref{eq:zpwapmdef}) and (\ref{eq:zmwapmdef}). One particular expression it is of interest to consider is thus
\[
\left(\begin{array}{c} e^{A_{\ron,+}t}\mfg z_{+,\wa}^{+} \\ -ie^{A_{\ron,+}t}\mfg z_{+,\wa}^{+}\end{array}\right)
=e^{i(\omega_{\rosh}-\varphi_{\rotot})}
\left(\begin{array}{c} e^{R_{\ron,+}^{-}t}\zeta_{+}^{+} \\ -ie^{R_{\ron,+}^{-}t}\zeta_{+}^{+}\end{array}\right).
\]
Combining this observation with similar reformulations of the expressions appearing on the left hand side of (\ref{eq:gzwazwadotfirstapp}) yields
\begin{equation}\label{eq:gzwazwadotfirstappsv}
\begin{split}
 & \left|\left(\begin{array}{c} \mfg z_{\wa} \\ \dot{z}_{\wa}\end{array}\right)-
e^{i(\omega_{\rosh}-\varphi_{\rotot})}\left(\begin{array}{c} e^{R_{\ron,+}^{-}t}\zeta_{+}^{+}+e^{R_{\ron,+}^{+}t}\zeta_{+}^{-} \\ 
-ie^{R_{\ron,+}^{-}t}\zeta_{+}^{+}-ie^{R_{\ron,+}^{+}t}\zeta_{+}^{-}\end{array}\right)\right.\\
 & \left.\phantom{\|}
-e^{i(\omega_{\rosh}+\varphi_{\rotot})}\left(\begin{array}{c} e^{R_{\ron,+}^{-}t}\zeta_{-}^{-}+e^{R_{\ron,+}^{+}t}\zeta_{-}^{+} \\ 
ie^{R_{\ron,+}^{-}t}\zeta_{-}^{-}+ie^{R_{\ron,+}^{+}t}\zeta_{-}^{+}\end{array}\right)\right|\\ 
\leq &  C_{B}\ldr{t}^{d_{\ron,+}-1}e^{(\kappa_{\ron,+}-\eta_{B})t}\ldr{\nu}^{s_{E}}\textstyle{\sum}_{\pm}(|\zeta_{\pm}^{+}|+|\zeta_{\pm}^{-}|),
\end{split}
\end{equation}
for all $t\geq 0$, where $C_{B}$ has the same dependence as in the case of (\ref{eq:uuweststatementnoise}).

\textbf{Comparing the different perspectives.} Comparing (\ref{eq:expasalongatimesequence}) and (\ref{eq:gzwazwadotfirstappsv}), it is clear that we 
want the following relations to hold:
\begin{equation}\label{eq:psipmpmitozetapmpmdefnoise}
\psi_{+}^{\pm}(\indexnot)=\zeta_{+}^{\pm}(\indexnot),\ \ \
\psi_{-}^{\pm}(\indexnot)=i\zeta_{-}^{\pm}(\indexnot).
\end{equation}
For a given $\indexnot\in\EFnindexset$, it is thus clear how to specify $\psi_{\infty}$ in terms of the $\zeta_{+}^{\pm}$ and $\zeta_{-}^{\pm}$. It is 
also important to note that, due to the properties of the right hand sides of (\ref{eq:psipmpmitozetapmpmdefnoise}), $\psi_{\infty}(\indexnot)$ is 
an element of the first generalised eigenspace of the $\eta_{B}$, $R_{\ron}(\indexnot)$-decomposition of $\cn{2m}$. Next, we can
then appeal to Lemma~\ref{lemma:asdataosccasenoise} with $\chi$ replaced by $\psi_{\infty}$. This yields initial data for (\ref{eq:fourierthesystemRge})
(with vanishing right hand side) at $T_{\ron}$. In fact, (\ref{eq:Psiinfchitoidnoise}) and (\ref{eq:Psiinfchitoidnoiseest}) hold. Thus
\begin{equation}\label{eq:dotzmfgzTronindnotnoise}
|\dot{z}(\indexnot,T_{\ron})|+\mfg(\indexnot,T_{\ron})|z(\indexnot,T_{\ron})|\leq C|e^{R_{\ron}(\indexnot)T_{\ron}}\psi_{\infty}(\indexnot)|
\leq C\ldr{T_{\ron}}^{d_{\ron,+}-1}e^{\kappa_{\ron,+}T_{\ron}}|\psi_{\infty}(\indexnot)|,
\end{equation}
where $C$ only depends on the coefficients of the equation (\ref{eq:thesystemRge}). Next, we 
need to derive an estimate for the energy of the mode at $t=0$. To this end, we combine (\ref{eq:dotzmfgzTronindnotnoise}) and 
Lemma~\ref{lemma:roughenestbalsetting} in order to obtain
\[
\me^{1/2}(\indexnot,0)\leq Ce^{(\kappa_{\ron,+}+1+\eta_{\robal})T_{\ron}}|\psi_{\infty}(\indexnot)|,
\]
where $C$ only depends on the coefficients of the 
equation (\ref{eq:thesystemRge}). Combining this estimate with an estimate for $T_{\ron}$ of the form (\ref{eq:Tronuppbdnoise}) yields 
\begin{equation}\label{eq:enindmodetimezeronoise}
\me^{1/2}(\indexnot,0)\leq C\ldr{\nu(\indexnot)}^{s_{a}}|\psi_{\infty}(\indexnot)|,
\end{equation}
where $C$ and $s_{a}$ only depend on the coefficients of the equation (\ref{eq:thesystemRge}). 

\textbf{Defining $\Phi_{\ron,\roT}$ and $\Phi_{\ron,\rog}$.} For reasons already mentioned, we can focus on the case of a dominant noisy spatial torus direction
here. Given $\chi_{\pm}\in \mW_{\ron}^{\pm}$, we calculate $\zeta_{\pm}^{+}$ and $\zeta_{\pm}^{-}$ as described above. Note also that (\ref{eq:zetapmpmbdspasnoise}) 
holds. Next, $\psi_{\infty}$ is defined by (\ref{eq:psiinfdecompnoise}) and (\ref{eq:psipmpmitozetapmpmdefnoise}). For a given $\indexnot\in\EFnindexset$, 
this yields $z(\indexnot,T_{\ron})$ and $\dot{z}(\indexnot,T_{\ron})$. Solving (\ref{eq:fourierthesystemRge}) backwards, this yields initial data for the 
$\indexnot$'th mode at $t=0$. Combining these maps for the individual modes yields the maps $\Phi_{\ron,\roT}$ and $\Phi_{\ron,\rog}$. Moreover, combining 
(\ref{eq:zetapmpmbdspasnoise}) and (\ref{eq:enindmodetimezeronoise}) yields (\ref{eq:Phironrogestnoise}) and (\ref{eq:PhironroTestnoise}). The injectivity
of the maps follows from the injectivity of the maps of which they are composed. Similarly, if $E_{\ron,+}=E_{\ron,-}=\cn{m}$, then the surjectivity statement
of Lemma~\ref{lemma:asdataosccasenoise} yields the desired surjectivity for fixed $\indexnot\in\EFnindexset$. 

\textbf{Asymptotics.} What remains to be verified is that (\ref{eq:mfesumuwestasspgen}) and (\ref{eq:mfesumuwestassptorus}) hold, and that surjectivity
holds if $E_{\ron,+}=E_{\ron,-}=\cn{m}$. Let $\chi_{\pm}\in \mW_{\ron}^{\pm}$ be given. Then we obtain initial data for (\ref{eq:thesystemRge}) (with $f=0$). 
Solving (\ref{eq:thesystemRge}) yields a solution $u$ such that $u=u_{\ron}$ (by construction). Moreover, if $z$ denotes the Fourier 
coefficient of $u$, then (\ref{eq:zzdottkestnoisestmt}) holds, where $\psi_{\infty}(\indexnot)$ is defined by (\ref{eq:psiinfdecompnoise}) and 
(\ref{eq:psipmpmitozetapmpmdefnoise}). Keeping (\ref{eq:TprekinvreplbyTpreinvnoisespeas}) in mind, the estimate
\begin{equation}\label{eq:unbarreduniqueseqest}
\left|\left(\begin{array}{c} \mfg(\indexnot,t_{k}) z(\indexnot,t_{k}) \\ \dot{z}(\indexnot,t_{k})\end{array}\right)
-e^{i\omega_{\rosh}(\indexnot,t_{k})}T_{\pre}^{-1}[D_{k}(\indexnot)]^{-1}e^{R_{\ron}(\indexnot)t_{k}}\psi_{\infty}(\indexnot)
\right|\leq C\ldr{t_{k}}^{N}e^{(\kappa_{\ron,+}-\eta_{B})t_{k}}
\end{equation}
holds for all $k\geq 0$, where $C$ is allowed to depend on the solution, $\indexnot$ etc. In what follows, it is important to keep in mind that 
$\psi_{\infty}(\indexnot)$ belongs to the first generalised eigenspace of the $\eta_{B}$, $R_{\ron}(\indexnot)$-decomposition of $\cn{2m}$.

Next, appealing to Remark~\ref{remark:specupmwanoise},
there are $\bze_{\pm}^{+}(\indexnot)$ and $\bze_{\pm}^{-}(\indexnot)$, $\indexnot\in\EFindexset$, which are uniquely associated with $u$ and satisfy
the conditions stated in Remark~\ref{remark:specupmwanoise}. Moreover, we obtain a function $\bu_{\wa}$ with the properties stated in 
Remark~\ref{remark:specupmwanoise}. In particular, (\ref{eq:uuweststatementnoisewa}) yields
\[
\mfe_{s}^{1/2}[u-\bu_{\wa}](t) \leq C_{B}\ldr{t}^{N}e^{(\kappa_{\ron,+}-\eta_{B})t}\mfe_{s+s_{B}}^{1/2}[u](0)
\]
for all $t\geq 0$ and all $s\in\ro$, so that, letting $\bz_{\wa}$ denote the Fourier coefficients of $\bu_{\wa}$,
\[
\left|\left(\begin{array}{c} \mfg(\indexnot,t) z(\indexnot,t) \\ \dot{z}(\indexnot,t)\end{array}\right)
-\left(\begin{array}{c} \mfg(\indexnot,t) \bz_{\wa}(\indexnot,t) \\ \dot{\bz}_{\wa}(\indexnot,t)\end{array}\right)\right|\leq C\ldr{t}^{N}e^{(\kappa_{\ron,+}-\eta_{B})t}
\]
for all $t\geq 0$, where $C$ is allowed to depend on the solution, $\indexnot$ etc. Next we can repeat, verbatim, the argument leading
up to (\ref{eq:gzwazwadotfirstappsv}). Thus (\ref{eq:gzwazwadotfirstappsv}) holds with $z_{\wa}$ replaced by $\bz_{\wa}$ and $\zeta_{\pm}^{+}$ and $\zeta_{\pm}^{-}$
replaced by the analogous barred quantities. Introducing $\bpsi_{\infty}(\indexnot)$ in analogy with (\ref{eq:psipmpmitozetapmpmdefnoise}) and 
(\ref{eq:psiinfdecompnoise}) (starting with $\bze_{\pm}^{+}(\indexnot)$ and $\bze_{\pm}^{-}(\indexnot)$), it follows that 
\begin{equation}\label{eq:barreduniqueseqest}
\left|\left(\begin{array}{c} \mfg(\indexnot,t_{k}) z(\indexnot,t_{k}) \\ \dot{z}(\indexnot,t_{k})\end{array}\right)
-e^{i\omega_{\rosh}(\indexnot,t_{k})}T_{\pre}^{-1}[D_{k}(\indexnot)]^{-1}e^{R_{\ron}(\indexnot)t_{k}}\bpsi_{\infty}(\indexnot)
\right|\leq C\ldr{t_{k}}^{N}e^{(\kappa_{\ron,+}-\eta_{B})t_{k}}
\end{equation}
for all $k\geq 0$, where $C$ is allowed to depend on the solution, $\indexnot$ etc., and we have appealed to the barred analogues of 
(\ref{eq:expasalongatimesequence}) and (\ref{eq:gzwazwadotfirstappsv}). Moreover, due to the properties of the $\bze_{\pm}^{+}(\indexnot)$ and 
$\bze_{\pm}^{-}(\indexnot)$, $\bpsi_{\infty}(\indexnot)$ belongs to the first generalised eigenspace of the $\eta_{B}$, $R_{\ron}(\indexnot)$-decomposition 
of $\cn{2m}$. Combining (\ref{eq:unbarreduniqueseqest}) and (\ref{eq:barreduniqueseqest}) with the fact that $\bpsi_{\infty}(\indexnot)$ and 
$\psi_{\infty}(\indexnot)$ belong to the first generalised eigenspace of the $\eta_{B}$, $R_{\ron}(\indexnot)$-decomposition of $\cn{2m}$ yields the 
conclusion that $\bpsi_{\infty}(\indexnot)=\psi_{\infty}(\indexnot)$. In particular, 
\begin{equation}\label{eq:zetapmpmeqbzetapmpm}
\bze_{\pm}^{+}(\indexnot)=\zeta_{\pm}^{+}(\indexnot),\ \ \
\bze_{\pm}^{-}(\indexnot)=\zeta_{\pm}^{-}(\indexnot)
\end{equation}
for all $\indexnot\in\EFindexset$. Next, we wish to use this information to relate $u_{\row}$, appearing in the statement of the lemma, with $\bu_{\wa}$. 
First of all, $u_{\row}$ is determined by $u_{\pm,\row}^{+}$ and $u_{\pm,\row}^{-}$ (the solutions to (\ref{eq:homtheeqnoise}) which are uniquely
determined by $\chi_{\pm}$). On the other hand, $u_{\pm,\row}^{+}$ and $u_{\pm,\row}^{-}$ correspond to functions $u_{\pm,\wa}^{+}$ and $u_{\pm,\wa}^{-}$
respectively, as described in Lemma~\ref{lemma:asymposccasescalareqnoise}. Moreover, the latter functions correspond to the $\zeta_{\pm}^{+}(\indexnot)$ and
$\zeta_{\pm}^{-}(\indexnot)$. Finally, each of these correspondences arise from a bijection. Turning to $\bu_{\wa}$, it can be written
\[
\bu_{\wa}=e^{A_{\ron,+}t}\bu_{+,\wa}+e^{A_{\ron,-}t}\bu_{-,\wa},
\]
where 
\[
\bu_{+,\wa}:=\bu_{+,\wa}^{+}+\bu_{-,\wa}^{-},\ \ \
\bu_{-,\wa}:=\bu_{+,\wa}^{-}+\bu_{-,\wa}^{+}
\]
and $\bu_{\tau,\wa}^{\upsilon}$, $\upsilon,\tau\in\{+,-\}$, are the functions uniquely determined by $\bze_{\pm}^{+}(\indexnot)$ and $\bze_{\pm}^{-}(\indexnot)$.
Due to (\ref{eq:zetapmpmeqbzetapmpm}) it follows that 
\[
u_{+,\wa}:=u_{+,\wa}^{+}+u_{-,\wa}^{-}=\bu_{+,\wa},\ \ \
u_{-,\wa}:=u_{-,\wa}^{+}+u_{+,\wa}^{-}=\bu_{-,\wa}.
\]
In particular, 
\[
u_{\wa}:=e^{A_{\ron,+}t}u_{+,\wa}+e^{A_{\ron,-}t}u_{-,\wa}=\bu_{\wa}.
\]
Combining (\ref{eq:mfesbdurowmuwa}), (\ref{eq:uuweststatementnoisewa}) (which we know to hold with $u_{\wa}$ replaced by $\bu_{\wa}$ and $f$ replaced
by $0$) and the fact that $u_{\wa}=\bu_{\wa}$ yields 
\begin{equation*}
\begin{split}
\mfe_{s}^{1/2}[u-u_{\row}](t) \leq & \mfe_{s}^{1/2}[u-u_{\wa}](t)+\mfe_{s}^{1/2}[u_{\wa}-u_{\row}](t)\\
 \leq & C_{B}\ldr{t}^{d_{\ron,+}-1}e^{(\kappa_{\ron,+}-\eta_{B})t}[\|\chi_{+}\|_{(s+s_{B})}+\|\chi_{-}\|_{(s+s_{B})}]\\
 & +C_{B}\ldr{t}^{N}e^{(\kappa_{\ron,+}-\eta_{B})t}\mfe_{s+s_{B}}^{1/2}[u](0)
\end{split}
\end{equation*}
for all $t\geq 0$ and all $s\in\ro$. Here $C_{B}>0$ and $s_{B}\geq 0$ only depend on $\eta_{B}$ and the coefficients of the equation (\ref{eq:thesystemRge}),
and $N$ only depends on $m$.
Moreover, $N=1$ in case the Jordan blocks of the matrices $A_{\ron,\pm}$ are trivial. In addition, if $E_{\ron,+}=E_{\ron,-}=\cn{m}$, then $N$ can be replaced by 
$d_{n,+}-1$. This is not exactly the estimate we want. However, combining it with (\ref{eq:PhironroTestnoise}) (which we already demonstrated to hold), 
it yields (\ref{eq:mfesumuwestasspgen}) and (\ref{eq:mfesumuwestassptorus}).

\textbf{Surjectivity.} In order to demonstrate surjectivity in case $E_{\ron,+}=E_{\ron,-}=\cn{m}$, let $u$ be a smooth solution to (\ref{eq:thesystemRge}) such
that $u=u_{\ron}$. Appealing to Lemma~\ref{lemma:asymposccaseitosoltowenoise} yields solutions $u_{\tau,\row}^{\upsilon}$, $\upsilon,\tau\in\{+,-\}$, and $u_{\pm,\row}$ 
to (\ref{eq:homtheeqnoise}) such that the conclusions of Lemma~\ref{lemma:asymposccaseitosoltowenoise} hold. Moreover, if $u_{\row}$ is defined by 
(\ref{eq:urowdefnoisetorus}), then 
\begin{equation}\label{eq:umurowsurjproof}
\mfe_{s}^{1/2}[u-u_{\row}](t) \leq C\ldr{t}^{N}e^{(\kappa_{\ron,+}-\eta_{B})t}
\end{equation}
for all $t\geq 0$ and $s\in\ro$, where $C$ is allowed to depend on the solution, $s$, etc. The initial data for $u_{\tau,\row}^{\upsilon}$ correspond to functions 
$\chi_{\pm}$ with properties stated in the present lemma. Letting $\bu$ be the solution corresponding to initial data $\Phi_{\ron,\roT}(\chi_{+},\chi_{-})$ at $t=0$, 
the earlier parts of the present proof guarantee that (\ref{eq:umurowsurjproof}) holds if we replace $u$ by $\bu$. Letting $v=u-\bu$ and $z$ denote the 
Fourier coefficients of $v$, we conclude that for each $\indexnot\in\EFnindexset$, 
\begin{equation}\label{eq:gztkzdottkdeglimitnoise}
\left|\left(\begin{array}{c} \mfg(\indexnot,t_{k}) z(\indexnot,t_{k}) \\ \dot{z}(\indexnot,t_{k})\end{array}\right)
\right|\leq C\ldr{t_{k}}^{N}e^{(\kappa_{\ron,+}-\eta_{B})t_{k}}
\end{equation}
for all $k\geq 0$. Assume now that the initial data for $z(\indexnot,\cdot)$ at $T_{\ron}$ are not zero. Then, due to the surjectivity of the map 
$\Psi_{\infty}$ introduced in Lemma~\ref{lemma:asdataosccasenoise}, there is a $0\neq \chi\in\cn{2m}$ such that (\ref{eq:Psiinfchitoidnoise}) holds. 
Moreover, Lemma~\ref{lemma:asdataosccasenoise} implies that (\ref{eq:zzdottkestnoisestmt}) holds with $\psi_{\infty}$ replaced by $\chi$. However, 
(\ref{eq:zzdottkestnoisestmt}) (with $\psi_{\infty}\neq 0$) is irreconcilable with (\ref{eq:gztkzdottkdeglimitnoise}). Thus $v=0$ and $u=\bu$. 
Thus the initial data of $u$ are in the image of $\Phi_{\ron,\roT}$. Since $u$ is an arbitrary smooth solution, the surjectivity of $\Phi_{\ron,\roT}$ follows.
\end{proof}

\part{Non-degenerate, diagonally dominated, convergent and 
balanced equations}\label{part:nondegcabeq}

\chapter[Convergent and balanced equations]{Equations with diagonally dominated, convergent and 
balanced coefficients}\label{chapter:diagdomconvbal}

\section{Introduction}

In Part~\ref{part:averaging}, we consider Fourier coefficients of solutions to the equation (\ref{eq:thesystemRge}) under circumstances
in which the behaviour is oscillatory. In particular, we derive formulae for how the solution evolves over one period of the oscillations;
cf., in particular, (\ref{eq:wprekpertransfin})--(\ref{eq:XiIappdefstfin}). This analysis is the basis for a further study of
the evolution of solutions. However, in order to make progress, we need to make additional assumptions. To be more specific, consider
the matrices $R^{ab}_{\pre,k}$ appearing on the right hand side of (\ref{eq:Xikpreappdefstfin}). These matrices are of central importance, since
they determine the leading order behaviour of the evolution. Returning to (\ref{eq:Rabprekdef}), it is clear that we need to study
$\a/\mfg$, $\dot{\ell}/\mfg$ and $X/\mfg$ in greater detail. Naively, we would like to think of the matrix $\Xi^{k,\pm}_{\fin,\app}$
as follows:
\begin{equation}\label{eq:Xikpmfinappapproxexpr}
\begin{split}
\Xi^{k,\pm}_{\fin,\app} = & \Id_{2m} \pm\frac{1}{2}\frac{2\pi}{\mfg(t_{k})}
\left(\begin{array}{cc} -\a(t_{k})+\dot{\ell}(t_{k})\Id_{m}+X(t_{k})
& 0 \\  0 & -\a(t_{k})+\dot{\ell}(t_{k})\Id_{m}-X(t_{k})\end{array}
\right)\\
 \approx & \Id_{2m} \pm\frac{1}{2}\int_{t_{k}}^{t_{k+1}}
\left(\begin{array}{cc} -\a(t)+\dot{\ell}(t)\Id_{m}+X(t)
& 0 \\  0 & -\a(t)+\dot{\ell}(t)\Id_{m}-X(t)\end{array}\right)dt\\
 \approx & \exp\left[\pm\frac{1}{2}\int_{t_{k}}^{t_{k+1}}
\left(\begin{array}{cc} -\a(t)+\dot{\ell}(t)\Id_{m}+X(t)
& 0 \\  0 & -\a(t)+\dot{\ell}(t)\Id_{m}-X(t)\end{array}\right)dt\right].
\end{split}
\end{equation}
The justification of the first approximate equality is based on (\ref{eq:tbmtagestge}) and the assumption that $\a$, $\dot{\ell}$ and 
$X$ do not vary too much over one period of the oscillations. The second approximate equality is based on the assumption that the
integral appearing in the final exponential expression is small. The advantage of the expression on the far right hand side of  
(\ref{eq:Xikpmfinappapproxexpr}) is that it behaves well under multiplication (at least in intervals in which $\a$ and $X$ are essentially
constant). It is thus tempting to hope that the product of the
$\Xi^{k,\pm}_{\fin,\app}$ can be approximated by the far right hand side of (\ref{eq:Xikpmfinappapproxexpr}), where the time interval
$[t_{k},t_{k+1}]$ is replaced by the union of the corresponding intervals. However, it is very optimistic to hope for this in general. 
The reason for this is that the resulting matrix can be expected to contract some directions more than others. Taking a vector corresponding
to a direction which is maximally contracted, the small errors associated with the approximations in (\ref{eq:Xikpmfinappapproxexpr})
can be expected to generate a small component in the direction which is maximally expanded. The contribution of this small error may then 
(over long periods of time) yield a bigger contribution to the end result than the original vector. This situation is reminiscent of the 
one considered already in Chapter~\ref{chapter:roughanalysisODEregion}. In fact, consider (\ref{eq:ODEregmod}) with $F=0$, where 
$A$ is constant and $A_{\rem}$ satisfies (\ref{eq:Aremestintro}). Naively, one might then expect $\exp(At)$ to be a good approximation
of the evolution. However, this expectation is not reasonable if the real eigenvalue spread of $A$, $\Rsp A$, exceeds the parameter
$\b_{\rem}>0$ appearing in (\ref{eq:Aremestintro}). To summarise the discussion, (\ref{eq:Xikpmfinappapproxexpr}) may give a 
rough idea of how solutions evolve, even when replacing the interval $[t_{k},t_{k+1}]$ by the union of several periods. However, as a guide
to the asymptotic behaviour, it should be treated with care. 

\textbf{Diagonally dominated, non-degenerate, convergent and balanced equations.}
Due to the above discussion, it is clear that we wish to eliminate the risk of a substantial error arising due to a vector corresponding 
to a contracting direction being rotated (by a small error) in such a way that it picks up a component corresponding to an expanding 
direction. In order to avoid problems of this type, it is convenient to keep track of which directions correspond to expansion and which
directions correspond to contraction. However, we would preferably like the matrices giving rise to these directions to (in the limit) be 
finite in number; otherwise error terms arising due to an unbounded number of algebraic manipulations could be significant. One way to 
fulfil these conditions is to consider situations in which the diagonal components of the metric 
are dominant and such that they converge after suitable rescaling by an exponential function. We also need to make corresponding assumptions
concerning $X^{l}$, $\a$ and $\zeta$. In order to develop some intuition, consider the model equation
\begin{equation}\label{eq:modeleqdiagdomconvbal}
u_{tt}-\textstyle{\sum}_{j=1}^{d}g^{jj}_{\infty}e^{2\b_{j}t}\d_{j}^{2}u+\sum_{j=1}^{d}X^{j}_{\infty}e^{\b_{j}t}\d_{j}u+\a_{\infty}u_{t}+\zeta_{\infty}u=0,
\end{equation}
where $0<g^{jj}_{\infty}\in\ro$, $X^{j}_{\infty},\a_{\infty},\zeta_{\infty}\in\Mn{m}{\co}$, $j=1,\dots,d$. Finally, we assume that 
$\b_{1}<\b_{2}<\cdots<\b_{d}$; this corresponds to the condition of non-degeneracy. For an equation of the form 
(\ref{eq:modeleqdiagdomconvbal}), $\mfg$ satisfies the relation
\[
\mfg^{2}(\indexnot,t)=\textstyle{\sum}_{j=1}^{d}g^{jj}_{\infty}e^{2\b_{j}t}n_{j}^{2}. 
\]
For a given $\indexnot=n\in\zn{d}$, it is then clear that the time interval $[0,\infty)$ can be divided into subintervals, which we refer to as 
\textit{eras},
in which one of the functions $g^{jj}_{\infty}e^{2\b_{j}t}n_{j}^{2}$, $j=1,\dots,d$, and $1$ dominates (if all the $\b_{j}$ are strictly negative, the function
$1$ will eventually dominate; we refer to the corresponding interval as the \textit{ODE era}). Let $I_{j}$ be an interval in which 
$g^{jj}_{\infty}e^{2\b_{j}t}n_{j}^{2}$ dominates. In $I_{j}$, $\mfg$ can then effectively be replaced by $(g^{jj}_{\infty})^{1/2}e^{\b_{j}t}|n_{j}|$ (the exact meaning 
of this statement, as well as a detailed justification of it, is provided in Lemmas~\ref{lemma:diagapprox} and \ref{lemma:Ijanalysis} below). Moreover, 
$X$ can be replaced by
\[
\frac{in_{j}X^{j}_{\infty}e^{\b_{j}t}}{(g^{jj}_{\infty})^{1/2}|n_{j}|e^{\b_{j}t}}=\pm i(g^{jj}_{\infty})^{-1/2}X^{j}_{\infty},
\]
where the sign equals the sign of $n_{j}$. Finally, $\dot{\ell}$ is effectively $\b_{j}$. Thus, for $t_{k}\in I_{j}$,
\begin{equation}\label{eq:Rabprekapprox}
R^{ab}_{\pre,k}\approx \frac{1}{(g^{jj}_{\infty})^{1/2}|n_{j}|e^{\b_{j}t_{k}}}[(-1)^{a}\a_{\infty}+\b_{j}\Id_{m}\pm i(-1)^{b}(g^{jj}_{\infty})^{-1/2}X^{j}_{\infty}],
\end{equation}
where the sign equals the sign of $n_{j}$. Note that the first factor roughly speaking corresponds to integration over the period of the oscillations
under consideration. The essential structure appearing on the right hand side of (\ref{eq:Rabprekapprox}) is thus captured by one element of
a finite set of matrices. Keeping track of expanding and contracting directions in $I_{j}$ is therefore straightforward. 
To summarise: for a fixed $n$, there are finitely many eras (never more than $d+1$); for each era (except for the ODE era, if it exists), 
the dominant part of the evolution over one period of the oscillations is determined by one matrix (chosen from a finite set of matrices); and the 
evolution in the ODE era (if it exists) can be analysed as described in Chapter~\ref{chapter:roughanalysisODEregion}. 

\textbf{Specifying the class of equations of interest.} With the above intuition in mind, we introduce the class of equations of interest in the 
present part of these notes. We do not demand that the equations be of the form (\ref{eq:modeleqdiagdomconvbal}). However, when $R=0$, we do 
demand that this be the dominant part asymptotically. On the other hand, we do not, at this stage, insist on non-degeneracy. For
this reason, the material of the present chapter also serves as a basis for the study of degenerate equations (a topic we do not address
in these notes). Moreover, as opposed to (\ref{eq:modeleqdiagdomconvbal}), we wish to allow $R>0$; cf. 
(\ref{eq:thesystemRge}). Giving a precise definition of the corresponding class of equations is the main purpose 
of Section~\ref{section:diagdomconvbal}. However, we also verify that this class is such that the results of Part~\ref{part:averaging} 
apply; cf. Lemma~\ref{lemma:consconvcoeff}. 

\textbf{Simplifying $R^{ab}_{\pre,k}$.} Our next goal is to simplify the expressions $\a/\mfg$, $\dot{\ell}/\mfg$ and $X/\mfg$ appearing on the 
right hand side of (\ref{eq:Rabprekdef}). We do this in two steps. In Section~\ref{section:simplmacoeff}, in particular in Lemma~\ref{lemma:diagapprox},
we estimate the error associated with ignoring the off-diagonal components of the matrix $g^{jl}$. In order to progress further, we 
need to divide the time interval $[0,\infty)$ into eras in which particular frequencies and exponential behaviour dominate. Defining such a 
division and deriving the associated simplified expressions for $\a/\mfg$, $\dot{\ell}/\mfg$ and $X/\mfg$ is the subject of 
Section~\ref{section:timeintfreeras}, in particular Lemma~\ref{lemma:Ijanalysis}. 

\textbf{Estimating sums.} The final observation in the present chapter concerns the estimation of sums that appear when calculating
matrix products of the type appearing in (\ref{eq:differenceeqrepres}). In practice, our ambition is not to calculate the product, but
rather the leading order behaviour. In order to be able to do so, we first have to divide the interval $[0,\infty)$ into frequency eras, 
as described above. In a given frequency era, we can then approximate $R^{ab}_{\pre,k}$ in such a way that the dominant behaviour is 
determined by one fixed matrix. Transforming the iteration so that this matrix is in Jordan normal form, the problem of estimating the 
evolution of solutions can then be reduced to estimating certain sums. In Section~\ref{section:estsumsfreera} we demonstrate how to estimate
the types of sums that appear; cf. Lemma~\ref{lemma:compsumswithintegrals}.

\section[Diagonally dominated, convergent and balanced equations]{Equations with diagonally dominant, convergent and balanced 
coefficients}\label{section:diagdomconvbal}

Let us return to the equation (\ref{eq:thesystemRge}) and impose the conditions of interest in the present part of these notes. 
The motivation for introducing these conditions is described in the previous section. 

\begin{definition}\label{def:convergeq}
Consider (\ref{eq:thesystemRge}). Assume the associated metric to be such that $(M,g)$ is a canonical separable cosmological model 
manifold. The coefficients $\a$, $X^{j}$ and $\zeta$, $j=1,\dots,d$, are referred to as the \textit{main coefficients} 
\index{Main!coefficients}%
\index{Coefficients!main}%
of the equation (\ref{eq:thesystemRge}). The coefficients $g^{0l}$, $g^{jj}$ and $g^{jl}$ ($j\neq l$), $j,l=1,\dots,d$, are referred to 
as the \textit{shift coefficients}, the \textit{diagonal coefficients} 
\index{Diagonal!coefficients}%
\index{Coefficients!diagonal}%
and the \textit{off-diagonal coefficients} 
\index{Off-diagonal!coefficients}%
\index{Coefficients!off-diagonal}%
of the equation (\ref{eq:thesystemRge}) respectively. The functions $a^{-2}_{r}$, $r=1,\dots,R$, are referred to as the \textit{auxiliary
diagonal coefficients}. 
\index{Auxiliary diagonal!coefficients}%
\index{Coefficients!auxiliary diagonal}%
The diagonal coefficients are said to be $C_{\exp}^{0}$-\textit{convergent and $C_{\exp}^{2}$-bounded} 
if there are real constants $\b_{j}$, strictly positive real numbers $g_{\infty}^{jj}$, $j=1,\dots,d$, and constants $\eta_{\rod},C_{\rod}>0$ 
such that 
\begin{equation}\label{eq:diagcoeffconv}
e^{\eta_{\rod}t}|e^{-2\b_{j}t}g^{jj}(t)-g_{\infty}^{jj}|+|e^{-2\b_{j}t}\dot{g}^{jj}(t)|+|e^{-2\b_{j}t}\ddot{g}^{jj}(t)|\leq C_{\rod}
\end{equation}
for all $t\geq 0$ and $j=1,\dots,d$. If, in addition, 
\begin{equation}\label{eq:diagcoeffcoconv}
|e^{-2\b_{j}t}\dot{g}^{jj}(t)-2\b_{j}g_{\infty}^{jj}|\leq C_{\rod}e^{-\eta_{\rod}t}
\end{equation}
for all $t\geq 0$ and $j=1,\dots,d$, the diagonal coefficients are said to be $C_{\exp}^{1}$-\textit{convergent and $C_{\exp}^{2}$-bounded}. 
Similarly, the auxiliary diagonal coefficients are said to be $C_{\exp}^{0}$-\textit{convergent and $C_{\exp}^{2}$-bounded} 
if there are real constants $\bRie{r}$, strictly positive real numbers $a_{r,\infty}$, $r=1,\dots,R$, and constants $\eta_{\road},
C_{\road}>0$ such that 
\begin{equation}\label{eq:adiagcoeffconv}
e^{\eta_{\road}t}|e^{\bRie{r}t}a_{r}(t)-a_{r,\infty}|+|e^{\bRie{r}t}\dot{a}_{r}(t)|+|e^{\bRie{r}t}\ddot{a}_{r}(t)|\leq C_{\road}
\end{equation}
for all $t\geq 0$ and $r=1,\dots,R$. If, in addition,
\begin{equation}\label{eq:adiagcoeffcoconv}
|e^{\bRie{r}t}\dot{a}_{r}(t)+\bRie{r}a_{r,\infty}|\leq C_{\road}e^{-\eta_{\road}t}
\end{equation}
for all $t\geq 0$ and $j=1,\dots,d$, the auxiliary diagonal coefficients are said to be $C_{\exp}^{1}$-\textit{convergent and 
$C_{\exp}^{2}$-bounded}. Assuming the diagonal coefficients are $C_{\exp}^{0}$-convergent and $C_{\exp}^{2}$-bounded, the main coefficients are 
said to be $C_{\exp}^{1}$-\textit{balanced and bounded} if there is a real constant $C_{\romn}>0$ such that 
\begin{equation}\label{eq:mainbd}
\begin{split}
e^{-\b_{j}t}\left(\|X^{j}(t)\|+\|\dot{X}^{j}(t)\|\right) 
+\|\a(t)\|+\|\dot{\a}(t)\|+\|\zeta(t)\|+\|\dot{\zeta}(t)\| & \leq C_{\romn}
\end{split}
\end{equation}
for all $t\geq 0$ and $j=1,\dots,d$. If, in addition, there are $\a_{\infty}, X^{j}_{\infty}, \zeta_{\infty}\in\Mn{m}{\co}$ as well 
as a real constant $\eta_{\romn}>0$ such that 
\begin{equation}\label{eq:mainconv}
\begin{split}
\|e^{-\b_{j}t}X^{j}(t)-X^{j}_{\infty}\|+\|\a(t)-\a_{\infty}\|+\|\zeta(t)-\zeta_{\infty}\|\leq C_{\romn}e^{-\eta_{\romn} t}
\end{split}
\end{equation}
for all $t\geq 0$ and $j=1,\dots,d$, the main coefficients are said to be $C^{0}_{\exp}$-\textit{balanced and convergent and} 
$C_{\exp}^{1}$-\textit{balanced and bounded}. The shift coefficients are said to be \textit{negligible} if there are real constants 
$C_{\rosh},\eta_{\rosh}>0$ such that 
\begin{equation}\label{eq:shtermneg}
\sup_{\xi\neq 0}\frac{|g^{0l}(t)\xi_{l}|}{[g^{jl}(t)\xi_{j}\xi_{l}]^{1/2}}+
\sup_{\xi\neq 0}\frac{|\dot{g}^{0l}(t)\xi_{l}|}{[g^{jl}(t)\xi_{j}\xi_{l}]^{1/2}}\leq C_{\rosh}e^{-\eta_{\rosh}t}
\end{equation}
for all $t\geq 0$, where Einstein's summation convention is enforced. The off-diagonal coefficients are said to be \textit{negligible}
if there are real constants $C_{\rood},\eta_{\rood}>0$ such that
\begin{equation}\label{eq:odtermneg}
\textstyle{\sum}_{j\neq l}[|g^{jl}(t)\xi_{j}\xi_{l}|+|\dot{g}^{jl}(t)\xi_{j}\xi_{l}|+|\ddot{g}^{jl}(t)\xi_{j}\xi_{l}|]
\leq C_{\rood}e^{-\eta_{\rood}t}\sum_{j}g^{jj}(t)\xi_{j}\xi_{j}
\end{equation}
for all $t\geq 0$ and $\xi\in\rn{d}$. If all the above conditions are satisfied, (\ref{eq:thesystemRge}) is said to be 
\textit{diagonally dominated, balanced and convergent}. 
\index{Diagonally dominated, balanced and convergent!equation}%
\index{Equation!diagonally dominated, balanced and convergent}%
\end{definition}
\begin{remark}
In (\ref{eq:odtermneg}), the sum is over all distinct $j$ and $l$, since neither $j$ nor $l$ is fixed from the context. Note, however, that if 
$l$ were fixed, then the sum appearing in (\ref{eq:odtermneg}) should have been interpreted as being over all $j$ different from $l$. 
\end{remark}

To begin with, it is of interest to verify that (\ref{eq:thesystemRge}) is oscillation adapted, cf. Definition~\ref{def:oscad}, when 
the requirements of the previous definition are fulfilled. 

\begin{lemma}\label{lemma:consconvcoeff}
Consider (\ref{eq:thesystemRge}). Assume that the diagonal and auxiliary diagonal coefficients are $C_{\exp}^{0}$-convergent and 
$C_{\exp}^{2}$-bounded; that the main coefficients are $C_{\exp}^{1}$-balanced and bounded; and that the shift coefficients and the off 
diagonal coefficients are negligible. Then there is a constant $C>1$ such that 
\begin{align}
C^{-1}g^{jl}(t)\xi_{j}\xi_{l} \leq & \textstyle{\sum}_{j}g^{jj}(t)\xi_{j}\xi_{j}\leq Cg^{jl}(t)\xi_{j}\xi_{l},\label{eq:geneqdiag}\\
C^{-1}g^{jl}(t)\xi_{j}\xi_{l} \leq & \textstyle{\sum}_{j}e^{2\b_{j}t}\xi_{j}\xi_{j}\leq Cg^{jl}(t)\xi_{j}\xi_{l}\label{eq:etbjdiag}
\end{align}
for all $t\geq 0$ and $\xi\in\rn{d}$. Moreover, 
\begin{equation}\label{eq:arnormeq}
C^{-1}\textstyle{\sum}_{r=1}^{R}a_{r}^{-2}(t)\eta_{r}^{2}\leq \sum_{r=1}^{R}e^{2\bRie{r}t}\eta_{r}^{2}
\leq C\sum_{r=1}^{R}a_{r}^{-2}(t)\eta_{r}^{2}
\end{equation}
for all $\eta\in\rn{R}$. Finally, choosing $\ellderbd$ and $C$ large enough and defining
\begin{equation}\label{eq:oscadddbac}
\mff_{\rosh}(t):=Ce^{-\eta_{\rosh}t},\ \ \
\mff_{X}(t):=C,\ \ \
\mff_{\roode}(t):=C,
\end{equation}
(\ref{eq:thesystemRge}) is oscillation adapted.
\end{lemma}
\begin{remark}
The constants $C$ and $\ellderbd$ appearing in the statement of the lemma only depend on the coefficients of the equation
(\ref{eq:thesystemRge}). 
\end{remark}
\begin{proof}
Let us begin by demonstrating the norm equivalences (\ref{eq:geneqdiag})--(\ref{eq:arnormeq}).

\textbf{Norm equivalences.} In order to prove that (\ref{eq:geneqdiag}) holds, estimate
\[
g^{jl}(t)\xi_{j}\xi_{l}=\textstyle{\sum}_{j\neq l}g^{jl}(t)\xi_{j}\xi_{l}+\sum_{j}g^{jj}(t)\xi_{j}\xi_{j}
\leq (1+C_{\rood}e^{-\eta_{\rood}t})\sum_{j}g^{jj}(t)\xi_{j}\xi_{j},
\]
where we used (\ref{eq:odtermneg}) in the last step. Thus there is a constant $C>1$ such that the first inequality in 
(\ref{eq:geneqdiag}) holds. Let $0\leq t_{\rood}<\infty$ be such that $C_{\rood}e^{-\eta_{\rood}t_{\rood}}\leq 1/2$. On $[0,t_{\rood}]$, it is 
clear that there is a constant $C>1$ such that the right inequality in (\ref{eq:geneqdiag}) holds; this is an immediate 
consequence of the fact that the $g^{jl}$ are smooth functions and the fact that $g^{jl}(t)$ are the components of a positive
definite matrix for every $t$. For $t\geq t_{\rood}$, 
\[
g^{jl}(t)\xi_{j}\xi_{l}=\textstyle{\sum}_{j\neq l}g^{jl}(t)\xi_{j}\xi_{l}+\sum_{j}g^{jj}(t)\xi_{j}\xi_{j}
\geq\sum_{j}g^{jj}(t)\xi_{j}\xi_{j}/2,
\]
where we used (\ref{eq:odtermneg}) and the fact that $C_{\rood}e^{-\eta_{\rood}t}\leq 1/2$ for $t\geq t_{\rood}$ in the last step. 
Thus (\ref{eq:geneqdiag}) holds. Using (\ref{eq:diagcoeffconv}), a similar argument yields the conclusion that there is a 
constant $C>1$ such that 
\[
C^{-1}\leq e^{-2\b_{j}t}g^{jj}(t)\leq C
\]
for all $t\geq 0$ and $\xi\in\rn{d}$. Combining this estimate with (\ref{eq:geneqdiag}) yields (\ref{eq:etbjdiag}).
Since $e^{\bRie{r}t}a_{r}(t)$ converges to a strictly positive number, there is similarly a constant $C>1$ (depending only on 
the coefficients of the equation), such that (\ref{eq:arnormeq}) holds.

\textbf{Bounding the derivatives of $\ell$.} Next, let us estimate
\begin{equation}\label{eq:dotgxijlest}
\begin{split}
|\dot{g}^{jl}\xi_{j}\xi_{l}| \leq & \textstyle{\sum}_{j\neq l}|\dot{g}^{jl}\xi_{j}\xi_{l}|
+\sum_{j}|\dot{g}^{jj}\xi_{j}\xi_{j}|\\
 \leq & Ce^{-\eta_{\rood}t}\textstyle{\sum}_{j}g^{jj}(t)\xi_{j}\xi_{j}+C\sum_{j}e^{2\b_{j}t}\xi_{j}\xi_{j}
\leq Cg^{jl}\xi_{j}\xi_{l}
\end{split}
\end{equation}
for all $t\geq 0$ and $\xi\in\rn{d}$, where we used (\ref{eq:odtermneg}) and the fact that $e^{-2\b_{j}t}\dot{g}^{jj}(t)$ is bounded
in the second step, and (\ref{eq:geneqdiag}) and (\ref{eq:etbjdiag}) in the third step. A similar argument concerning 
$\ddot{g}^{jl}$ yields the conclusion that there is a constant $C$ such that 
\begin{equation}\label{eq:ddotgxijlest}
|\ddot{g}^{jl}\xi_{j}\xi_{l}| \leq Cg^{jl}\xi_{j}\xi_{l}
\end{equation}
for all $t\geq 0$ and $\xi\in\rn{d}$. Appealing to (\ref{eq:adiagcoeffconv}) and (\ref{eq:arnormeq}), it can similarly be verified that 
\begin{equation}\label{eq:ddotarsumlest}
\left|\textstyle{\sum}_{r=1}^{R}\d_{t}(a_{r}^{-2})\eta_{r}^{2}\right|
+\left|\textstyle{\sum}_{r=1}^{R}\d_{t}^{2}(a_{r}^{-2})\eta_{r}^{2}\right|\leq C\textstyle{\sum}_{r=1}^{R}a_{r}^{-2}\eta_{r}^{2}
\end{equation}
for all $\eta\in\rn{R}$. Moreover, the constants $C$ appearing in (\ref{eq:dotgxijlest})--(\ref{eq:ddotarsumlest}) only depend on
the coefficients of the equation (\ref{eq:thesystemRge}). 

Let $0\neq \indexnot\in\EFindexset$, define $\mfg(\indexnot,t)$ by (\ref{eq:mfgnutdef}) and let $\ell(\indexnot,t):=\ln\mfg(\indexnot,t)$.
Compute
\begin{equation}\label{eq:dotellexpr}
\dot{\ell}(\indexnot,t)=\frac{1}{2\mfg^{2}(\indexnot,t)}\left(\dot{g}^{jl}(t)n_{j}n_{l}+\textstyle{\sum}_{r=1}^{R}\d_{t}(a_{r}^{-2})(t)\nu_{r,i_{r}}^{2}(\indexnot)\right)
\end{equation}
Due to (\ref{eq:dotgxijlest}) and (\ref{eq:ddotarsumlest}), we conclude that $|\dot{\ell}(\indexnot,t)|$ is bounded for $t\geq 0$. Since
\[
\ddot{\ell}(\indexnot,t)=\frac{1}{2\mfg^{2}(\indexnot,t)}\left(\ddot{g}^{jl}(t)n_{j}n_{l}
+\textstyle{\sum}_{r=1}^{R}\d_{t}^{2}(a_{r}^{-2})(t)\nu_{r,i_{r}}^{2}(\indexnot)\right)-2\dot{\ell}^{2}(\indexnot,t),
\]
(\ref{eq:ddotgxijlest}) and (\ref{eq:ddotarsumlest}) can similarly be used to yield the conclusion that $|\ddot{\ell}(\indexnot,t)|$ is bounded 
for $t\geq 0$. Thus (\ref{eq:estdtlnge}) holds for some constant $\ellderbd$ depending only on the coefficients of the equation 
(\ref{eq:thesystemRge}). 

\textbf{Bounding the remaining coefficients.} 
Next, let us consider (\ref{eq:shiftbd}). Let $0\neq \indexnot\in\EFindexset$ and let $\sigma$ be defined as in 
(\ref{eq:ellsigmaXgenRdef}). Due to (\ref{eq:shtermneg}) and the fact that $|\dot{\ell}|$ is bounded to the future,
it follows that (\ref{eq:shiftbd}) holds with $\mff_{\rosh}(t)=Ce^{-\eta_{\rosh}t}$, for some suitably big constant 
$C$, depending only on the coefficients of the equation (\ref{eq:thesystemRge}). 

Turning to (\ref{eq:Xbd}), note that if $X$ is 
defined as in (\ref{eq:ellsigmaXgenRdef}) and $n\neq 0$, then
\[
\|X\|\leq \frac{1}{\mfg}\textstyle{\sum}_{j}|n_{j}|\|X^{j}\|\leq C\left(\sum_{j}e^{2\b_{j}t}n_{j}^{2}\right)^{-1/2}\sum_{j}|n_{j}|e^{\b_{j}t}
\leq C
\]
for all $t\geq 0$, where we have used the fact that (\ref{eq:mainbd}) implies that $e^{-\b_{j}t}\|X^{j}\|$ is bounded, as well
as (\ref{eq:etbjdiag}). Since
\[
\dot{X}=\frac{n_{l}\dot{X}^{l}}{\mfg}-X\dot{\ell},
\]
a similar argument (using (\ref{eq:mainbd}) together with the bounds for $\|X\|$ and $|\dot{\ell}|$ we have 
already derived) yields $\|\dot{X}\|\leq C$ for all $t\geq 0$ and all $0\neq n\in\zn{d}$. Thus (\ref{eq:Xbd})
holds with $\mff_{X}=C$, where $C$ is a constant depending only on the coefficients of the equation (\ref{eq:thesystemRge}). 

Finally, (\ref{eq:mainbd}) implies that (\ref{eq:albd}) and (\ref{eq:zetabd}) hold for all $t\geq 0$, 
with $\mff_{\roode}(t)=C$, where $C$ has the same dependence as before. Since $\mff_{\rosh}$, $\mff_{X}$ and $\mff_{\roode}$ are all 
constants times exponential functions, it is clear that (\ref{eq:mffXdotbdetc}) holds for a suitably chosen $\ellderbd$. 
\end{proof}

\section{Simplifying the matrix coefficients}\label{section:simplmacoeff}

Consider the equation (\ref{eq:thesystemRge}). Assume that it is diagonally dominated, balanced and convergent; cf. 
Definition~\ref{def:convergeq}. Let $\baverage{1}<\baverage{2}<\dots<\baverage{Q}$ be the distinct elements of 
the union of the sets $\{\b_{1},\dots,\b_{d}\}$ and $\{\bRie{1},\dots,\bRie{R}\}$. For $1\leq q\leq Q$ and $\indexnot\in\EFindexset$, let 
\begin{equation}\label{eq:nuaveragedef}
\nuaverage{q}(\indexnot):=
\left(\textstyle{\sum}_{\{j:\b_{j}=\baverage{q}\}}g^{jj}_{\infty}n_{j}^{2}+\sum_{\{r:\bRie{r}=\baverage{q}\}}a_{r,\infty}^{-2}\nu_{r,i_{r}}^{2}(\indexnot)\right)^{1/2}.
\end{equation}
In the analysis we are about to carry out, it would be convenient to replace $\mfg(t)$ with
\begin{equation}\label{eq:mfgdiagdef}
\begin{split}
\mfg_{\diag}(\indexnot,t) := & \left(\textstyle{\sum}_{j=1}^{d}g^{jj}_{\infty}n_{j}^{2}e^{2\b_{j}t}+\sum_{r=1}^{R}a_{r,\infty}^{-2}\nu_{r,i_{r}}^{2}(\indexnot)
e^{2\bRie{r}t}\right)^{1/2}\\
 = & \left(\textstyle{\sum}_{q=1}^{Q}\nuaverage{q}^{2}(\indexnot)e^{2\baverage{q}t}\right)^{1/2}.
\end{split}
\end{equation}
It is therefore of interest to compare $\mfg(t)$ with $\mfg_{\diag}(t)$. We are also interested in approximating $X/\mfg$ and $\a/\mfg$ in terms 
of similar asymptotic data; cf. the introduction to the present chapter. 

\begin{lemma}\label{lemma:diagapprox}
Consider the equation (\ref{eq:thesystemRge}). Assume that it is diagonally dominated, balanced and convergent; cf. 
Definition~\ref{def:convergeq}. Then there is a constant $C>1$ such that 
\begin{equation}\label{eq:mfgmfgdiageq}
C^{-1}\mfg(\indexnot,t)\leq \mfg_{\diag}(\indexnot,t)\leq C\mfg(\indexnot,t)
\end{equation}
for all $t\geq 0$ and $\indexnot\in\EFindexset$, where $\mfg_{\diag}$ is defined in (\ref{eq:mfgdiagdef}). Moreover,
\begin{equation}\label{eq:mfgmmfgdiag}
|\mfg(\indexnot,t)-\mfg_{\diag}(\indexnot,t)| \leq Ce^{-\eta_{\diag}t}\mfg(\indexnot,t)
\end{equation}
for all $t\geq 0$ and $\indexnot\in\EFindexset$, where $\eta_{\diag}:=\min\{\eta_{\rood},\eta_{\rod},\eta_{\road}\}$. 
Letting $\ell_{\diag}:=\ln\mfg_{\diag}$,
\begin{equation}\label{eq:elltmelldiagtst}
\left|\frac{\dot{\ell}(\indexnot,t)}{\mfg(\indexnot,t)}-\frac{\dot{\ell}_{\diag}(\indexnot,t)}{\mfg_{\diag}(\indexnot,t)}\right|
\leq C\frac{e^{-\eta_{\diag}t}}{\mfg(\indexnot,t)}
\end{equation}
for all $t\geq 0$ and $0\neq\indexnot\in\EFindexset$. Letting 
\begin{equation}\label{eq:Xdiagdef}
X_{\diag}(\indexnot,t):=\frac{1}{\mfg_{\diag}(\indexnot,t)}\sum_{l}n_{l}e^{\b_{l}t}X^{l}_{\infty},
\end{equation}
the estimate
\begin{equation}\label{eq:XmXdiag}
\left\|\frac{X_{\diag}(\indexnot,t)}{\mfg_{\diag}(\indexnot,t)}-\frac{X(\indexnot,t)}{\mfg(\indexnot,t)}\right\| 
\leq C\frac{e^{-\eta_{X}t}}{\mfg(\indexnot,t)}
\end{equation}
holds for all $t\geq 0$ and $0\neq\indexnot\in\EFindexset$, where $\eta_{X}:=\min\{\eta_{\diag},\eta_{\romn}\}$. Moreover, 
\begin{equation}\label{eq:agagdiag}
\left\|\frac{\a_{\infty}}{\mfg_{\diag}(\indexnot,t)}-\frac{\a(t)}{\mfg(\indexnot,t)}\right\|\leq C\frac{e^{-\eta_{\a}t}}{\mfg(\indexnot,t)}
\end{equation}
for all $t\geq 0$ and $0\neq\indexnot\in\EFindexset$, where $\eta_{\a}:=\min\{\eta_{\diag},\eta_{\romn}\}$.
\end{lemma}
\begin{remark}
Many of the conclusions of the lemma can be obtained using weaker assumptions. 
\end{remark}
\begin{remark}
The constants $C$ appearing in the statement of the lemma only depend on the coefficients of the equation
(\ref{eq:thesystemRge}). 
\end{remark}
\begin{proof}
That there is a constant $C>1$, depending only on the coefficients of the equation (\ref{eq:thesystemRge}), such that (\ref{eq:mfgmfgdiageq}) 
holds for all $t\geq 0$ and $\indexnot\in\EFindexset$ is a consequence of (\ref{eq:etbjdiag}) and (\ref{eq:arnormeq}). Let us estimate
\begin{equation}\label{eq:ggdiagprel}
\begin{split}
 & |\mfg^{2}-\mfg^{2}_{\diag}|\\ 
\leq & \textstyle{\sum}_{j\neq l}\left|g^{jl}n_{j}n_{l}\right|+\sum_{j}e^{2\b_{j}t}|e^{-2\b_{j}t}g^{jj}-g^{jj}_{\infty}|n_{j}^{2}\\
 & +\textstyle{\sum}_{r=1}^{R}|e^{\bRie{r}t}a_{r}-a_{r,\infty}|(e^{\bRie{r}t}a_{r}+a_{r,\infty})[e^{\bRie{r}t}a_{r}]^{-2}a_{r,\infty}^{-2}\nu_{r,i_{r}}^{2}(\indexnot)
e^{2\bRie{r}t}\\
 \leq & Ce^{-\eta_{\rood}t}\mfg^{2}+Ce^{-\eta_{\rod}t}\mfg^{2}+Ce^{-\eta_{\road}t}\mfg^{2}\leq Ce^{-\eta_{\diag}t}\mfg^{2}
\end{split}
\end{equation}
for all $t\geq 0$ and $\indexnot\in\EFindexset$, where $C$ only depends on the coefficients of the equation (\ref{eq:thesystemRge})
and we have appealed to (\ref{eq:diagcoeffconv}), (\ref{eq:adiagcoeffconv}), (\ref{eq:odtermneg}), (\ref{eq:geneqdiag}), (\ref{eq:etbjdiag}) 
and (\ref{eq:arnormeq}). Assuming $\indexnot\neq 0$ and dividing (\ref{eq:ggdiagprel}) with $\mfg+\mfg_{\diag}$ yields (\ref{eq:mfgmmfgdiag}) 
(if $\indexnot=0$, this inequality holds trivially).

Let us turn to $\dot{\ell}/\mfg$. Recalling that $\dot{\ell}$ is given by (\ref{eq:dotellexpr}), 
\begin{equation}\label{eq:elltmelldiagt}
\begin{split}
\left|\frac{\dot{\ell}}{\mfg}-\frac{\dot{\ell}_{\diag}}{\mfg_{\diag}}\right| \leq & \frac{1}{2}\left|\frac{\dot{g}^{jl}n_{j}n_{l}}{\mfg^{3}}
-\frac{1}{\mfg_{\diag}^{3}}\sum_{j=1}^{d}2\b_{j}g^{jj}_{\infty}n_{j}^{2}e^{2\b_{j}t}\right|\\
 & + \frac{1}{2}\left|\sum_{r=1}^{R}\left(\frac{1}{\mfg^{3}}\d_{t}(a_{r}^{-2})-\frac{1}{\mfg_{\diag}^{3}}2\bRie{r}a_{r,\infty}^{-2}e^{2\bRie{r}t}\right)
\nu_{r,i_{r}}^{2}(\indexnot)\right| 
\end{split}
\end{equation}
The first term on the right hand side of (\ref{eq:elltmelldiagt}) can be estimated by 
\begin{equation}\label{eq:elltmelldiagtfirstterm}
\begin{split}
 & \frac{1}{2}\frac{1}{\mfg^{3}}\sum_{j\neq l}|\dot{g}^{jl}n_{j}n_{l}|
+\frac{1}{2}\frac{1}{\mfg^{3}}\sum_{j}e^{2\b_{j}t}|e^{-2\b_{j}t}\dot{g}^{jj}-2\b_{j}g^{jj}_{\infty}|n_{j}^{2}\\
 + & \frac{1}{2}\left|\frac{1}{\mfg^{3}}
-\frac{1}{\mfg_{\diag}^{3}}\right|\sum_{j=1}^{d}g^{jj}_{\infty}n_{j}^{2}2|\b_{j}|e^{2\b_{j}t}\leq C\frac{e^{-\eta_{\diag}t}}{\mfg}
\end{split}
\end{equation}
for all $t\geq 0$, where we appealed to (\ref{eq:diagcoeffcoconv}), (\ref{eq:odtermneg}), (\ref{eq:geneqdiag}), (\ref{eq:etbjdiag}), 
(\ref{eq:mfgmfgdiageq}) and (\ref{eq:mfgmmfgdiag}). Before turning to the second term on the right hand side of 
(\ref{eq:elltmelldiagt}), note that 
\begin{equation*}
\begin{split}
\d_{t}a_{r}^{-2}-2\bRie{r}a_{r,\infty}^{-2}e^{2\bRie{r}t} = & -2(a_{r}e^{\bRie{r}t})^{-3}(e^{\bRie{r}t}\dot{a}_{r}+\bRie{r}a_{r,\infty})e^{2\bRie{r}t}\\
 & +2\bRie{r}e^{2\bRie{r}t}a_{r,\infty}[(a_{r}e^{\bRie{r}t})^{-3}-a_{r,\infty}^{-3}].
\end{split}
\end{equation*}
Combining this observation with (\ref{eq:adiagcoeffconv}) and (\ref{eq:adiagcoeffcoconv}) yields
\[
|\d_{t}a_{r}^{-2}-2\bRie{r}a_{r,\infty}^{-2}e^{2\bRie{r}t}| \leq Ce^{-\eta_{\road}t}e^{2\bRie{r}t}
\]
for all $t\geq 0$, where $C$ only depends on the coefficients of the equation (\ref{eq:thesystemRge}). Due to this estimate, the
second term on the right hand side of (\ref{eq:elltmelldiagt}) can be estimated by 
\begin{equation*}
\begin{split}
 & \frac{1}{2}\sum_{r=1}^{R}\frac{1}{\mfg^{3}}|\d_{t}a_{r}^{-2}-2\bRie{r}a_{r,\infty}^{-2}e^{2\bRie{r}t}|\nu_{r,i_{r}}^{2}(\indexnot)\\
 & +\frac{1}{2}\sum_{r=1}^{R}\left|\frac{1}{\mfg^{3}}-\frac{1}{\mfg_{\diag}^{3}}\right|2|\bRie{r}|a_{r,\infty}^{-2}e^{2\bRie{r}t}
\nu_{r,i_{r}}^{2}(\indexnot)\leq C\frac{e^{-\eta_{\diag}t}}{\mfg}
\end{split}
\end{equation*}
for all $t\geq 0$, 
where we appealed to (\ref{eq:arnormeq}), (\ref{eq:mfgmfgdiageq}) and (\ref{eq:mfgmmfgdiag}), and $C$ only depends on the 
coefficients of the equation (\ref{eq:thesystemRge}). Combining this estimate with (\ref{eq:elltmelldiagt}) and 
(\ref{eq:elltmelldiagtfirstterm}) yields (\ref{eq:elltmelldiagtst}).

Next, let us turn to $X/\mfg$ and $\a/\mfg$. To begin with
\begin{equation*}
\begin{split}
\|X_{\diag}-X\| \leq & \frac{1}{\mfg_{\diag}}\sum_{l}|n_{l}|e^{\b_{l}t}\|e^{-\b_{l}t}X^{l}-X^{l}_{\infty}\|
+\left|\frac{1}{\mfg_{\diag}}-\frac{1}{\mfg}\right|\|n_{l}X^{l}\|\\
 \leq & Ce^{-\eta_{\romn}t}+Ce^{-\eta_{\diag}t}\|X\|\leq Ce^{-\eta_{X}t}
\end{split}
\end{equation*}
for all $t\geq 0$ and $\indexnot\in\EFindexset$, where we have appealed to  (\ref{eq:mainbd}), (\ref{eq:mainconv}), (\ref{eq:etbjdiag}), 
(\ref{eq:mfgmfgdiageq}) and (\ref{eq:mfgmmfgdiag}). Combining this estimate with (\ref{eq:mfgmfgdiageq}) and (\ref{eq:mfgmmfgdiag}) yields 
(\ref{eq:XmXdiag}). Moreover, $C$ only depends on the coefficients of the equation (\ref{eq:thesystemRge}). Estimate
\[
\left\|\frac{\a_{\infty}}{\mfg_{\diag}}-\frac{\a}{\mfg}\right\|\leq 
\frac{1}{\mfg_{\diag}}\|\a-\a_{\infty}\|
+\left|\frac{1}{\mfg_{\diag}}-\frac{1}{\mfg}\right|\|\a\|.
\]
Appealing to (\ref{eq:mainbd}), (\ref{eq:mainconv}), (\ref{eq:mfgmfgdiageq}) and (\ref{eq:mfgmmfgdiag}) yields
(\ref{eq:agagdiag}). 
\end{proof}

\section[Time intervals, frequency eras]{Choosing appropriate time intervals; defining the frequency eras}\label{section:timeintfreeras}

In order to be able to estimate the matrix products of interest, it is convenient to divide $[0,\infty)$ into 
intervals during which the coefficients of the iteration can be simplified. Specifying such intervals
and deriving preliminary information concerning the matrix coefficients in them is the main purpose of the next lemma. 

\begin{lemma}\label{lemma:Ijanalysis}
Consider the equation (\ref{eq:thesystemRge}). Assume that it is diagonally dominated, balanced and convergent; cf. 
Definition~\ref{def:convergeq}. Fix $0\neq \indexnot\in\EFindexset$. Using the notation introduced in connection with
(\ref{eq:nuaveragedef}), let $I_{q}$ denote 
the subset of $[0,\infty)$ such that 
\begin{equation}\label{eq:defprera}
\nuaverage{q}(\indexnot)e^{\baverage{q}t}\geq \nuaverage{p}(\indexnot)e^{\baverage{p}t}
\end{equation}
for all $p=1,\dots,Q$. Then $I_{q}$ is either empty or a closed interval. Moreover, if $q,p\in\{1,\dots,Q\}$ are such that 
$q\neq p$, then $I_{q}\cap I_{p}$ is either empty or consists of one point. Assuming $I_{q}$ to be non-empty, define
\begin{equation}\label{eq:mfhmfHjdef}
\mfH_{q}^{(s)}(\indexnot,t):=\sum_{p\neq q}\left(\frac{\nuaverage{p}(\indexnot)e^{\baverage{p}t}}{\nuaverage{q}(\indexnot)e^{\baverage{q}t}}\right)^{s},\ \ \
\mfH_{q}(\indexnot,t):=\mfH_{q}^{(2)}(\indexnot,t),
\end{equation}
\index{$\a$Aa@Notation!Auxiliary functions!$\mfH_{q}^{(s)}(\indexnot,t)$}%
\index{$\a$Aa@Notation!Auxiliary functions!$\mfH_{q}(\indexnot,t)$}%
where $0<s\in\ro$. Then there is a constant $C>1$ such that 
\begin{equation}\label{eq:njnogequ}
C^{-1}\mfg(\indexnot,t)\leq \nuaverage{q}(\indexnot)e^{\baverage{q}t}\leq C\mfg(\indexnot,t)
\end{equation}
for all $t\in I_{q}$. Moreover,
\begin{align}
\left|\frac{1}{\nuaverage{q}(\indexnot)e^{\baverage{q}t}}-\frac{1}{\mfg_{\diag}(\indexnot,t)}\right|
 \leq & \frac{C}{\mfg(\indexnot,t)}\mfH_{q}(\indexnot,t),\label{eq:ginvminnjvergest}\\
\left|\frac{1}{\nuaverage{q}^{2}(\indexnot)e^{2\baverage{q}t}}-\frac{1}{\mfg_{\diag}^{2}(\indexnot,t)}\right|
 \leq & \frac{C}{\mfg^{2}(\indexnot,t)}\mfH_{q}(\indexnot,t),\label{eq:gsqinvminnjvergest}\\
\left|\frac{\dot{\ell}_{\diag}(\indexnot,t)}{\mfg_{\diag}(\indexnot,t)}-\frac{\baverage{q}}{\nuaverage{q}(\indexnot)e^{\baverage{q}t}}\right| \leq &
 \frac{C}{\mfg(\indexnot,t)}\mfH_{q}(\indexnot,t)\label{eq:elldotgenestgest}
\end{align}
for all $t\in I_{q}$. Let 
\begin{equation}\label{eq:Xavdiagqindexnotdef}
\Xaverage{\diag}^{q}(\indexnot):=\frac{1}{\nuaverage{q}(\indexnot)}\sum_{\{j:\b_{j}=\baverage{q}\}}n_{j}X^{j}_{\infty};
\end{equation}
recall the notation introduced at the beginning of Section~\ref{section:simplmacoeff}. Then 
\begin{equation}\label{eq:XdiagbXdiagjest}
\left\|\frac{\Xaverage{\diag}^{q}(\indexnot)}{\nuaverage{q}(\indexnot)e^{\baverage{q}t}}-\frac{X_{\diag}(\indexnot,t)}{\mfg_{\diag}(\indexnot,t)}\right\|\leq 
\frac{C}{\mfg(\indexnot,t)}\mfH_{q}(\indexnot,t)+\frac{C}{\mfg^{2}(\indexnot,t)}
\sum_{\{j:\b_{j}\neq\baverage{q}\}}|n_{j}|e^{\b_{j}t}
\end{equation}
for $t\in I_{q}$, where $X_{\diag}$ is defined in (\ref{eq:Xdiagdef}). Moreover,
\begin{equation}\label{eq:adiagbadiagjest}
\left\|\frac{\a_{\infty}}{\nuaverage{q}(\indexnot)e^{\baverage{q}t}}-\frac{\a_{\infty}}{\mfg_{\diag}(\indexnot,t)}\right\|\leq 
\frac{C}{\mfg(\indexnot,t)}\mfH_{q}(\indexnot,t)
\end{equation}
for $t\in I_{q}$. 
\end{lemma}
\begin{remark}
The constants $C$ appearing in the statement of the lemma only depend on the coefficients of the equation
(\ref{eq:thesystemRge}). 
\end{remark}
\begin{proof}
Fix $q\in\{1,\dots,Q\}$ and assume $I_{q}$ to be non-empty. Let $t_{0},t_{1}\in I_{q}$ with $t_{0}\leq t_{1}$. Then 
$\nuaverage{q}(\indexnot)e^{\baverage{q}t}$ is greater 
than $\nuaverage{p}(\indexnot)e^{\baverage{p}t}$ for all $p<q$ and $t\geq t_{0}$. Moreover, $\nuaverage{q}(\indexnot)e^{\baverage{q}t}$ is greater than 
$\nuaverage{p}(\indexnot)e^{\baverage{p}t}$ for all $p>q$ and $t\leq t_{1}$. Thus the line connecting $t_{0}$ and $t_{1}$ is contained in $I_{q}$,
so that $I_{q}$ is connected. Since $I_{q}$ is closed by definition, we conclude that $I_{q}$ is a closed interval. By these arguments,
the statement concerning the intersection of the different $I_{q}$'s also follows. 

In $I_{q}$, it is of interest to write $\mfg_{\diag}$ as 
\[
\mfg_{\diag}(\indexnot,t)=\nuaverage{q}(\indexnot)e^{\baverage{q}t}\mfp_{q}(\indexnot,t);\ \ \ 
\mfp_{q}(\indexnot,t): = \left[1+\mfH_{q}(\indexnot,t)\right]^{1/2};
\]
cf. (\ref{eq:mfhmfHjdef}). Note that 
\[
\nuaverage{q}(\indexnot)e^{\baverage{q}t}\leq \mfg_{\diag}(\indexnot,t)\leq Q^{1/2} \nuaverage{q}(\indexnot)e^{\baverage{q}t}
\]
in $I_{q}$. Combining this estimate with (\ref{eq:mfgmfgdiageq}) it is clear that $\nuaverage{q}(\indexnot)e^{\baverage{q}t}$ and $\mfg(t)$
are equivalent in $I_{q}$. Thus (\ref{eq:njnogequ}) holds, and the constant $C$ only depends on the coefficients of the equation
(\ref{eq:thesystemRge}). Let us compute
\[
\frac{1}{\nuaverage{q}(\indexnot)e^{\baverage{q}t}}-\frac{1}{\mfg_{\diag}}
=\frac{1}{\nuaverage{q}(\indexnot)e^{\baverage{q}t}}\frac{\mfp_{q}-1}{\mfp_{q}}
=\frac{1}{\nuaverage{q}(\indexnot)e^{\baverage{q}t}}\frac{\mfp_{q}^{2}-1}{\mfp_{q}(\mfp_{q}+1)}.
\]
In particular, 
\begin{equation}\label{eq:ginvminnjverge}
\left|\frac{1}{\nuaverage{q}(\indexnot)e^{\baverage{q}t}}-\frac{1}{\mfg_{\diag}(\indexnot,t)}\right|
\leq\frac{1}{2}\frac{1}{\nuaverage{q}(\indexnot)e^{\baverage{q}t}}\mfH_{q}(\indexnot,t)
\end{equation}
for $t\in I_{q}$, so that (\ref{eq:ginvminnjvergest}) holds (where $C$ has the same dependence as before). As a consequence,
\begin{equation}\label{eq:gsqinvminnjverge}
\left|\frac{1}{\nuaverage{q}^{2}(\indexnot)e^{2\baverage{q}t}}-\frac{1}{\mfg_{\diag}^{2}(\indexnot,t)}\right|
\leq\frac{1}{\nuaverage{q}^{2}(\indexnot)e^{2\baverage{q}t}}\mfH_{q}(\indexnot,t)
\end{equation}
for $t\in I_{q}$, so that (\ref{eq:gsqinvminnjvergest}) holds. Again, the constant $C$ only depends on the coefficients of the equation
(\ref{eq:thesystemRge}). Next, let us turn to 
\begin{equation*}
\begin{split}
\dot{\ell}_{\diag} = & \frac{1}{\mfg_{\diag}^{2}}\sum_{p=1}^{Q}\baverage{p}\nuaverage{p}^{2}(\indexnot)e^{2\baverage{p}t}
 =  \baverage{q}+\frac{1}{\mfp_{q}^{2}}\sum_{p\neq q}(\baverage{p}-\baverage{q})
\left(\frac{\nuaverage{p}(\indexnot)e^{\baverage{p}t}}{\nuaverage{q}(\indexnot)e^{\baverage{q}t}}\right)^{2};
\end{split}
\end{equation*}
recall (\ref{eq:mfgdiagdef}). Thus
\[
\left|\dot{\ell}_{\diag}(\indexnot,t)-\baverage{q}\right|\leq\max_{p}|\baverage{p}-\baverage{q}|\cdot\mfH_{q}(\indexnot,t).
\]
Thus
\[
\left|\frac{\dot{\ell}_{\diag}(\indexnot,t)}{\mfg_{\diag}(\indexnot,t)}-\frac{\bbe_{q}}{\mfg_{\diag}(\indexnot,t)}\right|\leq
\frac{\max_{p}|\baverage{p}-\baverage{q}|}{\nuaverage{q}(\indexnot)e^{\baverage{q}t}}\mfH_{q}(\indexnot,t).
\]
Combining this estimate with (\ref{eq:ginvminnjverge}) yields
\begin{equation}\label{eq:elldotgenestge}
\left|\frac{\dot{\ell}_{\diag}(\indexnot,t)}{\mfg_{\diag}(\indexnot,t)}-\frac{\baverage{q}}{\nuaverage{q}(\indexnot)e^{\baverage{q}t}}\right|\leq
\frac{1}{\nuaverage{q}(\indexnot)e^{\baverage{q}t}}
\left(\frac{|\baverage{q}|}{2}+\max_{p}|\baverage{p}-\baverage{q}|\right)\mfH_{q}(\indexnot,t).
\end{equation}
Thus (\ref{eq:elldotgenestgest}) holds, where the constant $C$ only depends on the coefficients of the equation (\ref{eq:thesystemRge}). 

Next, let us turn to $X/\mfg$. Note that the norm of $X_{\diag}$ is bounded by a constant depending only on the coefficients of the equation 
(\ref{eq:thesystemRge}). Moreover, 
\begin{equation*}
\begin{split}
X_{\diag}(\indexnot,t)-\Xaverage{\diag}^{q}(\indexnot) = & \frac{1}{\mfg_{\diag}(\indexnot,t)}\sum_{l}n_{l}e^{\b_{l}t}X^{l}_{\infty}
-\frac{1}{\nuaverage{q}(\indexnot)}\sum_{\{j:\b_{j}=\baverage{q}\}}n_{j}X^{j}_{\infty}\\
 = & \left(\frac{1}{\mfg_{\diag}(\indexnot,t)}-\frac{1}{\nuaverage{q}(\indexnot)e^{\baverage{q}t}}\right)\sum_{l}n_{l}e^{\b_{l}t}X^{l}_{\infty}\\
 & +\frac{1}{\nuaverage{q}(\indexnot)e^{\baverage{q}t}}\sum_{\{j:\b_{j}\neq\baverage{q}\}}n_{j}e^{\b_{j}t}X^{j}_{\infty}.
\end{split}
\end{equation*}
Combining this equality with (\ref{eq:mfgmfgdiageq}), (\ref{eq:njnogequ}) and (\ref{eq:ginvminnjvergest}) yields
\begin{equation*}
\begin{split}
\left\|X_{\diag}(\indexnot,t)-\Xaverage{\diag}^{q}(\indexnot)\right\| \leq & C\mfH_{q}(\indexnot,t)
+\frac{C}{\mfg(\indexnot,t)}\sum_{\{j:\b_{j}\neq\baverage{q}\}}|n_{j}|e^{\b_{j}t}
\end{split}
\end{equation*}
for $t\in I_{q}$. Combining this estimate in its turn with (\ref{eq:njnogequ}), (\ref{eq:ginvminnjvergest}) and the previously mentioned bound 
on the norm of $X_{\diag}$ yields (\ref{eq:XdiagbXdiagjest}). Moreover, the constant $C$ only depends on the coefficients of the equation 
(\ref{eq:thesystemRge}). Finally, (\ref{eq:adiagbadiagjest}) is an immediate consequence of (\ref{eq:ginvminnjvergest}). The lemma follows. 
\end{proof}

The intervals $I_{q}$ defined in the previous lemma are important in the analysis to follow. It is therefore convenient to introduce
the following terminology. 

\begin{definition}\label{def:era}
Given assumptions and terminology as in the statement of Lemma~\ref{lemma:Ijanalysis}, let $I_{q}$, $q=1,\dots,Q$, be defined as in the statement
of that lemma. If $I_{q}$ is non-empty, then $I_{q}$ is referred to as a \textit{frequency era} 
\index{Frequency era}%
of the equation (\ref{eq:thesystemRge}) corresponding to $0\neq \indexnot\in\EFindexset$. If $\baverage{q}<0$,
then $I_{q}$ is called a \textit{silent era}; 
\index{Era!silent}%
\index{Silent!era}%
if $\baverage{q}=0$, then $I_{q}$ is called a \textit{transparent era}; 
\index{Era!transparent}%
\index{Transparent!era}%
and if $\baverage{q}>0$, then $I_{q}$ 
is called a \textit{noisy era}.
\index{Era!noisy}%
\index{Noisy!era}%
\end{definition}
\begin{remark}
If the equation and $\indexnot$ are understood from the context, we simply speak of a frequency era. 
\end{remark}

\section{Estimating sums in a frequency era}\label{section:estsumsfreera}

When calculating the matrix products appearing in (\ref{eq:differenceeqrepres}), it is convenient to restrict one's attention
to a frequency era; cf. Definition~\ref{def:era}. In that context, the following lemma is of interest. 

\begin{lemma}\label{lemma:compsumswithintegrals}
Consider the equation (\ref{eq:thesystemRge}). Assume that it is diagonally dominated, balanced and convergent; cf. 
Definition~\ref{def:convergeq}. Fix $0\neq \indexnot\in\EFindexset$ and let $I_{q}$ be a corresponding frequency era; 
cf. Definition~\ref{def:era}. Consider a sequence $\{t_{k}\}$ defined as in Definition~\ref{def:tkdefge},
assume that there are elements of the sequence belonging to $I_{q}$, let $k_{q,1}$ be the smallest and $k_{q,2}$ the largest $k$ 
such that $t_{k}\in I_{q}$, and assume that $k_{q,1}<k_{q,2}$ (if there is no upper bound on the $k$'s for which $t_{k}\in I_{q}$,
$k_{q,2}$ is defined to equal $\infty$). Let $k_{0}$ and $k_{1}$ be such that $k_{q,1}\leq k_{0}\leq k_{1}<k_{q,2}$ and define $\bt_{0}$
and $\bt_{1}$ by $\bt_{0}=t_{k_{0}}$ and $\bt_{1}=t_{k_{1}+1}$. Then 
\begin{equation}\label{eq:sumapproxint}
\begin{split}
 & \left|\sum_{k=k_{0}}^{k_{1}}\frac{2\pi e^{\lambda t_{k}}}{(\nuaverage{q}(\indexnot)e^{\baverage{q}t_{k}})^{a}}
-\int_{\bt_{0}}^{\bt_{1}}\frac{e^{\lambda t}}{(\nuaverage{q}(\indexnot)e^{\baverage{q}t})^{a-1}}dt\right|\\
 \leq & \int_{\bt_{0}}^{\bt_{1}}\left(\frac{Ce^{\lambda t}}{(\nuaverage{q}(\indexnot)e^{\baverage{q}t})^{a-1}}\left[e^{-\eta_{\diag}t}
+\mfH_{q}(\indexnot,t)\right]+\frac{Ce^{\lambda t}}{(\nuaverage{q}(\indexnot)e^{\baverage{q}t})^{a}}\right)dt,
\end{split}
\end{equation}
where $a$ and $\lambda$ are real numbers and the constant $C$ only depends on the coefficients of the equation (\ref{eq:thesystemRge}) and the 
numbers $\lambda$ and $a$. Moreover, 
\begin{align}
\sum_{k=k_{0}}^{k_{1}}\frac{2\pi e^{\lambda t_{k}}}{(\nuaverage{q}(\indexnot)e^{\baverage{q}t_{k}})^{a}} \leq & C
\int_{\bt_{0}}^{\bt_{1}}\frac{e^{\lambda t}}{(\nuaverage{q}(\indexnot)e^{\baverage{q}t})^{a-1}}dt,\label{eq:sumestint}\\
\sum_{k=k_{0}}^{k_{1}}\frac{2\pi e^{\lambda t_{k}}}{(\nuaverage{q}(\indexnot)e^{\baverage{q}t_{k}})^{a}}\mfH_{q}^{(r)}(\indexnot,t_{k}) \leq & C
\int_{\bt_{0}}^{\bt_{1}}\frac{e^{\lambda t}}{(\nuaverage{q}(\indexnot)e^{\baverage{q}t})^{a-1}}\mfH_{q}^{(r)}(\indexnot,t) dt,\label{eq:sumestmfHint}
\end{align}
where $a$, $\lambda$ and $r>0$ are real numbers and the constants $C$ only depend on the coefficients of the equation (\ref{eq:thesystemRge}) 
and the numbers $\lambda$, $a$ and $r$.
\end{lemma}
\begin{proof}
Assume that $[t_{k},t_{k+1}]\subseteq I_{q}$. Let $\lambda\in\ro$ and estimate
\begin{equation}\label{eq:elatdiff}
|e^{\lambda t}-e^{\lambda s}|\leq e^{\lambda u}|e^{\lambda(t-u)}-e^{\lambda(s-u)}|\leq C\frac{e^{\lambda u}}{\nuaverage{q}(\indexnot)e^{\baverage{q}u}}
\end{equation}
for $s,t,u\in [t_{k},t_{k+1}]$, where we appealed to (\ref{eq:gttaroughge}), (\ref{eq:tatbroughge}), (\ref{eq:njnogequ}) and the 
fact that $\mfg(\indexnot,t_{k})\geq 1$. Note also that $C$ only depends on $\lambda$ and the coefficients of the equation 
(\ref{eq:thesystemRge}). By a similar argument, there is a constant $C_{\lambda}>1$, depending only on $\lambda$, such that 
\begin{equation}\label{eq:elambdastequiv}
C_{\lambda}^{-1}e^{\lambda s}\leq e^{\lambda t}\leq C_{\lambda}e^{\lambda s}
\end{equation}
for $s,t\in [t_{k},t_{k+1}]$. In addition, (\ref{eq:elatdiff}) yields
\[
\left|\nuaverage{q}(\indexnot)e^{\baverage{q}t}-\nuaverage{q}(\indexnot)e^{\baverage{q}t_{k}}\right|\leq C
\]
for $t\in [t_{k},t_{k+1}]$, where $C$ only depends on the coefficients of the equation (\ref{eq:thesystemRge}). By a similar argument,
\begin{equation}\label{eq:mfHjdiff}
|\mfH_{q}(\indexnot,t)-\mfH_{q}(\indexnot,s)|\leq \frac{C}{\nuaverage{q}(\indexnot)e^{\baverage{q}u}}\mfH_{q}(\indexnot,u)
\end{equation}
for all $s,t,u\in [t_{k},t_{k+1}]$, where $C$ only depends on the coefficients of the equation (\ref{eq:thesystemRge}). Note, in particular, 
that $\mfH_{q}(\indexnot,t)$ and $\mfH_{q}(\indexnot,s)$ are equivalent for 
$s,t\in [t_{k},t_{k+1}]$; recall that $\nuaverage{q}(\indexnot)e^{\baverage{q}t}$ and $\mfg(\indexnot,t)$ are equivalent due to (\ref{eq:njnogequ})
and that $\mfg(\indexnot,t)\geq 1$ in $[t_{k},t_{k+1}]$ due to Lemma~\ref{lemma:gaintvarge}. Due to (\ref{eq:gttaroughge}), (\ref{eq:oogdiffge}), 
(\ref{eq:mfgmmfgdiag}), (\ref{eq:ginvminnjvergest}) and the fact that 
$\mfg(\indexnot,t)$, $\mfg_{\diag}(\indexnot,t)$ and $\nuaverage{q}(\indexnot)e^{\baverage{q}t}$ are equivalent in $I_{q}$ 
\begin{equation}\label{eq:mfgnjnoinf}
\left|\mfg(\indexnot,t)-\nuaverage{q}(\indexnot)e^{\baverage{q}t_{k}}\right|\leq C\mfg(\indexnot,t_{k})[e^{-\eta_{\diag}t_{k}}
+\mfH_{q}(\indexnot,t_{k})]+C
\end{equation}
for all $t\in [t_{k},t_{k+1}]$, where $C$ only depends on the coefficients of the equation (\ref{eq:thesystemRge}). For future reference, it is of 
interest to keep in mind that we could replace the $t_{k}$ appearing on the right hand side of (\ref{eq:mfgnjnoinf}) with any $s\in [t_{k},t_{k+1}]$; 
this is a consequence of (\ref{eq:gttaroughge}), (\ref{eq:elambdastequiv}) and (\ref{eq:mfHjdiff}). On the other hand, 
\begin{equation}\label{eq:intinteofnjnoinf}
\begin{split}
2\pi = & \int_{t_{k}}^{t_{k+1}}\mfg(\indexnot,t)dt=\int_{t_{k}}^{t_{k+1}}[\mfg(\indexnot,t)-\nuaverage{q}(\indexnot)e^{\baverage{q}t_{k}}]dt
 +(t_{k+1}-t_{k})\nuaverage{q}(\indexnot)e^{\baverage{q}t_{k}}.
\end{split}
\end{equation}
Combining (\ref{eq:tatbroughge}), (\ref{eq:njnogequ}), (\ref{eq:mfgnjnoinf}) and (\ref{eq:intinteofnjnoinf}) yields
\begin{equation}\label{eq:tkpomtkest}
\left| t_{k+1}-t_{k}-\frac{2\pi}{\nuaverage{q}(\indexnot)e^{\baverage{q}t_{k}}}\right|
\leq \frac{C}{\mfg(\indexnot,t_{k})}\left[e^{-\eta_{\diag}t_{k}}+\mfH_{q}(\indexnot,t_{k})\right]+\frac{C}{\mfg^{2}(\indexnot,t_{k})},
\end{equation}
where $C$ has the same dependence as before. For future reference, it is of interest to note that $t_{k}$ can be replaced by 
$t_{k+1}$ on the right hand side and in the third term inside the absolute value sign on the left hand side. 
One particular consequence of (\ref{eq:tkpomtkest}) is that 
\begin{equation*}
\begin{split}
 & \left|\frac{2\pi e^{\lambda t_{k}}}{(\nuaverage{q}(\indexnot)e^{\baverage{q}t_{k}})^{a}}-
\frac{e^{\lambda t_{k}}}{(\nuaverage{q}(\indexnot)e^{\baverage{q}t_{k}})^{a-1}}(t_{k+1}-t_{k})\right|\\
 \leq & \frac{Ce^{\lambda t_{k}}}{(\nuaverage{q}(\indexnot)e^{\baverage{q}t_{k}})^{a}}\left[e^{-\eta_{\diag}t_{k}}+\mfH_{q}(\indexnot,t_{k})\right]
+\frac{Ce^{\lambda t_{k}}}{(\nuaverage{q}(\indexnot)e^{\baverage{q}t_{k}})^{a+1}},
\end{split}
\end{equation*}
where we appealed to (\ref{eq:njnogequ}) and $C$ only depends on the coefficients of the equation (\ref{eq:thesystemRge}).
On the other hand,
\begin{equation*}
\begin{split}
 & \left|\int_{t_{k}}^{t_{k+1}}\frac{e^{\lambda t}}{(\nuaverage{q}(\indexnot)e^{\baverage{q}t})^{a-1}}dt-
\frac{e^{\lambda t_{k}}}{(\nuaverage{q}(\indexnot)e^{\baverage{q}t_{k}})^{a-1}}(t_{k+1}-t_{k})\right|
 \leq  \frac{Ce^{\lambda t_{k}}}{(\nuaverage{q}(\indexnot)e^{\baverage{q}t_{k}})^{a+1}},
\end{split}
\end{equation*}
where we appealed to (\ref{eq:tatbroughge}), (\ref{eq:njnogequ}) and (\ref{eq:elatdiff}), and $C$ only depends on $a$, $\lambda$ and the 
coefficients of the equation (\ref{eq:thesystemRge}). Combining these two estimates with (\ref{eq:tatbroughge}), (\ref{eq:njnogequ}) and the 
previously mentioned equivalences yields
\begin{equation*}
\begin{split}
 & \left|\frac{2\pi e^{\lambda t_{k}}}{(\nuaverage{q}(\indexnot)e^{\baverage{q}t_{k}})^{a}}-
\int_{t_{k}}^{t_{k+1}}\frac{e^{\lambda t}}{(\nuaverage{q}(\indexnot)e^{\baverage{q}t})^{a-1}}dt\right|\\
 \leq & \int_{t_{k}}^{t_{k+1}}\left(\frac{Ce^{\lambda t}}{(\nuaverage{q}(\indexnot)e^{\baverage{q}t})^{a-1}}\left[e^{-\eta_{\diag}t}
+\mfH_{q}(\indexnot,t)\right]+\frac{Ce^{\lambda t}}{(\nuaverage{q}(\indexnot)e^{\baverage{q}t})^{a}}\right)dt,
\end{split}
\end{equation*}
an estimate which implies (\ref{eq:sumapproxint}). Note also that $C$ only depends on $a$, $\lambda$ and the coefficients of the equation 
(\ref{eq:thesystemRge}). The estimates (\ref{eq:sumestint}) and (\ref{eq:sumestmfHint}) follow by similar, but simpler, arguments. 
\end{proof}

Next, it is of interest to compute the integrals appearing in (\ref{eq:sumapproxint}). Before doing so, let us introduce the 
following terminology. 

\begin{definition}\label{def:laajIajdef}
Consider the equation (\ref{eq:thesystemRge}). Assume that it is diagonally dominated, balanced and convergent; cf. 
Definition~\ref{def:convergeq}. Fix $0\neq \indexnot\in\EFindexset$ and let $I_{q}$ be a corresponding frequency era; 
cf. Definition~\ref{def:era}. Let $\baverage{q}$ and $\nuaverage{q}(\indexnot)$ be defined as at the beginning of 
Section~\ref{section:simplmacoeff}. Given $\lambda,a\in\ro$ and $t\in \ro$, define $\lambda_{a,q}$ and $I_{a,q}(\lambda,t)$ by 
\[
\lambda_{a,q}:=\lambda-(a-1)\baverage{q},\ \ \
I_{a,q}(\lambda,t):=\frac{1}{\nuaverage{q}^{a-1}(\indexnot)}\left\{\begin{array}{cc} t & \lambda_{a,q}=0,\\ 
\lambda_{a,q}^{-1}e^{\lambda_{a,q} t} & \lambda_{a,q}\neq 0.\end{array}
\right.
\]
\end{definition}
\textbf{Computing the integrals.} With the notation of Definition~\ref{def:laajIajdef}, 
\[
\int_{\bt_{0}}^{\bt_{1}}\frac{e^{\lambda t}}{(\nuaverage{q}(\indexnot)e^{\baverage{q}t})^{a-1}}dt=I_{a,q}(\lambda,\bt_{1})-I_{a,q}(\lambda,\bt_{0}).
\]
Most of the integrals appearing in Lemma~\ref{lemma:compsumswithintegrals} can be computed directly using this formula (with various values of
$\lambda$ and $a$). However, it is of interest to consider a generalised version of the integral involving $\mfH_{q}$. 
Let $0<r\in\ro$ and recall the notation introduced in (\ref{eq:mfhmfHjdef}). Then
\begin{equation}\label{eq:Hjrint}
\begin{split}
 & \int_{\bt_{0}}^{\bt_{1}}\frac{e^{\lambda t}}{(\nuaverage{q}(\indexnot)e^{\baverage{q}t})^{a-1}}\mfH_{q}^{(r)}(t)dt\\
 = & \sum_{p\neq q}\nuaverage{p}^{r}(\indexnot)\left[I_{a+r,q}(\lambda+r\baverage{p},\bt_{1})-I_{a+r,q}(\lambda+r\baverage{p},\bt_{0})\right].
\end{split}
\end{equation}
Assume first that $(\lambda+r\baverage{p})_{a+r,q}\neq 0$ for all $p\neq q$. Then
\begin{equation}\label{eq:Hjamointnd}
\begin{split}
\sum_{p\neq q}\nuaverage{p}^{r}(\indexnot)I_{a+r,q}(\lambda+r\baverage{p},t)
 = & \sum_{p\neq q}\frac{1}{(\lambda+r\baverage{p})_{a+r,q}}\frac{\nuaverage{p}^{r}(\indexnot)}{\nuaverage{q}^{r}(\indexnot)}
\frac{1}{\nuaverage{q}^{a-1}(\indexnot)}e^{(\lambda+r\baverage{p})_{a+r,q} t}\\
 = & \sum_{p\neq q}\frac{1}{(\lambda+r\baverage{p})_{a+r,q}}\frac{\nuaverage{p}^{r}(\indexnot)e^{r\baverage{p}t}}{\nuaverage{q}^{r}(\indexnot)e^{r\baverage{q}t}}
\frac{1}{(\nuaverage{q}(\indexnot)e^{\baverage{q}t})^{a-1}}e^{\lambda t}.
\end{split}
\end{equation}
Second, assume that there is one $p$, say $s$, such that $(\lambda+r\baverage{p})_{a+r,q}=0$ (note that there cannot be more than one such 
$p$, since $(\lambda+r\baverage{p})_{a+r,q}$ is strictly increasing with $p$). Then
\begin{equation}\label{eq:Hjamointd}
\begin{split}
\sum_{p\neq q}\nuaverage{p}^{r}(\indexnot)I_{a+r,q}(\lambda+r\baverage{p},t)
 = & \sum_{p\neq q,s}\frac{1}{(\lambda+r\baverage{p})_{a+r,q}}\frac{\nuaverage{p}^{r}(\indexnot)e^{r\baverage{p}t}}{\nuaverage{q}^{r}(\indexnot)e^{r\baverage{q}t}}
\frac{1}{(\nuaverage{q}(\indexnot)e^{\baverage{q}t})^{a-1}}e^{\lambda t}\\
 & +\frac{\nuaverage{s}^{r}(\indexnot)e^{r\baverage{s}t}}{\nuaverage{q}^{r}(\indexnot)e^{r\baverage{q}t}}
\frac{1}{(\nuaverage{q}(\indexnot)e^{\baverage{q}t})^{a-1}}te^{\lambda t}.
\end{split}
\end{equation}

\chapter[Non-degenerate equations]{Non-degenerate, diagonally dominated, convergent and balanced 
equations}\label{chapter:ndddconandbalancedeq}

\section{Introduction}

In the present chapter, we consider non-degenerate, diagonally dominated, convergent and balanced equations. The notion of a diagonally dominated, 
balanced and convergent equation is introduced in Definition~\ref{def:convergeq}. The requirement of non-degeneracy amounts to 
demanding that the set consisting of the $\b_{j}$, $j=1,\dots,d$, and the $\bRie{r}$, $r=1,\dots,R$, contain $d+R$ distinct elements; cf. 
Definition~\ref{def:nondegconvabal} below. For solutions to such equations, we are here interested in estimating how the energies of the associated
modes evolve over time; in later chapters, we estimate the energy of the entire solution. However, even when considering a single mode, there are normally
several frequency eras to take into consideration; cf. Definition~\ref{def:era}. Moreover, how the energy evolves typically depends on the frequency 
era. This leads us to divide the analysis into the following parts. 

\textbf{Estimating the energy of a mode during one frequency era.} The basic tool
for estimating the energy associated with a mode is the equality (\ref{eq:wprekpertransvarpi}). Iterating this equality leads to matrix products
of the type appearing in (\ref{eq:differenceeqrepres}). Our first goal is to estimate such matrix products. Since the result depends on the 
frequency era to which the corresponding elements of the time sequence belong, we start, in Section~\ref{section:modeonefreqera}, to fix one frequency
era, say $I_{q}$, and to estimate the norm of matrix products such that each factor corresponds to an element in the time sequence belonging to $I_{q}$. 
The relevant result is Lemma~\ref{lemma:maprnoes}. In the end, it turns out that, assuming $\baverage{q}\neq 0$, the estimates we obtain are
essentially optimal; we demonstrate this in Lemma~\ref{lemma:lbmatprod}. However, when $\baverage{q}=0$, there is room for improvement. 
We therefore treat this case separately in Section~\ref{section:refmodeanaltranspera}. When $\baverage{q}=0$, it is natural to divide the analysis into
two cases. First, there is the case that $\nuaverage{q}(\indexnot)$ is large. Then the conclusions of Lemma~\ref{lemma:maprnoes} are 
satisfactory. When $\nuaverage{q}(\indexnot)$ is small, basing an analysis on equalities such as (\ref{eq:wprekpertransvarpi}) is, however, not a 
good idea. In fact, it is better to consider the relevant system of differential equations directly. We derive the corresponding estimates in 
Lemma~\ref{lemma:Iqestbavqeqzero} below. 

\textbf{Estimating the energy of a mode during several frequency eras.}
Given the results of Lemma~\ref{lemma:maprnoes} and the refined analysis in the case of transparent eras, it is possible to draw conclusions concerning 
how the energy evolves over several frequency eras. This is the subject of Section~\ref{section:analysisforonemode}. In fact, the main result of the 
present chapter is Corollary~\ref{cor:solnoest} of Section~\ref{section:analysisforonemode}. Even though the corresponding estimates are applicable to 
time intervals containing several frequency eras, they are, however, still restricted to intervals corresponding to a time sequence of 
the type introduced in Definition~\ref{def:tkdefge}. In other words, the estimates only apply in situations where the behaviour is oscillatory. On the
other hand, the non-oscillatory setting can be dealt with using estimates of the type derived in Chapters~\ref{chapter:roughanalysisODEregion} and
\ref{chapter:ODEtransp}. In later chapters, we combine the results relevant in the oscillatory setting with the ones relevant in the non-oscillatory
setting in order to obtain a complete picture for one fixed mode $\indexnot$ and all future times. Combining the corresponding estimates then yields 
a complete picture for solutions to (\ref{eq:thesystemRge}). 

\textbf{Improved estimates for unbounded frequency eras.} For $\baverage{q}\neq 0$, the estimates derived in Lemma~\ref{lemma:maprnoes} are essentially
optimal. In fact, the only problem that might occur is that there might be non-trivial Jordan blocks of certain matrices; this leads to an $\e$-loss in the 
estimates. However, for frequency eras of the form $I_{q}=[t_{0},\infty)$, this loss can be removed even in the presence of non-trivial Jordan blocks. 
This is the subject of Section~\ref{section:thecaseofunbdfreeras}. 

\textbf{Optimality.} Finally, in Section~\ref{ssection:optimality}, we turn to the question of optimality. In case $\baverage{q}=0$, this 
issue is settled in Chapters~\ref{chapter:ODEtransp} and \ref{chapter:asympttranspcase}. In case $\baverage{q}\neq 0$, we demonstrate that 
the estimates derived in Lemma~\ref{lemma:maprnoes} concerning the matrix products are optimal.

\section{Mode analysis in one frequency era}\label{section:modeonefreqera}

In the present chapter, we focus on non-degenerate equations, a notion we now define. 

\begin{definition}\label{def:nondegconvabal}
Consider the equation (\ref{eq:thesystemRge}). Assume that it is diagonally dominated, balanced and convergent; cf. 
Definition~\ref{def:convergeq}. If the set consisting of the $\b_{j}$, $j=1,\dots,d$, and the $\bRie{r}$, $r=1,\dots,R$,
contains $d+R$ distinct elements, then (\ref{eq:thesystemRge}) is said to be \textit{non-degenerate, diagonally dominated, balanced and 
convergent}. 
\index{Non-degenerate, diagonally dominated, balanced and convergent!equation}%
\index{Equation!non-degenerate, diagonally dominated, balanced and convergent}%
\end{definition}

The following objects will play a central role in the analysis to follow.

\begin{definition}\label{def:tXqinfRqpmkappaqpm}
If (\ref{eq:thesystemRge}) is non-degenerate, diagonally dominated, balanced and convergent, then $Q=d+R$, using the notation introduced
at the beginning of Section~\ref{section:simplmacoeff}. Given $q\in \{1,\dots,Q\}$, define $\tX^{q}_{\infty}$ as follows. If $\baverage{q}=\b_{j}$
for some $j\in \{1,\dots,d\}$, then $j$ is uniquely determined by $q$ and
\begin{equation}\label{eq:tXqinfdef}
\tX^{q}_{\infty}:=(g^{jj}_{\infty})^{-1/2}X^{j}_{\infty}.
\end{equation}
If $\baverage{q}=\bRie{r}$ for some $r\in \{1,\dots,R\}$, then $r$ is uniquely determined by $q$ and $\tX^{q}_{\infty}:=0$. For $q\in\{1,\dots,Q\}$,
let 
\begin{equation}\label{eq:matrixdetgrowth}
R_{q,\pm}^{+}:=\pm\frac{1}{2}(-\a_{\infty}+\baverage{q}\Id_{m}+\tX^{q}_{\infty}),\ \  \
R_{q,\pm}^{-}:=\pm\frac{1}{2}(-\a_{\infty}+\baverage{q}\Id_{m}-\tX^{q}_{\infty}),
\end{equation}
and let
\begin{align}
\kappa_{q} := & \Rsp\left(\diag\{R_{q,+}^{+},R_{q,+}^{-}\}\right),\ \ \
\kappa_{q,\pm}:=\max\{\kappa_{\max}(R_{q,\pm}^{+}),\kappa_{\max}(R_{q,\pm}^{-})\},\label{eq:kappaqpmdef}\\
d_{q,\pm} := & d_{\max}(\diag\{R_{q,\pm}^{+},R_{q,\pm}^{-}\},\kappa_{q,\pm});\label{eq:dqpmdef}
\end{align}
\index{$\a$Aa@Notation!Matrix notation!kappaq@$\kappa_{q}$}%
cf. Definition~\ref{def:SpRspdef}. Finally, let 
\begin{equation}\label{eq:kappapmdef}
\kappa_{\rsp}:=\max_{q}\kappa_{q},\ \ \
\kappa_{\pm}:=\max_{q}\kappa_{q,\pm},
\end{equation}
\index{$\a$Aa@Notation!Matrix notation!kappa@$\kappa$}%
\end{definition}
\begin{remark}
It is of interest to keep in mind that $\kappa_{q}=\kappa_{q,+}+\kappa_{q,-}$. 
\end{remark}

Time sequences of the type introduced in Definition~\ref{def:tkdefge} will play a central role in the analysis to follow. However, for
technical reasons we need to restrict them slightly. 

\begin{definition}\label{def:tkdefgeaddbd}
Consider the equation (\ref{eq:thesystemRge}). Assume that it is non-degenerate, diagonally dominated, balanced and convergent; cf. 
Definition~\ref{def:nondegconvabal}. Let $0\neq \indexnot\in\EFindexset$ and assume $t_{0}\in [0,\infty)$ to 
be such that (\ref{eq:gtalbge}) holds with $t_{a}$ replaced by $t_{0}$. Assume, moreover, that 
\begin{equation}\label{eq:maxqdiaglbitokappaq}
\max_{q}\nuaverage{q}(\indexnot)e^{\baverage{q}t_{0}}\geq 2\pi (\kappa_{\rsp}+1),
\end{equation}
where $\kappa_{\rsp}$ is defined by (\ref{eq:kappapmdef}). Assume that $t_{k}\geq t_{0}$ has been defined for some $0\leq k\in\zo$ and that 
(\ref{eq:gtalbge}) and (\ref{eq:maxqdiaglbitokappaq}) hold with $t_{a}$ and $t_{0}$ replaced by $t_{k}$. Then $t_{k+1}$ is defined by the 
condition that (\ref{eq:inttktkpomfgcond}) holds.
If there is a final $k$ such that (\ref{eq:gtalbge}) and (\ref{eq:maxqdiaglbitokappaq}) hold with $t_{a}$ and $t_{0}$ replaced by $t_{k}$, 
then this $k$ is denoted by $k_{\fin,\ror}$. Moreover, $t_{\fin,\ror}:=t_{k_{\fin,\ror}}$. If there is no such final $k$, then $k_{\fin,\ror}:=\infty$ 
and $t_{\fin,\ror}:=\infty$. If, in addition to the above, $t_{0}\geq 2\pi$, then 
$t_{-1}$ is defined by the condition that (\ref{eq:inttktkpomfgcond}) holds with $k=-1$. Assume $t_{k}\leq t_{0}$ has been defined for some 
$0\geq k\in\zo$; that (\ref{eq:gtalbge}) and (\ref{eq:maxqdiaglbitokappaq}) hold with $t_{0}$ and $t_{a}$ replaced by $t_{k}$; and that 
$t_{k}\geq 2\pi$. Then $t_{k-1}$ is defined by the 
condition that (\ref{eq:inttktkpomfgcond}) holds with $k$ replaced by $k-1$. Let $k_{\roini,\ror}$ be the first $0\geq k\in\zo$ such that either 
$0\leq t_{k}<2\pi$ or one of the inequalities (\ref{eq:gtalbge}) and (\ref{eq:maxqdiaglbitokappaq}) (with $t_{0}$ and $t_{a}$ replaced by $t_{k}$)
is violated. Finally, $t_{\roini,\ror}:=t_{k_{\roini,\ror}}$. 
\end{definition}
\begin{remark}
Note that since the equation is diagonally dominated, balanced and convergent, it is also oscillation adapted; cf. Lemma~\ref{lemma:consconvcoeff}.
Given that $t_{0}$ satisfies the conditions of the definition, the sequence $t_{k}$ is thus simply a restriction of the sequence defined in 
Definition~\ref{def:tkdefge}. The conclusions derived concerning time sequences of the type introduced in Definition~\ref{def:tkdefge} thus 
apply to the time sequences defined here.
\end{remark}
\begin{remark}
Due to (\ref{eq:mfgdiagdef}), it is clear that 
\[
\max_{q}\nuaverage{q}(\indexnot)e^{\baverage{q}t}\leq\mfg_{\diag}(\indexnot,t)\leq \sqrt{Q}\max_{q}\nuaverage{q}(\indexnot)e^{\baverage{q}t}.
\]
Combining this estimate with (\ref{eq:mfgmfgdiageq}), it is clear that the estimate (\ref{eq:maxqdiaglbitokappaq}) is equivalent to 
a lower bound on $\mfg$ which only depends on the coefficients of the equation (\ref{eq:thesystemRge}).
\end{remark}

Next we turn to the question of how solutions develop in one frequency era. Here we focus on obtaining an upper bound on the norms of matrix products, 
but we return to the issue of optimality later. 

\begin{lemma}\label{lemma:maprnoes}
Consider the equation (\ref{eq:thesystemRge}). Assume that it is non-degenerate, diagonally dominated, balanced and convergent; cf. 
Definition~\ref{def:nondegconvabal}. Fix $0\neq \indexnot\in\EFindexset$ and let $I_{q}$ be a corresponding frequency era; cf. 
Definition~\ref{def:era}. Let $\{t_{k}\}$ be a sequence defined as in Definition~\ref{def:tkdefgeaddbd} and let $R_{q,\pm}^{+}$, $R_{q,\pm}^{-}$ and
$\kappa_{q,\pm}$ be given by Definition~\ref{def:tXqinfRqpmkappaqpm}. Then the following holds.
\begin{itemize}
\item Assume that the Jordan blocks of $R_{q,+}^{+}$ and $R_{q,+}^{-}$ corresponding to the eigenvalues with 
real part $\kappa_{q,+}$ are trivial. Then there are constants $C_{q,+}$ and $C_{\deg,q,+}$, where $C_{\deg,q,+}=0$ if $\baverage{q}\neq 0$,
depending only on the coefficients of the equation (\ref{eq:thesystemRge}), such that the following holds. If $t_{k}\in I_{q}$ for 
$k_{0}-1\leq k\leq k_{1}+1$, where $k_{0}\leq k_{1}$, then 
\begin{equation}\label{eq:Akprodndegplus}
\|A_{k_{1}}^{+}\cdots A_{k_{0}}^{+}\|\leq C_{q,+}\exp\left[\kappa_{q,+}(\bt_{1,+}-\bt_{0,+})
+\frac{C_{\deg,q,+}}{\nuaverage{q}(\indexnot)}(\bt_{1,+}-\bt_{0,+})\right],
\end{equation}
where $\bt_{0,+}=t_{k_{0}}$, $\bt_{1,+}=t_{k_{1}+1}$ and $A_{k}^{\pm}:=\Xi^{k,\pm}_{\fin}$; cf. Lemma~\ref{lemma:wkfinlemma}.
\item Assume that the Jordan blocks of $R_{q,-}^{+}$ and $R_{q,-}^{-}$ corresponding to the eigenvalues with 
real part $\kappa_{q,-}$ are trivial. Then there are constants $C_{q,-}$ and $C_{\deg,q,-}$, where $C_{\deg,q,-}=0$ if $\baverage{q}\neq 0$,
depending only on the coefficients of the equation (\ref{eq:thesystemRge}), such that the following holds. If $t_{k}\in I_{q}$ for 
$k_{0}-1\leq k\leq k_{1}+1$, where $k_{0}\leq k_{1}$, then 
\begin{equation}\label{eq:Akprodndegminus}
\|A_{k_{0}}^{-}\cdots A_{k_{1}}^{-}\|\leq C_{q,-}\exp\left[\kappa_{q,-}(\bt_{1,-}-\bt_{0,-})
+\frac{C_{\deg,q,-}}{\nuaverage{q}(\indexnot)}(\bt_{1,-}-\bt_{0,-})\right],
\end{equation}
where $\bt_{0,-}:=t_{k_{0}-1}$, $\bt_{1,-}:=t_{k_{1}}$ and $A_{k}^{\pm}$ is defined as above.
\item Assume that there is a non-trivial Jordan block of $R_{q,+}^{+}$ and $R_{q,+}^{-}$ corresponding to an eigenvalue with 
real part $\kappa_{q,+}$. For every $\e>0$, there are then constants $C_{q,\e,+}$ and $C_{\deg,q,+}$, where $C_{\deg,q,+}=0$ if 
$\baverage{q}\neq 0$, depending only on the coefficients of the equation (\ref{eq:thesystemRge}) and $\e$, such that the following
holds. If $t_{k}\in I_{q}$ for $k_{0}-1\leq k\leq k_{1}+1$, where $k_{0}\leq k_{1}$, then 
\begin{equation}\label{eq:Akproddegplus}
\|A_{k_{1}}^{+}\cdots A_{k_{0}}^{+}\|\leq C_{q,\e,+}
\exp\left[(\kappa_{q,+}+\e)(\bt_{1,+}-\bt_{0,+})+\frac{C_{\deg,q,+}}{\nuaverage{q}(\indexnot)}(\bt_{1,+}-\bt_{0,+})\right],
\end{equation}
where $\bt_{0,+}$, $\bt_{1,+}$ and $A_{k}^{+}$ are defined as above.
\item Assume that there is a non-trivial Jordan block of $R_{q,-}^{+}$ and $R_{q,-}^{-}$ corresponding to an eigenvalue with 
real part $\kappa_{q,-}$. For every $\e>0$, there are then constants $C_{q,\e,-}$ and $C_{\deg,q,-}$, where $C_{\deg,q,-}=0$ if 
$\baverage{q}\neq 0$, depending only on the coefficients of the equation (\ref{eq:thesystemRge}) and $\e$, such that the following
holds. If $t_{k}\in I_{q}$ for $k_{0}-1\leq k\leq k_{1}+1$, where $k_{0}\leq k_{1}$, then 
\begin{equation}\label{eq:Akproddegminus}
\|A_{k_{0}}^{-}\cdots A_{k_{1}}^{-}\|\leq C_{q,\e,-}
\exp\left[(\kappa_{q,-}+\e)(\bt_{1,-}-\bt_{0,-})+\frac{C_{\deg,q,-}}{\nuaverage{q}(\indexnot)}(\bt_{1,-}-\bt_{0,-})\right],
\end{equation}
where $\bt_{0,-}$, $\bt_{1,-}$ and $A_{k}^{-}$ are defined as above.
\end{itemize}
\end{lemma}
\begin{remark}
For a given $0\neq\indexnot\in\EFindexset$, it could be that there is no sequence defined as in Definition~\ref{def:tkdefgeaddbd}.
In that case, the lemma does not yield any conclusions. 
\end{remark}
\begin{remark}
It is important to keep in mind that the orders of the matrices appearing on the left hand sides of (\ref{eq:Akprodndegplus}) and 
(\ref{eq:Akprodndegminus}) are different; cf. also (\ref{eq:Akproddegplus}) and (\ref{eq:Akproddegminus}). This is due to the fact 
that in the case of the plus sign, we are starting at $t_{k_{0}}$ and evolving forwards to $t_{k_{1}}$. In the case of the minus sign
we are starting at $t_{k_{1}}$ and evolving backwards to $t_{k_{0}}$. 
\end{remark}
\begin{remark}\label{remark:kzkkonesuff}
If $k_{0}>k_{\roini,\ror}$ and $k_{1}<k_{\fin,\ror}$, it is sufficient to demand that $t_{k}\in I_{q}$ for $k_{0}\leq k\leq k_{1}$.
\end{remark}
\begin{proof}
We begin by estimating $A^{\pm}_{k}$ for $t_{k}\in I_{q}$. 

\textbf{Estimating $A^{\pm}_{k}$.} Since $A_{k}^{\pm}=\Xi^{k,\pm}_{\fin}$ and (\ref{eq:XikXikIpreestfin}) holds, it is of interest to simplify 
the matrix defined in (\ref{eq:Xikpreappdefstfin}). We therefore begin by simplifying $R^{ab}_{\pre,k}$, given by (\ref{eq:Rabprekdef}). 
Using the notation introduced in (\ref{eq:Xavdiagqindexnotdef}), 
\begin{equation}\label{eq:bXjinfndef}
\Xaverage{\diag}^{q}(\indexnot)=\pm\tX^{q}_{\infty},
\end{equation}
where $\tX^{q}_{\infty}$ is introduced in Definition~\ref{def:tXqinfRqpmkappaqpm}. In order to determine the sign, note that the right hand 
side of (\ref{eq:bXjinfndef}) is only non-zero when $q$ corresponds to a $j\in\{1,\dots,d\}$. Let us therefore assume this to be the 
case. In order for $I_{q}$ to be non-empty, it is then necessary for $n_{j}$ to be non-zero, and
\[
\Xaverage{\diag}^{q}(\indexnot)=\frac{n_{j}}{|n_{j}|}\tX^{q}_{\infty};
\]
cf. (\ref{eq:Xavdiagqindexnotdef}). Assume that $t_{k}\in I_{q}$. Then (\ref{eq:agagdiag}), (\ref{eq:njnogequ}) and  (\ref{eq:adiagbadiagjest}) 
yield 
\begin{equation}\label{eq:almfgapprox}
\left\|\frac{\a(t_{k})}{\mfg(\indexnot,t_{k})}-\frac{\a_{\infty}}{\nuaverage{q}(\indexnot)e^{\baverage{q}t_{k}}}\right\|
\leq \frac{C}{\nuaverage{q}(\indexnot)e^{\baverage{q}t_{k}}}\left[e^{-\eta_{\a}t_{k}}+\mfH_{q}(\indexnot,t_{k})\right],
\end{equation}
where $C$ only depends on the coefficients of (\ref{eq:thesystemRge}). Similarly, (\ref{eq:XmXdiag}), (\ref{eq:njnogequ}) and  
(\ref{eq:XdiagbXdiagjest}) yield
\begin{equation}\label{eq:Xmfgapprox}
\left\|
\frac{X(\indexnot,t_{k})}{\mfg(\indexnot,t_{k})}-\frac{\Xaverage{\diag}^{q}(\indexnot)}{\nuaverage{q}(\indexnot)e^{\baverage{q}t_{k}}}
\right\|\leq \frac{C}{\nuaverage{q}(\indexnot)e^{\baverage{q}t_{k}}}\left[e^{-\eta_{X}t_{k}}
+\sum_{s=1}^{2}\mfH_{q}^{(s)}(\indexnot,t_{k})\right],
\end{equation}
where $C$ only depends on the coefficients of (\ref{eq:thesystemRge}). Finally, (\ref{eq:elltmelldiagtst}), (\ref{eq:njnogequ}) and  
(\ref{eq:elldotgenestgest}) yield
\begin{equation}\label{eq:elldmfgapprox}
\left|
\frac{\dot{\ell}(\indexnot,t_{k})}{\mfg(\indexnot,t_{k})}-\frac{\baverage{q}}{\nuaverage{q}(\indexnot)e^{\baverage{q}t_{k}}}
\right|\leq \frac{C}{\nuaverage{q}(\indexnot)e^{\baverage{q}t_{k}}}
\left[e^{-\eta_{\diag}t_{k}}+\mfH_{q}(\indexnot,t_{k})\right],
\end{equation}
where $C$ only depends on the coefficients of (\ref{eq:thesystemRge}). Combining (\ref{eq:almfgapprox}), (\ref{eq:Xmfgapprox}) and 
(\ref{eq:elldmfgapprox}) yields
\begin{equation}\label{eq:Rabapproxest}
\begin{split}
 & \left\| R^{ab}_{\pre,k}-\frac{1}{\nuaverage{q}(\indexnot)e^{\baverage{q}t_{k}}}R^{ab}_{\infty,q}(\indexnot)\right\|
\leq \frac{C}{\nuaverage{q}(\indexnot)e^{\baverage{q}t_{k}}}\left[e^{-\eta_{R}t_{k}}
+\sum_{s=1}^{2}\mfH_{q}^{(s)}(\indexnot,t_{k})\right],
\end{split}
\end{equation}
where $C$ only depends on the coefficients of (\ref{eq:thesystemRge}), $R^{ab}_{\pre,k}$ is defined in (\ref{eq:Rabprekdef}),
\[
R^{ab}_{\infty,q}(\indexnot):=(-1)^{a}\a_{\infty}+\baverage{q}\Id_{m}+(-1)^{b}\Xaverage{\diag}^{q}(\indexnot)
\]
and $\eta_{R}:=\eta_{X}=\eta_{\a}\leq \eta_{\diag}$. Defining $A^{\app}_{k,q,\pm}$ by
\begin{equation}\label{eq:Rinfjn}
A^{\app}_{k,q,\pm}:=\Id_{2m}\pm\frac{2\pi}{\nuaverage{q}(\indexnot)e^{\baverage{q}t_{k}}}R_{\infty,q}(\indexnot),\ \ \
R_{\infty,q}(\indexnot):=\frac{1}{2}
\left(\begin{array}{cc} R^{10}_{\infty,q}(\indexnot) & 0 \\ 0 & R^{11}_{\infty,q}(\indexnot)\end{array}\right),
\end{equation}
Lemma~\ref{lemma:wkfinlemma} and (\ref{eq:Rabapproxest}) imply that 
\[
\|A_{k}^{\pm}-A^{\app}_{k,q,\pm}\|\leq \frac{C}{\nuaverage{q}(\indexnot)e^{\baverage{q}t_{k}}}\left[e^{-\eta_{A}t_{k}}
+\sum_{s=1}^{2}\mfH_{q}^{(s)}(\indexnot,t_{k})+\frac{1}{\nuaverage{q}(\indexnot)e^{\baverage{q}t_{k}}}\right],
\]
where $A_{k}^{\pm}=\Xi^{k,\pm}_{\fin}$, $\eta_{A}:=\min\{\eta_{R},\eta_{\rosh}\}$ and $C$ only depends on the coefficients of (\ref{eq:thesystemRge}). 
Note, however, that this estimate is only valid for $t_{k}\in I_{q}$. It is also of interest to note that the set 
$\{R^{+}_{q,+},R^{-}_{q,+}\}$ coincides with the set $\{R^{10}_{\infty,q}(\indexnot)/2,R^{11}_{\infty,q}(\indexnot)/2\}$. In this sense, the sign
issue arising in (\ref{eq:bXjinfndef}) does not affect the end result. For future reference, it is of interest to keep in mind that 
\begin{equation}\label{eq:kappaqpmrel}
\kappa_{q,+}=\kappa_{\max}[R_{\infty,q}(\indexnot)],\ \ \
\kappa_{q,-}=-\kappa_{\min}[R_{\infty,q}(\indexnot)];
\end{equation}
cf. Definition~\ref{def:SpRspdef}.

\textbf{Reformulating the matrix products.} As a next step, it is convenient to reformulate the matrix products by conjugating the
factors with appropriate matrices. To be more specific, fix $\e>0$ and let $T_{q,\e}$ be such that 
\[
J_{q,\e}:=T_{q,\e}R_{\infty,q}(\indexnot)T_{q,\e}^{-1}
\]
is in generalised Jordan normal form with the non-zero off-diagonal components equal to $\e$; cf. Remark~\ref{remark:genJordblock}. For the 
sake of brevity, we suppress the dependence of 
$J_{q,\e}$ and $T_{q,\e}$ on $\indexnot$; note, however, that the image of $R_{\infty,q}(\indexnot)$, considered as a function of $\indexnot$,
consists of at most two matrices, so that the constants appearing in the estimates below can be chosen to be independent of $\indexnot$.
Note also that 
\begin{align}
T_{q,\e}A_{k_{1}}^{+}\cdots A_{k_{0}}^{+}T_{q,\e}^{-1} = & A_{k_{1},q,\e}^{+}\cdots A_{k_{0},q,\e}^{+},\label{eq:Akdiagprodplus}\\
T_{q,\e}A_{k_{0}}^{-}\cdots A_{k_{1}}^{-}T_{q,\e}^{-1} = & A_{k_{0},q,\e}^{-}\cdots A_{k_{1},q,\e}^{-},\label{eq:Akdiagprodminus}
\end{align}
where
\begin{equation}\label{eq:Akjendef}
A_{k,q,\e}^{\pm}:=T_{q,\e}A_{k}^{\pm}T_{q,\e}^{-1}.
\end{equation}
Moreover, 
\[
\|A_{k,q,\e}^{\pm}-A^{\app,\pm}_{k,q,\e}\|\leq \frac{C_{\e}}{\nuaverage{q}(\indexnot)e^{\baverage{q}t_{k}}}\left[e^{-\eta_{A}t_{k}}
+\sum_{s=1}^{2}\mfH_{q}^{(s)}(\indexnot,t_{k})+\frac{1}{\nuaverage{q}(\indexnot)e^{\baverage{q}t_{k}}}\right],
\]
where $C_{\e}$ only depends on the coefficients of (\ref{eq:thesystemRge}) and $\e$, and 
\[
A^{\app,\pm}_{k,q,\e}:=\Id_{2m}\pm\frac{2\pi}{\nuaverage{q}(\indexnot)e^{\baverage{q}t_{k}}}J_{q,\e}.
\]
In order to obtain an upper bound on the norms of the matrix products on the right hand sides of (\ref{eq:Akdiagprodplus}) and
(\ref{eq:Akdiagprodminus}), it is convenient to consider
\begin{align}
e^{-\lambda (\bt_{1,+}-\bt_{0,+})}A_{k_{1},q,\e}^{+}\cdots A_{k_{0},q,\e}^{+} = & \hA_{k_{1},q,\e}^{+}\cdots \hA_{k_{0},q,\e}^{+},\label{eq:Akprodscaleredplus}\\
e^{-\lambda (\bt_{0,-}-\bt_{1,-})}A_{k_{0},q,\e}^{-}\cdots A_{k_{1},q,\e}^{-} = & \hA_{k_{0},q,\e}^{-}\cdots \hA_{k_{1},q,\e}^{-},\label{eq:Akprodscaleredminus}
\end{align}
where $\lambda$ is a real number and 
\begin{equation}\label{eq:hAkjendef}
\hA_{k,q,\e}^{\pm}:=e^{-\lambda (t_{k\pm 1}-t_{k})}A_{k,q,\e}^{\pm}.
\end{equation}
Note that 
\[
\left|e^{-\lambda (t_{k\pm 1}-t_{k})}-1+\lambda (t_{k\pm 1}-t_{k})\right|\leq\frac{C_{\lambda}}{\mfg^{2}(\indexnot,t_{k})}
\]
where the constant $C_{\lambda}$ only depends on $\lambda$ and we appealed to (\ref{eq:tatbroughge}); for future reference, 
it is of interest to keep in mind that a similar estimate holds when $\lambda$ is complex. Combining this estimate with 
(\ref{eq:tkpomtkest}) (and the comment made in connection with it) yields 
\begin{equation}\label{eq:elambdadelttest}
\left|e^{-\lambda (t_{k\pm 1}-t_{k})}-1\pm\frac{2\pi\lambda}{\nuaverage{q}(\indexnot)e^{\baverage{q}t_{k}}}\right|\leq
\frac{C_{\lambda}}{\mfg(\indexnot,t_{k})}\left[e^{-\eta_{\diag}t_{k}}+\mfH_{q}(\indexnot,t_{k})\right]+\frac{C_{\lambda}}{\mfg^{2}(\indexnot,t_{k})},
\end{equation}
where $C_{\lambda}$ only depends on the coefficients of (\ref{eq:thesystemRge}) and $\lambda$; again, a similar estimate
holds when $\lambda$ is complex. Thus
\[
\|\hA_{k,q,\e}^{\pm}-\hA^{\app,\pm}_{k,q,\e}\|\leq \frac{C_{\e,\lambda}}{\nuaverage{q}(\indexnot)e^{\baverage{q}t_{k}}}\left[e^{-\eta_{A}t_{k}}
+\sum_{s=1}^{2}\mfH_{q}^{(s)}(\indexnot,t_{k})+\frac{1}{\nuaverage{q}(\indexnot)e^{\baverage{q}t_{k}}}\right],
\]
where
\[
\hA^{\app,\pm}_{k,q,\e}:=\Id_{2m}\pm\frac{2\pi}{\nuaverage{q}(\indexnot)e^{\baverage{q}t_{k}}}[J_{q,\e}-\lambda\Id_{2m}].
\]
Moreover, $C_{\e,\lambda}$ only depends on the coefficients of (\ref{eq:thesystemRge}), $\e$ and $\lambda$.  

Before proceeding, it is of interest to eliminate the imaginary parts of the eigenvalues of 
$J_{q,\e}$. Note that $J_{q,\e}-\lambda\Id_{2m}$ consists of generalised Jordan blocks. Assume that 
the blocks have dimension $l_{1},\dots,l_{D}$. Let $\zeta_{r}$ be the imaginary part 
of the eigenvalue corresponding to the $r$'th generalised Jordan block. Define $S_{k}$ and $\chA_{k,q,\e}^{\pm}$ via
\begin{equation}\label{eq:ChAkjedef}
S_{k}:=\mathrm{diag}\{e^{-i\zeta_{1}t_{k}}\Id_{l_{1}},\dots,e^{-i\zeta_{D}t_{k}}\Id_{l_{D}}\},\ \ \
\chA_{k,q,\e}^{\pm}:=S_{k\pm 1}\hA_{k,q,\e}^{\pm}S_{k}^{-1}.
\end{equation}
Note that $\|S_{k}\|=\|S_{k}^{-1}\|=1$ and that 
\begin{align}
S_{k_{1}+1}\hA_{k_{1},q,\e}^{+}\cdots \hA_{k_{0},q,\e}^{+}S_{k_{0}}^{-1} = & \chA_{k_{1},q,\e}^{+}\cdots \chA_{k_{0},q,\e}^{+},\label{eq:AkprodSkredplus}\\
S_{k_{0}-1}\hA_{k_{0},q,\e}^{-}\cdots \hA_{k_{1},q,\e}^{-}S_{k_{1}}^{-1} = & \chA_{k_{0},q,\e}^{-}\cdots \chA_{k_{1},q,\e}^{-}.\label{eq:AkprodSkredminus}
\end{align}
Moreover, 
\begin{equation}\label{eq:chAkmchAkapp}
\|\chA_{k,q,\e}^{\pm}-\chA^{\app,\pm}_{k,q,\e}\|\leq \frac{C_{\e,\lambda}}{\nuaverage{q}(\indexnot)e^{\baverage{q}t_{k}}}\left[e^{-\eta_{A}t_{k}}
+\sum_{s=1}^{2}\mfH_{q}^{(s)}(\indexnot,t_{k})+\frac{1}{\nuaverage{q}(\indexnot)e^{\baverage{q}t_{k}}}\right],
\end{equation}
where
\begin{equation}\label{eq:chAkjedef}
\chA^{\app,\pm}_{k,q,\e}:=\Id_{2m}\pm\frac{2\pi}{\nuaverage{q}(\indexnot)e^{\baverage{q}t_{k}}}[\mathrm{Re}\{J_{q,\e}\}-\lambda\Id_{2m}];
\end{equation}
this is a consequence of an estimate similar to (\ref{eq:elambdadelttest}). Finally, the constant $C_{\e,\lambda}$ appearing in 
(\ref{eq:chAkmchAkapp}) only depends on $\e$, $\lambda$ and the coefficients of the equation (\ref{eq:thesystemRge}).

\textbf{Estimating the matrix products.} Let $\kappa_{q,+}$ be defined by (\ref{eq:kappaqpmdef}) and fix a real number $\lambda$ 
such that $\kappa_{q,+}<\lambda\leq\kappa_{q,+}+1/2$. By choosing $\e>0$ small enough (depending only on $\lambda$ and the coefficients 
of the equation), it can then be ensured that $\|\chA^{\app,+}_{k,q,\e}\|\leq 1$. The reason for this is the following. Since $t_{k}$ is 
an element of the sequence introduced in Definition~\ref{def:tkdefgeaddbd} and since $t_{k}\in I_{q}$, we know that 
\[
\nuaverage{q}(\indexnot)e^{\baverage{q}t_{k}}=\max_{\rho}\nuaverage{\rho}(\indexnot)e^{\baverage{\rho}t_{k}}\geq 2\pi (\kappa+1).
\]
Thus the diagonal components of the second term on the right hand side of (\ref{eq:chAkjedef}) are bounded in absolute value by
$(\kappa+1/2)/(\kappa+1)$; recall that (\ref{eq:kappaqpmrel}) holds. On the other hand, the diagonal components of 
$\mathrm{Re}\{J_{q,\e}\}-\lambda\Id_{2m}$ are bounded from
above by $\kappa_{q,+}-\lambda<0$. Choosing $\e>0$ small enough (the bound depending only on the coefficients of the equation and $\lambda$),
it can then be ensured that in the case of the $+$-sign, the second term on the right hand side of (\ref{eq:chAkjedef}) is negative 
definite and has a norm bounded from above by $1$. Moreover, $\|\chA^{\app,+}_{k,q,\e}\|\leq 1$. In case all the generalised Jordan blocks corresponding
to eigenvalues with real part $\kappa_{q,+}$ are trivial, $\lambda$ can be chosen to equal $\kappa_{q,+}$ and $\e>0$ can be chosen to be a 
fixed number depending only on the difference between $\kappa_{q,+}$ and the second largest real part of an eigenvalue of $J_{q,\e}$ (in 
case there is no second largest real part, there are no non-trivial generalised Jordan blocks, and then there is no need for $\e$).

By a similar argument, if we fix $\lambda$ so that $-\kappa_{q,-}-1/2\leq\lambda<-\kappa_{q,-}$, then there is an $\e>0$ (depending only
on $\lambda$ and the coefficients of the equation (\ref{eq:thesystemRge})) so that $\|\chA^{\app,-}_{k,q,\e}\|\leq 1$. Again, in case all the 
generalised Jordan blocks corresponding
to eigenvalues with real part $\kappa_{q,-}$ are trivial, $\lambda$ can be chosen to equal $-\kappa_{q,-}$ and $\e>0$ can be chosen to be a 
fixed number depending only on the difference between $-\kappa_{q,-}$ and the second smallest real part of an eigenvalue of $J_{q,\e}$.

With $\lambda$ and $\e$ chosen as above,
\begin{align*}
\|\chA_{k_{1},q,\e}^{+}\cdots \chA_{k_{0},q,\e}^{+}\|
 \leq &  \left(1+\|\chA_{k_{1},q,\e}^{+}-\chA^{\app,+}_{k_{1},q,\e}\|\right)
\cdots \left(1+\|\chA_{k_{0},q,\e}^{+}-\chA^{\app,+}_{k_{0},q,\e}\|\right),\\
 \|\chA_{k_{0},q,\e}^{-}\cdots \chA_{k_{1},q,\e}^{-}\|
 \leq & \left(1+\|\chA_{k_{1},q,\e}^{-}-\chA^{\app,-}_{k_{1},q,\e}\|\right)
\cdots \left(1+\|\chA_{k_{0},q,\e}^{-}-\chA^{\app,-}_{k_{0},q,\e}\|\right).
\end{align*}
Since $\ln (1+x)\leq x$ for $x\geq 0$, we conclude that the logarithms of the right hand sides are bounded from above by
\begin{equation}\label{eq:finsumtobeest}
\textstyle{\sum}_{k=k_{0}}^{k_{1}}\|\chA_{k,q,\e}^{\pm}-\chA^{\app,\pm}_{k,q,\e}\|.
\end{equation}
Due to (\ref{eq:chAkmchAkapp}), it is thus of interest to estimate the sum of the right hand side of (\ref{eq:chAkmchAkapp}).
Let us begin with 
\[
\sum_{k=k_{0}}^{k_{1}}\frac{1}{\nuaverage{q}(\indexnot)e^{\baverage{q}t_{k}}}e^{-\eta_{A}t_{k}}\leq C\int_{\bt_{0,+}}^{\bt_{1,+}}e^{-\eta_{A}t}dt\leq C,
\]
where we appealed to (\ref{eq:sumestint}) and the constant $C$ only depends on the coefficients of the equation (\ref{eq:thesystemRge}). 
Moreover, (\ref{eq:sumestint}) implies that
\[
\sum_{k=k_{0}}^{k_{1}}\frac{1}{\nuaverage{q}^{2}(\indexnot)e^{2\baverage{q}t_{k}}}
\leq C\int_{\bt_{0,+}}^{\bt_{1,+}}\frac{1}{\nuaverage{q}(\indexnot)e^{\baverage{q}t}}dt,
\]
where $C$ has the same dependence as before. If $\baverage{q}\neq 0$, the right hand side of this inequality is bounded. However, if $\baverage{q}=0$, 
we only have the bound
\begin{equation}\label{eq:sumbjeqz}
\sum_{k=k_{0}}^{k_{1}}\frac{1}{\nuaverage{q}^{2}(\indexnot)e^{2\baverage{q}t_{k}}}\leq \frac{C}{\nuaverage{q}(\indexnot)}(\bt_{1,+}-\bt_{0,+}).
\end{equation}
On the other hand, the constant $C$ only depends on the coefficients of the equation (\ref{eq:thesystemRge}). Finally
\[
\sum_{k=k_{0}}^{k_{1}}\frac{1}{\nuaverage{q}(\indexnot)e^{\baverage{q}t_{k}}}\mfH_{q}^{(s)}(\indexnot,t_{k})
\leq C\int_{\bt_{0,+}}^{\bt_{1,+}}\mfH_{q}^{(s)}(\indexnot,t)dt
\]
for $s=1,2$, where we appealed to (\ref{eq:sumestmfHint}) and $C$ has the same dependence as before. Note that the right 
hand side can be computed using (\ref{eq:Hjrint}). Since $\lambda=0$ and $a=1$, we, moreover, have 
$(\lambda+s\baverage{p})_{a+s,q}=s(\baverage{p}-\baverage{q})\neq 0$ (since $\baverage{p}\neq \baverage{q}$ for 
$p\neq q$ and $s>0$). We can therefore combine (\ref{eq:Hjrint}) and (\ref{eq:Hjamointnd}) in order to conclude that 
\[
\int_{\bt_{0,+}}^{\bt_{1,+}}\mfH_{q}^{(s)}(\indexnot,t)dt=\sum_{p\neq q}\frac{1}{s(\baverage{p}-\baverage{q})}
\left(\frac{\nuaverage{p}^{s}(\indexnot)e^{s\baverage{p}\bt_{1,+}}}{\nuaverage{q}^{s}(\indexnot)e^{s\baverage{q}\bt_{1,+}}}
-\frac{\nuaverage{p}^{s}(\indexnot)e^{s\baverage{p}\bt_{0,+}}}{\nuaverage{q}^{s}(\indexnot)e^{s\baverage{q}\bt_{0,+}}}\right)\leq C
\]
for $s=1,2$, where $C$ has the same dependence as before. To conclude, the sum appearing in (\ref{eq:finsumtobeest}) is bounded if 
$\baverage{q}\neq 0$ and is bounded by a constant plus a constant times the right hand side of (\ref{eq:sumbjeqz}) in case $\baverage{q}=0$. 
In both cases, the constant only depends on $\e$, $\lambda$ and the coefficients of the equation. Thus 
$\|\chA_{k_{1},q,\e}^{+}\cdots \chA_{k_{0},q,\e}^{+}\|$ and 
$\|\chA_{k_{0},q,\e}^{-}\cdots \chA_{k_{1},q,\e}^{-}\|$ are bounded if $\baverage{q}\neq 0$. In case $\baverage{q}=0$, it is necessary to include a 
factor of the form 
\[
\exp\left[\frac{C_{\deg,q,\pm}}{\nuaverage{q}(\indexnot)}(\bt_{1,+}-\bt_{0,+})\right].
\]
Combining these observations with (\ref{eq:Akdiagprodplus}), (\ref{eq:Akprodscaleredplus}) and (\ref{eq:AkprodSkredplus}) yields
\begin{equation*}
\begin{split}
\|A_{k_{1}}^{+}\cdots A_{k_{0}}^{+}\| \leq & C_{\e}\|A_{k_{1},q,\e}^{+}\cdots A_{k_{0},q,\e}^{+}\|
 \leq C_{\e}e^{\lambda (\bt_{1,+}-\bt_{0,+})}\|\hA_{k_{1},q,\e}^{+}\cdots \hA_{k_{0},q,\e}^{+}\|\\
 \leq & C_{\e}e^{\lambda (\bt_{1,+}-\bt_{0,+})}\|\chA_{k_{1},q,\e}\cdots \chA_{k_{0},q,\e}\|\\
 \leq & C_{\e,\lambda}\exp\left[\lambda (\bt_{1,+}-\bt_{0,+})+\frac{C_{\deg,q,+}}{\nuaverage{q}(\indexnot)}(\bt_{1,+}-\bt_{0,+})\right].
\end{split}
\end{equation*}
Here $C_{\e,\lambda}$ only depends on $\e$, $\lambda$ and the coefficients of the equation (\ref{eq:thesystemRge}). Moreover, $C_{\deg,q,+}$ only depends on 
$\e$, $\lambda$ and the coefficients of the equation (\ref{eq:thesystemRge}), and it vanishes if $\baverage{q}\neq 0$. A similar argument yields
\[
\|A_{k_{0}}^{-}\cdots A_{k_{1}}^{-}\| \leq 
C_{\e,\lambda}\exp\left[\lambda (\bt_{0,-}-\bt_{1,-})+\frac{C_{\deg,q,-}}{\nuaverage{q}(\indexnot)}(\bt_{1,-}-\bt_{0,-})\right],
\]
where the constants have the same dependence and properties as in the previous estimate. The appearance of $\bt_{1,-}-\bt_{0,-}$ in the last term 
inside the exponential requires some justification. However, $|\bt_{1,+}-\bt_{1,-}|$ and $|\bt_{0,+}-\bt_{0,-}|$ are bounded by numerical constants. 
Changing $\bt_{1,+}$ to $\bt_{1,-}$ and $\bt_{0,+}$ to $\bt_{0,-}$ thus only amounts to a change of the constant $C_{\e,\lambda}$. 
In the case of trivial generalised Jordan blocks we can choose $\lambda=\kappa_{q,+}$ and $\lambda=-\kappa_{q,-}$ for the respective cases. Moreover, in those 
cases, $\e$ can be chosen to depend only on the coefficients of the equation, as described above. Finally, if the relevant generalised Jordan blocks are 
non-trivial, we need to include a margin in the form of the $\e>0$ appearing in the statement of the lemma. The lemma follows. 

In order to justify Remark~\ref{remark:kzkkonesuff}, note that $\|A^{\pm}_{k}\|$ is bounded by a constant depending only on the 
coefficients of the equation. 
\end{proof}

\section{Refined estimates in transparent eras}\label{section:refmodeanaltranspera}

When $\baverage{q}\neq 0$, the estimates derived in Lemma~\ref{lemma:maprnoes} are satisfactory. When $\baverage{q}=0$ and $\nuaverage{q}(\indexnot)$ is 
large, they are also satisfactory. However, when $\baverage{q}=0$ and $\nuaverage{q}(\indexnot)$ is not large, there is room for improvement. 
We therefore treat this situation separately. In order to develop a feeling concerning how to approximate the solution in this setting, let 
\begin{equation}\label{eq:txqtyqvarpidef}
\left(\begin{array}{c} \tx_{q}(\indexnot,t) \\ \ty_{q}(\indexnot,t) \end{array}\right)
:=\left(\begin{array}{c} \nuaverage{q}(\indexnot)z(\indexnot,t) \\ \dot{z}(\indexnot,t) \end{array}\right),\ \ \
\varpi_{q}:=\left(\begin{array}{c} \tx_{q} \\ \ty_{q} \end{array}\right).
\end{equation}
It can then be calculated that 
\begin{equation}\label{eq:varpiqeq}
\dot{\varpi}_{q}(\indexnot,t)=A_{q,\infty}(\indexnot)\varpi_{q}(\indexnot,t)+A_{q,\rem}(\indexnot,t)\varpi_{q}(\indexnot,t)
+F(\indexnot,t),
\end{equation}
where
\begin{align}
A_{q,\infty}(\indexnot) := & \left(\begin{array}{cc} 0 & \nuaverage{q}(\indexnot)\Id_{m} \\ 
-\nuaverage{q}(\indexnot)\Id_{m}-i\Xaverage{\diag}^{q}(\indexnot)-[\nuaverage{q}(\indexnot)]^{-1}\zeta_{\infty} & -\a_{\infty}\end{array}\right),\label{eq:Aqinfinddef}\\
F(\indexnot,t) := & \left(\begin{array}{c} 0 \\ \hf(\indexnot,t) \end{array}\right).\label{eq:Findtdefdeg}
\end{align}
In order for this observation to be of interest, it is, however, necessary to estimate the norm of the matrix $A_{q,\rem}(\indexnot,t)$ appearing
in (\ref{eq:varpiqeq}). We do so in the proof of Lemma~\ref{lemma:Iqestbavqeqzero} below. Since $A_{q,\infty}(\indexnot)$ is independent of $t$, 
approximating the behaviour of solutions should be straightforward. Moreover, assuming $A_{q,\rem}(\indexnot,t)$ to be small, the quantities
\begin{equation}\label{eq:kapqindnotpm}
\kappa_{q,\indexnot,\pm}:=\kappa_{\max}[\pm A_{q,\infty}(\indexnot)]
\end{equation}
can be expected to correspond to the dominant behaviour solutions; $\kappa_{q,\indexnot,+}$ for increasing 
$t$ and $\kappa_{q,\indexnot,-}$ for decreasing $t$.

\begin{lemma}\label{lemma:Iqestbavqeqzero}
Consider the equation (\ref{eq:thesystemRge}). Assume that it is non-degenerate, diagonally dominated, balanced and convergent; cf. 
Definition~\ref{def:nondegconvabal}. Consider $0\neq \indexnot\in\EFindexset$ such that there is a frequency era $I_{q}$ with $\baverage{q}=0$; cf. 
Definition~\ref{def:era}.
Fix $\e>0$ and $K_{\roco}>0$. Then there is a constant $C_{\rocsp,\e}$ depending only on an upper bound on $K_{\roco}$, the spectrum of the Riemannian 
manifold corresponding to $q$ (if $q$ corresponds to a $j\in \{1,\dots,d\}$, then the relevant Riemannian manifold is $\so$ with the standard Euclidean 
metric), $\e$ and the coefficients of the equation (\ref{eq:thesystemRge}) such that
\begin{equation}\label{eq:westdegcaselowfre}
|w(\indexnot,t)|\leq C_{\rocsp,\e}e^{(\kappa_{q,\indexnot,+}+\e)(t-t_{-})}|w(\indexnot,t_{-})|
+C_{\rocsp,\e}\int_{t_{-}}^{t}e^{(\kappa_{q,\indexnot,+}+\e)(t-t')}|\hf(\indexnot,t')|dt'
\end{equation}
for $\indexnot\in\EFindexset$ such that $I_{q}\neq\varnothing$ and $\nuaverage{q}(\indexnot)\leq K_{\roco}$; $t_{-}\in I_{q}$; $t_{-}\leq t\in I_{q}$; and
$z$ a solution to (\ref{eq:fourierthesystemRge}). Here $w$ is defined in terms of $z$ as described in (\ref{eq:xyFodefshiftge}) and 
(\ref{eq:wdefshiftge}). Moreover, $\kappa_{q,\indexnot,+}$ is defined in (\ref{eq:kapqindnotpm}) and the $\e$ can be removed in case the Jordan 
blocks of $A_{q,\infty}(\indexnot)$ corresponding to eigenvalues with real parts equal to $\kappa_{q,\indexnot,+}$ are trivial; cf. (\ref{eq:Aqinfinddef}) 
for a definition of $A_{q,\infty}(\indexnot)$. Moreover,
\[
|w(\indexnot,t)|\leq C_{\rocsp,\e}e^{-(\kappa_{q,\indexnot,-}+\e)(t-t_{+})}|w(\indexnot,t_{+})|
+C_{\rocsp,\e}\int_{t}^{t_{+}}e^{-(\kappa_{q,\indexnot,-}+\e)(t-t')}|\hf(\indexnot,t')|dt'
\] 
for $\indexnot\in\EFindexset$ such that $I_{q}\neq\varnothing$ and $\nuaverage{q}(\indexnot)\leq K_{\roco}$; $t_{+}\in I_{q}$; and 
$t_{+}\geq t\in I_{q}$. Here $\kappa_{q,\indexnot,-}$ is defined in (\ref{eq:kapqindnotpm}) and the $\e$ can be removed in case the Jordan blocks of 
$-A_{q,\infty}(\indexnot)$ corresponding to eigenvalues with real parts equal to $\kappa_{q,\indexnot,-}$ are trivial. 
\end{lemma}
\begin{remark}
When applying this lemma, it is of interest to keep the following in mind. First, for a given $0< K_{\roco}\in\ro$, there are only finitely many
matrices of the form $A_{q,\infty}(\indexnot)$ for $\indexnot\in\EFindexset$ such that $\nuaverage{q}(\indexnot)\neq 0$ and 
$\nuaverage{q}(\indexnot)\leq K_{\roco}$; this is a consequence of the fact that the eigenvalues of the relevant Riemannian manifolds converge to infinity. 
Second, the matrices $A_{q,\infty}(\indexnot)$ are of the form (\ref{eq:Amudef}). Due to Lemma~\ref{lemma:Nmueigenvapprox}, the set of real parts of the
eigenvalues of $A_{q,\infty}(\indexnot)$ for $\indexnot\in\EFindexset$ such that $\nuaverage{q}(\indexnot)\neq 0$ is thus bounded. 
\end{remark}
\begin{remark}
Note that the set of matrices $A_{q,\infty}(\indexnot)$ for $\indexnot\in\EFindexset$ such that $\nuaverage{q}(\indexnot)\neq 0$ and 
$\nuaverage{q}(\indexnot)\leq K_{\roco}$ is determined by three things: $K_{\roco}$; the spectrum of the Riemannian 
manifold corresponding to $q$; and the coefficients of the equation (\ref{eq:thesystemRge}). That is why the constants appearing in
the estimates have the stated dependence. 
\end{remark}
\begin{proof}
As described in the statement of the lemma, we are here interested in situations such that $\nuaverage{q}(\indexnot)\leq K_{\roco}$ for some constant 
$K_{\roco}>0$. Moreover, we below allow the constants appearing in the estimates to depend on an upper bound on $K_{\roco}$. 

\textbf{Estimating the norm of $A_{q,\rem}(\indexnot,t)$.} The proof of the lemma 
is based on an analysis of solutions to (\ref{eq:varpiqeq}), where the constituents, except for $A_{q,\rem}(\indexnot,t)$, are defined in 
(\ref{eq:txqtyqvarpidef}), (\ref{eq:Aqinfinddef}) and (\ref{eq:Findtdefdeg}). As a first step, we need to estimate the norm of $A_{q,\rem}(\indexnot,t)$.
Note, to this end, that 
\begin{equation}\label{eq:mfgsqnuavapprox}
|\mfg^{2}(\indexnot,t)-\nuaverage{q}^{2}(\indexnot)|\leq C_{\roco}\nuaverage{q}(\indexnot)[\mfH_{q}(\indexnot,t)+e^{-\eta_{\diag}t}]
\end{equation}
for all $t\in I_{q}$, where $C_{\roco}$ depends only on an upper bound on $K_{\roco}$ and the coefficients of the equation (\ref{eq:thesystemRge}). In order to obtain 
this estimate, we have appealed to (\ref{eq:mfgmfgdiageq}), (\ref{eq:mfgmmfgdiag}), (\ref{eq:njnogequ}) and (\ref{eq:ginvminnjvergest}). Note also that 
(\ref{eq:njnogequ}) and the 
present assumptions imply that $\mfg(\indexnot,t)\leq C_{\roco}$ for $t\in I_{q}$, where $C_{\roco}$ depends only on an upper bound on $K_{\roco}$ and the coefficients 
of the equation (\ref{eq:thesystemRge}). Similarly, 
\begin{equation}\label{eq:sigmamfgprodest}
|2i\sigma(\indexnot,t)\mfg(\indexnot,t)|\leq C_{\roco}e^{-\eta_{\rosh}t}
\end{equation}
for all $t\in I_{q}$, where $C_{\roco}$ has the same dependence as before; recall that the equation is oscillation adapted with $\mff_{\rosh}(t)=Ce^{-\eta_{\rosh}t}$
due to Lemma~\ref{lemma:consconvcoeff}. Finally, let us estimate
\begin{equation}\label{eq:nlXlmnuavXdiagest}
\begin{split}
\|n_{l}X^{l}(t)-\nuaverage{q}(\indexnot)\Xaverage{\diag}^{q}(\indexnot)\| = & \left\|n_{l}X^{l}(t)-\textstyle{\sum}_{\{j:\b_{j}=\baverage{q}\}}n_{j}X^{j}_{\infty}\right\|\\
 \leq & C\mfH^{(1)}_{q}(\indexnot,t)\nuaverage{q}(\indexnot)+Ce^{-\eta_{\romn}t}\nuaverage{q}(\indexnot)
\end{split}
\end{equation}
for all $t\in I_{q}$, 
where we appealed to (\ref{eq:mainbd}), (\ref{eq:mainconv}) and (\ref{eq:Xavdiagqindexnotdef}), and the constants $C$ only depend on the coefficients of the 
equation (\ref{eq:thesystemRge}). Combining (\ref{eq:mainconv}) with (\ref{eq:mfgsqnuavapprox}), (\ref{eq:sigmamfgprodest}) and (\ref{eq:nlXlmnuavXdiagest})
yields the conclusion that (\ref{eq:varpiqeq}) holds, where $A_{q,\infty}(\indexnot)$ and $F(\indexnot,t)$ are defined by (\ref{eq:Aqinfinddef}) and 
(\ref{eq:Findtdefdeg}) respectively. Moreover, 
\[
\|A_{q,\rem}(\indexnot,t)\|\leq C_{\roco}\left(\textstyle{\sum}_{s=1}^{2}\mfH_{q}^{(s)}(\indexnot,t)+e^{-\eta_{\roB}t}+[\nuaverage{q}(\indexnot)]^{-1}e^{-\eta_{\romn}t}\right)
\]
for all $t\in I_{q}$, 
where $\eta_{\roB}:=\min\{\eta_{\rosh},\eta_{\romn},\eta_{\diag}\}$ and $C_{\roco}$ depends only on an upper bound on $K_{\roco}$ and the coefficients 
of the equation (\ref{eq:thesystemRge}). 

\textbf{Reformulating the equations.} Fix $\e>0$ and let $T_{q,\indexnot,\e}$ be such that 
\[
J_{q,\indexnot,\e}:=T_{q,\indexnot,\e}^{-1}A_{q,\infty}(\indexnot)T_{q,\indexnot,\e}
\]
consists of generalised Jordan blocks with the non-zero off-diagonal terms equal to $\e/2$; cf. Remark~\ref{remark:genJordblock}. Let 
\[
R_{q,\indexnot}(t):=\exp[-i\mathrm{Im}\{J_{q,\indexnot,\e}\}t];
\]
note that the imaginary part of $J_{q,\indexnot,\e}$ is independent of $\e$. Finally, let $\kappa_{q,\indexnot,\pm}$ be defined by (\ref{eq:kapqindnotpm})
and
\begin{equation}\label{eq:xiqindpm}
\xi_{q,\indexnot}^{\pm}(t):=\exp[\mp(\kappa_{q,\indexnot,\pm}+\eta)\bt_{\pm}]R_{q,\indexnot}(\bt_{\pm})T_{q,\indexnot,\e}^{-1}\varpi_{q}(\indexnot,t),
\end{equation}
where $\bt_{\pm}:=t-t_{\mp}$; $t_{-},t_{+}\in I_{q}$; and $\eta\geq 0$ remains to be determined. Then
\begin{equation}\label{eq:xiqindpmev}
\begin{split}
\dot{\xi}_{q,\indexnot}^{\pm}(t) = & [\mathrm{Re}\{J_{q,\indexnot,\e}\}\mp (\kappa_{q,\indexnot,\pm}+\eta)\Id_{2m}]\xi_{q,\indexnot}^{\pm}(t)
+A_{q,\rem,\e}^{\pm}(\indexnot,t)\xi_{q,\indexnot}^{\pm}(t)+G_{q,\e}^{\pm}(\indexnot,t),
\end{split}
\end{equation}
where
\begin{align*}
A_{q,\rem,\e}^{\pm}(\indexnot,t) := & R_{q,\indexnot}(\bt_{\pm})T_{q,\indexnot,\e}^{-1}A_{q,\rem}(\indexnot,t)T_{q,\indexnot,\e}[R_{q,\indexnot}(\bt_{\pm})]^{-1},\\
G_{q,\e}^{\pm}(\indexnot,t) := & \exp[\mp(\kappa_{q,\indexnot,\pm}+\eta)\bt_{\pm}]R_{q,\indexnot}(\bt_{\pm})T_{q,\indexnot,\e}^{-1}F(\indexnot,t).
\end{align*}
For future reference, it is of interest to note that 
\begin{equation}\label{eq:Aqremepest}
\|A_{q,\rem,\e}^{\pm}(\indexnot,t)\|\leq C_{\rocsp,\e}\left(\textstyle{\sum}_{s=1}^{2}\mfH_{q}^{(s)}(\indexnot,t)+e^{-\eta_{\roB}t}\right)
\end{equation}
for all $t\in I_{q}$, 
where $C_{\rocsp,\e}$ only depends on an upper bound on $K_{\roco}$, the spectrum of the Riemannian manifold corresponding to $q$ (if $q$ corresponds to 
a $j\in \{1,\dots,d\}$, then the relevant Riemannian manifold is $\so$ with the standard Euclidean metric), $\e$ and the coefficients of the equation
(\ref{eq:thesystemRge}). In what follows, we wish to choose $\e$ and $\eta$ in such a way that 
\begin{equation}\label{eq:Jqindenormposnegsd}
\mathrm{Re}\{J_{q,\indexnot,\e}\}-(\kappa_{q,\indexnot,+}+\eta)\Id_{2m},\ \ \
\mathrm{Re}\{J_{q,\indexnot,\e}\}+(\kappa_{q,\indexnot,-}+\eta)\Id_{2m}
\end{equation}
are negative semi-definite and positive semi-definite respectively; here we say that $A\in\Mn{k}{\co}$ is negative semi-definite if 
$\mathrm{Re}\{\ldr{w,Aw}\}\leq 0$ for all $w\in\cn{k}$ etc. In case the Jordan blocks of $A_{q,\infty}(\indexnot)$ corresponding to eigenvalues 
with real parts equal to $\kappa_{q,\indexnot,+}$ are all trivial, we choose $\eta=0$ and $\e$ small enough (the bound depending only on an upper bound on 
$K_{\roco}$, the spectrum of the Riemannian manifold corresponding to $q$ and the coefficients of the equation (\ref{eq:thesystemRge})). Under these
conditions, the first matrix appearing in (\ref{eq:Jqindenormposnegsd}) is negative semi-definite. In case there is a non-trivial Jordan block of 
$A_{q,\infty}(\indexnot)$ corresponding to an eigenvalue with real part equal to $\kappa_{q,\indexnot,+}$, we choose $\eta=\e$. Then the first matrix 
appearing in (\ref{eq:Jqindenormposnegsd}) is negative definite. The choices in the case of the second matrix appearing in (\ref{eq:Jqindenormposnegsd}) 
are similar. 

\textbf{Estimating the energy of the solution.} Given the above choices of $\e$ and $\eta$, 
\[
\frac{d}{dt}|\xi_{q,\indexnot}^{+}|^{2}\leq 2\|A_{q,\rem,\e}^{+}(\indexnot,\cdot)\| |\xi_{q,\indexnot}^{+}|^{2}+2|G_{q,\e}^{+}(\indexnot,\cdot)||\xi_{q,\indexnot}^{+}|
\]
for all $t\in I_{q}$, where we appealed to (\ref{eq:xiqindpmev}). Let 
\begin{equation}\label{eq:Lqepdef}
L_{q,\e}^{+}(\indexnot,t):=-\int_{t_{-}}^{t}\|A_{q,\rem,\e}^{+}(\indexnot,t')\|dt'.
\end{equation}
Then
\[
\frac{d}{dt}\left(\exp\left[2L_{q,\e}^{+}(\indexnot,\cdot)\right]|\xi_{q,\indexnot}^{+}|^{2}\right)\leq 
2\exp\left[2L_{q,\e}^{+}(\indexnot,\cdot)\right]|G_{q,\e}^{+}(\indexnot,\cdot)||\xi_{q,\indexnot}^{+}|.
\]
This estimate can be used to deduce that for $t_{-}\leq t\in I_{q}$,
\begin{equation}\label{eq:eLqexiqindpest}
\exp\left[L_{q,\e}^{+}(\indexnot,t)\right]|\xi_{q,\indexnot}^{+}(t)|\leq |\xi_{q,\indexnot}^{+}(t_{-})|+
\int_{t_{-}}^{t}\exp\left[L_{q,\e}^{+}(\indexnot,t')\right]|G_{q,\e}^{+}(\indexnot,t')|dt'.
\end{equation}
In order to estimate the right hand side of (\ref{eq:Lqepdef}), note that (\ref{eq:Aqremepest}) holds. Combining this estimate with (\ref{eq:Hjrint}) 
and (\ref{eq:Hjamointnd}) yields 
\[
|L_{q,\e}^{+}(\indexnot,t)|\leq C_{\rocsp,\e}
\]
for all $t\in I_{q}$, where $C_{\rocsp,\e}$ has the same dependence as in the case of (\ref{eq:Aqremepest}). Combining this estimate with 
(\ref{eq:eLqexiqindpest}) and recalling the definitions of the constituents of (\ref{eq:eLqexiqindpest}) yields
\[
|\varpi_{q}(\indexnot,t)|\leq C_{\rocsp,\e}e^{(\kappa_{q,\indexnot,+}+\eta)\bt_{+}}|\varpi_{q}(\indexnot,t_{-})|
+C_{\rocsp,\e}\int_{t_{-}}^{t}e^{(\kappa_{q,\indexnot,+}+\eta)(t-t')}|F(\indexnot,t')|dt'
\]
for $t_{-}\leq t\in I_{q}$, where $C_{\rocsp,\e}$ has the same dependence as in the case of (\ref{eq:Aqremepest}). A similar argument yields
\[
|\varpi_{q}(\indexnot,t)|\leq C_{\rocsp,\e}e^{-(\kappa_{q,\indexnot,-}+\eta)\bt_{-}}|\varpi_{q}(\indexnot,t_{+})|
+C_{\rocsp,\e}\int_{t}^{t_{+}}e^{-(\kappa_{q,\indexnot,-}+\eta)(t-t')}|F(\indexnot,t')|dt'
\] 
for $t_{+}\geq t\in I_{q}$, where $C_{\rocsp,\e}$ has the same dependence as in the case of (\ref{eq:Aqremepest}). Since $|\varpi_{q}(\indexnot,t)|$
and $|w(\indexnot,t)|$ are equivalent for $t\in I_{q}$, where the constant of equivalence only depends on the coefficients of (\ref{eq:thesystemRge}),
the lemma follows. 
\end{proof}

\section{Analysis for one mode}\label{section:analysisforonemode}

In Sections~\ref{section:modeonefreqera} and \ref{section:refmodeanaltranspera} we analyse how the energy of one mode evolves over one frequency
era. The purpose of the present section is to combine the corresponding results in order to obtain an estimate over intervals that include several
frequency eras. Before we state the corresponding result, we, however, need to introduce additional terminology. 

\begin{definition}\label{def:kappaqtrpmetc}
Consider the equation (\ref{eq:thesystemRge}). Assume that it is non-degenerate, diagonally dominated, balanced and convergent; cf. 
Definition~\ref{def:nondegconvabal}. If there is a $q\in \{1,\dots,Q\}$ such that $\baverage{q}=0$, say $q_{\trs}$, define the sets $\mK_{q_{\trs},\pm}$ by 
\[
\mK_{q_{\trs},\pm}:=\left\{\kappa_{q_{\trs},\indexnot,\pm}\ |\ \indexnot\in\EFindexset,\ \nuaverage{q_{\trs}}(\indexnot)\neq 0\right\},
\]
where $\kappa_{q_{\trs},\indexnot,\pm}$ is defined in (\ref{eq:kapqindnotpm}). Define $\bka_{q_{\trs},\pm}$ to be the limit of $\kappa_{q_{\trs},\indexnot,\pm}$ as 
$\nuaverage{q_{\trs}}(\indexnot)\rightarrow\infty$ and define
\[
\kappa_{q_{\trs},\trs,\pm}:=\sup\mK_{q_{\trs},\pm}.
\]
\end{definition}
\begin{remark}\label{remark:limitofrealparttranseras}
That the limits of $\kappa_{q_{\trs},\indexnot,\pm}$ as $\nuaverage{q_{\trs}}(\indexnot)\rightarrow\infty$ exist is a consequence of the fact that 
the matrices $A_{q_{\trs},\infty}(\indexnot)$ are of the form (\ref{eq:Amudef}); Lemma~\ref{lemma:Nmueigenvapprox}; the fact that (\ref{eq:bXjinfndef}) 
holds; and the fact that replacing $V$ by $-V$ in (\ref{eq:Amudef}) does not affect the asymptotic values of the real parts of the eigenvalues 
of $A_{\mu}$ (cf. Lemma~\ref{lemma:Nmueigenvapprox}). In fact, Lemma~\ref{lemma:Nmueigenvapprox} implies that 
\begin{align*}
\bka_{q_{\trs},+} = & \frac{1}{2}\max\left\{\kappa_{\max}\left(-\a_{\infty}+\tX^{q_{\trs}}_{\infty}\right),
\kappa_{\max}\left(-\a_{\infty}-\tX^{q_{\trs}}_{\infty}\right)\right\},\\
\bka_{q_{\trs},-} = & -\frac{1}{2}\min\left\{\kappa_{\min}\left(-\a_{\infty}+\tX^{q_{\trs}}_{\infty}\right),
\kappa_{\min}\left(-\a_{\infty}-\tX^{q_{\trs}}_{\infty}\right)\right\},
\end{align*}
where we appealed to (\ref{eq:bXjinfndef}) and used the terminology introduced in Definition~\ref{def:SpRspdef}. In particular, we thus know
that $\bka_{q_{\trs},\pm}=\kappa_{q_{\trs},\pm}$; cf. Definition~\ref{def:tXqinfRqpmkappaqpm}.
\end{remark}
\begin{remark}
Due to the fact that the limits $\bka_{q_{\trs},\pm}$ exist and the fact that, given $0<K\in\ro$, there are only finitely many matrices $A_{q_{\trs},\infty}(\indexnot)$ 
for $0<\nuaverage{q_{\trs}}(\indexnot)\leq K$, the sets $\mK_{q_{\trs},\pm}$ are bounded. Thus $\kappa_{q_{\trs},\trs,\pm}$ are finite numbers. 
\end{remark}

With the notation introduced in Definition~\ref{def:kappaqtrpmetc}, Lemmas~\ref{lemma:maprnoes} and \ref{lemma:Iqestbavqeqzero} yield the following corollary.

\begin{cor}\label{cor:solnoest}
Consider the equation (\ref{eq:thesystemRge}). Assume that it is non-degenerate, diagonally dominated, balanced and convergent; cf. 
Definition~\ref{def:nondegconvabal}. Given a $0\neq \indexnot\in\EFindexset$ such that there is a sequence $\{t_{k}\}$ of the type described in 
Definition~\ref{def:tkdefgeaddbd}, let $I_{\ror}:=[t_{\roini,\ror},t_{\fin,\ror}]$ (or $I_{\ror}:=[t_{\roini,\ror},\infty)$ in case $t_{\fin,\ror}=\infty$), where 
$t_{\roini,\ror}$ and $t_{\fin,\ror}$ are 
introduced in Definition~\ref{def:tkdefgeaddbd}. Define $R_{q,\pm}^{+}$ and $R_{q,\pm}^{-}$ by (\ref{eq:matrixdetgrowth}) for $q\in \{1,\dots,Q\}$, and define 
$\kappa_{\pm}$ by (\ref{eq:kappapmdef}). Then the following holds. 
\begin{enumerate}
\item Assume that $\baverage{q}\neq 0$ for all $q\in \{1,\dots,Q\}$ and let $\e>0$. Then there is a constant $C_{+,\e}$ depending only on $\e$ and 
the coefficients of the equation (\ref{eq:thesystemRge}), such that 
\begin{equation}\label{eq:westplus}
|w(\indexnot,t_{b})|\leq C_{+,\e}e^{(\kappa_{+}+\e)(t_{b}-t_{a})}|w(\indexnot,t_{a})|+C_{+,\e}\int_{t_{a}}^{t_{b}}e^{(\kappa_{+}+\e)(t_{b}-t')}|\hf(\indexnot,t')|dt'
\end{equation}
for $I_{a}:=[t_{a},t_{b}]\subseteq I_{\ror}$ and every solution $z$ to (\ref{eq:fourierthesystemRge}) corresponding to $\indexnot$. Here 
$w$ is defined in terms of $z$ as described in (\ref{eq:xyFodefshiftge}) and (\ref{eq:wdefshiftge}). Moreover, the following improvements hold:
\begin{itemize}
\item If $q_{i}\in \{1,\dots,Q\}$, $i=1,\dots,l$, are the only elements $q\in\{1,\dots,Q\}$ such that $I_{q}\cap I_{a}\neq 
\varnothing$, then $\kappa_{+}$ appearing in (\ref{eq:westplus}) can be replaced by $\kappa_{+,I_{a}}$, defined by
\begin{equation}\label{eq:kappapmIadef}
\kappa_{\pm,I_{a}}:=\max_{1\leq i\leq l}\kappa_{q_{i},\pm},
\end{equation}
where the dependence of $\kappa_{\pm,I_{a}}$ on $\indexnot$ has been suppressed for the sake of brevity.
\item If, in addition, the Jordan blocks of $R_{q_{i},+}^{+}$ and $R_{q_{i},+}^{-}$ corresponding to the eigenvalues with real part $\kappa_{+,I_{a}}$ are 
trivial for all $i=1,\dots,l$, then $\kappa_{+}$ appearing in (\ref{eq:westplus}) can be replaced by $\kappa_{+,I_{a}}$; the parameter $\e$ can be 
replaced by zero; and $C_{+,\e}$ can be replaced by a constant depending only on the coefficients of the equation (\ref{eq:thesystemRge}). 
\end{itemize}
\item Assume that $\baverage{q}\neq 0$ for all $q\in \{1,\dots,Q\}$ and let $\e>0$. Then there is a constant $C_{-,\e}$ depending only on $\e$ and 
the coefficients of the equation (\ref{eq:thesystemRge}), such that 
\begin{equation}\label{eq:westminus}
|w(\indexnot,t_{a})|\leq C_{-,\e}e^{-(\kappa_{-}+\e)(t_{a}-t_{b})}|w(\indexnot,t_{b})|+C_{-,\e}\int_{t_{a}}^{t_{b}}e^{-(\kappa_{-}+\e)(t_{a}-t')}|\hf(\indexnot,t')|dt'
\end{equation}
for $I_{a}:=[t_{a},t_{b}]\subseteq I_{\ror}$ and every solution $z$ to (\ref{eq:fourierthesystemRge}) corresponding to $\indexnot$. Moreover,
the following improvements hold:
\begin{itemize}
\item If $q_{i}\in \{1,\dots,Q\}$, $i=1,\dots,l$, are the only elements $q\in\{1,\dots,Q\}$ such that $I_{q}\cap I_{a}\neq 
\varnothing$, then $\kappa_{-}$ appearing in (\ref{eq:westminus}) can be replaced by $\kappa_{-,I_{a}}$ introduced in (\ref{eq:kappapmIadef}).
\item If, in addition, the Jordan blocks of $R_{q_{i},-}^{+}$ and $R_{q_{i},-}^{-}$ corresponding to the eigenvalues with real part $\kappa_{-,I_{a}}$ are 
trivial for all $i=1,\dots,l$, then $\kappa_{-}$ appearing in (\ref{eq:westplus}) can be replaced by $\kappa_{-,I_{a}}$; the parameter $\e$ can be 
replaced by zero; and $C_{-,\e}$ can be replaced by a constant depending only on the coefficients of the equation (\ref{eq:thesystemRge}). 
\end{itemize}
\item Assume that there is one $q\in \{1,\dots,Q\}$ such that $\baverage{q}=0$, say $q_{\trs}$, and let $\e>0$. Then there is a constant $C_{+,\e}$ 
depending only on $\e$, the spectrum of the Riemannian manifold corresponding to $q_{\trs}$ and the coefficients of the equation (\ref{eq:thesystemRge}), 
such that 
\begin{equation}\label{eq:westplusdeg}
\begin{split}
|w(\indexnot,t_{b})| \leq & C_{+,\e}e^{(\kappa_{\trs,+}+\e)(t_{b}-t_{a})}|w(\indexnot,t_{a})|
+C_{+,\e}\int_{t_{a}}^{t_{b}}e^{(\kappa_{\trs,+}+\e)(t_{b}-t')}|\hf(\indexnot,t')|dt'
\end{split}
\end{equation}
for $I_{a}:=[t_{a},t_{b}]\subseteq I_{\ror}$ and every solution $z$ to (\ref{eq:fourierthesystemRge}) corresponding to $\indexnot$, where 
\begin{equation}\label{eq:kappatrpmdef}
\kappa_{\trs,\pm}:=\max\{\kappa_{\pm},\kappa_{q_{\trs},\trs,\pm}\}
\end{equation}
and the $\kappa_{q,\trs,\pm}$ are introduced in Definition~\ref{def:kappaqtrpmetc}. If $I_{q_{\trs}}\cap I_{a}=\varnothing$, then the 
conclusions stated in case $1$ above hold (including the improvements, and with the same dependence of the constants as in case 1). If 
$I_{q_{\trs}}\cap I_{a}\neq\varnothing$ and $\kappa_{q_{\trs},+}<\kappa_{\trs,+}$, then the $\e$ appearing in (\ref{eq:westplusdeg}) can be removed if 
\begin{itemize}
\item the Jordan blocks of $R_{q,+}^{+}$ and $R_{q,+}^{-}$ corresponding to the eigenvalues with real part $\kappa_{\trs,+}$ are trivial for all 
$q\in \{1,\dots,Q\}$, and 
\item the Jordan blocks of $A_{q_{\trs},\infty}(\indexnottwo)$ corresponding to the eigenvalues with real part $\kappa_{\trs,+}$ are trivial for all 
$\indexnottwo\in\EFindexset$ such that $\nuaverage{q_{\trs}}(\indexnottwo)\neq 0$.
\end{itemize}
When $\e$ can be removed, $C_{+,\e}$ can be replaced by a constant depending only on the spectrum of the Riemannian manifold corresponding 
to $q_{\trs}$ and the coefficients of the equation (\ref{eq:thesystemRge}). 
\item Assume that there is one $q\in \{1,\dots,Q\}$ such that $\baverage{q}=0$, say $q_{\trs}$, and let $\e>0$. Then there is a constant $C_{-,\e}$ 
depending only on $\e$, the spectrum of the Riemannian manifold corresponding to $q_{\trs}$ and the coefficients of the equation (\ref{eq:thesystemRge}) 
such that
\begin{equation}\label{eq:westminusdeg}
\begin{split}
|w(\indexnot,t_{a})| \leq & C_{-,\e}e^{-(\kappa_{\trs,-}+\e)(t_{a}-t_{b})}|w(\indexnot,t_{b})|
+C_{-,\e}\int_{t_{a}}^{t_{b}}e^{-(\kappa_{\trs,-}+\e)(t_{a}-t')}|\hf(\indexnot,t')|dt'
\end{split}
\end{equation}
for $I_{a}:=[t_{a},t_{b}]\subseteq I_{\ror}$ and every solution $z$ to (\ref{eq:fourierthesystemRge}) corresponding to $\indexnot$, where $\kappa_{\trs,-}$
is defined by (\ref{eq:kappatrpmdef}). If $I_{q_{\trs}}\cap I_{a}=\varnothing$, then the conclusions stated in case $2$ above hold (including the 
improvements, and with the same dependence of the constants as in case 2). If $I_{q_{\trs}}\cap I_{a}\neq\varnothing$ and $\kappa_{q_{\trs},-}<\kappa_{\trs,-}$, then 
the $\e$ appearing in (\ref{eq:westminusdeg}) can be removed if 
\begin{itemize}
\item the Jordan blocks of $R_{q,-}^{+}$ and $R_{q,-}^{-}$ corresponding to the eigenvalues with real part $\kappa_{\trs,-}$ are trivial for all 
$q\in \{1,\dots,Q\}$, and 
\item the Jordan blocks of $-A_{q_{\trs},\infty}(\indexnottwo)$ corresponding to the eigenvalues with real part $\kappa_{\trs,-}$ are trivial for all 
$\indexnottwo\in\EFindexset$ such that $\nuaverage{q_{\trs}}(\indexnottwo)\neq 0$.
\end{itemize}
When $\e$ can be removed, $C_{-,\e}$ can be replaced by a constant depending only on the spectrum of the Riemannian manifold corresponding 
to $q_{\trs}$ and the coefficients of the equation (\ref{eq:thesystemRge}). 
\end{enumerate}
\end{cor}
\begin{remark}
In cases $3$ and $4$, the $\kappa_{\pm}$ which enters the definition of $\kappa_{\trs,\pm}$, cf. (\ref{eq:kappatrpmdef}), can be replaced by 
$\kappa_{\pm,I_{a}}$; cf. the improvements stated in cases $1$ and $2$. See also Corollary~\ref{cor:westvarrhocase} for further improvements.
\end{remark}
\begin{remark}
It is important to note that the constants appearing in the corollary depend neither on $\indexnot$ nor on the solution $z$. 
\end{remark} 
\begin{remark}\label{remark:abswsqinter}
Recall that $|w(\indexnot,t)|^{2}$ is given by 
\begin{equation}\label{eq:wintermsofzetaetc}
|w(\indexnot,t)|^{2}=|x_{1}(\indexnot,t)|^{2}+|y_{1}(\indexnot,t)|^{2}=|\dot{z}(\indexnot,t)|^{2}+\mfg^{2}(\indexnot,t)|z(\indexnot,t)|^{2},
\end{equation}
where $z$ is a solution to (\ref{eq:fourierthesystemRge}). In this sense, $|w(\indexnot,t)|^{2}$ represents the energy content of the 
$\indexnot$'th mode at time $t$. 
\end{remark}
\begin{proof}
We begin by proving (\ref{eq:westplus}) in the case that both $t_{a}$ and $t_{b}$ are elements of the sequence $\{t_{k}\}$. Without loss of generality, 
we can then assume that $t_{a}=t_{0}$ and that $t_{b}=t_{l_{b}+1}$ for some $0\leq l_{b}\in\zo$ (this can be achieved by simply relabelling the sequence). 
Let us begin by estimating
\begin{align}
|w(t_{k})| \leq & \|T_{\pre,k}^{-1}\|\cdot |w_{\pre,k}|\leq C|w_{\fin,k}|=C|\psi_{k}|,\label{eq:wtkpoprelest}\\
|\psi_{k}| = & |w_{\pre,k}|\leq \|T_{\pre,k}\|\cdot|w(t_{k})|\leq C|w(t_{k})|,\label{eq:invwtkpoprelest}
\end{align}
where $C$ only depends on the coefficients of the equation (\ref{eq:thesystemRge}). In order to arrive at this conclusion, we used the fact that 
$\psi_{k}=w_{\fin,k}$, (\ref{eq:wkpredef}), (\ref{eq:wfinkdef}), the fact that $D_{k}$ is an isometry and the fact that the norms of $T_{\pre,k}$ and its
inverse are bounded by a constant depending only on the coefficients of the equation (\ref{eq:thesystemRge}); cf. the comments made in connection with 
(\ref{eq:Tprekage}) and (\ref{eq:Tprekainvge}). On the other hand, recalling (\ref{eq:differenceeqrepres}), 
\begin{equation}\label{eq:psikponoest}
\begin{split}
|\psi_{k+1}| \leq & \|A_{k}^{+}\cdots A_{0}^{+}\|\cdot |\psi_{0}|+\|A_{k}^{+}\|\int_{t_{k}}^{t_{k+1}}|\hF_{k}(t)|dt\\
 & +\cdots+\|A_{k}^{+}\cdots A_{0}^{+}\|\int_{t_{0}}^{t_{1}}|\hF_{0}(t)|dt. 
\end{split}
\end{equation}
\textbf{Estimating the norm of the matrix products in the non-transparent setting.} In order to proceed, we need to estimate $\|A_{k_{1}}^{+}\cdots A_{k_{0}}^{+}\|$ 
in cases where $t_{k}$ does not necessarily belong to the same frequency era for all $k_{0}\leq k\leq k_{1}$. For simplicity, we start by considering the case
that there are no transparent eras. Given $k_{0},k_{1}$ such that $0\leq k_{0}\leq k_{1}\leq l_{b}<k_{\fin,\ror}$, there are, say, $l$ subintervals, 
denoted $[k_{0,j},k_{1,j}]$, $j=1,\dots,l$, such that $t_{k}\in I_{q_{j}}$ for $k\in [k_{0,j},k_{1,j}]$ and some choice of $q_{j}$. Moreover, $k_{0,1}=k_{0}$,
$k_{1,l}=k_{1}$ and $k_{1,j}+1=k_{0,j+1}$. We also assume that none of the intervals $I_{q_{j}}$ is a transparent era. Fix $\e>0$. Since $l\leq Q$, the following 
estimate holds:
\begin{equation}\label{eq:Akprodnoestgc}
\|A_{k_{1}}^{+}\cdots A_{k_{0}}^{+}\|\leq \|A_{k_{1,l}}^{+}\cdots A_{k_{0,l}}^{+}\|\cdots \|A_{k_{1,1}}^{+}\cdots A_{k_{0,1}}^{+}\|\leq C_{+,\e}e^{(\kappa_{+}+\e)(\bt_{1}-\bt_{0})},
\end{equation}
where $\bt_{1}=t_{k_{1}+1}$, $\bt_{0}=t_{k_{0}}$ and the constant $C_{+,\e}$ depends only on $\e$ and the coefficients of the equation (\ref{eq:thesystemRge}).
In order to obtain the last inequality in (\ref{eq:Akprodnoestgc}), we appealed to (\ref{eq:Akproddegplus}) $l\leq Q$ times, as well as 
Remark~\ref{remark:kzkkonesuff}; note that 
$C_{\deg,q_{j},+}=0$ for all $j=1,\dots,l$, using the notation of (\ref{eq:Akproddegplus}). Assume now that there are no non-trivial Jordan blocks of 
$R_{q,+}^{+}$ and $R_{q,+}^{-}$, $q=q_{j}$, $j=1,\dots,l$, corresponding to eigenvalues with real part $\kappa_{+}$. For a given $j$, this situation can arise
for two reasons. The first possibility is that $\kappa_{+}=\kappa_{q_{j},+}$, but the Jordan blocks of $R_{q_{j},+}^{+}$ and $R_{q_{j},+}^{-}$
corresponding to eigenvalues with real part $\kappa_{q_{j},+}$ are trivial. For such $j$, we can appeal to (\ref{eq:Akprodndegplus}) with
$C_{\deg,q_{j},+}=0$. The second possibility is that $\kappa_{+}>\kappa_{q_{j},+}$. In that case, we can appeal to (\ref{eq:Akproddegplus}) with 
$\e=\kappa_{+}-\kappa_{q_{j},+}$ and $C_{\deg,q_{j},+}=0$. Regardless, it is clear that when there are no non-trivial Jordan blocks of $R_{q,+}^{+}$ and 
$R_{q,+}^{-}$, $q=q_{j}$, $j=1,\dots,l$, corresponding to eigenvalues with real part $\kappa_{+}$, then the $\e$ appearing in (\ref{eq:Akprodnoestgc}) can be 
removed and $C_{+,\e}$ can be replaced by a constant $C_{+}$ depending only on the coefficients of the equation (\ref{eq:thesystemRge}).
From the above arguments, it is also clear that $\kappa_{+}$ can be replaced by $\kappa_{+,I_{k_{0},k_{1}}}$, where $I_{k_{0},k_{1}}:=[\bt_{0},\bt_{1}]$.

\textbf{Deriving energy estimates over intervals defined by the sequence.} Returning to Lemma~\ref{lemma:wkfinlemma}, in particular (\ref{eq:Fkfindef}), 
it is clear that 
\[
|\hF_{l}(t)|\leq C|\hf(t)|
\]
in the interval $[t_{l},t_{l+1}]$, where $C$ only depends on the coefficients of the equation (\ref{eq:thesystemRge}). Combining this estimate with 
(\ref{eq:wtkpoprelest}), (\ref{eq:psikponoest}) and (\ref{eq:Akprodnoestgc}) yields
\begin{equation}\label{eq:wtkpoprelestint}
\begin{split}
|w(t_{k+1})| \leq & C_{+,\e}e^{(\kappa_{+}+\e)(t_{k+1}-t_{0})}|\psi_{0}|+C_{+,\e}e^{(\kappa_{+}+\e)(t_{k+1}-t_{k})}\int_{t_{k}}^{t_{k+1}}|\hf(t)|dt\\
 & +\cdots+C_{+,\e}e^{(\kappa_{+}+\e)(t_{k+1}-t_{0})}\int_{t_{0}}^{t_{1}}|\hf(t)|dt\\
 \leq & C_{+,\e}e^{(\kappa_{+}+\e)(t_{k+1}-t_{0})}|\psi_{0}|+C_{+,\e}\int_{t_{0}}^{t_{k+1}}e^{(\kappa_{+}+\e)(t_{k+1}-t)}|\hf(t)|dt,
\end{split}
\end{equation}
where $C_{+,\e}$ depends only on $\e$ and the coefficients of the equation (\ref{eq:thesystemRge}). In order to obtain this estimate, we 
used the fact that $e^{(\kappa_{+}+\e)(t-t_{k})}\leq C_{\e}$ for $t\in [t_{k},t_{k+1}]$, where $C_{\e}$ depends only on $\e$ and the coefficients of the 
equation (\ref{eq:thesystemRge}). Combining (\ref{eq:invwtkpoprelest}) with (\ref{eq:wtkpoprelestint}) yields the conclusion that (\ref{eq:westplus}) 
holds for $t_{a}=t_{k_{0}}$ and $t_{b}=t_{k_{1}+1}$, where $k_{\roini,\ror}<k_{0}\leq k_{1}<k_{\fin,\ror}$. Moreover, if there are no non-trivial Jordan 
blocks of $R_{q,+}^{+}$ and $R_{q,+}^{-}$, $q=1,\dots,Q$, corresponding to eigenvalues with real part $\kappa_{+}$, then the $\e$ appearing in 
(\ref{eq:westplus}) can be removed and $C_{+,\e}$ can be replaced by a constant $C_{+}$ depending only on the coefficients of the equation 
(\ref{eq:thesystemRge}). Finally, the stated improvements also hold. 

\textbf{Extending the energy estimate beyond the time sequence.} In general, we can assume that $t_{k_{0}-1}<t_{a}\leq t_{k_{0}}$ and that 
$t_{k_{1}+1}\leq t_{b}<t_{k_{1}+2}$, where $k_{1}+2\leq k_{\fin,\ror}$. We then need to estimate the change of the solution in the interval $[t_{a},t_{k_{0}}]$ 
and in the interval $[t_{k_{1}+1},t_{b}]$. In order to obtain such estimates, it is sufficient to appeal to Lemma~\ref{lemma:roughenestbalsetting}. In fact, 
due to Lemma~\ref{lemma:consconvcoeff}, we know that (\ref{eq:thesystemRge}) is oscillation adapted with $\mff_{\rosh}$, $\mff_{X}$ and $\mff_{\roode}$ 
given by (\ref{eq:oscadddbac}). Moreover, in (\ref{eq:oscadddbac}), $C$ only depends on the coefficients of the equation (\ref{eq:thesystemRge}). Thus 
Lemma~\ref{lemma:roughenestbalsetting} applies (with $\betafun=0$). Appealing to (\ref{eq:meestroughbalset}) with 
$t_{0}$ replaced by $t_{k_{1}+1}$ and $t_{1}$ replaced by $t_{b}$ yields
\begin{equation}\label{eq:westlessthanper}
|w(t_{b})|\leq C|w(t_{k_{1}+1})|+C\int_{t_{k_{1}+1}}^{t_{b}}|\hf(t')|dt',
\end{equation}
where $C$ only depends on the coefficients of the equation (\ref{eq:thesystemRge}). In order to obtain this conclusion, we used the fact that 
$|w|$ and $\me^{1/2}$ are equivalent in the interval of interest, the constant of equivalence being numerical; cf. 
(\ref{eq:medefroughbalset}), (\ref{eq:wintermsofzetaetc}) and Definition~\ref{def:tkdefgeaddbd}. A similar estimate holds in the interval $[t_{a},t_{k_{0}}]$.
Combining these estimates  with (\ref{eq:wtkpoprelestint}) and the fact that $|\psi_{k_{0}}|$ can be bounded by a constant times $|w(t_{k_{0}})|$ yields 
(\ref{eq:westplus}). Moreover, the improvements stated in case $1$ also hold. 

\textbf{Energy estimates in the presence of transparent eras.} Assume that (\ref{eq:thesystemRge}) is such that there is a $q\in \{1,\dots,Q\}$
with the property that $\baverage{q}=0$, say $q_{\trs}$. Let $I_{a}$ be as in the statement of the corollary. If $I_{q_{\trs}}\cap I_{a}=\varnothing$,
then the previous analysis applies, so that the conclusions of case $1$ hold. In particular, (\ref{eq:westplusdeg}) holds. When 
$I_{q_{\trs}}\cap I_{a}\neq\varnothing$, either (\ref{eq:Akprodndegplus}) or (\ref{eq:Akproddegplus}) holds in intervals of the form $[k_{0,j},k_{1,j}]$;
cf. the division made in connection with (\ref{eq:Akprodnoestgc}). In the case of (\ref{eq:Akproddegplus}), we are free to replace $\e$ by $\e/2$, and 
we do so in what follows. In both (\ref{eq:Akprodndegplus}) and (\ref{eq:Akproddegplus}), the constant $C_{\deg,q_{\trs},+}$ appears (which only depends 
on the coefficients of (\ref{eq:thesystemRge}) and $\e$). This constant gives rise to a ``cut off'' in $\nuaverage{q_{\trs}}(\indexnot)$. In fact, define
\begin{equation}\label{eq:Krocoep}
K_{\roco,\e,+}:=2\e^{-1}C_{\deg,q_{\trs},+}.
\end{equation}
If $\nuaverage{q_{\trs}}(\indexnot)\geq K_{\roco,\e,+}$, then (\ref{eq:Akprodndegplus}) and (\ref{eq:Akproddegplus}) yield estimates which are good enough
that the previous analysis can be applied. Summing up, there are, given $\indexnot$ and $\{t_{k}\}$ as above, three cases to consider. Either there 
is no transparent era corresponding to $\indexnot$ (or $I_{q_{\trs}}\cap I_{a}=\varnothing$). In that case, the analysis in the absence of transparent
eras applies. The second possibility is that $I_{q_{\trs}}\cap I_{a}\neq\varnothing$, but that $\nuaverage{q_{\trs}}(\indexnot)\geq K_{\roco,\e,+}$. In that
case, (\ref{eq:Akprodndegplus}) and (\ref{eq:Akproddegplus}) yield estimates which are good enough that the analysis in the absence of transparent
eras applies. The third possibility is that $I_{q_{\trs}}\cap I_{a}\neq\varnothing$ and $\nuaverage{q_{\trs}}(\indexnot)\leq K_{\roco,\e,+}$. This situation
is one we need to consider in greater detail. 

\textbf{Analysis for low frequency transparent eras.} Assume that $I_{q_{\trs}}\cap I_{a}\neq\varnothing$ and 
$\nuaverage{q_{\trs}}(\indexnot)\leq K_{\roco,\e,+}$. Divide $I_{a}$ into three parts: $I_{a,1}:=[t_{a},\bt_{0}]$, $I_{a,2}:=[\bt_{0},\bt_{1}]$ and 
$I_{a,3}:=[\bt_{1},t_{b}]$, where $\bt_{0}$ is the infimum of $I_{q_{\trs}}\cap I_{a}$ and $\bt_{1}$ is the supremum of $I_{q_{\trs}}\cap I_{a}$. It could 
be that one or two of the intervals $I_{a,j}$ consists of one point. In that case, the corresponding interval(s) can be ignored in the analysis. Here 
we assume that all the intervals $I_{a,j}$ have non-empty interior, and leave the remaining cases to the reader. In the interval $I_{a,1}$, the 
analysis in the absence of transparent eras applies. Thus (\ref{eq:westplus}) (including the improvements) holds when $t_{b}$ is replaced by 
$\bt_{0}$; i.e.
\begin{equation}\label{eq:westplusbtz}
|w(\indexnot,\bt_{0})|\leq C_{+,\e}e^{(\kappa_{+}+\e)(\bt_{0}-t_{a})}|w(\indexnot,t_{a})|+C_{+,\e}\int_{t_{a}}^{\bt_{0}}e^{(\kappa_{+}+\e)(\bt_{0}-t')}|\hf(\indexnot,t')|dt',
\end{equation}
where $C_{+,\e}$ only depends on $\e$ and the coefficients of the equation (\ref{eq:thesystemRge}). In order to estimate $|w(\indexnot,t)|$ for
$t\in I_{a,2}$, we appeal to Lemma~\ref{lemma:Iqestbavqeqzero} with $K_{\roco}$ replaced by $K_{\roco,\e,+}$ defined in (\ref{eq:Krocoep}). 
Then the constant $C_{\rocsp,\e}$ appearing in (\ref{eq:westdegcaselowfre}) only depends on $\e$, the spectrum of the Riemannian manifold corresponding
to $q_{\trs}$ and the coefficients of the equation (\ref{eq:thesystemRge}). Combining (\ref{eq:westplusbtz}) and (\ref{eq:westdegcaselowfre}) (in the 
latter estimate, we replace $t_{-}$ with $\bt_{0}$) yields
\begin{equation*}
\begin{split}
|w(\indexnot,t)| \leq & C_{\rocsp,\e}e^{(\kappa_{q_{\trs},\indexnot,+}+\e)(t-\bt_{0})}|w(\indexnot,\bt_{0})|
+C_{\rocsp,\e}\int_{\bt_{0}}^{t}e^{(\kappa_{q_{\trs},\indexnot,+}+\e)(t-t')}|\hf(\indexnot,t')|dt'\\
 \leq & C_{+,\e}e^{(\kappa_{\trs,+}+\e)(t-t_{a})}|w(\indexnot,t_{a})|+C_{+,\e}\int_{t_{a}}^{t}e^{(\kappa_{\trs,+}+\e)(t-t')}|\hf(\indexnot,t')|dt'
\end{split}
\end{equation*}
for $t\in I_{a,2}$, where $C_{+,\e}$ only depends on $\e$, the spectrum of the Riemannian manifold corresponding to $q$ and the coefficients of the 
equation (\ref{eq:thesystemRge}). Thus (\ref{eq:westplusdeg}) holds for $t\in I_{a,1}\cup I_{a,2}$. In order to extend this estimate to all of 
$I_{a}$, it is sufficient to appeal to (\ref{eq:westplus}) in $I_{a,3}$. Thus (\ref{eq:westplusdeg}) holds. 

\textbf{Removing the $\e$.} What remains to be verified is that the $\e$ appearing in (\ref{eq:westplusdeg}) can be removed under the circumstances 
mentioned in the statement of the corollary. Due to the assumption that the Jordan blocks of $R_{q,+}^{+}$ and $R_{q,+}^{-}$ corresponding 
to the eigenvalues with real part $\kappa_{\trs,+}$ are trivial for all $q\in \{1,\dots,Q\}$, it is clear that the above analysis in the absence of 
transparent eras can be modified in that $\kappa_{+}$ can be replaced by $\kappa_{\trs,+}$ and the $\e$ can be removed (the constants are in that case
also independent of $\e$). In the presence of transparent eras, we may need to appeal to (\ref{eq:Akprodndegplus}) or (\ref{eq:Akproddegplus}).
However, we only appeal to these estimates when $\nuaverage{q_{\trs}}(\indexnot)$ is large. Moreover, how to make a choice of what ``large'' means 
is clear in the present setting. In fact, since $\kappa_{q_{\trs},+}<\kappa_{\trs,+}$ by assumption, we first fix an 
$0<\e<(\kappa_{\trs,+}-\kappa_{q_{\trs},+})/2$ (when necessary) and then choose $K_{\roco,+}$ according to 
\begin{equation}\label{ref:Kcoplusdef}
K_{\roco,+}:=\frac{2C_{\deg,q_{\trs},+}}{\kappa_{\trs,+}-\kappa_{q_{\trs},+}}.
\end{equation}
Then, for $\nuaverage{q_{\trs}}(\indexnot)\geq K_{\roco,+}$, the exponential factors appearing on the right hand sides of (\ref{eq:Akprodndegplus}) and 
(\ref{eq:Akproddegplus}) can be replaced by $\exp[\kappa_{\trs,+}(\bt_{1,+}-\bt_{0,+})]$. What remains to be taken into consideration is the finite
number of matrices of the form (\ref{eq:Aqinfinddef}) that occur for $\nuaverage{q_{\trs}}(\indexnot)\leq K_{\roco,+}$. For those with the property
that $\kappa_{q_{\trs},\indexnot,+}<\kappa_{\trs,+}$, it is sufficient to choose $\e=\kappa_{\trs,+}-\kappa_{q_{\trs},\indexnot,+}$ in (\ref{eq:westdegcaselowfre}).
Then the expression $\kappa_{q_{\trs},\indexnot,+}+\e$ appearing (twice) on the right hand side of (\ref{eq:westdegcaselowfre}) can be replaced by 
$\kappa_{\trs,+}$. For those matrices $A_{q_{\trs},\infty}(\indexnottwo)$ for which $\kappa_{q_{\trs},\indexnot,+}=\kappa_{\trs,+}$, the Jordan blocks of 
$A_{q_{\trs},\infty}(\indexnottwo)$ corresponding to the eigenvalues with real part $\kappa_{\trs,+}$ are trivial. In such cases, the $\e$ appearing on 
the right hand side of (\ref{eq:westdegcaselowfre}) can thus be removed. To summarise, in all the estimates needed to obtain the desired 
conclusion, the $\e$ can be removed. All the statements of the corollary concerning cases $1$ and $3$ follow. 

\textbf{Analysis in the past direction.} The analysis in the past direction is analogous to the above. We therefore leave the 
details of the proof to the reader. 
\end{proof}

\subsection{Refinements}\label{ssection:refinementsenest}

Even though Corollary~\ref{cor:solnoest} is sufficient for many applications, more refined estimates are sometimes needed. In preparation
for the statement of such results, it is convenient to introduce the following notation. 

\begin{definition}\label{def:varrhoeppl}
Consider the equation (\ref{eq:thesystemRge}). Assume that it is non-degenerate, diagonally dominated, balanced and convergent; cf. 
Definition~\ref{def:nondegconvabal}. Fix $\e>0$ and $K>0$. Given $0\neq\indexnot\in\EFindexset$, consider an interval $J=[s,t]$, where 
$0\leq s\leq t$. If $J$ is a subset of one frequency era, say $I_{q_{1}}$, let $k:=0$ and $I_{1}:=J$. Otherwise, let $s_{1}$ be the supremum of 
the set of $t'\in [s,t]$ such that $s$ and $t'$ belong to the same frequency era, say $I_{q_{1}}$. If $[s_{1},t]$ is contained in one frequency era, say $I_{q_{2}}$,
let $k:=1$, $I_{1}:=[s,s_{1}]$ and $I_{2}:=[s_{1},t]$. If $[s_{1},t]$ is not contained in one frequency era, let $k\geq 2$ and $s_{j}$, $j=2,\dots,k$, 
be defined by the conditions that $s\leq s_{1}\leq\cdots\leq s_{k}\leq t$; that for each $j=1,\dots,k-1$, $I_{j+1}:=[s_{j},s_{j+1}]$ 
equals a frequency era $I_{q_{j+1}}$ for some $q_{j+1}\in\{1,\dots,Q\}$; that $I_{1}:=[s,s_{1}]$ is a subset of a frequency era $I_{q_{1}}$; 
and that $I_{k+1}:=[s_{k},t]$ is a subset of a frequency era $I_{q_{k+1}}$. Define
\[
\varrho_{\e,K,+}(t,s):=\textstyle{\sum}_{j=1}^{k+1}\mu_{j,+}|I_{j}|,
\]
where $\mu_{j,\pm}:=\kappa_{q_{j},\pm}$ in case $q_{j}\neq 0$ and the Jordan blocks of $R_{q_{j},\pm}^{+}$ and $R_{q_{j},\pm}^{-}$ corresponding to eigenvalues 
with real part $\kappa_{q_{j},\pm}$ are trivial; $\mu_{j,\pm}:=\kappa_{q_{j},\pm}+\e$ in case $q_{j}\neq 0$ and there is a non-trivial Jordan block of 
either $R_{q_{j},\pm}^{+}$ or $R_{q_{j},\pm}^{-}$ corresponding to an eigenvalue with real part $\kappa_{q_{j},\pm}$; $\mu_{j,\pm}:=\kappa_{q_{j},\indexnot,\pm}$
if $q_{j}=0$, $\nuaverage{q_{j}}(\indexnot)\leq K$ and the Jordan blocks of $\pm A_{q_{j},\infty}(\indexnot)$ corresponding to eigenvalues 
with real part $\kappa_{q_{j},\indexnot,\pm}$ are trivial; and $\mu_{j,\pm}:=\kappa_{q_{j},\indexnot,\pm}+\e$ otherwise. Similarly, 
\[
\varrho_{\e,K,-}(s,t):=\textstyle{\sum}_{j=1}^{k+1}\mu_{j,-}|I_{j}|.
\]
\end{definition}
\begin{remark}
If $s=t$ and $t$ belongs to two frequency eras, some parts of the definition are ambiguous. However, this does not affect the end result. 
\end{remark}
\begin{remark}
When there are no transparent eras, the constant $K$ is irrelevant, and the definition can be simplified. In this case, we also use the
notation $\varrho_{\e,\pm}$. 
\end{remark}
\begin{remark}
The functions $\varrho_{\e,K,\pm}$ depend on $\indexnot\neq 0$. However, for the sake of brevity, we omit reference to this dependence. 
Note also that we only define $\varrho_{\e,K,+}(t,s)$ for $0\leq s\leq t$ and $\varrho_{\e,K,-}(t,s)$ for $0\leq t\leq s$. 
\end{remark}

With this notation, Corollary~\ref{cor:solnoest} can be improved as follows. 

\begin{cor}\label{cor:westvarrhocase}
Consider the equation (\ref{eq:thesystemRge}). Assume that it is non-degenerate, diagonally dominated, balanced and convergent; cf. 
Definition~\ref{def:nondegconvabal}. Given $\e>0$ and $K>0$, there is then a constant $C_{\e,K}$, depending only on $\e$, $K$,
the spectrum of the Riemannian manifold corresponding to the $q_{\trs}\in\{1,\dots,Q\}$ such that $\baverage{q_{\trs}}=0$ (in case there 
is such a $q_{\trs}$) and the coefficients of 
the equation, such that the following holds. Given a $0\neq \indexnot\in\EFindexset$ such that there is a sequence $\{t_{k}\}$ of the type described in 
Definition~\ref{def:tkdefgeaddbd}, let $I_{\ror}:=[t_{\roini,\ror},t_{\fin,\ror}]$ (or $I_{\ror}:=[t_{\roini,\ror},\infty)$ in case $t_{\fin,\ror}=\infty$), where 
$t_{\roini,\ror}$ and $t_{\fin,\ror}$ are introduced in 
Definition~\ref{def:tkdefgeaddbd}. Let $0\leq t_{a}\leq t_{b}$ be such that $I_{a}:=[t_{a},t_{b}]\subseteq I_{\ror}$. Then
\begin{align}
|w(\indexnot,t_{b})| \leq & C_{\e,K}e^{\varrho_{\e,K,+}(t_{b},t_{a})}|w(\indexnot,t_{a})|+C_{\e,K}\int_{t_{a}}^{t_{b}}e^{\varrho_{\e,K,+}(t_{b},t')}|\hf(\indexnot,t')|dt',
\label{eq:wplvarrhoest}\\
|w(\indexnot,t_{a})| \leq & C_{\e,K}e^{\varrho_{\e,K,-}(t_{a},t_{b})}|w(\indexnot,t_{b})|+C_{\e,K}\int_{t_{a}}^{t_{b}}e^{\varrho_{\e,K,-}(t_{a},t')}|\hf(\indexnot,t')|dt'
\label{eq:wmlvarrhoest}
\end{align}
for all solutions $z$ to (\ref{eq:fourierthesystemRge}). 
\end{cor}
\begin{remark}
The constant $K$ should be thought of as a frequency cut off. Considering
matrices of the form (\ref{eq:Aqinfinddef}), we can only carry out a detailed analysis (transforming the matrix to Jordan normal form etc.) for a finite
number of matrices. Here this finite number is determined by the bound $\nuaverage{q_{\trs}}(\indexnot)\leq K$ (where $q_{\trs}$ is defined by the condition that 
$\baverage{q_{\trs}}=0$); cf. Definition~\ref{def:varrhoeppl}.
\end{remark}
\begin{remark}\label{remark:removingbothepsilon}
In case there is no transparent era with non-empty intersection with $I_{a}$, then $C_{\e,K}$ can be replaced by a constant depending 
only on $\e$ and the coefficients of the equation (\ref{eq:thesystemRge}). In that setting, we also use the notation $\varrho_{\e,\pm}$ instead of 
$\varrho_{\e,K,\pm}$. 
\end{remark}
\begin{remark}\label{remark:removingbothepsilonandK}
When the Jordan blocks of $R_{q,+}^{+}$ and $R_{q,+}^{-}$ corresponding to eigenvalues with real part $\kappa_{q,+}$ are trivial for all
$q\in \{1,\dots,Q\}$ such that $I_{q}\cap I_{a}\neq\varnothing$, and there is no transparent era with non-empty intersection with $I_{a}$, then the
$C_{\e,K}$ appearing in (\ref{eq:wplvarrhoest}) can be replaced by a constant depending only on the coefficients of the equation (\ref{eq:thesystemRge}). 
A similar statement concerning (\ref{eq:wmlvarrhoest}) holds. 
\end{remark}
\begin{proof}
Let us begin by proving (\ref{eq:wplvarrhoest}) in the case that there is no transparent era with non-empty intersection with $I_{a}$. Given $t_{a}$ 
and $t_{b}$ as in the statement of the lemma, define the intervals $I_{j}$ as in Definition~\ref{def:varrhoeppl}, where $s$ and $t$ have 
been replaced by $t_{a}$ and $t_{b}$ respectively. Let, moreover, $s_{j}$, $j=1,\dots,k$, be defined as in Definition~\ref{def:varrhoeppl}. If $k=0$, 
i.e., $I_{1}=I_{a}$, then Corollary~\ref{cor:solnoest} yields
\begin{equation}\label{eq:westplusappliedpre}
|w(t_{b})|\leq C_{\e}e^{(\kappa_{q_{1},+}+\e)(t_{b}-t_{a})}|w(t_{a})|+C_{\e}\int_{t_{a}}^{t_{b}}e^{(\kappa_{q_{1},+}+\e)(t_{b}-t')}|\hf(t')|dt',
\end{equation}
where $C_{\e}$ only depends on $\e$ and the coefficients of the equation (\ref{eq:thesystemRge}). Moreover, in case the Jordan blocks of $R_{q_{1},+}^{+}$ and 
$R_{q_{1},+}^{-}$ corresponding to the eigenvalues with real part 
$\kappa_{q_{1},+}$ are trivial, the parameter $\e$ can be replaced by zero, and $C_{\e}$ can be replaced by a constant independent of $\e$. Thus 
(\ref{eq:wplvarrhoest}) holds if $k=0$, i.e., if $I_{1}=I_{a}$. Let us now assume, inductively, that (\ref{eq:wplvarrhoest}) holds if the number of intervals
$I_{j}$ into which $I_{a}$ is divided (when appealing to Definition~\ref{def:varrhoeppl}) is bounded from above by $k$. Assume now that $I_{a}$ is divided
into $k+1$ intervals. Then
\begin{equation*}
\begin{split}
|w(t_{b})| \leq & C_{\e}e^{(\kappa_{q_{k+1},+}+\e)(t_{b}-s_{k})}|w(s_{k})|+C_{\e}\int_{s_{k}}^{t_{b}}e^{(\kappa_{q_{k+1},+}+\e)(t_{b}-t')}|\hf(t')|dt'\\
 \leq & C_{\e}\exp\left[(\kappa_{q_{k+1},+}+\e)(t_{b}-s_{k})+\varrho_{\e,+}(s_{k},t_{a})\right]|w(t_{a})|\\
 & +C_{\e}\int_{t_{a}}^{s_{k}}\exp\left[(\kappa_{q_{k+1},+}+\e)(t_{b}-s_{k})+\varrho_{\e,+}(s_{k},t')\right]|\hf(t')|dt'\\
 & +C_{\e}\int_{s_{k}}^{t_{b}}e^{(\kappa_{q_{k+1},+}+\e)(t_{b}-t')}|\hf(t')|dt',
\end{split}
\end{equation*}
where we appealed to Corollary~\ref{cor:solnoest} in the first step and the inductive hypothesis in the second step. Moreover, the $\e$ can be removed
under circumstances analogous to those described in connection with (\ref{eq:westplusappliedpre}). By induction, we conclude that (\ref{eq:wplvarrhoest})
holds in the absence of transparent eras. Moreover, the $\e$ and $K$ can both be removed under the circumstances described in 
Remark~\ref{remark:removingbothepsilonandK}. In each step of the induction process, the constant appearing in the estimate grows. However, this does not 
lead to problems, since the number of steps is bounded by $Q$. 

\textbf{Presence of a transparent era.} In case there is a transparent era, we first assume $K$ to be larger than the $K_{\roco,\e,+}$ 
appearing in (\ref{eq:Krocoep}). Moreover, we assume that 
\[
\left|\kappa_{q_{\trs},\indexnot,+}-\kappa_{q_{\trs},+}\right|\leq \e/4
\]
for $\indexnot$ such that $\nuaverage{q_{\trs}}(\indexnot)\geq K$. Note that we can make these assumptions without loss of generality. Next we wish to use an 
argument similar to the one presented in connection with (\ref{eq:Krocoep}). In particular, we wish to appeal to (\ref{eq:Akprodndegplus}) or (\ref{eq:Akproddegplus}).
When doing so, we here assume the $\e$ appearing in (\ref{eq:Akproddegplus}) to have been replaced by $\e/4$. Given these assumptions, an argument similar
to the one presented in connection with (\ref{eq:Krocoep}) yields an estimate for transparent eras such that $\nuaverage{q_{\trs}}(\indexnot)\geq K$ which is 
good enough that we can argue as in the absence of transparent eras in order to obtain the desired conclusion. What remains is to consider transparent eras
such that $\nuaverage{q_{\trs}}(\indexnot)\leq K$. However, the relevant estimate is in that case supplied by Lemma~\ref{lemma:Iqestbavqeqzero}. This finishes
the proof of (\ref{eq:wplvarrhoest}), including the improvements described in Remarks~\ref{remark:removingbothepsilon} and \ref{remark:removingbothepsilonandK}.

The proof of (\ref{eq:wmlvarrhoest}) is similar, and we leave the details to the reader. 
\end{proof}

\section{The case of unbounded frequency eras}\label{section:thecaseofunbdfreeras}

Even though Lemma~\ref{lemma:maprnoes} yields an essentially optimal estimate (see Subsection~\ref{ssection:optimality} below), there is room for 
improvement in the case that there is an unbounded non-transparent era $I_{q}$. In fact, we have the following result.

\begin{lemma}\label{lemma:westunbfre}
Consider the equation (\ref{eq:thesystemRge}). Assume that it is non-degenerate, diagonally dominated, balanced and convergent; cf. 
Definition~\ref{def:nondegconvabal}. Fix $0\neq \indexnot\in\EFindexset$ and assume that $I_{q}$ is an unbounded and non-transparent era
containing $[t_{0},\infty)$ for some $q\in \{1,\dots,Q\}$ and $t_{0}\geq 0$. Assume, moreover, that (\ref{eq:gtalbge}) holds with $t_{a}$ replaced by 
$t_{0}$ and that (\ref{eq:maxqdiaglbitokappaq}) holds. Assume, finally, that if $\{t_{k}\}$ is the sequence given by
Definition~\ref{def:tkdefgeaddbd} and starting at $t_{0}$, then $t_{\fin,\ror}=\infty$. 
Define $R_{q,\pm}^{+}$ and $R_{q,\pm}^{-}$ by (\ref{eq:matrixdetgrowth}); define $\kappa_{q,\pm}$ by 
(\ref{eq:kappaqpmdef}); and define $d_{q,\pm}$ by (\ref{eq:dqpmdef}). Then there is a constant $C$ depending only on the spectrum of the Riemannian 
manifold corresponding to $q$ and the coefficients of the equation (\ref{eq:thesystemRge}), such that 
\begin{equation}\label{eq:westfininfnontrsera}
\begin{split}
|w(\indexnot,t)| \leq & Ce^{\kappa_{q,+}\bt}\ldr{\bt}^{d_{q,+}-1}|w(\indexnot,t_{a})|
+C\int_{t_{a}}^{t}\ldr{t-t'}^{d_{q,+}-1}e^{\kappa_{q,+}(t-t')}|\hf(\indexnot,t')|dt'
\end{split}
\end{equation}
for all $t_{0}\leq t_{a}\leq t$ and all solutions $z$ to (\ref{eq:fourierthesystemRge}), where $\bt=t-t_{a}$.
\end{lemma}
\begin{remark}
If $\baverage{q}>0$, it should be possible to remove the following assumptions: that (\ref{eq:gtalbge}) holds with $t_{a}$ replaced by $t_{0}$; that 
(\ref{eq:maxqdiaglbitokappaq}) 
holds; and that $t_{\fin,\ror}=\infty$. The reason for this is that $\baverage{q}>0$ implies that $\mfg(\indexnot,t)\rightarrow\infty$ as $t\rightarrow\infty$. 
Replacing $t_{0}$ by a later time, say $t_{b}\geq t_{0}$, all of the mentioned assumptions are therefore satisfied. On the other hand, the length of the 
interval $[t_{0},t_{b}]$ is bounded by 
a constant depending only on the spectrum of the Riemannian manifold corresponding to $q$ and the coefficients of the equation (\ref{eq:thesystemRge}).
Deriving a bound for the interval $[t_{0},t_{b}]$ should therefore not be a problem; cf. Lemma~\ref{lemma:roughenestbalsetting}. 
\end{remark}
\begin{proof}
Keeping the observations made in Subsection~\ref{ssection:estsolinhomeqprelobs}, in particular (\ref{eq:winhomformredtoPhiest}), in mind, 
it is clear that in order to obtain (\ref{eq:westfininfnontrsera}), it is sufficient to prove that 
\begin{equation}\label{eq:Phiindnottmsestfinnoisy}
\|\Phi_{\indexnot}(t;t')\|\leq C\ldr{t-t'}^{d_{q,+}-1}e^{\kappa_{q,+}(t-t')}
\end{equation}
for all $t_{0}\leq t'\leq t$, where $C$ depends only on the spectrum of the Riemannian manifold corresponding to $q$ and the coefficients of the equation 
(\ref{eq:thesystemRge}). On the other hand, in order to prove (\ref{eq:Phiindnottmsestfinnoisy}), it is sufficient to prove that 
\begin{equation}\label{eq:westhomverofeq}
|w(\indexnot,t)|\leq C\ldr{t-t'}^{d_{q,+}-1}e^{\kappa_{q,+}(t-t')}|w(\indexnot,t')|
\end{equation}
for all $t_{0}\leq t'\leq t$ and all solutions $w$ to the homogeneous version of (\ref{eq:wmamfprel}). Moreover, $C$ should have the same dependence as 
in the case of (\ref{eq:Phiindnottmsestfinnoisy}). The remainder of the proof is devoted to demonstrating (\ref{eq:westhomverofeq}).

\textbf{Preliminaries.} To begin with, note that since the frequency era $I_{q}$ contains $[t_{0},\infty)$ and since $t_{\fin,\ror}=\infty$, it is clear that 
$\baverage{q}>0$ and that $\baverage{q}>\baverage{p}$ for all $p\neq q$ such that $\nuaverage{p}(\indexnot)\neq 0$. Let $\{t_{k}\}$ be a sequence of the type 
described in Definition~\ref{def:tkdefgeaddbd}, starting at $t_{0}$. 

\textbf{Improving the matrix estimates.} Consider the proof of Lemma~\ref{lemma:maprnoes}. Since the assumptions of the present lemma imply 
that the assumptions of Lemma~\ref{lemma:maprnoes} are satisfied, we know that (\ref{eq:chAkmchAkapp}) and (\ref{eq:chAkjedef}) hold. Consider the right hand 
side of (\ref{eq:chAkmchAkapp}). The first and the last term inside the parenthesis decay exponentially (since $\baverage{q}>0$). Let us consider the middle 
term. Due to (\ref{eq:mfhmfHjdef}),
\begin{equation}\label{eq:mfHqsestdisttobdry}
\mfH_{q}^{(s)}(\indexnot,t_{k})=\sum_{p\neq q}\frac{\nuaverage{p}^{s}(\indexnot)e^{s\baverage{p}t_{k}}}{\nuaverage{q}^{s}(\indexnot)e^{s\baverage{q}t_{k}}}
=\sum_{p\neq q}\frac{\nuaverage{p}^{s}(\indexnot)e^{s\baverage{p}t_{0}}}{\nuaverage{q}^{s}(\indexnot)e^{s\baverage{q}t_{0}}}e^{-s(\baverage{q}-\baverage{p})(t_{k}-t_{0})}
\leq (Q-1)e^{-s\b_{\rodm}\bt_{k}}
\end{equation}
for $t_{k}\geq t_{0}$, where $s>0$; $\b_{\rodm}$ is the difference between $\baverage{q}$ and the second largest $\baverage{p}$ for which 
$\nuaverage{p}(\indexnot)\neq 0$; $\bt_{k}=t_{k}-t_{0}$; and we have used the fact that (by the definition of $t_{0}$) 
$\nuaverage{p}(\indexnot)e^{\baverage{p}t_{0}}\leq\nuaverage{q}(\indexnot)e^{\baverage{q}t_{0}}$ for all $p\in \{1,\dots,Q\}$. Thus (\ref{eq:chAkmchAkapp}) yields
\begin{equation}\label{eq:chAkmchAkappinfIj}
\|\chA_{k,q,\e}^{\pm}-\chA^{\app,\pm}_{k,q,\e}\|\leq \frac{C_{\e,\lambda}}{\nuaverage{q}(\indexnot)e^{\baverage{q}t_{k}}}e^{-\eta_{\rodm}\bt_{k}}
\end{equation}
for $t_{k}\geq t_{0}$, where the constant $C_{\e,\lambda}$ only depends on $\e$, $\lambda$, the spectrum of the Riemannian manifold corresponding to $q$ and the 
coefficients of the equation (\ref{eq:thesystemRge});
$\eta_{\rodm}:=\min\{\eta_{A},\b_{\rodm},\baverage{q}\}$; $\chA^{\app,\pm}_{k,q,\e}$ is given by (\ref{eq:chAkjedef}); and $\chA_{k,q,\e}^{\pm}$ is defined via 
(\ref{eq:Akjendef}), (\ref{eq:hAkjendef}) and (\ref{eq:ChAkjedef}), where $\lambda$ remains to be determined. 
Considering the proof of Lemma~\ref{lemma:maprnoes}, it is clear that by choosing $\lambda=\kappa_{q,+}$ and $\e$ small enough (the bound depending
only on the difference between $\lambda=\kappa_{q,+}$ and the second largest real part of an eigenvalue of the matrix $R_{\infty,q}(\indexnot)$ introduced 
in (\ref{eq:Rinfjn})), we can write (\ref{eq:chAkjedef}) as 
\begin{equation}\label{eq:chAkj}
\chA^{\app,+}_{k,q,\e}:=\Id_{2m}+\frac{2\pi}{\nuaverage{q}(\indexnot)e^{\baverage{q}t_{k}}}J_{R},
\end{equation}
where $J_{R}\in\Mn{2m}{\ro}$ consists of generalised Jordan blocks (cf. the terminology introduced in Remark~\ref{remark:genJordblock}) such that the blocks 
corresponding to strictly negative eigenvalues are negative definite. Since this choice corresponds to specific values of $\lambda$ and $\e$, depending only on 
the coefficients of the equation (\ref{eq:thesystemRge}), we can omit reference to $\e$ and $\lambda$ in what follows. Note also that $J_{R}$ can be chosen to
be independent of $\indexnot$. The reason for this is that there are only two possibilities of what $R_{\infty,q}(\indexnot)$ can be, and the difference 
corresponds to interchanging the blocks in the matrix appearing on the right hand side of the second equality in (\ref{eq:Rinfjn}). 

\textbf{Estimating the evolution along the time sequence.} Fix an integer $k_{0}\geq 0$. To begin with, we are interested in deriving estimates
for $k\geq k_{0}$. Consider (\ref{eq:wprekpertransvarpi}), where $\psi_{k}=w_{\fin,k}$, $A_{k}^{\pm}=\Xi^{k,\pm}_{\fin}$ and $\hF_{k}$ is replaced by $0$. 
This equality implies that
\begin{equation}\label{eq:rhokpoit}
\rho_{k+ 1}=\bA_{k,q,\e}^{+}\rho_{k},
\end{equation}
where 
\begin{align}
\rho_{k} := & e^{-\kappa_{q,+}t_{k}}e^{-J_{\rodiff}\tildt_{k}}S_{k}T_{q,\e}\psi_{k},\label{eq:rhokbAkdef}\\
\bA_{k,q,\e}^{+} := & e^{-J_{\rodiff}\tildt_{k+1}}\chA_{k,q,\e}^{+}e^{J_{\rodiff}\tildt_{k}},\nonumber
\end{align}
$\tildt_{k}:=t_{k}-t_{k_{0}}$ and $S_{k}$ and $T_{q,\e}$ are defined in the statement of Lemma~\ref{lemma:maprnoes}. Moreover, $J_{\tr}$ is defined to 
be the matrix obtained from $J_{R}$ by setting the Jordan blocks with zero eigenvalue to zero, and $J_{\rodiff}:=J_{R}-J_{\tr}$. Thus, if $k\geq k_{0}$, 
\begin{equation}\label{eq:rhokpointfor}
\begin{split}
\rho_{k+1} = & \bA_{k,q,\e}^{+}\cdots \bA_{k_{0},q,\e}^{+}\rho_{k_{0}}.
\end{split}
\end{equation}
Compute
\begin{equation}\label{eq:bAauxexp}
\bA_{k,q,\e}^{+}=e^{-J_{\rodiff}\tildt_{k+1}}(\chA_{k,q,\e}^{+}-\chA^{\app,+}_{k,q,\e})e^{J_{\rodiff}\tildt_{k}}
+e^{-J_{\rodiff}(t_{k+1}-t_{k})}\chA^{\app,+}_{k,q,\e};
\end{equation}
note that $J_{\rodiff}$ and $\chA^{\app,+}_{k,q,\e}$ commute. On the other hand, 
\begin{equation*}
\begin{split}
e^{-J_{\rodiff}(t_{k+1}-t_{k})}\chA^{\app,+}_{k,q,\e}
 = & \chA^{\app,+}_{k,q,\e}-J_{\rodiff}(t_{k+1}-t_{k})+\frac{2\pi}{\nuaverage{q}(\indexnot)e^{\baverage{q}t_{k}}}J_{R}[e^{-J_{\rodiff}(t_{k+1}-t_{k})}-\Id_{2m}]\\
 & +[e^{-J_{\rodiff}(t_{k+1}-t_{k})}-\Id_{2m}+J_{\rodiff}(t_{k+1}-t_{k})].
\end{split}
\end{equation*}
Introducing 
\begin{equation}\label{eq:bAkjappdef}
\bA^{\app,+}_{k,q}:=\Id_{2m}+\frac{2\pi}{\nuaverage{q}(\indexnot)e^{\baverage{q}t_{k}}}J_{\tr}
\end{equation}
and appealing to (\ref{eq:tkpomtkest}) yields 
\[
\|e^{-J_{\rodiff}(t_{k+1}-t_{k})}\chA^{\app,+}_{k,q,\e}-\bA^{\app,+}_{k,q}\|\leq \frac{C}{\nuaverage{q}(\indexnot)e^{\baverage{q}t_{k}}}e^{-\eta_{\rodm}\bt_{k}}
\]
for all $k\geq k_{0}$, where $C$ only depends on the spectrum of the Riemannian manifold corresponding to $q$ and the coefficients of the equation 
(\ref{eq:thesystemRge}); note that $\eta_{A}\leq \eta_{\diag}$. Combining this estimate with (\ref{eq:chAkmchAkappinfIj}) and (\ref{eq:bAauxexp}) yields
\[
\|\bA_{k,q,\e}^{+}-\bA^{\app,+}_{k,q}\|\leq \frac{C}{\nuaverage{q}(\indexnot)e^{\baverage{q}t_{k}}}\ldr{\tildt_{k}}^{N}e^{-\eta_{\rodm}\bt_{k}}
\]
for all $k\geq k_{0}$, where $C$ only depends on the spectrum of the Riemannian manifold corresponding to $q$ and the coefficients of the equation 
(\ref{eq:thesystemRge}), and $N$ is a non-negative integer depending only on $m$. Note also that $\e$ has been defined so that $\|\bA^{\app,+}_{k,q}\|\leq 1$. 
In particular, for $k\geq k_{0}$,
\begin{equation}\label{eq:bAkprod}
\|\bA_{k,q,\e}^{+}\cdots\bA_{k_{0},q,\e}^{+}\|\leq C
\end{equation}
for a constant $C$ depending only on the spectrum of the Riemannian manifold corresponding to $q$ and the coefficients of the equation 
(\ref{eq:thesystemRge}); the proof of this statement is similar to the argument presented in the proof of Lemma~\ref{lemma:maprnoes}. Combining this 
estimate with (\ref{eq:rhokpointfor}) yields
\begin{equation}\label{eq:rhokpobd}
|\rho_{k+1}|\leq C|\rho_{k_{0}}|
\end{equation}
for $k\geq k_{0}$, where the constant $C$ has the same dependence as in the case of (\ref{eq:bAkprod}). Returning to the definition of $\rho_{k}$ in 
terms of $\psi_{k}$, cf. (\ref{eq:rhokbAkdef}), and the definition of $\psi_{k}=w_{\fin,k}$, cf. (\ref{eq:wfinkdef}), this estimate yields
\begin{equation}\label{eq:wkposharpestimate}
\begin{split}
|w(\indexnot,t_{k+1})| \leq & Ce^{\kappa_{q,+}\tildt_{k+1}}\ldr{\tildt_{k+1}}^{d_{q,+}-1}|w(\indexnot,t_{k_{0}})|,
\end{split}
\end{equation}
where the constant $C$ has the same dependence as in the case of (\ref{eq:bAkprod}). Thus (\ref{eq:westhomverofeq}) holds if $t'=t_{k_{0}}$ and 
$t=t_{k}$. In general, $s\in [t_{k_{0}-1},t_{k_{0}}]$ and $t\in [t_{k},t_{k+1}]$ for some $0<k_{0}\leq k$. Appealing to 
Lemma~\ref{lemma:roughenestbalsetting} in order to estimate the evolution between $t'$ and $t_{k_{0}}$ and between $t_{k}$ and $t$ yields
the conclusion that (\ref{eq:westhomverofeq}) holds for all $t_{0}\leq t'\leq t$. Moreover, $C$ has the dependence stated in connection with 
(\ref{eq:Phiindnottmsestfinnoisy}). The lemma follows. 
\end{proof}

\section{Optimality}\label{ssection:optimality}

It is of interest to determine to what extent the estimates (\ref{eq:Akprodndegplus})--(\ref{eq:Akproddegminus}) are optimal. 

\begin{lemma}\label{lemma:lbmatprod}
Consider the equation (\ref{eq:thesystemRge}). Assume that it is non-degenerate, diagonally dominated, balanced and convergent; cf. 
Definition~\ref{def:nondegconvabal}. Fix $q\in \{1,\dots,Q\}$ and assume that $\baverage{q}\neq 0$. Let $R_{q,\pm}^{+}$, $R_{q,\pm}^{-}$ 
and $\kappa_{q,\pm}$ be given by Definition~\ref{def:tXqinfRqpmkappaqpm}. For every $\e>0$, there are then constants $C_{\e}>0$, 
$\mfg_{\e,\min}>0$, $t_{\roas,\e}\geq 0$ and $t_{\romar,\e}\geq 0$, depending only on 
the coefficients of the equation (\ref{eq:thesystemRge}) and $\e$, such that the following holds. Fix $0\neq\indexnot\in\EFindexset$; let $I_{q}$ be 
the frequency era corresponding to $\indexnot$ and $q$ (cf. Definition~\ref{def:era}); and define a sequence $\{t_{k}\}$ as in 
Definition~\ref{def:tkdefgeaddbd}. Then, if $t_{k}\in I_{q}$ and $\mfg(\indexnot,t_{k})\geq \mfg_{\e,\min}$ for $k_{-}\leq k\leq k_{+}$; 
$t_{k_{-}}\geq t_{\roas,\e}$; and the distance from $t_{k_{\pm}}$ to the boundary of $I_{q}$ is greater than $t_{\romar,\e}$, 
\begin{equation}\label{eq:Akmaprlb}
\|A_{k_{\pm}\mp 1}^{\pm}\cdots A_{k_{\mp}}^{\pm}\|\geq C_{\e}\exp\left[\pm(\kappa_{q,\pm}-\e)(t_{k_{\pm}}-t_{k_{\mp}})\right].
\end{equation}
In the case of the plus sign, the estimate can be improved under the following circumstances. 
If the Jordan blocks of $R_{q,+}^{+}$ and $R_{q,+}^{-}$ corresponding to the eigenvalues with real part $\kappa_{q,+}$ are trivial, then the constants 
$C_{\e}>0$, $\mfg_{\e,\min}>0$, $t_{\roas,\e}\geq 0$ and $t_{\romar,\e}\geq 0$ can be chosen to be independent of $\e$ and the right hand side of 
(\ref{eq:Akmaprlb}) can be replaced by $C\exp\left[\kappa_{q,+}(t_{k_{+}}-t_{k_{-}})\right]$ for a constant $C>0$ depending only on the coefficients of 
the equation (\ref{eq:thesystemRge}). The corresponding statement in the case of the minus sign is the following. If the Jordan blocks of $R_{q,-}^{+}$ 
and $R_{q,-}^{-}$ corresponding to the eigenvalues with real part $\kappa_{q,-}$ are trivial, then the constants 
$C_{\e}>0$, $\mfg_{\e,\min}>0$, $t_{\roas,\e}\geq 0$ and $t_{\romar,\e}\geq 0$ can be chosen to be independent of $\e$ and the right hand side of 
(\ref{eq:Akmaprlb}) can be replaced by $C\exp\left[-\kappa_{q,-}(t_{k_{-}}-t_{k_{+}})\right]$ for a constant $C>0$ depending only on the coefficients of 
the equation (\ref{eq:thesystemRge}).
\end{lemma}
\begin{remark}
Due to (\ref{eq:Akmaprlb}), the estimates (\ref{eq:Akprodndegplus}) and (\ref{eq:Akprodndegminus}) (with $C_{\deg,q,\pm}=0$) are optimal in terms of 
growth/decay in the case of trivial Jordan blocks. In the presence of non-trivial Jordan blocks, there is an $\e$-loss; cf. 
(\ref{eq:Akproddegplus}) and (\ref{eq:Akproddegminus}) (with $C_{\deg,q,\pm}=0$). 
\end{remark}
\begin{remark}
In case $\baverage{q}=0$, optimality results can be derived using Proposition~\ref{prop:spasdatrs}; cf. the proof of Theorem~\ref{thm:mainoptthm} 
below.
\end{remark}
\begin{remark}
The main requirements needed in order to obtain (\ref{eq:Akmaprlb}) have the following intuitive interpretation:
\begin{itemize}
\item The inequality $t_{k_{-}}\geq t_{\roas,\e}$ corresponds to the requirement that we are in the asymptotic region. Since we do not know much 
concerning the coefficients before we are in the asymptotic region, such an assumption seems natural. 
\item That the distance from $t_{k_{\pm}}$ to the boundary of $I_{q}$ is greater than $t_{\romar,\e}$ means that we are sufficiently far away 
from the boundary of the frequency era $I_{q}$. Since the behaviour close to another frequency era, say $I_{p}$, might as well be approximated by 
the relevant matrices in $I_{p}$, this is a natural assumption. 
\item The inequality $\mfg(\indexnot,t)\geq \mfg_{\e,\min}$ corresponds to the requirement that we are well in the oscillatory regime. Since the estimate
(\ref{eq:Akmaprlb}) is based on the assumption that there are oscillations and since the corresponding approximations improve with the frequency of the 
oscillations, this is also a natural assumption.
\end{itemize}
\end{remark}
\begin{remark}\label{remark:optsol}
The proof also yields a lower bound on the growth of solutions. In fact, define $w(\indexnot,t_{k})$ by 
\[
w(\indexnot,t_{k})=T_{\pre,k}^{-1}[D_{k}(\indexnot)]^{-1}\zeta_{k},
\]
where $\zeta_{k}$ is defined in (\ref{eq:psikdef}) below and $\xi(k)$ is defined via the iteration (\ref{eq:xikit}) and a suitable choice 
of $\xi(k_{a})$ (in the case of the plus sign, $k_{a}$ equals $k_{-}$ and in the case of the minus sign, $k_{a}$ equals $k_{+}$, and the choice
of $\xi(t_{k_{a}})$ depends on the sign ($\pm$)). Then (\ref{Akprodnorlb}) yields
\begin{equation}\label{eq:wtkpolb}
\frac{|w(\indexnot,t_{k_{\pm}})|}{|w(\indexnot,t_{k_{\mp}})|}\geq C_{\e}\exp[\pm(\kappa_{q,\pm}-\e)(t_{k_{\pm}}-t_{k_{\mp}})],
\end{equation}
where $C_{\e}>0$ depends only on $\e$ and the coefficients of the equation (\ref{eq:thesystemRge}). 
Note that there is a solution to the homogeneous version of the equation (\ref{eq:fourierthesystemRge}) such that $w(\indexnot,t_{k})$ 
corresponds to this solution, evaluated at $t_{k}$. Again, in the case of trivial Jordan blocks, the $\e$ can be removed, so that 
the growth/decay estimates derived in Corollary~\ref{cor:solnoest} are optimal in that case (for homogeneous equations and $\kappa_{q,+}=\kappa_{+}$
and $\kappa_{q,-}=\kappa_{-}$ respectively). 
\end{remark}
\begin{remark}
The optimality statement of Remark~\ref{remark:optsol} is limited in the following sense. For every choice of time interval $[t_{k_{-}},t_{k_{+}}]$ 
(satisfying the assumptions of the lemma, where $q$ is such that $\kappa_{q,+}=\kappa_{+}$, or such that $\kappa_{q,-}=\kappa_{-}$), we know that 
there is a solution such that a lower bound similar to the upper bound in (\ref{eq:westplus}), or (\ref{eq:westminus}), holds on that time interval 
(in the case of homogeneous equations). However, it is not clear that there is a solution such that the lower bound holds on a semi-infinite interval. 
Nevertheless, if $\baverage{q}>0$ and $\indexnot$ is such that the only non-zero $\nuaverage{p}(\indexnot)$ is $\nuaverage{q}(\indexnot)$, then $I_{q}$ is 
semi-infinite and we obtain optimality in the sense that there is a solution to the homogeneous version of the equation such that a lower bound of the 
desired form holds for all future times. In case $\baverage{q}=0$, optimality results are obtained by appealing to Proposition~\ref{prop:spasdatrs}.
However, in case $\baverage{q}<0$ and the spatially homogeneous growth rate is not the largest growth rate, it is not so clear that optimality holds
in the sense that we have the desired lower bound for all future times. 
\end{remark}
\begin{proof}
Consider the proof of Lemma~\ref{lemma:maprnoes}. Since the assumptions of the present lemma imply that the assumptions of Lemma~\ref{lemma:maprnoes} 
are satisfied, we know that (\ref{eq:chAkmchAkapp}) and (\ref{eq:chAkjedef}) hold. These estimates are the starting point for the analysis. However,
in order to be able to demonstrate the desired lower bounds, we need to isolate the component of the solution that corresponds to maximal growth
(minimal decay); we need 
to distinguish between the maximally growing part and the remainder; we need to obtain a lower bound for the growth of the maximally growing component; 
and, finally, we need to translate the corresponding lower bound to a lower bound on the norm of the matrix products. 

\textbf{Dividing solution vectors into a maximally growing component and a remainder.} Assume the top left $m_{+}\times m_{+}$-block of 
$\mathrm{Re}\{J_{q,\e}\}$ to correspond to the generalised Jordan blocks with eigenvalue $\kappa_{q,+}$; the bottom right $m_{-}\times m_{-}$-block of 
$\mathrm{Re}\{J_{q,\e}\}$ to correspond to the generalised Jordan blocks with eigenvalue $-\kappa_{q,-}$; and collect the remaining blocks in the middle
in an $m_{0}\times m_{0}$-matrix. Note that even if $J_{q,\e}$ depends on $\indexnot$, the collection of Jordan blocks of $J_{q,\e}$ does not depend on 
$\indexnot$. Note, moreover, that the eigenvalues of $J_{q,\e}$ depend neither on $\indexnot$ nor on $\e$. Divide vectors $\xi\in\cn{2m}$ into  
components $\xi_{1,\pm}$ and $\xi_{2,\pm}$, where $\xi_{1,+}$ consists of the first $m_{+}$ elements of $\xi$; $\xi_{1,-}$ consists of the 
last $m_{-}$ elements of $\xi$;  $\xi_{2,+}$ consists of the last $m_{0}+m_{-}$ elements of $\xi$; and $\xi_{2,-}$ consists of the first $m_{+}+m_{0}$ 
elements of $\xi$. Let $\xi(k_{a})\in\cn{2m}$ for some $k_{-}\leq k_{a}\leq k_{+}$, and define $\xi(k\pm 1)$ by 
\begin{equation}\label{eq:xikit}
\xi(k\pm 1):=\chA_{k,q,\e}^{\pm}\xi(k),
\end{equation}
where $\chA_{k,q,\e}^{\pm}$ is defined in (\ref{eq:ChAkjedef}). We write the components of $\xi(k)$ (according to the above division) $\xi_{1,\pm}(k)$ and 
$\xi_{2,\pm}(k)$. Then appealing to (\ref{eq:chAkmchAkapp}) and (\ref{eq:chAkjedef}) with $\lambda=0$ yields
\begin{equation}\label{eq:wokpolb}
\begin{split}
 & |\xi_{1,\pm}(k\pm 1)| \geq  \left(1+\frac{2\pi}{\nuaverage{q}(\indexnot)e^{\baverage{q}t_{k}}}(\kappa_{q,\pm}-\e)\right)|\xi_{1,\pm}(k)|\\
 & -\frac{C_{\e}}{\nuaverage{q}(\indexnot)e^{\baverage{q}t_{k}}}\left[e^{-\eta_{A}t_{k}}
+\sum_{s=1}^{2}\mfH_{q}^{(s)}(\indexnot,t_{k})+\frac{1}{\nuaverage{q}(\indexnot)e^{\baverage{q}t_{k}}}\right][|\xi_{1,\pm}(k)|+|\xi_{2,\pm}(k)|],
\end{split}
\end{equation}
where $C_{\e}$ depends only on $\e$ and the coefficients of the equation (\ref{eq:thesystemRge}). Let us consider the relative sizes 
of the terms on the right hand side. To begin with, note that, due to (\ref{eq:njnogequ}) and the fact that $t_{k}\in I_{q}$,
\begin{equation}\label{eq:nuavqebcoarseest}
\frac{1}{\nuaverage{q}(\indexnot)e^{\baverage{q}t_{k}}}\leq \frac{C}{\mfg(\indexnot,t_{k})}\leq C\mfg^{-1}_{\e,\min},
\end{equation}
where $C$ depends only on the coefficients of the equation (\ref{eq:thesystemRge}). By assuming $\mfg_{\e,\min}$ to be sufficiently large (depending only 
on the coefficients of the equation), it is thus clear that we can make the left hand side of (\ref{eq:nuavqebcoarseest}) as small as we wish; from now on
we, for the sake of convenience, always assume it to be less than or equal to $1$. Thus we can 
make the first factor in the first term on the right hand side of (\ref{eq:wokpolb}) strictly positive by assuming $\mfg_{\e,\min}$ to be sufficiently large,
the lower bound depending only on $\e$ and the coefficients of the equation (\ref{eq:thesystemRge}) (assuming $0<\e<1$, which we do below, the lower bound
in fact only depends on the coefficients of the equation (\ref{eq:thesystemRge})). Consider the second factor in the second 
term on the right hand side of (\ref{eq:wokpolb}). The third term in this factor can be assumed to be as small as we want by demanding $\mfg_{\e,\min}$ 
to be sufficiently large. Moreover, by assuming $t_{\roas,\e}$ to be large enough, we can assume $e^{-\eta_{A}t_{k}}$ to be as small as we wish. Finally, 
note that 
\[
\textstyle{\sum}_{s=1}^{2}\mfH_{q}^{(s)}(\indexnot,t_{k})\leq (Q-1)(e^{-\eta_{q}t_{\romar,\e}}+e^{-2\eta_{q}t_{\romar,\e}})
\]
for $k_{-}\leq k\leq k_{+}$, where $\eta_{q}=|\baverage{q}-\baverage{p}|$, where $\baverage{p}$ is the $\baverage{q}$ which is closest to $\baverage{q}$; 
cf. (\ref{eq:mfHqsestdisttobdry}). Again, by assuming $t_{\romar,\e}$ to be large enough, we can assume this expression to be as small as we wish. 

\textbf{Preliminary conclusions.} Due to the above observations, it is clear that by choosing $t_{\roas,\e}$, $t_{\romar,\e}$ and $\mfg_{\e,\min}$ to be large 
enough (the bound depending only on the coefficients of the equation and $\e$), the following holds. If $|\xi_{1,\pm}(k)|>0$ and $|\xi_{1,\pm}(k)|\geq |\xi_{2,\pm}(k)|$, 
then the right and side of (\ref{eq:wokpolb}) is strictly positive. Note also that we can remove the $-\e$ appearing in the first 
row (and replace $C_{\e}$ by a constant independent of $\e$ in the second row) in case the relevant generalised Jordan blocks are trivial. 

\textbf{Controlling the relative sizes of the components.} Let $\varkappa_{q,\pm}<\kappa_{q,\pm}$ be the second largest real part of an eigenvalue of 
$\pm J_{q,\e}$ (it could be that there is no such $\varkappa_{q,\pm}$, but for that to happen the diagonal components of $\mathrm{Re}\{J_{q,\e}\}$ all
have to agree, a situation which is much easier to deal with; in that case, there is no need to control the relative sizes of the components, since
there is only one component). Then, assuming $\mfg_{\e,\min}$ to be large enough (the lower bound depending only on the coefficients of the equation),
\begin{equation}\label{eq:xitwopmest}
\begin{split}
& |\xi_{2,\pm}(k\pm 1)| \leq  \left(1+\frac{2\pi}{\nuaverage{q}(\indexnot)e^{\baverage{q}t_{k}}}(\varkappa_{q,\pm}+\e)\right)|\xi_{2,\pm}(k)|\\
 & +\frac{C_{\e}}{\nuaverage{q}(\indexnot)e^{\baverage{q}t_{k}}}\left[e^{-\eta_{A}t_{k}}
+\sum_{s=1}^{2}\mfH_{q}^{(s)}(\indexnot,t_{k})+\frac{1}{\nuaverage{q}(\indexnot)e^{\baverage{q}t_{k}}}\right][|\xi_{1,\pm}(k)|+|\xi_{2,\pm}(k)|],
\end{split}
\end{equation}
where $C_{\e}$ depends only on $\e$ and the coefficients of the equation (\ref{eq:thesystemRge}). Let us now assume that $\xi(k_{a})\neq 0$ and that
$|\xi_{2,\pm}(k_{a})|/|\xi_{1,\pm}(k_{a})|\leq \de$ for some $0<\de\leq 1$; recall that $k_{a}$ is introduced in connection with (\ref{eq:xikit}). 
Moreover, make the inductive assumption that $|\xi_{2,\pm}(k)|/|\xi_{1,\pm}(k)|\leq 2\de$. Then (\ref{eq:wokpolb}) yields
\begin{equation}\label{eq:xionepmquotest}
\begin{split}
\frac{|\xi_{1,\pm}(k\pm 1)|}{|\xi_{1,\pm}(k)|} \geq & 1+\frac{2\pi}{\nuaverage{q}(\indexnot)e^{\baverage{q}t_{k}}}(\kappa_{q,\pm}-\e)\\
 & -\frac{C_{\e}}{\nuaverage{q}(\indexnot)e^{\baverage{q}t_{k}}}\left[e^{-\eta_{A}t_{k}}
+\sum_{s=1}^{2}\mfH_{q}^{(s)}(\indexnot,t_{k})+\frac{1}{\nuaverage{q}(\indexnot)e^{\baverage{q}t_{k}}}\right],
\end{split}
\end{equation}
where $C_{\e}$ depends only on $\e$ and the coefficients of the equation (\ref{eq:thesystemRge}); note that for general $\de$, the
constant would also depend on $\de$, but the restriction that $\de\leq 1$ allows us to avoid such a dependence. In what follows, we assume $t_{\roas,\e}$, 
$t_{\romar,\e}$ and $\mfg_{\e,\min}$ to be large enough that the second and third terms on the right hand side of (\ref{eq:xionepmquotest}) are $\ll 1$.
For our purposes, it is in fact sufficient to demand that the sum of the second and third terms are, in absolute value, bounded from above by $1/2$. 
The corresponding bounds on $t_{\roas,\e}$, $t_{\romar,\e}$ and $\mfg_{\e,\min}$ depend only on $\e$ and the coefficients of the equation (\ref{eq:thesystemRge}).
Due to (\ref{eq:xitwopmest}), there is an upper bound on $|\xi_{2,\pm}(k\pm 1)|/|\xi_{1,\pm}(k)|$ which is similar to (\ref{eq:xionepmquotest}). 
Combining this estimate with (\ref{eq:xionepmquotest}) yields
\begin{equation}\label{eq:wtnodbwono}
\begin{split}
\frac{|\xi_{2,\pm}(k\pm 1)|}{|\xi_{1,\pm}(k\pm 1)|} \leq & \frac{|\xi_{2,\pm}(k)|}{|\xi_{1,\pm}(k)|}
+\frac{2\pi}{\nuaverage{q}(\indexnot)e^{\baverage{q}t_{k}}}(\varkappa_{q,\pm}-\kappa_{q,\pm}+2\e)\frac{|\xi_{2,\pm}(k)|}{|\xi_{1,\pm}(k)|}\\
 & +\frac{C_{\e}}{\nuaverage{q}(\indexnot)e^{\baverage{q}t_{k}}}\left[e^{-\eta_{A}t_{k}}
+\sum_{s=1}^{2}\mfH_{q}^{(s)}(\indexnot,t_{k})+\frac{1}{\nuaverage{q}(\indexnot)e^{\baverage{q}t_{k}}}\right],
\end{split}
\end{equation}
where $C_{\e}$ has the same dependence as in the case of (\ref{eq:xionepmquotest}). In the end, we would like to argue that if 
$|\xi_{2,\pm}(k)|/|\xi_{1,\pm}(k)|$ starts out bounded by $\de$, then it cannot become larger than $2\de$. In order for us to obtain such a conclusion 
using (\ref{eq:wtnodbwono}), it is necessary for the last factor in the last term on the right hand side of (\ref{eq:wtnodbwono}) to be small. Moreover, 
$\e$ has to be small enough that $\varkappa_{q,\pm}-\kappa_{q,\pm}+2\e$ is negative. Thus, 
let us fix 
\[
0<\e\leq \frac{1}{6}\min\{\kappa_{q,+}-\varkappa_{q,+},\kappa_{q,-}-\varkappa_{q,-}\}
\]
and $0<\de\leq 1$. In order for the sum of the last two terms on the right hand side of (\ref{eq:wtnodbwono}) to be strictly negative when 
$|\xi_{2,\pm}(k)|/|\xi_{1,\pm}(k)|\geq \de$, it is sufficient to assume $t_{\roas,\e}$, $t_{\romar,\e}$ and $\mfg_{\e,\min}$ to be large enough, where the lower 
bounds in this case depend only on the coefficients of the equation (\ref{eq:thesystemRge}), $\e$ and $\de$. Moreover, again assuming $t_{\roas,\e}$, 
$t_{\romar,\e}$ and $\mfg_{\e,\min}$ to be large enough (the constants involved in the restrictions having the same dependence), if $k_{-}\leq k\leq k_{+}$ and 
$|\xi_{2,\pm}(k)|\leq \de |\xi_{1,\pm}(k)|$, then 
\[
\frac{|\xi_{2,\pm}(k\pm 1)|}{|\xi_{1,\pm}(k\pm 1)|}\leq \de+\frac{2\pi(\kappa_{q,\pm}-\varkappa_{q,\pm})}{\nuaverage{q}(\indexnot)e^{\baverage{q}t_{k}}}\de.
\]
Assuming $\mfg_{\e,\min}$ to be large enough, we can assume the second term on the right hand side to be less than $\de$, so that 
$|\xi_{2,\pm}(k\pm 1)|/|\xi_{1,\pm}(k\pm 1)|\leq 2\de$. Combining the above arguments, it is clear that if $|\xi_{2,\pm}(k_{a})|\leq \de |\xi_{1,\pm}(k_{a})|$,
then the quotient $|\xi_{2,\pm}(k)|/|\xi_{1,\pm}(k)|$ might reach a value in the interval $[\de,2\de]$, but if it does, it will decrease until it is smaller 
than $\de$ etc. In short, an inductive argument demonstrates that $|\xi_{2,+}(k)|/|\xi_{1,+}(k)|\leq 2\de$ for all $k_{a}\leq k\leq k_{+}$ and that 
$|\xi_{2,-}(k)|/|\xi_{1,-}(k)|\leq 2\de$ for all $k_{-}\leq k\leq k_{a}$ (though, needless to say, the initial conditions (for $k=k_{a}$) yielding the first 
conclusion typically differ from the initial conditions yielding the second conclusion). In the remaining arguments, the particular value of $\de$ is not
important. We therefore fix $\de:=1/4$ from now on. 

\textbf{Deriving a lower bound on the size of the dominant component.} Returning to (\ref{eq:wokpolb}) with the above information concerning the 
relative sizes of the components in mind, it is clear that 
\begin{equation}\label{eq:wokpoit}
\begin{split}
|\xi_{1,\pm}(k\pm 1)| \geq & \left(1+\frac{2\pi}{\nuaverage{q}(\indexnot)e^{\baverage{q}t_{k}}}(\kappa_{q,\pm}-\e)\right.\\
 & \left. -\frac{C_{\e}}{\nuaverage{q}(\indexnot)e^{\baverage{q}t_{k}}}\left[e^{-\eta_{A}t_{k}}
+\sum_{s=1}^{2}\mfH_{q}^{(s)}(\indexnot,t_{k})+\frac{1}{\nuaverage{q}(\indexnot)e^{\baverage{q}t_{k}}}\right]\right)|\xi_{1,\pm}(k)|.
\end{split}
\end{equation}
In the case of trivial generalised Jordan blocks, the $\e$ appearing in the first row can be removed. Moreover, in order to control the relative sizes of the 
components, it is sufficient to fix an $\e$ depending only on the coefficients of the equation. For that reason, $C_{\e}$ appearing on the right hand 
side can be replaced by a constant $C$ depending only on the coefficients of the equation. 
In order to estimate $|\xi_{1,\pm}(k_{\pm})|$ from below in terms of $|\xi_{1,\pm}(k_{a})|$, it is of interest to estimate the sum of the logarithms of the 
first factor on the right hand side (for $k=k_{a},\dots,k_{+}-1$ and $k=k_{-}+1,\dots,k_{a}$ respectively). In the case of the plus sign, it can be bounded 
from below by
\begin{equation*}
\begin{split}
 & \sum_{k=k_{a}}^{k_{+}-1}\left(\frac{2\pi}{\nuaverage{q}(\indexnot)e^{\baverage{q}t_{k}}}(\kappa_{q,+}-\e)\right.\\
 & \left. -\frac{C_{\e}}{\nuaverage{q}(\indexnot)e^{\baverage{q}t_{k}}}\left[e^{-\eta_{A}t_{k}}
+\sum_{s=1}^{2}\mfH_{q}^{(s)}(\indexnot,t_{k})+\frac{1}{\nuaverage{q}(\indexnot)e^{\baverage{q}t_{k}}}\right]\right)\\
 \geq & (\kappa_{q,+}-\e)(t_{k_{+}}-t_{k_{a}})-C_{\e},
\end{split}
\end{equation*}
assuming $\baverage{q}\neq 0$, where we have appealed to (\ref{eq:sumapproxint}), (\ref{eq:sumestint}) and (\ref{eq:sumestmfHint}). Moreover, $C_{\e}$ only 
depends on $\e$ and the coefficients of the equation (\ref{eq:thesystemRge}). Combining this estimate with (\ref{eq:wokpoit}) yields
\begin{equation}\label{eq:wokolb}
|\xi_{1,+}(k_{+})|\geq C_{\e}\exp[(\kappa_{q,+}-\e)(t_{k_{+}}-t_{k_{a}})]|\xi_{1,+}(k_{a})|,
\end{equation}
where $C_{\e}>0$ only depends on $\e$ and the coefficients of the equation (\ref{eq:thesystemRge}). A similar argument in the case of the minus sign
yields
\begin{equation}\label{eq:wokolbminus}
|\xi_{1,-}(k_{-})|\geq C_{\e}\exp[(-\kappa_{q,-}+\e)(t_{k_{-}}-t_{k_{a}})]|\xi_{1,-}(k_{a})|,
\end{equation}
where the constant has the same dependence as in the case of (\ref{eq:wokolb}). Note, moreover, that the $\e$ appearing in (\ref{eq:wokolb}) and 
(\ref{eq:wokolbminus}) can be removed in case the relevant Jordan blocks are trivial. One consequence of (\ref{eq:wokolb}) and (\ref{eq:wokolbminus})
is that 
\begin{equation}\label{eq:xikpmlowbd}
|\xi(k_{\pm})|\geq C_{\e}\exp[(\pm\kappa_{q,\pm}\mp\e)(t_{k_{\pm}}-t_{k_{a}})]|\xi(k_{a})|.
\end{equation}
Here $C_{\e}>0$ only depends on $\e$ and the coefficients of the equation (\ref{eq:thesystemRge}), and it is important to keep in mind that what 
$\xi(k_{a})$ is depends on the sign ($\pm$). 

\textbf{Deriving a lower bound on the norm of the matrix products.} Let us now translate the estimate (\ref{eq:xikpmlowbd})
to a lower bound on the norm of the matrix products. By definition, 
\begin{equation}\label{eq:xikitreform}
\begin{split}
\xi(k_{\pm}) = & \chA_{k_{\pm}\mp 1,q,\e}^{\pm}\cdots \chA_{k_{a},q,\e}^{\pm}\xi(k_{a})
=S_{k_{\pm}}\hA_{k_{\pm}\mp 1,q,\e}^{\pm}\cdots \hA_{k_{a},q,\e}^{\pm}S_{k_{a}}^{-1}\xi(k_{a})\\
 = & S_{k_{\pm}}A_{k_{\pm}\mp 1,q,\e}^{\pm}\cdots A_{k_{a},q,\e}^{\pm}S_{k_{a}}^{-1}\xi(k_{a})\\
 = & S_{k_{\pm}}T_{q,\e}A_{k_{\pm}\mp 1}^{\pm}\cdots A_{k_{a}}^{\pm}T_{q,\e}^{-1}S_{k_{a}}^{-1}\xi(k_{a})
\end{split}
\end{equation}
where we have appealed to (\ref{eq:Akjendef}), (\ref{eq:hAkjendef}) and (\ref{eq:ChAkjedef}); recall also that $\lambda=0$ in the present setting.  
In the arguments below, it is, however, important to keep in mind that $\xi(k_{a})$ depends on the sign ($\pm$). Letting
\begin{equation}\label{eq:psikdef}
\zeta_{k}:=T_{q,\e}^{-1}S_{k}^{-1}\xi(k),
\end{equation}
(\ref{eq:xikitreform}) can be written
\[
\zeta_{k_{\pm}}=A_{k_{\pm}\mp 1}^{\pm}\cdots A_{k_{a}}^{\pm}\zeta_{k_{a}}.
\]
In particular, (\ref{eq:xikpmlowbd}) yields
\begin{equation}\label{Akprodnorlb}
\|A_{k_{\pm}\mp 1}^{\pm}\cdots A_{k_{a}}^{\pm}\|\geq \frac{|\zeta_{k_{\pm}}|}{|\zeta_{k_{a}}|}\geq C_{\e}\frac{|\xi(k_{\pm})|}{|\xi(k_{a})|}
\geq C_{\e}\exp[(\pm\kappa_{q,\pm}\mp\e)(t_{k_{\pm}}-t_{k_{a}})],
\end{equation}
where the $\e$ can be removed in case the relevant Jordan blocks are trivial. The estimate for the matrix products stated in (\ref{eq:Akmaprlb}) follows
and the constants only depend on $\e$ and the coefficients of the equation; in the case of the plus sign, we can choose $k_{a}=k_{-}$ and in the case of 
the minus sign, we can choose $k_{a}=k_{+}$. The lemma follows. 
\end{proof}



\chapter{Estimating the Sobolev norms of solutions}\label{chapter:estsobnormsolbodoftext}

\section{Introduction}

Our next goal is to estimate how the Sobolev norms of solutions to (\ref{eq:thesystemRge}) evolve with time. To be more specific, consider a 
solution $u$ to (\ref{eq:thesystemRge}) and define $\mfe_{s}[u]$ by (\ref{eq:mfedef}), where $s\in\ro$ and $z(\indexnot,t)$ is given by 
(\ref{eq:znutdef}). Then we want to estimate $\mfe_{s}[u](t)$, $t\geq 0$, in terms of $\mfe_{s}[u](0)$ and integrals of appropriate norms of $f$. 
However, in the case of homogeneous equations, we are also interested in estimating $\mfe_{s_{1}}[u](0)$ in terms of $\mfe_{s_{2}}[u](T)$ for 
appropriate choices of $s_{1}$, $s_{2}$ and $T>0$. The reason we take an interest in the latter type of estimates is that they are needed 
when specifying the asymptotics; cf. the proof of Lemma~\ref{lemma:spasda}.
In order to obtain estimates, it is natural to consider each mode separately. However, even when considering only one mode, the time intervals
under consideration have to be divided into pieces. Due to Corollary~\ref{cor:solnoest}, we can estimate how the energy of one mode evolves in the 
interval $[t_{\roini,\ror},t_{\fin,\ror}]$ (or $[t_{\roini,\ror},t_{\fin,\ror})$ in case $t_{\fin,\ror}=\infty$); here $\{t_{k}\}$ is a time sequence and $t_{\roini,\ror}$
and $t_{\fin,\ror}$ are the corresponding numbers introduced in Definition~\ref{def:tkdefgeaddbd}. However, we need to proceed differently on 
$[t_{\fin,\ror},\infty)$ in case $t_{\fin,\ror}<\infty$ and on $[0,t_{\roini,\ror}]$ in case the behaviour is not initially oscillatory. We start by discussing 
the latter two situations in Section~\ref{section:ODEregime}. We then return to the problem of estimating the energy of the solution in 
Section~\ref{section:estsobnorm}. Finally, in Section~\ref{section:optsobest}, we verify the optimality of the results.

\section{Mode estimates for the entire future}\label{section:ODEregime}

When estimating how the energy of a mode develops over time, our main tools are Lemma~\ref{lemma:oderegest}, Lemma~\ref{lemma:roughenestbalsetting}
and Corollary~\ref{cor:solnoest}. The subintervals of $[0,\infty)$ to which we apply these results depend on the $\indexnot$ of the mode
under consideration. It is therefore convenient to introduce the following classification of the elements of $\EFindexset$. 

\begin{definition}\label{def:frequencysectors}
Consider the equation (\ref{eq:thesystemRge}). Assume that it is non-degenerate, diagonally dominated, balanced and convergent; cf. 
Definition~\ref{def:nondegconvabal}. Then the elements $\indexnot$ of $\EFindexset$ are classified as follows:
\begin{itemize}
\item If 
\begin{itemize}
\item the inequality $\baverage{q}<0$ holds for all $q\in\{1,\dots,Q\}$ such that $\nuaverage{q}(\indexnot)\neq 0$, and
\item either the inequality (\ref{eq:maxqdiaglbitokappaq}) with $t_{0}$ replaced by $0$ is violated, or the inequality (\ref{eq:gtalbge}) with 
$t_{a}$ replaced by $0$ is violated, 
\end{itemize}
then $\indexnot$ is said to belong to the \textit{low frequency silent sector}.
\index{Low frequency!silent sector}%
\index{Silent sector!low frequency}%
\index{Silent!sector}%
\index{Sector!silent}%
\item If
\begin{itemize}
\item the inequality $\baverage{q}<0$ holds for all $q\in\{1,\dots,Q\}$ such that $\nuaverage{q}(\indexnot)\neq 0$, 
\item the inequality (\ref{eq:gtalbge}) with $t_{a}$ replaced by $0$ is satisfied, and
\item the inequality (\ref{eq:maxqdiaglbitokappaq}) with $t_{0}$ replaced by $0$ is satisfied,
\end{itemize}
then $\indexnot$ is said to belong to the \textit{high frequency silent sector}.
\index{High frequency!silent sector}%
\index{Silent sector!high frequency}%
\index{Silent!sector}%
\index{Sector!silent}%
\item If 
\begin{itemize}
\item the inequality $\baverage{q}\leq 0$ holds for all $q\in\{1,\dots,Q\}$ such that $\nuaverage{q}(\indexnot)\neq 0$, 
\item either the inequality (\ref{eq:maxqdiaglbitokappaq}) with $t_{0}$ replaced by $0$ is violated, or the inequality (\ref{eq:gtalbge}) with 
$t_{a}$ replaced by $0$ is violated, and
\item there is one $q\in\{1,\dots,Q\}$ such that $\nuaverage{q}(\indexnot)\neq 0$ and $\baverage{q}=0$, 
\end{itemize}
then $\indexnot$ is said to belong to the \textit{low frequency transparent sector}.
\index{Low frequency!transparent sector}%
\index{Transparent sector!low frequency}%
\index{Transparent!sector}%
\index{Sector!transparent}%
\item If
\begin{itemize}
\item the inequality $\baverage{q}\leq 0$ holds for all $q\in\{1,\dots,Q\}$ such that $\nuaverage{q}(\indexnot)\neq 0$, 
\item the inequality (\ref{eq:gtalbge}) holds with $t_{a}$ replaced by $0$, 
\item the inequality (\ref{eq:maxqdiaglbitokappaq}) with $t_{0}$ replaced by $0$ is satisfied, and
\item there is one $q\in\{1,\dots,Q\}$ such that $\nuaverage{q}(\indexnot)\neq 0$ and $\baverage{q}=0$, 
\end{itemize}
then $\indexnot$ is said to belong to the \textit{high frequency transparent sector}.
\index{High frequency!transparent sector}%
\index{Transparent sector!high frequency}%
\index{Transparent!sector}%
\index{Sector!transparent}%
\item If there is one $q\in\{1,\dots,Q\}$ such that $\nuaverage{q}(\indexnot)\neq 0$ and $\baverage{q}>0$, 
then $\indexnot$ is said to belong to the \textit{noisy sector}.
\index{Noisy!sector}%
\index{Sector!noisy}%
\end{itemize}
\end{definition}
\begin{remark}
Since the equation is non-degenerate, diagonally dominated, balanced and convergent, Lemma~\ref{lemma:consconvcoeff} implies that it is 
oscillation adapted. Thus (\ref{eq:estdtlnge}) and (\ref{eq:mffXdotbdetc}) hold for all $0\neq\indexnot\in\EFindexset$, and $\ellderbd$ 
is the constant appearing in these estimates. 
\end{remark}
\begin{remark}
The element $\indexnot=0$ of $\EFindexset$ belongs to the low frequency silent sector. 
\end{remark}

In preparation for the next proposition, it is also convenient to introduce the following terminology. 

\begin{definition}\label{def:subdomnondeg}
Consider the equation (\ref{eq:thesystemRge}). Assume that it is non-degenerate, diagonally dominated, balanced and convergent; cf. 
Definition~\ref{def:nondegconvabal}. If there is a $q_{\trs}\in \{1,\dots,Q\}$ such that $\baverage{q_{\trs}}=0$, let 
$\kappa_{\rotot,+}:=\max\{\kappa_{\trs,+},\kappa_{\rosil,+}\}$, where $\kappa_{\trs,+}$ is defined in (\ref{eq:kappatrpmdef}); 
$\kappa_{\rosil,+}:=\kappa_{\max}(A_{\infty})$; and $A_{\infty}$ is defined in (\ref{eq:vAhFdef}). If there is no such $q_{\trs}$, let 
$\kappa_{\rotot,+}:=\max\{\kappa_{+},\kappa_{\rosil,+}\}$, where $\kappa_{\rosil,+}$ is defined as before and $\kappa_{+}$ is defined by 
(\ref{eq:kappapmdef}). In addition, (\ref{eq:thesystemRge}) is said to exhibit \textit{subdominant block non-degeneracy} if the
following holds: 
\begin{itemize}
\item if there is a $q_{\trs}\in \{1,\dots,Q\}$ such that $\baverage{q_{\trs}}=0$, then $\bka_{q_{\trs},+}<\kappa_{\rotot,+}$, 
\item for $q\in \{1,\dots,Q\}$ such that $\baverage{q}<\max\{0,\baverage{Q}\}$, all the Jordan blocks of $R_{q,+}^{+}$ and $R_{q,+}^{-}$ corresponding 
to the eigenvalues with real part $\kappa_{\rotot,+}$ are trivial (recall that $\baverage{1}<\cdots<\baverage{Q}$), and
\item if $\baverage{Q}>0$ and there is a $q_{\trs}\in \{1,\dots,Q\}$ such that $\baverage{q_{\trs}}=0$, then the Jordan blocks of 
$A_{q_{\trs},\infty}(\indexnottwo)$ corresponding to the eigenvalues with real part $\kappa_{\rotot,+}$ are trivial for all 
$\indexnottwo\in\EFindexset$ such that $\nuaverage{q_{\trs}}(\indexnottwo)\neq 0$.
\end{itemize}
\end{definition}
\begin{remark}\label{remark:subdomblockdegbody}
The motivation for the introduction of the terminology ``subdominant block non-degeneracy'' is that, when deriving estimates not involving the
loss of an $\e>0$, we wish to exclude the occurrence of frequency eras 
\begin{itemize}
\item in which the relevant matrices have an eigenvalue with real part $\kappa_{\rotot,+}$ and a corresponding non-trivial Jordan block, and
\item which are not of the form $[t_{0},\infty)$.
\end{itemize}
The reason we wish to exclude such frequency eras is that the polynomial factors in the estimates cause problems at the transition when a new 
frequency era begins; cf. the estimates appearing in the proof of Lemma~\ref{lemma:maprnoes}. We also wish to exclude $\bka_{q_{\trs},+}=\kappa_{\rotot,+}$, 
since the value $\bka_{q_{\trs},+}$ arises in the limit as $\nuaverage{q_{\trs}}(\indexnot)\rightarrow\infty$; returning to the analysis in 
Chapter~\ref{chapter:ODEtransp}, it is clear that there is then a tension between the following two goals: that of decomposing the matrices
into Jordan normal form; and that of controlling the norms of the corresponding transformation matrices.
\end{remark}

Given the terminology introduced in Definition~\ref{def:subdomnondeg}, we are in a position to state the main estimates concerning the individual modes 
of solutions to (\ref{eq:thesystemRge}).

\begin{prop}\label{prop:modeestentfuture}
Consider the equation (\ref{eq:thesystemRge}). Assume that it is non-degenerate, diagonally dominated, balanced and convergent; cf. 
Definition~\ref{def:nondegconvabal}. Fix $\e>0$. Then there is a constant $0<C_{\e}\in\ro$, depending only on $\e$, the spectra of 
$(M_{r},g_{r})$, $r\in \{1,\dots,R\}$, and the coefficients of the equation (\ref{eq:thesystemRge}), such that 
\begin{equation}\label{eq:megenestlspresob}
\me^{1/2}(\indexnot,t) \leq  C_{\e}e^{(\kappa_{\rotot,+}+\e)t}\me^{1/2}(\indexnot,0)+C_{\e}\int_{0}^{t}e^{(\kappa_{\rotot,+}+\e)(t-t')}|\hf(\indexnot,t')|dt'
\end{equation}
for all $t\geq 0$, all solutions $z$ to (\ref{eq:fourierthesystemRge}) and all $\indexnot\in\EFindexset$, where $\kappa_{\rotot,+}$ is given by 
Definition~\ref{def:subdomnondeg}. If, in addition to the above, the equation (\ref{eq:thesystemRge}) exhibits 
subdominant block non-degeneracy, there are the following improvements of (\ref{eq:megenestlspresob}):
\begin{itemize}
\item If $\baverage{Q}<0$, then 
\begin{equation}\label{eq:megenestlspresobsilcase}
\me^{1/2}(\indexnot,t) \leq  C\ldr{t}^{d_{\ros,+}-1}e^{\kappa_{\rotot,+}t}\me^{1/2}(\indexnot,0)
+C\int_{0}^{t}\ldr{t-t'}^{d_{\ros,+}-1}e^{\kappa_{\rotot,+}(t-t')}|\hf(\indexnot,t')|dt'
\end{equation}
holds for all $t\geq 0$, all solutions $z$ to (\ref{eq:fourierthesystemRge}) and all $\indexnot\in\EFindexset$, 
where $d_{\ros,+}:=d_{\max}(A_{\infty},\kappa_{\rotot,+})$; 
$A_{\infty}$ is defined in (\ref{eq:vAhFdef}); and the constant $C$ only depends on the spectra of $(M_{r},g_{r})$, $r\in \{1,\dots,R\}$, and 
the coefficients of the equation (\ref{eq:thesystemRge}).
\item If $\baverage{Q}=0$, then 
\begin{equation}\label{eq:megenestlspresobstscase}
\me^{1/2}(\indexnot,t) \leq  C\ldr{t}^{d_{\rosts,+}-1}e^{\kappa_{\rotot,+}t}\me^{1/2}(\indexnot,0)
+C\int_{0}^{t}\ldr{t-t'}^{d_{\rosts,+}-1}e^{\kappa_{\rotot,+}(t-t')}|\hf(\indexnot,t')|dt'
\end{equation}
holds for all $t\geq 0$, all solutions $z$ to (\ref{eq:fourierthesystemRge}) and all $\indexnot\in\EFindexset$, where 
$d_{\rosts,+}:=\max\{d_{\ros,+},d_{\rot,+}\}$; $d_{\ros,+}$ is defined as in the case of (\ref{eq:megenestlspresobsilcase});
$d_{\rot,+}$ is the largest dimension of a non-trivial Jordan block of 
a matrix of the form $A_{Q,\infty}(\indexnottwo)$ (where $\indexnottwo\in\EFindexset$ is such that $\nuaverage{Q}(\indexnottwo)\neq 0$) corresponding 
to an eigenvalue with real part $\kappa_{\rotot,+}$; and the constant $C$ only depends on the spectra of $(M_{r},g_{r})$, $r\in \{1,\dots,R\}$,
and the coefficients of the equation (\ref{eq:thesystemRge}). 
\item If $\baverage{Q}>0$, then 
\begin{equation}\label{eq:megenestlspresobsncase}
\me^{1/2}(\indexnot,t) \leq  C\ldr{t}^{d_{\rotot,+}-1}e^{\kappa_{\rotot,+}t}\me^{1/2}(\indexnot,0)
+C\int_{0}^{t}\ldr{t-t'}^{d_{\rotot,+}-1}e^{\kappa_{\rotot,+}(t-t')}|\hf(\indexnot,t')|dt'
\end{equation}
holds for all $t\geq 0$, all solutions $z$ to (\ref{eq:fourierthesystemRge}) and all $\indexnot\in\EFindexset$, where $d_{\rotot,+}:=\max\{d_{\ros,+},d_{\ron,+}\}$; 
$d_{\ros,+}$ is defined as in the case of (\ref{eq:megenestlspresobsilcase});
\[
d_{\ron,+}:=\max\{d_{\max}(R_{Q,+}^{+},\kappa_{\rotot,+}),d_{\max}(R_{Q,+}^{-},\kappa_{\rotot,+})\};
\]
and the constant $C$ only depends on the spectra of $(M_{r},g_{r})$, $r\in \{1,\dots,R\}$, and the coefficients of the equation 
(\ref{eq:thesystemRge}). 
\end{itemize}
\end{prop}
\begin{remark}
If there is no element of $\Spe A_{\infty}$ with real part $\kappa_{\rotot,+}$, then the $d_{\ros,+}$ appearing in (\ref{eq:megenestlspresobsilcase}) equals $1$, so
that the polynomial expressions appearing in (\ref{eq:megenestlspresobsilcase}) can be omitted; cf. Definition~\ref{def:SpRspdef}. Similar remarks can 
be made concerning (\ref{eq:megenestlspresobstscase}) and (\ref{eq:megenestlspresobsncase}). 
\end{remark}
\begin{proof}
We divide the proof according to the different cases identified in Definition~\ref{def:frequencysectors}. 

\textbf{The low frequency silent sector.} Assume that $\indexnot$ belongs to the low frequency silent sector. Then either the inequality 
(\ref{eq:maxqdiaglbitokappaq}) with $t_{0}$ replaced by $0$ is violated, or the inequality (\ref{eq:gtalbge}) with $t_{a}$ replaced by $0$ 
is violated. Regardless of which is the case, there is a constant $C$, depending only on the coefficients of the equation, such that
that $\nuaverage{q}(\indexnot)\leq C$ for all $q$. Combining this observation with (\ref{eq:mfgdiagdef}) and (\ref{eq:mfgmfgdiageq}) yields 
\[
\mfg(\indexnot,t)\leq Ce^{\b_{\min,-}t}
\]
for all $t\geq 0$, where $C$ only depends on the coefficients of the equation (\ref{eq:thesystemRge}) and $\b_{\min,-}$ is the largest $\baverage{q}$
which is strictly smaller than zero. In order to obtain this conclusion, we used the fact that $\baverage{q}<0$ holds for all 
$q\in\{1,\dots,Q\}$ such that $\nuaverage{q}(\indexnot)\neq 0$. In analogy with the arguments presented at the beginning of 
Section~\ref{section:roughODEbaenest}, it is then clear that (\ref{eq:fourierthesystemRge}) can be written as (\ref{eq:dotvroughODE}), where 
$v$, $A_{\infty}$ and $F$ are as defined in (\ref{eq:vAhFdef}). Moreover, $A_{\rem}$ is a smooth function taking its values in $\Mn{2m}{\co}$ and satisfying
(\ref{eq:Aremestgensys}), where, in the present setting, $C_{\rem}$ only depends on the coefficients of the equation (\ref{eq:thesystemRge});
$\b_{\rem}:=\min\{-\b_{\min,-},\eta_{\romn}\}$; and $T_{\roode}:=0$. Appealing to Lemma~\ref{lemma:oderegest} yields 
\begin{equation}\label{eq:meestlowfresilsect}
\me^{1/2}(t)\leq C\ldr{t}^{d_{\rosil,+}-1}e^{\kappa_{\rosil,+}t}\me^{1/2}(0)
+C\int_{0}^{t}\ldr{t-t'}^{d_{\rosil,+}-1}e^{\kappa_{\rosil,+}(t-t')}|\hf(t')|dt'
\end{equation}
for all $t\geq 0$, 
where $C$ only depends on the coefficients of the equation (\ref{eq:thesystemRge}); $\kappa_{\rosil,+}:=\kappa_{\max}(A_{\infty})$; and 
$d_{\rosil,+}:=d_{\max}(A_{\infty},\kappa_{\rosil,+})$. In order to obtain this conclusion, we used the fact that $|v|$ and $\me^{1/2}$ are equivalent for 
$t\geq 0$. In particular, it is clear that (\ref{eq:megenestlspresob}) holds.

\textbf{The high frequency silent sector.} Assume that $\indexnot$ belongs to the high frequency silent sector. Then we can define a time sequence
$\{t_{k}\}$ of the type introduced in Definition~\ref{def:tkdefgeaddbd}, where $t_{0}=0$. Since $\baverage{q}<0$ for all $q\in\{1,\dots,Q\}$ such that 
$\nuaverage{q}(\indexnot)\neq 0$, it is clear that $t_{\fin,\ror}<\infty$. Moreover, Corollary~\ref{cor:solnoest} applies to the interval 
$[0,t_{\fin,\ror}]$. This yields, for any $\e>0$,
\begin{equation}\label{eq:westplusapplied}
\me^{1/2}(t)\leq C_{+,\e}e^{(\kappa_{+}+\e)t}\me^{1/2}(0)+C_{+,\e}\int_{0}^{t}e^{(\kappa_{+}+\e)(t-t')}|\hf(t')|dt'
\end{equation}
for all $t\in [0,t_{\fin,\ror}]$, where $C_{+,\e}$ only depends on $\e$ and the coefficients of the equation (\ref{eq:thesystemRge}) and $\kappa_{+}$ is 
defined by (\ref{eq:kappapmdef}) (in fact, it is sufficient to compute the supremum over the $q$'s such that $\baverage{q}<0$). Moreover,
in case the Jordan blocks of $R_{q,+}^{+}$ and $R_{q,+}^{-}$ corresponding to the eigenvalues with real part $\kappa_{+}$ are trivial for all 
$q\in \{1,\dots,Q\}$, the parameter $\e$ can be replaced by zero, and $C_{+,\e}$ can be replaced by a constant $C_{+}$ 
independent of $\e$. In order to obtain (\ref{eq:westplusapplied}), we also used the fact that $|w|$ and $\me^{1/2}$ are equivalent on 
$[0,t_{\fin,\ror}]$. For $t\geq t_{\fin,\ror}$, an argument similar to the proof of (\ref{eq:meestlowfresilsect}) yields
\begin{equation}\label{eq:meestlowfresilsectfinint}
\begin{split}
\me^{1/2}(t) \leq & C\ldr{\bt}^{d_{\rosil,+}-1}e^{\kappa_{\rosil,+}\bt}\me^{1/2}(t_{\fin,\ror})
+C\int_{t_{\fin,\ror}}^{t}\ldr{t-t'}^{d_{\rosil,+}-1}e^{\kappa_{\rosil,+}(t-t')}|\hf(t')|dt'
\end{split}
\end{equation}
for all $t\geq t_{\fin,\ror}$, where $\bt:=t-t_{\fin,\ror}$; $C$ has the same dependence as in the case of 
(\ref{eq:meestlowfresilsect}); and $\kappa_{\rosil,+}$ and $d_{\rosil,+}$ are defined as in the case of (\ref{eq:meestlowfresilsect}). 
When combining (\ref{eq:westplusapplied}) and (\ref{eq:meestlowfresilsectfinint}), there are four different cases to consider. If
$\kappa_{+}<\kappa_{\rosil,+}$, then (\ref{eq:meestlowfresilsect}) holds for all $t\geq 0$. If $\kappa_{+}=\kappa_{\rosil,+}$ and the $\e$
appearing in (\ref{eq:westplusapplied}) can be removed (due to the relevant Jordan blocks being trivial), then (\ref{eq:meestlowfresilsect}) 
holds for all $t\geq 0$. If $\kappa_{+}=\kappa_{\rosil,+}$ and the $\e$ appearing in (\ref{eq:westplusapplied}) cannot be removed, then 
(\ref{eq:westplusapplied}) holds for all $t\geq 0$. If $\kappa_{+}>\kappa_{\rosil,+}$, then (\ref{eq:westplusapplied}) holds for all $t\geq 0$.
Moreover, in case the Jordan blocks of $R_{q,+}^{+}$ and $R_{q,+}^{-}$ corresponding to the eigenvalues with real part $\kappa_{+}$ are trivial 
for all $q\in \{1,\dots,Q\}$, the parameter $\e$ can be replaced by zero, and $C_{+,\e}$ can be replaced by a constant $C_{+}$ independent of 
$\e$. Again, it is clear that (\ref{eq:megenestlspresob}) holds.

At this stage, it is convenient to prove (\ref{eq:megenestlspresobsilcase}) in case $\baverage{Q}<0$ and the equation exhibits subdominant 
block non-degeneracy. Returning to Definition~\ref{def:subdomnondeg}, it is clear that the $\e$ appearing in (\ref{eq:westplusapplied}) can be 
removed, if we replace $\kappa_{+}$ by $\kappa_{\rotot,+}$. Moreover, if $\kappa_{\rosil,+}<\kappa_{\rotot,+}$, then $d_{\rosil,+}$ can be set equal to 
$1$ in (\ref{eq:meestlowfresilsect}) and (\ref{eq:meestlowfresilsectfinint}), if we replace $\kappa_{\rosil,+}$ in these estimates with $\kappa_{\rotot,+}$.
Considering the above arguments again, it is thus clear that (\ref{eq:megenestlspresobsilcase}) holds. 

\textbf{The transparent sectors.} Before treating the high/low frequency transparent sectors separately, it is convenient to make some general remarks
which apply to both cases. Assume that there are constants $0<K_{\roco}\in\ro$ and $0\leq t_{1}\in\ro$ such that $\mfg(\indexnot,t)\leq K_{\roco}$, 
for $t\geq t_{1}$, where $\indexnot$ is in the high or low frequency transparent sector. In this case, there is one element of $\{1,\dots,Q\}$, say 
$q_{\trs}$, such that $\baverage{q_{\trs}}=0$ and $\nuaverage{q_{\trs}}(\indexnot)\neq 0$. Moreover, there is a constant $C_{\rolb}>0$,
depending only on the coefficients of the equation (\ref{eq:thesystemRge}) and the spectrum of the Riemannian manifold corresponding to 
$q_{\trs}$, such that $\mfg(\indexnot,t)\geq C_{\rolb}^{-1}$ for all $t\geq t_{1}$. Keeping this observation in mind, and appealing to 
(\ref{eq:mfgdiagdef}) and (\ref{eq:mfgmmfgdiag}) yields 
\begin{equation}\label{eq:mfgapproxlts}
|\mfg(\indexnot,t)-\nuaverage{q_{\trs}}(\indexnot)|\leq Ce^{-\b_{\rolts,1}\bt}
\end{equation}
for all $t\geq t_{1}$, where $\bt:=t-t_{1}$, $\b_{\rolts,1}:=\min\{-2\b_{\min,-},\eta_{\diag}\}$ and $C$ only depends on $K_{\roco}$, the coefficients of the equation 
(\ref{eq:thesystemRge}) and the spectrum of the Riemannian manifold corresponding to $q_{\trs}$. Before reformulating the equations as in the beginning of 
Section~\ref{section:refmodeanaltranspera}, note that 
\begin{equation*}
\begin{split}
\|n_{l}X^{l}(t)-\nuaverage{q_{\trs}}(\indexnot)\Xaverage{\diag}^{q_{\trs}}(\indexnot)\| = &
\left\|n_{l}X^{l}(t)-\textstyle{\sum}_{\{j:\b_{j}=0\}}n_{j}X^{j}_{\infty}\right\|\\
 \leq & Ce^{\b_{\min,-}\bt}+\left\|\textstyle{\sum}_{\{j:\b_{j}=0\}}n_{j}[X^{j}(t)-X^{j}_{\infty}]\right\|
\end{split}
\end{equation*}
for all $t\geq t_{1}$, where the constant only depends on $K_{\roco}$ and the coefficients of the equation (\ref{eq:thesystemRge}). In order to obtain this 
conclusion, we appealed to (\ref{eq:mainbd}), (\ref{eq:mfgdiagdef}), (\ref{eq:mfgmfgdiageq}), (\ref{eq:Xavdiagqindexnotdef}) 
and the fact that $\mfg(\indexnot,t)\leq K_{\roco}$ for $t\geq t_{1}$. In particular, 
\begin{equation}\label{eq:nlXlapproxlts}
\|n_{l}X^{l}(t)-\nuaverage{q_{\trs}}(\indexnot)\Xaverage{\diag}^{q_{\trs}}(\indexnot)\|\leq Ce^{-\b_{\rolts,2}\bt}
\end{equation}
for all $t\geq t_{1}$, where $\b_{\rolts,2}:=\min\{-\b_{\min,-},\eta_{\romn}\}$ and $C$ only depends on $K_{\roco}$ and the coefficients of the equation 
(\ref{eq:thesystemRge}). Combining (\ref{eq:mainconv}), (\ref{eq:mfgapproxlts}) and (\ref{eq:nlXlapproxlts}) with the fact that 
the equation is oscillation adapted with $\mff_{\rosh}(t)=Ce^{-\eta_{\rosh}t}$ yields the conclusion that (\ref{eq:fourierthesystemRge})
takes the form (\ref{eq:varpiqeq}) (with $q=q_{\trs}$), where $\varpi_{q}$ is defined by (\ref{eq:txqtyqvarpidef}); $A_{q,\infty}(\indexnot)$
is defined by (\ref{eq:Aqinfinddef}); $F(\indexnot,t)$ is defined by (\ref{eq:Findtdefdeg}); and $A_{q,\rem}(\indexnot,t)$
satisfies the estimate
\[
\|A_{q_{\trs},\rem}(\indexnot,t)\|\leq Ce^{-\b_{\rolts}\bt}
\]
for all $t\geq t_{1}$, where 
\[
\b_{\rolts}:=\min\{-\b_{\min,-},\eta_{\romn},\eta_{\rosh},\eta_{\rood},\eta_{\rod},\eta_{\road}\}
\]
(recall that $\eta_{\diag}:=\min\{\eta_{\rood},\eta_{\rod},\eta_{\road}\}$) and $C$ only depends on $K_{\roco}$, the coefficients of the equation 
(\ref{eq:thesystemRge}) and the spectrum of the Riemannian manifold corresponding to $q_{\trs}$. Introducing
\[
\kappa^{+}_{\trs,\indexnot}:=\kappa_{\max}[A_{q_{\trs},\infty}(\indexnot)],\ \ \
d^{+}_{\trs,\indexnot}:=d_{\max}[A_{q_{\trs},\infty}(\indexnot),\kappa^{+}_{\trs,\indexnot}],
\]
Lemma~\ref{lemma:oderegest} yields the conclusion that 
\begin{equation}\label{eq:varpiestlftrssector}
|\varpi_{q_{\trs}}(\indexnot,t)|\leq C\ldr{\bt}^{d^{+}_{\trs,\indexnot}-1}e^{\kappa^{+}_{\trs,\indexnot}\bt}
|\varpi_{q_{\trs}}(\indexnot,t_{1})|+C\int_{t_{1}}^{t}\ldr{t-t'}^{d^{+}_{\trs,\indexnot}-1}e^{\kappa^{+}_{\trs,\indexnot}(t-t')}|\hf(\indexnot,t')|dt'
\end{equation}
for all $t\geq t_{1}$. However, the constant $C$ in this case depends on $A_{q_{\trs},\infty}(\indexnot)$ as well as $K_{\roco}$, the coefficients of the 
equation (\ref{eq:thesystemRge}) and the spectrum of the Riemannian manifold corresponding to $q_{\trs}$. On the other hand, since 
$\nuaverage{q_{\trs}}(\indexnot)\leq C$ for some $C$ depending only on $K_{\roco}$ and the coefficients of the equation (\ref{eq:thesystemRge}), we 
need only take a finite number of matrices of the form $A_{q_{\trs},\infty}(\indexnot)$ into account. For this reason, the constant $C$ appearing in 
(\ref{eq:varpiestlftrssector}) can be assumed to only depend on $K_{\roco}$, the coefficients of the equation (\ref{eq:thesystemRge}) and the spectrum of 
the Riemannian manifold corresponding to $q_{\trs}$. Since $|\varpi_{q_{\trs}}(\indexnot,t)|$ and $\me^{1/2}(\indexnot,t)$ are equivalent for $t\geq t_{1}$ in 
the present setting, (\ref{eq:varpiestlftrssector}) implies that 
\begin{equation}\label{eq:meestlowfretrssect}
\me^{1/2}(\indexnot,t)\leq C\ldr{\bt}^{d^{+}_{\trs,\indexnot}-1}e^{\kappa^{+}_{\trs,\indexnot}\bt}
\me^{1/2}(\indexnot,t_{1})+C\int_{t_{1}}^{t}\ldr{t-t'}^{d^{+}_{\trs,\indexnot}-1}e^{\kappa^{+}_{\trs,\indexnot}(t-t')}|\hf(\indexnot,t')|dt'
\end{equation}
for all $t\geq t_{1}$, where $C$ only depends on $K_{\roco}$, the coefficients of the equation (\ref{eq:thesystemRge}) and the spectrum of the Riemannian 
manifold corresponding to $q_{\trs}$. 

\textbf{The low frequency transparent sector.} 
Assume that $\indexnot$ belongs to the low frequency transparent sector. In this case, there
is one element of $\{1,\dots,Q\}$, say $q_{\trs}$, such that $\baverage{q_{\trs}}=0$ and $\nuaverage{q_{\trs}}(\indexnot)\neq 0$. Moreover, since 
either the inequality (\ref{eq:maxqdiaglbitokappaq}) with $t_{0}$ replaced by $0$ is violated, or the inequality (\ref{eq:gtalbge}) with 
$t_{a}$ replaced by $0$ is violated, it is clear that there is a constant $C$, depending only on the coefficients of the equation
(\ref{eq:thesystemRge}) such that $\mfg(\indexnot,t)\leq C$ for all $t\geq 0$. In order to obtain this conclusion, we also used the 
fact that $\baverage{q}\leq 0$ for all $q\in \{1,\dots,Q\}$ such that $\nuaverage{q}(\indexnot)\neq 0$. Thus (\ref{eq:meestlowfretrssect})
holds with $t_{1}=0$, and the constant $C$ only depends on the coefficients of the equation (\ref{eq:thesystemRge}) and the spectrum of the Riemannian 
manifold corresponding to $q_{\trs}$. In particular, it is clear that (\ref{eq:megenestlspresob}) holds.

\textbf{The high frequency transparent sector.} Assume that $\indexnot$ belongs to the high frequency transparent sector. Then we can construct
a time sequence of the type introduced in Definition~\ref{def:tkdefgeaddbd}, starting at $t_{0}=0$. Moreover, there are two possibilities: either
$t_{\fin,\ror}=\infty$ or $t_{\fin,\ror}<\infty$. If $t_{\fin,\ror}=\infty$, we can appeal to Corollary~\ref{cor:solnoest} for all $t\geq 0$. This yields 
the conclusion that for every $\e>0$, 
\begin{equation}\label{eq:meesthighfretrssect}
\me^{1/2}(\indexnot,t) \leq  C_{\e}e^{(\kappa_{\trs,+}+\e)t}\me^{1/2}(\indexnot,0)+C_{\e}\int_{0}^{t}e^{(\kappa_{\trs,+}+\e)(t-t')}|\hf(\indexnot,t')|dt'
\end{equation}
for all $t\geq 0$, where $C_{\e}$ only depends on $\e$, the coefficients of the equation (\ref{eq:thesystemRge}) and the spectrum of the Riemannian manifold 
corresponding to $q_{\trs}$ (where $q_{\trs}$ is defined by the condition that $\baverage{q_{\trs}}=0$). Moreover, $\kappa_{\trs,+}$ is defined by 
(\ref{eq:kappatrpmdef}), and we used the fact that $\me^{1/2}$ and $|w|$ are equivalent for $t\geq 0$. Note, finally, that the $\e$ appearing in 
(\ref{eq:meesthighfretrssect}) can be removed under certain circumstances. Since these circumstances are described in the statement of 
Corollary~\ref{cor:solnoest}, we do not repeat them here. 

Assume now that $t_{\fin,\ror}<\infty$. Then Corollary~\ref{cor:solnoest} applies in 
$[0,t_{\fin,\ror}]$, so that (\ref{eq:meesthighfretrssect}) holds for $t\in [0,t_{\fin,\ror}]$. As in the low frequency case, since either the inequality 
(\ref{eq:maxqdiaglbitokappaq}) with $t_{0}$ replaced by $t_{k_{\fin,\ror}+1}$ is violated, or the inequality (\ref{eq:gtalbge}) with 
$t_{a}$ replaced by $t_{k_{\fin,\ror}+1}$ is violated, it is clear that there is a constant $C$, depending only on the coefficients of the equation
(\ref{eq:thesystemRge}), such that $\mfg(\indexnot,t)\leq C$ for all $t\geq t_{\fin,\ror}$. Moreover, (\ref{eq:meestlowfretrssect}) yields the conclusion
that 
\[
\me^{1/2}(\indexnot,t)\leq C\ldr{\bt}^{d^{+}_{\trs,\indexnot}-1}e^{\kappa^{+}_{\trs,\indexnot}\bt}
\me^{1/2}(\indexnot,t_{\fin,\ror})+C\int_{t_{\fin,\ror}}^{t}\ldr{t-t'}^{d^{+}_{\trs,\indexnot}-1}e^{\kappa^{+}_{\trs,\indexnot}(t-t')}|\hf(\indexnot,t')|dt'
\]
for all $t\geq t_{\fin,\ror}$, where $\bt:=t-t_{\fin,\ror}$ and $C$ only depends on the coefficients of the equation (\ref{eq:thesystemRge}) and the spectrum of 
the Riemannian manifold corresponding to $q_{\trs}$. Since $\kappa^{+}_{\trs,\indexnot}\leq\kappa_{\trs,+}$, it is thus clear that (\ref{eq:meesthighfretrssect})
holds for all $t\geq 0$, so that (\ref{eq:megenestlspresob}) holds. 

Let us now prove that (\ref{eq:megenestlspresobstscase}) holds in case $\baverage{Q}=0$ and the equation exhibits subdominant block non-degeneracy. 
Due to the analysis in the two silent sectors, we know that the desired conclusion holds if $\indexnot$ belongs to either the low or the high frequency silent
sector. Due to the analysis presented above for $\indexnot$ belonging to the low frequency transparent sector, it also holds for such $\indexnot$. Let us 
therefore focus on the high frequency transparent sector. By assumption, $\kappa_{Q,+}<\kappa_{\rotot,+}$. Choose a $K_{\roco}$, depending only on the coefficients
of the equation (\ref{eq:thesystemRge}), such that if $\nuaverage{Q}(\indexnot)\geq K_{\roco}$, then (\ref{eq:gtalbge}) and 
(\ref{eq:maxqdiaglbitokappaq}) hold for all $t_{a}\geq 0$ and $t_{0}\geq 0$ respectively. Defining a time sequence as in Definition~\ref{def:tkdefgeaddbd}
with $t_{0}=0$, it then follows that $t_{\fin,\ror}=\infty$. We are thus allowed to apply conclusion 3 of Corollary~\ref{cor:solnoest} with $t_{a}=0$ and 
$t_{b}\geq 0$ arbitrary. If $\kappa_{\trs,+}<\kappa_{\rotot,+}$, then (\ref{eq:westplusdeg}) implies (\ref{eq:megenestlspresobstscase}). If 
$\kappa_{\trs,+}=\kappa_{\rotot,+}$, then $\kappa_{Q,+}=\bar{\kappa}_{Q,+}<\kappa_{\rotot,+}=\kappa_{\trs,+}$, where the strict inequality is due to the fact
that the equation exhibits subdominant block non-degeneracy. Moreover, the Jordan blocks of $R_{q,+}^{+}$ and $R_{q,+}^{-}$ corresponding to the eigenvalues
with real part $\kappa_{\trs,+}$ are trivial for all $q\in \{1,\dots,Q\}$. Considering conclusion 3 of Corollary~\ref{cor:solnoest}, the only obstruction
to removing the $\e$ appearing in (\ref{eq:westplusdeg}) is that there might be non-trivial Jordan blocks of $A_{Q,\infty}(\indexnottwo)$ 
corresponding to eigenvalues with real part $\kappa_{\trs,+}$ for some $\indexnottwo\in\EFindexset$ such that $\nuaverage{Q}(\indexnottwo)\neq 0$. 
Considering the proof of Corollary~\ref{cor:solnoest}, in particular the text adjacent to (\ref{ref:Kcoplusdef}), this obstruction can be eliminated if
we demand that $K_{\roco}$ be large enough, the bound depending only on the coefficients of the equation (\ref{eq:thesystemRge}) and the spectrum of 
the Riemannian manifold corresponding to $Q$. What remains to be considered is thus the case that $\nuaverage{q_{\trs}}(\indexnot)\leq K_{\roco}$.
In case $\indexnot$ belongs to the low frequency transparent sector, (\ref{eq:meestlowfretrssect}) with $t_{1}$ replaced by $0$ yields the desired 
conclusion. In case $\indexnot$ belongs to the high frequency transparent sector, define a time sequence as in Definition~\ref{def:tkdefgeaddbd},
starting at $t_{0}=0$. If $t_{\fin,\ror}<\infty$, then a combination of Corollary~\ref{cor:solnoest} and (\ref{eq:meestlowfretrssect}) yields the desired
conclusion. If $t_{\fin,\ror}=\infty$, then there is a $0\leq t_{1}<\infty$ such that $\mfg(\indexnot,t)\leq CK_{\roco}$ for all $t\geq t_{1}$, where the 
constant $C$ only depends on the coefficients of the equation (\ref{eq:thesystemRge}). Applying Corollary~\ref{cor:solnoest} in $[0,t_{1}]$ and 
(\ref{eq:meestlowfretrssect}) in $[t_{1},\infty)$ yields the desired conclusion. To conclude, (\ref{eq:megenestlspresobstscase}) holds.

\textbf{The noisy sector.} Assume that $\indexnot$ belongs to the noisy sector. Then there is a $q_{\ron}\in \{1,\dots,Q\}$ such that 
$\baverage{\ron}:=\baverage{q_{\ron}}>0$ and $\nuaverage{q_{\ron}}(\indexnot)\neq 0$. In particular, there is thus a constant $C>0$, depending only 
on the coefficients of the equation (\ref{eq:thesystemRge}), such that $\mfg(\indexnot,t)\geq C\nuaverage{q_{\ron}}(\indexnot)e^{\baverage{\ron}t}$ for 
all $t\geq 0$. In particular, there is thus a $t_{0}'\geq 0$ such that (\ref{eq:gtalbge}) and (\ref{eq:maxqdiaglbitokappaq}) hold for all $t_{a}\geq t_{0}'$ 
and $t_{0}\geq t_{0}'$ respectively. Moreover, $t_{0}'$ only depends on the coefficients of the equation (\ref{eq:thesystemRge}) and the spectrum of 
the Riemannian manifold corresponding to $q_{\ron}$. Define a time sequence as in Definition~\ref{def:tkdefgeaddbd}, starting at $t_{0}'$. Then 
$t_{\fin,\ror}=\infty$. If $t_{0}'=0$, we can appeal to Corollary~\ref{cor:solnoest} with $t_{a}=0$ and $t_{b}\geq 0$ arbitrary. If $t_{0}'>0$, 
appealing to Lemma~\ref{lemma:roughenestbalsetting} yields
\begin{equation}\label{eq:noisyinitiallylowg}
\me^{1/2}(\indexnot,t)\leq C\me^{1/2}(\indexnot,0)+C\int_{0}^{t}|\hf(\indexnot,t')|dt'
\end{equation}
for all $t\in [0,t_{0}']$, where $C$ only depends on the coefficients of the equation (\ref{eq:thesystemRge}) and the spectrum of the Riemannian 
manifold corresponding to $q_{\ron}$. Combining this estimate with Corollary~\ref{cor:solnoest} yields the conclusion that 
(\ref{eq:megenestlspresob}) holds. 

Finally, we wish to prove that (\ref{eq:megenestlspresobsncase}) holds in case $\baverage{Q}>0$ and the equation exhibits subdominant block non-degeneracy. 
For $\indexnot$ in one of the silent sectors, the stated estimate 
follows from earlier arguments. Assume now that there is a $q\in \{1,\dots,Q\}$ such that $\baverage{q}=0$ and $\nuaverage{q}(\indexnot)\neq 0$, 
say $q_{\trs}$. Assume, moreover, that there is no $q\in\{1,\dots,Q\}$ such that $\baverage{q}>0$ and $\nuaverage{q}(\indexnot)\neq 0$. As in the 
transparent setting, there is a $K_{\roco}>0$, depending only on the coefficients of the equation (\ref{eq:thesystemRge}), such that if 
$\nuaverage{q_{\trs}}(\indexnot)\geq K_{\roco}$, then (\ref{eq:gtalbge}) and (\ref{eq:maxqdiaglbitokappaq}) hold for all $t_{a}\geq 0$ and $t_{0}\geq 0$ 
respectively. Defining a time sequence as in Definition~\ref{def:tkdefgeaddbd} with $t_{0}=0$, it then follows that $t_{\fin,\ror}=\infty$. Moreover,
there is a $t_{1}\geq 0$ such that $I_{q_{\trs}}=[t_{1},\infty)$. For any $\e>0$, we can appeal to conclusion 1 of Corollary~\ref{cor:solnoest} on 
$[0,t_{1}-\e]$. Due to the assumption of subdominant block non-degeneracy, this yields an estimate of the form
\begin{equation}\label{eq:wevolutoptnoishighfretransp}
|w(\indexnot,t)|\leq Ce^{\kappa_{\rotot,+}t}|w(\indexnot,0)|+C\int_{0}^{t}e^{\kappa_{\rotot,+}(t-t')}|\hf(\indexnot,t')|dt'
\end{equation}
for all $t\in [0,t_{1}-\e]$. Moreover, the constant $C$ only depends on the coefficients of the equation (\ref{eq:thesystemRge}). In particular, 
it is independent of $\e$, so that (\ref{eq:wevolutoptnoishighfretransp}) holds on $[0,t_{1}]$. On the interval $[t_{1},\infty)$, we need to proceed
differently. However, assuming $K_{\roco}$ to be large enough, the bound depending only on the coefficients of the equation (\ref{eq:thesystemRge}) and 
the spectrum of the Riemannian manifold corresponding to $q_{\trs}$, an argument similar to the one presented in the text adjacent to (\ref{ref:Kcoplusdef})
yields the conclusion that  
\[
|w(\indexnot,t)|\leq Ce^{\kappa_{\rotot,+}(t-t_{1})}|w(\indexnot,t_{1})|+C\int_{t_{1}}^{t}e^{\kappa_{\rotot,+}(t-t')}|\hf(\indexnot,t')|dt'
\]
for all $t\geq t_{1}$, where $C$ only depends on the coefficients of the equation (\ref{eq:thesystemRge}) and the spectrum of the Riemannian manifold 
corresponding to $q_{\trs}$. Combining this estimate with (\ref{eq:wevolutoptnoishighfretransp}) yields (\ref{eq:megenestlspresobsncase}). Next we need 
to consider the case that $0<\nuaverage{q_{\trs}}(\indexnot)<K_{\roco}$. If $\indexnot$ is in the low frequency transparent sector, (\ref{eq:meestlowfretrssect})
holds with $t_{1}=0$, and the desired conclusion follows. If $\indexnot$ is in the high frequency transparent sector, there is a $t_{1}$ such that 
(\ref{eq:wevolutoptnoishighfretransp}) holds on $[0,t_{1}]$ and such that (\ref{eq:meestlowfretrssect}) holds for all $t\geq t_{1}$. Keeping in mind that the
assumption of subdominant block non-degeneracy is fulfilled, this yields (\ref{eq:megenestlspresobsncase}).

What remains to be considered is the $\indexnot\in\EFindexset$ such that $\nuaverage{q}(\indexnot)\neq 0$ for some $q\in \{1,\dots,Q\}$ with the property 
that $\baverage{q}>0$. Consider such a $\indexnot$ and assume that $q<Q$ and that $\nuaverage{p}(\indexnot)=0$ for $p>q$. Then there is a $t_{1}\geq 0$
such that $I_{q}=[t_{1},\infty)$. If there is no transparent era, we can appeal to conclusion 1 of Corollary~\ref{cor:solnoest} combined, possibly, with
an estimate of the form (\ref{eq:noisyinitiallylowg}) in order to obtain (\ref{eq:megenestlspresobsncase}). If there is a transparent era, we have to 
divide the analysis into the two cases $0<\nuaverage{q_{\trs}}(\indexnot)<K_{\roco}$ and $\nuaverage{q_{\trs}}(\indexnot)\geq K_{\roco}$. If $K_{\roco}$ is
large enough, the bound depending only on the coefficients of the equation (\ref{eq:thesystemRge}) and the spectrum of the Riemannian manifold 
corresponding to $q_{\trs}$, then the arguments presented in the text adjacent to (\ref{eq:Krocoep}) yield the conclusion that 
(\ref{eq:wevolutoptnoishighfretransp}) holds for all $t\geq 0$. If there is a transparent era with $0<\nuaverage{q_{\trs}}(\indexnot)<K_{\roco}$, then 
there is a $t_{0}'\geq 0$, depending only on the coefficients of the equation (\ref{eq:thesystemRge}) and the spectra of $(M_{r},g_{r})$, $r\in \{1,\dots,R\}$,
such that on $[t_{0}',\infty)$, there is a sequence as in Definition~\ref{def:tkdefgeaddbd}. Moreover, only noisy eras have a non-empty intersection with 
$[t_{0}',\infty)$. As a consequence, we can appeal to (\ref{eq:noisyinitiallylowg}) on $[0,t_{0}']$ and to conclusion 1 of Corollary~\ref{cor:solnoest}
on $[t_{0}',\infty)$. Assume, finally, that $\nuaverage{Q}(\indexnot)\neq 0$. Then there is a $t_{1}\geq 0$ such that $I_{Q}=[t_{1},\infty)$. In 
$[0,t_{1}]$ we can proceed as above in order to obtain the estimate
\[
\me^{1/2}(\indexnot,t) \leq  Ce^{\kappa_{\rotot,+}t}\me^{1/2}(\indexnot,0)+C\int_{0}^{t}e^{\kappa_{\rotot,+}(t-t')}|\hf(\indexnot,t')|dt'
\]
for all $0\leq t\leq t_{1}$. Finally, on $[t_{1},\infty)$, we can appeal to Lemma~\ref{lemma:westunbfre}.
Combining these estimates yields the desired conclusion. The proposition follows. 
\end{proof}

\section{Energy estimates}\label{section:estsobnorm}

Given the estimates derived in Proposition~\ref{prop:modeestentfuture}, it is straightforward to derive Sobolev estimates.

\begin{thm}\label{thm:sobestndddbaconv}
Consider the equation (\ref{eq:thesystemRge}). Assume that it is non-degenerate, diagonally dominated, balanced and convergent; cf. 
Definition~\ref{def:nondegconvabal}. Fix $\e>0$. Then there is a constant $0<C_{\e}\in\ro$, depending only on $\e$, the spectra of 
$(M_{r},g_{r})$, $r\in \{1,\dots,R\}$, and the coefficients of the equation (\ref{eq:thesystemRge}), such that 
\begin{equation}\label{eq:megenestlspresobsob}
\mfe_{s}^{1/2}[u](t) \leq  C_{\e}e^{(\kappa_{\rotot,+}+\e)t}\mfe^{1/2}_{s}[u](0)+C_{\e}\int_{0}^{t}e^{(\kappa_{\rotot,+}+\e)(t-t')}\|f(\cdot,t')\|_{(s)}dt'
\end{equation}
for all $t\geq 0$, all solutions $u$ to (\ref{eq:thesystemRge}) and all $s\in\ro$, where $\kappa_{\rotot,+}$ is given by Definition~\ref{def:subdomnondeg}. 
If, in addition to the above, the equation (\ref{eq:thesystemRge}) exhibits subdominant block non-degeneracy, there are the following improvements of 
(\ref{eq:megenestlspresobsob}):
\begin{itemize}
\item If $\baverage{Q}<0$, then 
\begin{equation}\label{eq:megenestlspresobsilcasesob}
\mfe^{1/2}_{s}[u](t) \leq  C\ldr{t}^{d_{\ros,+}-1}e^{\kappa_{\rotot,+}t}\mfe^{1/2}_{s}[u](0)
+C\int_{0}^{t}\ldr{t-t'}^{d_{\ros,+}-1}e^{\kappa_{\rotot,+}(t-t')}\|f(\cdot,t')\|_{(s)}dt'
\end{equation}
holds for all $t\geq 0$, all solutions $u$ to (\ref{eq:thesystemRge}) and all $s\in\ro$. Here $d_{\ros,+}=d_{\max}(A_{\infty},\kappa_{\rotot,+})$; $A_{\infty}$ is defined in 
(\ref{eq:vAhFdef}); and the constant $C$ only depends on the spectra of $(M_{r},g_{r})$, $r\in \{1,\dots,R\}$, and the coefficients 
of the equation (\ref{eq:thesystemRge}).
\item If $\baverage{Q}=0$, then 
\begin{equation}\label{eq:megenestlspresobstscasesob}
\mfe^{1/2}_{s}[u](t) \leq  C\ldr{t}^{d_{\rosts,+}-1}e^{\kappa_{\rotot,+}t}\mfe^{1/2}_{s}[u](0)
+C\int_{0}^{t}\ldr{t-t'}^{d_{\rosts,+}-1}e^{\kappa_{\rotot,+}(t-t')}\|f(\cdot,t')\|_{(s)}dt'
\end{equation}
holds for all $t\geq 0$, all solutions $u$ to (\ref{eq:thesystemRge}) and all $s\in\ro$. Here $d_{\rosts,+}:=\max\{d_{\ros,+},d_{\rot,+}\}$; $d_{\ros,+}$ is defined as 
in the case of (\ref{eq:megenestlspresobsilcasesob}); $d_{\rot,+}$ is the largest dimension of a non-trivial Jordan block of 
a matrix of the form $A_{Q,\infty}(\indexnottwo)$ (where $\indexnottwo\in\EFindexset$ is such that $\nuaverage{Q}(\indexnottwo)\neq 0$) corresponding 
to an eigenvalue with real part $\kappa_{\rotot,+}$; and the constant $C$ only depends on the spectra of $(M_{r},g_{r})$, $r\in \{1,\dots,R\}$, and the 
coefficients of the equation (\ref{eq:thesystemRge}). 
\item If $\baverage{Q}>0$, then 
\begin{equation}\label{eq:megenestlspresobsncasesob}
\mfe^{1/2}_{s}[u](t) \leq  C\ldr{t}^{d_{\rotot,+}-1}e^{\kappa_{\rotot,+}t}\mfe^{1/2}_{s}[u](0)
+C\int_{0}^{t}\ldr{t-t'}^{d_{\rotot,+}-1}e^{\kappa_{\rotot,+}(t-t')}\|f(\cdot,t')\|_{(s)}dt'
\end{equation}
holds for all $t\geq 0$, all solutions $u$ to (\ref{eq:thesystemRge}) and all $s\in\ro$. Here $d_{\rotot,+}:=\max\{d_{\ros,+},d_{\ron,+}\}$; $d_{\ros,+}$ is defined as 
in the case of (\ref{eq:megenestlspresobsilcasesob});
\[
d_{\ron,+}:=\max\{d_{\max}(R_{Q,+}^{+},\kappa_{\rotot,+}),d_{\max}(R_{Q,+}^{-},\kappa_{\rotot,+})\};
\]
and the constant $C$ only depends on the spectra of $(M_{r},g_{r})$, $r\in \{1,\dots,R\}$, and the coefficients of the equation 
(\ref{eq:thesystemRge}). 
\end{itemize}
\end{thm}
\begin{remark}
The objects $\mfe_{s}[u]$ and $\|f(\cdot,t)\|_{(s)}$ are defined in (\ref{eq:mfedefextra}) and (\ref{eq:fcdottHsnorm}) respectively.
\end{remark}
\begin{remark}\label{remark:maintheoremgeomcond}
If (\ref{eq:thesystemRge}) is geometrically non-degenerate, diagonally dominated, balanced and convergent, then it is non-degenerate, diagonally dominated, 
balanced and convergent; cf. Remark~\ref{remark:geomecharofdiagdombalaconv}.
\end{remark}
\begin{proof}
The statement is an immediate consequence of Proposition~\ref{prop:modeestentfuture} and Minkowski's inequality.
\end{proof}

\section{Optimality}\label{section:optsobest}

It is of interest to know if the estimates derived in Theorem~\ref{thm:sobestndddbaconv} are optimal in the case of homogeneous equations (i.e., when $f=0$).
There are various ways to define optimality, but here we are mainly interested in comparing $\kappa_{\rotot,+}$, introduced in Definition~\ref{def:subdomnondeg},
with $\kappa_{\roopt,+}$, defined as follows.

\begin{definition}\label{def:roopotplusdef}
Consider the equation (\ref{eq:thesystemRge}) with $f=0$. Assume that it is non-degenerate, diagonally dominated, balanced and convergent; cf. 
Definition~\ref{def:nondegconvabal}. Let $\ma$ be the set of $\kappa\in\ro$ such that there is a constant $0<C\in\ro$ with the property that 
\[
\mfe^{1/2}[u](t)\leq Ce^{\kappa t}\mfe^{1/2}[u](0)
\]
for all $t\geq 0$ and all solutions $u$ to (\ref{eq:thesystemRge}) with $f=0$; here $\mfe=\mfe_{0}$. Then $\kappa_{\roopt,+}$ is defined to be the infimum 
of $\ma$. 
\end{definition}
\begin{remark}
Due to Theorem~\ref{thm:sobestndddbaconv}, we know that $\ma$ is non-empty. Moreover, Lemma~\ref{lemma:roughenestbalsettingfullsol} implies that 
$\ma$ has a lower bound, so that $\kappa_{\roopt,+}\in\ro$.
\end{remark}

The main result concerning optimality is the following. 

\begin{thm}\label{thm:mainoptthm}
Consider the equation (\ref{eq:thesystemRge}) with $f=0$. Assume that it is non-degenerate, diagonally dominated, balanced and convergent; cf. 
Definition~\ref{def:nondegconvabal}.  Then $\kappa_{\roopt,+}=\kappa_{\rotot,+}$. Assume, in addition to the above, that the equation exhibits subdominant 
block non-degeneracy. If $\baverage{Q}<0$ and $\kappa_{\rosil,+}=\kappa_{\rotot,+}$, there is then a constant $0<C\in\ro$ and a solution $u\neq 0$ to 
(\ref{eq:thesystemRge}) with $f=0$ such that 
\begin{equation}\label{eq:mfeoptlbsilsett}
\mfe^{1/2}[u](t)\geq C\ldr{t}^{d_{\ros,+}-1}e^{\kappa_{\rotot,+}t}\mfe^{1/2}[u](0)
\end{equation}
for all $t\geq 0$. In this setting, the estimate 
(\ref{eq:megenestlspresobsilcasesob}) is thus optimal for $f=0$. If $\baverage{Q}=0$ and either $\kappa_{\rosil,+}=\kappa_{\rotot,+}$ or 
$\kappa_{Q,\trs,+}=\kappa_{\rotot,+}$ holds, then there is a constant $0<C\in\ro$ and a solution $u\neq 0$ to (\ref{eq:thesystemRge}) with $f=0$ such that 
\begin{equation}\label{eq:mfeoptlbtrssett}
\mfe^{1/2}[u](t)\geq C\ldr{t}^{d_{\rosts,+}-1}e^{\kappa_{\rotot,+}t}\mfe^{1/2}[u](0)
\end{equation}
for all $t\geq 0$. In this setting, the estimate 
(\ref{eq:megenestlspresobstscasesob}) is thus optimal for $f=0$. Finally, if $\baverage{Q}>0$ and either $\kappa_{\rosil,+}=\kappa_{\rotot,+}$ or 
$\kappa_{Q,+}=\kappa_{\rotot,+}$ holds, then there is a constant $0<C\in\ro$ and a solution $u\neq 0$ to (\ref{eq:thesystemRge}) 
with $f=0$ such that 
\begin{equation}\label{eq:mfeoptlbnsett}
\mfe^{1/2}[u](t)\geq C\ldr{t}^{d_{\rotot,+}-1}e^{\kappa_{\rotot,+}t}\mfe^{1/2}[u](0)
\end{equation}
for all $t\geq 0$. In this setting, the estimate (\ref{eq:megenestlspresobsncasesob}) is thus optimal for $f=0$.
\end{thm}
\begin{remark}
The notation $\kappa_{\rosil,+}$ and $\kappa_{\rotot,+}$ is introduced in Definition~\ref{def:subdomnondeg}; the notation $\kappa_{q,\trs,+}$ is introduced
in Definition~\ref{def:kappaqtrpmetc}; and the notation $\kappa_{q,+}$ is introduced in Definition~\ref{def:tXqinfRqpmkappaqpm}.
The notation $d_{\ros,+}$, $d_{\rosts,+}$ and $d_{\rotot,+}$ is introduced in connection with (\ref{eq:megenestlspresobsilcasesob}), 
(\ref{eq:megenestlspresobstscasesob}) and (\ref{eq:megenestlspresobsncasesob}) respectively.
\end{remark}
\begin{remark}
Note that in the present setting, the $\kappa_{\roopt,+}$ introduced in Definition~\ref{def:roopotplusdef} equals the $\nolossrate$ introduced in 
Definition~\ref{eq:defdecayrates}. In order to justify this statement, note, first of all, that $\mfe=\mfe_{\robas}$; cf. the comment made in connection
with (\ref{eq:mfedef}). Thus $\ma_{\ronl}$ introduced in Definition~\ref{eq:defdecayrates} equals $\ma$ introduced in Definition~\ref{def:roopotplusdef}.
In particular, $\ma_{\ronl}$ is non-empty and bounded from below, so that $\nolossrate$ is well defined and equals $\kappa_{\roopt,+}$. 
\end{remark}
\begin{proof}
Let us begin by proving that $\kappa_{\roopt,+}=\kappa_{\rotot,+}$. Due to Theorem~\ref{thm:sobestndddbaconv} and Definition~\ref{def:roopotplusdef}, it is clear 
that $\kappa_{\roopt,+}\leq \kappa_{\rotot,+}$. What we need to prove is the opposite inequality. 

Due to Definition~\ref{def:subdomnondeg} and (\ref{eq:kappatrpmdef}), we need to consider the following three cases: $\kappa_{\rotot,+}=\kappa_{\rosil,+}$;
$\kappa_{\rotot,+}=\kappa_{q_{\trs},\trs,+}$ (assuming there is a $q\in \{1,\dots, Q\}$ such that $\baverage{q}=0$, say $q_{\trs}$); and $\kappa_{\rotot,+}=\kappa_{q,+}$
for some $q\in \{1,\dots,Q\}$ such that $\baverage{q}\neq 0$. In the first case, the inequality $\kappa_{\roopt,+}\geq \kappa_{\rotot,+}$ follows from 
Lemma~\ref{lemma:spasODEsett}; it is sufficient to consider spatially homogeneous solutions to the equations. In the second case, the desired conclusion
follows from Proposition~\ref{prop:spasdatrs}. What remains to be considered is the case that there is a $q\in \{1,\dots,Q\}$ such that $\baverage{q}\neq 0$
and $\kappa_{q,+}=\kappa_{\rotot,+}$. Fix one such $q$ and denote it by $q_{0}$. In this case, we wish to appeal to Lemma~\ref{lemma:lbmatprod}. From now on, we 
focus on $\indexnot\in\EFindexset$ such that $\nuaverage{q}(\indexnot)=0$ for all $q\neq q_{0}$. In order to prove the desired lower bound on 
$\kappa_{\roopt,+}$, it is convenient to assume that $\kappa_{\roopt,+}<\kappa_{\rotot,+}=\kappa_{q_{0},+}$. Let $\de:=\kappa_{q_{0},+}-\kappa_{\roopt,+}>0$. Then there
is a constant $0<K_{\de}\in\ro$ such that 
\begin{equation}\label{eq:mfedelossest}
\mfe^{1/2}[u](t)\leq K_{\de}e^{(\kappa_{\roopt,+}+\de/2)t}\mfe^{1/2}[u](0)
\end{equation}
for all $t\geq 0$ and all solutions $u$ to (\ref{eq:thesystemRge}) with $f=0$; this is a direct consequence of the definition of $\kappa_{\roopt,+}$. Choose 
$\e:=\de/4$ and $T>0$. For $\indexnot$ such that 
$\nuaverage{q_{0}}(\indexnot)$ is large enough (the lower bound depending only on the coefficients of the equation (\ref{eq:thesystemRge}), $\e$ and $T$), 
we can define a sequence as in Definition~\ref{def:tkdefgeaddbd}, and the sequence satisfies $t_{0}=0$ and $t_{\fin,\ror}\geq T$. Moreover 
$[0,T]\subseteq I_{q_{0}}$; 
$\mfg(\indexnot,t)\geq\mfg_{\e,\min}$ for $t\in [0,T]$, where $\mfg_{\e,\min}$ is the constant appearing in the statement of Lemma~\ref{lemma:lbmatprod};
and $T\geq t_{\roas,\e}+2t_{\romar,\e}$, where $t_{\roas,\e}$ and $t_{\romar,\e}$ are the constants appearing in the statement of Lemma~\ref{lemma:lbmatprod}. 
Given $\indexnot$ and $\{t_{k}\}$ as above, 
let $k_{-}$ be the smallest $k$ such that $t_{k}\geq \max\{t_{\roas,\e},t_{\romar,\e}\}$ and let $k_{+}$ be the largest $k$ such that $t_{k}\leq T-t_{\romar,\e}$.
Then there is a solution $z$ to (\ref{eq:fourierthesystemRge}) such that 
\[
|w(\indexnot,t_{+})|\geq C_{\de}\exp[(\kappa_{q_{0},+}-\de/4)(t_{+}-t_{-})]|w(\indexnot,t_{-})|,
\] 
where $t_{\pm}:=t_{k_{\pm}}$ and $C_{\de}>0$ only depends on $\de$ and the coefficients of the equation (\ref{eq:thesystemRge}); this is a consequence of
(\ref{eq:wprekpertransvarpi}) and Lemma~\ref{lemma:lbmatprod}. Combining this estimate with (\ref{eq:meestroughbalset}), it is clear that 
\begin{equation}\label{eq:windtplindlb}
|w(\indexnot,t_{+})|\geq C_{\de}\exp[(\kappa_{q_{0},+}-\de/4)t_{+}]|w(\indexnot,0)|,
\end{equation}
where $C_{\de}>0$ only depends on $\de$ and the coefficients of the equation (\ref{eq:thesystemRge}); note
that $t_{-}$ is bounded by a number depending only on $\de$ and the coefficients of the equation (\ref{eq:thesystemRge}). For $T$ large enough,
it is clear that (\ref{eq:mfedelossest}) and (\ref{eq:windtplindlb}) are incompatible. 

Turning to the estimates (\ref{eq:mfeoptlbsilsett})--(\ref{eq:mfeoptlbnsett}), the first two follow from Lemma~\ref{lemma:spasODEsett} and
Proposition~\ref{prop:spasdatrs}. The last one is a consequence of the results of Section~\ref{section:spedataatinf}. The theorem follows.
\end{proof}

\chapter{The silent setting}

\section{Introduction}

In Chapter~\ref{chapter:weaksil}, we consider weakly silent, balanced and convergent equations. In particular, we derive detailed asymptotics
in Section~\ref{section:roughODEfutas}, and we demonstrate that it is possible to specify the asymptotics in Section~\ref{section:roughODEspecas}.
Even though the corresponding results are of interest, they suffer from one deficiency: it is not clear how to compute the particular values of 
the constants $s_{\rohom}$ and $s_{\roih}$ appearing in the statement of Lemma~\ref{lemma:roughas}; nor is it clear how to compute the value of 
$s_{\infty}$ appearing in Lemma~\ref{lemma:spasda}. Our goal in the present chapter is to use the improved estimates derived in 
Chapter~\ref{chapter:ndddconandbalancedeq} to arrive at specific values for these constants. In order to do so, we here need to make stronger 
assumptions than in Chapter~\ref{chapter:weaksil}. In fact, in all the results, we consider equations of the form (\ref{eq:thesystemRge}) which are
non-degenerate, diagonally dominated, balanced and convergent; cf. Definition~\ref{def:nondegconvabal}. Moreover, to ensure that we are in the 
silent setting, we assume that $\baverage{Q}<0$.

It would be of interest to demonstrate that the numbers $s_{\roh,\b,+}$, $s_{\roih,\b}$, $s_{\roh,\b}$ and $s_{\roh,-}$ appearing in Propositions~\ref{prop:roughasex} 
and \ref{prop:spasdaex} below are optimal. However, we do not attempt to do so here. Nevertheless, in the case of $s_{\roh,\b}$ and $s_{\roh,-}$, we 
at least conjecture them to be optimal. We also expect methods similar to those used in the proof of Theorem~\ref{thm:mainoptthm}
to be useful in arriving at such a conclusion. However, there might be a subtlety involved in the transition between the interval in which time sequences 
of the type introduced in Definition~\ref{def:tkdefgeaddbd} are well defined and the remaining interval, say $[T_{\roode},\infty)$, in which the 
ODE behaviour dominates. 

\textbf{Rough idea of proofs.}
The rough structures of the proofs of Propositions~\ref{prop:roughasex} and \ref{prop:spasdaex} below are similar to those of 
Lemmas~\ref{lemma:roughas} and \ref{lemma:spasda}. In particular, we consider the different modes $\indexnot\in\EFindexset$ one by one, and for each
mode, we divide the time interval $[0,\infty)$ into the subintervals $[0,T_{\roode}]$ and $[T_{\roode},\infty)$. However, as opposed to the division
made in the proofs of Lemmas~\ref{lemma:roughas} and \ref{lemma:spasda}, the time interval $[0,T_{\roode}]$ here equals $[0,t_{\fin,\ror}]$, where 
$t_{\fin,\ror}$ is introduced in Definition~\ref{def:tkdefgeaddbd}. In other words, there is a time sequence of the type introduced in 
Definition~\ref{def:tkdefgeaddbd} stretching from $t=0$ to $t=T_{\roode}$. In the interval $[T_{\roode},\infty)$, we appeal to Lemmas~\ref{lemma:ODEasymp} 
and \ref{lemma:spasODEsett}, as before. However, in the interval $[0,T_{\roode}]$ we appeal neither to Lemma~\ref{lemma:oscroughODE}, nor to the analogous
estimate in the other time direction (cf. the proof of Lemma~\ref{lemma:spasda}). Instead, we make use of the improved knowledge contained in the 
statement of Corollary~\ref{cor:westvarrhocase}. Considering Definition~\ref{def:varrhoeppl} and Corollary~\ref{cor:westvarrhocase}, it is clear that we
need to estimate expressions of the form 
\begin{equation}\label{eq:protosumrefsilas}
\textstyle{\sum}_{j=1}^{k+1}\rho_{j}|I_{j}|,
\end{equation}
where $\rho_{j}\in\ro$, $j=1,\dots,k+1$, are given numbers and $I_{j}$ are the intervals into which $[t',t]\subseteq [0,T_{\roode}]$ is divided when appealing to 
Definition~\ref{def:varrhoeppl}. In order to obtain
useful conclusions, we need to estimate expressions of this type from above. Using the definition of $T_{\roode}$ and the definition of the frequency 
eras (which enter into the definition of the $I_{j}$), an upper bound on (\ref{eq:protosumrefsilas}) can be derived. It is of the form
$s_{\mathrm{pro}}\ln\ldr{\nu}+C$, where $\nu=\nu(\indexnot)$, $C$ is a constant independent of $\indexnot$ and $s_{\mathrm{pro}}$ is a prototype value 
for the number of derivatives lost in the particular situation under consideration. Moreover, $s_{\mathrm{pro}}$ can be computed in terms of the 
$\rho_{j}$'s and the $\baverage{q}$'s. In the situations of interest here, $0\leq t'\leq t\leq T_{\roode}$,
and we obtain general upper bounds on (\ref{eq:protosumrefsilas}) for $t',t$ in this range. Since $t'$ could potentially equal $t$, it is clear that 
$s_{\mathrm{pro}}\geq 0$. However, in the case that $t'=0$ and $t=T_{\roode}$, it is sometimes possible to derive bounds of the form $s_{\mathrm{pro}}\ln\ldr{\nu}+C$,
where $s_{\mathrm{pro}}<0$. This can happen, in particular, when estimating the asymptotic functions $V_{\infty}$ appearing in Lemma~\ref{lemma:roughas} (as
well as Proposition~\ref{prop:roughasex} below), and signals a \textit{gain} in derivatives. 

\section{Future asymptotics, silent setting}\label{section:roughODEfutasex}

As a next step, we wish to strengthen Lemma~\ref{lemma:roughas} by improving the conclusions concerning the loss of derivatives. 
In particular, we wish to replace the constants $s_{\rohom}$ and $s_{\roih}$ appearing in (\ref{eq:uudothsest}) and (\ref{eq:uinfudinfHsest})
by particular values; cf. (\ref{eq:shbpdef}), (\ref{eq:sihbdef}) and (\ref{eq:shbdef}) below. 

\begin{prop}\label{prop:roughasex}
Consider the equation (\ref{eq:thesystemRge}). Assume that it is non-degenerate, diagonally dominated, balanced and convergent; cf. 
Definition~\ref{def:nondegconvabal}. Assume that $\baverage{Q}<0$ and that $f$ is a smooth function such that for every $s\in\ro$, 
\begin{equation}\label{eq:fnosilsdef}
\|f\|_{\rosil,s}:=\int_{0}^{\infty}e^{-\kappa_{\rosil,+}t}\|f(\cdot,t)\|_{(s)}dt<\infty,
\end{equation}
where $\kappa_{\rosil,+}:=\kappa_{\max}(A_{\infty})$ and $A_{\infty}$ is defined by (\ref{eq:vAhFdef}).
Let $\b_{\rem}:=\min\{-\baverage{Q},\eta_{\romn}\}$, where $\eta_{\romn}$ 
is the constant appearing in (\ref{eq:mainconv}). Fix $0<\b\leq\b_{\rem}$ and $\e>0$. Let $E_{a}$ be the first generalised eigenspace in the 
$\b,A_{\infty}$-decomposition 
of $\cn{2m}$; cf. Definition~\ref{def:defofgeneigenspintro}. Then there are constants $C_{\e,\b}$ and $N$, where $C_{\e,\b}$ 
only depends on $\e$, $\b$, the coefficients of the equation (\ref{eq:thesystemRge}) and the spectra of $(M_{r},g_{r})$, $r\in\{1,\dots,R\}$; 
and $N$ only depends on $m$, such that the following holds. Given a smooth solution $u$ to 
(\ref{eq:thesystemRge}), there is a $V_{\infty}\in C^{\infty}(\bM,E_{a})$ such that 
\begin{equation}\label{eq:uudothsestex}
\begin{split}
 & \left\|\left(\begin{array}{c} u(\cdot,t) \\ u_{t}(\cdot,t)\end{array}\right)
-e^{A_{\infty}t}V_{\infty}
-\int_{0}^{t}e^{A_{\infty}(t-t')}\left(\begin{array}{c} 0 \\ f(\cdot,t')\end{array}\right)dt'\right\|_{(s)} \\
\leq & C_{\e,\b}\ldr{t}^{N}e^{(\kappa_{\rosil,+}-\b)t}\left(\|u_{t}(\cdot,0)\|_{(s+s_{\roh,\b,+}+\e)}+\|u(\cdot,0)\|_{(s+s_{\roh,\b,+}+\e+1)}
+\|f\|_{\rosil,s+s_{\roih,\b}+\e}\right)
\end{split}
\end{equation}
holds for all $t\geq 0$ and all $s\in\ro$; recall that $\bM$ is given by (\ref{eq:bMdef}). Here
\begin{align}
s_{\roh,\b,+} := & \max_{1\leq q\leq Q}\max\left\{0,-\frac{\b+\kappa_{q,+}-\kappa_{\rosil,+}}{\baverage{q}}\right\},\label{eq:shbpdef}\\
s_{\roih,\b} := & \max\left\{-\frac{\kappa_{\roode}}{\baverage{Q}},
-\frac{\b}{\baverage{Q}}+\max_{1\leq q\leq Q}\max\left\{0,-\frac{\kappa_{q,+}-\kappa_{\rosil,+}}{\baverage{q}}\right\}\right\},\label{eq:sihbdef}
\end{align}
where $\kappa_{\roode}:=\Rsp A_{\infty}$, cf. Definition~\ref{def:SpRspdef}, and $\kappa_{q,+}$ is defined by (\ref{eq:kappaqpmdef}). Moreover, 
\begin{equation}\label{eq:uinfudinfHsestex}
\|V_{\infty}\|_{(s)}\leq C_{\e,\b}\left(\|u_{t}(\cdot,0)\|_{(s+s_{\roh,\b}+\e)}+\|u(\cdot,0)\|_{(s+s_{\roh,\b}+\e+1)}+\|f\|_{\rosil,s+s_{\roih,\b}+\e}\right),
\end{equation}
where 
\begin{equation}\label{eq:shbdef}
s_{\roh,\b}:=\max_{1\leq q\leq Q}\left(-\frac{\b+\kappa_{q,+}-\kappa_{\rosil,+}}{\baverage{q}}\right).
\end{equation}
\end{prop}
\begin{remark}\label{remark:optimalasest}
If the assumptions of Proposition~\ref{prop:roughasexintro} are satisfied, then the assumptions of the proposition are satisfied; 
cf. Remark~\ref{remark:maintheoremgeomcond}.
\end{remark}
\begin{remark}
The constants $\baverage{q}$, $q\in \{1,\dots,Q\}$, are introduced at the beginning of Section~\ref{section:simplmacoeff}.
\end{remark}
\begin{remark}
The function $V_{\infty}$ is uniquely determined by the fact that it satisfies the estimate (\ref{eq:uudothsestex}). 
\end{remark}
\begin{remark}
Both $s_{\roh,\b,+}$ and $s_{\roih,\b}$ are non-negative. However, $s_{\roh,\b}$ could be strictly negative. In this sense, the function $V_{\infty}$
could be more regular than the initial data. 
\end{remark}
\begin{remark}
As opposed to (\ref{eq:uudothsest}), $\b$ need not equal $\b_{\rem}$ in (\ref{eq:uudothsestex}). The reason for allowing $\b$ to vary in the 
range $(0,\b_{\rem}]$ is that a higher $\b$ yields more detailed asymptotics, but a lower $\b$ yields a lower loss of derivatives. 
\end{remark}
\begin{remark}\label{remark:improvedbetarem}
For particular equations, the value of $\b_{\rem}$ can be improved. In fact, if there are constants $\g_{i}\geq 0$, $i=1,2$, such that
\begin{equation}\label{eq:sigmagammaoneXgammatwo}
|\sigma(\indexnot,t)|\leq Ce^{-\g_{1} t},\ \ \
\|X(\indexnot,t)\|\leq Ce^{-\g_{2}t}
\end{equation}
for all $t\geq 0$ and all $0\neq\indexnot\in\EFindexset$, where $C$ only depends on the coefficients of the equation (\ref{eq:thesystemRge})
and $\sigma$ and $X$ are defined by (\ref{eq:ellsigmaXgenRdef}), then $\b_{\rem}$ can be replaced by 
\begin{equation}\label{eq:bremimprovement}
\b_{\rem}:=\min\{-2\baverage{Q},-\baverage{Q}+\g_{1},-\baverage{Q}+\g_{2},\eta_{\romn}\}.
\end{equation}
\end{remark}
\begin{remark}\label{remark:homimprsilasest}
In the case of homogeneous equations (i.e., when $f=0$), the $\e$ appearing in (\ref{eq:uudothsestex}) can be removed under the following circumstances.
If, in addition to the assumptions of the proposition, all Jordan blocks of $R_{q,+}^{+}$ and $R_{q,+}^{-}$ corresponding to eigenvalues such that
\begin{itemize}
\item the real part of the eigenvalue equals $\kappa_{q,+}$, and
\item $\kappa_{q,+}$ satisfies
\begin{equation}\label{eq:kappaqpyieldetarohbremark}
-\frac{\b+\kappa_{q,+}-\kappa_{\rosil,+}}{\baverage{q}}=s_{\roh,\b,+},
\end{equation}
\end{itemize}
are trivial for $q=1,\dots,Q$, then 
\begin{equation}\label{eq:uudothsestexhomimpr}
\begin{split}
 & \left\|\left(\begin{array}{c} u(\cdot,t) \\ u_{t}(\cdot,t)\end{array}\right)
-e^{A_{\infty}t}V_{\infty}\right\|_{(s)} \\
\leq & C_{\b}\ldr{t}^{N}e^{(\kappa_{\rosil,+}-\b)t}\left(\|u_{t}(\cdot,0)\|_{(s+s_{\roh,\b,+})}+\|u(\cdot,0)\|_{(s+s_{\roh,\b,+}+1)}\right)
\end{split}
\end{equation}
holds for all $t\geq 0$ and all $s\in\ro$, where $C_{\b}$ only depends on $\b$, the coefficients of the equation (\ref{eq:thesystemRge}) and the 
spectra of $(M_{r},g_{r})$, $r\in\{1,\dots,R\}$; and $N$ only depends on $m$. Note also that if there are no 
eigenvalues of $R_{q,+}^{+}$ or $R_{q,+}^{-}$ with real part equalling $\kappa_{q,+}$ and satisfying (\ref{eq:kappaqpyieldetarohbremark}), $q\in\{1,\dots,Q\}$, 
then the conclusion holds. 
\end{remark}
\begin{remark}\label{remark:homimprsilasfunest}
Again, in the case of homogeneous equations (i.e., when $f=0$), the $\e$ appearing in (\ref{eq:uinfudinfHsestex}) can be removed under the following 
circumstances. If, in addition to the assumptions of the proposition, all Jordan blocks of $R_{q,+}^{+}$ and $R_{q,+}^{-}$ corresponding to eigenvalues 
such that
\begin{itemize}
\item the real part of the eigenvalue equals $\kappa_{q,+}$, and
\item $\kappa_{q,+}$ satisfies
\begin{equation}\label{eq:kappaqpyieldetarohbremarktwo}
-\frac{\b+\kappa_{q,+}-\kappa_{\rosil,+}}{\baverage{q}}=s_{\roh,\b},
\end{equation}
\end{itemize}
are trivial for $q=1,\dots,Q$, then 
\begin{equation}\label{eq:uinfudinfHsestexhomimpr}
\|V_{\infty}\|_{(s)}\leq C_{\b}\left(\|u_{t}(\cdot,0)\|_{(s+s_{\roh,\b})}+\|u(\cdot,0)\|_{(s+s_{\roh,\b}+1)}\right),
\end{equation}
holds for all $s\in\ro$, where $C_{\b}$ only depends on $\b$, the coefficients of the equation (\ref{eq:thesystemRge}) and the 
spectra of $(M_{r},g_{r})$, $r\in\{1,\dots,R\}$. 
\end{remark}
\begin{proof}
Roughly speaking, the proof is quite similar to that of Lemma~\ref{lemma:roughas}. First, for a given $\indexnot\in\EFindexset$, we divide
the time interval $[0,\infty)$ into $[0,T_{\roode}]$ and $[T_{\roode},\infty)$. Moreover, we verify that (\ref{eq:fourierthesystemRge}) 
can be written as in (\ref{eq:dotvroughODE}), where the constituents are such that the conditions of Subsection~\ref{ssection:ODEreg} are satisfied
on $[T_{\roode},\infty)$. In the interval $[T_{\roode},\infty)$ we can appeal to Lemma~\ref{lemma:ODEasymp}. However, we need to reformulate this estimate
so that $T_{\roode}$ does not appear, and we need to derive estimates in the interval $[0,T_{\roode}]$. It is at this point that the proof diverges from 
that of Lemma~\ref{lemma:roughas}; even though the quantities to be estimated are the same as in the proof of Lemma~\ref{lemma:roughas}, we here
derive more detailed estimates by appealing to Corollary~\ref{cor:westvarrhocase} instead of to Lemma~\ref{lemma:oscroughODE}.

\textbf{Terminology and preliminary estimates.}
Given $\indexnot\in\EFindexset$, it is convenient to divide $[0,\infty)$ into two subintervals, say $[0,T_{\roode}]$ and 
$[T_{\roode},\infty)$; cf. the analysis presented in Section~\ref{section:roughODEfutas}. If $\indexnot\in\EFindexset$ is in the low frequency
silent sector, cf. Definition~\ref{def:frequencysectors}, we let $T_{\roode}:=0$. If $\indexnot\in\EFindexset$ is in the high frequency
silent sector, we let $T_{\roode}:=t_{\fin,\ror}$, where $t_{\fin,\ror}$ results from appealing to Definition~\ref{def:tkdefgeaddbd} with
$t_{0}:=0$. With these definitions, 
\begin{equation}\label{eq:mfgestsilnondeglatetimes}
\mfg(\indexnot,t)\leq Ce^{\baverage{Q}\bt}
\end{equation}
for all $t\geq T_{\roode}$, 
where $\bt:=t-T_{\roode}$ and $C$ only depends on the coefficients of the equation (\ref{eq:thesystemRge}). Combining this estimate with 
(\ref{eq:shtermneg}) yields the conclusion that 
\[
|g^{0l}(t)n_{l}|\leq Ce^{(\baverage{Q}-\eta_{\rosh})\bt}
\]
for all $t\geq T_{\roode}$, where $C$ only depends on the coefficients of the equation (\ref{eq:thesystemRge}). Combining (\ref{eq:mainbd}) and
(\ref{eq:mfgestsilnondeglatetimes}) yields
\[
\|n_{l}X^{l}(t)\|\leq Ce^{\baverage{Q}\bt}
\]
for all $t\geq T_{\roode}$, where $C$ only depends on the coefficients of the equation (\ref{eq:thesystemRge}). Finally, recall that 
(\ref{eq:mainconv}) holds. This yields estimates for $\a-\a_{\infty}$ and $\zeta-\zeta_{\infty}$. Combining these observations yields the conclusion
that the equation (\ref{eq:fourierthesystemRge}) can be written as in (\ref{eq:dotvroughODE}), where $v$, $A_{\infty}$ and $F$ are defined by 
(\ref{eq:vAhFdef}). Moreover, $A_{\rem}$ satisfies the estimate (\ref{eq:Aremestgensys}) for all $t\geq T_{\roode}$, where 
$\b_{\rem}:=\min\{-\baverage{Q},\eta_{\romn}\}$ and the constant $C_{\rem}$ only depends on the coefficients of the equation (\ref{eq:thesystemRge}).
Note also that if (\ref{eq:sigmagammaoneXgammatwo}) is satisfied, then (\ref{eq:Aremestgensys}) holds with $\b_{\rem}$ given by (\ref{eq:bremimprovement}).

\textbf{The expressions to be estimated.}
Fix $0<\b\leq\b_{\rem}$. For every $\indexnot\in\EFindexset$, we can appeal to Lemma~\ref{lemma:ODEasymp} for $t\geq T_{\roode}$, with this 
choice of $\b$; that we are allowed to do so is due to the above observations and the assumption that (\ref{eq:fnosilsdef}) holds. However, just 
as in the proof of Lemma~\ref{lemma:roughas}, we need to reformulate the estimate so that $T_{\roode}$ does not 
appear and to estimate the error in the interval $[0,T_{\roode}]$. Introduce, just as in the proof of Lemma~\ref{lemma:roughas},
\begin{equation}\label{eq:vinfmoddefex}
v_{\infty,\romod}:=v_{\infty}-\int_{0}^{T_{\roode}}e^{-A_{\infty}t'}F(t')dt',
\end{equation}
where we suppress the dependence of the constituents on $\indexnot$. Then (\ref{eq:vinfasres}) can be written
\begin{equation}\label{eq:vasadalltex}
\begin{split}
 & \left|v(t)-e^{A_{\infty}t}v_{\infty,\romod}-\int_{0}^{t}e^{A_{\infty}(t-t')}F(t')dt'\right|\\
 \leq & C\ldr{t}^{N}e^{(\kappa_{\rosil,+}-\b)t}e^{\b T_{\roode}}\left[e^{-\kappa_{\rosil,+}T_{\roode}}|v(T_{\roode})|+\|F\|_{\rosil}\right];
\end{split}
\end{equation}
recall that this estimate holds on $[T_{\roode},\infty)$. Here we use the notation
\begin{equation}\label{eq:Fnormrosildef}
\|F\|_{\rosil}:=\int_{0}^{\infty}e^{-\kappa_{\rosil,+}t}|F(t)|dt.
\end{equation}
Note also that the constant
$C$ only depends on $C_{\rem}$, $\b_{\rem}$ and $A_{\infty}$; i.e., only on the coefficients of the equation (\ref{eq:thesystemRge}). Similarly to the proof of 
Lemma~\ref{lemma:roughas}, it is of interest to estimate
\begin{equation}\label{eq:zTodetermsex}
e^{(\b-\kappa_{\rosil,+})t}|v(t)|,\ \ \
e^{(\b-\kappa_{\rosil,+})t}|e^{A_{\infty}t}v_{\infty}|,\ \ \
\left|e^{(\b-\kappa_{\rosil,+})t}\int_{t}^{T_{\roode}}e^{A_{\infty}(t-t')}F(t')dt'\right|
\end{equation}
for $t\in [0,T_{\roode}]$; this yields the desired estimate on the interval $[0,T_{\roode}]$. In order to obtain a good estimate for $t\geq T_{\roode}$, we need
to estimate 
\begin{equation}\label{eq:Todeinftytermsex}
e^{(\b-\kappa_{\rosil,+})T_{\roode}}|v(T_{\roode})|,\ \ \
e^{\b T_{\roode}}\|F\|_{\rosil}.
\end{equation}
Finally, we wish to estimate $v_{\infty,\romod}$ introduced in (\ref{eq:vinfmoddefex}).

\textbf{Preliminaries, the high frequency case.} Assume that $T_{\roode}>0$. Then we can appeal to Corollary~\ref{cor:westvarrhocase} in $[0,T_{\roode}]$. Moreover,
since there are no transparent eras, we can ignore the parameter $K$; cf. Remark~\ref{remark:removingbothepsilon}. For a given $\e>0$, there is thus 
a constant $C_{\e}$, depending only on $\e$ and the coefficients of the equation, such that the following holds. If $0\leq t_{a}\leq t_{b}$ are such that 
$I_{a}:=[t_{a},t_{b}]\subseteq [0,T_{\roode}]$, then
\[
|w(t_{b})| \leq C_{\e}e^{\varrho_{\e,+}(t_{b},t_{a})}|w(t_{a})|+C_{\e}\int_{t_{a}}^{t_{b}}e^{\varrho_{\e,+}(t_{b},t')}|\hf(t')|dt'
\]
for all solutions $z$ to (\ref{eq:fourierthesystemRge}), where we have omitted reference to $\indexnot$ for the sake of brevity. Here $\varrho_{\e,+}$ is 
introduced in Definition~\ref{def:varrhoeppl} (and the omission of the $K$ in the subscript is due to the absence of transparent eras). In particular, 
since $|v(t)|\leq C|w(t)|$ for $t\in [0,T_{\roode}]$ and a constant $C$ only depending on the coefficients of the equation, 
\begin{equation}\label{eq:prelestexptimesv}
\begin{split}
e^{(\b-\kappa_{\rosil,+})t}|v(t)|\leq & C_{\e}e^{(\b-\kappa_{\rosil,+})t+\varrho_{\e,+}(t,0)}\me^{1/2}(0)\\
 & +C_{\e}e^{\b t}\int_{0}^{t}e^{-\kappa_{\rosil,+}(t-t')+\varrho_{\e,+}(t,t')}e^{-\kappa_{\rosil,+}t'}|\hf(t')|dt'
\end{split}
\end{equation}
for all $t\in [0,T_{\roode}]$ and all solutions $z$ to (\ref{eq:fourierthesystemRge}), where we have used the fact that $|w(t)|\leq \sqrt{2}\me^{1/2}(t)$
and $C_{\e}$ depends only on $\e$ and the coefficients of the equation (\ref{eq:thesystemRge}). 

\textbf{Deriving estimates for the exponentials.} Due to (\ref{eq:prelestexptimesv}), it is clearly of interest to derive upper bounds on 
\begin{equation}\label{eq:mainsumstobeestimatedvarrhoep}
(\b-\kappa_{\rosil,+})t+\varrho_{\e,+}(t,0),\ \ \
\b t-\kappa_{\rosil,+}(t-t')+\varrho_{\e,+}(t,t'),
\end{equation}
where $0\leq t'\leq t\leq T_{\roode}$. Given $[t',t]\subseteq [0,T_{\roode}]$, divide it into intervals $I_{1},\dots,I_{k+1}$ as in
Definition~\ref{def:varrhoeppl}. Then, e.g.,
\begin{equation}\label{eq:mkappaoneplusvarrhoexpr}
\begin{split}
-\kappa_{\rosil,+}(t-t')+\varrho_{\e,+}(t,t') = & \textstyle{\sum}_{j=1}^{k+1}(\mu_{j,+}-\kappa_{\rosil,+})|I_{j}|,
\end{split}
\end{equation}
where the $\mu_{j,+}$ are introduced in Definition~\ref{def:varrhoeppl}. Clearly, it is of interest to estimate the right hand side of 
(\ref{eq:mkappaoneplusvarrhoexpr}). Other expressions appearing in (\ref{eq:mainsumstobeestimatedvarrhoep}) can be reformulated similarly. In general,
it is of interest to estimate
\[
\varsigma(t,t';\rho):=\textstyle{\sum}_{j=1}^{k+1}\rho_{j}|I_{j}|,
\]
where $\rho_{j}\in\ro$, $j=1,\dots,k+1$, and $\rho:=(\rho_{1},\dots,\rho_{k+1})$. Clearly, 
\[
\varsigma(t,t';\rho)\leq \rho_{\max}(t-t'),
\]
where $\rho_{\max}$ is the largest of the $\rho_{j}$. However, this estimate is too crude for most applications. In order to refine it, note that 
if $t'\leq s_{1}\leq\cdots\leq s_{k}\leq t$ corresponds to the division into the intervals $I_{1},\dots,I_{k+1}$, then 
\begin{equation}\label{eq:freqeratelescope}
\frac{\nuaverage{q_{1}}e^{\baverage{q_{1}}t'}}{\nuaverage{q_{k+1}}e^{\baverage{q_{k+1}}t}} =  
\frac{\nuaverage{q_{1}}e^{\baverage{q_{1}}t'}}{\nuaverage{q_{1}}e^{\baverage{q_{1}}s_{1}}}
\frac{\nuaverage{q_{2}}e^{\baverage{q_{2}}s_{1}}}{\nuaverage{q_{2}}e^{\baverage{q_{2}}s_{2}}}\cdots
\frac{\nuaverage{q_{k+1}}e^{\baverage{q_{k+1}}s_{k}}}{\nuaverage{q_{k+1}}e^{\baverage{q_{k+1}}t}}
 =  \exp\left[-\textstyle{\sum}_{j=1}^{k+1}\baverage{q_{j}}|I_{j}|\right], 
\end{equation}
where we suppressed the dependence on $\indexnot$ and the numbers $q_{j}$ are introduced in Definition~\ref{def:varrhoeppl}; this equality is a 
consequence of the definition of the frequency eras. Moreover, 
\begin{equation}\label{eq:qlfrepartbd}
\nuaverage{q_{k+1}}e^{\baverage{q_{k+1}}t}\geq 2\pi,\ \ \
\nuaverage{q_{1}}e^{\baverage{q_{1}}t'}\leq C\ldr{\nu},
\end{equation}
where $C$ only depends on the coefficients of the equation (\ref{eq:thesystemRge}). Combining (\ref{eq:freqeratelescope}) and (\ref{eq:qlfrepartbd})
yields
\begin{equation}\label{eq:bavqjtaujsumest}
-\textstyle{\sum}_{j=1}^{k+1}\baverage{q_{j}}|I_{j}|\leq\ln\ldr{\nu}+c_{0},
\end{equation}
where $c_{0}$ only depends on the coefficients of the equation (\ref{eq:thesystemRge}). Thus $\varsigma(t,t';\rho)$ can be estimated by 
\begin{equation}\label{eq:varsigmaprelest}
\varsigma(t,t';\rho) = \sum_{j=1}^{k+1}-\frac{\rho_{j}}{\baverage{q_{j}}}(-\baverage{q_{j}}|I_{j}|)
 \leq \max_{1\leq j\leq k+1}\max\left\{0,-\frac{\rho_{j}}{\baverage{q_{j}}}\right\}\ln\ldr{\nu}+C,
\end{equation}
where $C$ only depends on $\rho$ and the coefficients of the equation (\ref{eq:thesystemRge}). In particular, the following estimates hold:
\begin{align}
(\b-\kappa_{\rosil,+})t+\varrho_{\e,+}(t,0) \leq & \max_{1\leq j\leq k+1}\max\left\{0,-\frac{\b+\mu_{j,+}-\kappa_{\rosil,+}}{\baverage{q_{j}}}\right\}\ln\ldr{\nu}+C,\label{eq:vzTodeest}\\
-\kappa_{\rosil,+}(t-t')+\varrho_{\e,+}(t,t') \leq & 
\max_{1\leq j\leq k+1}\max\left\{0,-\frac{\mu_{j,+}-\kappa_{\rosil,+}}{\baverage{q_{j}}}\right\}\ln\ldr{\nu}+C,\label{eq:kaovarrhotms}\\
\b t \leq & -\frac{\b}{\baverage{Q}}\ln\ldr{\nu}+C\label{eq:betatestvzTode}
\end{align}
for all $0\leq t'\leq t\leq T_{\roode}$, where $C$ only depends on the coefficients of the equation (\ref{eq:thesystemRge}) (that the constant does not depend on $\e$
is ensured by assuming that $\e\leq 1$ and that it does not depend on $\b$ is due to the fact that $\b\leq\b_{\rem}$). It is of interest to improve the estimate 
(\ref{eq:varsigmaprelest}) when $t'=0$ and $t=T_{\roode}$. Note, to this end, that when $t'=0$, then 
\[
\nuaverage{q_{1}}\geq C\ldr{\nu}
\]
where $C$ only depends on the coefficients of the equation (\ref{eq:thesystemRge}) and the spectrum of the Riemannian manifold corresponding to $q_{1}$.
Moreover, when $t=T_{\roode}$, 
\[
\nuaverage{q_{k+1}}e^{\baverage{q_{k+1}}t}\leq C,
\]
where $C\geq 1$ only depends on the coefficients of the equation (\ref{eq:thesystemRge}). Combining these two estimates with (\ref{eq:freqeratelescope})
yields
\begin{equation}\label{eq:sumbaverageqisisumlb}
-\textstyle{\sum}_{j=1}^{k+1}\baverage{q_{j}}|I_{j}|\geq\ln\ldr{\nu}-c_{1},
\end{equation}
where $c_{1}$ only depends on the coefficients of the equation (\ref{eq:thesystemRge}) and the spectrum of the Riemannian manifold corresponding to $q_{1}$. 
For this reason, the estimate (\ref{eq:varsigmaprelest}) can be improved to 
\begin{equation}\label{eq:varsigmaToderhoest}
\varsigma(T_{\roode},0;\rho) \leq \max_{1\leq j\leq k+1}\left(-\frac{\rho_{j}}{\baverage{q_{j}}}\right)\ln\ldr{\nu}+C
\end{equation}
when $t'=0$ and $t=T_{\roode}$, where $C$ only depends on the coefficients of the equation (\ref{eq:thesystemRge}), $\rho$ and the spectrum of the Riemannian 
manifold corresponding to $q_{1}$. In particular, 
\begin{equation}\label{eq:varsigmaToderhoestspecase}
(\b-\kappa_{\rosil,+})T_{\roode}+\varrho_{\e,+}(T_{\roode},0)\leq \max_{1\leq j\leq k+1}\left(-\frac{\b+\mu_{j,+}-\kappa_{\rosil,+}}{\baverage{q_{j}}}\right)\ln\ldr{\nu}+C,
\end{equation}
where $C$ only depends on the coefficients of the equation (\ref{eq:thesystemRge}) and the spectrum of the Riemannian manifold 
corresponding to $q_{1}$.

\textbf{Estimating $v$.} Combining (\ref{eq:prelestexptimesv}) with (\ref{eq:vzTodeest})--(\ref{eq:betatestvzTode}) yields
\begin{equation}\label{eq:ebmkappaovldrnuest}
e^{(\b-\kappa_{\rosil,+})t}|v(t)|\leq C_{\e}\ldr{\nu}^{s_{\roh,\b,+}+\e}\me^{1/2}(0)+C_{\e}\ldr{\nu}^{s_{\roih,\b,1}+\e}\int_{0}^{t}e^{-\kappa_{\rosil,+}t'}|\hf(t')|dt'
\end{equation}
for all $t\in [0,T_{\roode}]$, 
where $C_{\e}$ only depends on $\e$ and the coefficients of the equation (\ref{eq:thesystemRge}). Here $s_{\roh,\b,+}$ is defined by (\ref{eq:shbpdef}) and
\[
s_{\roih,\b,1} := -\frac{\b}{\baverage{Q}}+\max_{1\leq q\leq Q}\max\left\{0,-\frac{\kappa_{q,+}-\kappa_{\rosil,+}}{\baverage{q}}\right\};
\]
note that $s_{\roih,\b,1}\leq s_{\roih,\b}$, where $s_{\roih,\b}$ is introduced in (\ref{eq:sihbdef}).
In some situations, the $\e$ appearing in (\ref{eq:ebmkappaovldrnuest}) can be removed. Due to complications that arise later in the proof, we here only
consider the case of homogeneous equations. If $f=0$ and the criteria written down in Remark~\ref{remark:homimprsilasest} are satisfied, then the $\e$ 
appearing in (\ref{eq:ebmkappaovldrnuest}) can be removed and the constant $C_{\e}$ can be replaced by one depending only on $\b$ and the coefficients of 
the equation (\ref{eq:thesystemRge}).

Due to (\ref{eq:varsigmaToderhoest}), the estimate (\ref{eq:ebmkappaovldrnuest}) can be improved in case $t=T_{\roode}$. In fact, 
(\ref{eq:prelestexptimesv}) and (\ref{eq:varsigmaToderhoestspecase}) yield
\begin{equation}\label{eq:ebmkappaovldrnuestimpTode}
e^{(\b-\kappa_{\rosil,+})T_{\roode}}|v(T_{\roode})|\leq C_{\e}\ldr{\nu}^{s_{\roh,\b}+\e}\me^{1/2}(0)
+C_{\e}\ldr{\nu}^{s_{\roih,\b,1}+\e}\int_{0}^{T_{\roode}}e^{-\kappa_{\rosil,+}t'}|\hf(t')|dt',
\end{equation}
where $s_{\roh,\b}$ is given by (\ref{eq:shbdef}) and $C_{\e}$ only depends on $\e$, the coefficients of the equation (\ref{eq:thesystemRge}) and the 
spectra of $(M_{r},g_{r})$, $r\in\{1,\dots,R\}$. Moreover, in the case of homogeneous equations, the criterion for removing the $\e$ appearing in 
(\ref{eq:ebmkappaovldrnuestimpTode}) is the same as the one stated in Remark~\ref{remark:homimprsilasfunest}. In that case, $C_{\e}$ is replaced by 
a constant $C_{\b}$ depending only on $\b$, the coefficients of the equation (\ref{eq:thesystemRge}) and the 
spectra of $(M_{r},g_{r})$, $r\in\{1,\dots,R\}$.

\textbf{Estimating $v_{\infty}$.} Let us consider the vector $v_{\infty}$ appearing in (\ref{eq:vinfasres}) in greater detail. To begin with, it is given by 
(\ref{eq:vinfform}), where $u_{\infty}$ satisfies the estimate (\ref{eq:winfTodeest}). Moreover, the constant $C$ appearing in (\ref{eq:winfTodeest}) only 
depends on the coefficients of the equation (\ref{eq:thesystemRge}). Due to the fact that $u_{\infty}\in E_{a}$, 
\begin{equation*}
\begin{split}
e^{(\b-\kappa_{\rosil,+})t}|e^{A_{\infty}t}v_{\infty}| = & e^{(\b-\kappa_{\rosil,+})t}|e^{A_{\infty}(t-T_{\roode})}u_{\infty}|\leq C_{\b}e^{(\b-\kappa_{\rosil,+})t}e^{-(\kappa_{\rosil,+}-\b)(T_{\roode}-t)}|u_{\infty}|\\
 \leq & C_{\b}e^{-(\kappa_{\rosil,+}-\b)T_{\roode}}|u_{\infty}|\leq C_{\b}[e^{(\b-\kappa_{\rosil,+})T_{\roode}}|v(T_{\roode})|+e^{\b T_{\roode}}\|\hf\|_{\rosil}]
\end{split}
\end{equation*}
for all $0\leq t\leq T_{\roode}$, where $C_{\b}$ only depends on $\b$ and the coefficients of the equation (\ref{eq:thesystemRge}); and we use the notation
introduced in (\ref{eq:Fnormrosildef}). Combining this estimate with 
(\ref{eq:betatestvzTode}) and (\ref{eq:ebmkappaovldrnuestimpTode}) yields
\begin{equation}\label{eq:vinftyesteb}
e^{(\b-\kappa_{\rosil,+})t}|e^{A_{\infty}t}v_{\infty}|\leq  C_{\e,\b}\ldr{\nu}^{s_{\roh,\b}+\e}\me^{1/2}(0)+C_{\e,\b}\ldr{\nu}^{s_{\roih,\b,1}+\e}\|\hf\|_{\rosil}
\end{equation}
for all $0\leq t\leq T_{\roode}$,
where $C_{\e,\b}$ only depends on $\e$, $\b$, the coefficients of the equation (\ref{eq:thesystemRge}) and the spectra of $(M_{r},g_{r})$, $r\in\{1,\dots,R\}$. 
Moreover, in the case of homogeneous equations, the criterion for removing the $\e$ appearing in (\ref{eq:vinftyesteb}) 
is the same as the one stated in Remark~\ref{remark:homimprsilasfunest}.
Note also that in the process of deriving the estimate (\ref{eq:vinftyesteb}), we derived an estimate for 
the right hand side of (\ref{eq:vasadalltex}). Note, finally, that setting $t$ to $0$ in (\ref{eq:vinftyesteb}) yields an estimate for $v_{\infty}$.

\textbf{Estimating the contribution from the inhomogeneity.} Let us turn to 
\begin{equation}\label{eq:FconzTodepartex}
\begin{split}
 & e^{(\b-\kappa_{\rosil,+})t}\int_{t}^{T_{\roode}}e^{A_{\infty}(t-t')}F(t')dt'\\
 = & \int_{t}^{T_{\roode}}e^{\b t'}e^{(A_{\infty}-\kappa_{\rosil,+}\Id_{2m}+\b\Id_{2m})(t-t')}e^{-\kappa_{\rosil,+}t'}F(t')dt'.
\end{split}
\end{equation}
When estimating the right hand side of (\ref{eq:FconzTodepartex}), we consider the cases $\b-\kappa_{\roode}>0$ and $\b-\kappa_{\roode}\leq 0$ separately; here 
$\kappa_{\roode}:=\Rsp A_{\infty}$ and the notation $\Rsp A$ is introduced in Definition~\ref{def:SpRspdef}. In the second case, there is potentially a polynomial loss, 
which we estimate crudely by $e^{-\e\baverage{Q} T_{\roode}}$. This yields
\begin{equation}\label{eq:Fintfinestex}
\sup_{t\in [0,T_{\roode}]}\left|e^{(\b-\kappa_{\rosil,+})t}\int_{t}^{T_{\roode}}e^{A_{\infty}(t-t')}F(t')dt'\right|
\leq C_{\e,\b}\ldr{\nu}^{s_{\roih,\b,2}+\e}\|\hf\|_{\rosil},
\end{equation}
where 
\[
s_{\roih,\b,2}:=-\max\{\b,\kappa_{\roode}\}/\baverage{Q}
\]
and $C_{\e,\b}$ only depends on $\e$, $\b$ and the coefficients of the equation (\ref{eq:thesystemRge}). In order to obtain this conclusion, we appealed 
to estimates such as (\ref{eq:betatestvzTode}).

\textbf{Estimating $v_{\infty,\romod}$.} Combining (\ref{eq:vinfmoddefex}), (\ref{eq:vinftyesteb}) (with $t$ set to $0$) and (\ref{eq:Fintfinestex}) yields
\begin{equation}\label{eq:vinfromodestex}
|v_{\infty,\romod}|\leq C_{\e,\b}\ldr{\nu}^{s_{\roh,\b}+\e}\me^{1/2}(0)+C_{\e,\b}\ldr{\nu}^{s_{\roih,\b}+\e}\|\hf\|_{\rosil}
\end{equation}
where $C_{\e,\b}$ only depends on $\e$, $\b$, the coefficients of the equation (\ref{eq:thesystemRge}) and the spectra of $(M_{r},g_{r})$, $r\in\{1,\dots,R\}$. 
Moreover, in the case of homogeneous equations, the criterion for removing the $\e$ appearing in (\ref{eq:vinfromodestex}) 
is the same as the one stated in Remark~\ref{remark:homimprsilasfunest}.

\textbf{Analysis for one mode.} Consider (\ref{eq:vasadalltex}). This estimate is valid on $[T_{\roode},\infty)$. In order to derive an estimate which is 
valid for all $t\geq 0$, it is of interest to note that
\begin{equation}\label{eq:mainmodestsilsettingasearlyt}
\begin{split}
 & e^{-(\kappa_{\rosil,+}-\b)t}\left|v(t)-e^{A_{\infty}t}v_{\infty,\romod}-\int_{0}^{t}e^{A_{\infty}(t-t')}F(t')dt'\right|\\
 \leq & C_{\e,\b}\ldr{\nu}^{s_{\roh,\b,+}+\e}\me^{1/2}(0)+C_{\e,\b}\ldr{\nu}^{s_{\roih,\b}+\e}\|\hf\|_{\rosil}
\end{split}
\end{equation}
for all $t\in [0,T_{\roode}]$. In order to obtain this estimate, we appealed to (\ref{eq:ebmkappaovldrnuest}), (\ref{eq:vinftyesteb}) and (\ref{eq:Fintfinestex}).
Moreover, the constant $C_{\e,\b}$ only depends on $\e$, $\b$, the coefficients of the equation (\ref{eq:thesystemRge}) and the spectra of 
$(M_{r},g_{r})$, $r\in\{1,\dots,R\}$. In the case of homogeneous equations, the criterion for removing the $\e$ is the same as the one 
stated in Remark~\ref{remark:homimprsilasest}. Combining (\ref{eq:vasadalltex}), (\ref{eq:betatestvzTode}) and (\ref{eq:ebmkappaovldrnuest}) yields
\begin{equation}\label{eq:mainmodestsilsettingaslatet}
\begin{split}
 & e^{-(\kappa_{\rosil,+}-\b)t}\left|v(t)-e^{A_{\infty}t}v_{\infty,\romod}-\int_{0}^{t}e^{A_{\infty}(t-t')}F(t')dt'\right|\\
 \leq & C_{\e}\ldr{t}^{N}\ldr{\nu}^{s_{\roh,\b,+}+\e}\me^{1/2}(0)+C_{\e}\ldr{t}^{N}\ldr{\nu}^{s_{\roih,\b,1}+\e}\|\hf\|_{\rosil}
\end{split}
\end{equation}
for all $t\geq T_{\roode}$, where $C_{\e}$ only depends on $\e$ and the coefficients of the equation (\ref{eq:thesystemRge}). Moreover, $N$ only depends 
on $m$. In the case of homogeneous equations, the criterion for removing the $\e$ is the same as the one stated in Remark~\ref{remark:homimprsilasest}. 
However, $C_{\e}$ is then replaced by a constant depending only on $\b$ and the coefficients of the equation (\ref{eq:thesystemRge}).
Combining (\ref{eq:mainmodestsilsettingasearlyt}) and 
(\ref{eq:mainmodestsilsettingaslatet}) yields
\begin{equation}\label{eq:mainmodestsilsettingasallt}
\begin{split}
 & e^{-(\kappa_{\rosil,+}-\b)t}\left|v(t)-e^{A_{\infty}t}v_{\infty,\romod}-\int_{0}^{t}e^{A_{\infty}(t-t')}F(t')dt'\right|\\
 \leq & C_{\e,\b}\ldr{t}^{N}\ldr{\nu}^{s_{\roh,\b,+}+\e}\me^{1/2}(0)+C_{\e,\b}\ldr{t}^{N}\ldr{\nu}^{s_{\roih,\b}+\e}\|\hf\|_{\rosil}
\end{split}
\end{equation}
for all $t\geq 0$, where $C_{\e,\b}$ only depends on $\e$, $\b$, the coefficients of the equation (\ref{eq:thesystemRge}) and the spectra of 
$(M_{r},g_{r})$, $r\in\{1,\dots,R\}$. Again, in the case of homogeneous equations, the criterion for removing 
the $\e$ is the same as the one stated in Remark~\ref{remark:homimprsilasest}. In case $T_{\roode}=0$, (\ref{eq:vasadalltex}) immediately yields the 
conclusion that (\ref{eq:mainmodestsilsettingasallt}) holds (with the factors involving $\ldr{\nu}$ removed). 

\textbf{Projecting the data at infinity.} At this stage we would like to define $V_{\infty}$ by the requirement that its $\indexnot$'th Fourier 
coefficient be given by $v_{\infty,\romod}(\indexnot)$. However, the resulting function need not take its values in $E_{a}$. Just as in the proof of
Lemma~\ref{lemma:roughas}, we therefore first need to project $v_{\infty,\romod}(\indexnot)$ to the right subspace and to estimate the corresponding 
error. In particular, we need to estimate 
\[
e^{-(\kappa_{\rosil,+}-\b)t}\left|e^{A_{\infty}t}\Pi_{b}\left(\int_{0}^{T_{\roode}}e^{-A_{\infty}t'}F(t')dt'\right)\right|,
\]
where $\Pi_{b}$ denotes the projection to the second generalised eigenspace in the $\b$, $A_{\infty}$-decomposition of $\cn{2m}$; 
cf. Lemma~\ref{lemma:eAtPibnormest}. However, appealing to (\ref{eq:Fintfinestex}) and Lemma~\ref{lemma:eAtPibnormest} yields
\begin{equation*}
\begin{split}
 & e^{-(\kappa_{\rosil,+}-\b)t}\left|e^{A_{\infty}t}\Pi_{b}\left(\int_{0}^{T_{\roode}}e^{-A_{\infty}t'}F(t')dt'\right)\right| \\
\leq & C\ldr{t}^{d_{b}-1}e^{\kappa_{b}t}
e^{-(\kappa_{\rosil,+}-\b)t}\left|\int_{0}^{T_{\roode}}e^{-A_{\infty}t'}F(t')dt'\right|\\
\leq & C_{\e,\b}\ldr{t}^{d_{b}-1}\ldr{\nu}^{s_{\roih,\b,2}+\e}\|\hf\|_{\rosil},
\end{split}
\end{equation*}
where $d_{b}$ and $\kappa_{b}$ are given in the statement of Lemma~\ref{lemma:eAtPibnormest} and $C_{\e,\b}$ only depends on $\e$, $\b$ and the 
coefficients of the equation (\ref{eq:thesystemRge}). Combining this estimate with (\ref{eq:mainmodestsilsettingasallt}) yields
\begin{equation}\label{eq:mainmodestsilsettingasalltproj}
\begin{split}
 & e^{-(\kappa_{\rosil,+}-\b)t}\left|v(t)-e^{A_{\infty}t}\Pi_{a}v_{\infty,\romod}-\int_{0}^{t}e^{A_{\infty}(t-t')}F(t')dt'\right|\\
 \leq & C_{\e,\b}\ldr{t}^{N}\ldr{\nu}^{s_{\roh,\b,+}+\e}\me^{1/2}(0)+C_{\e,\b}\ldr{t}^{N}\ldr{\nu}^{s_{\roih,\b}+\e}\|\hf\|_{\rosil}
\end{split}
\end{equation}
for all $t\geq 0$, where $C_{\e,\b}$ only depends on $\e$, $\b$, the coefficients of the equation (\ref{eq:thesystemRge}) and the spectra of 
$(M_{r},g_{r})$, $r\in\{1,\dots,R\}$. Here $\Pi_{a}$ denotes the projection to the first generalised eigenspace in the $\b$, $A_{\infty}$-decomposition 
of $\cn{2m}$; cf. Lemma~\ref{lemma:eAtPibnormest}. Finally, in the case of homogeneous equations, the criterion for removing 
the $\e$ in (\ref{eq:mainmodestsilsettingasalltproj}) is the same as the one stated in Remark~\ref{remark:homimprsilasest}.

\textbf{Summing up.} Given $p=(x,p_{1},\dots,p_{R})\in\bM$, where $\bM$ is given by (\ref{eq:bMdef}), define $V_{\infty}(p)$ by
\[
V_{\infty}(p):=\textstyle{\sum}_{\indexnot\in\EFindexset}\Pi_{a}[v_{\infty,\romod}(\indexnot)]\varphi_{\indexnot}(p),
\]
where $\varphi_{\indexnot}$ is given by (\ref{eq:varphinudef}). In order to justify that the sum on the right hand side makes sense and yields a smooth 
function, it is sufficient to appeal to the fact that $u$ is smooth; the assumption that (\ref{eq:fnosilsdef}) holds for all $s$; 
(\ref{eq:vinfromodestex}); and the Minkowski inequality. Then (\ref{eq:mainmodestsilsettingasalltproj}) implies that
\begin{equation*}
\begin{split}
 & \left\|\left(\begin{array}{c} u(\cdot,t) \\ u_{t}(\cdot,t)\end{array}\right)
-e^{A_{\infty}t}V_{\infty}
-\int_{0}^{t}e^{A_{\infty}(t-t')}\left(\begin{array}{c} 0 \\ f(\cdot,t')\end{array}\right)dt'\right\|_{(s)} \\
\leq & C_{\e,\b}\ldr{t}^{N}e^{(\kappa_{\rosil,+}-\b)t}\left(\mfe_{s+s_{\roh,\b,+}+\e}^{1/2}[u](0)+\|f\|_{\rosil,s+s_{\roih,\b}+\e}\right)
\end{split}
\end{equation*}
for all $t\geq 0$, where we appealed to the Minkowski inequality and $C_{\e,\b}$ and $N$ have the same dependence as in the case of 
(\ref{eq:mainmodestsilsettingasalltproj}). Thus (\ref{eq:uudothsestex}) holds. Moreover, Remark~\ref{remark:homimprsilasest} holds. Note also that, due 
to (\ref{eq:vinfromodestex}), 
\begin{equation}\label{eq:VinfHsestpf}
\|V_{\infty}\|_{(s)}\leq C_{\e,\b}\left(\mfe_{s+s_{\roh,\b}+\e}^{1/2}[u](0)
+\|f\|_{\rosil,s+s_{\roih,\b}+\e}\right),
\end{equation}
where $C_{\e,\b}$ has the same dependence as in the case of (\ref{eq:mainmodestsilsettingasallt}). Thus (\ref{eq:uinfudinfHsestex}) holds. 
Moreover, Remark~\ref{remark:homimprsilasfunest} holds. The proposition follows. 
\end{proof}

\section{Specifying the asymptotics}\label{section:roughODEspecasex}

As a next step, it is of interest to improve Lemma~\ref{lemma:spasda} by demonstrating that the conclusions hold for a specific value of 
$s_{\infty}$ that can be calculated in terms of the coefficients of the equation. 

\begin{prop}\label{prop:spasdaex}
Consider the equation (\ref{eq:thesystemRge}) with $f=0$. Assume that it is non-degenerate, diagonally dominated, balanced and convergent; cf. 
Definition~\ref{def:nondegconvabal}. Assume that $\baverage{Q}<0$ and let $\b_{\rem}:=\min\{-\baverage{Q},\eta_{\romn}\}$, where $\eta_{\romn}$ 
is the constant appearing in (\ref{eq:mainconv}). Fix $0<\b\leq\b_{\rem}$ and $\e>0$. Let $E_{a}$ be the first generalised eigenspace in the 
$\b,A_{\infty}$-decomposition of $\cn{2m}$, where $A_{\infty}$ is defined by (\ref{eq:vAhFdef}); cf. Definition~\ref{def:defofgeneigenspintro}. 
Then there is a linear injective map
\[
\Phi_{\b,\infty}:C^{\infty}(\bM,E_{a})\rightarrow C^{\infty}(\bM,\cn{2m})
\]
such that the following holds. First, if $\Phi^{j}_{\b,\infty}:C^{\infty}(\bM,E_{a})\rightarrow C^{\infty}(\bM,\cn{m})$, $j=1,2$, are defined by 
the condition that 
\[
\Phi_{\b,\infty}(\chi)=\left(\begin{array}{c} \Phi^{1}_{\b,\infty}(\chi) \\ \Phi^{2}_{\b,\infty}(\chi)\end{array}\right)
\]
for all $\chi\in C^{\infty}(\bM,E_{a})$, then 
\begin{equation}\label{eq:Phiinfnobdex}
\|\Phi_{\b,\infty}^{1}(\chi)\|_{(s+1)}+\|\Phi_{\b,\infty}^{2}(\chi)\|_{(s)}\leq C_{\e}\|\chi\|_{(s+s_{\roh,-}+\e)}
\end{equation}
for all $s\in\ro$ and all $\chi\in C^{\infty}(\bM,E_{a})$, where $C_{\e}$ only depends on $\e$, the coefficients of the equation 
(\ref{eq:thesystemRge}) and the spectra of $(M_{r},g_{r})$, $r\in\{1,\dots,R\}$, and 
\begin{equation}\label{eq:srohmdef}
s_{\roh,-}:=\max_{1\leq q\leq Q}\left(-\frac{\kappa_{q,-}+\kappa_{\rosil,+}}{\baverage{q}}\right).
\end{equation}
Secondly, if $\chi\in C^{\infty}(\bM,E_{a})$ and $u$ is the solution to (\ref{eq:thesystemRge}) (with $f=0$) such that 
\begin{equation}\label{eq:uuditoPhiinfchiex}
\left(\begin{array}{c} u(\cdot,0) \\ u_{t}(\cdot,0)\end{array}\right)=\Phi_{\b,\infty}(\chi),
\end{equation}
then 
\begin{equation}\label{eq:estspecasdataex}
\begin{split}
 & \left\|\left(\begin{array}{c} u(\cdot,t) \\ u_{t}(\cdot,t)\end{array}\right)
-e^{A_{\infty}t}\chi\right\|_{(s)} \\
\leq & C_{\e,\b}\ldr{t}^{N}e^{(\kappa_{\rosil,+}-\b)t}\left(\|u_{t}(\cdot,0)\|_{(s+s_{\roh,\b,+}+\e)}+\|u(\cdot,0)\|_{(s+s_{\roh,\b,+}+\e+1)}\right)
\end{split}
\end{equation}
for all $t\geq 0$ and all $s\in\ro$, 
where $C_{\e,\b}$ only depends on $\e$, $\b$, the coefficients of the equation (\ref{eq:thesystemRge}) and the spectra of 
$(M_{r},g_{r})$, $r\in\{1,\dots,R\}$, and $N$ only depends on $m$. Finally, if $E_{a}=\cn{2m}$ (i.e., if $\b>\Rsp A_{\infty}$; cf. 
Definition~\ref{def:SpRspdef}), then $\Phi_{\b,\infty}$ is surjective. 
\end{prop}
\begin{remark}\label{remark:optimalasspecest}
If the assumptions of Proposition~\ref{prop:spasdaexintro} are satisfied, then the assumptions of the proposition are satisfied; 
cf. Remark~\ref{remark:maintheoremgeomcond}.
\end{remark}
\begin{remark}\label{remark:removingepsilonbackwest}
If the Jordan blocks of $A_{\infty}$ corresponding to eigenvalues with real part $\kappa_{\rosil,+}$ are trivial, and if all Jordan blocks of $R_{q,-}^{+}$ and 
$R_{q,-}^{-}$ corresponding to eigenvalues such that
\begin{itemize}
\item the real part of the eigenvalue equals $\kappa_{q,-}$, and
\item $\kappa_{q,-}$ satisfies
\begin{equation}\label{eq:kappaqmsrohmrelstmt}
-\frac{\kappa_{q,-}+\kappa_{\rosil,+}}{\baverage{q}}=s_{\roh,-},
\end{equation}
\end{itemize}
are trivial for $q=1,\dots,Q$, then the $\e$ appearing in (\ref{eq:Phiinfnobdex}) can be removed, so that
\[
\|\Phi_{\b,\infty}^{1}(\chi)\|_{(s+1)}+\|\Phi_{\b,\infty}^{2}(\chi)\|_{(s)}\leq C\|\chi\|_{(s+s_{\roh,-})},
\]
where $C$ only depends on the coefficients of the equation (\ref{eq:thesystemRge}) and the spectra of 
$(M_{r},g_{r})$, $r\in\{1,\dots,R\}$.
\end{remark}
\begin{remark}
By combining (\ref{eq:Phiinfnobdex}), (\ref{eq:uuditoPhiinfchiex}) and (\ref{eq:estspecasdataex}), the norms of $u(\cdot,0)$ and $u_{t}(\cdot,0)$ 
appearing on the right hand side of (\ref{eq:estspecasdataex}) can be replaced by a suitable Sobolev norm of $\chi$. 
\end{remark}
\begin{remark}
In order to obtain a similar result in the case of inhomogeneous equations, it is sufficient to combine Propositions~\ref{prop:roughasex} and
\ref{prop:spasdaex}; cf. Remark~\ref{remark:inhomaschar}. 
\end{remark}
\begin{remark}
The same criteria and conclusions concerning improvements of $\b_{\rem}$ that are stated in Remark~\ref{remark:improvedbetarem} hold in the present 
proposition.
\end{remark}
\begin{proof}
Let $\chi\in C^{\infty}(\bM,E_{a})$, $\indexnot\in \EFindexset$ and $\hchi(\indexnot):=\ldrbox{\chi,\varphi_{\indexnot}}\in E_{a}$; cf. the notation 
introduced in Subsection~\ref{ssection:specprodset}. Let $u$ be a solution to (\ref{eq:thesystemRge}) and let $z(\indexnot,t)$ be given by 
(\ref{eq:znutdef}). Note that the $\indexnot$'th Fourier coefficient of the expression inside the norm on the left hand side of (\ref{eq:estspecasdataex}) 
can be written
\begin{equation}\label{eq:vasitochihex}
v-e^{A_{\infty}t}\hchi,
\end{equation}
where we use the notation (\ref{eq:vAhFdef}). On the other hand, $v$ satisfies (\ref{eq:dotvroughODE}) with $F=0$, and $A_{\rem}$ satisfies 
(\ref{eq:Aremestgensys}) for $t\geq T_{\roode}$ (where $T_{\roode}$ is defined as in the proof of Proposition~\ref{prop:roughasex}). Moreover,
$\b_{\rem}$ has the value given in the statement of Proposition~\ref{prop:roughasex} (or in Remark~\ref{remark:improvedbetarem}, if the conditions
of Remark~\ref{remark:improvedbetarem} apply) and $C_{\rem}$ only depends on the coefficients of the equation (\ref{eq:thesystemRge}). In particular, 
the conditions needed to apply Lemma~\ref{lemma:spasODEsett} are satisfied. In fact, specifying
$v$ by imposing the initial condition $v(T_{\roode}):=\Psi_{\infty}[\hchi(\indexnot)]$ yields the conclusion that (\ref{eq:vasitochihex}) is small 
asymptotically; 
here $\Psi_{\infty}$ is the map constructed in Lemma~\ref{lemma:spasODEsett}. Note that $\Psi_{\infty}$ depends on $\indexnot$. However, the estimates
involving $\Psi_{\infty}$ that we need do not. In order to estimate $|v(T_{\roode})|$ in terms of $|\hchi(\indexnot)|$, define $u_{\infty}\in E_{a}$ by 
\[
e^{-A_{\infty}T_{\roode}}u_{\infty}=\hchi(\indexnot).
\]
Thus $|u_{\infty}|=|e^{A_{\infty}T_{\roode}}\hchi(\indexnot)|$, so that if $\e>0$ is fixed, then (\ref{eq:Psiinfnorm}) yields the estimate
\begin{equation}\label{eq:vTodeabsestasdataex}
\begin{split}
|v(T_{\roode})| = & |\Psi_{\infty}(e^{-A_{\infty}T_{\roode}}u_{\infty})|\leq C|u_{\infty}|=C|e^{A_{\infty}T_{\roode}}\hchi(\indexnot)|\\
 \leq & C_{\e}e^{(\kappa_{\rosil,+}+\e)T_{\roode}}|\hchi(\indexnot)|
\end{split}
\end{equation}
where $C_{\e}$ only depends on $\e$ and the coefficients of the equation (\ref{eq:thesystemRge}). Moreover, the $\e$ can be removed if the Jordan blocks
of $A_{\infty}$ corresponding to eigenvalues with real part $\kappa_{\rosil,+}$ are trivial. In order to estimate $\me_{s}(0)$, it is natural to 
divide the analysis into two cases. To begin with, assume 
that $T_{\roode}=0$. Then 
\begin{equation}\label{eq:mesestTodeezex}
\begin{split}
\me^{1/2}(0)\leq & C|v(0)| = C|v(T_{\roode})|\leq C|\hchi(\indexnot)|,
\end{split}
\end{equation}
where $C$ only depends on the coefficients of the equation (\ref{eq:thesystemRge}) and we have appealed to (\ref{eq:vTodeabsestasdataex}) 
and the fact that $T_{\roode}=0$. 

\textbf{The high frequency case.} In case $T_{\roode}>0$, we appeal to Corollary~\ref{cor:westvarrhocase} with $f=0$. For $\e_{1}>0$, this yields
\begin{equation}\label{eq:mesqrtzprimest}
\me^{1/2}(0)\leq C|w(0)| \leq C_{\e_{1}}e^{\varrho_{\e_{1},-}(0,T_{\roode})}|w(T_{\roode})|,
\end{equation}
where $C_{\e_{1}}$ only depends on $\e_{1}$ and the coefficients of the equation (\ref{eq:thesystemRge}), and we have omitted the subscript $K$ due to 
the absence of transparent eras. Due to the definition of $T_{\roode}$, the expression $|w(T_{\roode})|$ is bounded by $C|v(T_{\roode})|$, where $C$
only depends on the coefficients of the equation (\ref{eq:thesystemRge}). Combining this observation with (\ref{eq:vTodeabsestasdataex}) and
(\ref{eq:mesqrtzprimest}) yields
\begin{equation}\label{eq:mesqrtzprelestCeetc}
\me^{1/2}(0)\leq C_{\e,\e_{1}}e^{\varrho_{\e_{1},-}(0,T_{\roode})}e^{(\kappa_{\rosil,+}+\e)T_{\roode}}|\hchi(\indexnot)|,
\end{equation}
where $C_{\e,\e_{1}}$ only depends on $\e$, $\e_{1}$ and the coefficients of the equation (\ref{eq:thesystemRge}). Moreover, the $\e$ can be removed 
in case the Jordan blocks of $A_{\infty}$ corresponding to eigenvalues with real part $\kappa_{\rosil,+}$ are trivial. Keeping 
Definition~\ref{def:varrhoeppl} in mind, 
\begin{equation}\label{eq:varrhoempkappaoeTode}
\varrho_{\e_{1},-}(0,T_{\roode})+(\kappa_{\rosil,+}+\e)T_{\roode}=\textstyle{\sum}_{j=1}^{k+1}(\mu_{j,-}+\kappa_{\rosil,+}+\e)|I_{j}|,
\end{equation}
where the intervals $I_{j}$, $j\in\{1,\dots,k+1\}$, are the ones into which $[0,T_{\roode}]$ is divided when appealing to Definition~\ref{def:varrhoeppl} and 
the numbers $\mu_{j,-}$ are introduced in Definition~\ref{def:varrhoeppl}. Rewriting the right hand side of (\ref{eq:varrhoempkappaoeTode})
as in the right hand side of the first line of (\ref{eq:varsigmaprelest}) and recalling (\ref{eq:bavqjtaujsumest}) (with $t'=0$ and $t=T_{\roode}$) and 
(\ref{eq:sumbaverageqisisumlb}) yields
\begin{equation}\label{eq:varrhoempkappaoeTodeimpest}
\varrho_{\e_{1},-}(0,T_{\roode})+(\kappa_{\rosil,+}+\e)T_{\roode}\leq \max_{1\leq j\leq k+1}\left(-\frac{\mu_{j,-}+\kappa_{\rosil,+}+\e}{\baverage{q_{j}}}\right)\ln\ldr{\nu}+C,
\end{equation}
where $C$ only depends on the coefficients of the equation (\ref{eq:thesystemRge}) and the spectra of $(M_{r},g_{r})$, $r\in\{1,\dots,R\}$ (in order to obtain 
this conclusion we assume that $\e,\e_{1}\leq 1$). Note that the $\e$ appearing explicitly on the right hand side of 
(\ref{eq:varrhoempkappaoeTodeimpest}) can be removed in case the Jordan blocks of $A_{\infty}$ corresponding to eigenvalues with real part $\kappa_{\rosil,+}$ are 
trivial. Let $s_{\roh,-}$ be defined by (\ref{eq:srohmdef}). Then (\ref{eq:varrhoempkappaoeTodeimpest}) implies
\[
\varrho_{\e_{1},-}(0,T_{\roode})+(\kappa_{\rosil,+}+\e)T_{\roode}\leq (s_{\roh,-}+\e_{2})\ln\ldr{\nu}+C,
\]
which, combined with (\ref{eq:mesqrtzprelestCeetc}), yields
\begin{equation}\label{eq:mesqrtmainestbackw}
\me_{s}^{1/2}(0)\leq C_{\e,\e_{1}}\ldr{\nu}^{s+s_{\roh,-}+\e_{2}}|\hchi(\indexnot)|.
\end{equation}
Here $\e_{2}$ is a linear combination of $\e$ and $\e_{1}$ (where the coefficients only depend on the coefficients of the equation (\ref{eq:thesystemRge})).
Moreover, $C_{\e,\e_{1}}$ only depends on $\e$ and $\e_{1}$, the coefficients of the equation (\ref{eq:thesystemRge}) and the spectra of 
$(M_{r},g_{r})$, $r\in\{1,\dots,R\}$. In addition, if the criteria stated in Remark~\ref{remark:removingepsilonbackwest} are satisfied, then $\e$, $\e_{1}$
and $\e_{2}$ appearing in (\ref{eq:mesqrtmainestbackw}) can be removed, so that
\begin{equation}\label{eq:mesqrtmainestbackwimproved}
\me_{s}^{1/2}(0)\leq C\ldr{\nu}^{s+s_{\roh,-}}|\hchi(\indexnot)|,
\end{equation}
where $C$ only depends on the coefficients of the equation (\ref{eq:thesystemRge}) and the spectra of $(M_{r},g_{r})$, $r\in\{1,\dots,R\}$. Combining 
(\ref{eq:mesestTodeezex}) and (\ref{eq:mesqrtmainestbackw}) yields the conclusion that (\ref{eq:mesqrtmainestbackw})
holds regardless of whether $T_{\roode}$ equals zero or not. Moreover, the improvement (\ref{eq:mesqrtmainestbackwimproved}) holds in case the conditions
stated in Remark~\ref{remark:removingepsilonbackwest} are fulfilled. 

Let $\Phi_{\b,\infty}$ be the map taking $\chi\in C^{\infty}(\bM,E_{a})$ to the element of $C^{\infty}(\bM,\cn{2m})$ whose $\indexnot$'th mode is given by $v(0)$, 
where $v$ is constructed as above. Then (\ref{eq:mesqrtmainestbackw}) implies that (\ref{eq:Phiinfnobdex}) holds; note that this implies that $\Phi_{\b,\infty}$ 
maps $\chi\in C^{\infty}(\bM,E_{a})$ to an element of $C^{\infty}(\bM,\cn{2m})$. Moreover, Remark~\ref{remark:removingepsilonbackwest} holds. 
Since $\Phi_{\b,\infty}$ is injective on the level of Fourier coefficients (this is a consequence of Lemma~\ref{lemma:spasODEsett}), 
it follows that $\Phi_{\b,\infty}$ is injective. Next, let $\chi\in C^{\infty}(\bM,E_{a})$ and let $u$ be the solution to (\ref{eq:thesystemRge}) (with $f=0$) satisfying 
(\ref{eq:uuditoPhiinfchiex}). Due to Proposition~\ref{prop:roughasex}, we know that (\ref{eq:estspecasdataex}) holds with $\chi$ replaced by some 
$V_{\infty}\in C^{\infty}(\bM,E_{a})$. Moreover, the constants $C_{\e,\b}$ and $N$ have the dependence stated in Proposition~\ref{prop:roughasex}. In order to determine 
$V_{\infty}$ in terms of $\chi$, note that (\ref{eq:estspecasdataex}) (with $\chi$ replaced by $V_{\infty}$) implies that 
\begin{equation}\label{eq:vprelaspfex}
\left|v(t)-e^{A_{\infty}t}v_{\infty}\right| \leq C\ldr{t}^{N}e^{(\kappa_{\rosil,+}-\b)t}
\end{equation}
for all $t\geq 0$, where $C$ depends on the initial data and $\indexnot$; $v$ is defined as in (\ref{eq:vAhFdef}) where $z$ is the $\indexnot$'th mode of $u$; 
and $v_{\infty}$ is the $\indexnot$'th mode of $V_{\infty}$. Due to the construction of $\Phi_{\b,\infty}$, we know that the estimate (\ref{eq:vprelaspfex}) holds 
with $v_{\infty}$ replaced by $\hchi(\indexnot)$. Thus
\[
\left|e^{A_{\infty}t}[\hchi(\indexnot)-v_{\infty}]\right| \leq C\ldr{t}^{N}e^{(\kappa_{\rosil,+}-\b)t}
\]
for all $t\geq 0$. Since $\hchi(\indexnot)-v_{\infty}\in E_{a}$, this estimate implies that $\hchi(\indexnot)-v_{\infty}=0$. Thus $V_{\infty}=\chi$, and 
the lemma follows, except for the statement concerning surjectivity. 

Assume that $E_{a}=\cn{2m}$ and let $\psi\in C^{\infty}(\bM,\cn{2m})$. We wish to demonstrate that $\psi$ is in the image of $\Phi_{\b,\infty}$. Let 
$u$ be the solution to (\ref{eq:thesystemRge}) with initial data given by (\ref{eq:uuditoPhiinfchiex}), where the right hand side has been replaced by 
$\psi$. Appealing to Proposition~\ref{prop:roughasex} yields a $V_{\infty}\in C^{\infty}(\bM,\cn{2m})$ such that 
(\ref{eq:uudothsestex}) holds (with $f=0$). Let $\bu$ be the solution to (\ref{eq:thesystemRge}) with initial data given by (\ref{eq:uuditoPhiinfchiex}),
where the right hand side has been replaced by $\Phi_{\b,\infty}(V_{\infty})$. Then, by the above arguments, (\ref{eq:uudothsestex}) holds (with $f=0$)
and $u$ replaced by $\bu$. In particular, there is thus a constant, depending on $u$, $\bu$ etc., such that 
\begin{equation}\label{eq:uidminusbuid}
\left\|\left(\begin{array}{c} u(\cdot,t) \\ u_{t}(\cdot,t)\end{array}\right)
-\left(\begin{array}{c} \bu(\cdot,t) \\ \bu_{t}(\cdot,t)\end{array}\right)\right\|_{(s)} 
\leq C_{\e,\b}\ldr{t}^{N}e^{(\kappa_{\rosil,+}-\b)t}
\end{equation}
for all $t\geq 0$ and all $s\in\ro$. Let $z$ and $\bz$ be the Fourier coefficients of $u$ and $\bu$ respectively, let $v$ and $\bv$ be defined in
analogy with (\ref{eq:vAhFdef}), starting with $z$ and $\bz$ respectively. Finally, let $V=v-\bv$. If $V(\indexnot,0)=0$ for all $\indexnot\in\EFindexset$, 
then $u=\bu$ (since $V(\indexnot,\cdot)$ solves a homogeneous equation, $V(\indexnot,0)=0$ for all $\indexnot\in\EFindexset$ implies that $V(\indexnot,t)=0$ 
for all $\indexnot\in\EFindexset$ and all $t\in I$), so that $\Phi_{\b,\infty}(V_{\infty})=\psi$. Thus $\psi$ is in the image of $\Phi_{\b,\infty}$. Assume now that 
there is a $\indexnot\in\EFindexset$ such that $V(\indexnot,0)\neq 0$. Note that $V(\indexnot,\cdot)$ is a solution to an equation to which 
Lemma~\ref{lemma:spasODEsett} applies. Moreover, applying Lemma~\ref{lemma:spasODEsett} yields a $\Psi_{\infty}$ which is linear and bijective. Since
$V(\indexnot,0)\neq 0$, there is a $0\neq\chi\in\cn{2m}$ such that $\Psi_{\infty}(\chi)=V(\indexnot,T_{\roode})$, where $T_{\roode}$ is determined by the 
$\indexnot$ under consideration. This means that 
\[
|V(\indexnot,t)-e^{A_{\infty}t}\chi|\leq C\ldr{t}^{N}e^{(\kappa_{\rosil,+}-\b)t}
\]
for all $t\geq T_{\roode}$ and a constant $C$ depending on $V$, $\indexnot$ etc. On the other hand, (\ref{eq:uidminusbuid}) implies that the same estimate holds
with $\chi$ set to zero. Due to the fact that $\Rsp A_{\infty}<\b$, these estimates are contradictory. To conclude, every element of $C^{\infty}(\bM,\cn{2m})$ is 
in the image of $\Phi_{\b,\infty}$. The lemma follows. 
\end{proof}

\part{Appendices}\label{part:appendices}

\chapter{Spectral aspects of differential operators and matrices}

The main object of study in these notes is the equation (\ref{eq:thesystemRge}). This equation is separable, and the main purpose of 
the present chapter is to derive the equations satisfied by the modes. We provide the relevant material in Section~\ref{section:divintomodes}.
To begin with, we recall the spectral properties of the Laplace-Beltrami operator on a closed Riemannian manifold in 
Subsection~\ref{ssection:laplacebeltramibasics}. We then turn to the product setting in Subsection~\ref{ssection:specprodset}; this is 
the setting relevant in the study of (\ref{eq:thesystemRge}). With this background material available, we are in a position to separate
the equation. This is the topic of Subsection~\ref{ssection:fourdecompmaineq}. Finally, in Subsection~\ref{ssection:highordenergies}, we 
express the energy in terms of the modes. 

The second purpose of the present chapter is to recall some of the basic properties of the Jordan normal form of a matrix, as well as of 
matrix exponentials of matrices. This is the subject of Section~\ref{section:jordannormalform}.

\section{Dividing the solution into modes}\label{section:divintomodes}

The equation (\ref{eq:thesystemRge}) is separable. It is therefore possible to analyse the behaviour of solutions by considering the individual
modes. In fact, the arguments presented in these notes are based on this perspective, as are the statements of several of the results. 
Before proceeding, we therefore here present the background relevant to separating (\ref{eq:thesystemRge}). As a starting point, let us
recall some basic facts concerning the Laplace-Beltrami operator 
\index{Laplace-Beltrami operator}%
on a closed Riemannian manifold.

\subsection{Spectral properties of the Laplace-Beltrami operator on a 
closed Riemannian manifold}\label{ssection:laplacebeltramibasics}

To begin with, let us recall the following result. 
\begin{thm}\label{thm:specclriem}
Let $(M,g)$ be a closed Riemannian manifold and $\Delta_{g}$ be the associated Laplace-Beltrami operator. Then the eigenvalues of $\Delta_{g}$ 
consist of a sequence $\lambda_{i}$, $0\leq i\in\zo$, such that $0=\lambda_{0}>\lambda_{1}>\cdots$ and $\lambda_{i}\rightarrow -\infty$ as 
$i\rightarrow\infty$. Moreover, if $\me_{i}$ is the eigenspace corresponding to $\lambda_{i}$, then $\me_{i}$ is finite dimensional and consists 
of smooth functions. In particular $\me_{0}$ is the set of constant functions. Finally, the set of eigenfunctions of $\Delta_{g}$ is a basis 
for $L^{2}(M)$. 
\end{thm}
\begin{remark}\label{remark:riemettomeasure}
Let $(M,g)$ be an oriented Riemannian manifold. There is an associated volume form $\mu_{g}$. This volume form defines a linear functional $\Lambda$
on $C_{0}(M,\co)$ (the set of complex valued continuous functions on $M$ with compact support):
\[
\Lambda(f)=\int_{M}f\mu_{g}.
\]
Using this linear functional, we can appeal to the Riesz Representation Theorem; cf., e.g., \cite[Theorem~2.14, p.~40]{rudin}. This yields the existence
of a positive measure $\lambda_{g}$ and a $\sigma$-algebra $\ma$ of subsets of $M$, such that $(M,\ma,\lambda_{g})$ is a complete measure space. Moreover, 
$\ma$ contains all the Borel sets and 
\[
\int_{M}f\lambda_{g}=\int_{M}f\mu_{g}
\]
for all $f\in C_{0}(M,\co)$. Given the measure $\lambda_{g}$, we can define $L^{p}(M)$. When we do so, the relevant underlying Riemannian metric should be
understood from the context, and we take it for granted that the functions in question are complex-valued. 
\end{remark}
\begin{remark}\label{remark:eigfeigvlab}
Since each of the eigenspaces is finite dimensional, 
\index{Eigenvalues!Laplace-Beltrami operator}%
\index{Eigenspaces!Laplace-Beltrami operator}%
\index{Laplace-Beltrami operator!Eigenvalues}%
\index{Laplace-Beltrami operator!Eigenspaces}%
there is a sequence $0=\bla_{0}\geq\bla_{1}\geq \cdots$, for $0\leq i\in\zo$, and corresponding 
orthonormal eigenfunctions $\varphi_{i}\in C^{\infty}(M)$, such that the sequence $\varphi_{i}$ is a basis for $L^{2}(M)$. Moreover, there is a sequence 
$0\leq \nu_{0}\leq \nu_{1}\leq\cdots$ such that $\bla_{i}=-\nu^{2}_{i}$. Note that $\nu_{i}\rightarrow\infty$ as $i\rightarrow\infty$. In what follows, we 
assume all closed manifolds to be connected. In that setting $\nu_{0}=0$ and $\nu_{1}>0$. 
\end{remark}
Let $(M,g)$ be a closed (connected and oriented) Riemannian manifold and $\varphi_{i}$ and $\nu_{i}$ be defined as in 
Remark~\ref{remark:eigfeigvlab}. Let $\mu_{g}$ be the volume form associated with $g$ and $\lambda_{g}$ be the measure defined by 
Remark~\ref{remark:riemettomeasure}. Define the inner product
\[
\ldr{u,v}_{g}:=\int_{M}uv^{*}\lambda_{g}
\]
\index{$\a$Aa@Notation!Bilinear forms!$\ldr{\cdot,\cdot}_{g}$}%
for $u,v\in L^{2}(M)$, where $*$ denotes complex conjugation; this turns $L^{2}(M)$ into a complex Hilbert space. For $u\in L^{2}(M)$ and $0\leq i\in\zo$, 
we also let $\hu(i):=\ldr{u,\varphi_{i}}_{g}$. The statement that $\varphi_{i}$ is a basis of $L^{2}(M)$, cf. Theorem~\ref{thm:specclriem}, is equivalent to 
the statement that 
\begin{equation}\label{eq:ultdecomp}
\int_{M}|u|^{2}\lambda_{g}=\sum_{i=0}^{\infty}|\hu(i)|^{2}
\end{equation}
for all $u\in L^{2}(M)$; cf. \cite[Theorem~4.18, p.~85]{rudin}. If $u\in C^{\infty}(M,\co)$ and $s\in\ro$, let
\begin{equation}\label{eq:Hsnormdef}
\|u\|_{(s)}:=\left(\sum_{i=0}^{\infty}\ldr{\nu_{i}}^{2s}|\hu(i)|^{2}\right)^{1/2}.
\end{equation}
If $s=k$, where $0\leq k\in\zo$, then this norm is equivalent to the standard $H^{k}(M)$ Sobolev norm associated with the metric $g$. From now on, we 
therefore loosely refer to the 
$\|\cdot\|_{(s)}$ as Sobolev norms. Moreover, in these notes, $H^{s}(M)$ is defined to be the set of complex valued distributions such that the right
hand side of (\ref{eq:Hsnormdef}) is finite, where the Riemannian metric should be understood from the context. 
\index{$\a$Aa@Notation!Function spaces!$H^{s}(M)$}%
We use notation similar to (\ref{eq:Hsnormdef}) if $u\in C^{\infty}(M,\cn{n})$; in that case, the $\hu(i)$ are computed component-wise.

\subsection{Spectral properties, the product setting}\label{ssection:specprodset}

Consider (\ref{eq:thesystemRge}), where the coefficients have the properties stated in connection with (\ref{eq:thesystemRge}). 
For $r=1,\dots,R$, we apply Remark~\ref{remark:eigfeigvlab} to the Laplace-Beltrami operator $\Delta_{g_{r}}$ on $(M_{r},g_{r})$. This yields 
orthonormal eigenfunctions $\varphi_{r,i}$, $0\leq i\in\zo$, of $\Delta_{g_{r}}$ and corresponding eigenvalues $-\nu_{r,i}^{2}$. Let $n\in\zn{d}$, 
$0\leq i_{r}\in\zo$, $r=1,\dots,R$, and 
\begin{equation}\label{eq:IgenReqdef}
\indexnot:=(n,i_{1},\dots,i_{R});
\end{equation}
\index{$\a$Aa@Notation!Frequencies!$\indexnot$}%
we denote the set of such $\indexnot$ by $\EFindexset$. 
\index{$\a$Aa@Notation!Frequency sets!$\EFindexset$}%
For each $\indexnot\in\EFindexset$, there is a uniquely associated $\nu$ given by 
\begin{equation}\label{eq:nugenReqdef}
\nu(\indexnot):=(n,\nu_{1,i_{1}},\dots,\nu_{R,i_{R}}).
\end{equation}
\index{$\a$Aa@Notation!Eigenvalues!$\nu(\indexnot)$}%
In case (\ref{eq:nugenReqdef}) holds, we also write, by slight abuse of notation, 
\begin{equation}\label{eq:nuroTetcdef}
\nu_{\roT}(\indexnot)=n,\ \ \
\nu_{\roT,j}(\indexnot)=n_{j},\ \ \
\nu_{r,i_{r}}(\indexnot)=\nu_{r,i_{r}}. 
\end{equation}
\index{$\a$Aa@Notation!Eigenvalues!$\nu_{\roT}(\indexnot)$}%
\index{$\a$Aa@Notation!Eigenvalues!$\nu_{\roT,j}(\indexnot)$}%
\index{$\a$Aa@Notation!Eigenvalues!$\nu_{r,i_{r}}(\indexnot)$}%
Let $x\in \tn{d}$ (where we think of $\tn{d}$ as $[0,2\pi]^{d}$ with the ends identified),
$p_{r}\in M_{r}$, $r=1,\dots,R$, and $p=(x,p_{1},\dots,p_{R})$. Given $\indexnot\in\EFindexset$, define $\varphi_{\indexnot}$ by
\begin{equation}\label{eq:varphinudef}
\varphi_{\indexnot}(p):=(2\pi)^{-d/2}e^{in\cdot x}\varphi_{1,i_{1}}(p_{1})\cdots \varphi_{R,i_{R}}(p_{R}).
\end{equation}
\index{$\a$Aa@Notation!Eigenfunctions!$\varphi_{\indexnot}$}%
Note that $\varphi_{\indexnot}$ is an eigenfunction of the operator $\Delta_{\roT}+\Delta_{1}+\dots+\Delta_{R}$, and the eigenvalue 
corresponding to $\varphi_{\indexnot}$ is $-|\nu(\indexnot)|^{2}$; here $\Delta_{\roT}$ is the standard Laplace operator on $\tn{d}$. Introduce the 
volume form $\mubox$ on $\bM$ by (\ref{eq:muboxdef}), where $\bM$ is given by (\ref{eq:bMdef}). Let $\lambdabox$ be obtained from $\mubox$ in 
the same way that $\lambda_{g}$ is obtained from $\mu_{g}$ in Remark~\ref{remark:riemettomeasure}. Using the measure $\lambdabox$, we can think
of $L^{2}(\bM)$ as a complex Hilbert space, in analogy with the above. We denote the corresponding inner product by
\begin{equation}\label{eq:ldrboxdef}
\ldrbox{u,v}:=\int_{\bM}uv^{*}\lambdabox
\end{equation}
\index{$\a$Aa@Notation!Bilinear forms!$\ldrbox{\cdot,\cdot}$}%
for $u,v\in L^{2}(\bM)$, where the star denotes complex conjugation. If $\indexnot_{a},\indexnot_{b}\in\EFindexset$, it is clear that 
$\ldrbox{\varphi_{\indexnot_{a}} ,\varphi_{\indexnot_{b}}}=0$ unless $\indexnot_{a}=\indexnot_{b}$. Moreover, 
$\ldrbox{\varphi_{\indexnot} ,\varphi_{\indexnot}}=1$ 
for $\indexnot\in\EFindexset$. Let $\indexnot$ be given by (\ref{eq:IgenReqdef}) and 
$u\in L^{2}(\bM)$. Define $\hu(\indexnot)$ by
\begin{equation}\label{eq:huindexnotdef}
\hu(\indexnot):=\ldrbox{u,\varphi_{\indexnot}}. 
\end{equation}
\index{$\a$Aa@Notation!Fourier coefficients!$\hu(\indexnot)$}%
Appealing successively to the fact that $e^{in\cdot x}$, $n\in\zn{d}$, is a basis of $L^{2}(\tn{d})$ and the fact that $\varphi_{r,i}$ is a basis of 
$L^{2}(M_{r})$, $r=1,\dots,R$, it can be verified that if $u\in C^{\infty}(\bM,\co)$, 
\begin{equation}\label{eq:parsevalonbM}
\int_{\bM}|u|^{2}\mubox=\sum_{\indexnot\in\EFindexset}|\hu(\indexnot)|^{2}.
\end{equation}
In particular, it is thus clear that $\varphi_{\indexnot}$, $\indexnot\in\EFindexset$, is a basis for $L^{2}(\bM)$. If $s\in\ro$ and 
$u\in C^{\infty}(\bM,\co)$, define
\begin{equation}\label{eq:HsnormonbM}
\|u\|_{(s)}:=\left(\sum_{\indexnot\in\EFindexset}\ldr{\nu(\indexnot)}^{2s}|\hu(\indexnot)|^{2}\right)^{1/2}.
\end{equation}
\index{$\a$Aa@Notation!Norms!$\normHs$}%
If $0\leq k\in\zo$, $\|\cdot\|_{(k)}$ is a norm which is equivalent to the standard Sobolev $H^{k}$-norms, and we therefore refer to the $\|\cdot\|_{(s)}$
as Sobolev norms or $H^{s}$-norms. We also define $H^{s}(\bM)$ as at the end of Subsection~\ref{ssection:laplacebeltramibasics}. We use a notation similar 
to (\ref{eq:HsnormonbM}) for $u\in C^{\infty}(\bM,\cn{n})$. It is also worth noting that a combination of (\ref{eq:parsevalonbM}) and polarisation type 
identities yields
\begin{equation}\label{eq:ldruvparsival}
\ldrbox{u,v}=\sum_{\indexnot\in\EFindexset}\hu(\indexnot)\hv(\indexnot)^{*};
\end{equation}
cf. \cite[Theorem~4.18, p.~85]{rudin}.

\subsection{Fourier decomposition of the main equation}\label{ssection:fourdecompmaineq}

Consider (\ref{eq:thesystemRge}). Assume the associated Lorentz manifold $(M,g)$ to be a canonical separable cosmological model manifold. 
Taking the inner product of (\ref{eq:thesystemRge}) with $\varphi_{\indexnot}$ (with respect to $\ldrbox{\cdot,\cdot}$ introduced in 
(\ref{eq:ldrboxdef})) yields
\begin{equation}\label{eq:fourierthesystemRgelong}
\begin{split}
\ddot{z}(\indexnot,t)+\mfg^{2}(\indexnot,t)z(\indexnot,t)-2\textstyle{\sum}_{l=1}^{d}in_{l}g^{0l}(t)\dot{z}(\indexnot,t) & \\
+\a(t)\dot{z}(\indexnot,t)+\textstyle{\sum}_{l=1}^{d}in_{l}X^{l}(t)z(\indexnot,t)+\zeta(t)z(\indexnot,t) & =\hf(\indexnot,t),
\end{split}
\end{equation}
where $n_{j}:=\nu_{\roT,j}(\indexnot)$, 
\begin{align}
\mfg(\indexnot,t) := & \left(\textstyle{\sum}_{j,l=1}^{d}g^{jl}(t)n_{j}n_{l}+\sum_{r=1}^{R}a_{r}^{-2}(t)\nu_{r,i_{r}}^{2}(\indexnot)\right)^{1/2}\label{eq:mfgnutdef}\\
z(\indexnot,t) := & \ldrbox{u(\cdot,t),\varphi_{\indexnot}},\label{eq:znutdef}\\
\hf(\indexnot,t) := & \ldrbox{f(\cdot,t),\varphi_{\indexnot}}.\label{eq:hfnutdef}
\end{align}
\index{$\a$Aa@Notation!Coefficients, Fourier side!$\mfg(\indexnot,t)$}%
\index{$\a$Aa@Notation!Fourier coefficients!$z(\indexnot,t)$}%
\index{$\a$Aa@Notation!Fourier coefficients!$\hf(\indexnot,t)$}%
For the sake of brevity, it is convenient to omit reference to the arguments $\indexnot$ and $t$ in (\ref{eq:fourierthesystemRgelong}), and
to use the Einstein summation convention for sums from $1$ to $d$. Then (\ref{eq:fourierthesystemRgelong}) can be written
\begin{equation}\label{eq:fourierthesystemRge}
\begin{split}
\ddot{z}+\mfg^{2}z-2in_{l}g^{0l}\dot{z} +\a\dot{z}+in_{l}X^{l}z+\zeta z & =\hf.
\end{split}
\end{equation}
Most of the arguments in these notes are devoted to a study of this equation. For $0\neq\indexnot\in\EFindexset$, it is often convenient to 
introduce the notation
\begin{equation}\label{eq:sigmaXdefintro}
\sigma(\indexnot,t):=\frac{n_{l}g^{0l}(t)}{\mfg(\indexnot,t)},\ \ \
X(\indexnot,t):=\frac{n_{l}X^{l}(t)}{\mfg(\indexnot,t)}.
\end{equation}
\index{$\a$Aa@Notation!Coefficients, Fourier side!$\sigma(\indexnot,t)$}%
\index{$\a$Aa@Notation!Coefficients, Fourier side!$X(\indexnot,t)$}%
Then (\ref{eq:fourierthesystemRge}) can be written
\begin{equation}\label{eq:fourierthesystemRgesigmaandX}
\ddot{z}+\mfg^{2}z-2i\sigma\mfg\dot{z} +\a\dot{z}+iX\mfg z+\zeta z=\hf.
\end{equation}

\subsection{Higher order energies}\label{ssection:highordenergies}

In Section~\ref{section:Questions} we introduce the basic energies of interest in these notes. However, it is also of interest to 
introduce higher order energies. In particular, we define
\begin{equation}\label{eq:mfedef}
\mfe_{s}[u](t):=\frac{1}{2}\sum_{\indexnot\in\EFindexset}\ldr{\nu(\indexnot)}^{2s}\left[|\dot{z}(\indexnot,t)|^{2}
+\mfg^{2}(\indexnot,t)|z(\indexnot,t)|^{2}+|z(\indexnot,t)|^{2}\right].
\end{equation}
\index{$\a$Aa@Notation!Energies!$\mfe_{s}$}%
Here $u\in C^{\infty}(M,\cn{m})$ and $z$ is defined by (\ref{eq:znutdef}). In what follows, we also use the notation $\mfe:=\mfe_{0}$.
Note that $\mfe=\mfe_{\robas}$, where $\mfe_{\robas}$ is defined in (\ref{eq:mfebasdef}). 

\section{The Jordan normal form}\label{section:jordannormalform}

Let $A\in\Mn{n}{\co}$. Then $A$ can be decomposed into Jordan normal form. 
\index{Jordan!normal form}%
This means that there is a matrix $T\in\Gl{n}{\co}$ such that $J:=T^{-1}AT$
is a block diagonal matrix. Moreover, the matrices on the diagonal are Jordan blocks; 
\index{Jordan!block}%
i.e., they are square matrices of the form 
\[
J_{\lambda,d}=\left(\begin{array}{cccccc} \lambda & 1 & 0 & \cdots & 0 & 0\\ 0 & \lambda & 1 & \cdots & 0 & 0\\ 
\vdots & \vdots & \vdots & \vdots & \vdots & \vdots\\
 0 & 0 & 0 & \cdots & \lambda & 1\\
 0 & 0 & 0 & \cdots & 0 & \lambda \end{array}\right).
\]
Equivalently, all the elements on the diagonal of $J_{\lambda,d}$ are $\lambda$, all the elements immediately above the diagonal are $1$, and all the remaining 
elements are $0$. Moreover, $d$ is the dimension of the Jordan block; i.e., $J_{\lambda,d}\in\Mn{d}{\co}$. Finally, $\lambda\in \Spe(A)$, where we use the notation
introduced in Definition~\ref{def:SpRspdef}. 

\textbf{Properties of Jordan blocks.} Before considering the matrix $A$ itself, it is useful to describe the properties of isolated Jordan blocks. Let us begin 
by computing the generalised eigenspace of $B:=J_{\lambda,d}$. Using the notation of Definition~\ref{def:defofgeneigenspintro}, it is clear that $n_{\lambda}=d$. 
Moreover $(B-\lambda\Id_{d})^{n_{\lambda}}=0$. This means that the generalised eigenspace of $B$, i.e. $E_{\lambda}$, equals $\cn{d}$. Next, $e^{Bt}$ is given by 
\begin{equation}\label{eq:matrixexponentialofJordanblock}
\exp(J_{\lambda,d}t)=e^{\lambda t}\exp[J_{0,d}t]=e^{\lambda t}\sum_{k=0}^{\infty}\frac{1}{k!}J_{0,d}^{k}t^{k}
=e^{\lambda t}\sum_{k=0}^{d-1}\frac{1}{k!}J_{0,d}^{k}t^{k};
\end{equation}
note that $J_{0,d}=J_{\lambda,d}-\lambda \Id_{d}$; that $J_{0,d}^{d-1}\neq 0$; that $J_{0,d}^{d}=0$; and that $J_{\lambda,d}$ commutes with $\lambda\Id_{d}$. In fact, 
the $k$'th power of the matrix $J_{0,d}$ has ones $k$ steps above the diagonal, and all the other elements are zero. In particular, for generic 
$v\in\cn{d}$, the leading order term in $\exp(J_{\lambda,d}t)v$ (as $t\rightarrow\infty$) has time dependence of the form $t^{d-1}\exp(\mathrm{Re}\{\lambda\}t)$. 

\textbf{Generalised eigenspaces of $A$.} In general, $J$ is of the form
\[
J=\diag\{J_{\lambda_{1},d_{1}},\dots,J_{\lambda_{m},d_{m}}\}.
\]
Note that the $\lambda_{i}$ need not be distinct. It is convenient to order the blocks. From now on, we assume that 
$\mathrm{Re}\{\lambda_{i}\}\geq \mathrm{Re}\{\lambda_{i+1}\}$ and that $d_{i}\geq d_{i+1}$ in case 
$\mathrm{Re}\{\lambda_{i}\}=\mathrm{Re}\{\lambda_{i+1}\}$. Assume now that $\lambda$ is an eigenvalue of $A$ and that $\lambda_{l},\dots,\lambda_{k}$ are all the 
$\lambda_{i}$ that equal $\lambda$, where we can assume that $1\leq l\leq k\leq m$. In that case, it is clear that $n_{\lambda}=d_{l}+\dots +d_{k}$.
Moreover, the generalised eigenspace of $J$ corresponding to $\lambda$ equals $\{0\}^{m_{a}}\times\cn{n_{\lambda}}\times\{0\}^{m_{b}}$, where 
$m_{a}=d_{1}+\dots +d_{l-1}$ and $m_{b}=d_{k+1}+\dots +d_{m}$. On the other hand, $v\in\cn{n}$ belongs to the generalised eigenspace of 
$J$ corresponding to $\lambda$ if and only if $Tv$ belongs to the generalised eigenspace of $A$ corresponding to $\lambda$. This allows us to calculate the 
generalised eigenspace of $A$ corresponding to $\lambda$. 

\textbf{Matrix exponentials.} In order to calculate $e^{At}$, given the above decomposition into Jordan normal form, it is sufficient to note that 
\[
e^{At}=Te^{Jt}T^{-1}=T\diag\{\exp(J_{\lambda_{1},d_{1}}t),\dots,\exp(J_{\lambda_{m},d_{m}}t)\}T^{-1}.
\]
Here $\exp(J_{\lambda_{i},d_{i}}t)$ can be calculated as in (\ref{eq:matrixexponentialofJordanblock}). In particular, it is clear that for generic 
$v\in\cn{d}$, the leading order term in $e^{At}v$ (as $t\rightarrow\infty$) has a time dependence of the form $t^{d_{1}-1}\exp(\mathrm{Re}\{\lambda_{1}\}t)$.
This motivates the introduction of $\kappa_{\max}$ and $\lambda_{\max}$ in Definition~\ref{def:SpRspdef}. In fact, 
$\kappa_{\max}(A)=\mathrm{Re}\{\lambda_{1}\}$ and $d_{\max}(A,\kappa_{\max})=d_{1}$ by construction. 

\textbf{First generalised eigenspace in the $\b$, $A$-decomposition of $\cn{n}$.} In these notes, we often want to solve a system of ODE's for which 
there is a leading order behaviour corresponding to a model equation $\dot{x}=Ax$. However, there are also error terms. As a consequence, it is
natural to divide the solutions to the model equation into a leading order part, which can be distinguished from the error terms, and a remainder, which 
cannot. This division can be made already on the level of the generalised eigenspaces of $A$, and this is the purpose of 
Definition~\ref{def:defofgeneigenspintro}. We already know that the leading order behaviour of solutions is $t^{d_{1}-1}\exp(\mathrm{Re}\{\lambda_{1}\}t)$.
The cut-off, distinguishing between the leading order part and the remainder, is characterised by a positive real number, say $\b>0$. Roughly speaking
we want to keep the part of the solution corresponding to eigenvalues with real part strictly larger than $\mathrm{Re}\{\lambda_{1}\}-\beta$. The remaining
part of the solution we do not want to keep. On the transformed, Jordan block, side, we thus want to focus on vectors $v\in\cn{n}$ such that the components
of $v$ corresponding to the Jordan blocks $J_{\lambda_{i},d_{i}}$ vanish if $\mathrm{Re}\{\lambda_{i}\}\leq \mathrm{Re}\{\lambda_{1}\}-\beta$. On the original, 
untransformed, side, this corresponds to focusing on $v$ in the first generalised eigenspace in the $\b$, $A$-decomposition of $\cn{n}$. On the Jordan block 
side, what remains, after removing the leading order part, is vectors $v\in\cn{n}$ such that the components of $v$ corresponding to the Jordan blocks 
$J_{\lambda_{i},d_{i}}$ vanish if $\mathrm{Re}\{\lambda_{i}\}>\mathrm{Re}\{\lambda_{1}\}-\beta$. On the original, untransformed, side, this corresponds to focusing 
on $v$ in the second generalised eigenspace in the $\b$, $A$-decomposition of $\cn{n}$; cf. Definition~\ref{def:fssubspetc}.

\chapter{Geometric preliminaries}\label{chapter:geometry}

The purpose of the present part of these notes is to derive basic conclusions concerning 
the geometry that arises in the study of equations of the form (\ref{eq:thesystemRge}). Moreover, since the results of 
Parts~\ref{part:roughansiltrs}--\ref{part:nondegcabeq} are based on analytical conditions, we need to relate these analytical conditions with the 
geometric conditions of Part~\ref{part:introduction}.

\section{Metrics of interest, lapse and shift decomposition}\label{section:basgeomglobhyp}

Consider (\ref{eq:thesystemRge}). By assumption, $g^{00}(t)<0$ and $g^{ij}(t)$, $i,j\in \{1,\dots,d\}$, are the components of a positive definite matrix 
for all $t\in I$. As a consequence, there are smooth functions $g_{\a\b}$, $\a,\b\in\{0,\dots,d\}$, from $I$ to $\ro$ such that $g_{\a\b}(t)$ are the 
components of the inverse of the matrix with components $g^{\a\b}(t)$. Moreover, $g_{00}(t)<0$ and $g_{ij}(t)$ are the components of a positive definite 
matrix for all $t$. These statements are a consequence of \cite[Lemma~8.5, p.~72]{minbok}. In \cite[Section~8.1]{minbok}, it is also demonstrated that 
if $g_{00}(t)<0$ and $g_{ij}(t)$ are the components of a positive definite matrix, then the matrix with components $g_{\a\b}(t)$ has one strictly negative and 
$d$ strictly positive eigenvalues. Starting with the equation (\ref{eq:thesystemRge}), we thus arrive at the Lorentz metric (\ref{eq:gfromthesystemRgeintro}). 
Next, let us verify that $(M,g)$ is a globally hyperbolic Lorentz manifold. 

\begin{lemma}\label{lemma:globallyhyperbol}
Let $(M,g)$ be a separable cosmological model manifold in the sense of Definition~\ref{def:sepcosmmodmanintro}. Then 
$(M,g)$ is a globally hyperbolic Lorentz manifold, and each hypersurface $\bM_{t}$, defined by (\ref{eq:bMtintro}) for $t\in I$, is a Cauchy hypersurface. 
\end{lemma}
\begin{proof}
It is sufficient to prove that the $\bM_{t}$, $t\in I$, are Cauchy hypersurfaces. Let $\g$ be an inextendible causal curve and let $\g^{0}$ denote the 
time component of $\g$. Then, letting $\pi_{0}:M\rightarrow I$ denote projection onto the last factor,
\begin{equation}\label{eq:dgzdtpfhyp}
\frac{d\g^{0}}{ds}=\frac{d}{ds}\pi_{0}\circ\g=\g' (\pi_{0})=d\pi_{0}(\g')=\ldr{\grad\pi_{0},\g'}.
\end{equation}
However, 
\[
\ldr{\grad\pi_{0},\grad\pi_{0}}=g_{\a\b}g^{0\a}g^{0\b}=g^{00}<0,
\]
so that $\grad\pi_{0}$ is timelike. For this reason, the right hand side of (\ref{eq:dgzdtpfhyp}) is either always strictly positive or always strictly
negative. In particular, $\g^{0}$ is strictly monotone, so that $\g$ can intersect a given $\bM_{t}$ at most once. Moreover, there are two possibilities: 
either the range of $\g^{0}$ equals $I$, or it does not. In order to demonstrate that the second case leads to a contradiction, assume that the range of 
$\g^{0}$ is contained in $(-\infty,t_{0})$, $t_{0}\in I$, and that $\g^{0}$ is strictly increasing (there are of course other cases to consider, but 
they can be dealt with in a similar fashion). If the domain of $\g$ is $J=(s_{-},s_{+})$, then $\g^{0}(s)$ converges to some $t_{1}$ as $s\rightarrow s_{+}$.
Moreover, $t_{1}\in I$. Let $\pi_{1}:M\rightarrow\bM$ denote the projection onto the first factor and $\bga:=\pi_{1}\circ\g$. If we could prove that $\bga(s)$
converges as $s\rightarrow s_{+}$, then we would know that $\g$ converges as $s\rightarrow s_{+}$. This would imply that $\g$ is extendible and contradict the 
assumptions (and thereby finish the proof). In order to prove that $\bga$ converges, note that due to the causality of $\g$, 
\[
g_{00}\dot{\g}^{0}\dot{\g}^{0}+2g_{0i}\dot{\g}^{0}\dot{\g}^{i}+g_{ij}\dot{\g}^{i}\dot{\g}^{j}+\textstyle{\sum}_{r}a_{r}^{2}g_{r}(\dot{\g}_{r},\dot{\g}_{r})\leq 0,
\]
where $\g_{r}$ denotes the $M_{r}$-component of $\g$. Using this estimate; the fact that for $s_{0}\in J$, $\g^{0}(s)$ is contained in a compact subset of $I$
for $s\in [s_{0},s_{+})$; and the fact that $\g^{0}$ converges, it can be argued that for any $s_{n}\rightarrow s_{+}$, $\bga(s_{n})$ is a Cauchy sequence with 
respect to the metric induced by
\begin{equation}\label{eq:metricconvglobhyp}
\textstyle{\sum}_{i,j}\de_{ij}dx^{i}\otimes dx^{j}+\sum_{r}g_{r}.
\end{equation}
Since (\ref{eq:metricconvglobhyp}) is a complete metric on $\bM$, we conclude that $\bga$ converges. This yields the desired contradiction. Thus the range of $\g^{0}$ 
equals $I$, and $\g^{0}$ is strictly increasing. In particular, given an inextendible causal curve $\g$, each $\bM_{t}$ is intersected once and only once by $\g$, so 
that each $\bM_{t}$ is a Cauchy hypersurface. 
\end{proof}

\subsection{Lapse and shift}\label{ssection:lapseandshift}
Let $(M,g)$ be a separable cosmological model manifold in the sense of Definition~\ref{def:sepcosmmodmanintro}. The first four terms on the right hand side of 
(\ref{eq:sepcosmodmeintro}) can be written
\begin{equation}\label{eq:hdefinclash}
\begin{split}
h := & g_{00}dt\otimes dt+g_{0i}dt\otimes dx^{i}+g_{i0}dx^{i}\otimes dt+g_{ij}dx^{i}\otimes dx^{j}\\
 = & -N^{2}dt\otimes dt+g_{ij}(\chi^{i}dt+dx^{i})\otimes (\chi^{j}dt+dx^{j}),
\end{split}
\end{equation}
where the $g_{ij}$ are the same functions as in (\ref{eq:sepcosmodmeintro}) and $\chi^{i}$ and $N>0$ are determined by the relations 
\begin{equation}\label{eq:shiftlapsedef}
g_{ij}\chi^{j}=g_{0i},\ \ \
N:=(-g_{00}+g_{ij}\chi^{i}\chi^{j})^{1/2}.
\end{equation}
Since the $g_{ij}$ are the components of a positive definite matrix, the first equation appearing in (\ref{eq:shiftlapsedef}) can be solved
in order to yield smooth functions $\chi^{i}$. Given $\chi^{i}$, the second equality in (\ref{eq:shiftlapsedef}) defines the strictly positive
function $N$. Note that the $\chi^{i}$ can be thought of as the components of a vector field $\chi=\chi^{i}\d_{i}$ which we refer to as the 
\textit{shift vector field}. The function $N$ is referred to as the \textit{lapse function}. On the other hand, if we start with a lapse
function, a shift vector field and the $g_{ij}$'s (which we assume to be the components of a positive definite matrix for every $t\in I$),
then we have to assume that 
\begin{equation}\label{eq:lapseshiftineq}
g_{ij}\chi^{i}\chi^{j}<N^{2}
\end{equation}
in order for $g_{00}$ (now considered to be defined by the second equality in (\ref{eq:hdefinclash})) to satisfy $g_{00}<0$.

\section{The components of the inverse}\label{section:confrescmetriccomofinv}

Let $(M,g)$ be a separable cosmological model manifold in the sense of Definition~\ref{def:sepcosmmodmanintro}, and define $h$ by 
(\ref{eq:hdefinclash}). Given $t\in I$, we can consider $h_{\a\b}(t)$, $\a,\b=0,\dots,d$, to be the components of an element of $\Mn{d+1}{\ro}$.
It is of interest to calculate the components of the inverse in terms of the the lapse function, the shift vector field and $h_{ij}$; 
considering (\ref{eq:hdefinclash}), we can consider these three objects to define $h$. Defining $\bh$ by 
\begin{equation}\label{eq:bhdef}
\bh:=h_{ij}dx^{i}\otimes dx^{j},
\end{equation}
we denote the components of $\bh$ by $\bh_{ij}=h_{ij}=g_{ij}$; $i,j=1,\dots,d$. For $t\in I$, $\bh_{ij}(t)$, $i,j=1,\dots,d$, are the components of a 
positive definite, symmetric element of $\Mn{d}{\ro}$. We denote the components of the inverse by $\bh^{ij}(t)$, $i,j=1,\dots,d$. 

\begin{lemma}\label{lemma:invitoshiftetc}
Let $(M,g)$ be a separable cosmological model manifold in the sense of Definition~\ref{def:sepcosmmodmanintro}, and define $h$ by 
(\ref{eq:hdefinclash}). Fix $t\in I$ and consider $h_{\a\b}(t)$, $\a,\b=0,\dots,d$, to be the components of an element of $\Mn{d+1}{\ro}$. 
This element is invertible. Denote the components of the inverse by $h^{\a\b}(t)$, $\a,\b=0,\dots,d$. Then, if the lapse function $N$ 
satisfies $N=1$,
\begin{equation}\label{eq:hinvcomform}
h^{00}=-1,\ \ \
h^{0i}=\chi^{i},\ \ \
h^{ij}=\bh^{ij}-\chi^{i}\chi^{j},
\end{equation}
where $\bh^{ij}$ is defined immediately prior to the statement of the lemma. Moreover, for $t\in I$, $h^{ij}(t)$, $i,j=1,\dots,d$, are
the components of a positive definite matrix. Finally, 
\[
\det h=-\det\bh,
\]
where $h$ and $\bh$ are considered to be matrix valued functions taking $t\in I$ to the elements of $\Mn{d+1}{\ro}$ and $\Mn{d}{\ro}$ 
with components $h_{\a\b}(t)$, $\a,\b=0,\dots,d$, and $\bh_{ij}(t)$, $i,j=1,\dots,d$, respectively. 
\end{lemma}
\begin{remark}\label{remark:huzzformgenerallapse}
In case $N\neq 1$, we can apply the lemma to the metric $\hh:=N^{-2}h$. This yields
\begin{equation}\label{eq:hinvcomformnonnorm}
h^{00}=-N^{-2},\ \ \
h^{0i}=N^{-2}\chi^{i},\ \ \
h^{ij}=\bh^{ij}-N^{-2}\chi^{i}\chi^{j},\ \ \
\det h=-N^{2}\det\bh.
\end{equation}
\end{remark}
\begin{proof}
The invertibility follows from \cite[Lemma~8.5, p.~72]{minbok}. 
Fix $t\in I$ and let $a:=h_{00}(t)$, $v\in\rn{d}$ be the vector with components $v_{i}=h_{0i}(t)$ and $\bH\in\Mn{d}{\ro}$ be the matrix with 
components $h_{ij}(t)$, $i,j=1,\dots,d$. Then the matrix with components $h_{\a\b}(t)$, $\a,\b=0,\dots,d$, can be written
\[
H:=\left(\begin{array}{cc}a & v^{t} \\ v & \bH \end{array}\right). 
\]
Thus
\begin{equation}\label{eq:Hfactorisation}
\left(\begin{array}{cc} 1 & -v^{t}\bH^{-1} \\ 0 & \Id_{d} \end{array}\right)H\left(\begin{array}{cc} 1 & 0 \\ -\bH^{-1}v & \Id_{d} \end{array}\right)
=\left(\begin{array}{cc}a-v^{t}\bH^{-1}v & 0 \\ 0 & \bH \end{array}\right).
\end{equation}
Note that the components of $\bH^{-1}v$ are given by 
\[
\bh^{ij}(t)h_{0j}(t)=\bh^{ij}(t)g_{jl}(t)\chi^{l}(t)=\chi^{i}(t),
\]
and that 
\begin{equation}\label{eq:auxshiftform}
\begin{split}
a-v^{t}\bH^{-1}v = & h_{00}(t)-\bh^{ij}(t)h_{0i}(t)h_{0j}(t)\\
 = & h_{00}(t)-\bh_{ij}(t)\chi^{i}(t)\chi^{j}(t)=-1;
\end{split}
\end{equation}
in the derivation of these equalities, we appealed to (\ref{eq:hdefinclash}) and the fact that $N=1$. Finally, the components of $\bH^{-1}v v^{t}\bH^{-1}$
are given by 
\[
\bh^{il}(t)v_{l}v_{k}\bh^{jk}(t)=\chi^{i}(t)\chi^{j}(t).
\]
Combining these observations with (\ref{eq:Hfactorisation}) yields (\ref{eq:hinvcomform}).
In order to prove that $h^{ij}(t)$ are the components of a positive definite matrix, let $A$ be a positive definite symmetric matrix such that 
$A^{2}=\bH$, let $x\in\rn{d}$ be such that $x^{i}=\chi^{i}(t)$ and let $w\in\rn{d}$. Then 
\begin{equation}\label{eq:chiiviabsest}
|\chi^{i}(t)w_{i}|=|x\cdot w|=|A^{-1}w\cdot Ax|\leq |A^{-1}w||Ax|=|x|_{\bH}|w|_{\bH^{-1}},
\end{equation}
where $|w|_{B}=(B_{ij}w^{i}w^{j})^{1/2}$ for $w\in\rn{d}$ and $B\in\Mn{d}{\ro}$ symmetric and positive semi-definite. Moreover, $|x|_{\bH}<1$, since
$h_{00}<0$ (by assumption) and $N=1$. Thus, if $w\neq 0$ and $x\neq 0$, 
\begin{equation}\label{eq:hinvijposdef}
\chi^{i}(t)\chi^{j}(t)w_{i}w_{j}\leq |x|_{\bH}^{2}|w|_{\bH^{-1}}^{2}<\bh^{lm}(t)w_{l}w_{m}.
\end{equation}
Combining this estimate with (\ref{eq:hinvcomform}), it is clear that $h^{ij}(t)$ are the components of a positive definite matrix. The last statement of 
the lemma is an immediate consequence of (\ref{eq:Hfactorisation}) and (\ref{eq:auxshiftform}). 
\end{proof}

\section{The second fundamental form}

The main purpose of the present section is to derive some basic properties of the second fundamental form. Before proceeding, let us introduce coordinates 
adapted to the structure of the metric. 
\begin{definition}\label{def:cancoord}
Let $(\mO_{\roT},\bx_{\roT})$ be standard local coordinates on $\tn{d}$, $(\mO_{r},\bx_{r})$ be local coordinates on $M_{r}$ and 
$\mO:=\mO_{\roT}\times \mO_{1}\times\cdots\times \mO_{R}\times I$. Let $x$ be the coordinates mapping
$p=(\xi_{\roT},\xi_{1},\dots,\xi_{R},t)\in \mO$ to $[t,\bx_{\roT}(\xi_{\roT}),\bx_{1}(\xi_{1}),\dots,\bx_{R}(\xi_{R})]$. Then $(\mO,x)$ are 
referred to as \textit{canonical local coordinates} 
\index{Canonical!local coordinates}%
\index{Local coordinates!canonical}%
on $M$. 
\end{definition}
\begin{remarks}
By the statement that $(\mO_{\roT},\bx_{\roT})$ are standard local coordinates on $\tn{d}$, we mean that $\d/\d\bx_{\roT}^{1},\dots,\d/\d\bx_{\roT}^{d}$ are the standard 
vector fields on $\tn{d}$. The components of $x$ are labelled $x^{\a}$, where $\a=0,\dots,D$ and $D$ is the dimension of the manifold $\bM$ defined
by (\ref{eq:bMdef}). In particular, $\d_{x^{i}}$, $i=1,\dots,d$, are tangent to the torus. 
\end{remarks}
Turning to the second fundamental form, note that 
\[
U:=-(-g^{00})^{-1/2}g^{0\mu}\d_{\mu}=\frac{1}{N}(\d_{t}-\chi^{i}\d_{i})
\]
is the future directed unit normal to the hypersurfaces $\bM_{t}$. If $\bk$ denotes the 
second fundamental form of $\bM_{t}$, $(\mO,x)$ are canonical local coordinates on $M$ (cf. Definition~\ref{def:cancoord}) and $\d_{i},\d_{j}$ 
denote coordinate vector fields corresponding to $x$ which are tangent to $\bM_{t}$, then 
\begin{equation}\label{eq:bkcomp}
\begin{split}
\bk_{ij} := & \bk(\d_{i},\d_{j})=\ldr{\nabla_{\d_{i}}U,\d_{j}}=-(-g^{00})^{-1/2}g^{0\mu}\ldr{\nabla_{\d_{i}}\d_{\mu},\d_{j}}\\
 = & -(-g^{00})^{-1/2}g^{0\mu}\frac{1}{2}(\d_{i}g_{\mu j}+\d_{\mu}g_{ij}-\d_{j}g_{i\mu})=\frac{1}{2}(-g^{00})^{1/2}\d_{t}g_{ij}.
\end{split}
\end{equation}
In the last step, we used the following facts: if $\a=0,\dots,d$, then $\d_{i}g_{\a\mu}=0$ for all $i=1,\dots,D$ and all $\mu=0,\dots,D$; 
$g^{0i}=g_{0i}=0$ if $i=d+1,\dots,D$; and $\d_{i}g_{\mu\nu}=0$ for all $i=1,\dots,d$ and all $\mu,\nu=0,\dots,D$. Combining (\ref{eq:hinvcomformnonnorm}) and 
(\ref{eq:bkcomp}) yields
\begin{equation}\label{eq:bkijform}
N\bk_{ij}=\frac{1}{2}\d_{t}g_{ij}.
\end{equation}
Due to this equality, it follows that 
\begin{equation}\label{eq:bkblock}
N\bk=\frac{1}{2}\sum_{i,j=1}^{d}\d_{t}\bh_{ij}dx^{i}\otimes dx^{j}+\sum_{r=1}^{R}\frac{\dot{a}_{r}}{a_{r}}a_{r}^{2}\pi_{r}^{*}g_{r},
\end{equation}
where $\pi_{r}:\bM_{t}\rightarrow M_{r}$ denotes the projection onto $M_{r}$; note that the metric $\bge$ induced on $\bM_{t}$ by $g$ can be written
\begin{equation}\label{eq:bgblock}
\bge=\textstyle{\sum}_{i,j=1}^{d}\bh_{ij}dx^{i}\otimes dx^{j}+\sum_{r=1}^{R}a_{r}^{2}\pi_{r}^{*}g_{r}.
\end{equation}
Due to (\ref{eq:bkblock}) and (\ref{eq:bgblock}), it is clear that $\bk$ has a block structure, and that this block structure corresponds to a block 
structure in $\bge$. It is therefore natural to introduce 
\[
\bk_{\roT}:=\frac{1}{2N}\sum_{i,j=1}^{d}\d_{t}\bh_{ij}dx^{i}\otimes dx^{j},\ \ \
\bk_{r}:=\frac{1}{N}\frac{\dot{a}_{r}}{a_{r}}a_{r}^{2}\pi_{r}^{*}g_{r},\ \ \
\bk_{\roRie}:=\sum_{r=1}^{R}\bk_{r}. 
\]
Taking the trace of these tensor fields with respect to $\bge$ yields 
\begin{equation}\label{eq:thetaroTrdef}
\theta_{\roT}:=\tr_{\bge}\bk_{\roT}=\frac{1}{2N}\bh^{ij}\d_{t}\bh_{ij},\ \ \
\theta_{r}:=\tr_{\bge}\bk_{r}=\frac{D_{r}}{N}\frac{\dot{a}_{r}}{a_{r}},
\end{equation}
where $D_{r}$ is the dimension of $M_{r}$. For future reference, it is convenient to note that 
\begin{equation}\label{eq:dtlndetbh}
\d_{t}\ln\det\bh=\bh^{ij}\d_{t}\bh_{ij}=2N\tr_{\bge}\bk_{\roT}=2N\theta_{\roT}.
\end{equation}
Summing up the above observations yields
\begin{equation}\label{eq:thetadecomp}
\theta:=\tr_{\bge}\bk=\frac{1}{N}\d_{t}\ln\sqrt{\det\bh}+\frac{D_{1}}{N}\frac{\dot{a}_{1}}{a_{1}}+\dots+\frac{D_{R}}{N}\frac{\dot{a}_{R}}{a_{R}}.
\end{equation}
For future reference, it is also of interest to note that 
\begin{equation}\label{eq:bknormbgsq}
|\bk|_{\bge}^{2}=|\bk_{\roT}|_{\bh}^{2}+|\bk_{\roRie}|_{\bge}^{2}=|\bk_{\roT}|_{\bh}^{2}+\sum_{r=1}^{R}\frac{D_{r}}{N^{2}}\left(\frac{\dot{a}_{r}}{a_{r}}\right)^{2}.
\end{equation}
Similarly, if $N=1$, 
\begin{equation}\label{eq:mlUbknormbgsq}
|\ml_{U}\bk|_{\bge}^{2}=|(\ml_{U}\bk)_{\roT}|_{\bh}^{2}+\sum_{r=1}^{R}D_{r}
\left[\frac{\ddot{a}_{r}}{a_{r}}+\left(\frac{\dot{a}_{r}}{a_{r}}\right)^{2}\right]^{2}.
\end{equation}

\section{Geometric convergence}\label{section:geometricconvergence}

Let $(M,g)$ be a separable cosmological model manifold in the sense of Definition~\ref{def:sepcosmmodmanintro}. In what follows, we 
use the terminology introduced in Definition~\ref{def:bkbgeV}. Moreover, we focus on manifolds such that Definitions~\ref{def:futureexpetcintro}
and \ref{definition:futbdfutconvgeometryintro} are satisfied. The purpose of the present section is to prove Lemma~\ref{lemma:bdandconvtocanonicalform}.
There is a similar result in the case of big crunch asymptotics. However, we leave the statement and proof of the corresponding result to the 
reader. 

\begin{proof}[Lemma~\ref{lemma:bdandconvtocanonicalform}]
To begin with, we rescale $g$ by $e^{2\lambda}$, for a function $\lambda$ (depending only on $t$) yet to be determined. This yields $\hg=e^{2\lambda}g$. Letting
$\hU:=e^{-\lambda}U$, we then have
\[
\hk_{ij}=\frac{1}{2}\hU \hg_{ij}=\frac{1}{2}e^{-\lambda}U(e^{2\lambda}g_{ij})=\hU(\lambda)\hg_{ij}+e^{\lambda}\bk_{ij},
\]
where $\hk$ is the second fundamental form induced on $\bM_{t}$ by $\hg$. Raising the indices of this equality with $\chg$ (the metric induced on
$\bM_{t}$ by $\hg$) yields
\begin{equation}\label{eq:hkuidj}
\hk^{i}_{\phantom{i}j}=\hU(\lambda)\de^{i}_{j}+e^{-\lambda}\bk^{i}_{\phantom{i}j}.
\end{equation}
Define $\lambda$ by $e^{-\lambda}\tr_{\bge}\bk=1$. Then, 
\[
\hU(\lambda)=e^{-\lambda}U(\lambda)=-U(e^{-\lambda})=-U[(\tr_{\bge}\bk)^{-1}],
\]
which is bounded by assumption. Thus $|\hk|_{\chg}$ is future bounded, as stated. 
Turning to the time coordinate $\tau$, it is clear by the assumptions that the interval $[t_{0},t_{+})$ corresponds to $[0,\infty)$
in $\tau$-time. Moreover, due to (\ref{eq:UitoNchiintro}) and the adjacent comments, 
\[
e^{\lambda}=\tr_{\bge}\bk=U(\tau)=N^{-1}\d_{t}\tau.
\]
In particular, $Ne^{\lambda}dt=d\tau$, and $\hg$ takes the form (\ref{eq:hgitotauintro}).
Combining the assumption of convergence with (\ref{eq:hkuidj}) and the fact that $e^{\lambda}=\tr_{\bge}\bk$ yields the conclusion 
that there is a $2$-tensor field $\hA$ of mixed type such that 
\[
|\hK-\hA|_{\varrho}\leq C\exp[-\eta |\ln V(t)|]
\]
for all $t\geq t_{0}$. Due to the definition of $\tau$, this estimate implies (\ref{eq:bkdbtrkconvwithratehverintro}) and yields the conclusion
of the lemma. 
\end{proof}

Due to this result, we, from now on, focus on separable cosmological model manifolds such that the lapse function equals $1$ and 
$I$ contains $[0,\infty)$.

\section{Norms}\label{section:normstensorsapp}

Before proceeding, it is convenient to make some observations of a purely linear algebra nature.

\begin{lemma}\label{lemma:normsofmultilinearfun}
Let $V$ be a finite dimensional real vector space; let $V^{*}$ denote the dual; and let $g$ be an inner product
on $V$. For $v\in V$, let $|v|_{g}:=[g(v,v)]^{1/2}$. Let $0\leq k,l\in\zo$, $k+l\geq 1$, and 
\[
T:V^{k}\times (V^{*})^{l}\rightarrow\ro
\]
be a multilinear map. Let $\{e_{i}\}$ be an orthonormal basis of $V$ and let $\{e^{i}\}$ denote the dual basis. Define $|T|_{g}$ by
\begin{equation}\label{eq:TnormgdefONver}
|T|_{g}:=\left(\textstyle{\sum}|T(e_{i_{1}},\dots,e_{i_{k}},e^{j_{1}},\dots,e^{j_{l}})|^{2}\right)^{1/2}.
\end{equation}
Then $|T|_{g}$ is independent of the choice of orthonormal basis. Moreover, if $v_{i}\in V$, $i=1,\dots,k$, and $\eta^{i}\in V^{*}$,
$i=1,\dots,l$, then 
\begin{equation}\label{eq:Tvodotvletcest}
|T(v_{1},\dots,v_{k},\eta^{1},\dots,\eta^{l})|\leq |T|_{g}|v_{1}|_{g}\cdots|v_{k}|_{g}|\eta^{1}|_{g}\cdots|\eta^{l}|_{g}.
\end{equation}
\end{lemma}
\begin{remark}
By an \textit{inner product on} $V$, we mean a positive definite symmetric bilinear form on $V$.
\end{remark}
\begin{remark}
Given $v\in V$, $v^{\flat}\in V^{*}$ is defined by $v^{\flat}(w)=g(v,w)$ for all $w\in V$. With this notation, if $\{e_{i}\}$ is an orthonormal
basis, then $e_{i}^{\flat}=e^{i}$ for all $i$. Note that $\flat$ is an isomorphism from $V$ to $V^{*}$. The inverse is denoted by $\sharp$; i.e.,
$(v^{\flat})^{\sharp}=v$. Note, also, that $g$ induces an inner product on $V^{*}$, denoted $g^{\sharp}$ and defined by 
$g^{\sharp}(\eta,\xi):=g(\eta^{\sharp},\xi^{\sharp})$ for all $\eta,\xi\in V^{*}$. In (\ref{eq:Tvodotvletcest}), $|\eta^{i}|_{g}$ should be thought of 
as being defined by (\ref{eq:TnormgdefONver}). Note, however, that if $\{e_{i}\}$ is an ON basis, $\{e^{i}\}$ is the dual basis and $\eta_{i}e^{i}=\eta\in V^{*}$, 
then $\eta(e_{i})=\eta_{i}$, $\{e^{i}\}$ is an ON-basis of $V^{*}$ with respect to $g^{\sharp}$, and 
\[
g^{\sharp}(\eta,\eta)=\eta_{i}\eta_{j}g^{\sharp}(e^{i},e^{j})=\textstyle{\sum}\eta_{i}^{2}=\textstyle{\sum}|\eta(e_{i})|^{2}=|\eta|_{g}^{2},
\] 
so that $|\eta|_{g}=[g^{\sharp}(\eta,\eta)]^{1/2}$. 
\end{remark}
\begin{proof}
We leave it to the reader to verify that $|T|_{g}$ is independent of the choice of ON basis. In order to 
prove (\ref{eq:Tvodotvletcest}) in the case that $k=2$ and $l=0$, let $v,w\in V$; $\{e_{i}\}$ be an ON basis; and and define $v^{i},w^{i}$
by $v=v^{i}e_{i}$ and $w=w^{i}e_{i}$. Then
\[
|v|_{g}=[g(v,v)]^{1/2}=\left(\textstyle{\sum}_{i}(v^{i})^{2}\right)^{1/2}.
\]
Moreover,
\begin{equation}
\begin{split}
|T(v,w)| = & \left|\textstyle{\sum}_{i,j}T(e_{i},e_{j})v^{i}w^{j}\right|
 \leq \left(\textstyle{\sum}_{j}\left[\textstyle{\sum}_{i}T(e_{i},e_{j})v^{i}\right]^{2}\right)^{1/2}|w|_{g}\\
 \leq & \left(\textstyle{\sum}_{i,j}[T(e_{i},e_{j})]^{2}\right)^{1/2}|v|_{g}|w|_{g}=|T|_{g}|v|_{g}|w|_{g}. 
\end{split}
\end{equation}
The proof of the general case is similar and left to the reader. 
\end{proof}

Under the assumptions of Lemma~\ref{lemma:normsofmultilinearfun}, let $\{e_{i}\}$ be a basis of $V$ which is not necessarily orthonormal. 
Let $\{e^{i}\}$ be the dual basis; $g_{ij}:=g(e_{i},e_{j})$; and $g^{ij}:=g^{\sharp}(e^{i},e^{j})$. Let $G^{ij}$ be the components of the inverse 
of the metric with components $g_{ij}$. Note that there are $a^{ij}\in\ro$ such that $(e^{i})^{\sharp}=a^{ij}e_{j}$. On the other hand, 
\begin{align*}
g[(e^{i})^{\sharp},e_{j}] = & [(e^{i})^{\sharp}]^{\flat}(e_{j})=e^{i}(e_{j})=\de^{i}_{j},\\
g[(e^{i})^{\sharp},e_{j}] = & g(a^{ik}e_{k},e_{j})=a^{ik}g_{kj}.
\end{align*}
Thus $a^{ij}=G^{ij}$. Moreover, 
\[
g^{ij}=g^{\sharp}(e^{i},e^{j})=g[(e^{i})^{\sharp},(e^{j})^{\sharp}]=G^{ik}G^{jl}g_{kl}=G^{ij}. 
\]
Let $A$ and $B$ be the symmetric positive definite matrices defined by the conditions that 
\[
g_{ij}=\textstyle{\sum}_{l}B_{il}B_{lj},\ \ \
g^{ij}=\textstyle{\sum}_{l}A^{il}A^{lj}. 
\]
Then $A$ is the inverse of $B$ and $f_{i}:=A^{ij}e_{j}$ defines an orthonormal basis whose dual basis is given by $f^{i}=B_{ik}e^{k}$ (we leave 
the verification of these statements to the reader). Let now $T$ have the properties stated in Lemma~\ref{lemma:normsofmultilinearfun}.
Introduce the notation 
\[
T_{i_{1}\cdots i_{k}}^{j_{1}\cdots j_{l}}=T(e_{i_{1}},\dots,e_{i_{k}},e^{j_{1}},\dots,e^{j_{l}}).
\]
Then
\begin{equation*}
\begin{split}
|T|_{g}^{2} = & \textstyle{\sum}|T(f_{i_{1}},\dots,f_{i_{k}},f^{j_{1}},\dots,f^{j_{l}})|^{2}\\
 = & \textstyle{\sum}A^{i_{1}n_{1}}\cdots A^{i_{k}n_{k}}A^{i_{1}r_{1}}\cdots A^{i_{k}r_{k}}
B_{j_{1}m_{1}}\cdots B_{j_{l}m_{l}}B_{j_{1}s_{1}}\cdots B_{j_{l}s_{l}}
T_{n_{1}\cdots n_{k}}^{m_{1}\cdots m_{l}}T_{r_{1}\cdots r_{k}}^{s_{1}\cdots s_{l}}\\
 = & \textstyle{\sum}g^{n_{1}r_{1}} \cdots g^{n_{k}r_{k}}g_{m_{1}s_{1}}\cdots g_{m_{l}s_{l}} 
T_{n_{1}\cdots n_{k}}^{m_{1}\cdots m_{l}}T_{r_{1}\cdots r_{k}}^{s_{1}\cdots s_{l}}. 
\end{split}
\end{equation*}
An alternate way to define the norm $|T|_{g}$ is thus via
\begin{equation}\label{eq:Tnormmultilinaltchar}
|T|_{g}=\left(\textstyle{\sum}g^{n_{1}r_{1}} \cdots g^{n_{k}r_{k}}g_{m_{1}s_{1}}\cdots g_{m_{l}s_{l}} 
T_{n_{1}\cdots n_{k}}^{m_{1}\cdots m_{l}}T_{r_{1}\cdots r_{k}}^{s_{1}\cdots s_{l}}\right)^{1/2}. 
\end{equation}

\subsection{Tensor estimates}

The above analysis yields estimates that are of importance in what follows. In particular, if $\lambda^{i},
\kappa_{i}\in\ro$, $i=1,\dots,d$, then 
\begin{equation}\label{eq:lambdakappacontrest}
|\lambda^{i}\kappa_{i}|\leq (\bh_{ij}\lambda^{i}\lambda^{j})^{1/2}(\bh^{kl}\kappa_{k}\kappa_{l})^{1/2}. 
\end{equation}
Moreover, if $\kappa$ is a symmetric covariant $2$-tensor field on $\bM$, then 
\begin{equation}\label{eq:kappavwcontrest}
|\kappa(v,w)|\leq |\kappa|_{\bge}|v|_{\bge}|w|_{\bge}
\end{equation}
for all $v,w\in T\bM$ with the same base point. 

On a slightly more technical note, let $\kappa$ be a symmetric covariant $2$-tensor field on $\tn{d}$. Define a $2$-tensor field $T$
by 
\[
T_{ij}:=\bh^{kl}\kappa_{ik}\kappa_{jl}.
\]
Then
\begin{equation}\label{eq:Tbhnormkappasq}
|T|_{\bh}\leq |\kappa|_{\bh}^{2}.
\end{equation}
In order to prove this statement, note that since $\kappa$ is symmetric, we can choose an orthonormal basis, say $\{e_{i}\}$, such that 
$\kappa$ is diagonal, with diagonal components $\kappa(e_{i},e_{i})=\kappa_{i}$. Then
\[
T(e_{i},e_{j})=\de^{kl}\kappa_{ik}\kappa_{jl}=\textstyle{\sum}_{l}\kappa_{il}\kappa_{jl}=\kappa_{i}^{2}\de_{ij}
\]
(no summation on $i$). Thus
\begin{equation}\label{eq:Tbhnormkappasqprel}
|T|_{\bh}^{2}=\textstyle{\sum}_{i,j}|T(e_{i},e_{j})|^{2}=\sum_{i,j}\kappa_{i}^{4}\de_{ij}=\sum_{i}\kappa_{i}^{4}\leq
\left(\sum_{i}\kappa_{i}^{2}\right)^{2}.
\end{equation}
On the other hand, 
\begin{equation}\label{eq:kappasqdiagexpr}
|\kappa|_{\bh}^{2}=\textstyle{\sum}_{i,j}\kappa_{ij}^{2}=\sum_{i,j}\kappa_{i}^{2}\de_{ij}=\sum_{i}\kappa_{i}^{2}.
\end{equation}
Combining (\ref{eq:Tbhnormkappasqprel}) and (\ref{eq:kappasqdiagexpr}) yields (\ref{eq:Tbhnormkappasq}).

\section{Quantitative bounds on the shift vector field}

Due to (\ref{eq:lapseshiftineq}), there is a bound on the shift vector field in terms of the lapse function. Even though this estimate is of 
interest, stronger assumptions lead to more useful conclusions. Drawing some such conclusions is the subject of the next lemma. 

\begin{lemma}\label{lemma:unifshiftbd}
Let $(M,g)$ be a canonical separable cosmological model manifold. Define $h$ by 
(\ref{eq:hdefinclash}). Assume that there is an $\shiftrb\in\ro$, satisfying $0<\shiftrb\leq 1$, such that 
\begin{equation}\label{eq:shiftbddhhdt}
h(\d_{t},\d_{t})\leq -\shiftrb^{2}
\end{equation}
for all $t\geq 0$. Then
\begin{equation}\label{eq:hinvbhinvequ}
\shiftrb^{2}\bh^{ij}v_{i}v_{j}\leq h^{ij}v_{i}v_{j}\leq \bh^{ij}v_{i}v_{j}
\end{equation}
for all $v\in\rn{d}$ and all $t\geq 0$. Moreover, 
\begin{equation}\label{eq:quotnormcharshift}
\sup_{t\geq 0}\sup_{0\neq v\in\rn{d}}\frac{|\chi^{i}v_{i}|}{(h^{ij}v_{i}v_{j})^{1/2}}\leq (\shiftrb^{-2}-1)^{1/2}.
\end{equation}
Similarly, if, in addition, $\lambda^{i}\in \ro$, $i=1,\dots,d$, then
\begin{equation}\label{eq:genshiftestreform}
\sup_{0\neq v\in\rn{d}}\frac{|\lambda^{i}v_{i}|}{(h^{ij}v_{i}v_{j})^{1/2}}\leq \shiftrb^{-1}(\bh_{ij}\lambda^{i}\lambda^{j})^{1/2}
\end{equation}
for all $t\geq 0$. In particular, if there is a constant $0<\kappa\in\ro$ such that 
\begin{equation}\label{eq:shiftdotbdd}
\bh_{ij}\lambda^{i}\lambda^{j}\leq \kappa^{2}
\end{equation}
for all $t\geq 0$, then 
\begin{equation}\label{eq:quotnormchargen}
\sup_{t\geq 0}\sup_{0\neq v\in\rn{d}}\frac{|\lambda^{i}v_{i}|}{(h^{ij}v_{i}v_{j})^{1/2}}\leq \frac{\kappa}{\shiftrb}.
\end{equation}
\end{lemma}
\begin{remark}
The estimate (\ref{eq:shiftbddhhdt}) is equivalent to 
\begin{equation}\label{eq:shiftbdd}
\bh_{ij}\chi^{i}\chi^{j}\leq 1-\shiftrb^{2}.
\end{equation}
\end{remark}
\begin{proof}
Combining (\ref{eq:hinvcomform}), (\ref{eq:chiiviabsest}) and (\ref{eq:shiftbdd}) yields
\begin{equation*}
\begin{split}
h^{ij}v_{i}v_{j} = & \bh^{ij}v_{i}v_{j}-\chi^{i}v_{i}\chi^{j}v_{j}\\
 \geq & \bh^{ij}v_{i}v_{j}-\bh_{ij}\chi^{i}\chi^{j}\bh^{lm}v_{l}v_{m}
\geq \shiftrb^{2}\bh^{ij}v_{i}v_{j}.
\end{split}
\end{equation*}
Thus (\ref{eq:hinvbhinvequ}) holds. One particular consequence of this estimate is that if $v\neq 0$, then 
\begin{equation*}
\begin{split}
\frac{|\chi^{i}v_{i}|}{(h^{ij}v_{i}v_{j})^{1/2}} \leq &  
\frac{(\bh_{ij}\chi^{i}\chi^{j})^{1/2}(\bh^{lm}v_{l}v_{m})^{1/2}}{(\shiftrb^{2}\bh^{ij}v_{i}v_{j})^{1/2}}
\leq (\shiftrb^{-2}-1)^{1/2}.
\end{split}
\end{equation*}
Thus (\ref{eq:quotnormcharshift}) holds. The arguments justifying the remaining statements of the lemma are similar. 
\end{proof}

\section{From geometric to analytical conditions, part I}

The conditions appearing in Parts~\ref{part:roughansiltrs}--\ref{part:nondegcabeq} are phrased in terms of 
$\ell$, $\sigma$, $X$ etc.; cf. (\ref{eq:ellsigmaXgenRdef}). However, the conditions appearing in Part~\ref{part:introduction}
are, largely, formulated in terms of the geometry. The purpose of the present part of these notes is to relate the 
two perspectives. In this section, we begin by taking the step from conditions on the second fundamental form and the shift
vector field to conditions on $\ell$; recall that $\mfg(\indexnot,t)$ is defined by (\ref{eq:mfgnutdef}) and that $\ell$ is 
defined by 
\begin{equation}\label{eq:elldefappendix}
\ell(\indexnot,t):=\ln\mfg(\indexnot,t)
\end{equation}
for $0\neq\indexnot\in\EFindexset$. In what follows, it is useful to divide $\mfg^{2}$ into two parts. We therefore introduce the 
notation 
\begin{equation}\label{eq:mfgTdef}
\mfg_{\roT}(\indexnot,t):=[g^{ij}(t)n_{i}n_{j}]^{1/2}=[h^{ij}(t)n_{i}n_{j}]^{1/2}
\end{equation}
for $\indexnot\in\EFindexset$, where $n_{j}=\nu_{\roT,j}(\indexnot)$, cf. (\ref{eq:nuroTetcdef}), and $h$ is defined by 
(\ref{eq:hdefinclash}).

\begin{lemma}\label{lemma:condyieldellderbd}
Let $(M,g)$ be a canonical separable cosmological model manifold. Assume that there is an 
$0<\shiftbd{0}\leq 1$ such that (\ref{eq:ddtuniftimelike}) holds for all $t\geq 0$; that there is an $0<\shiftbd{1}\in\ro$ such that (\ref{eq:LieDerchibd}) 
holds with $k=1$; and that there is a $0<C_{0}\in\ro$ such that $|\bk|_{\bge}\leq C_{0}$. Define $\ell$ by (\ref{eq:elldefappendix}). Then 
there is a constant $0<C\in\ro$, depending only on $C_{0}$, and $\shiftbd{k}$, $k=0,1$, such that 
\begin{equation}\label{eq:elldotbditosffandshift}
|\dot{\ell}(\indexnot,t)|\leq C
\end{equation}
for all $t\geq 0$ and all $0\neq \indexnot\in\EFindexset$. Assume, in addition to the above, that there is an $0<\shiftbd{2}\in\ro$ such that 
(\ref{eq:LieDerchibd}) holds with $k=2$ and that there is a $0<C_{1}\in\ro$ such that $|\ml_{U}\bk|_{\bge}\leq C_{1}$. Then there is a constant 
$0<C\in\ro$, depending only on $C_{0}$, $C_{1}$, and $\shiftbd{k}$, $k=0,1,2$, such that 
\begin{equation}\label{eq:ellddotbditosffandshift}
|\ddot{\ell}(\indexnot,t)|\leq C
\end{equation}
for all $t\geq 0$ and all $0\neq \indexnot\in\EFindexset$.
\end{lemma}
\begin{proof}
Due to Lemma~\ref{lemma:unifshiftbd}, we know that (\ref{eq:hinvbhinvequ}) holds. Moreover, (\ref{eq:hinvcomform}) yields $h^{ij}=\bh^{ij}-\chi^{i}\chi^{j}$.
Fixing $n\in\zn{d}$ and defining $\mfg_{\roT}$ according to (\ref{eq:mfgTdef}), it can thus be computed that 
\begin{equation}\label{eq:dotmfgroTprelapp}
2\mfg_{\roT}\dot{\mfg}_{\roT}=-\bh^{il}\bh^{jm}\d_{t}\bh_{lm}n_{i}n_{j}-2\chi^{i}n_{i}\dot{\chi}^{j}n_{j}.
\end{equation}
Due to (\ref{eq:lambdakappacontrest}), it can be estimated that
\begin{equation}\label{eq:chiuinlibasest}
|\chi^{i}n_{i}|\leq (\bh_{ij}\chi^{i}\chi^{j})^{1/2}(\bh^{ij}n_{i}n_{j})^{1/2}=|\chi|_{\bge}(\bh^{ij}n_{i}n_{j})^{1/2}.
\end{equation}
Combining this estimate with a similar estimate for $\dot{\chi}^{j}n_{j}$ yields
\begin{equation}\label{eq:shiftmfgtestapp}
|\chi^{i}n_{i}\dot{\chi}^{j}n_{j}|\leq |\chi|_{\bge}|\dot{\chi}|_{\bge}\bh^{ij}n_{i}n_{j}
\leq \shiftrb^{-2}|\chi|_{\bge}|\dot{\chi}|_{\bge}\mfg_{\roT}^{2},
\end{equation}
where we appealed to (\ref{eq:hinvbhinvequ}). Note that $2\bk_{\roT,lm}=\d_{t}\bh_{lm}$, so that the 
first term on the right hand side of (\ref{eq:dotmfgroTprelapp}) can be estimated by
\[
2|\bh^{il}\bh^{jm}\bk_{\roT,lm}n_{i}n_{j}|\leq 2|\bk_{\roT}|_{\bh}|v|_{\bh}^{2},
\]
where $v^{i}=\bh^{il}n_{l}$ and we appealed to Lemma~\ref{lemma:normsofmultilinearfun}. Note that 
\[
|v|_{\bh}^{2}=\bh_{ij}\bh^{il}\bh^{jm}n_{l}n_{m}=\bh^{ij}n_{i}n_{j}\leq\shiftrb^{-2}\mfg_{\roT}^{2},
\]
where we appealed to (\ref{eq:hinvbhinvequ}). Summing up yields
\begin{equation}\label{eq:dthlmninjabsest}
|\bh^{il}\bh^{jm}\d_{t}h_{lm}n_{i}n_{j}|\leq 2\shiftrb^{-2}|\bk_{\roT}|_{\bh}\mfg_{\roT}^{2}.
\end{equation}
Combining (\ref{eq:dotmfgroTprelapp}), (\ref{eq:shiftmfgtestapp}) and (\ref{eq:dthlmninjabsest}) yields
\begin{equation}\label{eq:dotmfgroTest}
|\dot{\mfg}_{\roT}|\leq \shiftrb^{-2}(|\bk_{\roT}|_{\bh}+|\chi|_{\bge}|\dot{\chi}|_{\bge})\mfg_{\roT}.
\end{equation}
Keeping (\ref{eq:bknormbgsq}) in mind, 
\begin{equation}\label{eq:twomfgmfgdotestapp}
\begin{split}
|2\mfg\dot{\mfg}| = & \left|2\mfg_{\roT}\dot{\mfg}_{\roT}-2\sum_{r=1}^{R}\frac{\dot{a}_{r}}{a_{r}}a_{r}^{-2}\nu_{r,i_{r}}^{2}\right|\\
 \leq & 2[\shiftrb^{-2}(|\bk_{\roT}|_{\bh}+|\chi|_{\bge}\cdot|\dot{\chi}|_{\bge})+|\bk_{\roRie}|_{\bge}]\mfg^{2}.
\end{split}
\end{equation}
Thus (\ref{eq:elldotbditosffandshift}) holds. 

Turning to the second derivative, note that differentiating (\ref{eq:dotmfgroTprelapp}) and 
carrying out estimates similar to the above yields
\begin{equation}\label{eq:twodotmfgTsqpetc}
|2\dot{\mfg}_{\roT}^{2}+2\mfg_{\roT}\ddot{\mfg}_{\roT}|\leq 2\shiftrb^{-2}(|(\ml_{U}\bk)_{\roT}|_{\bh}+4|\bk_{\roT}|_{\bh}^{2}
+|\chi|_{\bge}\cdot|\ddot{\chi}|_{\bge}+|\dot{\chi}|_{\bge}^{2})\mfg_{\roT}^{2},
\end{equation}
where we also appealed to (\ref{eq:Tbhnormkappasq}). On the other hand, 
\[
2\dot{\mfg}^{2}+2\mfg\ddot{\mfg}=2\dot{\mfg}_{\roT}^{2}+2\mfg_{\roT}\ddot{\mfg}_{\roT}
-2\sum_{r=1}^{R}\frac{\ddot{a}_{r}}{a_{r}}a_{r}^{-2}\nu_{r,i_{r}}^{2}
+6\sum_{r=1}^{R}\left(\frac{\dot{a}_{r}}{a_{r}}\right)^{2}a_{r}^{-2}\nu_{r,i_{r}}^{2}.
\]
The sum of the first two terms can, in absolute value, be bounded by the right hand side of (\ref{eq:twodotmfgTsqpetc}). Keeping 
(\ref{eq:bknormbgsq}) and (\ref{eq:mlUbknormbgsq}) in mind, the sum of the last two terms can be bounded in absolute value by 
\[
2\left(|(\ml_{U}\bk)_{\roRie}|_{\bge}+4|\bk_{\roRie}|_{\bge}^{2}\right)\textstyle{\sum}_{r=1}^{R}a_{r}^{-2}\nu_{r,i_{r}}^{2}.
\]
Adding up yields
\begin{equation*}
\begin{split}
|\mfg\ddot{\mfg}+\dot{\mfg}^{2}| \leq & 
\shiftrb^{-2}\left(|(\ml_{U}\bk)_{\roT}|_{\bh}+4|\bk_{\roT}|_{\bh}^{2}+|\chi|_{\bge}\cdot|\ddot{\chi}|_{\bge}+|\dot{\chi}|_{\bge}^{2}\right)\mfg^{2}\\
 & +\left(|(\ml_{U}\bk)_{\roRie}|_{\bge}+4|\bk_{\roRie}|_{\bge}^{2}\right)\mfg^{2}.
\end{split}
\end{equation*}
Combining this estimate with (\ref{eq:twomfgmfgdotestapp}) yields the conclusion of the lemma. 
\end{proof}

\subsection{Consequences of bounds on the second fundamental form}

In Chapter~\ref{chapter:silentequations}, in particular in Definition~\ref{definition:Cosilenceintro}, we define the notion of silence
in terms of a lower bound on the second fundamental form. However, in Chapter~\ref{chapter:weaksil}, in particular in 
Definition~\ref{def:roughODEtermo}, we define the notion of silence in terms of an upper bound on $\dot{\ell}$. In the present 
subsection, we relate the two perspectives. 

\begin{lemma}\label{lemma:condwsil}
Let $(M,g)$ be a canonical separable cosmological model manifold. Assume, moreover, that 
there is an $0<\shiftbd{0}\leq 1$ such that (\ref{eq:ddtuniftimelike}) holds for all $t\geq 0$ and a $0<\mu\in\ro$ and a continuous non-negative function 
$\betafun\in L^{1}([0,\infty))$ such that 
\begin{align}
\bk \geq & (\mu-\betafun)\bge,\label{eq:hhdotest}\\
|\chi|_{\bge}\cdot|\dot{\chi}|_{\bge} \leq & \betafun\label{eq:chichidotnormbdbybetafun}
\end{align}
for all $t\geq 0$. Then there is a continuous 
non-negative function $\betafun_{\mfg}\in L^{1}([0,\infty))$ such that 
\begin{equation}\label{eq:mfgdotestbetafun}
\dot{\ell}(\indexnot,t)\leq -\mu+\betafun_{\mfg}(t)
\end{equation}
for all $t\geq 0$ and all $0\neq\indexnot\in\EFindexset$. Moreover, $\betafun_{\mfg}:=2\shiftbd{0}^{-2}\betafun$.
\end{lemma}
\begin{remark}\label{remark:bkuppbdest}
The statement that $\bk\geq\eta\bge$ for all $t\geq 0$ is equivalent to the statement that $\bk(\xi,\xi)\geq\eta\bge(\xi,\xi)$ for all 
$\xi\in T\bM$ and all $t\geq 0$, where we think of $\bge$ and $\bk$ as being $2$-tensor fields on $\bM$. 
\end{remark}
\begin{remark}\label{remark:bkesttoelldotestuppbd}
If the estimate (\ref{eq:hhdotest}) is replaced with 
\[
\bk \leq (\mu_{+}+\betafun_{+})\bge,
\]
where $0<\mu_{+}\in\ro$ and $\betafun_{+}\in L^{1}([0,\infty))$ is a continuous non-negative function, then the conclusion 
(\ref{eq:mfgdotestbetafun}) is replaced by
\begin{equation}\label{eq:mfgdotestbetafunlow}
\dot{\ell}(\indexnot,t)\geq -\mu_{\low}-\betafun_{\low}(t)
\end{equation}
for all $t\geq 0$ and all $0\neq\indexnot\in\EFindexset$, where $\betafun_{\low}\in L^{1}([0,\infty))$ is a continuous non-negative function
(depending only on $\betafun$, $\betafun_{+}$ and $\shiftbd{0}$) and $0<\mu_{\low}\in\ro$ depends only on $\mu_{+}$ and $\shiftbd{0}$.
\end{remark}
\begin{proof}
The first steps of the proof of Lemma~\ref{lemma:condyieldellderbd} still apply. In particular, the estimates up to, and including, 
(\ref{eq:shiftmfgtestapp}) still hold. On the other hand, appealing to (\ref{eq:hhdotest}) and the fact that (\ref{eq:bkblock}) holds yields
\begin{equation}\label{eq:mainmfgtest}
\bh^{il}\bh^{jm}\d_{t}\bh_{lm}n_{i}n_{j}\geq 2(\mu-\betafun)\bh^{ij}n_{i}n_{j}\geq 2(\mu-\shiftbd{0}^{-2}\betafun)\mfg_{\roT}^{2}
\end{equation}
for all $t\geq 0$; in the last step, we also appealed to (\ref{eq:hinvbhinvequ}). Combining (\ref{eq:dotmfgroTprelapp}), 
(\ref{eq:shiftmfgtestapp}), (\ref{eq:chichidotnormbdbybetafun}) and (\ref{eq:mainmfgtest}) yields 
\begin{equation}\label{eq:dotmfgroTbetafunmfg}
\dot{\mfg}_{\roT}\leq -(\mu-2\shiftbd{0}^{-2}\betafun)\mfg_{\roT}
\end{equation}
for all $t\geq 0$. On the other hand, combining (\ref{eq:bkblock}), (\ref{eq:bgblock}) and (\ref{eq:hhdotest}) yields 
$\dot{a}_{r}/a_{r}\geq\mu-\betafun$ for all $t\geq 0$. Thus
\[
2\mfg\dot{\mfg}=2\mfg_{\roT}\dot{\mfg}_{\roT}-2\sum_{r=1}^{R}\frac{\dot{a}_{r}}{a_{r}}a_{r}^{-2}\nu_{r,i_{r}}^{2}\leq -2(\mu-\betafun_{\mfg})\mfg^{2}
\]
for all $t\geq 0$ and all $0\neq \indexnot\in\EFindexset$, where $\betafun_{\mfg}:=2\shiftbd{0}^{-2}\betafun$. The lemma follows.
\end{proof}

It is also of interest to record the following criteria, ensuring that (\ref{eq:elldotbdgeneral}) holds. 

\begin{lemma}\label{lemma:condbalweak}
Let $(M,g)$ be a canonical separable cosmological model manifold. Assume, moreover, that 
there is an $0<\shiftbd{0}\leq 1$ such that (\ref{eq:ddtuniftimelike}) holds for all $t\geq 0$ and a constant $0<C_{a}\in\ro$ and a continuous non-negative 
function $\betafun\in L^{1}([0,\infty))$ such that 
\begin{align}
|\bk|_{\bge}+|\chi|_{\bge}\cdot|\dot{\chi}|_{\bge} \leq & C_{a}+\betafun\label{eq:bkandshiftestbalweak}
\end{align}
for all $t\geq 0$. Then there is a continuous non-negative function $\betafun_{\robal}\in L^{1}([0,\infty))$ and a $0<C_{\ell}\in\ro$ such that 
\begin{equation}\label{eq:abselldotestbetafunbalweak}
|\dot{\ell}(\indexnot,t)|\leq C_{\ell}+\betafun_{\robal}(t)
\end{equation}
for all $t\geq 0$ and all $0\neq\indexnot\in\EFindexset$.
\end{lemma}
\begin{remark}
In the statement of the lemma, 
\[
C_{\ell}=(1+\shiftbd{0}^{-2})C_{a},\ \ \
\betafun_{\robal}=(1+\shiftbd{0}^{-2})\betafun. 
\]
\end{remark}
\begin{proof}
The statement follows from the assumptions and (\ref{eq:twomfgmfgdotestapp}).
\end{proof}

\subsection{Bounding $\sigma$ and $X$}

Recall that $\sigma$ and $X$ are defined by (\ref{eq:sigmaXdefintro}). It is of interest to bound the norms of these functions (and their
time derivatives) in terms of the norms of $\chi$, $\mcX$ (recall that $\mcX$ is introduced in (\ref{eq:mcXdef})) and their time derivatives. 

\begin{lemma}\label{lemma:sigmaXbdsandderbds}
Consider (\ref{eq:thesystemRge}). Assume the associated metric to be such that $(M,g)$ is a canonical separable cosmological model 
manifold. Define $\bh$ by (\ref{eq:bhdef}). Assume, moreover, that there is an $0<\shiftbd{0}\leq 1$ such that (\ref{eq:ddtuniftimelike}) holds 
for all $t\geq 0$. Then 
\begin{align}
|\sigma(\indexnot,t)| \leq & \shiftbd{0}^{-1}|\chi(t)|_{\bge},\label{eq:sigmabditochi}\\
\|X(\indexnot,t)\| \leq & \shiftbd{0}^{-1}|\mcX(t)|_{\bh},\label{eq:XbditomcX}\\
|\dot{\sigma}(\indexnot,t)| \leq & \shiftbd{0}^{-1}|\dot{\chi}(t)|_{\bge}+\shiftbd{0}^{-1}|\chi(t)|_{\bge}|\dot{\ell}(\indexnot,t)|,\label{eq:sigmadotbditochi}\\
\|\dot{X}(\indexnot,t)\| \leq & \shiftbd{0}^{-1}|\dot{\mcX}(t)|_{\bh}+\shiftbd{0}^{-1}|\mcX(t)|_{\bh}|\dot{\ell}(\indexnot,t)|\label{eq:XdotbditomcX}
\end{align}
for all $t\geq 0$ and all $0\neq \indexnot\in\EFindexset$. 
\end{lemma}
\begin{remark}
Recall that $|\mcX(t)|_{\bh}$ is defined by (\ref{eq:mcXnorm}).
\end{remark}
\begin{proof}
Estimate
\[
|\sigma(\indexnot,t)|=\frac{|n_{l}\chi^{l}(t)|}{\mfg(\indexnot,t)}\leq\shiftbd{0}^{-1}\frac{|\chi(t)|_{\bge}
\mfg_{\roT}(\indexnot,t)}{\mfg(\indexnot,t)}\leq \shiftbd{0}^{-1}|\chi(t)|_{\bge}
\]
for all $t\geq 0$ and all $0\neq \indexnot\in\EFindexset$, where we appealed to (\ref{eq:lambdakappacontrest})
and (\ref{eq:hinvbhinvequ}). Thus (\ref{eq:sigmabditochi}) holds. Estimate
\begin{equation}\label{eq:signsubtlmcXnorm}
\begin{split}
\|n_{l}X^{l}(t)\| \leq & \textstyle{\sum}_{l}|n_{l}|\cdot \|X^{l}(t)\|=\sum_{l}n_{l}\left(\frac{|n_{l}|}{n_{l}}\|X^{l}(t)\|\right)\\
 \leq & (\bh^{ij}(t)n_{i}n_{j})^{1/2}\left(\textstyle{\sum}_{l,k}\bh_{lk}(t)\frac{|n_{l}|}{n_{l}}\|X^{l}(t)\|\frac{|n_{k}|}{n_{k}}\|X^{k}(t)\|\right)^{1/2}\\
 \leq & \eta_{\rosh,0}^{-1}(h^{ij}(t)n_{i}n_{j})^{1/2}|\mcX(t)|_{\bh}\leq \eta_{\rosh,0}^{-1}\mfg_{\roT}(\indexnot,t)|\mcX(t)|_{\bh}
\end{split}
\end{equation}
for all $t\geq 0$, where $|n_{l}|/n_{l}$ should be thought of as equalling $1$ when $n_{l}=0$. Thus
\[
\|X(\indexnot,t)\|\leq \frac{\|n_{l}X^{l}(t)\|}{\mfg(\indexnot,t)}\leq \eta_{\rosh,0}^{-1}|\mcX(t)|_{\bh}
\]
and (\ref{eq:XbditomcX}) follows. The proofs of the remaining estimates are similar. 
\end{proof}

\subsection{Geometric criteria in the unbalanced setting}

Let us return to the conditions stated in Sections~\ref{section:PDEunbalterm}--\ref{section:bapropofiterseg}, in particular 
in Lemma~\ref{lemma:mainassumpubcase} and Assumptions~\ref{assumption:guooass} and \ref{assumption:Yoconvub}. 

\begin{lemma}\label{lemma:geomcritunbalset}
Consider (\ref{eq:thesystemRge}). Assume the associated metric to be such that $(M,g)$ is a canonical separable cosmological model 
manifold. Assume, moreover, that $d=1$, that $R=0$ and that there is an 
$0<\shiftbd{0}\leq 1$ such that (\ref{eq:ddtuniftimelike}) holds for all $t\geq 0$.  Assume that there are constants $K_{\rosh},K_{\bk},K_{\roode},\bK_{X}>0$;
$\b_{\rosh}<0$; $\b_{\roode}$; and $\b_{X}\geq 0$ such that 
\begin{align}
|\chi(t)|_{\bge}+|\dot{\chi}(t)|_{\bge} \leq & K_{\rosh}e^{\b_{\rosh}t},\label{eq:shiftexpdecbdsupexpgr}\\
|\ddot{\chi}(t)|_{\bge}+|\bk(t)|_{\bge}+|\ml_{U}\bk(t)|_{\bge} \leq & K_{\bk},\label{eq:shiftddbkzadsupexpgr}\\
\|\zeta(t)\|^{1/2}+\|\dot{\zeta}(t)\|^{1/2}+\|\a(t)\|+\|\dot{\a}(t)\| \leq & K_{\roode}e^{\b_{\roode}t},\label{eq:zetaalphaanddersupexpgr}\\
|\mcX(t)|_{\bh}+|\dot{\mcX}(t)|_{\bh} \leq & \bK_{X}e^{\b_{X}t}\label{eq:mcXamcXdsupexpgr}
\end{align}
for all $t\geq 0$. Assume, finally, that $\b_{X}-\b_{\roode}>0$. Then the assumptions of Lemma~\ref{lemma:mainassumpubcase} are 
satisfied. 

Assume, in addition to the above, that $\b_{X}>0$. Let $e_{1}$ be the unit vector field which is a positive multiple of $\d_{1}$, define 
$\mcY^{1}$ to be the matrix valued function such that $\mcY^{1}e_{1}=\mcX$ and assume that there is a matrix $\mcY^{1}_{\infty}\in \Mn{m}{\co}$ and 
constants $\etab_{X},L_{X}>0$ such that 
\begin{equation}\label{eq:mcYomcYoinfest}
\|e^{-\b_{X}t}\mcY^{1}(t)-\mcY^{1}_{\infty}\|\leq L_{X}e^{-\etab_{X}t}
\end{equation}
for all $t\geq 0$. Then, defining $Y^{1}$ by (\ref{eq:mflYovarsdefintro}), there is a constant $K_{X}$ such that 
\begin{equation}\label{eq:Yolimestimate}
\|e^{-\b_{X}t}Y^{1}(t)-\mcY^{1}_{\infty}\|\leq K_{X}e^{-\eta_{X}t}
\end{equation}
for all $t\geq 0$, where $K_{X}$ only depends on $\shiftbd{0}$, $K_{\rosh}$, $\bK_{X}$, $K_{\bk}$ and $L_{X}$. Moreover, 
$\eta_{X}:=\min\{-2\b_{\rosh},\etab_{X}\}$.
In particular, Assumptions~\ref{assumption:Yoconvub} are satisfied with $Y^{1}_{\infty}=\mcY^{1}_{\infty}$. 
\end{lemma}
\begin{remark}\label{remark:geomcritunbsetintrover}
If the conditions of the lemma are satisfied and $\mcY^{1}_{\infty}$ has an eigenvalue with a non-zero real part, then 
Assumptions~\ref{ass:mainassumpubcaseintro} and \ref{assumption:Yoconvubintro} are satisfied with $Y^{1}_{\infty}=\mcY^{1}_{\infty}$; this is a 
consequence of the fact that the assumptions of Lemma~\ref{lemma:mainassumpubcase} are satisfied and the fact that 
Assumptions~\ref{assumption:Yoconvub} hold. 
\end{remark}
\begin{proof}
Due to (\ref{eq:shiftexpdecbdsupexpgr}), (\ref{eq:shiftddbkzadsupexpgr}) and the fact that there is an $0<\shiftbd{0}\leq 1$ such that 
(\ref{eq:ddtuniftimelike}) holds for all $t\geq 0$, Lemma~\ref{lemma:condyieldellderbd} is applicable and yields the conclusion that there is a constant
$C_{\mfl}$ such that 
\begin{equation}\label{eq:dotmflddotmflest}
|\dot{\mfl}(t)|+|\ddot{\mfl}(t)|\leq C_{\mfl}
\end{equation}
for all $t\geq 0$; note that $\dot{\mfl}(t)=\dot{\ell}(\indexnot,t)$ in the present context. Moreover, $C_{\mfl}$ only depends on $K_{\bk}$, 
$\shiftbd{0}$ and $K_{\rosh}$. Combining (\ref{eq:dotmflddotmflest}) with the assumptions of the lemma as well as with 
Lemma~\ref{lemma:sigmaXbdsandderbds} yields the conclusion that there are constants $C_{\rosh}$ and $C_{X}$ such that 
\begin{equation}\label{eq:varsigmaYobdfromgeombd}
|\varsigma(t)|+|\dot{\varsigma}(t)|\leq C_{\rosh}e^{\b_{\rosh}t},\ \ \
\|Y^{1}(t)\|+\|\dot{Y}^{1}(t)\| \leq C_{X}e^{\b_{X}t}
\end{equation}
for all $t\geq 0$; note that $\|X(\indexnot,t)\|=\|Y^{1}(t)\|$, $|\sigma(\indexnot,t)|=|\varsigma(t)|$ etc. in the present context. In these 
estimates, $C_{\rosh}$ only depends on 
$K_{\rosh}$, $\shiftbd{0}$ and $K_{\bk}$; and $C_{X}$ only depends on $\bK_{X}$, $K_{\rosh}$, $\shiftbd{0}$ and $K_{\bk}$. Returning to the statement of 
Proposition~\ref{lemma:mainassumpubcase}, note that by defining $\eta_{\roode}:=\b_{X}-\b_{\roode}$, it is clear that $\eta_{\roode}>0$. 
Thus (\ref{eq:etaroodedef}) holds. Due to the fact that the left hand sides of (\ref{eq:betalowerbd}) and (\ref{eq:mfldaddbd}) are 
bounded, it is clear that we can find a constant $\ellderbd>0$ such that (\ref{eq:betalowerbd}) and (\ref{eq:mfldaddbd}) hold. 
Due to (\ref{eq:varsigmaYobdfromgeombd}), it is clear that (\ref{eq:shiftbdub}) and (\ref{eq:Xbdub}) hold. Finally, 
(\ref{eq:albdub}) and (\ref{eq:zetabdub}) follow from (\ref{eq:zetaalphaanddersupexpgr}). Thus the assumptions of 
Proposition~\ref{lemma:mainassumpubcase} are satisfied. 

Next, it is of interest to compare $\mcY^{1}$ with $Y^{1}$ introduced in (\ref{eq:mflYovarsdefintro}). By definition, $e_{1}=(g_{11})^{-1/2}\d_{1}$.
For this reason, 
\begin{equation}\label{eq:YomcYocomp}
Y^{1}=\frac{X^{1}}{(g^{11})^{1/2}}=\frac{\mcY^{1}}{(g_{11}g^{11})^{1/2}}.
\end{equation}
Due to (\ref{eq:hinvcomform}), we know that 
\[
g_{11}g^{11}=g_{11}(g_{11})^{-1}-g_{11}\chi^{1}\chi^{1}=1-g_{11}\chi^{1}\chi^{1}.
\]
Combining this observation with (\ref{eq:ddtuniftimelike}) yields $\shiftbd{0}^{2}\leq g_{11}g^{11}\leq 1$. In particular, 
\begin{equation}\label{eq:mcYoest}
\|\mcY^{1}(t)\|\leq \|Y^{1}(t)\|\leq C_{X}e^{\b_{X}t}
\end{equation}
for all $t\geq 0$, where we appealed to (\ref{eq:varsigmaYobdfromgeombd}). Moreover,
\begin{equation*}
\begin{split}
\left|1-\frac{1}{(g_{11}g^{11})^{1/2}}\right| = & 
\frac{|g^{11}g_{11}-1|}{(g_{11}g^{11})^{1/2}[1+(g_{11}g^{11})^{1/2}]}\\
 \leq & \shiftbd{0}^{-1}K_{\rosh}^{2}e^{2\b_{\rosh}t}
\end{split}
\end{equation*}
for all $t\geq 0$. Combining this estimate with (\ref{eq:YomcYocomp}) and (\ref{eq:mcYoest}) yields
\[
\|Y^{1}(t)-\mcY^{1}(t)\|\leq \shiftbd{0}^{-1}K_{\rosh}^{2}C_{X}e^{(\b_{X}+2\b_{\rosh})t}
\]
for all $t\geq 0$. Combining this estimate with (\ref{eq:mcYomcYoinfest}) yields (\ref{eq:Yolimestimate}). 
The lemma follows. 
\end{proof}

\begin{lemma}\label{lemma:guaranteeingbdsonguoounbset}
Let $(M,g)$ be a canonical separable cosmological model manifold. Assume, moreover, that $d=1$, 
that $R=0$ and that there is an $0<\shiftbd{0}\leq 1$ such that (\ref{eq:ddtuniftimelike}) holds for all $t\geq 0$. Assume, finally, that there are 
$\bbe_{j}\in\ro$, $j=1,2$, and a continuous function $0\leq\betafun\in L^{1}([0,\infty))$ such that 
\begin{equation}\label{eq:bkbobtwobdellderunbd}
(\bbe_{2}-\betafun)\bge\leq \bk\leq(\bbe_{1}+\betafun)\bge
\end{equation}
for all $t\geq 0$. Then there are constants $C_{i}>0$, $i=1,2$, depending only on $g_{11}(0)$, $C_{\betafun}:=\|\betafun\|_{1}$ and $\shiftbd{0}$
such that 
\[
C_{1}e^{-2\bbe_{1}t}\leq g^{11}(t)\leq C_{2}e^{-2\bbe_{2}t}
\]
for all $t\geq 0$. 
\end{lemma}
\begin{remark}
If (\ref{eq:thesystemRge}) is such that the the associated Lorentz manifold satisfies the assumptions of 
Lemma~\ref{lemma:guaranteeingbdsonguoounbset}, then Assumption~\ref{assumption:guooass} holds with $\b_{i}=-\bbe_{i}$. 
\end{remark}
\begin{proof}
Note that (\ref{eq:bkbobtwobdellderunbd}) is equivalent to
\[
\bbe_{2}-\betafun\leq \d_{t}\ln (g_{11})^{1/2}\leq\bbe_{1}+\betafun.
\]
Integrating this estimate yields
\begin{equation}\label{eq:goobbeotwoest}
g_{11}(0)\exp(2\bbe_{2}t-2C_{\betafun})\leq g_{11}(t)\leq g_{11}(0)\exp(2\bbe_{1}t+2C_{\betafun})
\end{equation}
for all $t\geq 0$, where $C_{\betafun}:=\|\betafun\|_{1}$. Note that $\shiftbd{0}^{2}\leq g_{11}(t)g^{11}(t)\leq 1$ for all $t\geq 0$.
This is a consequence of (\ref{eq:hinvcomform}) and (\ref{eq:ddtuniftimelike}); cf. the proof of Lemma~\ref{lemma:geomcritunbalset}.
Combining this estimate with (\ref{eq:goobbeotwoest}) yields 
\[
\frac{\shiftbd{0}^{2}}{g_{11}(0)}\exp(-2\bbe_{1}t-2C_{\betafun})\leq g^{11}(t)\leq 
\frac{1}{g_{11}(0)}\exp(-2\bbe_{2}t+2C_{\betafun})
\]
for all $t\geq 0$. The lemma follows. 
\end{proof}

\section{Implications concerning the causal structure}

In Definition~\ref{definition:Cosilenceintro}, we introduce the notion of $C^{1}$-silence. The purpose of the present section is to justify
the use of the word silent. 

\begin{lemma}\label{lemma:silenceconsforcauscurves}
Let $(M,g)$ be a canonical separable cosmological model manifold. Assume that there is a $0<\mu\in\ro$ and a continuous non-negative function 
$\betafun\in L^{1}([0,\infty))$ such that 
\begin{equation}\label{eq:hhdotestcausalityest}
\bk \geq (\mu-\betafun)\bge
\end{equation}
for all $t\geq 0$. Let $\bge_{\refer}$ be a fixed Riemannian metric on $\bM$. Then there is 
a constant $0<C\in\ro$, depending only on  $\|\betafun\|_{1}$, $a_{r}(0)$ ($r=1,\dots,R$), $g_{ij}(0)$ ($i,j=1,\dots,d$) and the 
reference metric $\bge_{\refer}$ such that if $\g:[0,\infty)\rightarrow M$ is a causal curve in $(M,g)$ with $\g(t)=[\bga(t),t]$, then 
\begin{equation}\label{eq:dbgaest}
|\dot{\bga}(t)|_{\bge_{\refer}}\leq Ce^{-\mu t}
\end{equation}
for all $t\geq 0$.
\end{lemma}
\begin{remark}\label{remark:silenceconsforcauscurves}
Due to (\ref{eq:dbgaest}), it is clear that $\bga(t)$ converges exponentially to a point, say $\bp[\g]\in\bM$, as $t\rightarrow\infty$. Moreover, if 
$\g_{i}:[0,\infty)\rightarrow M$, $i=1,2$, are two future oriented causal curves such that $\g_{i}(t)=[\bga_{i}(t),t]$ and $\bp[\g_{1}]\neq \bp[\g_{2}]$, 
then the observers $\g_{1}$ and $\g_{2}$ sooner or later lose the ability to communicate. In fact, there are $t_{i}\in [0,\infty)$, $i=1,2$, such that 
\[
J^{+}[\g_{1}(t_{1})]\cap J^{+}[\g_{2}(t_{2})]=\varnothing.
\] 
In this sense, there is silence asymptotically. 
\end{remark}
\begin{proof}
Given $\xi\in\rn{d}$, let 
\[
\mfN(t,\xi):=\bh_{ij}(t)\xi^{i}\xi^{j}.
\]
Then
\[
\dot{\mfN}(t,\xi)=\d_{t}\bh_{ij}\xi^{i}\xi^{j}=2\bk_{\roT,ij}\xi^{i}\xi^{j}\geq 2[\mu-\betafun(t)]\mfN(t,\xi)
\]
for all $t\geq 0$, where we appealed to (\ref{eq:bkblock}) and (\ref{eq:hhdotestcausalityest}). Integrating this estimate yields
\begin{equation}\label{eq:bhihexpgrowthest}
\bh_{ij}(t)\xi^{i}\xi^{j}\geq C|\xi|^{2}e^{2\mu t}
\end{equation}
for all $t\geq 0$ and all $\xi\in\rn{d}$, where $C>0$ only depends on $\|\betafun\|_{1}$ and $g_{ij}(0)$ ($i,j=1,\dots,d$). Similarly, 
(\ref{eq:bkblock}), (\ref{eq:bgblock}) and (\ref{eq:hhdotestcausalityest}) yield $a_{r}(t)\geq Ce^{\mu t}$ for all $t\geq 0$, where the 
constant $C>0$ only depends on $\|\betafun\|_{1}$ and $a_{r}(0)$. Let $\g:[0,\infty)\rightarrow M$ be a causal curve in $(M,g)$ with
\[
\g(t)=[\bga(t),t]=[\bga_{\roT}(t),\bga_{1}(t),\dots,\bga_{R}(t),t],
\]
where $\bga_{\roT}$ takes its values in $\tn{d}$ and $\bga_{r}$ takes its values in $M_{r}$, $r=1,\dots,R$. Then 
\[
-1+\bh_{ij}\chi^{i}\chi^{j}+2\bh_{ij}\chi^{i}\dot{\bga}_{\roT}^{j}
+\bh_{ij}\dot{\bga}_{\roT}^{i}\dot{\bga}_{\roT}^{j}+\textstyle{\sum}_{r=1}^{R}a_{r}^{2}|\dot{\bga}_{r}|_{g_{r}}^{2}\leq 0
\]
for all $t\geq 0$. Thus
\begin{equation*}
\begin{split}
|\dot{\bga}_{\roT}|_{\bh}^{2}+\textstyle{\sum}_{r=1}^{R}a_{r}^{2}|\dot{\bga}_{r}|_{g_{r}}^{2} \leq & 1+2|\chi|_{\bge}|\dot{\bga}_{\roT}|_{\bh}-|\chi|_{\bge}^{2}
\leq 1+|\chi|_{\bge}^{2}+\frac{1}{2}|\dot{\bga}_{\roT}|_{\bh}^{2}\\
 \leq & 2+\frac{1}{2}|\dot{\bga}_{\roT}|_{\bh}^{2},
\end{split}
\end{equation*}
where we used the fact that $|\chi|_{\bge}\leq 1$; recall that (\ref{eq:lapseshiftineq}) holds and that $N=1$. In particular,  
\[
|\dot{\bga}_{\roT}|_{\bh}^{2}+\textstyle{\sum}_{r=1}^{R}a_{r}^{2}|\dot{\bga}_{r}|_{g_{r}}^{2}\leq 4
\]
for all $t\geq 0$. Due to (\ref{eq:bhihexpgrowthest}) and the analogous estimate for the $a_{r}$, we conclude that there is a $0<C\in\ro$, 
depending only on $\|\betafun\|_{1}$, $a_{r}(0)$ ($r=1,\dots,R$), and $g_{ij}(0)$ ($i,j=1,\dots,d$), such that 
\[
|\dot{\bga}_{\roT}(t)|^{2}+\textstyle{\sum}_{r=1}^{R}|\dot{\bga}_{r}(t)|_{g_{r}}^{2}\leq Ce^{-2\mu t}
\]
for all $t\geq 0$. The lemma follows.
\end{proof}

\chapter{Relating the analytic and geometric conditions}

In Part~\ref{part:introduction}, we formulate the conditions geometrically. However, in 
Parts~\ref{part:roughansiltrs}--\ref{part:nondegcabeq}, the conditions are formulated 
analytically. The purpose of the present chapter is to relate the two perspectives. 

\section{Bounds on the coefficients in the silent setting}

Let us begin by deriving conclusions concerning the coefficients from the conditions appearing in 
Definition~\ref{definition:Cosilenceintro}. 

\begin{lemma}\label{lemma:coeffofspderdecexpsil}
Given that the assumptions of Lemma~\ref{lemma:condwsil} are fulfilled, the norm of the matrix with components $g^{ij}(t)$, $i,j=1,\dots,d$, is 
bounded by $Ce^{-2\mu t}$ for all $t\geq 0$, where $C$ only depends on $\shiftbd{0}$, $\|\betafun\|_{1}$ and $g^{ij}(0)$ ($i,j=1,\dots,d$). Moreover, 
$a_{r}^{-2}(t)\leq Ce^{-2\mu t}$ for all $t\geq 0$, where $C$ only depends on $\|\betafun\|_{1}$ and $a_{r}(0)$.
If, in addition, $\mcX$ is $C^{0}$-future bounded, then $\|X^{j}(t)\|\leq Ce^{-\mu t}$ for all $t\geq 0$ and all $j\in\{1,\dots,d\}$, where $C$ 
only depends on $\|\betafun\|_{1}$, $g_{ij}(0)$ ($i,j=1,\dots,d$) and the future bound on $|\mcX|_{\bh}$. Finally, 
$|\chi(t)|\leq Ce^{-\mu t}$ for all $t\geq 0$, where $C$ only depends on $\|\betafun\|_{1}$ and $g_{ij}(0)$ ($i,j=1,\dots,d$) 
(note that $|\chi(t)|$ denotes the Euclidean norm of the vector with components $\chi^{i}(t)$). 
\end{lemma}
\begin{proof}
Consider the estimate (\ref{eq:dotmfgroTbetafunmfg}). If we would replace
$n\in\zn{d}$ in the definition of $\mfg_{\roT}$ by $\xi\in\rn{d}$, the result would be the same. In particular, there is thus a constant $0<C\in\ro$,
depending only on $\shiftbd{0}$ and $\|\betafun\|_{1}$, such that 
\begin{equation}\label{eq:guijexpdecayestimateremark}
g^{ij}(t)\xi_{i}\xi_{j}\leq Cg^{ij}(0)\xi_{i}\xi_{j}e^{-2\mu t}
\end{equation}
for all $t\geq 0$ and all $\xi\in\rn{d}$. Thus the first statement of the lemma  holds. Note that the conditions of 
Lemma~\ref{lemma:silenceconsforcauscurves} are fulfilled. In what follows, we therefore use the conclusions of this 
lemma freely. The second statement follows from lower bound on $a_{r}(t)$ stated in connection with (\ref{eq:bhihexpgrowthest}). The statement 
concerning $\mcX$ follows from (\ref{eq:bhihexpgrowthest}) and the assumption that $\mcX$ is $C^{0}$-future bounded. The justification of the bound 
on $|\chi(t)|$ is similar. 
\end{proof}

\section{Conditions ensuring weak transparency}

Let us now, starting with the conditions formulated in Section~\ref{section:introtransintro}, derive conclusions concerning the coefficients
of the equation. 

\begin{lemma}\label{lemma:geometrictoanalytictrans}
Let $(M,g)$ be a canonical separable cosmological model manifold. 
Assume that there is a division of the cotangent space into transparent and silent subsets, as described in 
Definition~\ref{def:transpandsilentdivision}. Assume, moreover, that Definitions~\ref{def:expandconvdir} and 
\ref{def:shiftnegligibletrs} are fulfilled. Given $\indexnot\in\EFindexset$, let $\indexnot_{\trs}$ and $\indexnot_{\rosil}$ 
be given by Definition~\ref{def:indexnottrssilintro}. Then there are constants $C_{A}$ and $C_{B}$ such that 
\begin{align}
|\dot{\ell}(\indexnot_{\trs},t)| \leq & C_{A}e^{-\eta_{\trs}t},\label{eq:dotelltrsestfinal}\\
|\sigma(\indexnot_{\trs},t)| \leq & C_{B}e^{-\eta_{\trs}t}\label{eq:sigmatrsestfinal}
\end{align}
for all $t\geq 0$ and all $\indexnot\in\EFindexset$ such that $\indexnot_{\trs}\neq 0$. Moreover, 
\begin{equation}\label{eq:dotellsilestfinal}
\dot{\ell}(\indexnot_{\rosil},t)\leq -\b_{\rosil}+\shiftbd{0}^{-2}\betafun_{\rosil}(t)+\shiftbd{0}^{-2}\betafun(t)
\end{equation}
for all $t\geq 0$ and all $\indexnot\in\EFindexset$ such that $\indexnot_{\rosil}\neq 0$. 

Concerning the metric coefficients, there are constants $g^{j_{k}j_{l}}_{\infty}$, $k,l\in\{1,\dots,d_{\trs}\}$, and $q_{\infty,r_{j}}$,
$j=1,\dots,R_{\trs}$, such that  
\begin{equation}\label{eq:gacoefftrsdir}
|g^{j_{k}j_{l}}(t)-g^{j_{k}j_{l}}_{\infty}|\leq Ce^{-\eta_{\trs}t},\ \ \
|a_{r_{j}}^{-2}(t)-q_{\infty,r_{j}}|\leq Ce^{-\eta_{\trs}t}
\end{equation}
for all $t\geq 0$, where $C$ only depends on $C_{A}$, $\eta_{\trs}$, $g^{ij}(0)$ and $a_{r}(0)$. Moreover, $g^{j_{k}j_{l}}_{\infty}$ are the components 
of a positive definite matrix, and $q_{\infty,r_{j}}>0$. In addition, 
\begin{equation}\label{eq:gacoeffsildir}
|g^{\bj_{k}\bj_{l}}(t)|\leq Ce^{-2\b_{\rosil}t},\ \ \
a_{\bre_{j}}^{-2}(t)\leq Ce^{-2\b_{\rosil}t}
\end{equation}
for all $t\geq 0$, $k,l\in\{1,\dots,d_{\rosil}\}$ and $j=1,\dots,R_{\rosil}$, where $C$ only depends on 
$g^{ij}(0)$, $a_{r}(0)$, $\shiftbd{0}$, $c_{\rosil}:=\|\betafun_{\rosil}\|_{1}$ and $c_{\betafun}:=\|\betafun\|$. Moreover,
\begin{equation}\label{eq:gjkbjlest}
|g^{j_{k}\bj_{l}}(t)|\leq Ce^{-\b_{\rosil}t}
\end{equation}
for all $t\geq 0$, all $k\in\{1,\dots,d_{\trs}\}$ and all $l\in\{1,\dots,d_{\rosil}\}$, where $C$ only depends on $C_{A}$, $\eta_{\trs}$, $g^{ij}(0)$, 
$a_{r}(0)$, $\shiftbd{0}$, $c_{\rosil}$ and $c_{\betafun}$. Finally, 
\begin{equation}\label{eq:shiftestparttrscase}
|g^{0j_{k}}(t)|\leq C_{1}e^{-\eta_{\trs}t},\ \ \
|g^{0\bj_{l}}(t)|\leq C_{2}e^{-\b_{\rosil}t}
\end{equation}
for all $t\geq 0$, all $k\in\{1,\dots,d_{\trs}\}$ and all $l\in\{1,\dots,d_{\rosil}\}$, where $C_{1}$ only depends on $C_{A}$, $C_{B}$, $\eta_{\trs}$
and $g^{ij}(0)$ and $C_{2}$ only depends on $g^{ij}(0)$, $\shiftbd{0}$, $c_{\rosil}$ and $c_{\betafun}$.  
\end{lemma}
\begin{remark}
The constant $C_{A}$ appearing in (\ref{eq:dotelltrsestfinal}) only depends on $C_{a}$, $C_{b}$ and $\shiftbd{k}$, $k=0,1$. The constant 
$C_{B}$ appearing in (\ref{eq:sigmatrsestfinal}) only depends on $C_{b}$ and $\shiftbd{0}$. 
\end{remark}
\begin{remark}\label{remark:Xnormbdsiltrans}
If, in addition to the assumptions of the lemma, $\mcX$ is $C^{0}$-future bounded, then
\[
\|X^{\bj_{k}}(t)\|\leq Ce^{-\b_{\rosil}t}
\]
for all $t\geq 0$ and all $k\in \{1,\dots,d_{\rosil}\}$, where $C$ only depends on $g^{ij}(0)$, $C_{0}$, $\shiftbd{0}$, $c_{\rosil}$ and 
$c_{\betafun}$. The reason for this is that if $n\in\zn{d}$ corresponds to an element $\indexnot\in\EFindexset$ such that 
$\indexnot=\indexnot_{\rosil}$, then 
\[
\| n_{j}X^{j}(t)\|\leq \shiftbd{0}^{-1}|\mcX(t)|_{\bh}\mfg_{\roT}(\indexnot_{\rosil},t)\leq Ce^{-\b_{\rosil}t}\mfg_{\roT}(\indexnot_{\rosil},0)
\]
for all $t\geq 0$, where $C$ only depends on $C_{0}$, $\shiftbd{0}$, $c_{\rosil}$ and $c_{\betafun}$. Moreover, we appealed to 
(\ref{eq:lambdakappacontrest}) in the first step and to (\ref{eq:mfgsqrosilexpdecest}) below in the second step. 
\end{remark}
\begin{proof}
\textbf{The transparent directions.} Compute
\begin{equation}\label{eq:elldottransprel}
\textstyle{2}\mfg(\indexnot_{\trs},t)\dot{\mfg}(\indexnot_{\trs},t)=2\mfg_{\roT}(\indexnot_{\trs},t)\dot{\mfg}_{\roT}(\indexnot_{\trs},t)
-2\sum_{j=1}^{R_{\trs}}\frac{\dot{a}_{r_{j}}(t)}{a_{r_{j}}(t)}a_{r_{j}}^{-2}(t)\nu_{r_{j},i_{r_{j}}}^{2}(\indexnot_{\trs}).
\end{equation}
Let us begin by considering the first term on the right hand side. It can be computed as in (\ref{eq:dotmfgroTprelapp}). Consider the 
first term on the right hand side of (\ref{eq:dotmfgroTprelapp}). Since $\indexnot=\indexnot_{\trs}$ in the case of interest here, 
$v:=n_{j}dx^{j}\in T_{\trs}^{*}\bM$. Moreover, $\bh^{il}\bh^{jp}\d_{t}\bh_{lp}$ is the $ij$'th component of $2\bk^{\sharp}$, $i,j\in \{1,\dots,d\}$. 
Combining these observations with (\ref{eq:bkbdtrssettingintrocotan}) yields
\begin{equation}\label{eq:estsffconttomfgT}
\begin{split}
|\bh^{il}\bh^{jp}\d_{t}\bh_{lp}n_{i}n_{j}| = & 2|\bk^{\sharp}(v,v)|\leq 2C_{a}e^{-\eta_{\trs}t}|\bh^{ij}n_{i}n_{j}|
\leq 2\shiftbd{0}^{-2}C_{a}e^{-\eta_{\trs}t}\mfg_{\roT}^{2}
\end{split}
\end{equation}
for all $t\geq 0$, where we also appealed to (\ref{eq:hinvbhinvequ}). Consider the second term on the right hand side of 
(\ref{eq:dotmfgroTprelapp}). Note that
\begin{equation}\label{eq:sigmatrsprelest}
|\chi^{i}n_{i}|=|v(\chi)|\leq C_{b}e^{-\eta_{\trs}t}|v|_{\bge}\leq \shiftbd{0}^{-1}C_{b}e^{-\eta_{\trs}t}\mfg_{\roT}
\end{equation}
for all $t\geq 0$, 
where we appealed to (\ref{eq:hinvbhinvequ}) and (\ref{eq:transpshiftcondintro}). Since $\chi$ is $C^{1}$-bounded, we thus obtain 
\begin{equation}\label{eq:estshiftconttomfgT}
|\chi^{i}n_{i}\dot{\chi}^{j}n_{j}|\leq C_{b}\shiftbd{0}^{-2}\shiftbd{1}e^{-\eta_{\trs}t}\mfg_{\roT}^{2}.
\end{equation}
Combining (\ref{eq:estsffconttomfgT}) and (\ref{eq:estshiftconttomfgT}) yields 
\begin{equation}\label{eq:mfgTdotmfgesttrs}
2|\mfg_{\roT}(\indexnot_{\trs},t)\dot{\mfg}_{\roT}(\indexnot_{\trs},t)|
\leq 2\shiftbd{0}^{-2}(C_{a}+\shiftbd{1}C_{b})e^{-\eta_{\trs}t}\mfg_{\roT}^{2}(\indexnot_{\trs},t)
\end{equation}
for all $t\geq 0$. Let us now turn to the second term on the right hand side of (\ref{eq:elldottransprel}). Let $j\in \{1,\dots,R_{\trs}\}$,
$t\in [0,\infty)$ and $\xi$ be a unit cotangent vector (with respect to $\bge$) such that $\xi^{\sharp}$ is tangent to the $M_{r_{j}}$-factor 
in $\bM$. Then (\ref{eq:bkbdtrssettingintrocotan}) yields
\begin{equation}\label{eq:ardotthrartranstransest}
\left|\textstyle{\frac{\dot{a}_{r_{j}}(t)}{a_{r_{j}}(t)}}\right|\leq C_{a}e^{-\eta_{\trs}t};
\end{equation}
note that this estimate holds for all $t\geq 0$ and all $j\in \{1,\dots,R_{\trs}\}$. Thus the second term on the right hand side of 
(\ref{eq:elldottransprel}) can be estimated by 
\begin{equation}\label{eq:Riemmanfdesttrs}
2\left|\textstyle{\sum}_{j=1}^{R_{\trs}}\frac{\dot{a}_{r_{j}}(t)}{a_{r_{j}}(t)}a_{r_{j}}^{-2}(t)\nu_{r_{j},i_{r_{j}}}^{2}(\indexnot_{\trs})\right|
\leq 2C_{a}e^{-\eta_{\trs}t}\textstyle{\sum}_{j=1}^{R_{\trs}}a_{r_{j}}^{-2}(t)\nu_{r_{j},i_{r_{j}}}^{2}(\indexnot_{\trs})
\end{equation}
for all $t\geq 0$. Combining (\ref{eq:elldottransprel}), (\ref{eq:mfgTdotmfgesttrs}) and (\ref{eq:Riemmanfdesttrs}) yields
(\ref{eq:dotelltrsestfinal}) for all $t\geq 0$. Note also that (\ref{eq:sigmatrsprelest}) implies that (\ref{eq:sigmatrsestfinal}) holds 
for all $t\geq 0$.

\textbf{The silent directions.} Next, let us consider (\ref{eq:elldottransprel}) in the case that $\indexnot_{\trs}$ is replaced by 
$\indexnot_{\rosil}$. Since $\indexnot=\indexnot_{\rosil}$ in the case of interest here, $v:=n_{i}dx^{i}\in T^{*}_{\rosil}\bM$. The 
first term on the right hand side of (\ref{eq:elldottransprel}) can be expressed as in (\ref{eq:dotmfgroTprelapp}). Consider the first 
term on the right hand side of (\ref{eq:dotmfgroTprelapp}). Due to (\ref{eq:bkbdtrssettingsilentpartintrocotan}) this term can be estimated by 
\begin{equation}\label{eq:transsilentfirstpart}
\begin{split}
-\bh^{il}\bh^{jp}\d_{t}\bh_{lp}n_{i}n_{j} = & -2\bk^{\sharp}(v,v)\leq -2(\b_{\rosil}-\betafun_{\rosil})\bh^{ij}n_{i}n_{j}\\
 \leq & -2(\b_{\rosil}-\shiftbd{0}^{-2}\betafun_{\rosil})\mfg_{\roT}^{2},
\end{split}
\end{equation}
where we appealed to (\ref{eq:bkblock}) in the first step and (\ref{eq:hinvbhinvequ}) in the last step. Consider the second term on 
the right hand side of (\ref{eq:dotmfgroTprelapp}). It can be estimated by 
\begin{equation}\label{eq:transsilentsecondpart}
2\shiftbd{0}^{-2}|\chi|_{\bge}|\dot{\chi}|_{\bge}\mfg_{\roT}^{2}\leq 2\shiftbd{0}^{-2}\betafun \mfg_{\roT}^{2},
\end{equation}
where we appealed to (\ref{eq:chichidottrsest}). Finally, due to (\ref{eq:bkblock}) and 
(\ref{eq:bkbdtrssettingsilentpartintrocotan}),
\begin{equation}\label{eq:transsilentthirdpart}
-2\textstyle{\sum}_{j=1}^{R_{\rosil}}\frac{\dot{a}_{\bre_{j}}(t)}{a_{\bre_{j}}(t)}a_{\bre_{j}}^{-2}(t)\nu_{\bre_{j},i_{\bre_{j}}}^{2}(\indexnot_{\rosil})
\leq -2[\b_{\rosil}-\betafun_{\rosil}(t)]\textstyle{\sum}_{j=1}^{R_{\rosil}}a_{\bre_{j}}^{-2}(t)\nu_{\bre_{j},i_{\bre_{j}}}^{2}(\indexnot_{\rosil}).
\end{equation}
Combining (\ref{eq:transsilentfirstpart}), (\ref{eq:transsilentsecondpart}) and (\ref{eq:transsilentthirdpart}) yields
(\ref{eq:dotellsilestfinal}) for all $t\geq 0$.

\textbf{The coefficients of the equation.}
Next, let us turn to the coefficients $g^{ij}$ and $a_{r}$. Note, first of all, that the estimate (\ref{eq:dotelltrsestfinal}) implies that 
$\mfg(\indexnot_{\trs},t)$ converges exponentially to a limit, say $\mfg_{\infty}(\indexnot_{\trs})$, and that 
\begin{equation}\label{eq:mfgsqindtrslimexpest}
\left|\frac{\mfg^{2}(\indexnot_{\trs},t)}{\mfg_{\infty}^{2}(\indexnot_{\trs})}-1\right|+
\left|\frac{\mfg_{\infty}^{2}(\indexnot_{\trs})}{\mfg^{2}(\indexnot_{\trs},t)}-1\right|\leq Ce^{-\eta_{\trs}t}
\end{equation}
for all $t\geq 0$ and all $\indexnot\in\EFindexset$ such that $\indexnot_{\trs}\neq 0$, where $C$ only depends on $C_{A}$ and $\eta_{\trs}$. This 
estimate implies the existence of a constant $C\geq 1$, depending only on $C_{A}$ and $\eta_{\trs}$, such that 
\begin{equation}\label{eq:mfgsqindtrsestzinfcomp}
C^{-1}\mfg^{2}(\indexnot_{\trs},0)\leq \mfg^{2}_{\infty}(\indexnot_{\trs})\leq C\mfg^{2}(\indexnot_{\trs},0)
\end{equation}
for all $\indexnot\in\EFindexset$. Due to this estimate, $\mfg_{\infty}^{2}(\indexnot_{\trs})$, considered as a quadratic form, is 
positive definite. Combining (\ref{eq:mfgsqindtrslimexpest}) and (\ref{eq:mfgsqindtrsestzinfcomp}) yields
\[
|\mfg^{2}(\indexnot_{\trs},t)-\mfg^{2}_{\infty}(\indexnot_{\trs})|\leq Ce^{-\eta_{\trs}t}\mfg^{2}(\indexnot_{\trs},0)
\]
for all $t\geq 0$ and all $\indexnot\in\EFindexset$, where $C$ only depends on $C_{A}$ and $\eta_{\trs}$. This estimate 
implies that there are $g^{j_{k}j_{l}}_{\infty}$, $k,l\in\{1,\dots,d_{\trs}\}$, and $q_{\infty,r_{j}}$, $j=1,\dots,R_{\trs}$, such that (\ref{eq:gacoefftrsdir})
holds for all $t\geq 0$, where $C$ only depends on $C_{A}$, $\eta_{\trs}$, $g^{ij}(0)$ and $a_{r}(0)$. Note also that 
\[
\textstyle{\mfg}_{\infty}^{2}(\indexnot_{\trs})=g^{j_{k}j_{l}}_{\infty}n_{j_{k}}n_{j_{l}}
+\sum_{j=1}^{R_{\trs}}q_{\infty,r_{j}}\nu_{r_{j},i_{r_{j}}}^{2}(\indexnot_{\trs}).
\]
Since the left hand side of this equality is a positive definite quadratic form, $g^{j_{k}j_{l}}_{\infty}$ are the components of a positive 
definite matrix, and $q_{\infty,r_{j}}>0$. 

Turning to the silent directions, note that (\ref{eq:dotellsilestfinal}) yields
\begin{equation}\label{eq:mfgsqrosilexpdecest}
\mfg^{2}(\indexnot_{\rosil},t)\leq Ce^{-2\b_{\rosil}t}\mfg^{2}(\indexnot_{\rosil},0)
\end{equation}
for all $t\geq 0$, where $C$ only depends on $\shiftbd{0}$, $c_{\rosil}:=\|\betafun_{\rosil}\|_{1}$ and $c_{\betafun}:=\|\betafun\|$. Thus
(\ref{eq:gacoeffsildir}) holds for all $t\geq 0$, $k,l\in\{1,\dots,d_{\rosil}\}$ and $j=1,\dots,R_{\rosil}$, where $C$ only depends on 
$g^{ij}(0)$, $a_{r}(0)$, $\shiftbd{0}$, $c_{\rosil}$ and $c_{\betafun}$. Next, note that the Cauchy-Schwarz inequality yields
\begin{equation}\label{eq:guppmixsiltrsest}
|g^{j_{k}\bj_{l}}(t)|\leq [g^{j_{k}j_{k}}(t)]^{1/2}[g^{\bj_{l}\bj_{l}}(t)]^{1/2}\leq Ce^{-\b_{\rosil}t}
\end{equation}
for all $t\geq 0$, where $C$ only depends on $C_{A}$, $\eta_{\trs}$, $g^{ij}(0)$, $a_{r}(0)$, $\shiftbd{0}$, $c_{\rosil}$ 
and $c_{\betafun}$. Thus (\ref{eq:gjkbjlest}) holds. 

Next let us turn to the shift vector field $\chi^{l}=g^{0l}$. Note that 
\[
\sigma(\indexnot_{\trs},t)=\frac{n_{j_{i}}g^{0j_{i}}(t)}{\mfg(\indexnot_{\trs},t)}.
\]
Combining this equality with (\ref{eq:sigmatrsestfinal}), (\ref{eq:mfgsqindtrslimexpest}) and (\ref{eq:mfgsqindtrsestzinfcomp}) yields
\[
|n_{j_{i}}g^{0j_{i}}(t)|\leq Ce^{-\eta_{\trs}t}\mfg(\indexnot_{\trs},0)
\]
for all $t\geq 0$, where $C$ only depends on $C_{A}$, $C_{B}$ and $\eta_{\trs}$. Thus the first estimate appearing in (\ref{eq:shiftestparttrscase}) 
holds. Finally, if $n\in\zn{d}$ corresponds to a $\indexnot=\indexnot_{\rosil}$, then 
\[
|\chi^{i}(t)n_{i}|\leq \shiftbd{0}^{-1}(1-\shiftbd{0}^{2})^{1/2}\mfg_{\roT}(\indexnot_{\rosil},t)
\]
for all $t\geq 0$. Combining this estimate with (\ref{eq:mfgsqrosilexpdecest}) yields the second part of (\ref{eq:shiftestparttrscase}), where
$C_{2}$ only depends on $g^{ij}(0)$, $\shiftbd{0}$, $c_{\rosil}$ and $c_{\betafun}$. The lemma follows. 
\end{proof}

\section[A dominant noisy spatial direction]{Conditions ensuring the existence of a dominant noisy 
spatial direction}
 
The purpose of the present section is to relate the conditions appearing in Chapter~\ref{chapter:domnoisspdirintro}
with the conditions appearing in Part~\ref{part:dominantnoisyspdirection}.

\begin{lemma}\label{lemma:geometrictoanalyticnoise}
Let $(M,g)$ be a canonical separable cosmological model manifold. Assume that there is 
an $l\in \{1,\dots,d\}$ such that $g$ has a geometric dominant noisy spatial $\so$-direction corresponding to $l$; 
cf. Definition~\ref{def:domnoisyspasodir}. Assume, moreover, that there is an $0<\shiftbd{0}\leq 1$ such that (\ref{eq:ddtuniftimelike}) holds for all $t\geq 0$
and constants $C_{\rosh},\eta_{\rosh}>0$ such that 
\begin{equation}\label{eq:chichidotbdnoise}
|\chi(t)|_{\bge}+|\dot{\chi}(t)|_{\bge}\leq C_{\rosh}e^{-\eta_{\rosh}t}
\end{equation}
for all $t\geq 0$. Finally, assume that $\bk$ is $C^{0}$-future bounded; i.e., that there is a constant $C_{0}\geq 0$ such that 
\begin{equation}\label{eq:futureczbkbd}
|\bk|_{\bge}\leq C_{0}
\end{equation}
for all $t\geq 0$. Then there are constants $h^{ll}_{\infty},C>0$ such that 
\begin{equation}\label{eq:hlldthllestnoise}
|e^{-2\b_{\ron}t}h^{ll}(t)-h^{ll}_{\infty}|+|e^{-2\b_{\ron}t}\d_{t}h^{ll}(t)-2\b_{\ron}h^{ll}_{\infty}| \leq Ce^{-\eta_{\romar}t}
\end{equation}
for all $t\geq 0$, where $\eta_{\romar}:=\min\{2\eta_{\rosh},\eta_{\ron}\}$. Moreover, 
\begin{equation}\label{eq:gupsubnoiseest}
e^{-2\b_{\ron}t}\textstyle{\sum}_{j,k\neq l}|g^{jk}(t)|\leq Ce^{-2\eta_{\ron}t},\ \ \
e^{-2\b_{\ron}t}\sum_{r=1}^{R}a_{r}^{-2}(t)\leq Ce^{-2\eta_{\ron}t}
\end{equation}
for all $t\geq 0$. In addition,
\begin{equation}\label{eq:gujlnormexpdec}
e^{-2\b_{\ron}t}\textstyle{\sum}_{j\neq l}|g^{jl}(t)|\leq Ce^{-\eta_{\ron}t}
\end{equation}
for all $t\geq 0$.

Turning to the time derivatives, 
\begin{equation}\label{eq:dthuijnotbothlest}
e^{-2\b_{\ron}t}\textstyle{\sum}_{i,j\neq l}|\d_{t}h^{ij}(t)| \leq Ce^{-2\eta_{\ron}t},\ \ \
e^{-2\b_{\ron}t}\sum_{j\neq l}|\d_{t}h^{jl}(t)| \leq Ce^{-\eta_{\ron}t}
\end{equation}
for all $t\geq 0$. Finally, 
\begin{equation}\label{eq:dtarmtnormest}
e^{-2\b_{\ron}t}|\d_{t}a_{r}^{-2}(t)|\leq Ce^{-2\eta_{\ron}t}
\end{equation}
for all $t\geq 0$.
\end{lemma}
\begin{remark}\label{remark:constCdepnoisysetting}
The constant $C$ only depends on $g^{ij}(0)$, $\bge^{ij}(0)$, $a_{r}(0)$, $\|\betafun\|_{1}$, $\shiftbd{0}$, $\eta_{\rosh}$, 
$C_{\rosh}$, $C_{\ron}$, $C_{0}$, $\eta_{\ron}$ and $\b_{\ron}$
\end{remark}
\begin{remark}\label{remark:domnoisspdirlowbdonbk}
If, in addition to the assumptions of the lemma, there is a non-negative continuous function $\betafun_{\low}\in L^{1}([0,\infty))$ such
that (\ref{eq:bklowbddomnoisspdir}) holds for all $t\geq 0$, then there is a constant $C>0$ (depending only on $\b_{\ron}$, $\|\betafun_{\low}\|_{1}$, 
$\shiftbd{0}$, $C_{\rosh}$ and $\eta_{\rosh}$) such that $e^{-\b_{\ron}t}\mfg(\indexnot,t)\geq C\nu_{\ron}(\indexnot)$ for all $t\geq 0$ and all
$\indexnot\in\EFindexset$. Here $\nu_{\ron}(\indexnot)$ is defined by (\ref{eq:nurondefintro}). In order to justify that the limit appearing
in (\ref{eq:nurondefintro}) exists, note that 
\begin{equation*}
\begin{split}
e^{-2\b_{\ron}t}\mfg^{2}(\indexnot,t) = & \textstyle{\sum}_{i,j\neq l}e^{-2\b_{\ron}t}g^{ij}(t)n_{i}n_{j}
+2\textstyle{\sum}_{j\neq l}e^{-2\b_{\ron}t}g^{jl}(t)n_{j}n_{l}\\
 & +e^{-2\b_{\ron}t}g^{ll}(t)n_{l}^{2}+
\textstyle{\sum}_{r=1}^{R}e^{-2\b_{\ron}t}a_{r}^{-2}(t)\nu_{r,i_{r}}^{2}(\indexnot)\rightarrow h^{ll}_{\infty}n_{l}^{2},
\end{split}
\end{equation*}
due to the conclusions of the lemma. In order to justify the lower bound $e^{-\b_{\ron}t}\mfg(\indexnot,t)\geq C\nu_{\ron}(\indexnot)$, note that an 
argument similar to the beginning of the proof of Lemma~\ref{lemma:condwsil} yields
\begin{equation*}
\begin{split}
2\mfg_{\roT}\dot{\mfg}_{\roT} \leq & 2(\b_{\ron}+\betafun_{\low})\bh^{ij}n_{i}n_{j}-2\chi^{i}\dot{\chi}^{j}n_{i}n_{j}\\
 \leq & 2\b_{\ron}\mfg_{\roT}^{2}+2\b_{\ron}\chi^{i}\chi^{j}n_{i}n_{j}+2\shiftbd{0}^{-2}\betafun_{\low}\mfg_{\roT}^{2}
+2\shiftbd{0}^{-2}|\chi|_{\bge}|\dot{\chi}|_{\bge}\mfg_{\roT}^{2}.
\end{split}
\end{equation*}
The remaining part of the time derivative of $\mfg^{2}$ can be estimated more easily. Appealing to (\ref{eq:chichidotbdnoise})
yields $\dot{\ell}(\indexnot,t)\leq\b_{\ron}+\betafun_{2}(t)$ for all $t\geq 0$ and $0\neq \indexnot\in\EFindexset$. Here 
$\betafun_{2}\in L^{1}([0,\infty))$ is a non-negative continuous function whose $L^{1}$-norm only depends on $\b_{\ron}$, $\|\betafun_{\low}\|_{1}$, 
$\shiftbd{0}$, $C_{\rosh}$ and $\eta_{\rosh}$. Thus
\[
\d_{t}\ln[e^{-\b_{\ron}t}\mfg(\indexnot,t)]\leq \betafun_{2}(t)
\]
for all $t\geq 0$ and all $0\neq\indexnot\in\EFindexset$. Integrating this estimate from $t$ to $\infty$ yields 
$e^{-\b_{\ron}t}\mfg(\indexnot,t)\geq e^{-\|\betafun_{2}\|_{1}}\nu_{\ron}(\indexnot)$ for all $t\geq 0$ and all $0\neq\indexnot\in\EFindexset$.
\end{remark}
\begin{proof}
\textbf{The components of the inverse metric in the dominant noisy direction.} Note that $dx^{l}\in T^{*}_{\ron}\bM$. Appealing
to (\ref{eq:condonkdomnoisdirintro}) with $\xi:=dx^{l}$ thus yields
\begin{equation}\label{eq:dtexpmtbronbhhllestprel}
\left|\frac{1}{2}\bh^{il}\bh^{jl}\d_{t}\bh_{ij}+\b_{\ron}\bh^{ll}\right|\leq C_{\ron}e^{-\eta_{\ron}t}\bh^{ll}
\end{equation}
(no summation on $l$) for all $t\geq 0$. This estimate is, in its turn, equivalent to 
\begin{equation}\label{eq:dtexpmtbronbhhllest}
|\d_{t}(e^{-2\b_{\ron}t}\bh^{ll})|\leq 2C_{\ron}e^{-\eta_{\ron}t}e^{-2\b_{\ron}t}\bh^{ll}
\end{equation}
for all $t\geq 0$. Thus
\begin{equation}\label{eq:roughestexpmtbronbhhll}
\exp(-2C_{\ron}/\eta_{\ron})\leq\frac{e^{-2\b_{\ron}t}\bh^{ll}(t)}{\bh^{ll}(0)}\leq \exp(2C_{\ron}/\eta_{\ron})
\end{equation}
for all $t\geq 0$. Combining (\ref{eq:dtexpmtbronbhhllest}) and (\ref{eq:roughestexpmtbronbhhll}) yields the conclusion that 
$e^{-2\b_{\ron}t}\bh^{ll}(t)$ converges to, say, $\bh^{ll}_{\infty}>0$. Moreover,
\begin{equation}\label{eq:bhullasest}
|e^{-2\b_{\ron}t}\bh^{ll}(t)-\bh^{ll}_{\infty}|\leq C\bh^{ll}(0)e^{-\eta_{\ron}t}
\end{equation}
for all $t\geq 0$, where $C$ only depends on $C_{\ron}$ and $\eta_{\ron}$. Combining this estimate with (\ref{eq:dtexpmtbronbhhllestprel})
and (\ref{eq:roughestexpmtbronbhhll}) yields
\begin{equation}\label{eq:dtbhullasest}
|e^{-2\b_{\ron}t}\d_{t}\bh^{ll}(t)-2\b_{\ron}\bh^{ll}_{\infty}|\leq C\bh^{ll}(0)e^{-\eta_{\ron}t}
\end{equation}
for all $t\geq 0$, where $C$ only depends on $C_{\ron}$, $\b_{\ron}$ and $\eta_{\ron}$. In order to estimate $h^{ll}(t)$, we also need to 
estimate $\chi^{l}$ and $\dot{\chi}^{l}$. However, combining (\ref{eq:chichidotbdnoise}) with estimates of the form (\ref{eq:chiuinlibasest}) 
(with $n_{i}=\de_{il}$) yields
\[
e^{-\b_{\ron}t}|\chi^{l}(t)|+e^{-\b_{\ron}t}|\dot{\chi}^{l}(t)|\leq C[\bh^{ll}(0)]^{1/2}e^{-\eta_{\rosh}t} 
\]
for all $t\geq 0$, where we appealed to (\ref{eq:roughestexpmtbronbhhll}) and 
$C$ only depends on $C_{\rosh}$, $C_{\ron}$ and $\eta_{\ron}$. Combining this estimate with (\ref{eq:hinvcomform}),
(\ref{eq:bhullasest}) and (\ref{eq:dtbhullasest}) yields
\begin{align}
|e^{-2\b_{\ron}t}h^{ll}(t)-\bh^{ll}_{\infty}| \leq & C\bh^{ll}(0)e^{-\eta_{\romar}t},\label{eq:hullestnoisefinpf}\\
|e^{-2\b_{\ron}t}\d_{t}h^{ll}(t)-2\b_{\ron}\bh^{ll}_{\infty}| \leq & C\bh^{ll}(0)e^{-\eta_{\romar}t}\label{eq:dthullestnoisefinpf}
\end{align}
for all $t\geq 0$, where $\eta_{\romar}:=\min\{2\eta_{\rosh},\eta_{\ron}\}$ and $C$ only depends on $C_{\rosh}$, $C_{\ron}$, $\eta_{\ron}$
and $\b_{\ron}$. Thus (\ref{eq:hlldthllestnoise}) holds. 

\textbf{The subdominant directions.} 
Let us consider the remaining directions. In order to do so, let $\indexnot\in\EFindexset$ be such that $\nu_{\roT,l}(\indexnot)=0$.
If $n=\nu_{\roT}(\indexnot)$, we then have $v:=n_{i}dx^{i}\in T^{*}_{\rosub}\bM$. 
In analogy with the proof of Lemma~\ref{lemma:geometrictoanalytictrans}, let us consider the time derivative of $\mfg^{2}(\indexnot,\cdot)$.
As before, this derivative can be divided into several different parts. Let us start by considering the first term on the right hand 
side of (\ref{eq:dotmfgroTprelapp}). It can be estimated by
\begin{equation}\label{eq:dtbhjknjnksildomnoise}
\begin{split}
\d_{t}\bh^{jk}n_{j}n_{k} = & -\bh^{ji}\bh^{kp}\d_{t}\bh_{ip}n_{j}n_{k}=-2\bk^{\sharp}(v,v)\\
 \leq & 2(\b_{\ron}-\eta_{\ron}+\betafun)\bh^{jk}n_{j}n_{k}
\end{split}
\end{equation}
for all $t\geq 0$, where we appealed to (\ref{eq:bkblock}) and (\ref{eq:bkwwlbsubtmintro}). On the other hand, combining (\ref{eq:hinvcomform}),
(\ref{eq:lambdakappacontrest}), (\ref{eq:hinvbhinvequ}) and (\ref{eq:chichidotbdnoise}) yields
\[
|\bh^{jk}(t)n_{j}n_{k}-h^{jk}(t)n_{j}n_{k}|\leq Ce^{-2\eta_{\rosh}t}\mfg_{\roT}^{2}(\indexnot,t)
\]
for all $t\geq 0$, 
where $C$ only depends on $\shiftbd{0}$ and $C_{\rosh}$. Combining this estimate with (\ref{eq:hinvbhinvequ}) and 
(\ref{eq:dtbhjknjnksildomnoise}) yields
\begin{equation}\label{eq:dtbhjknjnksildomnoiseimprhs}
\d_{t}\bh^{jk}(t)n_{j}n_{k} \leq 2(\b_{\ron}-\eta_{\ron})\mfg_{\roT}^{2}(\indexnot,t)+C(\betafun+e^{-2\eta_{\rosh}t})\mfg_{\roT}^{2}(\indexnot,t)
\end{equation}
for all $t\geq 0$, 
where $C$ only depends on $\eta_{\ron}$, $\b_{\ron}$, $\shiftbd{0}$ and $C_{\rosh}$. Next, the second term on the right hand side of 
(\ref{eq:dotmfgroTprelapp}) can be estimated using (\ref{eq:shiftmfgtestapp}). In fact, combining (\ref{eq:shiftmfgtestapp}) with 
(\ref{eq:chichidotbdnoise}) yields
\[
2|\chi^{i}(t)n_{i}\dot{\chi}^{j}(t)n_{j}|\leq Ce^{-2\eta_{\rosh}t}\mfg_{\roT}^{2}(\indexnot,t)
\]
for all $t\geq 0$, where $C$ only depends on $\shiftbd{0}$ and $C_{\rosh}$. Combining this estimate with (\ref{eq:dtbhjknjnksildomnoiseimprhs})
yields
\[
2\mfg_{\roT}\dot{\mfg}_{\roT}\leq 2(\b_{\ron}-\eta_{\ron}+\betafun_{1})\mfg_{\roT}^{2}
\]
for all $t\geq 0$, where
\[
\betafun_{1}(t)=C[\betafun(t)+e^{-2\eta_{\rosh}t}]
\]
and $C\geq 1$ only depends on $\eta_{\ron}$, $\b_{\ron}$, $\shiftbd{0}$ and $C_{\rosh}$. Finally, a similar estimate holds for the time derivative of the remaining part
of $\mfg^{2}$; cf. the arguments presented in connection with (\ref{eq:transsilentthirdpart}). Summing up,
\[
2\mfg\dot{\mfg}\leq 2(\b_{\ron}-\eta_{\ron}+\betafun_{1})\mfg^{2}
\]
for all $t\geq 0$. Introducing $\mfh(\indexnot,t):=e^{-(\b_{\ron}-\eta_{\ron})t}\mfg(\indexnot,t)$, this estimate can be written
\[
\d_{t}\mfh^{2}\leq 2\betafun_{1}\mfh^{2}
\]
for all $t\geq 0$. Thus
\[
e^{-2(\b_{\ron}-\eta_{\ron})t}\mfg^{2}(\indexnot,t)\leq e^{2c_{1}}\mfg^{2}(\indexnot,0)
\]
for all $t\geq 0$, where $c_{1}:=\|\betafun_{1}\|_{1}$ only depends on $\|\betafun\|_{1}$, $\shiftbd{0}$, $\eta_{\ron}$, $\b_{\ron}$, $\eta_{\rosh}$ and 
$C_{\rosh}$. This estimate yields (\ref{eq:gupsubnoiseest}). 
Finally, similarly to the proof of the first estimate in (\ref{eq:guppmixsiltrsest}), $j\neq l$ yields
\[
e^{-2\b_{\ron}t}|g^{jl}(t)|\leq [e^{-2\b_{\ron}t}g^{jj}(t)]^{1/2}[e^{-2\b_{\ron}t}g^{ll}(t)]^{1/2}\leq Ce^{-\eta_{\ron}t}
\]
for all $t\geq 0$, where we appealed to (\ref{eq:hinvbhinvequ}), (\ref{eq:roughestexpmtbronbhhll}) and (\ref{eq:gupsubnoiseest}), and
$C$ only depends on the constants stated in Remark~\ref{remark:constCdepnoisysetting}. 

\textbf{The time derivatives in the subdominant directions.} Due to (\ref{eq:kappavwcontrest}),  
\[
|\d_{t}\bh^{ij}|=|\bh^{ip}\bh^{jq}\d_{t}\bh_{pq}|=2|\bk_{pq}\bh^{ip}\bh^{jq}|\leq 2|\bk|_{\bge}(\bh^{ii})^{1/2}(\bh^{jj})^{1/2}
\]
(no summation on $i$ or $j$) for all $t\geq 0$. Keeping (\ref{eq:hinvcomform}) and (\ref{eq:hinvbhinvequ}) in mind, this estimate implies
\begin{equation*}
\begin{split}
|\d_{t}h^{ij}| \leq & |\d_{t}\bh^{ij}|+|\dot{\chi}^{i}||\chi^{j}|+|\dot{\chi}^{j}||\chi^{i}|\\
 \leq & 2\shiftbd{0}^{-2}[|\bk|_{\bge}+|\chi|_{\bge}\cdot|\dot{\chi}|_{\bge}](h^{ii})^{1/2}(h^{jj})^{1/2}
\end{split}
\end{equation*}
for all $t\geq 0$, 
where we appealed to (\ref{eq:lambdakappacontrest}) in the second step. Keeping (\ref{eq:hinvbhinvequ}), (\ref{eq:chichidotbdnoise}), (\ref{eq:futureczbkbd}),
(\ref{eq:gupsubnoiseest}) and (\ref{eq:roughestexpmtbronbhhll}) in mind, this estimate implies that (\ref{eq:dthuijnotbothlest}) holds. Finally, note that 
\[
e^{-2\b_{\ron}t}|\d_{t}a_{r}^{-2}|=2e^{-2\b_{\ron}t}a_{r}^{-2}|a_{r}^{-1}\d_{t}a_{r}|\leq 2C_{0}e^{-2\b_{\ron}t}a_{r}^{-2}\leq Ce^{-2\eta_{\ron}t}
\]
for all $t\geq 0$, where $C$ only depends on the constants stated in Remark~\ref{remark:constCdepnoisysetting}. Thus (\ref{eq:dtarmtnormest}) holds, 
and the lemma follows. 
\end{proof}

In the case that $g$ has a geometric dominant noisy spatial direction corresponding to $M_{r}$, there is an analogous result.

\begin{lemma}\label{lemma:geometrictoanalyticnoiseRie}
Let $(M,g)$ be a canonical separable cosmological model manifold. Assume that there is 
an $r_{\ron}\in \{1,\dots,R\}$ such that $g$ has a geometric dominant noisy spatial direction corresponding to $M_{r_{\ron}}$; 
cf. Definition~\ref{def:domnoisyspaMrdir}.
Assume, moreover, that there are constants $C_{\rosh},C_{0},\eta_{\rosh}>0$ such that (\ref{eq:chichidotbdnoise}) and 
(\ref{eq:futureczbkbd}) hold for all $t\geq 0$. Then there are constants $q_{\infty,\ron},C>0$ such that 
\begin{equation}\label{eq:qinftronasnoise}
|e^{-2\b_{\ron}t}a_{r_{\ron}}^{-2}(t)-q_{\infty,\ron}|+|e^{-2\b_{\ron}t}\d_{t}a_{r_{\ron}}^{-2}(t)-2\b_{\ron}q_{\infty,\ron}| \leq Ce^{-\eta_{\ron}t},
\end{equation}
for all $t\geq 0$. Moreover, 
\begin{equation}\label{eq:gupsubnoiseestRie}
e^{-2\b_{\ron}t}\textstyle{\sum}_{j,k}|g^{jk}(t)|\leq Ce^{-2\eta_{\ron}t},\ \ \
e^{-2\b_{\ron}t}\sum_{r\neq r_{\ron}}a_{r}^{-2}(t)\leq Ce^{-2\eta_{\ron}t}
\end{equation}
for all $t\geq 0$. Finally,
\begin{align}
e^{-2\b_{\ron}t}\textstyle{\sum}_{i,j}|\d_{t}h^{ij}(t)| \leq & Ce^{-2\eta_{\ron}t},\label{eq:dthuijestRie}\\
e^{-2\b_{\ron}t}\textstyle{\sum}_{r\neq r_{\ron}}|\d_{t}a_{r}^{-2}(t)| \leq & Ce^{-2\eta_{\ron}t}\label{eq:dtarmtnormestRie}
\end{align}
for all $t\geq 0$.
\end{lemma}
\begin{remark}
The constant $C$ appearing in the estimates has the dependence described in Remark~\ref{remark:constCdepnoisysetting}.
\end{remark}
\begin{remark}\label{remark:domnoisspdirlowbdonbkgendir}
If, in addition to the assumptions of the lemma, there is a non-negative continuous function $\betafun_{\low}\in L^{1}([0,\infty))$ such
that (\ref{eq:bklowbddomnoisspdir}) holds for all $t\geq 0$, then there is a constant $C>0$ (depending only on $\b_{\ron}$, $\|\betafun_{\low}\|_{1}$, 
$\shiftbd{0}$, $C_{\rosh}$ and $\eta_{\rosh}$) such that $e^{-\b_{\ron}t}\mfg(\indexnot,t)\geq C\nu_{\ron}(\indexnot)$ for all $t\geq 0$ and all
$\indexnot\in\EFindexset$. The justification of this statement is similar to the one given in Remark~\ref{remark:domnoisspdirlowbdonbk}. 
\end{remark}
\begin{proof}
The statements concerning the subdominant directions are a consequence of the arguments presented in the proof of 
Lemma~\ref{lemma:geometrictoanalyticnoise}. In order to prove (\ref{eq:qinftronasnoise}), note that since
(\ref{eq:condonkdomnoisdirintro}) holds for all $v\in T_{\ron}^{*}\bM$, 
\[
|\dot{a}_{r_{\ron}}/a_{r_{\ron}}+\b_{\ron}|\leq C_{\ron}e^{-\eta_{\ron}t}
\]
for all $t\geq 0$. Thus
\[
|\d_{t}(e^{-2\b_{\ron}t}a_{r_{\ron}}^{-2})|=|e^{-2\b_{\ron}t}[-2\dot{a}_{r_{\ron}}/a_{r_{\ron}}-2\b_{\ron}]a_{r_{\ron}}^{-2}|\leq 2C_{\ron}e^{-\eta_{\ron}t}\cdot e^{-2\b_{\ron}t}a_{r_{\ron}}^{-2}
\]
for all $t\geq 0$. In particular, there is a $q_{\infty,\ron}>0$ such that 
\[
\left|\ln \frac{q_{\infty,\ron}}{e^{-2\b_{\ron}t}a_{r_{\ron}}^{-2}(t)}\right|\leq 2C_{\ron}\eta_{\ron}^{-1}e^{-\eta_{\ron}t}
\]
for all $t\geq 0$. Using this estimate, it can be verified that (\ref{eq:qinftronasnoise}) holds. The lemma follows. 
\end{proof}

\section{The asymptotically diagonal setting}

In the present section, we relate the conditions stated in Section~\ref{section:asdiageqintronoise} with those appearing in 
Part~\ref{part:nondegcabeq}. We begin by deriving conclusions concerning $\bge_{ij}$, given the assumption that $\bge$ is 
$C^{0}$-asymptotically diagonal; cf. Definition~\ref{def:diagconvgeometricdefintro}. 

\begin{lemma}
Let $(M,g)$ be a canonical separable cosmological model manifold. Assume that the corresponding 
metric $\bge$ is $C^{0}$-asymptotically diagonal; cf. Definition~\ref{def:diagconvgeometricdefintro}. Then there is a constant $C>1$, 
depending only on the metric $g$, such that 
\begin{equation}\label{eq:bgeuiiinvequivtobgeii}
C^{-1}\leq \bge^{ii}(t)\bge_{ii}(t)\leq C
\end{equation}
for all $t\geq 0$ (no summation on $i$). Moreover, there is a constant $C>0$, depending only on the metric $g$, such that for 
$i\neq j$, $i,j\in\{1,\dots,d\}$, 
\begin{equation}\label{eq:bgeineqjbgeuiijjest}
|[\bge^{ii}(t)]^{1/2}\bge_{ij}(t)[\bge^{jj}(t)]^{1/2}|\leq Ce^{-\kappa_{\rood}t}
\end{equation}
for all $t\geq 0$ (no summation on $i$ or $j$).
\end{lemma}
\begin{proof}
Let $A\in\Mn{d}{\ro}$ be the matrix with components $\bge^{ij}$; let $D$ be the matrix obtained from $A$ by setting the off-diagonal components to zero; let $R$ be 
the remainder (so that $A=D+R$); and let $S:=-D^{-1/2}RD^{-1/2}$. If $S^{ij}$ denotes the $ij$'th element of $S$, note that $S^{ii}=0$ for $i\in\{1,\dots,d\}$ 
(no summation on $i$) and that if $i\neq j$, $i,j\in\{1,\dots,d\}$, then
\[
|S^{ij}|=|(\bge^{ii})^{-1/2}\bge^{ij}(\bge^{jj})^{-1/2}|\leq C_{\rood}e^{-\kappa_{\rood}t}
\]
(no summation on $i$ or $j$) for all $t\geq 0$, where we appealed to (\ref{eq:metricupoffdiaggeomintro}). In particular, 
$\|S\|\leq d\cdot C_{\rood}e^{-\kappa_{\rood}t}$ for all $t\geq 0$. On the other hand, 
\[
A=D+R=D^{1/2}(\Id_{d}-S)D^{1/2},
\]
so that 
\[
D^{1/2}A^{-1}D^{1/2}=(\Id_{d}-S)^{-1}=\textstyle{\sum}_{l=0}^{\infty}S^{l},
\]
assuming $t$ to be large enough that $\|S\|\leq 1/2$. In particular,
\[
\|D^{1/2}A^{-1}D^{1/2}-\Id_{d}\|\leq 2d\cdot C_{\rood}e^{-\kappa_{\rood}t}
\]
for all $t$ large enough that $\|S\|\leq 1/2$. Since the left hand side of this inequality is always bounded, there is a constant $C>0$ such that
\begin{equation}\label{eq:DohAinvAohmIdest}
\|D^{1/2}A^{-1}D^{1/2}-\Id_{d}\|\leq Ce^{-\kappa_{\rood}t}
\end{equation}
for all $t\geq 0$, where $C$ only depends on the metric $g$. One particular consequence of this estimate is that 
\[
|\bge^{ii}(t)\bge_{ii}(t)-1|\leq Ce^{-\kappa_{\rood}t}
\]
for all $t\geq 0$ (no summation on $i$), where $C$ only depends on the metric $g$. As a consequence of this 
estimate, there is a constant $C>1$ such that (\ref{eq:bgeuiiinvequivtobgeii}) holds. Finally, 
(\ref{eq:DohAinvAohmIdest}) implies that for $i\neq j$, $i,j\in\{1,\dots,d\}$, (\ref{eq:bgeineqjbgeuiijjest}) holds. 
\end{proof}

Next, we turn to the problem of verifying that there are geometric conditions ensuring that the (non-geometric) assumptions of 
Definition~\ref{def:convergeq} are satisfied. 

\begin{lemma}\label{lemma:geomecharofdiagdombalaconv}
Consider (\ref{eq:thesystemRge}). Assume the associated metric to be such that $(M,g)$ is a canonical separable cosmological model manifold. 
Assume (\ref{eq:thesystemRge}) to be $C^{2}$-balanced; cf. Definition~\ref{definition:Cobal}. Assume,
moreover, the shift vector field to be negligible, $\bk$ to be diagonally convergent and $\bge$ to be $C^{2}$-asymptotically diagonal; cf. 
Definition~\ref{def:diagconvgeometricdefintro}. Finally, assume (\ref{eq:thesystemRge}) to be such that the main coefficients are convergent; cf. 
Definition~\ref{def:maincoeffconvasdiagsetintro}. Then, for each $j\in\{1,\dots,d\}$, there are constants $g^{jj}_{\infty},C>0$ such that
\begin{align}
|e^{-2\b_{j}t}g^{jj}(t)-g^{jj}_{\infty}| \leq & Ce^{-\kappa_{\romar}t},\label{eq:gujjgeomconv}\\
|e^{-2\b_{j}t}\d_{t}g^{jj}(t)-2\b_{j}g^{jj}_{\infty}| \leq & Ce^{-\kappa_{\romar}t}\label{eq:dtgujjgeomconv}
\end{align}
for all $t\geq 0$, where $\kappa_{\romar}:=\min\{2\kappa_{\rosh},\kappa_{\rod}\}$ and $C$ only depends on $C_{\rosh}$, $C_{\rod}$, $\kappa_{\rod}$, $\b_{j}$,
$\chi^{j}(0)$ and $g^{jj}(0)$. Moreover, for each $r\in\{1,\dots,R\}$, there are constants $a_{r,\infty},C>0$ such that
\begin{align}
|e^{\bRie{r}t}a_{r}(t)-a_{r,\infty}| \leq & Ce^{-\kappa_{\rod}t},\label{eq:argeomconv}\\
|e^{\bRie{r}t}\dot{a}_{r}(t)+\bRie{r}a_{r,\infty}| \leq & Ce^{-\kappa_{\rod}t}\label{eq:dtargeomconv}
\end{align}
for all $t\geq 0$, where $C$ only depends on $C_{\rod}$, $\kappa_{\rod}$, $\bRie{r}$ and $a_{r}(0)$. There is also a constant $C>0$ such that 
\begin{equation}\label{eq:dotgujjddotugjjbdgeometrset}
e^{-2\b_{j}t}|\dot{g}^{jj}(t)|+e^{-2\b_{j}t}|\ddot{g}^{jj}(t)|\leq C
\end{equation}
for all $t\geq 0$ and all $j\in\{1,\dots,d\}$, where $C$ only depends on the constants appearing in the definition of $C^{2}$-balance, 
$C_{\rosh}$, $C_{\rod}$, $\kappa_{\rod}$, $\b_{j}$, $\chi^{j}(0)$ and $g^{jj}(0)$. Similarly, there is a constant $C>0$ such that 
\begin{equation}\label{eq:dotarddotarbdgeometrset}
e^{\bRie{r}t}|\dot{a}_{r}(t)|+e^{\bRie{r}t}|\ddot{a}_{r}(t)|\leq C
\end{equation}
for all $t\geq 0$ and all $r\in\{1,\dots,R\}$, where $C$ only depends on the constants appearing in the definition of $C^{2}$-balance, 
$C_{\rod}$, $\kappa_{\rod}$, $\bRie{r}$ and $a_{r}(0)$. There is also a constant $C>0$ such that 
\begin{equation}\label{eq:XjdotXjbd}
e^{-\b_{j}t}\|X^{j}(t)\|+e^{-\b_{j}t}\|\dot{X}^{j}(t)\|\leq C
\end{equation}
for all $t\geq 0$, where $C$ only depends on the constants appearing in the definition of $C^{2}$-balance, $C_{\rosh}$, $C_{\rod}$, $\kappa_{\rod}$, 
$\b_{j}$, $\chi^{j}(0)$ and $g^{jj}(0)$. Turning to the shift vector field, 
\begin{equation}\label{eq:shiftbditonormgeometr}
\sup_{0\neq \xi\in\rn{d}}\frac{|g^{0l}(t)\xi_{l}|}{[g^{jl}(t)\xi_{j}\xi_{l}]^{1/2}}+
\sup_{0\neq \xi\in\rn{d}}\frac{|\dot{g}^{0l}(t)\xi_{l}|}{[g^{jl}(t)\xi_{j}\xi_{l}]^{1/2}}\leq \shiftrb^{-1}C_{\rosh}e^{-\kappa_{\rosh}t}
\end{equation}
for all $t\geq 0$. Finally, there are constants $C,\eta_{\rood}>0$ such that
\begin{equation}\label{eq:odtermneggeomtoanal}
\textstyle{\sum}_{j\neq l}[|g^{jl}(t)\xi_{j}\xi_{l}|+|\dot{g}^{jl}(t)\xi_{j}\xi_{l}|+|\ddot{g}^{jl}(t)\xi_{j}\xi_{l}|]
\leq Ce^{-\eta_{\rood}t}\sum_{j}g^{jj}(t)\xi_{j}\xi_{j}
\end{equation}
for all $t\geq 0$ and $\xi\in\rn{d}$, where $C$ only depends on the metric $g$ and $\eta_{\rood}:=\min\{\kappa_{\rosh},\kappa_{\rood}\}$.
\end{lemma}
\begin{remark}\label{remark:geomecharofdiagdombalaconv}
If the assumptions of the lemma hold, then (\ref{eq:thesystemRge}) is diagonally dominated, balanced and convergent; cf. 
Definition~\ref{def:convergeq}. 
\end{remark}
\begin{remark}
If we, in addition to the assumptions of the lemma, require $|\ddot{\chi}|_{\bge}$ to be bounded by the right hand side of 
(\ref{eq:chidotchinegligibleintro}), then $\eta_{\rood}$ appearing in (\ref{eq:odtermneggeomtoanal}) can be improved to 
$\eta_{\rood}:=\min\{2\kappa_{\rosh},\kappa_{\rood}\}$.
\end{remark}
\begin{proof}
That (\ref{eq:gujjgeomconv}) and (\ref{eq:dtgujjgeomconv}) hold follows by an argument similar to the beginning of the proof of 
Lemma~\ref{lemma:geometrictoanalyticnoise}. That (\ref{eq:argeomconv}) and (\ref{eq:dtargeomconv}) hold is a consequence of
an argument similar to the beginning of the proof of Lemma~\ref{lemma:geometrictoanalyticnoiseRie}.

\textbf{The time derivatives of the diagonal components of the metric.}
In order to estimate the first and second time derivatives of $g^{jj}$, note that $g^{jj}(t)=\mfg^{2}_{\roT}(\indexnot,t)$, where $\indexnot$
has been chosen so that $\nu_{\roT,j}(\indexnot)=1$ and all the other components of $\indexnot$ vanish. On the other hand, 
(\ref{eq:dotmfgroTest}) yields
\[
|\mfg_{\roT}\dot{\mfg}_{\roT}|\leq \shiftrb^{-2}(|\bk_{\roT}|_{\bge}+|\chi|_{\bge}\cdot|\dot{\chi}|_{\bge})\mfg_{\roT}^{2}
\]
for all $t\geq 0$. Thus
\begin{equation}\label{eq:dotgujjestgeom}
e^{-2\b_{j}t}|\dot{g}^{jj}|\leq Ce^{-2\b_{j}t}g^{jj}\leq C
\end{equation}
for all $t\geq 0$, where $C>0$ only depends on the constants appearing in the definition of $C^{2}$-balance, $C_{\rosh}$, $C_{\rod}$, $\kappa_{\rod}$, $\b_{j}$,
$\chi^{j}(0)$ and $g^{jj}(0)$. Turning to $\ddot{g}^{jj}$, note that it can be written
\[
\ddot{g}^{jj}=2\mfg_{\roT}\ddot{\mfg}_{\roT}+2\dot{\mfg}_{\roT}\dot{\mfg}_{\roT}.
\]
On the other hand, the right hand side can be estimated by (\ref{eq:twodotmfgTsqpetc}). Thus $e^{-2\b_{j}t}|\ddot{g}^{jj}|\leq C$ for all $t\geq 0$,
where $C>0$ has the same dependence as in the case of (\ref{eq:dotgujjestgeom}). To conclude, (\ref{eq:dotgujjddotugjjbdgeometrset}) holds.
Due to (\ref{eq:bknormbgsq}), (\ref{eq:mlUbknormbgsq}), (\ref{eq:argeomconv}) and the fact that the equation is $C^{2}$-balanced, it is also clear 
that (\ref{eq:dotarddotarbdgeometrset}) holds. 

\textbf{The main coefficients.}
The estimate (\ref{eq:XjdotXjbd}) follows from the fact that $\mcX$ is $C^{1}$-future bounded; (\ref{eq:lambdakappacontrest}); 
(\ref{eq:hinvbhinvequ}); and (\ref{eq:gujjgeomconv}).

\textbf{The shift vector field.}
Turning to the shift vector field, note that if $\xi\in\rn{d}$, then (\ref{eq:lambdakappacontrest}) yields
\[
|g^{0l}\xi_{l}|\leq |\chi|_{\bge}(\bge^{ij}\xi_{i}\xi_{j})^{1/2}.
\]
Combining this estimate with (\ref{eq:hinvbhinvequ}) and (\ref{eq:chidotchinegligibleintro}) yields
\[
\frac{|g^{0l}(t)\xi_{l}|}{[g^{ij}(t)\xi_{i}\xi_{j}]^{1/2}}\leq \shiftrb^{-1}C_{\rosh}e^{-\kappa_{\rosh}t}
\]
for all $0\neq \xi\in\rn{d}$ and all $t\geq 0$. A similar estimate holds for $\dot{\chi}$. In fact, (\ref{eq:shiftbditonormgeometr}) holds. 

\textbf{The off-diagonal terms.}
Finally, let us consider the off-diagonal terms. Due to (\ref{eq:metricupoffdiaggeomintro}), we know that if $i\neq j$, $i,j\in\{1,\dots,d\}$,
then
\[
|\bge^{ij}(t)|=|\bge^{\sharp}(dx^{i},dx^{j})|\leq C_{\rood}e^{-\kappa_{\rood}t}[\bge^{ii}(t)]^{1/2}[\bge^{jj}(t)]^{1/2}
\]
for all $t\geq 0$ (no summation on $i$ or $j$). Thus
\begin{equation}\label{eq:bgeuijxixjdiffest}
\begin{split}
\textstyle{\sum}_{i\neq j}|\bge^{ij}(t)\xi_{i}\xi_{j}| \leq & \textstyle{\sum}_{i\neq j}C_{\rood}e^{-\kappa_{\rood}t}[\bge^{ii}(t)\xi_{i}^{2}]^{1/2}
[\bge^{jj}(t)\xi_{j}^{2}]^{1/2}\\
 \leq & d\cdot \shiftrb^{-2}C_{\rood}e^{-\kappa_{\rood}t}\textstyle{\sum}_{j}g^{jj}(t)\xi_{j}^{2}
\end{split}
\end{equation}
for all $t\geq 0$, where we appealed to (\ref{eq:hinvbhinvequ}). Before deriving a similar estimate with $\bge^{ij}$ on the left hand
side replaced by $g^{ij}$, note that 
\begin{equation}\label{eq:chiixiiestnoisegeom}
\begin{split}
|\chi^{i}\xi_{i}|^{2} \leq & |\chi|_{\bge}^{2}\bge^{ij}\xi_{i}\xi_{j}=|\chi|_{\bge}^{2}\left(\textstyle{\sum}_{j}\bge^{jj}\xi_{j}\xi_{j}
+\textstyle{\sum}_{i\neq j}\bge^{ij}\xi_{i}\xi_{j}\right)\\
 \leq & \shiftrb^{-2}C_{\rosh}^{2}e^{-2\kappa_{\rosh}t}(1+d\cdot C_{\rood}e^{-\kappa_{\rood}t})\textstyle{\sum}_{j}g^{jj}\xi_{j}\xi_{j}
\end{split}
\end{equation}
for all $t\geq 0$, where appeal to (\ref{eq:lambdakappacontrest}) in the first step; and (\ref{eq:hinvbhinvequ}),
(\ref{eq:chidotchinegligibleintro}) and (\ref{eq:bgeuijxixjdiffest}) in the third step. In short
\begin{equation}\label{eq:chiixiifinestnoisegeom}
|\chi^{i}\xi_{i}|\leq Ce^{-\kappa_{\rosh}t}\left(\textstyle{\sum}_{j}g^{jj}\xi_{j}\xi_{j}\right)^{1/2}
\end{equation}
for all $t\geq 0$, where $C$ only depends on $C_{\rosh}$, $C_{\rood}$ and the constants appearing in the definition of $C^{2}$-balance.
Combining this estimate with (\ref{eq:hinvcomform}) and (\ref{eq:bgeuijxixjdiffest}) yields
\begin{equation}\label{eq:guijxixjfinestinpf}
\textstyle{\sum}_{i\neq j}|g^{ij}(t)\xi_{i}\xi_{j}| \leq
Ce^{-\min\{\kappa_{\rood},2\kappa_{\rosh}\}t}\textstyle{\sum}_{j}g^{jj}(t)\xi_{j}\xi_{j}
\end{equation}
for all $t\geq 0$, where $C$ has the same dependence as in the case of (\ref{eq:chiixiifinestnoisegeom}). 

\textbf{The first time derivative of the off-diagonal terms.} Turning to the time derivative of 
$g^{ij}$, note that $\d_{t}\bge^{ij}=-2\bk^{ij}$. For $i\neq j$, $i,j\in \{1,\dots,d\}$, 
\[
|\d_{t}\bge^{ij}|=2|\bk^{ij}|=2|\bk^{\sharp}(dx^{i},dx^{j})|\leq 2C_{\rood,1}e^{-\kappa_{\rood}t}(\bge^{ii})^{1/2}(\bge^{jj})^{1/2}
\]
for all $t\geq 0$, where we used the fact that $\bge$ is $C^{2}$-asymptotically diagonal. Consequently, an argument similar to the proof of 
(\ref{eq:guijxixjfinestinpf}) yields
\begin{equation}\label{eq:dtguijxixjfinestinpf}
\textstyle{\sum}_{i\neq j}|\dot{g}^{ij}(t)\xi_{i}\xi_{j}| \leq
Ce^{-\min\{\kappa_{\rood},2\kappa_{\rosh}\}t}\textstyle{\sum}_{j}g^{jj}(t)\xi_{j}\xi_{j}
\end{equation}
for all $t\geq 0$, where $C$ only depends on $C_{\rosh}$, $C_{\rood}$, $C_{\rood,1}$ and the constants appearing in the definition of $C^{2}$-balance.

\textbf{The second time derivative of the off-diagonal terms.} Turning to the second time derivative of $\bge^{ij}$, note that 
\begin{equation}\label{eq:ddotbgeuijform}
\d_{t}^{2}\bge^{ij}=8\bk^{il}\bk_{l}^{\phantom{l}j}-2(\ml_{U}\bk)^{ij}=8\textstyle{\sum}_{l\neq i}\bk^{il}\bk_{l}^{\phantom{l}j}+8\bk^{ii}\bk_{i}^{\phantom{i}j}-2(\ml_{U}\bk)^{ij}
\end{equation}
(no summation on $i$ in the second term on the far right hand side). Let us estimate 
\begin{equation}\label{eq:bkllujest}
\begin{split}
|\bk_{l}^{\phantom{l}j}| \leq & |\bge_{ll}\bk^{lj}|+\textstyle{\sum}_{r\neq l}|\bge_{lr}\bk^{rj}|\leq C|\bk|_{\bge}(\bge^{ll})^{-1/2}(\bge^{jj})^{1/2}\\
 & +C\textstyle{\sum}_{r\neq l}e^{-\kappa_{\rood}t}(\bge^{ll})^{-1/2}(\bge^{rr})^{-1/2}|\bk|_{\bge}(\bge^{rr})^{1/2}(\bge^{jj})^{1/2}
\end{split}
\end{equation}
for all $t\geq 0$ (no summation on $l$), where we appeal to (\ref{eq:kappavwcontrest}), (\ref{eq:bgeuiiinvequivtobgeii}) and 
(\ref{eq:bgeineqjbgeuiijjest}). In particular, there is thus a constant $C>0$ such that 
\begin{equation}\label{eq:bkllujcrudeest}
|\bk_{l}^{\phantom{l}j}|\leq C(\bge^{ll})^{-1/2}(\bge^{jj})^{1/2}
\end{equation}
for all $t\geq 0$, where $C$ only depends on the the metric $g$. However, in the case that $l\neq j$, 
this estimate can be improved by, in the last step of (\ref{eq:bkllujest}), appealing to (\ref{eq:mlUbkvivlestintro}). Doing so yields
\begin{equation}\label{eq:bkllujrefinedljdifest}
|\bk_{l}^{\phantom{l}j}|\leq Ce^{-\kappa_{\rood}t}(\bge^{ll})^{-1/2}(\bge^{jj})^{1/2}
\end{equation}
for all $t\geq 0$ and $l\neq j$ ($l,j\in\{1,\dots,d\}$), where $C$ only depends on the metric $g$. Returning to (\ref{eq:ddotbgeuijform}), consider 
the first term on the far right hand side. Appealing to (\ref{eq:mlUbkvivlestintro}) and (\ref{eq:bkllujcrudeest}) yields
the existence of a constant $C>0$ such that 
\begin{equation}\label{eq:ddotbgeuijfirsttermest}
\textstyle{\sum}_{l\neq i}|8\bk^{il}\bk_{l}^{\phantom{l}j}|\leq Ce^{-\kappa_{\rood}t}(\bge^{ii})^{1/2}(\bge^{jj})^{1/2}
\end{equation}
for all $t\geq 0$, where $C$ only depends on the metric $g$. Concerning the second term on the far right
hand side of (\ref{eq:ddotbgeuijform}), appealing to (\ref{eq:kappavwcontrest}) and (\ref{eq:bkllujrefinedljdifest}) yields the existence of a constant 
$C>0$ such that 
\begin{equation}\label{eq:ddotbgeuijsecondtermest}
|8\bk^{ii}\bk_{i}^{\phantom{i}j}|\leq Ce^{-\kappa_{\rood}t}(\bge^{ii})^{1/2}(\bge^{jj})^{1/2}
\end{equation}
for all $t\geq 0$, where $C$ only depends on the metric $g$. Combining (\ref{eq:ddotbgeuijform}), (\ref{eq:ddotbgeuijfirsttermest}) and 
(\ref{eq:ddotbgeuijsecondtermest}) with (\ref{eq:mlUbkvivlestintro}) yields a constant $C>0$ such that 
\[
|\d_{t}^{2}\bge^{ij}|\leq Ce^{-\kappa_{\rood}t}(g^{ii})^{1/2}(g^{jj})^{1/2}
\]
for all $t\geq 0$ and $i\neq j$ ($i,j\in\{1,\dots,d\}$), where we also appealed to (\ref{eq:hinvbhinvequ}) and $C$ only depends on the 
metric $g$. In order to estimate the second time derivative of $g^{ij}$, we also need to estimate
$\dot{\chi}^{i}\dot{\chi}^{j}$ and $\ddot{\chi}^{i}\chi^{j}$. The first of these expressions can be estimated as before. However, in the case of 
the second expression, we can estimate $\chi^{i}$ as before, but in order to estimate $\ddot{\chi}^{j}$, we need to appeal to the fact that 
the shift vector field is $C^{2}$-future bounded. This yields a constant $C>0$ such that 
\[
|\ddot{\chi}^{i}\chi^{j}\xi_{i}\xi_{j}|\leq Ce^{-\kappa_{\rosh}t}\textstyle{\sum}_{j}g^{jj}\xi_{j}\xi_{j}
\]
for all $t\geq 0$, where $C$ only depends on the metric $g$. Summing up, (\ref{eq:odtermneggeomtoanal}) holds and the lemma follows. 
\end{proof}

\printindex

\end{document}